\pdfoutput=1

\documentclass[acmsmall,screen]{acmart}

\setcopyright{rightsretained}
\acmPrice{}
\acmDOI{10.1145/3341695}
\acmYear{2019}
\copyrightyear{2019}
\acmJournal{PACMPL}
\acmVolume{3}
\acmNumber{ICFP}
\acmArticle{91}
\acmMonth{8}

\usepackage{xcolor}

\usepackage{hyperref}
\usepackage{booktabs}   
\usepackage{subcaption} 
\usepackage[utf8]{inputenc}
\usepackage{graphicx}
\usepackage{amsmath}
\usepackage{nccmath}
\usepackage{amssymb}
\usepackage{amsthm}
\usepackage{fancyvrb}
\usepackage{mathpartir}
\usepackage{xspace}
\usepackage{paralist}
\usepackage{tikz}
\usetikzlibrary{shapes,arrows}

\usepackage{blindtext}
\usepackage{mathtools}
\usepackage{pdfpages}

\usepackage{longtable}
\usepackage{multirow}
\usepackage{framed}
\usepackage[all,cmtip]{xy}

\usepackage{ottalt}

\newcommand{\ottdrule}[4][]{{\displaystyle\frac{\begin{array}{l}#2\end{array}}{#3}\quad\ottdrulename{#4}}}

\newcommand{\ottpremise}[1]{ #1 \\}
\newenvironment{ottdefnblock}[3][]{ \framebox{\mbox{#2}} \quad #3 \\[0pt]}{}

\newcommand{\ottmv}[1]{\mathit{#1}}

\newcommand{\ottsym}[1]{#1}

\newcommand{\ottdrulename}[1]{\textsc{#1}}

\newcommand{\ottcomp}[2]{\overline{#1}^{\,#2}}

\newcommand{\ottafterlastrule}{\\}

\newcommand{\ottdruleiTmXXtrue}[1]{\ottdrule[#1]{%
\ottpremise{ \vdash_{ctx}  \intermed {  {\intermed \Sigma}  } ; \intermed {  {\intermed { \Gamma_C } }  } ; \intermed {  {\intermed \Gamma}  } }%
}{
 \intermed {  {\intermed \Sigma}  } ; \intermed {  {\intermed { \Gamma_C } }  } ; \intermed {  {\intermed \Gamma}  }  \vdash_{tm}  \intermed {  \mathit{\intermed {True} }  } : \intermed {  \mathit{\intermed {Bool} }  } \itot{ \rightsquigarrow  \target {  \mathit{\target {True} }  } } }{%
{\ottdrulename{iTm\_true}}{}%
}}

\newcommand{\ottdruleiTmXXfalse}[1]{\ottdrule[#1]{%
\ottpremise{ \vdash_{ctx}  \intermed {  {\intermed \Sigma}  } ; \intermed {  {\intermed { \Gamma_C } }  } ; \intermed {  {\intermed \Gamma}  } }%
}{
 \intermed {  {\intermed \Sigma}  } ; \intermed {  {\intermed { \Gamma_C } }  } ; \intermed {  {\intermed \Gamma}  }  \vdash_{tm}  \intermed {  \mathit{\intermed {False} }  } : \intermed {  \mathit{\intermed {Bool} }  } \itot{ \rightsquigarrow  \target {  \mathit{\target {False} }  } } }{%
{\ottdrulename{iTm\_false}}{}%
}}

\newcommand{\ottdruleiTmXXvar}[1]{\ottdrule[#1]{%
\ottpremise{ ( \intermed {  {\intermed x}  } : \intermed {  {\intermed \sigma}  } )  \in  \intermed {  {\intermed \Gamma}  } }%
\ottpremise{ \vdash_{ctx}  \intermed {  {\intermed \Sigma}  } ; \intermed {  {\intermed { \Gamma_C } }  } ; \intermed {  {\intermed \Gamma}  } }%
}{
 \intermed {  {\intermed \Sigma}  } ; \intermed {  {\intermed { \Gamma_C } }  } ; \intermed {  {\intermed \Gamma}  }  \vdash_{tm}  \intermed {  {\intermed x}  } : \intermed {  {\intermed \sigma}  } \itot{ \rightsquigarrow  \target {  {\target x}  } } }{%
{\ottdrulename{iTm\_var}}{}%
}}

\newcommand{\ottdruleiTmXXlet}[1]{\ottdrule[#1]{%
\ottpremise{ \intermed {  {\intermed \Sigma}  } ; \intermed {  {\intermed { \Gamma_C } }  } ; \intermed {  {\intermed \Gamma}  }  \vdash_{tm}  \intermed {  {\intermed e}_{{\mathrm{1}}}  } : \intermed {  {\intermed \sigma}_{{\mathrm{1}}}  } \itot{ \rightsquigarrow  \target {  {\target e }_{{\mathrm{1}}}  } } }%
\ottpremise{ \intermed {  {\intermed \Sigma}  } ; \intermed {  {\intermed { \Gamma_C } }  } ; \intermed {  {\intermed \Gamma}  \ottsym{,}  {\intermed x}  \ottsym{:}  {\intermed \sigma}_{{\mathrm{1}}}  }  \vdash_{tm}  \intermed {  {\intermed e}_{{\mathrm{2}}}  } : \intermed {  {\intermed \sigma}_{{\mathrm{2}}}  } \itot{ \rightsquigarrow  \target {  {\target e }_{{\mathrm{2}}}  } } }%
\ottpremise{ \intermed {  {\intermed { \Gamma_C } }  } ; \intermed {  {\intermed \Gamma}  }  \vdash_{ty}  \intermed {  {\intermed \sigma}_{{\mathrm{1}}}  } \itot{ \rightsquigarrow  \target {  {\target \sigma }_{{\mathrm{1}}}  } } }%
}{
 \intermed {  {\intermed \Sigma}  } ; \intermed {  {\intermed { \Gamma_C } }  } ; \intermed {  {\intermed \Gamma}  }  \vdash_{tm}  \intermed {  \textbf{\intermed {let} } \, {\intermed x}  \ottsym{:}  {\intermed \sigma}_{{\mathrm{1}}}  \ottsym{=}  {\intermed e}_{{\mathrm{1}}} \, \textbf{\intermed {in} } \, {\intermed e}_{{\mathrm{2}}}  } : \intermed {  {\intermed \sigma}_{{\mathrm{2}}}  } \itot{ \rightsquigarrow  \target {  \textbf{\target {let} } \, {\target x}  \ottsym{:}  {\target \sigma }_{{\mathrm{1}}}  \ottsym{=}  {\target e }_{{\mathrm{1}}} \, \textbf{\target {in} } \, {\target e }_{{\mathrm{2}}}  } } }{%
{\ottdrulename{iTm\_let}}{}%
}}

\newcommand{\ottdruleiTmXXmethod}[1]{\ottdrule[#1]{%
\ottpremise{ \intermed {  {\intermed \Sigma}  } ; \intermed {  {\intermed { \Gamma_C } }  } ; \intermed {  {\intermed \Gamma}  }  \vdash_{d}  \intermed {  {\intermed d}  } : \intermed {  {\intermed {\mathit{TC} } } \, {\intermed \sigma}  } \itot{\,  \rightsquigarrow  \target {  {\target e }  } } }%
\ottpremise{ ( \intermed {  {\intermed m}  } : \intermed {  {\intermed {\mathit{TC} } } \, {\intermed a }  } : \intermed {  {\intermed \sigma}'  } )  \in  \intermed {  {\intermed { \Gamma_C } }  } }%
}{
 \intermed {  {\intermed \Sigma}  } ; \intermed {  {\intermed { \Gamma_C } }  } ; \intermed {  {\intermed \Gamma}  }  \vdash_{tm}  \intermed {  {\intermed d}  \ottsym{.}  {\intermed m}  } : \intermed {  [  {\intermed \sigma}  /  {\intermed a }  ]  {\intermed \sigma}'  } \itot{ \rightsquigarrow  \target {  {\target e }  \ottsym{.}  {\target m}  } } }{%
{\ottdrulename{iTm\_method}}{}%
}}

\newcommand{\ottdruleiTmXXarrI}[1]{\ottdrule[#1]{%
\ottpremise{ \intermed {  {\intermed \Sigma}  } ; \intermed {  {\intermed { \Gamma_C } }  } ; \intermed {  {\intermed \Gamma}  \ottsym{,}  {\intermed x}  \ottsym{:}  {\intermed \sigma}_{{\mathrm{1}}}  }  \vdash_{tm}  \intermed {  {\intermed e}  } : \intermed {  {\intermed \sigma}_{{\mathrm{2}}}  } \itot{ \rightsquigarrow  \target {  {\target e }  } } }%
\ottpremise{ \intermed {  {\intermed { \Gamma_C } }  } ; \intermed {  {\intermed \Gamma}  }  \vdash_{ty}  \intermed {  {\intermed \sigma}_{{\mathrm{1}}}  } \itot{ \rightsquigarrow  \target {  {\target \sigma }_{{\mathrm{1}}}  } } }%
}{
 \intermed {  {\intermed \Sigma}  } ; \intermed {  {\intermed { \Gamma_C } }  } ; \intermed {  {\intermed \Gamma}  }  \vdash_{tm}  \intermed {  \lambda  {\intermed x}  \ottsym{:}  {\intermed \sigma}_{{\mathrm{1}}}  \ottsym{.}  {\intermed e}  } : \intermed {  {\intermed \sigma}_{{\mathrm{1}}}  \rightarrow  {\intermed \sigma}_{{\mathrm{2}}}  } \itot{ \rightsquigarrow  \target {  \lambda  {\target x}  \ottsym{:}  {\target \sigma }_{{\mathrm{1}}}  \ottsym{.}  {\target e }  } } }{%
{\ottdrulename{iTm\_arrI}}{}%
}}

\newcommand{\ottdruleiTmXXarrE}[1]{\ottdrule[#1]{%
\ottpremise{ \intermed {  {\intermed \Sigma}  } ; \intermed {  {\intermed { \Gamma_C } }  } ; \intermed {  {\intermed \Gamma}  }  \vdash_{tm}  \intermed {  {\intermed e}_{{\mathrm{1}}}  } : \intermed {  {\intermed \sigma}_{{\mathrm{1}}}  \rightarrow  {\intermed \sigma}_{{\mathrm{2}}}  } \itot{ \rightsquigarrow  \target {  {\target e }_{{\mathrm{1}}}  } } }%
\ottpremise{ \intermed {  {\intermed \Sigma}  } ; \intermed {  {\intermed { \Gamma_C } }  } ; \intermed {  {\intermed \Gamma}  }  \vdash_{tm}  \intermed {  {\intermed e}_{{\mathrm{2}}}  } : \intermed {  {\intermed \sigma}_{{\mathrm{1}}}  } \itot{ \rightsquigarrow  \target {  {\target e }_{{\mathrm{2}}}  } } }%
}{
 \intermed {  {\intermed \Sigma}  } ; \intermed {  {\intermed { \Gamma_C } }  } ; \intermed {  {\intermed \Gamma}  }  \vdash_{tm}  \intermed {  {\intermed e}_{{\mathrm{1}}} \, {\intermed e}_{{\mathrm{2}}}  } : \intermed {  {\intermed \sigma}_{{\mathrm{2}}}  } \itot{ \rightsquigarrow  \target {  {\target e }_{{\mathrm{1}}} \, {\target e }_{{\mathrm{2}}}  } } }{%
{\ottdrulename{iTm\_arrE}}{}%
}}

\newcommand{\ottdruleiTmXXconstrI}[1]{\ottdrule[#1]{%
\ottpremise{ \intermed {  {\intermed \Sigma}  } ; \intermed {  {\intermed { \Gamma_C } }  } ; \intermed {  {\intermed \Gamma}  \ottsym{,}  {\intermed \delta }  \ottsym{:}  {\intermed Q}  }  \vdash_{tm}  \intermed {  {\intermed e}  } : \intermed {  {\intermed \sigma}  } \itot{ \rightsquigarrow  \target {  {\target e }  } } }%
\ottpremise{ \intermed {  {\intermed { \Gamma_C } }  } ; \intermed {  {\intermed \Gamma}  }  \vdash_{\mathit {Q} }  \intermed {  {\intermed Q}  }\, \itot{ \rightsquigarrow  \target {  {\target \sigma }  } } }%
}{
 \intermed {  {\intermed \Sigma}  } ; \intermed {  {\intermed { \Gamma_C } }  } ; \intermed {  {\intermed \Gamma}  }  \vdash_{tm}  \intermed {  \lambda  {\intermed \delta }  \ottsym{:}  {\intermed Q}  \ottsym{.}  {\intermed e}  } : \intermed {   {\intermed Q}  \Rightarrow  {\intermed \sigma}   } \itot{ \rightsquigarrow  \target {  \lambda  {\target \delta }  \ottsym{:}  {\target \sigma }  \ottsym{.}  {\target e }  } } }{%
{\ottdrulename{iTm\_constrI}}{}%
}}

\newcommand{\ottdruleiTmXXconstrE}[1]{\ottdrule[#1]{%
\ottpremise{ \intermed {  {\intermed \Sigma}  } ; \intermed {  {\intermed { \Gamma_C } }  } ; \intermed {  {\intermed \Gamma}  }  \vdash_{tm}  \intermed {  {\intermed e}  } : \intermed {   {\intermed Q}  \Rightarrow  {\intermed \sigma}   } \itot{ \rightsquigarrow  \target {  {\target e }_{{\mathrm{1}}}  } } }%
\ottpremise{ \intermed {  {\intermed \Sigma}  } ; \intermed {  {\intermed { \Gamma_C } }  } ; \intermed {  {\intermed \Gamma}  }  \vdash_{d}  \intermed {  {\intermed d}  } : \intermed {  {\intermed Q}  } \itot{\,  \rightsquigarrow  \target {  {\target e }_{{\mathrm{2}}}  } } }%
}{
 \intermed {  {\intermed \Sigma}  } ; \intermed {  {\intermed { \Gamma_C } }  } ; \intermed {  {\intermed \Gamma}  }  \vdash_{tm}  \intermed {  {\intermed e} \, {\intermed d}  } : \intermed {  {\intermed \sigma}  } \itot{ \rightsquigarrow  \target {  {\target e }_{{\mathrm{1}}} \, {\target e }_{{\mathrm{2}}}  } } }{%
{\ottdrulename{iTm\_constrE}}{}%
}}

\newcommand{\ottdruleiTmXXforallI}[1]{\ottdrule[#1]{%
\ottpremise{ \intermed {  {\intermed \Sigma}  } ; \intermed {  {\intermed { \Gamma_C } }  } ; \intermed {  {\intermed \Gamma}  \ottsym{,}  {\intermed a }  }  \vdash_{tm}  \intermed {  {\intermed e}  } : \intermed {  {\intermed \sigma}  } \itot{ \rightsquigarrow  \target {  {\target e }  } } }%
}{
 \intermed {  {\intermed \Sigma}  } ; \intermed {  {\intermed { \Gamma_C } }  } ; \intermed {  {\intermed \Gamma}  }  \vdash_{tm}  \intermed {  \Lambda  {\intermed a }  \ottsym{.}  {\intermed e}  } : \intermed {  \forall  {\intermed a }  \ottsym{.}  {\intermed \sigma}  } \itot{ \rightsquigarrow  \target {  \Lambda  {\target a }  \ottsym{.}  {\target e }  } } }{%
{\ottdrulename{iTm\_forallI}}{}%
}}

\newcommand{\ottdruleiTmXXforallE}[1]{\ottdrule[#1]{%
\ottpremise{ \intermed {  {\intermed \Sigma}  } ; \intermed {  {\intermed { \Gamma_C } }  } ; \intermed {  {\intermed \Gamma}  }  \vdash_{tm}  \intermed {  {\intermed e}  } : \intermed {  \forall  {\intermed a }  \ottsym{.}  {\intermed \sigma}'  } \itot{ \rightsquigarrow  \target {  {\target e }  } } }%
\ottpremise{ \intermed {  {\intermed { \Gamma_C } }  } ; \intermed {  {\intermed \Gamma}  }  \vdash_{ty}  \intermed {  {\intermed \sigma}  } \itot{ \rightsquigarrow  \target {  {\target \sigma }  } } }%
}{
 \intermed {  {\intermed \Sigma}  } ; \intermed {  {\intermed { \Gamma_C } }  } ; \intermed {  {\intermed \Gamma}  }  \vdash_{tm}  \intermed {  {\intermed e} \, {\intermed \sigma}  } : \intermed {  [  {\intermed \sigma}  /  {\intermed a }  ]  {\intermed \sigma}'  } \itot{ \rightsquigarrow  \target {  {\target e } \, {\target \sigma }  } } }{%
{\ottdrulename{iTm\_forallE}}{}%
}}

\newcommand{\ottdruleDXXvar}[1]{\ottdrule[#1]{%
\ottpremise{ ( \intermed {  {\intermed \delta }  } : \intermed {  {\intermed Q}  } )  \in  \intermed {  {\intermed \Gamma}  } }%
\ottpremise{ \vdash_{ctx}  \intermed {  {\intermed \Sigma}  } ; \intermed {  {\intermed { \Gamma_C } }  } ; \intermed {  {\intermed \Gamma}  } }%
}{
 \intermed {  {\intermed \Sigma}  } ; \intermed {  {\intermed { \Gamma_C } }  } ; \intermed {  {\intermed \Gamma}  }  \vdash_{d}  \intermed {  {\intermed \delta }  } : \intermed {  {\intermed Q}  } \itot{\,  \rightsquigarrow  \target {  {\target \delta }  } } }{%
{\ottdrulename{D\_var}}{}%
}}

\newcommand{\ottdruleDXXcon}[1]{\ottdrule[#1]{%
\ottpremise{ \intermed {  {\intermed \Sigma}  } = \intermed {  {\intermed \Sigma}_{{\mathrm{1}}}  \ottsym{,}  \ottsym{(}  {\intermed D }  \ottsym{:}  \forall  {\overline {\intermed a} }_{\ottmv{j}}  \ottsym{.}  {\overline {\intermed Q} }_{\ottmv{i}}  \Rightarrow  {\intermed {\mathit{TC} } } \, {\intermed \sigma}_{\ottmv{q}}  \ottsym{)}  \ottsym{.}  {\intermed m}  \mapsto  \Lambda  {\overline {\intermed a} }_{\ottmv{j}}  \ottsym{.}  \lambda  {\overline {\intermed \delta} }_{\ottmv{i}}  \ottsym{:}  {\overline {\intermed Q} }_{\ottmv{i}}  \ottsym{.}  {\intermed e}  \ottsym{,}  {\intermed \Sigma}_{{\mathrm{2}}}  } }%
\ottpremise{ \vdash_{ctx}  \intermed {  {\intermed \Sigma}  } ; \intermed {  {\intermed { \Gamma_C } }  } ; \intermed {  {\intermed \Gamma}  } }%
\ottpremise{\ottcomp{ \intermed {  {\intermed { \Gamma_C } }  } ; \intermed {  \bullet  \ottsym{,}  {\overline {\intermed a} }_{\ottmv{j}}  }  \vdash_{\mathit {Q} }  \intermed {  {\intermed Q}_{\ottmv{i}}  }\, \itot{ \rightsquigarrow  \target {  {\target \sigma }'_{\ottmv{i}}  } } }{\ottmv{i}}}%
\ottpremise{\ottcomp{ \intermed {  {\intermed { \Gamma_C } }  } ; \intermed {  {\intermed \Gamma}  }  \vdash_{ty}  \intermed {  {\intermed \sigma}_{\ottmv{j}}  } \itot{ \rightsquigarrow  \target {  {\target \sigma }_{\ottmv{j}}  } } }{\ottmv{j}}}%
\ottpremise{ \intermed {  {\intermed \Sigma}_{{\mathrm{1}}}  } ; \intermed {  {\intermed { \Gamma_C } }  } ; \intermed {  \bullet  \ottsym{,}  {\overline {\intermed a} }_{\ottmv{j}}  \ottsym{,}  {\overline {\intermed \delta} }_{\ottmv{i}}  \ottsym{:}  {\overline {\intermed Q} }_{\ottmv{i}}  }  \vdash_{tm}  \intermed {  {\intermed e}  } : \intermed {  [  {\intermed \sigma}_{\ottmv{q}}  /  {\intermed a }  ]  {\intermed \sigma}_{\ottmv{m}}  } \itot{ \rightsquigarrow  \target {  {\target e }  } } }%
\ottpremise{\ottcomp{ \intermed {  {\intermed \Sigma}  } ; \intermed {  {\intermed { \Gamma_C } }  } ; \intermed {  {\intermed \Gamma}  }  \vdash_{d}  \intermed {  {\intermed d}_{\ottmv{i}}  } : \intermed {  [  \overline {\intermed \sigma}_{\ottmv{j}}  /  {\overline {\intermed a} }_{\ottmv{j}}  ]  {\intermed Q}_{\ottmv{i}}  } \itot{\,  \rightsquigarrow  \target {  {\target e }_{\ottmv{i}}  } } }{\ottmv{i}}}%
}{
 \intermed {  {\intermed \Sigma}  } ; \intermed {  {\intermed { \Gamma_C } }  } ; \intermed {  {\intermed \Gamma}  }  \vdash_{d}  \intermed {  {\intermed D } \, \overline {\intermed \sigma}_{\ottmv{j}} \, {\overline {\intermed d} }_{\ottmv{i}}  } : \intermed {  {\intermed {\mathit{TC} } } \, [  \overline {\intermed \sigma}_{\ottmv{j}}  /  {\overline {\intermed a} }_{\ottmv{j}}  ]  {\intermed \sigma}_{\ottmv{q}}  } \itot{\,  \rightsquigarrow  \target {  \ottsym{(}  \Lambda  {\overline {\target a} }_{\ottmv{j}}  \ottsym{.}  \lambda \, \ottcomp{{\target \delta }_{\ottmv{i}}  \ottsym{:}  {\target \sigma }'_{\ottmv{i}}}{\ottmv{i}} \, \ottsym{.}  \ottsym{\{}  {\target m}  \ottsym{=}  {\target e }  \ottsym{\}}  \ottsym{)} \, {\overline {\target \sigma} }_{\ottmv{j}} \, {\overline {\target e} }_{\ottmv{i}}  } } }{%
{\ottdrulename{D\_con}}{}%
}}

\newcommand{\ottdruleiEvalXXappAbs}[1]{\ottdrule[#1]{%
}{
 \intermed {  {\intermed \Sigma}  }  \vdash  \intermed {  \ottsym{(}  \lambda  {\intermed x}  \ottsym{:}  {\intermed \sigma}  \ottsym{.}  {\intermed e}_{{\mathrm{1}}}  \ottsym{)} \, {\intermed e}_{{\mathrm{2}}}  }  \longrightarrow  \intermed {  [  {\intermed e}_{{\mathrm{2}}}  /  {\intermed x}  ]  {\intermed e}_{{\mathrm{1}}}  } }{%
{\ottdrulename{iEval\_appAbs}}{}%
}}

  \renewottcommands[ott]

\usepackage{listings}

\usepackage{etoc} 
\usepackage{thmtools}

\renewcommand\ottaltinferrule[4]{
  \inferrule*[right=\scriptsize{#1}]
    {#3}
    {#4}
}

\definecolor{NavyBlue}{HTML}{006EB8}
\definecolor{OliveGreen}{HTML}{3C8031}
\definecolor{BrickRed}{HTML}{B6321C}

\newcommand{\boxedfigureH}[3]{
  \begin{figure}[H]
    \centering
    \mbox{
      \begin{minipage}{\textwidth - .5cm}
        {#1}
      \end{minipage}
    }
    \caption{#2}
    \label{#3}
  \end{figure}
}

\newcommand{\boxedfigure}[3]{
  \begin{figure}
    \centering
    \mbox{
      \begin{minipage}{\textwidth - .5cm}
        {#1}
      \end{minipage}
    }
    \caption{#2}
    \label{#3}
  \end{figure}
}

\newcommand{\repeated}[2]{( {#1} ~ , ~ \forall {#2} )}

\newcommand{\srcL}{$\textcolor{NavyBlue}{\boldsymbol{\lambda}_{\textbf{\textsc{TC}}}}$\xspace}
\newcommand{\interL}{$\textcolor{OliveGreen}{\boldsymbol{F}_{\textbf{\textsc{D}}}}$\xspace}
\newcommand{\targetL}{$\textcolor{BrickRed}{\boldsymbol{F}_{\textbf{\{\}}}}$\xspace}

\newcommand{\gsp}{\,\,}

\newcommand{\sourceC}  [1]{\begin{color}{NavyBlue}{#1}\end{color}}
\newcommand{\source}   [1]{\sourceC{{#1}}}
\newcommand{\intermedC}[1]{\begin{color}{OliveGreen}{#1}\end{color}}
\newcommand{\intermed} [1]{\intermedC{{#1}}}

\newcommand{\targetC}  [1]{\begin{color}{BrickRed}{#1}\end{color}}
\newcommand{\target}   [1]{\targetC{{#1}}}

\newcommand{\ruleform}[1]{\fbox{$#1$}}

\definecolor{superlightgray}{gray}{0.80}

\newcommand{\band}{\wedge}

\newcommand{\proofStep}[3]{{\small \ruleform{\textbf{#1}}} \quad ${#2}$ \\
                           \begin{changemargin}{2em}{0em}
                             {#3}
                           \end{changemargin}
                           }
\newcommand{\proofStepNL}[3]{{\small \ruleform{\textbf{#1}}}
  \begin{equation*}
    {#2}
  \end{equation*}
  \begin{changemargin}{2em}{0em}
    {#3}
  \end{changemargin}
  }

\newcommand{\namedRuleform}[2]{{\small $\ruleform{#1}$} \hspace*{\fill} ({#2})}

\usepackage{newfloat,caption,float}

\DeclareFloatingEnvironment[
  listname = Examples,
  name = Example,
  placement = H,
  within = none,
  ]{example}
\newtheoremstyle{defstyle}
  {}
  {}
  {\itshape}
  {1em}
  {\scshape}
  {.}
  { }
  {}

\newtheorem{theorem}{Theorem}
\theoremstyle{defstyle}
\newtheorem{definition}{Definition}

\newenvironment{theoremB}{\begin{theorem}}{\end{theorem}}

\newcommand\stoi[1]{#1}
\newcommand\stot[1]{#1}
\newcommand\itot[1]{#1}

\begin{document}

\title[]{Coherence of Type Class Resolution}         


\author{Gert-Jan Bottu}
\affiliation{
  \department{Department of Computer Science}              
  \institution{KU Leuven}            
  \country{Belgium}                    
}
\email{gertjan.bottu@kuleuven.be}          

\author{Ningning Xie}
\affiliation{
  \department{Department of Computer Science}             
  \institution{The University of Hong Kong}           
  \country{China}                   
}
\email{nnxie@cs.hku.hk}

\author{Koar Marntirosian}
\affiliation{
  \department{Department of Computer Science}             
  \institution{KU Leuven}           
  \country{Belgium}                   
}
\email{koar.marntirosian@kuleuven.be}   

\author{Tom Schrijvers}
\orcid{0000-0001-8771-5559}             
\affiliation{
  \department{Department of Computer Science}             
  \institution{KU Leuven}           
  \country{Belgium}                   
}
\email{tom.schrijvers@kuleuven.be}         

\begin{abstract}
Elaboration-based type class resolution, as found in languages like Haskell,
Mercury and PureScript, is generally nondeterministic: there can be
multiple ways to satisfy a wanted constraint in terms of global 
instances and locally given constraints.
Coherence is the key property that keeps this sane; it guarantees that, despite
the nondeterminism, programs still behave predictably. Even though
elaboration-based resolution is generally assumed coherent, as far as we know, there
is no formal proof of this property in the presence of sources of
nondeterminism, like superclasses and flexible contexts.

This paper provides a formal proof to remedy the situation. The proof is
non-trivial because the semantics elaborates resolution into a target language
where different elaborations can be distinguished by contexts that do not have a
source language counterpart. Inspired by the notion of full abstraction, we
present a two-step strategy that first elaborates nondeterministically into an
intermediate language that preserves contextual equivalence, and then
deterministically elaborates from there into the target language. We use an
approach based on logical relations to establish contextual equivalence and thus
coherence for the first step of elaboration, while the second step's determinism
straightforwardly preserves this coherence property.
\end{abstract}



\maketitle

\part*{}
\invisiblelocaltableofcontents\label{parttoc:1}
\vspace{-0.5cm}

As a guide to the reader, we present the source language
\srcL{} in \source{blue}, the intermediate language \interL{}
in \intermed{green} and the target language \targetL{} in \target{red}.
We thus encourage the reader to view / print this paper in color.

\section{Introduction}

Type classes were initially introduced in Haskell~\cite{haskell98}
by Wadler and Blott~\cite{AdHocPolymorphism} to
make ad-hoc overloading less ad hoc, and they have since become one of Haskell's
core abstraction features.
Moreover, their resounding success has spread far beyond Haskell: several
languages have adopted them  (e.g., Mercury~\cite{mercury},
Coq~\cite{Sozeau2008}, PureScript~\cite{purescript}, Lean~\citep{de2015lean}),
and they have inspired various alternative language features (e.g., Scala's
implicits~\cite{implicits,Simplicitly}, Rust's traits~\cite{rust}, C++'s
concepts~\cite{concepts-article}, Agda's instance
arguments~\cite{Devriese2011}).

Type classes have also received a lot of attention from researchers with many
proposals for extensions and improvements, including
functional dependencies~\cite{fundeps-original}, associated
types~\cite{AssociatedTypeSynonyms}, quantified
constraints~\cite{Bottu2017} among other extensions.

Given the extensive attention that type classes have received, it may be
surprising that the metatheory of their elaboration-based semantics~\cite{ClassElaboration} has not yet been exhaustively
studied. In particular, as far as we know, while there have been many
informal arguments, the formal notion of \emph{coherence} has never been proven.
\citet{Reynolds91coherence} has defined coherence as follows:
\begin{quote}\it
``When a programming language has a sufficiently rich type structure, there can
be more than one proof of the same typing judgment; potentially this can lead
to semantic ambiguity since the semantics of a typed language is a function of
such proofs. When no such ambiguity arises, we say that the language is
coherent.''
\end{quote}

Type classes give rise to two main (potential) sources of incoherence. The first
source are \emph{ambiguous type schemes}, such as that of the function \texttt{foo}:
\begin{verbatim}
    > foo :: (Show a, Read a) => String -> String
    > foo s = show (read s)
\end{verbatim}
The type scheme of \texttt{foo} requires that the type with which \texttt{a}
will be instantiated must have \texttt{Show} and \texttt{Read} instances. This
restriction alone is too permissive, because the type part
(\texttt{String -> String}) of \texttt{foo}’s type scheme is not
sufficient for a deterministic instantiation of \texttt{a} from the calling
context.
\texttt{a} can thus be instantiated arbitrarily to any
type with \texttt{Show} and \texttt{Read} instances. Yet, the choice of type
may lead to a different behavior of \texttt{show} and \texttt{read}, and thus
of \texttt{foo} as a whole. For instance, \texttt{foo "1"} yields \texttt{"1"}
when \texttt{a} is instantiated to \texttt{Int}, and \texttt{"1.0"} when it is
instantiated to \texttt{Float}. To rule out this source of incoherence,
\citet{jones} requires type schemes to be unambiguous and has formally proven
that, for his system, this guarantees coherence.

The second source of ambiguity arises from the type class resolution mechanism
itself. Such mechanisms check
whether a particular type class constraint holds.
Usually, they are styled after resolution-based proof
search in logic, where type class instances act as Horn clauses and type scheme
constraints as additional facts.
Generally, this process is nondeterministic, but languages like Haskell, Mercury
or PureScript contain it by
requiring that type class instances do not overlap with 
each other or with locally given constraints. 
Nevertheless, superclasses remain as a source of nondeterminism; indeed, a superclass
constraint can be resolved through any of its subclass constraints. Hence,
in the presence of superclasses, type class resolution should properly be
considered as a potential source for incoherence.  Moreover, overlap between
locally wanted constraints and global instances is often allowed (e.g., through
GHC's \texttt{FlexibleContexts} pragma), but a formal argument
for its harmlessness is also lacking. \citet{jones} considered neither of these
aspects and simply assumed the coherence of resolution as a given.
\citet{DBLP:conf/haskell/Morris14} side-steps these issues with a denotational
semantics that is disconnected from the original elaboration-based semantics 
and its implementations (e.g., Hugs and GHC).

This paper aims to fill this gap in the metatheory of programming languages
featuring type classes, including industrial grade languages such as Haskell, by formally
establishing that elaboration-based type class resolution is coherent in the presence of
superclasses and flexible contexts. The proof of this property is considerably complicated by the
indirect, elaboration-based approach that is used to give meaning to programs
with type classes.
Indeed, the meaning of such programs is commonly given in terms of their
translation to a core language~\cite{ClassElaboration}, like System F, the
meaning of which is defined in 
the form of an operational semantics.
In this translation process, type classes are elaborated into explicitly passed
function dictionaries.
These dictionaries can, however, often be constructed in more than one way,
resulting in multiple possible translations for a single program. 
The problem is that different
translations of the same source program actually may have different meanings
in the core language. The reason for this discrepancy is that the core language is more expressive
than the source language and admits programs --- that cannot be expressed in
the source language --- in which the different dictionaries can be distinguished.

We solve this problem with a new two-step approach that splits the
problem into two subproblems. The midway point is an intermediate language that
makes type class dictionaries explicit, but---inspired by fully abstract
compilation---cannot distinguish between different elaborations from the same
source language term~\cite{Abadi1999}. We use a logical-relations approach to show that the
nondeterministic elaboration from the source language to this intermediate
language is coherent. Showing coherence for the elaboration from the
intermediate language to the target language is much simpler, because we can
formulate it in a deterministic fashion.

In summary, the contributions of this work are:
\begin{itemize}
\item We present a simple calculus \srcL{} with full-blown type class resolution (incl.
  superclasses), which isolates nondeterministic resolution. Furthermore, we
  present an elaboration from \srcL{} to the target language \targetL{}, System F with
  records, which are used to encode dictionaries.
\item We present an intermediate language \interL{} with explicit dictionary-passing.
  This language enforces the uniqueness of dictionaries, which captures
  the intention of type class instances. We study its metatheory,
  and define a logical relation to prove contextual equivalence. 
\item We present elaborations from \srcL{} to \interL{} and from \interL{} to
  \targetL{}, and prove that a direct translation from \srcL{} to \targetL{} can
  always be decomposed into an equivalent translation through \interL{}. 
\item We prove coherence of the elaboration between \srcL{} and \interL{}, using
  logical relations.
\item We prove that coherence is also preserved through the elaboration from
  \interL{} to \targetL{}. As a consequence, by combining this with the previous
  result, we prove that the elaboration between \srcL{} and \targetL{} is
  coherent.
  The latter coherence result implies coherence of elaboration-based
  type class resolution in the presence of superclasses and flexible contexts.
\end{itemize}

The full formalization and coherence proof can be found in the accompanying 122-page appendix.

The purpose of our work is twofold:
\begin{inparaenum}[1)]
\item To develop a proof technique to establish coherence of type class
  resolution.
  Because this result is achieved on a minimal calculus, this work becomes a
  basis for researchers investigating type class extensions and larger
  languages, as well as their impact on coherence.
\item To present a formal proof of coherence for language designers considering
  to adopt type classes.
  In doing so, we show that the informally trivial argument for the coherence of
  type class resolution is surprisingly hard to formalize.
\end{inparaenum}

\section{Overview}

This section provides some background on dictionary-passing elaboration of type
class resolution and discusses the potential nondeterminism introduced by
superclasses and local constraints. We then briefly introduce our calculi and
discuss the key ideas of the coherence proof. Throughout the section we use
Haskell-like syntax as the source language for examples, and to simplify our
informal discussion we use the same syntax without type classes as the target
language.

\subsection{Dictionary-Passing Elaboration}

A program is coherent if it has exactly one meaning --- i.e., its semantics is
unambiguously determined. For type classes this is not as straightforward as it
seems, because their dynamic semantics are not expressed directly but rather by
type-directed elaboration into a simpler language without type classes
such as System F. Thus the dynamic semantics of type classes are given
indirectly as the dynamic semantics of their elaborated forms.

\begin{example}[tp]
\begin{verbatim}
    > class Eq a where
    >   (==) :: a -> a -> Bool
    >
    > instance Eq Int where
    >   (==) = primEqInt
    >
    > instance (Eq a, Eq b) => Eq (a, b) where
    >   (x1,y1) == (x2,y2)  =  x1 == x2 && y1 == y2
    >
    > refl :: Eq a => a -> Bool
    > refl x = x == x
    >
    > main :: Bool
    > main = refl (5,42)
\end{verbatim}
\caption{Program with type classes.}\label{fig:ex1}
\end{example}

\paragraph{\textbf{Basic Elaboration.}}

Consider the small program with type classes in Example~\ref{fig:ex1}. We declare
a type class \verb|Eq| and instances for the \verb|Int| and pair types. The
function \verb|refl| trivially tests whether an expression is equivalent to
itself, which is called in \verb|main|.

The dictionary-passing elaboration translates this program into a System F-like
core language that does not feature type classes. The main idea of the
elaboration is to map a type class declaration onto a datatype that contains the
method implementations, a so-called \emph{(function) dictionary}.

\begin{verbatim}
    > data EqD a = EqD { (==) :: a -> a -> Bool }
\end{verbatim}
Then simple instances give rise to dictionary values:
\begin{verbatim}
    > eqInt :: EqD Int
    > eqInt = EqD { (==) = primEqInt }
\end{verbatim}
Instances with a non-empty context are translated to functions that take context dictionaries
to the instance dictionary.
\begin{verbatim}
    > eqPair :: (EqD a, EqD b) -> Eq (a,b)
    > eqPair (da, db) = 
    >   EqD { (==) = \(x1,y1) (x2,y2) -> (==) da x1 x2 && (==) db y1 y2 }
\end{verbatim}
Functions with qualified types, like \texttt{refl}, are translated to
functions that take explicit dictionaries as arguments.
\begin{verbatim}
    > refl :: EqD a -> a -> Bool
    > refl d x = (==) d x x
\end{verbatim}
Finally, calls to functions with a qualified type are mapped to calls that
explicitly pass the appropriate dictionary.
\begin{verbatim}
    > main :: Bool
    > main = refl (eqPair eqInt eqInt) (5,42)
\end{verbatim}

\paragraph{\textbf{Elaboration of Superclasses.}}

Superclasses require a small extension to the above elaboration scheme.
Consider the small program in Example~\ref{fig:ex2} where \texttt{Sub1} is a
subclass of \texttt{Base}. The function \verb|test1| has \verb|Sub1 a| in the
context and calls \verb|sub1| and \verb|base| in its definition.

The standard approach to encode superclass is to embed the superclass dictionary
in that of the subclass. For this case, \verb|Sub1D a| contains a field
\verb|super1| that points to the superclass:
\begin{verbatim}
    > data BaseD a = BaseD { base :: a -> Bool }
    > data Sub1D a = Sub1D { super1 :: BaseD a
    >                      , sub1 :: a -> Bool }  
\end{verbatim}
This way we can extract the superclass from the subclass when needed. The
function \texttt{test1} is then encoded as:
\begin{verbatim}
    > test1 :: Sub1 a -> a -> Bool
    > test1 d x = sub1 d x && base (super1 d) x
\end{verbatim}

\begin{example}[tp]
\begin{verbatim}
    > class Base a where
    >   base :: a -> Bool
    >
    > class Base a => Sub1 a where
    >   sub1 :: a -> Bool
    >
    > test1 :: Sub1 a => a -> Bool
    > test1 x = sub1 x && base x
\end{verbatim}
\caption{Program with superclasses.}\label{fig:ex2}
\end{example}

\paragraph{\textbf{Resolution.}}

Calls to functions with a qualified type generate type class constraints.
The process for checking whether these constraints can be satisfied, is known as
\emph{resolution}. For the sake of dictionary-passing elaboration, this resolution
process is augmented with the construction of the appropriate dictionary that
witnesses the satisfiability of the constraint.

\subsection{Nondeterminism and Coherence}
\label{sec:nondet}

For Haskell'98 programs there is usually only one way to construct a dictionary
for a type class constraint. Yet, in the presence of superclasses,
there may be multiple ways. Suppose we extend Example~\ref{fig:ex2} with an
additional subclass and the following function:
\begin{verbatim}
    > class Base a => Sub2 a where
    >   sub2 :: a -> Bool
    >
    > test2 :: (Sub1 a, Sub2 a) => a -> Bool
    > test2 x = base x
\end{verbatim}
There are two possible ways to resolve the \texttt{Base a} constraint that
arises from the call to \texttt{base} in function \texttt{test2}, resulting in
the following two translations: we can either
establish the desired constraint as the superclass of the given \texttt{Sub1 a}
constraint or as the superclass of the given \texttt{Sub2 a} constraint.
\begin{verbatim}
    > test2a, test2b :: (Sub1D a, Sub2D a) -> a -> Bool
    > test2a (d1,d2) x = base (super1 d1) x
    > test2b (d1,d2) x = base (super2 d2) x
\end{verbatim}

Fortunately, this nondeterminism is harmless because the difference between the
two elaborations cannot be observed. Indeed, for any given type \texttt{A},
Haskell'98 only allows a single instance \texttt{Base A}, and it does not matter whether we
access its dictionary directly or through one of its subclass instances. More
generally, this suggests that type class resolution in Haskell'98 is coherent.

If we relax the Haskell'98 non-overlap condition for locally given constraints 
and adopt flexible contexts (allowing for arbitrary types in class constraints,
rather than simple type variables), another source of nondeterminism
arises. Consider:
\begin{verbatim}
    > isZero :: Eq Int => Int -> Bool
    > isZero n = n == 0
\end{verbatim}
There are two ways to resolve the wanted \texttt{Eq Int} constraint that arises
from the use of \texttt{(==)}. Either we use the global \texttt{Eq Int}
constraint (in \texttt{isZero1}),
or we use the locally given \texttt{Eq Int} constraint, passed as argument
\texttt{d} (in \texttt{isZero2}):
\begin{verbatim}
    > isZero1, isZero2 :: EqD Int -> Int -> Bool
    > isZero1 d n = (==) eqInt n 0
    > isZero2 d n = (==) d     n 0
\end{verbatim}

Haskell'98 does not allow the \texttt{Eq Int} constraint in \texttt{isZero}'s
signature, which overlaps with the global \texttt{Eq Int} instance; it only
allows constraints on type variables in function signatures. This prevents the
above nondeterminism in the elaboration. Yet, the nondeterminism is, once more,
harmless; there is no way that the supplied dictionary \texttt{d} can be
anything other than the global instance's dictionary \texttt{eqInt}.
Informally, resolution remains coherent in the presence of flexible contexts.

\subsection{Contextual Difference}\label{sec:overview:difference}

While it is easy to provide an informal argument for the coherence of type class
resolution, formally establishing the property is much harder.
The indirect, elaboration-based attribution of a dynamic semantics in
particular is a complicating factor, since it requires us to reason about two
languages simultaneously.
Unfortunately, there is another factor that further complicates the proof:
different elaborations of the same source program can actually be distinguished
in the target language. Consider, for instance, the target program below:
\begin{verbatim}
    > discern :: ((Sub1D (), Sub2D ()) -> () -> Bool) -> Bool
    > discern f = 
    >   let b1 = BaseD { base = \() -> True }
    >       b2 = BaseD { base = \() -> False }
    >       d1 = Sub1D { super1 = b1 }
    >       d2 = Sub2D { super2 = b2 }
    >   in f (d1,d2) ()
\end{verbatim}
We find that \texttt{discern test2a} evaluates to \texttt{True} and
\texttt{discern test2b} evaluates to \texttt{False}.
Hence, since \texttt{discern} can differentiate between them, \texttt{test2a} and
\texttt{test2b} clearly do not have the same meaning in the target language.

The dictionaries for \texttt{Sub1 ()} and
\texttt{Sub2 ()} have different implementations for their \texttt{Base ()}
superclass. The source language would never allow this, but the target language
has no notion of type classes and happily admits \texttt{discern}'s violation
of source language rules.

The problem is that the target language is more expressive than the source
language. While \texttt{test2a} and \texttt{test2b} cannot be distinguished in
any program context that arises from the source language, we can write target
programs like \texttt{discern} that are not the image of any source program and
thus do not have to play by the source language rules.

\subsection{Our Approach to Proving Coherence}

To avoid the problem with contextual difference in the target language, we
employ a novel two-step approach.
We prove that any elaboration from a source language program into a
dictionary-passing encoding in the target language, can be decomposed in two
separate elaborations through an intermediate language.
We thus obtain two simpler problems for proving coherence of type class resolution. 

The source language, \srcL{} (presented in \source{blue}), features full-fledged
type class resolution, and simplifies term typing with a bidirectional type
system (a technique popularized by \citet{Pierce:2000:LTI:345099.345100})
to not distract from the main objective of coherent resolution.

The intermediate language, \interL{} (presented in \intermed{green}), is an extension of System F that explicitly
passes type class dictionaries, and preserves the source language invariant
that there is at most one such dictionary value for any combination of class
and type. We show \interL{} is type-safe and strongly normalizing, and define a
logical relation that captures the contextual equivalence of two \interL{} terms.

The target language, \targetL{} (presented in \target{red}), is a different variant of System~F
without direct support for type class dictionaries; instead it features
records, which can be used to encode dictionaries, but does not enforce
uniqueness of instances.

\begin{figure}[htp!]
  \centering
  \tikzstyle{block} = [ellipse, node distance=2cm,
     text centered, rounded corners, minimum height=1cm]
\tikzstyle{line} = [draw, very thick, -latex']

\begin{tikzpicture}[->, auto]
  \node [block] (SRC) {
    \parbox{1cm}{\centering
      \srcL{}\\
    (Fig.~\ref{fig:source_syntax})}
    };
  \node [block, right of=SRC, below of=SRC, node distance=2.6cm] (MID) {
    \parbox{1cm}{\centering
      \interL{}\\
      (Fig.~\ref{fig:inter_syntax})
    }
  };
  \node [block, right of=SRC, node distance=5cm] (TGT) {
    \parbox{1cm}{\centering
      \targetL{}\\
      (Fig.~\ref{fig:target_syntax})
    }
  };

  \path
  (SRC) edge[line] node[above]{Thm.~\ref{thm:coh}}            (TGT)
  (SRC) edge[line] node[left]{Thm.~\ref{thm:coh_exprA}}       (MID)
  (MID) edge[line] node[right]{Thm.~\ref{thm:mid-tgt-deter}}  (TGT);
\end{tikzpicture}
  \caption{The different calculi with elaborations}
  \label{fig:lang_diag}
\end{figure}
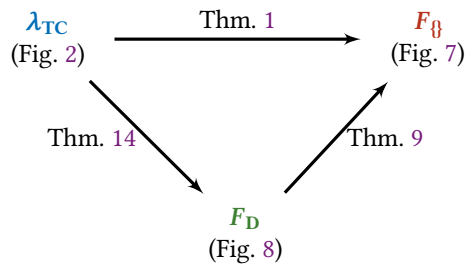

The different calculi are presented in Figure~\ref{fig:lang_diag}, where the
edges denote possible elaborations.

The coherence proof consists of two main parts:
\begin{description}
\item[Coherent Elaboration from \srcL{} to \interL{}.]
  Our elaboration from \srcL{} into \interL{} is nondeterministic, but type preserving.
  Furthermore, we show that any two \interL{} elaborations of the same \srcL{} term
  are logically related, and prove that this logical relation implies
  contextual equivalence. This establishes that the elaboration from \srcL{} to
  \interL{} is coherent.
\item[Deterministic Elaboration from \interL{} to \targetL{}.]
  Because of the syntactic similarity between \interL{} and
  \targetL{}, the elaboration from the former into the latter is a more
  straightforward affair. In addition to being type preserving, it is also
  deterministic, and preserves contextual equivalence.
\end{description}
These results are easily combined to show the coherence of the elaboration
from \srcL{} to \targetL{}, which implies coherence of elaboration-based type
class resolution.
The full proofs can be found in the appendix. Note that the proofs depend on a
number of standard boilerplate conjectures (e.g., substitution lemmas), which can be
found in Sections~\ref{app:src_conj} and~\ref{app:mid_conj} of the appendix.



\section{Source Language \srcL{}}\label{sec:source}

This section presents our source language \srcL{}, a basic calculus which
isolates nondeterministic resolution.
The calculus only supports features that are
essential for type class resolution and its coherence.

Consequently, the language is strongly normalizing, and thus does not support
recursive let expressions, mutual recursion or recursive methods.
This is a sensible choice, as recursion does not affect the fundamentals of the
coherence proof.
This work could include recursion through step
indexing~\citep{ahmed2006step}, a well-known technique,
but this would significantly clutter the proof.
Recursion is discussed in more detail in Section~\ref{sec:features}.

Furthermore, two notable design decisions were made in the support of
superclasses in \srcL{}.
Firstly, similar to GHC, \srcL{} derives all possible superclass constraints
from their subclass constraints in advance, instead of deriving them
``just-in-time'' during resolution. The advantage of this approach is that it
streamlines the actual resolution process.

Secondly, similar to Coq~\cite{Sozeau2008} and unlike \citet{AdHocPolymorphism}, we pass superclass
dictionaries alongside their subclass dictionaries, i.e., in a \emph{flattened}
form, instead of nesting them inside their subclass dictionaries.
As it is not too difficult to see that both approaches are isomorphic,
flattening the superclasses does not impact the coherence of resolution.
It does, however, considerably simplify the
proof, since this way neither our type class
resolution mechanism, nor the intermediate language \interL{}
(Section~\ref{sec:inter}) need to have any
support for superclasses and can treat them as regular local constraints.
A more structured representation would give rise to additional complexity, but
would not alter the essence of the proof.

\paragraph{\textbf{Syntax.}} Figure~\ref{fig:source_syntax} presents the, mostly
standard, syntax.
Programs consist of a number of class (with superclasses) and instance
declarations, and an expression.
For the sake of simplicity and well-foundedness, the declarations are ordered
and can only refer to previous declarations.

\boxedfigure
  {\input{figures/source_syntax.mng}}
  {\srcL{} syntax}
  {fig:source_syntax}

Following \citet{jones-94}'s qualified types framework, we distinguish between three
sorts of types: monotypes \({\source \tau }\), qualified types \({\source \rho}\) which include
constraints, and type schemes \({\source \sigma}\) which include type abstractions.
Constraints differentiate between full constraint schemes \({\source C}\)
and simple class constraints \({\source Q}\). Observe that we allow flexible contexts
in the qualified types; they are not restricted to constraints on type variables.

The definition of expressions \({\source e}\) is standard, but with a few notable exceptions.
Firstly, the language differentiates syntactically between regular
variables \({\source x}\) and method names \({\source m}\), which are introduced
in class declarations.
Secondly, type annotations \(\source{{\source e}  ::  {\source \tau }}\) allow the programmer to manually
assign a monotype to an expression. This is useful for resolving ambiguity---see
the \textit{Typing} paragraph below.
Finally, let bindings include type annotations with a type scheme \({\source \sigma}\),
allowing the programmer to introduce local constraints---also discussed
in the \textit{Typing} paragraph.
Note that we use Haskell syntax for class and instance declarations. 

There are three \srcL{} environments: two global ones and one local
environment.
Firstly, the global class environment \({\source { \Gamma_C } }\) stores all class declarations.
Each entry in \({\source { \Gamma_C } }\) contains the method name \({\source m}\), any superclasses \(\source{\overline {\source {\mathit{TC}_i~a } }}\), the
class \({\source {\mathit{TC} } } \, {\source a }\) itself and the corresponding method type \({\source \sigma}\).

Secondly, the global program context \({ \source P }\) contains all instance declarations.
Each entry in \({ \source P }\) consists of a unique dictionary constructor \({\source D }\), its corresponding
constraint \({\source C}\), the method name \({\source m}\) and its implementation
\({\source e}\), together with the context \({\source \Gamma}\) under which \({\source e}\) should be
interpreted. This context contains the local axioms available in this
instance declaration, as well as any axioms which explicitly annotate the method
type signature. 

Thirdly, the local typing environment \({\source \Gamma}\), besides containing the
default term and type variables \({\source x}\) and \({\source a }\), also stores any local
axioms \({\source Q}\). As opposed to the program context \({ \source P }\),
\({\source \Gamma}\) does not contain any type class instances. Instead, the (local) axioms
are associated with a dictionary variable \({\source \delta }\).
Sections~\ref{sec:inter} and \ref{sec:elabSI} explain the use of these dictionaries.

\paragraph{\textbf{Typing.}}
Our type system features two design choices to
eliminate the possibility of ambiguous type schemes.
This allows us to focus on the coherence of type class
resolution, by making our proof orthogonal to ambiguous type schemes, the source
of ambiguity which has already been studied by \citet{jones}.
We thus side-step an already solved problem and focus on tackling the full
problem of resolution coherence.

Firstly, we require type signatures to be unambiguous
(Figure~\ref{fig:source_typing_extra}, right-hand side)
to make sure that all newly introduced type variables
are bound in the head of the type (the remaining monotype after dropping all type
and constraint abstractions).
This prevents ambiguous expressions such as:
\begin{verbatim}
    > let f : forall a . Eq a => Int -> Int -- ambiguous
    >       = \ x . x + 1 in f 42
\end{verbatim}

Secondly, we use a bidirectional type system rather than
a fully declarative one. A bidirectional type system distinguishes between two
typing modes: \emph{inference} and \emph{check} mode.
The former synthesizes a type from the given expression, while the latter checks
whether a given expression is of a given type. 
Special in our setting is that
variables can only be typed in check mode, to ensure that only a single
instantiation exists.
This avoids the ambiguity that can arise when instantiating type
variables in inference mode.
Consider the following example:
\begin{verbatim}
    > let y : forall a . Eq a => a -> a = ...
    > in const 1 y
\end{verbatim}
where \texttt{const x} is the constant function, which evaluates to \texttt{x}
for any input.
The instantiation of \texttt{y}'s type scheme is not uniquely determined by the
context in which it is used.  In a declarative type system or in inference mode,
this ambiguity would result in multiple distinct typings and corresponding
elaborations. While this ambiguity is harmless, it is not the focus of this
work. Hence, to focus exclusively on the resolution, we use a bidirectional type
system with check mode for variables to eliminate this irrelevant source of ambiguity. 

Figure~\ref{fig:source_typing}~
\footnote{Note that lists, such as \(\source{{\overline {\source \tau} }_{\ottmv{i}}}\),
  are denoted by overlines, whereas collections of predicates are annotated by
  their range. For instance,
  \( \repeated{ \source {  {\source { \Gamma_C } }  } ; \source {  {\source \Gamma}  }  \vdash_{ty}  \source {  {\source \tau }_{\ottmv{i}}  }\,{ \stoi{ \rightsquigarrow  \target {  {\target \sigma }_{\ottmv{i}}  } } } }{i} \) iterates over \(i\).}
shows selected typing rules.
The full set of rules can be found in Section~\ref{app:src_tm_typing} of the appendix.
We ignore the red (elaboration-related) parts for now and explain them in
detail in Section~\ref{sec:elabST}.
\renewcommand\stoi[1]{}
\renewcommand\stot[1]{}
The judgments \( \source {  { \source P }  } ; \source {  {\source { \Gamma_C } }  } ; \source {  {\source \Gamma}  }  \vdash_{tm}  \source {  {\source e}  } \Rightarrow \source {  {\source \tau }  } \stot{ \rightsquigarrow  \target {  {\target e }  } } \) and
\( \source {  { \source P }  } ; \source {  {\source { \Gamma_C } }  } ; \source {  {\source \Gamma}  }  \vdash_{tm}  \source {  {\source e}  } \Leftarrow \source {  {\source \tau }  } \stot{ \rightsquigarrow  \target {  {\target e }  } } \) denote \emph{infer}ring a monotype 
\({\source \tau }\) for expression \({\source e}\) and \emph{check}ing \({\source e}\) to have a
monotype \({\source \tau }\) respectively, in environments \({ \source P }\), \({\source { \Gamma_C } }\) and
\({\source \Gamma}\).
Note that the constraint and type well-formedness relations \(\vdash_{Q}\) and
\(\vdash_{ty}\) are omitted, as they are standard
well-scopedness checks.
They can be found in Section~\ref{app:src_ty_wf} of the appendix.

\renewcommand\stoi[1]{{#1}}
\renewcommand\stot[1]{{#1}}

\newcommand{\vrulesep}{\unskip\ \vrule\ }
\newcommand{\hrulesep}{\hrule}
\boxedfigureH
  {
    \input{figures/source_typing.mng}
  }
  {\srcL{} typing, selected rules}
  {fig:source_typing}

\boxedfigureH
  {
    \begin{subfigure}[t]{0.49\textwidth}
      \input{figures/source_typing_extra.mng}
    \end{subfigure}
    \vrulesep
    \begin{subfigure}[t]{0.49\textwidth}
      \input{figures/source_typing_extra2.mng}
    \end{subfigure}
  }
  {Closure and unambiguity relations}
  {fig:source_typing_extra}

\boxedfigureH
  {\input{figures/source_resolution.mng}}
  {\srcL{} constraint entailment}
  {fig:source_entail}

\boxedfigureH
  {\input{figures/source_inst_typing.mng}}
  {\srcL{} instance declaration typing}
  {fig:source_inst}

Through a let binding (rule \textsc{sTm-infT-let}), the programmer provides a
type scheme for a variable, thus potentially introducing local constraints.
As explained above, the unambiguity check from
Figure~\ref{fig:source_typing_extra} (right-hand side) requires the
provided type scheme to be unambiguous.
In order to flatten the superclasses, the rule takes the closure
over the superclass relation (left-hand side of Figure~\ref{fig:source_typing_extra}) of the
user provided constraints \(\source{{\overline {\source Q} }_{\ottmv{i}}}\). It then adds the resulting set of constraints
\(\source{{\overline {\source Q} }_{\ottmv{k}}}\) to the typing environment, under which to typecheck \(\source{{\source e}_{{\mathrm{1}}}}\).
Finally, the type of \(\source{{\source e}_{{\mathrm{2}}}}\) is inferred under the extended environment. 

Rule~\textsc{sTm-checkT-meth} types a method call \({\source m}\) in
check mode, like regular variables, to avoid any ambiguity in the instantiation of the type variables in
the  method's type scheme.
This includes both the type variable \({\source a }\) from the class and any
additional free variables \(\source{{\overline {\source a} }_{\ottmv{j}}}\) in the method type.
Furthermore, the rules uses the unambig-relation to avoid ambiguity in the
method type scheme itself, by requiring that both sets
of type variables have to occur in the head of the method type.
The rule also checks that all required constraints
\(\source{{\overline {\source Q} }_{\ottmv{i}}}\) from the method type can be
entailed.

\renewcommand{\stoi}[1]{}
The instance typing rule can be found in
Figure~\ref{fig:source_inst}. The relation
\( \source {  { \source P }  } ; \source {  {\source { \Gamma_C } }  }  \vdash_{inst}  \source {  {\source {inst} }  } : \source {  { \source P }'  } \) denotes that an instance declaration
\({\source {inst} }\) results in a \srcL{} program context \(\source{{ \source P }'}\),
while being typed under environments \({ \source P }\) and \({\source { \Gamma_C } }\).
The unambig-relation for constraints
(Figure~\ref{fig:source_typing_extra}, bottom right),
similarly to the unambig-relation for types,
checks that all free type variables \(\source{{\overline {\source b} }_{\ottmv{k}}}\) in the instance context
occur in the instance type \({\source \tau }\) as well, in order to avoid ambiguity.
Like in the \textsc{sTm-infT-let} rule explained above, the
superclasses of the instance context \(\source{{\overline {\source Q} }_{\ottmv{p}}}\) are flattened into additional local
constraints \(\source{{\overline {\source Q} }_{\ottmv{q}}}\) and added to the environment.
The superclasses \(\source{{\overline {\source Q} }'_{\ottmv{i}}}\) of the instantiated type
class are then checked to be entailed under this extended environment.
The rule checks that no overlapping instance declarations \(\source{{\source D }'}\) have been defined.
Finally, the program context is extended with the new instance axiom \({\source D }\), consisting
of a constraint scheme that requires the full set of local constraints \(\source{{\overline {\source Q} }_{\ottmv{q}}}\).

\paragraph{\textbf{Type Class Resolution.}}
\renewcommand\stoi[1]{}
\renewcommand\stot[1]{}
The type class resolution rules can be found in
Figure~\ref{fig:source_entail}, where
\( \source {  { \source P }  } ; \source {  {\source { \Gamma_C } }  } ; \source {  {\source \Gamma}  }  \vDash  \source {  {\source Q}  } \stot{ \rightsquigarrow  \target {  {\target e }  } } \) denotes that a class constraint \({\source Q}\)
is entailed under the environments \({ \source P }\), \({\source { \Gamma_C } }\) and \({\source \Gamma}\).
A wanted constraint \({\source Q}\) can either be resolved
using a locally available constraint \({\source \delta }\) (\textsc{sEntailT-local}) or through a
global instance declaration \({\source D }\)
(\textsc{sEntailT-inst}). 
\renewcommand\stoi[1]{{#1}}
\renewcommand\stot[1]{{#1}}
The former is entirely straightforward. The latter is more involved as
an instance \({\source D }\) may have an instance context \(\source{{\overline {\source Q} }'_{\ottmv{i}}}\), which has to be
recursively resolved. Before resolving the context, the type variables
\(\source{{\overline {\source a} }_{\ottmv{j}}}\) are instantiated with the
corresponding concrete types \(\source{{\overline {\source \tau} }_{\ottmv{j}}}\), originating from the wanted
constraint \({\source Q}\).

Note that the type class resolution mechanism does not require any specific
support for superclasses, as these have all been flattened into regular local
constraints.

\section{Target Language \targetL{}}
\label{sec:target}

This section covers our target language \targetL{}, and the elaboration from
\srcL{} to \targetL{}. 

The target language is System F with records, which we consider a
reasonable subcalculus of those used by
Haskell compilers. Its syntax is shown in Figure~\ref{fig:target_syntax}.
We omit its standard typing rules and call-by-name operational semantics and refer the reader
to \citet[Chapter 23]{TAPL}, or Section~\ref{app:tgt} of the appendix.

\boxedfigure
  {\input{figures/target_syntax.mng}}
  {Target language syntax}
  {fig:target_syntax}

\subsection{Elaboration from \srcL{} to \targetL{}}
\label{sec:elabST}

The red aspects in Figure~\ref{fig:source_typing} denote the elaboration of
\srcL{} terms to \targetL{}. We have adopted the convention
that any red \targetL{} types are the elaborated forms of their
identically named blue \srcL{} counterparts. 
This elaboration maps most \srcL{} forms on identical \targetL{} terms,
with the exception of a few notable cases:
\begin{inparaenum}[(a)]
\item The interesting aspect of elaborating let expressions
  (\textsc{sTm-infT-let}) is that, as mentioned previously, superclasses are
  flattened into additional local constraints.
  The elaborated expression thus explicitly requires both the type
  variables and the full closure of the local constraints.
\item As opposed to \srcL{}, dictionary and type application are made explicit in
  \targetL{}. When elaborating variables \({\source x}\) and method references
  \({\source m}\) (\textsc{sTm-checkT-meth}), all previously substituted types
  \(\source{{\overline {\source \tau} }_{\ottmv{j}}}\) are now explicitly applied, together with the dictionary expressions \(\target{{\overline {\target e} }_{\ottmv{i}}}\).
  Furthermore, method names \({\source m}\) are elaborated to \targetL{} record
  labels \({\target m}\) and therefore cannot appear by themselves, but must be
  applied to a record expression \({\target e }\), which originates from resolving the
  class constraint.
\end{inparaenum}

Type class resolution (Figure~\ref{fig:source_entail}) of a \srcL{}
constraint \({\source Q}\) results in a \targetL{} expression \({\target e }\).
When resolving the wanted constraint using a locally available constraint \({\source \delta }\)
(\textsc{sEntailT-local}), this results in a regular term variable \({\target \delta }\)
(which keeps the name of its \srcL{} counterpart for readability).
On the other hand, when resolving with the use of a global instance declaration \({\source D }\) (\textsc{sEntailT-inst}), a record
expression is constructed, containing the method name \({\target m}\) and its corresponding
implementation \({\target e }\). This method implementation now explicitly abstracts over the
type variables \(\target{{\overline {\target b} }_{\ottmv{k}}}\) and term variables \(\target{{\overline {\target \delta} }_{\ottmv{h}}}\)
originating from the method types's class constraints \(\source{{\overline {\source Q} }_{\ottmv{h}}}\), which
annotate the class declaration.
Furthermore, the record expression is nested in abstractions
over the type variables \(\target{{\overline {\target a} }_{\ottmv{j}}}\) and term variables \(\target{{\overline {\target \delta} }'_{\ottmv{i}}}\) arising from the corresponding instance
constraints \(\source{{\overline {\source Q} }'_{\ottmv{i}}}\).
These abstractions are immediately instantiated by applying
\begin{inparaenum}[(a)]
\item the types \(\target{{\overline {\target \sigma} }_{\ottmv{j}}}\) needed for matching the wanted constraint \({\source Q}\) to the instance
  declaration \(\source{{\source Q}'}\) and
\item the expressions \(\target{{\overline {\target e} }_{\ottmv{i}}}\) constructed by resolving the
  instance context constraints \(\source{{\overline {\source Q} }'_{\ottmv{i}}}\).
\end{inparaenum}

\paragraph{\textbf{Example~\ref{fig:ex1} \srcL{} to \targetL{}.}}
Typing the Example~\ref{fig:ex1} program results in the following
environments:
\begin{fleqn}
  \begin{align*}
    \qquad {\source { \Gamma_C } } &= \source { {\source {(==)} }  \ottsym{:}  {\source {Eq} } \, {\source a }  \ottsym{:}  {\source a }  \rightarrow  {\source a }  \rightarrow  \mathit{\source {Bool} } }
  \end{align*}
\end{fleqn}
\begin{fleqn}
  \begin{align*}
    \qquad { \source P } =~&\source { \ottsym{(}  {\source D }_{{\mathrm{1}}}  \ottsym{:}  {\source {Eq} } \, { \source {Int} }  \ottsym{)}  \ottsym{.}  {\source {(==)} }  \mapsto  \bullet  \ottsym{:}   {\source {primEqInt} }  }
    \\ \source { , }~& \source { \ottsym{(}  {\source D }_{{\mathrm{2}}}  \ottsym{:}  \forall {\source a }  \ottsym{,}  {\source b }  \ottsym{.}  {\source {Eq} } \, {\source a }  \Rightarrow  {\source {Eq} } \, {\source b }  \Rightarrow  {\source {Eq} } \, \ottsym{(}  {\source a }  \ottsym{,}  {\source b }  \ottsym{)}  \ottsym{)}  \ottsym{.}  {\source {(==)} }  \mapsto {\source a }  \ottsym{,}  {\source b }  \ottsym{,}  {\source \delta }_{{\mathrm{1}}}  \ottsym{:}  {\source {Eq} } \, {\source a }  \ottsym{,}  {\source \delta }_{{\mathrm{2}}}  \ottsym{:}  {\source {Eq} } \, {\source b }  \ottsym{:} } \\
    &\quad \source { \lambda  \ottsym{(}  {\source x}_{{\mathrm{1}}}  \ottsym{,}  {\source y}_{{\mathrm{1}}}  \ottsym{)}  \ottsym{.}  \lambda  \ottsym{(}  {\source x}_{{\mathrm{2}}}  \ottsym{,}  {\source y}_{{\mathrm{2}}}  \ottsym{)}  \ottsym{.}~{\source {(\&\&)} }  \, \ottsym{(}  {\source {(==)} } \, {\source x}_{{\mathrm{1}}}  \, {\source x}_{{\mathrm{2}}}  \ottsym{)} \, \ottsym{(}  {\source {(==)} } \, {\source y}_{{\mathrm{1}}}  \, {\source y}_{{\mathrm{2}}}  \ottsym{)} }
  \end{align*}
\end{fleqn}
 
The \({\source {Eq} }\)~class straightforwardly gets stored in the class environment
\({\source { \Gamma_C } }\). Instances are stored in the \srcL{} program context \({ \source P }\)
(containing the dictionary constructor, the corresponding constraint, the method
implementation and the environment under which to interpret this expression).
Storing the instance declaration for \({\source {Eq} } \, { \source {Int} }\) is
clear-cut. The instance for tuples on the other hand is somewhat more complex, since it
requires an instance context, containing the local constraints \({\source {Eq} } \, {\source a }\) and
\({\source {Eq} } \, {\source b }\). These constraints are made explicit, that is, the
corresponding dictionaries are required by the elaborated
implementation.

Elaborating the \srcL{} expression results in the following \targetL{}
expression:
\begin{fleqn}
\begin{align*}
  \qquad &\target { \textbf{\target {let} } \, {\target {\mathit {refl} } } }
    \begin{aligned}[t]
      & \target { \ottsym{:}  \forall  {\target a }  \ottsym{.}  \ottsym{\{}  {\target {(==)} }  \ottsym{:}  {\target a }  \rightarrow  {\target a }  \rightarrow  \mathit{\target {Bool} }  \ottsym{\}}  \rightarrow  {\target a }  \rightarrow  \mathit{\target {Bool} } } \\
    & \target{ \ottsym{=}  \Lambda  {\target a }  \ottsym{.}  \lambda  {\target \delta }_{{\mathrm{3}}}  \ottsym{:}  \ottsym{\{}  {\target {(==)} }  \ottsym{:}  {\target a }  \rightarrow  {\target a }  \rightarrow  \mathit{\target {Bool} }  \ottsym{\}}  \ottsym{.}  \lambda  {\target x}  \ottsym{:}  {\target a }  \ottsym{.} {\target \delta }_{{\mathrm{3}}}  \ottsym{.}  {\target {(==)} }  \, {\target x} \, {\target x} \, }
    \end{aligned} \\
  & \target{ \textbf{\target {in} } \, \textbf{\target {let} } \, {\target {main} } }
    \begin{aligned}[t]
      & \target{ \ottsym{:}  \mathit{\target {Bool} } } \\
      & \target { \ottsym{=}  {\target {\mathit {refl} } } \, \ottsym{(}  { \target {Int} }  \ottsym{,}  { \target {Int} }  \ottsym{)} \, } \\
      &\quad \target { \ottsym{(}  \Lambda  {\target a }  \ottsym{.}  \Lambda  {\target b }  \ottsym{.}  \lambda  {\target \delta }_{{\mathrm{4}}}  \ottsym{:}  \ottsym{\{}  {\target {(==)} }  \ottsym{:}  {\target a }  \rightarrow  {\target a }  \rightarrow  \mathit{\target {Bool} }  \ottsym{\}}  \ottsym{.}  \lambda  {\target \delta }_{{\mathrm{5}}}  \ottsym{:}  \ottsym{\{}  {\target {(==)} }  \ottsym{:}  {\target b }  \rightarrow  {\target b }  \rightarrow  \mathit{\target {Bool} }  \ottsym{\}}  \ottsym{.} } \\
      &\qquad \target{ \ottsym{\{}  {\target {(==)} }  \ottsym{=}  \lambda  \ottsym{(}  {\target x}_{{\mathrm{1}}}  \ottsym{,}  {\target y}_{{\mathrm{1}}}  \ottsym{)}  \ottsym{:}  \ottsym{(}  {\target a }  \ottsym{,}  {\target b }  \ottsym{)}  \ottsym{.}  \lambda  \ottsym{(}  {\target x}_{{\mathrm{2}}}  \ottsym{,}  {\target y}_{{\mathrm{2}}}  \ottsym{)}  \ottsym{:}  \ottsym{(}  {\target a }  \ottsym{,}  {\target b }  \ottsym{)}  \ottsym{.} } \\
      &\qquad \quad \target{ {\target {(\&\&)} }  \, \ottsym{(}  {\target \delta }_{{\mathrm{4}}}  \ottsym{.}  {\target {(==)} }  \, {\target x}_{{\mathrm{1}}}  \, {\target x}_{{\mathrm{2}}} \ottsym{)} \, \ottsym{(}  {\target \delta }_{{\mathrm{5}}}  \ottsym{.}  {\target {(==)} }  \, {\target y}_{{\mathrm{1}}}  \, {\target y}_{{\mathrm{2}}}  \ottsym{)}  \ottsym{\}}  \ottsym{)} \, } \\
      &\quad \target{ { \target {Int} }  \, { \target {Int} }  \, \ottsym{\{}  {\target {(==)} }  \ottsym{=}   {\target {primEqInt} }   \ottsym{\}}  \, \ottsym{\{}  {\target {(==)} }  \ottsym{=}   {\target {primEqInt} }   \ottsym{\}}  \, \ottsym{(}  \ottsym{5}  \ottsym{,}  \ottsym{42}  \ottsym{)} \, }
    \end{aligned} \\
  & \target{ \textbf{\target {in} } \, {\target {main} } }
\end{align*}
\end{fleqn}
 
Note that the \({\source {Eq} } \, {\source a }\) constraint is made explicit in the implementation
of \({\target {\mathit {refl} } }\), by abstracting over the constraint
(elaborated to \targetL{} as the record type
\( \target {\ottsym{\{}  {\target {(==)} }  \ottsym{:}  {\target a }  \rightarrow  {\target a }  \rightarrow  \mathit{\target {Bool} }  \ottsym{\}}}\), which stores the method name and its
corresponding type)
with the use of the record variable \(\target{{\target \delta }_{{\mathrm{3}}}}\). 
When this function is called in \({\target {main} }\), both the type and the
dictionary variable are instantiated. The latter is performed by (recursively) constructing a
dictionary expression, using the type class resolution mechanism, as explained
above in Section~\ref{sec:elabST}.

\paragraph{\textbf{Example~\ref{fig:ex2} \srcL{} to \targetL{}.}}
Below is the environment generated by typing the Example~\ref{fig:ex2} program (including
the Section~\ref{sec:nondet} extension), which features superclasses.
\begin{fleqn}
  \begin{align*}
    \qquad {\source { \Gamma_C } } =~&\source { {\source {base} }  \ottsym{:}  {\source {Base} } \, {\source a }  \ottsym{:}  {\source a }  \rightarrow  \mathit{\source {Bool} } }
    \\ \source {,}~& \source{ {\source {sub} }_{{\mathrm{1}}}  \ottsym{:}  {\source {Base} } \, {\source a }  \Rightarrow  {\source {Sub} }_{{\mathrm{1}}} \, {\source a }  \ottsym{:}  {\source a }  \rightarrow  \mathit{\source {Bool} } }
    \\ \source {,}~& \source{ {\source {sub} }_{{\mathrm{2}}}  \ottsym{:}  {\source {Base} } \, {\source a }  \Rightarrow  {\source {Sub} }_{{\mathrm{2}}} \, {\source a }  \ottsym{:}  {\source a }  \rightarrow  \mathit{\source {Bool} } }
  \end{align*}
\end{fleqn}

The class environment \({\source { \Gamma_C } }\) contains three classes, two of which have superclasses.
However, since the example does not contain any instance declarations, the resulting
program context \({ \source P }\) is empty.

For space reasons, we focus solely on elaborating \texttt{test2}, which results in
the following \targetL{} expression:

\begin{fleqn}
\begin{align*}
  \qquad &\target{ \textbf{\target {let} } \, {\target {test} }_{{\mathrm{2}}}  }
    \begin{aligned}[t]
      &\target{ \ottsym{:} \forall  {\target a }  \ottsym{.}  \ottsym{\{}  {\target {base} }  \ottsym{:}  {\target a }  \rightarrow  \mathit{\target {Bool} }  \ottsym{\}}  \rightarrow  \ottsym{\{}  {\target {sub} }_{{\mathrm{1}}}  \ottsym{:}  {\target a }  \rightarrow  \mathit{\target {Bool} }  \ottsym{\}}  } \\
      &\target{ \quad \rightarrow  \ottsym{\{}  {\target {base} }  \ottsym{:}  {\target a }  \rightarrow  \mathit{\target {Bool} }  \ottsym{\}}  \rightarrow  \ottsym{\{}  {\target {sub} }_{{\mathrm{2}}}  \ottsym{:}  {\target a }  \rightarrow  \mathit{\target {Bool} }  \ottsym{\}}  \rightarrow  {\target a }  \rightarrow  \mathit{\target {Bool} } } \\
      &\target{ \ottsym{=}  \Lambda  {\target a }  \ottsym{.} \lambda  {\target \delta }_{{\mathrm{1}}}  \ottsym{:}  \ottsym{\{}  {\target {base} }  \ottsym{:}  {\target a }  \rightarrow  \mathit{\target {Bool} }  \ottsym{\}}  \ottsym{.}  \lambda  {\target \delta }_{{\mathrm{2}}}  \ottsym{:}  \ottsym{\{}  {\target {sub} }_{{\mathrm{1}}}  \ottsym{:}  {\target a }  \rightarrow  \mathit{\target {Bool} }  \ottsym{\}}  \ottsym{.} } \\
      &\target{ \quad \lambda  {\target \delta }_{{\mathrm{3}}}  \ottsym{:}  \ottsym{\{}  {\target {base} }  \ottsym{:}  {\target a }  \rightarrow  \mathit{\target {Bool} }  \ottsym{\}}  \ottsym{.}  \lambda  {\target \delta }_{{\mathrm{4}}}  \ottsym{:}  \ottsym{\{}  {\target {sub} }_{{\mathrm{2}}}  \ottsym{:}  {\target a }  \rightarrow  \mathit{\target {Bool} }  \ottsym{\}}  \ottsym{.} \lambda  {\target x}  \ottsym{:}  {\target a }  \ottsym{.}~{\target \delta }_{{\mathrm{1}}}  \ottsym{.}  {\target {base} } \, {\target x} \, } \\
    \end{aligned} \\
  &\target{ \textbf{\target {in} } \, \mathit{\target {True} } }
\end{align*}
\end{fleqn}

Note that the \srcL{} expression requires two local constraints:
\(\source{{\source {Sub} }_{{\mathrm{1}}} \, {\source a }}\) and \(\source{{\source {Sub} }_{{\mathrm{2}}} \, {\source a }}\).
However, after flattening the superclasses and adding them to the local
constraints, the elaborated \targetL{} expression requires (the elaborated form
of) the
\(\source{{\source {Base} } \, {\source a }}\), \(\source{{\source {Sub} }_{{\mathrm{1}}} \, {\source a }}\), \(\source{{\source {Base} } \, {\source a }}\) and
\(\source{{\source {Sub} }_{{\mathrm{2}}} \, {\source a }}\) constraints. Notice the duplicate \(\source{{\source {Base} } \, {\source a }}\) entry.
Either of these two entries can be used for calling
the method \({\target {base} }\). 
We have arbitrarily selected the first here. 
The next section proves that both options are equivalent
and can be used interchangeably.

\section{Coherence}
\label{sec:coherence}
\renewcommand\stoi[1]{{#1}} 
\renewcommand\itot[1]{}

This section provides an outline for our coherence proof, and defines the
required notions.
We first provide a definition of
\textit{contextual equivalence}~\citep{morris1969lambda}, which captures that
two expressions have the same meaning.

\subsection{Contextual Equivalence}

In order to formally discuss the concept of contextual equivalence, we first define the notion of an
\textit{expression context}.

\paragraph{\textbf{Expression Contexts.}}
An expression context ${\source M}$ is an expression
with a single hole, for which another expression \({\source e}\) can be filled in,
denoted as \(\source{{\source M} [ {\source e} ]}\).
The syntax can be found in Figure~\ref{fig:source_syntax}.

The typing judgment for an expression context \({\source M}\) is of the form
$ \source {  {\source M}  } : ( \source {  { \source P }  } ; \source {  {\source { \Gamma_C } }  } ; \source {  {\source \Gamma}  }  \Rightarrow  \source {  {\source \tau }  } )  \mapsto  ( \source {  { \source P }  } ; \source {  {\source { \Gamma_C } }  } ; \source {  {\source \Gamma}'  }  \Rightarrow  \source {  {\source \tau }'  } )  \rightsquigarrow  \target {  {\target M }  } $.
This means that for any
expression \({\source e}\) such that $ \source {  { \source P }  } ; \source {  {\source { \Gamma_C } }  } ; \source {  {\source \Gamma}  }  \vdash_{tm}  \source {  {\source e}  } \Rightarrow \source {  {\source \tau }  } \stot{ \rightsquigarrow  \target {  {\target e }  } } $,
we have $ \source {  { \source P }  } ; \source {  {\source { \Gamma_C } }  } ; \source {  {\source \Gamma}'  }  \vdash_{tm}  \source {   {\source M}  [  {\source e}  ]   } \Rightarrow \source {  {\source \tau }'  } \stot{ \rightsquigarrow  \target {  {\target e }'  } } $.
Following regular \srcL{} term typing, context typing spans all combinations
of type inference and checking mode:
$ \source {  {\source M}  } : ( \source {  { \source P }  } ; \source {  {\source { \Gamma_C } }  } ; \source {  {\source \Gamma}  }  \Rightarrow  \source {  {\source \tau }  } )  \mapsto  ( \source {  { \source P }  } ; \source {  {\source { \Gamma_C } }  } ; \source {  {\source \Gamma}'  }  \Leftarrow  \source {  {\source \tau }'  } )  \rightsquigarrow  \target {  {\target M }  } $, 
$ \source {  {\source M}  } : ( \source {  { \source P }  } ; \source {  {\source { \Gamma_C } }  } ; \source {  {\source \Gamma}  }  \Leftarrow  \source {  {\source \tau }  } )  \mapsto  ( \source {  { \source P }  } ; \source {  {\source { \Gamma_C } }  } ; \source {  {\source \Gamma}'  }  \Rightarrow  \source {  {\source \tau }'  } )  \rightsquigarrow  \target {  {\target M }  } $ and
$ \source {  {\source M}  } : ( \source {  { \source P }  } ; \source {  {\source { \Gamma_C } }  } ; \source {  {\source \Gamma}  }  \Leftarrow  \source {  {\source \tau }  } )  \mapsto  ( \source {  { \source P }  } ; \source {  {\source { \Gamma_C } }  } ; \source {  {\source \Gamma}'  }  \Leftarrow  \source {  {\source \tau }'  } )  \rightsquigarrow  \target {  {\target M }  } $.

For example, the simplest expression context is the
empty context $ \source {  \ottsym{[} \, \bullet \, \ottsym{]}  } : ( \source {  { \source P }  } ; \source {  {\source { \Gamma_C } }  } ; \source {  {\source \Gamma}  }  \Rightarrow  \source {  {\source \tau }  } )  \mapsto  ( \source {  { \source P }  } ; \source {  {\source { \Gamma_C } }  } ; \source {  {\source \Gamma}  }  \Rightarrow  \source {  {\source \tau }  } )  \rightsquigarrow  \target {  \ottsym{[} \, \target {\bullet} \, \ottsym{]}  } $.

Now we can formally define contextual equivalence. Note that the small step
operational semantics can be found in Section~\ref{app:tgt_op} of the appendix.
The environment and type well-formedness judgments can be found in
Sections~\ref{app:src_env_wf} and~\ref{app:src_ty_wf} of the appendix respectively.
\begin{definition}[Kleene Equivalence]\leavevmode \\
  Two \targetL{} expressions \gsp $\target{{\target e }_{{\mathrm{1}}}}$ \gsp and \gsp
  $\target{{\target e }_{{\mathrm{2}}}}$ \gsp are
  Kleene equivalent, written \gsp $ \target {  {\target e }_{{\mathrm{1}}}  }  \simeq  \target {  {\target e }_{{\mathrm{2}}}  } $, \\ if there exists 
    a value \gsp ${\target v}$ \gsp such that \gsp $ \target {  {\target e }_{{\mathrm{1}}}  }  \longrightarrow^{*}  \target {  {\target v}  } $, and \gsp $ \target {  {\target e }_{{\mathrm{2}}}  }  \longrightarrow^{*}  \target {  {\target v}  } $.
  \label{def:kleene_tgt}
\end{definition}

\begin{definition}[Contextual Equivalence]\leavevmode \\
  Two expressions \gsp
  $ \target {  {\target \Gamma }  }  \vdash_{tm}  \target {  {\target e }_{{\mathrm{1}}}  } : \target {  {\target \sigma }  } $ \gsp and \gsp $ \target {  {\target \Gamma }  }  \vdash_{tm}  \target {  {\target e }_{{\mathrm{2}}}  } : \target {  {\target \sigma }  } $, \\ where \gsp
  $ \vdash_{ctx}  \source {  { \source P }  } ; \source {  {\source { \Gamma_C } }  } ; \source {  {\source \Gamma}  }  \rightsquigarrow  \target {  {\target \Gamma }  } $ \gsp and \gsp $ \source {  {\source { \Gamma_C } }  } ; \source {  {\source \Gamma}  }  \vdash_{ty}  \source {  {\source \tau }  }\,{ \stoi{ \rightsquigarrow  \target {  {\target \sigma }  } } } $, \\
  are contextually equivalent, written \gsp
  $ \source {  { \source P }  } ; \source {  {\source { \Gamma_C } }  } ; \source {  {\source \Gamma}  }  \vdash  \target {  {\target e }_{{\mathrm{1}}}  }  \simeq_{ctx}  \target {  {\target e }_{{\mathrm{2}}}  } : \source {  {\source \tau }  } $, \\ if forall \gsp
  $ \source {  {\source M}  } : ( \source {  { \source P }  } ; \source {  {\source { \Gamma_C } }  } ; \source {  {\source \Gamma}  }  \Rightarrow  \source {  {\source \tau }  } )  \mapsto  ( \source {  { \source P }  } ; \source {  {\source { \Gamma_C } }  } ; \source {  \bullet  }  \Rightarrow  \source {  \mathit{\source {Bool} }  } )  \rightsquigarrow  \target {  {\target M }_{{\mathrm{1}}}  } $ \\
  and \gsp $ \source {  {\source M}  } : ( \source {  { \source P }  } ; \source {  {\source { \Gamma_C } }  } ; \source {  {\source \Gamma}  }  \Rightarrow  \source {  {\source \tau }  } )  \mapsto  ( \source {  { \source P }  } ; \source {  {\source { \Gamma_C } }  } ; \source {  \bullet  }  \Rightarrow  \source {  \mathit{\source {Bool} }  } )  \rightsquigarrow  \target {  {\target M }_{{\mathrm{2}}}  } $ \\
  implies \gsp $ \target {   {\target M }_{{\mathrm{1}}}  [  {\target e }_{{\mathrm{1}}}  ]   }  \simeq  \target {   {\target M }_{{\mathrm{2}}}  [  {\target e }_{{\mathrm{2}}}  ]   } $.
  \label{def:ctx_tgt}
\end{definition}

The definition is adapted from~\citet[Chapter 46]{PFPL46}.
Intuitively, contextual equivalence means that two open expressions are
observationally indistinguishable, when used in any program that instantiates
the expressions' free variables.

\subsection{Coherence}

We can now make a first attempt to prove 
that different translations of the same
source program are contextually equivalent. The program typing judgment
can be found in Section~\ref{app:src_tm_typing} of the appendix.

\begin{theoremB}[Coherence]\leavevmode \\
  If \gsp \( \source {  \bullet  } ; \source {  \bullet  }  \vdash_{pgm}  \source {  {\source {pgm} }  } : \source {  {\source \tau }  } ; \source {  { \source P }_{{\mathrm{1}}}  } ; \source {  {\source { \Gamma_C } }_{{\mathrm{1}}}  }  \rightsquigarrow  \target {  {\target e }_{{\mathrm{1}}}  } \) \gsp
  and \gsp \( \source {  \bullet  } ; \source {  \bullet  }  \vdash_{pgm}  \source {  {\source {pgm} }  } : \source {  {\source \tau }  } ; \source {  { \source P }_{{\mathrm{2}}}  } ; \source {  {\source { \Gamma_C } }_{{\mathrm{2}}}  }  \rightsquigarrow  \target {  {\target e }_{{\mathrm{2}}}  } \) \\
  then \gsp \( \source {  {\source { \Gamma_C } }_{{\mathrm{1}}}  } = \source {  {\source { \Gamma_C } }_{{\mathrm{2}}}  } \), \( \source {  { \source P }_{{\mathrm{1}}}  } = \source {  { \source P }_{{\mathrm{2}}}  } \) \gsp
  and \gsp \( \source {  { \source P }_{{\mathrm{1}}}  } ; \source {  {\source { \Gamma_C } }_{{\mathrm{1}}}  } ; \source {  \bullet  }  \vdash  \target {  {\target e }_{{\mathrm{1}}}  }  \simeq_{ctx}  \target {  {\target e }_{{\mathrm{2}}}  } : \source {  {\source \tau }  } \).
  \label{thm:coh}
\end{theoremB}

We first set out to prove the simpler variant, which only considers
expressions~\footnote{Theorem~\ref{thm:ecoh} also has a type checking mode
  counterpart, which has been omitted here for space reasons.}.

\begin{theoremB}[Expression Coherence]\leavevmode \\
  If \gsp \( \source {  { \source P }  } ; \source {  {\source { \Gamma_C } }  } ; \source {  {\source \Gamma}  }  \vdash_{tm}  \source {  {\source e}  } \Rightarrow \source {  {\source \tau }  } \stot{ \rightsquigarrow  \target {  {\target e }_{{\mathrm{1}}}  } } \) \gsp
  and \gsp \( \source {  { \source P }  } ; \source {  {\source { \Gamma_C } }  } ; \source {  {\source \Gamma}  }  \vdash_{tm}  \source {  {\source e}  } \Rightarrow \source {  {\source \tau }  } \stot{ \rightsquigarrow  \target {  {\target e }_{{\mathrm{2}}}  } } \) \\
  then \gsp \( \source {  { \source P }  } ; \source {  {\source { \Gamma_C } }  } ; \source {  {\source \Gamma}  }  \vdash  \target {  {\target e }_{{\mathrm{1}}}  }  \simeq_{ctx}  \target {  {\target e }_{{\mathrm{2}}}  } : \source {  {\source \tau }  } \).
  \label{thm:ecoh}
\end{theoremB}

The main requirement which makes type class resolution coherent is that type
class instances do not overlap. However, since \targetL{} uses records to encode
dictionaries, the \targetL{} language does not enforce this crucial uniqueness property.
In order to prove Theorem~\ref{thm:coh}, we introduce an
additional intermediate language \interL{}, which captures the invariant that
type class instances do not overlap, and makes it explicit. 

\renewcommand\itot[1]{#1}


\section{Intermediate Language \interL{}}\label{sec:inter}

This section presents our intermediate language \interL{}.
The language is modeled with three main goals in mind:
\begin{inparaenum}[(a)]
\item \interL{} should explicitly pass type class dictionaries, which are 
      implicit in \srcL;
\item the \interL{} type system should capture the uniqueness of
  dictionaries, thus enforcing the elaboration from
  \srcL{} to preserve full abstraction; and
\item \interL{} expressions should elaborate straightforwardly and
  deterministically to the target language \targetL{} (System F with records, see
  Section~\ref{sec:target}).
\end{inparaenum}

To this end, \interL{} is an extension of System F, with built-in support for
dictionaries. These dictionaries differ from those commonly used in
Haskell compilers in that they are special constants rather than a record of
method implementations. A separate global map \({\intermed \Sigma}\) from dictionaries
to method implementations gives access to the latter. Note that this setup does
not allow programs to introduce new (and possibly overlapping) dictionaries dynamically.
All dictionaries have to be provided upfront, where uniqueness is easily enforced.

\paragraph{\textbf{Syntax.}}
Figure~\ref{fig:inter_syntax} shows selected syntax of \interL{};
the basic System F constructs are omitted and can be found in 
Section~\ref{app:mid_syntax} of the appendix.

\boxedfigure
  {\input{figures/inter_syntax.mng}}
  {\interL{}, selected syntax}
  {fig:inter_syntax}

\interL{} introduces a new syntactic sort of 
dictionaries \({\intermed d}\) that can either be a dictionary variable
\({\intermed \delta }\) or a dictionary constructor \({\intermed D }\). A dictionary
constructor has a number (possibly zero) of type and dictionary parameters and always
appears in fully-applied form.
Each constructor corresponds to a unique instance declaration, and is
mapped to its method implementation by the global environment \({\intermed \Sigma}\).

Expressions have explicit application and abstraction forms for dictionaries.
Furthermore, similarly to \targetL{}, method names can no longer
be used on their own. Instead, they have to be applied explicitly to a
dictionary, in the form \({\intermed d}  \ottsym{.}  {\intermed m}\).

\interL{} types \({\intermed \sigma}\) or \({\intermed \tau}\) are identical to the well-known
System F types, with the addition of a special function type \(\intermed{ {\intermed Q}  \Rightarrow  {\intermed \sigma} }\)
for dictionary abstractions. 

Similarly to \srcL{}, \interL{} features two global
and a single local environment \({\intermed \Gamma}\). The latter is similar to the \srcL{}
typing environment \({\source \Gamma}\). However, there are notable differences between
the global environments. The \interL{} class environment \({\intermed { \Gamma_C } }\)
does not contain any superclass information.
The reason for this is that, as previously mentioned in Section~\ref{sec:source},
superclass constraints in the source language \srcL{} are flattened into local
constraints, and stored in the typing environment \({\source \Gamma}\).
The analog to the \srcL{} program context
\({ \source P }\) is the \interL{} method environment \({\intermed \Sigma}\), storing
information about all dictionary constructors \({\intermed D }\).
Each constructor corresponds to a unique instance declaration,
and stores the accompanying method implementations.

\paragraph{\textbf{Typing.}}
\renewcommand\itot[1]{} 
Figure~\ref{fig:inter_typing} (left-hand side) shows selected typing rules for
\interL{} expressions.
The red parts can be safely ignored
for now, as they will be explained in detail in Section~\ref{sec:elabIT}.
The judgment \( \intermed {  {\intermed \Sigma}  } ; \intermed {  {\intermed { \Gamma_C } }  } ; \intermed {  {\intermed \Gamma}  }  \vdash_{tm}  \intermed {  {\intermed e}  } : \intermed {  {\intermed \sigma}  } \itot{ \rightsquigarrow  \target {  {\target e }  } } \) expresses that the \interL{}
term \({\intermed e}\) of type \({\intermed \sigma}\) is well-typed under environments \({\intermed \Sigma}\),
\({\intermed { \Gamma_C } }\) and \({\intermed \Gamma}\).
As shown by rule~\textsc{iTm-method}, the type of a method variable applied to a
dictionary is simply the corresponding method type (as stored in the
static class environment), where the type variable has been substituted for the
corresponding dictionary type.
\renewcommand\itot[1]{#1} 

\boxedfigure
{
  \begin{subfigure}[t]{0.49\textwidth}
    \vskip 0pt
    \input{figures/inter_typing.mng}
  \end{subfigure}
  \vrulesep
  \begin{subfigure}[t]{0.49\textwidth}
    \vskip 0pt
    \input{figures/target_Q.mng}
    \hrulesep
    \input{figures/inter_eval.mng}
  \end{subfigure}
}
{\interL{} typing and operational semantics, selected rules}
{fig:inter_typing}

\boxedfigure
  {\input{figures/inter_envwf.mng}}
  {\interL{} environment well-formedness, selected rules}
  {fig:inter_env_wf}

\boxedfigure
  {\input{figures/inter_typing_dict.mng}}
  {\interL{} dictionary typing}
  {fig:inter_dict_typing}

\renewcommand\itot[1]{} 
Figure~\ref{fig:inter_dict_typing} shows the typing rules for
dictionaries. The relation
\( \intermed {  {\intermed \Sigma}  } ; \intermed {  {\intermed { \Gamma_C } }  } ; \intermed {  {\intermed \Gamma}  }  \vdash_{d}  \intermed {  {\intermed d}  } : \intermed {  {\intermed Q}  } \itot{\,  \rightsquigarrow  \target {  {\target e }  } } \) denotes that dictionary \({\intermed d}\) of
dictionary type \({\intermed Q}\) is well-formed
under environments \({\intermed \Sigma}\), \({\intermed { \Gamma_C } }\) and \({\intermed \Gamma}\).
Similarly to regular term variables \({\intermed x}\) (\textsc{iTm-var}),
the type of a dictionary variable \({\intermed \delta }\) (\textsc{D-var}) is obtained
from the typing environment \({\intermed \Gamma}\).
The type of a dictionary constructor \({\intermed D }\) (\textsc{D-con}), on the other
hand, is obtained by finding the corresponding entry in the method environment \({\intermed \Sigma}\) and
substituting any types \(\intermed{\overline {\intermed \sigma}_{\ottmv{j}}}\) applied to it in the
corresponding class constraint \(\intermed{{\intermed {\mathit{TC} } } \, {\intermed \sigma}_{\ottmv{q}}}\). All
applied dictionaries \(\intermed{{\overline {\intermed d} }_{\ottmv{i}}}\) have to be well-typed with the corresponding
constraint. Finally, the corresponding method implementation has to be
well-typed in the reduced method environment \(\intermed{{\intermed \Sigma}_{{\mathrm{1}}}}\), which only contains
the instances declared before \({\intermed D }\). As mentioned in
Section~\ref{sec:source}, this reduced environment disallows
recursive method implementations, as this would significantly clutter the
coherence proof while, as a feature, recursion is completely orthogonal to the desired property.
\renewcommand\itot[1]{#1} 

\paragraph{\textbf{Non-Overlapping Instances.}}
The main requirement for achieving coherence of type class resolution, is that
type class instances do not overlap. This requirement is common in Haskell and
is for example enforced in GHC (though strongly discouraged, 
the \texttt{OverlappingInstances} pragma disables it).
By storing all method implementations (with their corresponding instances) in a
single environment \({\intermed \Sigma}\), this invariant can easily be made explicit.

Figure~\ref{fig:inter_env_wf} shows the
environment well-formedness condition \( \vdash_{ctx}  \intermed {  {\intermed \Sigma}  } ; \intermed {  {\intermed { \Gamma_C } }  } ; \intermed {  {\intermed \Gamma}  } \) for the method environment.
Besides stating well-scopedness, it denotes that the method
environment \({\intermed \Sigma}\) cannot contain a second instance, for which the head of
the constraint overlaps with \(\intermed{{\intermed {\mathit{TC} } } \, {\intermed \sigma}}\), up to renaming.
This key property will be exploited in our coherence proof. 

\paragraph{\textbf{Operational Semantics.}}
As \interL{} is an extension of System F, its call-by-name operational
semantics  are mostly standard. The non-standard rules can be found
in Figure~\ref{fig:inter_typing} (bottom right), where \( \intermed {  {\intermed \Sigma}  }  \vdash  \intermed {  {\intermed e}  }  \longrightarrow  \intermed {  {\intermed e}'  } \) denotes
expression \({\intermed e}\) evaluating to \(\intermed{{\intermed e}'}\) in a single step, under method
environment \({\intermed \Sigma}\).

The evaluation rules for dictionary application (\textsc{iEval-DApp} and
\textsc{iEval-DAppAbs}) are identical to those for term and type application.
More interesting, however, is the evaluation for methods (\textsc{iEval-method}).
A method name applied to a dictionary evaluates in one step to the method
implementation, as stored in the environment \({\intermed \Sigma}\). 

\renewcommand\itot[1]{} 
\paragraph{\textbf{Metatheory.}}
\interL{} is type safe. That is, the
common progress and preservation properties hold:
\begin{theoremB}[Progress]\leavevmode \\
  If \gsp $ \intermed {  {\intermed \Sigma}  } ; \intermed {  \bullet  } ; \intermed {  \bullet  }  \vdash_{tm}  \intermed {  {\intermed e}  } : \intermed {  {\intermed \sigma}  } \itot{ \rightsquigarrow  \target {  {\target e }  } } $,
  then either \gsp ${\intermed e}$ \gsp is a value, 
  or there exists \gsp $\intermed{{\intermed e}'}$ \gsp such that \gsp $ \intermed {  {\intermed \Sigma}  }  \vdash  \intermed {  {\intermed e}  }  \longrightarrow  \intermed {  {\intermed e}'  } $.
\label{thm:progress_expressions}
\end{theoremB}
\begin{theoremB}[Preservation]\leavevmode \\
  If \gsp $ \intermed {  {\intermed \Sigma}  } ; \intermed {  {\intermed { \Gamma_C } }  } ; \intermed {  {\intermed \Gamma}  }  \vdash_{tm}  \intermed {  {\intermed e}  } : \intermed {  {\intermed \sigma}  } \itot{ \rightsquigarrow  \target {  {\target e }  } } $,
  and \gsp $ \intermed {  {\intermed \Sigma}  }  \vdash  \intermed {  {\intermed e}  }  \longrightarrow  \intermed {  {\intermed e}'  } $, 
  then \gsp $ \intermed {  {\intermed \Sigma}  } ; \intermed {  {\intermed { \Gamma_C } }  } ; \intermed {  {\intermed \Gamma}  }  \vdash_{tm}  \intermed {  {\intermed e}'  } : \intermed {  {\intermed \sigma}  } \itot{ \rightsquigarrow  \target {  {\target e }'  } } $.
  \label{thm:intermediate_preservation}
\end{theoremB}
Analogously to \srcL{}, \interL{}
rejects recursive expressions (including mutual recursion and recursive
methods). This allows for a normalizing language, that is,
any well-typed expression evaluates to a value, after a finite number of steps.
Note that since the small step operational semantics are deterministic,
normalization implies strong normalization.
\begin{theoremB}[Strong Normalization]\leavevmode \\
  If \( \intermed {  {\intermed \Sigma}  } ; \intermed {  {\intermed { \Gamma_C } }  } ; \intermed {  \bullet  }  \vdash_{tm}  \intermed {  {\intermed e}  } : \intermed {  {\intermed \sigma}  } \itot{ \rightsquigarrow  \target {  {\target e }  } } \)
  then all possible evaluation derivations for \({\intermed e}\) terminate :
  \( \exists {\intermed v} :  \intermed {  {\intermed \Sigma}  }  \vdash  \intermed {  {\intermed e}  }  \longrightarrow^{*}  \intermed {  {\intermed v}  } \).
  \label{theorem:inter-sn}
\end{theoremB}
The proof follows the familiar structure for proving normalization using logical
relations, as presented by Ahmed~\cite{Ahmed2015}, and can be found in
Section~\ref{app:sn} of the appendix.
\renewcommand\itot[1]{#1} 

\subsection{Elaboration from \srcL{} to \interL{}}
\label{sec:elabSI}

The green aspects in Figure~\ref{fig:sourceM_typing} denote the elaboration of
\srcL{} terms to \interL{}. Similarly to the elaboration from \srcL{} to
\targetL{}, we have adopted the convention
that any green \interL{} types or constraints are the elaborated forms of their
identically named blue \srcL{} counterparts.
This elaboration works analogously to the elaboration from \srcL{} to \targetL{},
as shown in Figure~\ref{fig:source_typing}.
The full set of rules can be found in Section~\ref{app:srcmid_tm_typing} of the appendix.

The only notable case is \textsc{sTm-check-meth}, where the entailment relation
for solving the type class constraint \({\source {\mathit{TC} } } \, {\source \tau }\) now results in a
dictionary \({\intermed d}\). As explained at the start of Section~\ref{sec:inter},
unlike \targetL{},
\interL{} differentiates syntactically between dictionaries and normal
expressions.

\boxedfigure
  {\input{figures/sourceM_typing.mng}}
  {\srcL{} typing with elaboration to \interL{}, selected rules}
  {fig:sourceM_typing}

\boxedfigure
  {\input{figures/sourceM_resolution.mng}}
  {\srcL{} constraint entailment with elaboration to \interL{}}
  {fig:sourceM_entail}

Type class resolution (Figure~\ref{fig:sourceM_entail}) of a \srcL{}
constraint \({\source Q}\) results in a \interL{} dictionary \({\intermed d}\).
When using a locally available constraint to resolve the wanted constraint
(\textsc{sEntail-local}), the corresponding dictionary variable \({\intermed \delta }\) is returned.
On the other hand, when resolving using a global instance declaration (\textsc{sEntail-inst}), a
dictionary is constructed by taking the corresponding constructor
\({\intermed D }\) and applying
\begin{inparaenum}[(a)]
\item the types \(\intermed{\overline {\intermed \sigma}_{\ottmv{j}}}\) needed for matching the wanted constraint to the instance
  declaration and
\item the dictionaries \(\intermed{{\overline {\intermed d} }_{\ottmv{i}}}\), constructed by resolving the
  instance context constraints.
\end{inparaenum}

\renewcommand\itot[1]{} 
\paragraph{\textbf{Metatheory.}}
We discuss the coherence of the elaboration from \srcL{} to \interL{} in detail
in Section~\ref{sec:coherence2}, and mention here
that it is type preserving:
\begin{theoremB}[Typing Preservation - Expressions]\leavevmode \\
If \gsp $ \source {  { \source P }  } ; \source {  {\source { \Gamma_C } }  } ; \source {  {\source \Gamma}  }  \vdash_{tm}^{\intermed {M} }  \source {  {\source e}  } \Rightarrow \source {  {\source \tau }  } \stoi{ \rightsquigarrow  \intermed {  {\intermed e}  } } $,
and \gsp $ \vdash_{ctx}^{\intermed {M} }  \source {  { \source P }  } ; \source {  {\source { \Gamma_C } }  } ; \source {  {\source \Gamma}  }  \rightsquigarrow  \intermed {  {\intermed \Sigma}  } ; \intermed {  {\intermed { \Gamma_C } }  } ; \intermed {  {\intermed \Gamma}  } $,
and \gsp $ \source {  {\source { \Gamma_C } }  } ; \source {  {\source \Gamma}  }  \vdash_{ty}^{\intermed {M} }  \source {  {\source \tau }  }\,{ \stoi{ \rightsquigarrow  \intermed {  {\intermed \sigma}  } } } $, \\
then \gsp $ \intermed {  {\intermed \Sigma}  } ; \intermed {  {\intermed { \Gamma_C } }  } ; \intermed {  {\intermed \Gamma}  }  \vdash_{tm}  \intermed {  {\intermed e}  } : \intermed {  {\intermed \sigma}  } \itot{ \rightsquigarrow  \target {  {\target e }  } } $.
\label{thm:typing_pres_expr}
\end{theoremB}
The same theorem holds for check mode, but is omitted for space
reasons. The full proofs can be found in Section~\ref{app:type_pres} of the appendix.
\renewcommand\itot[1]{#1} 

\paragraph{\textbf{Example~\ref{fig:ex1} \srcL{} to \interL{}.}}
Elaborating the \srcL{} environments that originate from Example~\ref{fig:ex1},
results in the following \interL{} environments:
\begin{fleqn}
  \begin{align*}
    \qquad {\intermed { \Gamma_C } } &= \intermed { {\intermed {(==)} }  \ottsym{:}  {\intermed {Eq} } \, {\intermed a }  \ottsym{:}  {\intermed a }  \rightarrow  {\intermed a }  \rightarrow  \mathit{\intermed {Bool} } }
  \end{align*}
\end{fleqn}
\begin{fleqn}
  \begin{align*}
    \qquad {\intermed \Sigma} =~&\intermed{ \ottsym{(}  {\intermed D }_{{\mathrm{1}}}  \ottsym{:}  {\intermed {Eq} } \, { \intermed {Int} }  \ottsym{)}  \ottsym{.}  {\intermed {(==)} }  \mapsto   {\intermed {primEqInt} } }
    \\ \intermed {,}~& \intermed{ \ottsym{(}  {\intermed D }_{{\mathrm{2}}}  \ottsym{:}  \forall  {\intermed a }  \ottsym{,}  {\intermed b }  \ottsym{.}    {\intermed {Eq} } \, {\intermed a }   \Rightarrow  {\intermed {Eq} } \, {\intermed b }   \Rightarrow  {\intermed {Eq} } \, \ottsym{(}  {\intermed a }  \ottsym{,}  {\intermed b }  \ottsym{)}  \ottsym{)}  \ottsym{.}  {\intermed {(==)} }  \mapsto } \\
                         &\quad \intermed { \Lambda  {\intermed a }  \ottsym{.}  \Lambda  {\intermed b }  \ottsym{.}  \lambda  {\intermed \delta }_{{\mathrm{1}}}  \ottsym{:}  {\intermed {Eq} } \, {\intermed a }  \ottsym{.}  \lambda  {\intermed \delta }_{{\mathrm{2}}}  \ottsym{:}  {\intermed {Eq} } \, {\intermed b }  \ottsym{.}  } \\
                         &\quad \intermed { \lambda  \ottsym{(}  {\intermed x}_{{\mathrm{1}}}  \ottsym{,}  {\intermed y}_{{\mathrm{1}}}  \ottsym{)}  \ottsym{:}  \ottsym{(}  {\intermed a }  \ottsym{,}  {\intermed b }  \ottsym{)}  \ottsym{.}  \lambda  \ottsym{(}  {\intermed x}_{{\mathrm{2}}}  \ottsym{,}  {\intermed y}_{{\mathrm{2}}}  \ottsym{)}  \ottsym{:}  \ottsym{(}  {\intermed a }  \ottsym{,}  {\intermed b }  \ottsym{)}  \ottsym{.} {\intermed {(\&\&)} }  \, \ottsym{(}  {\intermed \delta }_{{\mathrm{1}}}  \ottsym{.}  {\intermed {(==)} } \, {\intermed x}_{{\mathrm{1}}} \, {\intermed x}_{{\mathrm{2}}}  \ottsym{)} \, \ottsym{(}  {\intermed \delta }_{{\mathrm{2}}}  \ottsym{.}  {\intermed {(==)} }  \, {\intermed y}_{{\mathrm{1}}}  \, {\intermed y}_{{\mathrm{2}}}  \ottsym{)} }
  \end{align*}
\end{fleqn}
 
Note that both the class environment \({\intermed { \Gamma_C } }\) and the program context \({\intermed \Sigma}\)
are largely direct translations of their \srcL{} counterparts. One notable
difference is the fact that the environment \({\source \Gamma}\) under which to interpret
the \srcL{} method implementation is now explicitly abstracted over in the
\interL{} method implementation.
Consider for instance the case of \(\intermed{{\intermed D }_{{\mathrm{2}}}}\), where the variables
\({\intermed a }\), \({\intermed b }\), \(\intermed{{\intermed \delta }_{{\mathrm{1}}}}\) and \(\intermed{{\intermed \delta }_{{\mathrm{2}}}}\),
which in \srcL{} are implicitly provided by the typing environment, are now
explicit in the term level. 

Elaborating the \srcL{} expression results in the following \interL{}
expression:
\begin{fleqn}
\begin{align*}
  \qquad &\intermed { \textbf{\intermed {let} } \, {\intermed {\mathit{refl}} }  \ottsym{:}  \forall  {\intermed a }  \ottsym{.}  {\intermed {Eq} } \, {\intermed a }  \Rightarrow  {\intermed a }  \rightarrow  \mathit{\intermed {Bool} } }
  \\ &\qquad \quad \, \, \intermed{ \ottsym{=}  \Lambda  {\intermed a }  \ottsym{.}  \lambda  {\intermed \delta }_{{\mathrm{3}}}  \ottsym{:}  {\intermed {Eq} } \, {\intermed a }  \ottsym{.}  \lambda  {\intermed x}  \ottsym{:}  {\intermed a }  \ottsym{.}  {\intermed \delta }_{{\mathrm{3}}}  \ottsym{.}  {\intermed {(==)} }  \, {\intermed x}  \, {\intermed x} \, }
  \\ &\intermed{ \textbf{\intermed {in} } \, \textbf{\intermed {let} } \, {\intermed {main} }  \ottsym{:}  \mathit{\intermed {Bool} } }
     \\ &\qquad \quad \, \, \intermed{ \ottsym{=}  {\intermed {\mathit{refl}} } \, \ottsym{(}  { \intermed {Int} }  \ottsym{,}  { \intermed {Int} }  \ottsym{)}  \, \ottsym{(}  {\intermed D }_{{\mathrm{2}}} \, { \intermed {Int} } \, { \intermed {Int} }  \, {\intermed D }_{{\mathrm{1}}} \, {\intermed D }_{{\mathrm{1}}} \ottsym{)} \, \ottsym{(}  \ottsym{5}  \ottsym{,}  \ottsym{42}  \ottsym{)} \, }
     \\ &\intermed{ \textbf{\intermed {in} } \, {\intermed {main} } }
\end{align*}
\end{fleqn}

Unlike the corresponding \targetL{} expression, shown in Section~\ref{sec:elabST},
records storing the method types and implementations do not need to be passed
around explicitly.
In \interL{}, they are replaced by class constraints and dictionaries,
respectively. The construction of these dictionaries through type class
resolution is shown in Figure~\ref{fig:sourceM_entail}. 

\paragraph{\textbf{Example~\ref{fig:ex2} \srcL{} to \interL{}.}}
Elaborating Example~\ref{fig:ex2}, including the extension from
Section~\ref{sec:nondet}, results in the following \interL{} class environment
(since no instance declarations exist, the method environment \({\intermed \Sigma}\) remains empty):
\begin{fleqn}
  \begin{align*}
    \qquad {\intermed { \Gamma_C } } =~&\intermed{ {\intermed {base} }  \ottsym{:}  {\intermed {Base} } \, {\intermed a }  \ottsym{:}  {\intermed a }  \rightarrow  \mathit{\intermed {Bool} } }
    \\ \intermed{,}~&\intermed{ {\intermed {sub} }_{{\mathrm{1}}}  \ottsym{:}  {\intermed {Sub} }_{{\mathrm{1}}} \, {\intermed a }  \ottsym{:}  {\intermed a }  \rightarrow  \mathit{\intermed {Bool} } }
    \\ \intermed{,}~&\intermed{ {\intermed {sub} }_{{\mathrm{2}}}  \ottsym{:}  {\intermed {Sub} }_{{\mathrm{2}}} \, {\intermed a }  \ottsym{:}  {\intermed a }  \rightarrow  \mathit{\intermed {Bool} } }
  \end{align*}
\end{fleqn}
 
The \interL{} class environment no longer needs to
store superclasses, as these are all flattened into additional local
constraints during elaboration.

Similarly to Section~\ref{sec:elabST}, we focus solely on elaborating
\texttt{test2}, which results in the following \interL{} expression:

\begin{fleqn}
\begin{align*}
  \qquad \intermed{ \textbf{\intermed {let} } \, {\intermed {test} }_{{\mathrm{2}}} } & \intermed{\ottsym{:}  \forall  {\intermed a }  \ottsym{.}  {\intermed {Base} } \, {\intermed a }  \Rightarrow  {\intermed {Sub} }_{{\mathrm{1}}} \, {\intermed a }  \Rightarrow {\intermed {Base} } \, {\intermed a }  \Rightarrow {\intermed {Sub} }_{{\mathrm{2}}} \, {\intermed a }  \Rightarrow {\intermed a }  \rightarrow  \mathit{\intermed {Bool} } }
  \\ &\intermed{ \ottsym{=}  \Lambda  {\intermed a }  \ottsym{.}  \lambda  {\intermed \delta }_{{\mathrm{1}}}  \ottsym{:}  {\intermed {Base} } \, {\intermed a }  \ottsym{.}  \lambda  {\intermed \delta }_{{\mathrm{2}}}  \ottsym{:}  {\intermed {Sub} }_{{\mathrm{1}}} \, {\intermed a }  \ottsym{.} \lambda  {\intermed \delta }_{{\mathrm{3}}}  \ottsym{:}  {\intermed {Base} } \, {\intermed a }  \ottsym{.} \lambda  {\intermed \delta }_{{\mathrm{4}}}  \ottsym{:}  {\intermed {Sub} }_{{\mathrm{2}}} \, {\intermed a }  \ottsym{.} }
  \\ &\quad \intermed{ \lambda  {\intermed x}  \ottsym{:}  {\intermed a }  \ottsym{.} {\intermed \delta }_{{\mathrm{1}}}  \ottsym{.}  {\intermed {base} }  \, {\intermed x} \, }
  \\ \textbf{\intermed {in} } \, \mathit{\intermed {True} } \, \, &
\end{align*}
\end{fleqn}

The only difference with the \targetL{} elaboration is that we now use class
constraints instead of passing around a record type (storing the method types).

\subsection{Elaboration from \interL{} to \targetL{}}
\label{sec:elabIT}

As both \interL{} and \targetL{} are extensions of System F, the elaboration
from former to latter is mostly trivial, leaving common features unchanged. 
The mapping of \interL{} dictionaries into \targetL{}
records, however, is non-trivial.
Briefly, dictionary types are elaborated into record types, as shown in
Figure~\ref{fig:inter_typing} (top right), and dictionaries into record expressions,
possibly nested within type and term abstractions and applications, as shown in
Figure~\ref{fig:inter_dict_typing}.

In particular, a dictionary type, ${\intermed {\mathit{TC} } }$, which corresponds to a unique entry
$(\intermed{{\intermed m} : {\intermed {\mathit{TC} } } \, {\intermed a } : {\intermed \sigma}'})$ in the class environment ${\intermed { \Gamma_C } }$, elaborates to a record
type whose field has the same name as the dictionary type's method name, ${\intermed m}$, and the type of
that field is determined by the elaboration of $\intermed{{\intermed \sigma}'}$. A ${\intermed {\mathit{TC} } } \, {\intermed \sigma}$ dictionary
elaborates to a record expression which is surrounded, firstly, by abstractions over
type and term variables that arise from the method type's class constraints and, secondly,
by type and term applications that properly instantiate those abstractions.


\paragraph{\textbf{Metatheory.}}
The following theorems confirm that the \interL{}-to-\targetL{} elaboration is
indeed appropriate. 

The first theorem states that a well-typed \interL{} expression always elaborates to
a \targetL{} expression that is also well-typed in the translated context. 
\begin{theoremB}[Type Preservation]\leavevmode \\
  If \gsp $ \intermed {  {\intermed \Sigma}  } ; \intermed {  {\intermed { \Gamma_C } }  } ; \intermed {  {\intermed \Gamma}  }  \vdash_{tm}  \intermed {  {\intermed e}  } : \intermed {  {\intermed \sigma}  } \itot{ \rightsquigarrow  \target {  {\target e }  } } $, \\
  then there are unique \gsp ${\target \Gamma }$ \gsp and \gsp ${\target \sigma }$ \gsp such that \gsp
  $ \target {  {\target \Gamma }  }  \vdash_{tm}  \target {  {\target e }  } : \target {  {\target \sigma }  } $, \\
  where \gsp $ \intermed {  {\intermed { \Gamma_C } }  } ; \intermed {  {\intermed \Gamma}  }  \vdash_{ty}  \intermed {  {\intermed \sigma}  } \itot{ \rightsquigarrow  \target {  {\target \sigma }  } } $ \gsp and \gsp $ \intermed {  {\intermed { \Gamma_C } }  } ; \intermed {  {\intermed \Gamma}  }  \rightsquigarrow  \target {  {\target \Gamma }  } $.
\end{theoremB}
Secondly, and more importantly, the dynamic semantics is also preserved by the
elaboration.
\begin{theoremB}[Semantic Preservation]\leavevmode \\
  If \gsp \( \intermed {  {\intermed \Sigma}  } ; \intermed {  {\intermed { \Gamma_C } }  } ; \intermed {  \bullet  }  \vdash_{tm}  \intermed {  {\intermed e}  } : \intermed {  {\intermed \sigma}  } \itot{ \rightsquigarrow  \target {  {\target e }  } } \) \gsp
  and \gsp \( \intermed {  {\intermed \Sigma}  }  \vdash  \intermed {  {\intermed e}  }  \longrightarrow^{*}  \intermed {  {\intermed v}  } \) \\
  then there exists a \gsp \({\target v}\) \gsp such that \gsp \( \intermed {  {\intermed \Sigma}  } ; \intermed {  {\intermed { \Gamma_C } }  } ; \intermed {  \bullet  }  \vdash_{tm}  \intermed {  {\intermed v}  } : \intermed {  {\intermed \sigma}  } \itot{ \rightsquigarrow  \target {  {\target v}  } } \) \gsp
  and \gsp \( \target {  {\target e }  }  \simeq  \target {  {\target v}  } \).
\end{theoremB}
Thirdly, the elaboration is entirely deterministic.
\begin{theoremB}[Determinism]\leavevmode \\
  \label{thm:mid-tgt-deter}
    If \gsp $ \intermed {  {\intermed \Sigma}  } ; \intermed {  {\intermed { \Gamma_C } }  } ; \intermed {  {\intermed \Gamma}  }  \vdash_{tm}  \intermed {  {\intermed e}  } : \intermed {  {\intermed \sigma}  } \itot{ \rightsquigarrow  \target {  {\target e }_{{\mathrm{1}}}  } } $ \gsp and \gsp $ \intermed {  {\intermed \Sigma}  } ; \intermed {  {\intermed { \Gamma_C } }  } ; \intermed {  {\intermed \Gamma}  }  \vdash_{tm}  \intermed {  {\intermed e}  } : \intermed {  {\intermed \sigma}  } \itot{ \rightsquigarrow  \target {  {\target e }_{{\mathrm{2}}}  } } $,  
    then \gsp $ \target {  {\target e }_{{\mathrm{1}}}  } = \target {  {\target e }_{{\mathrm{2}}}  } $.
\end{theoremB}

\subsection{Elaboration Decomposition}
\renewcommand\itot[1]{{#1}}

An elaboration from \srcL{} to \targetL{} can always be decomposed into two
elaborations through \interL{}.
This intuition is formalized in
Theorems~\ref{thm:equiv_expr} and~\ref{thm:equiv_dict} respectively.

\begin{theoremB}[Elaboration Equivalence - Expressions]\leavevmode \\
  If \gsp \( \source {  { \source P }  } ; \source {  {\source { \Gamma_C } }  } ; \source {  {\source \Gamma}  }  \vdash_{tm}  \source {  {\source e}  } \Rightarrow \source {  {\source \tau }  } \stot{ \rightsquigarrow  \target {  {\target e }  } } \) \gsp
  and \gsp \( \vdash_{ctx}  \source {  { \source P }  } ; \source {  {\source { \Gamma_C } }  } ; \source {  {\source \Gamma}  }  \rightsquigarrow  \target {  {\target \Gamma }  } \) \\
  then \gsp \( \source {  { \source P }  } ; \source {  {\source { \Gamma_C } }  } ; \source {  {\source \Gamma}  }  \vdash_{tm}^{\intermed {M} }  \source {  {\source e}  } \Rightarrow \source {  {\source \tau }  } \stoi{ \rightsquigarrow  \intermed {  {\intermed e}  } } \) \gsp
  and \gsp \( \intermed {  {\intermed \Sigma}  } ; \intermed {  {\intermed { \Gamma_C } }  } ; \intermed {  {\intermed \Gamma}  }  \vdash_{tm}  \intermed {  {\intermed e}  } : \intermed {  {\intermed \sigma}  } \itot{ \rightsquigarrow  \target {  {\target e }  } } \) \\
  where \gsp \( \vdash_{ctx}^{\intermed {M} }  \source {  { \source P }  } ; \source {  {\source { \Gamma_C } }  } ; \source {  {\source \Gamma}  }  \rightsquigarrow  \intermed {  {\intermed \Sigma}  } ; \intermed {  {\intermed { \Gamma_C } }  } ; \intermed {  {\intermed \Gamma}  } \) \gsp
  and \gsp \( \intermed {  {\intermed { \Gamma_C } }  } ; \intermed {  {\intermed \Gamma}  }  \rightsquigarrow  \target {  {\target \Gamma }  } \) \gsp
  and \gsp \( \source {  {\source { \Gamma_C } }  } ; \source {  {\source \Gamma}  }  \vdash_{ty}^{\intermed {M} }  \source {  {\source \tau }  }\,{ \stoi{ \rightsquigarrow  \intermed {  {\intermed \sigma}  } } } \).
  \label{thm:equiv_expr}
\end{theoremB} 

\begin{theoremB}[Elaboration Equivalence - Dictionaries]\leavevmode \\
  If \gsp \( \source {  { \source P }  } ; \source {  {\source { \Gamma_C } }  } ; \source {  {\source \Gamma}  }  \vDash  \source {  {\source Q}  } \stot{ \rightsquigarrow  \target {  {\target e }  } } \) \gsp
  and \gsp \( \vdash_{ctx}  \source {  { \source P }  } ; \source {  {\source { \Gamma_C } }  } ; \source {  {\source \Gamma}  }  \rightsquigarrow  \target {  {\target \Gamma }  } \) \\
  then \gsp \( \source {  { \source P }  } ; \source {  {\source { \Gamma_C } }  } ; \source {  {\source \Gamma}  }  \vDash^{\intermed {M} }  \source {  {\source Q}  } \stoi{ \rightsquigarrow  \intermed {  {\intermed d}  } } \) \gsp
  and \gsp \( \intermed {  {\intermed \Sigma}  } ; \intermed {  {\intermed { \Gamma_C } }  } ; \intermed {  {\intermed \Gamma}  }  \vdash_{d}  \intermed {  {\intermed d}  } : \intermed {  {\intermed Q}  } \itot{\,  \rightsquigarrow  \target {  {\target e }  } } \) \\
  where \gsp \( \vdash_{ctx}^{\intermed {M} }  \source {  { \source P }  } ; \source {  {\source { \Gamma_C } }  } ; \source {  {\source \Gamma}  }  \rightsquigarrow  \intermed {  {\intermed \Sigma}  } ; \intermed {  {\intermed { \Gamma_C } }  } ; \intermed {  {\intermed \Gamma}  } \) \gsp
  and \gsp \( \intermed {  {\intermed { \Gamma_C } }  } ; \intermed {  {\intermed \Gamma}  }  \rightsquigarrow  \target {  {\target \Gamma }  } \).
  \label{thm:equiv_dict}
\end{theoremB} 

Theorem~\ref{thm:equiv_expr} also has a type checking mode counterpart, which
has been omitted for space reasons. The full proofs can be
found in Section~\ref{app:equiv} of the appendix.

\section{Coherence Revisited}
\label{sec:coherence2}
\renewcommand\stoi[1]{{#1}} 
\renewcommand\itot[1]{}

As mentioned previously in Section~\ref{sec:coherence}, the invariant that type
class instances do not overlap is crucial in proving Theorem~\ref{thm:coh}.
This uniqueness property is made explicit in \interL{}.
Our proof thus proceeds by elaborating the \srcL{} expression to two possibly
different \interL{} expressions and subsequently elaborating these \interL{}
expressions to \targetL{} expressions.
Consequently, the proof is split in two main steps.
The first part is the most involved, where we use a technique based on logical
relations to prove that any two \interL{} expressions originating from the same
\srcL{} expression are contextually equivalent.
The second part proves that the elaboration from \interL{} to
\targetL{} is contextual equivalence preserving. This step follows
straightforwardly from the fact that the \interL{}-to-\targetL{} elaboration is deterministic.
Together, these prove that the elaboration from \srcL{} to \targetL{} through
\interL{} is coherent. Theorem~\ref{thm:ecoh} follows from this result,
together with Theorem~\ref{thm:equiv_expr}. 

The remainder of this section explains the techniques we used to prove
Theorem~\ref{thm:coh} in detail.

\subsection{Coherent Elaboration from \srcL{} to \interL{}}

\subsubsection{\textbf{Logical Relations.}}
\renewcommand\itot[1]{}

Logical relations~\citep{plotkin1973lambda, statman1985logical,
tait1967intensional} are key to proving contextual equivalence. In
our type system, the logical relation for expressions is mostly standard,
though the relation for dictionaries is novel.

\paragraph{\textbf{Dictionaries.}}

The logical relation over two open dictionaries is defined by means of an
auxiliary relation on closed dictionaries.
We define this value relation for closed dictionaries as follows.
Note that from now on, we will omit elaborations when they are entirely
irrelevant. The appendix uses the same convention.

\begin{definition}[Value Relation for Dictionaries] \leavevmode \\
  The dictionary values \gsp
  $\intermed{{\intermed D } \, \overline {\intermed \sigma}_{\ottmv{j}} \, {\overline {\intermed { dv } } }_{{\mathrm{1}}\,\ottmv{i}}}$ \gsp
  and \gsp
  $\intermed{{\intermed D } \, \overline {\intermed \sigma}_{\ottmv{j}} \, {\overline {\intermed { dv } } }_{{\mathrm{2}}\,\ottmv{i}}}$ \gsp
  are in the value relation, defined as:

  \begin{flalign*}
    & ( \intermed {  {\intermed \Sigma}_{{\mathrm{1}}}  } : \intermed {  {\intermed D } \, \overline {\intermed \sigma}_{\ottmv{j}} \, {\overline {\intermed { dv } } }_{{\mathrm{1}}\,\ottmv{i}}  } , \intermed {  {\intermed \Sigma}_{{\mathrm{2}}}  } : \intermed {  {\intermed D } \, \overline {\intermed \sigma}_{\ottmv{j}} \, {\overline {\intermed { dv } } }_{{\mathrm{2}}\,\ottmv{i}}  } ) \in \mathcal{V} \llbracket \intermed {  {\intermed Q}  } \rrbracket ^{\intermed {  {\intermed { \Gamma_C } }  } }_{\intermed {  {\intermed R }  } }  \triangleq
    \repeated{ ( \intermed {  {\intermed \Sigma}_{{\mathrm{1}}}  } : \intermed {  {\intermed {dv} }_{{\mathrm{1}}\,\ottmv{i}}  } , \intermed {  {\intermed \Sigma}_{{\mathrm{2}}}  } : \intermed {  {\intermed {dv} }_{{\mathrm{2}}\,\ottmv{i}}  } ) \in \mathcal{V} \llbracket \intermed {  [  \overline {\intermed \sigma}_{\ottmv{j}}  /  {\overline {\intermed a} }_{\ottmv{j}}  ]  {\intermed Q}_{\ottmv{i}}  } \rrbracket ^{\intermed {  {\intermed { \Gamma_C } }  } }_{\intermed {  {\intermed R }  } } }{i} &\\
    &\qquad
      \begin{aligned}
        & \land \gsp  \intermed {  {\intermed \Sigma}_{{\mathrm{1}}}  } ; \intermed {  {\intermed { \Gamma_C } }  } ; \intermed {  \bullet  }  \vdash_{d}  \intermed {  {\intermed D } \, \overline {\intermed \sigma}_{\ottmv{j}} \, {\overline {\intermed { dv } } }_{{\mathrm{1}}\,\ottmv{i}}  } : \intermed {  {\intermed R }  \ottsym{(}  {\intermed Q}  \ottsym{)}  } \itot{\,  \rightsquigarrow  \target {  {\target e }_{{\mathrm{1}}}  } }  \gsp
        \land \gsp  \intermed {  {\intermed \Sigma}_{{\mathrm{2}}}  } ; \intermed {  {\intermed { \Gamma_C } }  } ; \intermed {  \bullet  }  \vdash_{d}  \intermed {  {\intermed D } \, \overline {\intermed \sigma}_{\ottmv{j}} \, {\overline {\intermed { dv } } }_{{\mathrm{2}}\,\ottmv{i}}  } : \intermed {  {\intermed R }  \ottsym{(}  {\intermed Q}  \ottsym{)}  } \itot{\,  \rightsquigarrow  \target {  {\target e }_{{\mathrm{2}}}  } } , \\
        & \textit{where }  ( \intermed {  {\intermed D }  } : \intermed {  \forall  {\overline {\intermed a} }_{\ottmv{j}}  \ottsym{.}  {\overline {\intermed Q} }_{\ottmv{i}}  \Rightarrow  {\intermed Q}'  } ) . \intermed {  {\intermed m}  }  \mapsto  \intermed {  {\intermed e}_{{\mathrm{1}}}  }  \in  \intermed {  {\intermed \Sigma}_{{\mathrm{1}}}  }  
        \land  \intermed {  {\intermed Q}  } = \intermed {  [  \overline {\intermed \sigma}_{\ottmv{j}}  /  {\overline {\intermed a} }_{\ottmv{j}}  ]  {\intermed Q}'  } 
      \end{aligned}
  \end{flalign*}
\end{definition}

The value relation is indexed by the dictionary type \({\intermed Q}\).
We require both dictionaries to be well-typed, and their dictionary arguments to
be in the value relation as well.
The relation has four additional parameters:
the contexts \(\intermed{{\intermed \Sigma}_{{\mathrm{1}}}}\) and \(\intermed{{\intermed \Sigma}_{{\mathrm{2}}}}\), which annotate the dictionaries, the
class environment \({\intermed { \Gamma_C } }\), used in the well-typing condition,
and the type substitution \({\intermed R }\).

In order to define logical equivalence between open dictionaries, we substitute
all free variables with closed terms, thus reducing them to closed dictionary
values. Three kinds of variables exist (term variables \({\intermed x}\), type
variables \({\intermed a }\) and dictionary variables \({\intermed \delta }\)). This results in three
separate semantic interpretations of the typing context \({\intermed \Gamma}\).
The type substitution \( \intermed {  {\intermed R }  } \in \mathcal{F} \llbracket \intermed {  {\intermed \Gamma}  } \rrbracket ^{\intermed {  {\intermed { \Gamma_C } }  } } \) maps all type variables
\( \intermed {  {\intermed a }  }  \in  \intermed {  {\intermed \Gamma}  } \) onto closed types.
\( \intermed {  {\intermed \phi }  } \in \mathcal{G} \llbracket \intermed {  {\intermed \Gamma}  } \rrbracket ^{\intermed {  {\intermed \Sigma}_{{\mathrm{1}}}  } , \intermed {  {\intermed \Sigma}_{{\mathrm{2}}}  } , \intermed {  {\intermed { \Gamma_C } }  } }_{\intermed {  {\intermed R }  } } \) maps each term
variable \( \intermed {  {\intermed x}  }  \in  \intermed {  {\intermed \Gamma}  } \) to two expressions \(\intermed{{\intermed e}_{{\mathrm{1}}}}\) and
\(\intermed{{\intermed e}_{{\mathrm{2}}}}\) that are in the
expression value relation (see Definition~\ref{def:expr_val_rel}),
and \( \intermed {  {\intermed \gamma }  } \in \mathcal{H} \llbracket \intermed {  {\intermed \Gamma}  } \rrbracket ^{\intermed {  {\intermed \Sigma}_{{\mathrm{1}}}  } , \intermed {  {\intermed \Sigma}_{{\mathrm{2}}}  } , \intermed {  {\intermed { \Gamma_C } }  } }_{\intermed {  {\intermed R }  } } \) maps each dictionary
variable \( \intermed {  {\intermed \delta }  }  \in  \intermed {  {\intermed \Gamma}  } \) to two logically related dictionary values.
We use \(\intermed{{\intermed \phi }_{{\mathrm{1}}}}\) and \(\intermed{{\intermed \phi }_{{\mathrm{2}}}}\) to denote the substitution for the first
and second expression, respectively.

\begin{definition}[Logical Equivalence for Dictionaries]\leavevmode \\
  Two dictionaries \gsp
  $\intermed{{\intermed d}_{{\mathrm{1}}}}$ \gsp
  and \gsp
  $\intermed{{\intermed d}_{{\mathrm{2}}}}$ \gsp
  are logically equivalent, defined as:
  \begin{flalign*}
    & \intermed {  {\intermed { \Gamma_C } }  } ; \intermed {  {\intermed \Gamma}  }  \vdash  \intermed {  {\intermed \Sigma}_{{\mathrm{1}}}  } : \intermed {  {\intermed d}_{{\mathrm{1}}}  }  \simeq_{log}  \intermed {  {\intermed \Sigma}_{{\mathrm{2}}}  } : \intermed {  {\intermed d}_{{\mathrm{2}}}  } : \intermed {  {\intermed Q}  } 
    \triangleq \gsp \forall  \intermed {  {\intermed R }  } \in \mathcal{F} \llbracket \intermed {  {\intermed \Gamma}  } \rrbracket ^{\intermed {  {\intermed { \Gamma_C } }  } } , \gsp
        \intermed {  {\intermed \phi }  } \in \mathcal{G} \llbracket \intermed {  {\intermed \Gamma}  } \rrbracket ^{\intermed {  {\intermed \Sigma}_{{\mathrm{1}}}  } , \intermed {  {\intermed \Sigma}_{{\mathrm{2}}}  } , \intermed {  {\intermed { \Gamma_C } }  } }_{\intermed {  {\intermed R }  } } , \gsp
        \intermed {  {\intermed \gamma }  } \in \mathcal{H} \llbracket \intermed {  {\intermed \Gamma}  } \rrbracket ^{\intermed {  {\intermed \Sigma}_{{\mathrm{1}}}  } , \intermed {  {\intermed \Sigma}_{{\mathrm{2}}}  } , \intermed {  {\intermed { \Gamma_C } }  } }_{\intermed {  {\intermed R }  } }  : &\\
    &\qquad  ( \intermed {  {\intermed \Sigma}_{{\mathrm{1}}}  } : \intermed {   {\intermed \gamma }_{{\mathrm{1}}}  ( {\intermed \phi }_{{\mathrm{1}}}  ( {\intermed R }  ( {\intermed d}_{{\mathrm{1}}} )))   } , \intermed {  {\intermed \Sigma}_{{\mathrm{2}}}  } : \intermed {   {\intermed \gamma }_{{\mathrm{2}}}  ( {\intermed \phi }_{{\mathrm{2}}}  ( {\intermed R }  ( {\intermed d}_{{\mathrm{2}}} )))   } ) \in \mathcal{V} \llbracket \intermed {  {\intermed Q}  } \rrbracket ^{\intermed {  {\intermed { \Gamma_C } }  } }_{\intermed {  {\intermed R }  } } 
  \end{flalign*}
\end{definition}

Two dictionaries \(\intermed{{\intermed d}_{{\mathrm{1}}}}\) and \(\intermed{{\intermed d}_{{\mathrm{2}}}}\) are logically
equivalent if any substitution of their free variables (with related expressions
/ dictionaries) results in related dictionary values. 

\paragraph{\textbf{Expressions.}}

The value relation for expressions is mostly standard, with two notable
deviations. Firstly, the relation is defined over two different method
environments \(\intermed{{\intermed \Sigma}_{{\mathrm{1}}}}\) and \(\intermed{{\intermed \Sigma}_{{\mathrm{2}}}}\).
Hence, both expressions are annotated with their respective
environment. Secondly, the dictionary abstraction case is novel.

\begin{definition}[Value Relation for Expressions]\leavevmode \\
  Two values \gsp
  $\intermed{{\intermed v}_{{\mathrm{1}}}}$ \gsp
  and \gsp
  $\intermed{{\intermed v}_{{\mathrm{2}}}}$ \gsp
  are in the value relation, defined as:
  \begin{flalign*}
    & ( \intermed {  {\intermed \Sigma}_{{\mathrm{1}}}  } : \intermed {  \mathit{\intermed {True} }  } , \intermed {  {\intermed \Sigma}_{{\mathrm{2}}}  } : \intermed {  \mathit{\intermed {True} }  } ) \in \mathcal{V} \llbracket \intermed {  \mathit{\intermed {Bool} }  } \rrbracket ^{\intermed {  {\intermed { \Gamma_C } }  } }_{\intermed {  {\intermed R }  } }  &\\
    & ( \intermed {  {\intermed \Sigma}_{{\mathrm{1}}}  } : \intermed {  \mathit{\intermed {False} }  } , \intermed {  {\intermed \Sigma}_{{\mathrm{2}}}  } : \intermed {  \mathit{\intermed {False} }  } ) \in \mathcal{V} \llbracket \intermed {  \mathit{\intermed {Bool} }  } \rrbracket ^{\intermed {  {\intermed { \Gamma_C } }  } }_{\intermed {  {\intermed R }  } }  &\\
    & ( \intermed {  {\intermed \Sigma}_{{\mathrm{1}}}  } : \intermed {  {\intermed v}_{{\mathrm{1}}}  } , \intermed {  {\intermed \Sigma}_{{\mathrm{2}}}  } : \intermed {  {\intermed v}_{{\mathrm{2}}}  } ) \in \mathcal{V} \llbracket \intermed {  {\intermed a }  } \rrbracket ^{\intermed {  {\intermed { \Gamma_C } }  } }_{\intermed {  {\intermed R }  } } 
    \triangleq &\\
    &\qquad  ( \intermed {  {\intermed a }  }  \mapsto  ( \intermed {  {\intermed \sigma}  } , \intermed {  { \intermed { \mathit{ r } } }  } ) )  \in  \intermed {  {\intermed R }  }  \gsp
    \land \gsp  ( \intermed {  {\intermed v}_{{\mathrm{1}}}  } , \intermed {  {\intermed v}_{{\mathrm{2}}}  } )  \in  \intermed {  { \intermed { \mathit{ r } } }  }  \gsp
    \land \gsp  \intermed {  {\intermed \Sigma}_{{\mathrm{1}}}  } ; \intermed {  {\intermed { \Gamma_C } }  } ; \intermed {  \bullet  }  \vdash_{tm}  \intermed {  {\intermed v}_{{\mathrm{1}}}  } : \intermed {  {\intermed \sigma}  } \itot{ \rightsquigarrow  \target {  {\target e }_{{\mathrm{1}}}  } }  \gsp
      \land \gsp  \intermed {  {\intermed \Sigma}_{{\mathrm{2}}}  } ; \intermed {  {\intermed { \Gamma_C } }  } ; \intermed {  \bullet  }  \vdash_{tm}  \intermed {  {\intermed v}_{{\mathrm{2}}}  } : \intermed {  {\intermed \sigma}  } \itot{ \rightsquigarrow  \target {  {\target e }_{{\mathrm{2}}}  } }  &\\
    & ( \intermed {  {\intermed \Sigma}_{{\mathrm{1}}}  } : \intermed {  \lambda  {\intermed x}  \ottsym{:}  {\intermed \sigma}_{{\mathrm{1}}}  \ottsym{.}  {\intermed e}_{{\mathrm{1}}}  } , \intermed {  {\intermed \Sigma}_{{\mathrm{2}}}  } : \intermed {  \lambda  {\intermed x}  \ottsym{:}  {\intermed \sigma}_{{\mathrm{1}}}  \ottsym{.}  {\intermed e}_{{\mathrm{2}}}  } ) \in \mathcal{V} \llbracket \intermed {  {\intermed \sigma}_{{\mathrm{1}}}  \rightarrow  {\intermed \sigma}_{{\mathrm{2}}}  } \rrbracket ^{\intermed {  {\intermed { \Gamma_C } }  } }_{\intermed {  {\intermed R }  } } 
    \triangleq &\\
    &\qquad  \intermed {  {\intermed \Sigma}_{{\mathrm{1}}}  } ; \intermed {  {\intermed { \Gamma_C } }  } ; \intermed {  \bullet  }  \vdash_{tm}  \intermed {  \lambda  {\intermed x}  \ottsym{:}  {\intermed \sigma}  \ottsym{.}  {\intermed e}_{{\mathrm{1}}}  } : \intermed {  {\intermed R }  \ottsym{(}  {\intermed \sigma}_{{\mathrm{1}}}  \rightarrow  {\intermed \sigma}_{{\mathrm{2}}}  \ottsym{)}  } \itot{ \rightsquigarrow  \target {  {\target e }_{{\mathrm{1}}}  } }  \gsp
    \land \gsp  \intermed {  {\intermed \Sigma}_{{\mathrm{2}}}  } ; \intermed {  {\intermed { \Gamma_C } }  } ; \intermed {  \bullet  }  \vdash_{tm}  \intermed {  \lambda  {\intermed x}  \ottsym{:}  {\intermed \sigma}  \ottsym{.}  {\intermed e}_{{\mathrm{2}}}  } : \intermed {  {\intermed R }  \ottsym{(}  {\intermed \sigma}_{{\mathrm{1}}}  \rightarrow  {\intermed \sigma}_{{\mathrm{2}}}  \ottsym{)}  } \itot{ \rightsquigarrow  \target {  {\target e }_{{\mathrm{2}}}  } }  &\\
    &\qquad \land \forall  ( \intermed {  {\intermed \Sigma}_{{\mathrm{1}}}  } : \intermed {  {\intermed e}_{{\mathrm{3}}}  } , \intermed {  {\intermed \Sigma}_{{\mathrm{2}}}  } : \intermed {  {\intermed e}_{{\mathrm{4}}}  } ) \in \mathcal{E} \llbracket \intermed {  {\intermed \sigma}_{{\mathrm{1}}}  } \rrbracket ^{\intermed {  {\intermed { \Gamma_C } }  } }_{\intermed {  {\intermed R }  } }  :
     ( \intermed {  {\intermed \Sigma}_{{\mathrm{1}}}  } : \intermed {  \ottsym{(}  \lambda  {\intermed x}  \ottsym{:}  {\intermed \sigma}  \ottsym{.}  {\intermed e}_{{\mathrm{1}}}  \ottsym{)} \, {\intermed e}_{{\mathrm{3}}}  } , \intermed {  {\intermed \Sigma}_{{\mathrm{2}}}  } : \intermed {  \ottsym{(}  \lambda  {\intermed x}  \ottsym{:}  {\intermed \sigma}  \ottsym{.}  {\intermed e}_{{\mathrm{2}}}  \ottsym{)} \, {\intermed e}_{{\mathrm{4}}}  } ) \in \mathcal{E} \llbracket \intermed {  {\intermed \sigma}_{{\mathrm{2}}}  } \rrbracket ^{\intermed {  {\intermed { \Gamma_C } }  } }_{\intermed {  {\intermed R }  } }  &\\
    & ( \intermed {  {\intermed \Sigma}_{{\mathrm{1}}}  } : \intermed {  \lambda  {\intermed \delta }  \ottsym{:}  {\intermed Q}  \ottsym{.}  {\intermed e}_{{\mathrm{1}}}  } , \intermed {  {\intermed \Sigma}_{{\mathrm{2}}}  } : \intermed {  \lambda  {\intermed \delta }  \ottsym{:}  {\intermed Q}  \ottsym{.}  {\intermed e}_{{\mathrm{2}}}  } ) \in \mathcal{V} \llbracket \intermed {   {\intermed Q}  \Rightarrow  {\intermed \sigma}   } \rrbracket ^{\intermed {  {\intermed { \Gamma_C } }  } }_{\intermed {  {\intermed R }  } } 
    \triangleq &\\
    &\qquad  \intermed {  {\intermed \Sigma}_{{\mathrm{1}}}  } ; \intermed {  {\intermed { \Gamma_C } }  } ; \intermed {  \bullet  }  \vdash_{tm}  \intermed {  \lambda  {\intermed \delta }  \ottsym{:}  {\intermed Q}  \ottsym{.}  {\intermed e}_{{\mathrm{1}}}  } : \intermed {  {\intermed R }  \ottsym{(}   {\intermed Q}  \Rightarrow  {\intermed \sigma}   \ottsym{)}  } \itot{ \rightsquigarrow  \target {  {\target e }_{{\mathrm{1}}}  } }  \gsp
    \land \gsp  \intermed {  {\intermed \Sigma}_{{\mathrm{2}}}  } ; \intermed {  {\intermed { \Gamma_C } }  } ; \intermed {  \bullet  }  \vdash_{tm}  \intermed {  \lambda  {\intermed \delta }  \ottsym{:}  {\intermed Q}  \ottsym{.}  {\intermed e}_{{\mathrm{2}}}  } : \intermed {  {\intermed R }  \ottsym{(}   {\intermed Q}  \Rightarrow  {\intermed \sigma}   \ottsym{)}  } \itot{ \rightsquigarrow  \target {  {\target e }_{{\mathrm{2}}}  } }  &\\
    &\qquad \land \forall  ( \intermed {  {\intermed \Sigma}_{{\mathrm{1}}}  } : \intermed {  {\intermed {dv} }_{{\mathrm{1}}}  } , \intermed {  {\intermed \Sigma}_{{\mathrm{2}}}  } : \intermed {  {\intermed {dv} }_{{\mathrm{2}}}  } ) \in \mathcal{V} \llbracket \intermed {  {\intermed Q}  } \rrbracket ^{\intermed {  {\intermed { \Gamma_C } }  } }_{\intermed {  {\intermed R }  } }  :
     ( \intermed {  {\intermed \Sigma}_{{\mathrm{1}}}  } : \intermed {  \ottsym{(}  \lambda  {\intermed \delta }  \ottsym{:}  {\intermed Q}  \ottsym{.}  {\intermed e}_{{\mathrm{1}}}  \ottsym{)} \, {\intermed {dv} }_{{\mathrm{1}}}  } , \intermed {  {\intermed \Sigma}_{{\mathrm{2}}}  } : \intermed {  \ottsym{(}  \lambda  {\intermed \delta }  \ottsym{:}  {\intermed Q}  \ottsym{.}  {\intermed e}_{{\mathrm{2}}}  \ottsym{)} \, {\intermed {dv} }_{{\mathrm{2}}}  } ) \in \mathcal{E} \llbracket \intermed {  {\intermed \sigma}  } \rrbracket ^{\intermed {  {\intermed { \Gamma_C } }  } }_{\intermed {  {\intermed R }  } }  &\\
    & ( \intermed {  {\intermed \Sigma}_{{\mathrm{1}}}  } : \intermed {  \Lambda  {\intermed a }  \ottsym{.}  {\intermed e}_{{\mathrm{1}}}  } , \intermed {  {\intermed \Sigma}_{{\mathrm{2}}}  } : \intermed {  \Lambda  {\intermed a }  \ottsym{.}  {\intermed e}_{{\mathrm{2}}}  } ) \in \mathcal{V} \llbracket \intermed {  \forall  {\intermed a }  \ottsym{.}  {\intermed \sigma}  } \rrbracket ^{\intermed {  {\intermed { \Gamma_C } }  } }_{\intermed {  {\intermed R }  } } 
    \triangleq &\\
    &\qquad  \intermed {  {\intermed \Sigma}_{{\mathrm{1}}}  } ; \intermed {  {\intermed { \Gamma_C } }  } ; \intermed {  \bullet  }  \vdash_{tm}  \intermed {  \Lambda  {\intermed a }  \ottsym{.}  {\intermed e}_{{\mathrm{1}}}  } : \intermed {  {\intermed R }  \ottsym{(}  \forall  {\intermed a }  \ottsym{.}  {\intermed \sigma}  \ottsym{)}  } \itot{ \rightsquigarrow  \target {  {\target e }_{{\mathrm{1}}}  } }  \gsp
    \land \gsp  \intermed {  {\intermed \Sigma}_{{\mathrm{2}}}  } ; \intermed {  {\intermed { \Gamma_C } }  } ; \intermed {  \bullet  }  \vdash_{tm}  \intermed {  \Lambda  {\intermed a }  \ottsym{.}  {\intermed e}_{{\mathrm{2}}}  } : \intermed {  {\intermed R }  \ottsym{(}  \forall  {\intermed a }  \ottsym{.}  {\intermed \sigma}  \ottsym{)}  } \itot{ \rightsquigarrow  \target {  {\target e }_{{\mathrm{2}}}  } }  &\\
    &\qquad \land \forall \intermed{{\intermed \sigma}'}, \forall  \intermed {  { \intermed { \mathit{ r } } }  } \in \mathit{Rel} [ \intermed {  {\intermed \sigma}'  } ]  : \gsp
       \intermed {  {\intermed { \Gamma_C } }  } ; \intermed {  \bullet  }  \vdash_{ty}  \intermed {  {\intermed \sigma}'  } \itot{ \rightsquigarrow  \target {  {\target \sigma }'  } }  \Rightarrow
       ( \intermed {  {\intermed \Sigma}_{{\mathrm{1}}}  } : \intermed {  \ottsym{(}  \Lambda  {\intermed a }  \ottsym{.}  {\intermed e}_{{\mathrm{1}}}  \ottsym{)} \, {\intermed \sigma}'  } , \intermed {  {\intermed \Sigma}_{{\mathrm{2}}}  } : \intermed {  \ottsym{(}  \Lambda  {\intermed a }  \ottsym{.}  {\intermed e}_{{\mathrm{2}}}  \ottsym{)} \, {\intermed \sigma}'  } ) \in \mathcal{E} \llbracket \intermed {  {\intermed \sigma}  } \rrbracket ^{\intermed {  {\intermed { \Gamma_C } }  } }_{\intermed {   {\intermed R }  ,  {\intermed a }  \mapsto (  {\intermed \sigma}' ,  { \intermed { \mathit{ r } } }  )   } } 
  \end{flalign*}
  \label{def:expr_val_rel}
\end{definition}

Consider the interesting case of dictionary abstraction. The relation
requires the terms to be
well-typed, and the applications for all related input dictionaries
to be in the expression relation \(\mathcal{E}\). The definition of this \(\mathcal{E}\) relation
is as follows:

\begin{definition}[Expression Relation]\leavevmode \\
  Two expressions \gsp
  $\intermed{{\intermed e}_{{\mathrm{1}}}}$ \gsp
  and \gsp
  $\intermed{{\intermed e}_{{\mathrm{2}}}}$ \gsp
  are in the expression relation, defined as:
  \begin{flalign*}
    & ( \intermed {  {\intermed \Sigma}_{{\mathrm{1}}}  } : \intermed {  {\intermed e}_{{\mathrm{1}}}  } , \intermed {  {\intermed \Sigma}_{{\mathrm{2}}}  } : \intermed {  {\intermed e}_{{\mathrm{2}}}  } ) \in \mathcal{E} \llbracket \intermed {  {\intermed \sigma}  } \rrbracket ^{\intermed {  {\intermed { \Gamma_C } }  } }_{\intermed {  {\intermed R }  } } 
    \triangleq  \intermed {  {\intermed \Sigma}_{{\mathrm{1}}}  } ; \intermed {  {\intermed { \Gamma_C } }  } ; \intermed {  \bullet  }  \vdash_{tm}  \intermed {  {\intermed e}_{{\mathrm{1}}}  } : \intermed {  {\intermed R }  \ottsym{(}  {\intermed \sigma}  \ottsym{)}  } \itot{ \rightsquigarrow  \target {  {\target e }_{{\mathrm{1}}}  } }  \gsp
    \land  \intermed {  {\intermed \Sigma}_{{\mathrm{2}}}  } ; \intermed {  {\intermed { \Gamma_C } }  } ; \intermed {  \bullet  }  \vdash_{tm}  \intermed {  {\intermed e}_{{\mathrm{2}}}  } : \intermed {  {\intermed R }  \ottsym{(}  {\intermed \sigma}  \ottsym{)}  } \itot{ \rightsquigarrow  \target {  {\target e }_{{\mathrm{2}}}  } }  &\\
    &\qquad \land \exists \intermed{{\intermed v}_{{\mathrm{1}}}} , \intermed{{\intermed v}_{{\mathrm{2}}}},
     \intermed {  {\intermed \Sigma}_{{\mathrm{1}}}  }  \vdash  \intermed {  {\intermed e}_{{\mathrm{1}}}  }  \longrightarrow^{*}  \intermed {  {\intermed v}_{{\mathrm{1}}}  } ,
     \intermed {  {\intermed \Sigma}_{{\mathrm{2}}}  }  \vdash  \intermed {  {\intermed e}_{{\mathrm{2}}}  }  \longrightarrow^{*}  \intermed {  {\intermed v}_{{\mathrm{2}}}  } ,
     ( \intermed {  {\intermed \Sigma}_{{\mathrm{1}}}  } : \intermed {  {\intermed v}_{{\mathrm{1}}}  } , \intermed {  {\intermed \Sigma}_{{\mathrm{2}}}  } : \intermed {  {\intermed v}_{{\mathrm{2}}}  } ) \in \mathcal{V} \llbracket \intermed {  {\intermed \sigma}  } \rrbracket ^{\intermed {  {\intermed { \Gamma_C } }  } }_{\intermed {  {\intermed R }  } } 
  \end{flalign*}
\end{definition}

In this definition, expressions are reduced to values and those values must be
in the value relation. This is well-defined because \interL{} is
strongly normalizing~(Theorem~\ref{theorem:inter-sn}).

Finally, we can give the definition of logical equivalence for open expressions:

\begin{definition}[Logical Equivalence for Expressions]\leavevmode \\
  Two expressions \gsp
  $\intermed{{\intermed e}_{{\mathrm{1}}}}$ \gsp
  and \gsp
  $\intermed{{\intermed e}_{{\mathrm{2}}}}$ \gsp
  are logically equivalent, defined as:
  \begin{flalign*}
    & \intermed {  {\intermed { \Gamma_C } }  } ; \intermed {  {\intermed \Gamma}  }  \vdash  \intermed {  {\intermed \Sigma}_{{\mathrm{1}}}  } : \intermed {  {\intermed e}_{{\mathrm{1}}}  }  \simeq_{log}  \intermed {  {\intermed \Sigma}_{{\mathrm{2}}}  } : \intermed {  {\intermed e}_{{\mathrm{2}}}  } : \intermed {  {\intermed \sigma}  } 
    \triangleq \forall  \intermed {  {\intermed R }  } \in \mathcal{F} \llbracket \intermed {  {\intermed \Gamma}  } \rrbracket ^{\intermed {  {\intermed { \Gamma_C } }  } } , \gsp
        \intermed {  {\intermed \phi }  } \in \mathcal{G} \llbracket \intermed {  {\intermed \Gamma}  } \rrbracket ^{\intermed {  {\intermed \Sigma}_{{\mathrm{1}}}  } , \intermed {  {\intermed \Sigma}_{{\mathrm{2}}}  } , \intermed {  {\intermed { \Gamma_C } }  } }_{\intermed {  {\intermed R }  } } , \gsp
        \intermed {  {\intermed \gamma }  } \in \mathcal{H} \llbracket \intermed {  {\intermed \Gamma}  } \rrbracket ^{\intermed {  {\intermed \Sigma}_{{\mathrm{1}}}  } , \intermed {  {\intermed \Sigma}_{{\mathrm{2}}}  } , \intermed {  {\intermed { \Gamma_C } }  } }_{\intermed {  {\intermed R }  } }  : &\\
    &\qquad   ( \intermed {  {\intermed \Sigma}_{{\mathrm{1}}}  } : \intermed {   {\intermed \gamma }_{{\mathrm{1}}}  ( {\intermed \phi }_{{\mathrm{1}}}  ( {\intermed R }  ( {\intermed e}_{{\mathrm{1}}} )))   } , \intermed {  {\intermed \Sigma}_{{\mathrm{2}}}  } : \intermed {   {\intermed \gamma }_{{\mathrm{2}}}  ( {\intermed \phi }_{{\mathrm{2}}}  ( {\intermed R }  ( {\intermed e}_{{\mathrm{2}}} )))   } ) \in \mathcal{E} \llbracket \intermed {  {\intermed \sigma}  } \rrbracket ^{\intermed {  {\intermed { \Gamma_C } }  } }_{\intermed {  {\intermed R }  } } 
  \end{flalign*}
\end{definition}

We also provide a definition of logical equivalence for contexts:

\begin{definition}[Logical Equivalence for Contexts]\leavevmode \\
  \label{def:log_M}
  Two contexts \gsp \(\intermed{{\intermed M}_{{\mathrm{1}}}}\) \gsp and \gsp \(\intermed{{\intermed M}_{{\mathrm{2}}}}\) \gsp
  are logically equivalent, defined as:
  \begin{flalign*}
    & \intermed {  {\intermed \Sigma}_{{\mathrm{1}}}  } : \intermed {  {\intermed M}_{{\mathrm{1}}}  }  \simeq_{log}  \intermed {  {\intermed \Sigma}_{{\mathrm{2}}}  } : \intermed {  {\intermed M}_{{\mathrm{2}}}  } : ( \intermed {  {\intermed { \Gamma_C } }  } ; \intermed {  {\intermed \Gamma}  }  \Rightarrow  \intermed {  {\intermed \sigma}  } )  \mapsto  ( \intermed {  {\intermed { \Gamma_C } }  } ; \intermed {  {\intermed \Gamma}'  }  \Rightarrow  \intermed {  {\intermed \sigma}'  } ) 
    \triangleq &\\
    &\qquad \forall \intermed{{\intermed e}_{{\mathrm{1}}}}, \intermed{{\intermed e}_{{\mathrm{2}}}}:
     \intermed {  {\intermed { \Gamma_C } }  } ; \intermed {  {\intermed \Gamma}  }  \vdash  \intermed {  {\intermed \Sigma}_{{\mathrm{1}}}  } : \intermed {  {\intermed e}_{{\mathrm{1}}}  }  \simeq_{log}  \intermed {  {\intermed \Sigma}_{{\mathrm{2}}}  } : \intermed {  {\intermed e}_{{\mathrm{2}}}  } : \intermed {  {\intermed \sigma}  }  \gsp \Rightarrow \gsp
     \intermed {  {\intermed { \Gamma_C } }  } ; \intermed {  {\intermed \Gamma}'  }  \vdash  \intermed {  {\intermed \Sigma}_{{\mathrm{1}}}  } : \intermed {   {\intermed M}_{{\mathrm{1}}}  [  {\intermed e}_{{\mathrm{1}}}  ]   }  \simeq_{log}  \intermed {  {\intermed \Sigma}_{{\mathrm{2}}}  } : \intermed {   {\intermed M}_{{\mathrm{2}}}  [  {\intermed e}_{{\mathrm{2}}}  ]   } : \intermed {  {\intermed \sigma}'  } 
  \end{flalign*}
\end{definition}

\subsubsection{\textbf{Proof of \srcL{}-to-\interL{} Coherence.}}

With the above definitions we are ready to formally state the metatheory and
establish the coherence theorems from \srcL{} to \interL{}.

\paragraph{\textbf{Design Principle of \interL.}} 

We emphasize that \interL{} captures the intention of type class
instances. Theorem~\ref{thm:dval_rel} states that any two dictionary
values for the same constraint are logically related:

\begin{theoremB}[Value Relation for Dictionary Values]\leavevmode \\
  \label{thm:dval_rel}
  If \gsp \( \intermed {  {\intermed \Sigma}_{{\mathrm{1}}}  } ; \intermed {  {\intermed { \Gamma_C } }  } ; \intermed {  \bullet  }  \vdash_{d}  \intermed {  {\intermed {dv} }_{{\mathrm{1}}}  } : \intermed {  {\intermed Q}  } \itot{\,  \rightsquigarrow  \target {  {\target e }_{{\mathrm{1}}}  } } \) \gsp and \gsp \( \intermed {  {\intermed \Sigma}_{{\mathrm{2}}}  } ; \intermed {  {\intermed { \Gamma_C } }  } ; \intermed {  \bullet  }  \vdash_{d}  \intermed {  {\intermed {dv} }_{{\mathrm{2}}}  } : \intermed {  {\intermed Q}  } \itot{\,  \rightsquigarrow  \target {  {\target e }_{{\mathrm{2}}}  } } \) \gsp
  and \gsp \( \intermed {  {\intermed { \Gamma_C } }  }  \vdash  \intermed {  {\intermed \Sigma}_{{\mathrm{1}}}  }  \simeq_{log}  \intermed {  {\intermed \Sigma}_{{\mathrm{2}}}  } \) \\
  then \gsp \( ( \intermed {  {\intermed \Sigma}_{{\mathrm{1}}}  } : \intermed {  {\intermed {dv} }_{{\mathrm{1}}}  } , \intermed {  {\intermed \Sigma}_{{\mathrm{2}}}  } : \intermed {  {\intermed {dv} }_{{\mathrm{2}}}  } ) \in \mathcal{V} \llbracket \intermed {  {\intermed Q}  } \rrbracket ^{\intermed {  {\intermed { \Gamma_C } }  } }_{\intermed {  \bullet  } } \).
\end{theoremB}

Note that two environments \(\intermed{{\intermed \Sigma}_{{\mathrm{1}}}}\) and \(\intermed{{\intermed \Sigma}_{{\mathrm{2}}}}\) are
logically equivalent under \({\intermed { \Gamma_C } }\), written \( \intermed {  {\intermed { \Gamma_C } }  }  \vdash  \intermed {  {\intermed \Sigma}_{{\mathrm{1}}}  }  \simeq_{log}  \intermed {  {\intermed \Sigma}_{{\mathrm{2}}}  } \),
when they contain the same dictionary constructors and the corresponding method
implementations are logically equivalent. The full definition can be found in
Section~\ref{app:log_env_rel} of the appendix.

\paragraph{\textbf{Coherent Resolution.}} 
\renewcommand\itot[1]{{#1}}

We now prove that constraint resolution is semantically
coherent, that is, if multiple resolutions of the same
constraint exist, they are logically equivalent.

\begin{theoremB}[Logical Coherence of Dictionary Resolution]\leavevmode \\
  If \gsp \( \source {  { \source P }  } ; \source {  {\source { \Gamma_C } }  } ; \source {  {\source \Gamma}  }  \vDash^{\intermed {M} }  \source {  {\source Q}  } \stoi{ \rightsquigarrow  \intermed {  {\intermed d}_{{\mathrm{1}}}  } } \) \gsp
  and \gsp \( \source {  { \source P }  } ; \source {  {\source { \Gamma_C } }  } ; \source {  {\source \Gamma}  }  \vDash^{\intermed {M} }  \source {  {\source Q}  } \stoi{ \rightsquigarrow  \intermed {  {\intermed d}_{{\mathrm{2}}}  } } \) \\
  and \gsp \( \vdash_{ctx}^{\intermed {M} }  \source {  { \source P }  } ; \source {  {\source { \Gamma_C } }  } ; \source {  {\source \Gamma}  }  \rightsquigarrow  \intermed {  {\intermed \Sigma}_{{\mathrm{1}}}  } ; \intermed {  {\intermed { \Gamma_C } }  } ; \intermed {  {\intermed \Gamma}  } \) \gsp
  and \gsp \( \vdash_{ctx}^{\intermed {M} }  \source {  { \source P }  } ; \source {  {\source { \Gamma_C } }  } ; \source {  {\source \Gamma}  }  \rightsquigarrow  \intermed {  {\intermed \Sigma}_{{\mathrm{2}}}  } ; \intermed {  {\intermed { \Gamma_C } }  } ; \intermed {  {\intermed \Gamma}  } \) \\
  then \gsp \( \intermed {  {\intermed { \Gamma_C } }  } ; \intermed {  {\intermed \Gamma}  }  \vdash  \intermed {  {\intermed \Sigma}_{{\mathrm{1}}}  } : \intermed {  {\intermed d}_{{\mathrm{1}}}  }  \simeq_{log}  \intermed {  {\intermed \Sigma}_{{\mathrm{2}}}  } : \intermed {  {\intermed d}_{{\mathrm{2}}}  } : \intermed {  {\intermed Q}  } \) \gsp
  where \gsp \( \source {  {\source { \Gamma_C } }  } ; \source {  {\source \Gamma}  }  \vdash_{\mathit {Q} }^{\intermed {M} }  \source {  {\source Q}  }  \rightsquigarrow  \intermed {  {\intermed Q}  } \).
  \label{thm:dict_cohA}
\end{theoremB}

\paragraph{\textbf{Coherent Elaboration.}} 

Furthermore, in order to prove that the elaboration from \srcL to \interL is coherent,
we show that all elaborations of the same expression are logically
equivalent~\footnote{Theorem~\ref{thm:coh_exprA} also has a type checking mode
  counterpart, which has been omitted here for space reasons.}.

\begin{theoremB}[Logical Coherence of Expression Elaboration]\leavevmode \\
  If \gsp \( \source {  { \source P }  } ; \source {  {\source { \Gamma_C } }  } ; \source {  {\source \Gamma}  }  \vdash_{tm}^{\intermed {M} }  \source {  {\source e}  } \Rightarrow \source {  {\source \tau }  } \stoi{ \rightsquigarrow  \intermed {  {\intermed e}_{{\mathrm{1}}}  } } \) \gsp
  and \gsp \( \source {  { \source P }  } ; \source {  {\source { \Gamma_C } }  } ; \source {  {\source \Gamma}  }  \vdash_{tm}^{\intermed {M} }  \source {  {\source e}  } \Rightarrow \source {  {\source \tau }  } \stoi{ \rightsquigarrow  \intermed {  {\intermed e}_{{\mathrm{2}}}  } } \) \\
  and \gsp \( \vdash_{ctx}^{\intermed {M} }  \source {  { \source P }  } ; \source {  {\source { \Gamma_C } }  } ; \source {  {\source \Gamma}  }  \rightsquigarrow  \intermed {  {\intermed \Sigma}_{{\mathrm{1}}}  } ; \intermed {  {\intermed { \Gamma_C } }  } ; \intermed {  {\intermed \Gamma}  } \) \gsp
  and \gsp \( \vdash_{ctx}^{\intermed {M} }  \source {  { \source P }  } ; \source {  {\source { \Gamma_C } }  } ; \source {  {\source \Gamma}  }  \rightsquigarrow  \intermed {  {\intermed \Sigma}_{{\mathrm{2}}}  } ; \intermed {  {\intermed { \Gamma_C } }  } ; \intermed {  {\intermed \Gamma}  } \) \\
  then \gsp \( \intermed {  {\intermed { \Gamma_C } }  } ; \intermed {  {\intermed \Gamma}  }  \vdash  \intermed {  {\intermed \Sigma}_{{\mathrm{1}}}  } : \intermed {  {\intermed e}_{{\mathrm{1}}}  }  \simeq_{log}  \intermed {  {\intermed \Sigma}_{{\mathrm{2}}}  } : \intermed {  {\intermed e}_{{\mathrm{2}}}  } : \intermed {  {\intermed \sigma}  } \) \gsp
  where \gsp \( \source {  {\source { \Gamma_C } }  } ; \source {  {\source \Gamma}  }  \vdash_{ty}^{\intermed {M} }  \source {  {\source \tau }  }\,{ \stoi{ \rightsquigarrow  \intermed {  {\intermed \sigma}  } } } \).
  \label{thm:coh_exprA}
\end{theoremB}

\paragraph{\textbf{Contextual Equivalence.}}
We prove that all logically equivalent expressions are contextually
equivalent. Together with Theorem~\ref{thm:coh_exprA}, this shows coherence of the
\srcL{}-to-\interL{} elaboration.

We first provide a formal definition of
contextual equivalence for \interL{} expressions.
Kleene equivalence for \interL{} is defined similarly to
Definition~\ref{def:kleene_tgt} and can be found in Section~\ref{app:log_kleene} of the appendix.

\renewcommand\itot[1]{}
\begin{definition}[Contextual Equivalence for \interL{} Expressions]\leavevmode \\
  Two expressions \gsp $ \intermed {  {\intermed \Sigma}_{{\mathrm{1}}}  } ; \intermed {  {\intermed { \Gamma_C } }  } ; \intermed {  {\intermed \Gamma}  }  \vdash_{tm}  \intermed {  {\intermed e}_{{\mathrm{1}}}  } : \intermed {  {\intermed \sigma}  } \itot{ \rightsquigarrow  \target {  {\target e }_{{\mathrm{1}}}  } } $ \gsp
  and \gsp $ \intermed {  {\intermed \Sigma}_{{\mathrm{2}}}  } ; \intermed {  {\intermed { \Gamma_C } }  } ; \intermed {  {\intermed \Gamma}  }  \vdash_{tm}  \intermed {  {\intermed e}_{{\mathrm{2}}}  } : \intermed {  {\intermed \sigma}  } \itot{ \rightsquigarrow  \target {  {\target e }_{{\mathrm{2}}}  } } $, \\
  are contextually equivalent, written 
  \gsp $ \intermed {  {\intermed { \Gamma_C } }  } ; \intermed {  {\intermed \Gamma}  }  \vdash  \intermed {  {\intermed \Sigma}_{{\mathrm{1}}}  } : \intermed {  {\intermed e}_{{\mathrm{1}}}  }  \simeq_{ctx}  \intermed {  {\intermed \Sigma}_{{\mathrm{2}}}  } : \intermed {  {\intermed e}_{{\mathrm{2}}}  } : \intermed {  {\intermed \sigma}  } $, \\
  if forall \gsp $ \intermed {  {\intermed M}_{{\mathrm{1}}}  } : ( \intermed {  {\intermed \Sigma}_{{\mathrm{1}}}  } ; \intermed {  {\intermed { \Gamma_C } }  } ; \intermed {  {\intermed \Gamma}  }  \Rightarrow  \intermed {  {\intermed \sigma}  } )  \mapsto  ( \intermed {  {\intermed \Sigma}_{{\mathrm{1}}}  } ; \intermed {  {\intermed { \Gamma_C } }  } ; \intermed {  \bullet  }  \Rightarrow  \intermed {  \mathit{\intermed {Bool} }  } ) \itot{ \rightsquigarrow  \target {  {\target M }_{{\mathrm{1}}}  } } $ \\
  and forall \gsp $ \intermed {  {\intermed M}_{{\mathrm{2}}}  } : ( \intermed {  {\intermed \Sigma}_{{\mathrm{2}}}  } ; \intermed {  {\intermed { \Gamma_C } }  } ; \intermed {  {\intermed \Gamma}  }  \Rightarrow  \intermed {  {\intermed \sigma}  } )  \mapsto  ( \intermed {  {\intermed \Sigma}_{{\mathrm{2}}}  } ; \intermed {  {\intermed { \Gamma_C } }  } ; \intermed {  \bullet  }  \Rightarrow  \intermed {  \mathit{\intermed {Bool} }  } ) \itot{ \rightsquigarrow  \target {  {\target M }_{{\mathrm{2}}}  } } $ \\
  where \gsp $ \intermed {  {\intermed \Sigma}_{{\mathrm{1}}}  } : \intermed {  {\intermed M}_{{\mathrm{1}}}  }  \simeq_{log}  \intermed {  {\intermed \Sigma}_{{\mathrm{2}}}  } : \intermed {  {\intermed M}_{{\mathrm{2}}}  } : ( \intermed {  {\intermed { \Gamma_C } }  } ; \intermed {  {\intermed \Gamma}  }  \Rightarrow  \intermed {  {\intermed \sigma}  } )  \mapsto  ( \intermed {  {\intermed { \Gamma_C } }  } ; \intermed {  \bullet  }  \Rightarrow  \intermed {  \mathit{\intermed {Bool} }  } ) $, \\
  we have that \gsp $ \intermed {  {\intermed \Sigma}_{{\mathrm{1}}}  } : \intermed {   {\intermed M}_{{\mathrm{1}}}  [  {\intermed e}_{{\mathrm{1}}}  ]   }  \simeq  \intermed {  {\intermed \Sigma}_{{\mathrm{2}}}  } : \intermed {   {\intermed M}_{{\mathrm{2}}}  [  {\intermed e}_{{\mathrm{2}}}  ]   } $.
  \label{def:ctx_mid}
\end{definition}
\renewcommand\itot[1]{#1}

\begin{theoremB}[Logical Equivalence implies Contextual Equivalence]\leavevmode \\
  \label{thm:coh_exprB}
  If \gsp \( \intermed {  {\intermed { \Gamma_C } }  } ; \intermed {  {\intermed \Gamma}  }  \vdash  \intermed {  {\intermed \Sigma}_{{\mathrm{1}}}  } : \intermed {  {\intermed e}_{{\mathrm{1}}}  }  \simeq_{log}  \intermed {  {\intermed \Sigma}_{{\mathrm{2}}}  } : \intermed {  {\intermed e}_{{\mathrm{2}}}  } : \intermed {  {\intermed \sigma}  } \) \gsp
  then \gsp \( \intermed {  {\intermed { \Gamma_C } }  } ; \intermed {  {\intermed \Gamma}  }  \vdash  \intermed {  {\intermed \Sigma}_{{\mathrm{1}}}  } : \intermed {  {\intermed e}_{{\mathrm{1}}}  }  \simeq_{ctx}  \intermed {  {\intermed \Sigma}_{{\mathrm{2}}}  } : \intermed {  {\intermed e}_{{\mathrm{2}}}  } : \intermed {  {\intermed \sigma}  } \).
\end{theoremB}

\subsection{Deterministic Elaboration from \interL{} to \targetL{}}
\subsubsection{\textbf{Contextual Equivalence.}}
\renewcommand\itot[1]{{#1}}

Similarly to expressions, the elaboration from a \srcL{} context \({\source M}\) to a
\targetL{} context \({\target M }\) can always be decomposed into two elaborations,
through a \interL{} context \({\intermed M}\).
The syntax and typing judgments can be found in Sections~\ref{app:syntax} and~\ref{app:context_typing} of
the appendix, respectively. 

We now formally define contextual equivalence for \targetL{} expressions through
\interL{} contexts.
\begin{definition}[Contextual Equivalence in \interL{} Context]\leavevmode \\
  Two expressions \gsp $ \target {  {\target \Gamma }  }  \vdash_{tm}  \target {  {\target e }_{{\mathrm{1}}}  } : \target {  {\target \sigma }  } $ \gsp and \gsp $ \target {  {\target \Gamma }  }  \vdash_{tm}  \target {  {\target e }_{{\mathrm{2}}}  } : \target {  {\target \sigma }  } $, \\
  where \gsp $ \vdash_{ctx}  \source {  { \source P }  } ; \source {  {\source { \Gamma_C } }  } ; \source {  {\source \Gamma}  }  \rightsquigarrow  \target {  {\target \Gamma }  } $ \gsp and \gsp $ \source {  {\source { \Gamma_C } }  } ; \source {  {\source \Gamma}  }  \vdash_{ty}  \source {  {\source \tau }  }\,{ \stoi{ \rightsquigarrow  \target {  {\target \sigma }  } } } $, \\
  are contextually equivalent, written \gsp $ \intermed {  {\intermed { \Gamma_C } }  } ; \intermed {  {\intermed \Gamma}  }  \vdash  \intermed {  {\intermed \Sigma}_{{\mathrm{1}}}  } : \target {  {\target e }_{{\mathrm{1}}}  }  \simeq_{ctx}  \intermed {  {\intermed \Sigma}_{{\mathrm{2}}}  } : \target {  {\target e }_{{\mathrm{2}}}  } : \intermed {  {\intermed \sigma}  } $, \\
  if forall \gsp $  \intermed {  {\intermed M}_{{\mathrm{1}}}  } : ( \intermed {  {\intermed \Sigma}_{{\mathrm{1}}}  } ; \intermed {  {\intermed { \Gamma_C } }  } ; \intermed {  {\intermed \Gamma}  }  \Rightarrow  \intermed {  {\intermed \sigma}  } )  \mapsto  ( \intermed {  {\intermed \Sigma}_{{\mathrm{1}}}  } ; \intermed {  {\intermed { \Gamma_C } }  } ; \intermed {  \bullet  }  \Rightarrow  \intermed {  \mathit{\intermed {Bool} }  } ) \itot{ \rightsquigarrow  \target {  {\target M }_{{\mathrm{1}}}  } } $ \\
  and forall \gsp $  \intermed {  {\intermed M}_{{\mathrm{2}}}  } : ( \intermed {  {\intermed \Sigma}_{{\mathrm{2}}}  } ; \intermed {  {\intermed { \Gamma_C } }  } ; \intermed {  {\intermed \Gamma}  }  \Rightarrow  \intermed {  {\intermed \sigma}  } )  \mapsto  ( \intermed {  {\intermed \Sigma}_{{\mathrm{2}}}  } ; \intermed {  {\intermed { \Gamma_C } }  } ; \intermed {  \bullet  }  \Rightarrow  \intermed {  \mathit{\intermed {Bool} }  } ) \itot{ \rightsquigarrow  \target {  {\target M }_{{\mathrm{2}}}  } } $ \\
  where \gsp $ \intermed {  {\intermed \Sigma}_{{\mathrm{1}}}  } : \intermed {  {\intermed M}_{{\mathrm{1}}}  }  \simeq_{log}  \intermed {  {\intermed \Sigma}_{{\mathrm{2}}}  } : \intermed {  {\intermed M}_{{\mathrm{2}}}  } : ( \intermed {  {\intermed { \Gamma_C } }  } ; \intermed {  {\intermed \Gamma}  }  \Rightarrow  \intermed {  {\intermed \sigma}  } )  \mapsto  ( \intermed {  {\intermed { \Gamma_C } }  } ; \intermed {  \bullet  }  \Rightarrow  \intermed {  \mathit{\intermed {Bool} }  } ) $, \\
  we have that \gsp $ \target {   {\target M }_{{\mathrm{1}}}  [  {\target e }_{{\mathrm{1}}}  ]   }  \simeq  \target {   {\target M }_{{\mathrm{2}}}  [  {\target e }_{{\mathrm{2}}}  ]   } $, \\
  where \gsp $ \vdash_{ctx}^{\intermed {M} }  \source {  { \source P }  } ; \source {  {\source { \Gamma_C } }  } ; \source {  {\source \Gamma}  }  \rightsquigarrow  \intermed {  {\intermed \Sigma}  } ; \intermed {  {\intermed { \Gamma_C } }  } ; \intermed {  {\intermed \Gamma}  } $ \gsp and \gsp $ \intermed {  {\intermed { \Gamma_C } }  } ; \intermed {  {\intermed \Gamma}  }  \rightsquigarrow  \target {  {\target \Gamma }  } $ \gsp and \gsp $ \source {  {\source { \Gamma_C } }  } ; \source {  {\source \Gamma}  }  \vdash_{ty}^{\intermed {M} }  \source {  {\source \tau }  }\,{ \stoi{ \rightsquigarrow  \intermed {  {\intermed \sigma}  } } } $ \gsp and \gsp $ \intermed {  {\intermed { \Gamma_C } }  } ; \intermed {  {\intermed \Gamma}  }  \vdash_{ty}  \intermed {  {\intermed \sigma}  } \itot{ \rightsquigarrow  \target {  {\target \sigma }  } } $. 
  \label{def:ctx_tgt_mid}
\end{definition}

\subsubsection{\textbf{Proof of \interL{}-to-\targetL{} Coherence.}}

We continue by proving that contextual equivalence is preserved by the
elaboration from \interL{} to \targetL{}:

\begin{theoremB}[Elaboration preserves Contextual Equivalence]\leavevmode \\
  If \gsp \( \intermed {  {\intermed { \Gamma_C } }  } ; \intermed {  {\intermed \Gamma}  }  \vdash  \intermed {  {\intermed \Sigma}_{{\mathrm{1}}}  } : \intermed {  {\intermed e}_{{\mathrm{1}}}  }  \simeq_{ctx}  \intermed {  {\intermed \Sigma}_{{\mathrm{2}}}  } : \intermed {  {\intermed e}_{{\mathrm{2}}}  } : \intermed {  {\intermed \sigma}  } \) \\
  and \gsp \( \intermed {  {\intermed \Sigma}_{{\mathrm{1}}}  } ; \intermed {  {\intermed { \Gamma_C } }  } ; \intermed {  {\intermed \Gamma}  }  \vdash_{tm}  \intermed {  {\intermed e}_{{\mathrm{1}}}  } : \intermed {  {\intermed \sigma}  } \itot{ \rightsquigarrow  \target {  {\target e }_{{\mathrm{1}}}  } } \) \gsp
  and \gsp \( \intermed {  {\intermed \Sigma}_{{\mathrm{2}}}  } ; \intermed {  {\intermed { \Gamma_C } }  } ; \intermed {  {\intermed \Gamma}  }  \vdash_{tm}  \intermed {  {\intermed e}_{{\mathrm{2}}}  } : \intermed {  {\intermed \sigma}  } \itot{ \rightsquigarrow  \target {  {\target e }_{{\mathrm{2}}}  } } \) \gsp
  and \gsp \( \intermed {  {\intermed { \Gamma_C } }  } ; \intermed {  {\intermed \Gamma}  }  \rightsquigarrow  \target {  {\target \Gamma }  } \) \\
  then \gsp \( \intermed {  {\intermed { \Gamma_C } }  } ; \intermed {  {\intermed \Gamma}  }  \vdash  \intermed {  {\intermed \Sigma}_{{\mathrm{1}}}  } : \target {  {\target e }_{{\mathrm{1}}}  }  \simeq_{ctx}  \intermed {  {\intermed \Sigma}_{{\mathrm{2}}}  } : \target {  {\target e }_{{\mathrm{2}}}  } : \intermed {  {\intermed \sigma}  } \).
  \label{thm:coh_exprC}
\end{theoremB}

\subsubsection{\textbf{Proof of \srcL{}-to-\targetL{} Coherence.}}

Finally, in order to link back to Theorem~\ref{thm:ecoh} (which has no notion of
\interL{}), we prove that contextual equivalence with \interL{} contexts implies
contextual equivalence with \srcL{} contexts:

\begin{theoremB}[Contextual Equivalence in \interL{} implies Contextual
    Equivalence in \srcL{}]\ \\
  If \gsp \( \intermed {  {\intermed { \Gamma_C } }  } ; \intermed {  {\intermed \Gamma}  }  \vdash  \intermed {  {\intermed \Sigma}_{{\mathrm{1}}}  } : \target {  {\target e }_{{\mathrm{1}}}  }  \simeq_{ctx}  \intermed {  {\intermed \Sigma}_{{\mathrm{2}}}  } : \target {  {\target e }_{{\mathrm{2}}}  } : \intermed {  {\intermed \sigma}  } \) \\
  and \gsp \( \vdash_{ctx}^{\intermed {M} }  \source {  { \source P }  } ; \source {  {\source { \Gamma_C } }  } ; \source {  {\source \Gamma}  }  \rightsquigarrow  \intermed {  {\intermed \Sigma}_{{\mathrm{1}}}  } ; \intermed {  {\intermed { \Gamma_C } }  } ; \intermed {  {\intermed \Gamma}  } \) \gsp
  and \gsp \( \vdash_{ctx}^{\intermed {M} }  \source {  { \source P }  } ; \source {  {\source { \Gamma_C } }  } ; \source {  {\source \Gamma}  }  \rightsquigarrow  \intermed {  {\intermed \Sigma}_{{\mathrm{2}}}  } ; \intermed {  {\intermed { \Gamma_C } }  } ; \intermed {  {\intermed \Gamma}  } \) \gsp
  and \gsp \( \source {  {\source { \Gamma_C } }  } ; \source {  {\source \Gamma}  }  \vdash_{ty}^{\intermed {M} }  \source {  {\source \tau }  }\,{ \stoi{ \rightsquigarrow  \intermed {  {\intermed \sigma}  } } } \) \\
  then \gsp \( \source {  { \source P }  } ; \source {  {\source { \Gamma_C } }  } ; \source {  {\source \Gamma}  }  \vdash  \target {  {\target e }_{{\mathrm{1}}}  }  \simeq_{ctx}  \target {  {\target e }_{{\mathrm{2}}}  } : \source {  {\source \tau }  } \).
  \label{thm:ctx_equiv_pres}
\end{theoremB}

For clarity, we restate coherence Theorems~\ref{thm:ecoh} and~\ref{thm:coh}:

\newtheorem*{thm:ecoh}{Theorem \ref*{thm:ecoh}}
\begin{thm:ecoh}[Expression Coherence - Restated]\leavevmode \\
  If \gsp \( \source {  { \source P }  } ; \source {  {\source { \Gamma_C } }  } ; \source {  {\source \Gamma}  }  \vdash_{tm}  \source {  {\source e}  } \Rightarrow \source {  {\source \tau }  } \stot{ \rightsquigarrow  \target {  {\target e }_{{\mathrm{1}}}  } } \) \gsp
  and \gsp \( \source {  { \source P }  } ; \source {  {\source { \Gamma_C } }  } ; \source {  {\source \Gamma}  }  \vdash_{tm}  \source {  {\source e}  } \Rightarrow \source {  {\source \tau }  } \stot{ \rightsquigarrow  \target {  {\target e }_{{\mathrm{2}}}  } } \) \\
  then \gsp \( \source {  { \source P }  } ; \source {  {\source { \Gamma_C } }  } ; \source {  {\source \Gamma}  }  \vdash  \target {  {\target e }_{{\mathrm{1}}}  }  \simeq_{ctx}  \target {  {\target e }_{{\mathrm{2}}}  } : \source {  {\source \tau }  } \).
\end{thm:ecoh}

Theorem~\ref{thm:ecoh} follows by combining
Theorems~\ref{thm:equiv_expr}, \ref{thm:coh_exprA}, \ref{thm:coh_exprB},
\ref{thm:coh_exprC} and~\ref{thm:ctx_equiv_pres}. 

\newtheorem*{thm:coh}{Theorem \ref*{thm:coh}}
\begin{thm:coh}[Coherence - Restated]\leavevmode \\
  If \gsp \( \source {  \bullet  } ; \source {  \bullet  }  \vdash_{pgm}  \source {  {\source {pgm} }  } : \source {  {\source \tau }  } ; \source {  { \source P }_{{\mathrm{1}}}  } ; \source {  {\source { \Gamma_C } }_{{\mathrm{1}}}  }  \rightsquigarrow  \target {  {\target e }_{{\mathrm{1}}}  } \) \gsp
  and \gsp \( \source {  \bullet  } ; \source {  \bullet  }  \vdash_{pgm}  \source {  {\source {pgm} }  } : \source {  {\source \tau }  } ; \source {  { \source P }_{{\mathrm{2}}}  } ; \source {  {\source { \Gamma_C } }_{{\mathrm{2}}}  }  \rightsquigarrow  \target {  {\target e }_{{\mathrm{2}}}  } \) \\
  then \gsp \( \source {  {\source { \Gamma_C } }_{{\mathrm{1}}}  } = \source {  {\source { \Gamma_C } }_{{\mathrm{2}}}  } \), \( \source {  { \source P }_{{\mathrm{1}}}  } = \source {  { \source P }_{{\mathrm{2}}}  } \) \gsp
  and \gsp \( \source {  { \source P }_{{\mathrm{1}}}  } ; \source {  {\source { \Gamma_C } }_{{\mathrm{1}}}  } ; \source {  \bullet  }  \vdash  \target {  {\target e }_{{\mathrm{1}}}  }  \simeq_{ctx}  \target {  {\target e }_{{\mathrm{2}}}  } : \source {  {\source \tau }  } \).
\end{thm:coh}

Finally, we show that Theorem~\ref{thm:coh} follows from
Theorem~\ref{thm:ecoh}.
The full proofs can be found in Section~\ref{app:coh} of the appendix.

\renewcommand\itot[1]{#1}


\section{Discussion of Possible Extensions}
\label{sec:features}

As the goal of this work was to find a proof technique to formally establish
coherence for type class resolution,
a stripped down source calculus was employed in order not to clutter the proof.
This section provides a brief discussion of extending our
coherence proof to support several mainstream language features.

\paragraph{\textbf{Ambiguous Type Schemes.}}

As mentioned previously, our work is
orthogonal to ambiguous type schemes, which have already been extensively studied
by \citet{jones}. We believe our work and the proof by Jones can be combined,
which would then relax the restriction of bidirectional type checking, and prove
coherence for both ambiguous type schemes and type class resolution.

\paragraph{\textbf{General Recursion.}}

Recursion is an important feature, present in any real world programming
language.
It is important to note that, while \srcL{} does not feature recursion on
the expression level (as it does not affect the essence of the coherence proof),
type class resolution itself is recursive.
Dictionary values are constructed dynamically from a statically given set of
dictionary constructors (one constructor per type class instance). The system
can thus recursively generate an arbitrary number of dictionaries from a finite
set of instances.

Our logical relations can be adapted to support general recursion, through
well-known techniques, such as step indexing~\cite{ahmed2006step}.
While this results in a 
significantly longer and more cluttered proof, we do not anticipate any
major complications.

\paragraph{\textbf{Multi-Parameter Type Classes.}}

Just like regular type class instances, multi-parameter instances (as supported
by GHC) are subject to
the no-overlap rule. Hence, they respect our main assumption. They may indeed
give rise to more ambiguity, but this is the kind of ambiguity that is studied
by \citet{jones}, not the kind that shows up during resolution.
Note that functional dependencies were originally introduced
by~\citet{fundeps-original} as a way to resolve the ambiguity caused by
multi-parameter instances.

\paragraph{\textbf{Dependent Types.}}

Dependently typed languages, e.g., Agda~\citep{Devriese2011} and
Idris~\citep{brady2013idris}, include language features that are inspired by
type classes. Proving resolution coherence in a dependently
typed setting requires significant extension of our calculi, as dependent types
collapse the term and type levels into a single level and thus enable more powerful
type signatures for classes and instances.
Furthermore, our logical relation needs to be
extended to support dependent types~\citep{bernardy2012proofs} as well.
Fortunately, the essence of our proof strategy still applies. That is, the
intermediate language incorporates separate binding structures for dictionaries,
and enforces the uniqueness of dictionaries. We thus believe a non-trivial
extension of our proof methodology can be used to prove coherence for type
class resolution in the setting of dependently typed languages.

\paragraph{\textbf{Non-overlapping Instances.}}

Our work is built on top of the assumption that type class instances do not
overlap. This is enforced during the type checking of instance declarations, and
made explicit in the intermediate language. Whether a constraint is entailed
directly from an instance, through user provided constraints in a type
annotation, or through local evidence, is not actually relevant, as all evidence
ultimately has to originate from a non-overlapping instance declaration.

Therefore, our work can be extended to include features where the assumption
holds true.
This includes, among others, GADT's~\cite{jones_gadts},
implication constraints~\cite{Bottu2017}, type constructors, higher
kinded types and constraint kinds~\cite{constraint_kinds},
e.g., \citet{Bottu2017} informally discuss the coherence of implication
constraints based on the same assumption.
These features are all included in GHC.

\paragraph{\textbf{Modules.}}

Modules, as supported by GHC, pose an interesting challenge, as they are known
to cause a form of
ambiguity.\footnote{\url{http://blog.ezyang.com/2014/07/type-classes-confluence-coherence-global-uniqueness/}}
GHC does not statically check the uniqueness of instances across modules, thus
indirectly allowing users to write overlapping instances, as long as no
ambiguity arises during resolution.
Adapting our global uniqueness assumption to accommodate this additional freedom
remains an interesting challenge.

\paragraph{\textbf{Laziness.}}

The operational semantics of the \interL{} and \targetL{} calculi in this work
are given through standard call-by-name semantics, in order to approximate 
Haskell's laziness. The system can easily be adapted to either call-by-value or
call-by-need, with little impact on the proofs.  

It is important to note though, that while expressions are evaluated lazily, type
class resolution itself is eager, and constructs the full dictionaries at
compile time. This complicates supporting certain GHC features that rely on laziness, like 
cyclic and infinite dictionaries. They could be supported
through loop detection and deferring the construction of
dictionaries to runtime, but these would nonetheless pose a significant challenge.

\section{Related Work}

\paragraph{\textbf{Type Classes.}}

\citet{jones,jones-94} formally proves coherence for the framework of qualified types,
which generalizes from type classes to arbitrary evidence-backed type
constraints. He focuses on nondeterminism in the typing derivation, and assumes
that resolution is coherent.

\citet{DBLP:conf/haskell/Morris14} presents an alternative, denotational
semantics for type classes (without superclasses) that avoids elaboration and
instead interprets qualified type schemes as the set of denotations of all its
monomorphic instantiations that satisfy the qualifiers. The nondeterminism of
resolution does not affect these semantics.

\citet{Kahl-Scheffczyk-2001} present named type class instances that are not
used during resolution, but can be explicitly passed to functions. Nevertheless,
they violate the uniqueness of instances, and give rise to incoherence of the
form illustrated by our \texttt{discern} function in Section~\ref{sec:overview:difference}.

Unlike most other languages with type classes (such as Haskell, Mercury or
PureScript) Coq~\cite{Sozeau2008} does not enforce the non-overlapping instances
condition. Consequently, coherence does not hold for type class resolution in
Coq. The reason for this alternative design choice is twofold:
\begin{inparaenum}[(a)]
\item Since Coq's type system is more complex than that found in regular
  programming languages, it is not always possible to decide whether two
  instances overlap~\cite[Chapter 2: Typeclasses]{SF4}.
\item Type class members in Coq are often proofs and, unlike for expressions,
  users are often indifferent to coherence in the presence of proofs (even
  though from a semantic point of view, Coq differentiates between them).
  This concept is known as
  ``proof irrelevance''~\cite{gilbert}, that is, as long as at least one proof
  exists, the concrete choice between these proofs is irrelevant.
\end{inparaenum}
Users can deal with this lack of coherence by either assigning priorities to
overlapping instances, or by manually curating the instance database and
locally removing specific instances.

\citet{winant} introduce explicit dictionary application to the Haskell
language, and prove coherence for this extended system.
Their proof is parametric in the constraint entailment judgment and thus assumes
that the constraint solver produces ``canonical'' evidence.
They proceed by introducing a disjointness condition to explicitly applied
dictionaries, in order to ensure that coherence is preserved by their extension.
Our paper proves their aforementioned assumption, by establishing coherence for
type class resolution.

\citet{Dreyer:2007:MTC:1190216.1190229} blend ML modules with Haskell type
class resolution. Unlike Haskell, they feature multiple global (or outer)
scopes; instances within one such global scope must not overlap. Moreover,
global instances are shadowed by those given through type signatures.  While
their language has been formalized, no formal proof of coherence is given.

\paragraph{\textbf{Implicits.}}

Cochis~\cite{cochis} is a calculus with highly expressive implicit resolution,
including local instances. It achieves coherence by imposing restrictions on
the implicit context and enforcing a deterministic resolution process. This allows
for a much simpler coherence proof.

OCaml's modular implicits~\cite{DBLP:journals/corr/WhiteBY15} do not enforce 
uniqueness of ``instances'' but dynamically ensure coherence by rejecting
programs where there are multiple possible resolution derivations. This
approach has not been formalized yet.

\paragraph{\textbf{Other.}}

Reynolds~\cite{Reynolds91coherence} introduced the notion of coherence in the
context of the Forsythe language's intersection types; he proved coherence
directly in terms of the denotational semantics of the language.

In contrast,
\citet{ecoop2018,disjoint} consider a setting where subtyping for intersection types is
elaborated to coercions. Inspired by \citet{biernacki2015logical}, they use an approach
based on contextual equivalence and logical relations, which has inspired us in turn.
However, they do not create an intermediate language to avoid the problem of a more
expressive target language. This leads to a notion of
contextual equivalence that straddles two languages and complicates their proofs.

\section{Conclusion}

We have formally proven that type class resolution is coherent by means of
logical relations and an
intermediate language with explicit dictionaries.
In future work we would like to mechanize the proof and adapt it to extensions
such as quantified class constraints and GADT's.

\begin{acks}                            
  This work would not have been possible without the enlightening discussions with
  Dominique Devriese and George Karachalias.
  Furthermore, we would like to thank Alexander Vandenbroucke, Ruben Pieters and Steven Keuchel,
  as well as the anonymous ICFP 2019 and Haskell Symposium 2018 reviewers,
  for their constructive feedback. 
  This research was partially supported by the Flemish Fund for Scientific
  Research (FWO) and the Hong Kong Research
  Grant Council projects number 17210617 and 17258816.
\end{acks}

\bibliography{paper}

\newpage
\appendix

\newtheorem{theoremA}{Theorem}
\newtheorem{lemmaA}{Lemma}
\newtheorem{definitionA}{Definition}
\newtheorem{corollaryA}{Corollary}

\renewenvironment{theoremB}{\begin{leftbar}\begin{theoremA}}{\end{theoremA}\end{leftbar}}
\newenvironment{lemmaB}{\begin{leftbar}\begin{lemmaA}}{\end{lemmaA}\end{leftbar}}
\newenvironment{corollaryB}{\begin{leftbar}\begin{corollaryA}}{\end{corollaryA}\end{leftbar}}

\renewcommand\stoi[1]{#1}
\renewcommand\itot[1]{#1}
\renewcommand\stot[1]{#1}

{\Huge \bf \centering Appendix\par}


\hypersetup{
  linkcolor=.
}

\part{}
\invisiblelocaltableofcontents\label{parttoc:2}

\begingroup\parindent 0pt \parfillskip 0pt \leftskip 0cm \rightskip 1cm
\newcommand*{\DotsAndPage}
{\nobreak\leaders\hbox{\normalsize\hbox to 1.5ex {\hss.\hss}}%
\hfill\nobreak
\rlap{\makebox[1cm]{\normalsize\etocpage}}\par}
\etocsetstyle {section}
{}
{\leavevmode\leftskip 0cm\relax}
{\bfseries\normalsize\makebox[.5cm][l]{\etocnumber.}%
\etocname\nobreak\hfill\nobreak
\rlap{\makebox[1cm]{\mdseries\etocpage}}\par}
{}
\etocsetstyle {subsection}
{}
{\leavevmode\leftskip .5cm\relax }
{\mdseries\normalsize\makebox[1cm][l]{\etocnumber}%
  \etocname
\DotsAndPage}
{}
\etocsetstyle {subsubsection}
{}
{\leavevmode\leftskip 1.5cm\relax }
{\mdseries\normalsize\makebox[1cm][l]{\etocnumber}%
\etocname\nobreak\hfill\nobreak
\rlap{\makebox[1cm]{\etocpage}}\par}
{}
\tableofcontents \ref {parttoc:2}
\endgroup

\newpage
\renewcommand{\listtheoremname}{List of Definitions}
\listoftheorems[ignoreall, show={definitionA}]
\renewcommand{\listtheoremname}{List of Lemmas and Theorems}
\listoftheorems[ignoreall, show={lemmaA,theoremA,corollaryA}]

\section*{Theorem Number Mapping}
\begin{table}[h!]
\centering
\label{tab:thm_mapping}
\begin{tabular}{@{}lll@{}}
\toprule
Theorem                                            & Paper & Appendix \\ \midrule
Coherence                                          & 1     & \ref{thmA:coh} \\
Expression Coherence                               & 2     & \ref{thmA:ecoh} \\
Progress                                           & 3     & \ref{thmA:progress_expressions}        \\
Preservation                                       & 4     & \ref{thmA:intermediate_preservation}    \\
Strong Normalization                               & 5     & \ref{thmA:strong_norm}       \\
Typing Preservation - Expressions                  & 6     & \ref{thmA:typing_pres_expr}        \\
Type Preservation                                  & 7     & \ref{thmA:tmelab}       \\
Semantic Preservation                              & 8     & \ref{lemma:val_sem_pres}       \\
Determinism                                        & 9     & \ref{thmA:tmelab:det}       \\ 
Elaboration Equivalence - Expressions              & 10    & \ref{thmA:equiv_expr} \\
Elaboration Equivalence - Dictionaries             & 11    & \ref{thmA:equiv_dict} \\
Value Relation for Dictionary Values               & 12    & \ref{thmA:dval_rel}       \\
Logical Coherence of Dictionary Resolution         & 13    & \ref{thmA:dict_cohA}       \\
Logical Coherence of Expression Elaboration        & 14    & \ref{thmA:coh_exprA}       \\
Logical implies Contextual Equivalence             & 15    & \ref{thmA:coh_exprB}       \\
Elaboration preserves Contextual Equivalence       & 16    & \ref{thmA:coh_exprC} \\
Contextual Equivalence in \interL{} implies Contextual Equivalence in \srcL{} & 17    & \ref{thmA:ctx_equiv_pres} \\
\bottomrule
\end{tabular}
\end{table}

\newpage
\input{doc/sections/typesystem.mng}
\newpage
\section{Logical Relations}
\label{sec:log_rel}
\renewcommand\itot[1]{}
In the definitions for the logical relations below, \( \intermed {  {\intermed \gamma }  } = \intermed {   {\intermed \gamma }' ,  {\intermed \delta }  \mapsto ( {\intermed {dv} }_{{\mathrm{1}}} ,  {\intermed {dv} }_{{\mathrm{2}}}  )   } \) and \( \intermed {  {\intermed \phi }  } = \intermed {   {\intermed \phi }' ,  {\intermed x}  \mapsto ( {\intermed e}_{{\mathrm{1}}} ,  {\intermed e}_{{\mathrm{2}}}  )   } \)
are substitutions which map all dictionary variables \( \intermed {  {\intermed \delta }  }  \in  \intermed {  {\intermed \Gamma}  } \) and term
variables \( \intermed {  {\intermed x}  }  \in  \intermed {  {\intermed \Gamma}  } \) onto two (possibly different) dictionary values and term
values respectively. Notation-wise, we adopted the convention that \({\intermed \gamma }_{{\mathrm{1}}}\)
maps the dictionary variable \({\intermed \delta }\) to the leftmost value \({\intermed {dv} }_{{\mathrm{1}}}\) and
\({\intermed \gamma }_{{\mathrm{2}}}\) substitutes \({\intermed \delta }\) for the rightmost value \({\intermed {dv} }_{{\mathrm{2}}}\).
Similarly for \({\intermed \phi }_{{\mathrm{1}}}\) and \({\intermed \phi }_{{\mathrm{2}}}\).

The third kind of substitution \( \intermed {  {\intermed R }  } = \intermed {   {\intermed R }'  ,  {\intermed a }  \mapsto (  {\intermed \sigma} ,  { \intermed { \mathit{ r } } }  )   } \) maps all type variables
\( \intermed {  {\intermed a }  }  \in  \intermed {  {\intermed \Gamma}  } \) onto closed types \({\intermed \sigma}\), while also storing a relation
\({ \intermed { \mathit{ r } } }\). This relation \({ \intermed { \mathit{ r } } }\) is an arbitrary member of the set of all
relations \( \mathit{Rel} [ \intermed {  {\intermed \sigma}  } ] \) which offer the following property:
\begin{equation*}
   \mathit{Rel} [ \intermed {  {\intermed \sigma}  } ]  = \{ { \intermed { \mathit{ r } } } \in \mathcal{P} ({\intermed v} \times {\intermed v})~|~\forall ({\intermed v}_{{\mathrm{1}}} , {\intermed v}_{{\mathrm{2}}}) \in { \intermed { \mathit{ r } } } :  \intermed {  {\intermed \Sigma}  } ; \intermed {  {\intermed { \Gamma_C } }  } ; \intermed {  \bullet  }  \vdash_{tm}  \intermed {  {\intermed v}_{{\mathrm{1}}}  } : \intermed {  {\intermed \sigma}  } \itot{ \rightsquigarrow  \target {  {\target e }_{{\mathrm{1}}}  } }  \band  \intermed {  {\intermed \Sigma}  } ; \intermed {  {\intermed { \Gamma_C } }  } ; \intermed {  \bullet  }  \vdash_{tm}  \intermed {  {\intermed v}_{{\mathrm{2}}}  } : \intermed {  {\intermed \sigma}  } \itot{ \rightsquigarrow  \target {  {\target e }_{{\mathrm{2}}}  } }  \}
\end{equation*}

\subsection{Dictionary Relation}
  \begin{flushleft}
    \namedRuleform{ ( \intermed {  {\intermed \Sigma}_{{\mathrm{1}}}  } : \intermed {  {\intermed {dv} }_{{\mathrm{1}}}  } , \intermed {  {\intermed \Sigma}_{{\mathrm{2}}}  } : \intermed {  {\intermed {dv} }_{{\mathrm{2}}}  } ) \in \mathcal{V} \llbracket \intermed {  {\intermed Q}  } \rrbracket ^{\intermed {  {\intermed { \Gamma_C } }  } }_{\intermed {  {\intermed R }  } } }
                  {Closed Dictionary Value Relation}
  \end{flushleft}
  \begin{align*}
     ( \intermed {  {\intermed \Sigma}_{{\mathrm{1}}}  } : \intermed {  {\intermed D } \, \overline {\intermed \sigma}_{\ottmv{j}} \, {\overline {\intermed { dv } } }_{{\mathrm{1}}\,\ottmv{i}}  } , \intermed {  {\intermed \Sigma}_{{\mathrm{2}}}  } : \intermed {  {\intermed D } \, \overline {\intermed \sigma}_{\ottmv{j}} \, {\overline {\intermed { dv } } }_{{\mathrm{2}}\,\ottmv{i}}  } ) \in \mathcal{V} \llbracket \intermed {  {\intermed Q}  } \rrbracket ^{\intermed {  {\intermed { \Gamma_C } }  } }_{\intermed {  {\intermed R }  } } 
    & \triangleq \ottcomp{ ( \intermed {  {\intermed \Sigma}_{{\mathrm{1}}}  } : \intermed {  {\intermed {dv} }_{{\mathrm{1}}\,\ottmv{i}}  } , \intermed {  {\intermed \Sigma}_{{\mathrm{2}}}  } : \intermed {  {\intermed {dv} }_{{\mathrm{2}}\,\ottmv{i}}  } ) \in \mathcal{V} \llbracket \intermed {  [  \overline {\intermed \sigma}_{\ottmv{j}}  /  {\overline {\intermed a} }_{\ottmv{j}}  ]  {\intermed Q}_{\ottmv{i}}  } \rrbracket ^{\intermed {  {\intermed { \Gamma_C } }  } }_{\intermed {  {\intermed R }  } } }{\ottmv{i}} \\
    & \land  \intermed {  {\intermed \Sigma}_{{\mathrm{1}}}  } ; \intermed {  {\intermed { \Gamma_C } }  } ; \intermed {  \bullet  }  \vdash_{d}  \intermed {  {\intermed D } \, \overline {\intermed \sigma}_{\ottmv{j}} \, {\overline {\intermed { dv } } }_{{\mathrm{1}}\,\ottmv{i}}  } : \intermed {  {\intermed R }  \ottsym{(}  {\intermed Q}  \ottsym{)}  } \itot{\,  \rightsquigarrow  \target {  {\target e }_{{\mathrm{1}}}  } }  \\
    & \land  \intermed {  {\intermed \Sigma}_{{\mathrm{2}}}  } ; \intermed {  {\intermed { \Gamma_C } }  } ; \intermed {  \bullet  }  \vdash_{d}  \intermed {  {\intermed D } \, \overline {\intermed \sigma}_{\ottmv{j}} \, {\overline {\intermed { dv } } }_{{\mathrm{2}}\,\ottmv{i}}  } : \intermed {  {\intermed R }  \ottsym{(}  {\intermed Q}  \ottsym{)}  } \itot{\,  \rightsquigarrow  \target {  {\target e }_{{\mathrm{2}}}  } }  \\
    & \textit{where }  ( \intermed {  {\intermed D }  } : \intermed {  \forall  {\overline {\intermed a} }_{\ottmv{j}}  \ottsym{.}  {\overline {\intermed Q} }_{\ottmv{i}}  \Rightarrow  {\intermed Q}'  } ) . \intermed {  {\intermed m}  }  \mapsto  \intermed {  {\intermed e}_{{\mathrm{1}}}  }  \in  \intermed {  {\intermed \Sigma}_{{\mathrm{1}}}  }  
     \land  \intermed {  {\intermed Q}  } = \intermed {  [  \overline {\intermed \sigma}_{\ottmv{j}}  /  {\overline {\intermed a} }_{\ottmv{j}}  ]  {\intermed Q}'  }  \\
  \end{align*}

  \begin{flushleft}
    \namedRuleform{ \intermed {  {\intermed { \Gamma_C } }  } ; \intermed {  {\intermed \Gamma}  }  \vdash  \intermed {  {\intermed \Sigma}_{{\mathrm{1}}}  } : \intermed {  {\intermed d}_{{\mathrm{1}}}  }  \simeq_{log}  \intermed {  {\intermed \Sigma}_{{\mathrm{2}}}  } : \intermed {  {\intermed d}_{{\mathrm{2}}}  } : \intermed {  {\intermed Q}  } }
                  {Logical Equivalence for Open Dictionaries}
  \end{flushleft}
  \begin{align*}
     \intermed {  {\intermed { \Gamma_C } }  } ; \intermed {  {\intermed \Gamma}  }  \vdash  \intermed {  {\intermed \Sigma}_{{\mathrm{1}}}  } : \intermed {  {\intermed d}_{{\mathrm{1}}}  }  \simeq_{log}  \intermed {  {\intermed \Sigma}_{{\mathrm{2}}}  } : \intermed {  {\intermed d}_{{\mathrm{2}}}  } : \intermed {  {\intermed Q}  } 
    & \triangleq \forall  \intermed {  {\intermed R }  } \in \mathcal{F} \llbracket \intermed {  {\intermed \Gamma}  } \rrbracket ^{\intermed {  {\intermed { \Gamma_C } }  } } ,\\
    &   \intermed {  {\intermed \phi }  } \in \mathcal{G} \llbracket \intermed {  {\intermed \Gamma}  } \rrbracket ^{\intermed {  {\intermed \Sigma}_{{\mathrm{1}}}  } , \intermed {  {\intermed \Sigma}_{{\mathrm{2}}}  } , \intermed {  {\intermed { \Gamma_C } }  } }_{\intermed {  {\intermed R }  } } , \\
    &   \intermed {  {\intermed \gamma }  } \in \mathcal{H} \llbracket \intermed {  {\intermed \Gamma}  } \rrbracket ^{\intermed {  {\intermed \Sigma}_{{\mathrm{1}}}  } , \intermed {  {\intermed \Sigma}_{{\mathrm{2}}}  } , \intermed {  {\intermed { \Gamma_C } }  } }_{\intermed {  {\intermed R }  } } , \\
    &   ( \intermed {  {\intermed \Sigma}_{{\mathrm{1}}}  } : \intermed {   {\intermed \gamma }_{{\mathrm{1}}}  ( {\intermed \phi }_{{\mathrm{1}}}  ( {\intermed R }  ( {\intermed d}_{{\mathrm{1}}} )))   } , \intermed {  {\intermed \Sigma}_{{\mathrm{2}}}  } : \intermed {   {\intermed \gamma }_{{\mathrm{2}}}  ( {\intermed \phi }_{{\mathrm{2}}}  ( {\intermed R }  ( {\intermed d}_{{\mathrm{2}}} )))   } ) \in \mathcal{V} \llbracket \intermed {  {\intermed Q}  } \rrbracket ^{\intermed {  {\intermed { \Gamma_C } }  } }_{\intermed {  {\intermed R }  } } 
  \end{align*}

\subsection{Expression Relation}
  \begin{flushleft}
  \namedRuleform{ ( \intermed {  {\intermed \Sigma}_{{\mathrm{1}}}  } : \intermed {  {\intermed v}_{{\mathrm{1}}}  } , \intermed {  {\intermed \Sigma}_{{\mathrm{2}}}  } : \intermed {  {\intermed v}_{{\mathrm{2}}}  } ) \in \mathcal{V} \llbracket \intermed {  {\intermed \sigma}  } \rrbracket ^{\intermed {  {\intermed { \Gamma_C } }  } }_{\intermed {  {\intermed R }  } } }
                {Closed Expression Value Relation}
  \end{flushleft}
  \begin{align*}
     ( \intermed {  {\intermed \Sigma}_{{\mathrm{1}}}  } : \intermed {  \mathit{\intermed {True} }  } , \intermed {  {\intermed \Sigma}_{{\mathrm{2}}}  } : \intermed {  \mathit{\intermed {True} }  } ) \in \mathcal{V} \llbracket \intermed {  \mathit{\intermed {Bool} }  } \rrbracket ^{\intermed {  {\intermed { \Gamma_C } }  } }_{\intermed {  {\intermed R }  } }  \\
     ( \intermed {  {\intermed \Sigma}_{{\mathrm{1}}}  } : \intermed {  \mathit{\intermed {False} }  } , \intermed {  {\intermed \Sigma}_{{\mathrm{2}}}  } : \intermed {  \mathit{\intermed {False} }  } ) \in \mathcal{V} \llbracket \intermed {  \mathit{\intermed {Bool} }  } \rrbracket ^{\intermed {  {\intermed { \Gamma_C } }  } }_{\intermed {  {\intermed R }  } }  \\
     ( \intermed {  {\intermed \Sigma}_{{\mathrm{1}}}  } : \intermed {  {\intermed v}_{{\mathrm{1}}}  } , \intermed {  {\intermed \Sigma}_{{\mathrm{2}}}  } : \intermed {  {\intermed v}_{{\mathrm{2}}}  } ) \in \mathcal{V} \llbracket \intermed {  {\intermed a }  } \rrbracket ^{\intermed {  {\intermed { \Gamma_C } }  } }_{\intermed {  {\intermed R }  } } 
    & \triangleq  ( \intermed {  {\intermed a }  }  \mapsto  ( \intermed {  {\intermed \sigma}  } , \intermed {  { \intermed { \mathit{ r } } }  } ) )  \in  \intermed {  {\intermed R }  }  \\
    & \land  \intermed {  {\intermed \Sigma}_{{\mathrm{1}}}  } ; \intermed {  {\intermed { \Gamma_C } }  } ; \intermed {  \bullet  }  \vdash_{tm}  \intermed {  {\intermed v}_{{\mathrm{1}}}  } : \intermed {  {\intermed \sigma}  } \itot{ \rightsquigarrow  \target {  {\target e }_{{\mathrm{1}}}  } }  \\
    & \land  \intermed {  {\intermed \Sigma}_{{\mathrm{2}}}  } ; \intermed {  {\intermed { \Gamma_C } }  } ; \intermed {  \bullet  }  \vdash_{tm}  \intermed {  {\intermed v}_{{\mathrm{2}}}  } : \intermed {  {\intermed \sigma}  } \itot{ \rightsquigarrow  \target {  {\target e }_{{\mathrm{2}}}  } }  \\
    & \land  ( \intermed {  {\intermed v}_{{\mathrm{1}}}  } , \intermed {  {\intermed v}_{{\mathrm{2}}}  } )  \in  \intermed {  { \intermed { \mathit{ r } } }  }  \\
     ( \intermed {  {\intermed \Sigma}_{{\mathrm{1}}}  } : \intermed {  \lambda  {\intermed x}  \ottsym{:}  {\intermed \sigma}_{{\mathrm{1}}}  \ottsym{.}  {\intermed e}_{{\mathrm{1}}}  } , \intermed {  {\intermed \Sigma}_{{\mathrm{2}}}  } : \intermed {  \lambda  {\intermed x}  \ottsym{:}  {\intermed \sigma}_{{\mathrm{1}}}  \ottsym{.}  {\intermed e}_{{\mathrm{2}}}  } ) \in \mathcal{V} \llbracket \intermed {  {\intermed \sigma}_{{\mathrm{1}}}  \rightarrow  {\intermed \sigma}_{{\mathrm{2}}}  } \rrbracket ^{\intermed {  {\intermed { \Gamma_C } }  } }_{\intermed {  {\intermed R }  } } 
    & \triangleq  \intermed {  {\intermed \Sigma}_{{\mathrm{1}}}  } ; \intermed {  {\intermed { \Gamma_C } }  } ; \intermed {  \bullet  }  \vdash_{tm}  \intermed {  \lambda  {\intermed x}  \ottsym{:}  {\intermed \sigma}  \ottsym{.}  {\intermed e}_{{\mathrm{1}}}  } : \intermed {  {\intermed R }  \ottsym{(}  {\intermed \sigma}_{{\mathrm{1}}}  \rightarrow  {\intermed \sigma}_{{\mathrm{2}}}  \ottsym{)}  } \itot{ \rightsquigarrow  \target {  {\target e }_{{\mathrm{1}}}  } }  \\
    & \land  \intermed {  {\intermed \Sigma}_{{\mathrm{2}}}  } ; \intermed {  {\intermed { \Gamma_C } }  } ; \intermed {  \bullet  }  \vdash_{tm}  \intermed {  \lambda  {\intermed x}  \ottsym{:}  {\intermed \sigma}  \ottsym{.}  {\intermed e}_{{\mathrm{2}}}  } : \intermed {  {\intermed R }  \ottsym{(}  {\intermed \sigma}_{{\mathrm{1}}}  \rightarrow  {\intermed \sigma}_{{\mathrm{2}}}  \ottsym{)}  } \itot{ \rightsquigarrow  \target {  {\target e }_{{\mathrm{2}}}  } }  \\
    & \land \forall  ( \intermed {  {\intermed \Sigma}_{{\mathrm{1}}}  } : \intermed {  {\intermed e}_{{\mathrm{3}}}  } , \intermed {  {\intermed \Sigma}_{{\mathrm{2}}}  } : \intermed {  {\intermed e}_{{\mathrm{4}}}  } ) \in \mathcal{E} \llbracket \intermed {  {\intermed \sigma}_{{\mathrm{1}}}  } \rrbracket ^{\intermed {  {\intermed { \Gamma_C } }  } }_{\intermed {  {\intermed R }  } }  : \\ 
    &  ( \intermed {  {\intermed \Sigma}_{{\mathrm{1}}}  } : \intermed {  \ottsym{(}  \lambda  {\intermed x}  \ottsym{:}  {\intermed \sigma}  \ottsym{.}  {\intermed e}_{{\mathrm{1}}}  \ottsym{)} \, {\intermed e}_{{\mathrm{3}}}  } , \intermed {  {\intermed \Sigma}_{{\mathrm{2}}}  } : \intermed {  \ottsym{(}  \lambda  {\intermed x}  \ottsym{:}  {\intermed \sigma}  \ottsym{.}  {\intermed e}_{{\mathrm{2}}}  \ottsym{)} \, {\intermed e}_{{\mathrm{4}}}  } ) \in \mathcal{E} \llbracket \intermed {  {\intermed \sigma}_{{\mathrm{2}}}  } \rrbracket ^{\intermed {  {\intermed { \Gamma_C } }  } }_{\intermed {  {\intermed R }  } }  \\
     ( \intermed {  {\intermed \Sigma}_{{\mathrm{1}}}  } : \intermed {  \lambda  {\intermed \delta }  \ottsym{:}  {\intermed Q}  \ottsym{.}  {\intermed e}_{{\mathrm{1}}}  } , \intermed {  {\intermed \Sigma}_{{\mathrm{2}}}  } : \intermed {  \lambda  {\intermed \delta }  \ottsym{:}  {\intermed Q}  \ottsym{.}  {\intermed e}_{{\mathrm{2}}}  } ) \in \mathcal{V} \llbracket \intermed {   {\intermed Q}  \Rightarrow  {\intermed \sigma}   } \rrbracket ^{\intermed {  {\intermed { \Gamma_C } }  } }_{\intermed {  {\intermed R }  } } 
    & \triangleq  \intermed {  {\intermed \Sigma}_{{\mathrm{1}}}  } ; \intermed {  {\intermed { \Gamma_C } }  } ; \intermed {  \bullet  }  \vdash_{tm}  \intermed {  \lambda  {\intermed \delta }  \ottsym{:}  {\intermed Q}  \ottsym{.}  {\intermed e}_{{\mathrm{1}}}  } : \intermed {  {\intermed R }  \ottsym{(}   {\intermed Q}  \Rightarrow  {\intermed \sigma}   \ottsym{)}  } \itot{ \rightsquigarrow  \target {  {\target e }_{{\mathrm{1}}}  } }  \\
    & \land  \intermed {  {\intermed \Sigma}_{{\mathrm{2}}}  } ; \intermed {  {\intermed { \Gamma_C } }  } ; \intermed {  \bullet  }  \vdash_{tm}  \intermed {  \lambda  {\intermed \delta }  \ottsym{:}  {\intermed Q}  \ottsym{.}  {\intermed e}_{{\mathrm{2}}}  } : \intermed {  {\intermed R }  \ottsym{(}   {\intermed Q}  \Rightarrow  {\intermed \sigma}   \ottsym{)}  } \itot{ \rightsquigarrow  \target {  {\target e }_{{\mathrm{2}}}  } }  \\
    & \land \forall  ( \intermed {  {\intermed \Sigma}_{{\mathrm{1}}}  } : \intermed {  {\intermed {dv} }_{{\mathrm{1}}}  } , \intermed {  {\intermed \Sigma}_{{\mathrm{2}}}  } : \intermed {  {\intermed {dv} }_{{\mathrm{2}}}  } ) \in \mathcal{V} \llbracket \intermed {  {\intermed Q}  } \rrbracket ^{\intermed {  {\intermed { \Gamma_C } }  } }_{\intermed {  {\intermed R }  } }  : \\
    &  ( \intermed {  {\intermed \Sigma}_{{\mathrm{1}}}  } : \intermed {  \ottsym{(}  \lambda  {\intermed \delta }  \ottsym{:}  {\intermed Q}  \ottsym{.}  {\intermed e}_{{\mathrm{1}}}  \ottsym{)} \, {\intermed {dv} }_{{\mathrm{1}}}  } , \intermed {  {\intermed \Sigma}_{{\mathrm{2}}}  } : \intermed {  \ottsym{(}  \lambda  {\intermed \delta }  \ottsym{:}  {\intermed Q}  \ottsym{.}  {\intermed e}_{{\mathrm{2}}}  \ottsym{)} \, {\intermed {dv} }_{{\mathrm{2}}}  } ) \in \mathcal{E} \llbracket \intermed {  {\intermed \sigma}  } \rrbracket ^{\intermed {  {\intermed { \Gamma_C } }  } }_{\intermed {  {\intermed R }  } }  \\
     ( \intermed {  {\intermed \Sigma}_{{\mathrm{1}}}  } : \intermed {  \Lambda  {\intermed a }  \ottsym{.}  {\intermed e}_{{\mathrm{1}}}  } , \intermed {  {\intermed \Sigma}_{{\mathrm{2}}}  } : \intermed {  \Lambda  {\intermed a }  \ottsym{.}  {\intermed e}_{{\mathrm{2}}}  } ) \in \mathcal{V} \llbracket \intermed {  \forall  {\intermed a }  \ottsym{.}  {\intermed \sigma}  } \rrbracket ^{\intermed {  {\intermed { \Gamma_C } }  } }_{\intermed {  {\intermed R }  } } 
    & \triangleq  \intermed {  {\intermed \Sigma}_{{\mathrm{1}}}  } ; \intermed {  {\intermed { \Gamma_C } }  } ; \intermed {  \bullet  }  \vdash_{tm}  \intermed {  \Lambda  {\intermed a }  \ottsym{.}  {\intermed e}_{{\mathrm{1}}}  } : \intermed {  {\intermed R }  \ottsym{(}  \forall  {\intermed a }  \ottsym{.}  {\intermed \sigma}  \ottsym{)}  } \itot{ \rightsquigarrow  \target {  {\target e }_{{\mathrm{1}}}  } }  \\
    & \land  \intermed {  {\intermed \Sigma}_{{\mathrm{2}}}  } ; \intermed {  {\intermed { \Gamma_C } }  } ; \intermed {  \bullet  }  \vdash_{tm}  \intermed {  \Lambda  {\intermed a }  \ottsym{.}  {\intermed e}_{{\mathrm{2}}}  } : \intermed {  {\intermed R }  \ottsym{(}  \forall  {\intermed a }  \ottsym{.}  {\intermed \sigma}  \ottsym{)}  } \itot{ \rightsquigarrow  \target {  {\target e }_{{\mathrm{2}}}  } }  \\
    & \land \forall \intermed{{\intermed \sigma}'}, \forall  \intermed {  { \intermed { \mathit{ r } } }  } \in \mathit{Rel} [ \intermed {  {\intermed \sigma}'  } ]  : \\
    &  \intermed {  {\intermed { \Gamma_C } }  } ; \intermed {  \bullet  }  \vdash_{ty}  \intermed {  {\intermed \sigma}'  } \itot{ \rightsquigarrow  \target {  {\target \sigma }'  } }  \Rightarrow \\
    &  ( \intermed {  {\intermed \Sigma}_{{\mathrm{1}}}  } : \intermed {  \ottsym{(}  \Lambda  {\intermed a }  \ottsym{.}  {\intermed e}_{{\mathrm{1}}}  \ottsym{)} \, {\intermed \sigma}'  } , \intermed {  {\intermed \Sigma}_{{\mathrm{2}}}  } : \intermed {  \ottsym{(}  \Lambda  {\intermed a }  \ottsym{.}  {\intermed e}_{{\mathrm{2}}}  \ottsym{)} \, {\intermed \sigma}'  } ) \in \mathcal{E} \llbracket \intermed {  {\intermed \sigma}  } \rrbracket ^{\intermed {  {\intermed { \Gamma_C } }  } }_{\intermed {   {\intermed R }  ,  {\intermed a }  \mapsto (  {\intermed \sigma}' ,  { \intermed { \mathit{ r } } }  )   } } 
  \end{align*}

  \begin{flushleft}
  \namedRuleform{ ( \intermed {  {\intermed \Sigma}_{{\mathrm{1}}}  } : \intermed {  {\intermed e}_{{\mathrm{1}}}  } , \intermed {  {\intermed \Sigma}_{{\mathrm{2}}}  } : \intermed {  {\intermed e}_{{\mathrm{2}}}  } ) \in \mathcal{E} \llbracket \intermed {  {\intermed \sigma}_{{\mathrm{1}}}  } \rrbracket ^{\intermed {  {\intermed { \Gamma_C } }  } }_{\intermed {  {\intermed R }  } } }
                {Closed Expression Relation}
  \end{flushleft}
  \begin{align*}
     ( \intermed {  {\intermed \Sigma}_{{\mathrm{1}}}  } : \intermed {  {\intermed e}_{{\mathrm{1}}}  } , \intermed {  {\intermed \Sigma}_{{\mathrm{2}}}  } : \intermed {  {\intermed e}_{{\mathrm{2}}}  } ) \in \mathcal{E} \llbracket \intermed {  {\intermed \sigma}  } \rrbracket ^{\intermed {  {\intermed { \Gamma_C } }  } }_{\intermed {  {\intermed R }  } } 
      & \triangleq  \intermed {  {\intermed \Sigma}_{{\mathrm{1}}}  } ; \intermed {  {\intermed { \Gamma_C } }  } ; \intermed {  \bullet  }  \vdash_{tm}  \intermed {  {\intermed e}_{{\mathrm{1}}}  } : \intermed {  {\intermed R }  \ottsym{(}  {\intermed \sigma}  \ottsym{)}  } \itot{ \rightsquigarrow  \target {  {\target e }_{{\mathrm{1}}}  } }  \\
      & \land  \intermed {  {\intermed \Sigma}_{{\mathrm{2}}}  } ; \intermed {  {\intermed { \Gamma_C } }  } ; \intermed {  \bullet  }  \vdash_{tm}  \intermed {  {\intermed e}_{{\mathrm{2}}}  } : \intermed {  {\intermed R }  \ottsym{(}  {\intermed \sigma}  \ottsym{)}  } \itot{ \rightsquigarrow  \target {  {\target e }_{{\mathrm{2}}}  } }  \\
      & \land \exists \intermed{{\intermed v}_{{\mathrm{1}}}} , \intermed{{\intermed v}_{{\mathrm{2}}}},
         \intermed {  {\intermed \Sigma}_{{\mathrm{1}}}  }  \vdash  \intermed {  {\intermed e}_{{\mathrm{1}}}  }  \longrightarrow^{*}  \intermed {  {\intermed v}_{{\mathrm{1}}}  } ,
         \intermed {  {\intermed \Sigma}_{{\mathrm{2}}}  }  \vdash  \intermed {  {\intermed e}_{{\mathrm{2}}}  }  \longrightarrow^{*}  \intermed {  {\intermed v}_{{\mathrm{2}}}  } ,
         ( \intermed {  {\intermed \Sigma}_{{\mathrm{1}}}  } : \intermed {  {\intermed v}_{{\mathrm{1}}}  } , \intermed {  {\intermed \Sigma}_{{\mathrm{2}}}  } : \intermed {  {\intermed v}_{{\mathrm{2}}}  } ) \in \mathcal{V} \llbracket \intermed {  {\intermed \sigma}  } \rrbracket ^{\intermed {  {\intermed { \Gamma_C } }  } }_{\intermed {  {\intermed R }  } } 
  \end{align*}

  \begin{flushleft}
    \namedRuleform{ \intermed {  {\intermed { \Gamma_C } }  } ; \intermed {  {\intermed \Gamma}  }  \vdash  \intermed {  {\intermed \Sigma}_{{\mathrm{1}}}  } : \intermed {  {\intermed e}_{{\mathrm{1}}}  }  \simeq_{log}  \intermed {  {\intermed \Sigma}_{{\mathrm{2}}}  } : \intermed {  {\intermed e}_{{\mathrm{2}}}  } : \intermed {  {\intermed \sigma}  } }
                  {Logical Equivalence for Open Expressions}
  \end{flushleft}
  \begin{align*}
     \intermed {  {\intermed { \Gamma_C } }  } ; \intermed {  {\intermed \Gamma}  }  \vdash  \intermed {  {\intermed \Sigma}_{{\mathrm{1}}}  } : \intermed {  {\intermed e}_{{\mathrm{1}}}  }  \simeq_{log}  \intermed {  {\intermed \Sigma}_{{\mathrm{2}}}  } : \intermed {  {\intermed e}_{{\mathrm{2}}}  } : \intermed {  {\intermed \sigma}  } 
    & \triangleq \forall  \intermed {  {\intermed R }  } \in \mathcal{F} \llbracket \intermed {  {\intermed \Gamma}  } \rrbracket ^{\intermed {  {\intermed { \Gamma_C } }  } } ,\\
    &   \intermed {  {\intermed \phi }  } \in \mathcal{G} \llbracket \intermed {  {\intermed \Gamma}  } \rrbracket ^{\intermed {  {\intermed \Sigma}_{{\mathrm{1}}}  } , \intermed {  {\intermed \Sigma}_{{\mathrm{2}}}  } , \intermed {  {\intermed { \Gamma_C } }  } }_{\intermed {  {\intermed R }  } } , \\
    &   \intermed {  {\intermed \gamma }  } \in \mathcal{H} \llbracket \intermed {  {\intermed \Gamma}  } \rrbracket ^{\intermed {  {\intermed \Sigma}_{{\mathrm{1}}}  } , \intermed {  {\intermed \Sigma}_{{\mathrm{2}}}  } , \intermed {  {\intermed { \Gamma_C } }  } }_{\intermed {  {\intermed R }  } } , \\
    &   ( \intermed {  {\intermed \Sigma}_{{\mathrm{1}}}  } : \intermed {   {\intermed \gamma }_{{\mathrm{1}}}  ( {\intermed \phi }_{{\mathrm{1}}}  ( {\intermed R }  ( {\intermed e}_{{\mathrm{1}}} )))   } , \intermed {  {\intermed \Sigma}_{{\mathrm{2}}}  } : \intermed {   {\intermed \gamma }_{{\mathrm{2}}}  ( {\intermed \phi }_{{\mathrm{2}}}  ( {\intermed R }  ( {\intermed e}_{{\mathrm{2}}} )))   } ) \in \mathcal{E} \llbracket \intermed {  {\intermed \sigma}  } \rrbracket ^{\intermed {  {\intermed { \Gamma_C } }  } }_{\intermed {  {\intermed R }  } } 
  \end{align*}

  \begin{flushleft}
    \namedRuleform{ \intermed {  {\intermed \Sigma}_{{\mathrm{1}}}  } : \intermed {  {\intermed M}_{{\mathrm{1}}}  }  \simeq_{log}  \intermed {  {\intermed \Sigma}_{{\mathrm{2}}}  } : \intermed {  {\intermed M}_{{\mathrm{2}}}  } : ( \intermed {  {\intermed { \Gamma_C } }  } ; \intermed {  {\intermed \Gamma}  }  \Rightarrow  \intermed {  {\intermed \sigma}  } )  \mapsto  ( \intermed {  {\intermed { \Gamma_C } }  } ; \intermed {  {\intermed \Gamma}'  }  \Rightarrow  \intermed {  {\intermed \sigma}'  } ) }
                  {Logical Equivalence for Contexts}
  \end{flushleft}
  \begin{align*}
     \intermed {  {\intermed \Sigma}_{{\mathrm{1}}}  } : \intermed {  {\intermed M}_{{\mathrm{1}}}  }  \simeq_{log}  \intermed {  {\intermed \Sigma}_{{\mathrm{2}}}  } : \intermed {  {\intermed M}_{{\mathrm{2}}}  } : ( \intermed {  {\intermed { \Gamma_C } }  } ; \intermed {  {\intermed \Gamma}  }  \Rightarrow  \intermed {  {\intermed \sigma}  } )  \mapsto  ( \intermed {  {\intermed { \Gamma_C } }  } ; \intermed {  {\intermed \Gamma}'  }  \Rightarrow  \intermed {  {\intermed \sigma}'  } ) 
    & \triangleq \forall \intermed{{\intermed e}_{{\mathrm{1}}}}, \intermed{{\intermed e}_{{\mathrm{2}}}} :
       \intermed {  {\intermed { \Gamma_C } }  } ; \intermed {  {\intermed \Gamma}  }  \vdash  \intermed {  {\intermed \Sigma}_{{\mathrm{1}}}  } : \intermed {  {\intermed e}_{{\mathrm{1}}}  }  \simeq_{log}  \intermed {  {\intermed \Sigma}_{{\mathrm{2}}}  } : \intermed {  {\intermed e}_{{\mathrm{2}}}  } : \intermed {  {\intermed \sigma}  }  \\
    & \Rightarrow  \intermed {  {\intermed { \Gamma_C } }  } ; \intermed {  {\intermed \Gamma}'  }  \vdash  \intermed {  {\intermed \Sigma}_{{\mathrm{1}}}  } : \intermed {   {\intermed M}_{{\mathrm{1}}}  [  {\intermed e}_{{\mathrm{1}}}  ]   }  \simeq_{log}  \intermed {  {\intermed \Sigma}_{{\mathrm{2}}}  } : \intermed {   {\intermed M}_{{\mathrm{2}}}  [  {\intermed e}_{{\mathrm{2}}}  ]   } : \intermed {  {\intermed \sigma}'  } 
  \end{align*}

\begin{definitionA}[Interpretation of type variables in type contexts] 
  \label{def:inter_ty}
  \begin{mathpar}
    \ottaltinferrule{}{}
    {  }
    {  \intermed {  \bullet  } \in \mathcal{F} \llbracket \intermed {  \bullet  } \rrbracket ^{\intermed {  {\intermed { \Gamma_C } }  } }  } \and

    \ottaltinferrule{}{}
    {  \intermed {  {\intermed R }  } \in \mathcal{F} \llbracket \intermed {  {\intermed \Gamma}  } \rrbracket ^{\intermed {  {\intermed { \Gamma_C } }  } }  }
    {  \intermed {  {\intermed R }  } \in \mathcal{F} \llbracket \intermed {  {\intermed \Gamma}  \ottsym{,}  {\intermed x}  \ottsym{:}  {\intermed \sigma}  } \rrbracket ^{\intermed {  {\intermed { \Gamma_C } }  } }   } \and

    \ottaltinferrule{}{}
    {  \intermed {  {\intermed R }  } \in \mathcal{F} \llbracket \intermed {  {\intermed \Gamma}  } \rrbracket ^{\intermed {  {\intermed { \Gamma_C } }  } } 
      \\  \intermed {  { \intermed { \mathit{ r } } }  } \in \mathit{Rel} [ \intermed {  {\intermed \sigma}  } ] 
      \\  \intermed {  {\intermed { \Gamma_C } }  } ; \intermed {  \bullet  }  \vdash_{ty}  \intermed {  {\intermed \sigma}  } \itot{ \rightsquigarrow  \target {  {\target \sigma }  } } 
    }
    {  \intermed {   {\intermed R }  ,  {\intermed a }  \mapsto (  {\intermed \sigma} ,  { \intermed { \mathit{ r } } }  )   } \in \mathcal{F} \llbracket \intermed {  {\intermed \Gamma}  \ottsym{,}  {\intermed a }  } \rrbracket ^{\intermed {  {\intermed { \Gamma_C } }  } }   } \and

    \ottaltinferrule{}{}
    {  \intermed {  {\intermed R }  } \in \mathcal{F} \llbracket \intermed {  {\intermed \Gamma}  } \rrbracket ^{\intermed {  {\intermed { \Gamma_C } }  } }  }
    {  \intermed {  {\intermed R }  } \in \mathcal{F} \llbracket \intermed {  {\intermed \Gamma}  \ottsym{,}  {\intermed \delta }  \ottsym{:}  {\intermed Q}  } \rrbracket ^{\intermed {  {\intermed { \Gamma_C } }  } }  }
  \end{mathpar}
\end{definitionA}

\begin{definitionA}[Interpretation of term variables in type contexts] 
  \label{def:inter_tm}
  \begin{mathpar}
    \ottaltinferrule{}{}
    {  }
    {  \intermed {  \bullet  } \in \mathcal{G} \llbracket \intermed {  \bullet  } \rrbracket ^{\intermed {  {\intermed \Sigma}_{{\mathrm{1}}}  } , \intermed {  {\intermed \Sigma}_{{\mathrm{2}}}  } , \intermed {  {\intermed { \Gamma_C } }  } }_{\intermed {  {\intermed R }  } }  } \and

    \ottaltinferrule{}{}
    {  \intermed {  {\intermed \phi }  } \in \mathcal{G} \llbracket \intermed {  {\intermed \Gamma}  } \rrbracket ^{\intermed {  {\intermed \Sigma}_{{\mathrm{1}}}  } , \intermed {  {\intermed \Sigma}_{{\mathrm{2}}}  } , \intermed {  {\intermed { \Gamma_C } }  } }_{\intermed {  {\intermed R }  } }  }
    {  \intermed {  {\intermed \phi }  } \in \mathcal{G} \llbracket \intermed {  {\intermed \Gamma}  \ottsym{,}  {\intermed a }  } \rrbracket ^{\intermed {  {\intermed \Sigma}_{{\mathrm{1}}}  } , \intermed {  {\intermed \Sigma}_{{\mathrm{2}}}  } , \intermed {  {\intermed { \Gamma_C } }  } }_{\intermed {  {\intermed R }  } }   } \and

    \ottaltinferrule{}{}
    {  \intermed {  {\intermed \phi }  } \in \mathcal{G} \llbracket \intermed {  {\intermed \Gamma}  } \rrbracket ^{\intermed {  {\intermed \Sigma}_{{\mathrm{1}}}  } , \intermed {  {\intermed \Sigma}_{{\mathrm{2}}}  } , \intermed {  {\intermed { \Gamma_C } }  } }_{\intermed {  {\intermed R }  } } 
      \\  ( \intermed {  {\intermed \Sigma}_{{\mathrm{1}}}  } : \intermed {  {\intermed e}_{{\mathrm{1}}}  } , \intermed {  {\intermed \Sigma}_{{\mathrm{2}}}  } : \intermed {  {\intermed e}_{{\mathrm{2}}}  } ) \in \mathcal{E} \llbracket \intermed {  {\intermed \sigma}  } \rrbracket ^{\intermed {  {\intermed { \Gamma_C } }  } }_{\intermed {  {\intermed R }  } } 
    }
    {  \intermed {   {\intermed \phi } ,  {\intermed x}  \mapsto ( {\intermed e}_{{\mathrm{1}}} ,  {\intermed e}_{{\mathrm{2}}}  )   } \in \mathcal{G} \llbracket \intermed {  {\intermed \Gamma}  \ottsym{,}  {\intermed x}  \ottsym{:}  {\intermed \sigma}  } \rrbracket ^{\intermed {  {\intermed \Sigma}_{{\mathrm{1}}}  } , \intermed {  {\intermed \Sigma}_{{\mathrm{2}}}  } , \intermed {  {\intermed { \Gamma_C } }  } }_{\intermed {  {\intermed R }  } }   } \and

    \ottaltinferrule{}{}
    {  \intermed {  {\intermed \phi }  } \in \mathcal{G} \llbracket \intermed {  {\intermed \Gamma}  } \rrbracket ^{\intermed {  {\intermed \Sigma}_{{\mathrm{1}}}  } , \intermed {  {\intermed \Sigma}_{{\mathrm{2}}}  } , \intermed {  {\intermed { \Gamma_C } }  } }_{\intermed {  {\intermed R }  } }  }
    {  \intermed {  {\intermed \phi }  } \in \mathcal{G} \llbracket \intermed {  {\intermed \Gamma}  \ottsym{,}  {\intermed \delta }  \ottsym{:}  {\intermed Q}  } \rrbracket ^{\intermed {  {\intermed \Sigma}_{{\mathrm{1}}}  } , \intermed {  {\intermed \Sigma}_{{\mathrm{2}}}  } , \intermed {  {\intermed { \Gamma_C } }  } }_{\intermed {  {\intermed R }  } }  }
  \end{mathpar}
\end{definitionA}

\begin{definitionA}[Interpretation of dictionary variables dictionary contexts] 
  \label{def:inter_d}
  \begin{mathpar}
    \ottaltinferrule{}{}
    {  }
    {  \intermed {  \bullet  } \in \mathcal{H} \llbracket \intermed {  \bullet  } \rrbracket ^{\intermed {  {\intermed \Sigma}_{{\mathrm{1}}}  } , \intermed {  {\intermed \Sigma}_{{\mathrm{2}}}  } , \intermed {  {\intermed { \Gamma_C } }  } }_{\intermed {  {\intermed R }  } }  } \and

    \ottaltinferrule{}{}
    {  \intermed {  {\intermed \gamma }  } \in \mathcal{H} \llbracket \intermed {  {\intermed \Gamma}  } \rrbracket ^{\intermed {  {\intermed \Sigma}_{{\mathrm{1}}}  } , \intermed {  {\intermed \Sigma}_{{\mathrm{2}}}  } , \intermed {  {\intermed { \Gamma_C } }  } }_{\intermed {  {\intermed R }  } }  }
    {  \intermed {  {\intermed \gamma }  } \in \mathcal{H} \llbracket \intermed {  {\intermed \Gamma}  \ottsym{,}  {\intermed a }  } \rrbracket ^{\intermed {  {\intermed \Sigma}_{{\mathrm{1}}}  } , \intermed {  {\intermed \Sigma}_{{\mathrm{2}}}  } , \intermed {  {\intermed { \Gamma_C } }  } }_{\intermed {  {\intermed R }  } }  } \and

    \ottaltinferrule{}{}
    {  \intermed {  {\intermed \gamma }  } \in \mathcal{H} \llbracket \intermed {  {\intermed \Gamma}  } \rrbracket ^{\intermed {  {\intermed \Sigma}_{{\mathrm{1}}}  } , \intermed {  {\intermed \Sigma}_{{\mathrm{2}}}  } , \intermed {  {\intermed { \Gamma_C } }  } }_{\intermed {  {\intermed R }  } }  }
    {  \intermed {  {\intermed \gamma }  } \in \mathcal{H} \llbracket \intermed {  {\intermed \Gamma}  \ottsym{,}  {\intermed x}  \ottsym{:}  {\intermed \sigma}  } \rrbracket ^{\intermed {  {\intermed \Sigma}_{{\mathrm{1}}}  } , \intermed {  {\intermed \Sigma}_{{\mathrm{2}}}  } , \intermed {  {\intermed { \Gamma_C } }  } }_{\intermed {  {\intermed R }  } }  } \and

    \ottaltinferrule{}{}
    {  \intermed {  {\intermed \gamma }  } \in \mathcal{H} \llbracket \intermed {  {\intermed \Gamma}  } \rrbracket ^{\intermed {  {\intermed \Sigma}_{{\mathrm{1}}}  } , \intermed {  {\intermed \Sigma}_{{\mathrm{2}}}  } , \intermed {  {\intermed { \Gamma_C } }  } }_{\intermed {  {\intermed R }  } } 
      \\  ( \intermed {  {\intermed \Sigma}_{{\mathrm{1}}}  } : \intermed {  {\intermed {dv} }_{{\mathrm{1}}}  } , \intermed {  {\intermed \Sigma}_{{\mathrm{2}}}  } : \intermed {  {\intermed {dv} }_{{\mathrm{2}}}  } ) \in \mathcal{V} \llbracket \intermed {  {\intermed Q}  } \rrbracket ^{\intermed {  {\intermed { \Gamma_C } }  } }_{\intermed {  {\intermed R }  } } 
    }
    {  \intermed {   {\intermed \gamma } ,  {\intermed \delta }  \mapsto ( {\intermed {dv} }_{{\mathrm{1}}} ,  {\intermed {dv} }_{{\mathrm{2}}}  )   } \in \mathcal{H} \llbracket \intermed {  {\intermed \Gamma}  \ottsym{,}  {\intermed \delta }  \ottsym{:}  {\intermed Q}  } \rrbracket ^{\intermed {  {\intermed \Sigma}_{{\mathrm{1}}}  } , \intermed {  {\intermed \Sigma}_{{\mathrm{2}}}  } , \intermed {  {\intermed { \Gamma_C } }  } }_{\intermed {  {\intermed R }  } }   }
  \end{mathpar}
\end{definitionA}

\subsection{Environment Relation}
\label{app:log_env_rel}

\drules[ctxLog]{$ \intermed {  {\intermed { \Gamma_C } }  }  \vdash  \intermed {  {\intermed \Sigma}_{{\mathrm{1}}}  }  \simeq_{log}  \intermed {  {\intermed \Sigma}_{{\mathrm{2}}}  } $}{Logical Equivalence for
  Environments}{empty, cons}

\newpage
\section{Strong Normalization Relations}
As opposed to Section~\ref{sec:log_rel}, the relations and substitutions in the
strong normalization relations described below, are unary.
The substitutions \( \intermed {  {\intermed { \gamma^{SN} } }  } = \intermed {   {\intermed { \gamma^{SN} } }' ,  {\intermed \delta }  \mapsto  {\intermed d}   } \) and
\( \intermed {  {\intermed { \phi^{SN} } }  } = \intermed {   {\intermed { \phi^{SN} } }' ,  {\intermed x}  \mapsto  {\intermed e}   } \) map all dictionary 
variables \( \intermed {  {\intermed \delta }  }  \in  \intermed {  {\intermed \Gamma}  } \) and term variables \( \intermed {  {\intermed x}  }  \in  \intermed {  {\intermed \Gamma}  } \) onto
well-typed dictionaries \({\intermed d}\) and expressions \( \intermed {  {\intermed e}  } \in \mathcal{SN} \llbracket \intermed {  {\intermed \sigma}  } \rrbracket ^{\intermed {  {\intermed \Sigma}  } , \intermed {  {\intermed { \Gamma_C } }  } }_{\intermed {  {\intermed { R^{SN} } }  } } \). The final kind of substitution \( \intermed {  {\intermed { R^{SN} } }  } = \intermed {   {\intermed { R^{SN} } }'  ,  {\intermed a }  \mapsto (  {\intermed \sigma} ,  { \intermed { \mathit{ r } } }  )   } \) maps all type variables \( \intermed {  {\intermed a }  }  \in  \intermed {  {\intermed \Gamma}  } \) onto closed types \({\intermed \sigma}\),
while also storing a relation \({ \intermed { \mathit{ r } } }\).
This relation \({ \intermed { \mathit{ r } } }\) is an arbitrary member of the
set of all relations \( \mathit{Rel} [ \intermed {  {\intermed \sigma}  } ] \) which offer the following property:
\begin{equation*}
   \mathit{Rel} [ \intermed {  {\intermed \sigma}  } ]  = \{ { \intermed { \mathit{ r } } } \in \mathcal{P} ({\intermed e})~|~\forall {\intermed e} \in { \intermed { \mathit{ r } } } :  \intermed {  {\intermed \Sigma}  } ; \intermed {  {\intermed { \Gamma_C } }  } ; \intermed {  \bullet  }  \vdash_{tm}  \intermed {  {\intermed e}  } : \intermed {  {\intermed \sigma}  } \itot{ \rightsquigarrow  \target {  {\target e }  } }  \}
\end{equation*}
We adopted the convention that \({\intermed { R^{SN} } }_{{\mathrm{1}}}  \ottsym{(}  {\intermed a }  \ottsym{)}\) maps the type variable
\({\intermed a }\) onto the closed type \({\intermed \sigma}\) and \({\intermed { R^{SN} } }_{{\mathrm{2}}}  \ottsym{(}  {\intermed a }  \ottsym{)}\) denotes the
contained set of expressions \({ \intermed { \mathit{ r } } }\).

  \begin{flushleft}
    \namedRuleform{ \intermed {  {\intermed e}  } \in \mathcal{SN} \llbracket \intermed {  {\intermed \sigma}  } \rrbracket ^{\intermed {  {\intermed \Sigma}  } , \intermed {  {\intermed { \Gamma_C } }  } }_{\intermed {  {\intermed { R^{SN} } }  } } }
                  {Strong Normalization Relation}
  \end{flushleft}
  \begin{align*}
     \intermed {  {\intermed e}  } \in \mathcal{SN} \llbracket \intermed {  \mathit{\intermed {Bool} }  } \rrbracket ^{\intermed {  {\intermed \Sigma}  } , \intermed {  {\intermed { \Gamma_C } }  } }_{\intermed {  {\intermed { R^{SN} } }  } } 
    & \triangleq  \intermed {  {\intermed \Sigma}  } ; \intermed {  {\intermed { \Gamma_C } }  } ; \intermed {  \bullet  }  \vdash_{tm}  \intermed {  {\intermed e}  } : \intermed {  \mathit{\intermed {Bool} }  } \itot{ \rightsquigarrow  \target {  {\target e }  } }  \\
    & \land \exists {\intermed v} :  \intermed {  {\intermed \Sigma}  }  \vdash  \intermed {  {\intermed e}  }  \longrightarrow^{*}  \intermed {  {\intermed v}  }  \\
     \intermed {  {\intermed e}  } \in \mathcal{SN} \llbracket \intermed {  {\intermed a }  } \rrbracket ^{\intermed {  {\intermed \Sigma}  } , \intermed {  {\intermed { \Gamma_C } }  } }_{\intermed {  {\intermed { R^{SN} } }  } } 
    & \triangleq  \intermed {  {\intermed \Sigma}  } ; \intermed {  {\intermed { \Gamma_C } }  } ; \intermed {  \bullet  }  \vdash_{tm}  \intermed {  {\intermed e}  } : \intermed {  {\intermed { R^{SN} } }_{{\mathrm{1}}}  \ottsym{(}  {\intermed a }  \ottsym{)}  } \itot{ \rightsquigarrow  \target {  {\target e }  } }  \\
    & \land \exists {\intermed v} :  \intermed {  {\intermed \Sigma}  }  \vdash  \intermed {  {\intermed e}  }  \longrightarrow^{*}  \intermed {  {\intermed v}  }  \\
    & \land  \intermed {  {\intermed v}  }  \in  \intermed {  {\intermed { R^{SN} } }_{{\mathrm{2}}}  } ( \intermed {  {\intermed a }  } )  \\
     \intermed {  {\intermed e}  } \in \mathcal{SN} \llbracket \intermed {  {\intermed \sigma}_{{\mathrm{1}}}  \rightarrow  {\intermed \sigma}_{{\mathrm{2}}}  } \rrbracket ^{\intermed {  {\intermed \Sigma}  } , \intermed {  {\intermed { \Gamma_C } }  } }_{\intermed {  {\intermed { R^{SN} } }  } } 
    & \triangleq  \intermed {  {\intermed \Sigma}  } ; \intermed {  {\intermed { \Gamma_C } }  } ; \intermed {  \bullet  }  \vdash_{tm}  \intermed {  {\intermed e}  } : \intermed {  {\intermed { R^{SN} } }_{{\mathrm{1}}}  \ottsym{(}  {\intermed \sigma}_{{\mathrm{1}}}  \rightarrow  {\intermed \sigma}_{{\mathrm{2}}}  \ottsym{)}  } \itot{ \rightsquigarrow  \target {  {\target e }  } }  \\
    & \land \exists {\intermed v} :  \intermed {  {\intermed \Sigma}  }  \vdash  \intermed {  {\intermed e}  }  \longrightarrow^{*}  \intermed {  {\intermed v}  }  \\
    & \land \forall {\intermed e}' :  \intermed {  {\intermed e}'  } \in \mathcal{SN} \llbracket \intermed {  {\intermed \sigma}_{{\mathrm{1}}}  } \rrbracket ^{\intermed {  {\intermed \Sigma}  } , \intermed {  {\intermed { \Gamma_C } }  } }_{\intermed {  {\intermed { R^{SN} } }  } } 
      \Rightarrow  \intermed {  {\intermed e} \, {\intermed e}'  } \in \mathcal{SN} \llbracket \intermed {  {\intermed \sigma}_{{\mathrm{2}}}  } \rrbracket ^{\intermed {  {\intermed \Sigma}  } , \intermed {  {\intermed { \Gamma_C } }  } }_{\intermed {  {\intermed { R^{SN} } }  } }  \\
     \intermed {  {\intermed e}  } \in \mathcal{SN} \llbracket \intermed {   {\intermed Q}  \Rightarrow  {\intermed \sigma}   } \rrbracket ^{\intermed {  {\intermed \Sigma}  } , \intermed {  {\intermed { \Gamma_C } }  } }_{\intermed {  {\intermed { R^{SN} } }  } } 
    & \triangleq  \intermed {  {\intermed \Sigma}  } ; \intermed {  {\intermed { \Gamma_C } }  } ; \intermed {  \bullet  }  \vdash_{tm}  \intermed {  {\intermed e}  } : \intermed {  {\intermed { R^{SN} } }_{{\mathrm{1}}}  \ottsym{(}   {\intermed Q}  \Rightarrow  {\intermed \sigma}   \ottsym{)}  } \itot{ \rightsquigarrow  \target {  {\target e }  } }  \\
    & \land \exists {\intermed v} :  \intermed {  {\intermed \Sigma}  }  \vdash  \intermed {  {\intermed e}  }  \longrightarrow^{*}  \intermed {  {\intermed v}  }  \\
    & \land \forall {\intermed d} :  \intermed {  {\intermed \Sigma}  } ; \intermed {  {\intermed { \Gamma_C } }  } ; \intermed {  \bullet  }  \vdash_{d}  \intermed {  {\intermed d}  } : \intermed {  {\intermed { R^{SN} } }_{{\mathrm{1}}}  \ottsym{(}  {\intermed Q}  \ottsym{)}  } \itot{\,  \rightsquigarrow  \target {  {\target e }  } } 
      \Rightarrow  \intermed {  {\intermed e} \, {\intermed d}  } \in \mathcal{SN} \llbracket \intermed {  {\intermed \sigma}  } \rrbracket ^{\intermed {  {\intermed \Sigma}  } , \intermed {  {\intermed { \Gamma_C } }  } }_{\intermed {  {\intermed { R^{SN} } }  } }  \\
     \intermed {  {\intermed e}  } \in \mathcal{SN} \llbracket \intermed {  \forall  {\intermed a }  \ottsym{.}  {\intermed \sigma}  } \rrbracket ^{\intermed {  {\intermed \Sigma}  } , \intermed {  {\intermed { \Gamma_C } }  } }_{\intermed {  {\intermed { R^{SN} } }  } } 
    & \triangleq  \intermed {  {\intermed \Sigma}  } ; \intermed {  {\intermed { \Gamma_C } }  } ; \intermed {  \bullet  }  \vdash_{tm}  \intermed {  {\intermed e}  } : \intermed {  {\intermed { R^{SN} } }_{{\mathrm{1}}}  \ottsym{(}  \forall  {\intermed a }  \ottsym{.}  {\intermed \sigma}  \ottsym{)}  } \itot{ \rightsquigarrow  \target {  {\target e }  } }  \\
    & \land \exists {\intermed v} :  \intermed {  {\intermed \Sigma}  }  \vdash  \intermed {  {\intermed e}  }  \longrightarrow^{*}  \intermed {  {\intermed v}  }  \\
    & \land \forall {\intermed \sigma}',  \intermed {  { \intermed { \mathit{ r } } }  } \in \mathit{Rel} [ \intermed {  {\intermed \sigma}'  } ]  :
       \intermed {  {\intermed e} \, {\intermed \sigma}'  } \in \mathcal{SN} \llbracket \intermed {  {\intermed \sigma}  } \rrbracket ^{\intermed {  {\intermed \Sigma}  } , \intermed {  {\intermed { \Gamma_C } }  } }_{\intermed {   {\intermed { R^{SN} } }  ,  {\intermed a }  \mapsto (  {\intermed \sigma}' ,  { \intermed { \mathit{ r } } }  )   } } 
  \end{align*}

\begin{definitionA}[Interpretation of type variables in type contexts for strong normalization] 
  \label{def:inter_sn_ty}
  \begin{mathpar}
    \ottaltinferrule{}{}
    {  }
    {  \intermed {  \bullet  } \in \mathcal{F^{SN} } \llbracket \intermed {  \bullet  } \rrbracket ^{\intermed {  {\intermed \Sigma}  }, \intermed {  {\intermed { \Gamma_C } }  } }  } \and

    \ottaltinferrule{}{}
    {  \intermed {  {\intermed { R^{SN} } }  } \in \mathcal{F^{SN} } \llbracket \intermed {  {\intermed \Gamma}  } \rrbracket ^{\intermed {  {\intermed \Sigma}  }, \intermed {  {\intermed { \Gamma_C } }  } }  }
    {  \intermed {  {\intermed { R^{SN} } }  } \in \mathcal{F^{SN} } \llbracket \intermed {  {\intermed \Gamma}  \ottsym{,}  {\intermed x}  \ottsym{:}  {\intermed \sigma}  } \rrbracket ^{\intermed {  {\intermed \Sigma}  }, \intermed {  {\intermed { \Gamma_C } }  } }   } \and

    \ottaltinferrule{}{}
    {  \intermed {  {\intermed { R^{SN} } }  } \in \mathcal{F^{SN} } \llbracket \intermed {  {\intermed \Gamma}  } \rrbracket ^{\intermed {  {\intermed \Sigma}  }, \intermed {  {\intermed { \Gamma_C } }  } } 
      \\  \intermed {  { \intermed { \mathit{ r } } }  } = \mathcal{SN} \llbracket \intermed {  {\intermed \sigma}  } \rrbracket ^{\intermed {  {\intermed \Sigma}  } , \intermed {  {\intermed { \Gamma_C } }  } }_{\intermed {  {\intermed { R^{SN} } }  } } 
      \\  \intermed {  {\intermed { \Gamma_C } }  } ; \intermed {  \bullet  }  \vdash_{ty}  \intermed {  {\intermed { R^{SN} } }_{{\mathrm{1}}}  \ottsym{(}  {\intermed \sigma}  \ottsym{)}  } \itot{ \rightsquigarrow  \target {  {\target \sigma }  } } 
    }
    {  \intermed {   {\intermed { R^{SN} } }  ,  {\intermed a }  \mapsto (  {\intermed { R^{SN} } }_{{\mathrm{1}}}  \ottsym{(}  {\intermed \sigma}  \ottsym{)} ,  { \intermed { \mathit{ r } } }  )   } \in \mathcal{F^{SN} } \llbracket \intermed {  {\intermed \Gamma}  \ottsym{,}  {\intermed a }  } \rrbracket ^{\intermed {  {\intermed \Sigma}  }, \intermed {  {\intermed { \Gamma_C } }  } }   } \and

    \ottaltinferrule{}{}
    {  \intermed {  {\intermed { R^{SN} } }  } \in \mathcal{F^{SN} } \llbracket \intermed {  {\intermed \Gamma}  } \rrbracket ^{\intermed {  {\intermed \Sigma}  }, \intermed {  {\intermed { \Gamma_C } }  } }  }
    {  \intermed {  {\intermed { R^{SN} } }  } \in \mathcal{F^{SN} } \llbracket \intermed {  {\intermed \Gamma}  \ottsym{,}  {\intermed \delta }  \ottsym{:}  {\intermed Q}  } \rrbracket ^{\intermed {  {\intermed \Sigma}  }, \intermed {  {\intermed { \Gamma_C } }  } }  }
  \end{mathpar}
\end{definitionA}

\begin{definitionA}[Interpretation of term variables in type contexts for strong normalization] 
  \label{def:inter_sn_tm}
  \begin{mathpar}
    \ottaltinferrule{}{}
    {  }
    {  \intermed {  \bullet  } \in \mathcal{G^{SN} } \llbracket \intermed {  \bullet  } \rrbracket ^{\intermed {  {\intermed \Sigma}  } , \intermed {  {\intermed { \Gamma_C } }  } }_{\intermed {  {\intermed { R^{SN} } }  } }  } \and

    \ottaltinferrule{}{}
    {  \intermed {  {\intermed { \phi^{SN} } }  } \in \mathcal{G^{SN} } \llbracket \intermed {  {\intermed \Gamma}  } \rrbracket ^{\intermed {  {\intermed \Sigma}  } , \intermed {  {\intermed { \Gamma_C } }  } }_{\intermed {  {\intermed { R^{SN} } }  } }  }
    {  \intermed {  {\intermed { \phi^{SN} } }  } \in \mathcal{G^{SN} } \llbracket \intermed {  {\intermed \Gamma}  \ottsym{,}  {\intermed a }  } \rrbracket ^{\intermed {  {\intermed \Sigma}  } , \intermed {  {\intermed { \Gamma_C } }  } }_{\intermed {  {\intermed { R^{SN} } }  } }   } \and

    \ottaltinferrule{}{}
    {  \intermed {  {\intermed { \phi^{SN} } }  } \in \mathcal{G^{SN} } \llbracket \intermed {  {\intermed \Gamma}  } \rrbracket ^{\intermed {  {\intermed \Sigma}  } , \intermed {  {\intermed { \Gamma_C } }  } }_{\intermed {  {\intermed { R^{SN} } }  } } 
      \\  \intermed {  {\intermed e}  } \in \mathcal{SN} \llbracket \intermed {  {\intermed \sigma}  } \rrbracket ^{\intermed {  {\intermed \Sigma}  } , \intermed {  {\intermed { \Gamma_C } }  } }_{\intermed {  {\intermed { R^{SN} } }  } } 
    }
    {  \intermed {   {\intermed { \phi^{SN} } } ,  {\intermed x}  \mapsto  {\intermed e}   } \in \mathcal{G^{SN} } \llbracket \intermed {  {\intermed \Gamma}  \ottsym{,}  {\intermed x}  \ottsym{:}  {\intermed \sigma}  } \rrbracket ^{\intermed {  {\intermed \Sigma}  } , \intermed {  {\intermed { \Gamma_C } }  } }_{\intermed {  {\intermed { R^{SN} } }  } }   } \and

    \ottaltinferrule{}{}
    {  \intermed {  {\intermed { \phi^{SN} } }  } \in \mathcal{G^{SN} } \llbracket \intermed {  {\intermed \Gamma}  } \rrbracket ^{\intermed {  {\intermed \Sigma}  } , \intermed {  {\intermed { \Gamma_C } }  } }_{\intermed {  {\intermed { R^{SN} } }  } }  }
    {  \intermed {  {\intermed { \phi^{SN} } }  } \in \mathcal{G^{SN} } \llbracket \intermed {  {\intermed \Gamma}  \ottsym{,}  {\intermed \delta }  \ottsym{:}  {\intermed Q}  } \rrbracket ^{\intermed {  {\intermed \Sigma}  } , \intermed {  {\intermed { \Gamma_C } }  } }_{\intermed {  {\intermed { R^{SN} } }  } }  }
  \end{mathpar}
\end{definitionA}

\begin{definitionA}[Interpretation of dictionary variables dictionary contexts for strong normalization] 
  \label{def:inter_sn_d}
  \begin{mathpar}
    \ottaltinferrule{}{}
    {  }
    {  \intermed {  \bullet  } \in \mathcal{H^{SN} } \llbracket \intermed {  \bullet  } \rrbracket ^{\intermed {  {\intermed \Sigma}  } , \intermed {  {\intermed { \Gamma_C } }  } }_{\intermed {  {\intermed { R^{SN} } }  } }  } \and

    \ottaltinferrule{}{}
    {  \intermed {  {\intermed { \gamma^{SN} } }  } \in \mathcal{H^{SN} } \llbracket \intermed {  {\intermed \Gamma}  } \rrbracket ^{\intermed {  {\intermed \Sigma}  } , \intermed {  {\intermed { \Gamma_C } }  } }_{\intermed {  {\intermed { R^{SN} } }  } }  }
    {  \intermed {  {\intermed { \gamma^{SN} } }  } \in \mathcal{H^{SN} } \llbracket \intermed {  {\intermed \Gamma}  \ottsym{,}  {\intermed a }  } \rrbracket ^{\intermed {  {\intermed \Sigma}  } , \intermed {  {\intermed { \Gamma_C } }  } }_{\intermed {  {\intermed { R^{SN} } }  } }  } \and

    \ottaltinferrule{}{}
    {  \intermed {  {\intermed { \gamma^{SN} } }  } \in \mathcal{H^{SN} } \llbracket \intermed {  {\intermed \Gamma}  } \rrbracket ^{\intermed {  {\intermed \Sigma}  } , \intermed {  {\intermed { \Gamma_C } }  } }_{\intermed {  {\intermed { R^{SN} } }  } }  }
    {  \intermed {  {\intermed { \gamma^{SN} } }  } \in \mathcal{H^{SN} } \llbracket \intermed {  {\intermed \Gamma}  \ottsym{,}  {\intermed x}  \ottsym{:}  {\intermed \sigma}  } \rrbracket ^{\intermed {  {\intermed \Sigma}  } , \intermed {  {\intermed { \Gamma_C } }  } }_{\intermed {  {\intermed { R^{SN} } }  } }  } \and

    \ottaltinferrule{}{}
    {  \intermed {  {\intermed { \gamma^{SN} } }  } \in \mathcal{H^{SN} } \llbracket \intermed {  {\intermed \Gamma}  } \rrbracket ^{\intermed {  {\intermed \Sigma}  } , \intermed {  {\intermed { \Gamma_C } }  } }_{\intermed {  {\intermed { R^{SN} } }  } } 
      \\  \intermed {  {\intermed \Sigma}  } ; \intermed {  {\intermed { \Gamma_C } }  } ; \intermed {  \bullet  }  \vdash_{d}  \intermed {  {\intermed d}  } : \intermed {  {\intermed { R^{SN} } }_{{\mathrm{1}}}  \ottsym{(}  {\intermed Q}  \ottsym{)}  } \itot{\,  \rightsquigarrow  \target {  {\target e }  } } 
    }
    {  \intermed {   {\intermed { \gamma^{SN} } } ,  {\intermed \delta }  \mapsto  {\intermed d}   } \in \mathcal{H^{SN} } \llbracket \intermed {  {\intermed \Gamma}  \ottsym{,}  {\intermed \delta }  \ottsym{:}  {\intermed Q}  } \rrbracket ^{\intermed {  {\intermed \Sigma}  } , \intermed {  {\intermed { \Gamma_C } }  } }_{\intermed {  {\intermed { R^{SN} } }  } }   }
  \end{mathpar}
\end{definitionA}

\section{Equivalence Relations}

\subsection{Kleene Equivalence Relations}
\label{app:log_kleene}

  \begin{flushleft}
    \namedRuleform{ \intermed {  {\intermed \Sigma}_{{\mathrm{1}}}  } : \intermed {  {\intermed e}_{{\mathrm{1}}}  }  \simeq  \intermed {  {\intermed \Sigma}_{{\mathrm{2}}}  } : \intermed {  {\intermed e}_{{\mathrm{2}}}  } }
                  {Kleene Equivalence for \interL{} Expressions}
  \end{flushleft}
  \begin{align*}
     \intermed {  {\intermed \Sigma}_{{\mathrm{1}}}  } : \intermed {  {\intermed e}_{{\mathrm{1}}}  }  \simeq  \intermed {  {\intermed \Sigma}_{{\mathrm{2}}}  } : \intermed {  {\intermed e}_{{\mathrm{2}}}  } 
    & \triangleq \exists {\intermed v} :  \intermed {  {\intermed \Sigma}_{{\mathrm{1}}}  }  \vdash  \intermed {  {\intermed e}_{{\mathrm{1}}}  }  \longrightarrow^{*}  \intermed {  {\intermed v}  } 
      \band  \intermed {  {\intermed \Sigma}_{{\mathrm{2}}}  }  \vdash  \intermed {  {\intermed e}_{{\mathrm{2}}}  }  \longrightarrow^{*}  \intermed {  {\intermed v}  } 
  \end{align*}

  \begin{flushleft}
    \namedRuleform{ \target {  {\target e }_{{\mathrm{1}}}  }  \simeq  \target {  {\target e }_{{\mathrm{2}}}  } }
                  {Kleene Equivalence for \targetL{} Expressions}
  \end{flushleft}
  \begin{align*}
     \target {  {\target e }_{{\mathrm{1}}}  }  \simeq  \target {  {\target e }_{{\mathrm{2}}}  } 
    & \triangleq \exists {\target v} :  \target {  {\target e }_{{\mathrm{1}}}  }  \longrightarrow^{*}  \target {  {\target v}  } 
      \band  \target {  {\target e }_{{\mathrm{2}}}  }  \longrightarrow^{*}  \target {  {\target v}  } 
  \end{align*}

\subsection{Contextual Equivalence Relations}

  \begin{flushleft}
    \namedRuleform{ \intermed {  {\intermed { \Gamma_C } }  } ; \intermed {  {\intermed \Gamma}  }  \vdash  \intermed {  {\intermed \Sigma}_{{\mathrm{1}}}  } : \intermed {  {\intermed e}_{{\mathrm{1}}}  }  \simeq_{ctx}  \intermed {  {\intermed \Sigma}_{{\mathrm{2}}}  } : \intermed {  {\intermed e}_{{\mathrm{2}}}  } : \intermed {  {\intermed \sigma}  } }
                  {Contextual Equivalence for \interL{} Expressions}
  \end{flushleft}
  \begin{align*}
     \intermed {  {\intermed { \Gamma_C } }  } ; \intermed {  {\intermed \Gamma}  }  \vdash  \intermed {  {\intermed \Sigma}_{{\mathrm{1}}}  } : \intermed {  {\intermed e}_{{\mathrm{1}}}  }  \simeq_{ctx}  \intermed {  {\intermed \Sigma}_{{\mathrm{2}}}  } : \intermed {  {\intermed e}_{{\mathrm{2}}}  } : \intermed {  {\intermed \sigma}  } 
    & \triangleq \forall  \intermed {  {\intermed M}_{{\mathrm{1}}}  } : ( \intermed {  {\intermed \Sigma}_{{\mathrm{1}}}  } ; \intermed {  {\intermed { \Gamma_C } }  } ; \intermed {  {\intermed \Gamma}  }  \Rightarrow  \intermed {  {\intermed \sigma}  } )  \mapsto  ( \intermed {  {\intermed \Sigma}_{{\mathrm{1}}}  } ; \intermed {  {\intermed { \Gamma_C } }  } ; \intermed {  \bullet  }  \Rightarrow  \intermed {  \mathit{\intermed {Bool} }  } ) \itot{ \rightsquigarrow  \target {  {\target M }_{{\mathrm{1}}}  } }  \\
    & \band \forall  \intermed {  {\intermed M}_{{\mathrm{2}}}  } : ( \intermed {  {\intermed \Sigma}_{{\mathrm{2}}}  } ; \intermed {  {\intermed { \Gamma_C } }  } ; \intermed {  {\intermed \Gamma}  }  \Rightarrow  \intermed {  {\intermed \sigma}  } )  \mapsto  ( \intermed {  {\intermed \Sigma}_{{\mathrm{2}}}  } ; \intermed {  {\intermed { \Gamma_C } }  } ; \intermed {  \bullet  }  \Rightarrow  \intermed {  \mathit{\intermed {Bool} }  } ) \itot{ \rightsquigarrow  \target {  {\target M }_{{\mathrm{2}}}  } }  \\
    & \band  \intermed {  {\intermed \Sigma}_{{\mathrm{1}}}  } : \intermed {  {\intermed M}_{{\mathrm{1}}}  }  \simeq_{log}  \intermed {  {\intermed \Sigma}_{{\mathrm{2}}}  } : \intermed {  {\intermed M}_{{\mathrm{2}}}  } : ( \intermed {  {\intermed { \Gamma_C } }  } ; \intermed {  {\intermed \Gamma}  }  \Rightarrow  \intermed {  {\intermed \sigma}  } )  \mapsto  ( \intermed {  {\intermed { \Gamma_C } }  } ; \intermed {  \bullet  }  \Rightarrow  \intermed {  \mathit{\intermed {Bool} }  } )  \\
    & \Rightarrow  \intermed {  {\intermed \Sigma}_{{\mathrm{1}}}  } : \intermed {   {\intermed M}_{{\mathrm{1}}}  [  {\intermed e}_{{\mathrm{1}}}  ]   }  \simeq  \intermed {  {\intermed \Sigma}_{{\mathrm{2}}}  } : \intermed {   {\intermed M}_{{\mathrm{2}}}  [  {\intermed e}_{{\mathrm{2}}}  ]   } 
  \end{align*}

  \begin{flushleft}
    \namedRuleform{ \source {  { \source P }  } ; \source {  {\source { \Gamma_C } }  } ; \source {  {\source \Gamma}  }  \vdash  \target {  {\target e }_{{\mathrm{1}}}  }  \simeq_{ctx}  \target {  {\target e }_{{\mathrm{2}}}  } : \source {  {\source \tau }  } }
                  {Contextual Equivalence for \targetL{} Expressions}
  \end{flushleft}
  \begin{align*}
     \source {  { \source P }  } ; \source {  {\source { \Gamma_C } }  } ; \source {  {\source \Gamma}  }  \vdash  \target {  {\target e }_{{\mathrm{1}}}  }  \simeq_{ctx}  \target {  {\target e }_{{\mathrm{2}}}  } : \source {  {\source \tau }  } 
    & \triangleq \forall  \source {  {\source M}  } : ( \source {  { \source P }  } ; \source {  {\source { \Gamma_C } }  } ; \source {  {\source \Gamma}  }  \Rightarrow  \source {  {\source \tau }  } )  \mapsto  ( \source {  { \source P }  } ; \source {  {\source { \Gamma_C } }  } ; \source {  \bullet  }  \Rightarrow  \source {  \mathit{\source {Bool} }  } )  \rightsquigarrow  \target {  {\target M }_{{\mathrm{1}}}  }  \\
    & \band  \source {  {\source M}  } : ( \source {  { \source P }  } ; \source {  {\source { \Gamma_C } }  } ; \source {  {\source \Gamma}  }  \Rightarrow  \source {  {\source \tau }  } )  \mapsto  ( \source {  { \source P }  } ; \source {  {\source { \Gamma_C } }  } ; \source {  \bullet  }  \Rightarrow  \source {  \mathit{\source {Bool} }  } )  \rightsquigarrow  \target {  {\target M }_{{\mathrm{2}}}  }  \\
    & \Rightarrow  \target {   {\target M }_{{\mathrm{1}}}  [  {\target e }_{{\mathrm{1}}}  ]   }  \simeq  \target {   {\target M }_{{\mathrm{2}}}  [  {\target e }_{{\mathrm{2}}}  ]   } 
  \end{align*}

  \renewcommand\itot[1]{#1}

  \begin{flushleft}
    \namedRuleform{ \intermed {  {\intermed { \Gamma_C } }  } ; \intermed {  {\intermed \Gamma}  }  \vdash  \intermed {  {\intermed \Sigma}_{{\mathrm{1}}}  } : \target {  {\target e }_{{\mathrm{1}}}  }  \simeq_{ctx}  \intermed {  {\intermed \Sigma}_{{\mathrm{2}}}  } : \target {  {\target e }_{{\mathrm{2}}}  } : \intermed {  {\intermed \sigma}  } }
                  {Contextual Equivalence for \targetL{} Expressions in
                    \interL{} context}
  \end{flushleft}
  \begin{align*}
     \intermed {  {\intermed { \Gamma_C } }  } ; \intermed {  {\intermed \Gamma}  }  \vdash  \intermed {  {\intermed \Sigma}_{{\mathrm{1}}}  } : \target {  {\target e }_{{\mathrm{1}}}  }  \simeq_{ctx}  \intermed {  {\intermed \Sigma}_{{\mathrm{2}}}  } : \target {  {\target e }_{{\mathrm{2}}}  } : \intermed {  {\intermed \sigma}  } 
    & \triangleq \forall  \intermed {  {\intermed M}_{{\mathrm{1}}}  } : ( \intermed {  {\intermed \Sigma}_{{\mathrm{1}}}  } ; \intermed {  {\intermed { \Gamma_C } }  } ; \intermed {  {\intermed \Gamma}  }  \Rightarrow  \intermed {  {\intermed \sigma}  } )  \mapsto  ( \intermed {  {\intermed \Sigma}_{{\mathrm{1}}}  } ; \intermed {  {\intermed { \Gamma_C } }  } ; \intermed {  \bullet  }  \Rightarrow  \intermed {  \mathit{\intermed {Bool} }  } ) \itot{ \rightsquigarrow  \target {  {\target M }_{{\mathrm{1}}}  } }  \\
    & \band \forall  \intermed {  {\intermed M}_{{\mathrm{2}}}  } : ( \intermed {  {\intermed \Sigma}_{{\mathrm{2}}}  } ; \intermed {  {\intermed { \Gamma_C } }  } ; \intermed {  {\intermed \Gamma}  }  \Rightarrow  \intermed {  {\intermed \sigma}  } )  \mapsto  ( \intermed {  {\intermed \Sigma}_{{\mathrm{2}}}  } ; \intermed {  {\intermed { \Gamma_C } }  } ; \intermed {  \bullet  }  \Rightarrow  \intermed {  \mathit{\intermed {Bool} }  } ) \itot{ \rightsquigarrow  \target {  {\target M }_{{\mathrm{2}}}  } }  \\
    & \band  \intermed {  {\intermed \Sigma}_{{\mathrm{1}}}  } : \intermed {  {\intermed M}_{{\mathrm{1}}}  }  \simeq_{log}  \intermed {  {\intermed \Sigma}_{{\mathrm{2}}}  } : \intermed {  {\intermed M}_{{\mathrm{2}}}  } : ( \intermed {  {\intermed { \Gamma_C } }  } ; \intermed {  {\intermed \Gamma}  }  \Rightarrow  \intermed {  {\intermed \sigma}  } )  \mapsto  ( \intermed {  {\intermed { \Gamma_C } }  } ; \intermed {  \bullet  }  \Rightarrow  \intermed {  \mathit{\intermed {Bool} }  } )  \\
    & \Rightarrow  \target {   {\target M }_{{\mathrm{1}}}  [  {\target e }_{{\mathrm{1}}}  ]   }  \simeq  \target {   {\target M }_{{\mathrm{2}}}  [  {\target e }_{{\mathrm{2}}}  ]   }  
  \end{align*}


\newpage

\renewcommand\itot[1]{}

\section{\srcL{} Theorems}

\subsection{Conjectures}
\label{app:src_conj}

We are confident that the following lemmas can be proven using well-known proof
techniques. 

\begin{lemmaB}[Type Variable Substitution in \srcL{} Constraint Typing]\ \\
  \renewcommand\itot[1]{}
  \label{lemma:tvar_subst_ctr}
  If \( \source {  {\source { \Gamma_C } }  } ; \source {  {\source \Gamma}_{{\mathrm{1}}}  \ottsym{,}  {\source a }  \ottsym{,}  {\source \Gamma}_{{\mathrm{2}}}  }  \vdash_{\mathit {Q} }^{\intermed {M} }  \source {  {\source Q}  }  \rightsquigarrow  \intermed {  {\intermed Q}  } \)
  and \( \source {  {\source { \Gamma_C } }  } ; \source {  {\source \Gamma}_{{\mathrm{1}}}  }  \vdash_{ty}^{\intermed {M} }  \source {  {\source \tau }  }\,{ \stoi{ \rightsquigarrow  \intermed {  {\intermed \sigma}  } } } \)
  then \( \source {  {\source { \Gamma_C } }  } ; \source {  {\source \Gamma}_{{\mathrm{1}}}  \ottsym{,}  [  {\source \tau }  /  {\source a }  ]  {\source \Gamma}_{{\mathrm{2}}}  }  \vdash_{\mathit {Q} }^{\intermed {M} }  \source {  [  {\source \tau }  /  {\source a }  ]  {\source Q}  }  \rightsquigarrow  \intermed {  [  {\intermed \sigma}  /  {\intermed a }  ]  {\intermed Q}  } \).
\end{lemmaB}

\begin{lemmaB}[Type Well-Formedness Environment Weakening]\ \\
  \renewcommand\itot[1]{}
  \label{lemma:type_weakening}
  If \( \source {  {\source { \Gamma_C } }_{{\mathrm{1}}}  } ; \source {  {\source \Gamma}_{{\mathrm{1}}}  }  \vdash_{ty}^{\intermed {M} }  \source {  {\source \sigma}  }\,{ \stoi{ \rightsquigarrow  \intermed {  {\intermed \sigma}  } } } \)
  and \( \vdash_{ctx}^{\intermed {M} }  \source {  \bullet  } ; \source {  {\source { \Gamma_C } }_{{\mathrm{1}}}  \ottsym{,}  {\source { \Gamma_C } }_{{\mathrm{2}}}  } ; \source {  {\source \Gamma}_{{\mathrm{1}}}  \ottsym{,}  {\source \Gamma}_{{\mathrm{2}}}  }  \rightsquigarrow  \intermed {  \bullet  } ; \intermed {  {\intermed { \Gamma_C } }  } ; \intermed {  {\intermed \Gamma}  } \)
  then \( \source {  {\source { \Gamma_C } }_{{\mathrm{1}}}  \ottsym{,}  {\source { \Gamma_C } }_{{\mathrm{2}}}  } ; \source {  {\source \Gamma}_{{\mathrm{1}}}  \ottsym{,}  {\source \Gamma}_{{\mathrm{2}}}  }  \vdash_{ty}^{\intermed {M} }  \source {  {\source \sigma}  }\,{ \stoi{ \rightsquigarrow  \intermed {  {\intermed \sigma}  } } } \).
\end{lemmaB}

\begin{lemmaB}[Class Constraint Well-Formedness Environment Weakening]\ \\
  \renewcommand\itot[1]{}
  \label{lemma:class_constraint_weakening}
  If \( \source {  {\source { \Gamma_C } }_{{\mathrm{1}}}  } ; \source {  {\source \Gamma}_{{\mathrm{1}}}  }  \vdash_{\mathit {Q} }^{\intermed {M} }  \source {  {\source Q}  }  \rightsquigarrow  \intermed {  {\intermed Q}'  } \)
  and \( \vdash_{ctx}^{\intermed {M} }  \source {  \bullet  } ; \source {  {\source { \Gamma_C } }_{{\mathrm{1}}}  \ottsym{,}  {\source { \Gamma_C } }_{{\mathrm{2}}}  } ; \source {  {\source \Gamma}_{{\mathrm{1}}}  \ottsym{,}  {\source \Gamma}_{{\mathrm{2}}}  }  \rightsquigarrow  \intermed {  \bullet  } ; \intermed {  {\intermed { \Gamma_C } }  } ; \intermed {  {\intermed \Gamma}  } \)
  then \( \source {  {\source { \Gamma_C } }_{{\mathrm{1}}}  \ottsym{,}  {\source { \Gamma_C } }_{{\mathrm{2}}}  } ; \source {  {\source \Gamma}_{{\mathrm{1}}}  \ottsym{,}  {\source \Gamma}_{{\mathrm{2}}}  }  \vdash_{\mathit {Q} }^{\intermed {M} }  \source {  {\source Q}  }  \rightsquigarrow  \intermed {  {\intermed Q}'  } \).
\end{lemmaB}

\begin{lemmaB}[Constraint Well-Formedness Environment Weakening]\ \\
  \renewcommand\itot[1]{}
  \label{lemma:constraint_weakening}
  If \( \source {  {\source { \Gamma_C } }_{{\mathrm{1}}}  } ; \source {  {\source \Gamma}_{{\mathrm{1}}}  }  \vdash_{\mathit {C} }^{\intermed {M} }  \source {  {\source C}  }  \rightsquigarrow  \intermed {  {\intermed C}'  } \)
  and \( \vdash_{ctx}^{\intermed {M} }  \source {  \bullet  } ; \source {  {\source { \Gamma_C } }_{{\mathrm{1}}}  \ottsym{,}  {\source { \Gamma_C } }_{{\mathrm{2}}}  } ; \source {  {\source \Gamma}_{{\mathrm{1}}}  \ottsym{,}  {\source \Gamma}_{{\mathrm{2}}}  }  \rightsquigarrow  \intermed {  \bullet  } ; \intermed {  {\intermed { \Gamma_C } }  } ; \intermed {  {\intermed \Gamma}  } \)
  then \( \source {  {\source { \Gamma_C } }_{{\mathrm{1}}}  \ottsym{,}  {\source { \Gamma_C } }_{{\mathrm{2}}}  } ; \source {  {\source \Gamma}_{{\mathrm{1}}}  \ottsym{,}  {\source \Gamma}_{{\mathrm{2}}}  }  \vdash_{\mathit {C} }^{\intermed {M} }  \source {  {\source C}  }  \rightsquigarrow  \intermed {  {\intermed C}'  } \).
\end{lemmaB}

\begin{lemmaB}[Context Well-Formedness Class Environment Weakening]\ \\
  \renewcommand\itot[1]{}
  \label{lemma:ctx_weakening_class}
  If \( \vdash_{ctx}^{\intermed {M} }  \source {  { \source P }  } ; \source {  {\source { \Gamma_C } }_{{\mathrm{1}}}  } ; \source {  {\source \Gamma}  }  \rightsquigarrow  \intermed {  {\intermed \Sigma}  } ; \intermed {  {\intermed { \Gamma_C } }_{{\mathrm{1}}}  } ; \intermed {  {\intermed \Gamma}  } \)
  and \( \vdash_{ctx}^{\intermed {M} }  \source {  \bullet  } ; \source {  {\source { \Gamma_C } }_{{\mathrm{1}}}  \ottsym{,}  {\source { \Gamma_C } }_{{\mathrm{2}}}  } ; \source {  {\source \Gamma}  }  \rightsquigarrow  \intermed {  \bullet  } ; \intermed {  {\intermed { \Gamma_C } }_{{\mathrm{1}}}  \ottsym{,}  {\intermed { \Gamma_C } }_{{\mathrm{2}}}  } ; \intermed {  {\intermed \Gamma}  } \)
  then \( \vdash_{ctx}^{\intermed {M} }  \source {  { \source P }  } ; \source {  {\source { \Gamma_C } }_{{\mathrm{1}}}  \ottsym{,}  {\source { \Gamma_C } }_{{\mathrm{2}}}  } ; \source {  {\source \Gamma}  }  \rightsquigarrow  \intermed {  {\intermed \Sigma}  } ; \intermed {  {\intermed { \Gamma_C } }_{{\mathrm{1}}}  \ottsym{,}  {\intermed { \Gamma_C } }_{{\mathrm{2}}}  } ; \intermed {  {\intermed \Gamma}  } \).
\end{lemmaB}

\begin{lemmaB}[Context Well-Formedness Typing Environment Weakening]\ \\
  \renewcommand\itot[1]{}
  \label{lemma:ctx_weakening_typing}
  If \( \vdash_{ctx}^{\intermed {M} }  \source {  { \source P }  } ; \source {  {\source { \Gamma_C } }  } ; \source {  {\source \Gamma}_{{\mathrm{1}}}  }  \rightsquigarrow  \intermed {  {\intermed \Sigma}  } ; \intermed {  {\intermed { \Gamma_C } }  } ; \intermed {  {\intermed \Gamma}_{{\mathrm{1}}}  } \)
  and \( \vdash_{ctx}^{\intermed {M} }  \source {  \bullet  } ; \source {  {\source { \Gamma_C } }  } ; \source {  {\source \Gamma}_{{\mathrm{1}}}  \ottsym{,}  {\source \Gamma}_{{\mathrm{2}}}  }  \rightsquigarrow  \intermed {  \bullet  } ; \intermed {  {\intermed { \Gamma_C } }  } ; \intermed {  {\intermed \Gamma}_{{\mathrm{1}}}  \ottsym{,}  {\intermed \Gamma}_{{\mathrm{2}}}  } \)
  then \( \vdash_{ctx}^{\intermed {M} }  \source {  { \source P }  } ; \source {  {\source { \Gamma_C } }  } ; \source {  {\source \Gamma}_{{\mathrm{1}}}  \ottsym{,}  {\source \Gamma}_{{\mathrm{2}}}  }  \rightsquigarrow  \intermed {  {\intermed \Sigma}  } ; \intermed {  {\intermed { \Gamma_C } }  } ; \intermed {  {\intermed \Gamma}_{{\mathrm{1}}}  \ottsym{,}  {\intermed \Gamma}_{{\mathrm{2}}}  } \).
\end{lemmaB}

\subsection{Lemmas}

\begin{lemmaB}[Determinism of Context Typing]\leavevmode
  \label{lemma:ctx_type_unique}
  \begin{itemize}
  \item If \( \source {  {\source M}  } : ( \source {  { \source P }  } ; \source {  {\source { \Gamma_C } }  } ; \source {  {\source \Gamma}  }  \Rightarrow  \source {  {\source \tau }  } )  \mapsto  ( \source {  { \source P }  } ; \source {  {\source { \Gamma_C } }  } ; \source {  {\source \Gamma}'  }  \Rightarrow  \source {  {\source \tau }_{{\mathrm{1}}}  } )  \rightsquigarrow  \intermed {  {\intermed M}_{{\mathrm{1}}}  } \)
       and \( \source {  {\source M}  } : ( \source {  { \source P }  } ; \source {  {\source { \Gamma_C } }  } ; \source {  {\source \Gamma}  }  \Rightarrow  \source {  {\source \tau }  } )  \mapsto  ( \source {  { \source P }  } ; \source {  {\source { \Gamma_C } }  } ; \source {  {\source \Gamma}'  }  \Rightarrow  \source {  {\source \tau }_{{\mathrm{2}}}  } )  \rightsquigarrow  \intermed {  {\intermed M}_{{\mathrm{2}}}  } \) \\
       then \( \source {  {\source \tau }_{{\mathrm{1}}}  } = \source {  {\source \tau }_{{\mathrm{2}}}  } \).
  \item If \( \source {  {\source M}  } : ( \source {  { \source P }  } ; \source {  {\source { \Gamma_C } }  } ; \source {  {\source \Gamma}  }  \Rightarrow  \source {  {\source \tau }  } )  \mapsto  ( \source {  { \source P }  } ; \source {  {\source { \Gamma_C } }  } ; \source {  {\source \Gamma}'  }  \Leftarrow  \source {  {\source \tau }_{{\mathrm{1}}}  } )  \rightsquigarrow  \intermed {  {\intermed M}_{{\mathrm{1}}}  } \)
       and \( \source {  {\source M}  } : ( \source {  { \source P }  } ; \source {  {\source { \Gamma_C } }  } ; \source {  {\source \Gamma}  }  \Rightarrow  \source {  {\source \tau }  } )  \mapsto  ( \source {  { \source P }  } ; \source {  {\source { \Gamma_C } }  } ; \source {  {\source \Gamma}'  }  \Leftarrow  \source {  {\source \tau }_{{\mathrm{2}}}  } )  \rightsquigarrow  \intermed {  {\intermed M}_{{\mathrm{2}}}  } \) \\
       then \( \source {  {\source \tau }_{{\mathrm{1}}}  } = \source {  {\source \tau }_{{\mathrm{2}}}  } \).
  \item If \( \source {  {\source M}  } : ( \source {  { \source P }  } ; \source {  {\source { \Gamma_C } }  } ; \source {  {\source \Gamma}  }  \Leftarrow  \source {  {\source \tau }  } )  \mapsto  ( \source {  { \source P }  } ; \source {  {\source { \Gamma_C } }  } ; \source {  {\source \Gamma}'  }  \Rightarrow  \source {  {\source \tau }_{{\mathrm{1}}}  } )  \rightsquigarrow  \intermed {  {\intermed M}_{{\mathrm{1}}}  } \)
       and \( \source {  {\source M}  } : ( \source {  { \source P }  } ; \source {  {\source { \Gamma_C } }  } ; \source {  {\source \Gamma}  }  \Leftarrow  \source {  {\source \tau }  } )  \mapsto  ( \source {  { \source P }  } ; \source {  {\source { \Gamma_C } }  } ; \source {  {\source \Gamma}'  }  \Rightarrow  \source {  {\source \tau }_{{\mathrm{2}}}  } )  \rightsquigarrow  \intermed {  {\intermed M}_{{\mathrm{2}}}  } \) \\
       then \( \source {  {\source \tau }_{{\mathrm{1}}}  } = \source {  {\source \tau }_{{\mathrm{2}}}  } \).
  \item If \( \source {  {\source M}  } : ( \source {  { \source P }  } ; \source {  {\source { \Gamma_C } }  } ; \source {  {\source \Gamma}  }  \Leftarrow  \source {  {\source \tau }  } )  \mapsto  ( \source {  { \source P }  } ; \source {  {\source { \Gamma_C } }  } ; \source {  {\source \Gamma}'  }  \Leftarrow  \source {  {\source \tau }_{{\mathrm{1}}}  } )  \rightsquigarrow  \intermed {  {\intermed M}_{{\mathrm{1}}}  } \)
       and \( \source {  {\source M}  } : ( \source {  { \source P }  } ; \source {  {\source { \Gamma_C } }  } ; \source {  {\source \Gamma}  }  \Leftarrow  \source {  {\source \tau }  } )  \mapsto  ( \source {  { \source P }  } ; \source {  {\source { \Gamma_C } }  } ; \source {  {\source \Gamma}'  }  \Leftarrow  \source {  {\source \tau }_{{\mathrm{2}}}  } )  \rightsquigarrow  \intermed {  {\intermed M}_{{\mathrm{2}}}  } \) \\
       then \( \source {  {\source \tau }_{{\mathrm{1}}}  } = \source {  {\source \tau }_{{\mathrm{2}}}  } \).
  \end{itemize}
\end{lemmaB}

\proof
By straightforward induction on the first typing derivation, in combination with
case analysis on the second derivation. \\
\qed

\begin{lemmaB}[Class Constraint Elaboration to \interL{} Uniqueness]\ \\
  \label{lemma:q_elab_unique}
  If \( \source {  {\source { \Gamma_C } }  } ; \source {  {\source \Gamma}  }  \vdash_{\mathit {Q} }^{\intermed {M} }  \source {  {\source Q}  }  \rightsquigarrow  \intermed {  {\intermed Q}_{{\mathrm{1}}}  } \) and \( \source {  {\source { \Gamma_C } }  } ; \source {  {\source \Gamma}  }  \vdash_{\mathit {Q} }^{\intermed {M} }  \source {  {\source Q}  }  \rightsquigarrow  \intermed {  {\intermed Q}_{{\mathrm{2}}}  } \),
  then \( \intermed {  {\intermed Q}_{{\mathrm{1}}}  } = \intermed {  {\intermed Q}_{{\mathrm{2}}}  } \).
\end{lemmaB}

\proof
By mutual induction on both well-formedness derivations, together with
Lemma~\ref{lemma:ty_elab_unique}. \\
\qed

\begin{lemmaB}[Type Elaboration to \interL{} Uniqueness]\ \\
  \label{lemma:ty_elab_unique}
  If \( \source {  {\source { \Gamma_C } }  } ; \source {  {\source \Gamma}  }  \vdash_{ty}^{\intermed {M} }  \source {  {\source \sigma}  }\,{ \stoi{ \rightsquigarrow  \intermed {  {\intermed \sigma}_{{\mathrm{1}}}  } } } \) and \( \source {  {\source { \Gamma_C } }  } ; \source {  {\source \Gamma}  }  \vdash_{ty}^{\intermed {M} }  \source {  {\source \sigma}  }\,{ \stoi{ \rightsquigarrow  \intermed {  {\intermed \sigma}_{{\mathrm{2}}}  } } } \),
  then \( \intermed {  {\intermed \sigma}_{{\mathrm{1}}}  } = \intermed {  {\intermed \sigma}_{{\mathrm{2}}}  } \).
\end{lemmaB}

\proof
By mutual induction on both well-formedness derivations, together with
Lemma~\ref{lemma:q_elab_unique}. \\
\qed

\begin{lemmaB}[Constraint Elaboration to \interL{} Uniqueness]\ \\
  \label{lemma:c_elab_unique}
  If \( \source {  {\source { \Gamma_C } }  } ; \source {  {\source \Gamma}  }  \vdash_{\mathit {C} }^{\intermed {M} }  \source {  {\source C}  }  \rightsquigarrow  \intermed {  {\intermed C}_{{\mathrm{1}}}  } \) and \( \source {  {\source { \Gamma_C } }  } ; \source {  {\source \Gamma}  }  \vdash_{\mathit {C} }^{\intermed {M} }  \source {  {\source C}  }  \rightsquigarrow  \intermed {  {\intermed C}_{{\mathrm{2}}}  } \),
  then \( \intermed {  {\intermed C}_{{\mathrm{1}}}  } = \intermed {  {\intermed C}_{{\mathrm{2}}}  } \).
\end{lemmaB}

\proof
By straightforward induction on both well-formedness derivations, in combination
with Lemma~\ref{lemma:q_elab_unique}. \\
\qed

\begin{lemmaB}[Environment Elaboration to \interL{} Uniqueness]\ \\
  \label{lemma:env_elab_unique}
  If \( \vdash_{ctx}^{\intermed {M} }  \source {  { \source P }  } ; \source {  {\source { \Gamma_C } }  } ; \source {  {\source \Gamma}  }  \rightsquigarrow  \intermed {  {\intermed \Sigma}_{{\mathrm{1}}}  } ; \intermed {  {\intermed { \Gamma_C } }_{{\mathrm{1}}}  } ; \intermed {  {\intermed \Gamma}_{{\mathrm{2}}}  } \)
  and \( \vdash_{ctx}^{\intermed {M} }  \source {  { \source P }  } ; \source {  {\source { \Gamma_C } }  } ; \source {  {\source \Gamma}  }  \rightsquigarrow  \intermed {  {\intermed \Sigma}_{{\mathrm{2}}}  } ; \intermed {  {\intermed { \Gamma_C } }_{{\mathrm{2}}}  } ; \intermed {  {\intermed \Gamma}_{{\mathrm{2}}}  } \),
  then \( \intermed {  {\intermed { \Gamma_C } }_{{\mathrm{1}}}  } = \intermed {  {\intermed { \Gamma_C } }_{{\mathrm{2}}}  } \) and \( \intermed {  {\intermed \Gamma}_{{\mathrm{1}}}  } = \intermed {  {\intermed \Gamma}_{{\mathrm{2}}}  } \).
\end{lemmaB}

\proof
By straightforward induction on both well-formedness derivations, in combination
with Lemmas~\ref{lemma:q_elab_unique}, \ref{lemma:ty_elab_unique} and~\ref{lemma:c_elab_unique}. \\
\qed

\begin{lemmaB}[Environment Well-Formedness of \srcL{} Typing]\leavevmode
  \label{lemma:ctx_well_source_typing}
  \begin{itemize}
  \item If \( \source {  { \source P }  } ; \source {  {\source { \Gamma_C } }  } ; \source {  {\source \Gamma}  }  \vdash_{tm}  \source {  {\source e}  } \Rightarrow \source {  {\source \tau }  } \stot{ \rightsquigarrow  \target {  {\target e }  } } \) then \( \vdash_{ctx}  \source {  { \source P }  } ; \source {  {\source { \Gamma_C } }  } ; \source {  {\source \Gamma}  }  \rightsquigarrow  \target {  {\target \Gamma }  } \).
  \item If \( \source {  { \source P }  } ; \source {  {\source { \Gamma_C } }  } ; \source {  {\source \Gamma}  }  \vdash_{tm}  \source {  {\source e}  } \Leftarrow \source {  {\source \tau }  } \stot{ \rightsquigarrow  \target {  {\target e }  } } \) then \( \vdash_{ctx}  \source {  { \source P }  } ; \source {  {\source { \Gamma_C } }  } ; \source {  {\source \Gamma}  }  \rightsquigarrow  \target {  {\target \Gamma }  } \).
  \end{itemize}
\end{lemmaB}

\proof
By straightforward induction on the typing derivation. \\
\qed

\begin{lemmaB}[Environment Well-Formedness of \srcL{} Typing through \interL{}]\leavevmode
  \label{lemma:ctx_well_source_mid_typing}
  \begin{itemize}
  \item If \( \source {  { \source P }  } ; \source {  {\source { \Gamma_C } }  } ; \source {  {\source \Gamma}  }  \vdash_{tm}^{\intermed {M} }  \source {  {\source e}  } \Rightarrow \source {  {\source \tau }  } \stoi{ \rightsquigarrow  \intermed {  {\intermed e}  } } \) then \( \vdash_{ctx}^{\intermed {M} }  \source {  { \source P }  } ; \source {  {\source { \Gamma_C } }  } ; \source {  {\source \Gamma}  }  \rightsquigarrow  \intermed {  {\intermed \Sigma}  } ; \intermed {  {\intermed { \Gamma_C } }  } ; \intermed {  {\intermed \Gamma}  } \).
  \item If \( \source {  { \source P }  } ; \source {  {\source { \Gamma_C } }  } ; \source {  {\source \Gamma}  }  \vdash_{tm}^{\intermed {M} }  \source {  {\source e}  } \Leftarrow \source {  {\source \tau }  } \stoi{ \rightsquigarrow  \intermed {  {\intermed e}  } } \) then \( \vdash_{ctx}^{\intermed {M} }  \source {  { \source P }  } ; \source {  {\source { \Gamma_C } }  } ; \source {  {\source \Gamma}  }  \rightsquigarrow  \intermed {  {\intermed \Sigma}  } ; \intermed {  {\intermed { \Gamma_C } }  } ; \intermed {  {\intermed \Gamma}  } \).
  \end{itemize}
\end{lemmaB}

\proof
By straightforward induction on the typing derivation. \\
\qed

\begin{lemmaB}[Well-Formedness of \srcL{} Typing Result] \leavevmode
  \label{lemma:type_well_source_typing}
  \begin{itemize}
  \item If \( \source {  { \source P }  } ; \source {  {\source { \Gamma_C } }  } ; \source {  {\source \Gamma}  }  \vdash_{tm}^{\intermed {M} }  \source {  {\source e}  } \Rightarrow \source {  {\source \tau }  } \stoi{ \rightsquigarrow  \intermed {  {\intermed e}  } } \)
    then \( \source {  {\source { \Gamma_C } }  } ; \source {  {\source \Gamma}  }  \vdash_{ty}^{\intermed {M} }  \source {  {\source \tau }  }\,{ \stoi{ \rightsquigarrow  \intermed {  {\intermed \sigma}  } } } \).
  \item If \( \source {  { \source P }  } ; \source {  {\source { \Gamma_C } }  } ; \source {  {\source \Gamma}  }  \vdash_{tm}^{\intermed {M} }  \source {  {\source e}  } \Leftarrow \source {  {\source \tau }  } \stoi{ \rightsquigarrow  \intermed {  {\intermed e}  } } \)
    then \( \source {  {\source { \Gamma_C } }  } ; \source {  {\source \Gamma}  }  \vdash_{ty}^{\intermed {M} }  \source {  {\source \tau }  }\,{ \stoi{ \rightsquigarrow  \intermed {  {\intermed \sigma}  } } } \).
  \end{itemize}
\end{lemmaB}

\proof
By straightforward induction on the typing derivation. \\
\qed

\begin{lemmaB}[Preservation of Environment Term Variables from \srcL{} to \interL{}]\leavevmode
  \label{lemma:pres_env_tm}
  \begin{itemize}
  \item If \( ( \source {  {\source x}  } : \source {  {\source \sigma}  } )  \in  \source {  {\source \Gamma}  } \)
    and \( \vdash_{ctx}^{\intermed {M} }  \source {  { \source P }  } ; \source {  {\source { \Gamma_C } }  } ; \source {  {\source \Gamma}  }  \rightsquigarrow  \intermed {  {\intermed \Sigma}  } ; \intermed {  {\intermed { \Gamma_C } }  } ; \intermed {  {\intermed \Gamma}  } \)
    then \( ( \intermed {  {\intermed x}  } : \intermed {  {\intermed \sigma}  } )  \in  \intermed {  {\intermed \Gamma}  } \)
    where \( \source {  {\source { \Gamma_C } }  } ; \source {  {\source \Gamma}  }  \vdash_{ty}^{\intermed {M} }  \source {  {\source \sigma}  }\,{ \stoi{ \rightsquigarrow  \intermed {  {\intermed \sigma}  } } } \).
  \item If \( \source {  {\source x}  }  \notin   \textbf {dom}  ( \source {  {\source \Gamma}  } ) \)
    and \( \vdash_{ctx}^{\intermed {M} }  \source {  { \source P }  } ; \source {  {\source { \Gamma_C } }  } ; \source {  {\source \Gamma}  }  \rightsquigarrow  \intermed {  {\intermed \Sigma}  } ; \intermed {  {\intermed { \Gamma_C } }  } ; \intermed {  {\intermed \Gamma}  } \)
    then \( \intermed {  {\intermed x}  }  \notin   \textbf {dom}  ( \intermed {  {\intermed \Gamma}  } ) \).
  \end{itemize}
\end{lemmaB}

\proof
By straightforward induction on the environment elaboration derivation. \\
\qed

\begin{lemmaB}[Preservation of Environment Type Variables from \srcL{} to \interL{}]\leavevmode
  \label{lemma:pres_env_ty}
  \begin{itemize}
  \item If \( \source {  {\source a }  }  \in  \source {  {\source \Gamma}  } \)
    and \( \vdash_{ctx}^{\intermed {M} }  \source {  { \source P }  } ; \source {  {\source { \Gamma_C } }  } ; \source {  {\source \Gamma}  }  \rightsquigarrow  \intermed {  {\intermed \Sigma}  } ; \intermed {  {\intermed { \Gamma_C } }  } ; \intermed {  {\intermed \Gamma}  } \)
    then \( \intermed {  {\intermed a }  }  \in  \intermed {  {\intermed \Gamma}  } \).
  \item If \( \source {  {\source a }  }  \notin  \source {  {\source \Gamma}  } \)
    and \( \vdash_{ctx}^{\intermed {M} }  \source {  { \source P }  } ; \source {  {\source { \Gamma_C } }  } ; \source {  {\source \Gamma}  }  \rightsquigarrow  \intermed {  {\intermed \Sigma}  } ; \intermed {  {\intermed { \Gamma_C } }  } ; \intermed {  {\intermed \Gamma}  } \)
    then \( \intermed {  {\intermed a }  }  \notin  \intermed {  {\intermed \Gamma}  } \).
  \end{itemize}
\end{lemmaB}

\proof
By straightforward induction on the environment elaboration derivation. \\
\qed

\begin{lemmaB}[Preservation of Environment Dictionary Variables from \srcL{} to \interL{}]\leavevmode
  \label{lemma:pres_env_d}
  \begin{itemize}
  \item If \( ( \source {  {\source \delta }  } : \source {  {\source Q}  } )  \in  \source {  {\source \Gamma}  } \)
    and \( \vdash_{ctx}^{\intermed {M} }  \source {  { \source P }  } ; \source {  {\source { \Gamma_C } }  } ; \source {  {\source \Gamma}  }  \rightsquigarrow  \intermed {  {\intermed \Sigma}  } ; \intermed {  {\intermed { \Gamma_C } }  } ; \intermed {  {\intermed \Gamma}  } \)
    then \( ( \intermed {  {\intermed \delta }  } : \intermed {  {\intermed Q}  } )  \in  \intermed {  {\intermed \Gamma}  } \)
    where \( \source {  {\source { \Gamma_C } }  } ; \source {  {\source \Gamma}  }  \vdash_{\mathit {Q} }^{\intermed {M} }  \source {  {\source Q}  }  \rightsquigarrow  \intermed {  {\intermed Q}  } \).
  \item If \( \source {  {\source \delta }  }  \notin   \textbf {dom}  ( \source {  {\source \Gamma}  } ) \)
    and \( \vdash_{ctx}^{\intermed {M} }  \source {  { \source P }  } ; \source {  {\source { \Gamma_C } }  } ; \source {  {\source \Gamma}  }  \rightsquigarrow  \intermed {  {\intermed \Sigma}  } ; \intermed {  {\intermed { \Gamma_C } }  } ; \intermed {  {\intermed \Gamma}  } \)
    then \( \intermed {  {\intermed \delta }  }  \notin   \textbf {dom}  ( \intermed {  {\intermed \Gamma}  } ) \).
  \end{itemize}
\end{lemmaB}

\proof
By straightforward induction on the environment elaboration derivation. \\
\qed

\begin{lemmaB}[Environment Well-Formedness Strengthening]\ \\
  \label{lemma:env_strengthening}
  If \( \vdash_{ctx}  \source {  { \source P }  } ; \source {  {\source { \Gamma_C } }  } ; \source {  {\source \Gamma}  }  \rightsquigarrow  \target {  {\target \Gamma }  } \)
  then \( \vdash_{ctx}  \source {  { \source P }  } ; \source {  {\source { \Gamma_C } }  } ; \source {  \bullet  }  \rightsquigarrow  \target {  \target {\bullet}  } \).
\end{lemmaB}

\proof
By case analysis on the hypothesis,
the last rules used to construct it must be (possibly zero)
consecutive applications of \textsc{sCtxT-pgmInst}. Revert those rules, to obtain
\( \vdash_{ctx}  \source {  \bullet  } ; \source {  {\source { \Gamma_C } }  } ; \source {  {\source \Gamma}  }  \rightsquigarrow  \target {  {\target \Gamma }  } \).
By further case analysis (\textsc{sCtxT-tyEnvTm}, \textsc{sCtxT-tyEnvTy} and
\textsc{sCtxT-tyEnvD}), we get \( \vdash_{ctx}  \source {  \bullet  } ; \source {  {\source { \Gamma_C } }  } ; \source {  \bullet  }  \rightsquigarrow  \target {  \target {\bullet}  } \).
The goal follows by consecutively
re-applying rule \textsc{sCtxT-pgmInst} with the appropriate premises. \\
\qed

\begin{lemmaB}[Environment Well-Formedness with \interL{} Elaboration
    Strengthening]\ \\
  \label{lemma:env_mid_strengthening}
  If \( \vdash_{ctx}^{\intermed {M} }  \source {  { \source P }  } ; \source {  {\source { \Gamma_C } }  } ; \source {  {\source \Gamma}  }  \rightsquigarrow  \intermed {  {\intermed \Sigma}  } ; \intermed {  {\intermed { \Gamma_C } }  } ; \intermed {  {\intermed \Gamma}  } \)
  then \( \vdash_{ctx}^{\intermed {M} }  \source {  { \source P }  } ; \source {  {\source { \Gamma_C } }  } ; \source {  \bullet  }  \rightsquigarrow  \intermed {  {\intermed \Sigma}  } ; \intermed {  {\intermed { \Gamma_C } }  } ; \intermed {  \bullet  } \).
\end{lemmaB}

\proof
By case analysis on the hypothesis,
the last rules used to construct it must be (possibly zero)
consecutive applications of \textsc{sCtx-pgmInst}. Revert those rules, to obtain
\( \vdash_{ctx}^{\intermed {M} }  \source {  \bullet  } ; \source {  {\source { \Gamma_C } }  } ; \source {  {\source \Gamma}  }  \rightsquigarrow  \intermed {  \bullet  } ; \intermed {  {\intermed { \Gamma_C } }  } ; \intermed {  {\intermed \Gamma}  } \).
By further case analysis (\textsc{sCtx-tyEnvTm}, \textsc{sCtx-tyEnvTy} and
\textsc{sCtx-tyEnvD}), we get \( \vdash_{ctx}^{\intermed {M} }  \source {  \bullet  } ; \source {  {\source { \Gamma_C } }  } ; \source {  \bullet  }  \rightsquigarrow  \intermed {  \bullet  } ; \intermed {  {\intermed { \Gamma_C } }  } ; \intermed {  \bullet  } \).
The goal follows by consecutively
re-applying rule \textsc{sCtx-pgmInst} with the appropriate premises. \\
\qed

\subsection{Typing Preservation}
\label{app:type_pres}

\begin{theoremB}[Typing Preservation - Expressions]\leavevmode
\label{thmA:typing_pres_expr}
  \begin{itemize}
  \item If $ \source {  { \source P }  } ; \source {  {\source { \Gamma_C } }  } ; \source {  {\source \Gamma}  }  \vdash_{tm}^{\intermed {M} }  \source {  {\source e}  } \Rightarrow \source {  {\source \tau }  } \stoi{ \rightsquigarrow  \intermed {  {\intermed e}  } } $,
    and $ \source {  {\source { \Gamma_C } }  } ; \source {  {\source \Gamma}  }  \vdash_{ty}^{\intermed {M} }  \source {  {\source \tau }  }\,{ \stoi{ \rightsquigarrow  \intermed {  {\intermed \sigma}  } } } $,
    then $ \vdash_{ctx}^{\intermed {M} }  \source {  { \source P }  } ; \source {  {\source { \Gamma_C } }  } ; \source {  {\source \Gamma}  }  \rightsquigarrow  \intermed {  {\intermed \Sigma}  } ; \intermed {  {\intermed { \Gamma_C } }  } ; \intermed {  {\intermed \Gamma}  } $,
    and $ \intermed {  {\intermed \Sigma}  } ; \intermed {  {\intermed { \Gamma_C } }  } ; \intermed {  {\intermed \Gamma}  }  \vdash_{tm}  \intermed {  {\intermed e}  } : \intermed {  {\intermed \sigma}  } \itot{ \rightsquigarrow  \target {  {\target e }  } } $.
  \item If $ \source {  { \source P }  } ; \source {  {\source { \Gamma_C } }  } ; \source {  {\source \Gamma}  }  \vdash_{tm}^{\intermed {M} }  \source {  {\source e}  } \Leftarrow \source {  {\source \tau }  } \stoi{ \rightsquigarrow  \intermed {  {\intermed e}  } } $,
    and $ \source {  {\source { \Gamma_C } }  } ; \source {  {\source \Gamma}  }  \vdash_{ty}^{\intermed {M} }  \source {  {\source \tau }  }\,{ \stoi{ \rightsquigarrow  \intermed {  {\intermed \sigma}  } } } $,
    then $ \vdash_{ctx}^{\intermed {M} }  \source {  { \source P }  } ; \source {  {\source { \Gamma_C } }  } ; \source {  {\source \Gamma}  }  \rightsquigarrow  \intermed {  {\intermed \Sigma}  } ; \intermed {  {\intermed { \Gamma_C } }  } ; \intermed {  {\intermed \Gamma}  } $,
    and $ \intermed {  {\intermed \Sigma}  } ; \intermed {  {\intermed { \Gamma_C } }  } ; \intermed {  {\intermed \Gamma}  }  \vdash_{tm}  \intermed {  {\intermed e}  } : \intermed {  {\intermed \sigma}  } \itot{ \rightsquigarrow  \target {  {\target e }  } } $.
  \end{itemize}
\end{theoremB}

\begin{figure}[htp]
  \centering

  \def\svgscale{0.75}
\begingroup%
  \makeatletter%
  \providecommand\color[2][]{%
    \errmessage{(Inkscape) Color is used for the text in Inkscape, but the package 'color.sty' is not loaded}%
    \renewcommand\color[2][]{}%
  }%
  \providecommand\transparent[1]{%
    \errmessage{(Inkscape) Transparency is used (non-zero) for the text in Inkscape, but the package 'transparent.sty' is not loaded}%
    \renewcommand\transparent[1]{}%
  }%
  \providecommand\rotatebox[2]{#2}%
  \newcommand*\fsize{\dimexpr\f@size pt\relax}%
  \newcommand*\lineheight[1]{\fontsize{\fsize}{#1\fsize}\selectfont}%
  \ifx\svgwidth\undefined%
    \setlength{\unitlength}{144.00000146bp}%
    \ifx\svgscale\undefined%
      \relax%
    \else%
      \setlength{\unitlength}{\unitlength * \real{\svgscale}}%
    \fi%
  \else%
    \setlength{\unitlength}{\svgwidth}%
  \fi%
  \global\let\svgwidth\undefined%
  \global\let\svgscale\undefined%
  \makeatother%
  \begin{picture}(1,1.30555556)%
    \lineheight{1}%
    \setlength\tabcolsep{0pt}%
    \put(0,0){\includegraphics[width=\unitlength,page=1]{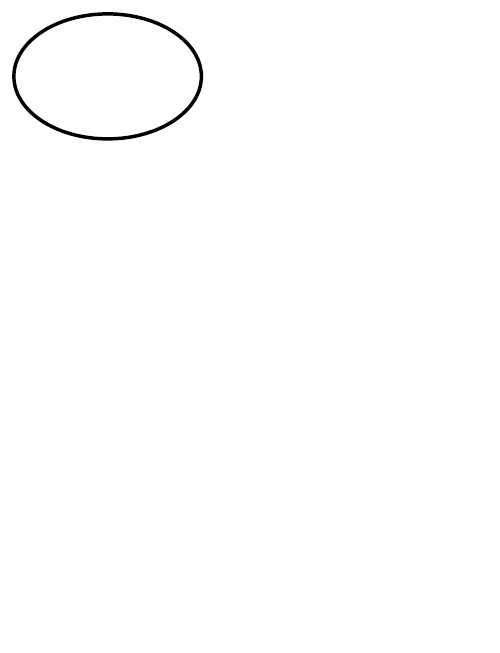}}%
    \put(0.21527778,1.12708335){\color[rgb]{0,0,0}\makebox(0,0)[t]{\lineheight{1.25}\smash{\begin{tabular}[t]{c}T1\end{tabular}}}}%
    \put(0,0){\includegraphics[width=\unitlength,page=2]{mut_dep_graph_pres.pdf}}%
    \put(0.21527778,0.12708333){\color[rgb]{0,0,0}\makebox(0,0)[t]{\lineheight{1.25}\smash{\begin{tabular}[t]{c}T7\end{tabular}}}}%
    \put(0,0){\includegraphics[width=\unitlength,page=3]{mut_dep_graph_pres.pdf}}%
    \put(0.65972222,0.62708335){\color[rgb]{0,0,0}\makebox(0,0)[t]{\lineheight{1.25}\smash{\begin{tabular}[t]{c}T6\end{tabular}}}}%
    \put(0,0){\includegraphics[width=\unitlength,page=4]{mut_dep_graph_pres.pdf}}%
  \end{picture}%
\endgroup%

  \caption{Dependency graph for Theorems~\ref{thmA:typing_pres_expr},
    \ref{thmA:typing_pres_constraint} and
    \ref{thmA:typing_pres_env}}
  \label{fig:dep_graph_pres}
\end{figure}

\proof
This theorem is mutually proven with Theorems~\ref{thmA:typing_pres_constraint}
and~\ref{thmA:typing_pres_env}. This mutual dependency is illustrated in
Figure~\ref{fig:dep_graph_pres}, where an arrow from A to B denotes A being
dependent on B. Note that at the dependency from
Theorem~\ref{thmA:typing_pres_env} to \ref{thmA:typing_pres_expr}, the size of
\({ \source P }\) is strictly decreasing, whereas \({ \source P }\) remains constant at every
other dependency. Consequently, the size of \({ \source P }\) is strictly decreasing in
every possible cycle. The induction thus remains well-founded.

By applying Lemma~\ref{lemma:ctx_well_source_mid_typing} to the first
hypothesis, we get:
\begin{equation}
   \vdash_{ctx}^{\intermed {M} }  \source {  { \source P }  } ; \source {  {\source { \Gamma_C } }  } ; \source {  {\source \Gamma}  }  \rightsquigarrow  \intermed {  {\intermed \Sigma}  } ; \intermed {  {\intermed { \Gamma_C } }  } ; \intermed {  {\intermed \Gamma}  } 
  \label{eq:typing_pres_expr0}
\end{equation}
We continue by induction on the lexicographic order of the tuple (size of the
expression, typing mode). Regarding typing mode, we define type checking to be
larger than type inference. In each mutual dependency, we know that the
tuple size decreases, meaning that the induction is well-founded.

\begin{description}
  \item[Part 1]~\\
\proofStep{\textsc{sTm-inf-true}}
  {
    \drule{sTm-inf-true}
  }
  {
    By \textsc{sTy-bool}, we know that:
    \begin{equation*}
       \source {  {\source { \Gamma_C } }  } ; \source {  {\source \Gamma}  }  \vdash_{ty}^{\intermed {M} }  \source {  \mathit{\source {Bool} }  }\,{ \stoi{ \rightsquigarrow  \intermed {  \mathit{\intermed {Bool} }  } } } 
    \end{equation*}
    The goal to be proven is the following:
    \begin{equation*}
       \intermed {  {\intermed \Sigma}  } ; \intermed {  {\intermed { \Gamma_C } }  } ; \intermed {  {\intermed \Gamma}  }  \vdash_{tm}  \intermed {  \mathit{\intermed {True} }  } : \intermed {  \mathit{\intermed {Bool} }  } \itot{ \rightsquigarrow  \target {  \mathit{\target {True} }  } } 
    \end{equation*}
    From Theorem~\ref{thmA:typing_pres_env}, we know that:
    \begin{equation*}
       \vdash_{ctx}  \intermed {  {\intermed \Sigma}  } ; \intermed {  {\intermed { \Gamma_C } }  } ; \intermed {  {\intermed \Gamma}  } 
    \end{equation*}
    The goal follows from \textsc{iTm-true}. \\
  }
\proofStep{\textsc{sTm-inf-false}}
  {
    \drule{sTm-inf-false}
  }
  { Similar to the \textsc{sTm-inf-true} case. \\
  }
\proofStep{\textsc{sTm-inf-let}}
  {}
  {
 \drule{sTm-inf-let}

    Given
    \begin{equation*}
     \source {  {\source { \Gamma_C } }  } ; \source {  {\source \Gamma}  }  \vdash_{ty}^{\intermed {M} }  \source {  {\source \tau }_{{\mathrm{2}}}  }\,{ \stoi{ \rightsquigarrow  \intermed {  {\intermed \sigma}_{{\mathrm{2}}}  } } } 
    \end{equation*}
    The goal to be proven is the following:
    \begin{equation*}
       \intermed {  {\intermed \Sigma}  } ; \intermed {  {\intermed { \Gamma_C } }  } ; \intermed {  {\intermed \Gamma}  }  \vdash_{tm}  \intermed {  \textbf{\intermed {let} } \, {\intermed x}  \ottsym{:}  \forall  {\overline {\intermed a} }_{\ottmv{j}}  \ottsym{.}   {\overline {\intermed Q} }_{\ottmv{k}}  \Rightarrow  {\intermed \sigma}   \ottsym{=}  \Lambda  {\overline {\intermed a} }_{\ottmv{j}}  \ottsym{.}  \lambda  {\overline {\intermed \delta} }_{\ottmv{k}}  \ottsym{:}  {\overline {\intermed Q} }_{\ottmv{k}}  \ottsym{.}  {\intermed e}_{{\mathrm{1}}} \, \textbf{\intermed {in} } \, {\intermed e}_{{\mathrm{2}}}  } : \intermed {  {\intermed \sigma}_{{\mathrm{2}}}  } \itot{ \rightsquigarrow  \target {  \textbf{\target {let} } \, {\target x}  \ottsym{:}  {\target \sigma }  \ottsym{=}  {\target e }_{{\mathrm{1}}} \, \textbf{\target {in} } \, {\target e }_{{\mathrm{2}}}  } } 
    \end{equation*}
    By case analysis on Equation~\ref{eq:typing_pres_expr0} (\textsc{sCtx-pgmInst}), we know:
    \begin{equation}
      \label{eq:preserve:let1}
       \vdash_{ctx}^{\intermed {M} }  \source {  \bullet  } ; \source {  {\source { \Gamma_C } }  } ; \source {  {\source \Gamma}  }  \rightsquigarrow  \intermed {  \bullet  } ; \intermed {  {\intermed { \Gamma_C } }  } ; \intermed {  {\intermed \Gamma}  } 
    \end{equation}
    From the rule premise we know that:
    \begin{equation}
      \label{eq:preserve:let2}
       \source {  {\source { \Gamma_C } }  } ; \source {  {\source \Gamma}  }  \vdash_{ty}^{\intermed {M} }  \source {  \forall  {\overline {\source a} }_{\ottmv{j}}  \ottsym{.}  {\overline {\source Q} }_{\ottmv{k}}  \Rightarrow  {\source \tau }_{{\mathrm{1}}}  }\,{ \stoi{ \rightsquigarrow  \intermed {  \forall  {\overline {\intermed a} }_{\ottmv{j}}  \ottsym{.}   {\overline {\intermed Q} }_{\ottmv{k}}  \Rightarrow  {\intermed \sigma}   } } } 
    \end{equation}
    Applying Theorem~\ref{thmA:typing_pres_types} to
    Equations~\ref{eq:preserve:let1} and \ref{eq:preserve:let2},
    we get that:
    \begin{equation}
      \label{eq:preserve:let3}
       \intermed {  {\intermed { \Gamma_C } }  } ; \intermed {  {\intermed \Gamma}  }  \vdash_{ty}  \intermed {  \forall  {\overline {\intermed a} }_{\ottmv{j}}  \ottsym{.}   {\overline {\intermed Q} }_{\ottmv{k}}  \Rightarrow  {\intermed \sigma}   } \itot{ \rightsquigarrow  \target {  {\target \sigma }  } } 
    \end{equation}
    By repeated case analysis on Equation~\ref{eq:preserve:let2} (\textsc{sTy-scheme} and \textsc{sTy-qual}), we get that:
    \begin{gather}
       \source {  {\overline {\source a} }_{\ottmv{j}}  }  \notin  \source {  {\source \Gamma}  } 
      \nonumber \\
      \ottcomp{ \source {  {\source { \Gamma_C } }  } ; \source {  {\source \Gamma}  \ottsym{,}  {\overline {\source a} }_{\ottmv{j}}  }  \vdash_{\mathit {Q} }^{\intermed {M} }  \source {  {\source Q}_{\ottmv{k}}  }  \rightsquigarrow  \intermed {  {\intermed Q}_{\ottmv{k}}  } }{\ottmv{k}}
      \nonumber
    \end{gather}
    Applying these results, together with Equation~\ref{eq:preserve:let1},
    to \textsc{sCtx-tyEnvTy} and \textsc{sCtx-tyEnvD}, we get:
    \begin{equation}
      \label{eq:preserve:let100}
       \vdash_{ctx}^{\intermed {M} }  \source {  \bullet  } ; \source {  {\source { \Gamma_C } }  } ; \source {  {\source \Gamma}  \ottsym{,}  {\overline {\source a} }_{\ottmv{j}}  \ottsym{,}  {\overline {\source \delta} }_{\ottmv{k}}  \ottsym{:}  {\overline {\source Q} }_{\ottmv{k}}  }  \rightsquigarrow  \intermed {  \bullet  } ; \intermed {  {\intermed { \Gamma_C } }  } ; \intermed {  {\intermed \Gamma}  \ottsym{,}  {\overline {\intermed a} }_{\ottmv{j}}  \ottsym{,}  {\overline {\intermed \delta} }_{\ottmv{k}}  \ottsym{:}  {\overline {\intermed Q} }_{\ottmv{k}}  } 
    \end{equation}
    By weakening (Lemma~\ref{lemma:ctx_weakening_typing}) on
    Equations~\ref{eq:typing_pres_expr0} and \ref{eq:preserve:let100}, we get:
    \begin{equation}
      \label{eq:preserve:let4}
       \vdash_{ctx}^{\intermed {M} }  \source {  { \source P }  } ; \source {  {\source { \Gamma_C } }  } ; \source {  {\source \Gamma}  \ottsym{,}  {\overline {\source a} }_{\ottmv{j}}  \ottsym{,}  {\overline {\source \delta} }_{\ottmv{k}}  \ottsym{:}  {\overline {\source Q} }_{\ottmv{k}}  }  \rightsquigarrow  \intermed {  {\intermed \Sigma}  } ; \intermed {  {\intermed { \Gamma_C } }  } ; \intermed {  {\intermed \Gamma}  \ottsym{,}  {\overline {\intermed a} }_{\ottmv{j}}  \ottsym{,}  {\overline {\intermed \delta} }_{\ottmv{k}}  \ottsym{:}  {\overline {\intermed Q} }_{\ottmv{k}}  } 
    \end{equation}
    The rule premise also gives us that:
    \begin{equation}
      \label{eq:preserve:let5}
       \source {  { \source P }  } ; \source {  {\source { \Gamma_C } }  } ; \source {  {\source \Gamma}  \ottsym{,}  {\overline {\source a} }_{\ottmv{j}}  \ottsym{,}  {\overline {\source \delta} }_{\ottmv{k}}  \ottsym{:}  {\overline {\source Q} }_{\ottmv{k}}  }  \vdash_{tm}^{\intermed {M} }  \source {  {\source e}_{{\mathrm{1}}}  } \Leftarrow \source {  {\source \tau }_{{\mathrm{1}}}  } \stoi{ \rightsquigarrow  \intermed {  {\intermed e}_{{\mathrm{1}}}  } } 
    \end{equation}
    By applying induction hypothesis with Equations~\ref{eq:preserve:let2} and
    \ref{eq:preserve:let5}, we get that:
    \begin{equation*}
       \intermed {  {\intermed \Sigma}  } ; \intermed {  {\intermed { \Gamma_C } }  } ; \intermed {  {\intermed \Gamma}  \ottsym{,}  {\overline {\intermed a} }_{\ottmv{j}}  \ottsym{,}  {\overline {\intermed \delta} }_{\ottmv{k}}  \ottsym{:}  {\overline {\intermed Q} }_{\ottmv{k}}  }  \vdash_{tm}  \intermed {  {\intermed e}_{{\mathrm{1}}}  } : \intermed {  {\intermed \sigma}  } \itot{ \rightsquigarrow  \target {  {\target e }_{{\mathrm{1}}}  } } 
    \end{equation*}
    Because of \textsc{iTm-constrI} and \textsc{iTm-forallI}, it is equivalent
    to say that:
    \begin{equation}
      \label{eq:preserve:let6}
       \intermed {  {\intermed \Sigma}  } ; \intermed {  {\intermed { \Gamma_C } }  } ; \intermed {  {\intermed \Gamma}  }  \vdash_{tm}  \intermed {  \Lambda  {\overline {\intermed a} }_{\ottmv{j}}  \ottsym{.}  \lambda  {\overline {\intermed \delta} }_{\ottmv{k}}  \ottsym{:}  {\overline {\intermed Q} }_{\ottmv{k}}  \ottsym{.}  {\intermed e}_{{\mathrm{1}}}  } : \intermed {  \forall  {\overline {\intermed a} }_{\ottmv{j}}  \ottsym{.}   {\overline {\intermed Q} }_{\ottmv{k}}  \Rightarrow  {\intermed \sigma}   } \itot{ \rightsquigarrow  \target {  {\target e }_{{\mathrm{1}}}  } } 
    \end{equation}
    Through a similar analysis, we get that:
    \begin{equation}
      \label{eq:preserve:let7}
       \intermed {  {\intermed \Sigma}  } ; \intermed {  {\intermed { \Gamma_C } }  } ; \intermed {  {\intermed \Gamma}  \ottsym{,}  {\intermed x}  \ottsym{:}  \forall  {\overline {\intermed a} }_{\ottmv{j}}  \ottsym{.}   {\overline {\intermed Q} }_{\ottmv{k}}  \Rightarrow  {\intermed \sigma}   }  \vdash_{tm}  \intermed {  {\intermed e}_{{\mathrm{2}}}  } : \intermed {  {\intermed \sigma}_{{\mathrm{2}}}  } \itot{ \rightsquigarrow  \target {  {\target e }_{{\mathrm{2}}}  } } 
    \end{equation}
    By \textsc{iTm-let}, in combination with Equations~\ref{eq:preserve:let3},
    \ref{eq:preserve:let6} and \ref{eq:preserve:let7}, the goal has been proven. \\
  }
\proofStep{\textsc{sTm-inf-ArrE}}
  {}
  { \drule{sTm-inf-ArrE}

    From the rule premise:
    \begin{gather}
      \label{eq:preserve:app1}
       \source {  { \source P }  } ; \source {  {\source { \Gamma_C } }  } ; \source {  {\source \Gamma}  }  \vdash_{tm}^{\intermed {M} }  \source {  {\source e}_{{\mathrm{1}}}  } \Rightarrow \source {  {\source \tau }_{{\mathrm{1}}}  \rightarrow  {\source \tau }_{{\mathrm{2}}}  } \stoi{ \rightsquigarrow  \intermed {  {\intermed e}_{{\mathrm{1}}}  } } \\
      \label{eq:preserve:app2}
       \source {  { \source P }  } ; \source {  {\source { \Gamma_C } }  } ; \source {  {\source \Gamma}  }  \vdash_{tm}^{\intermed {M} }  \source {  {\source e}_{{\mathrm{2}}}  } \Leftarrow \source {  {\source \tau }_{{\mathrm{1}}}  } \stoi{ \rightsquigarrow  \intermed {  {\intermed e}_{{\mathrm{2}}}  } } 
    \end{gather}
    The goal to be proven is the following:
    \begin{equation*}
       \intermed {  {\intermed \Sigma}  } ; \intermed {  {\intermed { \Gamma_C } }  } ; \intermed {  {\intermed \Gamma}  }  \vdash_{tm}  \intermed {  {\intermed e}_{{\mathrm{1}}} \, {\intermed e}_{{\mathrm{2}}}  } : \intermed {  {\intermed \sigma}_{{\mathrm{2}}}  } \itot{ \rightsquigarrow  \target {  {\target e }_{{\mathrm{1}}} \, {\target e }_{{\mathrm{2}}}  } } 
    \end{equation*}
    where $ \source {  {\source { \Gamma_C } }  } ; \source {  {\source \Gamma}  }  \vdash_{ty}^{\intermed {M} }  \source {  {\source \tau }_{{\mathrm{2}}}  }\,{ \stoi{ \rightsquigarrow  \intermed {  {\intermed \sigma}_{{\mathrm{2}}}  } } } $.

    Because the typing result is well-formed
    (Lemma~\ref{lemma:type_well_source_typing}), we know:
    \begin{equation*}
       \source {  {\source { \Gamma_C } }  } ; \source {  {\source \Gamma}  }  \vdash_{ty}^{\intermed {M} }  \source {  {\source \tau }_{{\mathrm{1}}}  \rightarrow  {\source \tau }_{{\mathrm{2}}}  }\,{ \stoi{ \rightsquigarrow  \intermed {  {\intermed \sigma}_{{\mathrm{1}}}  \rightarrow  {\intermed \sigma}_{{\mathrm{2}}}  } } } 
    \end{equation*}

    By applying the induction hypothesis on Equations~\ref{eq:preserve:app1} and
    \ref{eq:preserve:app2}, we know respectively:
    \begin{gather*}
       \intermed {  {\intermed \Sigma}  } ; \intermed {  {\intermed { \Gamma_C } }  } ; \intermed {  {\intermed \Gamma}  }  \vdash_{tm}  \intermed {  {\intermed e}_{{\mathrm{1}}}  } : \intermed {  {\intermed \sigma}_{{\mathrm{1}}}  \rightarrow  {\intermed \sigma}_{{\mathrm{2}}}  } \itot{ \rightsquigarrow  \target {  {\target e }_{{\mathrm{1}}}  } }  \\
       \intermed {  {\intermed \Sigma}  } ; \intermed {  {\intermed { \Gamma_C } }  } ; \intermed {  {\intermed \Gamma}  }  \vdash_{tm}  \intermed {  {\intermed e}_{{\mathrm{2}}}  } : \intermed {  {\intermed \sigma}_{{\mathrm{1}}}  } \itot{ \rightsquigarrow  \target {  {\target e }_{{\mathrm{2}}}  } } 
    \end{gather*}
    The goal follows from \textsc{iTm-arrE}. \\
  }
\proofStep{\textsc{sTm-inf-Ann}}
  { \drule{sTm-inf-Ann}}
  {
    Follows directly from the induction hypothesis. \\
  }
  
  \item[Part 2]~\\
\proofStep{\textsc{sTm-check-var}}
  {
  }
  {
    \drule{sTm-check-var}

    From the rule premise:
    \begin{equation*}
       ( \source {  {\source x}  } : \source {  \forall  {\overline {\source a} }_{\ottmv{j}}  \ottsym{.}  {\overline {\source Q} }_{\ottmv{i}}  \Rightarrow  {\source \tau }  } )  \in  \source {  {\source \Gamma}  } 
    \end{equation*}
    By repeated case analysis on Equation~\ref{eq:typing_pres_expr0} (\textsc{sCtx-pgmInst}), we get that:
    \begin{equation}
      \label{eq:preserve_exp:var1b}
       \vdash_{ctx}^{\intermed {M} }  \source {  \bullet  } ; \source {  {\source { \Gamma_C } }  } ; \source {  {\source \Gamma}  }  \rightsquigarrow  \intermed {  \bullet  } ; \intermed {  {\intermed { \Gamma_C } }  } ; \intermed {  {\intermed \Gamma}  } 
    \end{equation}
    By case analysis on Equation~\ref{eq:preserve_exp:var1b} (\textsc{sCtx-tyEnvTm}), we know:
    \begin{gather}
      \label{eq:preserve_exp:var1a}
       ( \source {  {\source x}  } : \source {  \forall  {\overline {\source a} }_{\ottmv{j}}  \ottsym{.}  {\overline {\source Q} }_{\ottmv{i}}  \Rightarrow  {\source \tau }  } )  \in  \source {  {\source \Gamma}  } \\ 
       \source {  {\source { \Gamma_C } }  } ; \source {  {\source \Gamma}_{{\mathrm{1}}}  }  \vdash_{ty}^{\intermed {M} }  \source {  \forall  {\overline {\source a} }_{\ottmv{j}}  \ottsym{.}  {\overline {\source Q} }_{\ottmv{i}}  \Rightarrow  {\source \tau }  }\,{ \stoi{ \rightsquigarrow  \intermed {  \forall  {\overline {\intermed a} }_{\ottmv{j}}  \ottsym{.}   {\overline {\intermed Q} }_{\ottmv{i}}  \Rightarrow  {\intermed \sigma}   } } } 
      \label{eq:preserve_exp:var1c}
    \end{gather}
    where \( \source {  {\source \Gamma}  } = \source {  {\source \Gamma}_{{\mathrm{1}}}  \ottsym{,}  {\source x}  \ottsym{:}  \forall  {\overline {\source a} }_{\ottmv{j}}  \ottsym{.}  {\overline {\source Q} }_{\ottmv{i}}  \Rightarrow  {\source \tau }  \ottsym{,}  {\source \Gamma}_{{\mathrm{2}}}  } \).

    By applying Lemma~\ref{lemma:pres_env_tm} to Equation~\ref{eq:preserve_exp:var1a}, we get:
    \begin{equation}
      \label{eq:preserve_exp:var1}
       ( \intermed {  {\intermed x}  } : \intermed {  \forall  {\overline {\intermed a} }_{\ottmv{j}}  \ottsym{.}   {\overline {\intermed Q} }_{\ottmv{i}}  \Rightarrow  {\intermed \sigma}   } )  \in  \intermed {  {\intermed \Gamma}  } 
    \end{equation}
    Furthermore, from the rule premise, we know that:
    \begin{equation}
      \ottcomp{ \source {  {\source { \Gamma_C } }  } ; \source {  {\source \Gamma}  }  \vdash_{ty}^{\intermed {M} }  \source {  {\source \tau }_{\ottmv{j}}  }\,{ \stoi{ \rightsquigarrow  \intermed {  {\intermed \sigma}_{\ottmv{j}}  } } } }{\ottmv{j}}
      \label{eq:preserve_exp:var2e}
    \end{equation}
    By Typing Preservation - Types (Theorem~\ref{thmA:typing_pres_types}),
    together with Equation~\ref{eq:preserve_exp:var1b}, we have:
    \begin{equation}
      \label{eq:preserve_exp:var2}
      \ottcomp{ \intermed {  {\intermed { \Gamma_C } }  } ; \intermed {  {\intermed \Gamma}  }  \vdash_{ty}  \intermed {  {\intermed \sigma}_{\ottmv{j}}  } \itot{ \rightsquigarrow  \target {  {\target \sigma }_{\ottmv{j}}  } } }{\ottmv{j}}
    \end{equation}
    Similarly, the rule premise tells us that:
    \begin{equation}
      \label{eq:preserve_exp:var2b}
      \ottcomp{ \source {  { \source P }  } ; \source {  {\source { \Gamma_C } }  } ; \source {  {\source \Gamma}  }  \vDash^{\intermed {M} }  \source {  [  {\overline {\source \tau} }_{\ottmv{j}}  /  {\overline {\source a} }_{\ottmv{j}}  ]  {\source Q}_{\ottmv{i}}  } \stoi{ \rightsquigarrow  \intermed {  {\intermed d}_{\ottmv{i}}  } } }{\ottmv{i}}
    \end{equation}
    By applying weakening (Lemma~\ref{lemma:type_weakening}) to Equation~\ref{eq:preserve_exp:var1c}, we get:
    \begin{equation}
      \label{eq:preserve_exp:var2d}
       \source {  {\source { \Gamma_C } }  } ; \source {  {\source \Gamma}  }  \vdash_{ty}^{\intermed {M} }  \source {  \forall  {\overline {\source a} }_{\ottmv{j}}  \ottsym{.}  {\overline {\source Q} }_{\ottmv{i}}  \Rightarrow  {\source \tau }  }\,{ \stoi{ \rightsquigarrow  \intermed {  \forall  {\overline {\intermed a} }_{\ottmv{j}}  \ottsym{.}   {\overline {\intermed Q} }_{\ottmv{i}}  \Rightarrow  {\intermed \sigma}   } } } 
    \end{equation}
    By repeated case analysis on Equation~\ref{eq:preserve_exp:var2d} (sTy-qual), we get that:
    \begin{equation}
      \label{eq:preserve_exp:var2c}
      \ottcomp{ \source {  {\source { \Gamma_C } }  } ; \source {  {\source \Gamma}  \ottsym{,}  {\overline {\source a} }_{\ottmv{j}}  }  \vdash_{\mathit {Q} }^{\intermed {M} }  \source {  {\source Q}_{\ottmv{i}}  }  \rightsquigarrow  \intermed {  {\intermed Q}_{\ottmv{i}}  } }{\ottmv{i}}
    \end{equation}
    By applying Lemma~\ref{lemma:tvar_subst_ctr} on Equations~\ref{eq:preserve_exp:var2c}
    and~\ref{eq:preserve_exp:var2e}, we get:
    \begin{equation}
      \label{eq:preserve_exp:var2f}
      \ottcomp{ \source {  {\source { \Gamma_C } }  } ; \source {  {\source \Gamma}  }  \vdash_{\mathit {Q} }^{\intermed {M} }  \source {  [  {\overline {\source \tau} }_{\ottmv{j}}  /  {\overline {\source a} }_{\ottmv{j}}  ]  {\source Q}_{\ottmv{i}}  }  \rightsquigarrow  \intermed {  [  \overline {\intermed \sigma}_{\ottmv{j}}  /  {\overline {\intermed a} }_{\ottmv{j}}  ]  {\intermed Q}_{\ottmv{i}}  } }{\ottmv{i}}
    \end{equation}
    By Typing Preservation - Constraints Proving (Theorem~\ref{thmA:typing_pres_constraint}),
    applied to Equations~\ref{eq:preserve_exp:var2b}, \ref{eq:typing_pres_expr0}
    and~\ref{eq:preserve_exp:var2f}, we have:
    \begin{equation}
      \label{eq:preserve_exp:var3}
      \ottcomp{ \intermed {  {\intermed \Sigma}  } ; \intermed {  {\intermed { \Gamma_C } }  } ; \intermed {  {\intermed \Gamma}  }  \vdash_{d}  \intermed {  {\intermed d}_{\ottmv{i}}  } : \intermed {  [  \overline {\intermed \sigma}_{\ottmv{j}}  /  {\overline {\intermed a} }_{\ottmv{j}}  ]  {\intermed Q}_{\ottmv{i}}  } \itot{\,  \rightsquigarrow  \target {  {\target e }_{\ottmv{i}}  } } }{\ottmv{i}}
    \end{equation}
    
    The goal to be proven is the following:
    \begin{equation*}
       \intermed {  {\intermed \Sigma}  } ; \intermed {  {\intermed { \Gamma_C } }  } ; \intermed {  {\intermed \Gamma}  }  \vdash_{tm}  \intermed {  {\intermed x} \, \overline {\intermed \sigma}_{\ottmv{j}} \, {\overline {\intermed d} }_{\ottmv{i}}  } : \intermed {  [  \overline {\intermed \sigma}_{\ottmv{j}}  /  {\overline {\intermed a} }_{\ottmv{j}}  ]  {\intermed \sigma}  } \itot{ \rightsquigarrow  \target {  {\target x} \, {\target \sigma }_{\ottmv{j}} \, {\target e }_{\ottmv{i}}  } } 
    \end{equation*}
    where $ \source {  {\source { \Gamma_C } }  } ; \source {  {\source \Gamma}  }  \vdash_{ty}^{\intermed {M} }  \source {  [  {\overline {\source \tau} }_{\ottmv{j}}  /  {\overline {\source a} }_{\ottmv{j}}  ]  {\source \tau }  }\,{ \stoi{ \rightsquigarrow  \intermed {  [  \overline {\intermed \sigma}_{\ottmv{j}}  /  {\overline {\intermed a} }_{\ottmv{j}}  ]  {\intermed \sigma}  } } } $.
    
    From Equation~\ref{eq:preserve_exp:var1}, by applying \textsc{iTm-var}, we get
    \begin{equation}
      \label{eq:preserve_exp:var4}
       \intermed {  {\intermed \Sigma}  } ; \intermed {  {\intermed { \Gamma_C } }  } ; \intermed {  {\intermed \Gamma}  }  \vdash_{tm}  \intermed {  {\intermed x}  } : \intermed {  \forall  {\overline {\intermed a} }_{\ottmv{j}}  \ottsym{.}   {\overline {\intermed Q} }_{\ottmv{i}}  \Rightarrow  {\intermed \sigma}   } \itot{ \rightsquigarrow  \target {  {\target x}  } } 
    \end{equation}
    By Equations~\ref{eq:preserve_exp:var2}, \ref{eq:preserve_exp:var4} and \textsc{iTm-forallE}, we get:
    \begin{equation}
      \label{eq:preserve_exp:var5}
       \intermed {  {\intermed \Sigma}  } ; \intermed {  {\intermed { \Gamma_C } }  } ; \intermed {  {\intermed \Gamma}  }  \vdash_{tm}  \intermed {  {\intermed x} \, \overline {\intermed \sigma}_{\ottmv{j}}  } : \intermed {  [  \overline {\intermed \sigma}_{\ottmv{j}}  /  {\overline {\intermed a} }_{\ottmv{j}}  ]   {\overline {\intermed Q} }_{\ottmv{i}}  \Rightarrow  [  \overline {\intermed \sigma}_{\ottmv{j}}  /  {\overline {\intermed a} }_{\ottmv{j}}  ]  {\intermed \sigma}   } \itot{ \rightsquigarrow  \target {  {\target x} \, {\target \sigma }_{\ottmv{j}}  } } 
    \end{equation}
    By Equations~\ref{eq:preserve_exp:var3} and \ref{eq:preserve_exp:var5},
    in combination with rule \textsc{iTm-constrE}, we get
    \begin{equation}
       \intermed {  {\intermed \Sigma}  } ; \intermed {  {\intermed { \Gamma_C } }  } ; \intermed {  {\intermed \Gamma}  }  \vdash_{tm}  \intermed {  {\intermed x} \, \overline {\intermed \sigma}_{\ottmv{j}} \, {\overline {\intermed d} }_{\ottmv{i}}  } : \intermed {  [  \overline {\intermed \sigma}_{\ottmv{j}}  /  {\overline {\intermed a} }_{\ottmv{j}}  ]  {\intermed \sigma}  } \itot{ \rightsquigarrow  \target {  {\target x} \, {\target \sigma }_{\ottmv{j}} \, {\target e }_{\ottmv{i}}  } } 
    \end{equation}
    which is exactly the goal. \\
  }
\proofStep{\textsc{sTm-check-meth}}
  {}
  {
    \drule{sTm-check-meth}

    The goal to be proven is the following:
    \begin{equation}
       \intermed {  {\intermed \Sigma}  } ; \intermed {  {\intermed { \Gamma_C } }  } ; \intermed {  {\intermed \Gamma}  }  \vdash_{tm}  \intermed {  {\intermed d}  \ottsym{.}  {\intermed m} \, \overline {\intermed \sigma}_{\ottmv{j}} \, {\overline {\intermed d} }_{\ottmv{i}}  } : \intermed {  [  \overline {\intermed \sigma}_{\ottmv{j}}  /  {\overline {\intermed a} }_{\ottmv{j}}  ]  [  {\intermed \sigma}  /  {\intermed a }  ]  {\intermed \sigma}'  } \itot{ \rightsquigarrow  \target {  {\target e }  } } 
      \label{eqg:preserve_exp:meth}
    \end{equation}
    where \( \source {  {\source { \Gamma_C } }  } ; \source {  \bullet  \ottsym{,}  {\overline {\source a} }_{\ottmv{j}}  \ottsym{,}  {\source a }  }  \vdash_{ty}^{\intermed {M} }  \source {  {\source \tau }'  }\,{ \stoi{ \rightsquigarrow  \intermed {  {\intermed \sigma}'  } } } \).

    From the rule premise, we get that:
    \begin{gather}
      \label{eq:preserve_exp:m:1}
       \source {  { \source P }  } ; \source {  {\source { \Gamma_C } }  } ; \source {  {\source \Gamma}  }  \vDash^{\intermed {M} }  \source {  {\source {\mathit{TC} } } \, {\source \tau }  } \stoi{ \rightsquigarrow  \intermed {  {\intermed d}  } } \\
      \label{eq:preserve_exp:m:2}
       \source {  {\source { \Gamma_C } }  } ; \source {  {\source \Gamma}  }  \vdash_{ty}^{\intermed {M} }  \source {  {\source \tau }  }\,{ \stoi{ \rightsquigarrow  \intermed {  {\intermed \sigma}  } } } 
    \end{gather}
    By repeated case analysis on Equation~\ref{eq:typing_pres_expr0} (\textsc{sCtx-clsEnv}),
    together with the first rule premise, we get:
    \begin{equation*}
       \source {  {\source { \Gamma_C } }_{{\mathrm{1}}}  } ; \source {  \bullet  \ottsym{,}  {\source a }  }  \vdash_{ty}^{\intermed {M} }  \source {  \forall  {\overline {\source a} }_{\ottmv{j}}  \ottsym{.}  {\overline {\source Q} }_{\ottmv{i}}  \Rightarrow  {\source \tau }'  }\,{ \stoi{ \rightsquigarrow  \intermed {  {\intermed \sigma}''  } } } 
    \end{equation*}
    where \( \source {  {\source { \Gamma_C } }  } = \source {  {\source { \Gamma_C } }_{{\mathrm{1}}}  }, \source {  {\source m}  \ottsym{:}  {\overline {\source Q} }'_{\ottmv{k}}  \Rightarrow  {\source {\mathit{TC} } } \, {\source a }  \ottsym{:}  \forall  {\overline {\source a} }_{\ottmv{j}}  \ottsym{.}  {\overline {\source Q} }_{\ottmv{i}}  \Rightarrow  {\source \tau }'  \ottsym{,}  {\source { \Gamma_C } }_{{\mathrm{2}}}  } \).

    Following \textsc{sQ-TC}, in combination with this result,
    Equation~\ref{eq:preserve_exp:m:2} and the first rule premise, we have:
    \begin{align}
      \label{eq:preserve_exp:m:3}
       \source {  {\source { \Gamma_C } }  } ; \source {  {\source \Gamma}  }  \vdash_{\mathit {Q} }^{\intermed {M} }  \source {  {\source {\mathit{TC} } } \, {\source \tau }  }  \rightsquigarrow  \intermed {  {\intermed {\mathit{TC} } } \, {\intermed \sigma}  } 
    \end{align}
    Applying Typing Preservation - Constraints Proving
    (Theorem~\ref{thmA:typing_pres_constraint}) on
    Equations~\ref{eq:preserve_exp:m:1} and \ref{eq:preserve_exp:m:3}, we get:
    \begin{align}
      \label{eq:preserve_exp:m:4}
        \intermed {  {\intermed \Sigma}  } ; \intermed {  {\intermed { \Gamma_C } }  } ; \intermed {  {\intermed \Gamma}  }  \vdash_{d}  \intermed {  {\intermed d}  } : \intermed {  {\intermed {\mathit{TC} } } \, {\intermed \sigma}  } \itot{\,  \rightsquigarrow  \target {  {\target e }  } } 
    \end{align}
    Furthermore, we know from the rule premise that:
    \begin{align*}
       ( \source {  {\source m}  } : \source {  {\overline {\source Q} }'_{\ottmv{k}}  }  \Rightarrow  \source {  {\source {\mathit{TC} } }  }~\source {  {\source a }  } : \source {  \forall  {\overline {\source a} }_{\ottmv{j}}  \ottsym{.}  {\overline {\source Q} }_{\ottmv{i}}  \Rightarrow  {\source \tau }'  } )  \in  \source {  {\source { \Gamma_C } }  } 
    \end{align*}
    Consequently, by repeated case analysis on Equation~\ref{eq:typing_pres_expr0}
    (\textsc{sCtx-clsEnv}), we know that:
    \begin{align}
      \label{eq:preserve_exp:m:5}
       ( \intermed {  {\intermed m}  } : \intermed {  {\intermed {\mathit{TC} } } \, {\intermed a }  } : \intermed {  {\intermed \sigma}''  } )  \in  \intermed {  {\intermed { \Gamma_C } }  } 
    \end{align}
    By Equations~\ref{eq:preserve_exp:m:4} and \ref{eq:preserve_exp:m:5},
    in combination with rule \textsc{iTm-method}, we get:
    \begin{align*}
       \intermed {  {\intermed \Sigma}  } ; \intermed {  {\intermed { \Gamma_C } }  } ; \intermed {  {\intermed \Gamma}  }  \vdash_{tm}  \intermed {  {\intermed d}  \ottsym{.}  {\intermed m}  } : \intermed {  [  {\intermed \sigma}  /  {\intermed a }  ]  {\intermed \sigma}''  } \itot{ \rightsquigarrow  \target {  {\target e }  \ottsym{.}  {\target m}  } } 
    \end{align*}

    The rest of the proof is similar to case \textsc{sTm-check-var}. \\
  }
\proofStep{\textsc{sTm-check-ArrI}}
  {}
  { \drule{sTm-check-ArrI}

    The second hypothesis is:
    \begin{equation*}
       \vdash_{ctx}^{\intermed {M} }  \source {  { \source P }  } ; \source {  {\source { \Gamma_C } }  } ; \source {  {\source \Gamma}  }  \rightsquigarrow  \intermed {  {\intermed \Sigma}  } ; \intermed {  {\intermed { \Gamma_C } }  } ; \intermed {  {\intermed \Gamma}  } 
    \end{equation*}
    It is easy to verify that 
    \begin{equation*}
       \vdash_{ctx}^{\intermed {M} }  \source {  { \source P }  } ; \source {  {\source { \Gamma_C } }  } ; \source {  {\source \Gamma}  \ottsym{,}  {\source x}  \ottsym{:}  {\source \tau }_{{\mathrm{1}}}  }  \rightsquigarrow  \intermed {  {\intermed \Sigma}  } ; \intermed {  {\intermed { \Gamma_C } }  } ; \intermed {  {\intermed \Gamma}  \ottsym{,}  {\intermed x}  \ottsym{:}  {\intermed \sigma}  } 
    \end{equation*}
    The goal follows directly by applying the induction hypothesis,
    in combination with rule \textsc{iTm-arrI}. \\
  }
\proofStep{\textsc{sTm-check-Inf}}
  { \drule{sTm-check-Inf}}
  {
    Follows directly from the induction hypothesis.
  }
\end{description}
\qed

\begin{theoremB}[Typing Preservation - Instance]\ \\
  If $ \source {  { \source P }  } ; \source {  {\source { \Gamma_C } }  }  \vdash_{inst}^{\intermed {M} }  \source {  {\source {inst} }  } : \source {  { \source P }'  } $,
  and $ \vdash_{ctx}^{\intermed {M} }  \source {  { \source P }  } ; \source {  {\source { \Gamma_C } }  } ; \source {  \bullet  }  \rightsquigarrow  \intermed {  {\intermed \Sigma}  } ; \intermed {  {\intermed { \Gamma_C } }  } ; \intermed {  \bullet  } $
  then we have $ \vdash_{ctx}^{\intermed {M} }  \source {  { \source P }  \ottsym{,}  { \source P }'  } ; \source {  {\source { \Gamma_C } }  } ; \source {  \bullet  }  \rightsquigarrow  \intermed {  {\intermed \Sigma}  \ottsym{,}  {\intermed \Sigma}'  } ; \intermed {  {\intermed { \Gamma_C } }  } ; \intermed {  \bullet  } $.
\label{thmA:typing_pres_instance}
\end{theoremB}
\proof

We restate the rule for typing instance declarations  for reference:

\[\drule{sInst-inst}\]
By inversion of rule \textsc{sInst-inst}, we know that:
\begin{gather*}
   \source {  { \source P }'  } = \source {  \ottsym{(}  {\source D }  \ottsym{:}  \forall  {\overline {\source b} }_{\ottmv{k}}  \ottsym{.}  {\overline {\source Q} }_{\ottmv{q}}  \Rightarrow  {\source {\mathit{TC} } } \, {\source \tau }  \ottsym{)}  \ottsym{.}  {\source m}  \mapsto  \bullet  \ottsym{,}  {\overline {\source b} }_{\ottmv{k}}  \ottsym{,}  {\overline {\source \delta} }_{\ottmv{q}}  \ottsym{:}  {\overline {\source Q} }_{\ottmv{q}}  \ottsym{,}  {\overline {\source a} }_{\ottmv{j}}  \ottsym{,}  {\overline {\source \delta} }_{\ottmv{h}}  \ottsym{:}  [  {\source \tau }  /  {\source a }  ]  {\overline {\source Q} }_{\ottmv{h}}  \ottsym{:}  {\source e}  } 
\end{gather*}
Therefore our goal is
\begin{equation}
  \label{eq:pres_inst:1}
   \vdash_{ctx}^{\intermed {M} }  \source {  { \source P }  \ottsym{,}  \ottsym{(}  {\source D }  \ottsym{:}  \forall  {\overline {\source b} }_{\ottmv{k}}  \ottsym{.}  {\overline {\source Q} }_{\ottmv{q}}  \Rightarrow  {\source {\mathit{TC} } } \, {\source \tau }  \ottsym{)}  \ottsym{.}  {\source m}  \mapsto  \bullet  \ottsym{,}  {\overline {\source b} }_{\ottmv{k}}  \ottsym{,}  {\overline {\source \delta} }_{\ottmv{q}}  \ottsym{:}  {\overline {\source Q} }_{\ottmv{q}}  \ottsym{,}  {\overline {\source a} }_{\ottmv{j}}  \ottsym{,}  {\overline {\source \delta} }_{\ottmv{h}}  \ottsym{:}  [  {\source \tau }  /  {\source a }  ]  {\overline {\source Q} }_{\ottmv{h}}  \ottsym{:}  {\source e}  } ; \source {  {\source { \Gamma_C } }  } ; \source {  \bullet  }  \rightsquigarrow  \intermed {  {\intermed \Sigma}  \ottsym{,}  {\intermed \Sigma}'  } ; \intermed {  {\intermed { \Gamma_C } }  } ; \intermed {  \bullet  } 
\end{equation}
From the hypothesis, we know that:
\begin{align}
  \label{eq:pres_inst:2}
   \vdash_{ctx}^{\intermed {M} }  \source {  { \source P }  } ; \source {  {\source { \Gamma_C } }  } ; \source {  \bullet  }  \rightsquigarrow  \intermed {  {\intermed \Sigma}  } ; \intermed {  {\intermed { \Gamma_C } }  } ; \intermed {  \bullet  } 
\end{align}
Goal~\ref{eq:pres_inst:1} follows directly from \textsc{sCtx-pgmInst} with
\( \intermed {  {\intermed \Sigma}'  } = \intermed {  \ottsym{(}  {\intermed D }  \ottsym{:}  \forall  {\overline {\intermed b} }_{\ottmv{k}}  \ottsym{.}  {\overline {\intermed Q} }_{\ottmv{q}}  \Rightarrow  {\intermed {\mathit{TC} } } \, {\intermed \sigma}  \ottsym{)}  \ottsym{.}  {\intermed m}  \mapsto  \Lambda  {\overline {\intermed b} }_{\ottmv{k}}  \ottsym{.}  \lambda  {\overline {\intermed \delta} }_{\ottmv{q}}  \ottsym{:}  {\overline {\intermed Q} }_{\ottmv{q}}  \ottsym{.}  \Lambda  {\overline {\intermed a} }_{\ottmv{j}}  \ottsym{.}  \lambda  {\overline {\intermed \delta} }_{\ottmv{h}}  \ottsym{:}  [  {\intermed \sigma}  /  {\intermed a }  ]  {\overline {\intermed Q} }_{\ottmv{h}}  \ottsym{.}  {\intermed e}  } \), if we can show the following:
\begin{gather}
  \label{eq:pres_inst:3a}
   \textbf {unambig}  ( \source {  \forall  {\overline {\source b} }_{\ottmv{k}}  \ottsym{.}  {\overline {\source Q} }_{\ottmv{q}}  \Rightarrow  {\source {\mathit{TC} } } \, {\source \tau }  } )  \\
  \label{eq:pres_inst:3}
   \source {  {\source { \Gamma_C } }  } ; \source {  \bullet  }  \vdash_{\mathit {C} }^{\intermed {M} }  \source {  \forall  {\overline {\source b} }_{\ottmv{k}}  \ottsym{.}  {\overline {\source Q} }_{\ottmv{q}}  \Rightarrow  {\source {\mathit{TC} } } \, {\source \tau }  }  \rightsquigarrow  \intermed {  \forall  {\overline {\intermed b} }_{\ottmv{k}}  \ottsym{.}  {\overline {\intermed Q} }_{\ottmv{q}}  \Rightarrow  {\intermed {\mathit{TC} } } \, {\intermed \sigma}  }  \\
  \label{eq:pres_inst:4}
   ( \source {  {\source m}  } : \source {  {\overline {\source Q} }'_{\ottmv{i}}  }  \Rightarrow  \source {  {\source {\mathit{TC} } }  }~\source {  {\source a }  } : \source {  \forall  {\overline {\source a} }_{\ottmv{j}}  \ottsym{.}  {\overline {\source Q} }_{\ottmv{h}}  \Rightarrow  {\source \tau }_{{\mathrm{1}}}  } )  \in  \source {  {\source { \Gamma_C } }  } \\
  \label{eq:pres_inst:5}
   \source {  { \source P }  } ; \source {  {\source { \Gamma_C } }  } ; \source {  \bullet  \ottsym{,}  {\overline {\source b} }_{\ottmv{k}}  \ottsym{,}  {\overline {\source \delta} }_{\ottmv{q}}  \ottsym{:}  {\overline {\source Q} }_{\ottmv{q}}  \ottsym{,}  {\overline {\source a} }_{\ottmv{j}}  \ottsym{,}  {\overline {\source \delta} }_{\ottmv{h}}  \ottsym{:}  [  {\source \tau }  /  {\source a }  ]  {\overline {\source Q} }_{\ottmv{h}}  }  \vdash_{tm}^{\intermed {M} }  \source {  {\source e}  } \Leftarrow \source {  [  {\source \tau }  /  {\source a }  ]  {\source \tau }_{{\mathrm{1}}}  } \stoi{ \rightsquigarrow  \intermed {  {\intermed e}  } }  \\
  \label{eq:pres_inst:6}
   \source {  {\source { \Gamma_C } }  } ; \source {  \bullet  \ottsym{,}  {\source a }  }  \vdash_{ty}^{\intermed {M} }  \source {  \forall  {\overline {\source a} }_{\ottmv{j}}  \ottsym{.}  {\overline {\source Q} }_{\ottmv{h}}  \Rightarrow  {\source \tau }_{{\mathrm{1}}}  }\,{ \stoi{ \rightsquigarrow  \intermed {  {\intermed \sigma}'  } } }  \\
  \label{eq:pres_inst:7}
   \source {  {\source D }  }  \notin   \textbf {dom}  ( \source {  { \source P }  } )  \\
  \label{eq:pres_inst:8}
   ( \source {  {\source D }'  } : \source {  \forall  {\overline {\source b} }'_{\ottmv{k}}  \ottsym{.}  {\overline {\source Q} }''_{\ottmv{h}}  \Rightarrow  {\source {\mathit{TC} } } \, {\source \tau }''  } ) . \source {  {\source m}'  }  \mapsto  \source {  {\source \Gamma}'  } : \source {  {\source e}'  }  \notin  \source {  { \source P }  }  \\
  \label{eq:pres_inst:9}
   \textbf {where}  \source {  [  {\overline {\source \tau} }_{\ottmv{j}}  /  {\overline {\source b} }_{\ottmv{j}}  ]  {\source \tau }  } = \source {  [  {\overline {\source \tau} }'_{\ottmv{k}}  /  {\overline {\source b} }'_{\ottmv{k}}  ]  {\source \tau }''  }  \\
  \label{eq:pres_inst:10}
   \vdash_{ctx}^{\intermed {M} }  \source {  { \source P }  } ; \source {  {\source { \Gamma_C } }  } ; \source {  \bullet  }  \rightsquigarrow  \intermed {  {\intermed \Sigma}  } ; \intermed {  {\intermed { \Gamma_C } }  } ; \intermed {  \bullet  }  
\end{gather}
Goal~\ref{eq:pres_inst:10} is exactly Equation~\ref{eq:pres_inst:2}, which we
already have. Goals~\ref{eq:pres_inst:3a} and \ref{eq:pres_inst:4} follow
directly from the premise of \textsc{sInst-inst}. The premise also tells us that
${\source D }$ is freshly generated, which satisfies Goal~\ref{eq:pres_inst:7}. Similarly
Goals~\ref{eq:pres_inst:5}, \ref{eq:pres_inst:8} and \ref{eq:pres_inst:9} can be proven directly from
the rule premise. 

From the premise, we know:
\begin{gather}
   \source {  {\source { \Gamma_C } }  } ; \source {  \bullet  \ottsym{,}  {\overline {\source b} }_{\ottmv{k}}  }  \vdash_{ty}^{\intermed {M} }  \source {  {\source \tau }  }\,{ \stoi{ \rightsquigarrow  \intermed {  {\intermed \sigma}  } } }  \\
  \ottcomp{ \source {  {\source { \Gamma_C } }  } ; \source {  \bullet  \ottsym{,}  {\overline {\source b} }_{\ottmv{k}}  }  \vdash_{\mathit {Q} }^{\intermed {M} }  \source {  {\source Q}_{\ottmv{q}}  }  \rightsquigarrow  \intermed {  {\intermed Q}_{\ottmv{q}}  } }{\ottmv{q}}
\end{gather}
Goal~\ref{eq:pres_inst:3} follows directly from the definition of
well-formedness of constraints and types.
Goals~\ref{eq:pres_inst:5} and
\ref{eq:pres_inst:6} remain to be proven.

From the rule premise, we know that
\begin{equation}
 ( \source {  {\source m}  } : \source {  {\overline {\source Q} }'_{\ottmv{i}}  }  \Rightarrow  \source {  {\source {\mathit{TC} } }  }~\source {  {\source a }  } : \source {  \forall  {\overline {\source a} }_{\ottmv{j}}  \ottsym{.}  {\overline {\source Q} }_{\ottmv{h}}  \Rightarrow  {\source \tau }_{{\mathrm{1}}}  } )  \in  \source {  {\source { \Gamma_C } }  } 
\end{equation}
From the definition of well-formedness of the source context, we know that:
\begin{gather*}
 \source {  {\source { \Gamma_C } }_{{\mathrm{1}}}  } ; \source {  \bullet  \ottsym{,}  {\source a }  }  \vdash_{ty}^{\intermed {M} }  \source {  \forall  {\overline {\source a} }_{\ottmv{j}}  \ottsym{.}  {\overline {\source Q} }_{\ottmv{h}}  \Rightarrow  {\source \tau }_{{\mathrm{1}}}  }\,{ \stoi{ \rightsquigarrow  \intermed {  {\intermed \sigma}'  } } }  \\
 \source {  {\source { \Gamma_C } }  } = \source {  {\source { \Gamma_C } }_{{\mathrm{1}}}  \ottsym{,}  {\source { \Gamma_C } }_{{\mathrm{2}}}  } 
\end{gather*}
By weakening of class environment (Lemma~\ref{lemma:type_weakening}), we
can prove Goal~\ref{eq:pres_inst:6}. \\
\qed

\begin{theoremB}[Typing Preservation - Classes]\ \\
  If $ \source {  {\source { \Gamma_C } }  }  \vdash_{cls}^{\intermed {M} }  \source {  {\source {cls} }  } : \source {  {\source { \Gamma_C } }'  } $,
  and $ \vdash_{ctx}^{\intermed {M} }  \source {  { \source P }  } ; \source {  {\source { \Gamma_C } }  } ; \source {  \bullet  }  \rightsquigarrow  \intermed {  {\intermed \Sigma}  } ; \intermed {  {\intermed { \Gamma_C } }  } ; \intermed {  \bullet  } $,
  then we have $ \vdash_{ctx}^{\intermed {M} }  \source {  { \source P }  } ; \source {  {\source { \Gamma_C } }  \ottsym{,}  {\source { \Gamma_C } }'  } ; \source {  \bullet  }  \rightsquigarrow  \intermed {  {\intermed \Sigma}  } ; \intermed {  {\intermed { \Gamma_C } }  \ottsym{,}  {\intermed { \Gamma_C } }'  } ; \intermed {  \bullet  } $.
\label{thmA:typing_pres_class}
\end{theoremB}
\proof

We restate the rule for class declaration typing for reference:
\[\drule{sCls-cls}\]
By case analysis, we know that $\source{{\source { \Gamma_C } }'}$ is of the form
\begin{equation*}
\source{\bullet  \ottsym{,}  {\source m}  \ottsym{:}  \overline {\source {\mathit{TC}_i~a } }  \Rightarrow  {\source {\mathit{TC} } } \, {\source a }  \ottsym{:}  \forall  {\overline {\source a} }_{\ottmv{j}}  \ottsym{.}  {\overline {\source Q} }_{\ottmv{p}}  \Rightarrow  {\source \tau }}
\end{equation*}
The goal to be proven is the following:
\begin{equation}
  \label{eq:pres_class:1}
 \vdash_{ctx}^{\intermed {M} }  \source {  { \source P }  } ; \source {  {\source { \Gamma_C } }  \ottsym{,}  {\source m}  \ottsym{:}  \overline {\source {\mathit{TC}_i~a } }  \Rightarrow  {\source {\mathit{TC} } } \, {\source a }  \ottsym{:}  \forall  {\overline {\source a} }_{\ottmv{j}}  \ottsym{.}  {\overline {\source Q} }_{\ottmv{p}}  \Rightarrow  {\source \tau }  } ; \source {  \bullet  }  \rightsquigarrow  \intermed {  {\intermed \Sigma}  } ; \intermed {  {\intermed { \Gamma_C } }  \ottsym{,}  {\intermed m}  \ottsym{:}  {\intermed {\mathit{TC} } } \, {\intermed a }  \ottsym{:}  {\intermed \sigma}  } ; \intermed {  \bullet  } 
\end{equation}
We can derive from \textsc{sCtx-clsEnv} that
\begin{equation}
  \label{eq:pres_class:8}
 \vdash_{ctx}^{\intermed {M} }  \source {  \bullet  } ; \source {  {\source { \Gamma_C } }  \ottsym{,}  {\source m}  \ottsym{:}  \overline {\source {\mathit{TC}_i~a } }  \Rightarrow  {\source {\mathit{TC} } } \, {\source a }  \ottsym{:}  \forall  {\overline {\source a} }_{\ottmv{j}}  \ottsym{.}  {\overline {\source Q} }_{\ottmv{p}}  \Rightarrow  {\source \tau }  } ; \source {  \bullet  }  \rightsquigarrow  \intermed {  \bullet  } ; \intermed {  {\intermed { \Gamma_C } }  \ottsym{,}  {\intermed m}  \ottsym{:}  {\intermed {\mathit{TC} } } \, {\intermed a }  \ottsym{:}  {\intermed \sigma}  } ; \intermed {  \bullet  } 
\end{equation}
assuming we can show that:
\begin{gather}
  \label{eq:pres_class:2}
 \source {  {\source { \Gamma_C } }  } ; \source {  \bullet  \ottsym{,}  {\source a }  }  \vdash_{ty}^{\intermed {M} }  \source {  \forall  {\overline {\source a} }_{\ottmv{k}}  \ottsym{.}  {\overline {\source Q} }_{\ottmv{p}}  \Rightarrow  {\source \tau }  }\,{ \stoi{ \rightsquigarrow  \intermed {  {\intermed \sigma}  } } }  \\
  \label{eq:pres_class:3}
 \source {  {\overline {\source a} }_{\ottmv{j}}  \ottsym{,}  {\source a }  } =  \textbf {fv}  ( \source {  {\source \tau }  } )  \\
  \label{eq:pres_class:4}
\ottcomp{ \source {  {\source { \Gamma_C } }  } ; \source {  \bullet  \ottsym{,}  {\source a }  }  \vdash_{\mathit {Q} }^{\intermed {M} }  \source {  {\source {\mathit{TC} } }_{\ottmv{i}} \, {\source a }  }  \rightsquigarrow  \intermed {  {\intermed Q}_{\ottmv{i}}  } }{\ottmv{i}} \\
  \label{eq:pres_class:5}
 \source {  {\source m}  }  \notin   \textbf {dom}  ( \source {  {\source { \Gamma_C } }  } )  \\
  \label{eq:pres_class:6}
 \source {  {\source {\mathit{TC} } } \, {\source b }  }  \notin   \textbf {dom}  ( \source {  {\source { \Gamma_C } }  } )  \\
  \label{eq:pres_class:7}
 \vdash_{ctx}^{\intermed {M} }  \source {  \bullet  } ; \source {  {\source { \Gamma_C } }  } ; \source {  \bullet  }  \rightsquigarrow  \intermed {  \bullet  } ; \intermed {  {\intermed { \Gamma_C } }  } ; \intermed {  \bullet  } 
\end{gather}
Goals~\ref{eq:pres_class:2} till \ref{eq:pres_class:6} follow directly from the
premises and from the hypothesis. Goal~\ref{eq:pres_class:7} follows by repeated
inversion on the second hypothesis.
Finally, Goal~\ref{eq:pres_class:1} follows from Equation~\ref{eq:pres_class:8}
by the definition of environment well-formedness and the second hypothesis. \\
\qed

\begin{theoremB}[Typing Preservation - Programs]\ \\
  If $ \source {  { \source P }  } ; \source {  {\source { \Gamma_C } }  }  \vdash_{pgm}^{\intermed {M} }  \source {  {\source {pgm} }  } : \source {  {\source \tau }  } ; \source {  { \source P }'  } ; \source {  {\source { \Gamma_C } }'  }  \rightsquigarrow  \intermed {  {\intermed e}  } $,
  and $ \vdash_{ctx}^{\intermed {M} }  \source {  { \source P }  } ; \source {  {\source { \Gamma_C } }  } ; \source {  \bullet  }  \rightsquigarrow  \intermed {  {\intermed \Sigma}  } ; \intermed {  {\intermed { \Gamma_C } }  } ; \intermed {  \bullet  } $,
  and $ \source {  {\source { \Gamma_C } }  \ottsym{,}  {\source { \Gamma_C } }'  } ; \source {  \bullet  }  \vdash_{ty}^{\intermed {M} }  \source {  {\source \tau }  }\,{ \stoi{ \rightsquigarrow  \intermed {  {\intermed \sigma}  } } } $
  then we have $ \vdash_{ctx}^{\intermed {M} }  \source {  { \source P }  \ottsym{,}  { \source P }'  } ; \source {  {\source { \Gamma_C } }  \ottsym{,}  {\source { \Gamma_C } }'  } ; \source {  \bullet  }  \rightsquigarrow  \intermed {  {\intermed \Sigma}  \ottsym{,}  {\intermed \Sigma}'  } ; \intermed {  {\intermed { \Gamma_C } }  \ottsym{,}  {\intermed { \Gamma_C } }'  } ; \intermed {  \bullet  } $,
  and we have $ \intermed {  {\intermed \Sigma}  \ottsym{,}  {\intermed \Sigma}'  } ; \intermed {  {\intermed { \Gamma_C } }  \ottsym{,}  {\intermed { \Gamma_C } }'  } ; \intermed {  \bullet  }  \vdash_{tm}  \intermed {  {\intermed e}  } : \intermed {  {\intermed \sigma}  } \itot{ \rightsquigarrow  \target {  {\target e }  } } $.
\label{thmA:typing_pres_pgm}
\end{theoremB}

\proof By structural induction on the typing derivation.\\
\proofStep{\textsc{sPgmCls}}
  {\drule{sPgm-cls}}
  {
    We know that
    \begin{equation*}
       \vdash_{ctx}^{\intermed {M} }  \source {  { \source P }  } ; \source {  {\source { \Gamma_C } }  } ; \source {  \bullet  }  \rightsquigarrow  \intermed {  {\intermed \Sigma}  } ; \intermed {  {\intermed { \Gamma_C } }  } ; \intermed {  \bullet  } 
    \end{equation*}
    By inversion it follows that:
    \begin{equation*}
       \vdash_{ctx}^{\intermed {M} }  \source {  \bullet  } ; \source {  {\source { \Gamma_C } }  } ; \source {  \bullet  }  \rightsquigarrow  \intermed {  \bullet  } ; \intermed {  {\intermed { \Gamma_C } }  } ; \intermed {  \bullet  } 
    \end{equation*}
    By Typing Preservation - Classes (Theorem~\ref{thmA:typing_pres_class}), we know
    \begin{equation*}
       \vdash_{ctx}^{\intermed {M} }  \source {  \bullet  } ; \source {  {\source { \Gamma_C } }  \ottsym{,}  {\source { \Gamma_C } }'  } ; \source {  \bullet  }  \rightsquigarrow  \intermed {  \bullet  } ; \intermed {  {\intermed { \Gamma_C } }  \ottsym{,}  {\intermed { \Gamma_C } }'  } ; \intermed {  \bullet  } 
    \end{equation*}
    Through weakening (Lemma~\ref{lemma:ctx_weakening_class}), we know that
    \begin{equation*}
       \vdash_{ctx}^{\intermed {M} }  \source {  { \source P }  } ; \source {  {\source { \Gamma_C } }  \ottsym{,}  {\source { \Gamma_C } }'  } ; \source {  \bullet  }  \rightsquigarrow  \intermed {  {\intermed \Sigma}  } ; \intermed {  {\intermed { \Gamma_C } }  \ottsym{,}  {\intermed { \Gamma_C } }'  } ; \intermed {  \bullet  } 
    \end{equation*}
    The goal follows directly from the induction hypothesis. \\
    }
\proofStep{\textsc{sPgm-Inst}}
  { \drule{sPgm-inst}}
  {
    We know that
    \begin{equation*}
       \vdash_{ctx}^{\intermed {M} }  \source {  { \source P }  } ; \source {  {\source { \Gamma_C } }  } ; \source {  \bullet  }  \rightsquigarrow  \intermed {  {\intermed \Sigma}  } ; \intermed {  {\intermed { \Gamma_C } }  } ; \intermed {  \bullet  } 
    \end{equation*}
    By Typing Preservation - Instance (Theorem~\ref{thmA:typing_pres_instance}),
    we know that 
    \begin{equation}
      \label{eq:pres_pgm:1}
       \vdash_{ctx}^{\intermed {M} }  \source {  { \source P }  \ottsym{,}  { \source P }'  } ; \source {  {\source { \Gamma_C } }  } ; \source {  \bullet  }  \rightsquigarrow  \intermed {  {\intermed \Sigma}  \ottsym{,}  {\intermed \Sigma}'  } ; \intermed {  {\intermed { \Gamma_C } }  } ; \intermed {  \bullet  } 
    \end{equation}
    The goal follows directly from the induction hypothesis. \\
  }
\proofStep{\textsc{sPgm-expr}}
  {\drule{sPgm-expr}}
  {
    Follows directly from Typing Preservation - Expressions
    (Theorem~\ref{thmA:typing_pres_expr}). 
  }
\qed

\begin{theoremB}[Typing Preservation - Types and Class Constraints]\leavevmode
  \begin{itemize}
    \item
      If \(  \source {  {\source { \Gamma_C } }  } ; \source {  {\source \Gamma}  }  \vdash_{ty}^{\intermed {M} }  \source {  {\source \sigma}  }\,{ \stoi{ \rightsquigarrow  \intermed {  {\intermed \sigma}  } } } \),
      and \(  \vdash_{ctx}^{\intermed {M} }  \source {  \bullet  } ; \source {  {\source { \Gamma_C } }  } ; \source {  {\source \Gamma}  }  \rightsquigarrow  \intermed {  \bullet  } ; \intermed {  {\intermed { \Gamma_C } }  } ; \intermed {  {\intermed \Gamma}  }  \),
      then \(  \intermed {  {\intermed { \Gamma_C } }  } ; \intermed {  {\intermed \Gamma}  }  \vdash_{ty}  \intermed {  {\intermed \sigma}  } \itot{ \rightsquigarrow  \target {  {\target \sigma }  } }  \).
    \item 
      If \(  \source {  {\source { \Gamma_C } }  } ; \source {  {\source \Gamma}  }  \vdash_{\mathit {Q} }^{\intermed {M} }  \source {  {\source Q}  }  \rightsquigarrow  \intermed {  {\intermed Q}  } \),
      and \(  \vdash_{ctx}^{\intermed {M} }  \source {  \bullet  } ; \source {  {\source { \Gamma_C } }  } ; \source {  {\source \Gamma}  }  \rightsquigarrow  \intermed {  \bullet  } ; \intermed {  {\intermed { \Gamma_C } }  } ; \intermed {  {\intermed \Gamma}  }  \),
      then \(  \intermed {  {\intermed { \Gamma_C } }  } ; \intermed {  {\intermed \Gamma}  }  \vdash_{\mathit {Q} }  \intermed {  {\intermed Q}  }\, \itot{ \rightsquigarrow  \target {  {\target \sigma }  } }  \).
  \end{itemize}
  \label{thmA:typing_pres_types}
\end{theoremB}

\proof By induction on the lexicographic order of the tuple (size of
${\source { \Gamma_C } }$, the derivation height of type well-formedness and the constraint
well-formedness). In each mutual dependency, the size of the tuple is
decreasing, so we know that the induction is well-founded.

\begin{description}
  \item[Part 1]~\\
    \proofStep{\textsc{sTy-arrow}}
    { \drule{sTy-bool} }
    { Follows directly by \textsc{iTy-bool}.\\}
    \proofStep{\textsc{sTy-var}}
    { \drule{sTy-var} }
    { It is easy to verify that for any environment for which $ \vdash_{ctx}^{\intermed {M} }  \source {  { \source P }  } ; \source {  {\source { \Gamma_C } }  } ; \source {  {\source \Gamma}  }  \rightsquigarrow  \intermed {  {\intermed \Sigma}  } ; \intermed {  {\intermed { \Gamma_C } }  } ; \intermed {  {\intermed \Gamma}  } $ holds,
      \( \source {  {\source a }  }  \in  \source {  {\source \Gamma}  } \) implies \( \intermed {  {\intermed a }  }  \in  \intermed {  {\intermed \Gamma}  } \). Therefore, the goal follows
      from \textsc{iTy-var}. \\}
    \proofStep{\textsc{sTy-arrow}}
    { \drule{sTy-arrow} }
    { By induction hypothesis, we get
        \begin{gather*}
          \intermed {  {\intermed { \Gamma_C } }  } ; \intermed {  {\intermed \Gamma}  }  \vdash_{ty}  \intermed {  {\intermed \sigma}_{{\mathrm{1}}}  } \itot{ \rightsquigarrow  \target {  {\target \sigma }_{{\mathrm{1}}}  } }  \\
          \intermed {  {\intermed { \Gamma_C } }  } ; \intermed {  {\intermed \Gamma}  }  \vdash_{ty}  \intermed {  {\intermed \sigma}_{{\mathrm{2}}}  } \itot{ \rightsquigarrow  \target {  {\target \sigma }_{{\mathrm{2}}}  } }  
        \end{gather*}
      The goal follows directly from \textsc{iTy-Arrow}.\\
    }
    \proofStep{\textsc{sTy-qual}}
    { \drule{sTy-qual} }
    { By induction hypothesis, we get
      \begin{align*}
          \intermed {  {\intermed { \Gamma_C } }  } ; \intermed {  {\intermed \Gamma}  }  \vdash_{ty}  \intermed {  {\intermed \sigma}  } \itot{ \rightsquigarrow  \target {  {\target \sigma }  } }  
      \end{align*}
      By Part 2 of this lemma, we get
      \begin{align*}
          \intermed {  {\intermed { \Gamma_C } }  } ; \intermed {  {\intermed \Gamma}  }  \vdash_{\mathit {Q} }  \intermed {  {\intermed Q}  }\, \itot{ \rightsquigarrow  \target {  {\target \sigma }'  } }  
      \end{align*}
      The goal follows directly from \textsc{iTy-Qual}.\\
    }
    \proofStep{\textsc{sTy-scheme}}
    { \drule{sTy-scheme} }
    {
      Given $ \vdash_{ctx}^{\intermed {M} }  \source {  \bullet  } ; \source {  {\source { \Gamma_C } }  } ; \source {  {\source \Gamma}  }  \rightsquigarrow  \intermed {  \bullet  } ; \intermed {  {\intermed { \Gamma_C } }  } ; \intermed {  {\intermed \Gamma}  } $, by \textsc{sCtx-EnvTy},
      we know that $ \vdash_{ctx}^{\intermed {M} }  \source {  \bullet  } ; \source {  {\source { \Gamma_C } }  } ; \source {  {\source \Gamma}  \ottsym{,}  {\source a }  }  \rightsquigarrow  \intermed {  \bullet  } ; \intermed {  {\intermed { \Gamma_C } }  } ; \intermed {  {\intermed \Gamma}  \ottsym{,}  {\intermed a }  } $.
      By induction hypothesis, we get
      \begin{align*}
          \intermed {  {\intermed { \Gamma_C } }  } ; \intermed {  {\intermed \Gamma}  \ottsym{,}  {\intermed a }  }  \vdash_{ty}  \intermed {  {\intermed \sigma}  } \itot{ \rightsquigarrow  \target {  {\target \sigma }  } }  
      \end{align*}
      The goal follows directly by \textsc{iTy-scheme}.
    }

  \item[Part 2]~\\
    \proofStep{\textsc{sQ-TC}}
    { \drule{sQ-TC}}
    {
      By Part 1 of this lemma, we know that
      \begin{align}
        \label{eq:preserve_type:1}
          \intermed {  {\intermed { \Gamma_C } }  } ; \intermed {  {\intermed \Gamma}  }  \vdash_{ty}  \intermed {  {\intermed \sigma}  } \itot{ \rightsquigarrow  \target {  {\target \sigma }  } }  
      \end{align}
      It is easy to verify that given any environment for which
      \begin{gather}
         \vdash_{ctx}^{\intermed {M} }  \source {  \bullet  } ; \source {  {\source { \Gamma_C } }  } ; \source {  {\source \Gamma}  }  \rightsquigarrow  \intermed {  \bullet  } ; \intermed {  {\intermed { \Gamma_C } }  } ; \intermed {  {\intermed \Gamma}  }  
        \nonumber \\
         \source {  {\source { \Gamma_C } }  } = \source {  {\source { \Gamma_C } }_{{\mathrm{1}}}  }, \source {  {\source m}  \ottsym{:}  {\overline {\source Q} }_{\ottmv{i}}  \Rightarrow  {\source {\mathit{TC} } } \, {\source a }  \ottsym{:}  {\source \sigma}  \ottsym{,}  {\source { \Gamma_C } }_{{\mathrm{2}}}  } 
        \label{eq:preserve_type:2}
      \end{gather}
      then
      \begin{gather*}
       \intermed {  {\intermed { \Gamma_C } }  } = \intermed {  {\intermed { \Gamma_C } }_{{\mathrm{1}}}  }, \intermed {  {\intermed m}  \ottsym{:}  {\intermed {\mathit{TC} } } \, {\intermed a }  \ottsym{:}  {\intermed \sigma}'  \ottsym{,}  {\intermed { \Gamma_C } }_{{\mathrm{2}}}  } \\
       \vdash_{ctx}^{\intermed {M} }  \source {  \bullet  } ; \source {  {\source { \Gamma_C } }_{{\mathrm{1}}}  } ; \source {  \bullet  }  \rightsquigarrow  \intermed {  \bullet  } ; \intermed {  {\intermed { \Gamma_C } }_{{\mathrm{1}}}  } ; \intermed {  \bullet  } \\
       \source {  {\source { \Gamma_C } }_{{\mathrm{1}}}  } ; \source {  \bullet  \ottsym{,}  {\source a }  }  \vdash_{ty}^{\intermed {M} }  \source {  {\source \sigma}  }\,{ \stoi{ \rightsquigarrow  \intermed {  {\intermed \sigma}'  } } } 
      \end{gather*}
      By \textsc{sCtx-TyEnvTy}, we get
      \begin{align*}
       \vdash_{ctx}^{\intermed {M} }  \source {  \bullet  } ; \source {  {\source { \Gamma_C } }_{{\mathrm{1}}}  } ; \source {  \bullet  \ottsym{,}  {\source a }  }  \rightsquigarrow  \intermed {  \bullet  } ; \intermed {  {\intermed { \Gamma_C } }_{{\mathrm{1}}}  } ; \intermed {  \bullet  \ottsym{,}  {\intermed a }  } 
      \end{align*}
      The size of ${\intermed { \Gamma_C } }_{{\mathrm{1}}}$ is trivially smaller than ${\intermed { \Gamma_C } }$. So by induction
      hypothesis, we have
      \begin{align}
        \label{eq:preserve_type:3}
          \intermed {  {\intermed { \Gamma_C } }  } ; \intermed {  \bullet  \ottsym{,}  {\intermed a }  }  \vdash_{ty}  \intermed {  {\intermed \sigma}'  } \itot{ \rightsquigarrow  \target {  {\target \sigma }'  } }  
      \end{align}
      The goal follows directly from \textsc{sQ-TC}, and Equations
      \ref{eq:preserve_type:1}, \ref{eq:preserve_type:2},
      \ref{eq:preserve_type:3}.
      
    }

\end{description}
\qed

\begin{theoremB}[Typing Preservation - Constraints Proving]\ \\
  If $ \source {  { \source P }  } ; \source {  {\source { \Gamma_C } }  } ; \source {  {\source \Gamma}  }  \vDash^{\intermed {M} }  \source {  {\source Q}  } \stoi{ \rightsquigarrow  \intermed {  {\intermed d}  } } $,
  and $  \vdash_{ctx}^{\intermed {M} }  \source {  { \source P }  } ; \source {  {\source { \Gamma_C } }  } ; \source {  {\source \Gamma}  }  \rightsquigarrow  \intermed {  {\intermed \Sigma}  } ; \intermed {  {\intermed { \Gamma_C } }  } ; \intermed {  {\intermed \Gamma}  }  $,
  and $  \source {  {\source { \Gamma_C } }  } ; \source {  {\source \Gamma}  }  \vdash_{\mathit {Q} }^{\intermed {M} }  \source {  {\source Q}  }  \rightsquigarrow  \intermed {  {\intermed Q}  } $,
  then $ \intermed {  {\intermed \Sigma}  } ; \intermed {  {\intermed { \Gamma_C } }  } ; \intermed {  {\intermed \Gamma}  }  \vdash_{d}  \intermed {  {\intermed d}  } : \intermed {  {\intermed Q}  } \itot{\,  \rightsquigarrow  \target {  {\target e }  } } $.
  \label{thmA:typing_pres_constraint}
\end{theoremB}

\proof By induction on the constraint resolution derivation tree.
This theorem is mutually proven with Theorems~\ref{thmA:typing_pres_expr}
and~\ref{thmA:typing_pres_env} (Figure~\ref{fig:dep_graph_pres}).
Note that at the dependency from
Theorem~\ref{thmA:typing_pres_env} to \ref{thmA:typing_pres_expr}, the size of
\({ \source P }\) is strictly decreasing, whereas \({ \source P }\) remains constant at every
other dependency. Consequently, the size of \({ \source P }\) is strictly decreasing in
every possible cycle. The induction thus remains well-founded.\\
\proofStep{\textsc{sEntail-local}}{\drule{sEntail-local}}
{
  It is easy to verify that given
  \begin{gather*}
     \vdash_{ctx}^{\intermed {M} }  \source {  { \source P }  } ; \source {  {\source { \Gamma_C } }  } ; \source {  {\source \Gamma}  }  \rightsquigarrow  \intermed {  {\intermed \Sigma}  } ; \intermed {  {\intermed { \Gamma_C } }  } ; \intermed {  {\intermed \Gamma}  }   \\
     ( \source {  {\source \delta }  } : \source {  {\source Q}  } )  \in  \source {  {\source \Gamma}  } 
  \end{gather*}
  we can derive
  \begin{gather*}
     ( \intermed {  {\intermed \delta }  } : \intermed {  {\intermed Q}'  } )  \in  \intermed {  {\intermed \Gamma}  } \\
     \source {  {\source { \Gamma_C } }  } ; \source {  {\source \Gamma}_{{\mathrm{1}}}  }  \vdash_{\mathit {Q} }^{\intermed {M} }  \source {  {\source Q}  }  \rightsquigarrow  \intermed {  {\intermed Q}'  }  \\
     \source {  {\source \Gamma}  } = \source {  {\source \Gamma}_{{\mathrm{1}}}  \ottsym{,}  {\source \delta }  \ottsym{:}  {\source Q}  \ottsym{,}  {\source \Gamma}_{{\mathrm{2}}}  } 
  \end{gather*}
  By weakening lemma (Lemma~\ref{lemma:class_constraint_weakening})
  \begin{align*}
   \source {  {\source { \Gamma_C } }  } ; \source {  {\source \Gamma}  }  \vdash_{\mathit {Q} }^{\intermed {M} }  \source {  {\source Q}  }  \rightsquigarrow  \intermed {  {\intermed Q}'  }  
  \end{align*}
  We already know from the hypothesis:
  \begin{align*}
   \source {  {\source { \Gamma_C } }  } ; \source {  {\source \Gamma}  }  \vdash_{\mathit {Q} }^{\intermed {M} }  \source {  {\source Q}  }  \rightsquigarrow  \intermed {  {\intermed Q}  } 
  \end{align*}
  Since the elaboration of $ \vdash_{\mathit {Q} } $ is deterministic (Lemma~\ref{lemma:q_elab_unique}), we know that ${\intermed Q} =
  {\intermed Q}'$.
  The goal follows directly by \textsc{D-var}. \\
}
\proofStep{\textsc{sEntail-inst}}
          {}
{
  \drule{sEntail-inst}

  The goal to be proven is the following:
  \begin{equation*}
     \intermed {  {\intermed \Sigma}  } ; \intermed {  {\intermed { \Gamma_C } }  } ; \intermed {  {\intermed \Gamma}  }  \vdash_{d}  \intermed {  {\intermed D } \, \overline {\intermed \sigma}_{\ottmv{j}} \, {\overline {\intermed d} }_{\ottmv{i}}  } : \intermed {  {\intermed Q}  } \itot{\,  \rightsquigarrow  \target {  {\target e }  } } ~
    \textit{where}~ \intermed {  {\intermed Q}  } = \intermed {  {\intermed {\mathit{TC} } } \, {\intermed \sigma}_{\ottmv{q}}  } 
  \end{equation*}
  Using \textsc{D-con}, proving this is equivalent to proving:
  \begin{gather}
    \label{eq:preserve_proof:1}
     ( \intermed {  {\intermed D }  } : \intermed {  \forall  {\overline {\intermed a} }_{\ottmv{j}}  \ottsym{.}  {\overline {\intermed Q} }_{\ottmv{i}}  \Rightarrow  {\intermed {\mathit{TC} } } \, {\intermed \sigma}'_{\ottmv{q}}  } ) . \intermed {  {\intermed m}  }  \mapsto  \intermed {  \Lambda  {\overline {\intermed a} }_{\ottmv{j}}  \ottsym{.}  \lambda  {\overline {\intermed \delta} }_{\ottmv{i}}  \ottsym{:}  {\overline {\intermed Q} }_{\ottmv{i}}  \ottsym{.}  {\intermed e}  }  \in  \intermed {  {\intermed \Sigma}  }  \\
    \textit{where}~ \intermed {  {\intermed \sigma}_{\ottmv{q}}  } = \intermed {  [  \overline {\intermed \sigma}_{\ottmv{j}}  /  {\overline {\intermed a} }_{\ottmv{j}}  ]  {\intermed \sigma}'_{\ottmv{q}}  }  \nonumber \\
    \label{eq:preserve_proof:2}
     \vdash_{ctx}  \intermed {  {\intermed \Sigma}  } ; \intermed {  {\intermed { \Gamma_C } }  } ; \intermed {  {\intermed \Gamma}  }  \\
    \label{eq:preserve_proof:3}
    \ottcomp{ \intermed {  {\intermed { \Gamma_C } }  } ; \intermed {  \bullet  \ottsym{,}  {\overline {\intermed a} }_{\ottmv{j}}  }  \vdash_{\mathit {Q} }  \intermed {  {\intermed Q}_{\ottmv{i}}  }\, \itot{ \rightsquigarrow  \target {  {\target \sigma }'_{\ottmv{i}}  } } }{\ottmv{i}} \\
    \label{eq:preserve_proof:4}
    \ottcomp{ \intermed {  {\intermed { \Gamma_C } }  } ; \intermed {  {\intermed \Gamma}  }  \vdash_{ty}  \intermed {  {\intermed \sigma}_{\ottmv{j}}  } \itot{ \rightsquigarrow  \target {  {\target \sigma }_{\ottmv{j}}  } } }{\ottmv{j}} \\
    \label{eq:preserve_proof:5}
     \intermed {  {\intermed \Sigma}_{{\mathrm{1}}}  } ; \intermed {  {\intermed { \Gamma_C } }  } ; \intermed {  \bullet  \ottsym{,}  {\overline {\intermed a} }_{\ottmv{j}}  \ottsym{,}  {\overline {\intermed \delta} }_{\ottmv{i}}  \ottsym{:}  {\overline {\intermed Q} }_{\ottmv{i}}  }  \vdash_{tm}  \intermed {  {\intermed e}  } : \intermed {  [  {\intermed \sigma}'_{\ottmv{q}}  /  {\intermed a }  ]  {\intermed \sigma}_{\ottmv{m}}  } \itot{ \rightsquigarrow  \target {  {\target e }'  } }  \\
    \textit{where}~ \intermed {  {\intermed \Sigma}  } = \intermed {  {\intermed \Sigma}_{{\mathrm{1}}}  \ottsym{,}  \ottsym{(}  {\intermed D }  \ottsym{:}  \forall  {\overline {\intermed a} }_{\ottmv{j}}  \ottsym{.}  {\overline {\intermed Q} }_{\ottmv{i}}  \Rightarrow  {\intermed Q}  \ottsym{)}  \ottsym{.}  {\intermed m}  \mapsto  {\intermed e}  \ottsym{,}  {\intermed \Sigma}_{{\mathrm{2}}}  }  \nonumber \\
    \label{eq:preserve_proof:6}
    \ottcomp{ \intermed {  {\intermed \Sigma}  } ; \intermed {  {\intermed { \Gamma_C } }  } ; \intermed {  {\intermed \Gamma}  }  \vdash_{d}  \intermed {  {\intermed d}_{\ottmv{i}}  } : \intermed {  [  \overline {\intermed \sigma}_{\ottmv{j}}  /  {\overline {\intermed a} }_{\ottmv{j}}  ]  {\intermed Q}_{\ottmv{i}}  } \itot{\,  \rightsquigarrow  \target {  {\target e }_{\ottmv{i}}  } } }{\ottmv{i}}
  \end{gather}
  The rule premise tells us, among other things, that:
  \begin{gather}
    \label{eq:preserve_proof:7}
     \source {  { \source P }  } = \source {  { \source P }_{{\mathrm{1}}}  }, \source {  \ottsym{(}  {\source D }  \ottsym{:}  \forall  {\overline {\source a} }_{\ottmv{j}}  \ottsym{.}  {\overline {\source Q} }_{\ottmv{i}}  \Rightarrow  {\source Q}'  \ottsym{)}  \ottsym{.}  {\source m}  \mapsto  \bullet  \ottsym{,}  {\overline {\source a} }_{\ottmv{j}}  \ottsym{,}  {\overline {\source \delta} }_{\ottmv{i}}  \ottsym{:}  {\overline {\source Q} }_{\ottmv{i}}  \ottsym{,}  {\overline {\source b} }_{\ottmv{k}}  \ottsym{,}  {\overline {\source \delta} }_{\ottmv{h}}  \ottsym{:}  {\overline {\source Q} }_{\ottmv{h}}  \ottsym{:}  {\source e}  \ottsym{,}  { \source P }_{{\mathrm{2}}}  }  \\
    \label{eq:preserve_proof:8}
     \vdash_{ctx}^{\intermed {M} }  \source {  { \source P }  } ; \source {  {\source { \Gamma_C } }  } ; \source {  {\source \Gamma}  }  \rightsquigarrow  \intermed {  {\intermed \Sigma}  } ; \intermed {  {\intermed { \Gamma_C } }  } ; \intermed {  {\intermed \Gamma}  }  \\
    \label{eq:preserve_proof:9}
    \ottcomp{ \source {  {\source { \Gamma_C } }  } ; \source {  {\source \Gamma}  }  \vdash_{ty}^{\intermed {M} }  \source {  {\source \tau }_{\ottmv{j}}  }\,{ \stoi{ \rightsquigarrow  \intermed {  {\intermed \sigma}_{\ottmv{j}}  } } } }{\ottmv{j}} \\
    \label{eq:preserve_proof:10}
    \ottcomp{ \source {  { \source P }  } ; \source {  {\source { \Gamma_C } }  } ; \source {  {\source \Gamma}  }  \vDash^{\intermed {M} }  \source {  [  {\overline {\source \tau} }_{\ottmv{j}}  /  {\overline {\source a} }_{\ottmv{j}}  ]  {\source Q}_{\ottmv{i}}  } \stoi{ \rightsquigarrow  \intermed {  {\intermed d}_{\ottmv{i}}  } } }{\ottmv{i}} 
  \end{gather}
  Goal~\ref{eq:preserve_proof:1} follows directly from
  Equations~\ref{eq:preserve_proof:7} and \ref{eq:preserve_proof:8}.
  Goal~\ref{eq:preserve_proof:2} follows by applying
  Equation~\ref{eq:preserve_proof:8} to preservation
  Theorem~\ref{thmA:typing_pres_env}.

  Consequently, from Equation~\ref{eq:preserve_proof:2}, in combination with
  rule \textsc{iCtx-MEnv}, we know that:
  \begin{equation}
     \intermed {  {\intermed { \Gamma_C } }  } ; \intermed {  \bullet  }  \vdash_{\mathit {C} }  \intermed {  \forall  {\overline {\intermed a} }_{\ottmv{j}}  \ottsym{.}  {\overline {\intermed Q} }_{\ottmv{i}}  \Rightarrow  {\intermed {\mathit{TC} } } \, {\intermed \sigma}'_{\ottmv{q}}  } 
    \label{eq:preserve_proof:11}
  \end{equation}
  Goal~\ref{eq:preserve_proof:3} follows from rule \textsc{iC-abs}, in
  combination with Equation~\ref{eq:preserve_proof:11}.

  By inversion on Equation~\ref{eq:preserve_proof:8}, we know that:
  \begin{equation}
     \vdash_{ctx}^{\intermed {M} }  \source {  \bullet  } ; \source {  {\source { \Gamma_C } }  } ; \source {  {\source \Gamma}  }  \rightsquigarrow  \intermed {  \bullet  } ; \intermed {  {\intermed { \Gamma_C } }  } ; \intermed {  {\intermed \Gamma}  } 
    \label{eq:preserve_proof:12}
  \end{equation}
  Goal~\ref{eq:preserve_proof:4} follows by applying
  Equations~\ref{eq:preserve_proof:9} and \ref{eq:preserve_proof:12} to
  Theorem~\ref{thmA:typing_pres_types}. 
  
  From rule \textsc{iCtx-MEnv}, in combination with
  Equations~\ref{eq:preserve_proof:1} and \ref{eq:preserve_proof:2}, we know
  that:
  \begin{equation}
     \intermed {  {\intermed \Sigma}_{{\mathrm{1}}}  } ; \intermed {  {\intermed { \Gamma_C } }  } ; \intermed {  \bullet  }  \vdash_{tm}  \intermed {  \Lambda  {\overline {\intermed a} }_{\ottmv{j}}  \ottsym{.}  \lambda  {\overline {\intermed \delta} }_{\ottmv{i}}  \ottsym{:}  {\overline {\intermed Q} }_{\ottmv{i}}  \ottsym{.}  {\intermed e}  } : \intermed {  \forall  {\overline {\intermed a} }_{\ottmv{j}}  \ottsym{.}   {\overline {\intermed Q} }_{\ottmv{i}}  \Rightarrow  [  {\intermed \sigma}'_{\ottmv{q}}  /  {\intermed a }  ]  {\intermed \sigma}_{\ottmv{m}}   } \itot{ \rightsquigarrow  \target {  {\target e }_{\ottmv{m}}  } } 
    \label{eq:preserve_proof:13}
  \end{equation}
  Goal~\ref{eq:preserve_proof:5} follows from inversion of rules
  \textsc{iTm-forallI} and \textsc{iTm-constrI}, on Equation~\ref{eq:preserve_proof:13}.

  Finally, Goal~\ref{eq:preserve_proof:6} follows by applying the induction
  hypothesis on Equation~\ref{eq:preserve_proof:10}, assuming we can show that:
  \begin{equation}
    \ottcomp{ \source {  {\source { \Gamma_C } }  } ; \source {  {\source \Gamma}  }  \vdash_{\mathit {Q} }^{\intermed {M} }  \source {  [  {\overline {\source \tau} }_{\ottmv{j}}  /  {\overline {\source a} }_{\ottmv{j}}  ]  {\source Q}_{\ottmv{i}}  }  \rightsquigarrow  \intermed {  [  \overline {\intermed \sigma}_{\ottmv{j}}  /  {\overline {\intermed a} }_{\ottmv{j}}  ]  {\intermed Q}_{\ottmv{i}}  } }{\ottmv{i}}
    \label{eq:preserve_proof:14}
  \end{equation}
  From rule \textsc{sCtx-pgmInst}, in combination with
  Equation~\ref{eq:preserve_proof:7}, we know that:
  \begin{equation}
     \source {  {\source { \Gamma_C } }  } ; \source {  \bullet  }  \vdash_{\mathit {C} }^{\intermed {M} }  \source {  \forall  {\overline {\source a} }_{\ottmv{j}}  \ottsym{.}  {\overline {\source Q} }_{\ottmv{i}}  \Rightarrow  {\source Q}'  }  \rightsquigarrow  \intermed {  \forall  {\overline {\intermed a} }_{\ottmv{j}}  \ottsym{.}  {\overline {\intermed Q} }_{\ottmv{i}}  \Rightarrow  {\intermed {\mathit{TC} } } \, {\intermed \sigma}'_{\ottmv{q}}  } 
    \label{eq:preserve_proof:15}
  \end{equation}
  From weakening Lemma~\ref{lemma:constraint_weakening}, we know that:
  \begin{equation}
     \source {  {\source { \Gamma_C } }  } ; \source {  {\source \Gamma}  }  \vdash_{\mathit {C} }^{\intermed {M} }  \source {  \forall  {\overline {\source a} }_{\ottmv{j}}  \ottsym{.}  {\overline {\source Q} }_{\ottmv{i}}  \Rightarrow  {\source Q}'  }  \rightsquigarrow  \intermed {  \forall  {\overline {\intermed a} }_{\ottmv{j}}  \ottsym{.}  {\overline {\intermed Q} }_{\ottmv{i}}  \Rightarrow  {\intermed {\mathit{TC} } } \, {\intermed \sigma}'_{\ottmv{q}}  } 
    \label{eq:preserve_proof:16}
  \end{equation}
  From rule \textsc{sQ-TC}, in combination with
  Equation~\ref{eq:preserve_proof:16}, we know that:
  \begin{equation}
    \ottcomp{ \source {  {\source { \Gamma_C } }  } ; \source {  {\source \Gamma}  \ottsym{,}  {\overline {\source a} }_{\ottmv{j}}  }  \vdash_{\mathit {Q} }^{\intermed {M} }  \source {  {\source Q}_{\ottmv{i}}  }  \rightsquigarrow  \intermed {  {\intermed Q}_{\ottmv{i}}  } }{\ottmv{i}}
    \label{eq:preserve_proof:17}
  \end{equation}
  Goal~\ref{eq:preserve_proof:14} follows by applying type substitution
  Lemma~\ref{lemma:tvar_subst_ctr} on Equations~\ref{eq:preserve_proof:17} and
  \ref{eq:preserve_proof:9} (in combination with weakening
  Lemma~\ref{lemma:type_weakening}). 
}
\qed

\begin{theoremB}[Typing Preservation - Environment Well-Formedness]\ \\
  If $  \vdash_{ctx}^{\intermed {M} }  \source {  { \source P }  } ; \source {  {\source { \Gamma_C } }  } ; \source {  {\source \Gamma}  }  \rightsquigarrow  \intermed {  {\intermed \Sigma}  } ; \intermed {  {\intermed { \Gamma_C } }  } ; \intermed {  {\intermed \Gamma}  }  $,
  then $ \vdash_{ctx}  \intermed {  {\intermed \Sigma}  } ; \intermed {  {\intermed { \Gamma_C } }  } ; \intermed {  {\intermed \Gamma}  } $.
  \label{thmA:typing_pres_env}
\end{theoremB}

\proof By induction on the well-formedness derivation.
This theorem is mutually proven with Theorems~\ref{thmA:typing_pres_expr}
and~\ref{thmA:typing_pres_constraint} (Figure~\ref{fig:dep_graph_pres}).
Note that at the dependency from
Theorem~\ref{thmA:typing_pres_env} to \ref{thmA:typing_pres_expr}, the size of
\({ \source P }\) is strictly decreasing, whereas \({ \source P }\) remains constant at every
other dependency. Consequently, the size of \({ \source P }\) is strictly decreasing in
every possible cycle. The induction thus remains well-founded.\\
\proofStep{\textsc{sCtx-empty}}{
  \drule{sCtx-empty}
}
{
Follows directly by \textsc{iCtx-empty}. \\
}
\proofStep{\textsc{sCtx-tyEnvTm}}{
  \drule{sCtx-tyEnvTm}
}
{
 By induction hypothesis, we know 
  \begin{equation*}
     \vdash_{ctx}  \intermed {  \bullet  } ; \intermed {  {\intermed { \Gamma_C } }  } ; \intermed {  {\intermed \Gamma}  } 
  \end{equation*}
  By Typing Preservation - Types and Class Constraints
  (Theorem~\ref{thmA:typing_pres_types}), we know
  \begin{equation*}
     \intermed {  {\intermed { \Gamma_C } }  } ; \intermed {  {\intermed \Gamma}  }  \vdash_{ty}  \intermed {  {\intermed \sigma}  } \itot{ \rightsquigarrow  \target {  {\target \sigma }  } } 
  \end{equation*}
  Since $ \source {  {\source x}  }  \notin   \textbf {dom}  ( \source {  {\source \Gamma}  } ) $, it is easy to verify that $ \intermed {  {\intermed x}  }  \notin   \textbf {dom}  ( \intermed {  {\intermed \Gamma}  } ) $.
  Therefore the goal follows directly by \textsc{iCtx-tyEnvTm}. \\
}
\proofStep{\textsc{sCtx-tyEnvTy}}{
  \drule{sCtx-tyEnvTy}
}
{
  By induction hypothesis, we know
  \begin{equation*}
     \vdash_{ctx}  \intermed {  \bullet  } ; \intermed {  {\intermed { \Gamma_C } }  } ; \intermed {  {\intermed \Gamma}  } 
  \end{equation*}
  Since we know $ \source {  {\source a }  }  \notin  \source {  {\source \Gamma}  } $, it is easy to verify that $ \intermed {  {\intermed a }  }  \notin  \intermed {  {\intermed \Gamma}  } $.
  Therefore the goal follows directly by \textsc{iCtx-tyEnvTy}. \\
}
\proofStep{\textsc{sCtx-tyEnvD}}{
  \drule{sCtx-tyEnvD}
}
{
  By induction hypothesis, we know
  \begin{equation*}
     \vdash_{ctx}  \intermed {  \bullet  } ; \intermed {  {\intermed { \Gamma_C } }  } ; \intermed {  {\intermed \Gamma}  } 
  \end{equation*}
  By Typing Preservation - Types and Class Constraints
  (Theorem~\ref{thmA:typing_pres_types}), we know
  \begin{equation*}
     \intermed {  {\intermed { \Gamma_C } }  } ; \intermed {  {\intermed \Gamma}  }  \vdash_{\mathit {Q} }  \intermed {  {\intermed Q}  }\, \itot{ \rightsquigarrow  \target {  {\target \sigma }  } } 
  \end{equation*}
  Since $ \source {  {\source \delta }  }  \notin   \textbf {dom}  ( \source {  {\source \Gamma}  } ) $, it is easy to verify that $ \intermed {  {\intermed \delta }  }  \notin   \textbf {dom}  ( \intermed {  {\intermed \Gamma}  } ) $.
  Therefore the goal follows directly by \textsc{iCtx-tyEnvD}. \\  
}
\proofStep{\textsc{sCtx-clsEnv}}{
}
{
  \drule{sCtx-clsEnv}

  By induction hypothesis, we know
  \begin{equation*}
     \vdash_{ctx}  \intermed {  \bullet  } ; \intermed {  {\intermed { \Gamma_C } }  } ; \intermed {  \bullet  } 
  \end{equation*}
  By \textsc{sCtx-tyEnvTy}, we know that 
  \begin{equation*}
     \vdash_{ctx}^{\intermed {M} }  \source {  \bullet  } ; \source {  {\source { \Gamma_C } }  } ; \source {  \bullet  \ottsym{,}  {\source a }  }  \rightsquigarrow  \intermed {  \bullet  } ; \intermed {  {\intermed { \Gamma_C } }  } ; \intermed {  \bullet  \ottsym{,}  {\intermed a }  } 
  \end{equation*}
  Then by Typing Preservation - Types and Class Constraints
  (Theorem~\ref{thmA:typing_pres_types}), we know
  \begin{equation*}
     \intermed {  {\intermed { \Gamma_C } }  } ; \intermed {  \bullet  \ottsym{,}  {\intermed a }  }  \vdash_{ty}  \intermed {  {\intermed \sigma}  } \itot{ \rightsquigarrow  \target {  {\target \sigma }  } } 
  \end{equation*}
  It is easy to verify that given $ \source {  {\source m}  }  \notin   \textbf {dom}  ( \source {  {\source { \Gamma_C } }  } ) ,  \source {  {\source {\mathit{TC} } } \, {\source b }  }  \notin   \textbf {dom}  ( \source {  {\source { \Gamma_C } }  } ) $, we can derive $ \intermed {  {\intermed m}  }  \notin   \textbf {dom}  ( \intermed {  {\intermed { \Gamma_C } }  } ) ,  \intermed {  {\intermed {\mathit{TC} } } \, {\intermed b }  }  \notin   \textbf {dom}  ( \intermed {  {\intermed { \Gamma_C } }  } ) $.

  The goal follows directly by \textsc{iCtx-ClsEnv}. \\
}
\proofStep{\textsc{sCtx-pgmInst}}{
}
{
  \drule{sCtx-pgmInst}

  By induction hypothesis, we know that
  \begin{equation}
    \label{eq:preserve_ctx:goal:0}
     \vdash_{ctx}  \intermed {  {\intermed \Sigma}  } ; \intermed {  {\intermed { \Gamma_C } }  } ; \intermed {  {\intermed \Gamma}  } 
  \end{equation}
  Also, since we know that
  \begin{equation*}
     \vdash_{ctx}^{\intermed {M} }  \source {  { \source P }  } ; \source {  {\source { \Gamma_C } }  } ; \source {  {\source \Gamma}  }  \rightsquigarrow  \intermed {  {\intermed \Sigma}  } ; \intermed {  {\intermed { \Gamma_C } }  } ; \intermed {  {\intermed \Gamma}  } 
  \end{equation*}
  by applying inversion, we get:
  \begin{equation}
    \label{eq:preserve_ctx:1}
     \vdash_{ctx}^{\intermed {M} }  \source {  \bullet  } ; \source {  {\source { \Gamma_C } }  } ; \source {  {\source \Gamma}  }  \rightsquigarrow  \intermed {  \bullet  } ; \intermed {  {\intermed { \Gamma_C } }  } ; \intermed {  {\intermed \Gamma}  } 
  \end{equation}
  From the premise, we already know
  \begin{align}
    \label{eq:preserve_ctx:2}
     \source {  {\source { \Gamma_C } }  } ; \source {  \bullet  }  \vdash_{\mathit {C} }^{\intermed {M} }  \source {  \forall  {\overline {\source b} }_{\ottmv{j}}  \ottsym{.}  {\overline {\source Q} }_{\ottmv{i}}  \Rightarrow  {\source {\mathit{TC} } } \, {\source \tau }  }  \rightsquigarrow  \intermed {  \forall  {\overline {\intermed b} }_{\ottmv{j}}  \ottsym{.}  {\overline {\intermed Q} }_{\ottmv{i}}  \Rightarrow  {\intermed {\mathit{TC} } } \, {\intermed \sigma}  } 
  \end{align}
  Then by Typing Preservation - Types and Class Constraints
  (Theorem~\ref{thmA:typing_pres_types}) on Equations~\ref{eq:preserve_ctx:1} and
  \ref{eq:preserve_ctx:2}, we know
  \begin{equation}
    \label{eq:preserve_ctx:goal:1}
     \intermed {  {\intermed { \Gamma_C } }  } ; \intermed {  \bullet  }  \vdash_{\mathit {C} }  \intermed {  \forall  {\overline {\intermed b} }_{\ottmv{j}}  \ottsym{.}  {\overline {\intermed Q} }_{\ottmv{i}}  \Rightarrow  {\intermed {\mathit{TC} } } \, {\intermed \sigma}  } 
  \end{equation}
  From the 3\textsuperscript{rd} rule premise, we know that
  \begin{equation}
    \label{eq:preserve_ctx:3}
     ( \source {  {\source m}  } : \source {  {\overline {\source Q} }'_{\ottmv{m}}  }  \Rightarrow  \source {  {\source {\mathit{TC} } }  }~\source {  {\source a }  } : \source {  \forall  {\overline {\source a} }_{\ottmv{k}}  \ottsym{.}  {\overline {\source Q} }_{\ottmv{h}}  \Rightarrow  {\source \tau }'  } )  \in  \source {  {\source { \Gamma_C } }  } 
  \end{equation}
  By inversion on Equation~\ref{eq:preserve_ctx:1} (\textsc{sCtx-clsEnv}),
  together with Equation~\ref{eq:preserve_ctx:3}, we get
  \begin{gather}
    \label{eq:preserve_ctx:41}
     ( \intermed {  {\intermed m}  } : \intermed {  {\intermed {\mathit{TC} } } \, {\intermed a }  } : \intermed {  {\intermed \sigma}_{{\mathrm{1}}}  } )  \in  \intermed {  {\intermed { \Gamma_C } }  }  \\
    \label{eq:preserve_ctx:4}
     \source {  {\source { \Gamma_C } }_{{\mathrm{1}}}  } ; \source {  \bullet  \ottsym{,}  {\source a }  }  \vdash_{ty}^{\intermed {M} }  \source {  \forall  {\overline {\source a} }_{\ottmv{k}}  \ottsym{.}  {\overline {\source Q} }_{\ottmv{h}}  \Rightarrow  {\source \tau }'  }\,{ \stoi{ \rightsquigarrow  \intermed {  {\intermed \sigma}_{{\mathrm{1}}}  } } }  \\
    \label{eq:preserve_ctx:5}
     \source {  {\source { \Gamma_C } }  } = \source {  {\source { \Gamma_C } }_{{\mathrm{1}}}  \ottsym{,}  {\source { \Gamma_C } }_{{\mathrm{2}}}  } 
  \end{gather}
  By applying weakening (Lemma~\ref{lemma:type_weakening}) on
  Equation~\ref{eq:preserve_ctx:4}, we have
  \begin{align}
    \label{eq:preserve_ctx:6}
     \source {  {\source { \Gamma_C } }  } ; \source {  \bullet  \ottsym{,}  {\source a }  }  \vdash_{ty}^{\intermed {M} }  \source {  \forall  {\overline {\source a} }_{\ottmv{k}}  \ottsym{.}  {\overline {\source Q} }_{\ottmv{h}}  \Rightarrow  {\source \tau }'  }\,{ \stoi{ \rightsquigarrow  \intermed {  {\intermed \sigma}_{{\mathrm{1}}}  } } }  
  \end{align}
  From the 5\textsuperscript{th} rule premise, we know that
  \begin{align}
    \label{eq:preserve_ctx:7}
     \source {  {\source { \Gamma_C } }  } ; \source {  \bullet  \ottsym{,}  {\source a }  }  \vdash_{ty}^{\intermed {M} }  \source {  \forall  {\overline {\source a} }_{\ottmv{k}}  \ottsym{.}  {\overline {\source Q} }_{\ottmv{h}}  \Rightarrow  {\source \tau }'  }\,{ \stoi{ \rightsquigarrow  \intermed {  \forall  {\overline {\intermed a} }_{\ottmv{k}}  \ottsym{.}   {\overline {\intermed Q} }_{\ottmv{h}}  \Rightarrow  {\intermed \sigma}'   } } } 
  \end{align}
  Because the elaboration of types is deterministic (Lemma~\ref{lemma:ty_elab_unique}), combining
  Equations~\ref{eq:preserve_ctx:6} and \ref{eq:preserve_ctx:7}, we know that
  $ \intermed {  {\intermed \sigma}_{{\mathrm{1}}}  } = \intermed {  \forall  {\overline {\intermed a} }_{\ottmv{k}}  \ottsym{.}   {\overline {\intermed Q} }_{\ottmv{h}}  \Rightarrow  {\intermed \sigma}'   } $. By rewriting Equation~\ref{eq:preserve_ctx:41}, we get
  \begin{align}
    \label{eq:preserve_ctx:goal:2}
     ( \intermed {  {\intermed m}  } : \intermed {  {\intermed {\mathit{TC} } } \, {\intermed a }  } : \intermed {  \forall  {\overline {\intermed a} }_{\ottmv{k}}  \ottsym{.}   {\overline {\intermed Q} }_{\ottmv{h}}  \Rightarrow  {\intermed \sigma}'   } )  \in  \intermed {  {\intermed { \Gamma_C } }  }  
  \end{align}
  By applying Theorem~\ref{thmA:typing_pres_expr} to the 4\textsuperscript{th} rule premise, we get:
  \begin{equation}
     \intermed {  {\intermed \Sigma}  } ; \intermed {  {\intermed { \Gamma_C } }  } ; \intermed {  \bullet  \ottsym{,}  {\overline {\intermed b} }_{\ottmv{j}}  \ottsym{,}  {\overline {\intermed \delta} }_{\ottmv{i}}  \ottsym{:}  {\overline {\intermed Q} }_{\ottmv{i}}  \ottsym{,}  {\overline {\intermed a} }_{\ottmv{k}}  \ottsym{,}  {\overline {\intermed \delta} }_{\ottmv{h}}  \ottsym{:}  [  {\intermed \sigma}  /  {\intermed a }  ]  {\overline {\intermed Q} }_{\ottmv{h}}  }  \vdash_{tm}  \intermed {  {\intermed e}  } : \intermed {  [  {\intermed \sigma}  /  {\intermed a }  ]  {\intermed \sigma}'  } \itot{ \rightsquigarrow  \target {  {\target e }  } }  
    \label{eq:preserve_ctx:8}
  \end{equation}
  Lemma~\ref{lemma:ctx_well_inter_typing}, applied to this result, gives us:
  \begin{equation}
     \vdash_{ctx}  \intermed {  {\intermed \Sigma}  } ; \intermed {  {\intermed { \Gamma_C } }  } ; \intermed {  \bullet  \ottsym{,}  {\overline {\intermed b} }_{\ottmv{j}}  \ottsym{,}  {\overline {\intermed \delta} }_{\ottmv{i}}  \ottsym{:}  {\overline {\intermed Q} }_{\ottmv{i}}  \ottsym{,}  {\overline {\intermed a} }_{\ottmv{k}}  \ottsym{,}  {\overline {\intermed \delta} }_{\ottmv{h}}  \ottsym{:}  [  {\intermed \sigma}  /  {\intermed a }  ]  {\overline {\intermed Q} }_{\ottmv{h}}  } 
    \label{eq:preserve_ctx:9}
  \end{equation}
  Furthermore, by applying \textsc{iTm-constrI} and \textsc{iTm-forallI} to
  Equation~\ref{eq:preserve_ctx:8}, in combination with inversion on
  Equation~\ref{eq:preserve_ctx:9}, we get
  \begin{equation*}
     \intermed {  {\intermed \Sigma}  } ; \intermed {  {\intermed { \Gamma_C } }  } ; \intermed {  \bullet  }  \vdash_{tm}  \intermed {  \Lambda  {\overline {\intermed b} }_{\ottmv{j}}  \ottsym{.}  \lambda  {\overline {\intermed \delta} }_{\ottmv{i}}  \ottsym{:}  {\overline {\intermed Q} }_{\ottmv{i}}  \ottsym{.}  \Lambda  {\overline {\intermed a} }_{\ottmv{k}}  \ottsym{.}  \lambda  {\overline {\intermed \delta} }_{\ottmv{h}}  \ottsym{:}  [  {\intermed \sigma}  /  {\intermed a }  ]  {\overline {\intermed Q} }_{\ottmv{h}}  \ottsym{.}  {\intermed e}  } : \intermed {  \forall  {\overline {\intermed b} }_{\ottmv{j}}  \ottsym{.}   {\overline {\intermed Q} }_{\ottmv{i}}  \Rightarrow  \forall  {\overline {\intermed a} }_{\ottmv{k}}  \ottsym{.}   [  {\intermed \sigma}  /  {\intermed a }  ]  {\overline {\intermed Q} }_{\ottmv{h}}  \Rightarrow  [  {\intermed \sigma}  /  {\intermed a }  ]  {\intermed \sigma}'    } \itot{ \rightsquigarrow  \target {  {\target e }'  } }  
  \end{equation*}
  which is equivalent to
  \begin{equation}
     \intermed {  {\intermed \Sigma}  } ; \intermed {  {\intermed { \Gamma_C } }  } ; \intermed {  \bullet  }  \vdash_{tm}  \intermed {  \Lambda  {\overline {\intermed b} }_{\ottmv{j}}  \ottsym{.}  \lambda  {\overline {\intermed \delta} }_{\ottmv{i}}  \ottsym{:}  {\overline {\intermed Q} }_{\ottmv{i}}  \ottsym{.}  \Lambda  {\overline {\intermed a} }_{\ottmv{k}}  \ottsym{.}  \lambda  {\overline {\intermed \delta} }_{\ottmv{h}}  \ottsym{:}  [  {\intermed \sigma}  /  {\intermed a }  ]  {\overline {\intermed Q} }_{\ottmv{h}}  \ottsym{.}  {\intermed e}  } : \intermed {  \forall  {\overline {\intermed b} }_{\ottmv{j}}  \ottsym{.}   {\overline {\intermed Q} }_{\ottmv{i}}  \Rightarrow  [  {\intermed \sigma}  /  {\intermed a }  ]  \ottsym{(}  \forall  {\overline {\intermed a} }_{\ottmv{k}}  \ottsym{.}   {\overline {\intermed Q} }_{\ottmv{h}}  \Rightarrow  {\intermed \sigma}'   \ottsym{)}   } \itot{ \rightsquigarrow  \target {  {\target e }'  } }  
    \label{eq:preserve_ctx:10}
  \end{equation}
  Given
  \begin{gather*}
     \textbf {unambig}  ( \source {  \forall  {\overline {\source b} }_{\ottmv{j}}  \ottsym{.}  {\overline {\source Q} }_{\ottmv{i}}  \Rightarrow  {\source {\mathit{TC} } } \, {\source \tau }  } )  \\
     \source {  {\source D }  }  \notin   \textbf {dom}  ( \source {  { \source P }  } )  \\
     ( \source {  {\source D }'  } : \source {  \forall  {\overline {\source b} }'_{\ottmv{k}}  \ottsym{.}  {\overline {\source Q} }''_{\ottmv{h}}  \Rightarrow  {\source {\mathit{TC} } } \, {\source \tau }''  } ) . \source {  {\source m}'  }  \mapsto  \source {  {\source \Gamma}'  } : \source {  {\source e}'  }  \notin  \source {  { \source P }  } \\
     \textbf {where}  \source {  [  {\overline {\source \tau} }_{\ottmv{j}}  /  {\overline {\source b} }_{\ottmv{j}}  ]  {\source \tau }  } = \source {  [  {\overline {\source \tau} }'_{\ottmv{k}}  /  {\overline {\source b} }'_{\ottmv{k}}  ]  {\source \tau }''  } 
  \end{gather*}
  It is easy to verify that 
  \begin{gather}
    \label{eq:preserve_ctx:goal:7}
     \textbf {unambig}  ( \intermed {  \forall  {\overline {\intermed b} }_{\ottmv{j}}  \ottsym{.}  {\overline {\intermed Q} }_{\ottmv{i}}  \Rightarrow  {\intermed {\mathit{TC} } } \, {\intermed \sigma}  } )  \\
    \label{eq:preserve_ctx:goal:4}
     \intermed {  {\intermed D }  }  \notin   \textbf {dom}  ( \intermed {  {\intermed \Sigma}  } )  \\
    \label{eq:preserve_ctx:goal:5}
     ( \intermed {  {\intermed D }'  } : \intermed {  \forall  {\overline {\intermed a} }'_{\ottmv{m}}  \ottsym{.}  {\overline {\intermed Q} }''_{\ottmv{n}}  \Rightarrow  {\intermed {\mathit{TC} } } \, {\intermed \sigma}''  } ) . \intermed {  {\intermed m}'  }  \mapsto  \intermed {  {\intermed e}'  }  \notin  \intermed {  {\intermed \Sigma}  }  \\
    \label{eq:preserve_ctx:goal:6}
     \textbf {where}  \intermed {  [  \overline {\intermed \sigma}_{\ottmv{j}}  /  {\overline {\intermed a} }_{\ottmv{j}}  ]  {\intermed \sigma}  } = \intermed {  [  \overline {\intermed \sigma}'_{\ottmv{m}}  /  {\overline {\intermed a} }'_{\ottmv{m}}  ]  {\intermed \sigma}''  }  
  \end{gather}
  The goal follows by
  combining Equations~\ref{eq:preserve_ctx:goal:0}, \ref{eq:preserve_ctx:goal:1},
  \ref{eq:preserve_ctx:goal:2}, \ref{eq:preserve_ctx:10},
  \ref{eq:preserve_ctx:goal:7}, 
  \ref{eq:preserve_ctx:goal:4}, \ref{eq:preserve_ctx:goal:5},
  \ref{eq:preserve_ctx:goal:6}, and the rule \textsc{iCtx-MEnv}.
}
\qed

\newpage
\section{\interL{} Theorems}

\subsection{Conjectures}
\label{app:mid_conj}

We are confident that the following lemmas can be proven using well-known proof
techniques. 

\begin{lemmaB}[Type Variable Substitution in Types]\ \\
  \renewcommand\itot[1]{#1}
  \label{lemma:tvar_tsubst}
  If \( \intermed {  {\intermed { \Gamma_C } }  } ; \intermed {  {\intermed \Gamma}_{{\mathrm{1}}}  \ottsym{,}  {\intermed a }  \ottsym{,}  {\intermed \Gamma}_{{\mathrm{2}}}  }  \vdash_{ty}  \intermed {  {\intermed \sigma}_{{\mathrm{1}}}  } \itot{ \rightsquigarrow  \target {  {\target \sigma }_{{\mathrm{1}}}  } } \)
  and \( \intermed {  {\intermed { \Gamma_C } }  } ; \intermed {  {\intermed \Gamma}_{{\mathrm{1}}}  }  \vdash_{ty}  \intermed {  {\intermed \sigma}_{{\mathrm{2}}}  } \itot{ \rightsquigarrow  \target {  {\target \sigma }_{{\mathrm{2}}}  } } \)
  then \( \intermed {  {\intermed { \Gamma_C } }  } ; \intermed {  {\intermed \Gamma}_{{\mathrm{1}}}  \ottsym{,}  [  {\intermed \sigma}_{{\mathrm{2}}}  /  {\intermed a }  ]  {\intermed \Gamma}_{{\mathrm{2}}}  }  \vdash_{ty}  \intermed {  [  {\intermed \sigma}_{{\mathrm{2}}}  /  {\intermed a }  ]  {\intermed \sigma}_{{\mathrm{1}}}  } \itot{ \rightsquigarrow  \target {  [  {\target \sigma }_{{\mathrm{2}}}  /  {\target a }  ]  {\target \sigma }_{{\mathrm{1}}}  } } \).
\end{lemmaB}

\begin{lemmaB}[Type Variable Substitution in Dictionaries]\ \\
  \renewcommand\itot[1]{#1}
  \label{lemma:tvar_tsubst_iq}
  If \( \intermed {  {\intermed { \Gamma_C } }  } ; \intermed {  {\intermed \Gamma}_{{\mathrm{1}}}  \ottsym{,}  {\intermed a }  \ottsym{,}  {\intermed \Gamma}_{{\mathrm{2}}}  }  \vdash_{\mathit {Q} }  \intermed {  {\intermed Q}  }\, \itot{ \rightsquigarrow  \target {  {\target \sigma }  } } \)
  and \( \intermed {  {\intermed { \Gamma_C } }  } ; \intermed {  {\intermed \Gamma}_{{\mathrm{1}}}  }  \vdash_{ty}  \intermed {  {\intermed \sigma}'  } \itot{ \rightsquigarrow  \target {  {\target \sigma }'  } } \)
  then \( \intermed {  {\intermed { \Gamma_C } }  } ; \intermed {  {\intermed \Gamma}_{{\mathrm{1}}}  \ottsym{,}  [  {\intermed \sigma}'  /  {\intermed a }  ]  {\intermed \Gamma}_{{\mathrm{2}}}  }  \vdash_{\mathit {Q} }  \intermed {  [  {\intermed \sigma}'  /  {\intermed a }  ]  {\intermed Q}  }\, \itot{ \rightsquigarrow  \target {  [  {\target \sigma }'  /  {\target a }  ]  {\target \sigma }  } } \).
\end{lemmaB}

\begin{lemmaB}[Variable Substitution]\ \\
  \renewcommand\itot[1]{#1}
  \label{lemma:var_subst}
  If \( \intermed {  {\intermed \Sigma}  } ; \intermed {  {\intermed { \Gamma_C } }  } ; \intermed {  {\intermed \Gamma}_{{\mathrm{1}}}  \ottsym{,}  {\intermed x}  \ottsym{:}  {\intermed \sigma}_{{\mathrm{2}}}  \ottsym{,}  {\intermed \Gamma}_{{\mathrm{2}}}  }  \vdash_{tm}  \intermed {  {\intermed e}_{{\mathrm{1}}}  } : \intermed {  {\intermed \sigma}_{{\mathrm{1}}}  } \itot{ \rightsquigarrow  \target {  {\target e }_{{\mathrm{1}}}  } } \)
  and \( \intermed {  {\intermed \Sigma}  } ; \intermed {  {\intermed { \Gamma_C } }  } ; \intermed {  {\intermed \Gamma}_{{\mathrm{1}}}  \ottsym{,}  {\intermed \Gamma}_{{\mathrm{2}}}  }  \vdash_{tm}  \intermed {  {\intermed e}_{{\mathrm{2}}}  } : \intermed {  {\intermed \sigma}_{{\mathrm{2}}}  } \itot{ \rightsquigarrow  \target {  {\target e }_{{\mathrm{2}}}  } } \)
  then \( \intermed {  {\intermed \Sigma}  } ; \intermed {  {\intermed { \Gamma_C } }  } ; \intermed {  {\intermed \Gamma}_{{\mathrm{1}}}  \ottsym{,}  {\intermed \Gamma}_{{\mathrm{2}}}  }  \vdash_{tm}  \intermed {  [  {\intermed e}_{{\mathrm{2}}}  /  {\intermed x}  ]  {\intermed e}_{{\mathrm{1}}}  } : \intermed {  {\intermed \sigma}_{{\mathrm{1}}}  } \itot{ \rightsquigarrow  \target {  [  {\target e }_{{\mathrm{2}}}  /  {\target x}  ]  {\target e }_{{\mathrm{1}}}  } } \).
\end{lemmaB}

\begin{lemmaB}[Reverse Variable Substitution]\ \\
  \renewcommand\itot[1]{}
  \label{lemma:rev_var_subst}
  If \( \intermed {  {\intermed \Sigma}  } ; \intermed {  {\intermed { \Gamma_C } }  } ; \intermed {  {\intermed \Gamma}  }  \vdash_{tm}  \intermed {  [  {\intermed e}_{{\mathrm{2}}}  /  {\intermed x}  ]  {\intermed e}_{{\mathrm{1}}}  } : \intermed {  {\intermed \sigma}_{{\mathrm{1}}}  } \itot{ \rightsquigarrow  \target {  {\target e }_{{\mathrm{3}}}  } } \)
  and \( \intermed {  {\intermed \Sigma}  } ; \intermed {  {\intermed { \Gamma_C } }  } ; \intermed {  {\intermed \Gamma}  }  \vdash_{tm}  \intermed {  {\intermed e}_{{\mathrm{2}}}  } : \intermed {  {\intermed \sigma}_{{\mathrm{2}}}  } \itot{ \rightsquigarrow  \target {  {\target e }_{{\mathrm{2}}}  } } \)
  then \( \intermed {  {\intermed \Sigma}  } ; \intermed {  {\intermed { \Gamma_C } }  } ; \intermed {  {\intermed \Gamma}  \ottsym{,}  {\intermed x}  \ottsym{:}  {\intermed \sigma}_{{\mathrm{2}}}  }  \vdash_{tm}  \intermed {  {\intermed e}_{{\mathrm{1}}}  } : \intermed {  {\intermed \sigma}_{{\mathrm{1}}}  } \itot{ \rightsquigarrow  \target {  {\target e }_{{\mathrm{1}}}  } } \).
\end{lemmaB}

\begin{lemmaB}[Dictionary Variable Substitution]\ \\
  \renewcommand\itot[1]{#1}
  \label{lemma:dvar_subst}
  If \( \intermed {  {\intermed \Sigma}  } ; \intermed {  {\intermed { \Gamma_C } }  } ; \intermed {  {\intermed \Gamma}_{{\mathrm{1}}}  \ottsym{,}  {\intermed \delta }  \ottsym{:}  {\intermed Q}  \ottsym{,}  {\intermed \Gamma}_{{\mathrm{2}}}  }  \vdash_{tm}  \intermed {  {\intermed e}  } : \intermed {  {\intermed \sigma}  } \itot{ \rightsquigarrow  \target {  {\target e }  } } \)
  and \( \intermed {  {\intermed \Sigma}  } ; \intermed {  {\intermed { \Gamma_C } }  } ; \intermed {  {\intermed \Gamma}_{{\mathrm{1}}}  \ottsym{,}  {\intermed \Gamma}_{{\mathrm{2}}}  }  \vdash_{d}  \intermed {  {\intermed d}  } : \intermed {  {\intermed Q}  } \itot{\,  \rightsquigarrow  \target {  {\target e }'  } } \)
  then \( \intermed {  {\intermed \Sigma}  } ; \intermed {  {\intermed { \Gamma_C } }  } ; \intermed {  {\intermed \Gamma}_{{\mathrm{1}}}  \ottsym{,}  {\intermed \Gamma}_{{\mathrm{2}}}  }  \vdash_{tm}  \intermed {  [  {\intermed d}  /  {\intermed \delta }  ]  {\intermed e}  } : \intermed {  {\intermed \sigma}  } \itot{ \rightsquigarrow  \target {  [  {\target e }'  /  {\target \delta }  ]  {\target e }  } } \).
\end{lemmaB}

\begin{lemmaB}[Reverse Dictionary Variable Substitution]\ \\
  \renewcommand\itot[1]{}
  \label{lemma:rev_dvar_subst}
  If \( \intermed {  {\intermed \Sigma}  } ; \intermed {  {\intermed { \Gamma_C } }  } ; \intermed {  {\intermed \Gamma}  }  \vdash_{tm}  \intermed {  [  {\intermed d}  /  {\intermed \delta }  ]  {\intermed e}  } : \intermed {  {\intermed \sigma}  } \itot{ \rightsquigarrow  \target {  {\target e }_{{\mathrm{1}}}  } } \)
  and \( \intermed {  {\intermed \Sigma}  } ; \intermed {  {\intermed { \Gamma_C } }  } ; \intermed {  {\intermed \Gamma}  }  \vdash_{d}  \intermed {  {\intermed d}  } : \intermed {  {\intermed Q}  } \itot{\,  \rightsquigarrow  \target {  {\target e }_{{\mathrm{2}}}  } } \)
  then \( \intermed {  {\intermed \Sigma}  } ; \intermed {  {\intermed { \Gamma_C } }  } ; \intermed {  {\intermed \Gamma}  \ottsym{,}  {\intermed \delta }  \ottsym{:}  {\intermed Q}  }  \vdash_{tm}  \intermed {  {\intermed e}  } : \intermed {  {\intermed \sigma}  } \itot{ \rightsquigarrow  \target {  {\target e }  } } \).
\end{lemmaB}

\begin{lemmaB}[Type Variable Substitution]\ \\
  \renewcommand\itot[1]{#1}
  \label{lemma:tvar_subst}
  If \( \intermed {  {\intermed \Sigma}  } ; \intermed {  {\intermed { \Gamma_C } }  } ; \intermed {  {\intermed \Gamma}_{{\mathrm{1}}}  \ottsym{,}  {\intermed a }  \ottsym{,}  {\intermed \Gamma}_{{\mathrm{2}}}  }  \vdash_{tm}  \intermed {  {\intermed e}  } : \intermed {  {\intermed \sigma}  } \itot{ \rightsquigarrow  \target {  {\target e }  } } \)
  and \( \intermed {  {\intermed { \Gamma_C } }  } ; \intermed {  {\intermed \Gamma}_{{\mathrm{1}}}  }  \vdash_{ty}  \intermed {  {\intermed \sigma}'  } \itot{ \rightsquigarrow  \target {  {\target \sigma }'  } } \)
  then \( \intermed {  {\intermed \Sigma}  } ; \intermed {  {\intermed { \Gamma_C } }  } ; \intermed {  {\intermed \Gamma}_{{\mathrm{1}}}  \ottsym{,}  [  {\intermed \sigma}'  /  {\intermed a }  ]  {\intermed \Gamma}_{{\mathrm{2}}}  }  \vdash_{tm}  \intermed {  [  {\intermed \sigma}'  /  {\intermed a }  ]  {\intermed e}  } : \intermed {  [  {\intermed \sigma}'  /  {\intermed a }  ]  {\intermed \sigma}  } \itot{ \rightsquigarrow  \target {  [  {\target \sigma }'  /  {\target a }  ]  {\target e }  } } \).
\end{lemmaB}

\begin{lemmaB}[Reverse Type Variable Substitution]\ \\
  \renewcommand\itot[1]{}
  \label{lemma:rev_tvar_subst}
  If \( \intermed {  {\intermed \Sigma}  } ; \intermed {  {\intermed { \Gamma_C } }  } ; \intermed {  {\intermed \Gamma}  }  \vdash_{tm}  \intermed {  [  {\intermed \sigma}'  /  {\intermed a }  ]  {\intermed e}  } : \intermed {  [  {\intermed \sigma}'  /  {\intermed a }  ]  {\intermed \sigma}  } \itot{ \rightsquigarrow  \target {  [  {\target \sigma }'  /  {\target a }  ]  {\target e }  } } \)
  then \( \intermed {  {\intermed \Sigma}  } ; \intermed {  {\intermed { \Gamma_C } }  } ; \intermed {  {\intermed \Gamma}  \ottsym{,}  {\intermed a }  }  \vdash_{tm}  \intermed {  {\intermed e}  } : \intermed {  {\intermed \sigma}  } \itot{ \rightsquigarrow  \target {  {\target e }  } } \).
\end{lemmaB}

\begin{lemmaB}[Dictionary Variable Substitution in Dictionaries]\ \\
  If \( \intermed {  {\intermed \Sigma}  } ; \intermed {  {\intermed { \Gamma_C } }  } ; \intermed {  {\intermed \Gamma}_{{\mathrm{1}}}  \ottsym{,}  {\intermed \delta }  \ottsym{:}  {\intermed Q}'  \ottsym{,}  {\intermed \Gamma}_{{\mathrm{2}}}  }  \vdash_{d}  \intermed {  {\intermed D } \, \overline {\intermed \sigma}_{\ottmv{j}} \, {\overline {\intermed d} }_{\ottmv{i}}  } : \intermed {  {\intermed Q}  } \itot{\,  \rightsquigarrow  \target {  {\target e }  } } \)
  and \( \intermed {  {\intermed \Sigma}  } ; \intermed {  {\intermed { \Gamma_C } }  } ; \intermed {  {\intermed \Gamma}_{{\mathrm{1}}}  \ottsym{,}  {\intermed \Gamma}_{{\mathrm{2}}}  }  \vdash_{d}  \intermed {  {\intermed d}  } : \intermed {  {\intermed Q}'  } \itot{\,  \rightsquigarrow  \target {  {\target e }'  } } \)
  then \( \intermed {  {\intermed \Sigma}  } ; \intermed {  {\intermed { \Gamma_C } }  } ; \intermed {  {\intermed \Gamma}_{{\mathrm{1}}}  \ottsym{,}  {\intermed \Gamma}_{{\mathrm{2}}}  }  \vdash_{d}  \intermed {  {\intermed D } \, \overline {\intermed \sigma}_{\ottmv{j}} \, [  {\intermed d}  /  {\intermed \delta }  ]  {\overline {\intermed d} }_{\ottmv{i}}  } : \intermed {  {\intermed Q}  } \itot{\,  \rightsquigarrow  \target {  [  {\target e }'  /  {\target \delta }  ]  {\target e }  } } \).
  \label{lemma:dvar_subst_dict}
\end{lemmaB}

\begin{lemmaB}[Type Variable Substitution in Dictionaries]\ \\
  If \( \intermed {  {\intermed \Sigma}  } ; \intermed {  {\intermed { \Gamma_C } }  } ; \intermed {  {\intermed \Gamma}_{{\mathrm{1}}}  \ottsym{,}  {\intermed a }  \ottsym{,}  {\intermed \Gamma}_{{\mathrm{2}}}  }  \vdash_{d}  \intermed {  {\intermed D } \, \overline {\intermed \sigma}_{\ottmv{j}} \, {\overline {\intermed d} }_{\ottmv{i}}  } : \intermed {  {\intermed Q}  } \itot{\,  \rightsquigarrow  \target {  {\target e }  } } \)
  and \( \intermed {  {\intermed { \Gamma_C } }  } ; \intermed {  {\intermed \Gamma}_{{\mathrm{1}}}  }  \vdash_{ty}  \intermed {  {\intermed \sigma}  } \itot{ \rightsquigarrow  \target {  {\target \sigma }  } } \)
  then \( \intermed {  {\intermed \Sigma}  } ; \intermed {  {\intermed { \Gamma_C } }  } ; \intermed {  {\intermed \Gamma}_{{\mathrm{1}}}  \ottsym{,}  [  {\intermed \sigma}  /  {\intermed a }  ]  {\intermed \Gamma}_{{\mathrm{2}}}  }  \vdash_{d}  \intermed {  {\intermed D } \, [  {\intermed \sigma}  /  {\intermed a }  ]  \overline {\intermed \sigma}_{\ottmv{j}} \, [  {\intermed \sigma}  /  {\intermed a }  ]  {\overline {\intermed d} }_{\ottmv{i}}  } : \intermed {  [  {\intermed \sigma}  /  {\intermed a }  ]  {\intermed Q}  } \itot{\,  \rightsquigarrow  \target {  [  {\target \sigma }  /  {\target a }  ]  {\target e }  } } \).
  \label{lemma:tvar_subst_dict}
\end{lemmaB}

\begin{lemmaB}[Expression Well-Typed Method Environment Weakening]\ \\
  \renewcommand\itot[1]{}
  \label{lemma:exp_weakening_method}
  If \( \intermed {  {\intermed \Sigma}_{{\mathrm{1}}}  } ; \intermed {  {\intermed { \Gamma_C } }  } ; \intermed {  {\intermed \Gamma}  }  \vdash_{tm}  \intermed {  {\intermed e}  } : \intermed {  {\intermed \sigma}  } \itot{ \rightsquigarrow  \target {  {\target e }  } } \)
  and \( \vdash_{ctx}  \intermed {  {\intermed \Sigma}_{{\mathrm{1}}}  \ottsym{,}  {\intermed \Sigma}_{{\mathrm{2}}}  } ; \intermed {  {\intermed { \Gamma_C } }  } ; \intermed {  {\intermed \Gamma}  } \)
  then \( \intermed {  {\intermed \Sigma}_{{\mathrm{1}}}  \ottsym{,}  {\intermed \Sigma}_{{\mathrm{2}}}  } ; \intermed {  {\intermed { \Gamma_C } }  } ; \intermed {  {\intermed \Gamma}  }  \vdash_{tm}  \intermed {  {\intermed e}  } : \intermed {  {\intermed \sigma}  } \itot{ \rightsquigarrow  \target {  {\target e }  } } \).
\end{lemmaB}

\begin{lemmaB}[Type Well-Formedness Dictionary Environment Weakening]\ \\
  \renewcommand\itot[1]{#1}
  \label{lemma:ty_weakening}
  If \( \vdash_{ctx}  \intermed {  {\intermed \Sigma}  } ; \intermed {  {\intermed { \Gamma_C } }_{{\mathrm{1}}}  } ; \intermed {  {\intermed \Gamma}_{{\mathrm{1}}}  } \)
  and \( \intermed {  {\intermed { \Gamma_C } }_{{\mathrm{1}}}  } ; \intermed {  {\intermed \Gamma}_{{\mathrm{1}}}  }  \vdash_{ty}  \intermed {  {\intermed \sigma}  } \itot{ \rightsquigarrow  \target {  {\target \sigma }  } } \)
  and \( \vdash_{ctx}  \intermed {  {\intermed \Sigma}  } ; \intermed {  {\intermed { \Gamma_C } }_{{\mathrm{1}}}  \ottsym{,}  {\intermed { \Gamma_C } }_{{\mathrm{2}}}  } ; \intermed {  {\intermed \Gamma}_{{\mathrm{1}}}  \ottsym{,}  {\intermed \Gamma}_{{\mathrm{2}}}  } \),
  then \( \intermed {  {\intermed { \Gamma_C } }_{{\mathrm{1}}}  \ottsym{,}  {\intermed { \Gamma_C } }_{{\mathrm{2}}}  } ; \intermed {  {\intermed \Gamma}_{{\mathrm{1}}}  \ottsym{,}  {\intermed \Gamma}_{{\mathrm{2}}}  }  \vdash_{ty}  \intermed {  {\intermed \sigma}  } \itot{ \rightsquigarrow  \target {  {\target \sigma }  } } \).
\end{lemmaB}

\begin{lemmaB}[Class Constraint Well-Formedness Environment Weakening]\ \\
  \renewcommand\itot[1]{#1}
  \label{lemma:q_weakening}
  If \( \intermed {  {\intermed { \Gamma_C } }_{{\mathrm{1}}}  } ; \intermed {  {\intermed \Gamma}_{{\mathrm{1}}}  }  \vdash_{\mathit {Q} }  \intermed {  {\intermed Q}  }\, \itot{ \rightsquigarrow  \target {  {\target \sigma }  } } \)
  and \( \vdash_{ctx}  \intermed {  {\intermed \Sigma}  } ; \intermed {  {\intermed { \Gamma_C } }_{{\mathrm{1}}}  \ottsym{,}  {\intermed { \Gamma_C } }_{{\mathrm{2}}}  } ; \intermed {  {\intermed \Gamma}_{{\mathrm{1}}}  \ottsym{,}  {\intermed \Gamma}_{{\mathrm{2}}}  } \)
  then \( \intermed {  {\intermed { \Gamma_C } }_{{\mathrm{1}}}  \ottsym{,}  {\intermed { \Gamma_C } }_{{\mathrm{2}}}  } ; \intermed {  {\intermed \Gamma}_{{\mathrm{1}}}  \ottsym{,}  {\intermed \Gamma}_{{\mathrm{2}}}  }  \vdash_{\mathit {Q} }  \intermed {  {\intermed Q}  }\, \itot{ \rightsquigarrow  \target {  {\target \sigma }  } } \).
\end{lemmaB}

\begin{lemmaB}[Logical Equivalence Environment Weakening]\ \\
  \renewcommand\itot[1]{}
  \label{lemma:log_weakening}
  If \( \intermed {  {\intermed { \Gamma_C } }_{{\mathrm{1}}}  } ; \intermed {  {\intermed \Gamma}_{{\mathrm{1}}}  }  \vdash  \intermed {  {\intermed \Sigma}_{{\mathrm{1}}}  } : \intermed {  {\intermed e}_{{\mathrm{1}}}  }  \simeq_{log}  \intermed {  {\intermed \Sigma}_{{\mathrm{2}}}  } : \intermed {  {\intermed e}_{{\mathrm{2}}}  } : \intermed {  {\intermed \sigma}  } \)
  and \( \vdash_{ctx}  \intermed {  {\intermed \Sigma}_{{\mathrm{1}}}  \ottsym{,}  {\intermed \Sigma}'_{{\mathrm{1}}}  } ; \intermed {  {\intermed { \Gamma_C } }_{{\mathrm{1}}}  \ottsym{,}  {\intermed { \Gamma_C } }_{{\mathrm{2}}}  } ; \intermed {  {\intermed \Gamma}_{{\mathrm{1}}}  \ottsym{,}  {\intermed \Gamma}_{{\mathrm{2}}}  } \)
  and \( \vdash_{ctx}  \intermed {  {\intermed \Sigma}_{{\mathrm{2}}}  \ottsym{,}  {\intermed \Sigma}'_{{\mathrm{2}}}  } ; \intermed {  {\intermed { \Gamma_C } }_{{\mathrm{1}}}  \ottsym{,}  {\intermed { \Gamma_C } }_{{\mathrm{2}}}  } ; \intermed {  {\intermed \Gamma}_{{\mathrm{1}}}  \ottsym{,}  {\intermed \Gamma}_{{\mathrm{2}}}  } \), \\
  then \( \intermed {  {\intermed { \Gamma_C } }_{{\mathrm{1}}}  \ottsym{,}  {\intermed { \Gamma_C } }_{{\mathrm{2}}}  } ; \intermed {  {\intermed \Gamma}_{{\mathrm{1}}}  \ottsym{,}  {\intermed \Gamma}_{{\mathrm{2}}}  }  \vdash  \intermed {  {\intermed \Sigma}_{{\mathrm{1}}}  \ottsym{,}  {\intermed \Sigma}'_{{\mathrm{1}}}  } : \intermed {  {\intermed e}_{{\mathrm{1}}}  }  \simeq_{log}  \intermed {  {\intermed \Sigma}_{{\mathrm{2}}}  \ottsym{,}  {\intermed \Sigma}'_{{\mathrm{2}}}  } : \intermed {  {\intermed e}_{{\mathrm{2}}}  } : \intermed {  {\intermed \sigma}  } \).
\end{lemmaB}

\begin{lemmaB}[Strong Normalization Relation Method Environment Weakening]\ \\
  \renewcommand\itot[1]{}
  \label{lemma:sn_menv_weakening}
  If \( \intermed {  {\intermed e}  } \in \mathcal{SN} \llbracket \intermed {  {\intermed \sigma}  } \rrbracket ^{\intermed {  {\intermed \Sigma}_{{\mathrm{1}}}  } , \intermed {  {\intermed { \Gamma_C } }  } }_{\intermed {  {\intermed { R^{SN} } }  } } \)
  and \( \vdash_{ctx}  \intermed {  {\intermed \Sigma}_{{\mathrm{1}}}  \ottsym{,}  {\intermed \Sigma}_{{\mathrm{2}}}  } ; \intermed {  {\intermed { \Gamma_C } }  } ; \intermed {  {\intermed \Gamma}  } \)
  then \( \intermed {  {\intermed e}  } \in \mathcal{SN} \llbracket \intermed {  {\intermed \sigma}  } \rrbracket ^{\intermed {  {\intermed \Sigma}_{{\mathrm{1}}}  \ottsym{,}  {\intermed \Sigma}_{{\mathrm{2}}}  } , \intermed {  {\intermed { \Gamma_C } }  } }_{\intermed {  {\intermed { R^{SN} } }  } } \).
\end{lemmaB}

\begin{lemmaB}[Dictionary Value Relation Preserved under Substitution]\ \\
  \renewcommand\itot[1]{}
  \label{lemma:dict_Rsubst}
  \( ( \intermed {  {\intermed \Sigma}_{{\mathrm{1}}}  } : \intermed {  {\intermed {dv} }_{{\mathrm{1}}}  } , \intermed {  {\intermed \Sigma}_{{\mathrm{2}}}  } : \intermed {  {\intermed {dv} }_{{\mathrm{2}}}  } ) \in \mathcal{V} \llbracket \intermed {  {\intermed Q}  } \rrbracket ^{\intermed {  {\intermed { \Gamma_C } }  } }_{\intermed {  {\intermed R }_{{\mathrm{1}}}  \ottsym{,}  {\intermed R }_{{\mathrm{2}}}  } } \)
  if and only if
  \( ( \intermed {  {\intermed \Sigma}_{{\mathrm{1}}}  } : \intermed {  {\intermed {dv} }_{{\mathrm{1}}}  } , \intermed {  {\intermed \Sigma}_{{\mathrm{2}}}  } : \intermed {  {\intermed {dv} }_{{\mathrm{2}}}  } ) \in \mathcal{V} \llbracket \intermed {  {\intermed R }_{{\mathrm{2}}}  \ottsym{(}  {\intermed Q}  \ottsym{)}  } \rrbracket ^{\intermed {  {\intermed { \Gamma_C } }  } }_{\intermed {  {\intermed R }_{{\mathrm{1}}}  } } \).
\end{lemmaB}

\subsection{Lemmas}

\begin{lemmaB}[Environment Well-Formedness Strengthening]\ \\
  \label{lemma:env_strengthening_mid}
  If \( \vdash_{ctx}  \intermed {  {\intermed \Sigma}  } ; \intermed {  {\intermed { \Gamma_C } }  } ; \intermed {  {\intermed \Gamma}  } \)
  then \( \vdash_{ctx}  \intermed {  {\intermed \Sigma}  } ; \intermed {  {\intermed { \Gamma_C } }  } ; \intermed {  \bullet  } \).
\end{lemmaB}

\proof
By case analysis on the hypothesis,
the last rules used to construct it must be (possibly zero)
consecutive applications of \textsc{iCtx-MEnv}. Revert those rules, to obtain
\( \vdash_{ctx}  \intermed {  \bullet  } ; \intermed {  {\intermed { \Gamma_C } }  } ; \intermed {  {\intermed \Gamma}  } \).
By further case analysis (with \textsc{iCtx-tyEnvTm}, \textsc{iCtx-tyEnvTy} and
\textsc{iCtx-tyEnvD}), we get \( \vdash_{ctx}  \intermed {  \bullet  } ; \intermed {  {\intermed { \Gamma_C } }  } ; \intermed {  \bullet  } \).
The goal follows by consecutively
re-applying rule \textsc{iCtx-MEnv} with the appropriate premises. \\
\qed

\begin{lemmaB}[Variable Strengthening in Dictionaries]\ \\
  If \( \intermed {  {\intermed \Sigma}  } ; \intermed {  {\intermed { \Gamma_C } }  } ; \intermed {  {\intermed \Gamma}_{{\mathrm{1}}}  \ottsym{,}  {\intermed x}  \ottsym{:}  {\intermed \sigma}  \ottsym{,}  {\intermed \Gamma}_{{\mathrm{2}}}  }  \vdash_{d}  \intermed {  {\intermed D } \, \overline {\intermed \sigma}_{\ottmv{j}} \, {\overline {\intermed d} }_{\ottmv{i}}  } : \intermed {  {\intermed Q}  } \itot{\,  \rightsquigarrow  \target {  {\target e }  } } \)
  then \( \intermed {  {\intermed \Sigma}  } ; \intermed {  {\intermed { \Gamma_C } }  } ; \intermed {  {\intermed \Gamma}_{{\mathrm{1}}}  \ottsym{,}  {\intermed \Gamma}_{{\mathrm{2}}}  }  \vdash_{d}  \intermed {  {\intermed D } \, \overline {\intermed \sigma}_{\ottmv{j}} \, {\overline {\intermed d} }_{\ottmv{i}}  } : \intermed {  {\intermed Q}  } \itot{\,  \rightsquigarrow  \target {  {\target e }  } } \).
  \label{lemma:var_subst_dict}
\end{lemmaB}

\proof
By straightforward induction on the well-formedness derivation. \\
\qed

\begin{lemmaB}[Variable Strengthening in Types]\ \\
  If\, \( \intermed {  {\intermed { \Gamma_C } }  } ; \intermed {  {\intermed \Gamma}_{{\mathrm{1}}}  \ottsym{,}  {\intermed x}  \ottsym{:}  {\intermed \sigma}  \ottsym{,}  {\intermed \Gamma}_{{\mathrm{2}}}  }  \vdash_{ty}  \intermed {  {\intermed \sigma}'  } \itot{ \rightsquigarrow  \target {  {\target \sigma }'  } } \)
  then\, \( \intermed {  {\intermed { \Gamma_C } }  } ; \intermed {  {\intermed \Gamma}_{{\mathrm{1}}}  \ottsym{,}  {\intermed \Gamma}_{{\mathrm{2}}}  }  \vdash_{ty}  \intermed {  {\intermed \sigma}'  } \itot{ \rightsquigarrow  \target {  {\target \sigma }'  } } \).
  \label{lemma:var_subst_types}
\end{lemmaB}

\proof
By straightforward induction on the well-formedness derivation. \\
\qed

\begin{lemmaB}[Method Type Well-Formedness]\ \\
  \renewcommand\itot[1]{#1}
  \label{lemma:classmethodwty}
  If \( \vdash_{ctx}  \intermed {  {\intermed \Sigma}  } ; \intermed {  {\intermed { \Gamma_C } }  } ; \intermed {  {\intermed \Gamma}  } \)
  and \( \intermed {  {\intermed { \Gamma_C } }  } = \intermed {  {\intermed { \Gamma_C } }_{{\mathrm{1}}}  }, \intermed {  {\intermed m}  \ottsym{:}  {\intermed {\mathit{TC} } } \, {\intermed a }  \ottsym{:}  {\intermed \sigma}  \ottsym{,}  {\intermed { \Gamma_C } }_{{\mathrm{2}}}  } \)
  then there is a \({\target \sigma }\) such,
  that \( \intermed {  {\intermed { \Gamma_C } }_{{\mathrm{1}}}  } ; \intermed {  \bullet  \ottsym{,}  {\intermed a }  }  \vdash_{ty}  \intermed {  {\intermed \sigma}  } \itot{ \rightsquigarrow  \target {  {\target \sigma }  } } \).
\end{lemmaB}

\proof
By straightforward induction on the environment well-formedness derivation. \\
\qed

\begin{lemmaB}[Method Environment Well-Formedness]\ \\
  \renewcommand\itot[1]{#1}
  \label{lemma:inSS2C}
  If \( \vdash_{ctx}  \intermed {  {\intermed \Sigma}  } ; \intermed {  {\intermed { \Gamma_C } }  } ; \intermed {  {\intermed \Gamma}  } \)
  and \( ( \intermed {  {\intermed D }  } : \intermed {  \forall  {\overline {\intermed a} }_{\ottmv{j}}  \ottsym{.}  {\overline {\intermed Q} }_{\ottmv{i}}  \Rightarrow  {\intermed {\mathit{TC} } } \, {\intermed \sigma}  } ) . \intermed {  {\intermed m}  }  \mapsto  \intermed {  {\intermed e}  }  \in  \intermed {  {\intermed \Sigma}  } \), where $i = 1 \ldots n$,
  then there are unique ${\intermed a }$, $\intermed{{\intermed \sigma}_{\ottmv{m}}}$, $\target{{\target \sigma }_{\ottmv{m}}}$ and \(\target{{\overline {\target \sigma} }_{\ottmv{i}}}\), such
  that \( ( \intermed {  {\intermed m}  } : \intermed {  {\intermed {\mathit{TC} } } \, {\intermed a }  } : \intermed {  {\intermed \sigma}_{\ottmv{m}}  } )  \in  \intermed {  {\intermed { \Gamma_C } }  } \)
  and \( \intermed {  {\intermed { \Gamma_C } }  } ; \intermed {  \bullet  \ottsym{,}  {\intermed a }  }  \vdash_{ty}  \intermed {  {\intermed \sigma}_{\ottmv{m}}  } \itot{ \rightsquigarrow  \target {  {\target \sigma }_{\ottmv{m}}  } } \)
  and \( \intermed {  {\intermed \Sigma}  } ; \intermed {  {\intermed { \Gamma_C } }  } ; \intermed {  \bullet  }  \vdash_{tm}  \intermed {  {\intermed e}  } : \intermed {  \forall  {\overline {\intermed a} }_{\ottmv{j}}  \ottsym{.}   {\overline {\intermed Q} }_{\ottmv{i}}  \Rightarrow  [  {\intermed \sigma}  /  {\intermed a }  ]  {\intermed \sigma}_{\ottmv{m}}   } \itot{ \rightsquigarrow  \target {  {\target e }  } } \)
  and \(\ottcomp{ \intermed {  {\intermed { \Gamma_C } }  } ; \intermed {  \bullet  \ottsym{,}  {\overline {\intermed a} }_{\ottmv{j}}  }  \vdash_{\mathit {Q} }  \intermed {  {\intermed Q}_{\ottmv{i}}  }\, \itot{ \rightsquigarrow  \target {  {\target \sigma }_{\ottmv{i}}  } } }{\ottmv{i}}\).
  and \( \intermed {  {\intermed { \Gamma_C } }  } ; \intermed {  \bullet  \ottsym{,}  {\overline {\intermed a} }_{\ottmv{j}}  }  \vdash_{\mathit {Q} }  \intermed {  {\intermed {\mathit{TC} } } \, {\intermed \sigma}  }\, \itot{ \rightsquigarrow  \target {  [  {\target \sigma }  /  {\target a }  ]  \ottsym{\{}  {\target m}  \ottsym{:}  {\target \sigma }_{\ottmv{m}}  \ottsym{\}}  } } \)
  and \( \intermed {  {\intermed { \Gamma_C } }  } ; \intermed {  \bullet  \ottsym{,}  {\overline {\intermed a} }_{\ottmv{j}}  }  \vdash_{ty}  \intermed {  {\intermed \sigma}  } \itot{ \rightsquigarrow  \target {  {\target \sigma }  } } \).
\end{lemmaB}

\proof
By straightforward induction on the environment well-formedness derivation. \\
\qed

\begin{lemmaB}[Determinism of Evaluation]\ \\
  \renewcommand\itot[1]{}
  \label{lemma:eval_unique}
  If \( \intermed {  {\intermed \Sigma}  }  \vdash  \intermed {  {\intermed e}  }  \longrightarrow  \intermed {  {\intermed e}_{{\mathrm{1}}}  } \)
  and \( \intermed {  {\intermed \Sigma}  }  \vdash  \intermed {  {\intermed e}  }  \longrightarrow  \intermed {  {\intermed e}_{{\mathrm{2}}}  } \)
  then \( \intermed {  {\intermed e}_{{\mathrm{1}}}  } = \intermed {  {\intermed e}_{{\mathrm{2}}}  } \).
\end{lemmaB}

\proof
By straightforward induction on the evaluation derivation. \\
\qed

\begin{lemmaB}[Preservation of Environment Type Variables from \interL{} to \targetL{}]\leavevmode
  \begin{itemize}
  \item If \( \intermed {  {\intermed a }  }  \in  \intermed {  {\intermed \Gamma}  } \)
    and \( \intermed {  {\intermed { \Gamma_C } }  } ; \intermed {  {\intermed \Gamma}  }  \rightsquigarrow  \target {  {\target \Gamma }  } \)
    then \( \target {  {\target a }  }  \in  \target {  {\target \Gamma }  } \).
  \item If \( \intermed {  {\intermed a }  }  \notin  \intermed {  {\intermed \Gamma}  } \)
    and \( \intermed {  {\intermed { \Gamma_C } }  } ; \intermed {  {\intermed \Gamma}  }  \rightsquigarrow  \target {  {\target \Gamma }  } \)
    then \( \target {  {\target a }  }  \notin  \target {  {\target \Gamma }  } \).
  \end{itemize}
  \label{lemma:pres_env_tyT}
\end{lemmaB}

\proof
By straightforward induction on the environment elaboration derivation. \\
\qed

\begin{lemmaB}[Well-Formedness of \interL{} Typing Result]\ \\
  \renewcommand\itot[1]{{#1}}
  \label{lemma:type_well_inter_typing}
  If \( \intermed {  {\intermed \Sigma}  } ; \intermed {  {\intermed { \Gamma_C } }  } ; \intermed {  {\intermed \Gamma}  }  \vdash_{tm}  \intermed {  {\intermed e}  } : \intermed {  {\intermed \sigma}  } \itot{ \rightsquigarrow  \target {  {\target e }  } } \)
  then \( \intermed {  {\intermed { \Gamma_C } }  } ; \intermed {  {\intermed \Gamma}  }  \vdash_{ty}  \intermed {  {\intermed \sigma}  } \itot{ \rightsquigarrow  \target {  {\target \sigma }  } } \).
\end{lemmaB}

\proof
By straightforward induction on the typing derivation. \\
\qed

\begin{lemmaB}[Context Well-Formedness of \interL{} Typing]\ \\
  \renewcommand\itot[1]{}
  \label{lemma:ctx_well_inter_typing}
  If \( \intermed {  {\intermed \Sigma}  } ; \intermed {  {\intermed { \Gamma_C } }  } ; \intermed {  {\intermed \Gamma}  }  \vdash_{tm}  \intermed {  {\intermed e}  } : \intermed {  {\intermed \sigma}  } \itot{ \rightsquigarrow  \target {  {\target e }  } } \)
  then \( \vdash_{ctx}  \intermed {  {\intermed \Sigma}  } ; \intermed {  {\intermed { \Gamma_C } }  } ; \intermed {  {\intermed \Gamma}  } \).
\end{lemmaB}

\proof
By straightforward induction on the typing derivation. \\
\qed

\begin{lemmaB}[Context Well-Formedness of Dictionary Typing]\ \\
  \renewcommand\itot[1]{}
  \label{lemma:ctx_well_dict_typing}
  If \( \intermed {  {\intermed \Sigma}  } ; \intermed {  {\intermed { \Gamma_C } }  } ; \intermed {  {\intermed \Gamma}  }  \vdash_{d}  \intermed {  {\intermed d}  } : \intermed {  {\intermed Q}  } \itot{\,  \rightsquigarrow  \target {  {\target e }  } } \)
  then \( \vdash_{ctx}  \intermed {  {\intermed \Sigma}  } ; \intermed {  {\intermed { \Gamma_C } }  } ; \intermed {  {\intermed \Gamma}  } \).
\end{lemmaB}

\proof
By straightforward induction on the dictionary typing derivation. \\
\qed

\subsection{Type Safety}

\begin{theoremB}[Preservation]\ \\
  \label{thmA:intermediate_preservation}
  If $ \intermed {  {\intermed \Sigma}  } ; \intermed {  {\intermed { \Gamma_C } }  } ; \intermed {  {\intermed \Gamma}  }  \vdash_{tm}  \intermed {  {\intermed e}  } : \intermed {  {\intermed \sigma}  } \itot{ \rightsquigarrow  \target {  {\target e }  } } $,
  and $ \intermed {  {\intermed \Sigma}  }  \vdash  \intermed {  {\intermed e}  }  \longrightarrow  \intermed {  {\intermed e}'  } $,
  then $ \intermed {  {\intermed \Sigma}  } ; \intermed {  {\intermed { \Gamma_C } }  } ; \intermed {  {\intermed \Gamma}  }  \vdash_{tm}  \intermed {  {\intermed e}'  } : \intermed {  {\intermed \sigma}  } \itot{ \rightsquigarrow  \target {  {\target e }'  } } $.
\end{theoremB}

\proof
By induction on the typing derivation. \\
\proofStep{\textsc{iTm-true}}
{\drule{iTm-true}}
{ $\mathit{\intermed {True} }$ is already a value, so impossible case. \\
}
\proofStep{\textsc{iTm-false}}
{\drule{iTm-false}}
{ $\mathit{\intermed {False} }$ is already a value, so impossible case. \\
}
\proofStep{\textsc{iTm-var}}
{\drule{iTm-var}}
{ ${\intermed x}$ cannot be reduced, so impossible case. \\
}
\proofStep{\textsc{iTm-let}}
{\drule{iTm-let}}
{ By inversion on the evaluation (\textsc{iEval-let}), we get that
  \begin{equation*}
     \intermed {  {\intermed e}'  } = \intermed {  [  {\intermed e}_{{\mathrm{1}}}  /  {\intermed x}  ]  {\intermed e}_{{\mathrm{2}}}  } 
  \end{equation*}
  Given
  \begin{gather*}
     \intermed {  {\intermed \Sigma}  } ; \intermed {  {\intermed { \Gamma_C } }  } ; \intermed {  {\intermed \Gamma}  }  \vdash_{tm}  \intermed {  {\intermed e}_{{\mathrm{1}}}  } : \intermed {  {\intermed \sigma}_{{\mathrm{1}}}  } \itot{ \rightsquigarrow  \target {  {\target e }_{{\mathrm{1}}}  } } \\
     \intermed {  {\intermed \Sigma}  } ; \intermed {  {\intermed { \Gamma_C } }  } ; \intermed {  {\intermed \Gamma}  \ottsym{,}  {\intermed x}  \ottsym{:}  {\intermed \sigma}_{{\mathrm{1}}}  }  \vdash_{tm}  \intermed {  {\intermed e}_{{\mathrm{2}}}  } : \intermed {  {\intermed \sigma}_{{\mathrm{2}}}  } \itot{ \rightsquigarrow  \target {  {\target e }_{{\mathrm{2}}}  } } 
  \end{gather*}
  The goal follows directly from Lemma~\ref{lemma:var_subst}. \\
}
\proofStep{\textsc{iTm-method}}
{\drule{iTm-method}}
{ By inversion on the evaluation (\textsc{iEval-method}), we get that 
  \begin{gather}
    \label{eq:pres:method:1}
     \intermed {  {\intermed e}'  } = \intermed {  {\intermed e} \, \overline {\intermed \sigma}_{\ottmv{j}} \, {\overline {\intermed d} }_{\ottmv{i}}  }  \\
    \label{eq:pres:method:2}
     \intermed {  {\intermed d}  } = \intermed {  {\intermed D } \, \overline {\intermed \sigma}_{\ottmv{j}} \, {\overline {\intermed d} }_{\ottmv{i}}  }  
  \end{gather}
  The goal to be proven thus becomes:
  \begin{gather}
    \label{eq:pres:method:4}
     \intermed {  {\intermed \Sigma}  } ; \intermed {  {\intermed { \Gamma_C } }  } ; \intermed {  \bullet  }  \vdash_{tm}  \intermed {  {\intermed e} \, \overline {\intermed \sigma}_{\ottmv{j}} \, {\overline {\intermed d} }_{\ottmv{i}}  } : \intermed {  [  {\intermed \sigma}  /  {\intermed a }  ]  {\intermed \sigma}'  } \itot{ \rightsquigarrow  \target {  {\target e }'  } } 
  \end{gather}
  By substituting Equation~\ref{eq:pres:method:2} in the 1\textsuperscript{st} rule premise, we get that
  \begin{gather}
    \label{eq:pres:method:3}
     \intermed {  {\intermed \Sigma}  } ; \intermed {  {\intermed { \Gamma_C } }  } ; \intermed {  {\intermed \Gamma}  }  \vdash_{d}  \intermed {  {\intermed D } \, \overline {\intermed \sigma}_{\ottmv{j}} \, {\overline {\intermed d} }_{\ottmv{i}}  } : \intermed {  {\intermed {\mathit{TC} } } \, {\intermed \sigma}  } \itot{\,  \rightsquigarrow  \target {  {\target e }  } } 
  \end{gather}
  By inversion on Equation~\ref{eq:pres:method:3} (\textsc{D-con}), we know that
  \begin{gather}
    \label{eq:pres:method:5}
     ( \intermed {  {\intermed D }  } : \intermed {  \forall  {\overline {\intermed a} }_{\ottmv{j}}  \ottsym{.}  {\overline {\intermed Q} }_{\ottmv{i}}  \Rightarrow  {\intermed {\mathit{TC} } } \, {\intermed \sigma}_{{\mathrm{1}}}  } ) . \intermed {  {\intermed m}  }  \mapsto  \intermed {  \Lambda  {\overline {\intermed a} }_{\ottmv{j}}  \ottsym{.}  \lambda  {\overline {\intermed \delta} }_{\ottmv{i}}  \ottsym{:}  {\overline {\intermed Q} }_{\ottmv{i}}  \ottsym{.}  {\intermed e}  }  \in  \intermed {  {\intermed \Sigma}  } \\
    \label{eq:pres:method:6}
     \intermed {  {\intermed \sigma}  } = \intermed {  [  \overline {\intermed \sigma}_{\ottmv{j}}  /  {\overline {\intermed a} }_{\ottmv{j}}  ]  {\intermed \sigma}_{{\mathrm{1}}}  }  \\
    \label{eq:pres:method:7}
    \ottcomp{ \intermed {  {\intermed \Sigma}  } ; \intermed {  {\intermed { \Gamma_C } }  } ; \intermed {  {\intermed \Gamma}  }  \vdash_{d}  \intermed {  {\intermed d}_{\ottmv{i}}  } : \intermed {  [  \overline {\intermed \sigma}_{\ottmv{j}}  /  {\overline {\intermed a} }_{\ottmv{j}}  ]  {\intermed Q}_{\ottmv{i}}  } \itot{\,  \rightsquigarrow  \target {  {\target e }_{\ottmv{i}}  } } }{\ottmv{i}}\\
    \label{eq:pres:method:11}
    \ottcomp{ \intermed {  {\intermed { \Gamma_C } }  } ; \intermed {  {\intermed \Gamma}  }  \vdash_{ty}  \intermed {  {\intermed \sigma}_{\ottmv{j}}  } \itot{ \rightsquigarrow  \target {  {\target \sigma }_{\ottmv{j}}  } } }{\ottmv{j}} \\
    \label{eq:pres:method:12}
     \vdash_{ctx}  \intermed {  {\intermed \Sigma}  } ; \intermed {  {\intermed { \Gamma_C } }  } ; \intermed {  {\intermed \Gamma}  } 
  \end{gather}
  By combining this result with Equation~\ref{eq:pres:method:5}, through inversion
  (\textsc{iCtx-MEnv}), we know that
  \begin{gather}
    \label{eq:pres:method:8}
     \intermed {  {\intermed \Sigma}_{{\mathrm{1}}}  } ; \intermed {  {\intermed { \Gamma_C } }  } ; \intermed {  \bullet  }  \vdash_{tm}  \intermed {  \Lambda  {\overline {\intermed a} }_{\ottmv{j}}  \ottsym{.}  \lambda  {\overline {\intermed \delta} }_{\ottmv{i}}  \ottsym{:}  {\overline {\intermed Q} }_{\ottmv{i}}  \ottsym{.}  {\intermed e}  } : \intermed {  \forall  {\overline {\intermed a} }_{\ottmv{j}}  \ottsym{.}   {\overline {\intermed Q} }_{\ottmv{i}}  \Rightarrow  [  {\intermed \sigma}_{{\mathrm{1}}}  /  {\intermed a }  ]  {\intermed \sigma}'   } \itot{ \rightsquigarrow  \target {  {\target e }_{{\mathrm{1}}}  } } 
  \end{gather}
  where \( \intermed {  {\intermed \Sigma}  } = \intermed {  {\intermed \Sigma}_{{\mathrm{1}}}  \ottsym{,}  \ottsym{(}  {\intermed D }  \ottsym{:}  \forall  {\overline {\intermed a} }_{\ottmv{j}}  \ottsym{.}  {\overline {\intermed Q} }_{\ottmv{i}}  \Rightarrow  {\intermed {\mathit{TC} } } \, {\intermed \sigma}_{{\mathrm{1}}}  \ottsym{)}  \ottsym{.}  {\intermed m}  \mapsto  \Lambda  {\overline {\intermed a} }_{\ottmv{j}}  \ottsym{.}  \lambda  {\overline {\intermed \delta} }_{\ottmv{i}}  \ottsym{:}  {\overline {\intermed Q} }_{\ottmv{i}}  \ottsym{.}  {\intermed e}  \ottsym{,}  {\intermed \Sigma}_{{\mathrm{2}}}  } \).
  By \textsc{iTm-forallE} and Equations~\ref{eq:pres:method:8} and
  \ref{eq:pres:method:11}, we have
  \begin{gather}
    \label{eq:pres:method:9}
     \intermed {  {\intermed \Sigma}_{{\mathrm{1}}}  } ; \intermed {  {\intermed { \Gamma_C } }  } ; \intermed {  \bullet  }  \vdash_{tm}  \intermed {  {\intermed e} \, \overline {\intermed \sigma}_{\ottmv{j}}  } : \intermed {  [  \overline {\intermed \sigma}_{\ottmv{j}}  /  {\overline {\intermed a} }_{\ottmv{j}}  ]   {\overline {\intermed Q} }_{\ottmv{i}}  \Rightarrow  [  \overline {\intermed \sigma}_{\ottmv{j}}  /  {\overline {\intermed a} }_{\ottmv{j}}  ]  [  {\intermed \sigma}_{{\mathrm{1}}}  /  {\intermed a }  ]  {\intermed \sigma}'   } \itot{ \rightsquigarrow  \target {  {\target e }_{{\mathrm{1}}} \, {\overline {\target \sigma} }_{\ottmv{j}}  } } 
  \end{gather}
  By \textsc{iTm-constrE} and Equations~\ref{eq:pres:method:9} and \ref{eq:pres:method:7}, we have
  \begin{gather}
    \label{eq:pres:method:10}
     \intermed {  {\intermed \Sigma}_{{\mathrm{1}}}  } ; \intermed {  {\intermed { \Gamma_C } }  } ; \intermed {  \bullet  }  \vdash_{tm}  \intermed {  {\intermed e} \, \overline {\intermed \sigma}_{\ottmv{j}} \, {\overline {\intermed d} }_{\ottmv{i}}  } : \intermed {  [  \overline {\intermed \sigma}_{\ottmv{j}}  /  {\overline {\intermed a} }_{\ottmv{j}}  ]  [  {\intermed \sigma}_{{\mathrm{1}}}  /  {\intermed a }  ]  {\intermed \sigma}'  } \itot{ \rightsquigarrow  \target {  {\target e }_{{\mathrm{1}}} \, {\overline {\target \sigma} }_{\ottmv{j}} \, {\overline {\target e} }_{\ottmv{i}}  } } 
  \end{gather}
  From the 2\textsuperscript{nd} rule premise we know that
  \begin{gather}
     ( \intermed {  {\intermed m}  } : \intermed {  {\intermed {\mathit{TC} } } \, {\intermed a }  } : \intermed {  {\intermed \sigma}'  } )  \in  \intermed {  {\intermed { \Gamma_C } }  } 
  \end{gather}
  Combining this result with Equation~\ref{eq:pres:method:12}, by inversion
  (\textsc{iCtx-clsEnv}), we know
  \begin{gather}
     \intermed {  {\intermed { \Gamma_C } }_{{\mathrm{1}}}  } ; \intermed {  \bullet  \ottsym{,}  {\intermed a }  }  \vdash_{ty}  \intermed {  {\intermed \sigma}'  } \itot{ \rightsquigarrow  \target {  {\target \sigma }'  } }  \\
     \intermed {  {\intermed { \Gamma_C } }  } = \intermed {  {\intermed { \Gamma_C } }_{{\mathrm{1}}}  \ottsym{,}  {\intermed { \Gamma_C } }_{{\mathrm{2}}}  } 
  \end{gather}
  Therefore, $\intermed{\overline {\intermed \sigma}_{\ottmv{j}}}$ are not free variables in $\intermed{{\intermed \sigma}'}$.
  Equation~\ref{eq:pres:method:9} thus simplifies to
  \begin{gather}
    \label{eq:pres:method:13}
     \intermed {  {\intermed \Sigma}_{{\mathrm{1}}}  } ; \intermed {  {\intermed { \Gamma_C } }  } ; \intermed {  \bullet  }  \vdash_{tm}  \intermed {  {\intermed e} \, \overline {\intermed \sigma}_{\ottmv{j}} \, {\overline {\intermed d} }_{\ottmv{i}}  } : \intermed {  [  [  \overline {\intermed \sigma}_{\ottmv{j}}  /  {\overline {\intermed a} }_{\ottmv{j}}  ]  {\intermed \sigma}_{{\mathrm{1}}}  /  {\intermed a }  ]  {\intermed \sigma}'  } \itot{ \rightsquigarrow  \target {  {\target e }_{{\mathrm{1}}} \, {\overline {\target \sigma} }_{\ottmv{j}} \, {\overline {\target e} }_{\ottmv{i}}  } } 
  \end{gather}
  By applying Equation~\ref{eq:pres:method:13} to
  Lemma~\ref{lemma:exp_weakening_method}, in combination with
  Equation~\ref{eq:pres:method:12}, we get
  \begin{gather*}
     \intermed {  {\intermed \Sigma}  } ; \intermed {  {\intermed { \Gamma_C } }  } ; \intermed {  \bullet  }  \vdash_{tm}  \intermed {  {\intermed e} \, \overline {\intermed \sigma}_{\ottmv{j}} \, {\overline {\intermed d} }_{\ottmv{i}}  } : \intermed {  [  [  \overline {\intermed \sigma}_{\ottmv{j}}  /  {\overline {\intermed a} }_{\ottmv{j}}  ]  {\intermed \sigma}_{{\mathrm{1}}}  /  {\intermed a }  ]  {\intermed \sigma}'  } \itot{ \rightsquigarrow  \target {  {\target e }_{{\mathrm{1}}} \, {\overline {\target \sigma} }_{\ottmv{j}} \, {\overline {\target e} }_{\ottmv{i}}  } } 
  \end{gather*}
  Goal~\ref{eq:pres:method:4} follows by combining this result with
  Equation~\ref{eq:pres:method:6}. \\
}
\proofStep{\textsc{iTm-arrI}}
{\drule{iTm-arrI}}
{$\intermed{\lambda  {\intermed x}  \ottsym{:}  {\intermed \sigma}_{{\mathrm{1}}}  \ottsym{.}  {\intermed e}}$ is already a value, so impossible case. \\}
\proofStep{\textsc{iTm-arrE}}
{\drule{iTm-arrE}}
{
  By inversion on the evaluation, we have two possible cases:
    \begin{itemize}
    \item Case \textsc{iEval-app}.
      \drule{iEval-app}

      By induction hypothesis, we get
      \begin{gather*}
         \intermed {  {\intermed \Sigma}  } ; \intermed {  {\intermed { \Gamma_C } }  } ; \intermed {  {\intermed \Gamma}  }  \vdash_{tm}  \intermed {  {\intermed e}'_{{\mathrm{1}}}  } : \intermed {  {\intermed \sigma}_{{\mathrm{1}}}  \rightarrow  {\intermed \sigma}_{{\mathrm{2}}}  } \itot{ \rightsquigarrow  \target {  {\target e }'  } } 
      \end{gather*}
      By \textsc{iTm-arrE} we get
      \begin{gather*}
         \intermed {  {\intermed \Sigma}  } ; \intermed {  {\intermed { \Gamma_C } }  } ; \intermed {  {\intermed \Gamma}  }  \vdash_{tm}  \intermed {  {\intermed e}'_{{\mathrm{1}}} \, {\intermed e}_{{\mathrm{2}}}  } : \intermed {  {\intermed \sigma}_{{\mathrm{2}}}  } \itot{ \rightsquigarrow  \target {  {\target e }_{{\mathrm{1}}} \, {\target e }_{{\mathrm{2}}}  } } 
      \end{gather*}
    \item Case \textsc{iEval-appAbs}.
      $\ottdruleiEvalXXappAbs{}$

      From premise, we know
      \begin{gather*}
         \intermed {  {\intermed \Sigma}  } ; \intermed {  {\intermed { \Gamma_C } }  } ; \intermed {  {\intermed \Gamma}  }  \vdash_{tm}  \intermed {  \lambda  {\intermed x}  \ottsym{:}  {\intermed \sigma}  \ottsym{.}  {\intermed e}_{{\mathrm{1}}}  } : \intermed {  {\intermed \sigma}  \rightarrow  {\intermed \sigma}_{{\mathrm{2}}}  } \itot{ \rightsquigarrow  \target {  {\target e }_{{\mathrm{1}}}  } } 
      \end{gather*}
      By inversion (\textsc{iTm-arrI})
      \begin{gather*}
         \intermed {  {\intermed \Sigma}  } ; \intermed {  {\intermed { \Gamma_C } }  } ; \intermed {  {\intermed \Gamma}  \ottsym{,}  {\intermed x}  \ottsym{:}  {\intermed \sigma}  }  \vdash_{tm}  \intermed {  {\intermed e}_{{\mathrm{1}}}  } : \intermed {  {\intermed \sigma}_{{\mathrm{2}}}  } \itot{ \rightsquigarrow  \target {  {\target e }_{{\mathrm{1}}}  } } 
      \end{gather*}
      By substitution (Lemma~\ref{lemma:var_subst}) we get
      \begin{gather*}
         \intermed {  {\intermed \Sigma}  } ; \intermed {  {\intermed { \Gamma_C } }  } ; \intermed {  {\intermed \Gamma}  }  \vdash_{tm}  \intermed {  [  {\intermed e}_{{\mathrm{2}}}  /  {\intermed x}  ]  {\intermed e}_{{\mathrm{1}}}  } : \intermed {  {\intermed \sigma}_{{\mathrm{2}}}  } \itot{ \rightsquigarrow  \target {  [  {\target e }_{{\mathrm{2}}}  /  {\target x}  ]  {\target e }_{{\mathrm{1}}}  } } 
      \end{gather*}
    \end{itemize}
}
\proofStep{\textsc{iTm-constrI}}
{\drule{iTm-constrI}}
{$\intermed{\lambda  {\intermed \delta }  \ottsym{:}  {\intermed Q}  \ottsym{.}  {\intermed e}}$ is already a value, so impossible case. \\}
\proofStep{\textsc{iTm-constrE}}
{\drule{iTm-constrE}}
{Similar to case \textsc{iTm-arrE}. The only difference lies in applying
  Lemma~\ref{lemma:dvar_subst}. \\}
\proofStep{\textsc{iTm-forallI}}
{\drule{iTm-constrI}}
{$\intermed{\Lambda  {\intermed a }  \ottsym{.}  {\intermed e}}$ is already a value, so impossible case. \\}
\proofStep{\textsc{iTm-forallE}}
{\drule{iTm-forallE}}
{Similar to case \textsc{iTm-arrE}. The only difference lies in applying Lemma~\ref{lemma:tvar_subst}.}

\qed

\begin{theoremB}[Progress]\ \\
  If $ \intermed {  {\intermed \Sigma}  } ; \intermed {  \bullet  } ; \intermed {  \bullet  }  \vdash_{tm}  \intermed {  {\intermed e}  } : \intermed {  {\intermed \sigma}  } \itot{ \rightsquigarrow  \target {  {\target e }  } } $,
  then either ${\intermed e}$ is a value,
  or there exists $\intermed{{\intermed e}'}$ such that $ \intermed {  {\intermed \Sigma}  }  \vdash  \intermed {  {\intermed e}  }  \longrightarrow  \intermed {  {\intermed e}'  } $.
\label{thmA:progress_expressions}
\end{theoremB}

\proof By structural induction on the typing derivation. \\
\proofStep{\textsc{iTm-true}}
{ \drule{iTm-true}}
{
  with $ \intermed {  {\intermed { \Gamma_C } }  } = \intermed {  \bullet  } $, $ \intermed {  {\intermed \Gamma}  } = \intermed {  \bullet  } $.

  $\mathit{\intermed {True} }$ is a value. \\
}
\proofStep{\textsc{iTm-false}}
  { \drule{iTm-false} }
  { 
     with $ \intermed {  {\intermed { \Gamma_C } }  } = \intermed {  \bullet  } $, $ \intermed {  {\intermed \Gamma}  } = \intermed {  \bullet  } $.

     $\mathit{\intermed {False} }$ is a value. \\
  }
\proofStep{\textsc{iTm-var}}
  { \drule{iTm-var} }
  {
     with $ \intermed {  {\intermed { \Gamma_C } }  } = \intermed {  \bullet  } $, $ \intermed {  {\intermed \Gamma}  } = \intermed {  \bullet  } $.

    ${\intermed x}$ cannot be in an empty context. Impossible case. \\
  }
\proofStep{\textsc{iTm-let}}
  { \drule{iTm-let} }
  {
    with $ \intermed {  {\intermed { \Gamma_C } }  } = \intermed {  \bullet  } $, $ \intermed {  {\intermed \Gamma}  } = \intermed {  \bullet  } $.

    By \textsc{iEval-let}:
    \begin{equation*}
       \intermed {  {\intermed e}'  } = \intermed {  [  {\intermed e}_{{\mathrm{1}}}  /  {\intermed x}  ]  {\intermed e}_{{\mathrm{2}}}  } 
    \end{equation*}
  }
\proofStep{\textsc{iTm-method}}
  { \drule{iTm-method} }
  {
    with $ \intermed {  {\intermed { \Gamma_C } }  } = \intermed {  \bullet  } $, $ \intermed {  {\intermed \Gamma}  } = \intermed {  \bullet  } $.

    ${\intermed m}$ cannot be in an empty context. Impossible case. \\
  }
\proofStep{\textsc{iTm-arrI}}
  { \drule{iTm-arrI} }
  {
    with $ \intermed {  {\intermed { \Gamma_C } }  } = \intermed {  \bullet  } $, $ \intermed {  {\intermed \Gamma}  } = \intermed {  \bullet  } $.

    $\intermed{ \lambda  {\intermed x}  \ottsym{:}  {\intermed \sigma}_{{\mathrm{1}}}  \ottsym{.}  {\intermed e} }$ is a value. \\
  }
\proofStep{\textsc{iTm-arrE}}
  {\drule{iTm-arrE} }
  {

    with $ \intermed {  {\intermed { \Gamma_C } }  } = \intermed {  \bullet  } $, $ \intermed {  {\intermed \Gamma}  } = \intermed {  \bullet  } $.

    From the 1\textsuperscript{st} rule premise:
    \begin{equation}
       \intermed {  {\intermed \Sigma}  } ; \intermed {  {\intermed { \Gamma_C } }  } ; \intermed {  {\intermed \Gamma}  }  \vdash_{tm}  \intermed {  {\intermed e}_{{\mathrm{1}}}  } : \intermed {  {\intermed \sigma}_{{\mathrm{1}}}  \rightarrow  {\intermed \sigma}_{{\mathrm{2}}}  } \itot{ \rightsquigarrow  \target {  {\target e }_{{\mathrm{1}}}  } } 
      \label{proofeq:progress_app1}
    \end{equation}
    By applying the induction hypothesis on Equation~\ref{proofeq:progress_app1},
    we know that either:
    \begin{itemize}
    \item \( \intermed{ {\intermed e}_{{\mathrm{1}}} } \) is a value. Because it has an arrow type, we know:
      \begin{equation*}
         \intermed {  {\intermed e}_{{\mathrm{1}}}  } = \intermed {  \ottsym{(}  \lambda  {\intermed x}  \ottsym{:}  {\intermed \sigma}  \ottsym{.}  {\intermed e}_{{\mathrm{1}}}  \ottsym{)} \, {\intermed e}_{{\mathrm{2}}}  } 
      \end{equation*}
      By \textsc{iEval-appAbs}:
      \begin{equation*}
         \intermed {  {\intermed e}'  } = \intermed {  [  {\intermed e}_{{\mathrm{2}}}  /  {\intermed x}  ]  {\intermed e}_{{\mathrm{1}}}  } 
      \end{equation*}
    \item There exists an $ \intermed { {\intermed e}'_{{\mathrm{1}}} } $ where \(  \intermed {  {\intermed \Sigma}  }  \vdash  \intermed {  {\intermed e}_{{\mathrm{1}}}  }  \longrightarrow  \intermed {  {\intermed e}'_{{\mathrm{1}}}  }  \).
      By \textsc{iEval-app}:
      \begin{equation*}
         \intermed {  {\intermed e}'  } = \intermed {  {\intermed e}'_{{\mathrm{1}}} \, {\intermed e}_{{\mathrm{2}}}  } 
      \end{equation*}
    \end{itemize}
  }
\proofStep{\textsc{iTm-constrI}}
  {\drule{iTm-constrI}}
  {
    with $ \intermed {  {\intermed { \Gamma_C } }  } = \intermed {  \bullet  } $, $ \intermed {  {\intermed \Gamma}  } = \intermed {  \bullet  } $.

    \( \intermed{ \lambda  {\intermed \delta }  \ottsym{:}  {\intermed Q}  \ottsym{.}  {\intermed e} } \) is a value. \\
  }
\proofStep{\textsc{iTm-constrE}}
  {\drule{iTm-constrE}}
  {
    with $ \intermed {  {\intermed { \Gamma_C } }  } = \intermed {  \bullet  } $, $ \intermed {  {\intermed \Gamma}  } = \intermed {  \bullet  } $.

    From the 1\textsuperscript{st} rule premise:
    \begin{equation}
       \intermed {  {\intermed \Sigma}  } ; \intermed {  {\intermed { \Gamma_C } }  } ; \intermed {  {\intermed \Gamma}  }  \vdash_{tm}  \intermed {  {\intermed e}  } : \intermed {   {\intermed Q}  \Rightarrow  {\intermed \sigma}   } \itot{ \rightsquigarrow  \target {  {\target e }_{{\mathrm{1}}}  } } 
      \label{proofeq:progress_dapp1}
    \end{equation}
    By applying the induction hypothesis on
    Equation~\ref{proofeq:progress_dapp1}, we know that either:
    \begin{itemize}
    \item \( \intermed{ {\intermed e}_{{\mathrm{1}}} } \) is a value. By case analysis, we know:
      \begin{equation*}
         \intermed {  {\intermed e}_{{\mathrm{1}}}  } = \intermed {  \ottsym{(}  \lambda  {\intermed \delta }  \ottsym{:}  {\intermed Q}  \ottsym{.}  {\intermed e}  \ottsym{)}  } 
      \end{equation*}
      By \textsc{iEval-DAppAbs}:
      \begin{equation*}
         \intermed {  {\intermed e}'  } = \intermed {  [  {\intermed d}  /  {\intermed \delta }  ]  {\intermed e}  } 
      \end{equation*}
    \item There exists an \( \intermed { {\intermed e}' } \) where $ \intermed {  {\intermed \Sigma}  }  \vdash  \intermed {  {\intermed e}  }  \longrightarrow  \intermed {  {\intermed e}'  } $
      By \textsc{iEval-DApp}:
      \begin{equation*}
         \intermed {  {\intermed \Sigma}  }  \vdash  \intermed {  {\intermed e} \, {\intermed d}  }  \longrightarrow  \intermed {  {\intermed e}' \, {\intermed d}  } 
      \end{equation*}
    \end{itemize}
  }
\proofStep{\textsc{iTm-forallI}}
  {\drule{iTm-forallI}}
  {
    with $ \intermed {  {\intermed { \Gamma_C } }  } = \intermed {  \bullet  } $, $ \intermed {  {\intermed \Gamma}  } = \intermed {  \bullet  } $.

    $ \intermed { \Lambda  {\intermed a }  \ottsym{.}  {\intermed e} } $ is a value. \\
  }
\proofStep{\textsc{iTm-forallE}}
  {\drule{iTm-forallE}}
  {
    with $ \intermed {  {\intermed { \Gamma_C } }  } = \intermed {  \bullet  } $, $ \intermed {  {\intermed \Gamma}  } = \intermed {  \bullet  } $.

    From the 1\textsuperscript{st} rule premise:
    \begin{equation}
       \intermed {  {\intermed \Sigma}  } ; \intermed {  {\intermed { \Gamma_C } }  } ; \intermed {  {\intermed \Gamma}  }  \vdash_{tm}  \intermed {  {\intermed e}  } : \intermed {  \forall  {\intermed a }  \ottsym{.}  {\intermed \sigma}'  } \itot{ \rightsquigarrow  \target {  {\target e }  } } 
      \label{proofeq:progress_tyapp1}
    \end{equation}
    By applying the induction hypothesis on
    Equation~\ref{proofeq:progress_tyapp1}, we know that either:
    \begin{itemize}
    \item \( \intermed { {\intermed e}_{{\mathrm{1}}} } \) is a value. By case analysis, we know:
      \begin{equation*}
         \intermed {  {\intermed e}_{{\mathrm{1}}}  } = \intermed {  \ottsym{(}  \Lambda  {\intermed a }  \ottsym{.}  {\intermed e}  \ottsym{)}  } 
      \end{equation*}
      By \textsc{iEval-tyAppAbs}:
      \begin{equation*}
         \intermed {  {\intermed e}'  } = \intermed {  [  {\intermed \sigma}  /  {\intermed a }  ]  {\intermed e}  } 
      \end{equation*}
    \item There exists an \( \intermed{ {\intermed e}'_{{\mathrm{1}}} } \) where \(  \intermed {  {\intermed \Sigma}  }  \vdash  \intermed {  {\intermed e}  }  \longrightarrow  \intermed {  {\intermed e}'  }  \).
      By \textsc{iEval-tyApp}:
      \begin{equation*}
         \intermed {  {\intermed e}'  } = \intermed {  {\intermed e}'_{{\mathrm{1}}} \, {\intermed \sigma}  } 
      \end{equation*}
    \end{itemize}
  }
\qed

\subsection{Strong Normalization}
\label{app:sn}
\renewcommand\itot[1]{}

\begin{theoremB}[Strong Normalization]\ \\
  If \( \intermed {  {\intermed \Sigma}  } ; \intermed {  {\intermed { \Gamma_C } }  } ; \intermed {  \bullet  }  \vdash_{tm}  \intermed {  {\intermed e}  } : \intermed {  {\intermed \sigma}  } \itot{ \rightsquigarrow  \target {  {\target e }  } } \)
  then all possible evaluation derivations for \({\intermed e}\) terminate :
  \( \exists {\intermed v} :  \intermed {  {\intermed \Sigma}  }  \vdash  \intermed {  {\intermed e}  }  \longrightarrow^{*}  \intermed {  {\intermed v}  } \).
  \label{thmA:strong_norm}
\end{theoremB}
\proof By Theorem~\ref{thmA:sn:part:a} and \ref{thmA:sn:part:b}, with ${\intermed { R^{SN} } } =
\intermed{\bullet}$, $ \intermed {  {\intermed { \phi^{SN} } }  } = \intermed {  \bullet  } $, $ \intermed {  {\intermed { \gamma^{SN} } }  } = \intermed {  \bullet  } $, since $ \intermed {  {\intermed \Gamma}  } = \intermed {  \bullet  } $,
it follows that:
\begin{equation*}
  \exists {\intermed v} :  \intermed {  {\intermed \Sigma}  }  \vdash  \intermed {  {\intermed e}  }  \longrightarrow^{*}  \intermed {  {\intermed v}  } 
\end{equation*}
Furthermore, since evaluation in \interL{} is deterministic
(Lemma~\ref{lemma:eval_unique}), there is exactly 1
possible evaluation derivation. Consequently, all derivations terminate. \\
\qed


\begin{lemmaB}[Well Typedness from Strong Normalization]\ \\
  \label{lemma:sn_typing}
  $ \intermed {  {\intermed e}  } \in \mathcal{SN} \llbracket \intermed {  {\intermed \sigma}  } \rrbracket ^{\intermed {  {\intermed \Sigma}  } , \intermed {  {\intermed { \Gamma_C } }  } }_{\intermed {  {\intermed { R^{SN} } }  } } $,
  then $ \intermed {  {\intermed \Sigma}  } ; \intermed {  {\intermed { \Gamma_C } }  } ; \intermed {  \bullet  }  \vdash_{tm}  \intermed {  {\intermed e}  } : \intermed {  {\intermed { R^{SN} } }  \ottsym{(}  {\intermed \sigma}  \ottsym{)}  } \itot{ \rightsquigarrow  \target {  {\target e }  } } $
\end{lemmaB}
\proof The goal is baked into the relation. It follows by simple induction on
${\intermed \sigma}$. \\
\qed

\begin{lemmaB}[Strong Normalization preserved by forward/backward reduction]\ \\
  \label{lemma:sn_pres}
  Suppose $ \intermed {  {\intermed \Sigma}  } ; \intermed {  {\intermed { \Gamma_C } }  } ; \intermed {  \bullet  }  \vdash_{tm}  \intermed {  {\intermed e}_{{\mathrm{1}}}  } : \intermed {  {\intermed { R^{SN} } }  \ottsym{(}  {\intermed \sigma}  \ottsym{)}  } \itot{ \rightsquigarrow  \target {  {\target e }_{{\mathrm{1}}}  } } $,
  and $ \intermed {  {\intermed \Sigma}  }  \vdash  \intermed {  {\intermed e}_{{\mathrm{1}}}  }  \longrightarrow  \intermed {  {\intermed e}_{{\mathrm{2}}}  } $, then
  \begin{itemize}
    \item If $ \intermed {  {\intermed e}_{{\mathrm{1}}}  } \in \mathcal{SN} \llbracket \intermed {  {\intermed \sigma}  } \rrbracket ^{\intermed {  {\intermed \Sigma}  } , \intermed {  {\intermed { \Gamma_C } }  } }_{\intermed {  {\intermed { R^{SN} } }  } } $,
      then $ \intermed {  {\intermed e}_{{\mathrm{2}}}  } \in \mathcal{SN} \llbracket \intermed {  {\intermed \sigma}  } \rrbracket ^{\intermed {  {\intermed \Sigma}  } , \intermed {  {\intermed { \Gamma_C } }  } }_{\intermed {  {\intermed { R^{SN} } }  } } $.
    \item If $ \intermed {  {\intermed e}_{{\mathrm{2}}}  } \in \mathcal{SN} \llbracket \intermed {  {\intermed \sigma}  } \rrbracket ^{\intermed {  {\intermed \Sigma}  } , \intermed {  {\intermed { \Gamma_C } }  } }_{\intermed {  {\intermed { R^{SN} } }  } } $,
      then $ \intermed {  {\intermed e}_{{\mathrm{1}}}  } \in \mathcal{SN} \llbracket \intermed {  {\intermed \sigma}  } \rrbracket ^{\intermed {  {\intermed \Sigma}  } , \intermed {  {\intermed { \Gamma_C } }  } }_{\intermed {  {\intermed { R^{SN} } }  } } $.
  \end{itemize}
\end{lemmaB}
\proof

\begin{description}
  \item[Part 1]
    By induction on ${\intermed \sigma}$. \\
    \proofStep{Bool}
    {
      \begin{aligned}
     \intermed {  {\intermed e}_{{\mathrm{1}}}  } \in \mathcal{SN} \llbracket \intermed {  \mathit{\intermed {Bool} }  } \rrbracket ^{\intermed {  {\intermed \Sigma}  } , \intermed {  {\intermed { \Gamma_C } }  } }_{\intermed {  {\intermed { R^{SN} } }  } } 
    & \triangleq  \intermed {  {\intermed \Sigma}  } ; \intermed {  {\intermed { \Gamma_C } }  } ; \intermed {  \bullet  }  \vdash_{tm}  \intermed {  {\intermed e}_{{\mathrm{1}}}  } : \intermed {  \mathit{\intermed {Bool} }  } \itot{ \rightsquigarrow  \target {  {\target e }  } }  \\
    & \land \exists {\intermed v} :  \intermed {  {\intermed \Sigma}  }  \vdash  \intermed {  {\intermed e}_{{\mathrm{1}}}  }  \longrightarrow^{*}  \intermed {  {\intermed v}  }  \\
      \end{aligned}
    }
    {
      By Preservation (Theorem~\ref{thmA:intermediate_preservation}), we know
      that $ \intermed {  {\intermed \Sigma}  } ; \intermed {  {\intermed { \Gamma_C } }  } ; \intermed {  \bullet  }  \vdash_{tm}  \intermed {  {\intermed e}_{{\mathrm{2}}}  } : \intermed {  \mathit{\intermed {Bool} }  } \itot{ \rightsquigarrow  \target {  {\target e }  } } $.
      Because the evaluation in \interL{} is deterministic (Lemma~\ref{lemma:eval_unique}), given $ \intermed {  {\intermed \Sigma}  }  \vdash  \intermed {  {\intermed e}_{{\mathrm{1}}}  }  \longrightarrow^{*}  \intermed {  {\intermed v}  } $,
      we have $ \intermed {  {\intermed \Sigma}  }  \vdash  \intermed {  {\intermed e}_{{\mathrm{2}}}  }  \longrightarrow^{*}  \intermed {  {\intermed v}  } $. \\
    }
    \proofStep{Type variable}
    {
      \begin{aligned}
     \intermed {  {\intermed e}_{{\mathrm{1}}}  } \in \mathcal{SN} \llbracket \intermed {  {\intermed a }  } \rrbracket ^{\intermed {  {\intermed \Sigma}  } , \intermed {  {\intermed { \Gamma_C } }  } }_{\intermed {  {\intermed { R^{SN} } }  } } 
    & \triangleq  \intermed {  {\intermed \Sigma}  } ; \intermed {  {\intermed { \Gamma_C } }  } ; \intermed {  \bullet  }  \vdash_{tm}  \intermed {  {\intermed e}_{{\mathrm{1}}}  } : \intermed {  {\intermed { R^{SN} } }_{{\mathrm{1}}}  \ottsym{(}  {\intermed a }  \ottsym{)}  } \itot{ \rightsquigarrow  \target {  {\target e }  } }  \\
    & \land \exists {\intermed v} :  \intermed {  {\intermed \Sigma}  }  \vdash  \intermed {  {\intermed e}_{{\mathrm{1}}}  }  \longrightarrow^{*}  \intermed {  {\intermed v}  }  \\
    & \land  \intermed {  {\intermed v}  }  \in  \intermed {  {\intermed { R^{SN} } }_{{\mathrm{2}}}  } ( \intermed {  {\intermed a }  } )  \\
      \end{aligned}
    }
    {
      Similar to Bool case. \\}
    \proofStep{Function}
    {
      \begin{aligned}
     \intermed {  {\intermed e}_{{\mathrm{1}}}  } \in \mathcal{SN} \llbracket \intermed {  {\intermed \sigma}_{{\mathrm{1}}}  \rightarrow  {\intermed \sigma}_{{\mathrm{2}}}  } \rrbracket ^{\intermed {  {\intermed \Sigma}  } , \intermed {  {\intermed { \Gamma_C } }  } }_{\intermed {  {\intermed { R^{SN} } }  } } 
    & \triangleq  \intermed {  {\intermed \Sigma}  } ; \intermed {  {\intermed { \Gamma_C } }  } ; \intermed {  \bullet  }  \vdash_{tm}  \intermed {  {\intermed e}_{{\mathrm{1}}}  } : \intermed {  {\intermed { R^{SN} } }_{{\mathrm{1}}}  \ottsym{(}  {\intermed \sigma}_{{\mathrm{1}}}  \rightarrow  {\intermed \sigma}_{{\mathrm{2}}}  \ottsym{)}  } \itot{ \rightsquigarrow  \target {  {\target e }  } }  \\
    & \land \exists {\intermed v} :  \intermed {  {\intermed \Sigma}  }  \vdash  \intermed {  {\intermed e}_{{\mathrm{1}}}  }  \longrightarrow^{*}  \intermed {  {\intermed v}  }  \\
    & \land \forall {\intermed e}' :  \intermed {  {\intermed e}'  } \in \mathcal{SN} \llbracket \intermed {  {\intermed \sigma}_{{\mathrm{1}}}  } \rrbracket ^{\intermed {  {\intermed \Sigma}  } , \intermed {  {\intermed { \Gamma_C } }  } }_{\intermed {  {\intermed { R^{SN} } }  } } 
      \Rightarrow  \intermed {  {\intermed e}_{{\mathrm{1}}} \, {\intermed e}'  } \in \mathcal{SN} \llbracket \intermed {  {\intermed \sigma}_{{\mathrm{2}}}  } \rrbracket ^{\intermed {  {\intermed \Sigma}  } , \intermed {  {\intermed { \Gamma_C } }  } }_{\intermed {  {\intermed { R^{SN} } }  } }  \\
      \end{aligned}
    }
    {
      By Preservation (Theorem~\ref{thmA:intermediate_preservation}), we know
      that $ \intermed {  {\intermed \Sigma}  } ; \intermed {  {\intermed { \Gamma_C } }  } ; \intermed {  \bullet  }  \vdash_{tm}  \intermed {  {\intermed e}_{{\mathrm{2}}}  } : \intermed {  {\intermed { R^{SN} } }_{{\mathrm{1}}}  \ottsym{(}  {\intermed \sigma}_{{\mathrm{1}}}  \rightarrow  {\intermed \sigma}_{{\mathrm{2}}}  \ottsym{)}  } \itot{ \rightsquigarrow  \target {  {\target e }  } } $. Because
      the evaluation
      in \interL{} is deterministic (Lemma~\ref{lemma:eval_unique}), given $ \intermed {  {\intermed \Sigma}  }  \vdash  \intermed {  {\intermed e}_{{\mathrm{1}}}  }  \longrightarrow^{*}  \intermed {  {\intermed v}  } $, we have $ \intermed {  {\intermed \Sigma}  }  \vdash  \intermed {  {\intermed e}_{{\mathrm{2}}}  }  \longrightarrow^{*}  \intermed {  {\intermed v}  } $. Given any ${\intermed e}' :  \intermed {  {\intermed e}'  } \in \mathcal{SN} \llbracket \intermed {  {\intermed \sigma}_{{\mathrm{1}}}  } \rrbracket ^{\intermed {  {\intermed \Sigma}  } , \intermed {  {\intermed { \Gamma_C } }  } }_{\intermed {  {\intermed { R^{SN} } }  } } $, we know that $ \intermed {  {\intermed \Sigma}  }  \vdash  \intermed {  {\intermed e}_{{\mathrm{1}}}  }  \longrightarrow  \intermed {  {\intermed e}_{{\mathrm{2}}}  } $, so $ \intermed {  {\intermed \Sigma}  }  \vdash  \intermed {  {\intermed e}_{{\mathrm{1}}} \, {\intermed e}'  }  \longrightarrow  \intermed {  {\intermed e}_{{\mathrm{2}}} \, {\intermed e}'  } $. By induction hypothesis, we get $  \intermed {  {\intermed e}_{{\mathrm{2}}} \, {\intermed e}'  } \in \mathcal{SN} \llbracket \intermed {  {\intermed \sigma}_{{\mathrm{2}}}  } \rrbracket ^{\intermed {  {\intermed \Sigma}  } , \intermed {  {\intermed { \Gamma_C } }  } }_{\intermed {  {\intermed { R^{SN} } }  } }  $. \\
    }
    \proofStep{Function over constraint}
    { \intermed {  {\intermed e}  } \in \mathcal{SN} \llbracket \intermed {   {\intermed Q}  \Rightarrow  {\intermed \sigma}   } \rrbracket ^{\intermed {  {\intermed \Sigma}  } , \intermed {  {\intermed { \Gamma_C } }  } }_{\intermed {  {\intermed { R^{SN} } }  } } }
    {
      Similar to Function case. \\
    }
    \proofStep{Polymorphic type}
    { \intermed {  {\intermed e}  } \in \mathcal{SN} \llbracket \intermed {  \forall  {\intermed a }  \ottsym{.}  {\intermed \sigma}  } \rrbracket ^{\intermed {  {\intermed \Sigma}  } , \intermed {  {\intermed { \Gamma_C } }  } }_{\intermed {  {\intermed { R^{SN} } }  } } }
    {
      Similar to Function case.
    }

    \item[Part 2]
      Similar to Part 1.
\end{description}
\qed

\begin{lemmaB}[Substitution for Context Interpretation]\leavevmode
  \label{lemma:sn_ctx_inter}
  \begin{itemize}
    \item
  If \( \intermed {  {\intermed \Sigma}  } ; \intermed {  {\intermed { \Gamma_C } }  } ; \intermed {  {\intermed \Gamma}  }  \vdash_{tm}  \intermed {  {\intermed e}  } : \intermed {  {\intermed \sigma}  } \itot{ \rightsquigarrow  \target {  {\target e }_{{\mathrm{1}}}  } } \)
  then \( \forall  \intermed {  {\intermed { R^{SN} } }  } \in \mathcal{F^{SN} } \llbracket \intermed {  {\intermed \Gamma}  } \rrbracket ^{\intermed {  {\intermed \Sigma}  }, \intermed {  {\intermed { \Gamma_C } }  } } \),
  \( \intermed {  {\intermed { \phi^{SN} } }  } \in \mathcal{G^{SN} } \llbracket \intermed {  {\intermed \Gamma}  } \rrbracket ^{\intermed {  {\intermed \Sigma}  } , \intermed {  {\intermed { \Gamma_C } }  } }_{\intermed {  {\intermed { R^{SN} } }  } } \) and
  \( \intermed {  {\intermed { \gamma^{SN} } }  } \in \mathcal{H^{SN} } \llbracket \intermed {  {\intermed \Gamma}  } \rrbracket ^{\intermed {  {\intermed \Sigma}  } , \intermed {  {\intermed { \Gamma_C } }  } }_{\intermed {  {\intermed { R^{SN} } }  } } \), \\
  we have $ \intermed {  {\intermed \Sigma}  } ; \intermed {  {\intermed { \Gamma_C } }  } ; \intermed {  \bullet  }  \vdash_{tm}  \intermed {   {\intermed { \gamma^{SN} } }  ( {\intermed { \phi^{SN} } }  ( {\intermed { R^{SN} } }  ( {\intermed e} )))   } : \intermed {  {\intermed { R^{SN} } }  \ottsym{(}  {\intermed \sigma}  \ottsym{)}  } \itot{ \rightsquigarrow  \target {  {\target e }_{{\mathrm{2}}}  } } $.
    \item
  If \( \intermed {  {\intermed \Sigma}  } ; \intermed {  {\intermed { \Gamma_C } }  } ; \intermed {  {\intermed \Gamma}  }  \vdash_{d}  \intermed {  {\intermed d}  } : \intermed {  {\intermed Q}  } \itot{\,  \rightsquigarrow  \target {  {\target e }_{{\mathrm{1}}}  } } \)
  then \( \forall  \intermed {  {\intermed { R^{SN} } }  } \in \mathcal{F^{SN} } \llbracket \intermed {  {\intermed \Gamma}  } \rrbracket ^{\intermed {  {\intermed \Sigma}  }, \intermed {  {\intermed { \Gamma_C } }  } } \),
  \( \intermed {  {\intermed { \phi^{SN} } }  } \in \mathcal{G^{SN} } \llbracket \intermed {  {\intermed \Gamma}  } \rrbracket ^{\intermed {  {\intermed \Sigma}  } , \intermed {  {\intermed { \Gamma_C } }  } }_{\intermed {  {\intermed { R^{SN} } }  } } \) and
  \( \intermed {  {\intermed { \gamma^{SN} } }  } \in \mathcal{H^{SN} } \llbracket \intermed {  {\intermed \Gamma}  } \rrbracket ^{\intermed {  {\intermed \Sigma}  } , \intermed {  {\intermed { \Gamma_C } }  } }_{\intermed {  {\intermed { R^{SN} } }  } } \), \\
  we have $ \intermed {  {\intermed \Sigma}  } ; \intermed {  {\intermed { \Gamma_C } }  } ; \intermed {  \bullet  }  \vdash_{d}  \intermed {   {\intermed { \gamma^{SN} } }  ( {\intermed { \phi^{SN} } }  ( {\intermed { R^{SN} } }  ( {\intermed d} )))   } : \intermed {  {\intermed { R^{SN} } }  \ottsym{(}  {\intermed Q}  \ottsym{)}  } \itot{\,  \rightsquigarrow  \target {  {\target e }_{{\mathrm{2}}}  } } $.
    \item
  If \( \intermed {  {\intermed { \Gamma_C } }  } ; \intermed {  {\intermed \Gamma}  }  \vdash_{ty}  \intermed {  {\intermed \sigma}  } \itot{ \rightsquigarrow  \target {  {\target \sigma }  } } \)
  then \( \forall  \intermed {  {\intermed { R^{SN} } }  } \in \mathcal{F^{SN} } \llbracket \intermed {  {\intermed \Gamma}  } \rrbracket ^{\intermed {  {\intermed \Sigma}  }, \intermed {  {\intermed { \Gamma_C } }  } } \),
  \( \intermed {  {\intermed { \phi^{SN} } }  } \in \mathcal{G^{SN} } \llbracket \intermed {  {\intermed \Gamma}  } \rrbracket ^{\intermed {  {\intermed \Sigma}  } , \intermed {  {\intermed { \Gamma_C } }  } }_{\intermed {  {\intermed { R^{SN} } }  } } \) and
  \( \intermed {  {\intermed { \gamma^{SN} } }  } \in \mathcal{H^{SN} } \llbracket \intermed {  {\intermed \Gamma}  } \rrbracket ^{\intermed {  {\intermed \Sigma}  } , \intermed {  {\intermed { \Gamma_C } }  } }_{\intermed {  {\intermed { R^{SN} } }  } } \), \\
  we have $ \intermed {  {\intermed { \Gamma_C } }  } ; \intermed {  \bullet  }  \vdash_{ty}  \intermed {  {\intermed { R^{SN} } }  \ottsym{(}  {\intermed \sigma}  \ottsym{)}  } \itot{ \rightsquigarrow  \target {  {\target \sigma }_{{\mathrm{2}}}  } } $.
  \item
  If \( \intermed {  {\intermed { \Gamma_C } }  } ; \intermed {  {\intermed \Gamma}  }  \vdash_{\mathit {Q} }  \intermed {  {\intermed Q}  }\, \itot{ \rightsquigarrow  \target {  {\target \sigma }  } } \)
  then \( \forall  \intermed {  {\intermed { R^{SN} } }  } \in \mathcal{F^{SN} } \llbracket \intermed {  {\intermed \Gamma}  } \rrbracket ^{\intermed {  {\intermed \Sigma}  }, \intermed {  {\intermed { \Gamma_C } }  } } \),
  \( \intermed {  {\intermed { \phi^{SN} } }  } \in \mathcal{G^{SN} } \llbracket \intermed {  {\intermed \Gamma}  } \rrbracket ^{\intermed {  {\intermed \Sigma}  } , \intermed {  {\intermed { \Gamma_C } }  } }_{\intermed {  {\intermed { R^{SN} } }  } } \) and
  \( \intermed {  {\intermed { \gamma^{SN} } }  } \in \mathcal{H^{SN} } \llbracket \intermed {  {\intermed \Gamma}  } \rrbracket ^{\intermed {  {\intermed \Sigma}  } , \intermed {  {\intermed { \Gamma_C } }  } }_{\intermed {  {\intermed { R^{SN} } }  } } \), \\
  we have $ \intermed {  {\intermed { \Gamma_C } }  } ; \intermed {  \bullet  }  \vdash_{\mathit {Q} }  \intermed {  {\intermed { R^{SN} } }  \ottsym{(}  {\intermed Q}  \ottsym{)}  }\, \itot{ \rightsquigarrow  \target {  {\target \sigma }_{{\mathrm{2}}}  } } $.
  \end{itemize}
\end{lemmaB}

\proof

By induction on ${\intermed e}$, ${\intermed d}$ ${\intermed \sigma}$, and ${\intermed Q}$
respectively. The goal follows from Definitions~\ref{def:inter_ty},
\ref{def:inter_tm} and \ref{def:inter_d}.

\qed

\begin{lemmaB}[Compositionality for Strong Normalization]\ \\
  \label{lemma:comp_sn}
  Let ${ \intermed { \mathit{ r } } }$ = $ \mathcal{SN} \llbracket \intermed {  {\intermed \sigma}_{{\mathrm{2}}}  } \rrbracket ^{\intermed {  {\intermed \Sigma}  } , \intermed {  {\intermed { \Gamma_C } }  } }_{\intermed {  {\intermed { R^{SN} } }  } } $,
  then $ \intermed {  {\intermed e}  } \in \mathcal{SN} \llbracket \intermed {  {\intermed \sigma}_{{\mathrm{1}}}  } \rrbracket ^{\intermed {  {\intermed \Sigma}  } , \intermed {  {\intermed { \Gamma_C } }  } }_{\intermed {   {\intermed { R^{SN} } }  ,  {\intermed a }  \mapsto (  {\intermed { R^{SN} } }  \ottsym{(}  {\intermed \sigma}_{{\mathrm{2}}}  \ottsym{)} ,  { \intermed { \mathit{ r } } }  )   } } $
  if and only if $ \intermed {  {\intermed e}  } \in \mathcal{SN} \llbracket \intermed {  [  {\intermed \sigma}_{{\mathrm{2}}}  /  {\intermed a }  ]  {\intermed \sigma}_{{\mathrm{1}}}  } \rrbracket ^{\intermed {  {\intermed \Sigma}  } , \intermed {  {\intermed { \Gamma_C } }  } }_{\intermed {  {\intermed { R^{SN} } }  } } $.
\end{lemmaB}

\proof

By induction on ${\intermed \sigma}_{{\mathrm{1}}}$. \\
\proofStep{Bool}
{ \intermed {  {\intermed \sigma}_{{\mathrm{1}}}  } = \intermed {  \mathit{\intermed {Bool} }  } }
{
  Since $ \intermed {  [  {\intermed \sigma}_{{\mathrm{2}}}  /  {\intermed a }  ]  \mathit{\intermed {Bool} }  } = \intermed {  \mathit{\intermed {Bool} }  } $, the goal follows directly. \\
}
\proofStep{Type variable}
{ \intermed {  {\intermed \sigma}_{{\mathrm{1}}}  } = \intermed {  {\intermed b }  } }
{
  Depending on whether ${\intermed b }$ is the same variable as ${\intermed a }$, we have two cases:
  \begin{itemize}
    \item If $ \intermed {  {\intermed b }  } = \intermed {  {\intermed a }  } $, then $ \intermed {  [  {\intermed \sigma}_{{\mathrm{2}}}  /  {\intermed a }  ]  {\intermed a }  } = \intermed {  {\intermed \sigma}_{{\mathrm{2}}}  } $.
      \begin{description}
        \item[Part 1: From left to right.]
      If $ \intermed {  {\intermed e}  } \in \mathcal{SN} \llbracket \intermed {  {\intermed a }  } \rrbracket ^{\intermed {  {\intermed \Sigma}  } , \intermed {  {\intermed { \Gamma_C } }  } }_{\intermed {   {\intermed { R^{SN} } }  ,  {\intermed a }  \mapsto (  {\intermed { R^{SN} } }  \ottsym{(}  {\intermed \sigma}_{{\mathrm{2}}}  \ottsym{)} ,  { \intermed { \mathit{ r } } }  )   } } $, it
      means that:
      \begin{gather}
        \label{eq:comp:var:1}
        \exists {\intermed v} :  \intermed {  {\intermed \Sigma}  }  \vdash  \intermed {  {\intermed e}  }  \longrightarrow^{*}  \intermed {  {\intermed v}  }   \\
        \label{eq:comp:var:2}
         \intermed {  {\intermed v}  } \in \mathcal{SN} \llbracket \intermed {  {\intermed \sigma}_{{\mathrm{2}}}  } \rrbracket ^{\intermed {  {\intermed \Sigma}  } , \intermed {  {\intermed { \Gamma_C } }  } }_{\intermed {  {\intermed { R^{SN} } }  } } 
      \end{gather}
      Combining with Strong Normalization preserved by forward/backward
      reduction (Lemma~\ref{lemma:sn_pres}), the goal is proven by
      Equations~\ref{eq:comp:var:1} and \ref{eq:comp:var:2}:
      \begin{gather}
         \intermed {  {\intermed e}  } \in \mathcal{SN} \llbracket \intermed {  {\intermed \sigma}_{{\mathrm{2}}}  } \rrbracket ^{\intermed {  {\intermed \Sigma}  } , \intermed {  {\intermed { \Gamma_C } }  } }_{\intermed {  {\intermed { R^{SN} } }  } } 
      \end{gather}
        \item[Part 2: From right to left.]
          From the hypothesis, we know that:
          \begin{gather}
            \label{eq:comp:var:3}
             \intermed {  {\intermed e}  } \in \mathcal{SN} \llbracket \intermed {  {\intermed \sigma}_{{\mathrm{2}}}  } \rrbracket ^{\intermed {  {\intermed \Sigma}  } , \intermed {  {\intermed { \Gamma_C } }  } }_{\intermed {  {\intermed { R^{SN} } }  } } 
          \end{gather}
          We want to
          prove that $ \intermed {  {\intermed e}  } \in \mathcal{SN} \llbracket \intermed {  {\intermed a }  } \rrbracket ^{\intermed {  {\intermed \Sigma}  } , \intermed {  {\intermed { \Gamma_C } }  } }_{\intermed {   {\intermed { R^{SN} } }  ,  {\intermed a }  \mapsto (  {\intermed { R^{SN} } }  \ottsym{(}  {\intermed \sigma}_{{\mathrm{2}}}  \ottsym{)} ,  { \intermed { \mathit{ r } } }  )   } } $. By the definition, this goal is equivalent to:
          \begin{gather}
            \label{eq:comp:var:4}
             \intermed {  {\intermed \Sigma}  } ; \intermed {  {\intermed { \Gamma_C } }  } ; \intermed {  \bullet  }  \vdash_{tm}  \intermed {  {\intermed e}  } : \intermed {  \ottsym{(}   {\intermed { R^{SN} } }  ,  {\intermed a }  \mapsto (  {\intermed { R^{SN} } }  \ottsym{(}  {\intermed \sigma}_{{\mathrm{2}}}  \ottsym{)} ,  { \intermed { \mathit{ r } } }  )   \ottsym{)}  \ottsym{(}  {\intermed a }  \ottsym{)}  } \itot{ \rightsquigarrow  \target {  {\target e }  } }  \\
            \label{eq:comp:var:5}
            \exists {\intermed v} :  \intermed {  {\intermed \Sigma}  }  \vdash  \intermed {  {\intermed e}  }  \longrightarrow^{*}  \intermed {  {\intermed v}  }   \\
            \label{eq:comp:var:6}
             \intermed {  {\intermed v}  }  \in  \intermed {  \ottsym{(}   {\intermed { R^{SN} } }  ,  {\intermed a }  \mapsto (  {\intermed { R^{SN} } }  \ottsym{(}  {\intermed \sigma}_{{\mathrm{2}}}  \ottsym{)} ,  { \intermed { \mathit{ r } } }  )   \ottsym{)}  } _2~\ottsym{(}  {\intermed a }  \ottsym{)}
          \end{gather}
        Equation~\ref{eq:comp:var:4} simplifies to:
          \begin{gather}
            \label{eq:comp:var:7}
             \intermed {  {\intermed \Sigma}  } ; \intermed {  {\intermed { \Gamma_C } }  } ; \intermed {  \bullet  }  \vdash_{tm}  \intermed {  {\intermed e}  } : \intermed {  {\intermed { R^{SN} } }  \ottsym{(}  {\intermed \sigma}_{{\mathrm{2}}}  \ottsym{)}  } \itot{ \rightsquigarrow  \target {  {\target e }  } }  
          \end{gather}
        By Well-Typedness from Strong Normalization
        (Lemma~\ref{lemma:sn_typing}), Equation~\ref{eq:comp:var:3} proves
        \ref{eq:comp:var:7}.
        By  Strong Normalization - Part B (Theorem~\ref{thmA:sn:part:b}),
        Equation~\ref{eq:comp:var:3} proves \ref{eq:comp:var:5}.

        We already know that $\ottsym{(}   {\intermed { R^{SN} } }  ,  {\intermed a }  \mapsto (  {\intermed { R^{SN} } }  \ottsym{(}  {\intermed \sigma}_{{\mathrm{2}}}  \ottsym{)} ,  { \intermed { \mathit{ r } } }  )   \ottsym{)}_2~\ottsym{(}  {\intermed a }  \ottsym{)} = { \intermed { \mathit{ r } } } =
         \mathcal{SN} \llbracket \intermed {  {\intermed \sigma}_{{\mathrm{2}}}  } \rrbracket ^{\intermed {  {\intermed \Sigma}  } , \intermed {  {\intermed { \Gamma_C } }  } }_{\intermed {  {\intermed { R^{SN} } }  } } $, so we simplify
        Equation~\ref{eq:comp:var:6} to get
          \begin{gather}
            \label{eq:comp:var:8}
             \intermed {  {\intermed v}  } \in \mathcal{SN} \llbracket \intermed {  {\intermed \sigma}_{{\mathrm{2}}}  } \rrbracket ^{\intermed {  {\intermed \Sigma}  } , \intermed {  {\intermed { \Gamma_C } }  } }_{\intermed {  {\intermed { R^{SN} } }  } } 
          \end{gather}
          From Equation~\ref{eq:comp:var:3} and Strong Normalization preserved
          by forward/backward reduction (Lemma~\ref{lemma:sn_pres}), we can
          prove Goal~\ref{eq:comp:var:8}.
        \end{description}
        
    \item If ${\intermed b } \neq {\intermed a }$, since $ \intermed {  [  {\intermed \sigma}_{{\mathrm{2}}}  /  {\intermed a }  ]  {\intermed b }  } = \intermed {  {\intermed b }  } $, the goal
      follows directly.
  \end{itemize}
}
\proofStep{Function}
{ \intermed {  {\intermed \sigma}_{{\mathrm{1}}}  } = \intermed {  {\intermed \sigma}_{{\mathrm{11}}}  \rightarrow  {\intermed \sigma}_{{\mathrm{12}}}  } }
{
  \begin{description}
    \item[Part 1: From left to right.]

  From the hypothesis, we get:
  \begin{equation*}
     \intermed {  {\intermed e}  } \in \mathcal{SN} \llbracket \intermed {  {\intermed \sigma}_{{\mathrm{11}}}  \rightarrow  {\intermed \sigma}_{{\mathrm{12}}}  } \rrbracket ^{\intermed {  {\intermed \Sigma}  } , \intermed {  {\intermed { \Gamma_C } }  } }_{\intermed {   {\intermed { R^{SN} } }  ,  {\intermed a }  \mapsto (  {\intermed { R^{SN} } }  \ottsym{(}  {\intermed \sigma}_{{\mathrm{2}}}  \ottsym{)} ,  { \intermed { \mathit{ r } } }  )   } } 
  \end{equation*}
  We thus know that:
  \begin{gather}
    \label{eq:comp:arr:1}
     \intermed {  {\intermed \Sigma}  } ; \intermed {  {\intermed { \Gamma_C } }  } ; \intermed {  \bullet  }  \vdash_{tm}  \intermed {  {\intermed e}  } : \intermed {  \ottsym{(}   {\intermed { R^{SN} } }  ,  {\intermed a }  \mapsto (  {\intermed { R^{SN} } }  \ottsym{(}  {\intermed \sigma}_{{\mathrm{2}}}  \ottsym{)} ,  { \intermed { \mathit{ r } } }  )   \ottsym{)}  \ottsym{(}  {\intermed \sigma}_{{\mathrm{11}}}  \rightarrow  {\intermed \sigma}_{{\mathrm{12}}}  \ottsym{)}  } \itot{ \rightsquigarrow  \target {  {\target e }  } }  \\
    \label{eq:comp:arr:2}
     \exists {\intermed v} :  \intermed {  {\intermed \Sigma}  }  \vdash  \intermed {  {\intermed e}  }  \longrightarrow^{*}  \intermed {  {\intermed v}  }  \\
    \label{eq:comp:arr:3}
     \forall {\intermed e}' :  \intermed {  {\intermed e}'  } \in \mathcal{SN} \llbracket \intermed {  {\intermed \sigma}_{{\mathrm{11}}}  } \rrbracket ^{\intermed {  {\intermed \Sigma}  } , \intermed {  {\intermed { \Gamma_C } }  } }_{\intermed {   {\intermed { R^{SN} } }  ,  {\intermed a }  \mapsto (  {\intermed { R^{SN} } }  \ottsym{(}  {\intermed \sigma}_{{\mathrm{2}}}  \ottsym{)} ,  { \intermed { \mathit{ r } } }  )   } } 
      \Rightarrow  \intermed {  {\intermed e} \, {\intermed e}'  } \in \mathcal{SN} \llbracket \intermed {  {\intermed \sigma}_{{\mathrm{12}}}  } \rrbracket ^{\intermed {  {\intermed \Sigma}  } , \intermed {  {\intermed { \Gamma_C } }  } }_{\intermed {   {\intermed { R^{SN} } }  ,  {\intermed a }  \mapsto (  {\intermed { R^{SN} } }  \ottsym{(}  {\intermed \sigma}_{{\mathrm{2}}}  \ottsym{)} ,  { \intermed { \mathit{ r } } }  )   } }  
  \end{gather}
  Our goal is to prove that $ \intermed {  {\intermed e}  } \in \mathcal{SN} \llbracket \intermed {  [  {\intermed \sigma}_{{\mathrm{2}}}  /  {\intermed a }  ]  {\intermed \sigma}_{{\mathrm{11}}}  \rightarrow  [  {\intermed \sigma}_{{\mathrm{2}}}  /  {\intermed a }  ]  {\intermed \sigma}_{{\mathrm{12}}}  } \rrbracket ^{\intermed {  {\intermed \Sigma}  } , \intermed {  {\intermed { \Gamma_C } }  } }_{\intermed {  {\intermed { R^{SN} } }  } } $. This is equivalent to proving:
  \begin{gather}
    \label{eq:comp:arr:4}
     \intermed {  {\intermed \Sigma}  } ; \intermed {  {\intermed { \Gamma_C } }  } ; \intermed {  \bullet  }  \vdash_{tm}  \intermed {  {\intermed e}  } : \intermed {  {\intermed { R^{SN} } }  \ottsym{(}  [  {\intermed \sigma}_{{\mathrm{2}}}  /  {\intermed a }  ]  {\intermed \sigma}_{{\mathrm{11}}}  \rightarrow  [  {\intermed \sigma}_{{\mathrm{2}}}  /  {\intermed a }  ]  {\intermed \sigma}_{{\mathrm{12}}}  \ottsym{)}  } \itot{ \rightsquigarrow  \target {  {\target e }  } }  \\
    \label{eq:comp:arr:5}
     \exists {\intermed v} :  \intermed {  {\intermed \Sigma}  }  \vdash  \intermed {  {\intermed e}  }  \longrightarrow^{*}  \intermed {  {\intermed v}  }  \\
    \label{eq:comp:arr:6}
     \forall {\intermed e}' :  \intermed {  {\intermed e}'  } \in \mathcal{SN} \llbracket \intermed {  [  {\intermed \sigma}_{{\mathrm{2}}}  /  {\intermed a }  ]  {\intermed \sigma}_{{\mathrm{11}}}  } \rrbracket ^{\intermed {  {\intermed \Sigma}  } , \intermed {  {\intermed { \Gamma_C } }  } }_{\intermed {  {\intermed { R^{SN} } }  } } 
      \Rightarrow  \intermed {  {\intermed e} \, {\intermed e}'  } \in \mathcal{SN} \llbracket \intermed {  [  {\intermed \sigma}_{{\mathrm{2}}}  /  {\intermed a }  ]  {\intermed \sigma}_{{\mathrm{12}}}  } \rrbracket ^{\intermed {  {\intermed \Sigma}  } , \intermed {  {\intermed { \Gamma_C } }  } }_{\intermed {  {\intermed { R^{SN} } }  } }  
  \end{gather}
  Goal~\ref{eq:comp:arr:4} and \ref{eq:comp:arr:5} are proven directly by
  Equations~\ref{eq:comp:arr:1} and \ref{eq:comp:arr:2}. Only
  Goal~\ref{eq:comp:arr:6} remains to be proven.
  
  Given  $ \intermed {  {\intermed e}'  } \in \mathcal{SN} \llbracket \intermed {  [  {\intermed \sigma}_{{\mathrm{2}}}  /  {\intermed a }  ]  {\intermed \sigma}_{{\mathrm{11}}}  } \rrbracket ^{\intermed {  {\intermed \Sigma}  } , \intermed {  {\intermed { \Gamma_C } }  } }_{\intermed {  {\intermed { R^{SN} } }  } } $, the induction
  hypothesis tells us that $ \intermed {  {\intermed e}'  } \in \mathcal{SN} \llbracket \intermed {  {\intermed \sigma}_{{\mathrm{11}}}  } \rrbracket ^{\intermed {  {\intermed \Sigma}  } , \intermed {  {\intermed { \Gamma_C } }  } }_{\intermed {   {\intermed { R^{SN} } }  ,  {\intermed a }  \mapsto (  {\intermed { R^{SN} } }  \ottsym{(}  {\intermed \sigma}_{{\mathrm{2}}}  \ottsym{)} ,  { \intermed { \mathit{ r } } }  )   } } $. In combination with equation Equation~\ref{eq:comp:arr:3}, we get
  \begin{align}
       \intermed {  {\intermed e} \, {\intermed e}'  } \in \mathcal{SN} \llbracket \intermed {  {\intermed \sigma}_{{\mathrm{12}}}  } \rrbracket ^{\intermed {  {\intermed \Sigma}  } , \intermed {  {\intermed { \Gamma_C } }  } }_{\intermed {   {\intermed { R^{SN} } }  ,  {\intermed a }  \mapsto (  {\intermed { R^{SN} } }  \ottsym{(}  {\intermed \sigma}_{{\mathrm{2}}}  \ottsym{)} ,  { \intermed { \mathit{ r } } }  )   } }  
  \end{align}
  By induction hypothesis, we get:
  \begin{align}
       \intermed {  {\intermed e} \, {\intermed e}'  } \in \mathcal{SN} \llbracket \intermed {  [  {\intermed \sigma}_{{\mathrm{2}}}  /  {\intermed a }  ]  {\intermed \sigma}_{{\mathrm{12}}}  } \rrbracket ^{\intermed {  {\intermed \Sigma}  } , \intermed {  {\intermed { \Gamma_C } }  } }_{\intermed {  {\intermed { R^{SN} } }  } }  
  \end{align}
  The goal has been proven.
    \item[Part 2: From right to left.] Similar to Part 1. \\
      \end{description} 
}
\proofStep{Function over constraints}
{ \intermed {  {\intermed \sigma}_{{\mathrm{1}}}  } = \intermed {   {\intermed Q}  \Rightarrow  {\intermed \sigma}_{{\mathrm{12}}}   } }
{
  \begin{description}
    \item[Part 1: From left to right.]

  We know from the hypothesis that:
  \begin{equation*}
     \intermed {  {\intermed e}  } \in \mathcal{SN} \llbracket \intermed {   {\intermed Q}  \Rightarrow  {\intermed \sigma}_{{\mathrm{12}}}   } \rrbracket ^{\intermed {  {\intermed \Sigma}  } , \intermed {  {\intermed { \Gamma_C } }  } }_{\intermed {   {\intermed { R^{SN} } }  ,  {\intermed a }  \mapsto (  {\intermed { R^{SN} } }  \ottsym{(}  {\intermed \sigma}_{{\mathrm{2}}}  \ottsym{)} ,  { \intermed { \mathit{ r } } }  )   } } 
  \end{equation*}
  It follows that:
  \begin{gather}
    \label{eq:comp:arrc:1}
     \intermed {  {\intermed \Sigma}  } ; \intermed {  {\intermed { \Gamma_C } }  } ; \intermed {  \bullet  }  \vdash_{tm}  \intermed {  {\intermed e}  } : \intermed {  \ottsym{(}   {\intermed { R^{SN} } }  ,  {\intermed a }  \mapsto (  {\intermed { R^{SN} } }  \ottsym{(}  {\intermed \sigma}_{{\mathrm{2}}}  \ottsym{)} ,  { \intermed { \mathit{ r } } }  )   \ottsym{)}  \ottsym{(}   {\intermed Q}  \Rightarrow  {\intermed \sigma}_{{\mathrm{12}}}   \ottsym{)}  } \itot{ \rightsquigarrow  \target {  {\target e }  } }  \\
    \label{eq:comp:arrc:2}
     \exists {\intermed v} :  \intermed {  {\intermed \Sigma}  }  \vdash  \intermed {  {\intermed e}  }  \longrightarrow^{*}  \intermed {  {\intermed v}  }  \\
    \label{eq:comp:arrc:3}
      \forall {\intermed d} :  \intermed {  {\intermed \Sigma}  } ; \intermed {  {\intermed { \Gamma_C } }  } ; \intermed {  \bullet  }  \vdash_{d}  \intermed {  {\intermed d}  } : \intermed {  \ottsym{(}   {\intermed { R^{SN} } }  ,  {\intermed a }  \mapsto (  {\intermed { R^{SN} } }  \ottsym{(}  {\intermed \sigma}_{{\mathrm{2}}}  \ottsym{)} ,  { \intermed { \mathit{ r } } }  )   \ottsym{)}  \ottsym{(}  {\intermed Q}  \ottsym{)}  } \itot{\,  \rightsquigarrow  \target {  {\target e }  } } 
      \Rightarrow  \intermed {  {\intermed e} \, {\intermed d}  } \in \mathcal{SN} \llbracket \intermed {  {\intermed \sigma}_{{\mathrm{12}}}  } \rrbracket ^{\intermed {  {\intermed \Sigma}  } , \intermed {  {\intermed { \Gamma_C } }  } }_{\intermed {   {\intermed { R^{SN} } }  ,  {\intermed a }  \mapsto (  {\intermed { R^{SN} } }  \ottsym{(}  {\intermed \sigma}_{{\mathrm{2}}}  \ottsym{)} ,  { \intermed { \mathit{ r } } }  )   } }  
  \end{gather}
  Our goal is to prove that $ \intermed {  {\intermed e}  } \in \mathcal{SN} \llbracket \intermed {   [  {\intermed \sigma}_{{\mathrm{2}}}  /  {\intermed a }  ]  {\intermed Q}  \Rightarrow  [  {\intermed \sigma}_{{\mathrm{2}}}  /  {\intermed a }  ]  {\intermed \sigma}_{{\mathrm{12}}}   } \rrbracket ^{\intermed {  {\intermed \Sigma}  } , \intermed {  {\intermed { \Gamma_C } }  } }_{\intermed {  {\intermed { R^{SN} } }  } } $. This is equivalent to proving:
  \begin{gather}
    \label{eq:comp:arrc:4}
     \intermed {  {\intermed \Sigma}  } ; \intermed {  {\intermed { \Gamma_C } }  } ; \intermed {  \bullet  }  \vdash_{tm}  \intermed {  {\intermed e}  } : \intermed {  {\intermed { R^{SN} } }  \ottsym{(}   [  {\intermed \sigma}_{{\mathrm{2}}}  /  {\intermed a }  ]  {\intermed Q}  \Rightarrow  [  {\intermed \sigma}_{{\mathrm{2}}}  /  {\intermed a }  ]  {\intermed \sigma}_{{\mathrm{12}}}   \ottsym{)}  } \itot{ \rightsquigarrow  \target {  {\target e }  } }  \\
    \label{eq:comp:arrc:5}
     \exists {\intermed v} :  \intermed {  {\intermed \Sigma}  }  \vdash  \intermed {  {\intermed e}  }  \longrightarrow^{*}  \intermed {  {\intermed v}  }  \\
    \label{eq:comp:arrc:6}
     \forall {\intermed d}' :  \intermed {  {\intermed \Sigma}  } ; \intermed {  {\intermed { \Gamma_C } }  } ; \intermed {  \bullet  }  \vdash_{d}  \intermed {  {\intermed d}'  } : \intermed {  {\intermed { R^{SN} } }_{{\mathrm{1}}}  \ottsym{(}  [  {\intermed \sigma}_{{\mathrm{2}}}  /  {\intermed a }  ]  {\intermed Q}  \ottsym{)}  } \itot{\,  \rightsquigarrow  \target {  {\target e }  } } 
      \Rightarrow  \intermed {  {\intermed e} \, {\intermed d}'  } \in \mathcal{SN} \llbracket \intermed {  [  {\intermed \sigma}_{{\mathrm{2}}}  /  {\intermed a }  ]  {\intermed \sigma}_{{\mathrm{12}}}  } \rrbracket ^{\intermed {  {\intermed \Sigma}  } , \intermed {  {\intermed { \Gamma_C } }  } }_{\intermed {  {\intermed { R^{SN} } }  } }  
  \end{gather}
  Goals~\ref{eq:comp:arrc:4} and \ref{eq:comp:arrc:5} are proven directly by
  Equations~\ref{eq:comp:arrc:1} and \ref{eq:comp:arrc:2}.
  Only Goal~\ref{eq:comp:arrc:6} remains to be proven.
  Given  $ \intermed {  {\intermed \Sigma}  } ; \intermed {  {\intermed { \Gamma_C } }  } ; \intermed {  \bullet  }  \vdash_{d}  \intermed {  {\intermed d}'  } : \intermed {  {\intermed { R^{SN} } }_{{\mathrm{1}}}  \ottsym{(}  [  {\intermed \sigma}_{{\mathrm{2}}}  /  {\intermed a }  ]  {\intermed Q}  \ottsym{)}  } \itot{\,  \rightsquigarrow  \target {  {\target e }  } } $,
  in combination with Equation~\ref{eq:comp:arrc:3}, we get that:
  \begin{gather}
       \intermed {  {\intermed e} \, {\intermed d}'  } \in \mathcal{SN} \llbracket \intermed {  {\intermed \sigma}_{{\mathrm{12}}}  } \rrbracket ^{\intermed {  {\intermed \Sigma}  } , \intermed {  {\intermed { \Gamma_C } }  } }_{\intermed {   {\intermed { R^{SN} } }  ,  {\intermed a }  \mapsto (  {\intermed { R^{SN} } }  \ottsym{(}  {\intermed \sigma}_{{\mathrm{2}}}  \ottsym{)} ,  { \intermed { \mathit{ r } } }  )   } }  
  \end{gather}
  By induction hypothesis, we get:
  \begin{gather}
       \intermed {  {\intermed e} \, {\intermed d}'  } \in \mathcal{SN} \llbracket \intermed {  [  {\intermed \sigma}_{{\mathrm{2}}}  /  {\intermed a }  ]  {\intermed \sigma}_{{\mathrm{12}}}  } \rrbracket ^{\intermed {  {\intermed \Sigma}  } , \intermed {  {\intermed { \Gamma_C } }  } }_{\intermed {  {\intermed { R^{SN} } }  } }  
  \end{gather}
  The goal has been proven.
    \item[Part 2: From right to left.] Similar to Part 1. \\
      \end{description}
}
\proofStep{Polymorphic types}
{  \intermed {  {\intermed \sigma}_{{\mathrm{1}}}  } = \intermed {  \forall  {\intermed b }  \ottsym{.}  {\intermed \sigma}_{{\mathrm{12}}}  }  }
{
  \begin{description}
    \item[Part 1: From left to right.]

  We know from the hypothesis that:
  \begin{equation*}
     \intermed {  {\intermed e}  } \in \mathcal{SN} \llbracket \intermed {  \forall  {\intermed b }  \ottsym{.}  {\intermed \sigma}_{{\mathrm{12}}}  } \rrbracket ^{\intermed {  {\intermed \Sigma}  } , \intermed {  {\intermed { \Gamma_C } }  } }_{\intermed {   {\intermed { R^{SN} } }  ,  {\intermed a }  \mapsto (  {\intermed { R^{SN} } }  \ottsym{(}  {\intermed \sigma}_{{\mathrm{2}}}  \ottsym{)} ,  { \intermed { \mathit{ r } } }  )   } } 
  \end{equation*}
  This implies that:
  \begin{gather}
    \label{eq:comp:all:1}
     \intermed {  {\intermed \Sigma}  } ; \intermed {  {\intermed { \Gamma_C } }  } ; \intermed {  \bullet  }  \vdash_{tm}  \intermed {  {\intermed e}  } : \intermed {  \ottsym{(}   {\intermed { R^{SN} } }  ,  {\intermed a }  \mapsto (  {\intermed { R^{SN} } }  \ottsym{(}  {\intermed \sigma}_{{\mathrm{2}}}  \ottsym{)} ,  { \intermed { \mathit{ r } } }  )   \ottsym{)}  \ottsym{(}  \forall  {\intermed b }  \ottsym{.}  {\intermed \sigma}_{{\mathrm{12}}}  \ottsym{)}  } \itot{ \rightsquigarrow  \target {  {\target e }  } }  \\
    \label{eq:comp:all:2}
     \exists {\intermed v} :  \intermed {  {\intermed \Sigma}  }  \vdash  \intermed {  {\intermed e}  }  \longrightarrow^{*}  \intermed {  {\intermed v}  }  \\
    \label{eq:comp:all:3}
      \forall {\intermed \sigma}' ,  \intermed {  { \intermed { \mathit{ r } } }  } = \mathcal{SN} \llbracket \intermed {  {\intermed \sigma}'  } \rrbracket ^{\intermed {  {\intermed \Sigma}  } , \intermed {  {\intermed { \Gamma_C } }  } }_{\intermed {  {\intermed { R^{SN} } }  } }  :
       \intermed {  {\intermed e} \, {\intermed \sigma}'  } \in \mathcal{SN} \llbracket \intermed {  {\intermed \sigma}_{{\mathrm{12}}}  } \rrbracket ^{\intermed {  {\intermed \Sigma}  } , \intermed {  {\intermed { \Gamma_C } }  } }_{\intermed {   {\intermed { R^{SN} } }  ,  {\intermed a }  \mapsto (  {\intermed { R^{SN} } }  \ottsym{(}  {\intermed \sigma}_{{\mathrm{2}}}  \ottsym{)} ,  { \intermed { \mathit{ r } } }  )   } }  
  \end{gather}
  Our goal is to prove that $ \intermed {  {\intermed e}  } \in \mathcal{SN} \llbracket \intermed {  \forall  {\intermed b }  \ottsym{.}  [  {\intermed \sigma}_{{\mathrm{2}}}  /  {\intermed a }  ]  {\intermed \sigma}_{{\mathrm{12}}}  } \rrbracket ^{\intermed {  {\intermed \Sigma}  } , \intermed {  {\intermed { \Gamma_C } }  } }_{\intermed {  {\intermed { R^{SN} } }  } } $. This is equivalent to proving:
  \begin{gather}
    \label{eq:comp:all:4}
     \intermed {  {\intermed \Sigma}  } ; \intermed {  {\intermed { \Gamma_C } }  } ; \intermed {  \bullet  }  \vdash_{tm}  \intermed {  {\intermed e}  } : \intermed {  {\intermed { R^{SN} } }  \ottsym{(}  \forall  {\intermed b }  \ottsym{.}  [  {\intermed \sigma}_{{\mathrm{2}}}  /  {\intermed a }  ]  {\intermed \sigma}_{{\mathrm{12}}}  \ottsym{)}  } \itot{ \rightsquigarrow  \target {  {\target e }  } }  \\
    \label{eq:comp:all:5}
     \exists {\intermed v} :  \intermed {  {\intermed \Sigma}  }  \vdash  \intermed {  {\intermed e}  }  \longrightarrow^{*}  \intermed {  {\intermed v}  }  \\
    \label{eq:comp:all:6}
     \forall {\intermed \sigma}' ,  \intermed {  { \intermed { \mathit{ r } } }  } = \mathcal{SN} \llbracket \intermed {  {\intermed \sigma}'  } \rrbracket ^{\intermed {  {\intermed \Sigma}  } , \intermed {  {\intermed { \Gamma_C } }  } }_{\intermed {  {\intermed { R^{SN} } }  } } 
      \Rightarrow  \intermed {  {\intermed e} \, {\intermed \sigma}'  } \in \mathcal{SN} \llbracket \intermed {  [  {\intermed \sigma}_{{\mathrm{2}}}  /  {\intermed a }  ]  {\intermed \sigma}_{{\mathrm{12}}}  } \rrbracket ^{\intermed {  {\intermed \Sigma}  } , \intermed {  {\intermed { \Gamma_C } }  } }_{\intermed {  {\intermed { R^{SN} } }  } }  
  \end{gather}
  Goals~\ref{eq:comp:all:4} and \ref{eq:comp:all:5} are directly proven by
  Equations~\ref{eq:comp:all:1} and \ref{eq:comp:all:2}.
  Only Goal~\ref{eq:comp:all:6} remains to be proven.
  Given  ${\intermed \sigma}'$ and $ \intermed {  { \intermed { \mathit{ r } } }  } = \mathcal{SN} \llbracket \intermed {  {\intermed \sigma}'  } \rrbracket ^{\intermed {  {\intermed \Sigma}  } , \intermed {  {\intermed { \Gamma_C } }  } }_{\intermed {  {\intermed { R^{SN} } }  } } $,
  by feeding it to Equation~\ref{eq:comp:all:3}, we get that:
  \begin{gather}
       \intermed {  {\intermed e} \, {\intermed \sigma}'  } \in \mathcal{SN} \llbracket \intermed {  {\intermed \sigma}_{{\mathrm{12}}}  } \rrbracket ^{\intermed {  {\intermed \Sigma}  } , \intermed {  {\intermed { \Gamma_C } }  } }_{\intermed {   {\intermed { R^{SN} } }  ,  {\intermed a }  \mapsto (  {\intermed { R^{SN} } }  \ottsym{(}  {\intermed \sigma}_{{\mathrm{2}}}  \ottsym{)} ,  { \intermed { \mathit{ r } } }  )   } }  
  \end{gather}
  By induction hypothesis, we get:
  \begin{gather}
       \intermed {  {\intermed e} \, {\intermed \sigma}'  } \in \mathcal{SN} \llbracket \intermed {  [  {\intermed \sigma}_{{\mathrm{2}}}  /  {\intermed a }  ]  {\intermed \sigma}_{{\mathrm{12}}}  } \rrbracket ^{\intermed {  {\intermed \Sigma}  } , \intermed {  {\intermed { \Gamma_C } }  } }_{\intermed {  {\intermed { R^{SN} } }  } }  
  \end{gather}
  The goal has been proven.
    \item[Part 2: From right to left.] Similar to Part 1.
      \end{description}
}
\qed

\begin{corollaryB}[Compositionality for Strong Normalization (Context Interpretation)]
  \label{cor:comp_sn_ctx}
  Suppose $ \intermed {  {\intermed { R^{SN} } }  } \in \mathcal{F^{SN} } \llbracket \intermed {  {\intermed \Gamma}  } \rrbracket ^{\intermed {  {\intermed \Sigma}  }, \intermed {  {\intermed { \Gamma_C } }  } } $,
  then $ \intermed {  {\intermed e}  } \in \mathcal{SN} \llbracket \intermed {  {\intermed \sigma}  } \rrbracket ^{\intermed {  {\intermed \Sigma}  } , \intermed {  {\intermed { \Gamma_C } }  } }_{\intermed {  {\intermed { R^{SN} } }  } } $
  if and only if $ \intermed {  {\intermed e}  } \in \mathcal{SN} \llbracket \intermed {  {\intermed { R^{SN} } }  \ottsym{(}  {\intermed \sigma}  \ottsym{)}  } \rrbracket ^{\intermed {  {\intermed \Sigma}  } , \intermed {  {\intermed { \Gamma_C } }  } }_{\intermed {  \bullet  } } $.
\end{corollaryB}

\proof

The choices from $ \intermed {  {\intermed { R^{SN} } }  } \in \mathcal{F^{SN} } \llbracket \intermed {  {\intermed \Gamma}  } \rrbracket ^{\intermed {  {\intermed \Sigma}  }, \intermed {  {\intermed { \Gamma_C } }  } } $ always satisfy the
precondition of Compositionality for Strong Normalization
(Lemma~\ref{lemma:comp_sn}). Therefore,
we can do
induction on ${\intermed \Gamma}$ and apply Compositionality for Strong Normalization
(Lemma~\ref{lemma:comp_sn}), in combination with the induction hypothesis,
to prove the goal.

\qed

\begin{theoremB}[Strong Normalization - Part A]\ \\
  \label{thmA:sn:part:a}
  If \( \intermed {  {\intermed \Sigma}  } ; \intermed {  {\intermed { \Gamma_C } }  } ; \intermed {  {\intermed \Gamma}  }  \vdash_{tm}  \intermed {  {\intermed e}  } : \intermed {  {\intermed \sigma}  } \itot{ \rightsquigarrow  \target {  {\target e }  } } \)
  then \( \forall  \intermed {  {\intermed { R^{SN} } }  } \in \mathcal{F^{SN} } \llbracket \intermed {  {\intermed \Gamma}  } \rrbracket ^{\intermed {  {\intermed \Sigma}  }, \intermed {  {\intermed { \Gamma_C } }  } } \),
  \( \intermed {  {\intermed { \phi^{SN} } }  } \in \mathcal{G^{SN} } \llbracket \intermed {  {\intermed \Gamma}  } \rrbracket ^{\intermed {  {\intermed \Sigma}  } , \intermed {  {\intermed { \Gamma_C } }  } }_{\intermed {  {\intermed { R^{SN} } }  } } \) and
  \( \intermed {  {\intermed { \gamma^{SN} } }  } \in \mathcal{H^{SN} } \llbracket \intermed {  {\intermed \Gamma}  } \rrbracket ^{\intermed {  {\intermed \Sigma}  } , \intermed {  {\intermed { \Gamma_C } }  } }_{\intermed {  {\intermed { R^{SN} } }  } } \), \\
  it holds that \( \intermed {   {\intermed { \gamma^{SN} } }  ( {\intermed { \phi^{SN} } }  ( {\intermed { R^{SN} } }  ( {\intermed e} )))   } \in \mathcal{SN} \llbracket \intermed {  {\intermed \sigma}  } \rrbracket ^{\intermed {  {\intermed \Sigma}  } , \intermed {  {\intermed { \Gamma_C } }  } }_{\intermed {  {\intermed { R^{SN} } }  } } \).
\end{theoremB}

\proof By induction on the first hypothesis of the theorem.\\
\proofStep{\textsc{iTm-true}}
{\drule{iTm-true}}
{We know $ \intermed {   {\intermed { \gamma^{SN} } }  ( {\intermed { \phi^{SN} } }  ( {\intermed { R^{SN} } }  ( \mathit{\intermed {True} } )))   } = \intermed {  \mathit{\intermed {True} }  } $. So the goal is
  $ \intermed {  \mathit{\intermed {True} }  } \in \mathcal{SN} \llbracket \intermed {  {\intermed \sigma}  } \rrbracket ^{\intermed {  {\intermed \Sigma}  } , \intermed {  {\intermed { \Gamma_C } }  } }_{\intermed {  {\intermed { R^{SN} } }  } } $. The goal follows directly since
  $ \intermed {  {\intermed \Sigma}  } ; \intermed {  {\intermed { \Gamma_C } }  } ; \intermed {  \bullet  }  \vdash_{tm}  \intermed {  \mathit{\intermed {True} }  } : \intermed {  \mathit{\intermed {Bool} }  } \itot{ \rightsquigarrow  \target {  \mathit{\target {True} }  } } $ and $ \intermed {  {\intermed \Sigma}  }  \vdash  \intermed {  \mathit{\intermed {True} }  }  \longrightarrow^{*}  \intermed {  \mathit{\intermed {True} }  } $. \\
}
\proofStep{\textsc{iTm-false}}
{\drule{iTm-false}}
{ Similar to the \textsc{iTm-true} case. \\
}
\proofStep{\textsc{iTm-var}}
{\drule{iTm-var}}
{
  We know that $ \intermed {  {\intermed { \phi^{SN} } }  } \in \mathcal{G^{SN} } \llbracket \intermed {  {\intermed \Gamma}  } \rrbracket ^{\intermed {  {\intermed \Sigma}  } , \intermed {  {\intermed { \Gamma_C } }  } }_{\intermed {  {\intermed { R^{SN} } }  } } $ and $ ( \intermed {  {\intermed x}  } : \intermed {  {\intermed \sigma}  } )  \in  \intermed {  {\intermed \Gamma}  } $, so we know that $ \intermed {  {\intermed x}  }  \mapsto  \intermed {  {\intermed e}  }  \in  \intermed {  {\intermed { \phi^{SN} } }  } $ for some ${\intermed e}$ with $ \intermed {  {\intermed e}  } \in \mathcal{SN} \llbracket \intermed {  {\intermed \sigma}  } \rrbracket ^{\intermed {  {\intermed \Sigma}  } , \intermed {  {\intermed { \Gamma_C } }  } }_{\intermed {  {\intermed { R^{SN} } }  } } $. Since $ \intermed {  {\intermed e}  } \in \mathcal{SN} \llbracket \intermed {  {\intermed \sigma}  } \rrbracket ^{\intermed {  {\intermed \Sigma}  } , \intermed {  {\intermed { \Gamma_C } }  } }_{\intermed {  {\intermed { R^{SN} } }  } } $, we know from the definition of the relation that ${\intermed e}$ does not
   contain any free variables in ${\intermed \Gamma}$. Consequently, $\intermed{ {\intermed { \gamma^{SN} } }  ( {\intermed { \phi^{SN} } }  ( {\intermed { R^{SN} } }  ( {\intermed e} ))) } =
   {\intermed e}$. Therefore, $ \intermed {   {\intermed { \gamma^{SN} } }  ( {\intermed { \phi^{SN} } }  ( {\intermed { R^{SN} } }  ( {\intermed x} )))   } = \intermed {  {\intermed e}  } $. Now our goal
   becomes $ \intermed {  {\intermed e}  } \in \mathcal{SN} \llbracket \intermed {  {\intermed \sigma}  } \rrbracket ^{\intermed {  {\intermed \Sigma}  } , \intermed {  {\intermed { \Gamma_C } }  } }_{\intermed {  {\intermed { R^{SN} } }  } } $, which we already know. \\
}
\proofStep{\textsc{iTm-let}}
{\drule{iTm-let}}
{
  By induction hypothesis, we know
  \begin{gather}
    \label{eq:sn:let1}
     \intermed {   {\intermed { \gamma^{SN} } }  ( {\intermed { \phi^{SN} } }  ( {\intermed { R^{SN} } }  ( {\intermed e}_{{\mathrm{1}}} )))   } \in \mathcal{SN} \llbracket \intermed {  {\intermed \sigma}_{{\mathrm{1}}}  } \rrbracket ^{\intermed {  {\intermed \Sigma}  } , \intermed {  {\intermed { \Gamma_C } }  } }_{\intermed {  {\intermed { R^{SN} } }  } }  \\
    \label{eq:sn:let1b}
     \intermed {   {\intermed { \gamma^{SN} } }  ( {\intermed { \phi^{SN} } }_{{\mathrm{2}}}  ( {\intermed { R^{SN} } }  ( {\intermed e}_{{\mathrm{2}}} )))   } \in \mathcal{SN} \llbracket \intermed {  {\intermed \sigma}_{{\mathrm{2}}}  } \rrbracket ^{\intermed {  {\intermed \Sigma}  } , \intermed {  {\intermed { \Gamma_C } }  } }_{\intermed {  {\intermed { R^{SN} } }  } } 
  \end{gather}
  Given Equation~\ref{eq:sn:let1}, we can choose $ \intermed {  {\intermed { \phi^{SN} } }_{{\mathrm{2}}}  } = \intermed {   {\intermed { \phi^{SN} } } ,  {\intermed x}  \mapsto   {\intermed { \gamma^{SN} } }  ( {\intermed { \phi^{SN} } }  ( {\intermed { R^{SN} } }  ( {\intermed e}_{{\mathrm{1}}} )))    } $. Equation~\ref{eq:sn:let1b} thus reduces to:
  \begin{gather}
    \label{eq:sn:let2}
     \intermed {   {\intermed { \gamma^{SN} } }  (  {\intermed { \phi^{SN} } } ,  {\intermed x}  \mapsto   {\intermed { \gamma^{SN} } }  ( {\intermed { \phi^{SN} } }  ( {\intermed { R^{SN} } }  ( {\intermed e}_{{\mathrm{1}}} )))    ( {\intermed { R^{SN} } }  ( {\intermed e}_{{\mathrm{2}}} )))   } \in \mathcal{SN} \llbracket \intermed {  {\intermed \sigma}_{{\mathrm{2}}}  } \rrbracket ^{\intermed {  {\intermed \Sigma}  } , \intermed {  {\intermed { \Gamma_C } }  } }_{\intermed {  {\intermed { R^{SN} } }  } } 
  \end{gather}
  Simplifying Equation~\ref{eq:sn:let2} results in:
  \begin{gather}
    \label{eq:sn:let3}
     \intermed {  [   {\intermed { \gamma^{SN} } }  ( {\intermed { \phi^{SN} } }  ( {\intermed { R^{SN} } }  ( {\intermed e}_{{\mathrm{1}}} )))   /  {\intermed x}  ]  \ottsym{(}   {\intermed { \gamma^{SN} } }  ( {\intermed { \phi^{SN} } }  ( {\intermed { R^{SN} } }  ( {\intermed e}_{{\mathrm{2}}} )))   \ottsym{)}  } \in \mathcal{SN} \llbracket \intermed {  {\intermed \sigma}_{{\mathrm{2}}}  } \rrbracket ^{\intermed {  {\intermed \Sigma}  } , \intermed {  {\intermed { \Gamma_C } }  } }_{\intermed {  {\intermed { R^{SN} } }  } } 
  \end{gather}
  Our goal is $ \intermed {   {\intermed { \gamma^{SN} } }  ( {\intermed { \phi^{SN} } }  ( {\intermed { R^{SN} } }  ( \textbf{\intermed {let} } \, {\intermed x}  \ottsym{:}  {\intermed \sigma}_{{\mathrm{1}}}  \ottsym{=}  {\intermed e}_{{\mathrm{1}}} \, \textbf{\intermed {in} } \, {\intermed e}_{{\mathrm{2}}} )))   } \in \mathcal{SN} \llbracket \intermed {  {\intermed \sigma}_{{\mathrm{2}}}  } \rrbracket ^{\intermed {  {\intermed \Sigma}  } , \intermed {  {\intermed { \Gamma_C } }  } }_{\intermed {  {\intermed { R^{SN} } }  } } $.

  We know that
  \begin{gather}
    \intermed{ {\intermed { \gamma^{SN} } }  ( {\intermed { \phi^{SN} } }  ( {\intermed { R^{SN} } }  ( \textbf{\intermed {let} } \, {\intermed x}  \ottsym{:}  {\intermed \sigma}_{{\mathrm{1}}}  \ottsym{=}  {\intermed e}_{{\mathrm{1}}} \, \textbf{\intermed {in} } \, {\intermed e}_{{\mathrm{2}}} ))) } \nonumber \\
    \label{eq:sn:let4}
    = \intermed{\textbf{\intermed {let} } \, {\intermed x}  \ottsym{:}  {\intermed \sigma}_{{\mathrm{1}}}  \ottsym{=}  \ottsym{(}   {\intermed { \gamma^{SN} } }  ( {\intermed { \phi^{SN} } }  ( {\intermed { R^{SN} } }  ( {\intermed e}_{{\mathrm{1}}} )))   \ottsym{)} \, \textbf{\intermed {in} } \, \ottsym{(}   {\intermed { \gamma^{SN} } }  ( {\intermed { \phi^{SN} } }  ( {\intermed { R^{SN} } }  ( {\intermed e}_{{\mathrm{2}}} )))   \ottsym{)}} \\
    {\intermed \Sigma}  \vdash  \intermed{\textbf{\intermed {let} } \, {\intermed x}  \ottsym{:}  {\intermed \sigma}_{{\mathrm{1}}}  \ottsym{=}  \ottsym{(}   {\intermed { \gamma^{SN} } }  ( {\intermed { \phi^{SN} } }  ( {\intermed { R^{SN} } }  ( {\intermed e}_{{\mathrm{1}}} )))   \ottsym{)} \, \textbf{\intermed {in} } \, \ottsym{(}   {\intermed { \gamma^{SN} } }  ( {\intermed { \phi^{SN} } }  ( {\intermed { R^{SN} } }  ( {\intermed e}_{{\mathrm{2}}} )))   \ottsym{)}} \nonumber \\
    \label{eq:sn:let5}
     \longrightarrow  \intermed{[   {\intermed { \gamma^{SN} } }  ( {\intermed { \phi^{SN} } }  ( {\intermed { R^{SN} } }  ( {\intermed e}_{{\mathrm{1}}} )))   /  {\intermed x}  ]  \ottsym{(}   {\intermed { \gamma^{SN} } }  ( {\intermed { \phi^{SN} } }  ( {\intermed { R^{SN} } }  ( {\intermed e}_{{\mathrm{2}}} )))   \ottsym{)}}
  \end{gather}
  Consequently, by Substitution for Context Interpretation (Lemma~\ref{lemma:sn_ctx_inter}), we
  know that 
  \begin{gather}
    \label{eq:sn:let6}
     \intermed {  {\intermed \Sigma}  } ; \intermed {  {\intermed { \Gamma_C } }  } ; \intermed {  \bullet  }  \vdash_{tm}  \intermed {  \textbf{\intermed {let} } \, {\intermed x}  \ottsym{:}  {\intermed \sigma}_{{\mathrm{1}}}  \ottsym{=}  \ottsym{(}   {\intermed { \gamma^{SN} } }  ( {\intermed { \phi^{SN} } }  ( {\intermed { R^{SN} } }  ( {\intermed e}_{{\mathrm{1}}} )))   \ottsym{)} \, \textbf{\intermed {in} } \, \ottsym{(}   {\intermed { \gamma^{SN} } }  ( {\intermed { \phi^{SN} } }  ( {\intermed { R^{SN} } }  ( {\intermed e}_{{\mathrm{2}}} )))   \ottsym{)}  } : \intermed {  {\intermed { R^{SN} } }  \ottsym{(}  {\intermed \sigma}_{{\mathrm{2}}}  \ottsym{)}  } \itot{ \rightsquigarrow  \target {  {\target e }  } } 
  \end{gather}
  By Equations~\ref{eq:sn:let3}, \ref{eq:sn:let5} and \ref{eq:sn:let6}, in
  combination with Strong
  Normalization preserved by forward/backward reduction
  (Lemma~\ref{lemma:sn_pres}), we get that
  \begin{gather}
    \label{eq:sn:let7}
     \intermed {  \textbf{\intermed {let} } \, {\intermed x}  \ottsym{:}  {\intermed \sigma}_{{\mathrm{1}}}  \ottsym{=}  \ottsym{(}   {\intermed { \gamma^{SN} } }  ( {\intermed { \phi^{SN} } }  ( {\intermed { R^{SN} } }  ( {\intermed e}_{{\mathrm{1}}} )))   \ottsym{)} \, \textbf{\intermed {in} } \, \ottsym{(}   {\intermed { \gamma^{SN} } }  ( {\intermed { \phi^{SN} } }  ( {\intermed { R^{SN} } }  ( {\intermed e}_{{\mathrm{2}}} )))   \ottsym{)}  } \in \mathcal{SN} \llbracket \intermed {  {\intermed \sigma}_{{\mathrm{2}}}  } \rrbracket ^{\intermed {  {\intermed \Sigma}  } , \intermed {  {\intermed { \Gamma_C } }  } }_{\intermed {  {\intermed { R^{SN} } }  } } 
  \end{gather}
  The goal follows from Equations~\ref{eq:sn:let4} and \ref{eq:sn:let7}. \\
}
\proofStep{\textsc{iTm-method}}
{\drule{iTm-method}}
{
  By inversion on the dictionary typing, we have two cases:

  \begin{itemize}
    \item $\drule{D-var}$
      with $ \intermed {  {\intermed Q}  } = \intermed {  {\intermed {\mathit{TC} } } \, {\intermed \sigma}  } $.
      
      Given $ \intermed {  {\intermed { \gamma^{SN} } }  } \in \mathcal{H^{SN} } \llbracket \intermed {  {\intermed \Gamma}  } \rrbracket ^{\intermed {  {\intermed \Sigma}  } , \intermed {  {\intermed { \Gamma_C } }  } }_{\intermed {  {\intermed { R^{SN} } }  } } $, we know that there
      exists a ${\intermed {dv} }$ in ${\intermed { \gamma^{SN} } }$ such that
      \begin{gather}
        \label{eq:sn:method1}
         \intermed {  {\intermed \Sigma}  } ; \intermed {  {\intermed { \Gamma_C } }  } ; \intermed {  \bullet  }  \vdash_{d}  \intermed {  {\intermed {dv} }  } : \intermed {  {\intermed {\mathit{TC} } } \, {\intermed { R^{SN} } }  \ottsym{(}  {\intermed \sigma}  \ottsym{)}  } \itot{\,  \rightsquigarrow  \target {  {\target e }  } } .
      \end{gather}
      Without loss of generality, suppose
      \begin{gather}
        \label{eq:sn:method3}
       \intermed {  {\intermed {dv} }  } = \intermed {  {\intermed D } \, \overline {\intermed \sigma}_{\ottmv{j}} \, {\overline {\intermed { dv } } }_{\ottmv{j}}  } .
      \end{gather}
      Therefore
      \begin{gather}
        \label{eq:sn:method2}
       \intermed {   {\intermed { \gamma^{SN} } }  ( {\intermed { \phi^{SN} } }  ( {\intermed { R^{SN} } }  ( {\intermed d}  \ottsym{.}  {\intermed m} )))   } = \intermed {  \ottsym{(}  {\intermed D } \, \overline {\intermed \sigma}_{\ottmv{j}} \, {\overline {\intermed { dv } } }_{\ottmv{j}}  \ottsym{)}  \ottsym{.}  {\intermed m}  } .
      \end{gather}
      Now our goal is to prove that:
      \begin{gather}
        \label{eq:sn:method21}
       \intermed {  \ottsym{(}  {\intermed D } \, \overline {\intermed \sigma}_{\ottmv{j}} \, {\overline {\intermed { dv } } }_{\ottmv{j}}  \ottsym{)}  \ottsym{.}  {\intermed m}  } \in \mathcal{SN} \llbracket \intermed {  [  {\intermed \sigma}  /  {\intermed a }  ]  {\intermed \sigma}'  } \rrbracket ^{\intermed {  {\intermed \Sigma}  } , \intermed {  {\intermed { \Gamma_C } }  } }_{\intermed {  {\intermed { R^{SN} } }  } } .
      \end{gather}
      Substituting Equation~\ref{eq:sn:method3} in Equation~\ref{eq:sn:method1}
      results in:
      \begin{gather}
        \label{eq:sn:method4}
         \intermed {  {\intermed \Sigma}  } ; \intermed {  {\intermed { \Gamma_C } }  } ; \intermed {  \bullet  }  \vdash_{d}  \intermed {  {\intermed D } \, \overline {\intermed \sigma}_{\ottmv{j}} \, {\overline {\intermed { dv } } }_{\ottmv{j}}  } : \intermed {  {\intermed {\mathit{TC} } } \, {\intermed { R^{SN} } }  \ottsym{(}  {\intermed \sigma}  \ottsym{)}  } \itot{\,  \rightsquigarrow  \target {  {\target e }  } } .
      \end{gather}
      By inversion, we have
      \begin{gather}
        \label{eq:sn:method5}
         ( \intermed {  {\intermed D }  } : \intermed {  \forall  {\overline {\intermed a} }_{\ottmv{j}}  \ottsym{.}  {\overline {\intermed Q} }_{\ottmv{i}}  \Rightarrow  {\intermed {\mathit{TC} } } \, {\intermed \sigma}_{{\mathrm{1}}}  } ) . \intermed {  {\intermed m}  }  \mapsto  \intermed {  \Lambda  {\overline {\intermed a} }_{\ottmv{j}}  \ottsym{.}  \lambda  {\overline {\intermed \delta} }_{\ottmv{i}}  \ottsym{:}  {\overline {\intermed Q} }_{\ottmv{i}}  \ottsym{.}  {\intermed e}  }  \in  \intermed {  {\intermed \Sigma}  } \\
        \label{eq:sn:method51}
        \ottcomp{ \intermed {  {\intermed \Sigma}  } ; \intermed {  {\intermed { \Gamma_C } }  } ; \intermed {  \bullet  }  \vdash_{d}  \intermed {  {\intermed {dv} }_{\ottmv{i}}  } : \intermed {  [  \overline {\intermed \sigma}_{\ottmv{j}}  /  {\overline {\intermed a} }_{\ottmv{j}}  ]  {\intermed Q}_{\ottmv{i}}  } \itot{\,  \rightsquigarrow  \target {  {\target e }_{\ottmv{i}}  } } }{\ottmv{i}} \\
        \label{eq:sn:method52}
         \intermed {  {\intermed { R^{SN} } }  \ottsym{(}  {\intermed \sigma}  \ottsym{)}  } = \intermed {  [  \overline {\intermed \sigma}_{\ottmv{j}}  /  {\overline {\intermed a} }_{\ottmv{j}}  ]  {\intermed \sigma}_{{\mathrm{1}}}  }  \\
        \label{eq:sn:method53}
         \vdash_{ctx}  \intermed {  {\intermed \Sigma}  } ; \intermed {  {\intermed { \Gamma_C } }  } ; \intermed {  {\intermed \Gamma}  }  \\
        \label{eq:sn:method53b}
        \ottcomp{ \intermed {  {\intermed { \Gamma_C } }  } ; \intermed {  \bullet  \ottsym{,}  {\overline {\intermed a} }_{\ottmv{j}}  }  \vdash_{\mathit {Q} }  \intermed {  {\intermed Q}_{\ottmv{i}}  }\, \itot{ \rightsquigarrow  \target {  {\target \sigma }  } } }{\ottmv{i}} \\
        \label{eq:sn:method53d}
        \ottcomp{ \intermed {  {\intermed { \Gamma_C } }  } ; \intermed {  {\intermed \Gamma}  }  \vdash_{ty}  \intermed {  {\intermed \sigma}_{\ottmv{j}}  } \itot{ \rightsquigarrow  \target {  {\target \sigma }_{\ottmv{j}}  } } }{\ottmv{j}} \\
        \label{eq:sn:method54}
         \intermed {  {\intermed \Sigma}_{{\mathrm{1}}}  } ; \intermed {  {\intermed { \Gamma_C } }  } ; \intermed {  \bullet  \ottsym{,}  {\overline {\intermed a} }_{\ottmv{j}}  \ottsym{,}  {\overline {\intermed \delta} }_{\ottmv{i}}  \ottsym{:}  {\overline {\intermed Q} }_{\ottmv{i}}  }  \vdash_{tm}  \intermed {  {\intermed e}  } : \intermed {  [  {\intermed \sigma}_{{\mathrm{1}}}  /  {\intermed a }  ]  {\intermed \sigma}'  } \itot{ \rightsquigarrow  \target {  {\target e }  } } 
        \\
        \nonumber
        \textit{where}~ \intermed {  {\intermed \Sigma}  } = \intermed {  {\intermed \Sigma}_{{\mathrm{1}}}  \ottsym{,}  \ottsym{(}  {\intermed D }  \ottsym{:}  \forall  {\overline {\intermed a} }_{\ottmv{j}}  \ottsym{.}  {\overline {\intermed Q} }_{\ottmv{i}}  \Rightarrow  {\intermed {\mathit{TC} } } \, {\intermed \sigma}_{{\mathrm{1}}}  \ottsym{)}  \ottsym{.}  {\intermed m}  \mapsto  \Lambda  {\overline {\intermed a} }_{\ottmv{j}}  \ottsym{.}  \lambda  {\overline {\intermed \delta} }_{\ottmv{i}}  \ottsym{:}  {\overline {\intermed Q} }_{\ottmv{i}}  \ottsym{.}  {\intermed e}  \ottsym{,}  {\intermed \Sigma}_{{\mathrm{2}}}  } 
      \end{gather}
      By evaluation rule (\textsc{iEval}), together with Equation~\ref{eq:sn:method5}, we
      know that:
      \begin{gather}
        \label{eq:sn:method10}
         \intermed {  {\intermed \Sigma}  }  \vdash  \intermed {  \ottsym{(}  {\intermed D } \, \overline {\intermed \sigma}_{\ottmv{j}} \, {\overline {\intermed { dv } } }_{\ottmv{j}}  \ottsym{)}  \ottsym{.}  {\intermed m}  }  \longrightarrow  \intermed {  \ottsym{(}  \Lambda  {\overline {\intermed a} }_{\ottmv{j}}  \ottsym{.}  \lambda  {\overline {\intermed \delta} }_{\ottmv{i}}  \ottsym{:}  {\overline {\intermed Q} }_{\ottmv{i}}  \ottsym{.}  {\intermed e}  \ottsym{)} \, \overline {\intermed \sigma}_{\ottmv{j}} \, {\overline {\intermed { dv } } }_{\ottmv{j}}  } 
      \end{gather}
      By Strong Normalization preserved by forward/backward reduction
      (Lemma~\ref{lemma:sn_pres}), Goal~\ref{eq:sn:method21} becomes:
      \begin{gather}
        \label{eq:sn:method11}
       \intermed {  {\intermed e} \, \overline {\intermed \sigma}_{\ottmv{j}} \, {\overline {\intermed { dv } } }_{\ottmv{j}}  } \in \mathcal{SN} \llbracket \intermed {  [  {\intermed \sigma}  /  {\intermed a }  ]  {\intermed \sigma}'  } \rrbracket ^{\intermed {  {\intermed \Sigma}  } , \intermed {  {\intermed { \Gamma_C } }  } }_{\intermed {  {\intermed { R^{SN} } }  } } .
      \end{gather}
      By applying weakening Lemma~\ref{lemma:q_weakening} on
      Equation~\ref{eq:sn:method53b}, we get:
      \begin{gather}
        \ottcomp{ \intermed {  {\intermed { \Gamma_C } }  } ; \intermed {  {\intermed \Gamma}  }  \vdash_{\mathit {Q} }  \intermed {  {\intermed Q}_{\ottmv{i}}  }\, \itot{ \rightsquigarrow  \target {  {\target \sigma }  } } }{\ottmv{i}} \\
        \label{eq:sn:method53c}
      \end{gather}
      By applying rules \textsc{iTm-forallI} and \textsc{iTm-constrI} (in
      combination with Equation~\ref{eq:sn:method53c}) on
      Equation~\ref{eq:sn:method54}, we get that:
      \begin{gather}
        \label{eq:sn:method55}
         \intermed {  {\intermed \Sigma}_{{\mathrm{1}}}  } ; \intermed {  {\intermed { \Gamma_C } }  } ; \intermed {  \bullet  }  \vdash_{tm}  \intermed {  \Lambda  {\overline {\intermed a} }_{\ottmv{j}}  \ottsym{.}  \lambda  {\overline {\intermed \delta} }  \ottsym{:}  {\overline {\intermed Q} }_{\ottmv{i}}  \ottsym{.}  {\intermed e}  } : \intermed {  \forall  {\overline {\intermed a} }_{\ottmv{j}}  \ottsym{.}   {\overline {\intermed Q} }_{\ottmv{i}}  \Rightarrow  [  {\intermed \sigma}_{{\mathrm{1}}}  /  {\intermed a }  ]  {\intermed \sigma}'   } \itot{ \rightsquigarrow  \target {  {\target e }  } } 
      \end{gather}
      By applying rules \textsc{iTm-forallE} (in combination with
      Equation~\ref{eq:sn:method53d}) and \textsc{iTm-constrE} (in combination
      with Equation~\ref{eq:sn:method51}) on Equation~\ref{eq:sn:method55}, we
      get that:
      \begin{gather}
        \label{eq:sn:method301}
         \intermed {  {\intermed \Sigma}_{{\mathrm{1}}}  } ; \intermed {  {\intermed { \Gamma_C } }  } ; \intermed {  \bullet  }  \vdash_{tm}  \intermed {  {\intermed e} \, \overline {\intermed \sigma}_{\ottmv{j}} \, {\overline {\intermed { dv } } }_{\ottmv{i}}  } : \intermed {  [  \overline {\intermed \sigma}_{\ottmv{j}}  /  {\overline {\intermed a} }_{\ottmv{j}}  ]  [  {\intermed \sigma}_{{\mathrm{1}}}  /  {\intermed a }  ]  {\intermed \sigma}'  } \itot{ \rightsquigarrow  \target {  {\target e }  } }  
      \end{gather}
      From \textsc{iCtx-clsEnv}, we know that $\intermed{{\intermed \sigma}'}$ only contains one free
      variable ${\intermed a }$. We can thus simplify Equation~\ref{eq:sn:method301} to:
      \begin{gather}
        \label{eq:sn:method303}
         \intermed {  {\intermed \Sigma}_{{\mathrm{1}}}  } ; \intermed {  {\intermed { \Gamma_C } }  } ; \intermed {  \bullet  }  \vdash_{tm}  \intermed {  {\intermed e} \, \overline {\intermed \sigma}_{\ottmv{j}} \, {\overline {\intermed { dv } } }_{\ottmv{i}}  } : \intermed {  [  [  \overline {\intermed \sigma}_{\ottmv{j}}  /  {\overline {\intermed a} }_{\ottmv{j}}  ]  {\intermed \sigma}_{{\mathrm{1}}}  /  {\intermed a }  ]  {\intermed \sigma}'  } \itot{ \rightsquigarrow  \target {  {\target e }  } }  
      \end{gather}
      And by substituting Equation~\ref{eq:sn:method52}:
      \begin{gather}
        \label{eq:sn:method304}
         \intermed {  {\intermed \Sigma}_{{\mathrm{1}}}  } ; \intermed {  {\intermed { \Gamma_C } }  } ; \intermed {  \bullet  }  \vdash_{tm}  \intermed {  {\intermed e} \, \overline {\intermed \sigma}_{\ottmv{j}} \, {\overline {\intermed { dv } } }_{\ottmv{i}}  } : \intermed {  [  {\intermed { R^{SN} } }  \ottsym{(}  {\intermed \sigma}  \ottsym{)}  /  {\intermed a }  ]  {\intermed \sigma}'  } \itot{ \rightsquigarrow  \target {  {\target e }  } }  
      \end{gather}
      Again, since $\intermed{{\intermed \sigma}'}$ contains only one free variable ${\intermed a }$, we can
      further rewrite this result to:
      \begin{gather}
        \label{eq:sn:method305}
         \intermed {  {\intermed \Sigma}_{{\mathrm{1}}}  } ; \intermed {  {\intermed { \Gamma_C } }  } ; \intermed {  \bullet  }  \vdash_{tm}  \intermed {  {\intermed e} \, \overline {\intermed \sigma}_{\ottmv{j}} \, {\overline {\intermed { dv } } }_{\ottmv{i}}  } : \intermed {  {\intermed { R^{SN} } }  \ottsym{(}  [  {\intermed \sigma}  /  {\intermed a }  ]  {\intermed \sigma}'  \ottsym{)}  } \itot{ \rightsquigarrow  \target {  {\target e }  } }  
      \end{gather}
      By induction hypothesis on Equation~\ref{eq:sn:method305}, we get:
      \begin{gather}
         \intermed {  {\intermed e} \, \overline {\intermed \sigma}_{\ottmv{j}} \, {\overline {\intermed { dv } } }_{\ottmv{i}}  } \in \mathcal{SN} \llbracket \intermed {  {\intermed { R^{SN} } }  \ottsym{(}  [  {\intermed \sigma}  /  {\intermed a }  ]  {\intermed \sigma}'  \ottsym{)}  } \rrbracket ^{\intermed {  {\intermed \Sigma}'  } , \intermed {  {\intermed { \Gamma_C } }  } }_{\intermed {  \bullet  } } 
      \end{gather}
      By weakening over $\intermed{{\intermed \Sigma}'}$ (Lemma~\ref{lemma:sn_menv_weakening}), we get:
      \begin{gather}
        \label{eq:sn:method306}
         \intermed {  {\intermed e} \, \overline {\intermed \sigma}_{\ottmv{j}} \, {\overline {\intermed { dv } } }_{\ottmv{i}}  } \in \mathcal{SN} \llbracket \intermed {  {\intermed { R^{SN} } }  \ottsym{(}  [  {\intermed \sigma}  /  {\intermed a }  ]  {\intermed \sigma}'  \ottsym{)}  } \rrbracket ^{\intermed {  {\intermed \Sigma}  } , \intermed {  {\intermed { \Gamma_C } }  } }_{\intermed {  \bullet  } } 
      \end{gather}
      Because we know that $ \intermed {  {\intermed { R^{SN} } }  } \in \mathcal{F^{SN} } \llbracket \intermed {  {\intermed \Gamma}  } \rrbracket ^{\intermed {  {\intermed \Sigma}  }, \intermed {  {\intermed { \Gamma_C } }  } } $, by
      Compositionality for Strong Normalization (Context Interpretation)
      (Corollary~\ref{cor:comp_sn_ctx}), Equation~\ref{eq:sn:method306} proves
      our Goal~\ref{eq:sn:method11}.
      
    \item $\drule{D-con}$ \\
      The goal to be proven is the following:
      \begin{gather}
        \label{eq:sn:method:con:g1}
         \intermed {   {\intermed { \gamma^{SN} } }  ( {\intermed { \phi^{SN} } }  ( {\intermed { R^{SN} } }  ( {\intermed d}  \ottsym{.}  {\intermed m} )))   } \in \mathcal{SN} \llbracket \intermed {  [  {\intermed \sigma}  /  {\intermed a }  ]  {\intermed \sigma}'  } \rrbracket ^{\intermed {  {\intermed \Sigma}  } , \intermed {  {\intermed { \Gamma_C } }  } }_{\intermed {  {\intermed { R^{SN} } }  } }  \\
        \label{eq:sn:method:con:g2}
        \textit{where}~ \intermed {  {\intermed \sigma}  } = \intermed {  [  \overline {\intermed \sigma}_{\ottmv{j}}  /  {\overline {\intermed a} }_{\ottmv{j}}  ]  {\intermed \sigma}_{\ottmv{q}}  } 
      \end{gather}
      From the rule premise, we know that:
      \begin{gather}
        \label{eq:sn:method:con:g6}
         ( \intermed {  {\intermed D }  } : \intermed {  \forall  {\overline {\intermed a} }_{\ottmv{j}}  \ottsym{.}  {\overline {\intermed Q} }_{\ottmv{i}}  \Rightarrow  {\intermed {\mathit{TC} } } \, {\intermed \sigma}_{\ottmv{q}}  } ) . \intermed {  {\intermed m}  }  \mapsto  \intermed {  \Lambda  {\overline {\intermed a} }_{\ottmv{j}}  \ottsym{.}  \lambda  {\overline {\intermed \delta} }_{\ottmv{i}}  \ottsym{:}  {\overline {\intermed Q} }_{\ottmv{i}}  \ottsym{.}  {\intermed e}  }  \in  \intermed {  {\intermed \Sigma}  }  \\
        \label{eq:sn:method:con:g5}
        \ottcomp{ \intermed {  {\intermed { \Gamma_C } }  } ; \intermed {  \bullet  \ottsym{,}  {\overline {\intermed a} }_{\ottmv{j}}  }  \vdash_{\mathit {Q} }  \intermed {  {\intermed Q}_{\ottmv{i}}  }\, \itot{ \rightsquigarrow  \target {  {\target \sigma }_{\ottmv{i}}  } } }{\ottmv{i}} \\
        \label{eq:sn:method:con:g7}
        \ottcomp{ \intermed {  {\intermed { \Gamma_C } }  } ; \intermed {  {\intermed \Gamma}  }  \vdash_{ty}  \intermed {  {\intermed \sigma}_{\ottmv{j}}  } \itot{ \rightsquigarrow  \target {  {\target \sigma }_{\ottmv{j}}  } } }{\ottmv{j}} \\
        \label{eq:sn:method:con:g3}
         \intermed {  {\intermed \Sigma}_{{\mathrm{1}}}  } ; \intermed {  {\intermed { \Gamma_C } }  } ; \intermed {  \bullet  \ottsym{,}  {\overline {\intermed a} }_{\ottmv{j}}  \ottsym{,}  {\overline {\intermed \delta} }_{\ottmv{i}}  \ottsym{:}  {\overline {\intermed Q} }_{\ottmv{i}}  }  \vdash_{tm}  \intermed {  {\intermed e}  } : \intermed {  [  {\intermed \sigma}_{\ottmv{q}}  /  {\intermed a }  ]  {\intermed \sigma}_{\ottmv{m}}  } \itot{ \rightsquigarrow  \target {  {\target e }  } }  \\
        \nonumber
        \textit{where}~ \intermed {  {\intermed \Sigma}  } = \intermed {  {\intermed \Sigma}_{{\mathrm{1}}}  \ottsym{,}  \ottsym{(}  {\intermed D }  \ottsym{:}  \forall  {\overline {\intermed a} }_{\ottmv{j}}  \ottsym{.}  {\overline {\intermed Q} }_{\ottmv{i}}  \Rightarrow  {\intermed {\mathit{TC} } } \, {\intermed \sigma}_{\ottmv{q}}  \ottsym{)}  \ottsym{.}  {\intermed m}  \mapsto  \Lambda  {\overline {\intermed a} }_{\ottmv{j}}  \ottsym{.}  \lambda  {\overline {\intermed \delta} }_{\ottmv{i}}  \ottsym{:}  {\overline {\intermed Q} }_{\ottmv{i}}  \ottsym{.}  {\intermed e}  \ottsym{,}  {\intermed \Sigma}_{{\mathrm{2}}}  } 
      \end{gather}

      By evaluation rule (\textsc{iEval}), we know that:
      \begin{gather}
        \label{eq:sn:method:con1}
         \intermed {  {\intermed \Sigma}  }  \vdash  \intermed {  \ottsym{(}  {\intermed D } \, \overline {\intermed \sigma}_{\ottmv{j}} \, {\overline {\intermed d} }_{\ottmv{i}}  \ottsym{)}  \ottsym{.}  {\intermed m}  }  \longrightarrow  \intermed {  \ottsym{(}  \Lambda  {\overline {\intermed a} }_{\ottmv{j}}  \ottsym{.}  \lambda  {\overline {\intermed \delta} }_{\ottmv{i}}  \ottsym{:}  {\overline {\intermed Q} }_{\ottmv{i}}  \ottsym{.}  {\intermed e}  \ottsym{)} \, \overline {\intermed \sigma}_{\ottmv{j}} \, {\overline {\intermed d} }_{\ottmv{i}}  } 
      \end{gather}
      By applying the substitutions, we can verify that:
      \begin{gather*}
         \intermed {  {\intermed \Sigma}  }  \vdash  \intermed {   {\intermed { \gamma^{SN} } }  ( {\intermed { \phi^{SN} } }  ( {\intermed { R^{SN} } }  ( \ottsym{(}  {\intermed D } \, \overline {\intermed \sigma}_{\ottmv{j}} \, {\overline {\intermed d} }_{\ottmv{i}}  \ottsym{)}  \ottsym{.}  {\intermed m} )))   }  \longrightarrow  \intermed {  \ottsym{(}  \Lambda  {\overline {\intermed a} }_{\ottmv{j}}  \ottsym{.}  \lambda  {\overline {\intermed \delta} }_{\ottmv{i}}  \ottsym{:}  {\overline {\intermed Q} }_{\ottmv{i}}  \ottsym{.}  {\intermed e}  \ottsym{)} \, {\intermed { R^{SN} } }  \ottsym{(}  \overline {\intermed \sigma}_{\ottmv{j}}  \ottsym{)} \, {\intermed { R^{SN} } }  \ottsym{(}  {\intermed { \gamma^{SN} } }  \ottsym{(}  {\overline {\intermed d} }_{\ottmv{i}}  \ottsym{)}  \ottsym{)}  } 
      \end{gather*}
      By Strong Normalization preserved by forward/backward reduction
      (Lemma~\ref{lemma:sn_pres}), our goal becomes:
      \begin{gather}
        \label{eq:sn:method:con2}
       \intermed {  \ottsym{(}  \Lambda  {\overline {\intermed a} }_{\ottmv{j}}  \ottsym{.}  \lambda  {\overline {\intermed \delta} }_{\ottmv{i}}  \ottsym{:}  {\overline {\intermed Q} }_{\ottmv{i}}  \ottsym{.}  {\intermed e}  \ottsym{)} \, {\intermed { R^{SN} } }  \ottsym{(}  \overline {\intermed \sigma}_{\ottmv{j}}  \ottsym{)} \, {\intermed { R^{SN} } }  \ottsym{(}  {\intermed { \gamma^{SN} } }  \ottsym{(}  {\overline {\intermed d} }_{\ottmv{i}}  \ottsym{)}  \ottsym{)}  } \in \mathcal{SN} \llbracket \intermed {  [  {\intermed \sigma}  /  {\intermed a }  ]  {\intermed \sigma}'  } \rrbracket ^{\intermed {  {\intermed \Sigma}  } , \intermed {  {\intermed { \Gamma_C } }  } }_{\intermed {  {\intermed { R^{SN} } }  } } .
      \end{gather}
      By applying rules \textsc{iTm-constrI} (in combination with
      Equation~\ref{eq:sn:method:con:g5}) and \textsc{iTm-forallI},
      Equation~\ref{eq:sn:method:con:g3} reduces to:
      \begin{gather}
        \label{eq:sn:method:con:g4}
         \intermed {  {\intermed \Sigma}_{{\mathrm{1}}}  } ; \intermed {  {\intermed { \Gamma_C } }  } ; \intermed {  \bullet  }  \vdash_{tm}  \intermed {  \Lambda  {\overline {\intermed a} }_{\ottmv{j}}  \ottsym{.}  \lambda  {\overline {\intermed \delta} }_{\ottmv{i}}  \ottsym{:}  {\overline {\intermed Q} }_{\ottmv{i}}  \ottsym{.}  {\intermed e}  } : \intermed {  \forall  {\overline {\intermed a} }_{\ottmv{j}}  \ottsym{.}   {\overline {\intermed Q} }_{\ottmv{i}}  \Rightarrow  [  {\intermed \sigma}_{\ottmv{q}}  /  {\intermed a }  ]  {\intermed \sigma}_{\ottmv{m}}   } \itot{ \rightsquigarrow  \target {  {\target e }  } } 
      \end{gather}
      By applying this to \textsc{iTm-forallE}, in combination with
      Equation~\ref{eq:sn:method:con:g7}, we get: 
      \begin{gather*}
         \intermed {  {\intermed \Sigma}_{{\mathrm{1}}}  } ; \intermed {  {\intermed { \Gamma_C } }  } ; \intermed {  \bullet  }  \vdash_{tm}  \intermed {  \ottsym{(}  \Lambda  {\overline {\intermed a} }_{\ottmv{j}}  \ottsym{.}  \lambda  {\overline {\intermed \delta} }_{\ottmv{i}}  \ottsym{:}  {\overline {\intermed Q} }_{\ottmv{i}}  \ottsym{.}  {\intermed e}  \ottsym{)} \, {\intermed { R^{SN} } }  \ottsym{(}  \overline {\intermed \sigma}_{\ottmv{j}}  \ottsym{)}  } : \intermed {  [  {\intermed { R^{SN} } }  \ottsym{(}  \overline {\intermed \sigma}_{\ottmv{j}}  \ottsym{)}  /  {\overline {\intermed a} }_{\ottmv{j}}  ]   {\overline {\intermed Q} }_{\ottmv{i}}  \Rightarrow  [  {\intermed { R^{SN} } }  \ottsym{(}  \overline {\intermed \sigma}_{\ottmv{j}}  \ottsym{)}  /  {\overline {\intermed a} }_{\ottmv{j}}  ]  [  {\intermed \sigma}_{\ottmv{q}}  /  {\intermed a }  ]  {\intermed \sigma}_{\ottmv{m}}   } \itot{ \rightsquigarrow  \target {  {\target e }_{{\mathrm{1}}}  } } 
      \end{gather*}
      Through weakening Lemma~\ref{lemma:exp_weakening_method}, we know this is
      equivalent to:
      \begin{gather}
        \label{eq:sn:method:con:g4p}
         \intermed {  {\intermed \Sigma}  } ; \intermed {  {\intermed { \Gamma_C } }  } ; \intermed {  \bullet  }  \vdash_{tm}  \intermed {  \ottsym{(}  \Lambda  {\overline {\intermed a} }_{\ottmv{j}}  \ottsym{.}  \lambda  {\overline {\intermed \delta} }_{\ottmv{i}}  \ottsym{:}  {\overline {\intermed Q} }_{\ottmv{i}}  \ottsym{.}  {\intermed e}  \ottsym{)} \, {\intermed { R^{SN} } }  \ottsym{(}  \overline {\intermed \sigma}_{\ottmv{j}}  \ottsym{)}  } : \intermed {  [  {\intermed { R^{SN} } }  \ottsym{(}  \overline {\intermed \sigma}_{\ottmv{j}}  \ottsym{)}  /  {\overline {\intermed a} }_{\ottmv{j}}  ]   {\overline {\intermed Q} }_{\ottmv{i}}  \Rightarrow  [  {\intermed { R^{SN} } }  \ottsym{(}  \overline {\intermed \sigma}_{\ottmv{j}}  \ottsym{)}  /  {\overline {\intermed a} }_{\ottmv{j}}  ]  [  {\intermed \sigma}_{\ottmv{q}}  /  {\intermed a }  ]  {\intermed \sigma}_{\ottmv{m}}   } \itot{ \rightsquigarrow  \target {  {\target e }_{{\mathrm{1}}}  } } 
      \end{gather}
      The 6\textsuperscript{th} rule premise tells us that:
      \begin{gather*}
      \ottcomp{ \intermed {  {\intermed \Sigma}  } ; \intermed {  {\intermed { \Gamma_C } }  } ; \intermed {  {\intermed \Gamma}  }  \vdash_{d}  \intermed {  {\intermed d}_{\ottmv{i}}  } : \intermed {  [  \overline {\intermed \sigma}_{\ottmv{j}}  /  {\overline {\intermed a} }_{\ottmv{j}}  ]  {\intermed Q}_{\ottmv{i}}  } \itot{\,  \rightsquigarrow  \target {  {\target e }_{\ottmv{i}}  } } }{\ottmv{i}}
      \end{gather*}
      From Equation~\ref{eq:sn:method:con:g5}, we know that $\intermed{{\overline {\intermed Q} }_{\ottmv{j}}}$ only
      contains free variables $\intermed{{\overline {\intermed a} }_{\ottmv{j}}}$. By applying the substition, we can thus
      verify that:
      \begin{gather*}
      \ottcomp{ \intermed {  {\intermed \Sigma}  } ; \intermed {  {\intermed { \Gamma_C } }  } ; \intermed {  \bullet  }  \vdash_{d}  \intermed {  {\intermed { R^{SN} } }  \ottsym{(}  {\intermed { \gamma^{SN} } }  \ottsym{(}  {\intermed d}_{\ottmv{i}}  \ottsym{)}  \ottsym{)}  } : \intermed {  [  {\intermed { R^{SN} } }  \ottsym{(}  \overline {\intermed \sigma}_{\ottmv{j}}  \ottsym{)}  /  {\overline {\intermed a} }_{\ottmv{j}}  ]  {\intermed Q}_{\ottmv{i}}  } \itot{\,  \rightsquigarrow  \target {  {\target e }'_{\ottmv{i}}  } } }{\ottmv{i}}
      \end{gather*}
      Therefore, by \textsc{iTm-constrE},
      Equation~\ref{eq:sn:method:con:g4p} is equivalent to:
      \begin{gather}
        \label{eq:sn:method:con:g4b}
         \intermed {  {\intermed \Sigma}  } ; \intermed {  {\intermed { \Gamma_C } }  } ; \intermed {  \bullet  }  \vdash_{tm}  \intermed {  \ottsym{(}  \Lambda  {\overline {\intermed a} }_{\ottmv{j}}  \ottsym{.}  \lambda  {\overline {\intermed \delta} }_{\ottmv{i}}  \ottsym{:}  {\overline {\intermed Q} }_{\ottmv{i}}  \ottsym{.}  {\intermed e}  \ottsym{)} \, {\intermed { R^{SN} } }  \ottsym{(}  \overline {\intermed \sigma}_{\ottmv{j}}  \ottsym{)} \, {\intermed { R^{SN} } }  \ottsym{(}  {\intermed { \gamma^{SN} } }  \ottsym{(}  {\overline {\intermed d} }_{\ottmv{i}}  \ottsym{)}  \ottsym{)}  } : \intermed {  [  {\intermed { R^{SN} } }  \ottsym{(}  \overline {\intermed \sigma}_{\ottmv{j}}  \ottsym{)}  /  {\overline {\intermed a} }_{\ottmv{j}}  ]  [  {\intermed \sigma}_{\ottmv{q}}  /  {\intermed a }  ]  {\intermed \sigma}_{\ottmv{m}}  } \itot{ \rightsquigarrow  \target {  {\target e }_{{\mathrm{1}}}  } } 
      \end{gather}
      From Lemma~\ref{lemma:type_well_inter_typing}, in combination with
      Equation~\ref{eq:sn:method:con:g3}, we know that $\intermed{[  {\intermed \sigma}_{\ottmv{q}}  /  {\intermed a }  ]  {\intermed \sigma}_{\ottmv{m}}}$
      only contains free variables $\intermed{{\overline {\intermed a} }_{\ottmv{j}}}$. we can thus rewrite
      Equation~\ref{eq:sn:method:con:g4b} to:
      \begin{gather}
        \label{eq:sn:method:con3}
         \intermed {  {\intermed \Sigma}  } ; \intermed {  {\intermed { \Gamma_C } }  } ; \intermed {  \bullet  }  \vdash_{tm}  \intermed {  \ottsym{(}  \Lambda  {\overline {\intermed a} }_{\ottmv{j}}  \ottsym{.}  \lambda  {\overline {\intermed \delta} }_{\ottmv{i}}  \ottsym{:}  {\overline {\intermed Q} }_{\ottmv{i}}  \ottsym{.}  {\intermed e}  \ottsym{)} \, {\intermed { R^{SN} } }  \ottsym{(}  \overline {\intermed \sigma}_{\ottmv{j}}  \ottsym{)} \, {\intermed { R^{SN} } }  \ottsym{(}  {\intermed { \gamma^{SN} } }  \ottsym{(}  {\overline {\intermed d} }_{\ottmv{i}}  \ottsym{)}  \ottsym{)}  } : \intermed {  {\intermed { R^{SN} } }  \ottsym{(}  [  \overline {\intermed \sigma}_{\ottmv{j}}  /  {\overline {\intermed a} }_{\ottmv{j}}  ]  [  {\intermed \sigma}_{\ottmv{q}}  /  {\intermed a }  ]  {\intermed \sigma}_{\ottmv{m}}  \ottsym{)}  } \itot{ \rightsquigarrow  \target {  {\target e }_{{\mathrm{1}}}  } } 
      \end{gather}
      From the environment well-formedness we know that \( \intermed {  {\intermed \sigma}'  } = \intermed {  {\intermed \sigma}_{\ottmv{m}}  } \) and
      that $\intermed{{\intermed \sigma}'}$ only contains free variable $\intermed{{\intermed a }}$. We can thus rewrite
      Equation~\ref{eq:sn:method:con3} to:
      \begin{gather}
        \label{eq:sn:method:con5}
         \intermed {  {\intermed \Sigma}  } ; \intermed {  {\intermed { \Gamma_C } }  } ; \intermed {  \bullet  }  \vdash_{tm}  \intermed {  \ottsym{(}  \Lambda  {\overline {\intermed a} }_{\ottmv{j}}  \ottsym{.}  \lambda  {\overline {\intermed \delta} }_{\ottmv{i}}  \ottsym{:}  {\overline {\intermed Q} }_{\ottmv{i}}  \ottsym{.}  {\intermed e}  \ottsym{)} \, {\intermed { R^{SN} } }  \ottsym{(}  \overline {\intermed \sigma}_{\ottmv{j}}  \ottsym{)} \, {\intermed { R^{SN} } }  \ottsym{(}  {\intermed { \gamma^{SN} } }  \ottsym{(}  {\overline {\intermed d} }_{\ottmv{i}}  \ottsym{)}  \ottsym{)}  } : \intermed {  {\intermed { R^{SN} } }  \ottsym{(}  [  [  \overline {\intermed \sigma}_{\ottmv{j}}  /  {\overline {\intermed a} }_{\ottmv{j}}  ]  {\intermed \sigma}_{{\mathrm{2}}}  /  {\intermed a }  ]  {\intermed \sigma}'  \ottsym{)}  } \itot{ \rightsquigarrow  \target {  {\target e }_{{\mathrm{1}}}  } } 
      \end{gather}
      By substituting Equation~\ref{eq:sn:method:con:g2}, we have:
      \begin{gather}
        \label{eq:sn:method:con6}
         \intermed {  {\intermed \Sigma}  } ; \intermed {  {\intermed { \Gamma_C } }  } ; \intermed {  \bullet  }  \vdash_{tm}  \intermed {  \ottsym{(}  \Lambda  {\overline {\intermed a} }_{\ottmv{j}}  \ottsym{.}  \lambda  {\overline {\intermed \delta} }_{\ottmv{i}}  \ottsym{:}  {\overline {\intermed Q} }_{\ottmv{i}}  \ottsym{.}  {\intermed e}  \ottsym{)} \, {\intermed { R^{SN} } }  \ottsym{(}  \overline {\intermed \sigma}_{\ottmv{j}}  \ottsym{)} \, {\intermed { R^{SN} } }  \ottsym{(}  {\intermed { \gamma^{SN} } }  \ottsym{(}  {\overline {\intermed d} }_{\ottmv{i}}  \ottsym{)}  \ottsym{)}  } : \intermed {  {\intermed { R^{SN} } }  \ottsym{(}  [  {\intermed \sigma}  /  {\intermed a }  ]  {\intermed \sigma}'  \ottsym{)}  } \itot{ \rightsquigarrow  \target {  {\target e }_{{\mathrm{1}}}  } } 
      \end{gather}
      By induction hypothesis on Equation~\ref{eq:sn:method:con6}, we get:
      \begin{gather}
        \label{eq:sn:method:con7}
       \intermed {  \ottsym{(}  \Lambda  {\overline {\intermed a} }_{\ottmv{j}}  \ottsym{.}  \lambda  {\overline {\intermed \delta} }_{\ottmv{i}}  \ottsym{:}  {\overline {\intermed Q} }_{\ottmv{i}}  \ottsym{.}  {\intermed e}  \ottsym{)} \, {\intermed { R^{SN} } }  \ottsym{(}  \overline {\intermed \sigma}_{\ottmv{j}}  \ottsym{)} \, {\intermed { R^{SN} } }  \ottsym{(}  {\intermed { \gamma^{SN} } }  \ottsym{(}  {\overline {\intermed d} }_{\ottmv{i}}  \ottsym{)}  \ottsym{)}  } \in \mathcal{SN} \llbracket \intermed {  {\intermed { R^{SN} } }  \ottsym{(}  [  {\intermed \sigma}  /  {\intermed a }  ]  {\intermed \sigma}'  \ottsym{)}  } \rrbracket ^{\intermed {  {\intermed \Sigma}  } , \intermed {  {\intermed { \Gamma_C } }  } }_{\intermed {  \bullet  } } .
      \end{gather}
      Because we know that $ \intermed {  {\intermed { R^{SN} } }  } \in \mathcal{F^{SN} } \llbracket \intermed {  {\intermed \Gamma}  } \rrbracket ^{\intermed {  {\intermed \Sigma}  }, \intermed {  {\intermed { \Gamma_C } }  } } $, by
      Compositionality for Strong Normalization (Context Interpretation)
      (Corollary~\ref{cor:comp_sn_ctx}), Equation~\ref{eq:sn:method:con7}
      proves our Goal~\ref{eq:sn:method:con2}.
  \end{itemize}
}
\proofStep{\textsc{iTm-arrI}}
{\drule{iTm-arrI}}
{
  Because $ \intermed {   {\intermed { \gamma^{SN} } }  ( {\intermed { \phi^{SN} } }  ( {\intermed { R^{SN} } }  ( \ottsym{(}  \lambda  {\intermed x}  \ottsym{:}  {\intermed \sigma}_{{\mathrm{1}}}  \ottsym{.}  {\intermed e}  \ottsym{)} )))   } = \intermed {  \lambda  {\intermed x}  \ottsym{:}  {\intermed { R^{SN} } }  \ottsym{(}  {\intermed \sigma}_{{\mathrm{1}}}  \ottsym{)}  \ottsym{.}  \ottsym{(}   {\intermed { \gamma^{SN} } }  ( {\intermed { \phi^{SN} } }  ( {\intermed { R^{SN} } }  ( {\intermed e} )))   \ottsym{)}  } $,
  our goal is to show that:
  \begin{gather}
     \intermed {  \lambda  {\intermed x}  \ottsym{:}  {\intermed { R^{SN} } }  \ottsym{(}  {\intermed \sigma}_{{\mathrm{1}}}  \ottsym{)}  \ottsym{.}  \ottsym{(}   {\intermed { \gamma^{SN} } }  ( {\intermed { \phi^{SN} } }  ( {\intermed { R^{SN} } }  ( {\intermed e} )))   \ottsym{)}  } \in \mathcal{SN} \llbracket \intermed {  {\intermed \sigma}_{{\mathrm{1}}}  \rightarrow  {\intermed \sigma}_{{\mathrm{2}}}  } \rrbracket ^{\intermed {  {\intermed \Sigma}  } , \intermed {  {\intermed { \Gamma_C } }  } }_{\intermed {  {\intermed { R^{SN} } }  } } 
  \end{gather}
  By definition, we need to prove the following goals:
  \begin{gather}
        \label{eq:sn:arri:1}
     \intermed {  {\intermed \Sigma}  } ; \intermed {  {\intermed { \Gamma_C } }  } ; \intermed {  \bullet  }  \vdash_{tm}  \intermed {  \lambda  {\intermed x}  \ottsym{:}  {\intermed { R^{SN} } }  \ottsym{(}  {\intermed \sigma}_{{\mathrm{1}}}  \ottsym{)}  \ottsym{.}  \ottsym{(}   {\intermed { \gamma^{SN} } }  ( {\intermed { \phi^{SN} } }  ( {\intermed { R^{SN} } }  ( {\intermed e} )))   \ottsym{)}  } : \intermed {  {\intermed { R^{SN} } }_{{\mathrm{1}}}  \ottsym{(}  {\intermed \sigma}_{{\mathrm{1}}}  \rightarrow  {\intermed \sigma}_{{\mathrm{2}}}  \ottsym{)}  } \itot{ \rightsquigarrow  \target {  {\target e }  } } \\
        \label{eq:sn:arri:2}
    \exists {\intermed v} :  \intermed {  {\intermed \Sigma}  }  \vdash  \intermed {  \lambda  {\intermed x}  \ottsym{:}  {\intermed { R^{SN} } }  \ottsym{(}  {\intermed \sigma}_{{\mathrm{1}}}  \ottsym{)}  \ottsym{.}  \ottsym{(}   {\intermed { \gamma^{SN} } }  ( {\intermed { \phi^{SN} } }  ( {\intermed { R^{SN} } }  ( {\intermed e} )))   \ottsym{)}  }  \longrightarrow^{*}  \intermed {  {\intermed v}  }  \\
        \label{eq:sn:arri:3}
        \forall \intermed{{\intermed e}'} :  \intermed {  {\intermed e}'  } \in \mathcal{SN} \llbracket \intermed {  {\intermed \sigma}_{{\mathrm{1}}}  } \rrbracket ^{\intermed {  {\intermed \Sigma}  } , \intermed {  {\intermed { \Gamma_C } }  } }_{\intermed {  {\intermed { R^{SN} } }  } } 
      \Rightarrow  \intermed {  \ottsym{(}  \lambda  {\intermed x}  \ottsym{:}  {\intermed { R^{SN} } }  \ottsym{(}  {\intermed \sigma}_{{\mathrm{1}}}  \ottsym{)}  \ottsym{.}  \ottsym{(}   {\intermed { \gamma^{SN} } }  ( {\intermed { \phi^{SN} } }  ( {\intermed { R^{SN} } }  ( {\intermed e} )))   \ottsym{)}  \ottsym{)} \, {\intermed e}'  } \in \mathcal{SN} \llbracket \intermed {  {\intermed \sigma}_{{\mathrm{2}}}  } \rrbracket ^{\intermed {  {\intermed \Sigma}  } , \intermed {  {\intermed { \Gamma_C } }  } }_{\intermed {  {\intermed { R^{SN} } }  } }  
  \end{gather}
  By Substitution for Context Interpretation (Lemma~\ref{lemma:sn_ctx_inter}),
  we can easily prove Equation~\ref{eq:sn:arri:1}.
  
  Furthermore,
  $\intermed{\lambda  {\intermed x}  \ottsym{:}  {\intermed { R^{SN} } }  \ottsym{(}  {\intermed \sigma}_{{\mathrm{1}}}  \ottsym{)}  \ottsym{.}  \ottsym{(}   {\intermed { \gamma^{SN} } }  ( {\intermed { \phi^{SN} } }  ( {\intermed { R^{SN} } }  ( {\intermed e} )))   \ottsym{)}}$ is already a value,
  which proves Equation~\ref{eq:sn:arri:2}. Now given
  \begin{gather}
        \label{eq:sn:arri:4}
        \forall \intermed{{\intermed e}'} :  \intermed {  {\intermed e}'  } \in \mathcal{SN} \llbracket \intermed {  {\intermed \sigma}_{{\mathrm{1}}}  } \rrbracket ^{\intermed {  {\intermed \Sigma}  } , \intermed {  {\intermed { \Gamma_C } }  } }_{\intermed {  {\intermed { R^{SN} } }  } } 
  \end{gather}
  We need to show
  \begin{gather}
        \label{eq:sn:arri:5}
        \intermed {  \ottsym{(}  \lambda  {\intermed x}  \ottsym{:}  {\intermed { R^{SN} } }  \ottsym{(}  {\intermed \sigma}_{{\mathrm{1}}}  \ottsym{)}  \ottsym{.}  \ottsym{(}   {\intermed { \gamma^{SN} } }  ( {\intermed { \phi^{SN} } }  ( {\intermed { R^{SN} } }  ( {\intermed e} )))   \ottsym{)}  \ottsym{)} \, {\intermed e}'  } \in \mathcal{SN} \llbracket \intermed {  {\intermed \sigma}_{{\mathrm{2}}}  } \rrbracket ^{\intermed {  {\intermed \Sigma}  } , \intermed {  {\intermed { \Gamma_C } }  } }_{\intermed {  {\intermed { R^{SN} } }  } }  
  \end{gather}
  Let $ \intermed {  {\intermed { \phi^{SN} } }'  } = \intermed {   {\intermed { \phi^{SN} } } ,  {\intermed x}  \mapsto  {\intermed e}'   } $. By induction hypothesis, we have
  \begin{gather}
        \label{eq:sn:arri:6}
     \intermed {   {\intermed { \gamma^{SN} } }  (  {\intermed { \phi^{SN} } } ,  {\intermed x}  \mapsto  {\intermed e}'   ( {\intermed { R^{SN} } }  ( {\intermed e} )))   } \in \mathcal{SN} \llbracket \intermed {  {\intermed \sigma}_{{\mathrm{2}}}  } \rrbracket ^{\intermed {  {\intermed \Sigma}  } , \intermed {  {\intermed { \Gamma_C } }  } }_{\intermed {  {\intermed { R^{SN} } }  } } 
  \end{gather}
  We know that:
  \begin{gather*}
    \intermed{\ottsym{(}  \lambda  {\intermed x}  \ottsym{:}  {\intermed { R^{SN} } }  \ottsym{(}  {\intermed \sigma}_{{\mathrm{1}}}  \ottsym{)}  \ottsym{.}  \ottsym{(}   {\intermed { \gamma^{SN} } }  ( {\intermed { \phi^{SN} } }  ( {\intermed { R^{SN} } }  ( {\intermed e} )))   \ottsym{)}  \ottsym{)} \, {\intermed e}'} \\
     \longrightarrow  \intermed{[  {\intermed e}'  /  {\intermed x}  ]  \ottsym{(}   {\intermed { \gamma^{SN} } }  ( {\intermed { \phi^{SN} } }  ( {\intermed { R^{SN} } }  ( {\intermed e} )))   \ottsym{)}} \\
    = \intermed{ {\intermed { \gamma^{SN} } }  (  {\intermed { \phi^{SN} } } ,  {\intermed x}  \mapsto  {\intermed e}'   ( {\intermed { R^{SN} } }  ( {\intermed e} ))) }
  \end{gather*}
  Consequently, by Strong Normalization preserved by forward/backward
  reduction(Lemma~\ref{lemma:sn_pres}), Equation~\ref{eq:sn:arri:6} proves
  \ref{eq:sn:arri:5}. \\
}
\proofStep{\textsc{iTm-arrE}}
{\drule{iTm-arrE}}
{
  By induction hypothesis, we have:
  \begin{gather}
    \label{eq:sn:arre:1}
     \intermed {   {\intermed { \gamma^{SN} } }  ( {\intermed { \phi^{SN} } }  ( {\intermed { R^{SN} } }  ( {\intermed e}_{{\mathrm{1}}} )))   } \in \mathcal{SN} \llbracket \intermed {  {\intermed \sigma}_{{\mathrm{1}}}  \rightarrow  {\intermed \sigma}_{{\mathrm{2}}}  } \rrbracket ^{\intermed {  {\intermed \Sigma}  } , \intermed {  {\intermed { \Gamma_C } }  } }_{\intermed {  {\intermed { R^{SN} } }  } }  \\
    \label{eq:sn:arre:2}
     \intermed {   {\intermed { \gamma^{SN} } }  ( {\intermed { \phi^{SN} } }  ( {\intermed { R^{SN} } }  ( {\intermed e}_{{\mathrm{2}}} )))   } \in \mathcal{SN} \llbracket \intermed {  {\intermed \sigma}_{{\mathrm{1}}}  } \rrbracket ^{\intermed {  {\intermed \Sigma}  } , \intermed {  {\intermed { \Gamma_C } }  } }_{\intermed {  {\intermed { R^{SN} } }  } }  
  \end{gather}
  By the definition of Equation~\ref{eq:sn:arre:1}, applying
  Equation~\ref{eq:sn:arre:2} results in:
  \begin{gather*}
     \intermed {  \ottsym{(}   {\intermed { \gamma^{SN} } }  ( {\intermed { \phi^{SN} } }  ( {\intermed { R^{SN} } }  ( {\intermed e}_{{\mathrm{1}}} )))   \ottsym{)} \, \ottsym{(}   {\intermed { \gamma^{SN} } }  ( {\intermed { \phi^{SN} } }  ( {\intermed { R^{SN} } }  ( {\intermed e}_{{\mathrm{2}}} )))   \ottsym{)}  } \in \mathcal{SN} \llbracket \intermed {  {\intermed \sigma}_{{\mathrm{2}}}  } \rrbracket ^{\intermed {  {\intermed \Sigma}  } , \intermed {  {\intermed { \Gamma_C } }  } }_{\intermed {  {\intermed { R^{SN} } }  } } 
  \end{gather*}
  which is exactly our goal since
  \begin{gather*}
  \intermed{ {\intermed { \gamma^{SN} } }  ( {\intermed { \phi^{SN} } }  ( {\intermed { R^{SN} } }  ( \ottsym{(}  {\intermed e}_{{\mathrm{1}}} \, {\intermed e}_{{\mathrm{2}}}  \ottsym{)} ))) }
  = \intermed{\ottsym{(}   {\intermed { \gamma^{SN} } }  ( {\intermed { \phi^{SN} } }  ( {\intermed { R^{SN} } }  ( {\intermed e}_{{\mathrm{1}}} )))   \ottsym{)} \, \ottsym{(}   {\intermed { \gamma^{SN} } }  ( {\intermed { \phi^{SN} } }  ( {\intermed { R^{SN} } }  ( {\intermed e}_{{\mathrm{2}}} )))   \ottsym{)}}
  \end{gather*}
}
\proofStep{\textsc{iTm-constrI}}
{\drule{iTm-constrI}}
{
  Because $ \intermed {   {\intermed { \gamma^{SN} } }  ( {\intermed { \phi^{SN} } }  ( {\intermed { R^{SN} } }  ( \ottsym{(}  \lambda  {\intermed \delta }  \ottsym{:}  {\intermed Q}  \ottsym{.}  {\intermed e}  \ottsym{)} )))   } = \intermed {  \lambda  {\intermed \delta }  \ottsym{:}  {\intermed { R^{SN} } }  \ottsym{(}  {\intermed Q}  \ottsym{)}  \ottsym{.}  \ottsym{(}   {\intermed { \gamma^{SN} } }  ( {\intermed { \phi^{SN} } }  ( {\intermed { R^{SN} } }  ( {\intermed e} )))   \ottsym{)}  } $,
  our goal is to show that:
  \begin{gather}
     \intermed {  \lambda  {\intermed \delta }  \ottsym{:}  {\intermed { R^{SN} } }  \ottsym{(}  {\intermed Q}  \ottsym{)}  \ottsym{.}  \ottsym{(}   {\intermed { \gamma^{SN} } }  ( {\intermed { \phi^{SN} } }  ( {\intermed { R^{SN} } }  ( {\intermed e} )))   \ottsym{)}  } \in \mathcal{SN} \llbracket \intermed {   {\intermed Q}  \Rightarrow  {\intermed \sigma}   } \rrbracket ^{\intermed {  {\intermed \Sigma}  } , \intermed {  {\intermed { \Gamma_C } }  } }_{\intermed {  {\intermed { R^{SN} } }  } } 
  \end{gather}
  By definition, we need to prove the following goals:
  \begin{gather}
        \label{eq:sn:di:1}
     \intermed {  {\intermed \Sigma}  } ; \intermed {  {\intermed { \Gamma_C } }  } ; \intermed {  \bullet  }  \vdash_{tm}  \intermed {  \lambda  {\intermed \delta }  \ottsym{:}  {\intermed { R^{SN} } }  \ottsym{(}  {\intermed Q}  \ottsym{)}  \ottsym{.}  \ottsym{(}   {\intermed { \gamma^{SN} } }  ( {\intermed { \phi^{SN} } }  ( {\intermed { R^{SN} } }  ( {\intermed e} )))   \ottsym{)}  } : \intermed {  {\intermed { R^{SN} } }_{{\mathrm{1}}}  \ottsym{(}   {\intermed Q}  \Rightarrow  {\intermed \sigma}   \ottsym{)}  } \itot{ \rightsquigarrow  \target {  {\target e }  } } \\
        \label{eq:sn:di:2}
    \exists {\intermed v} :  \intermed {  {\intermed \Sigma}  }  \vdash  \intermed {  \lambda  {\intermed \delta }  \ottsym{:}  {\intermed { R^{SN} } }  \ottsym{(}  {\intermed Q}  \ottsym{)}  \ottsym{.}  \ottsym{(}   {\intermed { \gamma^{SN} } }  ( {\intermed { \phi^{SN} } }  ( {\intermed { R^{SN} } }  ( {\intermed e} )))   \ottsym{)}  }  \longrightarrow^{*}  \intermed {  {\intermed v}  }  \\
        \label{eq:sn:di:3}
        \forall {\intermed d} :  \intermed {  {\intermed \Sigma}  } ; \intermed {  {\intermed { \Gamma_C } }  } ; \intermed {  \bullet  }  \vdash_{d}  \intermed {  {\intermed d}  } : \intermed {  {\intermed { R^{SN} } }_{{\mathrm{1}}}  \ottsym{(}  {\intermed Q}  \ottsym{)}  } \itot{\,  \rightsquigarrow  \target {  {\target e }  } } 
      \Rightarrow  \intermed {  \ottsym{(}  \lambda  {\intermed \delta }  \ottsym{:}  {\intermed { R^{SN} } }  \ottsym{(}  {\intermed Q}  \ottsym{)}  \ottsym{.}  \ottsym{(}   {\intermed { \gamma^{SN} } }  ( {\intermed { \phi^{SN} } }  ( {\intermed { R^{SN} } }  ( {\intermed e} )))   \ottsym{)}  \ottsym{)} \, {\intermed d}  } \in \mathcal{SN} \llbracket \intermed {  {\intermed \sigma}  } \rrbracket ^{\intermed {  {\intermed \Sigma}  } , \intermed {  {\intermed { \Gamma_C } }  } }_{\intermed {  {\intermed { R^{SN} } }  } }  
  \end{gather}
  By Substitution for Context Interpretation (Lemma~\ref{lemma:sn_ctx_inter}),
  we can easily prove Equation~\ref{eq:sn:di:1}.

  Furthermore,
  $\intermed{\lambda  {\intermed \delta }  \ottsym{:}  {\intermed { R^{SN} } }  \ottsym{(}  {\intermed Q}  \ottsym{)}  \ottsym{.}  \ottsym{(}   {\intermed { \gamma^{SN} } }  ( {\intermed { \phi^{SN} } }  ( {\intermed { R^{SN} } }  ( {\intermed e} )))   \ottsym{)}}$ is already a value,
  which proves Equation~\ref{eq:sn:di:2}. Now given
  \begin{gather}
        \label{eq:sn:di:4}
        \forall {\intermed d} :  \intermed {  {\intermed \Sigma}  } ; \intermed {  {\intermed { \Gamma_C } }  } ; \intermed {  \bullet  }  \vdash_{d}  \intermed {  {\intermed d}  } : \intermed {  {\intermed { R^{SN} } }_{{\mathrm{1}}}  \ottsym{(}  {\intermed Q}  \ottsym{)}  } \itot{\,  \rightsquigarrow  \target {  {\target e }  } } 
  \end{gather}
  We need to show
  \begin{gather}
        \label{eq:sn:di:5}
        \intermed {  \ottsym{(}  \lambda  {\intermed \delta }  \ottsym{:}  {\intermed { R^{SN} } }  \ottsym{(}  {\intermed Q}  \ottsym{)}  \ottsym{.}  \ottsym{(}   {\intermed { \gamma^{SN} } }  ( {\intermed { \phi^{SN} } }  ( {\intermed { R^{SN} } }  ( {\intermed e} )))   \ottsym{)}  \ottsym{)} \, {\intermed d}  } \in \mathcal{SN} \llbracket \intermed {  {\intermed \sigma}  } \rrbracket ^{\intermed {  {\intermed \Sigma}  } , \intermed {  {\intermed { \Gamma_C } }  } }_{\intermed {  {\intermed { R^{SN} } }  } }  
  \end{gather}
  Let $ \intermed {  {\intermed { \gamma^{SN} } }'  } = \intermed {   {\intermed { \gamma^{SN} } } ,  {\intermed \delta }  \mapsto  {\intermed d}   } $. By induction hypothesis, we have
  \begin{gather}
    \label{eq:sn:di:6}
     \intermed {    {\intermed { \gamma^{SN} } } ,  {\intermed \delta }  \mapsto  {\intermed d}   ( {\intermed { \phi^{SN} } }  ( {\intermed { R^{SN} } }  ( {\intermed e} )))   } \in \mathcal{SN} \llbracket \intermed {  {\intermed \sigma}  } \rrbracket ^{\intermed {  {\intermed \Sigma}  } , \intermed {  {\intermed { \Gamma_C } }  } }_{\intermed {  {\intermed { R^{SN} } }  } } 
  \end{gather}
  We know that:
  \begin{gather*}
    \intermed{\ottsym{(}  \lambda  {\intermed \delta }  \ottsym{:}  {\intermed { R^{SN} } }  \ottsym{(}  {\intermed Q}  \ottsym{)}  \ottsym{.}  \ottsym{(}   {\intermed { \gamma^{SN} } }  ( {\intermed { \phi^{SN} } }  ( {\intermed { R^{SN} } }  ( {\intermed e} )))   \ottsym{)}  \ottsym{)} \, {\intermed d}} \\
     \longrightarrow  \intermed{[  {\intermed d}  /  {\intermed \delta }  ]  \ottsym{(}   {\intermed { \gamma^{SN} } }  ( {\intermed { \phi^{SN} } }  ( {\intermed { R^{SN} } }  ( {\intermed e} )))   \ottsym{)}} \\
    = \intermed{  {\intermed { \gamma^{SN} } } ,  {\intermed \delta }  \mapsto  {\intermed d}   ( {\intermed { \phi^{SN} } }  ( {\intermed { R^{SN} } }  ( {\intermed e} ))) }
  \end{gather*}
  Consequently, by Strong Normalization preserved by forward/backward
  reduction(Lemma~\ref{lemma:sn_pres}), Equation~\ref{eq:sn:di:6} proves
  \ref{eq:sn:di:5}. \\
}
\proofStep{\textsc{iTm-constrE}}
{\drule{iTm-constrE}}
{
  By induction hypothesis, we have
  \begin{gather}
    \label{eq:sn:de:1}
     \intermed {   {\intermed { \gamma^{SN} } }  ( {\intermed { \phi^{SN} } }  ( {\intermed { R^{SN} } }  ( {\intermed e} )))   } \in \mathcal{SN} \llbracket \intermed {   {\intermed Q}  \Rightarrow  {\intermed \sigma}   } \rrbracket ^{\intermed {  {\intermed \Sigma}  } , \intermed {  {\intermed { \Gamma_C } }  } }_{\intermed {  {\intermed { R^{SN} } }  } }  
  \end{gather}
  By Substitution for Context Interpretation (Lemma~\ref{lemma:sn_ctx_inter}),
  we know that:
  \begin{gather}
    \label{eq:sn:de:2}
     \intermed {  {\intermed \Sigma}  } ; \intermed {  {\intermed { \Gamma_C } }  } ; \intermed {  \bullet  }  \vdash_{d}  \intermed {   {\intermed { \gamma^{SN} } }  ( {\intermed { \phi^{SN} } }  ( {\intermed { R^{SN} } }  ( {\intermed d} )))   } : \intermed {  {\intermed { R^{SN} } }  \ottsym{(}  {\intermed Q}  \ottsym{)}  } \itot{\,  \rightsquigarrow  \target {  {\target e }_{{\mathrm{2}}}  } }  
  \end{gather}
  By the definition of Equation~\ref{eq:sn:de:1}, applying
  Equation~\ref{eq:sn:de:2} results in:
  \begin{gather*}
     \intermed {  \ottsym{(}   {\intermed { \gamma^{SN} } }  ( {\intermed { \phi^{SN} } }  ( {\intermed { R^{SN} } }  ( {\intermed e} )))   \ottsym{)} \, \ottsym{(}   {\intermed { \gamma^{SN} } }  ( {\intermed { \phi^{SN} } }  ( {\intermed { R^{SN} } }  ( {\intermed d} )))   \ottsym{)}  } \in \mathcal{SN} \llbracket \intermed {  {\intermed \sigma}  } \rrbracket ^{\intermed {  {\intermed \Sigma}  } , \intermed {  {\intermed { \Gamma_C } }  } }_{\intermed {  {\intermed { R^{SN} } }  } } 
  \end{gather*}
  which is exactly our goal since
  \begin{equation*}
   \intermed {   {\intermed { \gamma^{SN} } }  ( {\intermed { \phi^{SN} } }  ( {\intermed { R^{SN} } }  ( \ottsym{(}  {\intermed e} \, {\intermed d}  \ottsym{)} )))   } = \intermed {  \ottsym{(}   {\intermed { \gamma^{SN} } }  ( {\intermed { \phi^{SN} } }  ( {\intermed { R^{SN} } }  ( {\intermed e} )))   \ottsym{)} \, \ottsym{(}   {\intermed { \gamma^{SN} } }  ( {\intermed { \phi^{SN} } }  ( {\intermed { R^{SN} } }  ( {\intermed d} )))   \ottsym{)}  } 
  \end{equation*}
}
\proofStep{\textsc{iTm-forallI}}
{\drule{iTm-forallI}}
{
  Because $ \intermed {   {\intermed { \gamma^{SN} } }  ( {\intermed { \phi^{SN} } }  ( {\intermed { R^{SN} } }  ( \ottsym{(}  \Lambda  {\intermed a }  \ottsym{.}  {\intermed e}  \ottsym{)} )))   } = \intermed {  \Lambda  {\intermed a }  \ottsym{.}  \ottsym{(}   {\intermed { \gamma^{SN} } }  ( {\intermed { \phi^{SN} } }  ( {\intermed { R^{SN} } }  ( {\intermed e} )))   \ottsym{)}  } $,
  our goal is to show that:
  \begin{gather}
     \intermed {  \Lambda  {\intermed a }  \ottsym{.}  \ottsym{(}   {\intermed { \gamma^{SN} } }  ( {\intermed { \phi^{SN} } }  ( {\intermed { R^{SN} } }  ( {\intermed e} )))   \ottsym{)}  } \in \mathcal{SN} \llbracket \intermed {  \forall  {\intermed a }  \ottsym{.}  {\intermed \sigma}  } \rrbracket ^{\intermed {  {\intermed \Sigma}  } , \intermed {  {\intermed { \Gamma_C } }  } }_{\intermed {  {\intermed { R^{SN} } }  } } 
  \end{gather}
  By definition, we need to prove the following goals:
  \begin{gather}
        \label{eq:sn:alli:1}
     \intermed {  {\intermed \Sigma}  } ; \intermed {  {\intermed { \Gamma_C } }  } ; \intermed {  \bullet  }  \vdash_{tm}  \intermed {  \Lambda  {\intermed a }  \ottsym{.}  \ottsym{(}   {\intermed { \gamma^{SN} } }  ( {\intermed { \phi^{SN} } }  ( {\intermed { R^{SN} } }  ( {\intermed e} )))   \ottsym{)}  } : \intermed {  {\intermed { R^{SN} } }_{{\mathrm{1}}}  \ottsym{(}  \forall  {\intermed a }  \ottsym{.}  {\intermed \sigma}  \ottsym{)}  } \itot{ \rightsquigarrow  \target {  {\target e }  } } \\
        \label{eq:sn:alli:2}
    \exists {\intermed v} :  \intermed {  {\intermed \Sigma}  }  \vdash  \intermed {  \Lambda  {\intermed a }  \ottsym{.}  \ottsym{(}   {\intermed { \gamma^{SN} } }  ( {\intermed { \phi^{SN} } }  ( {\intermed { R^{SN} } }  ( {\intermed e} )))   \ottsym{)}  }  \longrightarrow^{*}  \intermed {  {\intermed v}  }  \\
        \label{eq:sn:alli:3}
        \forall \intermed{{\intermed \sigma}'} ,  \intermed {  { \intermed { \mathit{ r } } }  } = \mathcal{SN} \llbracket \intermed {  {\intermed \sigma}'  } \rrbracket ^{\intermed {  {\intermed \Sigma}  } , \intermed {  {\intermed { \Gamma_C } }  } }_{\intermed {  {\intermed { R^{SN} } }  } } 
      \Rightarrow  \intermed {  \ottsym{(}  \Lambda  {\intermed a }  \ottsym{.}  \ottsym{(}   {\intermed { \gamma^{SN} } }  ( {\intermed { \phi^{SN} } }  ( {\intermed { R^{SN} } }  ( {\intermed e} )))   \ottsym{)}  \ottsym{)} \, {\intermed \sigma}'  } \in \mathcal{SN} \llbracket \intermed {  {\intermed \sigma}  } \rrbracket ^{\intermed {  {\intermed \Sigma}  } , \intermed {  {\intermed { \Gamma_C } }  } }_{\intermed {   {\intermed { R^{SN} } }  ,  {\intermed a }  \mapsto (  {\intermed \sigma}' ,  { \intermed { \mathit{ r } } }  )   } }  
  \end{gather}
  By Substitution for Context Interpretation (Lemma~\ref{lemma:sn_ctx_inter}),
  we can easily prove Equation~\ref{eq:sn:alli:1}.

  Furthermore,
  $ \intermed {  {\intermed \Sigma}  }  \vdash  \intermed {  \Lambda  {\intermed a }  \ottsym{.}  \ottsym{(}   {\intermed { \gamma^{SN} } }  ( {\intermed { \phi^{SN} } }  ( {\intermed { R^{SN} } }  ( {\intermed e} )))   \ottsym{)}  }  \longrightarrow^{*}  \intermed {  {\intermed v}  } $ is already a value,
  which proves Equation~\ref{eq:sn:alli:2}. Now given
  \begin{gather}
        \label{eq:sn:alli:4}
        \forall \intermed{{\intermed \sigma}'} ,  \intermed {  { \intermed { \mathit{ r } } }  } = \mathcal{SN} \llbracket \intermed {  {\intermed \sigma}'  } \rrbracket ^{\intermed {  {\intermed \Sigma}  } , \intermed {  {\intermed { \Gamma_C } }  } }_{\intermed {  {\intermed { R^{SN} } }  } } 
  \end{gather}
  We need to show
  \begin{gather}
        \label{eq:sn:alli:5}
       \intermed {  \ottsym{(}  \Lambda  {\intermed a }  \ottsym{.}  \ottsym{(}   {\intermed { \gamma^{SN} } }  ( {\intermed { \phi^{SN} } }  ( {\intermed { R^{SN} } }  ( {\intermed e} )))   \ottsym{)}  \ottsym{)} \, {\intermed \sigma}'  } \in \mathcal{SN} \llbracket \intermed {  {\intermed \sigma}  } \rrbracket ^{\intermed {  {\intermed \Sigma}  } , \intermed {  {\intermed { \Gamma_C } }  } }_{\intermed {   {\intermed { R^{SN} } }  ,  {\intermed a }  \mapsto (  {\intermed \sigma}' ,  { \intermed { \mathit{ r } } }  )   } }  
  \end{gather}
  Let $ \intermed {  {\intermed { R^{SN} } }'  } = \intermed {   {\intermed { R^{SN} } }  ,  {\intermed a }  \mapsto (  {\intermed \sigma}' ,  { \intermed { \mathit{ r } } }  )   } $. By induction hypothesis, we have:
  \begin{gather}
        \label{eq:sn:alli:6}
     \intermed {   {\intermed { \gamma^{SN} } }  ( {\intermed { \phi^{SN} } }  (  {\intermed { R^{SN} } }  ,  {\intermed a }  \mapsto (  {\intermed \sigma}' ,  { \intermed { \mathit{ r } } }  )   ( {\intermed e} )))   } \in \mathcal{SN} \llbracket \intermed {  {\intermed \sigma}  } \rrbracket ^{\intermed {  {\intermed \Sigma}  } , \intermed {  {\intermed { \Gamma_C } }  } }_{\intermed {   {\intermed { R^{SN} } }  ,  {\intermed a }  \mapsto (  {\intermed \sigma}' ,  { \intermed { \mathit{ r } } }  )   } } 
  \end{gather}
  We know that:
  \begin{gather*}
    \intermed{\ottsym{(}  \Lambda  {\intermed a }  \ottsym{.}  \ottsym{(}   {\intermed { \gamma^{SN} } }  ( {\intermed { \phi^{SN} } }  ( {\intermed { R^{SN} } }  ( {\intermed e} )))   \ottsym{)}  \ottsym{)} \, {\intermed \sigma}'} \\
     \longrightarrow  \intermed{[  {\intermed \sigma}'  /  {\intermed a }  ]  \ottsym{(}   {\intermed { \gamma^{SN} } }  ( {\intermed { \phi^{SN} } }  ( {\intermed { R^{SN} } }  ( {\intermed e} )))   \ottsym{)}} \\
    = \intermed{ {\intermed { \gamma^{SN} } }  ( {\intermed { \phi^{SN} } }  (  {\intermed { R^{SN} } }  ,  {\intermed a }  \mapsto (  {\intermed \sigma}' ,  { \intermed { \mathit{ r } } }  )   ( {\intermed e} ))) }
  \end{gather*}
  Consequently, by Strong Normalization preserved by forward/backward
  reduction(Lemma~\ref{lemma:sn_pres}), Equation~\ref{eq:sn:alli:6} proves
  \ref{eq:sn:alli:5}. \\
}
\proofStep{\textsc{iTm-forallE}}
{\drule{iTm-forallE}}
{
  By induction hypothesis, we have:
  \begin{gather}
    \label{eq:sn:alle:1}
     \intermed {   {\intermed { \gamma^{SN} } }  ( {\intermed { \phi^{SN} } }  ( {\intermed { R^{SN} } }  ( {\intermed e} )))   } \in \mathcal{SN} \llbracket \intermed {  \forall  {\intermed a }  \ottsym{.}  {\intermed \sigma}'  } \rrbracket ^{\intermed {  {\intermed \Sigma}  } , \intermed {  {\intermed { \Gamma_C } }  } }_{\intermed {  {\intermed { R^{SN} } }  } }  
  \end{gather}
  By Substitution for Context Interpretation (Lemma~\ref{lemma:sn_ctx_inter}),
  we know:
  \begin{gather}
    \label{eq:sn:alle:2}
     \intermed {  {\intermed { \Gamma_C } }  } ; \intermed {  \bullet  }  \vdash_{ty}  \intermed {  {\intermed { R^{SN} } }  \ottsym{(}  {\intermed \sigma}  \ottsym{)}  } \itot{ \rightsquigarrow  \target {  {\target \sigma }  } }  
  \end{gather}
  Choose $ \intermed {  { \intermed { \mathit{ r } } }  } = \mathcal{SN} \llbracket \intermed {  {\intermed \sigma}  } \rrbracket ^{\intermed {  {\intermed \Sigma}  } , \intermed {  {\intermed { \Gamma_C } }  } }_{\intermed {  {\intermed { R^{SN} } }  } } $.
  By the definition of Equation~\ref{eq:sn:alle:1}, applying
  Equation~\ref{eq:sn:alle:2} results in:
  \begin{gather*}
     \intermed {  \ottsym{(}   {\intermed { \gamma^{SN} } }  ( {\intermed { \phi^{SN} } }  ( {\intermed { R^{SN} } }  ( {\intermed e} )))   \ottsym{)} \, \ottsym{(}  {\intermed { R^{SN} } }  \ottsym{(}  {\intermed \sigma}  \ottsym{)}  \ottsym{)}  } \in \mathcal{SN} \llbracket \intermed {  {\intermed \sigma}'  } \rrbracket ^{\intermed {  {\intermed \Sigma}  } , \intermed {  {\intermed { \Gamma_C } }  } }_{\intermed {   {\intermed { R^{SN} } }  ,  {\intermed a }  \mapsto (  {\intermed { R^{SN} } }  \ottsym{(}  {\intermed \sigma}  \ottsym{)} ,  { \intermed { \mathit{ r } } }  )   } } 
  \end{gather*}
  By Compositionality for Strong Normalization (Lemma~\ref{lemma:comp_sn}), we get:
  \begin{gather*}
     \intermed {  \ottsym{(}   {\intermed { \gamma^{SN} } }  ( {\intermed { \phi^{SN} } }  ( {\intermed { R^{SN} } }  ( {\intermed e} )))   \ottsym{)} \, \ottsym{(}  {\intermed { R^{SN} } }  \ottsym{(}  {\intermed \sigma}  \ottsym{)}  \ottsym{)}  } \in \mathcal{SN} \llbracket \intermed {  [  {\intermed \sigma}  /  {\intermed a }  ]  {\intermed \sigma}'  } \rrbracket ^{\intermed {  {\intermed \Sigma}  } , \intermed {  {\intermed { \Gamma_C } }  } }_{\intermed {   {\intermed { R^{SN} } }  ,  {\intermed a }  \mapsto (  {\intermed \sigma} ,  { \intermed { \mathit{ r } } }  )   } } 
  \end{gather*}
  which is exactly our goal since
  \begin{gather*}
   \intermed {   {\intermed { \gamma^{SN} } }  ( {\intermed { \phi^{SN} } }  ( {\intermed { R^{SN} } }  ( \ottsym{(}  {\intermed e} \, {\intermed \sigma}  \ottsym{)} )))   } = \intermed {  \ottsym{(}   {\intermed { \gamma^{SN} } }  ( {\intermed { \phi^{SN} } }  ( {\intermed { R^{SN} } }  ( {\intermed e} )))   \ottsym{)} \, \ottsym{(}  {\intermed { R^{SN} } }  \ottsym{(}  {\intermed \sigma}  \ottsym{)}  \ottsym{)}  } 
  \end{gather*}
}
\qed

\begin{theoremB}[Strong Normalization - Part B]\ \\
  \label{thmA:sn:part:b}
  If \( \intermed {  {\intermed e}  } \in \mathcal{SN} \llbracket \intermed {  {\intermed \sigma}  } \rrbracket ^{\intermed {  {\intermed \Sigma}  } , \intermed {  {\intermed { \Gamma_C } }  } }_{\intermed {  {\intermed { R^{SN} } }  } } \)
  then \( \exists {\intermed v} :  \intermed {  {\intermed \Sigma}  }  \vdash  \intermed {  {\intermed e}  }  \longrightarrow^{*}  \intermed {  {\intermed v}  } \).
\end{theoremB}

\proof

This goal is baked into the relation. It follows by straightforward induction on
${\intermed \sigma}$. \\
\qed

\newpage
\section{Elaboration Equivalence Theorems}
\label{app:equiv}
\renewcommand\itot[1]{#1}

\begin{theoremB}[Equivalence - Environments]\leavevmode
  If \( \vdash_{ctx}  \source {  { \source P }  } ; \source {  {\source { \Gamma_C } }  } ; \source {  {\source \Gamma}  }  \rightsquigarrow  \target {  {\target \Gamma }  } \) \\
  then \( \vdash_{ctx}^{\intermed {M} }  \source {  { \source P }  } ; \source {  {\source { \Gamma_C } }  } ; \source {  {\source \Gamma}  }  \rightsquigarrow  \intermed {  {\intermed \Sigma}  } ; \intermed {  {\intermed { \Gamma_C } }  } ; \intermed {  {\intermed \Gamma}  } \)
  and \( \intermed {  {\intermed { \Gamma_C } }  } ; \intermed {  {\intermed \Gamma}  }  \rightsquigarrow  \target {  {\target \Gamma }  } \).
  \label{thmA:equiv_env}
\end{theoremB}

\begin{figure}[htp]
  \centering

  \def\svgscale{0.75}
\begingroup%
  \makeatletter%
  \providecommand\color[2][]{%
    \errmessage{(Inkscape) Color is used for the text in Inkscape, but the package 'color.sty' is not loaded}%
    \renewcommand\color[2][]{}%
  }%
  \providecommand\transparent[1]{%
    \errmessage{(Inkscape) Transparency is used (non-zero) for the text in Inkscape, but the package 'transparent.sty' is not loaded}%
    \renewcommand\transparent[1]{}%
  }%
  \providecommand\rotatebox[2]{#2}%
  \newcommand*\fsize{\dimexpr\f@size pt\relax}%
  \newcommand*\lineheight[1]{\fontsize{\fsize}{#1\fsize}\selectfont}%
  \ifx\svgwidth\undefined%
    \setlength{\unitlength}{93.5922019bp}%
    \ifx\svgscale\undefined%
      \relax%
    \else%
      \setlength{\unitlength}{\unitlength * \real{\svgscale}}%
    \fi%
  \else%
    \setlength{\unitlength}{\svgwidth}%
  \fi%
  \global\let\svgwidth\undefined%
  \global\let\svgscale\undefined%
  \makeatother%
  \begin{picture}(1,2.0087144)%
    \lineheight{1}%
    \setlength\tabcolsep{0pt}%
    \put(0,0){\includegraphics[width=\unitlength,page=1]{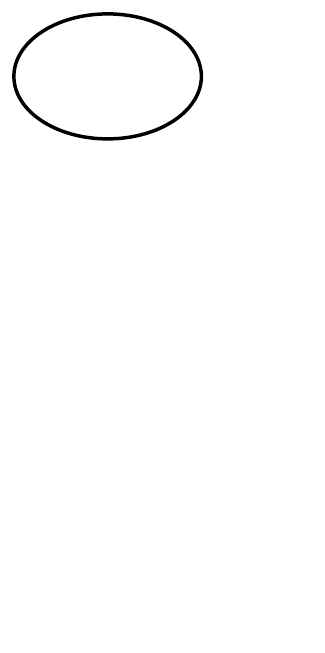}}%
    \put(0.33122418,1.7341189){\color[rgb]{0,0,0}\makebox(0,0)[t]{\lineheight{1.25}\smash{\begin{tabular}[t]{c}T13\end{tabular}}}}%
    \put(0,0){\includegraphics[width=\unitlength,page=2]{mut_dep_graph_equiv.pdf}}%
    \put(0.33122418,0.96482402){\color[rgb]{0,0,0}\makebox(0,0)[t]{\lineheight{1.25}\smash{\begin{tabular}[t]{c}T16\end{tabular}}}}%
    \put(0,0){\includegraphics[width=\unitlength,page=3]{mut_dep_graph_equiv.pdf}}%
    \put(0.33122418,0.19552912){\color[rgb]{0,0,0}\makebox(0,0)[t]{\lineheight{1.25}\smash{\begin{tabular}[t]{c}T15\end{tabular}}}}%
    \put(0,0){\includegraphics[width=\unitlength,page=4]{mut_dep_graph_equiv.pdf}}%
  \end{picture}%
\endgroup%

  \caption{Dependency graph for Theorems~\ref{thmA:equiv_env},
    \ref{thmA:equiv_dict} and
    \ref{thmA:equiv_expr}}
  \label{fig:dep_graph_equiv}
\end{figure}

\proof
By induction on the environment well-formedness relation.
This theorem is mutually proven with Theorems~\ref{thmA:equiv_dict}
and~\ref{thmA:equiv_expr} (Figure~\ref{fig:dep_graph_equiv}).
Note that at the dependencies between Theorem~\ref{thmA:equiv_env}
and~\ref{thmA:equiv_expr} and between Theorem~\ref{thmA:equiv_dict}
and~\ref{thmA:equiv_expr}, the size of \({ \source P }\) is strictly decreasing, whereas
\({ \source P }\) remains constant at every other dependency. Consequenty, the size of
\({ \source P }\) is strictly decreasing at every cycle and the induction remains well-founded.\\
\proofStep{\textsc{sCtxT-empty}}
{
   \vdash_{ctx}  \source {  \bullet  } ; \source {  \bullet  } ; \source {  \bullet  }  \rightsquigarrow  \target {  \target {\bullet}  } 
}
{
  The goal follows directly from \textsc{sCtx-empty} and \textsc{Ctx-Empty}. \\
}
\proofStep{\textsc{sCtxT-clsEnv}}
{
   \vdash_{ctx}  \source {  \bullet  } ; \source {  {\source { \Gamma_C } }  \ottsym{,}  {\source m}  \ottsym{:}  \overline {\source {\mathit{TC}_i~a } }  \Rightarrow  {\source {\mathit{TC} } } \, {\source a }  \ottsym{:}  \forall  {\overline {\source a} }_{\ottmv{j}}  \ottsym{.}  \ottcomp{{\source {\mathit{TC} } }_{\ottmv{i}} \, {\source a }}{\ottmv{i}}  \Rightarrow  {\source \tau }  } ; \source {  \bullet  }  \rightsquigarrow  \target {  \target {\bullet}  } 
}
{
  The goal to be proven is the following:
  \begin{gather}
     \vdash_{ctx}^{\intermed {M} }  \source {  \bullet  } ; \source {  {\source { \Gamma_C } }  \ottsym{,}  {\source m}  \ottsym{:}  \overline {\source {\mathit{TC}_i~a } }  \Rightarrow  {\source {\mathit{TC} } } \, {\source a }  \ottsym{:}  \forall  {\overline {\source a} }_{\ottmv{j}}  \ottsym{.}  \ottcomp{{\source {\mathit{TC} } }_{\ottmv{i}} \, {\source a }}{\ottmv{i}}  \Rightarrow  {\source \tau }  } ; \source {  \bullet  }  \rightsquigarrow  \intermed {  {\intermed \Sigma}  } ; \intermed {  {\intermed { \Gamma_C } }  } ; \intermed {  {\intermed \Gamma}  } 
    \label{eqg:eq_env1} \\
     \intermed {  {\intermed { \Gamma_C } }  } ; \intermed {  {\intermed \Gamma}  }  \rightsquigarrow  \target {  \target {\bullet}  } 
    \label{eqg:eq_env2}
  \end{gather}
  From the rule premise we get that:
  \begin{gather}
     \source {  {\source { \Gamma_C } }  } ; \source {  \bullet  \ottsym{,}  {\source a }  }  \vdash_{ty}  \source {  \forall  {\overline {\source a} }_{\ottmv{j}}  \ottsym{.}  \ottcomp{{\source {\mathit{TC} } }_{\ottmv{i}} \, {\source a }}{\ottmv{i}}  \Rightarrow  {\source \tau }  }\,{ \stoi{ \rightsquigarrow  \target {  {\target \sigma }  } } } 
    \label{eq:eq_env1} \\
    \ottcomp{ \source {  {\source { \Gamma_C } }  } ; \source {  \bullet  \ottsym{,}  {\source a }  }  \vdash_{\mathit {Q} }  \source {  {\source {\mathit{TC} } }_{\ottmv{i}} \, {\source a }  }  \rightsquigarrow  \target {  {\target \sigma }_{\ottmv{i}}  } }{\ottmv{i}}
    \label{eq:eq_env2} \\
     \vdash_{ctx}  \source {  \bullet  } ; \source {  {\source { \Gamma_C } }  } ; \source {  \bullet  }  \rightsquigarrow  \target {  \target {\bullet}  } 
    \label{eq:eq_env3}
  \end{gather}
  By applying the induction hypothesis to Equation~\ref{eq:eq_env3}, we get:
  \begin{gather}
     \vdash_{ctx}^{\intermed {M} }  \source {  \bullet  } ; \source {  {\source { \Gamma_C } }  } ; \source {  \bullet  }  \rightsquigarrow  \intermed {  \bullet  } ; \intermed {  {\intermed { \Gamma_C } }'  } ; \intermed {  \bullet  } 
    \label{eq:eq_env4} \\
     \intermed {  {\intermed { \Gamma_C } }'  } ; \intermed {  \bullet  }  \rightsquigarrow  \target {  \target {\bullet}  } 
    \label{eq:eq_env5} 
  \end{gather}
  From \textsc{sCtxT-tyEnvTy}, \textsc{sCtx-tyEnvTy} and \textsc{Ctx-TVar}, in combination with
  Equation~\ref{eq:eq_env3}, \ref{eq:eq_env4} and~\ref{eq:eq_env5}, respectively, we get:
  \begin{gather}
     \vdash_{ctx}  \source {  \bullet  } ; \source {  {\source { \Gamma_C } }  } ; \source {  \bullet  \ottsym{,}  {\source a }  }  \rightsquigarrow  \target {  \target {\bullet}  \ottsym{,}  {\target a }  } 
    \nonumber \\
     \vdash_{ctx}^{\intermed {M} }  \source {  \bullet  } ; \source {  {\source { \Gamma_C } }  } ; \source {  \bullet  \ottsym{,}  {\source a }  }  \rightsquigarrow  \intermed {  \bullet  } ; \intermed {  {\intermed { \Gamma_C } }'  } ; \intermed {  \bullet  \ottsym{,}  {\intermed a }  } 
    \nonumber \\
     \intermed {  {\intermed { \Gamma_C } }'  } ; \intermed {  \bullet  \ottsym{,}  {\intermed a }  }  \rightsquigarrow  \target {  \target {\bullet}  \ottsym{,}  {\target a }  } 
    \nonumber 
  \end{gather}
  Applying type and constraint equivalence (Theorem~\ref{thmA:equiv_ty}) to
  Equations~\ref{eq:eq_env1} and~\ref{eq:eq_env2}, together with these results, gives us:
  \begin{gather}
     \source {  {\source { \Gamma_C } }  } ; \source {  \bullet  \ottsym{,}  {\source a }  }  \vdash_{ty}^{\intermed {M} }  \source {  \forall  {\overline {\source a} }_{\ottmv{j}}  \ottsym{.}  \ottcomp{{\source {\mathit{TC} } }_{\ottmv{i}} \, {\source a }}{\ottmv{i}}  \Rightarrow  {\source \tau }  }\,{ \stoi{ \rightsquigarrow  \intermed {  {\intermed \sigma}  } } } 
    \label{eq:eq_env6} \\
     \intermed {  {\intermed { \Gamma_C } }'  } ; \intermed {  \bullet  \ottsym{,}  {\intermed a }  }  \vdash_{ty}  \intermed {  {\intermed \sigma}  } \itot{ \rightsquigarrow  \target {  {\target \sigma }  } } 
    \label{eq:eq_env7} \\
    \ottcomp{ \source {  {\source { \Gamma_C } }  } ; \source {  \bullet  \ottsym{,}  {\source a }  }  \vdash_{\mathit {Q} }^{\intermed {M} }  \source {  {\source {\mathit{TC} } }_{\ottmv{i}} \, {\source a }  }  \rightsquigarrow  \intermed {  {\intermed Q}_{\ottmv{i}}  } }{\ottmv{i}}
    \label{eq:eq_env8} \\
    \ottcomp{ \intermed {  {\intermed { \Gamma_C } }'  } ; \intermed {  \bullet  \ottsym{,}  {\intermed a }  }  \vdash_{\mathit {Q} }  \intermed {  {\intermed Q}_{\ottmv{i}}  }\, \itot{ \rightsquigarrow  \target {  {\target \sigma }_{\ottmv{i}}  } } }{\ottmv{i}}
    \label{eq:eq_env9} 
  \end{gather}
  Goal~\ref{eqg:eq_env1} follows from \textsc{sCtx-clsEnv}, in combination with
  Equations~\ref{eq:eq_env4}, \ref{eq:eq_env6} and~\ref{eq:eq_env8}, with
  \( \intermed {  {\intermed \Sigma}  } = \intermed {  \bullet  } \), \( \intermed {  {\intermed { \Gamma_C } }  } = \intermed {  {\intermed { \Gamma_C } }'  }, \intermed {  {\intermed m}  \ottsym{:}  {\intermed {\mathit{TC} } } \, {\intermed a }  \ottsym{:}  {\intermed \sigma}  } \) and \( \intermed {  {\intermed \Gamma}  } = \intermed {  \bullet  } \).
  Consequently, Goal~\ref{eqg:eq_env2} follows from \textsc{Ctx-Empty}. \\
}
\proofStep{\textsc{sCtxT-tyEnvTm}}
{
   \vdash_{ctx}  \source {  \bullet  } ; \source {  {\source { \Gamma_C } }  } ; \source {  {\source \Gamma}  \ottsym{,}  {\source x}  \ottsym{:}  {\source \sigma}  }  \rightsquigarrow  \target {  {\target \Gamma }  \ottsym{,}  {\target x}  \ottsym{:}  {\target \sigma }  } 
}
{
  The goal to be proven is the following:
  \begin{gather}
     \vdash_{ctx}^{\intermed {M} }  \source {  \bullet  } ; \source {  {\source { \Gamma_C } }  } ; \source {  {\source \Gamma}  \ottsym{,}  {\source x}  \ottsym{:}  {\source \sigma}  }  \rightsquigarrow  \intermed {  {\intermed \Sigma}  } ; \intermed {  {\intermed { \Gamma_C } }  } ; \intermed {  {\intermed \Gamma}  } 
    \label{eqg:eq_env3} \\
     \intermed {  {\intermed { \Gamma_C } }  } ; \intermed {  {\intermed \Gamma}  }  \rightsquigarrow  \target {  {\target \Gamma }  \ottsym{,}  {\target x}  \ottsym{:}  {\target \sigma }  } 
    \label{eqg:eq_env4}
  \end{gather}
  From the rule premise we get that:
  \begin{gather}
     \source {  {\source { \Gamma_C } }  } ; \source {  {\source \Gamma}  }  \vdash_{ty}  \source {  {\source \sigma}  }\,{ \stoi{ \rightsquigarrow  \target {  {\target \sigma }  } } } 
    \label{eq:eq_env12} \\
     \source {  {\source x}  }  \notin   \textbf {dom}  ( \source {  {\source \Gamma}  } ) 
    \label{eq:eq_env13} \\
     \vdash_{ctx}  \source {  \bullet  } ; \source {  {\source { \Gamma_C } }  } ; \source {  {\source \Gamma}  }  \rightsquigarrow  \target {  {\target \Gamma }  } 
    \label{eq:eq_env14}
  \end{gather}
  By applying the induction hypothesis to Equation~\ref{eq:eq_env14}, we get:
  \begin{gather}
     \vdash_{ctx}^{\intermed {M} }  \source {  \bullet  } ; \source {  {\source { \Gamma_C } }  } ; \source {  {\source \Gamma}  }  \rightsquigarrow  \intermed {  \bullet  } ; \intermed {  {\intermed { \Gamma_C } }  } ; \intermed {  {\intermed \Gamma}'  } 
    \label{eq:eq_env15} \\
     \intermed {  {\intermed { \Gamma_C } }  } ; \intermed {  {\intermed \Gamma}'  }  \rightsquigarrow  \target {  {\target \Gamma }'  } 
    \label{eq:eq_env16}
  \end{gather}
  We know from type equivalence (Theorem~\ref{thmA:equiv_ty}), in combination with
  Equations~\ref{eq:eq_env12}, \ref{eq:eq_env15} and~\ref{eq:eq_env16}, that:
  \begin{gather}
     \source {  {\source { \Gamma_C } }  } ; \source {  {\source \Gamma}  }  \vdash_{ty}^{\intermed {M} }  \source {  {\source \sigma}  }\,{ \stoi{ \rightsquigarrow  \intermed {  {\intermed \sigma}  } } } 
    \label{eq:eq_env17} \\
     \intermed {  {\intermed { \Gamma_C } }  } ; \intermed {  {\intermed \Gamma}  }  \vdash_{ty}  \intermed {  {\intermed \sigma}  } \itot{ \rightsquigarrow  \target {  {\target \sigma }  } } 
    \label{eq:eq_env18}
  \end{gather}
  Goal~\ref{eqg:eq_env3} follows from \textsc{sCtx-tyEnvTm}, in combination with
  Equations~\ref{eq:eq_env13}, \ref{eq:eq_env15} and~\ref{eq:eq_env17}, with
  \( \intermed {  {\intermed \Sigma}  } = \intermed {  \bullet  } \) and \( \intermed {  {\intermed \Gamma}  } = \intermed {  {\intermed \Gamma}'  \ottsym{,}  {\intermed x}  \ottsym{:}  {\intermed \sigma}  } \).
  Consequently, Goal~\ref{eqg:eq_env4} follows from \textsc{Ctx-Var}, in
  combination with Equations~\ref{eq:eq_env16} and~\ref{eq:eq_env18}, with
  \( \target {  {\target \Gamma }  } = \target {  {\target \Gamma }'  } \). \\
}
\proofStep{\textsc{sCtxT-tyEnvTy}}
{
   \vdash_{ctx}  \source {  \bullet  } ; \source {  {\source { \Gamma_C } }  } ; \source {  {\source \Gamma}  \ottsym{,}  {\source a }  }  \rightsquigarrow  \target {  {\target \Gamma }  \ottsym{,}  {\target a }  } 
}
{
  Similar to the \textsc{sCtxT-tyEnvTm} case. \\
}
\proofStep{\textsc{sCtxT-tyEnvD}}
{
   \vdash_{ctx}  \source {  \bullet  } ; \source {  {\source { \Gamma_C } }  } ; \source {  {\source \Gamma}  \ottsym{,}  {\source \delta }  \ottsym{:}  {\source {\mathit{TC} } } \, {\source \tau }  }  \rightsquigarrow  \target {  {\target \Gamma }  \ottsym{,}  {\target \delta }  \ottsym{:}  {\target \sigma }  } 
}
{
  Similar to the \textsc{sCtxT-tyEnvTm} case. \\
}
\proofStep{\textsc{sCtxT-pgmInst}}
{
   \vdash_{ctx}  \source {  { \source P }  \ottsym{,}  \ottsym{(}  {\source D }  \ottsym{:}  \forall  {\overline {\source b} }_{\ottmv{j}}  \ottsym{.}  {\overline {\source Q} }_{\ottmv{i}}  \Rightarrow  {\source {\mathit{TC} } } \, {\source \tau }  \ottsym{)}  \ottsym{.}  {\source m}  \mapsto  \bullet  \ottsym{,}  {\overline {\source b} }_{\ottmv{j}}  \ottsym{,}  {\overline {\source \delta} }_{\ottmv{i}}  \ottsym{:}  {\overline {\source Q} }_{\ottmv{i}}  \ottsym{,}  {\overline {\source a} }_{\ottmv{k}}  \ottsym{,}  {\overline {\source \delta} }_{\ottmv{h}}  \ottsym{:}  [  {\source \tau }  /  {\source a }  ]  {\overline {\source Q} }'_{\ottmv{h}}  \ottsym{:}  {\source e}  } ; \source {  {\source { \Gamma_C } }  } ; \source {  {\source \Gamma}  }  \rightsquigarrow  \target {  {\target \Gamma }  } 
}
{
  The goal to be proven is the following:
  \begin{gather}
     \vdash_{ctx}^{\intermed {M} }  \source {  { \source P }  \ottsym{,}  \ottsym{(}  {\source D }  \ottsym{:}  \forall  {\overline {\source b} }_{\ottmv{j}}  \ottsym{.}  {\overline {\source Q} }_{\ottmv{i}}  \Rightarrow  {\source {\mathit{TC} } } \, {\source \tau }  \ottsym{)}  \ottsym{.}  {\source m}  \mapsto  \bullet  \ottsym{,}  {\overline {\source b} }_{\ottmv{j}}  \ottsym{,}  {\overline {\source \delta} }_{\ottmv{i}}  \ottsym{:}  {\overline {\source Q} }_{\ottmv{i}}  \ottsym{,}  {\overline {\source a} }_{\ottmv{k}}  \ottsym{,}  {\overline {\source \delta} }_{\ottmv{h}}  \ottsym{:}  [  {\source \tau }  /  {\source a }  ]  {\overline {\source Q} }'_{\ottmv{h}}  \ottsym{:}  {\source e}  } ; \source {  {\source { \Gamma_C } }  } ; \source {  {\source \Gamma}  }  \rightsquigarrow  \intermed {  {\intermed \Sigma}  } ; \intermed {  {\intermed { \Gamma_C } }  } ; \intermed {  {\intermed \Gamma}  } 
    \label{eqg:eq_env5} \\
     \intermed {  {\intermed { \Gamma_C } }  } ; \intermed {  {\intermed \Gamma}  }  \rightsquigarrow  \target {  {\target \Gamma }  } 
    \label{eqg:eq_env6}
  \end{gather}
  From the rule premise we get that:
  \begin{gather}
     \textbf {unambig}  ( \source {  \forall  {\overline {\source b} }_{\ottmv{j}}  \ottsym{.}  {\overline {\source Q} }_{\ottmv{i}}  \Rightarrow  {\source {\mathit{TC} } } \, {\source \tau }  } ) 
    \label{eq:eq_env19} \\
     \source {  {\source { \Gamma_C } }  } ; \source {  \bullet  }  \vdash_{\mathit {C} }  \source {  \forall  {\overline {\source b} }_{\ottmv{j}}  \ottsym{.}  {\overline {\source Q} }_{\ottmv{i}}  \Rightarrow  {\source {\mathit{TC} } } \, {\source \tau }  }  \rightsquigarrow  \target {  \forall  {\overline {\target b} }_{\ottmv{j}}  \ottsym{.}  {\overline {\target \sigma} }_{\ottmv{i}}  \rightarrow  [  {\target \sigma }  /  {\target a }  ]  \ottsym{\{}  {\target m}  \ottsym{:}  \forall  {\overline {\target a} }_{\ottmv{k}}  \ottsym{.}  {\overline {\target \sigma} }'_{\ottmv{h}}  \rightarrow  {\target \sigma }'  \ottsym{\}}  } 
    \label{eq:eq_env20} \\
     ( \source {  {\source m}  } : \source {  {\overline {\source Q} }'_{\ottmv{m}}  }  \Rightarrow  \source {  {\source {\mathit{TC} } }  }~\source {  {\source a }  } : \source {  \forall  {\overline {\source a} }_{\ottmv{k}}  \ottsym{.}  {\overline {\source Q} }'_{\ottmv{h}}  \Rightarrow  {\source \tau }'  } )  \in  \source {  {\source { \Gamma_C } }  } 
    \label{eq:eq_env21} \\
     \source {  {\source { \Gamma_C } }  } ; \source {  \bullet  \ottsym{,}  {\source a }  }  \vdash_{ty}  \source {  \forall  {\overline {\source a} }_{\ottmv{k}}  \ottsym{.}  {\overline {\source Q} }'_{\ottmv{h}}  \Rightarrow  {\source \tau }'  }\,{ \stoi{ \rightsquigarrow  \target {  \forall  {\overline {\target a} }_{\ottmv{k}}  \ottsym{.}  {\overline {\target \sigma} }'_{\ottmv{h}}  \rightarrow  {\target \sigma }'  } } } 
    \label{eq:eq_env22} \\
     \source {  {\source { \Gamma_C } }  } ; \source {  \bullet  \ottsym{,}  {\overline {\source b} }_{\ottmv{j}}  }  \vdash_{ty}  \source {  {\source \tau }  }\,{ \stoi{ \rightsquigarrow  \target {  {\target \sigma }  } } } 
    \label{eq:eq_env23} \\
     \source {  { \source P }  } ; \source {  {\source { \Gamma_C } }  } ; \source {  \bullet  \ottsym{,}  {\overline {\source b} }_{\ottmv{j}}  \ottsym{,}  {\overline {\source \delta} }_{\ottmv{i}}  \ottsym{:}  {\overline {\source Q} }_{\ottmv{i}}  \ottsym{,}  {\overline {\source a} }_{\ottmv{k}}  \ottsym{,}  {\overline {\source \delta} }_{\ottmv{h}}  \ottsym{:}  [  {\source \tau }  /  {\source a }  ]  {\overline {\source Q} }'_{\ottmv{h}}  }  \vdash_{tm}  \source {  {\source e}  } \Leftarrow \source {  [  {\source \tau }  /  {\source a }  ]  {\source \tau }'  } \stot{ \rightsquigarrow  \target {  {\target e }  } } 
    \label{eq:eq_env24} \\
     \source {  {\source D }  }  \notin   \textbf {dom}  ( \source {  { \source P }  } ) 
    \label{eq:eq_env25} \\
     ( \source {  {\source D }'  } : \source {  \forall  {\overline {\source b} }'_{\ottmv{k}}  \ottsym{.}  {\overline {\source Q} }''_{\ottmv{h}}  \Rightarrow  {\source {\mathit{TC} } } \, {\source \tau }''  } ) . \source {  {\source m}'  }  \mapsto  \source {  {\source \Gamma}'  } : \source {  {\source e}'  }  \notin  \source {  { \source P }  } 
     \textbf {where}  \source {  [  {\overline {\source \tau} }_{\ottmv{j}}  /  {\overline {\source b} }_{\ottmv{j}}  ]  {\source \tau }  } = \source {  [  {\overline {\source \tau} }'_{\ottmv{k}}  /  {\overline {\source b} }'_{\ottmv{k}}  ]  {\source \tau }''  } 
    \label{eq:eq_env26} \\
     \vdash_{ctx}  \source {  { \source P }  } ; \source {  {\source { \Gamma_C } }  } ; \source {  {\source \Gamma}  }  \rightsquigarrow  \target {  {\target \Gamma }  } 
    \label{eq:eq_env27}
  \end{gather}
  By applying the induction hypothesis to Equation~\ref{eq:eq_env27}, we get: 
  \begin{gather}
     \vdash_{ctx}^{\intermed {M} }  \source {  { \source P }  } ; \source {  {\source { \Gamma_C } }  } ; \source {  {\source \Gamma}  }  \rightsquigarrow  \intermed {  {\intermed \Sigma}'  } ; \intermed {  {\intermed { \Gamma_C } }  } ; \intermed {  {\intermed \Gamma}  } 
    \label{eq:eq_env29} \\
     \intermed {  {\intermed { \Gamma_C } }  } ; \intermed {  {\intermed \Gamma}  }  \rightsquigarrow  \target {  {\target \Gamma }  } 
    \label{eq:eq_env30}
  \end{gather}
  Goal~\ref{eqg:eq_env6} follows directly from Equation~\ref{eq:eq_env30}. 
  From type and constraint equivalence (Theorem~\ref{thmA:equiv_ty}, the required
  assumptions follow straightforwardly from \textsc{sCtxT-tyEnvTy},
  \textsc{sCtx-tyEnvTy} and \textsc{Ctx-TVar}, in combination with
  Equations~\ref{eq:eq_env27}, \ref{eq:eq_env29} and~\ref{eq:eq_env30}),
  together with Equations~\ref{eq:eq_env20}, \ref{eq:eq_env22} and~\ref{eq:eq_env23},
  we know: 
  \begin{gather}
     \source {  {\source { \Gamma_C } }  } ; \source {  \bullet  }  \vdash_{\mathit {C} }^{\intermed {M} }  \source {  \forall  {\overline {\source b} }_{\ottmv{j}}  \ottsym{.}  {\overline {\source Q} }_{\ottmv{i}}  \Rightarrow  {\source {\mathit{TC} } } \, {\source \tau }  }  \rightsquigarrow  \intermed {  \forall  {\overline {\intermed b} }_{\ottmv{j}}  \ottsym{.}  {\overline {\intermed Q} }_{\ottmv{i}}  \Rightarrow  {\intermed {\mathit{TC} } } \, {\intermed \sigma}  } 
    \label{eq:eq_env32} \\
     \source {  {\source { \Gamma_C } }  } ; \source {  \bullet  \ottsym{,}  {\source a }  }  \vdash_{ty}^{\intermed {M} }  \source {  \forall  {\overline {\source a} }_{\ottmv{k}}  \ottsym{.}  {\overline {\source Q} }'_{\ottmv{h}}  \Rightarrow  {\source \tau }'  }\,{ \stoi{ \rightsquigarrow  \intermed {  \forall  {\overline {\intermed a} }_{\ottmv{k}}  \ottsym{.}   {\overline {\intermed Q} }'_{\ottmv{h}}  \Rightarrow  {\intermed \sigma}'   } } } 
    \label{eq:eq_env33} \\
     \intermed {  {\intermed { \Gamma_C } }  } ; \intermed {  \bullet  \ottsym{,}  {\intermed a }  }  \vdash_{ty}  \intermed {  \forall  {\overline {\intermed a} }_{\ottmv{k}}  \ottsym{.}   {\overline {\intermed Q} }'_{\ottmv{h}}  \Rightarrow  {\intermed \sigma}'   } \itot{ \rightsquigarrow  \target {  \forall  {\overline {\target a} }_{\ottmv{k}}  \ottsym{.}  {\overline {\target \sigma} }'_{\ottmv{h}}  \rightarrow  {\target \sigma }'  } } 
    \label{eq:eq_env34} \\
     \source {  {\source { \Gamma_C } }  } ; \source {  \bullet  \ottsym{,}  {\overline {\source b} }_{\ottmv{j}}  }  \vdash_{ty}^{\intermed {M} }  \source {  {\source \tau }  }\,{ \stoi{ \rightsquigarrow  \intermed {  {\intermed \sigma}  } } } 
    \label{eq:eq_env35} \\
     \intermed {  {\intermed { \Gamma_C } }  } ; \intermed {  \bullet  \ottsym{,}  {\overline {\intermed b} }_{\ottmv{j}}  }  \vdash_{ty}  \intermed {  {\intermed \sigma}  } \itot{ \rightsquigarrow  \target {  {\target \sigma }  } } 
    \label{eq:eq_env36}
  \end{gather}
  Similarly, from expression equivalence (Theorem~\ref{thmA:equiv_expr}, the
  environment well-formedness assumption is constructed straightforwardly),
  together with Equation~\ref{eq:eq_env24}, we get:
  \begin{equation}
     \source {  { \source P }  } ; \source {  {\source { \Gamma_C } }  } ; \source {  \bullet  \ottsym{,}  {\overline {\source b} }_{\ottmv{j}}  \ottsym{,}  {\overline {\source \delta} }_{\ottmv{i}}  \ottsym{:}  {\overline {\source Q} }_{\ottmv{i}}  \ottsym{,}  {\overline {\source a} }_{\ottmv{k}}  \ottsym{,}  {\overline {\source \delta} }_{\ottmv{h}}  \ottsym{:}  [  {\source \tau }  /  {\source a }  ]  {\overline {\source Q} }'_{\ottmv{h}}  }  \vdash_{tm}^{\intermed {M} }  \source {  {\source e}  } \Leftarrow \source {  [  {\source \tau }  /  {\source a }  ]  {\source \tau }'  } \stoi{ \rightsquigarrow  \intermed {  {\intermed e}  } } 
    \label{eq:eq_env37}
  \end{equation}
  Goal~\ref{eqg:eq_env5} follows from \textsc{sCtx-pgmInst}, in combination with
  Equations~\ref{eq:eq_env19}, \ref{eq:eq_env32}, \ref{eq:eq_env21},
  \ref{eq:eq_env37}, \ref{eq:eq_env33}, \ref{eq:eq_env25}, \ref{eq:eq_env26}
  and~\ref{eq:eq_env29}, and with \( \intermed {  {\intermed \Sigma}  } = \intermed {  {\intermed \Sigma}'  \ottsym{,}  \ottsym{(}  {\intermed D }  \ottsym{:}  \forall  {\overline {\intermed b} }_{\ottmv{j}}  \ottsym{.}  {\overline {\intermed Q} }_{\ottmv{i}}  \Rightarrow  {\intermed {\mathit{TC} } } \, {\intermed \sigma}'  \ottsym{)}  \ottsym{.}  {\intermed m}  \mapsto  \Lambda  {\overline {\intermed b} }_{\ottmv{j}}  \ottsym{.}  \lambda  {\overline {\intermed \delta} }_{\ottmv{i}}  \ottsym{:}  {\overline {\intermed Q} }_{\ottmv{i}}  \ottsym{.}  \Lambda  {\overline {\intermed a} }_{\ottmv{k}}  \ottsym{.}  \lambda  {\overline {\intermed \delta} }_{\ottmv{h}}  \ottsym{:}  [  {\intermed \sigma}  /  {\intermed a }  ]  {\overline {\intermed Q} }'_{\ottmv{h}}  \ottsym{.}  {\intermed e}  } \).
}
\qed

\begin{theoremB}[Equivalence - Types and Constraints]\leavevmode
  \begin{itemize}
  \item
    If \( \source {  {\source { \Gamma_C } }  } ; \source {  {\source \Gamma}  }  \vdash_{ty}  \source {  {\source \sigma}  }\,{ \stoi{ \rightsquigarrow  \target {  {\target \sigma }  } } } \)
    and \( \vdash_{ctx}  \source {  { \source P }  } ; \source {  {\source { \Gamma_C } }  } ; \source {  {\source \Gamma}  }  \rightsquigarrow  \target {  {\target \Gamma }  } \) \\
    and \( \vdash_{ctx}^{\intermed {M} }  \source {  { \source P }  } ; \source {  {\source { \Gamma_C } }  } ; \source {  {\source \Gamma}  }  \rightsquigarrow  \intermed {  {\intermed \Sigma}  } ; \intermed {  {\intermed { \Gamma_C } }  } ; \intermed {  {\intermed \Gamma}  } \)
    and \( \intermed {  {\intermed { \Gamma_C } }  } ; \intermed {  {\intermed \Gamma}  }  \rightsquigarrow  \target {  {\target \Gamma }  } \) \\
    then \( \source {  {\source { \Gamma_C } }  } ; \source {  {\source \Gamma}  }  \vdash_{ty}^{\intermed {M} }  \source {  {\source \sigma}  }\,{ \stoi{ \rightsquigarrow  \intermed {  {\intermed \sigma}  } } } \)
    and \( \intermed {  {\intermed { \Gamma_C } }  } ; \intermed {  {\intermed \Gamma}  }  \vdash_{ty}  \intermed {  {\intermed \sigma}  } \itot{ \rightsquigarrow  \target {  {\target \sigma }  } } \).
  \item
    If \( \source {  {\source { \Gamma_C } }  } ; \source {  {\source \Gamma}  }  \vdash_{\mathit {Q} }  \source {  {\source Q}  }  \rightsquigarrow  \target {  {\target \sigma }  } \)
    and \( \vdash_{ctx}  \source {  { \source P }  } ; \source {  {\source { \Gamma_C } }  } ; \source {  {\source \Gamma}  }  \rightsquigarrow  \target {  {\target \Gamma }  } \) \\
    and \( \vdash_{ctx}^{\intermed {M} }  \source {  { \source P }  } ; \source {  {\source { \Gamma_C } }  } ; \source {  {\source \Gamma}  }  \rightsquigarrow  \intermed {  {\intermed \Sigma}  } ; \intermed {  {\intermed { \Gamma_C } }  } ; \intermed {  {\intermed \Gamma}  } \)
    and \( \intermed {  {\intermed { \Gamma_C } }  } ; \intermed {  {\intermed \Gamma}  }  \rightsquigarrow  \target {  {\target \Gamma }  } \) \\
    then \( \source {  {\source { \Gamma_C } }  } ; \source {  {\source \Gamma}  }  \vdash_{\mathit {Q} }^{\intermed {M} }  \source {  {\source Q}  }  \rightsquigarrow  \intermed {  {\intermed Q}  } \)
    and \( \intermed {  {\intermed { \Gamma_C } }  } ; \intermed {  {\intermed \Gamma}  }  \vdash_{\mathit {Q} }  \intermed {  {\intermed Q}  }\, \itot{ \rightsquigarrow  \target {  {\target \sigma }  } } \).
  \item
    If \( \source {  {\source { \Gamma_C } }  } ; \source {  {\source \Gamma}  }  \vdash_{\mathit {C} }  \source {  {\source C}  }  \rightsquigarrow  \target {  {\target \sigma }  } \)
    and \( \vdash_{ctx}  \source {  { \source P }  } ; \source {  {\source { \Gamma_C } }  } ; \source {  {\source \Gamma}  }  \rightsquigarrow  \target {  {\target \Gamma }  } \) \\
    and \( \vdash_{ctx}^{\intermed {M} }  \source {  { \source P }  } ; \source {  {\source { \Gamma_C } }  } ; \source {  {\source \Gamma}  }  \rightsquigarrow  \intermed {  {\intermed \Sigma}  } ; \intermed {  {\intermed { \Gamma_C } }  } ; \intermed {  {\intermed \Gamma}  } \)
    and \( \intermed {  {\intermed { \Gamma_C } }  } ; \intermed {  {\intermed \Gamma}  }  \rightsquigarrow  \target {  {\target \Gamma }  } \) \\
    then \( \source {  {\source { \Gamma_C } }  } ; \source {  {\source \Gamma}  }  \vdash_{\mathit {C} }^{\intermed {M} }  \source {  {\source C}  }  \rightsquigarrow  \intermed {  {\intermed C}  } \)
    and \( \intermed {  {\intermed { \Gamma_C } }  } ; \intermed {  {\intermed \Gamma}  }  \vdash_{\mathit {C} }  \intermed {  {\intermed C}  } \).
  \end{itemize}
  \label{thmA:equiv_ty}
\end{theoremB}

\proof
By induction on the size of the type \({\source \sigma}\), class constraint
\({\source Q}\) or constraint \({\source C}\).

\begin{description}
\item[Part 1] By case analysis on the type well-formedness derivation. \\
  \proofStep{\textsc{sTyT-bool}}
  {
     \source {  {\source { \Gamma_C } }  } ; \source {  {\source \Gamma}  }  \vdash_{ty}  \source {  \mathit{\source {Bool} }  }\,{ \stoi{ \rightsquigarrow  \target {  \mathit{\target {Bool} }  } } } 
  }
  {
    The goal to be proven is the following:
    \begin{gather}
       \source {  {\source { \Gamma_C } }  } ; \source {  {\source \Gamma}  }  \vdash_{ty}^{\intermed {M} }  \source {  \mathit{\source {Bool} }  }\,{ \stoi{ \rightsquigarrow  \intermed {  {\intermed \sigma}  } } } 
      \label{eqg:eq_ty1} \\
       \intermed {  {\intermed { \Gamma_C } }  } ; \intermed {  {\intermed \Gamma}  }  \vdash_{ty}  \intermed {  {\intermed \sigma}  } \itot{ \rightsquigarrow  \target {  \mathit{\target {Bool} }  } } 
      \label{eqg:eq_ty2} 
    \end{gather}
    Goals~\ref{eqg:eq_ty1} and~\ref{eqg:eq_ty2} follow directly from
    \textsc{sTy-bool} and \textsc{iTy-bool} respectively,
    with \( \intermed {  {\intermed \sigma}  } = \intermed {  \mathit{\intermed {Bool} }  } \). \\
  }
  \proofStep{\textsc{sTyT-var}}
  {
     \source {  {\source { \Gamma_C } }  } ; \source {  {\source \Gamma}  }  \vdash_{ty}  \source {  {\source a }  }\,{ \stoi{ \rightsquigarrow  \target {  {\target a }  } } } 
  }
  {
    The goal to be proven is the following:
    \begin{gather}
       \source {  {\source { \Gamma_C } }  } ; \source {  {\source \Gamma}  }  \vdash_{ty}^{\intermed {M} }  \source {  {\source a }  }\,{ \stoi{ \rightsquigarrow  \intermed {  {\intermed \sigma}  } } } 
      \label{eqg:eq_ty3} \\
       \intermed {  {\intermed { \Gamma_C } }  } ; \intermed {  {\intermed \Gamma}  }  \vdash_{ty}  \intermed {  {\intermed \sigma}  } \itot{ \rightsquigarrow  \target {  {\target a }  } } 
      \label{eqg:eq_ty4} 
    \end{gather}
    From the rule premise we get that:
    \begin{equation}
       \source {  {\source a }  }  \in  \source {  {\source \Gamma}  } 
      \label{eq:eq_ty1}
    \end{equation}
    By applying Lemmas~\ref{lemma:pres_env_ty} and~\ref{lemma:pres_env_tyT} to
    Equation~\ref{eq:eq_ty1}, we get:
    \begin{gather}
       \intermed {  {\intermed a }  }  \in  \intermed {  {\intermed \Gamma}  } 
      \label{eq:eq_ty1b} \\
       \target {  {\target a }  }  \in  \target {  {\target \Gamma }  } 
      \label{eq:eq_ty1c}
    \end{gather}
    Goal~\ref{eqg:eq_ty3} and~\ref{eqg:eq_ty4} follow directly from
    \textsc{sTy-var} and \textsc{iTy-var} respectively, in combination with
    Equations~\ref{eq:eq_ty1b} and~\ref{eq:eq_ty1c}, with \( \intermed {  {\intermed \sigma}  } = \intermed {  {\intermed a }  } \). \\
  }
  \proofStep{\textsc{sTyT-arrow}}
  {
     \source {  {\source { \Gamma_C } }  } ; \source {  {\source \Gamma}  }  \vdash_{ty}  \source {  {\source \tau }_{{\mathrm{1}}}  \rightarrow  {\source \tau }_{{\mathrm{2}}}  }\,{ \stoi{ \rightsquigarrow  \target {  {\target \sigma }_{{\mathrm{1}}}  \rightarrow  {\target \sigma }_{{\mathrm{2}}}  } } } 
  }
  {
    The goal to be proven is the following:
    \begin{gather}
       \source {  {\source { \Gamma_C } }  } ; \source {  {\source \Gamma}  }  \vdash_{ty}^{\intermed {M} }  \source {  {\source \tau }_{{\mathrm{1}}}  \rightarrow  {\source \tau }_{{\mathrm{2}}}  }\,{ \stoi{ \rightsquigarrow  \intermed {  {\intermed \sigma}  } } } 
      \label{eqg:eq_ty5} \\
       \intermed {  {\intermed { \Gamma_C } }  } ; \intermed {  {\intermed \Gamma}  }  \vdash_{ty}  \intermed {  {\intermed \sigma}  } \itot{ \rightsquigarrow  \target {  {\target \sigma }_{{\mathrm{1}}}  \rightarrow  {\target \sigma }_{{\mathrm{2}}}  } } 
      \label{eqg:eq_ty6} 
    \end{gather}
    From the rule premise we get that:
    \begin{gather}
       \source {  {\source { \Gamma_C } }  } ; \source {  {\source \Gamma}  }  \vdash_{ty}  \source {  {\source \tau }_{{\mathrm{1}}}  }\,{ \stoi{ \rightsquigarrow  \target {  {\target \sigma }_{{\mathrm{1}}}  } } } 
      \label{eq:eq_ty2} \\
       \source {  {\source { \Gamma_C } }  } ; \source {  {\source \Gamma}  }  \vdash_{ty}  \source {  {\source \tau }_{{\mathrm{2}}}  }\,{ \stoi{ \rightsquigarrow  \target {  {\target \sigma }_{{\mathrm{2}}}  } } } 
      \label{eq:eq_ty3}
    \end{gather}
    By applying the induction hypothesis on Equations~\ref{eq:eq_ty2}
    and~\ref{eq:eq_ty3}, we get:
    \begin{gather}
       \source {  {\source { \Gamma_C } }  } ; \source {  {\source \Gamma}  }  \vdash_{ty}^{\intermed {M} }  \source {  {\source \tau }_{{\mathrm{1}}}  }\,{ \stoi{ \rightsquigarrow  \intermed {  {\intermed \sigma}_{{\mathrm{1}}}  } } } 
      \label{eq:eq_ty4} \\
       \intermed {  {\intermed { \Gamma_C } }  } ; \intermed {  {\intermed \Gamma}  }  \vdash_{ty}  \intermed {  {\intermed \sigma}_{{\mathrm{1}}}  } \itot{ \rightsquigarrow  \target {  {\target \sigma }_{{\mathrm{1}}}  } } 
      \label{eq:eq_ty5} \\
       \source {  {\source { \Gamma_C } }  } ; \source {  {\source \Gamma}  }  \vdash_{ty}^{\intermed {M} }  \source {  {\source \tau }_{{\mathrm{2}}}  }\,{ \stoi{ \rightsquigarrow  \intermed {  {\intermed \sigma}_{{\mathrm{2}}}  } } } 
      \label{eq:eq_ty6} \\
       \intermed {  {\intermed { \Gamma_C } }  } ; \intermed {  {\intermed \Gamma}  }  \vdash_{ty}  \intermed {  {\intermed \sigma}_{{\mathrm{2}}}  } \itot{ \rightsquigarrow  \target {  {\target \sigma }_{{\mathrm{2}}}  } } 
      \label{eq:eq_ty7}
    \end{gather}
    Goals~\ref{eqg:eq_ty5} and~\ref{eqg:eq_ty6} follow directly from
    \textsc{sTy-arrow} and \textsc{iTy-arrow} respectively, in combination with
    Equations~\ref{eq:eq_ty4}, \ref{eq:eq_ty5}, \ref{eq:eq_ty6}
    and~\ref{eq:eq_ty7}, with \( \intermed {  {\intermed \sigma}  } = \intermed {  {\intermed \sigma}_{{\mathrm{1}}}  \rightarrow  {\intermed \sigma}_{{\mathrm{2}}}  } \). \\
  }
  \proofStep{\textsc{sTyT-qual}}
  {
     \source {  {\source { \Gamma_C } }  } ; \source {  {\source \Gamma}  }  \vdash_{ty}  \source {  {\source Q}  \Rightarrow  {\source \rho}  }\,{ \stoi{ \rightsquigarrow  \target {  {\target \sigma }_{{\mathrm{1}}}  \rightarrow  {\target \sigma }_{{\mathrm{2}}}  } } } 
  }
  {
    The goal to be proven is the following:
    \begin{gather}
       \source {  {\source { \Gamma_C } }  } ; \source {  {\source \Gamma}  }  \vdash_{ty}^{\intermed {M} }  \source {  {\source Q}  \Rightarrow  {\source \rho}  }\,{ \stoi{ \rightsquigarrow  \intermed {  {\intermed \sigma}  } } } 
      \label{eqg:eq_ty7} \\
       \intermed {  {\intermed { \Gamma_C } }  } ; \intermed {  {\intermed \Gamma}  }  \vdash_{ty}  \intermed {  {\intermed \sigma}  } \itot{ \rightsquigarrow  \target {  {\target \sigma }_{{\mathrm{1}}}  \rightarrow  {\target \sigma }_{{\mathrm{2}}}  } } 
      \label{eqg:eq_ty8} 
    \end{gather}
    From the rule premise we get that:
    \begin{gather}
       \source {  {\source { \Gamma_C } }  } ; \source {  {\source \Gamma}  }  \vdash_{\mathit {Q} }  \source {  {\source Q}  }  \rightsquigarrow  \target {  {\target \sigma }_{{\mathrm{1}}}  } 
      \label{eq:eq_ty8} \\
       \source {  {\source { \Gamma_C } }  } ; \source {  {\source \Gamma}  }  \vdash_{ty}  \source {  {\source \rho}  }\,{ \stoi{ \rightsquigarrow  \target {  {\target \sigma }_{{\mathrm{2}}}  } } } 
      \label{eq:eq_ty9}
    \end{gather}
    By applying the induction hypothesis on Equation~\ref{eq:eq_ty9}, we get:
    \begin{gather}
       \source {  {\source { \Gamma_C } }  } ; \source {  {\source \Gamma}  }  \vdash_{ty}^{\intermed {M} }  \source {  {\source \rho}  }\,{ \stoi{ \rightsquigarrow  \intermed {  {\intermed \sigma}_{{\mathrm{2}}}  } } } 
      \label{eq:eq_ty10} \\
       \intermed {  {\intermed { \Gamma_C } }  } ; \intermed {  {\intermed \Gamma}  }  \vdash_{ty}  \intermed {  {\intermed \sigma}_{{\mathrm{2}}}  } \itot{ \rightsquigarrow  \target {  {\target \sigma }_{{\mathrm{2}}}  } } 
      \label{eq:eq_ty11}
    \end{gather}
    By applying Part 2 of this theorem on Equation~\ref{eq:eq_ty8}, we get:
    \begin{gather}
       \source {  {\source { \Gamma_C } }  } ; \source {  {\source \Gamma}  }  \vdash_{\mathit {Q} }^{\intermed {M} }  \source {  {\source Q}  }  \rightsquigarrow  \intermed {  {\intermed Q}  } 
      \label{eq:eq_ty12} \\
       \intermed {  {\intermed { \Gamma_C } }  } ; \intermed {  {\intermed \Gamma}  }  \vdash_{\mathit {Q} }  \intermed {  {\intermed Q}  }\, \itot{ \rightsquigarrow  \target {  {\target \sigma }_{{\mathrm{1}}}  } } 
      \label{eq:eq_ty13}
    \end{gather}
    Goals~\ref{eqg:eq_ty7} and~\ref{eqg:eq_ty8} follow directly from
    \textsc{sTy-qual} and \textsc{iTy-qual} respectively, in combination with
    Equations~\ref{eq:eq_ty10}, \ref{eq:eq_ty11}, \ref{eq:eq_ty12}
    and~\ref{eq:eq_ty13}, with \( \intermed {  {\intermed \sigma}  } = \intermed {   {\intermed Q}  \Rightarrow  {\intermed \sigma}_{{\mathrm{2}}}   } \). \\
  }
  \proofStep{\textsc{sTyT-scheme}}
  {
     \source {  {\source { \Gamma_C } }  } ; \source {  {\source \Gamma}  }  \vdash_{ty}  \source {  \forall  {\source a }  \ottsym{.}  {\source \sigma}  }\,{ \stoi{ \rightsquigarrow  \target {  \forall  {\target a }  \ottsym{.}  {\target \sigma }  } } } 
  }
  {
    The goal to be proven is the following:
    \begin{gather}
       \source {  {\source { \Gamma_C } }  } ; \source {  {\source \Gamma}  }  \vdash_{ty}^{\intermed {M} }  \source {  \forall  {\source a }  \ottsym{.}  {\source \sigma}  }\,{ \stoi{ \rightsquigarrow  \intermed {  {\intermed \sigma}  } } } 
      \label{eqg:eq_ty9} \\
       \intermed {  {\intermed { \Gamma_C } }  } ; \intermed {  {\intermed \Gamma}  }  \vdash_{ty}  \intermed {  {\intermed \sigma}  } \itot{ \rightsquigarrow  \target {  \forall  {\target a }  \ottsym{.}  {\target \sigma }  } } 
      \label{eqg:eq_ty10} 
    \end{gather}
    From the rule premise we get that:
    \begin{gather}
       \source {  {\source a }  }  \notin  \source {  {\source \Gamma}  } 
      \label{eq:eq_ty14} \\
       \source {  {\source { \Gamma_C } }  } ; \source {  {\source \Gamma}  \ottsym{,}  {\source a }  }  \vdash_{ty}  \source {  {\source \sigma}  }\,{ \stoi{ \rightsquigarrow  \target {  {\target \sigma }  } } } 
      \label{eq:eq_ty15}
    \end{gather}
    By repeated case analysis on the 2\textsuperscript{nd} hypothesis (\textsc{sCtxT-pgmInst}),
    we get:
    \begin{equation}
       \vdash_{ctx}  \source {  \bullet  } ; \source {  {\source { \Gamma_C } }  } ; \source {  {\source \Gamma}  }  \rightsquigarrow  \target {  {\target \Gamma }  } 
      \label{eq:eq_ty16a}
    \end{equation}
    From \textsc{sCtxT-tyEnvTy}, in combination with
    Equations~\ref{eq:eq_ty16a} and~\ref{eq:eq_ty14}, we know that:
    \begin{equation}
       \vdash_{ctx}  \source {  \bullet  } ; \source {  {\source { \Gamma_C } }  } ; \source {  {\source \Gamma}  \ottsym{,}  {\source a }  }  \rightsquigarrow  \target {  {\target \Gamma }  \ottsym{,}  {\target a }  } 
      \label{eq:eq_ty16}
    \end{equation}
    Similarly, we get from \textsc{sCtx-tyEnvTy} and \textsc{ctx-TVar} that:
    \begin{gather}
       \vdash_{ctx}^{\intermed {M} }  \source {  \bullet  } ; \source {  {\source { \Gamma_C } }  } ; \source {  {\source \Gamma}  \ottsym{,}  {\source a }  }  \rightsquigarrow  \intermed {  \bullet  } ; \intermed {  {\intermed { \Gamma_C } }  } ; \intermed {  {\intermed \Gamma}  \ottsym{,}  {\intermed a }  } 
      \label{eq:eq_ty16b} \\
       \intermed {  {\intermed { \Gamma_C } }  } ; \intermed {  {\intermed \Gamma}  \ottsym{,}  {\intermed a }  }  \rightsquigarrow  \target {  {\target \Gamma }  \ottsym{,}  {\target a }  } 
      \label{eq:eq_ty16c}
    \end{gather}
    By applying the induction hypothesis on Equation~\ref{eq:eq_ty15}, together
    with Equations~\ref{eq:eq_ty16}, \ref{eq:eq_ty16b} and~\ref{eq:eq_ty16c}, we get:
    \begin{gather}
       \source {  {\source { \Gamma_C } }  } ; \source {  {\source \Gamma}  \ottsym{,}  {\source a }  }  \vdash_{ty}^{\intermed {M} }  \source {  {\source \sigma}  }\,{ \stoi{ \rightsquigarrow  \intermed {  {\intermed \sigma}'  } } } 
      \label{eq:eq_ty17} \\
       \intermed {  {\intermed { \Gamma_C } }  } ; \intermed {  {\intermed \Gamma}  \ottsym{,}  {\intermed a }  }  \vdash_{ty}  \intermed {  {\intermed \sigma}'  } \itot{ \rightsquigarrow  \target {  {\target \sigma }  } } 
      \label{eq:eq_ty18}
    \end{gather}
    Goals~\ref{eqg:eq_ty9} and~\ref{eqg:eq_ty10} follow directly from
    \textsc{sTy-scheme} and \textsc{iTy-scheme} respectively, in combination
    with Equations~\ref{eq:eq_ty14}, \ref{eq:eq_ty17} and~\ref{eq:eq_ty18}, 
    with \( \intermed {  {\intermed \sigma}  } = \intermed {  \forall  {\intermed a }  \ottsym{.}  {\intermed \sigma}'  } \). \\
  }
\item[Part 2] By case analysis on the class constraint well-formedness
  derivation. \\
  \proofStep{\textsc{sQT-TC}}
  {
     \source {  {\source { \Gamma_C } }  } ; \source {  {\source \Gamma}  }  \vdash_{\mathit {Q} }  \source {  {\source {\mathit{TC} } } \, {\source \tau }  }  \rightsquigarrow  \target {  [  {\target \sigma }'  /  {\target a }  ]  \ottsym{\{}  {\target m}  \ottsym{:}  {\target \sigma }  \ottsym{\}}  } 
  }
  {
    The goal to be proven is the following:
    \begin{gather}
       \source {  {\source { \Gamma_C } }  } ; \source {  {\source \Gamma}  }  \vdash_{\mathit {Q} }^{\intermed {M} }  \source {  {\source {\mathit{TC} } } \, {\source \tau }  }  \rightsquigarrow  \intermed {  {\intermed Q}  } 
      \label{eqg:eq_ty11} \\
       \intermed {  {\intermed { \Gamma_C } }  } ; \intermed {  {\intermed \Gamma}  }  \vdash_{\mathit {Q} }  \intermed {  {\intermed Q}  }\, \itot{ \rightsquigarrow  \target {  [  {\target \sigma }'  /  {\target a }  ]  \ottsym{\{}  {\target m}  \ottsym{:}  {\target \sigma }  \ottsym{\}}  } } 
      \label{eqg:eq_ty12} 
    \end{gather}
    From the rule premise we get that:
    \begin{gather}
       \source {  {\source { \Gamma_C } }  } ; \source {  {\source \Gamma}  }  \vdash_{ty}  \source {  {\source \tau }  }\,{ \stoi{ \rightsquigarrow  \target {  {\target \sigma }'  } } } 
      \label{eq:eq_ty19} \\
       \source {  {\source { \Gamma_C } }  } = \source {  {\source { \Gamma_C } }_{{\mathrm{1}}}  }, \source {  {\source m}  \ottsym{:}  {\overline {\source Q} }_{\ottmv{i}}  \Rightarrow  {\source {\mathit{TC} } } \, {\source a }  \ottsym{:}  {\source \sigma}  \ottsym{,}  {\source { \Gamma_C } }_{{\mathrm{2}}}  } 
      \label{eq:eq_ty20} \\
       \source {  {\source { \Gamma_C } }_{{\mathrm{1}}}  } ; \source {  \bullet  \ottsym{,}  {\source a }  }  \vdash_{ty}  \source {  {\source \sigma}  }\,{ \stoi{ \rightsquigarrow  \target {  {\target \sigma }  } } } 
      \label{eq:eq_ty21} 
    \end{gather}
    By repeated case analysis on the 2\textsuperscript{nd} hypothesis (\textsc{sCtxT-pgmInst}),
    we get:
    \begin{equation}
       \vdash_{ctx}  \source {  \bullet  } ; \source {  {\source { \Gamma_C } }  } ; \source {  {\source \Gamma}  }  \rightsquigarrow  \target {  {\target \Gamma }  } 
      \label{eq:eq_ty21a}
    \end{equation}
    From \textsc{sCtxT-tyEnvTy}, together with Equation~\ref{eq:eq_ty21a}
    and the fact that \( \source {  {\source a }  }  \notin  \source {  \bullet  } \), we know that:
    \begin{equation}
       \vdash_{ctx}  \source {  \bullet  } ; \source {  {\source { \Gamma_C } }_{{\mathrm{1}}}  } ; \source {  \bullet  \ottsym{,}  {\source a }  }  \rightsquigarrow  \target {  \target {\bullet}  \ottsym{,}  {\target a }  } 
      \label{eq:eq_ty34}
    \end{equation}
    Similarly, we get from \textsc{sCtx-tyEnvTy} and \textsc{ctx-TVar} that:
    \begin{gather}
       \vdash_{ctx}^{\intermed {M} }  \source {  \bullet  } ; \source {  {\source { \Gamma_C } }_{{\mathrm{1}}}  } ; \source {  \bullet  \ottsym{,}  {\source a }  }  \rightsquigarrow  \intermed {  \bullet  } ; \intermed {  {\intermed { \Gamma_C } }_{{\mathrm{1}}}  } ; \intermed {  \bullet  \ottsym{,}  {\intermed a }  } 
      \label{eq:eq_ty34b} \\
       \intermed {  {\intermed { \Gamma_C } }_{{\mathrm{1}}}  } ; \intermed {  \bullet  \ottsym{,}  {\intermed a }  }  \rightsquigarrow  \target {  {\target \Gamma }  \ottsym{,}  {\target a }  } 
      \label{eq:eq_ty34c}
    \end{gather}
    By applying Part 1 of this theorem to Equations~\ref{eq:eq_ty19}
    and~\ref{eq:eq_ty21}, together with Equations~\ref{eq:eq_ty34},
    \ref{eq:eq_ty34b} and~\ref{eq:eq_ty34c}, we get:
    \begin{gather}
       \source {  {\source { \Gamma_C } }  } ; \source {  {\source \Gamma}  }  \vdash_{ty}^{\intermed {M} }  \source {  {\source \tau }  }\,{ \stoi{ \rightsquigarrow  \intermed {  {\intermed \sigma}'  } } } 
      \label{eq:eq_ty24} \\
       \intermed {  {\intermed { \Gamma_C } }  } ; \intermed {  {\intermed \Gamma}  }  \vdash_{ty}  \intermed {  {\intermed \sigma}'  } \itot{ \rightsquigarrow  \target {  {\target \sigma }'  } } 
      \label{eq:eq_ty25} \\
       \source {  {\source { \Gamma_C } }_{{\mathrm{1}}}  } ; \source {  \bullet  \ottsym{,}  {\source a }  }  \vdash_{ty}^{\intermed {M} }  \source {  {\source \sigma}  }\,{ \stoi{ \rightsquigarrow  \intermed {  {\intermed \sigma}  } } } 
      \label{eq:eq_ty26} \\
       \intermed {  {\intermed { \Gamma_C } }_{{\mathrm{1}}}  } ; \intermed {  \bullet  \ottsym{,}  {\intermed a }  }  \vdash_{ty}  \intermed {  {\intermed \sigma}  } \itot{ \rightsquigarrow  \target {  {\target \sigma }  } } 
      \label{eq:eq_ty27}
    \end{gather}
    Goal~\ref{eqg:eq_ty11} follows from \textsc{sQ-TC}, together with
    Equations~\ref{eq:eq_ty24}, \ref{eq:eq_ty20} and~\ref{eq:eq_ty26},
    with \( \intermed {  {\intermed Q}  } = \intermed {  {\intermed {\mathit{TC} } } \, {\intermed \sigma}'  } \). 
    Consequently, Goal~\ref{eqg:eq_ty12} follows from \textsc{iQ-TC}, together
    with Equations~\ref{eq:eq_ty25}, \ref{eq:eq_ty20} and~\ref{eq:eq_ty27}. \\
  }
\item[Part 3] By case analysis on the constraint well-formedness derivation. \\
  \proofStep{\textsc{sCT-abs}}
  {
     \source {  {\source { \Gamma_C } }  } ; \source {  {\source \Gamma}  }  \vdash_{\mathit {C} }  \source {  \forall  {\overline {\source a} }_{\ottmv{j}}  \ottsym{.}  {\overline {\source Q} }_{\ottmv{i}}  \Rightarrow  {\source Q}  }  \rightsquigarrow  \target {  \forall  {\overline {\target a} }_{\ottmv{j}}  \ottsym{.}  {\overline {\target \sigma} }_{\ottmv{i}}  \rightarrow  {\target \sigma }  } 
  }
  {
    The goal to be proven is the following:
    \begin{gather}
       \source {  {\source { \Gamma_C } }  } ; \source {  {\source \Gamma}  }  \vdash_{\mathit {C} }^{\intermed {M} }  \source {  \forall  {\overline {\source a} }_{\ottmv{j}}  \ottsym{.}  {\overline {\source Q} }_{\ottmv{i}}  \Rightarrow  {\source Q}  }  \rightsquigarrow  \intermed {  {\intermed C}  } 
      \label{eqg:eq_ty13} \\
       \intermed {  {\intermed { \Gamma_C } }  } ; \intermed {  {\intermed \Gamma}  }  \vdash_{\mathit {C} }  \intermed {  {\intermed C}  } 
      \label{eqg:eq_ty14}
    \end{gather}
    From the rule premise we get that:
    \begin{gather}
      \ottcomp{ \source {  {\source { \Gamma_C } }  } ; \source {  {\source \Gamma}  \ottsym{,}  {\overline {\source a} }_{\ottmv{j}}  }  \vdash_{\mathit {Q} }  \source {  {\source Q}_{\ottmv{i}}  }  \rightsquigarrow  \target {  {\target \sigma }_{\ottmv{i}}  } }{\ottmv{i}}
      \label{eq:eq_ty22} \\
       \source {  {\source { \Gamma_C } }  } ; \source {  {\source \Gamma}  \ottsym{,}  {\overline {\source a} }_{\ottmv{j}}  }  \vdash_{\mathit {Q} }  \source {  {\source Q}  }  \rightsquigarrow  \target {  {\target \sigma }  } 
      \label{eq:eq_ty23} \\
       \source {  {\overline {\source a} }_{\ottmv{j}}  }  \notin  \source {  {\source \Gamma}  } 
      \label{eq:eq_ty28}
    \end{gather}
    By repeated case analysis on the 2\textsuperscript{nd} hypothesis (\textsc{sCtxT-pgmInst}),
    we get:
    \begin{equation}
       \vdash_{ctx}  \source {  \bullet  } ; \source {  {\source { \Gamma_C } }  } ; \source {  {\source \Gamma}  }  \rightsquigarrow  \target {  {\target \Gamma }  } 
      \label{eq:eq_ty28a}
    \end{equation}
    By \textsc{sCtxT-tyEnvTy}, it follows from Equations~\ref{eq:eq_ty28} and~\ref{eq:eq_ty28a} that:
    \begin{equation}
       \vdash_{ctx}  \source {  \bullet  } ; \source {  {\source { \Gamma_C } }  } ; \source {  {\source \Gamma}  \ottsym{,}  {\overline {\source a} }_{\ottmv{j}}  }  \rightsquigarrow  \target {  {\target \Gamma }  \ottsym{,}  {\overline {\target a} }_{\ottmv{j}}  } 
      \label{eq:eq_ty29}
    \end{equation}
    Similarly, we get from \textsc{sCtx-tyEnvTy} and \textsc{ctx-TVar} that:
    \begin{gather}
       \vdash_{ctx}^{\intermed {M} }  \source {  \bullet  } ; \source {  {\source { \Gamma_C } }  } ; \source {  {\source \Gamma}  \ottsym{,}  {\overline {\source a} }_{\ottmv{j}}  }  \rightsquigarrow  \intermed {  \bullet  } ; \intermed {  {\intermed { \Gamma_C } }  } ; \intermed {  {\intermed \Gamma}  \ottsym{,}  {\overline {\intermed a} }_{\ottmv{j}}  } 
      \label{eq:eq_ty29b} \\
       \intermed {  {\intermed { \Gamma_C } }  } ; \intermed {  {\intermed \Gamma}  \ottsym{,}  {\overline {\intermed a} }_{\ottmv{j}}  }  \rightsquigarrow  \target {  {\target \Gamma }  \ottsym{,}  {\overline {\target a} }_{\ottmv{j}}  } 
      \label{eq:eq_ty29c}
    \end{gather}
    By applying Part 2 of this theorem to Equations~\ref{eq:eq_ty22}
    and~\ref{eq:eq_ty23}, together with Equations~\ref{eq:eq_ty29},
    \ref{eq:eq_ty29b} and~\ref{eq:eq_ty29c}, we get:
    \begin{gather}
      \ottcomp{ \source {  {\source { \Gamma_C } }  } ; \source {  {\source \Gamma}  \ottsym{,}  {\overline {\source a} }_{\ottmv{j}}  }  \vdash_{\mathit {Q} }^{\intermed {M} }  \source {  {\source Q}_{\ottmv{i}}  }  \rightsquigarrow  \intermed {  {\intermed Q}_{\ottmv{i}}  } }{\ottmv{i}}
      \label{eq:eq_ty30} \\
      \ottcomp{ \intermed {  {\intermed { \Gamma_C } }  } ; \intermed {  {\intermed \Gamma}  \ottsym{,}  {\overline {\intermed a} }_{\ottmv{j}}  }  \vdash_{\mathit {Q} }  \intermed {  {\intermed Q}_{\ottmv{i}}  }\, \itot{ \rightsquigarrow  \target {  {\target \sigma }_{\ottmv{i}}  } } }{\ottmv{i}}
      \label{eq:eq_ty31} \\
       \source {  {\source { \Gamma_C } }  } ; \source {  {\source \Gamma}  \ottsym{,}  {\overline {\source a} }_{\ottmv{j}}  }  \vdash_{\mathit {Q} }^{\intermed {M} }  \source {  {\source Q}  }  \rightsquigarrow  \intermed {  {\intermed Q}  } 
      \label{eq:eq_ty32} \\
       \intermed {  {\intermed { \Gamma_C } }  } ; \intermed {  {\intermed \Gamma}  \ottsym{,}  {\overline {\intermed a} }_{\ottmv{j}}  }  \vdash_{\mathit {Q} }  \intermed {  {\intermed Q}  }\, \itot{ \rightsquigarrow  \target {  {\target \sigma }  } } 
      \label{eq:eq_ty33}
    \end{gather}
    Goal~\ref{eqg:eq_ty13} follows from \textsc{sC-abs}, together with
    Equations~\ref{eq:eq_ty30} and~\ref{eq:eq_ty32},
    with \( \intermed {  {\intermed C}  } = \intermed {  \forall  {\overline {\intermed a} }_{\ottmv{j}}  \ottsym{.}  {\overline {\intermed Q} }_{\ottmv{i}}  \Rightarrow  {\intermed Q}  } \).
    Consequently, Goal~\ref{eqg:eq_ty14} follows from \textsc{iC-abs}, together
    with Equations~\ref{eq:eq_ty31} and~\ref{eq:eq_ty33}. \\
  }
\end{description}
\qed

\begin{theoremB}[Equivalence - Dictionaries]\leavevmode
  If \( \source {  { \source P }  } ; \source {  {\source { \Gamma_C } }  } ; \source {  {\source \Gamma}  }  \vDash  \source {  {\source Q}  } \stot{ \rightsquigarrow  \target {  {\target e }  } } \)
  and \( \vdash_{ctx}  \source {  { \source P }  } ; \source {  {\source { \Gamma_C } }  } ; \source {  {\source \Gamma}  }  \rightsquigarrow  \target {  {\target \Gamma }  } \) \\
  then \( \source {  { \source P }  } ; \source {  {\source { \Gamma_C } }  } ; \source {  {\source \Gamma}  }  \vDash^{\intermed {M} }  \source {  {\source Q}  } \stoi{ \rightsquigarrow  \intermed {  {\intermed d}  } } \)
  and \( \intermed {  {\intermed \Sigma}  } ; \intermed {  {\intermed { \Gamma_C } }  } ; \intermed {  {\intermed \Gamma}  }  \vdash_{d}  \intermed {  {\intermed d}  } : \intermed {  {\intermed Q}  } \itot{\,  \rightsquigarrow  \target {  {\target e }  } } \) \\
  where \( \source {  {\source { \Gamma_C } }  } ; \source {  {\source \Gamma}  }  \vdash_{\mathit {Q} }^{\intermed {M} }  \source {  {\source Q}  }  \rightsquigarrow  \intermed {  {\intermed Q}  } \)
  and \( \vdash_{ctx}^{\intermed {M} }  \source {  { \source P }  } ; \source {  {\source { \Gamma_C } }  } ; \source {  {\source \Gamma}  }  \rightsquigarrow  \intermed {  {\intermed \Sigma}  } ; \intermed {  {\intermed { \Gamma_C } }  } ; \intermed {  {\intermed \Gamma}  } \)
  and \( \intermed {  {\intermed { \Gamma_C } }  } ; \intermed {  {\intermed \Gamma}  }  \rightsquigarrow  \target {  {\target \Gamma }  } \).
  \label{thmA:equiv_dict}
\end{theoremB} 

\proof
By induction on the entailment derivation, and mutually proven with
Theorems~\ref{thmA:equiv_env}
and~\ref{thmA:equiv_expr} (Figure~\ref{fig:dep_graph_equiv}).
Note that at the dependencies between Theorem~\ref{thmA:equiv_env}
and~\ref{thmA:equiv_expr} and between Theorem~\ref{thmA:equiv_dict}
and~\ref{thmA:equiv_expr}, the size of \({ \source P }\) is strictly decreasing, whereas
\({ \source P }\) remains constant at every other dependency. Consequenty, the size of
\({ \source P }\) is strictly decreasing at every cycle and the induction remains well-founded.

From environment equivalence (Theorem~\ref{thmA:equiv_env}), in combination with
the 2\textsuperscript{nd} hypothesis, we derive that:
\begin{gather}
     \vdash_{ctx}^{\intermed {M} }  \source {  { \source P }  } ; \source {  {\source { \Gamma_C } }  } ; \source {  {\source \Gamma}  }  \rightsquigarrow  \intermed {  {\intermed \Sigma}  } ; \intermed {  {\intermed { \Gamma_C } }  } ; \intermed {  {\intermed \Gamma}  } 
    \label{eq:eq_d0a} \\
     \intermed {  {\intermed { \Gamma_C } }  } ; \intermed {  {\intermed \Gamma}  }  \rightsquigarrow  \target {  {\target \Gamma }  } 
    \label{eq:eq_d0b}
\end{gather}
\proofStep{\textsc{sEntailT-local}}
{
   \source {  { \source P }  } ; \source {  {\source { \Gamma_C } }  } ; \source {  {\source \Gamma}  }  \vDash  \source {  {\source Q}  } \stot{ \rightsquigarrow  \target {  {\target \delta }  } } 
}
{
  The goal to be proven is the following:
  \begin{gather}
     \source {  { \source P }  } ; \source {  {\source { \Gamma_C } }  } ; \source {  {\source \Gamma}  }  \vDash^{\intermed {M} }  \source {  {\source Q}  } \stoi{ \rightsquigarrow  \intermed {  {\intermed d}  } } 
    \label{eqg:eq_d1} \\
     \intermed {  {\intermed \Sigma}  } ; \intermed {  {\intermed { \Gamma_C } }  } ; \intermed {  {\intermed \Gamma}  }  \vdash_{d}  \intermed {  {\intermed d}  } : \intermed {  {\intermed Q}  } \itot{\,  \rightsquigarrow  \target {  {\target \delta }  } } 
    \label{eqg:eq_d2} \\
     \source {  {\source { \Gamma_C } }  } ; \source {  {\source \Gamma}  }  \vdash_{\mathit {Q} }^{\intermed {M} }  \source {  {\source Q}  }  \rightsquigarrow  \intermed {  {\intermed Q}  } 
    \label{eqg:eq_d2b}
  \end{gather}
  We know from the rule premise that:
  \begin{equation*}
     ( \source {  {\source \delta }  } : \source {  {\source Q}  } )  \in  \source {  {\source \Gamma}  } 
  \end{equation*} 
  Consequently, Goal~\ref{eqg:eq_d2b} follows by repeated case analysis on
  Equation~\ref{eq:eq_d0a} (\textsc{sCtx-tyEnvD}).
  Furthermore, we know from Lemma~\ref{lemma:pres_env_d} that:
  \begin{equation*}
     ( \intermed {  {\intermed \delta }  } : \intermed {  {\intermed Q}  } )  \in  \intermed {  {\intermed \Gamma}  } 
  \end{equation*}
  Goal~\ref{eqg:eq_d2} follows by \textsc{D-var}.
  Goal~\ref{eqg:eq_d1} follows by \textsc{sEntail-local}, with \( \intermed {  {\intermed d}  } = \intermed {  {\intermed \delta }  } \). \\
}
\proofStep{\textsc{sEntailT-inst}}
{
   \source {  { \source P }  } ; \source {  {\source { \Gamma_C } }  } ; \source {  {\source \Gamma}  }  \vDash  \source {  {\source Q}  } \stot{ \rightsquigarrow  \target {  \ottsym{(}  \Lambda  {\overline {\target a} }_{\ottmv{j}}  \ottsym{.}  \lambda \, \ottcomp{{\target \delta }_{\ottmv{i}}  \ottsym{:}  {\target \sigma }'_{\ottmv{i}}}{\ottmv{i}} \, \ottsym{.}  \ottsym{\{}  {\target m}  \ottsym{=}  \Lambda  {\overline {\target b} }_{\ottmv{k}}  \ottsym{.}  \lambda \, \ottcomp{{\target \delta }'_{\ottmv{h}}  \ottsym{:}  {\target \sigma }''_{\ottmv{h}}}{\ottmv{h}} \, \ottsym{.}  {\target e }  \ottsym{\}}  \ottsym{)} \, {\overline {\target \sigma} }_{\ottmv{j}} \, {\overline {\target e} }_{\ottmv{i}}  } } 
}
{
  The goal to be proven is the following:
  \begin{gather}
     \source {  { \source P }  } ; \source {  {\source { \Gamma_C } }  } ; \source {  {\source \Gamma}  }  \vDash^{\intermed {M} }  \source {  {\source Q}  } \stoi{ \rightsquigarrow  \intermed {  {\intermed d}  } } 
    \label{eqg:eq_d3} \\
     \intermed {  {\intermed \Sigma}  } ; \intermed {  {\intermed { \Gamma_C } }  } ; \intermed {  {\intermed \Gamma}  }  \vdash_{d}  \intermed {  {\intermed d}  } : \intermed {  {\intermed Q}  } \itot{\,  \rightsquigarrow  \target {  \ottsym{(}  \Lambda  {\overline {\target a} }_{\ottmv{j}}  \ottsym{.}  \lambda \, \ottcomp{{\target \delta }_{\ottmv{i}}  \ottsym{:}  {\target \sigma }'_{\ottmv{i}}}{\ottmv{i}} \, \ottsym{.}  \ottsym{\{}  {\target m}  \ottsym{=}  \Lambda  {\overline {\target b} }_{\ottmv{k}}  \ottsym{.}  \lambda \, \ottcomp{{\target \delta }'_{\ottmv{h}}  \ottsym{:}  {\target \sigma }''_{\ottmv{h}}}{\ottmv{h}} \, \ottsym{.}  {\target e }  \ottsym{\}}  \ottsym{)} \, {\overline {\target \sigma} }_{\ottmv{j}} \, {\overline {\target e} }_{\ottmv{i}}  } } 
    \label{eqg:eq_d4} \\
     \source {  {\source { \Gamma_C } }  } ; \source {  {\source \Gamma}  }  \vdash_{\mathit {Q} }^{\intermed {M} }  \source {  {\source Q}  }  \rightsquigarrow  \intermed {  {\intermed Q}  } 
    \label{eqg:eq_d4b}
  \end{gather}
  From the rule premise we get that:
  \begin{gather}
     \source {  { \source P }  } = \source {  { \source P }_{{\mathrm{1}}}  }, \source {  \ottsym{(}  {\source D }  \ottsym{:}  \forall  {\overline {\source a} }_{\ottmv{j}}  \ottsym{.}  {\overline {\source Q} }'_{\ottmv{i}}  \Rightarrow  {\source Q}'  \ottsym{)}  \ottsym{.}  {\source m}  \mapsto  \bullet  \ottsym{,}  {\overline {\source a} }_{\ottmv{j}}  \ottsym{,}  {\overline {\source \delta} }_{\ottmv{i}}  \ottsym{:}  {\overline {\source Q} }'_{\ottmv{i}}  \ottsym{,}  {\overline {\source b} }_{\ottmv{k}}  \ottsym{,}  {\overline {\source \delta} }_{\ottmv{h}}  \ottsym{:}  {\overline {\source Q} }_{\ottmv{h}}  \ottsym{:}  {\source e}  \ottsym{,}  { \source P }_{{\mathrm{2}}}  } 
    \label{eq:eq_d1} \\
     \source {  {\source Q}  } = \source {  [  {\overline {\source \tau} }_{\ottmv{j}}  /  {\overline {\source a} }_{\ottmv{j}}  ]  {\source Q}'  } 
    \label{eq:eq_d2} \\
     \source {  { \source P }_{{\mathrm{1}}}  } ; \source {  {\source { \Gamma_C } }  } ; \source {  \bullet  \ottsym{,}  {\overline {\source a} }_{\ottmv{j}}  \ottsym{,}  {\overline {\source \delta} }_{\ottmv{i}}  \ottsym{:}  {\overline {\source Q} }'_{\ottmv{i}}  \ottsym{,}  {\overline {\source b} }_{\ottmv{k}}  \ottsym{,}  {\overline {\source \delta} }_{\ottmv{h}}  \ottsym{:}  {\overline {\source Q} }_{\ottmv{h}}  }  \vdash_{tm}  \source {  {\source e}  } \Rightarrow \source {  {\source \tau }  } \stot{ \rightsquigarrow  \target {  {\target e }  } } 
    \label{eq:eq_d2b} \\
     \vdash_{ctx}  \source {  { \source P }  } ; \source {  {\source { \Gamma_C } }  } ; \source {  {\source \Gamma}  }  \rightsquigarrow  \target {  {\target \Gamma }  } 
    \label{eq:eq_d3} \\
    \ottcomp{ \source {  {\source { \Gamma_C } }  } ; \source {  {\source \Gamma}  }  \vdash_{ty}  \source {  {\source \tau }_{\ottmv{j}}  }\,{ \stoi{ \rightsquigarrow  \target {  {\target \sigma }_{\ottmv{j}}  } } } }{\ottmv{j}}
    \label{eq:eq_d4} \\
    \ottcomp{ \source {  {\source { \Gamma_C } }  } ; \source {  \bullet  \ottsym{,}  {\overline {\source a} }_{\ottmv{j}}  }  \vdash_{\mathit {Q} }  \source {  {\source Q}'_{\ottmv{i}}  }  \rightsquigarrow  \target {  {\target \sigma }'_{\ottmv{i}}  } }{\ottmv{i}}
    \label{eq:eq_d5} \\
    \ottcomp{ \source {  {\source { \Gamma_C } }  } ; \source {  \bullet  \ottsym{,}  {\overline {\source a} }_{\ottmv{j}}  \ottsym{,}  {\overline {\source b} }_{\ottmv{k}}  }  \vdash_{\mathit {Q} }  \source {  {\source Q}_{\ottmv{h}}  }  \rightsquigarrow  \target {  {\target \sigma }''_{\ottmv{h}}  } }{\ottmv{h}}
    \label{eq:eq_d5b} \\
    \ottcomp{ \source {  { \source P }  } ; \source {  {\source { \Gamma_C } }  } ; \source {  {\source \Gamma}  }  \vDash  \source {  [  {\overline {\source \tau} }_{\ottmv{j}}  /  {\overline {\source a} }_{\ottmv{j}}  ]  {\source Q}'_{\ottmv{i}}  } \stot{ \rightsquigarrow  \target {  {\target e }_{\ottmv{i}}  } } }{\ottmv{i}}
    \label{eq:eq_d6} 
  \end{gather}
  By applying the induction hypothesis to Equation~\ref{eq:eq_d6}, we get:
  \begin{gather}
    \ottcomp{ \source {  { \source P }  } ; \source {  {\source { \Gamma_C } }  } ; \source {  {\source \Gamma}  }  \vDash^{\intermed {M} }  \source {  [  {\overline {\source \tau} }_{\ottmv{j}}  /  {\overline {\source a} }_{\ottmv{j}}  ]  {\source Q}'_{\ottmv{i}}  } \stoi{ \rightsquigarrow  \intermed {  {\intermed d}_{\ottmv{i}}  } } }{\ottmv{i}}
    \label{eq:eq_d7} \\
    \ottcomp{ \intermed {  {\intermed \Sigma}  } ; \intermed {  {\intermed { \Gamma_C } }  } ; \intermed {  {\intermed \Gamma}  }  \vdash_{d}  \intermed {  {\intermed d}_{\ottmv{i}}  } : \intermed {  [  \overline {\intermed \sigma}_{\ottmv{j}}  /  {\overline {\intermed a} }_{\ottmv{j}}  ]  {\intermed Q}'_{\ottmv{i}}  } \itot{\,  \rightsquigarrow  \target {  {\target e }_{\ottmv{i}}  } } }{\ottmv{i}}
    \label{eq:eq_d8}
  \end{gather}
  By repeated case analysis on the 2\textsuperscript{nd} hypothesis, we know that:
  \begin{equation*}
     \vdash_{ctx}  \source {  \bullet  } ; \source {  {\source { \Gamma_C } }  } ; \source {  \bullet  }  \rightsquigarrow  \target {  \target {\bullet}  } 
  \end{equation*}
  From \textsc{sCtxT-tyEnvTy}, we get that:
  \begin{equation*}
     \vdash_{ctx}  \source {  \bullet  } ; \source {  {\source { \Gamma_C } }  } ; \source {  \bullet  \ottsym{,}  {\overline {\source a} }_{\ottmv{j}}  }  \rightsquigarrow  \target {  \target {\bullet}  \ottsym{,}  {\overline {\target a} }_{\ottmv{j}}  } 
  \end{equation*}
  Similarly, we can derive from Equations~\ref{eq:eq_d0a} and~\ref{eq:eq_d0b} that:
  \begin{gather}
     \vdash_{ctx}^{\intermed {M} }  \source {  \bullet  } ; \source {  {\source { \Gamma_C } }  } ; \source {  \bullet  \ottsym{,}  {\overline {\source a} }_{\ottmv{j}}  }  \rightsquigarrow  \intermed {  \bullet  } ; \intermed {  {\intermed { \Gamma_C } }  } ; \intermed {  \bullet  \ottsym{,}  {\overline {\intermed a} }_{\ottmv{j}}  } 
    \nonumber \\
     \intermed {  {\intermed { \Gamma_C } }  } ; \intermed {  \bullet  \ottsym{,}  {\overline {\intermed a} }_{\ottmv{j}}  }  \rightsquigarrow  \target {  \target {\bullet}  \ottsym{,}  {\overline {\target a} }_{\ottmv{j}}  } 
    \nonumber
  \end{gather}
  It follows from type and constraint equivalence (Theorem~\ref{thmA:equiv_ty}), in
  combination with Equations~\ref{eq:eq_d4}, \ref{eq:eq_d5}, \ref{eq:eq_d0a} and~\ref{eq:eq_d0b} that:
  \begin{gather}
    \ottcomp{ \source {  {\source { \Gamma_C } }  } ; \source {  {\source \Gamma}  }  \vdash_{ty}^{\intermed {M} }  \source {  {\source \tau }_{\ottmv{j}}  }\,{ \stoi{ \rightsquigarrow  \intermed {  {\intermed \sigma}_{\ottmv{j}}  } } } }{\ottmv{j}}
    \label{eq:eq_d11} \\
    \ottcomp{ \intermed {  {\intermed { \Gamma_C } }  } ; \intermed {  {\intermed \Gamma}  }  \vdash_{ty}  \intermed {  {\intermed \sigma}_{\ottmv{j}}  } \itot{ \rightsquigarrow  \target {  {\target \sigma }_{\ottmv{j}}  } } }{\ottmv{j}}
    \label{eq:eq_d12} \\
    \ottcomp{ \source {  {\source { \Gamma_C } }  } ; \source {  \bullet  \ottsym{,}  {\overline {\source a} }_{\ottmv{j}}  }  \vdash_{\mathit {Q} }^{\intermed {M} }  \source {  {\source Q}'_{\ottmv{i}}  }  \rightsquigarrow  \intermed {  {\intermed Q}'_{\ottmv{i}}  } }{\ottmv{i}}
    \label{eq:eq_d13} \\
    \ottcomp{ \intermed {  {\intermed { \Gamma_C } }  } ; \intermed {  \bullet  \ottsym{,}  {\overline {\intermed a} }_{\ottmv{j}}  }  \vdash_{\mathit {Q} }  \intermed {  {\intermed Q}'_{\ottmv{i}}  }\, \itot{ \rightsquigarrow  \target {  {\target \sigma }'_{\ottmv{i}}  } } }{\ottmv{i}}
    \label{eq:eq_d14}
  \end{gather}
  By repeated case analysis on Equation~\ref{eq:eq_d0a} (\textsc{sCtx-pgmInst}),
  together with Equation~\ref{eq:eq_d1}, we know that:
  \begin{gather}
     \source {  {\source { \Gamma_C } }  } ; \source {  \bullet  }  \vdash_{\mathit {C} }^{\intermed {M} }  \source {  \forall  {\overline {\source a} }_{\ottmv{j}}  \ottsym{.}  {\overline {\source Q} }'_{\ottmv{i}}  \Rightarrow  {\source Q}'  }  \rightsquigarrow  \intermed {  \forall  {\overline {\intermed a} }_{\ottmv{j}}  \ottsym{.}  {\overline {\intermed Q} }'_{\ottmv{i}}  \Rightarrow  {\intermed Q}'  } 
    \label{eq:eq_d6b} \\
     \vdash_{ctx}^{\intermed {M} }  \source {  \bullet  } ; \source {  {\source { \Gamma_C } }  } ; \source {  {\source \Gamma}  }  \rightsquigarrow  \intermed {  \bullet  } ; \intermed {  {\intermed { \Gamma_C } }  } ; \intermed {  {\intermed \Gamma}  } 
    \label{eq:eq_d6e}
  \end{gather}
  By applying Lemma~\ref{lemma:constraint_weakening} to Equations~\ref{eq:eq_d6b}
  and~\ref{eq:eq_d6e}, we get:
  \begin{equation}
     \source {  {\source { \Gamma_C } }  } ; \source {  {\source \Gamma}  }  \vdash_{\mathit {C} }^{\intermed {M} }  \source {  \forall  {\overline {\source a} }_{\ottmv{j}}  \ottsym{.}  {\overline {\source Q} }'_{\ottmv{i}}  \Rightarrow  {\source Q}'  }  \rightsquigarrow  \intermed {  \forall  {\overline {\intermed a} }_{\ottmv{j}}  \ottsym{.}  {\overline {\intermed Q} }'_{\ottmv{i}}  \Rightarrow  {\intermed Q}'  } 
    \label{eq:eq_d6f} 
  \end{equation}
  By case analysis on Equation~\ref{eq:eq_d6b} (\textsc{sC-abs}), we get that:
  \begin{equation}
     \source {  {\source { \Gamma_C } }  } ; \source {  {\source \Gamma}  \ottsym{,}  {\overline {\source a} }_{\ottmv{j}}  }  \vdash_{\mathit {Q} }^{\intermed {M} }  \source {  {\source Q}'  }  \rightsquigarrow  \intermed {  {\intermed Q}'  } 
    \label{eq:eq_d6c} 
  \end{equation}
  It follows from Lemma~\ref{lemma:tvar_subst_ctr} and
  Equations~\ref{eq:eq_d6c} and~\ref{eq:eq_d11} that:
  \begin{equation}
     \source {  {\source { \Gamma_C } }  } ; \source {  {\source \Gamma}  }  \vdash_{\mathit {Q} }^{\intermed {M} }  \source {  [  {\overline {\source \tau} }_{\ottmv{j}}  /  {\overline {\source a} }_{\ottmv{j}}  ]  {\source Q}'  }  \rightsquigarrow  \intermed {  [  \overline {\intermed \sigma}_{\ottmv{j}}  /  {\overline {\intermed a} }_{\ottmv{j}}  ]  {\intermed Q}'  } 
    \label{eq:eq_d6d} 
  \end{equation}
  Goal~\ref{eqg:eq_d4b} follows directly from Equation~\ref{eq:eq_d6d},
  since we know that \( \source {  {\source Q}  } = \source {  [  {\overline {\source \tau} }_{\ottmv{j}}  /  {\overline {\source a} }_{\ottmv{j}}  ]  {\source Q}'  } \) (Equation~\ref{eq:eq_d2}).

  By applying Lemma~\ref{lemma:ctx_well_source_typing} to
  Equation~\ref{eq:eq_d2b}, we get that:
  \begin{equation}
     \vdash_{ctx}  \source {  { \source P }_{{\mathrm{1}}}  } ; \source {  {\source { \Gamma_C } }  } ; \source {  \bullet  \ottsym{,}  {\overline {\source a} }_{\ottmv{j}}  \ottsym{,}  {\overline {\source \delta} }_{\ottmv{i}}  \ottsym{:}  {\overline {\source Q} }'_{\ottmv{i}}  \ottsym{,}  {\overline {\source b} }_{\ottmv{k}}  \ottsym{,}  {\overline {\source \delta} }_{\ottmv{h}}  \ottsym{:}  {\overline {\source Q} }_{\ottmv{h}}  }  \rightsquigarrow  \target {  {\target \Gamma }'  } 
    \label{eq:eq_d14a}
  \end{equation}
  From expression equivalence (Theorem~\ref{thmA:equiv_expr}), in combination
  with Equations~\ref{eq:eq_d2b} and~\ref{eq:eq_d14a}, we get:
  \begin{gather}
     \source {  { \source P }_{{\mathrm{1}}}  } ; \source {  {\source { \Gamma_C } }  } ; \source {  \bullet  \ottsym{,}  {\overline {\source a} }_{\ottmv{j}}  \ottsym{,}  {\overline {\source \delta} }_{\ottmv{i}}  \ottsym{:}  {\overline {\source Q} }'_{\ottmv{i}}  \ottsym{,}  {\overline {\source b} }_{\ottmv{k}}  \ottsym{,}  {\overline {\source \delta} }_{\ottmv{h}}  \ottsym{:}  {\overline {\source Q} }_{\ottmv{h}}  }  \vdash_{tm}^{\intermed {M} }  \source {  {\source e}  } \Rightarrow \source {  {\source \tau }  } \stoi{ \rightsquigarrow  \intermed {  {\intermed e}  } } 
    \label{eq:eq_d14b} \\
     \intermed {  {\intermed \Sigma}_{{\mathrm{1}}}  } ; \intermed {  {\intermed { \Gamma_C } }  } ; \intermed {  \bullet  \ottsym{,}  {\overline {\intermed a} }_{\ottmv{j}}  \ottsym{,}  {\overline {\intermed \delta} }_{\ottmv{i}}  \ottsym{:}  {\overline {\intermed Q} }'_{\ottmv{i}}  \ottsym{,}  {\overline {\intermed b} }_{\ottmv{k}}  \ottsym{,}  {\overline {\intermed \delta} }_{\ottmv{h}}  \ottsym{:}  {\overline {\intermed Q} }_{\ottmv{h}}  }  \vdash_{tm}  \intermed {  {\intermed e}  } : \intermed {  {\intermed \sigma}  } \itot{ \rightsquigarrow  \target {  {\target e }  } } 
    \label{eq:eq_d14c}
  \end{gather}
  Goal~\ref{eqg:eq_d3} follows from \textsc{sEntail-inst}, in combination with
  Equations~\ref{eq:eq_d1}, \ref{eq:eq_d2}, \ref{eq:eq_d0a},
  \ref{eq:eq_d11} and~\ref{eq:eq_d7}, with \( \intermed {  {\intermed d}  } = \intermed {  {\intermed D } \, \overline {\intermed \sigma}_{\ottmv{j}} \, {\overline {\intermed d} }_{\ottmv{i}}  } \).

  By inversion on Equation~\ref{eq:eq_d0a}, in combination with
  Equation~\ref{eq:eq_d1}, we get:
  \begin{equation}
     \intermed {  {\intermed \Sigma}  } = \intermed {  {\intermed \Sigma}_{{\mathrm{1}}}  \ottsym{,}  \ottsym{(}  {\intermed D }  \ottsym{:}  \forall  {\overline {\intermed a} }_{\ottmv{j}}  \ottsym{.}  {\overline {\intermed Q} }'_{\ottmv{i}}  \Rightarrow  {\intermed Q}'  \ottsym{)}  \ottsym{.}  {\intermed m}  \mapsto  \Lambda  {\overline {\intermed a} }_{\ottmv{j}}  \ottsym{.}  \lambda  {\overline {\intermed \delta} }_{\ottmv{i}}  \ottsym{:}  {\overline {\intermed Q} }'_{\ottmv{i}}  \ottsym{.}  \Lambda  {\overline {\intermed b} }_{\ottmv{k}}  \ottsym{.}  \lambda  {\overline {\intermed \delta} }_{\ottmv{h}}  \ottsym{:}  {\overline {\intermed Q} }_{\ottmv{h}}  \ottsym{.}  {\intermed e}  \ottsym{,}  {\intermed \Sigma}_{{\mathrm{2}}}  } 
    \label{eq:eq_d14d}
  \end{equation}
  By applying preservation Theorem~\ref{thmA:typing_pres_env} to
  Equation~\ref{eq:eq_d0a}, we get:
  \begin{equation}
     \vdash_{ctx}  \intermed {  {\intermed \Sigma}  } ; \intermed {  {\intermed { \Gamma_C } }  } ; \intermed {  {\intermed \Gamma}  } 
    \label{eq:eq_d14e}
  \end{equation}
  Finally, Goal~\ref{eqg:eq_d4} follows from \textsc{D-con}, in combination with
  Equations~\ref{eq:eq_d14}, \ref{eq:eq_d12}, \ref{eq:eq_d8}, \ref{eq:eq_d14c},
  \ref{eq:eq_d14d} and~\ref{eq:eq_d14e}. \\
}
\qed

\begin{theoremB}[Equivalence - Expressions]\leavevmode
  \begin{itemize}
  \item If \( \source {  { \source P }  } ; \source {  {\source { \Gamma_C } }  } ; \source {  {\source \Gamma}  }  \vdash_{tm}  \source {  {\source e}  } \Rightarrow \source {  {\source \tau }  } \stot{ \rightsquigarrow  \target {  {\target e }  } } \)
    and \( \vdash_{ctx}  \source {  { \source P }  } ; \source {  {\source { \Gamma_C } }  } ; \source {  {\source \Gamma}  }  \rightsquigarrow  \target {  {\target \Gamma }  } \) \\
    then \( \source {  { \source P }  } ; \source {  {\source { \Gamma_C } }  } ; \source {  {\source \Gamma}  }  \vdash_{tm}^{\intermed {M} }  \source {  {\source e}  } \Rightarrow \source {  {\source \tau }  } \stoi{ \rightsquigarrow  \intermed {  {\intermed e}  } } \)
    and \( \intermed {  {\intermed \Sigma}  } ; \intermed {  {\intermed { \Gamma_C } }  } ; \intermed {  {\intermed \Gamma}  }  \vdash_{tm}  \intermed {  {\intermed e}  } : \intermed {  {\intermed \sigma}  } \itot{ \rightsquigarrow  \target {  {\target e }  } } \) \\
    where \( \vdash_{ctx}^{\intermed {M} }  \source {  { \source P }  } ; \source {  {\source { \Gamma_C } }  } ; \source {  {\source \Gamma}  }  \rightsquigarrow  \intermed {  {\intermed \Sigma}  } ; \intermed {  {\intermed { \Gamma_C } }  } ; \intermed {  {\intermed \Gamma}  } \)
    and \( \intermed {  {\intermed { \Gamma_C } }  } ; \intermed {  {\intermed \Gamma}  }  \rightsquigarrow  \target {  {\target \Gamma }  } \)
    and \( \source {  {\source { \Gamma_C } }  } ; \source {  {\source \Gamma}  }  \vdash_{ty}^{\intermed {M} }  \source {  {\source \tau }  }\,{ \stoi{ \rightsquigarrow  \intermed {  {\intermed \sigma}  } } } \).
  \item If \( \source {  { \source P }  } ; \source {  {\source { \Gamma_C } }  } ; \source {  {\source \Gamma}  }  \vdash_{tm}  \source {  {\source e}  } \Leftarrow \source {  {\source \tau }  } \stot{ \rightsquigarrow  \target {  {\target e }  } } \)
    and \( \vdash_{ctx}  \source {  { \source P }  } ; \source {  {\source { \Gamma_C } }  } ; \source {  {\source \Gamma}  }  \rightsquigarrow  \target {  {\target \Gamma }  } \) \\
    then \( \source {  { \source P }  } ; \source {  {\source { \Gamma_C } }  } ; \source {  {\source \Gamma}  }  \vdash_{tm}^{\intermed {M} }  \source {  {\source e}  } \Leftarrow \source {  {\source \tau }  } \stoi{ \rightsquigarrow  \intermed {  {\intermed e}  } } \)
    and \( \intermed {  {\intermed \Sigma}  } ; \intermed {  {\intermed { \Gamma_C } }  } ; \intermed {  {\intermed \Gamma}  }  \vdash_{tm}  \intermed {  {\intermed e}  } : \intermed {  {\intermed \sigma}  } \itot{ \rightsquigarrow  \target {  {\target e }  } } \) \\
    where \( \vdash_{ctx}^{\intermed {M} }  \source {  { \source P }  } ; \source {  {\source { \Gamma_C } }  } ; \source {  {\source \Gamma}  }  \rightsquigarrow  \intermed {  {\intermed \Sigma}  } ; \intermed {  {\intermed { \Gamma_C } }  } ; \intermed {  {\intermed \Gamma}  } \)
    and \( \intermed {  {\intermed { \Gamma_C } }  } ; \intermed {  {\intermed \Gamma}  }  \rightsquigarrow  \target {  {\target \Gamma }  } \)
    and \( \source {  {\source { \Gamma_C } }  } ; \source {  {\source \Gamma}  }  \vdash_{ty}^{\intermed {M} }  \source {  {\source \tau }  }\,{ \stoi{ \rightsquigarrow  \intermed {  {\intermed \sigma}  } } } \).
  \end{itemize}
  \label{thmA:equiv_expr}
\end{theoremB} 

\proof
By induction on the lexicographic order of the tuple (size of the expression
\({\source e}\), typing mode). Regarding typing mode, we define type checking to be
larger than type inference. In each mutual dependency, we know that the tuple
size decreases, meaning that the induction is well-founded. 

Furthermore, this theorem is mutually proven with
Theorems~\ref{thmA:equiv_env}
and~\ref{thmA:equiv_dict} (Figure~\ref{fig:dep_graph_equiv}).
Note that at the dependencies between Theorem~\ref{thmA:equiv_env}
and~\ref{thmA:equiv_expr} and between Theorem~\ref{thmA:equiv_dict}
and~\ref{thmA:equiv_expr}, the size of \({ \source P }\) is strictly decreasing, whereas
\({ \source P }\) remains constant at every other dependency. Consequenty, the size of
\({ \source P }\) is strictly decreasing at every cycle and the induction remains well-founded.

From environment equivalence (Theorem~\ref{thmA:equiv_env}), in combination with
the 2\textsuperscript{nd} hypothesis, we derive that:
\begin{gather}
     \vdash_{ctx}^{\intermed {M} }  \source {  { \source P }  } ; \source {  {\source { \Gamma_C } }  } ; \source {  {\source \Gamma}  }  \rightsquigarrow  \intermed {  {\intermed \Sigma}  } ; \intermed {  {\intermed { \Gamma_C } }  } ; \intermed {  {\intermed \Gamma}  } 
    \label{eq:eq_expr0a} \\
     \intermed {  {\intermed { \Gamma_C } }  } ; \intermed {  {\intermed \Gamma}  }  \rightsquigarrow  \target {  {\target \Gamma }  } 
    \label{eq:eq_expr0b}
\end{gather}
Consequently, by Theorem~\ref{thmA:typing_pres_env} we derive that
\begin{equation}
   \vdash_{ctx}  \intermed {  {\intermed \Sigma}  } ; \intermed {  {\intermed { \Gamma_C } }  } ; \intermed {  {\intermed \Gamma}  } 
  \label{eq:eq_expr0c}
\end{equation}

\begin{description}
\item[Part 1] 
  By case analysis on the typing derivation. \\
  \proofStep{\textsc{sTm-infT-true}}
  {
     \source {  { \source P }  } ; \source {  {\source { \Gamma_C } }  } ; \source {  {\source \Gamma}  }  \vdash_{tm}  \source {  \mathit{\source {True} }  } \Rightarrow \source {  \mathit{\source {Bool} }  } \stot{ \rightsquigarrow  \target {  \mathit{\target {True} }  } } 
  }
  {
    The goal follows by \textsc{sTm-inf-true}, \textsc{iTm-true}
    (in combination with Equation~\ref{eq:eq_expr0c})
    and \textsc{sTy-bool}, with \( \intermed {  {\intermed e}  } = \intermed {  \mathit{\intermed {True} }  } \). \\
  }
  \proofStep{\textsc{sTm-infT-false}}
  {
     \source {  { \source P }  } ; \source {  {\source { \Gamma_C } }  } ; \source {  {\source \Gamma}  }  \vdash_{tm}  \source {  \mathit{\source {False} }  } \Rightarrow \source {  \mathit{\source {Bool} }  } \stot{ \rightsquigarrow  \target {  \mathit{\target {False} }  } } 
  }
  {
    Similar to the \textsc{sTm-infT-true} case. \\
  }
  \proofStep{\textsc{sTm-infT-let}}
  {
     \source {  { \source P }  } ; \source {  {\source { \Gamma_C } }  } ; \source {  {\source \Gamma}  }  \vdash_{tm}  \source {  \textbf{\source {let} } \, {\source x}  \ottsym{:}  \forall  {\overline {\source a} }_{\ottmv{j}}  \ottsym{.}  {\overline {\source Q} }_{\ottmv{i}}  \Rightarrow  {\source \tau }_{{\mathrm{1}}}  \ottsym{=}  {\source e}_{{\mathrm{1}}} \, \textbf{\source {in} } \, {\source e}_{{\mathrm{2}}}  } \Rightarrow \source {  {\source \tau }_{{\mathrm{2}}}  } \stot{ \rightsquigarrow  \target {  \textbf{\target {let} } \, {\target x}  \ottsym{:}  \forall  {\overline {\target a} }_{\ottmv{j}}  \ottsym{.}  {\overline {\target \sigma} }_{\ottmv{k}}  \rightarrow  {\target \sigma }  \ottsym{=}  \Lambda  {\overline {\target a} }_{\ottmv{j}}  \ottsym{.}  \lambda \, \ottcomp{{\target \delta }_{\ottmv{k}}  \ottsym{:}  {\target \sigma }_{\ottmv{k}}}{\ottmv{k}} \, \ottsym{.}  {\target e }_{{\mathrm{1}}} \, \textbf{\target {in} } \, {\target e }_{{\mathrm{2}}}  } } 
  }
  {
    The goal to be proven is the following:
    \begin{gather}
       \source {  { \source P }  } ; \source {  {\source { \Gamma_C } }  } ; \source {  {\source \Gamma}  }  \vdash_{tm}^{\intermed {M} }  \source {  \textbf{\source {let} } \, {\source x}  \ottsym{:}  \forall  {\overline {\source a} }_{\ottmv{j}}  \ottsym{.}  {\overline {\source Q} }_{\ottmv{i}}  \Rightarrow  {\source \tau }_{{\mathrm{1}}}  \ottsym{=}  {\source e}_{{\mathrm{1}}} \, \textbf{\source {in} } \, {\source e}_{{\mathrm{2}}}  } \Rightarrow \source {  {\source \tau }_{{\mathrm{2}}}  } \stoi{ \rightsquigarrow  \intermed {  {\intermed e}  } } 
      \label{eqg:eq_expr1} \\
       \intermed {  {\intermed \Sigma}  } ; \intermed {  {\intermed { \Gamma_C } }  } ; \intermed {  {\intermed \Gamma}  }  \vdash_{tm}  \intermed {  {\intermed e}  } : \intermed {  {\intermed \sigma}_{{\mathrm{2}}}  } \itot{ \rightsquigarrow  \target {  \textbf{\target {let} } \, {\target x}  \ottsym{:}  \forall  {\overline {\target a} }_{\ottmv{j}}  \ottsym{.}  {\overline {\target \sigma} }_{\ottmv{k}}  \rightarrow  {\target \sigma }  \ottsym{=}  \Lambda  {\overline {\target a} }_{\ottmv{j}}  \ottsym{.}  \lambda \, \ottcomp{{\target \delta }_{\ottmv{k}}  \ottsym{:}  {\target \sigma }_{\ottmv{k}}}{\ottmv{k}} \, \ottsym{.}  {\target e }_{{\mathrm{1}}} \, \textbf{\target {in} } \, {\target e }_{{\mathrm{2}}}  } } 
      \label{eqg:eq_expr2} \\
       \source {  {\source { \Gamma_C } }  } ; \source {  {\source \Gamma}  }  \vdash_{ty}^{\intermed {M} }  \source {  {\source \tau }_{{\mathrm{2}}}  }\,{ \stoi{ \rightsquigarrow  \intermed {  {\intermed \sigma}_{{\mathrm{2}}}  } } } 
      \label{eqg:eq_expr2b}
    \end{gather}
    From the rule premise we know that:
    \begin{gather}
       \source {  {\source x}  }  \notin   \textbf {dom}  ( \source {  {\source \Gamma}  } ) 
      \label{eq:eq_expr1} \\
       \textbf {unambig}  ( \source {  \forall  {\overline {\source a} }_{\ottmv{j}}  \ottsym{.}  {\overline {\source Q} }_{\ottmv{i}}  \Rightarrow  {\source \tau }_{{\mathrm{1}}}  } ) 
      \label{eq:eq_expr2} \\
       \textbf {closure}  ( \source {  {\source { \Gamma_C } }  } ; \source {  {\overline {\source Q} }_{\ottmv{i}}  } ) = \source {  {\overline {\source Q} }_{\ottmv{k}}  } 
      \label{eq:eq_expr3} \\
      \ottcomp{ \source {  {\source { \Gamma_C } }  } ; \source {  {\source \Gamma}  }  \vdash_{\mathit {Q} }  \source {  {\source Q}_{\ottmv{k}}  }  \rightsquigarrow  \target {  {\target \sigma }_{\ottmv{k}}  } }{\ottmv{k}}
      \label{eq:eq_expr4} \\
       \source {  {\source { \Gamma_C } }  } ; \source {  {\source \Gamma}  }  \vdash_{ty}  \source {  \forall  {\overline {\source a} }_{\ottmv{j}}  \ottsym{.}  {\overline {\source Q} }_{\ottmv{k}}  \Rightarrow  {\source \tau }_{{\mathrm{1}}}  }\,{ \stoi{ \rightsquigarrow  \target {  \forall  {\overline {\target a} }_{\ottmv{j}}  \ottsym{.}  {\overline {\target \sigma} }_{\ottmv{k}}  \rightarrow  {\target \sigma }  } } } 
      \label{eq:eq_expr5} \\
       \source {  {\overline {\source \delta} }_{\ottmv{k}}  }~ \textbf {fresh} 
      \label{eq:eq_expr6} \\
       \source {  { \source P }  } ; \source {  {\source { \Gamma_C } }  } ; \source {  {\source \Gamma}  \ottsym{,}  {\overline {\source a} }_{\ottmv{j}}  \ottsym{,}  {\overline {\source \delta} }_{\ottmv{k}}  \ottsym{:}  {\overline {\source Q} }_{\ottmv{k}}  }  \vdash_{tm}  \source {  {\source e}_{{\mathrm{1}}}  } \Leftarrow \source {  {\source \tau }_{{\mathrm{1}}}  } \stot{ \rightsquigarrow  \target {  {\target e }_{{\mathrm{1}}}  } } 
      \label{eq:eq_expr7} \\
       \source {  { \source P }  } ; \source {  {\source { \Gamma_C } }  } ; \source {  {\source \Gamma}  \ottsym{,}  {\source x}  \ottsym{:}  \forall  {\overline {\source a} }_{\ottmv{j}}  \ottsym{.}  {\overline {\source Q} }_{\ottmv{k}}  \Rightarrow  {\source \tau }_{{\mathrm{1}}}  }  \vdash_{tm}  \source {  {\source e}_{{\mathrm{2}}}  } \Rightarrow \source {  {\source \tau }_{{\mathrm{2}}}  } \stot{ \rightsquigarrow  \target {  {\target e }_{{\mathrm{2}}}  } } 
      \label{eq:eq_expr8} 
    \end{gather}
    By applying Lemma~\ref{lemma:type_well_source_typing} to Equation~\ref{eq:eq_expr8},
    we get that:
    \begin{equation}
       \source {  {\source { \Gamma_C } }  } ; \source {  {\source \Gamma}  \ottsym{,}  {\source x}  \ottsym{:}  \forall  {\overline {\source a} }_{\ottmv{j}}  \ottsym{.}  {\overline {\source Q} }_{\ottmv{k}}  \Rightarrow  {\source \tau }_{{\mathrm{1}}}  }  \vdash_{ty}^{\intermed {M} }  \source {  {\source \tau }_{{\mathrm{2}}}  }\,{ \stoi{ \rightsquigarrow  \intermed {  {\intermed \sigma}_{{\mathrm{2}}}  } } } 
      \label{eq:eq_expr8b} 
    \end{equation}
    It is straightforward to see from the definition of type well-formedness,
    that Goal~\ref{eqg:eq_expr2b} follows from Equation~\ref{eq:eq_expr8b},
    since term variables in the environment are not relevant for type well-formedness.

    We know from the hypothesis that \( \vdash_{ctx}  \source {  { \source P }  } ; \source {  {\source { \Gamma_C } }  } ; \source {  {\source \Gamma}  }  \rightsquigarrow  \target {  {\target \Gamma }  } \).
    By repeated case analysis on this result (\textsc{sCtxT-pgmInst}), we get that
    \( \vdash_{ctx}  \source {  \bullet  } ; \source {  {\source { \Gamma_C } }  } ; \source {  {\source \Gamma}  }  \rightsquigarrow  \target {  {\target \Gamma }  } \).
    From \textsc{sCtxT-tyEnvTm}, \textsc{sCtxT-tyEnvTy} and \textsc{sCtxT-tyEnvD}, in
    combination with Equations~\ref{eq:eq_expr1}, \ref{eq:eq_expr4},
    \ref{eq:eq_expr5} and~\ref{eq:eq_expr6}, we know that:
    \begin{gather}
       \vdash_{ctx}  \source {  { \source P }  } ; \source {  {\source { \Gamma_C } }  } ; \source {  {\source \Gamma}  \ottsym{,}  {\overline {\source a} }_{\ottmv{j}}  \ottsym{,}  {\overline {\source \delta} }_{\ottmv{k}}  \ottsym{:}  {\overline {\source Q} }_{\ottmv{k}}  }  \rightsquigarrow  \target {  {\target \Gamma }  \ottsym{,}  {\overline {\target a} }_{\ottmv{j}}  \ottsym{,} \, \ottcomp{{\target \delta }_{\ottmv{k}}  \ottsym{:}  {\target \sigma }_{\ottmv{k}}}{\ottmv{k}}  } 
      \label{eq:eq_expr9} \\
       \vdash_{ctx}  \source {  { \source P }  } ; \source {  {\source { \Gamma_C } }  } ; \source {  {\source \Gamma}  \ottsym{,}  {\source x}  \ottsym{:}  \forall  {\overline {\source a} }_{\ottmv{j}}  \ottsym{.}  {\overline {\source Q} }_{\ottmv{k}}  \Rightarrow  {\source \tau }_{{\mathrm{1}}}  }  \rightsquigarrow  \target {  {\target \Gamma }  \ottsym{,}  {\target x}  \ottsym{:}  \forall  {\overline {\target a} }_{\ottmv{j}}  \ottsym{.}  {\overline {\target \sigma} }_{\ottmv{k}}  \rightarrow  {\target \sigma }  } 
      \label{eq:eq_expr10} 
    \end{gather}
    Applying the induction hypothesis on Equations~\ref{eq:eq_expr7}
    and~\ref{eq:eq_expr8}, in combination with Equations~\ref{eq:eq_expr9}
    and~\ref{eq:eq_expr10}, results in:
    \begin{gather}
       \source {  { \source P }  } ; \source {  {\source { \Gamma_C } }  } ; \source {  {\source \Gamma}  \ottsym{,}  {\overline {\source a} }_{\ottmv{j}}  \ottsym{,}  {\overline {\source \delta} }_{\ottmv{k}}  \ottsym{:}  {\overline {\source Q} }_{\ottmv{k}}  }  \vdash_{tm}^{\intermed {M} }  \source {  {\source e}_{{\mathrm{1}}}  } \Leftarrow \source {  {\source \tau }_{{\mathrm{1}}}  } \stoi{ \rightsquigarrow  \intermed {  {\intermed e}_{{\mathrm{1}}}  } } 
      \label{eq:eq_expr11} \\
       \intermed {  {\intermed \Sigma}  } ; \intermed {  {\intermed { \Gamma_C } }  } ; \intermed {  {\intermed \Gamma}  \ottsym{,}  {\overline {\intermed a} }_{\ottmv{j}}  \ottsym{,}  {\overline {\intermed \delta} }_{\ottmv{k}}  \ottsym{:}  {\overline {\intermed Q} }_{\ottmv{k}}  }  \vdash_{tm}  \intermed {  {\intermed e}_{{\mathrm{1}}}  } : \intermed {  {\intermed \sigma}_{{\mathrm{1}}}  } \itot{ \rightsquigarrow  \target {  {\target e }_{{\mathrm{1}}}  } } 
      \label{eq:eq_expr12} \\
       \source {  { \source P }  } ; \source {  {\source { \Gamma_C } }  } ; \source {  {\source \Gamma}  \ottsym{,}  {\source x}  \ottsym{:}  \forall  {\overline {\source a} }_{\ottmv{j}}  \ottsym{.}  {\overline {\source Q} }_{\ottmv{k}}  \Rightarrow  {\source \tau }_{{\mathrm{1}}}  }  \vdash_{tm}^{\intermed {M} }  \source {  {\source e}_{{\mathrm{2}}}  } \Rightarrow \source {  {\source \tau }_{{\mathrm{2}}}  } \stoi{ \rightsquigarrow  \intermed {  {\intermed e}_{{\mathrm{2}}}  } } 
      \label{eq:eq_expr13} \\
       \intermed {  {\intermed \Sigma}  } ; \intermed {  {\intermed { \Gamma_C } }  } ; \intermed {  {\intermed \Gamma}  \ottsym{,}  {\intermed x}  \ottsym{:}  \forall  {\overline {\intermed a} }_{\ottmv{j}}  \ottsym{.}   {\overline {\intermed Q} }_{\ottmv{k}}  \Rightarrow  {\intermed \sigma}_{{\mathrm{1}}}   }  \vdash_{tm}  \intermed {  {\intermed e}_{{\mathrm{2}}}  } : \intermed {  {\intermed \sigma}_{{\mathrm{2}}}  } \itot{ \rightsquigarrow  \target {  {\target e }_{{\mathrm{2}}}  } } 
      \label{eq:eq_expr14} 
    \end{gather}
    From constraint equivalence (Theorem~\ref{thmA:equiv_ty}), in combination with
    Equation~\ref{eq:eq_expr4}, we get:
    \begin{equation}
      \ottcomp{ \intermed {  {\intermed { \Gamma_C } }  } ; \intermed {  {\intermed \Gamma}  }  \vdash_{\mathit {Q} }  \intermed {  {\intermed Q}_{\ottmv{k}}  }\, \itot{ \rightsquigarrow  \target {  {\target \sigma }_{\ottmv{k}}  } } }{\ottmv{k}}
      \label{eq:eq_expr15} 
    \end{equation}
    By applying \textsc{iTm-forallI} and \textsc{iTm-constrI} to
    Equation~\ref{eq:eq_expr12}, together with Equation~\ref{eq:eq_expr15}, we get:
    \begin{equation}
       \intermed {  {\intermed \Sigma}  } ; \intermed {  {\intermed { \Gamma_C } }  } ; \intermed {  {\intermed \Gamma}  }  \vdash_{tm}  \intermed {  \Lambda  {\overline {\intermed a} }_{\ottmv{j}}  \ottsym{.}  \lambda  {\overline {\intermed \delta} }_{\ottmv{k}}  \ottsym{:}  {\overline {\intermed Q} }_{\ottmv{k}}  \ottsym{.}  {\intermed e}_{{\mathrm{1}}}  } : \intermed {  \forall  {\overline {\intermed a} }_{\ottmv{j}}  \ottsym{.}   {\overline {\intermed Q} }_{\ottmv{k}}  \Rightarrow  {\intermed \sigma}_{{\mathrm{1}}}   } \itot{ \rightsquigarrow  \target {  \Lambda  {\overline {\target a} }_{\ottmv{j}}  \ottsym{.}  \lambda \, \ottcomp{{\target \delta }_{\ottmv{k}}  \ottsym{:}  {\target \sigma }_{\ottmv{k}}}{\ottmv{k}} \, \ottsym{.}  {\target e }_{{\mathrm{1}}}  } } 
      \label{eq:eq_expr16} 
    \end{equation}
    From Lemma~\ref{lemma:type_well_inter_typing}, in combination with
    Equation~\ref{eq:eq_expr16}, we know that:
    \begin{equation}
       \intermed {  {\intermed { \Gamma_C } }  } ; \intermed {  {\intermed \Gamma}  }  \vdash_{ty}  \intermed {  \forall  {\overline {\intermed a} }_{\ottmv{j}}  \ottsym{.}   {\overline {\intermed Q} }_{\ottmv{k}}  \Rightarrow  {\intermed \sigma}_{{\mathrm{1}}}   } \itot{ \rightsquigarrow  \target {  \forall  {\overline {\target a} }_{\ottmv{j}}  \ottsym{.}  {\overline {\target \sigma} }_{\ottmv{k}}  \rightarrow  {\target \sigma }  } } 
      \label{eq:eq_expr16b} 
    \end{equation}
    Goals~\ref{eqg:eq_expr1} and~\ref{eqg:eq_expr2} follow from
    \textsc{sTm-inf-let} and \textsc{iTm-let} respectively, with
    \begin{equation*}
     \intermed {  {\intermed e}  } = \intermed {  \textbf{\intermed {let} } \, {\intermed x}  \ottsym{:}  \forall  {\overline {\intermed a} }_{\ottmv{j}}  \ottsym{.}   {\overline {\intermed Q} }_{\ottmv{k}}  \Rightarrow  {\intermed \sigma}_{{\mathrm{1}}}   \ottsym{=}  \Lambda  {\overline {\intermed a} }_{\ottmv{j}}  \ottsym{.}  \lambda  {\overline {\intermed \delta} }_{\ottmv{k}}  \ottsym{:}  {\overline {\intermed Q} }_{\ottmv{k}}  \ottsym{.}  {\intermed e}_{{\mathrm{1}}} \, \textbf{\intermed {in} } \, {\intermed e}_{{\mathrm{2}}}  } 
    \end{equation*}
  }
  \proofStep{\textsc{sTm-infT-ArrE}}
  {
     \source {  { \source P }  } ; \source {  {\source { \Gamma_C } }  } ; \source {  {\source \Gamma}  }  \vdash_{tm}  \source {  {\source e}_{{\mathrm{1}}} \, {\source e}_{{\mathrm{2}}}  } \Rightarrow \source {  {\source \tau }_{{\mathrm{2}}}  } \stot{ \rightsquigarrow  \target {  {\target e }_{{\mathrm{1}}} \, {\target e }_{{\mathrm{2}}}  } } 
  }
  {
    The goal to be proven is the following:
    \begin{gather}
       \source {  { \source P }  } ; \source {  {\source { \Gamma_C } }  } ; \source {  {\source \Gamma}  }  \vdash_{tm}^{\intermed {M} }  \source {  {\source e}_{{\mathrm{1}}} \, {\source e}_{{\mathrm{2}}}  } \Rightarrow \source {  {\source \tau }_{{\mathrm{2}}}  } \stoi{ \rightsquigarrow  \intermed {  {\intermed e}  } } 
      \label{eqg:eq_expr3} \\
       \intermed {  {\intermed \Sigma}  } ; \intermed {  {\intermed { \Gamma_C } }  } ; \intermed {  {\intermed \Gamma}  }  \vdash_{tm}  \intermed {  {\intermed e}  } : \intermed {  {\intermed \sigma}_{{\mathrm{2}}}  } \itot{ \rightsquigarrow  \target {  {\target e }_{{\mathrm{1}}} \, {\target e }_{{\mathrm{2}}}  } } 
      \label{eqg:eq_expr4} \\
       \source {  {\source { \Gamma_C } }  } ; \source {  {\source \Gamma}  }  \vdash_{ty}^{\intermed {M} }  \source {  {\source \tau }_{{\mathrm{2}}}  }\,{ \stoi{ \rightsquigarrow  \intermed {  {\intermed \sigma}_{{\mathrm{2}}}  } } } 
      \label{eqg:eq_expr4b}
    \end{gather}
    From the rule premise we know that:
    \begin{gather}
       \source {  { \source P }  } ; \source {  {\source { \Gamma_C } }  } ; \source {  {\source \Gamma}  }  \vdash_{tm}  \source {  {\source e}_{{\mathrm{1}}}  } \Rightarrow \source {  {\source \tau }_{{\mathrm{1}}}  \rightarrow  {\source \tau }_{{\mathrm{2}}}  } \stot{ \rightsquigarrow  \target {  {\target e }_{{\mathrm{1}}}  } } 
      \label{eq:eq_expr17} \\
       \source {  { \source P }  } ; \source {  {\source { \Gamma_C } }  } ; \source {  {\source \Gamma}  }  \vdash_{tm}  \source {  {\source e}_{{\mathrm{2}}}  } \Leftarrow \source {  {\source \tau }_{{\mathrm{1}}}  } \stot{ \rightsquigarrow  \target {  {\target e }_{{\mathrm{2}}}  } } 
      \label{eq:eq_expr18} 
    \end{gather}
    By applying the induction hypothesis to Equations~\ref{eq:eq_expr17}
    and~\ref{eq:eq_expr18}, we get:
    \begin{gather}
       \source {  { \source P }  } ; \source {  {\source { \Gamma_C } }  } ; \source {  {\source \Gamma}  }  \vdash_{tm}^{\intermed {M} }  \source {  {\source e}_{{\mathrm{1}}}  } \Rightarrow \source {  {\source \tau }_{{\mathrm{1}}}  \rightarrow  {\source \tau }_{{\mathrm{2}}}  } \stoi{ \rightsquigarrow  \intermed {  {\intermed e}_{{\mathrm{1}}}  } } 
      \label{eq:eq_expr19} \\
       \intermed {  {\intermed \Sigma}  } ; \intermed {  {\intermed { \Gamma_C } }  } ; \intermed {  {\intermed \Gamma}  }  \vdash_{tm}  \intermed {  {\intermed e}_{{\mathrm{1}}}  } : \intermed {  {\intermed \sigma}_{{\mathrm{1}}}  \rightarrow  {\intermed \sigma}_{{\mathrm{2}}}  } \itot{ \rightsquigarrow  \target {  {\target e }_{{\mathrm{1}}}  } } 
      \label{eq:eq_expr20} \\
       \source {  { \source P }  } ; \source {  {\source { \Gamma_C } }  } ; \source {  {\source \Gamma}  }  \vdash_{tm}^{\intermed {M} }  \source {  {\source e}_{{\mathrm{2}}}  } \Leftarrow \source {  {\source \tau }_{{\mathrm{1}}}  } \stoi{ \rightsquigarrow  \intermed {  {\intermed e}_{{\mathrm{2}}}  } } 
      \label{eq:eq_expr21} \\
       \intermed {  {\intermed \Sigma}  } ; \intermed {  {\intermed { \Gamma_C } }  } ; \intermed {  {\intermed \Gamma}  }  \vdash_{tm}  \intermed {  {\intermed e}_{{\mathrm{2}}}  } : \intermed {  {\intermed \sigma}_{{\mathrm{1}}}  } \itot{ \rightsquigarrow  \target {  {\target e }_{{\mathrm{2}}}  } } 
      \label{eq:eq_expr22} 
    \end{gather}
    Goals~\ref{eqg:eq_expr3} and~\ref{eqg:eq_expr4} follow from
    \textsc{sTm-inf-ArrE} and \textsc{iTm-arrE} respectively, in combination
    with Equations~\ref{eq:eq_expr19}, \ref{eq:eq_expr20}, \ref{eq:eq_expr21}
    and~\ref{eq:eq_expr22}. Goal~\ref{eqg:eq_expr4b} follows by applying
    Lemma~\ref{lemma:type_well_source_typing} to Equation~\ref{eqg:eq_expr3}. \\
  }
  \proofStep{\textsc{sTm-infT-Ann}}
  {
     \source {  { \source P }  } ; \source {  {\source { \Gamma_C } }  } ; \source {  {\source \Gamma}  }  \vdash_{tm}  \source {  {\source e}  ::  {\source \tau }  } \Rightarrow \source {  {\source \tau }  } \stot{ \rightsquigarrow  \target {  {\target e }  } } 
  }
  {
    The goal to be proven is the following:
    \begin{gather}
       \source {  { \source P }  } ; \source {  {\source { \Gamma_C } }  } ; \source {  {\source \Gamma}  }  \vdash_{tm}^{\intermed {M} }  \source {  {\source e}  ::  {\source \tau }  } \Rightarrow \source {  {\source \tau }  } \stoi{ \rightsquigarrow  \intermed {  {\intermed e}  } } 
      \label{eqg:eq_expr5} \\
       \intermed {  {\intermed \Sigma}  } ; \intermed {  {\intermed { \Gamma_C } }  } ; \intermed {  {\intermed \Gamma}  }  \vdash_{tm}  \intermed {  {\intermed e}  } : \intermed {  {\intermed \sigma}  } \itot{ \rightsquigarrow  \target {  {\target e }  } } 
      \label{eqg:eq_expr6} \\
       \source {  {\source { \Gamma_C } }  } ; \source {  {\source \Gamma}  }  \vdash_{ty}^{\intermed {M} }  \source {  {\source \tau }  }\,{ \stoi{ \rightsquigarrow  \intermed {  {\intermed \sigma}  } } } 
      \label{eqg:eq_expr6b}
    \end{gather}
    From the rule premise we know that:
    \begin{equation}
       \source {  { \source P }  } ; \source {  {\source { \Gamma_C } }  } ; \source {  {\source \Gamma}  }  \vdash_{tm}  \source {  {\source e}  } \Leftarrow \source {  {\source \tau }  } \stot{ \rightsquigarrow  \target {  {\target e }  } } 
      \label{eq:eq_expr23}
    \end{equation}
    Goals~\ref{eqg:eq_expr5}, \ref{eqg:eq_expr6} and~\ref{eqg:eq_expr6b} follow by applying Part
    2 of this theorem to Equation~\ref{eq:eq_expr23}. \\
  }
\item[Part 2] 
  By case analysis on the typing derivation. \\
  \proofStep{\textsc{sTm-checkT-var}}
  {
     \source {  { \source P }  } ; \source {  {\source { \Gamma_C } }  } ; \source {  {\source \Gamma}  }  \vdash_{tm}  \source {  {\source x}  } \Leftarrow \source {  [  {\overline {\source \tau} }_{\ottmv{j}}  /  {\overline {\source a} }_{\ottmv{j}}  ]  {\source \tau }  } \stot{ \rightsquigarrow  \target {  {\target x} \, {\overline {\target \sigma} }_{\ottmv{j}} \, {\overline {\target e} }_{\ottmv{i}}  } } 
  }
  {
    The goal to be proven is the following:
    \begin{gather}
       \source {  { \source P }  } ; \source {  {\source { \Gamma_C } }  } ; \source {  {\source \Gamma}  }  \vdash_{tm}^{\intermed {M} }  \source {  {\source x}  } \Leftarrow \source {  [  {\overline {\source \tau} }_{\ottmv{j}}  /  {\overline {\source a} }_{\ottmv{j}}  ]  {\source \tau }  } \stoi{ \rightsquigarrow  \intermed {  {\intermed e}  } } 
      \label{eqg:eq_expr7} \\
       \intermed {  {\intermed \Sigma}  } ; \intermed {  {\intermed { \Gamma_C } }  } ; \intermed {  {\intermed \Gamma}  }  \vdash_{tm}  \intermed {  {\intermed e}  } : \intermed {  {\intermed \sigma}'  } \itot{ \rightsquigarrow  \target {  {\target x} \, {\overline {\target \sigma} }_{\ottmv{j}} \, {\overline {\target e} }_{\ottmv{i}}  } } 
      \label{eqg:eq_expr8} \\
       \source {  {\source { \Gamma_C } }  } ; \source {  {\source \Gamma}  }  \vdash_{ty}^{\intermed {M} }  \source {  [  {\overline {\source \tau} }_{\ottmv{j}}  /  {\overline {\source a} }_{\ottmv{j}}  ]  {\source \tau }  }\,{ \stoi{ \rightsquigarrow  \intermed {  {\intermed \sigma}'  } } } 
      \label{eqg:eq_expr8b} 
    \end{gather}
    From the rule premise we know that:
    \begin{gather}
       ( \source {  {\source x}  } : \source {  \forall  {\overline {\source a} }_{\ottmv{j}}  \ottsym{.}  {\overline {\source Q} }_{\ottmv{i}}  \Rightarrow  {\source \tau }  } )  \in  \source {  {\source \Gamma}  } 
      \label{eq:eq_expr24} \\
       \textbf {unambig}  ( \source {  \forall  {\overline {\source a} }_{\ottmv{j}}  \ottsym{.}  {\overline {\source Q} }_{\ottmv{i}}  \Rightarrow  {\source \tau }  } ) 
      \label{eq:eq_expr25} \\
      \ottcomp{ \source {  { \source P }  } ; \source {  {\source { \Gamma_C } }  } ; \source {  {\source \Gamma}  }  \vDash  \source {  [  {\overline {\source \tau} }_{\ottmv{j}}  /  {\overline {\source a} }_{\ottmv{j}}  ]  {\source Q}_{\ottmv{i}}  } \stot{ \rightsquigarrow  \target {  {\target e }_{\ottmv{i}}  } } }{\ottmv{i}}
      \label{eq:eq_expr26} \\
      \ottcomp{ \source {  {\source { \Gamma_C } }  } ; \source {  {\source \Gamma}  }  \vdash_{ty}  \source {  {\source \tau }_{\ottmv{j}}  }\,{ \stoi{ \rightsquigarrow  \target {  {\target \sigma }_{\ottmv{j}}  } } } }{\ottmv{j}}
      \label{eq:eq_expr27} \\
       \vdash_{ctx}  \source {  { \source P }  } ; \source {  {\source { \Gamma_C } }  } ; \source {  {\source \Gamma}  }  \rightsquigarrow  \target {  {\target \Gamma }  } 
      \label{eq:eq_expr28} 
    \end{gather}
    We know from Lemma~\ref{lemma:pres_env_tm}, in combination with Equations~\ref{eq:eq_expr24}
    and~\ref{eq:eq_expr0a}, that:
    \begin{equation}
       ( \intermed {  {\intermed x}  } : \intermed {  \forall  {\overline {\intermed a} }_{\ottmv{j}}  \ottsym{.}   {\overline {\intermed Q} }_{\ottmv{i}}  \Rightarrow  {\intermed \sigma}   } )  \in  \intermed {  {\intermed \Gamma}  } 
      \label{eq:eq_expr24b}
    \end{equation}
    By applying type equivalence (Theorem~\ref{thmA:equiv_ty}) to
    Equation~\ref{eq:eq_expr27}, we get:
    \begin{gather}
      \ottcomp{ \source {  {\source { \Gamma_C } }  } ; \source {  {\source \Gamma}  }  \vdash_{ty}^{\intermed {M} }  \source {  {\source \tau }_{\ottmv{j}}  }\,{ \stoi{ \rightsquigarrow  \intermed {  {\intermed \sigma}_{\ottmv{j}}  } } } }{\ottmv{j}}
      \label{eq:eq_expr31} \\
      \ottcomp{ \intermed {  {\intermed { \Gamma_C } }  } ; \intermed {  {\intermed \Gamma}  }  \vdash_{ty}  \intermed {  {\intermed \sigma}_{\ottmv{j}}  } \itot{ \rightsquigarrow  \target {  {\target \sigma }_{\ottmv{j}}  } } }{\ottmv{j}}
      \label{eq:eq_expr32} 
    \end{gather}
    By applying dictionary equivalence (Theorem~\ref{thmA:equiv_dict}) to
    Equation~\ref{eq:eq_expr26}, we get:
    \begin{gather}
      \ottcomp{ \source {  { \source P }  } ; \source {  {\source { \Gamma_C } }  } ; \source {  {\source \Gamma}  }  \vDash^{\intermed {M} }  \source {  [  {\overline {\source \tau} }_{\ottmv{j}}  /  {\overline {\source a} }_{\ottmv{j}}  ]  {\source Q}_{\ottmv{i}}  } \stoi{ \rightsquigarrow  \intermed {  {\intermed d}_{\ottmv{i}}  } } }{\ottmv{i}}
      \label{eq:eq_expr29} \\
      \ottcomp{ \intermed {  {\intermed \Sigma}  } ; \intermed {  {\intermed { \Gamma_C } }  } ; \intermed {  {\intermed \Gamma}  }  \vdash_{d}  \intermed {  {\intermed d}_{\ottmv{i}}  } : \intermed {  [  \overline {\intermed \sigma}_{\ottmv{j}}  /  {\overline {\intermed a} }_{\ottmv{j}}  ]  {\intermed Q}_{\ottmv{i}}  } \itot{\,  \rightsquigarrow  \target {  {\target e }_{\ottmv{i}}  } } }{\ottmv{i}}
      \label{eq:eq_expr30} \\
      \ottcomp{ \source {  {\source { \Gamma_C } }  } ; \source {  {\source \Gamma}  }  \vdash_{\mathit {Q} }^{\intermed {M} }  \source {  [  {\overline {\source \tau} }_{\ottmv{j}}  /  {\overline {\source a} }_{\ottmv{j}}  ]  {\source Q}_{\ottmv{i}}  }  \rightsquigarrow  \intermed {  [  \overline {\intermed \sigma}_{\ottmv{j}}  /  {\overline {\intermed a} }_{\ottmv{j}}  ]  {\intermed Q}_{\ottmv{i}}  } }{\ottmv{i}}
      \label{eq:eq_expr30b} 
    \end{gather}
    Goal~\ref{eqg:eq_expr7} follows from \textsc{sTm-check-var}, in combination
    with Equations~\ref{eq:eq_expr24}, \ref{eq:eq_expr25}, \ref{eq:eq_expr29},
    \ref{eq:eq_expr31} and~\ref{eq:eq_expr0a}, with
    \( \intermed {  {\intermed e}  } = \intermed {  {\intermed x} \, \overline {\intermed \sigma}_{\ottmv{j}} \, {\overline {\intermed d} }_{\ottmv{i}}  } \).
    Goal~\ref{eqg:eq_expr8} follows from \textsc{iTm-var}, \textsc{iTm-forallE}
    and \textsc{iTm-constrE}, in combination with Equations~\ref{eq:eq_expr0c}, \ref{eq:eq_expr24b},
    \ref{eq:eq_expr32} and~\ref{eq:eq_expr30}, with \( \intermed {  {\intermed \sigma}'  } = \intermed {  [  \overline {\intermed \sigma}_{\ottmv{j}}  /  {\overline {\intermed a} }_{\ottmv{j}}  ]  {\intermed \sigma}  } \).
    Goal~\ref{eqg:eq_expr8b} follows
    by applying Lemma~\ref{lemma:type_well_source_typing} to Equation~\ref{eqg:eq_expr7}. \\
  }
  \proofStep{\textsc{sTm-checkT-meth}}
  {
     \source {  { \source P }  } ; \source {  {\source { \Gamma_C } }  } ; \source {  {\source \Gamma}  }  \vdash_{tm}  \source {  {\source m}  } \Leftarrow \source {  [  {\overline {\source \tau} }_{\ottmv{j}}  /  {\overline {\source a} }_{\ottmv{j}}  ]  [  {\source \tau }  /  {\source a }  ]  {\source \tau }'  } \stot{ \rightsquigarrow  \target {  {\target e }  \ottsym{.}  {\target m} \, {\overline {\target \sigma} }_{\ottmv{j}} \, {\overline {\target e} }_{\ottmv{i}}  } } 
  }
  {
    The goal to be proven is the following:
    \begin{gather}
       \source {  { \source P }  } ; \source {  {\source { \Gamma_C } }  } ; \source {  {\source \Gamma}  }  \vdash_{tm}^{\intermed {M} }  \source {  {\source m}  } \Leftarrow \source {  [  {\overline {\source \tau} }_{\ottmv{j}}  /  {\overline {\source a} }_{\ottmv{j}}  ]  [  {\source \tau }  /  {\source a }  ]  {\source \tau }'  } \stoi{ \rightsquigarrow  \intermed {  {\intermed e}  } } 
      \label{eqg:eq_expr9} \\
       \intermed {  {\intermed \Sigma}  } ; \intermed {  {\intermed { \Gamma_C } }  } ; \intermed {  {\intermed \Gamma}  }  \vdash_{tm}  \intermed {  {\intermed e}  } : \intermed {  {\intermed \sigma}_{{\mathrm{0}}}  } \itot{ \rightsquigarrow  \target {  {\target e }  \ottsym{.}  {\target m} \, {\overline {\target \sigma} }_{\ottmv{j}} \, {\overline {\target e} }_{\ottmv{i}}  } } 
      \label{eqg:eq_expr10} \\
       \source {  {\source { \Gamma_C } }  } ; \source {  {\source \Gamma}  }  \vdash_{ty}^{\intermed {M} }  \source {  [  {\overline {\source \tau} }_{\ottmv{j}}  /  {\overline {\source a} }_{\ottmv{j}}  ]  [  {\source \tau }  /  {\source a }  ]  {\source \tau }'  }\,{ \stoi{ \rightsquigarrow  \intermed {  {\intermed \sigma}_{{\mathrm{0}}}  } } } 
      \label{eqg:eq_expr10b} 
    \end{gather}
    From the rule premise we know that:
    \begin{gather}
       ( \source {  {\source m}  } : \source {  {\overline {\source Q} }'_{\ottmv{k}}  }  \Rightarrow  \source {  {\source {\mathit{TC} } }  }~\source {  {\source a }  } : \source {  \forall  {\overline {\source a} }_{\ottmv{j}}  \ottsym{.}  {\overline {\source Q} }_{\ottmv{i}}  \Rightarrow  {\source \tau }'  } )  \in  \source {  {\source { \Gamma_C } }  } 
      \label{eq:eq_expr33} \\
       \textbf {unambig}  ( \source {  \forall  {\overline {\source a} }_{\ottmv{j}}  \ottsym{,}  {\source a }  \ottsym{.}  {\overline {\source Q} }_{\ottmv{i}}  \Rightarrow  {\source \tau }'  } ) 
      \label{eq:eq_expr34} \\
       \source {  { \source P }  } ; \source {  {\source { \Gamma_C } }  } ; \source {  {\source \Gamma}  }  \vDash  \source {  {\source {\mathit{TC} } } \, {\source \tau }  } \stot{ \rightsquigarrow  \target {  {\target e }  } } 
      \label{eq:eq_expr35} \\
       \source {  {\source { \Gamma_C } }  } ; \source {  {\source \Gamma}  }  \vdash_{ty}  \source {  {\source \tau }  }\,{ \stoi{ \rightsquigarrow  \target {  {\target \sigma }  } } } 
      \label{eq:eq_expr36} \\
      \ottcomp{ \source {  { \source P }  } ; \source {  {\source { \Gamma_C } }  } ; \source {  {\source \Gamma}  }  \vDash  \source {  [  {\overline {\source \tau} }_{\ottmv{j}}  /  {\overline {\source a} }_{\ottmv{j}}  ]  [  {\source \tau }  /  {\source a }  ]  {\source Q}_{\ottmv{i}}  } \stot{ \rightsquigarrow  \target {  {\target e }_{\ottmv{i}}  } } }{\ottmv{i}}
      \label{eq:eq_expr37} \\
      \ottcomp{ \source {  {\source { \Gamma_C } }  } ; \source {  {\source \Gamma}  }  \vdash_{ty}  \source {  {\source \tau }_{\ottmv{j}}  }\,{ \stoi{ \rightsquigarrow  \target {  {\target \sigma }_{\ottmv{j}}  } } } }{\ottmv{j}}
      \label{eq:eq_expr38} \\
       \vdash_{ctx}  \source {  { \source P }  } ; \source {  {\source { \Gamma_C } }  } ; \source {  {\source \Gamma}  }  \rightsquigarrow  \target {  {\target \Gamma }  } 
      \label{eq:eq_expr39}
    \end{gather}
    By repeated case analysis on Equation~\ref{eq:eq_expr0a} (\textsc{sCtx-clsEnv}),
    together with Equation~\ref{eq:eq_expr33}, we know that:
    \begin{equation}
       ( \intermed {  {\intermed m}  } : \intermed {  {\intermed {\mathit{TC} } } \, {\intermed a }  } : \intermed {  \forall  {\overline {\intermed a} }_{\ottmv{j}}  \ottsym{.}   {\overline {\intermed Q} }_{\ottmv{i}}  \Rightarrow  {\intermed \sigma}'   } )  \in  \intermed {  {\intermed { \Gamma_C } }  } 
      \label{eq:eq_expr33b}
    \end{equation}
    By applying type equivalence (Theorem~\ref{thmA:equiv_ty}) to
    Equations~\ref{eq:eq_expr36} and~\ref{eq:eq_expr38}, we get:
    \begin{gather}
       \source {  {\source { \Gamma_C } }  } ; \source {  {\source \Gamma}  }  \vdash_{ty}^{\intermed {M} }  \source {  {\source \tau }  }\,{ \stoi{ \rightsquigarrow  \intermed {  {\intermed \sigma}  } } } 
      \label{eq:eq_expr44} \\
       \intermed {  {\intermed { \Gamma_C } }  } ; \intermed {  {\intermed \Gamma}  }  \vdash_{ty}  \intermed {  {\intermed \sigma}  } \itot{ \rightsquigarrow  \target {  {\target \sigma }  } } 
      \label{eq:eq_expr45} \\
      \ottcomp{ \source {  {\source { \Gamma_C } }  } ; \source {  {\source \Gamma}  }  \vdash_{ty}^{\intermed {M} }  \source {  {\source \tau }_{\ottmv{j}}  }\,{ \stoi{ \rightsquigarrow  \intermed {  {\intermed \sigma}_{\ottmv{j}}  } } } }{\ottmv{j}}
      \label{eq:eq_expr46} \\
      \ottcomp{ \intermed {  {\intermed { \Gamma_C } }  } ; \intermed {  {\intermed \Gamma}  }  \vdash_{ty}  \intermed {  {\intermed \sigma}_{\ottmv{j}}  } \itot{ \rightsquigarrow  \target {  {\target \sigma }_{\ottmv{j}}  } } }{\ottmv{j}}
      \label{eq:eq_expr47}
    \end{gather}
    By applying dictionary equivalence (Theorem~\ref{thmA:equiv_dict}) to
    Equations~\ref{eq:eq_expr35} and~\ref{eq:eq_expr37}, we get:
    \begin{gather}
       \source {  { \source P }  } ; \source {  {\source { \Gamma_C } }  } ; \source {  {\source \Gamma}  }  \vDash^{\intermed {M} }  \source {  {\source {\mathit{TC} } } \, {\source \tau }  } \stoi{ \rightsquigarrow  \intermed {  {\intermed d}  } } 
      \label{eq:eq_expr40} \\
       \intermed {  {\intermed \Sigma}  } ; \intermed {  {\intermed { \Gamma_C } }  } ; \intermed {  {\intermed \Gamma}  }  \vdash_{d}  \intermed {  {\intermed d}  } : \intermed {  {\intermed {\mathit{TC} } } \, {\intermed \sigma}  } \itot{\,  \rightsquigarrow  \target {  {\target e }  } } 
      \label{eq:eq_expr41} \\
      \ottcomp{ \source {  { \source P }  } ; \source {  {\source { \Gamma_C } }  } ; \source {  {\source \Gamma}  }  \vDash^{\intermed {M} }  \source {  [  {\overline {\source \tau} }_{\ottmv{j}}  /  {\overline {\source a} }_{\ottmv{j}}  ]  [  {\source \tau }  /  {\source a }  ]  {\source Q}_{\ottmv{i}}  } \stoi{ \rightsquigarrow  \intermed {  {\intermed d}_{\ottmv{i}}  } } }{\ottmv{i}}
      \label{eq:eq_expr42} \\
      \ottcomp{ \intermed {  {\intermed \Sigma}  } ; \intermed {  {\intermed { \Gamma_C } }  } ; \intermed {  {\intermed \Gamma}  }  \vdash_{d}  \intermed {  {\intermed d}_{\ottmv{i}}  } : \intermed {  [  \overline {\intermed \sigma}_{\ottmv{j}}  /  {\overline {\intermed a} }_{\ottmv{j}}  ]  [  {\intermed \sigma}  /  {\intermed a }  ]  {\intermed Q}_{\ottmv{i}}  } \itot{\,  \rightsquigarrow  \target {  {\target e }_{\ottmv{i}}  } } }{\ottmv{i}}
      \label{eq:eq_expr43}
    \end{gather}
    Goal~\ref{eqg:eq_expr9} follows from \textsc{sTm-check-meth}, in combination
    with Equations~\ref{eq:eq_expr33}, \ref{eq:eq_expr34}, \ref{eq:eq_expr40},
    \ref{eq:eq_expr44}, \ref{eq:eq_expr42}, \ref{eq:eq_expr46}
    and~\ref{eq:eq_expr0a},
    with \( \intermed {  {\intermed e}  } = \intermed {  {\intermed d}  \ottsym{.}  {\intermed m} \, \overline {\intermed \sigma}_{\ottmv{j}} \, {\overline {\intermed d} }_{\ottmv{i}}  } \).
    Consequently, Goal~\ref{eqg:eq_expr10} follows from \textsc{iTm-method},
    \textsc{iTm-forallE} and \textsc{iTm-constrE}, in combination with
    Equations~\ref{eq:eq_expr41}, \ref{eq:eq_expr33b}, \ref{eq:eq_expr47}
    and~\ref{eq:eq_expr43}, with \( \intermed {  {\intermed \sigma}_{{\mathrm{0}}}  } = \intermed {  [  \overline {\intermed \sigma}_{\ottmv{j}}  /  {\overline {\intermed a} }_{\ottmv{j}}  ]  [  {\intermed \sigma}  /  {\intermed a }  ]  {\intermed \sigma}'  } \).
    Goal~\ref{eqg:eq_expr10b} follows
    by applying Lemma~\ref{lemma:type_well_source_typing} to Equation~\ref{eqg:eq_expr9}. \\
  }
  \proofStep{\textsc{sTm-checkT-ArrI}}
  {
     \source {  { \source P }  } ; \source {  {\source { \Gamma_C } }  } ; \source {  {\source \Gamma}  }  \vdash_{tm}  \source {  \lambda  {\source x}  \ottsym{.}  {\source e}  } \Leftarrow \source {  {\source \tau }_{{\mathrm{1}}}  \rightarrow  {\source \tau }_{{\mathrm{2}}}  } \stot{ \rightsquigarrow  \target {  \lambda  {\target x}  \ottsym{:}  {\target \sigma }  \ottsym{.}  {\target e }  } } 
  }
  {
    The goal to be proven is the following:
    \begin{gather}
       \source {  { \source P }  } ; \source {  {\source { \Gamma_C } }  } ; \source {  {\source \Gamma}  }  \vdash_{tm}^{\intermed {M} }  \source {  \lambda  {\source x}  \ottsym{.}  {\source e}  } \Leftarrow \source {  {\source \tau }_{{\mathrm{1}}}  \rightarrow  {\source \tau }_{{\mathrm{2}}}  } \stoi{ \rightsquigarrow  \intermed {  {\intermed e}  } } 
      \label{eqg:eq_expr11} \\
       \intermed {  {\intermed \Sigma}  } ; \intermed {  {\intermed { \Gamma_C } }  } ; \intermed {  {\intermed \Gamma}  }  \vdash_{tm}  \intermed {  {\intermed e}  } : \intermed {  {\intermed \sigma}_{{\mathrm{0}}}  } \itot{ \rightsquigarrow  \target {  \lambda  {\target x}  \ottsym{:}  {\target \sigma }  \ottsym{.}  {\target e }  } } 
      \label{eqg:eq_expr12} \\
       \source {  {\source { \Gamma_C } }  } ; \source {  {\source \Gamma}  }  \vdash_{ty}^{\intermed {M} }  \source {  {\source \tau }_{{\mathrm{1}}}  \rightarrow  {\source \tau }_{{\mathrm{2}}}  }\,{ \stoi{ \rightsquigarrow  \intermed {  {\intermed \sigma}_{{\mathrm{0}}}  } } } 
      \label{eqg:eq_expr12b}
    \end{gather}
    From the rule premise we know that:
    \begin{gather}
       \source {  {\source x}  }  \notin   \textbf {dom}  ( \source {  {\source \Gamma}  } ) 
      \label{eq:eq_expr48} \\
       \source {  { \source P }  } ; \source {  {\source { \Gamma_C } }  } ; \source {  {\source \Gamma}  \ottsym{,}  {\source x}  \ottsym{:}  {\source \tau }_{{\mathrm{1}}}  }  \vdash_{tm}  \source {  {\source e}  } \Leftarrow \source {  {\source \tau }_{{\mathrm{2}}}  } \stot{ \rightsquigarrow  \target {  {\target e }  } } 
      \label{eq:eq_expr49} \\
       \source {  {\source { \Gamma_C } }  } ; \source {  {\source \Gamma}  }  \vdash_{ty}  \source {  {\source \tau }_{{\mathrm{1}}}  }\,{ \stoi{ \rightsquigarrow  \target {  {\target \sigma }  } } } 
      \label{eq:eq_expr50}
    \end{gather}
    By applying type equivalence (Theorem~\ref{thmA:equiv_ty}) to
    Equation~\ref{eq:eq_expr50}, we get:
    \begin{gather}
       \source {  {\source { \Gamma_C } }  } ; \source {  {\source \Gamma}  }  \vdash_{ty}^{\intermed {M} }  \source {  {\source \tau }_{{\mathrm{1}}}  }\,{ \stoi{ \rightsquigarrow  \intermed {  {\intermed \sigma}  } } } 
      \label{eq:eq_expr52} \\
       \intermed {  {\intermed { \Gamma_C } }  } ; \intermed {  {\intermed \Gamma}  }  \vdash_{ty}  \intermed {  {\intermed \sigma}  } \itot{ \rightsquigarrow  \target {  {\target \sigma }  } } 
      \label{eq:eq_expr53}
    \end{gather}
    From \textsc{sCtxT-tyEnvTm}, together with the 2\textsuperscript{nd}
    hypothesis and Equation~\ref{eq:eq_expr50}, we know that:
    \begin{equation}
       \vdash_{ctx}  \source {  { \source P }  } ; \source {  {\source { \Gamma_C } }  } ; \source {  {\source \Gamma}  \ottsym{,}  {\source x}  \ottsym{:}  {\source \tau }_{{\mathrm{1}}}  }  \rightsquigarrow  \target {  {\target \Gamma }  \ottsym{,}  {\target x}  \ottsym{:}  {\target \sigma }  } 
      \label{eq:eq_expr54}
    \end{equation}
    By applying the induction hypothesis on Equation~\ref{eq:eq_expr49},
    together with Equation~\ref{eq:eq_expr54}, we get:
    \begin{gather}
       \source {  { \source P }  } ; \source {  {\source { \Gamma_C } }  } ; \source {  {\source \Gamma}  \ottsym{,}  {\source x}  \ottsym{:}  {\source \tau }_{{\mathrm{1}}}  }  \vdash_{tm}^{\intermed {M} }  \source {  {\source e}  } \Leftarrow \source {  {\source \tau }_{{\mathrm{2}}}  } \stoi{ \rightsquigarrow  \intermed {  {\intermed e}'  } } 
      \label{eq:eq_expr55} \\
       \intermed {  {\intermed \Sigma}  } ; \intermed {  {\intermed { \Gamma_C } }  } ; \intermed {  {\intermed \Gamma}  \ottsym{,}  {\intermed x}  \ottsym{:}  {\intermed \sigma}  }  \vdash_{tm}  \intermed {  {\intermed e}'  } : \intermed {  {\intermed \sigma}_{{\mathrm{2}}}  } \itot{ \rightsquigarrow  \target {  {\target e }  } } 
      \label{eq:eq_expr56} \\
       \source {  {\source { \Gamma_C } }  } ; \source {  {\source \Gamma}  \ottsym{,}  {\source x}  \ottsym{:}  {\source \tau }_{{\mathrm{1}}}  }  \vdash_{ty}^{\intermed {M} }  \source {  {\source \tau }_{{\mathrm{2}}}  }\,{ \stoi{ \rightsquigarrow  \intermed {  {\intermed \sigma}_{{\mathrm{2}}}  } } } 
      \label{eq:eq_expr56b} 
    \end{gather}
    Goal~\ref{eqg:eq_expr11} follows from \textsc{sTm-check-ArrI}, together with
    Equations~\ref{eq:eq_expr48}, \ref{eq:eq_expr55} and~\ref{eq:eq_expr52},
    with \( \intermed {  {\intermed e}  } = \intermed {  \lambda  {\intermed x}  \ottsym{:}  {\intermed \sigma}  \ottsym{.}  {\intermed e}'  } \).
    Consequently, Goal~\ref{eqg:eq_expr12} follows from \textsc{iTm-arrI},
    together with Equations~\ref{eq:eq_expr56} and~\ref{eq:eq_expr53}, with
    \( \intermed {  {\intermed \sigma}_{{\mathrm{0}}}  } = \intermed {  {\intermed \sigma}  \rightarrow  {\intermed \sigma}_{{\mathrm{2}}}  } \).
    Goal~\ref{eqg:eq_expr12b} follows
    by applying Lemma~\ref{lemma:type_well_source_typing} to Equation~\ref{eqg:eq_expr11}. \\
  }
  \proofStep{\textsc{sTm-checkT-Inf}}
  {
     \source {  { \source P }  } ; \source {  {\source { \Gamma_C } }  } ; \source {  {\source \Gamma}  }  \vdash_{tm}  \source {  {\source e}  } \Leftarrow \source {  {\source \tau }  } \stot{ \rightsquigarrow  \target {  {\target e }  } } 
  }
  {
    The goal to be proven is the following:
    \begin{gather}
       \source {  { \source P }  } ; \source {  {\source { \Gamma_C } }  } ; \source {  {\source \Gamma}  }  \vdash_{tm}^{\intermed {M} }  \source {  {\source e}  } \Leftarrow \source {  {\source \tau }  } \stoi{ \rightsquigarrow  \intermed {  {\intermed e}  } } 
      \label{eqg:eq_expr13} \\
       \intermed {  {\intermed \Sigma}  } ; \intermed {  {\intermed { \Gamma_C } }  } ; \intermed {  {\intermed \Gamma}  }  \vdash_{tm}  \intermed {  {\intermed e}  } : \intermed {  {\intermed \sigma}  } \itot{ \rightsquigarrow  \target {  {\target e }  } } 
      \label{eqg:eq_expr14} \\
       \source {  {\source { \Gamma_C } }  } ; \source {  {\source \Gamma}  }  \vdash_{ty}^{\intermed {M} }  \source {  {\source \tau }  }\,{ \stoi{ \rightsquigarrow  \intermed {  {\intermed \sigma}  } } } 
      \label{eqg:eq_expr14b}
    \end{gather}
    From the rule premise we know that:
    \begin{gather}
       \source {  { \source P }  } ; \source {  {\source { \Gamma_C } }  } ; \source {  {\source \Gamma}  }  \vdash_{tm}  \source {  {\source e}  } \Rightarrow \source {  {\source \tau }  } \stot{ \rightsquigarrow  \target {  {\target e }  } } 
      \label{eq:eq_expr51}
    \end{gather}
    Goals~\ref{eqg:eq_expr13}, \ref{eqg:eq_expr14} and~\ref{eqg:eq_expr14b} follow directly by
    applying Part 1 of this theorem to Equation~\ref{eq:eq_expr51}. 
  }
\end{description}
\qed

\begin{theoremB}[Equivalence - Contexts]\leavevmode
  \begin{itemize}
  \item
    If \( \source {  {\source M}  } : ( \source {  { \source P }  } ; \source {  {\source { \Gamma_C } }  } ; \source {  {\source \Gamma}  }  \Rightarrow  \source {  {\source \tau }  } )  \mapsto  ( \source {  { \source P }  } ; \source {  {\source { \Gamma_C } }  } ; \source {  {\source \Gamma}'  }  \Rightarrow  \source {  {\source \tau }'  } )  \rightsquigarrow  \target {  {\target M }  } \) \\
    and \( \vdash_{ctx}  \source {  { \source P }  } ; \source {  {\source { \Gamma_C } }  } ; \source {  {\source \Gamma}  }  \rightsquigarrow  \target {  {\target \Gamma }  } \)
    and \( \vdash_{ctx}  \source {  { \source P }  } ; \source {  {\source { \Gamma_C } }  } ; \source {  {\source \Gamma}'  }  \rightsquigarrow  \target {  {\target \Gamma }'  } \) \\
    then \( \source {  {\source M}  } : ( \source {  { \source P }  } ; \source {  {\source { \Gamma_C } }  } ; \source {  {\source \Gamma}  }  \Rightarrow  \source {  {\source \tau }  } )  \mapsto  ( \source {  { \source P }  } ; \source {  {\source { \Gamma_C } }  } ; \source {  {\source \Gamma}'  }  \Rightarrow  \source {  {\source \tau }'  } )  \rightsquigarrow  \intermed {  {\intermed M}  } \) \\
    and \( \intermed {  {\intermed M}  } : ( \intermed {  {\intermed \Sigma}  } ; \intermed {  {\intermed { \Gamma_C } }  } ; \intermed {  {\intermed \Gamma}  }  \Rightarrow  \intermed {  {\intermed \sigma}  } )  \mapsto  ( \intermed {  {\intermed \Sigma}  } ; \intermed {  {\intermed { \Gamma_C } }  } ; \intermed {  {\intermed \Gamma}'  }  \Rightarrow  \intermed {  {\intermed \sigma}'  } ) \itot{ \rightsquigarrow  \target {  {\target M }  } } \) \\
    where \( \vdash_{ctx}^{\intermed {M} }  \source {  { \source P }  } ; \source {  {\source { \Gamma_C } }  } ; \source {  {\source \Gamma}  }  \rightsquigarrow  \intermed {  {\intermed \Sigma}  } ; \intermed {  {\intermed { \Gamma_C } }  } ; \intermed {  {\intermed \Gamma}  } \)
    and \( \vdash_{ctx}^{\intermed {M} }  \source {  { \source P }  } ; \source {  {\source { \Gamma_C } }  } ; \source {  {\source \Gamma}'  }  \rightsquigarrow  \intermed {  {\intermed \Sigma}  } ; \intermed {  {\intermed { \Gamma_C } }  } ; \intermed {  {\intermed \Gamma}'  } \) \\
    and \( \source {  {\source { \Gamma_C } }  } ; \source {  {\source \Gamma}  }  \vdash_{ty}^{\intermed {M} }  \source {  {\source \tau }  }\,{ \stoi{ \rightsquigarrow  \intermed {  {\intermed \sigma}  } } } \)
    and \( \source {  {\source { \Gamma_C } }  } ; \source {  {\source \Gamma}'  }  \vdash_{ty}^{\intermed {M} }  \source {  {\source \tau }'  }\,{ \stoi{ \rightsquigarrow  \intermed {  {\intermed \sigma}'  } } } \).
  \item
    If \( \source {  {\source M}  } : ( \source {  { \source P }  } ; \source {  {\source { \Gamma_C } }  } ; \source {  {\source \Gamma}  }  \Rightarrow  \source {  {\source \tau }  } )  \mapsto  ( \source {  { \source P }  } ; \source {  {\source { \Gamma_C } }  } ; \source {  {\source \Gamma}'  }  \Leftarrow  \source {  {\source \tau }'  } )  \rightsquigarrow  \target {  {\target M }  } \) \\
    and \( \vdash_{ctx}  \source {  { \source P }  } ; \source {  {\source { \Gamma_C } }  } ; \source {  {\source \Gamma}  }  \rightsquigarrow  \target {  {\target \Gamma }  } \)
    and \( \vdash_{ctx}  \source {  { \source P }  } ; \source {  {\source { \Gamma_C } }  } ; \source {  {\source \Gamma}'  }  \rightsquigarrow  \target {  {\target \Gamma }'  } \) \\
    then \( \source {  {\source M}  } : ( \source {  { \source P }  } ; \source {  {\source { \Gamma_C } }  } ; \source {  {\source \Gamma}  }  \Rightarrow  \source {  {\source \tau }  } )  \mapsto  ( \source {  { \source P }  } ; \source {  {\source { \Gamma_C } }  } ; \source {  {\source \Gamma}'  }  \Leftarrow  \source {  {\source \tau }'  } )  \rightsquigarrow  \intermed {  {\intermed M}  } \) \\
    and \( \intermed {  {\intermed M}  } : ( \intermed {  {\intermed \Sigma}  } ; \intermed {  {\intermed { \Gamma_C } }  } ; \intermed {  {\intermed \Gamma}  }  \Rightarrow  \intermed {  {\intermed \sigma}  } )  \mapsto  ( \intermed {  {\intermed \Sigma}  } ; \intermed {  {\intermed { \Gamma_C } }  } ; \intermed {  {\intermed \Gamma}'  }  \Rightarrow  \intermed {  {\intermed \sigma}'  } ) \itot{ \rightsquigarrow  \target {  {\target M }  } } \) \\
    where \( \vdash_{ctx}^{\intermed {M} }  \source {  { \source P }  } ; \source {  {\source { \Gamma_C } }  } ; \source {  {\source \Gamma}  }  \rightsquigarrow  \intermed {  {\intermed \Sigma}  } ; \intermed {  {\intermed { \Gamma_C } }  } ; \intermed {  {\intermed \Gamma}  } \)
    and \( \vdash_{ctx}^{\intermed {M} }  \source {  { \source P }  } ; \source {  {\source { \Gamma_C } }  } ; \source {  {\source \Gamma}'  }  \rightsquigarrow  \intermed {  {\intermed \Sigma}  } ; \intermed {  {\intermed { \Gamma_C } }  } ; \intermed {  {\intermed \Gamma}'  } \) \\
    and \( \source {  {\source { \Gamma_C } }  } ; \source {  {\source \Gamma}  }  \vdash_{ty}^{\intermed {M} }  \source {  {\source \tau }  }\,{ \stoi{ \rightsquigarrow  \intermed {  {\intermed \sigma}  } } } \)
    and \( \source {  {\source { \Gamma_C } }  } ; \source {  {\source \Gamma}'  }  \vdash_{ty}^{\intermed {M} }  \source {  {\source \tau }'  }\,{ \stoi{ \rightsquigarrow  \intermed {  {\intermed \sigma}'  } } } \).
  \item
    If \( \source {  {\source M}  } : ( \source {  { \source P }  } ; \source {  {\source { \Gamma_C } }  } ; \source {  {\source \Gamma}  }  \Leftarrow  \source {  {\source \tau }  } )  \mapsto  ( \source {  { \source P }  } ; \source {  {\source { \Gamma_C } }  } ; \source {  {\source \Gamma}'  }  \Rightarrow  \source {  {\source \tau }'  } )  \rightsquigarrow  \target {  {\target M }  } \) \\
    and \( \vdash_{ctx}  \source {  { \source P }  } ; \source {  {\source { \Gamma_C } }  } ; \source {  {\source \Gamma}  }  \rightsquigarrow  \target {  {\target \Gamma }  } \)
    and \( \vdash_{ctx}  \source {  { \source P }  } ; \source {  {\source { \Gamma_C } }  } ; \source {  {\source \Gamma}'  }  \rightsquigarrow  \target {  {\target \Gamma }'  } \) \\
    then \( \source {  {\source M}  } : ( \source {  { \source P }  } ; \source {  {\source { \Gamma_C } }  } ; \source {  {\source \Gamma}  }  \Leftarrow  \source {  {\source \tau }  } )  \mapsto  ( \source {  { \source P }  } ; \source {  {\source { \Gamma_C } }  } ; \source {  {\source \Gamma}'  }  \Rightarrow  \source {  {\source \tau }'  } )  \rightsquigarrow  \intermed {  {\intermed M}  } \) \\
    and \( \intermed {  {\intermed M}  } : ( \intermed {  {\intermed \Sigma}  } ; \intermed {  {\intermed { \Gamma_C } }  } ; \intermed {  {\intermed \Gamma}  }  \Rightarrow  \intermed {  {\intermed \sigma}  } )  \mapsto  ( \intermed {  {\intermed \Sigma}  } ; \intermed {  {\intermed { \Gamma_C } }  } ; \intermed {  {\intermed \Gamma}'  }  \Rightarrow  \intermed {  {\intermed \sigma}'  } ) \itot{ \rightsquigarrow  \target {  {\target M }  } } \) \\
    where \( \vdash_{ctx}^{\intermed {M} }  \source {  { \source P }  } ; \source {  {\source { \Gamma_C } }  } ; \source {  {\source \Gamma}  }  \rightsquigarrow  \intermed {  {\intermed \Sigma}  } ; \intermed {  {\intermed { \Gamma_C } }  } ; \intermed {  {\intermed \Gamma}  } \)
    and \( \vdash_{ctx}^{\intermed {M} }  \source {  { \source P }  } ; \source {  {\source { \Gamma_C } }  } ; \source {  {\source \Gamma}'  }  \rightsquigarrow  \intermed {  {\intermed \Sigma}  } ; \intermed {  {\intermed { \Gamma_C } }  } ; \intermed {  {\intermed \Gamma}'  } \) \\
    and \( \source {  {\source { \Gamma_C } }  } ; \source {  {\source \Gamma}  }  \vdash_{ty}^{\intermed {M} }  \source {  {\source \tau }  }\,{ \stoi{ \rightsquigarrow  \intermed {  {\intermed \sigma}  } } } \)
    and \( \source {  {\source { \Gamma_C } }  } ; \source {  {\source \Gamma}'  }  \vdash_{ty}^{\intermed {M} }  \source {  {\source \tau }'  }\,{ \stoi{ \rightsquigarrow  \intermed {  {\intermed \sigma}'  } } } \).
  \item
    If \( \source {  {\source M}  } : ( \source {  { \source P }  } ; \source {  {\source { \Gamma_C } }  } ; \source {  {\source \Gamma}  }  \Leftarrow  \source {  {\source \tau }  } )  \mapsto  ( \source {  { \source P }  } ; \source {  {\source { \Gamma_C } }  } ; \source {  {\source \Gamma}'  }  \Leftarrow  \source {  {\source \tau }'  } )  \rightsquigarrow  \target {  {\target M }  } \) \\
    and \( \vdash_{ctx}  \source {  { \source P }  } ; \source {  {\source { \Gamma_C } }  } ; \source {  {\source \Gamma}  }  \rightsquigarrow  \target {  {\target \Gamma }  } \)
    and \( \vdash_{ctx}  \source {  { \source P }  } ; \source {  {\source { \Gamma_C } }  } ; \source {  {\source \Gamma}'  }  \rightsquigarrow  \target {  {\target \Gamma }'  } \) \\
    then \( \source {  {\source M}  } : ( \source {  { \source P }  } ; \source {  {\source { \Gamma_C } }  } ; \source {  {\source \Gamma}  }  \Leftarrow  \source {  {\source \tau }  } )  \mapsto  ( \source {  { \source P }  } ; \source {  {\source { \Gamma_C } }  } ; \source {  {\source \Gamma}'  }  \Leftarrow  \source {  {\source \tau }'  } )  \rightsquigarrow  \intermed {  {\intermed M}  } \) \\
    and \( \intermed {  {\intermed M}  } : ( \intermed {  {\intermed \Sigma}  } ; \intermed {  {\intermed { \Gamma_C } }  } ; \intermed {  {\intermed \Gamma}  }  \Rightarrow  \intermed {  {\intermed \sigma}  } )  \mapsto  ( \intermed {  {\intermed \Sigma}  } ; \intermed {  {\intermed { \Gamma_C } }  } ; \intermed {  {\intermed \Gamma}'  }  \Rightarrow  \intermed {  {\intermed \sigma}'  } ) \itot{ \rightsquigarrow  \target {  {\target M }  } } \) \\
    where \( \vdash_{ctx}^{\intermed {M} }  \source {  { \source P }  } ; \source {  {\source { \Gamma_C } }  } ; \source {  {\source \Gamma}  }  \rightsquigarrow  \intermed {  {\intermed \Sigma}  } ; \intermed {  {\intermed { \Gamma_C } }  } ; \intermed {  {\intermed \Gamma}  } \)
    and \( \vdash_{ctx}^{\intermed {M} }  \source {  { \source P }  } ; \source {  {\source { \Gamma_C } }  } ; \source {  {\source \Gamma}'  }  \rightsquigarrow  \intermed {  {\intermed \Sigma}  } ; \intermed {  {\intermed { \Gamma_C } }  } ; \intermed {  {\intermed \Gamma}'  } \) \\
    and \( \source {  {\source { \Gamma_C } }  } ; \source {  {\source \Gamma}  }  \vdash_{ty}^{\intermed {M} }  \source {  {\source \tau }  }\,{ \stoi{ \rightsquigarrow  \intermed {  {\intermed \sigma}  } } } \)
    and \( \source {  {\source { \Gamma_C } }  } ; \source {  {\source \Gamma}'  }  \vdash_{ty}^{\intermed {M} }  \source {  {\source \tau }'  }\,{ \stoi{ \rightsquigarrow  \intermed {  {\intermed \sigma}'  } } } \).
  \end{itemize}
  \label{thmA:equiv_ctx}
\end{theoremB}

\proof
By straightforward induction on the typing derivation. \\
\qed

\newpage
\section{Coherence Theorems}
\label{app:coh}
\renewcommand\itot[1]{}

\subsection{Compatibility Lemmas}

\begin{lemmaB}[Compatibility - Term Abstraction]\ \\
  \begin{mathpar}
    \ottaltinferrule{}{}
    {
       \intermed {  {\intermed { \Gamma_C } }  } ; \intermed {  {\intermed \Gamma}  \ottsym{,}  {\intermed x}  \ottsym{:}  {\intermed \sigma}_{{\mathrm{1}}}  }  \vdash  \intermed {  {\intermed \Sigma}_{{\mathrm{1}}}  } : \intermed {  {\intermed e}_{{\mathrm{1}}}  }  \simeq_{log}  \intermed {  {\intermed \Sigma}_{{\mathrm{2}}}  } : \intermed {  {\intermed e}_{{\mathrm{2}}}  } : \intermed {  {\intermed \sigma}_{{\mathrm{2}}}  } 
    }
    {  \intermed {  {\intermed { \Gamma_C } }  } ; \intermed {  {\intermed \Gamma}  }  \vdash  \intermed {  {\intermed \Sigma}_{{\mathrm{1}}}  } : \intermed {  \lambda  {\intermed x}  \ottsym{:}  {\intermed \sigma}_{{\mathrm{1}}}  \ottsym{.}  {\intermed e}_{{\mathrm{1}}}  }  \simeq_{log}  \intermed {  {\intermed \Sigma}_{{\mathrm{2}}}  } : \intermed {  \lambda  {\intermed x}  \ottsym{:}  {\intermed \sigma}_{{\mathrm{1}}}  \ottsym{.}  {\intermed e}_{{\mathrm{2}}}  } : \intermed {  {\intermed \sigma}_{{\mathrm{1}}}  \rightarrow  {\intermed \sigma}_{{\mathrm{2}}}  }  }
    \label{lemma:compat_abs}
  \end{mathpar}
\end{lemmaB}

\proof
By the definition of logical equivalence, suppose we have:
\begin{gather}
  \label{eq:compat_absR}
   \intermed {  {\intermed R }  } \in \mathcal{F} \llbracket \intermed {  {\intermed \Gamma}  } \rrbracket ^{\intermed {  {\intermed { \Gamma_C } }  } } \\
  \nonumber
   \intermed {  {\intermed \phi }  } \in \mathcal{G} \llbracket \intermed {  {\intermed \Gamma}  } \rrbracket ^{\intermed {  {\intermed \Sigma}_{{\mathrm{1}}}  } , \intermed {  {\intermed \Sigma}_{{\mathrm{2}}}  } , \intermed {  {\intermed { \Gamma_C } }  } }_{\intermed {  {\intermed R }  } } \\
  \label{eq:compat_absH}
   \intermed {  {\intermed \gamma }  } \in \mathcal{H} \llbracket \intermed {  {\intermed \Gamma}  } \rrbracket ^{\intermed {  {\intermed \Sigma}_{{\mathrm{1}}}  } , \intermed {  {\intermed \Sigma}_{{\mathrm{2}}}  } , \intermed {  {\intermed { \Gamma_C } }  } }_{\intermed {  {\intermed R }  } } 
\end{gather}
The goal to be proven is the following:
\begin{equation*}
   ( \intermed {  {\intermed \Sigma}_{{\mathrm{1}}}  } : \intermed {   {\intermed \gamma }_{{\mathrm{1}}}  ( {\intermed \phi }_{{\mathrm{1}}}  ( {\intermed R }  ( \lambda  {\intermed x}  \ottsym{:}  {\intermed \sigma}_{{\mathrm{1}}}  \ottsym{.}  {\intermed e}_{{\mathrm{1}}} )))   } , \intermed {  {\intermed \Sigma}_{{\mathrm{2}}}  } : \intermed {   {\intermed \gamma }_{{\mathrm{2}}}  ( {\intermed \phi }_{{\mathrm{2}}}  ( {\intermed R }  (  \lambda  {\intermed x}  \ottsym{:}  {\intermed \sigma}_{{\mathrm{1}}}  \ottsym{.}  {\intermed e}_{{\mathrm{2}}}  )))   } ) \in \mathcal{E} \llbracket \intermed {  {\intermed \sigma}_{{\mathrm{1}}}  \rightarrow  {\intermed \sigma}_{{\mathrm{2}}}  } \rrbracket ^{\intermed {  {\intermed { \Gamma_C } }  } }_{\intermed {  {\intermed R }  } } 
\end{equation*}
By the definition of the \( \mathcal{E} \) relation and the fact that term abstractions
are values, the goal reduces to:
\begin{gather*}
   \intermed {  {\intermed \Sigma}_{{\mathrm{1}}}  } ; \intermed {  {\intermed { \Gamma_C } }  } ; \intermed {  \bullet  }  \vdash_{tm}  \intermed {   {\intermed \gamma }_{{\mathrm{1}}}  ( {\intermed \phi }_{{\mathrm{1}}}  ( {\intermed R }  ( \lambda  {\intermed x}  \ottsym{:}  {\intermed \sigma}_{{\mathrm{1}}}  \ottsym{.}  {\intermed e}_{{\mathrm{1}}} )))   } : \intermed {  {\intermed R }  \ottsym{(}  {\intermed \sigma}_{{\mathrm{1}}}  \rightarrow  {\intermed \sigma}_{{\mathrm{2}}}  \ottsym{)}  } \itot{ \rightsquigarrow  \target {  {\target e }  } } \\
   \intermed {  {\intermed \Sigma}_{{\mathrm{2}}}  } ; \intermed {  {\intermed { \Gamma_C } }  } ; \intermed {  \bullet  }  \vdash_{tm}  \intermed {   {\intermed \gamma }_{{\mathrm{2}}}  ( {\intermed \phi }_{{\mathrm{2}}}  ( {\intermed R }  ( \lambda  {\intermed x}  \ottsym{:}  {\intermed \sigma}_{{\mathrm{1}}}  \ottsym{.}  {\intermed e}_{{\mathrm{2}}} )))   } : \intermed {  {\intermed R }  \ottsym{(}  {\intermed \sigma}_{{\mathrm{1}}}  \rightarrow  {\intermed \sigma}_{{\mathrm{2}}}  \ottsym{)}  } \itot{ \rightsquigarrow  \target {  {\target e }'  } } \\
   ( \intermed {  {\intermed \Sigma}_{{\mathrm{1}}}  } : \intermed {   {\intermed \gamma }_{{\mathrm{1}}}  ( {\intermed \phi }_{{\mathrm{1}}}  ( {\intermed R }  ( \lambda  {\intermed x}  \ottsym{:}  {\intermed \sigma}_{{\mathrm{1}}}  \ottsym{.}  {\intermed e}_{{\mathrm{1}}} )))   } , \intermed {  {\intermed \Sigma}_{{\mathrm{2}}}  } : \intermed {   {\intermed \gamma }_{{\mathrm{2}}}  ( {\intermed \phi }_{{\mathrm{2}}}  ( {\intermed R }  ( \lambda  {\intermed x}  \ottsym{:}  {\intermed \sigma}_{{\mathrm{1}}}  \ottsym{.}  {\intermed e}_{{\mathrm{2}}} )))   } ) \in \mathcal{V} \llbracket \intermed {  {\intermed \sigma}_{{\mathrm{1}}}  \rightarrow  {\intermed \sigma}_{{\mathrm{2}}}  } \rrbracket ^{\intermed {  {\intermed { \Gamma_C } }  } }_{\intermed {  {\intermed R }  } } 
\end{gather*}
By applying the substitutions, the goal simplifies to:
\begin{gather}
  \label{eq:compat_abs1}
   \intermed {  {\intermed \Sigma}_{{\mathrm{1}}}  } ; \intermed {  {\intermed { \Gamma_C } }  } ; \intermed {  \bullet  }  \vdash_{tm}  \intermed {  \lambda  {\intermed x}  \ottsym{:}  {\intermed R }  \ottsym{(}  {\intermed \sigma}_{{\mathrm{1}}}  \ottsym{)}  \ottsym{.}  \ottsym{(}   {\intermed \gamma }_{{\mathrm{1}}}  ( {\intermed \phi }_{{\mathrm{1}}}  ( {\intermed R }  ( {\intermed e}_{{\mathrm{1}}} )))   \ottsym{)}  } : \intermed {  {\intermed R }  \ottsym{(}  {\intermed \sigma}_{{\mathrm{1}}}  \rightarrow  {\intermed \sigma}_{{\mathrm{2}}}  \ottsym{)}  } \itot{ \rightsquigarrow  \target {  {\target e }  } } \\
  \label{eq:compat_abs2}
   \intermed {  {\intermed \Sigma}_{{\mathrm{2}}}  } ; \intermed {  {\intermed { \Gamma_C } }  } ; \intermed {  \bullet  }  \vdash_{tm}  \intermed {  \lambda  {\intermed x}  \ottsym{:}  {\intermed R }  \ottsym{(}  {\intermed \sigma}_{{\mathrm{1}}}  \ottsym{)}  \ottsym{.}  \ottsym{(}   {\intermed \gamma }_{{\mathrm{2}}}  ( {\intermed \phi }_{{\mathrm{2}}}  ( {\intermed R }  ( {\intermed e}_{{\mathrm{2}}} )))   \ottsym{)}  } : \intermed {  {\intermed R }  \ottsym{(}  {\intermed \sigma}_{{\mathrm{1}}}  \rightarrow  {\intermed \sigma}_{{\mathrm{2}}}  \ottsym{)}  } \itot{ \rightsquigarrow  \target {  {\target e }'  } } \\
  \label{eq:compat_abs3}
   ( \intermed {  {\intermed \Sigma}_{{\mathrm{1}}}  } : \intermed {  \lambda  {\intermed x}  \ottsym{:}  {\intermed R }  \ottsym{(}  {\intermed \sigma}_{{\mathrm{1}}}  \ottsym{)}  \ottsym{.}  \ottsym{(}   {\intermed \gamma }_{{\mathrm{1}}}  ( {\intermed \phi }_{{\mathrm{1}}}  ( {\intermed R }  ( {\intermed e}_{{\mathrm{1}}} )))   \ottsym{)}  } , \intermed {  {\intermed \Sigma}_{{\mathrm{2}}}  } : \intermed {  \lambda  {\intermed x}  \ottsym{:}  {\intermed R }  \ottsym{(}  {\intermed \sigma}_{{\mathrm{1}}}  \ottsym{)}  \ottsym{.}  \ottsym{(}   {\intermed \gamma }_{{\mathrm{2}}}  ( {\intermed \phi }_{{\mathrm{2}}}  ( {\intermed R }  ( {\intermed e}_{{\mathrm{2}}} )))   \ottsym{)}  } ) \in \mathcal{V} \llbracket \intermed {  {\intermed \sigma}_{{\mathrm{1}}}  \rightarrow  {\intermed \sigma}_{{\mathrm{2}}}  } \rrbracket ^{\intermed {  {\intermed { \Gamma_C } }  } }_{\intermed {  {\intermed R }  } } 
\end{gather}
By unfolding the definition of logical equivalence in the hypothesis of the theorem, we get:
\begin{equation}
   ( \intermed {  {\intermed \Sigma}_{{\mathrm{1}}}  } : \intermed {   {\intermed \gamma }'_{{\mathrm{1}}}  ( {\intermed \phi }'_{{\mathrm{1}}}  ( {\intermed R }'  ( {\intermed e}_{{\mathrm{1}}} )))   } , \intermed {  {\intermed \Sigma}_{{\mathrm{2}}}  } : \intermed {   {\intermed \gamma }'_{{\mathrm{2}}}  ( {\intermed \phi }'_{{\mathrm{2}}}  ( {\intermed R }'  ( {\intermed e}_{{\mathrm{2}}} )))   } ) \in \mathcal{E} \llbracket \intermed {  {\intermed \sigma}_{{\mathrm{2}}}  } \rrbracket ^{\intermed {  {\intermed { \Gamma_C } }  } }_{\intermed {  {\intermed R }'  } } 
  \label{eq:compat_abs4}
\end{equation}
for any \( \intermed {  {\intermed R }'  } \in \mathcal{F} \llbracket \intermed {  {\intermed \Gamma}  \ottsym{,}  {\intermed x}  \ottsym{:}  {\intermed \sigma}_{{\mathrm{1}}}  } \rrbracket ^{\intermed {  {\intermed { \Gamma_C } }  } } \),
\( \intermed {  {\intermed \phi }'  } \in \mathcal{G} \llbracket \intermed {  {\intermed \Gamma}  \ottsym{,}  {\intermed x}  \ottsym{:}  {\intermed \sigma}_{{\mathrm{1}}}  } \rrbracket ^{\intermed {  {\intermed \Sigma}_{{\mathrm{1}}}  } , \intermed {  {\intermed \Sigma}_{{\mathrm{2}}}  } , \intermed {  {\intermed { \Gamma_C } }  } }_{\intermed {  {\intermed R }'  } } \) and
\( \intermed {  {\intermed \gamma }'  } \in \mathcal{H} \llbracket \intermed {  {\intermed \Gamma}  \ottsym{,}  {\intermed x}  \ottsym{:}  {\intermed \sigma}_{{\mathrm{1}}}  } \rrbracket ^{\intermed {  {\intermed \Sigma}_{{\mathrm{1}}}  } , \intermed {  {\intermed \Sigma}_{{\mathrm{2}}}  } , \intermed {  {\intermed { \Gamma_C } }  } }_{\intermed {  {\intermed R }'  } } \).

By the definition of the \( \mathcal{F} \)-relation and from Equation~\ref{eq:compat_absR},
we have that \( \intermed {  {\intermed R }  } \in \mathcal{F} \llbracket \intermed {  {\intermed \Gamma}  \ottsym{,}  {\intermed x}  \ottsym{:}  {\intermed \sigma}_{{\mathrm{1}}}  } \rrbracket ^{\intermed {  {\intermed { \Gamma_C } }  } } \) and we choose \( \intermed {  {\intermed R }'  } = \intermed {  {\intermed R }  } \).
By case analysis on \(\intermed{{\intermed \phi }'}\), we know that
\( \intermed {  {\intermed \phi }'  } = \intermed {   {\intermed \phi }'' ,  {\intermed x}  \mapsto ( {\intermed e}_{{\mathrm{3}}} ,  {\intermed e}_{{\mathrm{4}}}  )   } \) for some
\( \intermed {  {\intermed \phi }''  } \in \mathcal{G} \llbracket \intermed {  {\intermed \Gamma}  } \rrbracket ^{\intermed {  {\intermed \Sigma}_{{\mathrm{1}}}  } , \intermed {  {\intermed \Sigma}_{{\mathrm{2}}}  } , \intermed {  {\intermed { \Gamma_C } }  } }_{\intermed {  {\intermed R }  } } \)
and expressions \(\intermed{{\intermed e}_{{\mathrm{3}}}}\) and \(\intermed{{\intermed e}_{{\mathrm{4}}}}\) such that
\begin{equation}
  \label{eq:compat_abs_ie34}
   ( \intermed {  {\intermed \Sigma}_{{\mathrm{1}}}  } : \intermed {  {\intermed e}_{{\mathrm{3}}}  } , \intermed {  {\intermed \Sigma}_{{\mathrm{2}}}  } : \intermed {  {\intermed e}_{{\mathrm{4}}}  } ) \in \mathcal{E} \llbracket \intermed {  {\intermed \sigma}_{{\mathrm{1}}}  } \rrbracket ^{\intermed {  {\intermed { \Gamma_C } }  } }_{\intermed {  {\intermed R }  } } 
\end{equation}
We choose \( \intermed {  {\intermed \phi }''  } = \intermed {  {\intermed \phi }  } \).
Lastly, by the definition of the \( \mathcal{H} \)-relation and from
Equation~\ref{eq:compat_absH}, we have that
\( \intermed {  {\intermed \gamma }  } \in \mathcal{H} \llbracket \intermed {  {\intermed \Gamma}  \ottsym{,}  {\intermed x}  \ottsym{:}  {\intermed \sigma}_{{\mathrm{1}}}  } \rrbracket ^{\intermed {  {\intermed \Sigma}_{{\mathrm{1}}}  } , \intermed {  {\intermed \Sigma}_{{\mathrm{2}}}  } , \intermed {  {\intermed { \Gamma_C } }  } }_{\intermed {  {\intermed R }  } } \) and we choose
\( \intermed {  {\intermed \gamma }'  } = \intermed {  {\intermed \gamma }  } \).

With the above mentioned choices for \(\intermed{{\intermed \gamma }'}\), \(\intermed{{\intermed \phi }'}\) and
\(\intermed{{\intermed R }'}\), unfolding the definition of the \( \mathcal{E} \)-relation in
Equation~\ref{eq:compat_abs4}, gives us:
\begin{gather}
  \label{eq:compat_abs5}
         \intermed {  {\intermed \Sigma}_{{\mathrm{1}}}  } ; \intermed {  {\intermed { \Gamma_C } }  } ; \intermed {  \bullet  }  \vdash_{tm}  \intermed {   {\intermed \gamma }_{{\mathrm{1}}}  ( \ottsym{(}   {\intermed \phi }_{{\mathrm{1}}} ,  {\intermed x}  \mapsto  {\intermed e}_{{\mathrm{3}}}   \ottsym{)}  ( {\intermed R }  ( {\intermed e}_{{\mathrm{1}}} )))   } : \intermed {  {\intermed R }  \ottsym{(}  {\intermed \sigma}_{{\mathrm{2}}}  \ottsym{)}  } \itot{ \rightsquigarrow  \target {  {\target e }_{{\mathrm{1}}}  } } \\
  \label{eq:compat_abs6}
         \intermed {  {\intermed \Sigma}_{{\mathrm{2}}}  } ; \intermed {  {\intermed { \Gamma_C } }  } ; \intermed {  \bullet  }  \vdash_{tm}  \intermed {   {\intermed \gamma }_{{\mathrm{2}}}  ( \ottsym{(}   {\intermed \phi }_{{\mathrm{2}}} ,  {\intermed x}  \mapsto  {\intermed e}_{{\mathrm{4}}}   \ottsym{)}  ( {\intermed R }  ( {\intermed e}_{{\mathrm{2}}} )))   } : \intermed {  {\intermed R }  \ottsym{(}  {\intermed \sigma}_{{\mathrm{2}}}  \ottsym{)}  } \itot{ \rightsquigarrow  \target {  {\target e }_{{\mathrm{2}}}  } } \\
        \label{eq:compat_abs7a}
\begin{split}    
  \exists \intermed{{\intermed v}_{{\mathrm{3}}}}, \intermed{{\intermed v}_{{\mathrm{4}}}} & :
     \intermed {  {\intermed \Sigma}_{{\mathrm{1}}}  }  \vdash  \intermed {   {\intermed \gamma }_{{\mathrm{1}}}  ( \ottsym{(}   {\intermed \phi }_{{\mathrm{1}}} ,  {\intermed x}  \mapsto  {\intermed e}_{{\mathrm{3}}}   \ottsym{)}  ( {\intermed R }  ( {\intermed e}_{{\mathrm{1}}} )))   }  \longrightarrow^{*}  \intermed {  {\intermed v}_{{\mathrm{3}}}  } \\
    & \band  \intermed {  {\intermed \Sigma}_{{\mathrm{2}}}  }  \vdash  \intermed {   {\intermed \gamma }_{{\mathrm{2}}}  ( \ottsym{(}   {\intermed \phi }_{{\mathrm{2}}} ,  {\intermed x}  \mapsto  {\intermed e}_{{\mathrm{4}}}   \ottsym{)}  ( {\intermed R }  ( {\intermed e}_{{\mathrm{2}}} )))   }  \longrightarrow^{*}  \intermed {  {\intermed v}_{{\mathrm{4}}}  } \\
      & \band  ( \intermed {  {\intermed \Sigma}_{{\mathrm{1}}}  } : \intermed {  {\intermed v}_{{\mathrm{3}}}  } , \intermed {  {\intermed \Sigma}_{{\mathrm{2}}}  } : \intermed {  {\intermed v}_{{\mathrm{4}}}  } ) \in \mathcal{V} \llbracket \intermed {  {\intermed \sigma}_{{\mathrm{2}}}  } \rrbracket ^{\intermed {  {\intermed { \Gamma_C } }  } }_{\intermed {  {\intermed R }  } } 
\end{split}
\end{gather}
By unfolding the definition of the \( \mathcal{E} \)-relation in
Equation~\ref{eq:compat_abs_ie34},
we know that:
\begin{gather}
   \intermed {  {\intermed \Sigma}_{{\mathrm{1}}}  } ; \intermed {  {\intermed { \Gamma_C } }  } ; \intermed {  \bullet  }  \vdash_{tm}  \intermed {  {\intermed e}_{{\mathrm{3}}}  } : \intermed {  {\intermed \sigma}_{{\mathrm{1}}}  } \itot{ \rightsquigarrow  \target {  {\target e }_{{\mathrm{3}}}  } } 
  \label{eq:compat_abs5c} \\
   \intermed {  {\intermed \Sigma}_{{\mathrm{2}}}  } ; \intermed {  {\intermed { \Gamma_C } }  } ; \intermed {  \bullet  }  \vdash_{tm}  \intermed {  {\intermed e}_{{\mathrm{4}}}  } : \intermed {  {\intermed \sigma}_{{\mathrm{1}}}  } \itot{ \rightsquigarrow  \target {  {\target e }_{{\mathrm{4}}}  } } 
  \label{eq:compat_abs6c}
\end{gather}
Note that neither \(\intermed{{\intermed e}_{{\mathrm{3}}}}\) nor \(\intermed{{\intermed e}_{{\mathrm{4}}}}\) contain any free
variables, thus \( \intermed {   {\intermed \gamma }_{{\mathrm{1}}}  ( {\intermed \phi }_{{\mathrm{1}}}  ( {\intermed R }  ( {\intermed e}_{{\mathrm{3}}} )))   } = \intermed {  {\intermed e}_{{\mathrm{3}}}  } \) and
\( \intermed {   {\intermed \gamma }_{{\mathrm{2}}}  ( {\intermed \phi }_{{\mathrm{2}}}  ( {\intermed R }  ( {\intermed e}_{{\mathrm{4}}} )))   } = \intermed {  {\intermed e}_{{\mathrm{4}}}  } \). Taking these equations into account, from the
definition of substitution, Equations~\ref{eq:compat_abs5} and \ref{eq:compat_abs6} are rewritten to:
\begin{gather}
  \label{eq:compat_abs5b}
   \intermed {  {\intermed \Sigma}_{{\mathrm{1}}}  } ; \intermed {  {\intermed { \Gamma_C } }  } ; \intermed {  \bullet  }  \vdash_{tm}  \intermed {  [  {\intermed e}_{{\mathrm{3}}}  /  {\intermed x}  ]  \ottsym{(}   {\intermed \gamma }_{{\mathrm{1}}}  ( {\intermed \phi }_{{\mathrm{1}}}  ( {\intermed R }  ( {\intermed e}_{{\mathrm{1}}} )))   \ottsym{)}  } : \intermed {  {\intermed R }  \ottsym{(}  {\intermed \sigma}_{{\mathrm{2}}}  \ottsym{)}  } \itot{ \rightsquigarrow  \target {  {\target e }_{{\mathrm{1}}}  } } \\
  \label{eq:compat_abs6b}
   \intermed {  {\intermed \Sigma}_{{\mathrm{2}}}  } ; \intermed {  {\intermed { \Gamma_C } }  } ; \intermed {  \bullet  }  \vdash_{tm}  \intermed {  [  {\intermed e}_{{\mathrm{4}}}  /  {\intermed x}  ]  \ottsym{(}   {\intermed \gamma }_{{\mathrm{2}}}  ( {\intermed \phi }_{{\mathrm{2}}}  ( {\intermed R }  ( {\intermed e}_{{\mathrm{2}}} )))   \ottsym{)}  } : \intermed {  {\intermed R }  \ottsym{(}  {\intermed \sigma}_{{\mathrm{2}}}  \ottsym{)}  } \itot{ \rightsquigarrow  \target {  {\target e }_{{\mathrm{2}}}  } } 
\end{gather}
By applying the substitution Lemma~\ref{lemma:rev_var_subst} on
Equations~\ref{eq:compat_abs5b} and \ref{eq:compat_abs6b} respectively, in
combination with Equations~\ref{eq:compat_abs5c} and \ref{eq:compat_abs6c}, we get:
\begin{gather}
  \label{eq:compat_abs13}
   \intermed {  {\intermed \Sigma}_{{\mathrm{1}}}  } ; \intermed {  {\intermed { \Gamma_C } }  } ; \intermed {  \bullet  \ottsym{,}  {\intermed x}  \ottsym{:}  {\intermed \sigma}_{{\mathrm{1}}}  }  \vdash_{tm}  \intermed {   {\intermed \gamma }_{{\mathrm{1}}}  ( {\intermed \phi }_{{\mathrm{1}}}  ( {\intermed R }  ( {\intermed e}_{{\mathrm{1}}} )))   } : \intermed {  {\intermed R }  \ottsym{(}  {\intermed \sigma}_{{\mathrm{2}}}  \ottsym{)}  } \itot{ \rightsquigarrow  \target {  {\target e }'_{{\mathrm{1}}}  } } \\
  \label{eq:compat_abs14}
   \intermed {  {\intermed \Sigma}_{{\mathrm{2}}}  } ; \intermed {  {\intermed { \Gamma_C } }  } ; \intermed {  \bullet  \ottsym{,}  {\intermed x}  \ottsym{:}  {\intermed \sigma}_{{\mathrm{1}}}  }  \vdash_{tm}  \intermed {   {\intermed \gamma }_{{\mathrm{2}}}  ( {\intermed \phi }_{{\mathrm{2}}}  ( {\intermed R }  ( {\intermed e}_{{\mathrm{2}}} )))   } : \intermed {  {\intermed R }  \ottsym{(}  {\intermed \sigma}_{{\mathrm{2}}}  \ottsym{)}  } \itot{ \rightsquigarrow  \target {  {\target e }'_{{\mathrm{2}}}  } } 
\end{gather}
By Lemma~\ref{lemma:ctx_well_inter_typing}, it follows from
Equation~\ref{eq:compat_abs13} that \( \vdash_{ctx}  \intermed {  {\intermed \Sigma}_{{\mathrm{1}}}  } ; \intermed {  {\intermed { \Gamma_C } }  } ; \intermed {  \bullet  \ottsym{,}  {\intermed x}  \ottsym{:}  {\intermed \sigma}_{{\mathrm{1}}}  } \).
By case analysis on this result, rule \textsc{iCtx-tyEnvTm} (the rule with which variable
\({\intermed x}\) wad inserted in the environment) tells us that:
\begin{equation}
  \label{eq:compat_abs13b}
   \intermed {  {\intermed { \Gamma_C } }  } ; \intermed {  \bullet  }  \vdash_{ty}  \intermed {  {\intermed \sigma}_{{\mathrm{1}}}  } \itot{ \rightsquigarrow  \target {  {\target \sigma }_{{\mathrm{1}}}  } } 
\end{equation}
Goals~\ref{eq:compat_abs1} and \ref{eq:compat_abs2} follow from applying the typing rule
\textsc{iTm-arrI} on Equations~\ref{eq:compat_abs13} and \ref{eq:compat_abs14} respectively,
together with Equation~\ref{eq:compat_abs13b}.

It remains to show Goal~\ref{eq:compat_abs3}. By unfolding the definition of the
\( \mathcal{V} \) relation, the goal simplifies to
\begin{align}
  \label{eq:compat_abs10}
  &\forall \intermed{{\intermed e}_{{\mathrm{5}}}}\,\intermed{{\intermed e}_{{\mathrm{6}}}},~\text{if}~
     ( \intermed {  {\intermed \Sigma}_{{\mathrm{1}}}  } : \intermed {  {\intermed e}_{{\mathrm{5}}}  } , \intermed {  {\intermed \Sigma}_{{\mathrm{2}}}  } : \intermed {  {\intermed e}_{{\mathrm{6}}}  } ) \in \mathcal{E} \llbracket \intermed {  {\intermed \sigma}_{{\mathrm{1}}}  } \rrbracket ^{\intermed {  {\intermed { \Gamma_C } }  } }_{\intermed {  {\intermed R }  } } ,\\
  \label{eq:compat_absglast}
  &\text{then}~ ( \intermed {  {\intermed \Sigma}_{{\mathrm{1}}}  } : \intermed {   \lambda  {\intermed x}  \ottsym{:}  {\intermed R }  \ottsym{(}  {\intermed \sigma}_{{\mathrm{1}}}  \ottsym{)}  \ottsym{.}  \ottsym{(}   {\intermed \gamma }_{{\mathrm{1}}}  ( {\intermed \phi }_{{\mathrm{1}}}  ( {\intermed R }  ( {\intermed e}_{{\mathrm{1}}} )))   \ottsym{)}  \, {\intermed e}_{{\mathrm{5}}}  } , \intermed {  {\intermed \Sigma}_{{\mathrm{2}}}  } : \intermed {   \lambda  {\intermed x}  \ottsym{:}  {\intermed R }  \ottsym{(}  {\intermed \sigma}_{{\mathrm{1}}}  \ottsym{)}  \ottsym{.}  \ottsym{(}   {\intermed \gamma }_{{\mathrm{2}}}  ( {\intermed \phi }_{{\mathrm{2}}}  ( {\intermed R }  ( {\intermed e}_{{\mathrm{2}}} )))   \ottsym{)}  \, {\intermed e}_{{\mathrm{6}}}  } ) \in \mathcal{E} \llbracket \intermed {  {\intermed \sigma}_{{\mathrm{2}}}  } \rrbracket ^{\intermed {  {\intermed { \Gamma_C } }  } }_{\intermed {  {\intermed R }  } } 
\end{align}
Then, suppose expressions \(\intermed{{\intermed e}_{{\mathrm{5}}}}\) and \(\intermed{{\intermed e}_{{\mathrm{6}}}}\), such that
Equation~\ref{eq:compat_abs10} holds. By unfolding the definition of the \(\mathcal{E}\)
relation in Equation~\ref{eq:compat_abs10}, we have
\begin{gather}
  \label{eq:compat_abs15}
   \intermed {  {\intermed \Sigma}_{{\mathrm{1}}}  } ; \intermed {  {\intermed { \Gamma_C } }  } ; \intermed {  \bullet  }  \vdash_{tm}  \intermed {  {\intermed e}_{{\mathrm{5}}}  } : \intermed {  {\intermed \sigma}_{{\mathrm{1}}}  } \itot{ \rightsquigarrow  \target {  {\target e }'_{{\mathrm{5}}}  } } \\
  \label{eq:compat_abs16}
   \intermed {  {\intermed \Sigma}_{{\mathrm{2}}}  } ; \intermed {  {\intermed { \Gamma_C } }  } ; \intermed {  \bullet  }  \vdash_{tm}  \intermed {  {\intermed e}_{{\mathrm{6}}}  } : \intermed {  {\intermed \sigma}_{{\mathrm{1}}}  } \itot{ \rightsquigarrow  \target {  {\target e }'_{{\mathrm{6}}}  } } 
\end{gather}
We also unfold the definition of the \( \mathcal{E} \) relation in
Goal~\ref{eq:compat_absglast},
to get:
\begin{gather}
  \label{eq:compat_abs8}
   \intermed {  {\intermed \Sigma}_{{\mathrm{1}}}  } ; \intermed {  {\intermed { \Gamma_C } }  } ; \intermed {  \bullet  }  \vdash_{tm}  \intermed {  \ottsym{(}  \lambda  {\intermed x}  \ottsym{:}  {\intermed R }  \ottsym{(}  {\intermed \sigma}_{{\mathrm{1}}}  \ottsym{)}  \ottsym{.}  \ottsym{(}   {\intermed \gamma }_{{\mathrm{1}}}  ( {\intermed \phi }_{{\mathrm{1}}}  ( {\intermed R }  ( {\intermed e}_{{\mathrm{1}}} )))   \ottsym{)}  \ottsym{)} \, {\intermed e}_{{\mathrm{5}}}  } : \intermed {  {\intermed \sigma}_{{\mathrm{2}}}  } \itot{ \rightsquigarrow  \target {  {\target e }_{{\mathrm{5}}}  } } \\
  \label{eq:compat_abs9}
   \intermed {  {\intermed \Sigma}_{{\mathrm{2}}}  } ; \intermed {  {\intermed { \Gamma_C } }  } ; \intermed {  \bullet  }  \vdash_{tm}  \intermed {  \ottsym{(}  \lambda  {\intermed x}  \ottsym{:}  {\intermed R }  \ottsym{(}  {\intermed \sigma}_{{\mathrm{1}}}  \ottsym{)}  \ottsym{.}  \ottsym{(}   {\intermed \gamma }_{{\mathrm{2}}}  ( {\intermed \phi }_{{\mathrm{2}}}  ( {\intermed R }  ( {\intermed e}_{{\mathrm{2}}} )))   \ottsym{)}  \ottsym{)} \, {\intermed e}_{{\mathrm{6}}}  } : \intermed {  {\intermed \sigma}_{{\mathrm{2}}}  } \itot{ \rightsquigarrow  \target {  {\target e }_{{\mathrm{6}}}  } } \\
\begin{split}
  \label{eq:compat_abs11}
  \exists \intermed{{\intermed v}_{{\mathrm{5}}}}, \intermed{{\intermed v}_{{\mathrm{6}}}} & :
     \intermed {  {\intermed \Sigma}_{{\mathrm{1}}}  }  \vdash  \intermed {  \ottsym{(}  \lambda  {\intermed x}  \ottsym{:}  {\intermed R }  \ottsym{(}  {\intermed \sigma}_{{\mathrm{1}}}  \ottsym{)}  \ottsym{.}  \ottsym{(}   {\intermed \gamma }_{{\mathrm{1}}}  ( {\intermed \phi }_{{\mathrm{1}}}  ( {\intermed R }  ( {\intermed e}_{{\mathrm{1}}} )))   \ottsym{)}  \ottsym{)} \, {\intermed e}_{{\mathrm{5}}}  }  \longrightarrow^{*}  \intermed {  {\intermed v}_{{\mathrm{5}}}  }  \\
    & \band  \intermed {  {\intermed \Sigma}_{{\mathrm{2}}}  }  \vdash  \intermed {  \ottsym{(}  \lambda  {\intermed x}  \ottsym{:}  {\intermed R }  \ottsym{(}  {\intermed \sigma}_{{\mathrm{1}}}  \ottsym{)}  \ottsym{.}  \ottsym{(}   {\intermed \gamma }_{{\mathrm{2}}}  ( {\intermed \phi }_{{\mathrm{2}}}  ( {\intermed R }  ( {\intermed e}_{{\mathrm{2}}} )))   \ottsym{)}  \ottsym{)} \, {\intermed e}_{{\mathrm{6}}}  }  \longrightarrow^{*}  \intermed {  {\intermed v}_{{\mathrm{6}}}  }  \\
    & \band  ( \intermed {  {\intermed \Sigma}_{{\mathrm{1}}}  } : \intermed {  {\intermed v}_{{\mathrm{5}}}  } , \intermed {  {\intermed \Sigma}_{{\mathrm{2}}}  } : \intermed {  {\intermed v}_{{\mathrm{6}}}  } ) \in \mathcal{V} \llbracket \intermed {  {\intermed \sigma}_{{\mathrm{2}}}  } \rrbracket ^{\intermed {  {\intermed { \Gamma_C } }  } }_{\intermed {  {\intermed R }  } } 
  \end{split}
\end{gather}
Goals~\ref{eq:compat_abs8} and \ref{eq:compat_abs9} follow by applying the \textsc{iTm-arrE}
typing rule once on Equations~\ref{eq:compat_abs13} and~\ref{eq:compat_abs15} and once more
on Equations~\ref{eq:compat_abs14} and~\ref{eq:compat_abs16}. Note that
Equations~\ref{eq:compat_abs13} and~\ref{eq:compat_abs14} have been already proven above.



By case analysis, it is easy to see that the first step of the evaluations in
Goal~\ref{eq:compat_abs11} is \textsc{iEval-appAbs}, reducing the goal to:
\begin{equation*}
  \begin{split}
  \exists \intermed{{\intermed v}_{{\mathrm{5}}}}, \intermed{{\intermed v}_{{\mathrm{6}}}} & :
     \intermed {  {\intermed \Sigma}_{{\mathrm{1}}}  }  \vdash  \intermed {   {\intermed \gamma }_{{\mathrm{1}}}  ( {\intermed \phi }_{{\mathrm{1}}}  ( {\intermed R }  ( [  {\intermed e}_{{\mathrm{5}}}  /  {\intermed x}  ]  {\intermed e}_{{\mathrm{1}}} )))   }  \longrightarrow^{*}  \intermed {  {\intermed v}_{{\mathrm{5}}}  }  \\
    & \band  \intermed {  {\intermed \Sigma}_{{\mathrm{2}}}  }  \vdash  \intermed {   {\intermed \gamma }_{{\mathrm{2}}}  ( {\intermed \phi }_{{\mathrm{2}}}  ( {\intermed R }  ( [  {\intermed e}_{{\mathrm{6}}}  /  {\intermed x}  ]  {\intermed e}_{{\mathrm{2}}} )))   }  \longrightarrow^{*}  \intermed {  {\intermed v}_{{\mathrm{6}}}  } \\
    & \band  ( \intermed {  {\intermed \Sigma}_{{\mathrm{1}}}  } : \intermed {  {\intermed v}_{{\mathrm{5}}}  } , \intermed {  {\intermed \Sigma}_{{\mathrm{2}}}  } : \intermed {  {\intermed v}_{{\mathrm{6}}}  } ) \in \mathcal{V} \llbracket \intermed {  {\intermed \sigma}_{{\mathrm{2}}}  } \rrbracket ^{\intermed {  {\intermed { \Gamma_C } }  } }_{\intermed {  {\intermed R }  } } 
  \end{split}
\end{equation*}
We choose \( \intermed {  {\intermed e}_{{\mathrm{3}}}  } = \intermed {  {\intermed e}_{{\mathrm{5}}}  } \) and \( \intermed {  {\intermed e}_{{\mathrm{4}}}  } = \intermed {  {\intermed e}_{{\mathrm{6}}}  } \) in Equation~\ref{eq:compat_abs_ie34}.
The goal follows by choosing \( \intermed {  {\intermed v}_{{\mathrm{5}}}  } = \intermed {  {\intermed v}_{{\mathrm{3}}}  } \) and \( \intermed {  {\intermed v}_{{\mathrm{6}}}  } = \intermed {  {\intermed v}_{{\mathrm{4}}}  } \) from
Equation~\ref{eq:compat_abs7a}. \\
\qed

\begin{lemmaB}[Compatibility - Term Application]\ \\
  \begin{mathpar}
    \ottaltinferrule{}{}
    {
       \intermed {  {\intermed { \Gamma_C } }  } ; \intermed {  {\intermed \Gamma}  }  \vdash  \intermed {  {\intermed \Sigma}_{{\mathrm{1}}}  } : \intermed {  {\intermed e}_{{\mathrm{1}}}  }  \simeq_{log}  \intermed {  {\intermed \Sigma}_{{\mathrm{2}}}  } : \intermed {  {\intermed e}'_{{\mathrm{1}}}  } : \intermed {  {\intermed \sigma}_{{\mathrm{1}}}  \rightarrow  {\intermed \sigma}_{{\mathrm{2}}}  } 
      \\  \intermed {  {\intermed { \Gamma_C } }  } ; \intermed {  {\intermed \Gamma}  }  \vdash  \intermed {  {\intermed \Sigma}_{{\mathrm{1}}}  } : \intermed {  {\intermed e}_{{\mathrm{2}}}  }  \simeq_{log}  \intermed {  {\intermed \Sigma}_{{\mathrm{2}}}  } : \intermed {  {\intermed e}'_{{\mathrm{2}}}  } : \intermed {  {\intermed \sigma}_{{\mathrm{1}}}  } 
    }
    {  \intermed {  {\intermed { \Gamma_C } }  } ; \intermed {  {\intermed \Gamma}  }  \vdash  \intermed {  {\intermed \Sigma}_{{\mathrm{1}}}  } : \intermed {  {\intermed e}_{{\mathrm{1}}} \, {\intermed e}_{{\mathrm{2}}}  }  \simeq_{log}  \intermed {  {\intermed \Sigma}_{{\mathrm{2}}}  } : \intermed {  {\intermed e}'_{{\mathrm{1}}} \, {\intermed e}'_{{\mathrm{2}}}  } : \intermed {  {\intermed \sigma}_{{\mathrm{2}}}  }  }
    \label{lemma:compat_app}
  \end{mathpar}
\end{lemmaB}

\proof By inlining the definition of logical equivalence, suppose we have:
\begin{gather*}
   \intermed {  {\intermed R }  } \in \mathcal{F} \llbracket \intermed {  {\intermed \Gamma}  } \rrbracket ^{\intermed {  {\intermed { \Gamma_C } }  } } \\
   \intermed {  {\intermed \phi }  } \in \mathcal{G} \llbracket \intermed {  {\intermed \Gamma}  } \rrbracket ^{\intermed {  {\intermed \Sigma}_{{\mathrm{1}}}  } , \intermed {  {\intermed \Sigma}_{{\mathrm{2}}}  } , \intermed {  {\intermed { \Gamma_C } }  } }_{\intermed {  {\intermed R }  } } \\
   \intermed {  {\intermed \gamma }  } \in \mathcal{H} \llbracket \intermed {  {\intermed \Gamma}  } \rrbracket ^{\intermed {  {\intermed \Sigma}_{{\mathrm{1}}}  } , \intermed {  {\intermed \Sigma}_{{\mathrm{2}}}  } , \intermed {  {\intermed { \Gamma_C } }  } }_{\intermed {  {\intermed R }  } } 
\end{gather*}
The goal to be proven is the following:
\begin{equation*}
   ( \intermed {  {\intermed \Sigma}_{{\mathrm{1}}}  } : \intermed {   {\intermed \gamma }_{{\mathrm{1}}}  ( {\intermed \phi }_{{\mathrm{1}}}  ( {\intermed R }  ( {\intermed e}_{{\mathrm{1}}} \, {\intermed e}_{{\mathrm{2}}} )))   } , \intermed {  {\intermed \Sigma}_{{\mathrm{2}}}  } : \intermed {   {\intermed \gamma }_{{\mathrm{2}}}  ( {\intermed \phi }_{{\mathrm{2}}}  ( {\intermed R }  (  {\intermed e}'_{{\mathrm{1}}} \, {\intermed e}'_{{\mathrm{2}}}  )))   } ) \in \mathcal{E} \llbracket \intermed {  {\intermed \sigma}_{{\mathrm{2}}}  } \rrbracket ^{\intermed {  {\intermed { \Gamma_C } }  } }_{\intermed {  {\intermed R }  } } 
\end{equation*}
By applying the definition of the \( \mathcal{E} \) relation, the goal reduces
to:
\begin{gather}
  \label{eq:compat_app1}
   \intermed {  {\intermed \Sigma}_{{\mathrm{1}}}  } ; \intermed {  {\intermed { \Gamma_C } }  } ; \intermed {  \bullet  }  \vdash_{tm}  \intermed {   {\intermed \gamma }_{{\mathrm{1}}}  ( {\intermed \phi }_{{\mathrm{1}}}  ( {\intermed R }  ( {\intermed e}_{{\mathrm{1}}} \, {\intermed e}_{{\mathrm{2}}} )))   } : \intermed {  {\intermed R }  \ottsym{(}  {\intermed \sigma}_{{\mathrm{2}}}  \ottsym{)}  } \itot{ \rightsquigarrow  \target {  {\target e }  } } \\
  \label{eq:compat_app2}
   \intermed {  {\intermed \Sigma}_{{\mathrm{2}}}  } ; \intermed {  {\intermed { \Gamma_C } }  } ; \intermed {  \bullet  }  \vdash_{tm}  \intermed {   {\intermed \gamma }_{{\mathrm{2}}}  ( {\intermed \phi }_{{\mathrm{2}}}  ( {\intermed R }  ( {\intermed e}'_{{\mathrm{1}}} \, {\intermed e}'_{{\mathrm{2}}} )))   } : \intermed {  {\intermed R }  \ottsym{(}  {\intermed \sigma}_{{\mathrm{2}}}  \ottsym{)}  } \itot{ \rightsquigarrow  \target {  {\target e }'  } } \\
  \label{eq:compat_app3}
\begin{split}
  \exists \intermed{{\intermed v}_{{\mathrm{1}}}}, \intermed{{\intermed v}_{{\mathrm{2}}}} & :
     \intermed {  {\intermed \Sigma}_{{\mathrm{1}}}  }  \vdash  \intermed {   {\intermed \gamma }_{{\mathrm{1}}}  ( {\intermed \phi }_{{\mathrm{1}}}  ( {\intermed R }  ( {\intermed e}_{{\mathrm{1}}} \, {\intermed e}_{{\mathrm{2}}} )))   }  \longrightarrow^{*}  \intermed {  {\intermed v}_{{\mathrm{1}}}  }  \\ 
    & \band  \intermed {  {\intermed \Sigma}_{{\mathrm{2}}}  }  \vdash  \intermed {   {\intermed \gamma }_{{\mathrm{2}}}  ( {\intermed \phi }_{{\mathrm{2}}}  ( {\intermed R }  ( {\intermed e}'_{{\mathrm{1}}} \, {\intermed e}'_{{\mathrm{2}}} )))   }  \longrightarrow^{*}  \intermed {  {\intermed v}_{{\mathrm{2}}}  } \\ 
    & \band  ( \intermed {  {\intermed \Sigma}_{{\mathrm{1}}}  } : \intermed {  {\intermed v}_{{\mathrm{1}}}  } , \intermed {  {\intermed \Sigma}_{{\mathrm{2}}}  } : \intermed {  {\intermed v}_{{\mathrm{2}}}  } ) \in \mathcal{V} \llbracket \intermed {  {\intermed \sigma}_{{\mathrm{2}}}  } \rrbracket ^{\intermed {  {\intermed { \Gamma_C } }  } }_{\intermed {  {\intermed R }  } } 
\end{split}
\end{gather}
By applying the substitutions in Goals~\ref{eq:compat_app1} and
\ref{eq:compat_app2}, they reduce to:
\begin{gather}
   \intermed {  {\intermed \Sigma}_{{\mathrm{1}}}  } ; \intermed {  {\intermed { \Gamma_C } }  } ; \intermed {  \bullet  }  \vdash_{tm}  \intermed {  \ottsym{(}   {\intermed \gamma }_{{\mathrm{1}}}  ( {\intermed \phi }_{{\mathrm{1}}}  ( {\intermed R }  ( {\intermed e}_{{\mathrm{1}}} )))   \ottsym{)} \, \ottsym{(}   {\intermed \gamma }_{{\mathrm{1}}}  ( {\intermed \phi }_{{\mathrm{1}}}  ( {\intermed R }  ( {\intermed e}_{{\mathrm{2}}} )))   \ottsym{)}  } : \intermed {  {\intermed R }  \ottsym{(}  {\intermed \sigma}_{{\mathrm{2}}}  \ottsym{)}  } \itot{ \rightsquigarrow  \target {  {\target e }  } } 
  \label{eq:compat_app10}\\
   \intermed {  {\intermed \Sigma}_{{\mathrm{2}}}  } ; \intermed {  {\intermed { \Gamma_C } }  } ; \intermed {  \bullet  }  \vdash_{tm}  \intermed {  \ottsym{(}   {\intermed \gamma }_{{\mathrm{2}}}  ( {\intermed \phi }_{{\mathrm{2}}}  ( {\intermed R }  ( {\intermed e}'_{{\mathrm{1}}} )))   \ottsym{)} \, \ottsym{(}   {\intermed \gamma }_{{\mathrm{2}}}  ( {\intermed \phi }_{{\mathrm{2}}}  ( {\intermed R }  ( {\intermed e}'_{{\mathrm{2}}} )))   \ottsym{)}  } : \intermed {  {\intermed R }  \ottsym{(}  {\intermed \sigma}_{{\mathrm{2}}}  \ottsym{)}  } \itot{ \rightsquigarrow  \target {  {\target e }'  } } 
  \label{eq:compat_app11}
\end{gather}
By inlining the definition of logical equivalence in the premise of the rule, we
get:
\begin{align}
   \intermed {  {\intermed \Sigma}_{{\mathrm{1}}}  } ; \intermed {  {\intermed { \Gamma_C } }  } ; \intermed {  \bullet  }  \vdash_{tm}  \intermed {   {\intermed \gamma }_{{\mathrm{1}}}  ( {\intermed \phi }_{{\mathrm{1}}}  ( {\intermed R }  ( {\intermed e}_{{\mathrm{1}}} )))   } : \intermed {  {\intermed R }  \ottsym{(}  {\intermed \sigma}_{{\mathrm{1}}}  \rightarrow  {\intermed \sigma}_{{\mathrm{2}}}  \ottsym{)}  } \itot{ \rightsquigarrow  \target {  {\target e }_{{\mathrm{1}}}  } } 
  \label{eq:compat_app4} \\
   \intermed {  {\intermed \Sigma}_{{\mathrm{2}}}  } ; \intermed {  {\intermed { \Gamma_C } }  } ; \intermed {  \bullet  }  \vdash_{tm}  \intermed {   {\intermed \gamma }_{{\mathrm{2}}}  ( {\intermed \phi }_{{\mathrm{2}}}  ( {\intermed R }  ( {\intermed e}'_{{\mathrm{1}}} )))   } : \intermed {  {\intermed R }  \ottsym{(}  {\intermed \sigma}_{{\mathrm{1}}}  \rightarrow  {\intermed \sigma}_{{\mathrm{2}}}  \ottsym{)}  } \itot{ \rightsquigarrow  \target {  {\target e }'_{{\mathrm{1}}}  } } 
  \label{eq:compat_app5}
\end{align}
\begin{align}
  \exists \intermed{{\intermed v}_{{\mathrm{3}}}}, \intermed{{\intermed v}_{{\mathrm{4}}}} & :
     \intermed {  {\intermed \Sigma}_{{\mathrm{1}}}  }  \vdash  \intermed {   {\intermed \gamma }_{{\mathrm{1}}}  ( {\intermed \phi }_{{\mathrm{1}}}  ( {\intermed R }  ( {\intermed e}_{{\mathrm{1}}} )))   }  \longrightarrow^{*}  \intermed {  {\intermed v}_{{\mathrm{3}}}  } 
    \label{eq:compat_app6a} \\
    & \band  \intermed {  {\intermed \Sigma}_{{\mathrm{2}}}  }  \vdash  \intermed {   {\intermed \gamma }_{{\mathrm{2}}}  ( {\intermed \phi }_{{\mathrm{2}}}  ( {\intermed R }  ( {\intermed e}'_{{\mathrm{1}}} )))   }  \longrightarrow^{*}  \intermed {  {\intermed v}_{{\mathrm{4}}}  } 
    \label{eq:compat_app6b} \\
    & \band  ( \intermed {  {\intermed \Sigma}_{{\mathrm{1}}}  } : \intermed {  {\intermed v}_{{\mathrm{3}}}  } , \intermed {  {\intermed \Sigma}_{{\mathrm{2}}}  } : \intermed {  {\intermed v}_{{\mathrm{4}}}  } ) \in \mathcal{V} \llbracket \intermed {  {\intermed \sigma}_{{\mathrm{1}}}  \rightarrow  {\intermed \sigma}_{{\mathrm{2}}}  } \rrbracket ^{\intermed {  {\intermed { \Gamma_C } }  } }_{\intermed {  {\intermed R }  } } 
  \label{eq:compat_app6c}
\end{align}
and
\begin{gather}
   \intermed {  {\intermed \Sigma}_{{\mathrm{1}}}  } ; \intermed {  {\intermed { \Gamma_C } }  } ; \intermed {  \bullet  }  \vdash_{tm}  \intermed {   {\intermed \gamma }_{{\mathrm{1}}}  ( {\intermed \phi }_{{\mathrm{1}}}  ( {\intermed R }  ( {\intermed e}_{{\mathrm{2}}} )))   } : \intermed {  {\intermed R }  \ottsym{(}  {\intermed \sigma}_{{\mathrm{1}}}  \ottsym{)}  } \itot{ \rightsquigarrow  \target {  {\target e }_{{\mathrm{2}}}  } } 
  \label{eq:compat_app7} \\
   \intermed {  {\intermed \Sigma}_{{\mathrm{2}}}  } ; \intermed {  {\intermed { \Gamma_C } }  } ; \intermed {  \bullet  }  \vdash_{tm}  \intermed {   {\intermed \gamma }_{{\mathrm{2}}}  ( {\intermed \phi }_{{\mathrm{2}}}  ( {\intermed R }  ( {\intermed e}'_{{\mathrm{2}}} )))   } : \intermed {  {\intermed R }  \ottsym{(}  {\intermed \sigma}_{{\mathrm{1}}}  \ottsym{)}  } \itot{ \rightsquigarrow  \target {  {\target e }'_{{\mathrm{2}}}  } } 
  \label{eq:compat_app8}\\
\begin{split}\nonumber
  \exists \intermed{{\intermed v}_{{\mathrm{5}}}}, \intermed{{\intermed v}_{{\mathrm{6}}}} & :
     \intermed {  {\intermed \Sigma}_{{\mathrm{1}}}  }  \vdash  \intermed {   {\intermed \gamma }_{{\mathrm{1}}}  ( {\intermed \phi }_{{\mathrm{1}}}  ( {\intermed R }  ( {\intermed e}_{{\mathrm{2}}} )))   }  \longrightarrow^{*}  \intermed {  {\intermed v}_{{\mathrm{5}}}  } \\
    & \band  \intermed {  {\intermed \Sigma}_{{\mathrm{2}}}  }  \vdash  \intermed {   {\intermed \gamma }_{{\mathrm{2}}}  ( {\intermed \phi }_{{\mathrm{2}}}  ( {\intermed R }  ( {\intermed e}'_{{\mathrm{2}}} )))   }  \longrightarrow^{*}  \intermed {  {\intermed v}_{{\mathrm{6}}}  } \\
    & \band  ( \intermed {  {\intermed \Sigma}_{{\mathrm{1}}}  } : \intermed {  {\intermed v}_{{\mathrm{5}}}  } , \intermed {  {\intermed \Sigma}_{{\mathrm{2}}}  } : \intermed {  {\intermed v}_{{\mathrm{6}}}  } ) \in \mathcal{V} \llbracket \intermed {  {\intermed \sigma}_{{\mathrm{1}}}  } \rrbracket ^{\intermed {  {\intermed { \Gamma_C } }  } }_{\intermed {  {\intermed R }  } } 
\end{split}
\end{gather}
By applying both Equation~\ref{eq:compat_app4} and \ref{eq:compat_app7} and both
Equation~\ref{eq:compat_app5} and \ref{eq:compat_app8} respectively to \textsc{iTm-arrE},
Goal~\ref{eq:compat_app10} and \ref{eq:compat_app11} are proven.

By applying the substitution, Goal~\ref{eq:compat_app3} reduces to:
\begin{align*}
  \begin{split}
  \exists \intermed{{\intermed v}_{{\mathrm{1}}}}, \intermed{{\intermed v}_{{\mathrm{2}}}} & :
     \intermed {  {\intermed \Sigma}_{{\mathrm{1}}}  }  \vdash  \intermed {  \ottsym{(}   {\intermed \gamma }_{{\mathrm{1}}}  ( {\intermed \phi }_{{\mathrm{1}}}  ( {\intermed R }  ( {\intermed e}_{{\mathrm{1}}} )))   \ottsym{)} \, \ottsym{(}   {\intermed \gamma }_{{\mathrm{1}}}  ( {\intermed \phi }_{{\mathrm{1}}}  ( {\intermed R }  ( {\intermed e}_{{\mathrm{2}}} )))   \ottsym{)}  }  \longrightarrow^{*}  \intermed {  {\intermed v}_{{\mathrm{1}}}  }  \\
    & \band  \intermed {  {\intermed \Sigma}_{{\mathrm{2}}}  }  \vdash  \intermed {  \ottsym{(}   {\intermed \gamma }_{{\mathrm{2}}}  ( {\intermed \phi }_{{\mathrm{2}}}  ( {\intermed R }  ( {\intermed e}'_{{\mathrm{1}}} )))   \ottsym{)} \, \ottsym{(}   {\intermed \gamma }_{{\mathrm{2}}}  ( {\intermed \phi }_{{\mathrm{2}}}  ( {\intermed R }  ( {\intermed e}'_{{\mathrm{2}}} )))   \ottsym{)}  }  \longrightarrow^{*}  \intermed {  {\intermed v}_{{\mathrm{2}}}  } \\
    & \band  ( \intermed {  {\intermed \Sigma}_{{\mathrm{1}}}  } : \intermed {  {\intermed v}_{{\mathrm{1}}}  } , \intermed {  {\intermed \Sigma}_{{\mathrm{2}}}  } : \intermed {  {\intermed v}_{{\mathrm{2}}}  } ) \in \mathcal{V} \llbracket \intermed {  {\intermed \sigma}_{{\mathrm{2}}}  } \rrbracket ^{\intermed {  {\intermed { \Gamma_C } }  } }_{\intermed {  {\intermed R }  } } 
  \end{split}
\end{align*}
Through case analysis, we see that we should first (repeatedly) apply
\textsc{iEval-app} and Equation~\ref{eq:compat_app6a} and \ref{eq:compat_app6b}
respectively. The goal reduces to:
\begin{align}
  \begin{split}
  \exists \intermed{{\intermed v}_{{\mathrm{1}}}}, \intermed{{\intermed v}_{{\mathrm{2}}}} & :
     \intermed {  {\intermed \Sigma}_{{\mathrm{1}}}  }  \vdash  \intermed {  {\intermed v}_{{\mathrm{3}}} \, \ottsym{(}   {\intermed \gamma }_{{\mathrm{1}}}  ( {\intermed \phi }_{{\mathrm{1}}}  ( {\intermed R }  ( {\intermed e}_{{\mathrm{2}}} )))   \ottsym{)}  }  \longrightarrow^{*}  \intermed {  {\intermed v}_{{\mathrm{1}}}  }  \\
    & \band  \intermed {  {\intermed \Sigma}_{{\mathrm{2}}}  }  \vdash  \intermed {  {\intermed v}_{{\mathrm{4}}} \, \ottsym{(}   {\intermed \gamma }_{{\mathrm{2}}}  ( {\intermed \phi }_{{\mathrm{2}}}  ( {\intermed R }  ( {\intermed e}'_{{\mathrm{2}}} )))   \ottsym{)}  }  \longrightarrow^{*}  \intermed {  {\intermed v}_{{\mathrm{2}}}  }  \\
    & \band  ( \intermed {  {\intermed \Sigma}_{{\mathrm{1}}}  } : \intermed {  {\intermed v}_{{\mathrm{1}}}  } , \intermed {  {\intermed \Sigma}_{{\mathrm{2}}}  } : \intermed {  {\intermed v}_{{\mathrm{2}}}  } ) \in \mathcal{V} \llbracket \intermed {  {\intermed \sigma}_{{\mathrm{2}}}  } \rrbracket ^{\intermed {  {\intermed { \Gamma_C } }  } }_{\intermed {  {\intermed R }  } } 
  \end{split}
  \label{eq:compat_app13}
\end{align}
By the definition of the \( \mathcal{V} \)-relation in Equation~\ref{eq:compat_app6c}, we
know that:
\begin{gather}
  \nonumber
   \intermed {  {\intermed \Sigma}_{{\mathrm{1}}}  } ; \intermed {  {\intermed { \Gamma_C } }  } ; \intermed {  \bullet  }  \vdash_{tm}  \intermed {  {\intermed v}_{{\mathrm{3}}}  } : \intermed {  {\intermed R }  \ottsym{(}  {\intermed \sigma}_{{\mathrm{1}}}  \rightarrow  {\intermed \sigma}_{{\mathrm{2}}}  \ottsym{)}  } \itot{ \rightsquigarrow  \target {  {\target e }_{{\mathrm{3}}}  } } \\
  \nonumber
   \intermed {  {\intermed \Sigma}_{{\mathrm{2}}}  } ; \intermed {  {\intermed { \Gamma_C } }  } ; \intermed {  \bullet  }  \vdash_{tm}  \intermed {  {\intermed v}_{{\mathrm{4}}}  } : \intermed {  {\intermed R }  \ottsym{(}  {\intermed \sigma}_{{\mathrm{1}}}  \rightarrow  {\intermed \sigma}_{{\mathrm{2}}}  \ottsym{)}  } \itot{ \rightsquigarrow  \target {  {\target e }_{{\mathrm{4}}}  } } \\
  \forall  ( \intermed {  {\intermed \Sigma}_{{\mathrm{1}}}  } : \intermed {  {\intermed e}_{{\mathrm{5}}}  } , \intermed {  {\intermed \Sigma}_{{\mathrm{2}}}  } : \intermed {  {\intermed e}_{{\mathrm{6}}}  } ) \in \mathcal{E} \llbracket \intermed {  {\intermed \sigma}_{{\mathrm{1}}}  } \rrbracket ^{\intermed {  {\intermed { \Gamma_C } }  } }_{\intermed {  {\intermed R }  } }  :
   ( \intermed {  {\intermed \Sigma}_{{\mathrm{1}}}  } : \intermed {  {\intermed v}_{{\mathrm{3}}} \, {\intermed e}_{{\mathrm{5}}}  } , \intermed {  {\intermed \Sigma}_{{\mathrm{2}}}  } : \intermed {  {\intermed v}_{{\mathrm{4}}} \, {\intermed e}_{{\mathrm{6}}}  } ) \in \mathcal{E} \llbracket \intermed {  {\intermed \sigma}_{{\mathrm{2}}}  } \rrbracket ^{\intermed {  {\intermed { \Gamma_C } }  } }_{\intermed {  {\intermed R }  } } 
  \label{eq:compat_app18}
\end{gather}
We choose \( \intermed {  {\intermed e}_{{\mathrm{5}}}  } = \intermed {   {\intermed \gamma }_{{\mathrm{1}}}  ( {\intermed \phi }_{{\mathrm{1}}}  ( {\intermed R }  ( {\intermed e}_{{\mathrm{2}}} )))   } \) and
\( \intermed {  {\intermed e}_{{\mathrm{6}}}  } = \intermed {   {\intermed \gamma }_{{\mathrm{2}}}  ( {\intermed \phi }_{{\mathrm{2}}}  ( {\intermed R }  ( {\intermed e}'_{{\mathrm{2}}} )))   } \). Goal~\ref{eq:compat_app13} now follows from
the definition of the \( \mathcal{E} \)-relation in Equation~\ref{eq:compat_app18}. \\
\qed

\begin{lemmaB}[Compatibility - Dictionary Abstraction]\ \\
  \begin{mathpar}
    \ottaltinferrule{}{}
    {
       \intermed {  {\intermed { \Gamma_C } }  } ; \intermed {  {\intermed \Gamma}  \ottsym{,}  {\intermed \delta }  \ottsym{:}  {\intermed Q}  }  \vdash  \intermed {  {\intermed \Sigma}_{{\mathrm{1}}}  } : \intermed {  {\intermed e}_{{\mathrm{1}}}  }  \simeq_{log}  \intermed {  {\intermed \Sigma}_{{\mathrm{2}}}  } : \intermed {  {\intermed e}_{{\mathrm{2}}}  } : \intermed {  {\intermed \sigma}  } 
    }
    {  \intermed {  {\intermed { \Gamma_C } }  } ; \intermed {  {\intermed \Gamma}  }  \vdash  \intermed {  {\intermed \Sigma}_{{\mathrm{1}}}  } : \intermed {  \lambda  {\intermed \delta }  \ottsym{:}  {\intermed Q}  \ottsym{.}  {\intermed e}_{{\mathrm{1}}}  }  \simeq_{log}  \intermed {  {\intermed \Sigma}_{{\mathrm{2}}}  } : \intermed {  \lambda  {\intermed \delta }  \ottsym{:}  {\intermed Q}  \ottsym{.}  {\intermed e}_{{\mathrm{2}}}  } : \intermed {   {\intermed Q}  \Rightarrow  {\intermed \sigma}   }  }
    \label{lemma:compat_dabs}
  \end{mathpar}
\end{lemmaB}

\proof
By unfolding the definition of logical equivalence in the conclusion of the lemma,
suppose we have:
\begin{gather}
  \label{eq:compat_dabsR}
   \intermed {  {\intermed R }  } \in \mathcal{F} \llbracket \intermed {  {\intermed \Gamma}  } \rrbracket ^{\intermed {  {\intermed { \Gamma_C } }  } } \\
  \label{eq:compat_dabsG}
   \intermed {  {\intermed \phi }  } \in \mathcal{G} \llbracket \intermed {  {\intermed \Gamma}  } \rrbracket ^{\intermed {  {\intermed \Sigma}_{{\mathrm{1}}}  } , \intermed {  {\intermed \Sigma}_{{\mathrm{2}}}  } , \intermed {  {\intermed { \Gamma_C } }  } }_{\intermed {  {\intermed R }  } } \\
  \nonumber
   \intermed {  {\intermed \gamma }  } \in \mathcal{H} \llbracket \intermed {  {\intermed \Gamma}  } \rrbracket ^{\intermed {  {\intermed \Sigma}_{{\mathrm{1}}}  } , \intermed {  {\intermed \Sigma}_{{\mathrm{2}}}  } , \intermed {  {\intermed { \Gamma_C } }  } }_{\intermed {  {\intermed R }  } } 
\end{gather}
The goal to be proven is the following:
\begin{equation*}
   ( \intermed {  {\intermed \Sigma}_{{\mathrm{1}}}  } : \intermed {   {\intermed \gamma }_{{\mathrm{1}}}  ( {\intermed \phi }_{{\mathrm{1}}}  ( {\intermed R }  ( \lambda  {\intermed \delta }  \ottsym{:}  {\intermed Q}  \ottsym{.}  {\intermed e}_{{\mathrm{1}}} )))   } , \intermed {  {\intermed \Sigma}_{{\mathrm{2}}}  } : \intermed {   {\intermed \gamma }_{{\mathrm{2}}}  ( {\intermed \phi }_{{\mathrm{2}}}  ( {\intermed R }  (  \lambda  {\intermed \delta }  \ottsym{:}  {\intermed Q}  \ottsym{.}  {\intermed e}_{{\mathrm{2}}}  )))   } ) \in \mathcal{E} \llbracket \intermed {   {\intermed Q}  \Rightarrow  {\intermed \sigma}   } \rrbracket ^{\intermed {  {\intermed { \Gamma_C } }  } }_{\intermed {  {\intermed R }  } } 
\end{equation*}
By applying the definition of the \( \mathcal{E} \) relation (taking into
account that \(\intermed{\lambda  {\intermed \delta }  \ottsym{:}  {\intermed Q}  \ottsym{.}  {\intermed e}}\) is a value) and partially applying
the substitutions, the goal reduces to:
\begin{gather}
   \intermed {  {\intermed \Sigma}_{{\mathrm{1}}}  } ; \intermed {  {\intermed { \Gamma_C } }  } ; \intermed {  \bullet  }  \vdash_{tm}  \intermed {  \lambda  {\intermed \delta }  \ottsym{:}  {\intermed R }  \ottsym{(}  {\intermed Q}  \ottsym{)}  \ottsym{.}  \ottsym{(}   {\intermed \gamma }_{{\mathrm{1}}}  ( {\intermed \phi }_{{\mathrm{1}}}  ( {\intermed R }  ( {\intermed e}_{{\mathrm{1}}} )))   \ottsym{)}  } : \intermed {  {\intermed R }  \ottsym{(}   {\intermed Q}  \Rightarrow  {\intermed \sigma}   \ottsym{)}  } \itot{ \rightsquigarrow  \target {  {\target e }  } } 
  \label{eq:compat_dabs1} \\
   \intermed {  {\intermed \Sigma}_{{\mathrm{2}}}  } ; \intermed {  {\intermed { \Gamma_C } }  } ; \intermed {  \bullet  }  \vdash_{tm}  \intermed {  \lambda  {\intermed \delta }  \ottsym{:}  {\intermed R }  \ottsym{(}  {\intermed Q}  \ottsym{)}  \ottsym{.}  \ottsym{(}   {\intermed \gamma }_{{\mathrm{2}}}  ( {\intermed \phi }_{{\mathrm{2}}}  ( {\intermed R }  ( {\intermed e}_{{\mathrm{2}}} )))   \ottsym{)}  } : \intermed {  {\intermed R }  \ottsym{(}   {\intermed Q}  \Rightarrow  {\intermed \sigma}   \ottsym{)}  } \itot{ \rightsquigarrow  \target {  {\target e }'  } } 
  \label{eq:compat_dabs2} \\
   ( \intermed {  {\intermed \Sigma}_{{\mathrm{1}}}  } : \intermed {  \lambda  {\intermed \delta }  \ottsym{:}  {\intermed R }  \ottsym{(}  {\intermed Q}  \ottsym{)}  \ottsym{.}  \ottsym{(}   {\intermed \gamma }_{{\mathrm{1}}}  ( {\intermed \phi }_{{\mathrm{1}}}  ( {\intermed R }  ( {\intermed e}_{{\mathrm{1}}} )))   \ottsym{)}  } , \intermed {  {\intermed \Sigma}_{{\mathrm{2}}}  } : \intermed {  \lambda  {\intermed \delta }  \ottsym{:}  {\intermed R }  \ottsym{(}  {\intermed Q}  \ottsym{)}  \ottsym{.}  \ottsym{(}   {\intermed \gamma }_{{\mathrm{2}}}  ( {\intermed \phi }_{{\mathrm{2}}}  ( {\intermed R }  ( {\intermed e}_{{\mathrm{2}}} )))   \ottsym{)}  } ) \in \mathcal{V} \llbracket \intermed {   {\intermed Q}  \Rightarrow  {\intermed \sigma}   } \rrbracket ^{\intermed {  {\intermed { \Gamma_C } }  } }_{\intermed {  {\intermed R }  } } 
  \label{eq:compat_dabs3}
\end{gather}
By unfolding the definition of logical equivalence in the premise of this lemma, we get:
\begin{equation}
   ( \intermed {  {\intermed \Sigma}_{{\mathrm{1}}}  } : \intermed {   {\intermed \gamma }'_{{\mathrm{1}}}  ( {\intermed \phi }'_{{\mathrm{1}}}  ( {\intermed R }'  ( {\intermed e}_{{\mathrm{1}}} )))   } , \intermed {  {\intermed \Sigma}_{{\mathrm{2}}}  } : \intermed {   {\intermed \gamma }'_{{\mathrm{2}}}  ( {\intermed \phi }'_{{\mathrm{2}}}  ( {\intermed R }'  ( {\intermed e}_{{\mathrm{2}}} )))   } ) \in \mathcal{E} \llbracket \intermed {  {\intermed \sigma}  } \rrbracket ^{\intermed {  {\intermed { \Gamma_C } }  } }_{\intermed {  {\intermed R }'  } } 
  \label{eq:compat_dabs4}
\end{equation}
for any \( \intermed {  {\intermed R }'  } \in \mathcal{F} \llbracket \intermed {  {\intermed \Gamma}  \ottsym{,}  {\intermed \delta }  \ottsym{:}  {\intermed Q}  } \rrbracket ^{\intermed {  {\intermed { \Gamma_C } }  } } \), \( \intermed {  {\intermed \phi }'  } \in \mathcal{G} \llbracket \intermed {  {\intermed \Gamma}  \ottsym{,}  {\intermed \delta }  \ottsym{:}  {\intermed Q}  } \rrbracket ^{\intermed {  {\intermed \Sigma}_{{\mathrm{1}}}  } , \intermed {  {\intermed \Sigma}_{{\mathrm{2}}}  } , \intermed {  {\intermed { \Gamma_C } }  } }_{\intermed {  {\intermed R }'  } } \) and \( \intermed {  {\intermed \gamma }'  } \in \mathcal{H} \llbracket \intermed {  {\intermed \Gamma}  \ottsym{,}  {\intermed \delta }  \ottsym{:}  {\intermed Q}  } \rrbracket ^{\intermed {  {\intermed \Sigma}_{{\mathrm{1}}}  } , \intermed {  {\intermed \Sigma}_{{\mathrm{2}}}  } , \intermed {  {\intermed { \Gamma_C } }  } }_{\intermed {  {\intermed R }'  } } \).

By the definition of the \( \mathcal{F} \) and the \( \mathcal{G} \) relations and from
Equations~\ref{eq:compat_dabsR} and~\ref{eq:compat_dabsG}, we have that
\( \intermed {  {\intermed R }  } \in \mathcal{F} \llbracket \intermed {  {\intermed \Gamma}  \ottsym{,}  {\intermed \delta }  \ottsym{:}  {\intermed Q}  } \rrbracket ^{\intermed {  {\intermed { \Gamma_C } }  } } \) and
\( \intermed {  {\intermed \phi }  } \in \mathcal{G} \llbracket \intermed {  {\intermed \Gamma}  \ottsym{,}  {\intermed \delta }  \ottsym{:}  {\intermed Q}  } \rrbracket ^{\intermed {  {\intermed \Sigma}_{{\mathrm{1}}}  } , \intermed {  {\intermed \Sigma}_{{\mathrm{2}}}  } , \intermed {  {\intermed { \Gamma_C } }  } }_{\intermed {  {\intermed R }'  } } \). We choose \( \intermed {  {\intermed R }'  } = \intermed {  {\intermed R }  } \)
and \( \intermed {  {\intermed \phi }'  } = \intermed {  {\intermed \phi }  } \). By case analysis on \(\intermed{{\intermed \gamma }'}\), we know that
\( \intermed {  {\intermed \gamma }'  } = \intermed {   {\intermed \gamma }'' ,  {\intermed \delta }  \mapsto ( {\intermed {dv} }_{{\mathrm{1}}} ,  {\intermed {dv} }_{{\mathrm{2}}}  )   } \) for some
\( \intermed {  {\intermed \gamma }''  } \in \mathcal{H} \llbracket \intermed {  {\intermed \Gamma}  } \rrbracket ^{\intermed {  {\intermed \Sigma}_{{\mathrm{1}}}  } , \intermed {  {\intermed \Sigma}_{{\mathrm{2}}}  } , \intermed {  {\intermed { \Gamma_C } }  } }_{\intermed {  {\intermed R }  } } \) and some dictionaries
\(\intermed{{\intermed {dv} }_{{\mathrm{1}}}}\) and \(\intermed{{\intermed {dv} }_{{\mathrm{2}}}}\) such that
\begin{equation}
  \label{eq:compat_dabs_dv12}
   ( \intermed {  {\intermed \Sigma}_{{\mathrm{1}}}  } : \intermed {  {\intermed {dv} }_{{\mathrm{1}}}  } , \intermed {  {\intermed \Sigma}_{{\mathrm{2}}}  } : \intermed {  {\intermed {dv} }_{{\mathrm{2}}}  } ) \in \mathcal{V} \llbracket \intermed {  {\intermed Q}  } \rrbracket ^{\intermed {  {\intermed { \Gamma_C } }  } }_{\intermed {  {\intermed R }  } } 
\end{equation}
We choose \( \intermed {  {\intermed \gamma }''  } = \intermed {  {\intermed \gamma }  } \).
Then, unfolding the definition of the \( \mathcal{E} \) relation in Equation~\ref{eq:compat_dabs4}
results in:
\begin{gather}
   \intermed {  {\intermed \Sigma}_{{\mathrm{1}}}  } ; \intermed {  {\intermed { \Gamma_C } }  } ; \intermed {  \bullet  }  \vdash_{tm}  \intermed {   \ottsym{(}   {\intermed \gamma }_{{\mathrm{1}}} ,  {\intermed \delta }  \mapsto  {\intermed {dv} }_{{\mathrm{1}}}   \ottsym{)}  ( {\intermed \phi }_{{\mathrm{1}}}  ( {\intermed R }  ( {\intermed e}_{{\mathrm{1}}} )))   } : \intermed {  {\intermed R }  \ottsym{(}  {\intermed \sigma}  \ottsym{)}  } \itot{ \rightsquigarrow  \target {  {\target e }_{{\mathrm{1}}}  } } 
  \label{eq:compat_dabs4b} \\
   \intermed {  {\intermed \Sigma}_{{\mathrm{2}}}  } ; \intermed {  {\intermed { \Gamma_C } }  } ; \intermed {  \bullet  }  \vdash_{tm}  \intermed {   \ottsym{(}   {\intermed \gamma }_{{\mathrm{2}}} ,  {\intermed \delta }  \mapsto  {\intermed {dv} }_{{\mathrm{2}}}   \ottsym{)}  ( {\intermed \phi }_{{\mathrm{2}}}  ( {\intermed R }  ( {\intermed e}_{{\mathrm{2}}} )))   } : \intermed {  {\intermed R }  \ottsym{(}  {\intermed \sigma}  \ottsym{)}  } \itot{ \rightsquigarrow  \target {  {\target e }_{{\mathrm{2}}}  } } 
  \label{eq:compat_dabs5} \\
\begin{split}
\label{eq:compat_dabs6}
  \exists \intermed{{\intermed v}_{{\mathrm{1}}}}, \intermed{{\intermed v}_{{\mathrm{2}}}} & :
     \intermed {  {\intermed \Sigma}_{{\mathrm{1}}}  }  \vdash  \intermed {   \ottsym{(}   {\intermed \gamma }_{{\mathrm{1}}} ,  {\intermed \delta }  \mapsto  {\intermed {dv} }_{{\mathrm{1}}}   \ottsym{)}  ( {\intermed \phi }_{{\mathrm{1}}}  ( {\intermed R }  ( {\intermed e}_{{\mathrm{1}}} )))   }  \longrightarrow^{*}  \intermed {  {\intermed v}_{{\mathrm{1}}}  } \\
    & \band  \intermed {  {\intermed \Sigma}_{{\mathrm{2}}}  }  \vdash  \intermed {   \ottsym{(}   {\intermed \gamma }_{{\mathrm{2}}} ,  {\intermed \delta }  \mapsto  {\intermed {dv} }_{{\mathrm{2}}}   \ottsym{)}  ( {\intermed \phi }_{{\mathrm{2}}}  ( {\intermed R }  ( {\intermed e}_{{\mathrm{2}}} )))   }  \longrightarrow^{*}  \intermed {  {\intermed v}_{{\mathrm{2}}}  } \\
    & \band  ( \intermed {  {\intermed \Sigma}_{{\mathrm{1}}}  } : \intermed {  {\intermed v}_{{\mathrm{1}}}  } , \intermed {  {\intermed \Sigma}_{{\mathrm{2}}}  } : \intermed {  {\intermed v}_{{\mathrm{2}}}  } ) \in \mathcal{V} \llbracket \intermed {  {\intermed \sigma}  } \rrbracket ^{\intermed {  {\intermed { \Gamma_C } }  } }_{\intermed {  {\intermed R }  } } 
\end{split}
\end{gather}
From the definition of the \( \mathcal{V} \) relation in Equation~\ref{eq:compat_dabs_dv12},
it follows that:
\begin{gather}
   \intermed {  {\intermed \Sigma}_{{\mathrm{1}}}  } ; \intermed {  {\intermed { \Gamma_C } }  } ; \intermed {  \bullet  }  \vdash_{d}  \intermed {  {\intermed {dv} }_{{\mathrm{1}}}  } : \intermed {  {\intermed Q}  } \itot{\,  \rightsquigarrow  \target {  {\target e }_{{\mathrm{3}}}  } } 
  \label{eq:compat_dabs4be} \\
   \intermed {  {\intermed \Sigma}_{{\mathrm{2}}}  } ; \intermed {  {\intermed { \Gamma_C } }  } ; \intermed {  \bullet  }  \vdash_{d}  \intermed {  {\intermed {dv} }_{{\mathrm{2}}}  } : \intermed {  {\intermed Q}  } \itot{\,  \rightsquigarrow  \target {  {\target e }_{{\mathrm{4}}}  } } 
  \label{eq:compat_dabs5be}
\end{gather}
Since neither \(\intermed{{\intermed {dv} }_{{\mathrm{1}}}}\) nor \(\intermed{{\intermed {dv} }_{{\mathrm{2}}}}\) contain any free variables, we know
that \( \intermed {  {\intermed \gamma }_{{\mathrm{1}}}  \ottsym{(}  {\intermed {dv} }_{{\mathrm{1}}}  \ottsym{)}  } = \intermed {  {\intermed {dv} }_{{\mathrm{1}}}  } \) and \( \intermed {  {\intermed \gamma }_{{\mathrm{2}}}  \ottsym{(}  {\intermed {dv} }_{{\mathrm{2}}}  \ottsym{)}  } = \intermed {  {\intermed {dv} }_{{\mathrm{2}}}  } \).
Consequently, Equations~\ref{eq:compat_dabs4b} and \ref{eq:compat_dabs5} are
equivalent to:
\begin{gather}
   \intermed {  {\intermed \Sigma}_{{\mathrm{1}}}  } ; \intermed {  {\intermed { \Gamma_C } }  } ; \intermed {  \bullet  }  \vdash_{tm}  \intermed {  [  {\intermed {dv} }_{{\mathrm{1}}}  /  {\intermed \delta }  ]  \ottsym{(}   {\intermed \gamma }_{{\mathrm{1}}}  ( {\intermed \phi }_{{\mathrm{1}}}  ( {\intermed R }  ( {\intermed e}_{{\mathrm{1}}} )))   \ottsym{)}  } : \intermed {  {\intermed R }  \ottsym{(}  {\intermed \sigma}  \ottsym{)}  } \itot{ \rightsquigarrow  \target {  {\target e }_{{\mathrm{1}}}  } } 
  \label{eq:compat_dabs4ce} \\
   \intermed {  {\intermed \Sigma}_{{\mathrm{2}}}  } ; \intermed {  {\intermed { \Gamma_C } }  } ; \intermed {  \bullet  }  \vdash_{tm}  \intermed {  [  {\intermed {dv} }_{{\mathrm{2}}}  /  {\intermed \delta }  ]  \ottsym{(}   {\intermed \gamma }_{{\mathrm{2}}}  ( {\intermed \phi }_{{\mathrm{2}}}  ( {\intermed R }  ( {\intermed e}_{{\mathrm{2}}} )))   \ottsym{)}  } : \intermed {  {\intermed R }  \ottsym{(}  {\intermed \sigma}  \ottsym{)}  } \itot{ \rightsquigarrow  \target {  {\target e }_{{\mathrm{2}}}  } } 
  \label{eq:compat_dabs5ce}
\end{gather}
By applying the substitution Lemma~\ref{lemma:rev_dvar_subst} on
Equations~\ref{eq:compat_dabs4ce} and \ref{eq:compat_dabs5ce} respectively, in
combination with Equations~\ref{eq:compat_dabs4be} and \ref{eq:compat_dabs5be}, we
find:
\begin{gather}
   \intermed {  {\intermed \Sigma}_{{\mathrm{1}}}  } ; \intermed {  {\intermed { \Gamma_C } }  } ; \intermed {  \bullet  \ottsym{,}  {\intermed \delta }  \ottsym{:}  {\intermed Q}  }  \vdash_{tm}  \intermed {   {\intermed \gamma }_{{\mathrm{1}}}  ( {\intermed \phi }_{{\mathrm{1}}}  ( {\intermed R }  ( {\intermed e}_{{\mathrm{1}}} )))   } : \intermed {  {\intermed R }  \ottsym{(}  {\intermed \sigma}  \ottsym{)}  } \itot{ \rightsquigarrow  \target {  {\target e }'_{{\mathrm{1}}}  } } 
  \label{eq:compat_dabs7} \\
   \intermed {  {\intermed \Sigma}_{{\mathrm{2}}}  } ; \intermed {  {\intermed { \Gamma_C } }  } ; \intermed {  \bullet  \ottsym{,}  {\intermed \delta }  \ottsym{:}  {\intermed Q}  }  \vdash_{tm}  \intermed {   {\intermed \gamma }_{{\mathrm{2}}}  ( {\intermed \phi }_{{\mathrm{2}}}  ( {\intermed R }  ( {\intermed e}_{{\mathrm{2}}} )))   } : \intermed {  {\intermed R }  \ottsym{(}  {\intermed \sigma}  \ottsym{)}  } \itot{ \rightsquigarrow  \target {  {\target e }'_{{\mathrm{2}}}  } } 
  \label{eq:compat_dabs8}
\end{gather}
By Lemma~\ref{lemma:ctx_well_inter_typing}, it follows from
Equation~\ref{eq:compat_dabs7} that \( \vdash_{ctx}  \intermed {  {\intermed \Sigma}_{{\mathrm{1}}}  } ; \intermed {  {\intermed { \Gamma_C } }  } ; \intermed {  \bullet  \ottsym{,}  {\intermed \delta }  \ottsym{:}  {\intermed Q}  } \).
Consequently, we know from \textsc{iCtx-tyEnvD} that:
\begin{equation}
  \label{eq:compat_dabs7b}
   \intermed {  {\intermed { \Gamma_C } }  } ; \intermed {  \bullet  }  \vdash_{\mathit {Q} }  \intermed {  {\intermed Q}  }\, \itot{ \rightsquigarrow  \target {  {\target \sigma }_{\ottmv{q}}  } } 
\end{equation}
Since \({\intermed Q}\) does not contain any free variables, it is straightforward to
see that \( \intermed {  {\intermed R }  \ottsym{(}  {\intermed Q}  \ottsym{)}  } = \intermed {  {\intermed Q}  } \).

Consequently, goals~\ref{eq:compat_dabs1} and \ref{eq:compat_dabs2} follow by applying
Equation~\ref{eq:compat_dabs7} and \ref{eq:compat_dabs8} respectively, in
combination with Equation~\ref{eq:compat_dabs7b}, to \textsc{iTm-constrI}.

By unfolding the definition of \( \mathcal{V} \),
Goal~\ref{eq:compat_dabs3} reduces to:
\begin{gather}
   \intermed {  {\intermed \Sigma}_{{\mathrm{1}}}  } ; \intermed {  {\intermed { \Gamma_C } }  } ; \intermed {  \bullet  }  \vdash_{tm}  \intermed {  \lambda  {\intermed \delta }  \ottsym{:}  {\intermed R }  \ottsym{(}  {\intermed Q}  \ottsym{)}  \ottsym{.}  \ottsym{(}   {\intermed \gamma }_{{\mathrm{1}}}  ( {\intermed \phi }_{{\mathrm{1}}}  ( {\intermed R }  ( {\intermed e}_{{\mathrm{1}}} )))   \ottsym{)}  } : \intermed {  {\intermed R }  \ottsym{(}   {\intermed Q}  \Rightarrow  {\intermed \sigma}   \ottsym{)}  } \itot{ \rightsquigarrow  \target {  {\target e }  } } 
  \label{eq:compat_dabs9} \\
   \intermed {  {\intermed \Sigma}_{{\mathrm{2}}}  } ; \intermed {  {\intermed { \Gamma_C } }  } ; \intermed {  \bullet  }  \vdash_{tm}  \intermed {  \lambda  {\intermed \delta }  \ottsym{:}  {\intermed R }  \ottsym{(}  {\intermed Q}  \ottsym{)}  \ottsym{.}  \ottsym{(}   {\intermed \gamma }_{{\mathrm{2}}}  ( {\intermed \phi }_{{\mathrm{2}}}  ( {\intermed R }  ( {\intermed e}_{{\mathrm{2}}} )))   \ottsym{)}  } : \intermed {  {\intermed R }  \ottsym{(}   {\intermed Q}  \Rightarrow  {\intermed \sigma}   \ottsym{)}  } \itot{ \rightsquigarrow  \target {  {\target e }'  } } 
  \label{eq:compat_dabs10} \\
  \forall \intermed{{\intermed {dv} }_{{\mathrm{3}}}}\,\intermed{{\intermed {dv} }_{{\mathrm{4}}}},~\text{if}~ ( \intermed {  {\intermed \Sigma}_{{\mathrm{1}}}  } : \intermed {  {\intermed {dv} }_{{\mathrm{3}}}  } , \intermed {  {\intermed \Sigma}_{{\mathrm{2}}}  } : \intermed {  {\intermed {dv} }_{{\mathrm{4}}}  } ) \in \mathcal{V} \llbracket \intermed {  {\intermed Q}  } \rrbracket ^{\intermed {  {\intermed { \Gamma_C } }  } }_{\intermed {  {\intermed R }  } }  :
  \label{eq:compat_dabs11'} \\
  \text{then}~ ( \intermed {  {\intermed \Sigma}_{{\mathrm{1}}}  } : \intermed {  \ottsym{(}  \lambda  {\intermed \delta }  \ottsym{:}  {\intermed R }  \ottsym{(}  {\intermed Q}  \ottsym{)}  \ottsym{.}  \ottsym{(}   {\intermed \gamma }_{{\mathrm{1}}}  ( {\intermed \phi }_{{\mathrm{1}}}  ( {\intermed R }  ( {\intermed e}_{{\mathrm{1}}} )))   \ottsym{)}  \ottsym{)} \, {\intermed {dv} }_{{\mathrm{3}}}  } , \intermed {  {\intermed \Sigma}_{{\mathrm{2}}}  } : \intermed {  \ottsym{(}  \lambda  {\intermed \delta }  \ottsym{:}  {\intermed R }  \ottsym{(}  {\intermed Q}  \ottsym{)}  \ottsym{.}  \ottsym{(}   {\intermed \gamma }_{{\mathrm{2}}}  ( {\intermed \phi }_{{\mathrm{2}}}  ( {\intermed R }  ( {\intermed e}_{{\mathrm{2}}} )))   \ottsym{)}  \ottsym{)} \, {\intermed {dv} }_{{\mathrm{4}}}  } ) \in \mathcal{E} \llbracket \intermed {  {\intermed \sigma}  } \rrbracket ^{\intermed {  {\intermed { \Gamma_C } }  } }_{\intermed {  {\intermed R }  } } 
  \label{eq:compat_dabs11}
\end{gather}
Goals~\ref{eq:compat_dabs9} and \ref{eq:compat_dabs10} are identical to
Goals~\ref{eq:compat_dabs1} and~\ref{eq:compat_dabs2}, which have been proven above.
For the final goal, suppose dictionaries \(\intermed{{\intermed {dv} }_{{\mathrm{3}}}}\) and
\(\intermed{{\intermed {dv} }_{{\mathrm{4}}}}\) such that Equation~\ref{eq:compat_dabs11'} holds. By the definition
of the \(\mathcal{V}\) relation in Equation~\ref{eq:compat_dabs11'}, we obtain:
\begin{gather}
  \label{eq:compat_dabs15}
   \intermed {  {\intermed \Sigma}_{{\mathrm{1}}}  } ; \intermed {  {\intermed { \Gamma_C } }  } ; \intermed {  \bullet  }  \vdash_{d}  \intermed {  {\intermed {dv} }_{{\mathrm{3}}}  } : \intermed {  {\intermed Q}  } \itot{\,  \rightsquigarrow  \target {  {\target e }'_{{\mathrm{3}}}  } } \\
  \label{eq:compat_dabs16}
   \intermed {  {\intermed \Sigma}_{{\mathrm{2}}}  } ; \intermed {  {\intermed { \Gamma_C } }  } ; \intermed {  \bullet  }  \vdash_{d}  \intermed {  {\intermed {dv} }_{{\mathrm{4}}}  } : \intermed {  {\intermed Q}  } \itot{\,  \rightsquigarrow  \target {  {\target e }'_{{\mathrm{4}}}  } } 
\end{gather}
We unfold
the definition of the \( \mathcal{E} \) relation in Goal~\ref{eq:compat_dabs11},
reducing it to:
\begin{gather}
   \intermed {  {\intermed \Sigma}_{{\mathrm{1}}}  } ; \intermed {  {\intermed { \Gamma_C } }  } ; \intermed {  \bullet  }  \vdash_{tm}  \intermed {  \ottsym{(}  \lambda  {\intermed \delta }  \ottsym{:}  {\intermed R }  \ottsym{(}  {\intermed Q}  \ottsym{)}  \ottsym{.}  \ottsym{(}   {\intermed \gamma }_{{\mathrm{1}}}  ( {\intermed \phi }_{{\mathrm{1}}}  ( {\intermed R }  ( {\intermed e}_{{\mathrm{1}}} )))   \ottsym{)}  \ottsym{)} \, {\intermed {dv} }_{{\mathrm{3}}}  } : \intermed {  {\intermed R }  \ottsym{(}  {\intermed \sigma}  \ottsym{)}  } \itot{ \rightsquigarrow  \target {  {\target e }_{{\mathrm{5}}}  } } 
  \label{eq:compat_dabs12}\\
   \intermed {  {\intermed \Sigma}_{{\mathrm{2}}}  } ; \intermed {  {\intermed { \Gamma_C } }  } ; \intermed {  \bullet  }  \vdash_{tm}  \intermed {  \ottsym{(}  \lambda  {\intermed \delta }  \ottsym{:}  {\intermed R }  \ottsym{(}  {\intermed Q}  \ottsym{)}  \ottsym{.}  \ottsym{(}   {\intermed \gamma }_{{\mathrm{2}}}  ( {\intermed \phi }_{{\mathrm{2}}}  ( {\intermed R }  ( {\intermed e}_{{\mathrm{2}}} )))   \ottsym{)}  \ottsym{)} \, {\intermed {dv} }_{{\mathrm{4}}}  } : \intermed {  {\intermed R }  \ottsym{(}  {\intermed \sigma}  \ottsym{)}  } \itot{ \rightsquigarrow  \target {  {\target e }_{{\mathrm{6}}}  } } 
  \label{eq:compat_dabs13}\\
\begin{split}
    \label{eq:compat_dabs14}
  \exists \intermed{{\intermed v}_{{\mathrm{3}}}}, \intermed{{\intermed v}_{{\mathrm{4}}}} & :
   \intermed {  {\intermed \Sigma}_{{\mathrm{1}}}  }  \vdash  \intermed {  \ottsym{(}  \lambda  {\intermed \delta }  \ottsym{:}  {\intermed R }  \ottsym{(}  {\intermed Q}  \ottsym{)}  \ottsym{.}  \ottsym{(}   {\intermed \gamma }_{{\mathrm{1}}}  ( {\intermed \phi }_{{\mathrm{1}}}  ( {\intermed R }  ( {\intermed e}_{{\mathrm{1}}} )))   \ottsym{)}  \ottsym{)} \, {\intermed {dv} }_{{\mathrm{3}}}  }  \longrightarrow^{*}  \intermed {  {\intermed v}_{{\mathrm{3}}}  } \\
  & \band  \intermed {  {\intermed \Sigma}_{{\mathrm{2}}}  }  \vdash  \intermed {  \ottsym{(}  \lambda  {\intermed \delta }  \ottsym{:}  {\intermed R }  \ottsym{(}  {\intermed Q}  \ottsym{)}  \ottsym{.}  \ottsym{(}   {\intermed \gamma }_{{\mathrm{2}}}  ( {\intermed \phi }_{{\mathrm{2}}}  ( {\intermed R }  ( {\intermed e}_{{\mathrm{2}}} )))   \ottsym{)}  \ottsym{)} \, {\intermed {dv} }_{{\mathrm{4}}}  }  \longrightarrow^{*}  \intermed {  {\intermed v}_{{\mathrm{4}}}  } \\
  & \band  ( \intermed {  {\intermed \Sigma}_{{\mathrm{1}}}  } : \intermed {  {\intermed v}_{{\mathrm{3}}}  } , \intermed {  {\intermed \Sigma}_{{\mathrm{2}}}  } : \intermed {  {\intermed v}_{{\mathrm{4}}}  } ) \in \mathcal{V} \llbracket \intermed {  {\intermed \sigma}  } \rrbracket ^{\intermed {  {\intermed { \Gamma_C } }  } }_{\intermed {  {\intermed R }  } } 
\end{split}
\end{gather}
Goals~\ref{eq:compat_dabs12} and \ref{eq:compat_dabs13} follow by applying the
\textsc{iTm-constrE} typing rule once on Equations~\ref{eq:compat_dabs9}
and~\ref{eq:compat_dabs15} and once on Equations~\ref{eq:compat_dabs10}
and~\ref{eq:compat_dabs16}.

Through case analysis, it is straightforward to note that the first step of
the evaluation paths in Equation~\ref{eq:compat_dabs14} should be by rule
\textsc{iEval-DAppAbs}. The goal reduces to:
\begin{align*}
  \exists \intermed{{\intermed v}_{{\mathrm{3}}}}, \intermed{{\intermed v}_{{\mathrm{4}}}} & :
   \intermed {  {\intermed \Sigma}_{{\mathrm{1}}}  }  \vdash  \intermed {  [  {\intermed {dv} }_{{\mathrm{3}}}  /  {\intermed \delta }  ]  \ottsym{(}   {\intermed \gamma }_{{\mathrm{1}}}  ( {\intermed \phi }_{{\mathrm{1}}}  ( {\intermed R }  ( {\intermed e}_{{\mathrm{1}}} )))   \ottsym{)}  }  \longrightarrow^{*}  \intermed {  {\intermed v}_{{\mathrm{3}}}  } \\
  & \band  \intermed {  {\intermed \Sigma}_{{\mathrm{2}}}  }  \vdash  \intermed {  [  {\intermed {dv} }_{{\mathrm{4}}}  /  {\intermed \delta }  ]  \ottsym{(}   {\intermed \gamma }_{{\mathrm{2}}}  ( {\intermed \phi }_{{\mathrm{2}}}  ( {\intermed R }  ( {\intermed e}_{{\mathrm{2}}} )))   \ottsym{)}  }  \longrightarrow^{*}  \intermed {  {\intermed v}_{{\mathrm{4}}}  } \\
  & \band  ( \intermed {  {\intermed \Sigma}_{{\mathrm{1}}}  } : \intermed {  {\intermed v}_{{\mathrm{3}}}  } , \intermed {  {\intermed \Sigma}_{{\mathrm{2}}}  } : \intermed {  {\intermed v}_{{\mathrm{4}}}  } ) \in \mathcal{V} \llbracket \intermed {  {\intermed \sigma}  } \rrbracket ^{\intermed {  {\intermed { \Gamma_C } }  } }_{\intermed {  {\intermed R }  } } 
\end{align*}
The above goal follows directly from Equation~\ref{eq:compat_dabs6}, by choosing \( \intermed {  {\intermed {dv} }_{{\mathrm{1}}}  } = \intermed {  {\intermed {dv} }_{{\mathrm{3}}}  } \), \( \intermed {  {\intermed {dv} }_{{\mathrm{2}}}  } = \intermed {  {\intermed {dv} }_{{\mathrm{4}}}  } \), \( \intermed {  {\intermed v}_{{\mathrm{3}}}  } = \intermed {  {\intermed v}_{{\mathrm{1}}}  } \) and \( \intermed {  {\intermed v}_{{\mathrm{4}}}  } = \intermed {  {\intermed v}_{{\mathrm{2}}}  } \). \\
\qed

\begin{lemmaB}[Compatibility - Dictionary Application]\ \\
  \begin{mathpar}
    \ottaltinferrule{}{}
    {
       \intermed {  {\intermed { \Gamma_C } }  } ; \intermed {  {\intermed \Gamma}  }  \vdash  \intermed {  {\intermed \Sigma}_{{\mathrm{1}}}  } : \intermed {  {\intermed e}_{{\mathrm{1}}}  }  \simeq_{log}  \intermed {  {\intermed \Sigma}_{{\mathrm{2}}}  } : \intermed {  {\intermed e}_{{\mathrm{2}}}  } : \intermed {   {\intermed Q}  \Rightarrow  {\intermed \sigma}   } 
      \\  \intermed {  {\intermed { \Gamma_C } }  } ; \intermed {  {\intermed \Gamma}  }  \vdash  \intermed {  {\intermed \Sigma}_{{\mathrm{1}}}  } : \intermed {  {\intermed d}_{{\mathrm{1}}}  }  \simeq_{log}  \intermed {  {\intermed \Sigma}_{{\mathrm{2}}}  } : \intermed {  {\intermed d}_{{\mathrm{2}}}  } : \intermed {  {\intermed Q}  } 
    }
    {  \intermed {  {\intermed { \Gamma_C } }  } ; \intermed {  {\intermed \Gamma}  }  \vdash  \intermed {  {\intermed \Sigma}_{{\mathrm{1}}}  } : \intermed {  {\intermed e}_{{\mathrm{1}}} \, {\intermed d}_{{\mathrm{1}}}  }  \simeq_{log}  \intermed {  {\intermed \Sigma}_{{\mathrm{2}}}  } : \intermed {  {\intermed e}_{{\mathrm{2}}} \, {\intermed d}_{{\mathrm{2}}}  } : \intermed {  {\intermed \sigma}  }  }
    \label{lemma:compat_dapp}
  \end{mathpar}
\end{lemmaB}

\proof
By inlining the definition of logical equivalence, suppose we have:
\begin{gather*}
   \intermed {  {\intermed R }  } \in \mathcal{F} \llbracket \intermed {  {\intermed \Gamma}  } \rrbracket ^{\intermed {  {\intermed { \Gamma_C } }  } } \\
   \intermed {  {\intermed \phi }  } \in \mathcal{G} \llbracket \intermed {  {\intermed \Gamma}  } \rrbracket ^{\intermed {  {\intermed \Sigma}_{{\mathrm{1}}}  } , \intermed {  {\intermed \Sigma}_{{\mathrm{2}}}  } , \intermed {  {\intermed { \Gamma_C } }  } }_{\intermed {  {\intermed R }  } } \\
   \intermed {  {\intermed \gamma }  } \in \mathcal{H} \llbracket \intermed {  {\intermed \Gamma}  } \rrbracket ^{\intermed {  {\intermed \Sigma}_{{\mathrm{1}}}  } , \intermed {  {\intermed \Sigma}_{{\mathrm{2}}}  } , \intermed {  {\intermed { \Gamma_C } }  } }_{\intermed {  {\intermed R }  } } 
\end{gather*}
The goal to be proven is the following:
\begin{equation*}
   ( \intermed {  {\intermed \Sigma}_{{\mathrm{1}}}  } : \intermed {   {\intermed \gamma }_{{\mathrm{1}}}  ( {\intermed \phi }_{{\mathrm{1}}}  ( {\intermed R }  (  {\intermed e}_{{\mathrm{1}}} \, {\intermed d}_{{\mathrm{1}}}  )))   } , \intermed {  {\intermed \Sigma}_{{\mathrm{2}}}  } : \intermed {   {\intermed \gamma }_{{\mathrm{2}}}  ( {\intermed \phi }_{{\mathrm{2}}}  ( {\intermed R }  (  {\intermed e}_{{\mathrm{2}}} \, {\intermed d}_{{\mathrm{2}}}  )))   } ) \in \mathcal{E} \llbracket \intermed {  {\intermed \sigma}  } \rrbracket ^{\intermed {  {\intermed { \Gamma_C } }  } }_{\intermed {  {\intermed R }  } } 
\end{equation*}
By unfolding the definition of the \( \mathcal{E} \) relation in the goal above, and by
simplifying the substitutions, the goal reduces to:
\begin{gather}
  \label{eq:compat_dapp4}
   \intermed {  {\intermed \Sigma}_{{\mathrm{1}}}  } ; \intermed {  {\intermed { \Gamma_C } }  } ; \intermed {  \bullet  }  \vdash_{tm}  \intermed {  \ottsym{(}   {\intermed \gamma }_{{\mathrm{1}}}  ( {\intermed \phi }_{{\mathrm{1}}}  ( {\intermed R }  ( {\intermed e}_{{\mathrm{1}}} )))   \ottsym{)} \, \ottsym{(}   {\intermed \gamma }_{{\mathrm{1}}}  ( {\intermed \phi }_{{\mathrm{1}}}  ( {\intermed R }  ( {\intermed d}_{{\mathrm{1}}} )))   \ottsym{)}  } : \intermed {  {\intermed R }  \ottsym{(}  {\intermed \sigma}  \ottsym{)}  } \itot{ \rightsquigarrow  \target {  {\target e }  } } \\
  \label{eq:compat_dapp5}
   \intermed {  {\intermed \Sigma}_{{\mathrm{2}}}  } ; \intermed {  {\intermed { \Gamma_C } }  } ; \intermed {  \bullet  }  \vdash_{tm}  \intermed {  \ottsym{(}   {\intermed \gamma }_{{\mathrm{2}}}  ( {\intermed \phi }_{{\mathrm{2}}}  ( {\intermed R }  ( {\intermed e}_{{\mathrm{2}}} )))   \ottsym{)} \, \ottsym{(}   {\intermed \gamma }_{{\mathrm{2}}}  ( {\intermed \phi }_{{\mathrm{2}}}  ( {\intermed R }  ( {\intermed d}_{{\mathrm{2}}} )))   \ottsym{)}  } : \intermed {  {\intermed R }  \ottsym{(}  {\intermed \sigma}  \ottsym{)}  } \itot{ \rightsquigarrow  \target {  {\target e }'  } } \\
  \label{eq:compat_dapp17}
\begin{split}
  \exists \intermed{{\intermed v}_{{\mathrm{1}}}}, \intermed{{\intermed v}_{{\mathrm{2}}}} & :
     \intermed {  {\intermed \Sigma}_{{\mathrm{1}}}  }  \vdash  \intermed {  \ottsym{(}   {\intermed \gamma }_{{\mathrm{1}}}  ( {\intermed \phi }_{{\mathrm{1}}}  ( {\intermed R }  ( {\intermed e}_{{\mathrm{1}}} )))   \ottsym{)} \, \ottsym{(}   {\intermed \gamma }_{{\mathrm{1}}}  ( {\intermed \phi }_{{\mathrm{1}}}  ( {\intermed R }  ( {\intermed d}_{{\mathrm{1}}} )))   \ottsym{)}  }  \longrightarrow^{*}  \intermed {  {\intermed v}_{{\mathrm{1}}}  } \\ 
    & \band  \intermed {  {\intermed \Sigma}_{{\mathrm{2}}}  }  \vdash  \intermed {  \ottsym{(}   {\intermed \gamma }_{{\mathrm{2}}}  ( {\intermed \phi }_{{\mathrm{2}}}  ( {\intermed R }  ( {\intermed e}_{{\mathrm{2}}} )))   \ottsym{)} \, \ottsym{(}   {\intermed \gamma }_{{\mathrm{2}}}  ( {\intermed \phi }_{{\mathrm{2}}}  ( {\intermed R }  ( {\intermed d}_{{\mathrm{2}}} )))   \ottsym{)}  }  \longrightarrow^{*}  \intermed {  {\intermed v}_{{\mathrm{2}}}  } \\ 
    & \band  ( \intermed {  {\intermed \Sigma}_{{\mathrm{1}}}  } : \intermed {  {\intermed v}_{{\mathrm{1}}}  } , \intermed {  {\intermed \Sigma}_{{\mathrm{2}}}  } : \intermed {  {\intermed v}_{{\mathrm{2}}}  } ) \in \mathcal{V} \llbracket \intermed {  {\intermed \sigma}  } \rrbracket ^{\intermed {  {\intermed { \Gamma_C } }  } }_{\intermed {  {\intermed R }  } } 
\end{split}
\end{gather}
By inlining the definitions of logical equivalence and the \(\mathcal{E}\)
relation in the first premise of this lemma, we get:
\begin{gather}
  \label{eq:compat_dapp6}
   \intermed {  {\intermed \Sigma}_{{\mathrm{1}}}  } ; \intermed {  {\intermed { \Gamma_C } }  } ; \intermed {  \bullet  }  \vdash_{tm}  \intermed {   {\intermed \gamma }_{{\mathrm{1}}}  ( {\intermed \phi }_{{\mathrm{1}}}  ( {\intermed R }  ( {\intermed e}_{{\mathrm{1}}} )))   } : \intermed {  {\intermed R }  \ottsym{(}   {\intermed Q}  \Rightarrow  {\intermed \sigma}   \ottsym{)}  } \itot{ \rightsquigarrow  \target {  {\target e }_{{\mathrm{1}}}  } } \\
  \label{eq:compat_dapp7}
   \intermed {  {\intermed \Sigma}_{{\mathrm{2}}}  } ; \intermed {  {\intermed { \Gamma_C } }  } ; \intermed {  \bullet  }  \vdash_{tm}  \intermed {   {\intermed \gamma }_{{\mathrm{2}}}  ( {\intermed \phi }_{{\mathrm{2}}}  ( {\intermed R }  ( {\intermed e}_{{\mathrm{2}}} )))   } : \intermed {  {\intermed R }  \ottsym{(}   {\intermed Q}  \Rightarrow  {\intermed \sigma}   \ottsym{)}  } \itot{ \rightsquigarrow  \target {  {\target e }_{{\mathrm{2}}}  } } \\
  \label{eq:compat_dapp8}
  \begin{split}
  \exists \intermed{{\intermed v}_{{\mathrm{3}}}}, \intermed{{\intermed v}_{{\mathrm{4}}}} & :
     \intermed {  {\intermed \Sigma}_{{\mathrm{1}}}  }  \vdash  \intermed {   {\intermed \gamma }_{{\mathrm{1}}}  ( {\intermed \phi }_{{\mathrm{1}}}  ( {\intermed R }  ( {\intermed e}_{{\mathrm{1}}} )))   }  \longrightarrow^{*}  \intermed {  {\intermed v}_{{\mathrm{3}}}  } \\
    & \band  \intermed {  {\intermed \Sigma}_{{\mathrm{2}}}  }  \vdash  \intermed {   {\intermed \gamma }_{{\mathrm{2}}}  ( {\intermed \phi }_{{\mathrm{2}}}  ( {\intermed R }  ( {\intermed e}_{{\mathrm{2}}} )))   }  \longrightarrow^{*}  \intermed {  {\intermed v}_{{\mathrm{4}}}  } \\
    & \band  ( \intermed {  {\intermed \Sigma}_{{\mathrm{1}}}  } : \intermed {  {\intermed v}_{{\mathrm{3}}}  } , \intermed {  {\intermed \Sigma}_{{\mathrm{2}}}  } : \intermed {  {\intermed v}_{{\mathrm{4}}}  } ) \in \mathcal{V} \llbracket \intermed {   {\intermed Q}  \Rightarrow  {\intermed \sigma}   } \rrbracket ^{\intermed {  {\intermed { \Gamma_C } }  } }_{\intermed {  {\intermed R }  } } 
  \end{split}
\end{gather}
Similarly, by unfolding the definition of logical equivalence 
in the second premise of the rule, we get:
\begin{equation}
   ( \intermed {  {\intermed \Sigma}_{{\mathrm{1}}}  } : \intermed {   {\intermed \gamma }_{{\mathrm{1}}}  ( {\intermed \phi }_{{\mathrm{1}}}  ( {\intermed R }  ( {\intermed d}_{{\mathrm{1}}} )))   } , \intermed {  {\intermed \Sigma}_{{\mathrm{2}}}  } : \intermed {   {\intermed \gamma }_{{\mathrm{2}}}  ( {\intermed \phi }_{{\mathrm{2}}}  ( {\intermed R }  ( {\intermed d}_{{\mathrm{2}}} )))   } ) \in \mathcal{V} \llbracket \intermed {  {\intermed Q}  } \rrbracket ^{\intermed {  {\intermed { \Gamma_C } }  } }_{\intermed {  {\intermed R }  } } 
  \label{eq:compat_dapp16}
\end{equation}
From the definition of the \(\mathcal{V}\) relation in
Equation~\ref{eq:compat_dapp16} we have:
\begin{gather}
  \label{eq:compat_dapp11}
   \intermed {  {\intermed \Sigma}_{{\mathrm{1}}}  } ; \intermed {  {\intermed { \Gamma_C } }  } ; \intermed {  \bullet  }  \vdash_{d}  \intermed {   {\intermed \gamma }_{{\mathrm{1}}}  ( {\intermed \phi }_{{\mathrm{1}}}  ( {\intermed R }  ( {\intermed d}_{{\mathrm{1}}} )))   } : \intermed {  {\intermed R }  \ottsym{(}  {\intermed Q}  \ottsym{)}  } \itot{\,  \rightsquigarrow  \target {  {\target e }_{{\mathrm{3}}}  } } \\
  \label{eq:compat_dapp12}
   \intermed {  {\intermed \Sigma}_{{\mathrm{2}}}  } ; \intermed {  {\intermed { \Gamma_C } }  } ; \intermed {  \bullet  }  \vdash_{d}  \intermed {   {\intermed \gamma }_{{\mathrm{2}}}  ( {\intermed \phi }_{{\mathrm{2}}}  ( {\intermed R }  ( {\intermed d}_{{\mathrm{2}}} )))   } : \intermed {  {\intermed R }  \ottsym{(}  {\intermed Q}  \ottsym{)}  } \itot{\,  \rightsquigarrow  \target {  {\target e }_{{\mathrm{4}}}  } } 
\end{gather}
Note that, by the definition of substitution, we have
\( \intermed {  {\intermed R }  \ottsym{(}   {\intermed Q}  \Rightarrow  {\intermed \sigma}   \ottsym{)}  } = \intermed {   {\intermed R }  \ottsym{(}  {\intermed Q}  \ottsym{)}  \Rightarrow  {\intermed R }  \ottsym{(}  {\intermed \sigma}  \ottsym{)}   } \). This allows the application of
the typing rule \textsc{iTm-constrE} once on
Equations~\ref{eq:compat_dapp6} and \ref{eq:compat_dapp11} and once more on
Equations~\ref{eq:compat_dapp7} and \ref{eq:compat_dapp12}, therefore proving
Goals~\ref{eq:compat_dapp4} and \ref{eq:compat_dapp5}.

Through application of the \textsc{iEval-DApp} evaluation rule on each step of the two
evaluation paths in Equation~\ref{eq:compat_dapp8},
Goal~\ref{eq:compat_dapp17} reduces to:
\begin{align}
\label{eq:compat_dapp18}
\begin{split}
  \exists \intermed{{\intermed v}_{{\mathrm{1}}}}, \intermed{{\intermed v}_{{\mathrm{2}}}} & :
     \intermed {  {\intermed \Sigma}_{{\mathrm{1}}}  }  \vdash  \intermed {  {\intermed v}_{{\mathrm{3}}} \, \ottsym{(}   {\intermed \gamma }_{{\mathrm{1}}}  ( {\intermed \phi }_{{\mathrm{1}}}  ( {\intermed R }  ( {\intermed d}_{{\mathrm{1}}} )))   \ottsym{)}  }  \longrightarrow^{*}  \intermed {  {\intermed v}_{{\mathrm{1}}}  } \\  
    & \band  \intermed {  {\intermed \Sigma}_{{\mathrm{2}}}  }  \vdash  \intermed {  {\intermed v}_{{\mathrm{4}}} \, \ottsym{(}   {\intermed \gamma }_{{\mathrm{2}}}  ( {\intermed \phi }_{{\mathrm{2}}}  ( {\intermed R }  ( {\intermed d}_{{\mathrm{2}}} )))   \ottsym{)}  }  \longrightarrow^{*}  \intermed {  {\intermed v}_{{\mathrm{2}}}  } \\ 
    & \band  ( \intermed {  {\intermed \Sigma}_{{\mathrm{1}}}  } : \intermed {  {\intermed v}_{{\mathrm{1}}}  } , \intermed {  {\intermed \Sigma}_{{\mathrm{2}}}  } : \intermed {  {\intermed v}_{{\mathrm{2}}}  } ) \in \mathcal{V} \llbracket \intermed {  {\intermed \sigma}  } \rrbracket ^{\intermed {  {\intermed { \Gamma_C } }  } }_{\intermed {  {\intermed R }  } } 
\end{split}
\end{align}
Unfolding the definition of the \(\mathcal{V}\) relation in
Equation~\ref{eq:compat_dapp8} results in:
\begin{gather}
  \nonumber
   \intermed {  {\intermed \Sigma}_{{\mathrm{1}}}  } ; \intermed {  {\intermed { \Gamma_C } }  } ; \intermed {  \bullet  }  \vdash_{tm}  \intermed {  {\intermed v}_{{\mathrm{3}}}  } : \intermed {  {\intermed R }  \ottsym{(}   {\intermed Q}  \Rightarrow  {\intermed \sigma}   \ottsym{)}  } \itot{ \rightsquigarrow  \target {  {\target e }_{{\mathrm{5}}}  } } \\
  \nonumber
   \intermed {  {\intermed \Sigma}_{{\mathrm{2}}}  } ; \intermed {  {\intermed { \Gamma_C } }  } ; \intermed {  \bullet  }  \vdash_{tm}  \intermed {  {\intermed v}_{{\mathrm{4}}}  } : \intermed {  {\intermed R }  \ottsym{(}   {\intermed Q}  \Rightarrow  {\intermed \sigma}   \ottsym{)}  } \itot{ \rightsquigarrow  \target {  {\target e }_{{\mathrm{6}}}  } } \\
  \label{eq:compat_dapp15}
  \forall  ( \intermed {  {\intermed \Sigma}_{{\mathrm{1}}}  } : \intermed {  {\intermed {dv} }_{{\mathrm{1}}}  } , \intermed {  {\intermed \Sigma}_{{\mathrm{2}}}  } : \intermed {  {\intermed {dv} }_{{\mathrm{2}}}  } ) \in \mathcal{V} \llbracket \intermed {  {\intermed Q}  } \rrbracket ^{\intermed {  {\intermed { \Gamma_C } }  } }_{\intermed {  {\intermed R }  } }  :
           ( \intermed {  {\intermed \Sigma}_{{\mathrm{1}}}  } : \intermed {  {\intermed v}_{{\mathrm{3}}} \, {\intermed {dv} }_{{\mathrm{1}}}  } , \intermed {  {\intermed \Sigma}_{{\mathrm{2}}}  } : \intermed {  {\intermed v}_{{\mathrm{4}}} \, {\intermed {dv} }_{{\mathrm{2}}}  } ) \in \mathcal{E} \llbracket \intermed {  {\intermed \sigma}  } \rrbracket ^{\intermed {  {\intermed { \Gamma_C } }  } }_{\intermed {  {\intermed R }  } }   
\end{gather}
We take \( \intermed {  {\intermed {dv} }_{{\mathrm{1}}}  } = \intermed {   {\intermed \gamma }_{{\mathrm{1}}}  ( {\intermed \phi }_{{\mathrm{1}}}  ( {\intermed R }  ( {\intermed d}_{{\mathrm{1}}} )))   } \) and
\( \intermed {  {\intermed {dv} }_{{\mathrm{2}}}  } = \intermed {   {\intermed \gamma }_{{\mathrm{2}}}  ( {\intermed \phi }_{{\mathrm{2}}}  ( {\intermed R }  ( {\intermed d}_{{\mathrm{2}}} )))   } \).
Goal~\ref{eq:compat_dapp18} follows from the definition of the \(\mathcal{E}\)
relation in Equation~\ref{eq:compat_dapp15}. \\
\qed

\begin{lemmaB}[Compatibility - Type Abstraction]\ \\
  \begin{mathpar}
    \ottaltinferrule{}{}
    {
       \intermed {  {\intermed { \Gamma_C } }  } ; \intermed {  {\intermed \Gamma}  \ottsym{,}  {\intermed a }  }  \vdash  \intermed {  {\intermed \Sigma}_{{\mathrm{1}}}  } : \intermed {  {\intermed e}_{{\mathrm{1}}}  }  \simeq_{log}  \intermed {  {\intermed \Sigma}_{{\mathrm{2}}}  } : \intermed {  {\intermed e}_{{\mathrm{2}}}  } : \intermed {  {\intermed \sigma}  } 
    }
    {  \intermed {  {\intermed { \Gamma_C } }  } ; \intermed {  {\intermed \Gamma}  }  \vdash  \intermed {  {\intermed \Sigma}_{{\mathrm{1}}}  } : \intermed {  \Lambda  {\intermed a }  \ottsym{.}  {\intermed e}_{{\mathrm{1}}}  }  \simeq_{log}  \intermed {  {\intermed \Sigma}_{{\mathrm{2}}}  } : \intermed {  \Lambda  {\intermed a }  \ottsym{.}  {\intermed e}_{{\mathrm{2}}}  } : \intermed {  \forall  {\intermed a }  \ottsym{.}  {\intermed \sigma}  }  }
    \label{lemma:compat_tyabs}
  \end{mathpar}
\end{lemmaB}

\proof
By unfolding the definition of logical equivalence, suppose we have:
\begin{gather}
  \label{eq:comat_tyabsR}
   \intermed {  {\intermed R }  } \in \mathcal{F} \llbracket \intermed {  {\intermed \Gamma}  } \rrbracket ^{\intermed {  {\intermed { \Gamma_C } }  } } \\
  \label{eq:comat_tyabsG}
   \intermed {  {\intermed \phi }  } \in \mathcal{G} \llbracket \intermed {  {\intermed \Gamma}  } \rrbracket ^{\intermed {  {\intermed \Sigma}_{{\mathrm{1}}}  } , \intermed {  {\intermed \Sigma}_{{\mathrm{2}}}  } , \intermed {  {\intermed { \Gamma_C } }  } }_{\intermed {  {\intermed R }  } } \\
  \label{eq:comat_tyabsH}
   \intermed {  {\intermed \gamma }  } \in \mathcal{H} \llbracket \intermed {  {\intermed \Gamma}  } \rrbracket ^{\intermed {  {\intermed \Sigma}_{{\mathrm{1}}}  } , \intermed {  {\intermed \Sigma}_{{\mathrm{2}}}  } , \intermed {  {\intermed { \Gamma_C } }  } }_{\intermed {  {\intermed R }  } } 
\end{gather}
The goal to be proven is the following:
\begin{equation*}
   ( \intermed {  {\intermed \Sigma}_{{\mathrm{1}}}  } : \intermed {   {\intermed \gamma }_{{\mathrm{1}}}  ( {\intermed \phi }_{{\mathrm{1}}}  ( {\intermed R }  ( \Lambda  {\intermed a }  \ottsym{.}  {\intermed e}_{{\mathrm{1}}} )))   } , \intermed {  {\intermed \Sigma}_{{\mathrm{2}}}  } : \intermed {   {\intermed \gamma }_{{\mathrm{2}}}  ( {\intermed \phi }_{{\mathrm{2}}}  ( {\intermed R }  (  \Lambda  {\intermed a }  \ottsym{.}  {\intermed e}_{{\mathrm{2}}}  )))   } ) \in \mathcal{E} \llbracket \intermed {  \forall  {\intermed a }  \ottsym{.}  {\intermed \sigma}  } \rrbracket ^{\intermed {  {\intermed { \Gamma_C } }  } }_{\intermed {  {\intermed R }  } } 
\end{equation*}
Because \( \intermed {  {\intermed a }  }  \notin  \intermed {  {\intermed \Gamma}  } \), from Equation~\ref{eq:comat_tyabsR} it follows that
\({\intermed a }\) is not in the domain of \({\intermed R }\). Furthermore, from
Equations~\ref{eq:comat_tyabsG} and~\ref{eq:comat_tyabsH} it follows that for every mapping
\( \intermed {  {\intermed x}  }  \mapsto  ( \intermed {  {\intermed e}'_{{\mathrm{1}}}  } , \intermed {  {\intermed e}'_{{\mathrm{2}}}  } )  \in  \intermed {  {\intermed \phi }  } \) and for every
mapping \( \intermed {  {\intermed \delta }  }  \mapsto  ( \intermed {  {\intermed {dv} }'_{{\mathrm{1}}}  } , \intermed {  {\intermed {dv} }'_{{\mathrm{2}}}  } )  \in  \intermed {  {\intermed \gamma }  } \), we have
\({\intermed a } \notin  \textbf {fv} (\intermed{{\intermed e}'_{\ottmv{i}}})\) and
\({\intermed a } \notin  \textbf {fv} (\intermed{{\intermed {dv} }'_{\ottmv{i}}})\), where \(\intermed{\ottmv{i}}\in\{1,2\}\).
Therefore, we obtain
\( \intermed {   {\intermed \gamma }_{\ottmv{i}}  ( {\intermed \phi }_{\ottmv{i}}  ( {\intermed R }  (  \Lambda  {\intermed a }  \ottsym{.}  {\intermed e}_{\ottmv{i}}  )))   } = \intermed {  \Lambda  {\intermed a }  \ottsym{.}    {\intermed \gamma }_{\ottmv{i}}  ( {\intermed \phi }_{\ottmv{i}}  ( {\intermed R }  ( {\intermed e}_{\ottmv{i}} )))    } \),
for \(\intermed{\ottmv{i}}\in\{1,2\}\). With these equations, the goal above reduces to
\begin{equation*}
   ( \intermed {  {\intermed \Sigma}_{{\mathrm{1}}}  } : \intermed {  \Lambda  {\intermed a }  \ottsym{.}    {\intermed \gamma }_{{\mathrm{1}}}  ( {\intermed \phi }_{{\mathrm{1}}}  ( {\intermed R }  ( {\intermed e}_{{\mathrm{1}}} )))    } , \intermed {  {\intermed \Sigma}_{{\mathrm{2}}}  } : \intermed {  \Lambda  {\intermed a }  \ottsym{.}    {\intermed \gamma }_{{\mathrm{2}}}  ( {\intermed \phi }_{{\mathrm{2}}}  ( {\intermed R }  ( {\intermed e}_{{\mathrm{2}}} )))    } ) \in \mathcal{E} \llbracket \intermed {  \forall  {\intermed a }  \ottsym{.}  {\intermed \sigma}  } \rrbracket ^{\intermed {  {\intermed { \Gamma_C } }  } }_{\intermed {  {\intermed R }  } } 
\end{equation*}
By applying the definition of the \( \mathcal{E} \) relation, taking into
account that expressions of the form \(\intermed{\Lambda  {\intermed a }  \ottsym{.}  {\intermed e}}\) are values, the goal
reduces to:
\begin{gather}
  \label{eq:compat_tyabs1}
   \intermed {  {\intermed \Sigma}_{{\mathrm{1}}}  } ; \intermed {  {\intermed { \Gamma_C } }  } ; \intermed {  \bullet  }  \vdash_{tm}  \intermed {  \Lambda  {\intermed a }  \ottsym{.}    {\intermed \gamma }_{{\mathrm{1}}}  ( {\intermed \phi }_{{\mathrm{1}}}  ( {\intermed R }  ( {\intermed e}_{{\mathrm{1}}} )))    } : \intermed {  {\intermed R }  \ottsym{(}  \forall  {\intermed a }  \ottsym{.}  {\intermed \sigma}  \ottsym{)}  } \itot{ \rightsquigarrow  \target {  {\target e }  } } \\
  \label{eq:compat_tyabs2}
   \intermed {  {\intermed \Sigma}_{{\mathrm{2}}}  } ; \intermed {  {\intermed { \Gamma_C } }  } ; \intermed {  \bullet  }  \vdash_{tm}  \intermed {  \Lambda  {\intermed a }  \ottsym{.}    {\intermed \gamma }_{{\mathrm{2}}}  ( {\intermed \phi }_{{\mathrm{2}}}  ( {\intermed R }  ( {\intermed e}_{{\mathrm{2}}} )))    } : \intermed {  {\intermed R }  \ottsym{(}  \forall  {\intermed a }  \ottsym{.}  {\intermed \sigma}  \ottsym{)}  } \itot{ \rightsquigarrow  \target {  {\target e }'  } } \\
  \label{eq:compat_tyabs3}
   ( \intermed {  {\intermed \Sigma}_{{\mathrm{1}}}  } : \intermed {  \Lambda  {\intermed a }  \ottsym{.}    {\intermed \gamma }_{{\mathrm{1}}}  ( {\intermed \phi }_{{\mathrm{1}}}  ( {\intermed R }  ( {\intermed e}_{{\mathrm{1}}} )))    } , \intermed {  {\intermed \Sigma}_{{\mathrm{2}}}  } : \intermed {  \Lambda  {\intermed a }  \ottsym{.}    {\intermed \gamma }_{{\mathrm{2}}}  ( {\intermed \phi }_{{\mathrm{2}}}  ( {\intermed R }  ( {\intermed e}_{{\mathrm{2}}} )))    } ) \in \mathcal{V} \llbracket \intermed {  \forall  {\intermed a }  \ottsym{.}  {\intermed \sigma}  } \rrbracket ^{\intermed {  {\intermed { \Gamma_C } }  } }_{\intermed {  {\intermed R }  } } 
\end{gather}
Suppose any \(\intermed{{\intermed \sigma}'}\) such that 
\begin{equation}
  \label{eq:compat_tyabs20}
   \intermed {  {\intermed { \Gamma_C } }  } ; \intermed {  \bullet  }  \vdash_{ty}  \intermed {  {\intermed \sigma}'  } \itot{ \rightsquigarrow  \target {  {\target \sigma }'  } } 
\end{equation}
and any \( \intermed {  { \intermed { \mathit{ r } } }  } \in \mathit{Rel} [ \intermed {  {\intermed \sigma}'  } ] \). Then, inlining the
definition of the \(\mathcal{V}\) relation in
Goal~\ref{eq:compat_tyabs3}, reduces it to:
\begin{equation}
  \label{eq:compat_tyabs8}
  ( \intermed {  {\intermed \Sigma}_{{\mathrm{1}}}  } : \intermed {  \ottsym{(}  \Lambda  {\intermed a }  \ottsym{.}    {\intermed \gamma }_{{\mathrm{1}}}  ( {\intermed \phi }_{{\mathrm{1}}}  ( {\intermed R }  ( {\intermed e}_{{\mathrm{1}}} )))    \ottsym{)} \, {\intermed \sigma}'  } , \intermed {  {\intermed \Sigma}_{{\mathrm{2}}}  } : \intermed {  \ottsym{(}  \Lambda  {\intermed a }  \ottsym{.}    {\intermed \gamma }_{{\mathrm{2}}}  ( {\intermed \phi }_{{\mathrm{2}}}  ( {\intermed R }  ( {\intermed e}_{{\mathrm{2}}} )))    \ottsym{)} \, {\intermed \sigma}'  } ) \in \mathcal{E} \llbracket \intermed {  {\intermed \sigma}  } \rrbracket ^{\intermed {  {\intermed { \Gamma_C } }  } }_{\intermed {   {\intermed R }  ,  {\intermed a }  \mapsto (  {\intermed \sigma}' ,  { \intermed { \mathit{ r } } }  )   } } 
\end{equation}

Unfolding the definition of logical equivalence in the premise of this lemma, gives us:
\begin{equation}
   ( \intermed {  {\intermed \Sigma}_{{\mathrm{1}}}  } : \intermed {   {\intermed \gamma }'_{{\mathrm{1}}}  ( {\intermed \phi }'_{{\mathrm{1}}}  ( {\intermed R }'  ( {\intermed e}_{{\mathrm{1}}} )))   } , \intermed {  {\intermed \Sigma}_{{\mathrm{2}}}  } : \intermed {   {\intermed \gamma }'_{{\mathrm{2}}}  ( {\intermed \phi }'_{{\mathrm{2}}}  ( {\intermed R }'  ( {\intermed e}_{{\mathrm{2}}} )))   } ) \in \mathcal{E} \llbracket \intermed {  {\intermed \sigma}  } \rrbracket ^{\intermed {  {\intermed { \Gamma_C } }  } }_{\intermed {  {\intermed R }'  } } 
  \label{eq:compat_tyabs4}
\end{equation}
for any \( \intermed {  {\intermed R }'  } \in \mathcal{F} \llbracket \intermed {  {\intermed \Gamma}  \ottsym{,}  {\intermed a }  } \rrbracket ^{\intermed {  {\intermed { \Gamma_C } }  } } \),
\( \intermed {  {\intermed \phi }'  } \in \mathcal{G} \llbracket \intermed {  {\intermed \Gamma}  \ottsym{,}  {\intermed a }  } \rrbracket ^{\intermed {  {\intermed \Sigma}_{{\mathrm{1}}}  } , \intermed {  {\intermed \Sigma}_{{\mathrm{2}}}  } , \intermed {  {\intermed { \Gamma_C } }  } }_{\intermed {  {\intermed R }'  } } \) and
\( \intermed {  {\intermed \gamma }'  } \in \mathcal{H} \llbracket \intermed {  {\intermed \Gamma}  \ottsym{,}  {\intermed a }  } \rrbracket ^{\intermed {  {\intermed \Sigma}_{{\mathrm{1}}}  } , \intermed {  {\intermed \Sigma}_{{\mathrm{2}}}  } , \intermed {  {\intermed { \Gamma_C } }  } }_{\intermed {  {\intermed R }'  } } \).

By the definition of the \( \mathcal{F} \) relation, we know that
\( \intermed {  {\intermed R }'  } = \intermed {   {\intermed R }''  ,  {\intermed a }  \mapsto (  {\intermed \sigma}'' ,  { \intermed { \mathit{ r } } }'  )   } \) for some \( \intermed {  {\intermed R }''  } \in \mathcal{F} \llbracket \intermed {  {\intermed \Gamma}  } \rrbracket ^{\intermed {  {\intermed { \Gamma_C } }  } } \) and
\(\intermed{{\intermed \sigma}''}\) such that
\( \intermed {  {\intermed { \Gamma_C } }  } ; \intermed {  \bullet  }  \vdash_{ty}  \intermed {  {\intermed \sigma}''  } \itot{ \rightsquigarrow  \target {  {\target \sigma }'  } } \) and \( \intermed {  { \intermed { \mathit{ r } } }'  } \in \mathit{Rel} [ \intermed {  {\intermed \sigma}''  } ] \).
We choose \( \intermed {  {\intermed R }''  } = \intermed {  {\intermed R }  } \), \( \intermed {  {\intermed \sigma}''  } = \intermed {  {\intermed \sigma}'  } \) and \( \intermed {  { \intermed { \mathit{ r } } }'  } = \intermed {  { \intermed { \mathit{ r } } }  } \).
By the definition of the \( \mathcal{G} \) and \( \mathcal{H} \) relations and from
Equations~\ref{eq:comat_tyabsG} and~\ref{eq:comat_tyabsH}, we have that
\( \intermed {  {\intermed \phi }  } \in \mathcal{G} \llbracket \intermed {  {\intermed \Gamma}  \ottsym{,}  {\intermed a }  } \rrbracket ^{\intermed {  {\intermed \Sigma}_{{\mathrm{1}}}  } , \intermed {  {\intermed \Sigma}_{{\mathrm{2}}}  } , \intermed {  {\intermed { \Gamma_C } }  } }_{\intermed {  {\intermed R }  } } \) and \( \intermed {  {\intermed \gamma }  } \in \mathcal{H} \llbracket \intermed {  {\intermed \Gamma}  \ottsym{,}  {\intermed a }  } \rrbracket ^{\intermed {  {\intermed \Sigma}_{{\mathrm{1}}}  } , \intermed {  {\intermed \Sigma}_{{\mathrm{2}}}  } , \intermed {  {\intermed { \Gamma_C } }  } }_{\intermed {  {\intermed R }  } } \). Then, we choose \( \intermed {  {\intermed \phi }'  } = \intermed {  {\intermed \phi }  } \) and
\( \intermed {  {\intermed \gamma }'  } = \intermed {  {\intermed \gamma }  } \).

Unfolding the definition of the \(\mathcal{E}\) relation in
Equation~\ref{eq:compat_tyabs4}, results in:
\begin{gather}
  \label{eq:compat_tyabs5}
   \intermed {  {\intermed \Sigma}_{{\mathrm{1}}}  } ; \intermed {  {\intermed { \Gamma_C } }  } ; \intermed {  \bullet  }  \vdash_{tm}  \intermed {   {\intermed \gamma }_{{\mathrm{1}}}  ( {\intermed \phi }_{{\mathrm{1}}}  (  {\intermed R }  ,  {\intermed a }  \mapsto (  {\intermed \sigma}' ,  { \intermed { \mathit{ r } } }  )   ( {\intermed e}_{{\mathrm{1}}} )))   } : \intermed {  \ottsym{(}   {\intermed R }  ,  {\intermed a }  \mapsto (  {\intermed \sigma}' ,  { \intermed { \mathit{ r } } }  )   \ottsym{)}  \ottsym{(}  {\intermed \sigma}  \ottsym{)}  } \itot{ \rightsquigarrow  \target {  {\target e }_{{\mathrm{1}}}  } } \\
  \label{eq:compat_tyabs6}
   \intermed {  {\intermed \Sigma}_{{\mathrm{2}}}  } ; \intermed {  {\intermed { \Gamma_C } }  } ; \intermed {  \bullet  }  \vdash_{tm}  \intermed {   {\intermed \gamma }_{{\mathrm{2}}}  ( {\intermed \phi }_{{\mathrm{2}}}  (  {\intermed R }  ,  {\intermed a }  \mapsto (  {\intermed \sigma}' ,  { \intermed { \mathit{ r } } }  )   ( {\intermed e}_{{\mathrm{2}}} )))   } : \intermed {  \ottsym{(}   {\intermed R }  ,  {\intermed a }  \mapsto (  {\intermed \sigma}' ,  { \intermed { \mathit{ r } } }  )   \ottsym{)}  \ottsym{(}  {\intermed \sigma}  \ottsym{)}  } \itot{ \rightsquigarrow  \target {  {\target e }_{{\mathrm{2}}}  } } \\
  \label{eq:compat_tyabs7}
\begin{split}
  \exists \intermed{{\intermed v}_{{\mathrm{3}}}}, \intermed{{\intermed v}_{{\mathrm{4}}}} & :
   \intermed {  {\intermed \Sigma}_{{\mathrm{1}}}  }  \vdash  \intermed {   {\intermed \gamma }_{{\mathrm{1}}}  ( {\intermed \phi }_{{\mathrm{1}}}  (  {\intermed R }  ,  {\intermed a }  \mapsto (  {\intermed \sigma}' ,  { \intermed { \mathit{ r } } }  )   ( {\intermed e}_{{\mathrm{1}}} )))   }  \longrightarrow^{*}  \intermed {  {\intermed v}_{{\mathrm{3}}}  } \\
  & \band  \intermed {  {\intermed \Sigma}_{{\mathrm{2}}}  }  \vdash  \intermed {   {\intermed \gamma }_{{\mathrm{2}}}  ( {\intermed \phi }_{{\mathrm{2}}}  (  {\intermed R }  ,  {\intermed a }  \mapsto (  {\intermed \sigma}' ,  { \intermed { \mathit{ r } } }  )   ( {\intermed e}_{{\mathrm{2}}} )))   }  \longrightarrow^{*}  \intermed {  {\intermed v}_{{\mathrm{4}}}  } \\
  & \band  ( \intermed {  {\intermed \Sigma}_{{\mathrm{1}}}  } : \intermed {  {\intermed v}_{{\mathrm{1}}}  } , \intermed {  {\intermed \Sigma}_{{\mathrm{2}}}  } : \intermed {  {\intermed v}_{{\mathrm{2}}}  } ) \in \mathcal{V} \llbracket \intermed {  {\intermed \sigma}  } \rrbracket ^{\intermed {  {\intermed { \Gamma_C } }  } }_{\intermed {  \ottsym{(}   {\intermed R }  ,  {\intermed a }  \mapsto (  {\intermed \sigma}' ,  { \intermed { \mathit{ r } } }  )   \ottsym{)}  } } 
\end{split}
\end{gather}
By the definition of substitution, and because \(\intermed{{\intermed \sigma}'}\) has no free
variables, it follows that \( \intermed {  \ottsym{(}   {\intermed R }  ,  {\intermed a }  \mapsto (  {\intermed \sigma}' ,  { \intermed { \mathit{ r } } }  )   \ottsym{)}  \ottsym{(}  {\intermed \sigma}  \ottsym{)}  } = \intermed {  [  {\intermed \sigma}'  /  {\intermed a }  ]  \ottsym{(}  {\intermed R }  \ottsym{(}  {\intermed \sigma}  \ottsym{)}  \ottsym{)}  } \).
Furthermode, because \( \intermed {  {\intermed a }  }  \notin  \intermed {  {\intermed \Gamma}  } \), from Equations~\ref{eq:comat_tyabsG}
and~\ref{eq:comat_tyabsH} it follows that \( \intermed {   {\intermed \gamma }_{\ottmv{i}}  ( {\intermed \phi }_{\ottmv{i}}  ( \ottsym{(}   {\intermed R }  ,  {\intermed a }  \mapsto (  {\intermed \sigma}' ,  { \intermed { \mathit{ r } } }  )   \ottsym{)}  ( {\intermed e} )))   } = \intermed {  [  {\intermed \sigma}'  /  {\intermed a }  ]  \ottsym{(}   {\intermed \gamma }_{\ottmv{i}}  ( {\intermed \phi }_{\ottmv{i}}  ( {\intermed R }  ( {\intermed e} )))   \ottsym{)}  } \),
for any expression \(\intermed{{\intermed e}}\) and \(\intermed{\ottmv{i}}\in\{1,2\}\). Taking these equalities into account, by
applying Equations~\ref{eq:compat_tyabs5} and \ref{eq:compat_tyabs6} to reverse
substitution Lemma~\ref{lemma:rev_tvar_subst} gives us:
\begin{align}
  \label{eq:compat_tyabs9}
   \intermed {  {\intermed \Sigma}_{{\mathrm{1}}}  } ; \intermed {  {\intermed { \Gamma_C } }  } ; \intermed {  \bullet  \ottsym{,}  {\intermed a }  }  \vdash_{tm}  \intermed {   {\intermed \gamma }_{{\mathrm{1}}}  ( {\intermed \phi }_{{\mathrm{1}}}  ( {\intermed R }  ( {\intermed e}_{{\mathrm{1}}} )))   } : \intermed {  {\intermed R }  \ottsym{(}  {\intermed \sigma}  \ottsym{)}  } \itot{ \rightsquigarrow  \target {  {\target e }'_{{\mathrm{1}}}  } } \\
  \label{eq:compat_tyabs10}
   \intermed {  {\intermed \Sigma}_{{\mathrm{2}}}  } ; \intermed {  {\intermed { \Gamma_C } }  } ; \intermed {  \bullet  \ottsym{,}  {\intermed a }  }  \vdash_{tm}  \intermed {   {\intermed \gamma }_{{\mathrm{2}}}  ( {\intermed \phi }_{{\mathrm{2}}}  ( {\intermed R }  ( {\intermed e}_{{\mathrm{2}}} )))   } : \intermed {  {\intermed R }  \ottsym{(}  {\intermed \sigma}  \ottsym{)}  } \itot{ \rightsquigarrow  \target {  {\target e }'_{{\mathrm{2}}}  } } 
\end{align}
Because \({\intermed a }\) is not in the domain of \({\intermed R }\), we have
\( \intermed {  {\intermed R }  \ottsym{(}  \forall  {\intermed a }  \ottsym{.}  {\intermed \sigma}  \ottsym{)}  } = \intermed {  \forall  {\intermed a }  \ottsym{.}  {\intermed R }  \ottsym{(}  {\intermed \sigma}  \ottsym{)}  } \). 
Hence, Goals~\ref{eq:compat_tyabs1} and \ref{eq:compat_tyabs2} follow by passing
Equations~\ref{eq:compat_tyabs9} and \ref{eq:compat_tyabs10} to
\textsc{iTm-forallI}, respectively.

Unfolding the definition of the \(\mathcal{E}\) relation in
Goal~\ref{eq:compat_tyabs8} and since \( \intermed {  \ottsym{(}   {\intermed R }  ,  {\intermed a }  \mapsto (  {\intermed \sigma}' ,  { \intermed { \mathit{ r } } }  )   \ottsym{)}  \ottsym{(}  {\intermed \sigma}  \ottsym{)}  } = \intermed {  [  {\intermed \sigma}'  /  {\intermed a }  ]  \ottsym{(}  {\intermed R }  \ottsym{(}  {\intermed \sigma}  \ottsym{)}  \ottsym{)}  } \), the goal
reduces to:
\begin{gather}
    \label{eq:compat_tyabs11a}
   \intermed {  {\intermed \Sigma}_{{\mathrm{1}}}  } ; \intermed {  {\intermed { \Gamma_C } }  } ; \intermed {  \bullet  }  \vdash_{tm}  \intermed {  \ottsym{(}  \Lambda  {\intermed a }  \ottsym{.}    {\intermed \gamma }_{{\mathrm{1}}}  ( {\intermed \phi }_{{\mathrm{1}}}  ( {\intermed R }  ( {\intermed e}_{{\mathrm{1}}} )))    \ottsym{)} \, {\intermed \sigma}'  } : \intermed {  [  {\intermed \sigma}'  /  {\intermed a }  ]  \ottsym{(}  {\intermed R }  \ottsym{(}  {\intermed \sigma}  \ottsym{)}  \ottsym{)}  } \itot{ \rightsquigarrow  \target {  {\target e }_{{\mathrm{3}}}  } } \\
    \label{eq:compat_tyabs11b}
   \intermed {  {\intermed \Sigma}_{{\mathrm{2}}}  } ; \intermed {  {\intermed { \Gamma_C } }  } ; \intermed {  \bullet  }  \vdash_{tm}  \intermed {  \ottsym{(}  \Lambda  {\intermed a }  \ottsym{.}    {\intermed \gamma }_{{\mathrm{2}}}  ( {\intermed \phi }_{{\mathrm{2}}}  ( {\intermed R }  ( {\intermed e}_{{\mathrm{2}}} )))    \ottsym{)} \, {\intermed \sigma}'  } : \intermed {  [  {\intermed \sigma}'  /  {\intermed a }  ]  \ottsym{(}  {\intermed R }  \ottsym{(}  {\intermed \sigma}  \ottsym{)}  \ottsym{)}  } \itot{ \rightsquigarrow  \target {  {\target e }_{{\mathrm{4}}}  } } \\
  \label{eq:compat_tyabs11c}
  \begin{split}
  \exists \intermed{{\intermed v}_{{\mathrm{5}}}}, \intermed{{\intermed v}_{{\mathrm{6}}}} :\, &  \intermed {  {\intermed \Sigma}_{{\mathrm{1}}}  }  \vdash  \intermed {  \ottsym{(}  \Lambda  {\intermed a }  \ottsym{.}    {\intermed \gamma }_{{\mathrm{1}}}  ( {\intermed \phi }_{{\mathrm{1}}}  ( {\intermed R }  ( {\intermed e}_{{\mathrm{1}}} )))    \ottsym{)} \, {\intermed \sigma}'  }  \longrightarrow^{*}  \intermed {  {\intermed v}_{{\mathrm{5}}}  } \\
   \band\, &  \intermed {  {\intermed \Sigma}_{{\mathrm{2}}}  }  \vdash  \intermed {  \ottsym{(}  \Lambda  {\intermed a }  \ottsym{.}    {\intermed \gamma }_{{\mathrm{2}}}  ( {\intermed \phi }_{{\mathrm{2}}}  ( {\intermed R }  ( {\intermed e}_{{\mathrm{2}}} )))    \ottsym{)} \, {\intermed \sigma}'  }  \longrightarrow^{*}  \intermed {  {\intermed v}_{{\mathrm{6}}}  } \\
   \band\, &  ( \intermed {  {\intermed \Sigma}_{{\mathrm{1}}}  } : \intermed {  {\intermed v}_{{\mathrm{5}}}  } , \intermed {  {\intermed \Sigma}_{{\mathrm{2}}}  } : \intermed {  {\intermed v}_{{\mathrm{6}}}  } ) \in \mathcal{V} \llbracket \intermed {  {\intermed \sigma}  } \rrbracket ^{\intermed {  {\intermed { \Gamma_C } }  } }_{\intermed {   {\intermed R }  ,  {\intermed a }  \mapsto (  {\intermed \sigma}' ,  { \intermed { \mathit{ r } } }  )   } } 
  \end{split}
\end{gather}
Goals~\ref{eq:compat_tyabs11a} and \ref{eq:compat_tyabs11b} follow by applying
Goals~\ref{eq:compat_tyabs1} and \ref{eq:compat_tyabs2} (which have previously
been proven) to \textsc{iTm-forallE}, respectively, together with
Equation~\ref{eq:compat_tyabs20}.
The first step of both evaluation paths in Equation \ref{eq:compat_tyabs11c} can only be taken by appropriate instantiations of rule \textsc{iEval-tyAppAbs}. With this,
Goal~\ref{eq:compat_tyabs11c} can be further reduced to
\begin{align*}
  \exists \intermed{{\intermed v}_{{\mathrm{5}}}}, \intermed{{\intermed v}_{{\mathrm{6}}}} & :  \intermed {  {\intermed \Sigma}_{{\mathrm{1}}}  }  \vdash  \intermed {  [  {\intermed \sigma}'  /  {\intermed a }  ]  \ottsym{(}   {\intermed \gamma }_{{\mathrm{1}}}  ( {\intermed \phi }_{{\mathrm{1}}}  ( {\intermed R }  ( {\intermed e}_{{\mathrm{1}}} )))   \ottsym{)}  }  \longrightarrow^{*}  \intermed {  {\intermed v}_{{\mathrm{5}}}  } \\
  & \band  \intermed {  {\intermed \Sigma}_{{\mathrm{2}}}  }  \vdash  \intermed {  [  {\intermed \sigma}'  /  {\intermed a }  ]  \ottsym{(}   {\intermed \gamma }_{{\mathrm{2}}}  ( {\intermed \phi }_{{\mathrm{2}}}  ( {\intermed R }  ( {\intermed e}_{{\mathrm{2}}} )))   \ottsym{)}  }  \longrightarrow^{*}  \intermed {  {\intermed v}_{{\mathrm{6}}}  } \\
  & \band  ( \intermed {  {\intermed \Sigma}_{{\mathrm{1}}}  } : \intermed {  {\intermed v}_{{\mathrm{5}}}  } , \intermed {  {\intermed \Sigma}_{{\mathrm{2}}}  } : \intermed {  {\intermed v}_{{\mathrm{6}}}  } ) \in \mathcal{V} \llbracket \intermed {  {\intermed \sigma}  } \rrbracket ^{\intermed {  {\intermed { \Gamma_C } }  } }_{\intermed {   {\intermed R }  ,  {\intermed a }  \mapsto (  {\intermed \sigma}' ,  { \intermed { \mathit{ r } } }  )   } } 
\end{align*}
which follows from Equation~\ref{eq:compat_tyabs7} by choosing \( \intermed {  {\intermed v}_{{\mathrm{5}}}  } = \intermed {  {\intermed v}_{{\mathrm{3}}}  } \)
and \( \intermed {  {\intermed v}_{{\mathrm{6}}}  } = \intermed {  {\intermed v}_{{\mathrm{4}}}  } \). \\
\qed

\begin{lemmaB}[Compatibility - Type Application]\ \\
  \begin{mathpar}
    \ottaltinferrule{}{}
    {
       \intermed {  {\intermed { \Gamma_C } }  } ; \intermed {  {\intermed \Gamma}  }  \vdash  \intermed {  {\intermed \Sigma}_{{\mathrm{1}}}  } : \intermed {  {\intermed e}_{{\mathrm{1}}}  }  \simeq_{log}  \intermed {  {\intermed \Sigma}_{{\mathrm{2}}}  } : \intermed {  {\intermed e}_{{\mathrm{2}}}  } : \intermed {  \forall  {\intermed a }  \ottsym{.}  {\intermed \sigma}'  } 
      \\  \intermed {  {\intermed { \Gamma_C } }  } ; \intermed {  {\intermed \Gamma}  }  \vdash_{ty}  \intermed {  {\intermed \sigma}  } \itot{ \rightsquigarrow  \target {  {\target \sigma }  } } 
    }
    {  \intermed {  {\intermed { \Gamma_C } }  } ; \intermed {  {\intermed \Gamma}  }  \vdash  \intermed {  {\intermed \Sigma}_{{\mathrm{1}}}  } : \intermed {  {\intermed e}_{{\mathrm{1}}} \, {\intermed \sigma}  }  \simeq_{log}  \intermed {  {\intermed \Sigma}_{{\mathrm{2}}}  } : \intermed {  {\intermed e}_{{\mathrm{2}}} \, {\intermed \sigma}  } : \intermed {  [  {\intermed \sigma}  /  {\intermed a }  ]  {\intermed \sigma}'  }  }
    \label{lemma:compat_tyapp}
  \end{mathpar}
\end{lemmaB}

\proof
By inlining the definition of logical equivalence, suppose we have:
\begin{gather*}
   \intermed {  {\intermed R }  } \in \mathcal{F} \llbracket \intermed {  {\intermed \Gamma}  } \rrbracket ^{\intermed {  {\intermed { \Gamma_C } }  } } \\
   \intermed {  {\intermed \phi }  } \in \mathcal{G} \llbracket \intermed {  {\intermed \Gamma}  } \rrbracket ^{\intermed {  {\intermed \Sigma}_{{\mathrm{1}}}  } , \intermed {  {\intermed \Sigma}_{{\mathrm{2}}}  } , \intermed {  {\intermed { \Gamma_C } }  } }_{\intermed {  {\intermed R }  } } \\
   \intermed {  {\intermed \gamma }  } \in \mathcal{H} \llbracket \intermed {  {\intermed \Gamma}  } \rrbracket ^{\intermed {  {\intermed \Sigma}_{{\mathrm{1}}}  } , \intermed {  {\intermed \Sigma}_{{\mathrm{2}}}  } , \intermed {  {\intermed { \Gamma_C } }  } }_{\intermed {  {\intermed R }  } } 
\end{gather*}
Note that, by the definition of the \( \mathcal{F} \) relation, \({\intermed a }\) is not in the domain of \({\intermed R }\), since \( \intermed {  {\intermed a }  }  \notin  \intermed {  {\intermed \Gamma}  } \).
The goal to be proven is the following:
\begin{equation}
  \label{eq:compat_tyapp21}
   ( \intermed {  {\intermed \Sigma}_{{\mathrm{1}}}  } : \intermed {   {\intermed \gamma }_{{\mathrm{1}}}  ( {\intermed \phi }_{{\mathrm{1}}}  ( {\intermed R }  ( {\intermed e}_{{\mathrm{1}}} \, {\intermed \sigma} )))   } , \intermed {  {\intermed \Sigma}_{{\mathrm{2}}}  } : \intermed {   {\intermed \gamma }_{{\mathrm{2}}}  ( {\intermed \phi }_{{\mathrm{2}}}  ( {\intermed R }  ( {\intermed e}_{{\mathrm{2}}} \, {\intermed \sigma} )))   } ) \in \mathcal{E} \llbracket \intermed {  [  {\intermed \sigma}  /  {\intermed a }  ]  {\intermed \sigma}'  } \rrbracket ^{\intermed {  {\intermed { \Gamma_C } }  } }_{\intermed {  {\intermed R }  } } 
\end{equation}
From the definition of substitution we have that
\begin{gather*}
   \intermed {   {\intermed \gamma }_{\ottmv{i}}  ( {\intermed \phi }_{\ottmv{i}}  ( {\intermed R }  ( {\intermed e}_{\ottmv{i}} \, {\intermed \sigma} )))   } = \intermed {    {\intermed \gamma }_{\ottmv{i}}  ( {\intermed \phi }_{\ottmv{i}}  ( {\intermed R }  ( {\intermed e}_{\ottmv{i}} )))   \, {\intermed R }  \ottsym{(}  {\intermed \sigma}  \ottsym{)}  } ,~
  \text{for}~ \intermed{i}\in\{1,2\}\\
  \text{and}~~ \intermed {  {\intermed R }  \ottsym{(}  [  {\intermed \sigma}  /  {\intermed a }  ]  {\intermed \sigma}'  \ottsym{)}  } = \intermed {  [  {\intermed R }  \ottsym{(}  {\intermed \sigma}  \ottsym{)}  /  {\intermed a }  ]  {\intermed R }  \ottsym{(}  {\intermed \sigma}'  \ottsym{)}  } 
\end{gather*}
Taking into account these equalities and
by unfolding the definition of the \( \mathcal{E} \) relation,
Goal~\ref{eq:compat_tyapp21} reduces to:
\begin{gather}
  \label{eq:compat_tyapp4}
   \intermed {  {\intermed \Sigma}_{{\mathrm{1}}}  } ; \intermed {  {\intermed { \Gamma_C } }  } ; \intermed {  \bullet  }  \vdash_{tm}  \intermed {    {\intermed \gamma }_{{\mathrm{1}}}  ( {\intermed \phi }_{{\mathrm{1}}}  ( {\intermed R }  ( {\intermed e}_{{\mathrm{1}}} )))   \, {\intermed R }  \ottsym{(}  {\intermed \sigma}  \ottsym{)}  } : \intermed {  [  {\intermed R }  \ottsym{(}  {\intermed \sigma}  \ottsym{)}  /  {\intermed a }  ]  {\intermed R }  \ottsym{(}  {\intermed \sigma}'  \ottsym{)}  } \itot{ \rightsquigarrow  \target {  {\target e }  } } \\
  \label{eq:compat_tyapp5}
   \intermed {  {\intermed \Sigma}_{{\mathrm{2}}}  } ; \intermed {  {\intermed { \Gamma_C } }  } ; \intermed {  \bullet  }  \vdash_{tm}  \intermed {    {\intermed \gamma }_{{\mathrm{2}}}  ( {\intermed \phi }_{{\mathrm{2}}}  ( {\intermed R }  ( {\intermed e}_{{\mathrm{2}}} )))   \, {\intermed R }  \ottsym{(}  {\intermed \sigma}  \ottsym{)}  } : \intermed {  [  {\intermed R }  \ottsym{(}  {\intermed \sigma}  \ottsym{)}  /  {\intermed a }  ]  {\intermed R }  \ottsym{(}  {\intermed \sigma}'  \ottsym{)}  } \itot{ \rightsquigarrow  \target {  {\target e }'  } } \\
\begin{split}\label{eq:compat_tyapp6}
  \exists \intermed{{\intermed v}_{{\mathrm{1}}}}, \intermed{{\intermed v}_{{\mathrm{2}}}} & :
     \intermed {  {\intermed \Sigma}_{{\mathrm{1}}}  }  \vdash  \intermed {  \ottsym{(}   {\intermed \gamma }_{{\mathrm{1}}}  ( {\intermed \phi }_{{\mathrm{1}}}  ( {\intermed R }  ( {\intermed e}_{{\mathrm{1}}} )))   \ottsym{)} \, {\intermed R }  \ottsym{(}  {\intermed \sigma}  \ottsym{)}  }  \longrightarrow^{*}  \intermed {  {\intermed v}_{{\mathrm{1}}}  }  \\
    & \band  \intermed {  {\intermed \Sigma}_{{\mathrm{2}}}  }  \vdash  \intermed {  \ottsym{(}   {\intermed \gamma }_{{\mathrm{2}}}  ( {\intermed \phi }_{{\mathrm{2}}}  ( {\intermed R }  ( {\intermed e}_{{\mathrm{2}}} )))   \ottsym{)} \, {\intermed R }  \ottsym{(}  {\intermed \sigma}  \ottsym{)}  }  \longrightarrow^{*}  \intermed {  {\intermed v}_{{\mathrm{2}}}  }  \\
    & \band  ( \intermed {  {\intermed \Sigma}_{{\mathrm{1}}}  } : \intermed {  {\intermed v}_{{\mathrm{1}}}  } , \intermed {  {\intermed \Sigma}_{{\mathrm{2}}}  } : \intermed {  {\intermed v}_{{\mathrm{2}}}  } ) \in \mathcal{V} \llbracket \intermed {  [  {\intermed \sigma}  /  {\intermed a }  ]  {\intermed \sigma}'  } \rrbracket ^{\intermed {  {\intermed { \Gamma_C } }  } }_{\intermed {  {\intermed R }  } } 
\end{split}
\end{gather}

By inlining the definition of logical equivalence in the first premise of this lemma, we
get
\[
   ( \intermed {  {\intermed \Sigma}_{{\mathrm{1}}}  } : \intermed {   {\intermed \gamma }_{{\mathrm{1}}}  ( {\intermed \phi }_{{\mathrm{1}}}  ( {\intermed R }  ( {\intermed e}_{{\mathrm{1}}} )))   } , \intermed {  {\intermed \Sigma}_{{\mathrm{2}}}  } : \intermed {   {\intermed \gamma }_{{\mathrm{2}}}  ( {\intermed \phi }_{{\mathrm{2}}}  ( {\intermed R }  ( {\intermed e}_{{\mathrm{2}}} )))   } ) \in \mathcal{E} \llbracket \intermed {  \forall  {\intermed a }  \ottsym{.}  {\intermed \sigma}'  } \rrbracket ^{\intermed {  {\intermed { \Gamma_C } }  } }_{\intermed {  {\intermed R }  } } 
\]
Unfolding the definition of the \(\mathcal{E}\) relation results in:
\begin{gather}
  \label{eq:compat_tyapp8}
   \intermed {  {\intermed \Sigma}_{{\mathrm{1}}}  } ; \intermed {  {\intermed { \Gamma_C } }  } ; \intermed {  \bullet  }  \vdash_{tm}  \intermed {   {\intermed \gamma }_{{\mathrm{1}}}  ( {\intermed \phi }_{{\mathrm{1}}}  ( {\intermed R }  ( {\intermed e}_{{\mathrm{1}}} )))   } : \intermed {  {\intermed R }  \ottsym{(}  \forall  {\intermed a }  \ottsym{.}  {\intermed \sigma}'  \ottsym{)}  } \itot{ \rightsquigarrow  \target {  {\target e }_{{\mathrm{1}}}  } } \\
  \label{eq:compat_tyapp9}
   \intermed {  {\intermed \Sigma}_{{\mathrm{2}}}  } ; \intermed {  {\intermed { \Gamma_C } }  } ; \intermed {  \bullet  }  \vdash_{tm}  \intermed {   {\intermed \gamma }_{{\mathrm{2}}}  ( {\intermed \phi }_{{\mathrm{2}}}  ( {\intermed R }  ( {\intermed e}_{{\mathrm{2}}} )))   } : \intermed {  {\intermed R }  \ottsym{(}  \forall  {\intermed a }  \ottsym{.}  {\intermed \sigma}'  \ottsym{)}  } \itot{ \rightsquigarrow  \target {  {\target e }_{{\mathrm{2}}}  } } \\
\begin{split}\label{eq:compat_tyapp10}
  \exists \intermed{{\intermed v}_{{\mathrm{3}}}}, \intermed{{\intermed v}_{{\mathrm{4}}}} & :
   \intermed {  {\intermed \Sigma}_{{\mathrm{1}}}  }  \vdash  \intermed {   {\intermed \gamma }_{{\mathrm{1}}}  ( {\intermed \phi }_{{\mathrm{1}}}  ( {\intermed R }  ( {\intermed e}_{{\mathrm{1}}} )))   }  \longrightarrow^{*}  \intermed {  {\intermed v}_{{\mathrm{3}}}  } \\
  & \band  \intermed {  {\intermed \Sigma}_{{\mathrm{2}}}  }  \vdash  \intermed {   {\intermed \gamma }_{{\mathrm{2}}}  ( {\intermed \phi }_{{\mathrm{2}}}  ( {\intermed R }  ( {\intermed e}_{{\mathrm{2}}} )))   }  \longrightarrow^{*}  \intermed {  {\intermed v}_{{\mathrm{4}}}  } \\
  & \band  ( \intermed {  {\intermed \Sigma}_{{\mathrm{1}}}  } : \intermed {  {\intermed v}_{{\mathrm{3}}}  } , \intermed {  {\intermed \Sigma}_{{\mathrm{2}}}  } : \intermed {  {\intermed v}_{{\mathrm{4}}}  } ) \in \mathcal{V} \llbracket \intermed {  \forall  {\intermed a }  \ottsym{.}  {\intermed \sigma}'  } \rrbracket ^{\intermed {  {\intermed { \Gamma_C } }  } }_{\intermed {  {\intermed R }  } } 
\end{split}
\end{gather}
Starting from the second premise of this lemma, by sequentially applying
Lemma~\ref{lemma:tvar_tsubst} with the substitutions of \({\intermed R }\) on \(\intermed{{\intermed \sigma}}\),
it follows that \( \intermed {  {\intermed { \Gamma_C } }  } ; \intermed {  {\intermed \Gamma}'  }  \vdash_{ty}  \intermed {  {\intermed R }  \ottsym{(}  {\intermed \sigma}  \ottsym{)}  } \itot{ \rightsquigarrow  \target {  {\target \sigma }  } } \), where \(\intermed{{\intermed \Gamma}'}\) only
contains term variables. Then, starting from this result, by sequentually applying
Lemma~\ref{lemma:var_subst_types}, we obtain
\begin{equation}
  \label{eq:compat_tyapp20}
   \intermed {  {\intermed { \Gamma_C } }  } ; \intermed {  \bullet  }  \vdash_{ty}  \intermed {  {\intermed R }  \ottsym{(}  {\intermed \sigma}  \ottsym{)}  } \itot{ \rightsquigarrow  \target {  {\target \sigma }  } } 
\end{equation}
Since \({\intermed a }\) is not in the domain of \({\intermed R }\), we have
\( \intermed {  {\intermed R }  \ottsym{(}  \forall  {\intermed a }  \ottsym{.}  {\intermed \sigma}  \ottsym{)}  } = \intermed {  \forall  {\intermed a }  \ottsym{.}  {\intermed R }  \ottsym{(}  {\intermed \sigma}  \ottsym{)}  } \). Consequently, Goals~\ref{eq:compat_tyapp4} and
\ref{eq:compat_tyapp5} follow by instantiating rule \textsc{iTm-forallE} with
Equations~\ref{eq:compat_tyapp8} and \ref{eq:compat_tyapp9}, respectively, together
with Equation~\ref{eq:compat_tyapp20}.

The definition of the \(\mathcal{V}\) relation in Equation~\ref{eq:compat_tyapp10}
tells us that:
\begin{equation}\label{eq:compat_tyapp11}
\begin{aligned}
  \forall \intermed{{\intermed \sigma}''},  \intermed {  { \intermed { \mathit{ r } } }  } \in \mathit{Rel} [ \intermed {  {\intermed \sigma}''  } ]  & :
   \intermed {  {\intermed { \Gamma_C } }  } ; \intermed {  \bullet  }  \vdash_{ty}  \intermed {  {\intermed \sigma}''  } \itot{ \rightsquigarrow  \target {  {\target \sigma }''  } } \\
  & \Rightarrow  ( \intermed {  {\intermed \Sigma}_{{\mathrm{1}}}  } : \intermed {  {\intermed v}_{{\mathrm{3}}} \, {\intermed \sigma}''  } , \intermed {  {\intermed \Sigma}_{{\mathrm{2}}}  } : \intermed {  {\intermed v}_{{\mathrm{4}}} \, {\intermed \sigma}''  } ) \in \mathcal{E} \llbracket \intermed {  {\intermed \sigma}'  } \rrbracket ^{\intermed {  {\intermed { \Gamma_C } }  } }_{\intermed {   {\intermed R }  ,  {\intermed a }  \mapsto (  {\intermed \sigma}'' ,  { \intermed { \mathit{ r } } }  )   } } 
\end{aligned}
\end{equation}

By repeatedly applying \textsc{iEval-tyApp} on each step of both evaluation paths in
Equation~\ref{eq:compat_tyapp10}, Goal~\ref{eq:compat_tyapp6} reduces to:
\begin{align*}
  \exists \intermed{{\intermed v}_{{\mathrm{1}}}}, \intermed{{\intermed v}_{{\mathrm{2}}}} & :
     \intermed {  {\intermed \Sigma}_{{\mathrm{1}}}  }  \vdash  \intermed {  {\intermed v}_{{\mathrm{3}}} \, {\intermed R }  \ottsym{(}  {\intermed \sigma}  \ottsym{)}  }  \longrightarrow^{*}  \intermed {  {\intermed v}_{{\mathrm{1}}}  } \\ 
    & \band  \intermed {  {\intermed \Sigma}_{{\mathrm{2}}}  }  \vdash  \intermed {  {\intermed v}_{{\mathrm{4}}} \, {\intermed R }  \ottsym{(}  {\intermed \sigma}  \ottsym{)}  }  \longrightarrow^{*}  \intermed {  {\intermed v}_{{\mathrm{2}}}  } \\ 
    & \band  ( \intermed {  {\intermed \Sigma}_{{\mathrm{1}}}  } : \intermed {  {\intermed v}_{{\mathrm{1}}}  } , \intermed {  {\intermed \Sigma}_{{\mathrm{2}}}  } : \intermed {  {\intermed v}_{{\mathrm{2}}}  } ) \in \mathcal{V} \llbracket \intermed {  [  {\intermed \sigma}  /  {\intermed a }  ]  {\intermed \sigma}'  } \rrbracket ^{\intermed {  {\intermed { \Gamma_C } }  } }_{\intermed {  {\intermed R }  } } 
\end{align*}
which follows directly from~\ref{eq:compat_tyapp11} by choosing \( \intermed {  {\intermed \sigma}''  } = \intermed {  {\intermed \sigma}  } \)
and unfolding the definition of the \(\mathcal{E}\) relation. \\
\qed

\begin{lemmaB}[Compatibility - Let Binding]\ \\
  \begin{mathpar}
    \ottaltinferrule{}{}
    {
       \intermed {  {\intermed { \Gamma_C } }  } ; \intermed {  {\intermed \Gamma}  }  \vdash  \intermed {  {\intermed \Sigma}_{{\mathrm{1}}}  } : \intermed {  {\intermed e}'_{{\mathrm{1}}}  }  \simeq_{log}  \intermed {  {\intermed \Sigma}_{{\mathrm{2}}}  } : \intermed {  {\intermed e}'_{{\mathrm{2}}}  } : \intermed {  {\intermed \sigma}_{{\mathrm{1}}}  } 
      \\  \intermed {  {\intermed { \Gamma_C } }  } ; \intermed {  {\intermed \Gamma}  \ottsym{,}  {\intermed x}  \ottsym{:}  {\intermed \sigma}_{{\mathrm{1}}}  }  \vdash  \intermed {  {\intermed \Sigma}_{{\mathrm{1}}}  } : \intermed {  {\intermed e}_{{\mathrm{1}}}  }  \simeq_{log}  \intermed {  {\intermed \Sigma}_{{\mathrm{2}}}  } : \intermed {  {\intermed e}_{{\mathrm{2}}}  } : \intermed {  {\intermed \sigma}_{{\mathrm{2}}}  } 
    }
    {  \intermed {  {\intermed { \Gamma_C } }  } ; \intermed {  {\intermed \Gamma}  }  \vdash  \intermed {  {\intermed \Sigma}_{{\mathrm{1}}}  } : \intermed {  \textbf{\intermed {let} } \, {\intermed x}  \ottsym{:}  {\intermed \sigma}_{{\mathrm{1}}}  \ottsym{=}  {\intermed e}'_{{\mathrm{1}}} \, \textbf{\intermed {in} } \, {\intermed e}_{{\mathrm{1}}}  }  \simeq_{log}  \intermed {  {\intermed \Sigma}_{{\mathrm{2}}}  } : \intermed {  \textbf{\intermed {let} } \, {\intermed x}  \ottsym{:}  {\intermed \sigma}_{{\mathrm{1}}}  \ottsym{=}  {\intermed e}'_{{\mathrm{2}}} \, \textbf{\intermed {in} } \, {\intermed e}_{{\mathrm{2}}}  } : \intermed {  {\intermed \sigma}_{{\mathrm{2}}}  }  }
    \label{lemma:compat_let}
  \end{mathpar}
\end{lemmaB}

\proof
By inlining the definition of logical equivalence, suppose we have:
\begin{equation}
  \label{eq:compat_let1}
  \begin{gathered}
     \intermed {  {\intermed R }  } \in \mathcal{F} \llbracket \intermed {  {\intermed \Gamma}  } \rrbracket ^{\intermed {  {\intermed { \Gamma_C } }  } } \\
     \intermed {  {\intermed \phi }  } \in \mathcal{G} \llbracket \intermed {  {\intermed \Gamma}  } \rrbracket ^{\intermed {  {\intermed \Sigma}_{{\mathrm{1}}}  } , \intermed {  {\intermed \Sigma}_{{\mathrm{2}}}  } , \intermed {  {\intermed { \Gamma_C } }  } }_{\intermed {  {\intermed R }  } } \\
     \intermed {  {\intermed \gamma }  } \in \mathcal{H} \llbracket \intermed {  {\intermed \Gamma}  } \rrbracket ^{\intermed {  {\intermed \Sigma}_{{\mathrm{1}}}  } , \intermed {  {\intermed \Sigma}_{{\mathrm{2}}}  } , \intermed {  {\intermed { \Gamma_C } }  } }_{\intermed {  {\intermed R }  } } 
  \end{gathered}
\end{equation}
The goal to be proven is the following:
\begin{equation}
  \label{eq:compat_let20}
   ( \intermed {  {\intermed \Sigma}_{{\mathrm{1}}}  } : \intermed {   {\intermed \gamma }_{{\mathrm{1}}}  ( {\intermed \phi }_{{\mathrm{1}}}  ( {\intermed R }  (  \textbf{\intermed {let} } \, {\intermed x}  \ottsym{:}  {\intermed \sigma}_{{\mathrm{1}}}  \ottsym{=}  {\intermed e}'_{{\mathrm{1}}} \, \textbf{\intermed {in} } \, {\intermed e}_{{\mathrm{1}}}  )))   } , \intermed {  {\intermed \Sigma}_{{\mathrm{2}}}  } : \intermed {   {\intermed \gamma }_{{\mathrm{2}}}  ( {\intermed \phi }_{{\mathrm{2}}}  ( {\intermed R }  (  \textbf{\intermed {let} } \, {\intermed x}  \ottsym{:}  {\intermed \sigma}_{{\mathrm{1}}}  \ottsym{=}  {\intermed e}'_{{\mathrm{2}}} \, \textbf{\intermed {in} } \, {\intermed e}_{{\mathrm{2}}}  )))   } ) \in \mathcal{E} \llbracket \intermed {  {\intermed \sigma}_{{\mathrm{2}}}  } \rrbracket ^{\intermed {  {\intermed { \Gamma_C } }  } }_{\intermed {  {\intermed R }  } } 
\end{equation}
From the definition of substitution, it follows that
\[
   \intermed {   {\intermed \gamma }_{\ottmv{i}}  ( {\intermed \phi }_{\ottmv{i}}  ( {\intermed R }  (  \textbf{\intermed {let} } \, {\intermed x}  \ottsym{:}  {\intermed \sigma}_{{\mathrm{1}}}  \ottsym{=}  {\intermed e}'_{\ottmv{i}} \, \textbf{\intermed {in} } \, {\intermed e}_{\ottmv{i}}  )))   } = \intermed {  \textbf{\intermed {let} } \, {\intermed x}  \ottsym{:}  {\intermed R }  \ottsym{(}  {\intermed \sigma}_{{\mathrm{1}}}  \ottsym{)}  \ottsym{=}  \ottsym{(}   {\intermed \gamma }_{\ottmv{i}}  ( {\intermed \phi }_{\ottmv{i}}  ( {\intermed R }  ( {\intermed e}'_{\ottmv{i}} )))   \ottsym{)} \, \textbf{\intermed {in} } \, \ottsym{(}   {\intermed \gamma }_{\ottmv{i}}  ( {\intermed \phi }_{\ottmv{i}}  ( {\intermed R }  ( {\intermed e}_{\ottmv{i}} )))   \ottsym{)}  } ,
  ~\text{for}~\intermed{i}\in\{1,2\}
\]
Taking into account this equality, by applying the definition of the
\( \mathcal{E} \) relation, Goal~\ref{eq:compat_let20} reduces
to:
\begin{gather}
  \label{eq:compat_let4}
   \intermed {  {\intermed \Sigma}_{{\mathrm{1}}}  } ; \intermed {  {\intermed { \Gamma_C } }  } ; \intermed {  \bullet  }  \vdash_{tm}  \intermed {  \textbf{\intermed {let} } \, {\intermed x}  \ottsym{:}  {\intermed R }  \ottsym{(}  {\intermed \sigma}_{{\mathrm{1}}}  \ottsym{)}  \ottsym{=}  \ottsym{(}   {\intermed \gamma }_{{\mathrm{1}}}  ( {\intermed \phi }_{{\mathrm{1}}}  ( {\intermed R }  ( {\intermed e}'_{{\mathrm{1}}} )))   \ottsym{)} \, \textbf{\intermed {in} } \, \ottsym{(}   {\intermed \gamma }_{{\mathrm{1}}}  ( {\intermed \phi }_{{\mathrm{1}}}  ( {\intermed R }  ( {\intermed e}_{{\mathrm{1}}} )))   \ottsym{)}  } : \intermed {  {\intermed R }  \ottsym{(}  {\intermed \sigma}_{{\mathrm{2}}}  \ottsym{)}  } \itot{ \rightsquigarrow  \target {  {\target e }  } } \\
  \label{eq:compat_let5}
   \intermed {  {\intermed \Sigma}_{{\mathrm{2}}}  } ; \intermed {  {\intermed { \Gamma_C } }  } ; \intermed {  \bullet  }  \vdash_{tm}  \intermed {  \textbf{\intermed {let} } \, {\intermed x}  \ottsym{:}  {\intermed R }  \ottsym{(}  {\intermed \sigma}_{{\mathrm{1}}}  \ottsym{)}  \ottsym{=}  \ottsym{(}   {\intermed \gamma }_{{\mathrm{2}}}  ( {\intermed \phi }_{{\mathrm{2}}}  ( {\intermed R }  ( {\intermed e}'_{{\mathrm{2}}} )))   \ottsym{)} \, \textbf{\intermed {in} } \, \ottsym{(}   {\intermed \gamma }_{{\mathrm{2}}}  ( {\intermed \phi }_{{\mathrm{2}}}  ( {\intermed R }  ( {\intermed e}_{{\mathrm{2}}} )))   \ottsym{)}  } : \intermed {  {\intermed R }  \ottsym{(}  {\intermed \sigma}_{{\mathrm{2}}}  \ottsym{)}  } \itot{ \rightsquigarrow  \target {  {\target e }'  } } \\
\begin{split}\label{eq:compat_let6}
  \exists \intermed{{\intermed v}_{{\mathrm{1}}}}, \intermed{{\intermed v}_{{\mathrm{2}}}} :\, &
     \intermed {  {\intermed \Sigma}_{{\mathrm{1}}}  }  \vdash  \intermed {  \textbf{\intermed {let} } \, {\intermed x}  \ottsym{:}  {\intermed R }  \ottsym{(}  {\intermed \sigma}_{{\mathrm{1}}}  \ottsym{)}  \ottsym{=}  \ottsym{(}   {\intermed \gamma }_{{\mathrm{1}}}  ( {\intermed \phi }_{{\mathrm{1}}}  ( {\intermed R }  ( {\intermed e}'_{{\mathrm{1}}} )))   \ottsym{)} \, \textbf{\intermed {in} } \, \ottsym{(}   {\intermed \gamma }_{{\mathrm{1}}}  ( {\intermed \phi }_{{\mathrm{1}}}  ( {\intermed R }  ( {\intermed e}_{{\mathrm{1}}} )))   \ottsym{)}  }  \longrightarrow^{*}  \intermed {  {\intermed v}_{{\mathrm{1}}}  }  \\
    \band\, &  \intermed {  {\intermed \Sigma}_{{\mathrm{2}}}  }  \vdash  \intermed {  \textbf{\intermed {let} } \, {\intermed x}  \ottsym{:}  {\intermed R }  \ottsym{(}  {\intermed \sigma}_{{\mathrm{1}}}  \ottsym{)}  \ottsym{=}  \ottsym{(}   {\intermed \gamma }_{{\mathrm{2}}}  ( {\intermed \phi }_{{\mathrm{2}}}  ( {\intermed R }  ( {\intermed e}'_{{\mathrm{2}}} )))   \ottsym{)} \, \textbf{\intermed {in} } \, \ottsym{(}   {\intermed \gamma }_{{\mathrm{2}}}  ( {\intermed \phi }_{{\mathrm{2}}}  ( {\intermed R }  ( {\intermed e}_{{\mathrm{2}}} )))   \ottsym{)}  }  \longrightarrow^{*}  \intermed {  {\intermed v}_{{\mathrm{2}}}  } \\
    \band\, &  ( \intermed {  {\intermed \Sigma}_{{\mathrm{1}}}  } : \intermed {  {\intermed v}_{{\mathrm{1}}}  } , \intermed {  {\intermed \Sigma}_{{\mathrm{2}}}  } : \intermed {  {\intermed v}_{{\mathrm{2}}}  } ) \in \mathcal{V} \llbracket \intermed {  {\intermed \sigma}_{{\mathrm{2}}}  } \rrbracket ^{\intermed {  {\intermed { \Gamma_C } }  } }_{\intermed {  {\intermed R }  } } 
\end{split}
\end{gather}
By inlining the definition of logical equivalence in the two hypotheses of this lemma, we
get:
\begin{gather}
  \label{eq:compat_let7}
   ( \intermed {  {\intermed \Sigma}_{{\mathrm{1}}}  } : \intermed {   {\intermed \gamma }_{{\mathrm{1}}}  ( {\intermed \phi }_{{\mathrm{1}}}  ( {\intermed R }  ( {\intermed e}'_{{\mathrm{1}}} )))   } , \intermed {  {\intermed \Sigma}_{{\mathrm{2}}}  } : \intermed {   {\intermed \gamma }_{{\mathrm{2}}}  ( {\intermed \phi }_{{\mathrm{2}}}  ( {\intermed R }  ( {\intermed e}'_{{\mathrm{2}}} )))   } ) \in \mathcal{E} \llbracket \intermed {  {\intermed \sigma}_{{\mathrm{1}}}  } \rrbracket ^{\intermed {  {\intermed { \Gamma_C } }  } }_{\intermed {  {\intermed R }  } } \\
  \label{eq:compat_let8}
   ( \intermed {  {\intermed \Sigma}_{{\mathrm{1}}}  } : \intermed {   {\intermed \gamma }'_{{\mathrm{1}}}  ( {\intermed \phi }'_{{\mathrm{1}}}  ( {\intermed R }'  ( {\intermed e}_{{\mathrm{1}}} )))   } , \intermed {  {\intermed \Sigma}_{{\mathrm{2}}}  } : \intermed {   {\intermed \gamma }'_{{\mathrm{2}}}  ( {\intermed \phi }'_{{\mathrm{2}}}  ( {\intermed R }'  ( {\intermed e}_{{\mathrm{2}}} )))   } ) \in \mathcal{E} \llbracket \intermed {  {\intermed \sigma}_{{\mathrm{2}}}  } \rrbracket ^{\intermed {  {\intermed { \Gamma_C } }  } }_{\intermed {  {\intermed R }'  } } 
\end{gather}
for any \( \intermed {  {\intermed R }'  } \in \mathcal{F} \llbracket \intermed {  {\intermed \Gamma}  \ottsym{,}  {\intermed x}  \ottsym{:}  {\intermed \sigma}_{{\mathrm{1}}}  } \rrbracket ^{\intermed {  {\intermed { \Gamma_C } }  } } \),
\( \intermed {  {\intermed \phi }'  } \in \mathcal{G} \llbracket \intermed {  {\intermed \Gamma}  \ottsym{,}  {\intermed x}  \ottsym{:}  {\intermed \sigma}_{{\mathrm{1}}}  } \rrbracket ^{\intermed {  {\intermed \Sigma}_{{\mathrm{1}}}  } , \intermed {  {\intermed \Sigma}_{{\mathrm{2}}}  } , \intermed {  {\intermed { \Gamma_C } }  } }_{\intermed {  {\intermed R }'  } } \) and
\( \intermed {  {\intermed \gamma }'  } \in \mathcal{H} \llbracket \intermed {  {\intermed \Gamma}  \ottsym{,}  {\intermed x}  \ottsym{:}  {\intermed \sigma}_{{\mathrm{1}}}  } \rrbracket ^{\intermed {  {\intermed \Sigma}_{{\mathrm{1}}}  } , \intermed {  {\intermed \Sigma}_{{\mathrm{2}}}  } , \intermed {  {\intermed { \Gamma_C } }  } }_{\intermed {  {\intermed R }'  } } \).
Note that in Equation~\ref{eq:compat_let7} we have already chosen the substitutions
\({\intermed R }\), \({\intermed \phi }\) and \({\intermed \gamma }\) from Equation~\ref{eq:compat_let1}.
By the definition of \( \mathcal{F} \), we obtain \( \intermed {  {\intermed R }  } \in \mathcal{F} \llbracket \intermed {  {\intermed \Gamma}  \ottsym{,}  {\intermed x}  \ottsym{:}  {\intermed \sigma}_{{\mathrm{1}}}  } \rrbracket ^{\intermed {  {\intermed { \Gamma_C } }  } } \).
Therefore, a valid choice for \(\intermed{{\intermed R }'}\) is \({\intermed R }\). From the definitions of
\( \mathcal{G} \) and \( \mathcal{H} \), it must hold that
\( \intermed {  {\intermed \phi }'  } = \intermed {   {\intermed \phi }'' ,  {\intermed x}  \mapsto ( {\intermed e}_{{\mathrm{3}}} ,  {\intermed e}_{{\mathrm{4}}}  )   } \) and
\( \intermed {  {\intermed \gamma }'  } = \intermed {  {\intermed \gamma }''  } \)
where
\( \intermed {  {\intermed \phi }''  } \in \mathcal{G} \llbracket \intermed {  {\intermed \Gamma}  } \rrbracket ^{\intermed {  {\intermed \Sigma}_{{\mathrm{1}}}  } , \intermed {  {\intermed \Sigma}_{{\mathrm{2}}}  } , \intermed {  {\intermed { \Gamma_C } }  } }_{\intermed {  {\intermed R }  } } \),
\( \intermed {  {\intermed \gamma }''  } \in \mathcal{H} \llbracket \intermed {  {\intermed \Gamma}  } \rrbracket ^{\intermed {  {\intermed \Sigma}_{{\mathrm{1}}}  } , \intermed {  {\intermed \Sigma}_{{\mathrm{2}}}  } , \intermed {  {\intermed { \Gamma_C } }  } }_{\intermed {  {\intermed R }  } } \) and
\( ( \intermed {  {\intermed \Sigma}_{{\mathrm{1}}}  } : \intermed {  {\intermed e}_{{\mathrm{3}}}  } , \intermed {  {\intermed \Sigma}_{{\mathrm{2}}}  } : \intermed {  {\intermed e}_{{\mathrm{4}}}  } ) \in \mathcal{E} \llbracket \intermed {  {\intermed \sigma}_{{\mathrm{1}}}  } \rrbracket ^{\intermed {  {\intermed { \Gamma_C } }  } }_{\intermed {  {\intermed R }  } } \).
We choose \( \intermed {  {\intermed \phi }''  } = \intermed {  {\intermed \phi }  } \) and \( \intermed {  {\intermed \gamma }''  } = \intermed {  {\intermed \gamma }  } \). It remains to
instantiate \(\intermed{{\intermed e}_{{\mathrm{3}}}}\) and \(\intermed{{\intermed e}_{{\mathrm{4}}}}\) with concrete choices.
For reasons of presentation, we defer this choice to the end of the proof.

From the definition of the \(\mathcal{E}\) relation in
Equations~\ref{eq:compat_let8}, we get:
\begin{gather}
  \label{eq:compat_let12}
   \intermed {  {\intermed \Sigma}_{{\mathrm{1}}}  } ; \intermed {  {\intermed { \Gamma_C } }  } ; \intermed {  \bullet  }  \vdash_{tm}  \intermed {   {\intermed \gamma }_{{\mathrm{1}}}  ( \ottsym{(}   {\intermed \phi }_{{\mathrm{1}}} ,  {\intermed x}  \mapsto  {\intermed e}_{{\mathrm{3}}}   \ottsym{)}  ( {\intermed R }  ( {\intermed e}_{{\mathrm{1}}} )))   } : \intermed {  {\intermed R }  \ottsym{(}  {\intermed \sigma}_{{\mathrm{2}}}  \ottsym{)}  } \itot{ \rightsquigarrow  \target {  {\target e }_{{\mathrm{2}}}  } } \\
  \label{eq:compat_let13}
   \intermed {  {\intermed \Sigma}_{{\mathrm{2}}}  } ; \intermed {  {\intermed { \Gamma_C } }  } ; \intermed {  \bullet  }  \vdash_{tm}  \intermed {   {\intermed \gamma }_{{\mathrm{2}}}  ( \ottsym{(}   {\intermed \phi }_{{\mathrm{2}}} ,  {\intermed x}  \mapsto  {\intermed e}_{{\mathrm{4}}}   \ottsym{)}  ( {\intermed R }  ( {\intermed e}_{{\mathrm{2}}} )))   } : \intermed {  {\intermed R }  \ottsym{(}  {\intermed \sigma}_{{\mathrm{2}}}  \ottsym{)}  } \itot{ \rightsquigarrow  \target {  {\target e }'_{{\mathrm{2}}}  } } \\
\begin{split}\label{eq:compat_let14}
  \exists \intermed{{\intermed v}_{{\mathrm{5}}}}, \intermed{{\intermed v}_{{\mathrm{6}}}} :\, &
   \intermed {  {\intermed \Sigma}_{{\mathrm{1}}}  }  \vdash  \intermed {   {\intermed \gamma }_{{\mathrm{1}}}  ( \ottsym{(}   {\intermed \phi }_{{\mathrm{1}}} ,  {\intermed x}  \mapsto  {\intermed e}_{{\mathrm{3}}}   \ottsym{)}  ( {\intermed R }  ( {\intermed e}_{{\mathrm{1}}} )))   }  \longrightarrow^{*}  \intermed {  {\intermed v}_{{\mathrm{5}}}  } \\
  \band\, &  \intermed {  {\intermed \Sigma}_{{\mathrm{2}}}  }  \vdash  \intermed {   {\intermed \gamma }_{{\mathrm{2}}}  ( \ottsym{(}   {\intermed \phi }_{{\mathrm{2}}} ,  {\intermed x}  \mapsto  {\intermed e}_{{\mathrm{4}}}   \ottsym{)}  ( {\intermed R }  ( {\intermed e}_{{\mathrm{2}}} )))   }  \longrightarrow^{*}  \intermed {  {\intermed v}_{{\mathrm{6}}}  } \\
  \band\, &  ( \intermed {  {\intermed \Sigma}_{{\mathrm{1}}}  } : \intermed {  {\intermed v}_{{\mathrm{5}}}  } , \intermed {  {\intermed \Sigma}_{{\mathrm{2}}}  } : \intermed {  {\intermed v}_{{\mathrm{6}}}  } ) \in \mathcal{V} \llbracket \intermed {  {\intermed \sigma}_{{\mathrm{2}}}  } \rrbracket ^{\intermed {  {\intermed { \Gamma_C } }  } }_{\intermed {  {\intermed R }  } } 
\end{split}
\end{gather}
Similarly, from Equation~\ref{eq:compat_let7} and from 
\( ( \intermed {  {\intermed \Sigma}_{{\mathrm{1}}}  } : \intermed {  {\intermed e}_{{\mathrm{3}}}  } , \intermed {  {\intermed \Sigma}_{{\mathrm{2}}}  } : \intermed {  {\intermed e}_{{\mathrm{4}}}  } ) \in \mathcal{E} \llbracket \intermed {  {\intermed \sigma}_{{\mathrm{1}}}  } \rrbracket ^{\intermed {  {\intermed { \Gamma_C } }  } }_{\intermed {  {\intermed R }  } } \), we get:
\begin{gather}
  \label{eq:compat_let9}
   \intermed {  {\intermed \Sigma}_{{\mathrm{1}}}  } ; \intermed {  {\intermed { \Gamma_C } }  } ; \intermed {  \bullet  }  \vdash_{tm}  \intermed {   {\intermed \gamma }_{{\mathrm{1}}}  ( {\intermed \phi }_{{\mathrm{1}}}  ( {\intermed R }  ( {\intermed e}'_{{\mathrm{1}}} )))   } : \intermed {  {\intermed R }  \ottsym{(}  {\intermed \sigma}_{{\mathrm{1}}}  \ottsym{)}  } \itot{ \rightsquigarrow  \target {  {\target e }_{{\mathrm{1}}}  } } \\
  \label{eq:compat_let10}
   \intermed {  {\intermed \Sigma}_{{\mathrm{2}}}  } ; \intermed {  {\intermed { \Gamma_C } }  } ; \intermed {  \bullet  }  \vdash_{tm}  \intermed {   {\intermed \gamma }_{{\mathrm{2}}}  ( {\intermed \phi }_{{\mathrm{2}}}  ( {\intermed R }  ( {\intermed e}'_{{\mathrm{2}}} )))   } : \intermed {  {\intermed R }  \ottsym{(}  {\intermed \sigma}_{{\mathrm{1}}}  \ottsym{)}  } \itot{ \rightsquigarrow  \target {  {\target e }'_{{\mathrm{1}}}  } } 
\end{gather}
and
\begin{gather}
   \intermed {  {\intermed \Sigma}_{{\mathrm{1}}}  } ; \intermed {  {\intermed { \Gamma_C } }  } ; \intermed {  \bullet  }  \vdash_{tm}  \intermed {  {\intermed e}_{{\mathrm{3}}}  } : \intermed {  {\intermed R }  \ottsym{(}  {\intermed \sigma}_{{\mathrm{1}}}  \ottsym{)}  } \itot{ \rightsquigarrow  \target {  {\target e }_{{\mathrm{3}}}  } } 
  \label{eq:compat_lete1} \\
   \intermed {  {\intermed \Sigma}_{{\mathrm{1}}}  } ; \intermed {  {\intermed { \Gamma_C } }  } ; \intermed {  \bullet  }  \vdash_{tm}  \intermed {  {\intermed e}_{{\mathrm{4}}}  } : \intermed {  {\intermed R }  \ottsym{(}  {\intermed \sigma}_{{\mathrm{1}}}  \ottsym{)}  } \itot{ \rightsquigarrow  \target {  {\target e }_{{\mathrm{4}}}  } } 
  \label{eq:compat_lete2}
\end{gather}
Note that from Equations~\ref{eq:compat_lete1} and~\ref{eq:compat_lete2} it is
evident that expressions \(\intermed{{\intermed e}_{{\mathrm{3}}}}\) and \(\intermed{{\intermed e}_{{\mathrm{4}}}}\)
contain no free variables. Therefore,
\begin{align*}
  &\intermed{ {\intermed \gamma }_{{\mathrm{1}}}  ( \ottsym{(}   {\intermed \phi }_{{\mathrm{1}}} ,  {\intermed x}  \mapsto  {\intermed e}_{{\mathrm{3}}}   \ottsym{)}  ( {\intermed R }  ( {\intermed e}_{{\mathrm{1}}} ))) }\\
  &~~~~= \intermed{{\intermed \gamma }_{{\mathrm{1}}}  \ottsym{(}  [  {\intermed e}_{{\mathrm{3}}}  /  {\intermed x}  ]  \ottsym{(}  {\intermed \phi }_{{\mathrm{1}}}  \ottsym{(}  {\intermed R }  \ottsym{(}  {\intermed e}_{{\mathrm{1}}}  \ottsym{)}  \ottsym{)}  \ottsym{)}  \ottsym{)}} && \text{(by definition)} \\
&~~~~= \intermed{[  {\intermed \gamma }_{{\mathrm{1}}}  \ottsym{(}  {\intermed e}_{{\mathrm{3}}}  \ottsym{)}  /  {\intermed x}  ]  \ottsym{(}   {\intermed \gamma }_{{\mathrm{1}}}  ( {\intermed \phi }_{{\mathrm{1}}}  ( {\intermed R }  ( {\intermed e}_{{\mathrm{1}}} )))   \ottsym{)}} && \text{(distributivity property)}\\
&~~~~= \intermed{[  {\intermed e}_{{\mathrm{3}}}  /  {\intermed x}  ]  \ottsym{(}   {\intermed \gamma }_{{\mathrm{1}}}  ( {\intermed \phi }_{{\mathrm{1}}}  ( {\intermed R }  ( {\intermed e}_{{\mathrm{1}}} )))   \ottsym{)}} && \text{(no free variables in \(\intermed{{\intermed e}_{{\mathrm{3}}}}\))}
\end{align*}
and similarly, \( \intermed {   {\intermed \gamma }_{{\mathrm{2}}}  ( \ottsym{(}   {\intermed \phi }_{{\mathrm{2}}} ,  {\intermed x}  \mapsto  {\intermed e}_{{\mathrm{4}}}   \ottsym{)}  ( {\intermed R }  ( {\intermed e}_{{\mathrm{2}}} )))   } = \intermed {  [  {\intermed e}_{{\mathrm{4}}}  /  {\intermed x}  ]    {\intermed \gamma }_{{\mathrm{2}}}  ( {\intermed \phi }_{{\mathrm{2}}}  ( {\intermed R }  ( {\intermed e}_{{\mathrm{2}}} )))    } \). Taking these equalities into account, we can apply the reverse
substitution Lemma~\ref{lemma:rev_var_subst} on
Equations~\ref{eq:compat_let12} and~\ref{eq:compat_let13}, in combination with
Equations~\ref{eq:compat_lete1} and~\ref{eq:compat_lete2}, respectively, to obtain:
\begin{gather}
  \label{eq:compat_let12b}
   \intermed {  {\intermed \Sigma}_{{\mathrm{1}}}  } ; \intermed {  {\intermed { \Gamma_C } }  } ; \intermed {  \bullet  \ottsym{,}  {\intermed x}  \ottsym{:}  {\intermed R }  \ottsym{(}  {\intermed \sigma}_{{\mathrm{1}}}  \ottsym{)}  }  \vdash_{tm}  \intermed {   {\intermed \gamma }_{{\mathrm{1}}}  ( {\intermed \phi }_{{\mathrm{1}}}  ( {\intermed R }  ( {\intermed e}_{{\mathrm{1}}} )))   } : \intermed {  {\intermed R }  \ottsym{(}  {\intermed \sigma}_{{\mathrm{2}}}  \ottsym{)}  } \itot{ \rightsquigarrow  \target {  {\target e }_{{\mathrm{2}}}  } } \\
  \label{eq:compat_let13b}
   \intermed {  {\intermed \Sigma}_{{\mathrm{2}}}  } ; \intermed {  {\intermed { \Gamma_C } }  } ; \intermed {  \bullet  \ottsym{,}  {\intermed x}  \ottsym{:}  {\intermed R }  \ottsym{(}  {\intermed \sigma}_{{\mathrm{1}}}  \ottsym{)}  }  \vdash_{tm}  \intermed {   {\intermed \gamma }_{{\mathrm{2}}}  ( {\intermed \phi }_{{\mathrm{2}}}  ( {\intermed R }  ( {\intermed e}_{{\mathrm{2}}} )))   } : \intermed {  {\intermed R }  \ottsym{(}  {\intermed \sigma}_{{\mathrm{2}}}  \ottsym{)}  } \itot{ \rightsquigarrow  \target {  {\target e }'_{{\mathrm{2}}}  } } 
\end{gather}
Combining Lemma~\ref{lemma:ctx_well_inter_typing} with
Equation~\ref{eq:compat_let12b}, yields
\( \vdash_{ctx}  \intermed {  {\intermed \Sigma}_{{\mathrm{1}}}  } ; \intermed {  {\intermed { \Gamma_C } }  } ; \intermed {  \bullet  \ottsym{,}  {\intermed x}  \ottsym{:}  {\intermed R }  \ottsym{(}  {\intermed \sigma}_{{\mathrm{1}}}  \ottsym{)}  } \) and, by case analysis
on this environment well-formedness judgment, it follows that
\( \intermed {  {\intermed { \Gamma_C } }  } ; \intermed {  \bullet  }  \vdash_{ty}  \intermed {  {\intermed R }  \ottsym{(}  {\intermed \sigma}_{{\mathrm{1}}}  \ottsym{)}  } \itot{ \rightsquigarrow  \target {  {\target \sigma }_{{\mathrm{1}}}  } } \).
Using this, Goals~\ref{eq:compat_let4} and \ref{eq:compat_let5} follow by
applying both Equations~\ref{eq:compat_let9} and \ref{eq:compat_let12b} and both
Equations~\ref{eq:compat_let10} and \ref{eq:compat_let13b} to \textsc{iTm-let},
respectively.

By case analysis, the first step of both evaluation paths in Equation~\ref{eq:compat_let6}
must be appropriate instantiations of rule \textsc{iEval-let}, according to which,
\begin{equation*}
   \intermed {  {\intermed \Sigma}_{\ottmv{i}}  }  \vdash  \intermed {  \textbf{\intermed {let} } \, {\intermed x}  \ottsym{:}  {\intermed R }  \ottsym{(}  {\intermed \sigma}_{{\mathrm{1}}}  \ottsym{)}  \ottsym{=}  \ottsym{(}   {\intermed \gamma }_{\ottmv{i}}  ( {\intermed \phi }_{\ottmv{i}}  ( {\intermed R }  ( {\intermed e}'_{\ottmv{i}} )))   \ottsym{)} \, \textbf{\intermed {in} } \, \ottsym{(}   {\intermed \gamma }_{\ottmv{i}}  ( {\intermed \phi }_{\ottmv{i}}  ( {\intermed R }  ( {\intermed e}_{\ottmv{i}} )))   \ottsym{)}  }  \longrightarrow  \intermed {  [   {\intermed \gamma }_{\ottmv{i}}  ( {\intermed \phi }_{\ottmv{i}}  ( {\intermed R }  ( {\intermed e}'_{\ottmv{i}} )))   /  {\intermed x}  ]  \ottsym{(}   {\intermed \gamma }_{\ottmv{i}}  ( {\intermed \phi }_{\ottmv{i}}  ( {\intermed R }  ( {\intermed e}_{\ottmv{i}} )))   \ottsym{)}  } 
\end{equation*}
for each \(\intermed{i}\in\{1,2\}\).
This simplifies Goal~\ref{eq:compat_let6} to:
\begin{equation*}
\begin{aligned}
  \exists \intermed{{\intermed v}_{{\mathrm{1}}}}, \intermed{{\intermed v}_{{\mathrm{2}}}} :\, &
     \intermed {  {\intermed \Sigma}_{{\mathrm{1}}}  }  \vdash  \intermed {  [   {\intermed \gamma }_{{\mathrm{1}}}  ( {\intermed \phi }_{{\mathrm{1}}}  ( {\intermed R }  ( {\intermed e}'_{{\mathrm{1}}} )))   /  {\intermed x}  ]  \ottsym{(}   {\intermed \gamma }_{{\mathrm{1}}}  ( {\intermed \phi }_{{\mathrm{1}}}  ( {\intermed R }  ( {\intermed e}_{{\mathrm{1}}} )))   \ottsym{)}  }  \longrightarrow^{*}  \intermed {  {\intermed v}_{{\mathrm{1}}}  }  \\
    \band\,&  \intermed {  {\intermed \Sigma}_{{\mathrm{2}}}  }  \vdash  \intermed {  [   {\intermed \gamma }_{{\mathrm{2}}}  ( {\intermed \phi }_{{\mathrm{2}}}  ( {\intermed R }  ( {\intermed e}'_{{\mathrm{2}}} )))   /  {\intermed x}  ]  \ottsym{(}   {\intermed \gamma }_{{\mathrm{2}}}  ( {\intermed \phi }_{{\mathrm{2}}}  ( {\intermed R }  ( {\intermed e}_{{\mathrm{2}}} )))   \ottsym{)}  }  \longrightarrow^{*}  \intermed {  {\intermed v}_{{\mathrm{2}}}  }  \\
    \band\,&  ( \intermed {  {\intermed \Sigma}_{{\mathrm{1}}}  } : \intermed {  {\intermed v}_{{\mathrm{1}}}  } , \intermed {  {\intermed \Sigma}_{{\mathrm{2}}}  } : \intermed {  {\intermed v}_{{\mathrm{2}}}  } ) \in \mathcal{V} \llbracket \intermed {  {\intermed \sigma}_{{\mathrm{2}}}  } \rrbracket ^{\intermed {  {\intermed { \Gamma_C } }  } }_{\intermed {  {\intermed R }  } } 
\end{aligned}
\end{equation*}

The goal follows from Equation~\ref{eq:compat_let14}. Because of
Equation~\ref{eq:compat_let7}, we can choose
\( \intermed {  {\intermed e}_{{\mathrm{3}}}  } = \intermed {   {\intermed \gamma }_{{\mathrm{1}}}  ( {\intermed \phi }_{{\mathrm{1}}}  ( {\intermed R }  ( {\intermed e}'_{{\mathrm{1}}} )))   } \)
and \( \intermed {  {\intermed e}_{{\mathrm{4}}}  } = \intermed {   {\intermed \gamma }_{{\mathrm{2}}}  ( {\intermed \phi }_{{\mathrm{2}}}  ( {\intermed R }  ( {\intermed e}'_{{\mathrm{2}}} )))   } \). \\
\qed

\begin{lemmaB}[Compatibility - Method]\ \\
  \begin{mathpar}
    \ottaltinferrule{}{}
    {
       \intermed {  {\intermed { \Gamma_C } }  } ; \intermed {  {\intermed \Gamma}  }  \vdash  \intermed {  {\intermed \Sigma}_{{\mathrm{1}}}  } : \intermed {  {\intermed d}_{{\mathrm{1}}}  }  \simeq_{log}  \intermed {  {\intermed \Sigma}_{{\mathrm{2}}}  } : \intermed {  {\intermed d}_{{\mathrm{2}}}  } : \intermed {  {\intermed {\mathit{TC} } } \, {\intermed \sigma}  } 
      \\  ( \intermed {  {\intermed m}  } : \intermed {  {\intermed {\mathit{TC} } } \, {\intermed a }  } : \intermed {  {\intermed \sigma}'  } )  \in  \intermed {  {\intermed { \Gamma_C } }  } 
      \\  \intermed {  {\intermed { \Gamma_C } }  }  \vdash  \intermed {  {\intermed \Sigma}_{{\mathrm{1}}}  }  \simeq_{log}  \intermed {  {\intermed \Sigma}_{{\mathrm{2}}}  } 
    }
    {  \intermed {  {\intermed { \Gamma_C } }  } ; \intermed {  {\intermed \Gamma}  }  \vdash  \intermed {  {\intermed \Sigma}_{{\mathrm{1}}}  } : \intermed {  {\intermed d}_{{\mathrm{1}}}  \ottsym{.}  {\intermed m}  }  \simeq_{log}  \intermed {  {\intermed \Sigma}_{{\mathrm{2}}}  } : \intermed {  {\intermed d}_{{\mathrm{2}}}  \ottsym{.}  {\intermed m}  } : \intermed {  [  {\intermed \sigma}  /  {\intermed a }  ]  {\intermed \sigma}'  }  }
    \label{lemma:compat_method}
  \end{mathpar}
\end{lemmaB}

\proof
By inlining the definition of logical equivalence, suppose we have:
\begin{equation}\label{eq:compat_method1}
\begin{gathered}
   \intermed {  {\intermed R }  } \in \mathcal{F} \llbracket \intermed {  {\intermed \Gamma}  } \rrbracket ^{\intermed {  {\intermed { \Gamma_C } }  } } \\
   \intermed {  {\intermed \phi }  } \in \mathcal{G} \llbracket \intermed {  {\intermed \Gamma}  } \rrbracket ^{\intermed {  {\intermed \Sigma}_{{\mathrm{1}}}  } , \intermed {  {\intermed \Sigma}_{{\mathrm{2}}}  } , \intermed {  {\intermed { \Gamma_C } }  } }_{\intermed {  {\intermed R }  } } \\
   \intermed {  {\intermed \gamma }  } \in \mathcal{H} \llbracket \intermed {  {\intermed \Gamma}  } \rrbracket ^{\intermed {  {\intermed \Sigma}_{{\mathrm{1}}}  } , \intermed {  {\intermed \Sigma}_{{\mathrm{2}}}  } , \intermed {  {\intermed { \Gamma_C } }  } }_{\intermed {  {\intermed R }  } } 
\end{gathered}
\end{equation}
The goal to be proven is the following:
\begin{equation*}
   ( \intermed {  {\intermed \Sigma}_{{\mathrm{1}}}  } : \intermed {   {\intermed \gamma }_{{\mathrm{1}}}  ( {\intermed \phi }_{{\mathrm{1}}}  ( {\intermed R }  ( {\intermed d}_{{\mathrm{1}}}  \ottsym{.}  {\intermed m} )))   } , \intermed {  {\intermed \Sigma}_{{\mathrm{2}}}  } : \intermed {   {\intermed \gamma }_{{\mathrm{2}}}  ( {\intermed \phi }_{{\mathrm{2}}}  ( {\intermed R }  ( {\intermed d}_{{\mathrm{2}}}  \ottsym{.}  {\intermed m} )))   } ) \in \mathcal{E} \llbracket \intermed {  [  {\intermed \sigma}  /  {\intermed a }  ]  {\intermed \sigma}'  } \rrbracket ^{\intermed {  {\intermed { \Gamma_C } }  } }_{\intermed {  {\intermed R }  } } 
\end{equation*}
By applying the definition of the \( \mathcal{E} \) relation, it reduces
to:
\begin{gather*}
   \intermed {  {\intermed \Sigma}_{{\mathrm{1}}}  } ; \intermed {  {\intermed { \Gamma_C } }  } ; \intermed {  \bullet  }  \vdash_{tm}  \intermed {   {\intermed \gamma }_{{\mathrm{1}}}  ( {\intermed \phi }_{{\mathrm{1}}}  ( {\intermed R }  ( {\intermed d}_{{\mathrm{1}}}  \ottsym{.}  {\intermed m} )))   } : \intermed {  {\intermed R }  \ottsym{(}  [  {\intermed \sigma}  /  {\intermed a }  ]  {\intermed \sigma}'  \ottsym{)}  } \itot{ \rightsquigarrow  \target {  {\target e }  } } \\
   \intermed {  {\intermed \Sigma}_{{\mathrm{2}}}  } ; \intermed {  {\intermed { \Gamma_C } }  } ; \intermed {  \bullet  }  \vdash_{tm}  \intermed {   {\intermed \gamma }_{{\mathrm{2}}}  ( {\intermed \phi }_{{\mathrm{2}}}  ( {\intermed R }  ( {\intermed d}_{{\mathrm{2}}}  \ottsym{.}  {\intermed m} )))   } : \intermed {  {\intermed R }  \ottsym{(}  [  {\intermed \sigma}  /  {\intermed a }  ]  {\intermed \sigma}'  \ottsym{)}  } \itot{ \rightsquigarrow  \target {  {\target e }'  } } \\
\begin{split}
  \exists \intermed{{\intermed v}_{{\mathrm{1}}}}, \intermed{{\intermed v}_{{\mathrm{2}}}} :\, &
     \intermed {  {\intermed \Sigma}_{{\mathrm{1}}}  }  \vdash  \intermed {   {\intermed \gamma }_{{\mathrm{1}}}  ( {\intermed \phi }_{{\mathrm{1}}}  ( {\intermed R }  ( {\intermed d}_{{\mathrm{1}}}  \ottsym{.}  {\intermed m} )))   }  \longrightarrow^{*}  \intermed {  {\intermed v}_{{\mathrm{1}}}  }  \\
    \band\, &  \intermed {  {\intermed \Sigma}_{{\mathrm{2}}}  }  \vdash  \intermed {   {\intermed \gamma }_{{\mathrm{2}}}  ( {\intermed \phi }_{{\mathrm{2}}}  ( {\intermed R }  ( {\intermed d}_{{\mathrm{2}}}  \ottsym{.}  {\intermed m} )))   }  \longrightarrow^{*}  \intermed {  {\intermed v}_{{\mathrm{2}}}  }  \\
    \band\, &  ( \intermed {  {\intermed \Sigma}_{{\mathrm{1}}}  } : \intermed {  {\intermed v}_{{\mathrm{1}}}  } , \intermed {  {\intermed \Sigma}_{{\mathrm{2}}}  } : \intermed {  {\intermed v}_{{\mathrm{2}}}  } ) \in \mathcal{V} \llbracket \intermed {  [  {\intermed \sigma}  /  {\intermed a }  ]  {\intermed \sigma}'  } \rrbracket ^{\intermed {  {\intermed { \Gamma_C } }  } }_{\intermed {  {\intermed R }  } } 
\end{split}
\end{gather*}
By applying the substitutions,
and because \( \intermed {  {\intermed R }  \ottsym{(}  [  {\intermed \sigma}  /  {\intermed a }  ]  {\intermed \sigma}'  \ottsym{)}  } = \intermed {  [  {\intermed R }  \ottsym{(}  {\intermed \sigma}  \ottsym{)}  /  {\intermed a }  ]  {\intermed R }  \ottsym{(}  {\intermed \sigma}'  \ottsym{)}  } \),
the goal further reduces to:
\begin{gather}
  \label{eq:compat_method4}
   \intermed {  {\intermed \Sigma}_{{\mathrm{1}}}  } ; \intermed {  {\intermed { \Gamma_C } }  } ; \intermed {  \bullet  }  \vdash_{tm}  \intermed {  \ottsym{(}   {\intermed \gamma }_{{\mathrm{1}}}  ( {\intermed \phi }_{{\mathrm{1}}}  ( {\intermed R }  ( {\intermed d}_{{\mathrm{1}}} )))   \ottsym{)}  \ottsym{.}  {\intermed m}  } : \intermed {  [  {\intermed R }  \ottsym{(}  {\intermed \sigma}  \ottsym{)}  /  {\intermed a }  ]  {\intermed R }  \ottsym{(}  {\intermed \sigma}'  \ottsym{)}  } \itot{ \rightsquigarrow  \target {  {\target e }  } } \\
  \label{eq:compat_method5}
   \intermed {  {\intermed \Sigma}_{{\mathrm{2}}}  } ; \intermed {  {\intermed { \Gamma_C } }  } ; \intermed {  \bullet  }  \vdash_{tm}  \intermed {  \ottsym{(}   {\intermed \gamma }_{{\mathrm{2}}}  ( {\intermed \phi }_{{\mathrm{2}}}  ( {\intermed R }  ( {\intermed d}_{{\mathrm{2}}} )))   \ottsym{)}  \ottsym{.}  {\intermed m}  } : \intermed {  [  {\intermed R }  \ottsym{(}  {\intermed \sigma}  \ottsym{)}  /  {\intermed a }  ]  {\intermed R }  \ottsym{(}  {\intermed \sigma}'  \ottsym{)}  } \itot{ \rightsquigarrow  \target {  {\target e }'  } } \\
\begin{split}\label{eq:compat_method6}
  \exists \intermed{{\intermed v}_{{\mathrm{1}}}}, \intermed{{\intermed v}_{{\mathrm{2}}}} :\ &
     \intermed {  {\intermed \Sigma}_{{\mathrm{1}}}  }  \vdash  \intermed {  \ottsym{(}   {\intermed \gamma }_{{\mathrm{1}}}  ( {\intermed \phi }_{{\mathrm{1}}}  ( {\intermed R }  ( {\intermed d}_{{\mathrm{1}}} )))   \ottsym{)}  \ottsym{.}  {\intermed m}  }  \longrightarrow^{*}  \intermed {  {\intermed v}_{{\mathrm{1}}}  }  \\
    \band\, &  \intermed {  {\intermed \Sigma}_{{\mathrm{2}}}  }  \vdash  \intermed {  \ottsym{(}   {\intermed \gamma }_{{\mathrm{2}}}  ( {\intermed \phi }_{{\mathrm{2}}}  ( {\intermed R }  ( {\intermed d}_{{\mathrm{2}}} )))   \ottsym{)}  \ottsym{.}  {\intermed m}  }  \longrightarrow^{*}  \intermed {  {\intermed v}_{{\mathrm{2}}}  }  \\
    \band\, &  ( \intermed {  {\intermed \Sigma}_{{\mathrm{1}}}  } : \intermed {  {\intermed v}_{{\mathrm{1}}}  } , \intermed {  {\intermed \Sigma}_{{\mathrm{2}}}  } : \intermed {  {\intermed v}_{{\mathrm{2}}}  } ) \in \mathcal{V} \llbracket \intermed {  [  {\intermed \sigma}  /  {\intermed a }  ]  {\intermed \sigma}'  } \rrbracket ^{\intermed {  {\intermed { \Gamma_C } }  } }_{\intermed {  {\intermed R }  } } 
\end{split}
\end{gather}
By inlining the definition of logical equivalence in the first hypothesis of this lemma
and choosing \({\intermed R }\), \({\intermed \phi }\) and \({\intermed \gamma }\) from Equation~\ref{eq:compat_method1}, we
get:
\begin{equation*}
   ( \intermed {  {\intermed \Sigma}_{{\mathrm{1}}}  } : \intermed {   {\intermed \gamma }_{{\mathrm{1}}}  ( {\intermed \phi }_{{\mathrm{1}}}  ( {\intermed R }  ( {\intermed d}_{{\mathrm{1}}} )))   } , \intermed {  {\intermed \Sigma}_{{\mathrm{2}}}  } : \intermed {   {\intermed \gamma }_{{\mathrm{2}}}  ( {\intermed \phi }_{{\mathrm{2}}}  ( {\intermed R }  ( {\intermed d}_{{\mathrm{2}}} )))   } ) \in \mathcal{V} \llbracket \intermed {  {\intermed {\mathit{TC} } } \, {\intermed \sigma}  } \rrbracket ^{\intermed {  {\intermed { \Gamma_C } }  } }_{\intermed {  {\intermed R }  } } 
  \label{eq:compat_method7}
\end{equation*}
Then, from the definition of \(\mathcal{V}\), we get:
\begin{gather}
  \nonumber
   \intermed {   {\intermed \gamma }_{{\mathrm{1}}}  ( {\intermed \phi }_{{\mathrm{1}}}  ( {\intermed R }  ( {\intermed d}_{{\mathrm{1}}} )))   } = \intermed {  {\intermed D } \, \overline {\intermed \sigma}_{\ottmv{j}} \, {\overline {\intermed { dv } } }_{{\mathrm{1}}\,\ottmv{i}}  } \\
  \nonumber
   \intermed {   {\intermed \gamma }_{{\mathrm{2}}}  ( {\intermed \phi }_{{\mathrm{2}}}  ( {\intermed R }  ( {\intermed d}_{{\mathrm{2}}} )))   } = \intermed {  {\intermed D } \, \overline {\intermed \sigma}_{\ottmv{j}} \, {\overline {\intermed { dv } } }_{{\mathrm{2}}\,\ottmv{i}}  } \\
  \label{eq:compat_method8}
   ( \intermed {  {\intermed D }  } : \intermed {  \forall  {\overline {\intermed a} }_{\ottmv{j}}  \ottsym{.}  {\overline {\intermed Q} }_{\ottmv{i}}  \Rightarrow  {\intermed {\mathit{TC} } } \, {\intermed \sigma}_{\ottmv{q}}  } ) . \intermed {  {\intermed m}  }  \mapsto  \intermed {  {\intermed e}'_{{\mathrm{1}}}  }  \in  \intermed {  {\intermed \Sigma}_{{\mathrm{1}}}  } \\
  \label{eq:compat_method9}
   \intermed {  {\intermed \Sigma}_{{\mathrm{1}}}  } ; \intermed {  {\intermed { \Gamma_C } }  } ; \intermed {  \bullet  }  \vdash_{d}  \intermed {  {\intermed D } \, \overline {\intermed \sigma}_{\ottmv{j}} \, {\overline {\intermed { dv } } }_{{\mathrm{1}}\,\ottmv{i}}  } : \intermed {  {\intermed {\mathit{TC} } } \, {\intermed R }  \ottsym{(}  {\intermed \sigma}  \ottsym{)}  } \itot{\,  \rightsquigarrow  \target {  {\target e }_{{\mathrm{1}}}  } } \\
  \label{eq:compat_method10}
   \intermed {  {\intermed \Sigma}_{{\mathrm{2}}}  } ; \intermed {  {\intermed { \Gamma_C } }  } ; \intermed {  \bullet  }  \vdash_{d}  \intermed {  {\intermed D } \, \overline {\intermed \sigma}_{\ottmv{j}} \, {\overline {\intermed { dv } } }_{{\mathrm{2}}\,\ottmv{i}}  } : \intermed {  {\intermed {\mathit{TC} } } \, {\intermed R }  \ottsym{(}  {\intermed \sigma}  \ottsym{)}  } \itot{\,  \rightsquigarrow  \target {  {\target e }_{{\mathrm{2}}}  } } \\
  \label{eq:compat_method11}
  \ottcomp{ ( \intermed {  {\intermed \Sigma}_{{\mathrm{1}}}  } : \intermed {  {\intermed {dv} }_{{\mathrm{1}}\,\ottmv{i}}  } , \intermed {  {\intermed \Sigma}_{{\mathrm{2}}}  } : \intermed {  {\intermed {dv} }_{{\mathrm{2}}\,\ottmv{i}}  } ) \in \mathcal{V} \llbracket \intermed {  [  \overline {\intermed \sigma}_{\ottmv{j}}  /  {\overline {\intermed a} }_{\ottmv{j}}  ]  {\intermed Q}_{\ottmv{i}}  } \rrbracket ^{\intermed {  {\intermed { \Gamma_C } }  } }_{\intermed {  {\intermed R }  } } }{\ottmv{i}}
\end{gather}
for some \(\intermed{\overline {\intermed \sigma}_{\ottmv{j}}}\), \(\intermed{{\overline {\intermed { dv } } }_{{\mathrm{1}}\,\ottmv{i}}}\), \(\intermed{{\overline {\intermed { dv } } }_{{\mathrm{2}}\,\ottmv{i}}}\), \(\intermed{{\overline {\intermed Q} }_{\ottmv{i}}}\),
\(\intermed{{\intermed e}'}\) and \(\intermed{{\intermed \sigma}_{\ottmv{q}}}\) such that \(\intermed{ \intermed {  {\intermed \sigma}  } = \intermed {  [  \overline {\intermed \sigma}_{\ottmv{j}}  /  {\overline {\intermed a} }_{\ottmv{j}}  ]  {\intermed \sigma}_{\ottmv{q}}  } }\).

Lemma~\ref{lemma:ctx_well_dict_typing}, applied on
Equations~\ref{eq:compat_method9} and~\ref{eq:compat_method10}, yields:
\begin{gather}
   \vdash_{ctx}  \intermed {  {\intermed \Sigma}_{{\mathrm{1}}}  } ; \intermed {  {\intermed { \Gamma_C } }  } ; \intermed {  \bullet  } 
  \label{eq:compat_method9b} \\
   \vdash_{ctx}  \intermed {  {\intermed \Sigma}_{{\mathrm{2}}}  } ; \intermed {  {\intermed { \Gamma_C } }  } ; \intermed {  \bullet  } 
  \label{eq:compat_method10b} 
\end{gather}
Also, from the second premise of this lemma's rule, there are
\(\intermed{{\intermed { \Gamma_C } }_{{\mathrm{1}}}}\) and \(\intermed{{\intermed { \Gamma_C } }_{{\mathrm{2}}}}\) such that
\( \intermed {  {\intermed { \Gamma_C } }  } = \intermed {  {\intermed { \Gamma_C } }_{{\mathrm{1}}}  }, \intermed {  {\intermed m}  \ottsym{:}  {\intermed {\mathit{TC} } } \, {\intermed a }  \ottsym{:}  {\intermed \sigma}'  \ottsym{,}  {\intermed { \Gamma_C } }_{{\mathrm{2}}}  } \). Then, from
Lemma~\ref{lemma:classmethodwty}, we get \( \intermed {  {\intermed { \Gamma_C } }_{{\mathrm{1}}}  } ; \intermed {  \bullet  \ottsym{,}  {\intermed a }  }  \vdash_{ty}  \intermed {  {\intermed \sigma}'  } \itot{ \rightsquigarrow  \target {  {\target \sigma }'  } } \),
which means that the only free variable appearing in \(\intermed{{\intermed \sigma}'}\) is the
fresh variable \({\intermed a }\). Then,
\begin{equation}\label{eq:compat_method2}
   \intermed {  {\intermed R }  \ottsym{(}  {\intermed \sigma}'  \ottsym{)}  } = \intermed {  {\intermed \sigma}'  } 
\end{equation}
Also, since the dictionary \(\intermed{{\intermed D } \, \overline {\intermed \sigma}_{\ottmv{j}} \, {\overline {\intermed { dv } } }_{{\mathrm{1}}\,\ottmv{i}}}\) is closed
(it is the result of applying the closing substitutions \({\intermed R }\),
\(\intermed{{\intermed \phi }_{{\mathrm{1}}}}\) and \(\intermed{{\intermed \gamma }_{{\mathrm{1}}}}\) on dictionary \(\intermed{{\intermed d}_{{\mathrm{1}}}}\)),
types \(\intermed{\overline {\intermed \sigma}_{\ottmv{j}}}\) can not contain any free variables. Hence,
\( \intermed {  {\intermed R }  \ottsym{(}  \overline {\intermed \sigma}_{\ottmv{j}}  \ottsym{)}  } = \intermed {  \overline {\intermed \sigma}_{\ottmv{j}}  } \).
In addition, from the last conclusion of Lemma~\ref{lemma:inSS2C},
supplied with \( \vdash_{ctx}  \intermed {  {\intermed \Sigma}_{{\mathrm{1}}}  } ; \intermed {  {\intermed { \Gamma_C } }  } ; \intermed {  \bullet  } \) and Equation~\ref{eq:compat_method8},
we have that \( \intermed {  {\intermed { \Gamma_C } }  } ; \intermed {  \bullet  \ottsym{,}  {\overline {\intermed a} }_{\ottmv{j}}  }  \vdash_{ty}  \intermed {  {\intermed \sigma}_{\ottmv{q}}  } \itot{ \rightsquigarrow  \target {  {\target \sigma }  } } \).
Because the type variables
\(\intermed{{\overline {\intermed a} }_{\ottmv{j}}}\) are not in \({\intermed \Gamma}\), they are not in the domain of \({\intermed R }\), thus \( \intermed {  {\intermed R }  \ottsym{(}  {\intermed \sigma}_{\ottmv{q}}  \ottsym{)}  } = \intermed {  {\intermed \sigma}_{\ottmv{q}}  } \).
Then,
\begin{equation}\label{eq:compat_method2'}
\begin{aligned}
  \intermed{{\intermed R }  \ottsym{(}  {\intermed \sigma}  \ottsym{)}} & = \intermed{{\intermed R }  \ottsym{(}  [  \overline {\intermed \sigma}_{\ottmv{j}}  /  {\overline {\intermed a} }_{\ottmv{j}}  ]  {\intermed \sigma}_{\ottmv{q}}  \ottsym{)}}\\
  & = \intermed{[  {\intermed R }  \ottsym{(}  \overline {\intermed \sigma}_{\ottmv{j}}  \ottsym{)}  /  {\overline {\intermed a} }_{\ottmv{j}}  ]  {\intermed R }  \ottsym{(}  {\intermed \sigma}_{\ottmv{q}}  \ottsym{)}}\\
  & =  \intermed {  [  \overline {\intermed \sigma}_{\ottmv{j}}  /  {\overline {\intermed a} }_{\ottmv{j}}  ]  {\intermed \sigma}_{\ottmv{q}}  } = \intermed {  {\intermed \sigma}  } 
\end{aligned}
\end{equation}
Equations~\ref{eq:compat_method9} and \ref{eq:compat_method10} can only stand
as conclusions of dictionary typing rule \textsc{D-con}. After rewriting them with
Equation~\ref{eq:compat_method2'} we can invert them and get their premises. 
Also because environment \(\intermed{{\intermed \Sigma}_{{\mathrm{1}}}}\) can only contain a unique entry
for each dictionary type (in this case, the one shown in Equation~\ref{eq:compat_method8}),
we finally conclude
\begin{gather}
  \label{eq:compat_method12}
  \ottcomp{ \intermed {  {\intermed { \Gamma_C } }  } ; \intermed {  \bullet  }  \vdash_{ty}  \intermed {  {\intermed \sigma}_{\ottmv{j}}  } \itot{ \rightsquigarrow  \target {  {\target \sigma }_{\ottmv{j}}  } } }{\ottmv{j}}\\
  \label{eq:compat_method13}
  \ottcomp{ \intermed {  {\intermed \Sigma}_{{\mathrm{1}}}  } ; \intermed {  {\intermed { \Gamma_C } }  } ; \intermed {  \bullet  }  \vdash_{d}  \intermed {  {\intermed {dv} }_{{\mathrm{1}}\,\ottmv{i}}  } : \intermed {  [  \overline {\intermed \sigma}_{\ottmv{j}}  /  {\overline {\intermed a} }_{\ottmv{j}}  ]  {\intermed Q}_{\ottmv{i}}  } \itot{\,  \rightsquigarrow  \target {  {\target e }_{{\mathrm{1}}\,\ottmv{i}}  } } }{\ottmv{i}}\\
  \label{eq:compat_method14}
  \ottcomp{ \intermed {  {\intermed \Sigma}_{{\mathrm{2}}}  } ; \intermed {  {\intermed { \Gamma_C } }  } ; \intermed {  \bullet  }  \vdash_{d}  \intermed {  {\intermed {dv} }_{{\mathrm{2}}\,\ottmv{i}}  } : \intermed {  [  \overline {\intermed \sigma}_{\ottmv{j}}  /  {\overline {\intermed a} }_{\ottmv{j}}  ]  {\intermed Q}_{\ottmv{i}}  } \itot{\,  \rightsquigarrow  \target {  {\target e }_{{\mathrm{2}}\,\ottmv{i}}  } } }{\ottmv{i}}\\
  \label{eq:compat_method15}
   \intermed {  {\intermed \Sigma}'_{{\mathrm{1}}}  } ; \intermed {  {\intermed { \Gamma_C } }  } ; \intermed {  \bullet  \ottsym{,}  {\overline {\intermed a} }_{\ottmv{j}}  \ottsym{,}  {\overline {\intermed \delta} }_{\ottmv{i}}  \ottsym{:}  {\overline {\intermed Q} }_{\ottmv{i}}  }  \vdash_{tm}  \intermed {  {\intermed e}_{{\mathrm{1}}}  } : \intermed {  [  {\intermed \sigma}_{\ottmv{q}}  /  {\intermed a }  ]  {\intermed \sigma}'  } \itot{ \rightsquigarrow  \target {  {\target e }_{{\mathrm{3}}}  } } \\
  \label{eq:compat_method15b}
   \intermed {  {\intermed \Sigma}'_{{\mathrm{2}}}  } ; \intermed {  {\intermed { \Gamma_C } }  } ; \intermed {  \bullet  \ottsym{,}  {\overline {\intermed a} }_{\ottmv{j}}  \ottsym{,}  {\overline {\intermed \delta} }_{\ottmv{i}}  \ottsym{:}  {\overline {\intermed Q} }_{\ottmv{i}}  }  \vdash_{tm}  \intermed {  {\intermed e}_{{\mathrm{2}}}  } : \intermed {  [  {\intermed \sigma}_{\ottmv{q}}  /  {\intermed a }  ]  {\intermed \sigma}'  } \itot{ \rightsquigarrow  \target {  {\target e }_{{\mathrm{4}}}  } } \\
  \label{eq:compat_method3c}
  \textit{where}~ \intermed {  {\intermed \Sigma}_{{\mathrm{1}}}  } = \intermed {  {\intermed \Sigma}'_{{\mathrm{1}}}  \ottsym{,}  \ottsym{(}  {\intermed D }  \ottsym{:}  \forall  {\overline {\intermed a} }_{\ottmv{j}}  \ottsym{.}  {\overline {\intermed Q} }_{\ottmv{i}}  \Rightarrow  {\intermed {\mathit{TC} } } \, {\intermed \sigma}_{\ottmv{q}}  \ottsym{)}  \ottsym{.}  {\intermed m}  \mapsto  {\intermed e}'_{{\mathrm{1}}}  \ottsym{,}  {\intermed \Sigma}''_{{\mathrm{1}}}  } \\
  \label{eq:compat_method3}
  \textit{and}~ \intermed {  {\intermed e}'_{{\mathrm{1}}}  } = \intermed {  \Lambda  {\overline {\intermed a} }_{\ottmv{j}}  \ottsym{.}  \lambda  {\overline {\intermed \delta} }_{\ottmv{i}}  \ottsym{:}  {\overline {\intermed Q} }_{\ottmv{i}}  \ottsym{.}  {\intermed e}_{{\mathrm{1}}}  }  \\
  \label{eq:compat_method3d}
  \textit{and}~ \intermed {  {\intermed \Sigma}_{{\mathrm{2}}}  } = \intermed {  {\intermed \Sigma}'_{{\mathrm{2}}}  \ottsym{,}  \ottsym{(}  {\intermed D }  \ottsym{:}  \forall  {\overline {\intermed a} }_{\ottmv{j}}  \ottsym{.}  {\overline {\intermed Q} }_{\ottmv{i}}  \Rightarrow  {\intermed {\mathit{TC} } } \, {\intermed \sigma}_{\ottmv{q}}  \ottsym{)}  \ottsym{.}  {\intermed m}  \mapsto  {\intermed e}'_{{\mathrm{2}}}  \ottsym{,}  {\intermed \Sigma}''_{{\mathrm{2}}}  } \\
  \label{eq:compat_method3b}
  \textit{and}~ \intermed {  {\intermed e}'_{{\mathrm{2}}}  } = \intermed {  \Lambda  {\overline {\intermed a} }_{\ottmv{j}}  \ottsym{.}  \lambda  {\overline {\intermed \delta} }_{\ottmv{i}}  \ottsym{:}  {\overline {\intermed Q} }_{\ottmv{i}}  \ottsym{.}  {\intermed e}_{{\mathrm{2}}}  } 
\end{gather}
With Equations~\ref{eq:compat_method2} and \ref{eq:compat_method2'},
Goals~\ref{eq:compat_method4} and \ref{eq:compat_method5} follow by using the
second premise of the theorem's rule and Equations~\ref{eq:compat_method9} and
\ref{eq:compat_method10}, respectively, in \textsc{iTm-method}.

Using Equations~\ref{eq:compat_method3c} and~\ref{eq:compat_method3d} in
\textsc{iEval-method}, results in:
\begin{gather*}
   \intermed {  {\intermed \Sigma}_{{\mathrm{1}}}  }  \vdash  \intermed {  \ottsym{(}  {\intermed D } \, \overline {\intermed \sigma}_{\ottmv{j}} \, {\overline {\intermed { dv } } }_{{\mathrm{1}}\,\ottmv{i}}  \ottsym{)}  \ottsym{.}  {\intermed m}  }  \longrightarrow  \intermed {  {\intermed e}'_{{\mathrm{1}}} \, \overline {\intermed \sigma}_{\ottmv{j}} \, {\overline {\intermed { dv } } }_{{\mathrm{1}}\,\ottmv{i}}  } \\
   \intermed {  {\intermed \Sigma}_{{\mathrm{2}}}  }  \vdash  \intermed {  \ottsym{(}  {\intermed D } \, \overline {\intermed \sigma}_{\ottmv{j}} \, {\overline {\intermed { dv } } }_{{\mathrm{2}}\,\ottmv{i}}  \ottsym{)}  \ottsym{.}  {\intermed m}  }  \longrightarrow  \intermed {  {\intermed e}'_{{\mathrm{2}}} \, \overline {\intermed \sigma}_{\ottmv{j}} \, {\overline {\intermed { dv } } }_{{\mathrm{2}}\,\ottmv{i}}  } 
\end{gather*}
This reduces Goal~\ref{eq:compat_method6} to:
\begin{equation}
\begin{aligned}\label{eq:compat_method16}
  \exists \intermed{{\intermed v}_{{\mathrm{1}}}}, \intermed{{\intermed v}_{{\mathrm{2}}}} :\, &
     \intermed {  {\intermed \Sigma}_{{\mathrm{1}}}  }  \vdash  \intermed {  {\intermed e}'_{{\mathrm{1}}} \, \overline {\intermed \sigma}_{\ottmv{j}} \, {\overline {\intermed { dv } } }_{{\mathrm{1}}\,\ottmv{i}}  }  \longrightarrow^{*}  \intermed {  {\intermed v}_{{\mathrm{1}}}  } \\
    \band\, &  \intermed {  {\intermed \Sigma}_{{\mathrm{2}}}  }  \vdash  \intermed {  {\intermed e}'_{{\mathrm{2}}} \, \overline {\intermed \sigma}_{\ottmv{j}} \, {\overline {\intermed { dv } } }_{{\mathrm{2}}\,\ottmv{i}}  }  \longrightarrow^{*}  \intermed {  {\intermed v}_{{\mathrm{2}}}  }  \\
    \band\, &  ( \intermed {  {\intermed \Sigma}_{{\mathrm{1}}}  } : \intermed {  {\intermed v}_{{\mathrm{1}}}  } , \intermed {  {\intermed \Sigma}_{{\mathrm{2}}}  } : \intermed {  {\intermed v}_{{\mathrm{2}}}  } ) \in \mathcal{V} \llbracket \intermed {  [  {\intermed \sigma}  /  {\intermed a }  ]  {\intermed \sigma}'  } \rrbracket ^{\intermed {  {\intermed { \Gamma_C } }  } }_{\intermed {  {\intermed R }  } } 
\end{aligned}
\end{equation}
From the definition of logical equivalence in the theorem's third hypothesis,
together with Equations~\ref{eq:compat_method3c} and~\ref{eq:compat_method3d},
we get that:
\begin{equation}
   \intermed {  {\intermed { \Gamma_C } }  } ; \intermed {  \bullet  }  \vdash  \intermed {  {\intermed \Sigma}'_{{\mathrm{1}}}  } : \intermed {  {\intermed e}'_{{\mathrm{1}}}  }  \simeq_{log}  \intermed {  {\intermed \Sigma}'_{{\mathrm{2}}}  } : \intermed {  {\intermed e}'_{{\mathrm{2}}}  } : \intermed {  \forall  {\overline {\intermed a} }_{\ottmv{j}}  \ottsym{.}  [  {\intermed \sigma}_{\ottmv{q}}  /  {\intermed a }  ]  {\intermed \sigma}'  } 
  \label{eq:compat_method17}
\end{equation}
Repeatedly applying compatibility
Lemma~\ref{lemma:compat_tyapp} to Equations~\ref{eq:compat_method17} and
\ref{eq:compat_method12}, results in:
\begin{equation}
   \intermed {  {\intermed { \Gamma_C } }  } ; \intermed {  \bullet  }  \vdash  \intermed {  {\intermed \Sigma}'_{{\mathrm{1}}}  } : \intermed {  {\intermed e}'_{{\mathrm{1}}} \, \overline {\intermed \sigma}_{\ottmv{j}}  }  \simeq_{log}  \intermed {  {\intermed \Sigma}'_{{\mathrm{2}}}  } : \intermed {  {\intermed e}'_{{\mathrm{2}}} \, \overline {\intermed \sigma}_{\ottmv{j}}  } : \intermed {  [  \overline {\intermed \sigma}_{\ottmv{j}}  /  {\overline {\intermed a} }_{\ottmv{j}}  ]   {\overline {\intermed Q} }_{\ottmv{i}}  \Rightarrow  [  \overline {\intermed \sigma}_{\ottmv{j}}  /  {\overline {\intermed a} }_{\ottmv{j}}  ]  [  {\intermed \sigma}_{\ottmv{q}}  /  {\intermed a }  ]  {\intermed \sigma}'   } 
  \label{eq:compat_method18}
\end{equation}
By applying weakening Lemma~\ref{lemma:log_weakening} on this result, in combination with
Equations~\ref{eq:compat_method9b} and~\ref{eq:compat_method10b}, we get:
\begin{equation}
   \intermed {  {\intermed { \Gamma_C } }  } ; \intermed {  \bullet  }  \vdash  \intermed {  {\intermed \Sigma}_{{\mathrm{1}}}  } : \intermed {  {\intermed e}'_{{\mathrm{1}}} \, \overline {\intermed \sigma}_{\ottmv{j}}  }  \simeq_{log}  \intermed {  {\intermed \Sigma}_{{\mathrm{2}}}  } : \intermed {  {\intermed e}'_{{\mathrm{2}}} \, \overline {\intermed \sigma}_{\ottmv{j}}  } : \intermed {  [  \overline {\intermed \sigma}_{\ottmv{j}}  /  {\overline {\intermed a} }_{\ottmv{j}}  ]   {\overline {\intermed Q} }_{\ottmv{i}}  \Rightarrow  [  \overline {\intermed \sigma}_{\ottmv{j}}  /  {\overline {\intermed a} }_{\ottmv{j}}  ]  [  {\intermed \sigma}_{\ottmv{q}}  /  {\intermed a }  ]  {\intermed \sigma}'   } 
  \label{eq:compat_method18b}
\end{equation}
From the definition of logical equivalence and
Equation~\ref{eq:compat_method11},
we can derive that:
\begin{equation*}
  \ottcomp{ \intermed {  {\intermed { \Gamma_C } }  } ; \intermed {  \bullet  }  \vdash  \intermed {  {\intermed \Sigma}_{{\mathrm{1}}}  } : \intermed {  {\intermed {dv} }_{{\mathrm{1}}\,\ottmv{i}}  }  \simeq_{log}  \intermed {  {\intermed \Sigma}_{{\mathrm{2}}}  } : \intermed {  {\intermed {dv} }_{{\mathrm{2}}\,\ottmv{i}}  } : \intermed {  [  \overline {\intermed \sigma}_{\ottmv{j}}  /  {\overline {\intermed a} }_{\ottmv{j}}  ]  {\intermed Q}_{\ottmv{i}}  } }{\ottmv{i}}
\end{equation*}
Repeatedly applying compatibility Lemma~\ref{lemma:compat_dapp} on
Equations~\ref{eq:compat_method18b}, together with
the above equation,
results in:
\begin{equation*}
   \intermed {  {\intermed { \Gamma_C } }  } ; \intermed {  \bullet  }  \vdash  \intermed {  {\intermed \Sigma}_{{\mathrm{1}}}  } : \intermed {  {\intermed e}'_{{\mathrm{1}}} \, \overline {\intermed \sigma}_{\ottmv{j}} \, {\overline {\intermed { dv } } }_{{\mathrm{1}}\,\ottmv{i}}  }  \simeq_{log}  \intermed {  {\intermed \Sigma}_{{\mathrm{2}}}  } : \intermed {  {\intermed e}'_{{\mathrm{2}}} \, \overline {\intermed \sigma}_{\ottmv{j}} \, {\overline {\intermed { dv } } }_{{\mathrm{2}}\,\ottmv{i}}  } : \intermed {  [  \overline {\intermed \sigma}_{\ottmv{j}}  /  {\overline {\intermed a} }_{\ottmv{j}}  ]  [  {\intermed \sigma}_{\ottmv{q}}  /  {\intermed a }  ]  {\intermed \sigma}'  } 
\end{equation*}
Since expressions \(\intermed{{\intermed e}'_{{\mathrm{1}}} \, \overline {\intermed \sigma}_{\ottmv{j}} \, {\overline {\intermed { dv } } }_{{\mathrm{1}}\,\ottmv{i}}}\) and \(\intermed{{\intermed e}'_{{\mathrm{2}}} \, \overline {\intermed \sigma}_{\ottmv{j}} \, {\overline {\intermed { dv } } }_{{\mathrm{2}}\,\ottmv{i}}}\) are closed, by the definition of the logical relation, applying
any substitutions
\( \intermed {  {\intermed R }  } \in \mathcal{F} \llbracket \intermed {  \bullet  } \rrbracket ^{\intermed {  {\intermed { \Gamma_C } }  } } \), \( \intermed {  {\intermed \phi }  } \in \mathcal{G} \llbracket \intermed {  \bullet  } \rrbracket ^{\intermed {  {\intermed \Sigma}_{{\mathrm{1}}}  } , \intermed {  {\intermed \Sigma}_{{\mathrm{2}}}  } , \intermed {  {\intermed { \Gamma_C } }  } }_{\intermed {  {\intermed R }  } } \)
and \( \intermed {  {\intermed \gamma }  } \in \mathcal{H} \llbracket \intermed {  \bullet  } \rrbracket ^{\intermed {  {\intermed \Sigma}_{{\mathrm{1}}}  } , \intermed {  {\intermed \Sigma}_{{\mathrm{2}}}  } , \intermed {  {\intermed { \Gamma_C } }  } }_{\intermed {  {\intermed R }  } } \) on both expressions should result in two terms that are related by the \(\mathcal{E}\) relation. By case analysis on \({\intermed R }\), \({\intermed \phi }\) and \({\intermed \gamma }\), only the empty substitutions are valid choices, returning exactly the same
expressions. Taking this into account, we have:
\begin{equation*}
   ( \intermed {  {\intermed \Sigma}_{{\mathrm{1}}}  } : \intermed {  {\intermed e}'_{{\mathrm{1}}} \, \overline {\intermed \sigma}_{\ottmv{j}} \, {\overline {\intermed { dv } } }_{{\mathrm{1}}\,\ottmv{i}}  } , \intermed {  {\intermed \Sigma}_{{\mathrm{2}}}  } : \intermed {  {\intermed e}'_{{\mathrm{2}}} \, \overline {\intermed \sigma}_{\ottmv{j}} \, {\overline {\intermed { dv } } }_{{\mathrm{2}}\,\ottmv{i}}  } ) \in \mathcal{E} \llbracket \intermed {  [  \overline {\intermed \sigma}_{\ottmv{j}}  /  {\overline {\intermed a} }_{\ottmv{j}}  ]  [  {\intermed \sigma}_{\ottmv{q}}  /  {\intermed a }  ]  {\intermed \sigma}'  } \rrbracket ^{\intermed {  {\intermed { \Gamma_C } }  } }_{\intermed {  {\intermed R }  } } 
\end{equation*}
In turn, unfolding the definition of the \(\mathcal{E}\) relation
results in:
\begin{equation}\label{eq:compat_method23}
\begin{aligned}
  \exists \intermed{{\intermed v}_{{\mathrm{3}}}}, \intermed{{\intermed v}_{{\mathrm{4}}}} :\, &
     \intermed {  {\intermed \Sigma}_{{\mathrm{1}}}  }  \vdash  \intermed {  {\intermed e}'_{{\mathrm{1}}} \, \overline {\intermed \sigma}_{\ottmv{j}} \, {\overline {\intermed { dv } } }_{{\mathrm{1}}\,\ottmv{i}}  }  \longrightarrow^{*}  \intermed {  {\intermed v}_{{\mathrm{3}}}  }  \\
    \band\, &  \intermed {  {\intermed \Sigma}_{{\mathrm{2}}}  }  \vdash  \intermed {  {\intermed e}'_{{\mathrm{2}}} \, \overline {\intermed \sigma}_{\ottmv{j}} \, {\overline {\intermed { dv } } }_{{\mathrm{2}}\,\ottmv{i}}  }  \longrightarrow^{*}  \intermed {  {\intermed v}_{{\mathrm{4}}}  } \\
    \band\, &  ( \intermed {  {\intermed \Sigma}_{{\mathrm{1}}}  } : \intermed {  {\intermed v}_{{\mathrm{3}}}  } , \intermed {  {\intermed \Sigma}_{{\mathrm{2}}}  } : \intermed {  {\intermed v}_{{\mathrm{4}}}  } ) \in \mathcal{V} \llbracket \intermed {  [  \overline {\intermed \sigma}_{\ottmv{j}}  /  {\overline {\intermed a} }_{\ottmv{j}}  ]  [  {\intermed \sigma}  /  {\intermed a }  ]  {\intermed \sigma}'  } \rrbracket ^{\intermed {  {\intermed { \Gamma_C } }  } }_{\intermed {  {\intermed R }  } } 
\end{aligned}
\end{equation}
Goal~\ref{eq:compat_method16} follows from Equation~\ref{eq:compat_method23} by
noting that \( \intermed {  [  \overline {\intermed \sigma}_{\ottmv{j}}  /  {\overline {\intermed a} }_{\ottmv{j}}  ]  [  {\intermed \sigma}  /  {\intermed a }  ]  {\intermed \sigma}'  } = \intermed {  [  [  \overline {\intermed \sigma}_{\ottmv{j}}  /  {\overline {\intermed a} }_{\ottmv{j}}  ]  {\intermed \sigma}  /  {\intermed a }  ]  [  \overline {\intermed \sigma}_{\ottmv{j}}  /  {\overline {\intermed a} }_{\ottmv{j}}  ]  {\intermed \sigma}'  } = \intermed {  [  [  \overline {\intermed \sigma}_{\ottmv{j}}  /  {\overline {\intermed a} }_{\ottmv{j}}  ]  {\intermed \sigma}  /  {\intermed a }  ]  {\intermed \sigma}'  }  \) and
taking \( \intermed {  {\intermed v}_{{\mathrm{3}}}  } = \intermed {  {\intermed v}_{{\mathrm{1}}}  } \) and \( \intermed {  {\intermed v}_{{\mathrm{4}}}  } = \intermed {  {\intermed v}_{{\mathrm{2}}}  } \). \\
\qed

\subsection{Helper Theorems}

\renewcommand\itot[1]{}

\begin{theoremB}[Congruence - Expressions]\ \\
  If \( \intermed {  {\intermed { \Gamma_C } }  } ; \intermed {  {\intermed \Gamma}  }  \vdash  \intermed {  {\intermed \Sigma}_{{\mathrm{1}}}  } : \intermed {  {\intermed e}_{{\mathrm{1}}}  }  \simeq_{log}  \intermed {  {\intermed \Sigma}_{{\mathrm{2}}}  } : \intermed {  {\intermed e}_{{\mathrm{2}}}  } : \intermed {  {\intermed \sigma}  } \) \\
  and \(  \intermed {  {\intermed M}_{{\mathrm{1}}}  } : ( \intermed {  {\intermed \Sigma}_{{\mathrm{1}}}  } ; \intermed {  {\intermed { \Gamma_C } }  } ; \intermed {  {\intermed \Gamma}  }  \Rightarrow  \intermed {  {\intermed \sigma}  } )  \mapsto  ( \intermed {  {\intermed \Sigma}_{{\mathrm{1}}}  } ; \intermed {  {\intermed { \Gamma_C } }  } ; \intermed {  {\intermed \Gamma}'  }  \Rightarrow  \intermed {  {\intermed \sigma}'  } ) \itot{ \rightsquigarrow  \target {  {\target M }_{{\mathrm{1}}}  } } \)
  and \(  \intermed {  {\intermed M}_{{\mathrm{2}}}  } : ( \intermed {  {\intermed \Sigma}_{{\mathrm{2}}}  } ; \intermed {  {\intermed { \Gamma_C } }  } ; \intermed {  {\intermed \Gamma}  }  \Rightarrow  \intermed {  {\intermed \sigma}  } )  \mapsto  ( \intermed {  {\intermed \Sigma}_{{\mathrm{2}}}  } ; \intermed {  {\intermed { \Gamma_C } }  } ; \intermed {  {\intermed \Gamma}'  }  \Rightarrow  \intermed {  {\intermed \sigma}'  } ) \itot{ \rightsquigarrow  \target {  {\target M }_{{\mathrm{2}}}  } } \) \\
  and \( \intermed {  {\intermed \Sigma}_{{\mathrm{1}}}  } : \intermed {  {\intermed M}_{{\mathrm{1}}}  }  \simeq_{log}  \intermed {  {\intermed \Sigma}_{{\mathrm{2}}}  } : \intermed {  {\intermed M}_{{\mathrm{2}}}  } : ( \intermed {  {\intermed { \Gamma_C } }  } ; \intermed {  {\intermed \Gamma}  }  \Rightarrow  \intermed {  {\intermed \sigma}  } )  \mapsto  ( \intermed {  {\intermed { \Gamma_C } }  } ; \intermed {  {\intermed \Gamma}'  }  \Rightarrow  \intermed {  {\intermed \sigma}'  } ) \) \\
  then \( \intermed {  {\intermed { \Gamma_C } }  } ; \intermed {  {\intermed \Gamma}'  }  \vdash  \intermed {  {\intermed \Sigma}_{{\mathrm{1}}}  } : \intermed {   {\intermed M}_{{\mathrm{1}}}  [  {\intermed e}_{{\mathrm{1}}}  ]   }  \simeq_{log}  \intermed {  {\intermed \Sigma}_{{\mathrm{2}}}  } : \intermed {   {\intermed M}_{{\mathrm{2}}}  [  {\intermed e}_{{\mathrm{2}}}  ]   } : \intermed {  {\intermed \sigma}'  } \).
  \label{thmA:congruence_expr}
\end{theoremB}

\proof
The goal follows directly from the definition of logical equivalence for
contexts. \\
\qed

\renewcommand\itot[1]{#1}

\begin{theoremB}[\targetL{} Context Preserved by Elaboration]\ \\
  If \( \intermed {  {\intermed \Sigma}  } ; \intermed {  {\intermed { \Gamma_C } }  } ; \intermed {  {\intermed \Gamma}  }  \vdash_{tm}  \intermed {  {\intermed e}  } : \intermed {  {\intermed \sigma}  } \itot{ \rightsquigarrow  \target {  {\target e }  } } \)
  and \( \intermed {  {\intermed M}  } : ( \intermed {  {\intermed \Sigma}  } ; \intermed {  {\intermed { \Gamma_C } }  } ; \intermed {  {\intermed \Gamma}  }  \Rightarrow  \intermed {  {\intermed \sigma}  } )  \mapsto  ( \intermed {  {\intermed \Sigma}  } ; \intermed {  {\intermed { \Gamma_C } }  } ; \intermed {  {\intermed \Gamma}'  }  \Rightarrow  \intermed {  {\intermed \sigma}'  } ) \itot{ \rightsquigarrow  \target {  {\target M }  } } \)
  then \( \intermed {  {\intermed \Sigma}  } ; \intermed {  {\intermed { \Gamma_C } }  } ; \intermed {  {\intermed \Gamma}'  }  \vdash_{tm}  \intermed {   {\intermed M}  [  {\intermed e}  ]   } : \intermed {  {\intermed \sigma}'  } \itot{ \rightsquigarrow  \target {   {\target M }  [  {\target e }  ]   } } \).
  \label{thmA:tgt_ctx_compat}
\end{theoremB}
\proof By structural induction on the typing derivation of \({\intermed M}\).\\
\proofStep{\textsc{iM-empty}}
{
   \intermed {  \ottsym{[} \, \bullet \, \ottsym{]}  } : ( \intermed {  {\intermed \Sigma}  } ; \intermed {  {\intermed { \Gamma_C } }  } ; \intermed {  {\intermed \Gamma}  }  \Rightarrow  \intermed {  {\intermed \sigma}  } )  \mapsto  ( \intermed {  {\intermed \Sigma}  } ; \intermed {  {\intermed { \Gamma_C } }  } ; \intermed {  {\intermed \Gamma}  }  \Rightarrow  \intermed {  {\intermed \sigma}  } ) \itot{ \rightsquigarrow  \target {  \ottsym{[} \, \bullet \, \ottsym{]}  } } 
}
{
  We need to show that:
  \begin{equation*}
     \intermed {  {\intermed \Sigma}  } ; \intermed {  {\intermed { \Gamma_C } }  } ; \intermed {  {\intermed \Gamma}  }  \vdash_{tm}  \intermed {  {\intermed e}  } : \intermed {  {\intermed \sigma}  } \itot{ \rightsquigarrow  \target {  {\target e }  } } 
  \end{equation*}
  which follows immediately from the first hypothesis of the theorem. \\
}
\proofStep{\textsc{iM-abs}}
{
   \intermed {  \lambda  {\intermed x}  \ottsym{:}  {\intermed \sigma}_{{\mathrm{1}}}  \ottsym{.}  {\intermed M}'  } : ( \intermed {  {\intermed \Sigma}  } ; \intermed {  {\intermed { \Gamma_C } }  } ; \intermed {  {\intermed \Gamma}  }  \Rightarrow  \intermed {  {\intermed \sigma}  } )  \mapsto  ( \intermed {  {\intermed \Sigma}  } ; \intermed {  {\intermed { \Gamma_C } }  } ; \intermed {  {\intermed \Gamma}'  }  \Rightarrow  \intermed {  {\intermed \sigma}_{{\mathrm{1}}}  \rightarrow  {\intermed \sigma}'  } ) \itot{ \rightsquigarrow  \target {  \lambda  {\target x}  \ottsym{:}  {\target \sigma }_{{\mathrm{1}}}  \ottsym{.}  {\target M }'  } } 
}
{
  We need to show the following.
  \begin{equation*}
     \intermed {  {\intermed \Sigma}  } ; \intermed {  {\intermed { \Gamma_C } }  } ; \intermed {  {\intermed \Gamma}'  }  \vdash_{tm}  \intermed {   \lambda  {\intermed x}  \ottsym{:}  {\intermed \sigma}_{{\mathrm{1}}}  \ottsym{.}  {\intermed M}'  [  {\intermed e}  ]   } : \intermed {  {\intermed \sigma}_{{\mathrm{1}}}  \rightarrow  {\intermed \sigma}'  } \itot{ \rightsquigarrow  \target {  \lambda  {\target x}  \ottsym{:}  {\target \sigma }_{{\mathrm{1}}}  \ottsym{.}   {\target M }'  [  {\target e }  ]   } } 
  \end{equation*}
  From the premises of rule \textsc{iM-abs}, we obtain:
  \begin{gather}
     \intermed {  {\intermed M}'  } : ( \intermed {  {\intermed \Sigma}  } ; \intermed {  {\intermed { \Gamma_C } }  } ; \intermed {  {\intermed \Gamma}  }  \Rightarrow  \intermed {  {\intermed \sigma}  } )  \mapsto  ( \intermed {  {\intermed \Sigma}  } ; \intermed {  {\intermed { \Gamma_C } }  } ; \intermed {  {\intermed \Gamma}'  \ottsym{,}  {\intermed x}  \ottsym{:}  {\intermed \sigma}_{{\mathrm{1}}}  }  \Rightarrow  \intermed {  {\intermed \sigma}'  } ) \itot{ \rightsquigarrow  \target {  {\target M }'  } } 
    \label{eq:tctx_abs1} \\
     \intermed {  {\intermed { \Gamma_C } }  } ; \intermed {  {\intermed \Gamma}'  }  \vdash_{ty}  \intermed {  {\intermed \sigma}_{{\mathrm{1}}}  } \itot{ \rightsquigarrow  \target {  {\target \sigma }_{{\mathrm{1}}}  } } 
    \label{eq:tctx_abs2}
  \end{gather}
  From the induction hypothesis applied on Equation~\ref{eq:tctx_abs1}, it follows that:
  \begin{equation}
     \intermed {  {\intermed \Sigma}  } ; \intermed {  {\intermed { \Gamma_C } }  } ; \intermed {  {\intermed \Gamma}'  \ottsym{,}  {\intermed x}  \ottsym{:}  {\intermed \sigma}_{{\mathrm{1}}}  }  \vdash_{tm}  \intermed {   {\intermed M}'  [  {\intermed e}  ]   } : \intermed {  {\intermed \sigma}'  } \itot{ \rightsquigarrow  \target {   {\target M }'  [  {\target e }  ]   } } 
    \label{eq:tctx_abs3}
  \end{equation}
  The goal follows from \textsc{iTm-arrI}, in combination with
  Equations~\ref{eq:tctx_abs2} and \ref{eq:tctx_abs3}. \\
}
\proofStep{\textsc{iM-appL}}
{
   \intermed {  {\intermed M}' \, {\intermed e}_{{\mathrm{2}}}  } : ( \intermed {  {\intermed \Sigma}  } ; \intermed {  {\intermed { \Gamma_C } }  } ; \intermed {  {\intermed \Gamma}  }  \Rightarrow  \intermed {  {\intermed \sigma}  } )  \mapsto  ( \intermed {  {\intermed \Sigma}  } ; \intermed {  {\intermed { \Gamma_C } }  } ; \intermed {  {\intermed \Gamma}'  }  \Rightarrow  \intermed {  {\intermed \sigma}'  } ) \itot{ \rightsquigarrow  \target {  {\target M }' \, {\target e }_{{\mathrm{2}}}  } } 
}
{
  The goal to be proven is the following:
  \begin{equation*}
     \intermed {  {\intermed \Sigma}  } ; \intermed {  {\intermed { \Gamma_C } }  } ; \intermed {  {\intermed \Gamma}'  }  \vdash_{tm}  \intermed {   {\intermed M}'  [  {\intermed e}  ]  \, {\intermed e}_{{\mathrm{2}}}  } : \intermed {  {\intermed \sigma}'  } \itot{ \rightsquigarrow  \target {   {\target M }'  [  {\target e }  ]  \, {\target e }_{{\mathrm{2}}}  } } 
  \end{equation*}
  From the premises of rule \textsc{iM-appL}, we obtain:
  \begin{gather}
     \intermed {  {\intermed M}'  } : ( \intermed {  {\intermed \Sigma}  } ; \intermed {  {\intermed { \Gamma_C } }  } ; \intermed {  {\intermed \Gamma}  }  \Rightarrow  \intermed {  {\intermed \sigma}  } )  \mapsto  ( \intermed {  {\intermed \Sigma}  } ; \intermed {  {\intermed { \Gamma_C } }  } ; \intermed {  {\intermed \Gamma}'  }  \Rightarrow  \intermed {  {\intermed \sigma}_{{\mathrm{1}}}  \rightarrow  {\intermed \sigma}'  } ) \itot{ \rightsquigarrow  \target {  {\target M }'  } } 
    \label{eq:tctx_appL1} \\
     \intermed {  {\intermed \Sigma}  } ; \intermed {  {\intermed { \Gamma_C } }  } ; \intermed {  {\intermed \Gamma}'  }  \vdash_{tm}  \intermed {  {\intermed e}_{{\mathrm{2}}}  } : \intermed {  {\intermed \sigma}_{{\mathrm{1}}}  } \itot{ \rightsquigarrow  \target {  {\target e }_{{\mathrm{2}}}  } } 
    \label{eq:tctx_appL2}
  \end{gather}
  From the induction hypothesis applied on Equation~\ref{eq:tctx_appL1}, it follows
  that:
  \begin{equation}
     \intermed {  {\intermed \Sigma}  } ; \intermed {  {\intermed { \Gamma_C } }  } ; \intermed {  {\intermed \Gamma}'  }  \vdash_{tm}  \intermed {   {\intermed M}'  [  {\intermed e}  ]   } : \intermed {  {\intermed \sigma}_{{\mathrm{1}}}  \rightarrow  {\intermed \sigma}'  } \itot{ \rightsquigarrow  \target {   {\target M }'  [  {\target e }  ]   } } 
    \label{eq:tctx_appL3}
  \end{equation}
  The goal follows from \textsc{iTm-arrE}, in combination with
  Equations~\ref{eq:tctx_appL2} and \ref{eq:tctx_appL3}. \\
}
\proofStep{\textsc{iM-appR}}
{
   \intermed {  {\intermed e}_{{\mathrm{1}}} \, {\intermed M}'  } : ( \intermed {  {\intermed \Sigma}  } ; \intermed {  {\intermed { \Gamma_C } }  } ; \intermed {  {\intermed \Gamma}  }  \Rightarrow  \intermed {  {\intermed \sigma}  } )  \mapsto  ( \intermed {  {\intermed \Sigma}  } ; \intermed {  {\intermed { \Gamma_C } }  } ; \intermed {  {\intermed \Gamma}'  }  \Rightarrow  \intermed {  {\intermed \sigma}'  } ) \itot{ \rightsquigarrow  \target {  {\target e }_{{\mathrm{1}}} \, {\target M }'  } } 
}
{
  The goal to be proven is the following:
  \begin{equation*}
     \intermed {  {\intermed \Sigma}  } ; \intermed {  {\intermed { \Gamma_C } }  } ; \intermed {  {\intermed \Gamma}'  }  \vdash_{tm}  \intermed {   {\intermed e}_{{\mathrm{1}}} \, {\intermed M}'  [  {\intermed e}  ]   } : \intermed {  {\intermed \sigma}'  } \itot{ \rightsquigarrow  \target {   {\target e }_{{\mathrm{1}}} \, {\target M }'  [  {\target e }  ]   } } 
  \end{equation*}
  From the premises of rule \textsc{iM-appR}, we obtain:
  \begin{gather}
     \intermed {  {\intermed M}'  } : ( \intermed {  {\intermed \Sigma}  } ; \intermed {  {\intermed { \Gamma_C } }  } ; \intermed {  {\intermed \Gamma}  }  \Rightarrow  \intermed {  {\intermed \sigma}  } )  \mapsto  ( \intermed {  {\intermed \Sigma}  } ; \intermed {  {\intermed { \Gamma_C } }  } ; \intermed {  {\intermed \Gamma}'  }  \Rightarrow  \intermed {  {\intermed \sigma}_{{\mathrm{1}}}  } ) \itot{ \rightsquigarrow  \target {  {\target M }'  } } 
    \label{eq:tctx_appR1} \\
     \intermed {  {\intermed \Sigma}  } ; \intermed {  {\intermed { \Gamma_C } }  } ; \intermed {  {\intermed \Gamma}'  }  \vdash_{tm}  \intermed {  {\intermed e}_{{\mathrm{1}}}  } : \intermed {  {\intermed \sigma}_{{\mathrm{1}}}  \rightarrow  {\intermed \sigma}'  } \itot{ \rightsquigarrow  \target {  {\target e }_{{\mathrm{1}}}  } } 
    \label{eq:tctx_appR2}
  \end{gather}
  From the induction hypothesis applied on Equation~\ref{eq:tctx_appR1}, it follows
  that:
  \begin{equation}
     \intermed {  {\intermed \Sigma}  } ; \intermed {  {\intermed { \Gamma_C } }  } ; \intermed {  {\intermed \Gamma}'  }  \vdash_{tm}  \intermed {   {\intermed M}'  [  {\intermed e}  ]   } : \intermed {  {\intermed \sigma}_{{\mathrm{1}}}  } \itot{ \rightsquigarrow  \target {   {\target M }'  [  {\target e }  ]   } } 
    \label{eq:tctx_appR3}
  \end{equation}
  The goal follows from \textsc{iTm-arrE}, in combination with
  Equations~\ref{eq:tctx_appR2} and \ref{eq:tctx_appR3}. \\
}
\proofStep{\textsc{iM-dictAbs}}
{
   \intermed {  \lambda  {\intermed \delta }  \ottsym{:}  {\intermed Q}  \ottsym{.}  {\intermed M}'  } : ( \intermed {  {\intermed \Sigma}  } ; \intermed {  {\intermed { \Gamma_C } }  } ; \intermed {  {\intermed \Gamma}  }  \Rightarrow  \intermed {  {\intermed \sigma}  } )  \mapsto  ( \intermed {  {\intermed \Sigma}  } ; \intermed {  {\intermed { \Gamma_C } }  } ; \intermed {  {\intermed \Gamma}'  }  \Rightarrow  \intermed {   {\intermed Q}  \Rightarrow  {\intermed \sigma}'   } ) \itot{ \rightsquigarrow  \target {  \lambda  {\target \delta }  \ottsym{:}  {\target \sigma }  \ottsym{.}  {\target M }'  } } 
}
{
  The goal to be proven is the following:
  \begin{equation*}
     \intermed {  {\intermed \Sigma}  } ; \intermed {  {\intermed { \Gamma_C } }  } ; \intermed {  {\intermed \Gamma}'  }  \vdash_{tm}  \intermed {   \lambda  {\intermed \delta }  \ottsym{:}  {\intermed Q}  \ottsym{.}  {\intermed M}'  [  {\intermed e}  ]   } : \intermed {   {\intermed Q}  \Rightarrow  {\intermed \sigma}'   } \itot{ \rightsquigarrow  \target {  \lambda  {\target \delta }  \ottsym{:}  {\target \sigma }  \ottsym{.}   {\target M }'  [  {\target e }  ]   } } 
  \end{equation*}
  From the premises of rule \textsc{iM-dictAbs}, we obtain:
  \begin{gather}
     \intermed {  {\intermed M}'  } : ( \intermed {  {\intermed \Sigma}  } ; \intermed {  {\intermed { \Gamma_C } }  } ; \intermed {  {\intermed \Gamma}  }  \Rightarrow  \intermed {  {\intermed \sigma}  } )  \mapsto  ( \intermed {  {\intermed \Sigma}  } ; \intermed {  {\intermed { \Gamma_C } }  } ; \intermed {  {\intermed \Gamma}'  \ottsym{,}  {\intermed \delta }  \ottsym{:}  {\intermed Q}  }  \Rightarrow  \intermed {  {\intermed \sigma}'  } ) \itot{ \rightsquigarrow  \target {  {\target M }'  } } 
    \label{eq:tctx_dabs1} \\
     \intermed {  {\intermed { \Gamma_C } }  } ; \intermed {  {\intermed \Gamma}'  }  \vdash_{\mathit {Q} }  \intermed {  {\intermed Q}  }\, \itot{ \rightsquigarrow  \target {  {\target \sigma }  } } 
    \label{eq:tctx_dabs2}
  \end{gather}
  From the induction hypothesis applied on Equation~\ref{eq:tctx_dabs1}, it follows
  that:
  \begin{equation}
     \intermed {  {\intermed \Sigma}  } ; \intermed {  {\intermed { \Gamma_C } }  } ; \intermed {  {\intermed \Gamma}'  \ottsym{,}  {\intermed \delta }  \ottsym{:}  {\intermed Q}  }  \vdash_{tm}  \intermed {   {\intermed M}'  [  {\intermed e}  ]   } : \intermed {  {\intermed \sigma}'  } \itot{ \rightsquigarrow  \target {   {\target M }'  [  {\target e }  ]   } } 
    \label{eq:tctx_dabs3}
  \end{equation}
  The goal follows from \textsc{iTm-constrI}, in combination with
  Equations~\ref{eq:tctx_dabs2} and \ref{eq:tctx_dabs3}. \\
}
\proofStep{\textsc{iM-dictApp}}
{
   \intermed {  {\intermed M}' \, {\intermed d}  } : ( \intermed {  {\intermed \Sigma}  } ; \intermed {  {\intermed { \Gamma_C } }  } ; \intermed {  {\intermed \Gamma}  }  \Rightarrow  \intermed {  {\intermed \sigma}  } )  \mapsto  ( \intermed {  {\intermed \Sigma}  } ; \intermed {  {\intermed { \Gamma_C } }  } ; \intermed {  {\intermed \Gamma}'  }  \Rightarrow  \intermed {  {\intermed \sigma}'  } ) \itot{ \rightsquigarrow  \target {  {\target M }' \, {\target e }  } } 
}
{
    The goal to be proven is the following.
  \begin{equation*}
     \intermed {  {\intermed \Sigma}  } ; \intermed {  {\intermed { \Gamma_C } }  } ; \intermed {  {\intermed \Gamma}'  }  \vdash_{tm}  \intermed {   {\intermed M}'  [  {\intermed e}  ]  \, {\intermed d}  } : \intermed {  {\intermed \sigma}'  } \itot{ \rightsquigarrow  \target {   {\target M }'  [  {\target e }  ]  \, {\target e }'  } } 
  \end{equation*}
  From the premises of rule \textsc{iM-dictApp}, we obtain:
  \begin{gather}
     \intermed {  {\intermed M}'  } : ( \intermed {  {\intermed \Sigma}  } ; \intermed {  {\intermed { \Gamma_C } }  } ; \intermed {  {\intermed \Gamma}  }  \Rightarrow  \intermed {  {\intermed \sigma}  } )  \mapsto  ( \intermed {  {\intermed \Sigma}  } ; \intermed {  {\intermed { \Gamma_C } }  } ; \intermed {  {\intermed \Gamma}'  }  \Rightarrow  \intermed {   {\intermed Q}  \Rightarrow  {\intermed \sigma}'   } ) \itot{ \rightsquigarrow  \target {  {\target M }'  } } 
    \label{eq:tctx_dapp1} \\
     \intermed {  {\intermed \Sigma}  } ; \intermed {  {\intermed { \Gamma_C } }  } ; \intermed {  {\intermed \Gamma}'  }  \vdash_{d}  \intermed {  {\intermed d}  } : \intermed {  {\intermed Q}  } \itot{\,  \rightsquigarrow  \target {  {\target e }'  } } 
    \label{eq:tctx_dapp2}
  \end{gather}
  From the induction hypothesis applied on Equation~\ref{eq:tctx_dapp1}, it follows
  that:
  \begin{equation}
     \intermed {  {\intermed \Sigma}  } ; \intermed {  {\intermed { \Gamma_C } }  } ; \intermed {  {\intermed \Gamma}'  }  \vdash_{tm}  \intermed {   {\intermed M}'  [  {\intermed e}  ]   } : \intermed {   {\intermed Q}  \Rightarrow  {\intermed \sigma}'   } \itot{ \rightsquigarrow  \target {   {\target M }'  [  {\target e }  ]   } } 
    \label{eq:tctx_dapp3}
  \end{equation}
  The goal follows from \textsc{iTm-constrE}, in combination with
  Equations~\ref{eq:tctx_dapp2} and \ref{eq:tctx_dapp3}. \\
}
\proofStep{\textsc{iM-tyAbs}}
{
   \intermed {  \Lambda  {\intermed a }  \ottsym{.}  {\intermed M}'  } : ( \intermed {  {\intermed \Sigma}  } ; \intermed {  {\intermed { \Gamma_C } }  } ; \intermed {  {\intermed \Gamma}  }  \Rightarrow  \intermed {  {\intermed \sigma}  } )  \mapsto  ( \intermed {  {\intermed \Sigma}  } ; \intermed {  {\intermed { \Gamma_C } }  } ; \intermed {  {\intermed \Gamma}'  }  \Rightarrow  \intermed {  \forall  {\intermed a }  \ottsym{.}  {\intermed \sigma}'  } ) \itot{ \rightsquigarrow  \target {  \Lambda  {\target a }  \ottsym{.}  {\target M }'  } } 
}
{
  The goal to be proven is the following:
  \begin{equation*}
     \intermed {  {\intermed \Sigma}  } ; \intermed {  {\intermed { \Gamma_C } }  } ; \intermed {  {\intermed \Gamma}'  }  \vdash_{tm}  \intermed {   \Lambda  {\intermed a }  \ottsym{.}  {\intermed M}'  [  {\intermed e}  ]   } : \intermed {  \forall  {\intermed a }  \ottsym{.}  {\intermed \sigma}'  } \itot{ \rightsquigarrow  \target {   \Lambda  {\target a }  \ottsym{.}  {\target M }'  [  {\target e }  ]   } } 
  \end{equation*}
  From the premises of rule \textsc{iM-tyAbs}, we obtain:
  \begin{gather*}
     \intermed {  {\intermed M}'  } : ( \intermed {  {\intermed \Sigma}  } ; \intermed {  {\intermed { \Gamma_C } }  } ; \intermed {  {\intermed \Gamma}  }  \Rightarrow  \intermed {  {\intermed \sigma}  } )  \mapsto  ( \intermed {  {\intermed \Sigma}  } ; \intermed {  {\intermed { \Gamma_C } }  } ; \intermed {  {\intermed \Gamma}'  \ottsym{,}  {\intermed a }  }  \Rightarrow  \intermed {  {\intermed \sigma}'  } ) \itot{ \rightsquigarrow  \target {  {\target M }'  } } 
    \label{eq:tctx_tabs1}
  \end{gather*}
  Applying the induction hypothesis on the above context typing, yields:
  \begin{equation*}
     \intermed {  {\intermed \Sigma}  } ; \intermed {  {\intermed { \Gamma_C } }  } ; \intermed {  {\intermed \Gamma}'  \ottsym{,}  {\intermed a }  }  \vdash_{tm}  \intermed {   {\intermed M}'  [  {\intermed e}  ]   } : \intermed {  {\intermed \sigma}'  } \itot{ \rightsquigarrow  \target {   {\target M }'  [  {\target e }  ]   } } 
    \label{eq:tctx_tabs2}
  \end{equation*}
  Using this result with rule \textsc{iTm-forallI}, we reach the goal.\\
}
\proofStep{\textsc{iM-tyApp}}
{
   \intermed {  {\intermed M}' \, {\intermed \sigma}''  } : ( \intermed {  {\intermed \Sigma}  } ; \intermed {  {\intermed { \Gamma_C } }  } ; \intermed {  {\intermed \Gamma}  }  \Rightarrow  \intermed {  {\intermed \sigma}  } )  \mapsto  ( \intermed {  {\intermed \Sigma}  } ; \intermed {  {\intermed { \Gamma_C } }  } ; \intermed {  {\intermed \Gamma}'  }  \Rightarrow  \intermed {  [  {\intermed \sigma}''  /  {\intermed a }  ]  {\intermed \sigma}'  } ) \itot{ \rightsquigarrow  \target {  {\target M }' \, {\target \sigma }''  } } 
}
{
  The goal to be proven is the following.
  \begin{equation*}
     \intermed {  {\intermed \Sigma}  } ; \intermed {  {\intermed { \Gamma_C } }  } ; \intermed {  {\intermed \Gamma}'  }  \vdash_{tm}  \intermed {   {\intermed M}'  [  {\intermed e}  ]  \, {\intermed \sigma}''  } : \intermed {  [  {\intermed \sigma}''  /  {\intermed a }  ]  {\intermed \sigma}'  } \itot{ \rightsquigarrow  \target {   {\target M }'  [  {\target e }  ]  \, {\target \sigma }''  } } 
  \end{equation*}
  From the premises of rule \textsc{iM-tyApp}, we obtain:
  \begin{gather}
     \intermed {  {\intermed M}'  } : ( \intermed {  {\intermed \Sigma}  } ; \intermed {  {\intermed { \Gamma_C } }  } ; \intermed {  {\intermed \Gamma}  }  \Rightarrow  \intermed {  {\intermed \sigma}  } )  \mapsto  ( \intermed {  {\intermed \Sigma}  } ; \intermed {  {\intermed { \Gamma_C } }  } ; \intermed {  {\intermed \Gamma}'  }  \Rightarrow  \intermed {  \forall  {\intermed a }  \ottsym{.}  {\intermed \sigma}'  } ) \itot{ \rightsquigarrow  \target {  {\target M }'  } } 
    \label{eq:tctx_tapp1} \\
     \intermed {  {\intermed { \Gamma_C } }  } ; \intermed {  {\intermed \Gamma}'  }  \vdash_{ty}  \intermed {  {\intermed \sigma}''  } \itot{ \rightsquigarrow  \target {  {\target \sigma }''  } } 
    \label{eq:tctx_tapp2}
  \end{gather}
  From the induction hypothesis applied on Equation~\ref{eq:tctx_tapp1}, we have
  that:
  \begin{equation}
     \intermed {  {\intermed \Sigma}  } ; \intermed {  {\intermed { \Gamma_C } }  } ; \intermed {  {\intermed \Gamma}'  }  \vdash_{tm}  \intermed {   {\intermed M}'  [  {\intermed e}  ]   } : \intermed {  \forall  {\intermed a }  \ottsym{.}  {\intermed \sigma}'  } \itot{ \rightsquigarrow  \target {   {\target M }'  [  {\target e }  ]   } } 
    \label{eq:tctx_tapp3}
  \end{equation}
  The goal follows from \textsc{iTm-forallE}, in combination with
  Equations~\ref{eq:tctx_tapp2} and \ref{eq:tctx_tapp3}. \\
}
\proofStep{\textsc{iM-letL}}
{
   \intermed {  \textbf{\intermed {let} } \, {\intermed x}  \ottsym{:}  {\intermed \sigma}_{{\mathrm{1}}}  \ottsym{=}  {\intermed M}' \, \textbf{\intermed {in} } \, {\intermed e}_{{\mathrm{2}}}  } : ( \intermed {  {\intermed \Sigma}  } ; \intermed {  {\intermed { \Gamma_C } }  } ; \intermed {  {\intermed \Gamma}  }  \Rightarrow  \intermed {  {\intermed \sigma}  } )  \mapsto  ( \intermed {  {\intermed \Sigma}  } ; \intermed {  {\intermed { \Gamma_C } }  } ; \intermed {  {\intermed \Gamma}'  }  \Rightarrow  \intermed {  {\intermed \sigma}'  } ) \itot{ \rightsquigarrow  \target {  \textbf{\target {let} } \, {\target x}  \ottsym{:}  {\target \sigma }_{{\mathrm{1}}}  \ottsym{=}  {\target M }' \, \textbf{\target {in} } \, {\target e }_{{\mathrm{2}}}  } } 
}
{
    The goal to be proven is the following:
  \begin{equation*}
     \intermed {  {\intermed \Sigma}  } ; \intermed {  {\intermed { \Gamma_C } }  } ; \intermed {  {\intermed \Gamma}'  }  \vdash_{tm}  \intermed {  \textbf{\intermed {let} } \, {\intermed x}  \ottsym{:}  {\intermed \sigma}_{{\mathrm{1}}}  \ottsym{=}   {\intermed M}'  [  {\intermed e}  ]  \, \textbf{\intermed {in} } \, {\intermed e}_{{\mathrm{2}}}  } : \intermed {  {\intermed \sigma}'  } \itot{ \rightsquigarrow  \target {  \textbf{\target {let} } \, {\target x}  \ottsym{:}  {\target \sigma }_{{\mathrm{1}}}  \ottsym{=}   {\target M }'  [  {\target e }  ]  \, \textbf{\target {in} } \, {\target e }_{{\mathrm{2}}}  } } 
  \end{equation*}
  From the premises of rule \textsc{iM-letL}, we obtain:
  \begin{gather}
     \intermed {  {\intermed M}'  } : ( \intermed {  {\intermed \Sigma}  } ; \intermed {  {\intermed { \Gamma_C } }  } ; \intermed {  {\intermed \Gamma}  }  \Rightarrow  \intermed {  {\intermed \sigma}  } )  \mapsto  ( \intermed {  {\intermed \Sigma}  } ; \intermed {  {\intermed { \Gamma_C } }  } ; \intermed {  {\intermed \Gamma}'  }  \Rightarrow  \intermed {  {\intermed \sigma}_{{\mathrm{1}}}  } ) \itot{ \rightsquigarrow  \target {  {\target M }'  } } 
    \label{eq:tctx_letL1} \\
     \intermed {  {\intermed \Sigma}  } ; \intermed {  {\intermed { \Gamma_C } }  } ; \intermed {  {\intermed \Gamma}'  \ottsym{,}  {\intermed x}  \ottsym{:}  {\intermed \sigma}_{{\mathrm{1}}}  }  \vdash_{tm}  \intermed {  {\intermed e}_{{\mathrm{2}}}  } : \intermed {  {\intermed \sigma}'  } \itot{ \rightsquigarrow  \target {  {\target e }_{{\mathrm{2}}}  } } 
    \label{eq:tctx_letL2} \\
     \intermed {  {\intermed { \Gamma_C } }  } ; \intermed {  {\intermed \Gamma}'  }  \vdash_{ty}  \intermed {  {\intermed \sigma}_{{\mathrm{1}}}  } \itot{ \rightsquigarrow  \target {  {\target \sigma }_{{\mathrm{1}}}  } } 
    \label{eq:tctx_letL3}
  \end{gather}
  From the induction hypothesis applied on Equation~\ref{eq:tctx_letL1}, we have
  that:
  \begin{equation}
     \intermed {  {\intermed \Sigma}  } ; \intermed {  {\intermed { \Gamma_C } }  } ; \intermed {  {\intermed \Gamma}'  }  \vdash_{tm}  \intermed {   {\intermed M}'  [  {\intermed e}  ]   } : \intermed {  {\intermed \sigma}_{{\mathrm{1}}}  } \itot{ \rightsquigarrow  \target {   {\target M }'  [  {\target e }  ]   } } 
    \label{eq:tctx_letL4}
  \end{equation}
  The goal follows from \textsc{iTm-let}, in combination with
  Equations~\ref{eq:tctx_letL2}, \ref{eq:tctx_letL3} and \ref{eq:tctx_letL4}. \\
}
\proofStep{\textsc{iM-letR}}
{
   \intermed {  \textbf{\intermed {let} } \, {\intermed x}  \ottsym{:}  {\intermed \sigma}_{{\mathrm{1}}}  \ottsym{=}  {\intermed e}_{{\mathrm{1}}} \, \textbf{\intermed {in} } \, {\intermed M}'  } : ( \intermed {  {\intermed \Sigma}  } ; \intermed {  {\intermed { \Gamma_C } }  } ; \intermed {  {\intermed \Gamma}  }  \Rightarrow  \intermed {  {\intermed \sigma}  } )  \mapsto  ( \intermed {  {\intermed \Sigma}  } ; \intermed {  {\intermed { \Gamma_C } }  } ; \intermed {  {\intermed \Gamma}'  }  \Rightarrow  \intermed {  {\intermed \sigma}'  } ) \itot{ \rightsquigarrow  \target {  \textbf{\target {let} } \, {\target x}  \ottsym{:}  {\target \sigma }_{{\mathrm{1}}}  \ottsym{=}  {\target e }_{{\mathrm{1}}} \, \textbf{\target {in} } \, {\target M }'  } } 
}
{
  The goal to be proven is the following:
  \begin{equation*}
     \intermed {  {\intermed \Sigma}  } ; \intermed {  {\intermed { \Gamma_C } }  } ; \intermed {  {\intermed \Gamma}'  }  \vdash_{tm}  \intermed {   \textbf{\intermed {let} } \, {\intermed x}  \ottsym{:}  {\intermed \sigma}_{{\mathrm{1}}}  \ottsym{=}  {\intermed e}_{{\mathrm{1}}} \, \textbf{\intermed {in} } \, {\intermed M}'  [  {\intermed e}  ]   } : \intermed {  {\intermed \sigma}'  } \itot{ \rightsquigarrow  \target {   \textbf{\target {let} } \, {\target x}  \ottsym{:}  {\target \sigma }_{{\mathrm{1}}}  \ottsym{=}  {\target e }_{{\mathrm{1}}} \, \textbf{\target {in} } \, {\target M }'  [  {\target e }  ]   } } 
  \end{equation*}
  From the rule premise:
  \begin{gather}
     \intermed {  {\intermed M}'  } : ( \intermed {  {\intermed \Sigma}  } ; \intermed {  {\intermed { \Gamma_C } }  } ; \intermed {  {\intermed \Gamma}  }  \Rightarrow  \intermed {  {\intermed \sigma}  } )  \mapsto  ( \intermed {  {\intermed \Sigma}  } ; \intermed {  {\intermed { \Gamma_C } }  } ; \intermed {  {\intermed \Gamma}'  \ottsym{,}  {\intermed x}  \ottsym{:}  {\intermed \sigma}_{{\mathrm{1}}}  }  \Rightarrow  \intermed {  {\intermed \sigma}'  } ) \itot{ \rightsquigarrow  \target {  {\target M }'  } } 
    \label{eq:tctx_letR1} \\
     \intermed {  {\intermed \Sigma}  } ; \intermed {  {\intermed { \Gamma_C } }  } ; \intermed {  {\intermed \Gamma}'  }  \vdash_{tm}  \intermed {  {\intermed e}_{{\mathrm{1}}}  } : \intermed {  {\intermed \sigma}_{{\mathrm{1}}}  } \itot{ \rightsquigarrow  \target {  {\target e }_{{\mathrm{1}}}  } } 
    \label{eq:tctx_letR2} \\
     \intermed {  {\intermed { \Gamma_C } }  } ; \intermed {  {\intermed \Gamma}'  }  \vdash_{ty}  \intermed {  {\intermed \sigma}_{{\mathrm{1}}}  } \itot{ \rightsquigarrow  \target {  {\target \sigma }_{{\mathrm{1}}}  } } 
    \label{eq:tctx_letR3}
  \end{gather}
  From the induction hypothesis applied on Equation~\ref{eq:tctx_letR1}, we have
  that:
  \begin{equation}
     \intermed {  {\intermed \Sigma}  } ; \intermed {  {\intermed { \Gamma_C } }  } ; \intermed {  {\intermed \Gamma}'  \ottsym{,}  {\intermed x}  \ottsym{:}  {\intermed \sigma}_{{\mathrm{1}}}  }  \vdash_{tm}  \intermed {   {\intermed M}'  [  {\intermed e}  ]   } : \intermed {  {\intermed \sigma}'  } \itot{ \rightsquigarrow  \target {   {\target M }'  [  {\target e }  ]   } } 
    \label{eq:tctx_letR4}
  \end{equation}
  The goal follows from \textsc{iTm-let}, in combination with
  Equations~\ref{eq:tctx_letR2}, \ref{eq:tctx_letR3} and \ref{eq:tctx_letR4}.
}
\qed

\renewcommand\itot[1]{}
\begin{theoremB}[Logical Equivalence Preserved by Forward/Backward Reduction]\ \\
  Given \( \intermed {  {\intermed \Sigma}_{{\mathrm{1}}}  }  \vdash  \intermed {  {\intermed e}_{{\mathrm{1}}}  }  \longrightarrow  \intermed {  {\intermed e}'_{{\mathrm{1}}}  } \) and \( \intermed {  {\intermed \Sigma}_{{\mathrm{2}}}  }  \vdash  \intermed {  {\intermed e}_{{\mathrm{2}}}  }  \longrightarrow  \intermed {  {\intermed e}'_{{\mathrm{2}}}  } \),
  \begin{itemize}
    \item If \( \intermed {  {\intermed { \Gamma_C } }  } ; \intermed {  {\intermed \Gamma}  }  \vdash  \intermed {  {\intermed \Sigma}_{{\mathrm{1}}}  } : \intermed {  {\intermed e}_{{\mathrm{1}}}  }  \simeq_{log}  \intermed {  {\intermed \Sigma}_{{\mathrm{2}}}  } : \intermed {  {\intermed e}_{{\mathrm{2}}}  } : \intermed {  {\intermed \sigma}  } \), then \( \intermed {  {\intermed { \Gamma_C } }  } ; \intermed {  {\intermed \Gamma}  }  \vdash  \intermed {  {\intermed \Sigma}_{{\mathrm{1}}}  } : \intermed {  {\intermed e}'_{{\mathrm{1}}}  }  \simeq_{log}  \intermed {  {\intermed \Sigma}_{{\mathrm{2}}}  } : \intermed {  {\intermed e}'_{{\mathrm{2}}}  } : \intermed {  {\intermed \sigma}  } \).
    \item If \( \intermed {  {\intermed { \Gamma_C } }  } ; \intermed {  {\intermed \Gamma}  }  \vdash  \intermed {  {\intermed \Sigma}_{{\mathrm{1}}}  } : \intermed {  {\intermed e}'_{{\mathrm{1}}}  }  \simeq_{log}  \intermed {  {\intermed \Sigma}_{{\mathrm{2}}}  } : \intermed {  {\intermed e}'_{{\mathrm{2}}}  } : \intermed {  {\intermed \sigma}  } \)
      and \( \intermed {  {\intermed \Sigma}_{{\mathrm{1}}}  } ; \intermed {  {\intermed { \Gamma_C } }  } ; \intermed {  {\intermed \Gamma}  }  \vdash_{tm}  \intermed {  {\intermed e}_{{\mathrm{1}}}  } : \intermed {  {\intermed \sigma}  } \itot{ \rightsquigarrow  \target {  {\target e }_{{\mathrm{1}}}  } } \)
      and \( \intermed {  {\intermed \Sigma}_{{\mathrm{2}}}  } ; \intermed {  {\intermed { \Gamma_C } }  } ; \intermed {  {\intermed \Gamma}  }  \vdash_{tm}  \intermed {  {\intermed e}_{{\mathrm{2}}}  } : \intermed {  {\intermed \sigma}  } \itot{ \rightsquigarrow  \target {  {\target e }_{{\mathrm{2}}}  } } \), \\
      then \( \intermed {  {\intermed { \Gamma_C } }  } ; \intermed {  {\intermed \Gamma}  }  \vdash  \intermed {  {\intermed \Sigma}_{{\mathrm{1}}}  } : \intermed {  {\intermed e}_{{\mathrm{1}}}  }  \simeq_{log}  \intermed {  {\intermed \Sigma}_{{\mathrm{2}}}  } : \intermed {  {\intermed e}_{{\mathrm{2}}}  } : \intermed {  {\intermed \sigma}  } \).
  \end{itemize}
  \label{thmA:log_eq_pres}
\end{theoremB}

\proof

\begin{description}
  \item[Part 1]
    By unfolding the definition of logical relation, we get:
    \begin{gather}
     \intermed {  {\intermed R }  } \in \mathcal{F} \llbracket \intermed {  {\intermed \Gamma}  } \rrbracket ^{\intermed {  {\intermed { \Gamma_C } }  } }  \\
     \intermed {  {\intermed \phi }  } \in \mathcal{G} \llbracket \intermed {  {\intermed \Gamma}  } \rrbracket ^{\intermed {  {\intermed \Sigma}_{{\mathrm{1}}}  } , \intermed {  {\intermed \Sigma}_{{\mathrm{2}}}  } , \intermed {  {\intermed { \Gamma_C } }  } }_{\intermed {  {\intermed R }  } }  \\
     \intermed {  {\intermed \gamma }  } \in \mathcal{H} \llbracket \intermed {  {\intermed \Gamma}  } \rrbracket ^{\intermed {  {\intermed \Sigma}_{{\mathrm{1}}}  } , \intermed {  {\intermed \Sigma}_{{\mathrm{2}}}  } , \intermed {  {\intermed { \Gamma_C } }  } }_{\intermed {  {\intermed R }  } }  \\
    \label{eq:logical:forward:1}
     ( \intermed {  {\intermed \Sigma}_{{\mathrm{1}}}  } : \intermed {   {\intermed \gamma }_{{\mathrm{1}}}  ( {\intermed \phi }_{{\mathrm{1}}}  ( {\intermed R }  ( {\intermed e}_{{\mathrm{1}}} )))   } , \intermed {  {\intermed \Sigma}_{{\mathrm{2}}}  } : \intermed {   {\intermed \gamma }_{{\mathrm{2}}}  ( {\intermed \phi }_{{\mathrm{2}}}  ( {\intermed R }  ( {\intermed e}_{{\mathrm{2}}} )))   } ) \in \mathcal{E} \llbracket \intermed {  {\intermed \sigma}  } \rrbracket ^{\intermed {  {\intermed { \Gamma_C } }  } }_{\intermed {  {\intermed R }  } } 
    \end{gather}
    Unfolding the definition of the closed expression relation in
    \ref{eq:logical:forward:1} results in:
    \begin{gather}
       \intermed {  {\intermed \Sigma}_{{\mathrm{1}}}  } ; \intermed {  {\intermed { \Gamma_C } }  } ; \intermed {  \bullet  }  \vdash_{tm}  \intermed {   {\intermed \gamma }_{{\mathrm{1}}}  ( {\intermed \phi }_{{\mathrm{1}}}  ( {\intermed R }  ( {\intermed e}_{{\mathrm{1}}} )))   } : \intermed {  {\intermed R }  \ottsym{(}  {\intermed \sigma}  \ottsym{)}  } \itot{ \rightsquigarrow  \target {  {\target e }_{{\mathrm{1}}}  } }  \\
       \intermed {  {\intermed \Sigma}_{{\mathrm{2}}}  } ; \intermed {  {\intermed { \Gamma_C } }  } ; \intermed {  \bullet  }  \vdash_{tm}  \intermed {   {\intermed \gamma }_{{\mathrm{2}}}  ( {\intermed \phi }_{{\mathrm{2}}}  ( {\intermed R }  ( {\intermed e}_{{\mathrm{2}}} )))   } : \intermed {  {\intermed R }  \ottsym{(}  {\intermed \sigma}  \ottsym{)}  } \itot{ \rightsquigarrow  \target {  {\target e }_{{\mathrm{2}}}  } }  \\
      \exists {\intermed v}_{{\mathrm{1}}}, {\intermed v}_{{\mathrm{2}}}, 
         \intermed {  {\intermed \Sigma}_{{\mathrm{1}}}  }  \vdash  \intermed {   {\intermed \gamma }_{{\mathrm{1}}}  ( {\intermed \phi }_{{\mathrm{1}}}  ( {\intermed R }  ( {\intermed e}_{{\mathrm{1}}} )))   }  \longrightarrow^{*}  \intermed {  {\intermed v}_{{\mathrm{1}}}  } ,\\
         \intermed {  {\intermed \Sigma}_{{\mathrm{2}}}  }  \vdash  \intermed {   {\intermed \gamma }_{{\mathrm{2}}}  ( {\intermed \phi }_{{\mathrm{2}}}  ( {\intermed R }  ( {\intermed e}_{{\mathrm{2}}} )))   }  \longrightarrow^{*}  \intermed {  {\intermed v}_{{\mathrm{2}}}  } ,\\
         ( \intermed {  {\intermed \Sigma}_{{\mathrm{1}}}  } : \intermed {  {\intermed v}_{{\mathrm{1}}}  } , \intermed {  {\intermed \Sigma}_{{\mathrm{2}}}  } : \intermed {  {\intermed v}_{{\mathrm{2}}}  } ) \in \mathcal{V} \llbracket \intermed {  {\intermed \sigma}  } \rrbracket ^{\intermed {  {\intermed { \Gamma_C } }  } }_{\intermed {  {\intermed R }  } } 
    \end{gather}
    By induction on ${\intermed e}_{{\mathrm{1}}}$, it is easy to verify that
    \begin{gather}
         \intermed {  {\intermed \Sigma}_{{\mathrm{1}}}  }  \vdash  \intermed {   {\intermed \gamma }_{{\mathrm{1}}}  ( {\intermed \phi }_{{\mathrm{1}}}  ( {\intermed R }  ( {\intermed e}_{{\mathrm{1}}} )))   }  \longrightarrow  \intermed {   {\intermed \gamma }_{{\mathrm{1}}}  ( {\intermed \phi }_{{\mathrm{1}}}  ( {\intermed R }  ( {\intermed e}'_{{\mathrm{1}}} )))   } 
    \end{gather}
    By preservation (Theorem~\ref{thmA:intermediate_preservation}) we have:
    \begin{gather}
       \intermed {  {\intermed \Sigma}_{{\mathrm{1}}}  } ; \intermed {  {\intermed { \Gamma_C } }  } ; \intermed {  \bullet  }  \vdash_{tm}  \intermed {   {\intermed \gamma }_{{\mathrm{1}}}  ( {\intermed \phi }_{{\mathrm{1}}}  ( {\intermed R }  ( {\intermed e}'_{{\mathrm{1}}} )))   } : \intermed {  {\intermed R }  \ottsym{(}  {\intermed \sigma}  \ottsym{)}  } \itot{ \rightsquigarrow  \target {  {\target e }_{{\mathrm{3}}}  } }  
    \end{gather}
    Because the evaluation in \interL{} is deterministic (Lemma~\ref{lemma:eval_unique}), we know that:
    \begin{gather}
         \intermed {  {\intermed \Sigma}_{{\mathrm{1}}}  }  \vdash  \intermed {   {\intermed \gamma }_{{\mathrm{1}}}  ( {\intermed \phi }_{{\mathrm{1}}}  ( {\intermed R }  ( {\intermed e}'_{{\mathrm{1}}} )))   }  \longrightarrow^{*}  \intermed {  {\intermed v}_{{\mathrm{1}}}  } 
    \end{gather}
    Similarly:
    \begin{gather}
       \intermed {  {\intermed \Sigma}_{{\mathrm{2}}}  } ; \intermed {  {\intermed { \Gamma_C } }  } ; \intermed {  \bullet  }  \vdash_{tm}  \intermed {   {\intermed \gamma }_{{\mathrm{2}}}  ( {\intermed \phi }_{{\mathrm{2}}}  ( {\intermed R }  ( {\intermed e}'_{{\mathrm{2}}} )))   } : \intermed {  {\intermed R }  \ottsym{(}  {\intermed \sigma}  \ottsym{)}  } \itot{ \rightsquigarrow  \target {  {\target e }_{{\mathrm{4}}}  } }  \\
       \intermed {  {\intermed \Sigma}_{{\mathrm{2}}}  }  \vdash  \intermed {   {\intermed \gamma }_{{\mathrm{2}}}  ( {\intermed \phi }_{{\mathrm{2}}}  ( {\intermed R }  ( {\intermed e}'_{{\mathrm{2}}} )))   }  \longrightarrow^{*}  \intermed {  {\intermed v}_{{\mathrm{2}}}  } 
    \end{gather}
    Combining those equations, results in:
    \begin{gather}
     ( \intermed {  {\intermed \Sigma}_{{\mathrm{1}}}  } : \intermed {   {\intermed \gamma }_{{\mathrm{1}}}  ( {\intermed \phi }_{{\mathrm{1}}}  ( {\intermed R }  ( {\intermed e}'_{{\mathrm{1}}} )))   } , \intermed {  {\intermed \Sigma}_{{\mathrm{2}}}  } : \intermed {   {\intermed \gamma }_{{\mathrm{2}}}  ( {\intermed \phi }_{{\mathrm{2}}}  ( {\intermed R }  ( {\intermed e}'_{{\mathrm{2}}} )))   } ) \in \mathcal{E} \llbracket \intermed {  {\intermed \sigma}  } \rrbracket ^{\intermed {  {\intermed { \Gamma_C } }  } }_{\intermed {  {\intermed R }  } } 
    \end{gather}
    The goal follows from the definition of logical equivalence.
  \item[Part 2]
    Similar to Part 1.
\end{description}
\qed

\begin{theoremB}[Dictionary Reflexivity]\ \\
  If \( \intermed {  {\intermed \Sigma}_{{\mathrm{1}}}  } ; \intermed {  {\intermed { \Gamma_C } }  } ; \intermed {  {\intermed \Gamma}  }  \vdash_{d}  \intermed {  {\intermed d}  } : \intermed {  {\intermed Q}  } \itot{\,  \rightsquigarrow  \target {  {\target e }  } } \)
  and \( \intermed {  {\intermed \Sigma}_{{\mathrm{2}}}  } ; \intermed {  {\intermed { \Gamma_C } }  } ; \intermed {  {\intermed \Gamma}  }  \vdash_{d}  \intermed {  {\intermed d}  } : \intermed {  {\intermed Q}  } \itot{\,  \rightsquigarrow  \target {  {\target e }  } } \)
  and \( \intermed {  {\intermed { \Gamma_C } }  }  \vdash  \intermed {  {\intermed \Sigma}_{{\mathrm{1}}}  }  \simeq_{log}  \intermed {  {\intermed \Sigma}_{{\mathrm{2}}}  } \),
  then \( \intermed {  {\intermed { \Gamma_C } }  } ; \intermed {  {\intermed \Gamma}  }  \vdash  \intermed {  {\intermed \Sigma}_{{\mathrm{1}}}  } : \intermed {  {\intermed d}  }  \simeq_{log}  \intermed {  {\intermed \Sigma}_{{\mathrm{2}}}  } : \intermed {  {\intermed d}  } : \intermed {  {\intermed Q}  } \).
  \label{thmA:dict_reflex}
\end{theoremB}

\proof
Proof by structural induction on the dictionary \({\intermed d}\) and consequently,
since \interL{} dictionary typing is syntax directed, on both typing derivations. \\
\proofStep{\( \intermed {  {\intermed d}  } = \intermed {  {\intermed \delta }  } \) (\textsc{D-var})}
{
   \intermed {  {\intermed \Sigma}_{{\mathrm{1}}}  } ; \intermed {  {\intermed { \Gamma_C } }  } ; \intermed {  {\intermed \Gamma}  }  \vdash_{d}  \intermed {  {\intermed \delta }  } : \intermed {  {\intermed Q}  } \itot{\,  \rightsquigarrow  \target {  {\target \delta }  } }  \band
   \intermed {  {\intermed \Sigma}_{{\mathrm{2}}}  } ; \intermed {  {\intermed { \Gamma_C } }  } ; \intermed {  {\intermed \Gamma}  }  \vdash_{d}  \intermed {  {\intermed \delta }  } : \intermed {  {\intermed Q}  } \itot{\,  \rightsquigarrow  \target {  {\target \delta }  } } 
}
{
  The goal to be proven is the following:
  \begin{equation*}
     \intermed {  {\intermed { \Gamma_C } }  } ; \intermed {  {\intermed \Gamma}  }  \vdash  \intermed {  {\intermed \Sigma}_{{\mathrm{1}}}  } : \intermed {  {\intermed \delta }  }  \simeq_{log}  \intermed {  {\intermed \Sigma}_{{\mathrm{2}}}  } : \intermed {  {\intermed \delta }  } : \intermed {  {\intermed Q}  } 
    \label{eq:drefl_var1}
  \end{equation*}
  By unfolding the definition of logical equivalence, the goal reduces to:
  \begin{equation*}
     ( \intermed {  {\intermed \Sigma}_{{\mathrm{1}}}  } : \intermed {   {\intermed \gamma }_{{\mathrm{1}}}  ( {\intermed \phi }_{{\mathrm{1}}}  ( {\intermed R }  ( {\intermed \delta } )))   } , \intermed {  {\intermed \Sigma}_{{\mathrm{2}}}  } : \intermed {   {\intermed \gamma }_{{\mathrm{2}}}  ( {\intermed \phi }_{{\mathrm{2}}}  ( {\intermed R }  ( {\intermed \delta } )))   } ) \in \mathcal{V} \llbracket \intermed {  {\intermed Q}  } \rrbracket ^{\intermed {  {\intermed { \Gamma_C } }  } }_{\intermed {  {\intermed R }  } } 
  \end{equation*}
  where \( \intermed {  {\intermed R }  } \in \mathcal{F} \llbracket \intermed {  {\intermed \Gamma}  } \rrbracket ^{\intermed {  {\intermed { \Gamma_C } }  } } \), \( \intermed {  {\intermed \phi }  } \in \mathcal{G} \llbracket \intermed {  {\intermed \Gamma}  } \rrbracket ^{\intermed {  {\intermed \Sigma}_{{\mathrm{1}}}  } , \intermed {  {\intermed \Sigma}_{{\mathrm{2}}}  } , \intermed {  {\intermed { \Gamma_C } }  } }_{\intermed {  {\intermed R }  } } \)
  and \( \intermed {  {\intermed \gamma }  } \in \mathcal{H} \llbracket \intermed {  {\intermed \Gamma}  } \rrbracket ^{\intermed {  {\intermed \Sigma}_{{\mathrm{1}}}  } , \intermed {  {\intermed \Sigma}_{{\mathrm{2}}}  } , \intermed {  {\intermed { \Gamma_C } }  } }_{\intermed {  {\intermed R }  } } \).

  From the given we know that \( ( \intermed {  {\intermed \delta }  } : \intermed {  {\intermed Q}  } )  \in  \intermed {  {\intermed \Gamma}  } \).
  Because of this, it follows from the definition of \(\mathcal{H}\) that:
  \begin{gather*}
     \intermed {   {\intermed \gamma }_{{\mathrm{1}}}  ( {\intermed \phi }_{{\mathrm{1}}}  ( {\intermed R }  ( {\intermed \delta } )))   } = \intermed {  {\intermed {dv} }_{{\mathrm{1}}}  }  \\
     \intermed {   {\intermed \gamma }_{{\mathrm{2}}}  ( {\intermed \phi }_{{\mathrm{2}}}  ( {\intermed R }  ( {\intermed \delta } )))   } = \intermed {  {\intermed {dv} }_{{\mathrm{2}}}  }  \\
     ( \intermed {  {\intermed \Sigma}_{{\mathrm{1}}}  } : \intermed {  {\intermed {dv} }_{{\mathrm{1}}}  } , \intermed {  {\intermed \Sigma}_{{\mathrm{2}}}  } : \intermed {  {\intermed {dv} }_{{\mathrm{2}}}  } ) \in \mathcal{V} \llbracket \intermed {  {\intermed Q}  } \rrbracket ^{\intermed {  {\intermed { \Gamma_C } }  } }_{\intermed {  {\intermed R }  } } 
  \end{gather*}
}
\proofStep{\( \intermed {  {\intermed d}  } = \intermed {  {\intermed D } \, \overline {\intermed \sigma}_{\ottmv{j}} \, {\overline {\intermed d} }_{\ottmv{i}}  } \) (\textsc{D-con})}
{
   \intermed {  {\intermed \Sigma}_{{\mathrm{1}}}  } ; \intermed {  {\intermed { \Gamma_C } }  } ; \intermed {  {\intermed \Gamma}  }  \vdash_{d}  \intermed {  {\intermed D } \, \overline {\intermed \sigma}_{\ottmv{j}} \, {\overline {\intermed d} }_{\ottmv{i}}  } : \intermed {  {\intermed {\mathit{TC} } } \, [  \overline {\intermed \sigma}_{\ottmv{j}}  /  {\overline {\intermed a} }_{\ottmv{j}}  ]  {\intermed \sigma}_{\ottmv{q}}  } \itot{\,  \rightsquigarrow  \target {  {\target e }  } }  \band
   \intermed {  {\intermed \Sigma}_{{\mathrm{2}}}  } ; \intermed {  {\intermed { \Gamma_C } }  } ; \intermed {  {\intermed \Gamma}  }  \vdash_{d}  \intermed {  {\intermed D } \, \overline {\intermed \sigma}_{\ottmv{j}} \, {\overline {\intermed d} }_{\ottmv{i}}  } : \intermed {  {\intermed {\mathit{TC} } } \, [  \overline {\intermed \sigma}_{\ottmv{j}}  /  {\overline {\intermed a} }_{\ottmv{j}}  ]  {\intermed \sigma}_{\ottmv{q}}  } \itot{\,  \rightsquigarrow  \target {  {\target e }  } } 
}
{
  The goal to be proven is the following:
  \begin{equation*}
     \intermed {  {\intermed { \Gamma_C } }  } ; \intermed {  {\intermed \Gamma}  }  \vdash  \intermed {  {\intermed \Sigma}_{{\mathrm{1}}}  } : \intermed {  {\intermed D } \, \overline {\intermed \sigma}_{\ottmv{j}} \, {\overline {\intermed d} }_{\ottmv{i}}  }  \simeq_{log}  \intermed {  {\intermed \Sigma}_{{\mathrm{2}}}  } : \intermed {  {\intermed D } \, \overline {\intermed \sigma}_{\ottmv{j}} \, {\overline {\intermed d} }_{\ottmv{i}}  } : \intermed {  {\intermed {\mathit{TC} } } \, [  \overline {\intermed \sigma}_{\ottmv{j}}  /  {\overline {\intermed a} }_{\ottmv{j}}  ]  {\intermed \sigma}_{\ottmv{q}}  } 
  \end{equation*}
  Unfolding the definition of logical equivalence in the goal results in:
  \begin{equation*}
     ( \intermed {  {\intermed \Sigma}_{{\mathrm{1}}}  } : \intermed {   {\intermed \gamma }_{{\mathrm{1}}}  ( {\intermed \phi }_{{\mathrm{1}}}  ( {\intermed R }  ( {\intermed D } \, \overline {\intermed \sigma}_{\ottmv{j}} \, {\overline {\intermed d} }_{\ottmv{i}} )))   } , \intermed {  {\intermed \Sigma}_{{\mathrm{2}}}  } : \intermed {   {\intermed \gamma }_{{\mathrm{2}}}  ( {\intermed \phi }_{{\mathrm{2}}}  ( {\intermed R }  ( {\intermed D } \, \overline {\intermed \sigma}_{\ottmv{j}} \, {\overline {\intermed d} }_{\ottmv{i}} )))   } ) \in \mathcal{V} \llbracket \intermed {  {\intermed {\mathit{TC} } } \, [  \overline {\intermed \sigma}_{\ottmv{j}}  /  {\overline {\intermed a} }_{\ottmv{j}}  ]  {\intermed \sigma}_{\ottmv{q}}  } \rrbracket ^{\intermed {  {\intermed { \Gamma_C } }  } }_{\intermed {  {\intermed R }  } } 
  \end{equation*}
  where \( \intermed {  {\intermed R }  } \in \mathcal{F} \llbracket \intermed {  {\intermed \Gamma}  } \rrbracket ^{\intermed {  {\intermed { \Gamma_C } }  } } \), \( \intermed {  {\intermed \phi }  } \in \mathcal{G} \llbracket \intermed {  {\intermed \Gamma}  } \rrbracket ^{\intermed {  {\intermed \Sigma}_{{\mathrm{1}}}  } , \intermed {  {\intermed \Sigma}_{{\mathrm{2}}}  } , \intermed {  {\intermed { \Gamma_C } }  } }_{\intermed {  {\intermed R }  } } \)
  and \( \intermed {  {\intermed \gamma }  } \in \mathcal{H} \llbracket \intermed {  {\intermed \Gamma}  } \rrbracket ^{\intermed {  {\intermed \Sigma}_{{\mathrm{1}}}  } , \intermed {  {\intermed \Sigma}_{{\mathrm{2}}}  } , \intermed {  {\intermed { \Gamma_C } }  } }_{\intermed {  {\intermed R }  } } \).
  
  By the definition of the \(\mathcal{V}\) relation and by distributivity of substitution
  over application, it suffices to show that:
  \begin{gather}
     ( \intermed {  {\intermed D }  } : \intermed {  \forall  {\overline {\intermed a} }_{\ottmv{j}}  \ottsym{.}  {\overline {\intermed Q} }_{\ottmv{i}}  \Rightarrow  {\intermed {\mathit{TC} } } \, {\intermed \sigma}_{\ottmv{q}}  } ) . \intermed {  {\intermed m}  }  \mapsto  \intermed {  {\intermed e}  }  \in  \intermed {  {\intermed \Sigma}_{{\mathrm{1}}}  } 
    \label{eq:drefl_con10} \\
    \ottcomp{ ( \intermed {  {\intermed \Sigma}_{{\mathrm{1}}}  } : \intermed {   {\intermed \gamma }_{{\mathrm{1}}}  ( {\intermed \phi }_{{\mathrm{1}}}  ( {\intermed R }  ( {\intermed d}_{\ottmv{i}} )))   } , \intermed {  {\intermed \Sigma}_{{\mathrm{2}}}  } : \intermed {   {\intermed \gamma }_{{\mathrm{2}}}  ( {\intermed \phi }_{{\mathrm{2}}}  ( {\intermed R }  ( {\intermed d}_{\ottmv{i}} )))   } ) \in \mathcal{V} \llbracket \intermed {  [  \overline {\intermed \sigma}_{\ottmv{j}}  /  {\overline {\intermed a} }_{\ottmv{j}}  ]  {\intermed Q}_{\ottmv{i}}  } \rrbracket ^{\intermed {  {\intermed { \Gamma_C } }  } }_{\intermed {  {\intermed R }  } } }{\ottmv{i}}
    \label{eq:drefl_con11} \\
     \intermed {  {\intermed \Sigma}_{{\mathrm{1}}}  } ; \intermed {  {\intermed { \Gamma_C } }  } ; \intermed {  \bullet  }  \vdash_{d}  \intermed {   {\intermed \gamma }_{{\mathrm{1}}}  ( {\intermed \phi }_{{\mathrm{1}}}  ( {\intermed R }  ( {\intermed D } \, \overline {\intermed \sigma}_{\ottmv{j}} \, {\overline {\intermed d} }_{\ottmv{i}} )))   } : \intermed {  {\intermed R }  \ottsym{(}  {\intermed {\mathit{TC} } } \, [  \overline {\intermed \sigma}_{\ottmv{j}}  /  {\overline {\intermed a} }_{\ottmv{j}}  ]  {\intermed \sigma}_{\ottmv{q}}  \ottsym{)}  } \itot{\,  \rightsquigarrow  \target {  {\target e }_{{\mathrm{1}}}  } } 
    \label{eq:drefl_con12} \\
     \intermed {  {\intermed \Sigma}_{{\mathrm{2}}}  } ; \intermed {  {\intermed { \Gamma_C } }  } ; \intermed {  \bullet  }  \vdash_{d}  \intermed {   {\intermed \gamma }_{{\mathrm{2}}}  ( {\intermed \phi }_{{\mathrm{2}}}  ( {\intermed R }  ( {\intermed D } \, \overline {\intermed \sigma}_{\ottmv{j}} \, {\overline {\intermed d} }_{\ottmv{i}} )))   } : \intermed {  {\intermed R }  \ottsym{(}  {\intermed {\mathit{TC} } } \, [  \overline {\intermed \sigma}_{\ottmv{j}}  /  {\overline {\intermed a} }_{\ottmv{j}}  ]  {\intermed \sigma}_{\ottmv{q}}  \ottsym{)}  } \itot{\,  \rightsquigarrow  \target {  {\target e }_{{\mathrm{2}}}  } } 
    \label{eq:drefl_con13}
  \end{gather}
  From the premises of the two \textsc{D-con} rules, we know that:
  \begin{gather}
     \vdash_{ctx}  \intermed {  {\intermed \Sigma}_{{\mathrm{1}}}  } ; \intermed {  {\intermed { \Gamma_C } }  } ; \intermed {  {\intermed \Gamma}  } 
    \label{eq:drefl_con3b} \\
     \vdash_{ctx}  \intermed {  {\intermed \Sigma}_{{\mathrm{2}}}  } ; \intermed {  {\intermed { \Gamma_C } }  } ; \intermed {  {\intermed \Gamma}  } 
    \label{eq:drefl_con3bb} \\
    \ottcomp{ \intermed {  {\intermed { \Gamma_C } }  } ; \intermed {  \bullet  \ottsym{,}  {\overline {\intermed a} }_{\ottmv{j}}  }  \vdash_{\mathit {Q} }  \intermed {  {\intermed Q}_{\ottmv{i}}  }\, \itot{ \rightsquigarrow  \target {  {\target \sigma }_{\ottmv{i}}  } } }{\ottmv{i}}
    \label{eq:drefl_con3c} \\
    \ottcomp{ \intermed {  {\intermed { \Gamma_C } }  } ; \intermed {  \bullet  \ottsym{,}  {\overline {\intermed a} }_{\ottmv{j}}  }  \vdash_{\mathit {Q} }  \intermed {  {\intermed Q}'_{\ottmv{i}}  }\, \itot{ \rightsquigarrow  \target {  {\target \sigma }_{\ottmv{i}}  } } }{\ottmv{i}}
    \label{eq:drefl_con3c'} \\
    \ottcomp{ \intermed {  {\intermed { \Gamma_C } }  } ; \intermed {  {\intermed \Gamma}  }  \vdash_{ty}  \intermed {  {\intermed \sigma}_{\ottmv{j}}  } \itot{ \rightsquigarrow  \target {  {\target \sigma }_{\ottmv{j}}  } } }{\ottmv{j}}
    \label{eq:drefl_con3d} \\
    \ottcomp{ \intermed {  {\intermed \Sigma}_{{\mathrm{1}}}  } ; \intermed {  {\intermed { \Gamma_C } }  } ; \intermed {  {\intermed \Gamma}  }  \vdash_{d}  \intermed {  {\intermed d}_{\ottmv{i}}  } : \intermed {  [  \overline {\intermed \sigma}_{\ottmv{j}}  /  {\overline {\intermed a} }_{\ottmv{j}}  ]  {\intermed Q}_{\ottmv{i}}  } \itot{\,  \rightsquigarrow  \target {  {\target e }_{\ottmv{i}}  } } }{\ottmv{i}}
    \label{eq:drefl_con3e} \\
    \ottcomp{ \intermed {  {\intermed \Sigma}_{{\mathrm{2}}}  } ; \intermed {  {\intermed { \Gamma_C } }  } ; \intermed {  {\intermed \Gamma}  }  \vdash_{d}  \intermed {  {\intermed d}_{\ottmv{i}}  } : \intermed {  [  \overline {\intermed \sigma}_{\ottmv{j}}  /  {\overline {\intermed a} }_{\ottmv{j}}  ]  {\intermed Q}'_{\ottmv{i}}  } \itot{\,  \rightsquigarrow  \target {  {\target e }_{\ottmv{i}}  } } }{\ottmv{i}}
    \label{eq:drefl_con3eb} \\
     \intermed {  {\intermed \Sigma}'_{{\mathrm{1}}}  } ; \intermed {  {\intermed { \Gamma_C } }  } ; \intermed {  \bullet  \ottsym{,}  {\overline {\intermed a} }_{\ottmv{j}}  \ottsym{,}  {\overline {\intermed \delta} }_{\ottmv{i}}  \ottsym{:}  {\overline {\intermed Q} }_{\ottmv{i}}  }  \vdash_{tm}  \intermed {  {\intermed e}_{{\mathrm{1}}}  } : \intermed {  [  \overline {\intermed \sigma}_{\ottmv{j}}  /  {\overline {\intermed a} }_{\ottmv{j}}  ]  {\intermed \sigma}_{\ottmv{m}}  } \itot{ \rightsquigarrow  \target {  {\target e }  } } 
    \label{eq:drefl_con3f} \\
     \intermed {  {\intermed \Sigma}'_{{\mathrm{2}}}  } ; \intermed {  {\intermed { \Gamma_C } }  } ; \intermed {  \bullet  \ottsym{,}  {\overline {\intermed a} }_{\ottmv{j}}  \ottsym{,}  {\overline {\intermed \delta} }_{\ottmv{i}}  \ottsym{:}  {\overline {\intermed Q} }'_{\ottmv{i}}  }  \vdash_{tm}  \intermed {  {\intermed e}_{{\mathrm{2}}}  } : \intermed {  [  \overline {\intermed \sigma}_{\ottmv{j}}  /  {\overline {\intermed a} }_{\ottmv{j}}  ]  {\intermed \sigma}_{\ottmv{m}}  } \itot{ \rightsquigarrow  \target {  {\target e }  } } 
    \label{eq:drefl_con3fb} \\
    \textit{where}~ \intermed {  {\intermed \Sigma}_{{\mathrm{1}}}  } = \intermed {  {\intermed \Sigma}'_{{\mathrm{1}}}  \ottsym{,}  \ottsym{(}  {\intermed D }  \ottsym{:}  \forall  {\overline {\intermed a} }_{\ottmv{j}}  \ottsym{.}  {\overline {\intermed Q} }_{\ottmv{i}}  \Rightarrow  {\intermed {\mathit{TC} } } \, {\intermed \sigma}_{\ottmv{q}}  \ottsym{)}  \ottsym{.}  {\intermed m}  \mapsto  \Lambda  {\overline {\intermed a} }_{\ottmv{j}}  \ottsym{.}  \lambda  {\overline {\intermed \delta} }_{\ottmv{i}}  \ottsym{:}  {\overline {\intermed Q} }_{\ottmv{i}}  \ottsym{.}  {\intermed e}_{{\mathrm{1}}}  \ottsym{,}  {\intermed \Sigma}''_{{\mathrm{1}}}  } 
    \label{eq:drefl_con3a} \\
    \textit{and}~\, \intermed {  {\intermed \Sigma}_{{\mathrm{2}}}  } = \intermed {  {\intermed \Sigma}'_{{\mathrm{2}}}  \ottsym{,}  \ottsym{(}  {\intermed D }  \ottsym{:}  \forall  {\overline {\intermed a} }_{\ottmv{j}}  \ottsym{.}  {\overline {\intermed Q} }'_{\ottmv{i}}  \Rightarrow  {\intermed {\mathit{TC} } }' \, {\intermed \sigma}'_{\ottmv{q}}  \ottsym{)}  \ottsym{.}  {\intermed m}'  \mapsto  \Lambda  {\overline {\intermed a} }_{\ottmv{j}}  \ottsym{.}  \lambda  {\overline {\intermed \delta} }_{\ottmv{i}}  \ottsym{:}  {\overline {\intermed Q} }'_{\ottmv{i}}  \ottsym{.}  {\intermed e}_{{\mathrm{2}}}  \ottsym{,}  {\intermed \Sigma}''_{{\mathrm{2}}}  } 
    \label{eq:drefl_con3ab}
  \end{gather}
  From the definition of logical equivalence in the third hypothesis of
  the theorem, we know that \( \intermed {  {\overline {\intermed Q} }_{\ottmv{i}}  } = \intermed {  {\overline {\intermed Q} }'_{\ottmv{i}}  } \),
  \( \intermed {  {\intermed {\mathit{TC} } }  } = \intermed {  {\intermed {\mathit{TC} } }'  } \), \( \intermed {  {\intermed \sigma}_{\ottmv{q}}  } = \intermed {  {\intermed \sigma}'_{\ottmv{q}}  } \), \( \intermed {  {\intermed m}  } = \intermed {  {\intermed m}'  } \).

  Goal~\ref{eq:drefl_con10} follows directly from Equation~\ref{eq:drefl_con3a} by
  setting \( \intermed {  {\intermed e}  } = \intermed {  \Lambda  {\overline {\intermed a} }_{\ottmv{j}}  \ottsym{.}  \lambda  {\overline {\intermed \delta} }_{\ottmv{i}}  \ottsym{:}  {\overline {\intermed Q} }'_{\ottmv{i}}  \ottsym{.}  {\intermed e}_{{\mathrm{2}}}  } \).

  By applying the induction hypothesis on Equations~\ref{eq:drefl_con3e}
  and~\ref{eq:drefl_con3eb}, we obtain:
  \begin{equation*}
    \ottcomp{ \intermed {  {\intermed { \Gamma_C } }  } ; \intermed {  {\intermed \Gamma}  }  \vdash  \intermed {  {\intermed \Sigma}_{{\mathrm{1}}}  } : \intermed {  {\intermed d}_{\ottmv{i}}  }  \simeq_{log}  \intermed {  {\intermed \Sigma}_{{\mathrm{2}}}  } : \intermed {  {\intermed d}_{\ottmv{i}}  } : \intermed {  [  \overline {\intermed \sigma}_{\ottmv{j}}  /  {\overline {\intermed a} }_{\ottmv{j}}  ]  {\intermed Q}_{\ottmv{i}}  } }{\ottmv{i}}
  \end{equation*}
  Unfolding the definition of logical equivalence in the above, we get:
  \begin{equation}
    \ottcomp{ ( \intermed {  {\intermed \Sigma}_{{\mathrm{1}}}  } : \intermed {  {\intermed {dv} }_{\ottmv{i}\,{\mathrm{1}}}  } , \intermed {  {\intermed \Sigma}_{{\mathrm{2}}}  } : \intermed {  {\intermed {dv} }_{\ottmv{i}\,{\mathrm{2}}}  } ) \in \mathcal{V} \llbracket \intermed {  [  \overline {\intermed \sigma}_{\ottmv{j}}  /  {\overline {\intermed a} }_{\ottmv{j}}  ]  {\intermed Q}_{\ottmv{i}}  } \rrbracket ^{\intermed {  {\intermed { \Gamma_C } }  } }_{\intermed {  {\intermed R }  } } }{\ottmv{i}}
    \label{eq:drefl_cone2}
  \end{equation}
  where \( \intermed {  {\intermed {dv} }_{\ottmv{i}\,{\mathrm{1}}}  } = \intermed {   {\intermed \gamma }_{{\mathrm{1}}}  ( {\intermed \phi }_{{\mathrm{1}}}  ( {\intermed R }  ( {\intermed d}_{\ottmv{i}} )))   } \) and
  \( \intermed {  {\intermed {dv} }_{\ottmv{i}\,{\mathrm{2}}}  } = \intermed {   {\intermed \gamma }_{{\mathrm{2}}}  ( {\intermed \phi }_{{\mathrm{2}}}  ( {\intermed R }  ( {\intermed d}_{\ottmv{i}} )))   } \).
  This proves Goal~\ref{eq:drefl_con11}.

  By applying Lemma~\ref{lemma:tvar_tsubst} in Equation~\ref{eq:drefl_con3d},
  there is an environment \({\intermed \Gamma}'\) (the resulting environment
  after applying on \({\intermed \Gamma}\) all type variable substitutions of \({\intermed R }\)) such that
  \begin{equation*}
    \ottcomp{ \intermed {  {\intermed { \Gamma_C } }  } ; \intermed {  {\intermed \Gamma}'  }  \vdash_{ty}  \intermed {  {\intermed R }  \ottsym{(}  {\intermed \sigma}_{\ottmv{j}}  \ottsym{)}  } \itot{ \rightsquigarrow  \target {  {\target \sigma }_{\ottmv{j}}  } } }{\ottmv{j}}
  \end{equation*}
  From Lemma~\ref{lemma:tvar_tsubst} and since the domain of \({\intermed R }\) contains all type
  variables of \({\intermed \Gamma}\), it is evident that all type variables are eliminated in
  \({\intermed \Gamma}'\). By consecutive applications of Lemma~\ref{lemma:var_subst_types} we have that
  \begin{equation}
    \ottcomp{ \intermed {  {\intermed { \Gamma_C } }  } ; \intermed {  \bullet  }  \vdash_{ty}  \intermed {  {\intermed R }  \ottsym{(}  {\intermed \sigma}_{\ottmv{j}}  \ottsym{)}  } \itot{ \rightsquigarrow  \target {  {\target \sigma }_{\ottmv{j}}  } } }{\ottmv{j}}
    \label{eq:drefl_con3d2}
  \end{equation}
  Furthermore, from the definiton of the \(\mathcal{V}\) relation in
  Equation~\ref{eq:drefl_cone2}, we obtain:
  \begin{gather*}
    \ottcomp{ \intermed {  {\intermed \Sigma}_{{\mathrm{1}}}  } ; \intermed {  {\intermed { \Gamma_C } }  } ; \intermed {  \bullet  }  \vdash_{d}  \intermed {  {\intermed {dv} }_{\ottmv{i}\,{\mathrm{1}}}  } : \intermed {  {\intermed R }  \ottsym{(}  [  \overline {\intermed \sigma}_{\ottmv{j}}  /  {\overline {\intermed a} }_{\ottmv{j}}  ]  {\intermed Q}_{\ottmv{i}}  \ottsym{)}  } \itot{\,  \rightsquigarrow  \target {  {\target e }_{\ottmv{i}\,{\mathrm{1}}}  } } }{\ottmv{i}}\\
    \ottcomp{ \intermed {  {\intermed \Sigma}_{{\mathrm{2}}}  } ; \intermed {  {\intermed { \Gamma_C } }  } ; \intermed {  \bullet  }  \vdash_{d}  \intermed {  {\intermed {dv} }_{\ottmv{i}\,{\mathrm{2}}}  } : \intermed {  {\intermed R }  \ottsym{(}  [  \overline {\intermed \sigma}_{\ottmv{j}}  /  {\overline {\intermed a} }_{\ottmv{j}}  ]  {\intermed Q}_{\ottmv{i}}  \ottsym{)}  } \itot{\,  \rightsquigarrow  \target {  {\target e }_{\ottmv{i}\,{\mathrm{2}}}  } } }{\ottmv{i}}
  \end{gather*}
  From Equation~\ref{eq:drefl_con3c}, we know that \(\intermed{{\intermed Q}_{\ottmv{i}}}\) only contains
  free variables \(\intermed{{\overline {\intermed a} }_{\ottmv{j}}}\). It follows that the above equations can be simplified to:
  \begin{gather}
    \ottcomp{ \intermed {  {\intermed \Sigma}_{{\mathrm{1}}}  } ; \intermed {  {\intermed { \Gamma_C } }  } ; \intermed {  \bullet  }  \vdash_{d}  \intermed {  {\intermed {dv} }_{\ottmv{i}\,{\mathrm{1}}}  } : \intermed {  [  {\intermed R }  \ottsym{(}  \overline {\intermed \sigma}_{\ottmv{j}}  \ottsym{)}  /  {\overline {\intermed a} }_{\ottmv{j}}  ]  {\intermed Q}_{\ottmv{i}}  } \itot{\,  \rightsquigarrow  \target {  {\target e }_{\ottmv{i}\,{\mathrm{1}}}  } } }{\ottmv{i}}
    \label{eq:drefl_con5b} \\
    \ottcomp{ \intermed {  {\intermed \Sigma}_{{\mathrm{2}}}  } ; \intermed {  {\intermed { \Gamma_C } }  } ; \intermed {  \bullet  }  \vdash_{d}  \intermed {  {\intermed {dv} }_{\ottmv{i}\,{\mathrm{2}}}  } : \intermed {  [  {\intermed R }  \ottsym{(}  \overline {\intermed \sigma}_{\ottmv{j}}  \ottsym{)}  /  {\overline {\intermed a} }_{\ottmv{j}}  ]  {\intermed Q}_{\ottmv{i}}  } \itot{\,  \rightsquigarrow  \target {  {\target e }_{\ottmv{i}\,{\mathrm{2}}}  } } }{\ottmv{i}}
    \label{eq:drefl_con6b}
  \end{gather}
  Applying the substitutions in Goals~\ref{eq:drefl_con12} and
  \ref{eq:drefl_con13},
  reduces them to:
  \begin{gather*}
     \intermed {  {\intermed \Sigma}_{{\mathrm{1}}}  } ; \intermed {  {\intermed { \Gamma_C } }  } ; \intermed {  \bullet  }  \vdash_{d}  \intermed {  {\intermed D } \, {\intermed R }  \ottsym{(}  \overline {\intermed \sigma}_{\ottmv{j}}  \ottsym{)} \, {\overline {\intermed { dv } } }_{\ottmv{i}\,{\mathrm{1}}}  } : \intermed {  {\intermed {\mathit{TC} } } \, {\intermed R }  \ottsym{(}  [  \overline {\intermed \sigma}_{\ottmv{j}}  /  {\overline {\intermed a} }_{\ottmv{j}}  ]  {\intermed \sigma}_{\ottmv{q}}  \ottsym{)}  } \itot{\,  \rightsquigarrow  \target {  {\target e }_{{\mathrm{1}}}  } } \\
     \intermed {  {\intermed \Sigma}_{{\mathrm{2}}}  } ; \intermed {  {\intermed { \Gamma_C } }  } ; \intermed {  \bullet  }  \vdash_{d}  \intermed {  {\intermed D } \, {\intermed R }  \ottsym{(}  \overline {\intermed \sigma}_{\ottmv{j}}  \ottsym{)} \, {\overline {\intermed { dv } } }_{\ottmv{i}\,{\mathrm{2}}}  } : \intermed {  {\intermed {\mathit{TC} } } \, {\intermed R }  \ottsym{(}  [  \overline {\intermed \sigma}_{\ottmv{j}}  /  {\overline {\intermed a} }_{\ottmv{j}}  ]  {\intermed \sigma}_{\ottmv{q}}  \ottsym{)}  } \itot{\,  \rightsquigarrow  \target {  {\target e }_{{\mathrm{2}}}  } } 
  \end{gather*}
  Similarly to \(\intermed{{\intermed Q}_{\ottmv{i}}}\), it follows from \textsc{iCtx-MEnv} that
  \(\intermed{{\intermed \sigma}_{\ottmv{q}}}\) only contains free variables \(\intermed{{\overline {\intermed a} }_{\ottmv{j}}}\). The above goals thus
  simplify to:
  \begin{gather}
     \intermed {  {\intermed \Sigma}_{{\mathrm{1}}}  } ; \intermed {  {\intermed { \Gamma_C } }  } ; \intermed {  \bullet  }  \vdash_{d}  \intermed {  {\intermed D } \, {\intermed R }  \ottsym{(}  \overline {\intermed \sigma}_{\ottmv{j}}  \ottsym{)} \, {\overline {\intermed { dv } } }_{\ottmv{i}\,{\mathrm{1}}}  } : \intermed {  {\intermed {\mathit{TC} } } \, [  {\intermed R }  \ottsym{(}  \overline {\intermed \sigma}_{\ottmv{j}}  \ottsym{)}  /  {\overline {\intermed a} }_{\ottmv{j}}  ]  {\intermed \sigma}_{\ottmv{q}}  } \itot{\,  \rightsquigarrow  \target {  {\target e }_{{\mathrm{1}}}  } } 
    \label{eq:drefl_con14b} \\
     \intermed {  {\intermed \Sigma}_{{\mathrm{2}}}  } ; \intermed {  {\intermed { \Gamma_C } }  } ; \intermed {  \bullet  }  \vdash_{d}  \intermed {  {\intermed D } \, {\intermed R }  \ottsym{(}  \overline {\intermed \sigma}_{\ottmv{j}}  \ottsym{)} \, {\overline {\intermed { dv } } }_{\ottmv{i}\,{\mathrm{2}}}  } : \intermed {  {\intermed {\mathit{TC} } } \, [  {\intermed R }  \ottsym{(}  \overline {\intermed \sigma}_{\ottmv{j}}  \ottsym{)}  /  {\overline {\intermed a} }_{\ottmv{j}}  ]  {\intermed \sigma}_{\ottmv{q}}  } \itot{\,  \rightsquigarrow  \target {  {\target e }_{{\mathrm{2}}}  } } 
    \label{eq:drefl_con15b}
  \end{gather}
  Goal~\ref{eq:drefl_con14b} follows from \textsc{D-con}, in combination with
  Equations~\ref{eq:drefl_con3a}, \ref{eq:drefl_con3b}, \ref{eq:drefl_con3c},
  \ref{eq:drefl_con3d2}, \ref{eq:drefl_con3f} and \ref{eq:drefl_con5b}.
  Goal~\ref{eq:drefl_con15b} follows from from \textsc{D-con}, in combination with
  Equations~\ref{eq:drefl_con3ab}, \ref{eq:drefl_con3bb}, \ref{eq:drefl_con3c},
  \ref{eq:drefl_con3d2}, \ref{eq:drefl_con3fb} and \ref{eq:drefl_con6b}.
  
}
\qed

\begin{theoremB}[Expression Reflexivity]\ \\
  If \( \intermed {  {\intermed \Sigma}_{{\mathrm{1}}}  } ; \intermed {  {\intermed { \Gamma_C } }  } ; \intermed {  {\intermed \Gamma}  }  \vdash_{tm}  \intermed {  {\intermed e}  } : \intermed {  {\intermed \sigma}  } \itot{ \rightsquigarrow  \target {  {\target e }  } } \)
  and \( \intermed {  {\intermed \Sigma}_{{\mathrm{2}}}  } ; \intermed {  {\intermed { \Gamma_C } }  } ; \intermed {  {\intermed \Gamma}  }  \vdash_{tm}  \intermed {  {\intermed e}  } : \intermed {  {\intermed \sigma}  } \itot{ \rightsquigarrow  \target {  {\target e }  } } \)
  and \( \intermed {  {\intermed { \Gamma_C } }  }  \vdash  \intermed {  {\intermed \Sigma}_{{\mathrm{1}}}  }  \simeq_{log}  \intermed {  {\intermed \Sigma}_{{\mathrm{2}}}  } \),
  then \( \intermed {  {\intermed { \Gamma_C } }  } ; \intermed {  {\intermed \Gamma}  }  \vdash  \intermed {  {\intermed \Sigma}_{{\mathrm{1}}}  } : \intermed {  {\intermed e}  }  \simeq_{log}  \intermed {  {\intermed \Sigma}_{{\mathrm{2}}}  } : \intermed {  {\intermed e}  } : \intermed {  {\intermed \sigma}  } \).
  \label{thmA:expr_reflex}
\end{theoremB}

\proof
The proof proceeds by induction on \({\intermed e}\) and consequently, since \interL{} term typing
is syntax directed, on both typing derivations. \\
\proofStep{\( \intermed {  {\intermed e}  } = \intermed {  \mathit{\intermed {True} }  } \) (\textsc{iTm-true})}
{
   \intermed {  {\intermed \Sigma}_{{\mathrm{1}}}  } ; \intermed {  {\intermed { \Gamma_C } }  } ; \intermed {  {\intermed \Gamma}  }  \vdash_{tm}  \intermed {  \mathit{\intermed {True} }  } : \intermed {  \mathit{\intermed {Bool} }  } \itot{ \rightsquigarrow  \target {  \mathit{\target {True} }  } }  \band
   \intermed {  {\intermed \Sigma}_{{\mathrm{2}}}  } ; \intermed {  {\intermed { \Gamma_C } }  } ; \intermed {  {\intermed \Gamma}  }  \vdash_{tm}  \intermed {  \mathit{\intermed {True} }  } : \intermed {  \mathit{\intermed {Bool} }  } \itot{ \rightsquigarrow  \target {  \mathit{\target {True} }  } } 
}
{
  The goal to prove is the following:
  \begin{equation*}
     \intermed {  {\intermed { \Gamma_C } }  } ; \intermed {  {\intermed \Gamma}  }  \vdash  \intermed {  {\intermed \Sigma}_{{\mathrm{1}}}  } : \intermed {  \mathit{\intermed {True} }  }  \simeq_{log}  \intermed {  {\intermed \Sigma}_{{\mathrm{2}}}  } : \intermed {  \mathit{\intermed {True} }  } : \intermed {  \mathit{\intermed {Bool} }  } 
  \end{equation*}
  Unfolding the definition of logical equivalence in the above, 
  results in the following goal:
  \begin{equation}
     ( \intermed {  {\intermed \Sigma}_{{\mathrm{1}}}  } : \intermed {   {\intermed \gamma }_{{\mathrm{1}}}  ( {\intermed \phi }_{{\mathrm{1}}}  ( {\intermed R }  ( \mathit{\intermed {True} } )))   } , \intermed {  {\intermed \Sigma}_{{\mathrm{2}}}  } : \intermed {   {\intermed \gamma }_{{\mathrm{2}}}  ( {\intermed \phi }_{{\mathrm{2}}}  ( {\intermed R }  ( \mathit{\intermed {True} } )))   } ) \in \mathcal{E} \llbracket \intermed {  \mathit{\intermed {Bool} }  } \rrbracket ^{\intermed {  {\intermed { \Gamma_C } }  } }_{\intermed {  {\intermed R }  } } 
    \label{eq:erefl_true2}
  \end{equation}
  where \( \intermed {  {\intermed R }  } \in \mathcal{F} \llbracket \intermed {  {\intermed \Gamma}  } \rrbracket ^{\intermed {  {\intermed { \Gamma_C } }  } } \), \( \intermed {  {\intermed \phi }  } \in \mathcal{G} \llbracket \intermed {  {\intermed \Gamma}  } \rrbracket ^{\intermed {  {\intermed \Sigma}_{{\mathrm{1}}}  } , \intermed {  {\intermed \Sigma}_{{\mathrm{2}}}  } , \intermed {  {\intermed { \Gamma_C } }  } }_{\intermed {  {\intermed R }  } } \)
  and \( \intermed {  {\intermed \gamma }  } \in \mathcal{H} \llbracket \intermed {  {\intermed \Gamma}  } \rrbracket ^{\intermed {  {\intermed \Sigma}_{{\mathrm{1}}}  } , \intermed {  {\intermed \Sigma}_{{\mathrm{2}}}  } , \intermed {  {\intermed { \Gamma_C } }  } }_{\intermed {  {\intermed R }  } } \). However, since \(\mathit{\intermed {True} }\)
  does not contain any free variables, we know that
  \( \intermed {   {\intermed \gamma }_{{\mathrm{1}}}  ( {\intermed \phi }_{{\mathrm{1}}}  ( {\intermed R }  ( \mathit{\intermed {True} } )))   } = \intermed {   {\intermed \gamma }_{{\mathrm{2}}}  ( {\intermed \phi }_{{\mathrm{2}}}  ( {\intermed R }  ( \mathit{\intermed {True} } )))   } = \intermed {  \mathit{\intermed {True} }  } \). Similarly, it follows that
  \( \intermed {  {\intermed R }  \ottsym{(}  \mathit{\intermed {Bool} }  \ottsym{)}  } = \intermed {  \mathit{\intermed {Bool} }  } \).

  Unfolding the definition of the \(\mathcal{E}\) relation in
  Goal~\ref{eq:erefl_true2}, reduces the goal to:
  \begin{gather}
     \intermed {  {\intermed \Sigma}_{{\mathrm{1}}}  } ; \intermed {  {\intermed { \Gamma_C } }  } ; \intermed {  \bullet  }  \vdash_{tm}  \intermed {  \mathit{\intermed {True} }  } : \intermed {  \mathit{\intermed {Bool} }  } \itot{ \rightsquigarrow  \target {  {\target e }  } } 
    \label{eq:erefl_true3} \\
     \intermed {  {\intermed \Sigma}_{{\mathrm{2}}}  } ; \intermed {  {\intermed { \Gamma_C } }  } ; \intermed {  \bullet  }  \vdash_{tm}  \intermed {  \mathit{\intermed {True} }  } : \intermed {  \mathit{\intermed {Bool} }  } \itot{ \rightsquigarrow  \target {  {\target e }  } } 
    \label{eq:erefl_true3b} \\
    \begin{aligned}
      \exists {\intermed v}_{{\mathrm{1}}}, {\intermed v}_{{\mathrm{2}}} :\, &
       \intermed {  {\intermed \Sigma}_{{\mathrm{1}}}  }  \vdash  \intermed {  \mathit{\intermed {True} }  }  \longrightarrow^{*}  \intermed {  {\intermed v}_{{\mathrm{1}}}  }   \\
      \band\, &  \intermed {  {\intermed \Sigma}_{{\mathrm{2}}}  }  \vdash  \intermed {  \mathit{\intermed {True} }  }  \longrightarrow^{*}  \intermed {  {\intermed v}_{{\mathrm{2}}}  }  \\
      \band\, &  ( \intermed {  {\intermed \Sigma}_{{\mathrm{1}}}  } : \intermed {  {\intermed v}_{{\mathrm{1}}}  } , \intermed {  {\intermed \Sigma}_{{\mathrm{2}}}  } : \intermed {  {\intermed v}_{{\mathrm{2}}}  } ) \in \mathcal{V} \llbracket \intermed {  \mathit{\intermed {Bool} }  } \rrbracket ^{\intermed {  {\intermed { \Gamma_C } }  } }_{\intermed {  {\intermed R }  } } 
    \end{aligned}
    \nonumber 
  \end{gather}
  Goals~\ref{eq:erefl_true3} and~\ref{eq:erefl_true3b} are satisified from the first and
  second hypotheses of the theorem.
  We set \( \intermed {  {\intermed v}_{{\mathrm{1}}}  } = \intermed {  {\intermed v}_{{\mathrm{2}}}  } = \intermed {  \mathit{\intermed {True} }  } \) and since \(\mathit{\intermed {True} }\) is a value, the term reductions
  above hold. Then, the last goal follows directly from the definition of the \(\mathcal{V}\)
  relation, according to which the following holds trivially.
  \begin{equation*}
     ( \intermed {  {\intermed \Sigma}_{{\mathrm{1}}}  } : \intermed {  \mathit{\intermed {True} }  } , \intermed {  {\intermed \Sigma}_{{\mathrm{2}}}  } : \intermed {  \mathit{\intermed {True} }  } ) \in \mathcal{V} \llbracket \intermed {  \mathit{\intermed {Bool} }  } \rrbracket ^{\intermed {  {\intermed { \Gamma_C } }  } }_{\intermed {  {\intermed R }  } } 
  \end{equation*}
}
\proofStep{\( \intermed {  {\intermed e}  } = \intermed {  \mathit{\intermed {False} }  } \) (\textsc{iTm-false})}
{
   \intermed {  {\intermed \Sigma}_{{\mathrm{1}}}  } ; \intermed {  {\intermed { \Gamma_C } }  } ; \intermed {  {\intermed \Gamma}  }  \vdash_{tm}  \intermed {  \mathit{\intermed {False} }  } : \intermed {  \mathit{\intermed {Bool} }  } \itot{ \rightsquigarrow  \target {  \mathit{\target {False} }  } }  \band
   \intermed {  {\intermed \Sigma}_{{\mathrm{2}}}  } ; \intermed {  {\intermed { \Gamma_C } }  } ; \intermed {  {\intermed \Gamma}  }  \vdash_{tm}  \intermed {  \mathit{\intermed {False} }  } : \intermed {  \mathit{\intermed {Bool} }  } \itot{ \rightsquigarrow  \target {  \mathit{\target {False} }  } } 
}
{
  The proof is similar to the \textsc{iTm-true} case. \\
}
\proofStep{\( \intermed {  {\intermed e}  } = \intermed {  {\intermed x}  } \) (\textsc{iTm-var})}
{
   \intermed {  {\intermed \Sigma}_{{\mathrm{1}}}  } ; \intermed {  {\intermed { \Gamma_C } }  } ; \intermed {  {\intermed \Gamma}  }  \vdash_{tm}  \intermed {  {\intermed x}  } : \intermed {  {\intermed \sigma}  } \itot{ \rightsquigarrow  \target {  {\target x}  } }  \band
   \intermed {  {\intermed \Sigma}_{{\mathrm{2}}}  } ; \intermed {  {\intermed { \Gamma_C } }  } ; \intermed {  {\intermed \Gamma}  }  \vdash_{tm}  \intermed {  {\intermed x}  } : \intermed {  {\intermed \sigma}  } \itot{ \rightsquigarrow  \target {  {\target x}  } } 
}
{
  The goal to be proven is the following:
  \begin{equation*}
     \intermed {  {\intermed { \Gamma_C } }  } ; \intermed {  {\intermed \Gamma}  }  \vdash  \intermed {  {\intermed \Sigma}_{{\mathrm{1}}}  } : \intermed {  {\intermed x}  }  \simeq_{log}  \intermed {  {\intermed \Sigma}_{{\mathrm{2}}}  } : \intermed {  {\intermed x}  } : \intermed {  {\intermed \sigma}  } 
  \end{equation*}
  By unfolding the definition of logical equivalence in the above, we have
  \begin{equation*}
     ( \intermed {  {\intermed \Sigma}_{{\mathrm{1}}}  } : \intermed {   {\intermed \gamma }_{{\mathrm{1}}}  ( {\intermed \phi }_{{\mathrm{1}}}  ( {\intermed R }  ( {\intermed x} )))   } , \intermed {  {\intermed \Sigma}_{{\mathrm{2}}}  } : \intermed {   {\intermed \gamma }_{{\mathrm{2}}}  ( {\intermed \phi }_{{\mathrm{2}}}  ( {\intermed R }  ( {\intermed x} )))   } ) \in \mathcal{E} \llbracket \intermed {  {\intermed \sigma}  } \rrbracket ^{\intermed {  {\intermed { \Gamma_C } }  } }_{\intermed {  {\intermed R }  } } 
  \end{equation*}
  for any \( \intermed {  {\intermed R }  } \in \mathcal{F} \llbracket \intermed {  {\intermed \Gamma}  } \rrbracket ^{\intermed {  {\intermed { \Gamma_C } }  } } \), \( \intermed {  {\intermed \phi }  } \in \mathcal{G} \llbracket \intermed {  {\intermed \Gamma}  } \rrbracket ^{\intermed {  {\intermed \Sigma}_{{\mathrm{1}}}  } , \intermed {  {\intermed \Sigma}_{{\mathrm{2}}}  } , \intermed {  {\intermed { \Gamma_C } }  } }_{\intermed {  {\intermed R }  } } \)
  and \( \intermed {  {\intermed \gamma }  } \in \mathcal{H} \llbracket \intermed {  {\intermed \Gamma}  } \rrbracket ^{\intermed {  {\intermed \Sigma}_{{\mathrm{1}}}  } , \intermed {  {\intermed \Sigma}_{{\mathrm{2}}}  } , \intermed {  {\intermed { \Gamma_C } }  } }_{\intermed {  {\intermed R }  } } \). 
  From the definition of the \(\mathcal{G}\) relation, we know that:
  \begin{gather}
     \intermed {   {\intermed \gamma }_{{\mathrm{1}}}  ( {\intermed \phi }_{{\mathrm{1}}}  ( {\intermed R }  ( {\intermed x} )))   } = \intermed {  {\intermed e}_{{\mathrm{1}}}  } 
    \nonumber \\
     \intermed {   {\intermed \gamma }_{{\mathrm{2}}}  ( {\intermed \phi }_{{\mathrm{2}}}  ( {\intermed R }  ( {\intermed x} )))   } = \intermed {  {\intermed e}_{{\mathrm{2}}}  } 
    \nonumber \\
     ( \intermed {  {\intermed \Sigma}_{{\mathrm{1}}}  } : \intermed {  {\intermed e}_{{\mathrm{1}}}  } , \intermed {  {\intermed \Sigma}_{{\mathrm{2}}}  } : \intermed {  {\intermed e}_{{\mathrm{2}}}  } ) \in \mathcal{E} \llbracket \intermed {  {\intermed \sigma}  } \rrbracket ^{\intermed {  {\intermed { \Gamma_C } }  } }_{\intermed {  {\intermed R }  } } 
    \label{eq:erefl_var3}
  \end{gather}
  The goal follows directly from Equation~\ref{eq:erefl_var3}. \\
}
\proofStep{\( \intermed {  {\intermed e}  } = \intermed {  \textbf{\intermed {let} } \, {\intermed x}  \ottsym{:}  {\intermed \sigma}_{{\mathrm{1}}}  \ottsym{=}  {\intermed e}_{{\mathrm{1}}} \, \textbf{\intermed {in} } \, {\intermed e}_{{\mathrm{2}}}  } \) (\textsc{iTm-let})}
{
   \intermed {  {\intermed \Sigma}_{{\mathrm{1}}}  } ; \intermed {  {\intermed { \Gamma_C } }  } ; \intermed {  {\intermed \Gamma}  }  \vdash_{tm}  \intermed {  \textbf{\intermed {let} } \, {\intermed x}  \ottsym{:}  {\intermed \sigma}_{{\mathrm{1}}}  \ottsym{=}  {\intermed e}_{{\mathrm{1}}} \, \textbf{\intermed {in} } \, {\intermed e}_{{\mathrm{2}}}  } : \intermed {  {\intermed \sigma}_{{\mathrm{2}}}  } \itot{ \rightsquigarrow  \target {  \textbf{\target {let} } \, {\target x}  \ottsym{:}  {\target \sigma }_{{\mathrm{1}}}  \ottsym{=}  {\target e }_{{\mathrm{1}}} \, \textbf{\target {in} } \, {\target e }_{{\mathrm{2}}}  } }  \band
   \intermed {  {\intermed \Sigma}_{{\mathrm{2}}}  } ; \intermed {  {\intermed { \Gamma_C } }  } ; \intermed {  {\intermed \Gamma}  }  \vdash_{tm}  \intermed {  \textbf{\intermed {let} } \, {\intermed x}  \ottsym{:}  {\intermed \sigma}_{{\mathrm{1}}}  \ottsym{=}  {\intermed e}_{{\mathrm{1}}} \, \textbf{\intermed {in} } \, {\intermed e}_{{\mathrm{2}}}  } : \intermed {  {\intermed \sigma}_{{\mathrm{2}}}  } \itot{ \rightsquigarrow  \target {  \textbf{\target {let} } \, {\target x}  \ottsym{:}  {\target \sigma }_{{\mathrm{1}}}  \ottsym{=}  {\target e }_{{\mathrm{1}}} \, \textbf{\target {in} } \, {\target e }_{{\mathrm{2}}}  } } 
}
{
  The goal to be proven is the following:
  \begin{equation*}
     \intermed {  {\intermed { \Gamma_C } }  } ; \intermed {  {\intermed \Gamma}  }  \vdash  \intermed {  {\intermed \Sigma}_{{\mathrm{1}}}  } : \intermed {  \textbf{\intermed {let} } \, {\intermed x}  \ottsym{:}  {\intermed \sigma}_{{\mathrm{1}}}  \ottsym{=}  {\intermed e}_{{\mathrm{1}}} \, \textbf{\intermed {in} } \, {\intermed e}_{{\mathrm{2}}}  }  \simeq_{log}  \intermed {  {\intermed \Sigma}_{{\mathrm{2}}}  } : \intermed {  \textbf{\intermed {let} } \, {\intermed x}  \ottsym{:}  {\intermed \sigma}_{{\mathrm{1}}}  \ottsym{=}  {\intermed e}_{{\mathrm{1}}} \, \textbf{\intermed {in} } \, {\intermed e}_{{\mathrm{2}}}  } : \intermed {  {\intermed \sigma}_{{\mathrm{2}}}  } 
  \end{equation*}
  By applying the induction hypothesis in the premises of the two \textsc{iTm-let} rules,
  we get:
  \begin{gather*}
     \intermed {  {\intermed { \Gamma_C } }  } ; \intermed {  {\intermed \Gamma}  }  \vdash  \intermed {  {\intermed \Sigma}_{{\mathrm{1}}}  } : \intermed {  {\intermed e}_{{\mathrm{1}}}  }  \simeq_{log}  \intermed {  {\intermed \Sigma}_{{\mathrm{2}}}  } : \intermed {  {\intermed e}_{{\mathrm{1}}}  } : \intermed {  {\intermed \sigma}_{{\mathrm{1}}}  } \\
     \intermed {  {\intermed { \Gamma_C } }  } ; \intermed {  {\intermed \Gamma}  \ottsym{,}  {\intermed x}  \ottsym{:}  {\intermed \sigma}_{{\mathrm{1}}}  }  \vdash  \intermed {  {\intermed \Sigma}_{{\mathrm{1}}}  } : \intermed {  {\intermed e}_{{\mathrm{2}}}  }  \simeq_{log}  \intermed {  {\intermed \Sigma}_{{\mathrm{2}}}  } : \intermed {  {\intermed e}_{{\mathrm{2}}}  } : \intermed {  {\intermed \sigma}_{{\mathrm{2}}}  } 
  \end{gather*}
  The goal follows directly by passing the above two Equations 
  to compatibility Lemma~\ref{lemma:compat_let}. \\
}
\proofStep{\( \intermed {  {\intermed e}  } = \intermed {  {\intermed d}  \ottsym{.}  {\intermed m}  } \) (\textsc{iTm-method})}
{
   \intermed {  {\intermed \Sigma}_{{\mathrm{1}}}  } ; \intermed {  {\intermed { \Gamma_C } }  } ; \intermed {  {\intermed \Gamma}  }  \vdash_{tm}  \intermed {  {\intermed d}  \ottsym{.}  {\intermed m}  } : \intermed {  [  {\intermed \sigma}  /  {\intermed a }  ]  {\intermed \sigma}'  } \itot{ \rightsquigarrow  \target {  {\target e }  \ottsym{.}  {\target m}  } }  \band
   \intermed {  {\intermed \Sigma}_{{\mathrm{2}}}  } ; \intermed {  {\intermed { \Gamma_C } }  } ; \intermed {  {\intermed \Gamma}  }  \vdash_{tm}  \intermed {  {\intermed d}  \ottsym{.}  {\intermed m}  } : \intermed {  [  {\intermed \sigma}  /  {\intermed a }  ]  {\intermed \sigma}'  } \itot{ \rightsquigarrow  \target {  {\target e }  \ottsym{.}  {\target m}  } } 
}
{
  The goal to be proven is the following:
  \begin{equation*}
     \intermed {  {\intermed { \Gamma_C } }  } ; \intermed {  {\intermed \Gamma}  }  \vdash  \intermed {  {\intermed \Sigma}_{{\mathrm{1}}}  } : \intermed {  {\intermed d}  \ottsym{.}  {\intermed m}  }  \simeq_{log}  \intermed {  {\intermed \Sigma}_{{\mathrm{2}}}  } : \intermed {  {\intermed d}  \ottsym{.}  {\intermed m}  } : \intermed {  [  {\intermed \sigma}  /  {\intermed a }  ]  {\intermed \sigma}'  } 
  \end{equation*}
  From the premises of rule \textsc{iTm-method} we have that
  \begin{gather}
     \intermed {  {\intermed \Sigma}_{{\mathrm{1}}}  } ; \intermed {  {\intermed { \Gamma_C } }  } ; \intermed {  {\intermed \Gamma}  }  \vdash_{d}  \intermed {  {\intermed d}  } : \intermed {  {\intermed {\mathit{TC} } } \, {\intermed \sigma}  } \itot{\,  \rightsquigarrow  \target {  {\target e }  } } 
    \label{eq:erefl_method2} \\
     \intermed {  {\intermed \Sigma}_{{\mathrm{2}}}  } ; \intermed {  {\intermed { \Gamma_C } }  } ; \intermed {  {\intermed \Gamma}  }  \vdash_{d}  \intermed {  {\intermed d}  } : \intermed {  {\intermed {\mathit{TC} } } \, {\intermed \sigma}  } \itot{\,  \rightsquigarrow  \target {  {\target e }  } } 
    \label{eq:erefl_method2b} \\
     ( \intermed {  {\intermed m}  } : \intermed {  {\intermed {\mathit{TC} } } \, {\intermed a }  } : \intermed {  {\intermed \sigma}'  } )  \in  \intermed {  {\intermed { \Gamma_C } }  } 
    \label{eq:erefl_method3}
  \end{gather}
  Applying the Dictionary Reflexivity (Theorem~\ref{thmA:dict_reflex}) to
  Equations~\ref{eq:erefl_method2}~and~\ref{eq:erefl_method2b}, in combination
  with the theorem's third hypothesis, results in:
  \begin{equation}
     \intermed {  {\intermed { \Gamma_C } }  } ; \intermed {  {\intermed \Gamma}  }  \vdash  \intermed {  {\intermed \Sigma}_{{\mathrm{1}}}  } : \intermed {  {\intermed d}  }  \simeq_{log}  \intermed {  {\intermed \Sigma}_{{\mathrm{2}}}  } : \intermed {  {\intermed d}  } : \intermed {  {\intermed {\mathit{TC} } } \, {\intermed \sigma}  } 
    \label{eq:erefl_method4}
  \end{equation}
  The goal follows directly from compatibility Lemma~\ref{lemma:compat_method}
  and Equations~\ref{eq:erefl_method3} and \ref{eq:erefl_method4}, in
  combination with the third hypothesis. \\
}
\proofStep{\( \intermed {  {\intermed e}  } = \intermed {  \lambda  {\intermed x}  \ottsym{:}  {\intermed \sigma}_{{\mathrm{1}}}  \ottsym{.}  {\intermed e}'  } \) (\textsc{iTm-arrI})}
{
   \intermed {  {\intermed \Sigma}_{{\mathrm{1}}}  } ; \intermed {  {\intermed { \Gamma_C } }  } ; \intermed {  {\intermed \Gamma}  }  \vdash_{tm}  \intermed {  \lambda  {\intermed x}  \ottsym{:}  {\intermed \sigma}_{{\mathrm{1}}}  \ottsym{.}  {\intermed e}'  } : \intermed {  {\intermed \sigma}_{{\mathrm{1}}}  \rightarrow  {\intermed \sigma}_{{\mathrm{2}}}  } \itot{ \rightsquigarrow  \target {  \lambda  {\target x}  \ottsym{:}  {\target \sigma }_{{\mathrm{1}}}  \ottsym{.}  {\target e }  } }  \band
   \intermed {  {\intermed \Sigma}_{{\mathrm{2}}}  } ; \intermed {  {\intermed { \Gamma_C } }  } ; \intermed {  {\intermed \Gamma}  }  \vdash_{tm}  \intermed {  \lambda  {\intermed x}  \ottsym{:}  {\intermed \sigma}_{{\mathrm{1}}}  \ottsym{.}  {\intermed e}'  } : \intermed {  {\intermed \sigma}_{{\mathrm{1}}}  \rightarrow  {\intermed \sigma}_{{\mathrm{2}}}  } \itot{ \rightsquigarrow  \target {  \lambda  {\target x}  \ottsym{:}  {\target \sigma }_{{\mathrm{1}}}  \ottsym{.}  {\target e }  } } 
}
{
  The goal to be proven is the following:
  \begin{equation*}
     \intermed {  {\intermed { \Gamma_C } }  } ; \intermed {  {\intermed \Gamma}  }  \vdash  \intermed {  {\intermed \Sigma}_{{\mathrm{1}}}  } : \intermed {  \lambda  {\intermed x}  \ottsym{:}  {\intermed \sigma}_{{\mathrm{1}}}  \ottsym{.}  {\intermed e}'  }  \simeq_{log}  \intermed {  {\intermed \Sigma}_{{\mathrm{2}}}  } : \intermed {  \lambda  {\intermed x}  \ottsym{:}  {\intermed \sigma}_{{\mathrm{1}}}  \ottsym{.}  {\intermed e}'  } : \intermed {  {\intermed \sigma}_{{\mathrm{1}}}  \rightarrow  {\intermed \sigma}_{{\mathrm{2}}}  } 
  \end{equation*}
  By applying the induction hypothesis to the premises of the two \textsc{iTm-arrI} rules,
  we get:
  \begin{equation*}
     \intermed {  {\intermed { \Gamma_C } }  } ; \intermed {  {\intermed \Gamma}  \ottsym{,}  {\intermed x}  \ottsym{:}  {\intermed \sigma}_{{\mathrm{1}}}  }  \vdash  \intermed {  {\intermed \Sigma}_{{\mathrm{1}}}  } : \intermed {  {\intermed e}'  }  \simeq_{log}  \intermed {  {\intermed \Sigma}_{{\mathrm{2}}}  } : \intermed {  {\intermed e}'  } : \intermed {  {\intermed \sigma}_{{\mathrm{2}}}  } 
  \end{equation*}
  The goal follows by applying the above 
  to compatibility Lemma~\ref{lemma:compat_abs}. \\
}
\proofStep{\( \intermed {  {\intermed e}  } = \intermed {  {\intermed e}_{{\mathrm{1}}} \, {\intermed e}_{{\mathrm{2}}}  } \) (\textsc{iTm-arrE})}
{
   \intermed {  {\intermed \Sigma}_{{\mathrm{1}}}  } ; \intermed {  {\intermed { \Gamma_C } }  } ; \intermed {  {\intermed \Gamma}  }  \vdash_{tm}  \intermed {  {\intermed e}_{{\mathrm{1}}} \, {\intermed e}_{{\mathrm{2}}}  } : \intermed {  {\intermed \sigma}_{{\mathrm{2}}}  } \itot{ \rightsquigarrow  \target {  {\target e }_{{\mathrm{1}}} \, {\target e }_{{\mathrm{2}}}  } }  \band
   \intermed {  {\intermed \Sigma}_{{\mathrm{2}}}  } ; \intermed {  {\intermed { \Gamma_C } }  } ; \intermed {  {\intermed \Gamma}  }  \vdash_{tm}  \intermed {  {\intermed e}_{{\mathrm{1}}} \, {\intermed e}_{{\mathrm{2}}}  } : \intermed {  {\intermed \sigma}_{{\mathrm{2}}}  } \itot{ \rightsquigarrow  \target {  {\target e }_{{\mathrm{1}}} \, {\target e }_{{\mathrm{2}}}  } } 
}
{
  The goal to be proven is the following:
  \begin{equation*}
     \intermed {  {\intermed { \Gamma_C } }  } ; \intermed {  {\intermed \Gamma}  }  \vdash  \intermed {  {\intermed \Sigma}_{{\mathrm{1}}}  } : \intermed {  {\intermed e}_{{\mathrm{1}}} \, {\intermed e}_{{\mathrm{2}}}  }  \simeq_{log}  \intermed {  {\intermed \Sigma}_{{\mathrm{2}}}  } : \intermed {  {\intermed e}_{{\mathrm{1}}} \, {\intermed e}_{{\mathrm{2}}}  } : \intermed {  {\intermed \sigma}_{{\mathrm{2}}}  } 
  \end{equation*}
  By applying the induction hypothesis to the premises of the two \textsc{iTm-arrE} rules,
  we get:
  \begin{gather*}
     \intermed {  {\intermed { \Gamma_C } }  } ; \intermed {  {\intermed \Gamma}  }  \vdash  \intermed {  {\intermed \Sigma}_{{\mathrm{1}}}  } : \intermed {  {\intermed e}_{{\mathrm{1}}}  }  \simeq_{log}  \intermed {  {\intermed \Sigma}_{{\mathrm{2}}}  } : \intermed {  {\intermed e}_{{\mathrm{1}}}  } : \intermed {  {\intermed \sigma}_{{\mathrm{1}}}  \rightarrow  {\intermed \sigma}_{{\mathrm{2}}}  } \\
     \intermed {  {\intermed { \Gamma_C } }  } ; \intermed {  {\intermed \Gamma}  }  \vdash  \intermed {  {\intermed \Sigma}_{{\mathrm{1}}}  } : \intermed {  {\intermed e}_{{\mathrm{2}}}  }  \simeq_{log}  \intermed {  {\intermed \Sigma}_{{\mathrm{2}}}  } : \intermed {  {\intermed e}_{{\mathrm{2}}}  } : \intermed {  {\intermed \sigma}_{{\mathrm{1}}}  } 
  \end{gather*}
  The goal follows by passing the above two equations to compatibility
  Lemma~\ref{lemma:compat_app}. \\
}
\proofStep{\( \intermed {  {\intermed e}  } = \intermed {  \lambda  {\intermed \delta }  \ottsym{:}  {\intermed Q}  \ottsym{.}  {\intermed e}'  } \) (\textsc{iTm-constrI})}
{
   \intermed {  {\intermed \Sigma}_{{\mathrm{1}}}  } ; \intermed {  {\intermed { \Gamma_C } }  } ; \intermed {  {\intermed \Gamma}  }  \vdash_{tm}  \intermed {  \lambda  {\intermed \delta }  \ottsym{:}  {\intermed Q}  \ottsym{.}  {\intermed e}'  } : \intermed {   {\intermed Q}  \Rightarrow  {\intermed \sigma}   } \itot{ \rightsquigarrow  \target {  \lambda  {\target \delta }  \ottsym{:}  {\target \sigma }  \ottsym{.}  {\target e }  } }  \band
   \intermed {  {\intermed \Sigma}_{{\mathrm{2}}}  } ; \intermed {  {\intermed { \Gamma_C } }  } ; \intermed {  {\intermed \Gamma}  }  \vdash_{tm}  \intermed {  \lambda  {\intermed \delta }  \ottsym{:}  {\intermed Q}  \ottsym{.}  {\intermed e}'  } : \intermed {   {\intermed Q}  \Rightarrow  {\intermed \sigma}   } \itot{ \rightsquigarrow  \target {  \lambda  {\target \delta }  \ottsym{:}  {\target \sigma }  \ottsym{.}  {\target e }  } } 
}
{
  The goal to be proven is the following:
  \begin{equation*}
     \intermed {  {\intermed { \Gamma_C } }  } ; \intermed {  {\intermed \Gamma}  }  \vdash  \intermed {  {\intermed \Sigma}_{{\mathrm{1}}}  } : \intermed {  \lambda  {\intermed \delta }  \ottsym{:}  {\intermed Q}  \ottsym{.}  {\intermed e}'  }  \simeq_{log}  \intermed {  {\intermed \Sigma}_{{\mathrm{2}}}  } : \intermed {  \lambda  {\intermed \delta }  \ottsym{:}  {\intermed Q}  \ottsym{.}  {\intermed e}'  } : \intermed {   {\intermed Q}  \Rightarrow  {\intermed \sigma}   } 
  \end{equation*}
  By applying the induction hypothesis to the premises of the two \textsc{iTm-constrI}
  rules, we get:
  \begin{equation*}
     \intermed {  {\intermed { \Gamma_C } }  } ; \intermed {  {\intermed \Gamma}  \ottsym{,}  {\intermed \delta }  \ottsym{:}  {\intermed Q}  }  \vdash  \intermed {  {\intermed \Sigma}_{{\mathrm{1}}}  } : \intermed {  {\intermed e}'  }  \simeq_{log}  \intermed {  {\intermed \Sigma}_{{\mathrm{2}}}  } : \intermed {  {\intermed e}'  } : \intermed {  {\intermed \sigma}  } 
  \end{equation*}
  The goal follows directly by passing the above equation 
  to compatibility Lemma~\ref{lemma:compat_dabs}. \\
}
\proofStep{\( \intermed {  {\intermed e}  } = \intermed {  {\intermed e}' \, {\intermed d}  } \) (\textsc{iTm-constrE})}
{
   \intermed {  {\intermed \Sigma}_{{\mathrm{1}}}  } ; \intermed {  {\intermed { \Gamma_C } }  } ; \intermed {  {\intermed \Gamma}  }  \vdash_{tm}  \intermed {  {\intermed e}' \, {\intermed d}  } : \intermed {  {\intermed \sigma}  } \itot{ \rightsquigarrow  \target {  {\target e }_{{\mathrm{1}}} \, {\target e }_{{\mathrm{2}}}  } }  \band
   \intermed {  {\intermed \Sigma}_{{\mathrm{2}}}  } ; \intermed {  {\intermed { \Gamma_C } }  } ; \intermed {  {\intermed \Gamma}  }  \vdash_{tm}  \intermed {  {\intermed e}' \, {\intermed d}  } : \intermed {  {\intermed \sigma}  } \itot{ \rightsquigarrow  \target {  {\target e }_{{\mathrm{1}}} \, {\target e }_{{\mathrm{2}}}  } } 
}
{
  The goal to be proven is the following:
  \begin{equation*}
     \intermed {  {\intermed { \Gamma_C } }  } ; \intermed {  {\intermed \Gamma}  }  \vdash  \intermed {  {\intermed \Sigma}_{{\mathrm{1}}}  } : \intermed {  {\intermed e}' \, {\intermed d}  }  \simeq_{log}  \intermed {  {\intermed \Sigma}_{{\mathrm{2}}}  } : \intermed {  {\intermed e}' \, {\intermed d}  } : \intermed {  {\intermed \sigma}  } 
  \end{equation*}
  By applying the induction hypothesis to the premises of the two \textsc{iTm-constrE} rules,
  we get:
  \begin{equation}
     \intermed {  {\intermed { \Gamma_C } }  } ; \intermed {  {\intermed \Gamma}  }  \vdash  \intermed {  {\intermed \Sigma}_{{\mathrm{1}}}  } : \intermed {  {\intermed e}'  }  \simeq_{log}  \intermed {  {\intermed \Sigma}_{{\mathrm{2}}}  } : \intermed {  {\intermed e}'  } : \intermed {   {\intermed Q}  \Rightarrow  {\intermed \sigma}   } 
    \label{eq:erefl_constrE2}
  \end{equation}
  Furthermore, applying Dictionary Reflexivity (Theorem~\ref{thmA:dict_reflex}) in the
  premises of the two \textsc{iTm-constrE} rules, in combination with the
  theorem's third
  hypothesis, results in:
  \begin{equation}
     \intermed {  {\intermed { \Gamma_C } }  } ; \intermed {  {\intermed \Gamma}  }  \vdash  \intermed {  {\intermed \Sigma}_{{\mathrm{1}}}  } : \intermed {  {\intermed d}  }  \simeq_{log}  \intermed {  {\intermed \Sigma}_{{\mathrm{2}}}  } : \intermed {  {\intermed d}  } : \intermed {  {\intermed Q}  } 
    \label{eq:erefl_constrE3}
  \end{equation}
  The goal follows from compatibility Lemma~\ref{lemma:compat_dapp}
  and Equations~\ref{eq:erefl_constrE2} and \ref{eq:erefl_constrE3}. \\
}
\proofStep{\( \intermed {  {\intermed e}  } = \intermed {  \Lambda  {\intermed a }  \ottsym{.}  {\intermed e}'  } \) (\textsc{iTm-forallI})}
{
   \intermed {  {\intermed \Sigma}_{{\mathrm{1}}}  } ; \intermed {  {\intermed { \Gamma_C } }  } ; \intermed {  {\intermed \Gamma}  }  \vdash_{tm}  \intermed {  \Lambda  {\intermed a }  \ottsym{.}  {\intermed e}'  } : \intermed {  \forall  {\intermed a }  \ottsym{.}  {\intermed \sigma}  } \itot{ \rightsquigarrow  \target {  \Lambda  {\target a }  \ottsym{.}  {\target e }  } }  \band
   \intermed {  {\intermed \Sigma}_{{\mathrm{2}}}  } ; \intermed {  {\intermed { \Gamma_C } }  } ; \intermed {  {\intermed \Gamma}  }  \vdash_{tm}  \intermed {  \Lambda  {\intermed a }  \ottsym{.}  {\intermed e}'  } : \intermed {  \forall  {\intermed a }  \ottsym{.}  {\intermed \sigma}  } \itot{ \rightsquigarrow  \target {  \Lambda  {\target a }  \ottsym{.}  {\target e }  } } 
}
{
  The goal to be proven is the following:
  \begin{equation*}
     \intermed {  {\intermed { \Gamma_C } }  } ; \intermed {  {\intermed \Gamma}  }  \vdash  \intermed {  {\intermed \Sigma}_{{\mathrm{1}}}  } : \intermed {  \Lambda  {\intermed a }  \ottsym{.}  {\intermed e}'  }  \simeq_{log}  \intermed {  {\intermed \Sigma}_{{\mathrm{2}}}  } : \intermed {  \Lambda  {\intermed a }  \ottsym{.}  {\intermed e}'  } : \intermed {  \forall  {\intermed a }  \ottsym{.}  {\intermed \sigma}  } 
  \end{equation*}
  By applying the induction hypothesis to the premises of the two \textsc{iTm-forallI}
  rules, we get:
  \begin{equation*}
     \intermed {  {\intermed { \Gamma_C } }  } ; \intermed {  {\intermed \Gamma}  \ottsym{,}  {\intermed a }  }  \vdash  \intermed {  {\intermed \Sigma}_{{\mathrm{1}}}  } : \intermed {  {\intermed e}'  }  \simeq_{log}  \intermed {  {\intermed \Sigma}_{{\mathrm{2}}}  } : \intermed {  {\intermed e}'  } : \intermed {  {\intermed \sigma}_{{\mathrm{1}}}  } 
  \end{equation*}
  The goal follows directly by applying the above equation 
  to compatibility Lemma~\ref{lemma:compat_tyabs}. \\
}
\proofStep{\( \intermed {  {\intermed e}  } = \intermed {  {\intermed e}' \, {\intermed \sigma}  } \) (\textsc{iTm-forallE})}
{
   \intermed {  {\intermed \Sigma}_{{\mathrm{1}}}  } ; \intermed {  {\intermed { \Gamma_C } }  } ; \intermed {  {\intermed \Gamma}  }  \vdash_{tm}  \intermed {  {\intermed e}' \, {\intermed \sigma}  } : \intermed {  [  {\intermed \sigma}  /  {\intermed a }  ]  {\intermed \sigma}'  } \itot{ \rightsquigarrow  \target {  {\target e } \, {\target \sigma }  } }  \band
   \intermed {  {\intermed \Sigma}_{{\mathrm{2}}}  } ; \intermed {  {\intermed { \Gamma_C } }  } ; \intermed {  {\intermed \Gamma}  }  \vdash_{tm}  \intermed {  {\intermed e}' \, {\intermed \sigma}  } : \intermed {  [  {\intermed \sigma}  /  {\intermed a }  ]  {\intermed \sigma}'  } \itot{ \rightsquigarrow  \target {  {\target e } \, {\target \sigma }  } } 
}
{
  The goal to be proven is the following:
  \begin{equation*}
     \intermed {  {\intermed { \Gamma_C } }  } ; \intermed {  {\intermed \Gamma}  }  \vdash  \intermed {  {\intermed \Sigma}_{{\mathrm{1}}}  } : \intermed {  {\intermed e}' \, {\intermed \sigma}  }  \simeq_{log}  \intermed {  {\intermed \Sigma}_{{\mathrm{2}}}  } : \intermed {  {\intermed e}' \, {\intermed \sigma}  } : \intermed {  [  {\intermed \sigma}  /  {\intermed a }  ]  {\intermed \sigma}'  } 
  \end{equation*}
  From the premises of \textsc{iTm-forallE}, we obtain
  \begin{equation}
     \intermed {  {\intermed { \Gamma_C } }  } ; \intermed {  {\intermed \Gamma}  }  \vdash_{ty}  \intermed {  {\intermed \sigma}  } \itot{ \rightsquigarrow  \target {  {\target \sigma }  } } 
    \label{eq:erefl_forallE_wt}
  \end{equation}
  By applying the induction hypothesis to the premises of the two \textsc{iTm-forallE}
  rules, we get:
  \begin{equation}
     \intermed {  {\intermed { \Gamma_C } }  } ; \intermed {  {\intermed \Gamma}  }  \vdash  \intermed {  {\intermed \Sigma}_{{\mathrm{1}}}  } : \intermed {  {\intermed e}'_{{\mathrm{1}}}  }  \simeq_{log}  \intermed {  {\intermed \Sigma}_{{\mathrm{2}}}  } : \intermed {  {\intermed e}'_{{\mathrm{1}}}  } : \intermed {  {\intermed \sigma}_{{\mathrm{1}}}  } 
    \label{eq:erefl_forallE2}
  \end{equation}
  The goal follows from compatibility Lemma~\ref{lemma:compat_tyapp}
  and Equations~\ref{eq:erefl_forallE2} and~\ref{eq:erefl_forallE_wt}.
}
\qed

\begin{theoremB}[Context Reflexivity]\ \\
  Suppose\, \( \vdash_{ctx}^{\intermed {M} }  \source {  { \source P }  } ; \source {  {\source { \Gamma_C } }  } ; \source {  {\source \Gamma}  }  \rightsquigarrow  \intermed {  {\intermed \Sigma}_{{\mathrm{1}}}  } ; \intermed {  {\intermed { \Gamma_C } }  } ; \intermed {  {\intermed \Gamma}  } \)
    and\, \( \vdash_{ctx}^{\intermed {M} }  \source {  { \source P }  } ; \source {  {\source { \Gamma_C } }  } ; \source {  {\source \Gamma}'  }  \rightsquigarrow  \intermed {  {\intermed \Sigma}_{{\mathrm{1}}}  } ; \intermed {  {\intermed { \Gamma_C } }  } ; \intermed {  {\intermed \Gamma}'  } \) \\
    and\, \( \vdash_{ctx}^{\intermed {M} }  \source {  { \source P }  } ; \source {  {\source { \Gamma_C } }  } ; \source {  {\source \Gamma}  }  \rightsquigarrow  \intermed {  {\intermed \Sigma}_{{\mathrm{2}}}  } ; \intermed {  {\intermed { \Gamma_C } }  } ; \intermed {  {\intermed \Gamma}  } \)
    and\, \( \vdash_{ctx}^{\intermed {M} }  \source {  { \source P }  } ; \source {  {\source { \Gamma_C } }  } ; \source {  {\source \Gamma}'  }  \rightsquigarrow  \intermed {  {\intermed \Sigma}_{{\mathrm{2}}}  } ; \intermed {  {\intermed { \Gamma_C } }  } ; \intermed {  {\intermed \Gamma}'  } \) \\
    and\, \( \source {  {\source { \Gamma_C } }  } ; \source {  {\source \Gamma}  }  \vdash_{ty}^{\intermed {M} }  \source {  {\source \tau }  }\,{ \stoi{ \rightsquigarrow  \intermed {  {\intermed \sigma}  } } } \)
    and\, \( \source {  {\source { \Gamma_C } }  } ; \source {  {\source \Gamma}'  }  \vdash_{ty}^{\intermed {M} }  \source {  {\source \tau }'  }\,{ \stoi{ \rightsquigarrow  \intermed {  {\intermed \sigma}'  } } } \),
  \begin{itemize}
  \item If \( \source {  {\source M}  } : ( \source {  { \source P }  } ; \source {  {\source { \Gamma_C } }  } ; \source {  {\source \Gamma}  }  \Rightarrow  \source {  {\source \tau }  } )  \mapsto  ( \source {  { \source P }  } ; \source {  {\source { \Gamma_C } }  } ; \source {  {\source \Gamma}'  }  \Rightarrow  \source {  {\source \tau }'  } )  \rightsquigarrow  \intermed {  {\intermed M}_{{\mathrm{1}}}  } \) \\
    and \( \source {  {\source M}  } : ( \source {  { \source P }  } ; \source {  {\source { \Gamma_C } }  } ; \source {  {\source \Gamma}  }  \Rightarrow  \source {  {\source \tau }  } )  \mapsto  ( \source {  { \source P }  } ; \source {  {\source { \Gamma_C } }  } ; \source {  {\source \Gamma}'  }  \Rightarrow  \source {  {\source \tau }'  } )  \rightsquigarrow  \intermed {  {\intermed M}_{{\mathrm{2}}}  } \) \\
    then \( \intermed {  {\intermed \Sigma}_{{\mathrm{1}}}  } : \intermed {  {\intermed M}_{{\mathrm{1}}}  }  \simeq_{log}  \intermed {  {\intermed \Sigma}_{{\mathrm{2}}}  } : \intermed {  {\intermed M}_{{\mathrm{2}}}  } : ( \intermed {  {\intermed { \Gamma_C } }  } ; \intermed {  {\intermed \Gamma}  }  \Rightarrow  \intermed {  {\intermed \sigma}  } )  \mapsto  ( \intermed {  {\intermed { \Gamma_C } }  } ; \intermed {  {\intermed \Gamma}'  }  \Rightarrow  \intermed {  {\intermed \sigma}'  } ) \).
  \item If \( \source {  {\source M}  } : ( \source {  { \source P }  } ; \source {  {\source { \Gamma_C } }  } ; \source {  {\source \Gamma}  }  \Rightarrow  \source {  {\source \tau }  } )  \mapsto  ( \source {  { \source P }  } ; \source {  {\source { \Gamma_C } }  } ; \source {  {\source \Gamma}'  }  \Leftarrow  \source {  {\source \tau }'  } )  \rightsquigarrow  \intermed {  {\intermed M}_{{\mathrm{1}}}  } \) \\
    and \( \source {  {\source M}  } : ( \source {  { \source P }  } ; \source {  {\source { \Gamma_C } }  } ; \source {  {\source \Gamma}  }  \Rightarrow  \source {  {\source \tau }  } )  \mapsto  ( \source {  { \source P }  } ; \source {  {\source { \Gamma_C } }  } ; \source {  {\source \Gamma}'  }  \Leftarrow  \source {  {\source \tau }'  } )  \rightsquigarrow  \intermed {  {\intermed M}_{{\mathrm{2}}}  } \) \\
    then \( \intermed {  {\intermed \Sigma}_{{\mathrm{1}}}  } : \intermed {  {\intermed M}_{{\mathrm{1}}}  }  \simeq_{log}  \intermed {  {\intermed \Sigma}_{{\mathrm{2}}}  } : \intermed {  {\intermed M}_{{\mathrm{2}}}  } : ( \intermed {  {\intermed { \Gamma_C } }  } ; \intermed {  {\intermed \Gamma}  }  \Rightarrow  \intermed {  {\intermed \sigma}  } )  \mapsto  ( \intermed {  {\intermed { \Gamma_C } }  } ; \intermed {  {\intermed \Gamma}'  }  \Rightarrow  \intermed {  {\intermed \sigma}'  } ) \).
  \item If \( \source {  {\source M}  } : ( \source {  { \source P }  } ; \source {  {\source { \Gamma_C } }  } ; \source {  {\source \Gamma}  }  \Leftarrow  \source {  {\source \tau }  } )  \mapsto  ( \source {  { \source P }  } ; \source {  {\source { \Gamma_C } }  } ; \source {  {\source \Gamma}'  }  \Rightarrow  \source {  {\source \tau }'  } )  \rightsquigarrow  \intermed {  {\intermed M}_{{\mathrm{1}}}  } \) \\
    and \( \source {  {\source M}  } : ( \source {  { \source P }  } ; \source {  {\source { \Gamma_C } }  } ; \source {  {\source \Gamma}  }  \Leftarrow  \source {  {\source \tau }  } )  \mapsto  ( \source {  { \source P }  } ; \source {  {\source { \Gamma_C } }  } ; \source {  {\source \Gamma}'  }  \Rightarrow  \source {  {\source \tau }'  } )  \rightsquigarrow  \intermed {  {\intermed M}_{{\mathrm{2}}}  } \) \\
    then \( \intermed {  {\intermed \Sigma}_{{\mathrm{1}}}  } : \intermed {  {\intermed M}_{{\mathrm{1}}}  }  \simeq_{log}  \intermed {  {\intermed \Sigma}_{{\mathrm{2}}}  } : \intermed {  {\intermed M}_{{\mathrm{2}}}  } : ( \intermed {  {\intermed { \Gamma_C } }  } ; \intermed {  {\intermed \Gamma}  }  \Rightarrow  \intermed {  {\intermed \sigma}  } )  \mapsto  ( \intermed {  {\intermed { \Gamma_C } }  } ; \intermed {  {\intermed \Gamma}'  }  \Rightarrow  \intermed {  {\intermed \sigma}'  } ) \).
  \item If \( \source {  {\source M}  } : ( \source {  { \source P }  } ; \source {  {\source { \Gamma_C } }  } ; \source {  {\source \Gamma}  }  \Leftarrow  \source {  {\source \tau }  } )  \mapsto  ( \source {  { \source P }  } ; \source {  {\source { \Gamma_C } }  } ; \source {  {\source \Gamma}'  }  \Leftarrow  \source {  {\source \tau }'  } )  \rightsquigarrow  \intermed {  {\intermed M}_{{\mathrm{1}}}  } \) \\
    and \( \source {  {\source M}  } : ( \source {  { \source P }  } ; \source {  {\source { \Gamma_C } }  } ; \source {  {\source \Gamma}  }  \Leftarrow  \source {  {\source \tau }  } )  \mapsto  ( \source {  { \source P }  } ; \source {  {\source { \Gamma_C } }  } ; \source {  {\source \Gamma}'  }  \Leftarrow  \source {  {\source \tau }'  } )  \rightsquigarrow  \intermed {  {\intermed M}_{{\mathrm{2}}}  } \) \\
    then \( \intermed {  {\intermed \Sigma}_{{\mathrm{1}}}  } : \intermed {  {\intermed M}_{{\mathrm{1}}}  }  \simeq_{log}  \intermed {  {\intermed \Sigma}_{{\mathrm{2}}}  } : \intermed {  {\intermed M}_{{\mathrm{2}}}  } : ( \intermed {  {\intermed { \Gamma_C } }  } ; \intermed {  {\intermed \Gamma}  }  \Rightarrow  \intermed {  {\intermed \sigma}  } )  \mapsto  ( \intermed {  {\intermed { \Gamma_C } }  } ; \intermed {  {\intermed \Gamma}'  }  \Rightarrow  \intermed {  {\intermed \sigma}'  } ) \).
  \end{itemize}
  \label{thmA:ctx_reflex}
\end{theoremB}

\proof
The theorem is stated in a nested fashion, where all common hypotheses are introduced in the
outer statement. Each of the four inner statements extends the outer statement with two
context-typing hypotheses, and sets the conclusion of the theorem, which is identical for each of
the four cases.

Suppose expressions \(\intermed{{\intermed e}_{{\mathrm{1}}}}\) and \(\intermed{{\intermed e}_{{\mathrm{2}}}}\) such that
\begin{equation}
  \label{eq:ctx_refl0}
   \intermed {  {\intermed { \Gamma_C } }  } ; \intermed {  {\intermed \Gamma}  }  \vdash  \intermed {  {\intermed \Sigma}_{{\mathrm{1}}}  } : \intermed {  {\intermed e}_{{\mathrm{1}}}  }  \simeq_{log}  \intermed {  {\intermed \Sigma}_{{\mathrm{2}}}  } : \intermed {  {\intermed e}_{{\mathrm{2}}}  } : \intermed {  {\intermed \sigma}  } 
\end{equation}
Then, by unfolding the defintion of logical equivalence in the goal of all four
sub-statements, it suffices to show
that
\begin{equation}
  \label{eqg:ctx_refl0}
   \intermed {  {\intermed { \Gamma_C } }  } ; \intermed {  {\intermed \Gamma}'  }  \vdash  \intermed {  {\intermed \Sigma}_{{\mathrm{1}}}  } : \intermed {   {\intermed M}_{{\mathrm{1}}}  [  {\intermed e}_{{\mathrm{1}}}  ]   }  \simeq_{log}  \intermed {  {\intermed \Sigma}_{{\mathrm{2}}}  } : \intermed {   {\intermed M}_{{\mathrm{2}}}  [  {\intermed e}_{{\mathrm{2}}}  ]   } : \intermed {  {\intermed \sigma}'  } 
\end{equation}
We assume all hypotheses of the outer statement and we proceed by mutual induction on the
first hypothesis of the nested statements. Note that
context typing derivations are of finite size, thus mutual induction over them is safe.

\begin{description}
\item[Part 1] ~ \\
  \proofStep{\textsc{sM-inf-inf-empty}}
  {
     \source {  \ottsym{[} \, \bullet \, \ottsym{]}  } : ( \source {  { \source P }  } ; \source {  {\source { \Gamma_C } }  } ; \source {  {\source \Gamma}  }  \Rightarrow  \source {  {\source \tau }  } )  \mapsto  ( \source {  { \source P }  } ; \source {  {\source { \Gamma_C } }  } ; \source {  {\source \Gamma}  }  \Rightarrow  \source {  {\source \tau }  } )  \rightsquigarrow  \intermed {  \ottsym{[} \, \bullet \, \ottsym{]}  } 
  }
  {
    By case analysis on the second hypothesis of the nested statement, its last context
    typing rule
    must be \textsc{sM-inf-inf-empty} as well. Therefore, the first and second hypotheses
    of the nested statement become
    \begin{gather*}
       \source {  \ottsym{[} \, \bullet \, \ottsym{]}  } : ( \source {  { \source P }  } ; \source {  {\source { \Gamma_C } }  } ; \source {  {\source \Gamma}  }  \Rightarrow  \source {  {\source \tau }  } )  \mapsto  ( \source {  { \source P }  } ; \source {  {\source { \Gamma_C } }  } ; \source {  {\source \Gamma}  }  \Rightarrow  \source {  {\source \tau }  } )  \rightsquigarrow  \intermed {  \ottsym{[} \, \bullet \, \ottsym{]}  } \\
       \source {  \ottsym{[} \, \bullet \, \ottsym{]}  } : ( \source {  { \source P }  } ; \source {  {\source { \Gamma_C } }  } ; \source {  {\source \Gamma}  }  \Rightarrow  \source {  {\source \tau }  } )  \mapsto  ( \source {  { \source P }  } ; \source {  {\source { \Gamma_C } }  } ; \source {  {\source \Gamma}  }  \Rightarrow  \source {  {\source \tau }  } )  \rightsquigarrow  \intermed {  \ottsym{[} \, \bullet \, \ottsym{]}  } 
    \end{gather*}
    and we have \( \intermed {  {\intermed M}_{{\mathrm{1}}}  } = \intermed {  {\intermed M}_{{\mathrm{2}}}  }  = \intermed{\ottsym{[} \, \bullet \, \ottsym{]}}\). Thus,
    Goal~\ref{eqg:ctx_refl0} becomes
    \begin{equation*}
       \intermed {  {\intermed { \Gamma_C } }  } ; \intermed {  {\intermed \Gamma}'  }  \vdash  \intermed {  {\intermed \Sigma}_{{\mathrm{1}}}  } : \intermed {  {\intermed e}_{{\mathrm{1}}}  }  \simeq_{log}  \intermed {  {\intermed \Sigma}_{{\mathrm{2}}}  } : \intermed {  {\intermed e}_{{\mathrm{2}}}  } : \intermed {  {\intermed \sigma}  } 
      \label{eqg:ctx_refl1}
    \end{equation*}
    The above logical equivalence follows directly from Equation~\ref{eq:ctx_refl0}. \\
  }
  \proofStep{\textsc{sM-inf-inf-appL}}
  {
     \source {  {\source M}' \, {\source e}'_{{\mathrm{2}}}  } : ( \source {  { \source P }  } ; \source {  {\source { \Gamma_C } }  } ; \source {  {\source \Gamma}  }  \Rightarrow  \source {  {\source \tau }  } )  \mapsto  ( \source {  { \source P }  } ; \source {  {\source { \Gamma_C } }  } ; \source {  {\source \Gamma}'  }  \Rightarrow  \source {  {\source \tau }_{{\mathrm{2}}}  } )  \rightsquigarrow  \intermed {  {\intermed M}'_{{\mathrm{1}}} \, {\intermed e}_{{\mathrm{21}}}  } 
  }
  {
    By case analysis on the second hypothesis of the nested statement, its last context
    typing rule
    must be \textsc{sM-inf-inf-appL} as well. Therefore, the first and second hypotheses
    of the nested statement become
    \begin{gather*}
       \source {  {\source M}' \, {\source e}'_{{\mathrm{2}}}  } : ( \source {  { \source P }  } ; \source {  {\source { \Gamma_C } }  } ; \source {  {\source \Gamma}  }  \Rightarrow  \source {  {\source \tau }  } )  \mapsto  ( \source {  { \source P }  } ; \source {  {\source { \Gamma_C } }  } ; \source {  {\source \Gamma}'  }  \Rightarrow  \source {  {\source \tau }_{{\mathrm{2}}}  } )  \rightsquigarrow  \intermed {  {\intermed M}'_{{\mathrm{1}}} \, {\intermed e}_{{\mathrm{21}}}  } \\
       \source {  {\source M}' \, {\source e}'_{{\mathrm{2}}}  } : ( \source {  { \source P }  } ; \source {  {\source { \Gamma_C } }  } ; \source {  {\source \Gamma}  }  \Rightarrow  \source {  {\source \tau }  } )  \mapsto  ( \source {  { \source P }  } ; \source {  {\source { \Gamma_C } }  } ; \source {  {\source \Gamma}'  }  \Rightarrow  \source {  {\source \tau }_{{\mathrm{2}}}  } )  \rightsquigarrow  \intermed {  {\intermed M}'_{{\mathrm{2}}} \, {\intermed e}_{{\mathrm{22}}}  } 
    \end{gather*}
    and Goal~\ref{eqg:ctx_refl0} becomes
    \begin{equation}
       \intermed {  {\intermed { \Gamma_C } }  } ; \intermed {  {\intermed \Gamma}'  }  \vdash  \intermed {  {\intermed \Sigma}_{{\mathrm{1}}}  } : \intermed {   {\intermed M}'_{{\mathrm{1}}}  [  {\intermed e}_{{\mathrm{1}}}  ]  \, {\intermed e}_{{\mathrm{21}}}  }  \simeq_{log}  \intermed {  {\intermed \Sigma}_{{\mathrm{2}}}  } : \intermed {   {\intermed M}'_{{\mathrm{2}}}  [  {\intermed e}_{{\mathrm{2}}}  ]  \, {\intermed e}_{{\mathrm{22}}}  } : \intermed {  {\intermed \sigma}_{{\mathrm{2}}}  } 
      \label{eqg:ctx_refl2}
    \end{equation}
    From the premises of the two \textsc{sM-inf-inf-appL} rules, we obtain:
    \begin{gather}
       \source {  {\source M}'  } : ( \source {  { \source P }  } ; \source {  {\source { \Gamma_C } }  } ; \source {  {\source \Gamma}  }  \Rightarrow  \source {  {\source \tau }  } )  \mapsto  ( \source {  { \source P }  } ; \source {  {\source { \Gamma_C } }  } ; \source {  {\source \Gamma}'  }  \Rightarrow  \source {  {\source \tau }_{{\mathrm{1}}}  \rightarrow  {\source \tau }_{{\mathrm{2}}}  } )  \rightsquigarrow  \intermed {  {\intermed M}'_{{\mathrm{1}}}  } 
      \label{eq:ctx_refl5} \\
       \source {  {\source M}'  } : ( \source {  { \source P }  } ; \source {  {\source { \Gamma_C } }  } ; \source {  {\source \Gamma}  }  \Rightarrow  \source {  {\source \tau }  } )  \mapsto  ( \source {  { \source P }  } ; \source {  {\source { \Gamma_C } }  } ; \source {  {\source \Gamma}'  }  \Rightarrow  \source {  {\source \tau }'_{{\mathrm{1}}}  \rightarrow  {\source \tau }_{{\mathrm{2}}}  } )  \rightsquigarrow  \intermed {  {\intermed M}'_{{\mathrm{2}}}  } 
      \label{eq:ctx_refl6} \\
       \source {  { \source P }  } ; \source {  {\source { \Gamma_C } }  } ; \source {  {\source \Gamma}'  }  \vdash_{tm}^{\intermed {M} }  \source {  {\source e}'_{{\mathrm{2}}}  } \Leftarrow \source {  {\source \tau }_{{\mathrm{1}}}  } \stoi{ \rightsquigarrow  \intermed {  {\intermed e}_{{\mathrm{21}}}  } } 
      \label{eq:ctx_refl7} \\
       \source {  { \source P }  } ; \source {  {\source { \Gamma_C } }  } ; \source {  {\source \Gamma}'  }  \vdash_{tm}^{\intermed {M} }  \source {  {\source e}'_{{\mathrm{2}}}  } \Leftarrow \source {  {\source \tau }'_{{\mathrm{1}}}  } \stoi{ \rightsquigarrow  \intermed {  {\intermed e}_{{\mathrm{22}}}  } } 
      \label{eq:ctx_refl8}
    \end{gather}
    By applying Lemma~\ref{lemma:ctx_type_unique} to
    Equations~\ref{eq:ctx_refl5} and~\ref{eq:ctx_refl6}, we know that
    \( \source {  {\source \tau }'_{{\mathrm{1}}}  \rightarrow  {\source \tau }_{{\mathrm{2}}}  } = \source {  {\source \tau }_{{\mathrm{1}}}  \rightarrow  {\source \tau }_{{\mathrm{2}}}  } \).

    Applying the induction hypothesis on Equations~\ref{eq:ctx_refl5}
    and~\ref{eq:ctx_refl6}, yields:
    \begin{equation*}
       \intermed {  {\intermed \Sigma}_{{\mathrm{1}}}  } : \intermed {  {\intermed M}'_{{\mathrm{1}}}  }  \simeq_{log}  \intermed {  {\intermed \Sigma}_{{\mathrm{2}}}  } : \intermed {  {\intermed M}'_{{\mathrm{2}}}  } : ( \intermed {  {\intermed { \Gamma_C } }  } ; \intermed {  {\intermed \Gamma}  }  \Rightarrow  \intermed {  {\intermed \sigma}  } )  \mapsto  ( \intermed {  {\intermed { \Gamma_C } }  } ; \intermed {  {\intermed \Gamma}'  }  \Rightarrow  \intermed {  {\intermed \sigma}_{{\mathrm{1}}}  \rightarrow  {\intermed \sigma}_{{\mathrm{2}}}  } ) 
    \end{equation*}
    By unfolding the definition of logical equivalence in the above equation
    and applying it on Equation~\ref{eq:ctx_refl0}, we get:
    \begin{equation}
       \intermed {  {\intermed { \Gamma_C } }  } ; \intermed {  {\intermed \Gamma}'  }  \vdash  \intermed {  {\intermed \Sigma}_{{\mathrm{1}}}  } : \intermed {   {\intermed M}'_{{\mathrm{1}}}  [  {\intermed e}_{{\mathrm{1}}}  ]   }  \simeq_{log}  \intermed {  {\intermed \Sigma}_{{\mathrm{2}}}  } : \intermed {   {\intermed M}'_{{\mathrm{2}}}  [  {\intermed e}_{{\mathrm{2}}}  ]   } : \intermed {  {\intermed \sigma}_{{\mathrm{1}}}  \rightarrow  {\intermed \sigma}_{{\mathrm{2}}}  } 
      \label{eq:ctx_refl12}
    \end{equation}
    By applying Expression Coherence Theorem A (Theorem~\ref{thmA:coh_exprA}) on
    Equations~\ref{eq:ctx_refl7} and~\ref{eq:ctx_refl8} (and on the second and fourth
    hypotheses of the outer statement), we get:
    \begin{equation}
       \intermed {  {\intermed { \Gamma_C } }  } ; \intermed {  {\intermed \Gamma}'  }  \vdash  \intermed {  {\intermed \Sigma}_{{\mathrm{1}}}  } : \intermed {  {\intermed e}_{{\mathrm{21}}}  }  \simeq_{log}  \intermed {  {\intermed \Sigma}_{{\mathrm{2}}}  } : \intermed {  {\intermed e}_{{\mathrm{22}}}  } : \intermed {  {\intermed \sigma}_{{\mathrm{1}}}  } 
      \label{eq:ctx_refl13}
    \end{equation}
    Goal~\ref{eqg:ctx_refl2} follows from compatibility of term applications
    (Lemma~\ref{lemma:compat_app}, together
    with Equations~\ref{eq:ctx_refl12} and~\ref{eq:ctx_refl13}). \\
  }
  \proofStep{\textsc{sM-inf-inf-appR}}
  {
     \source {  {\source e}'_{{\mathrm{1}}} \, {\source M}'  } : ( \source {  { \source P }  } ; \source {  {\source { \Gamma_C } }  } ; \source {  {\source \Gamma}  }  \Rightarrow  \source {  {\source \tau }  } )  \mapsto  ( \source {  { \source P }  } ; \source {  {\source { \Gamma_C } }  } ; \source {  {\source \Gamma}'  }  \Rightarrow  \source {  {\source \tau }_{{\mathrm{2}}}  } )  \rightsquigarrow  \intermed {  {\intermed e}_{{\mathrm{11}}} \, {\intermed M}'_{{\mathrm{1}}}  } 
  }
  {
    By case analysis on the second hypothesis of the nested statement, its last context
    typing rule
    must be \textsc{sM-inf-inf-appR} as well. Therefore, the first and second hypotheses
    of the nested statement become
    \begin{gather*}
       \source {  {\source e}'_{{\mathrm{1}}} \, {\source M}'  } : ( \source {  { \source P }  } ; \source {  {\source { \Gamma_C } }  } ; \source {  {\source \Gamma}  }  \Rightarrow  \source {  {\source \tau }  } )  \mapsto  ( \source {  { \source P }  } ; \source {  {\source { \Gamma_C } }  } ; \source {  {\source \Gamma}'  }  \Rightarrow  \source {  {\source \tau }_{{\mathrm{2}}}  } )  \rightsquigarrow  \intermed {  {\intermed e}_{{\mathrm{11}}} \, {\intermed M}'_{{\mathrm{1}}}  } \\
       \source {  {\source e}'_{{\mathrm{1}}} \, {\source M}'  } : ( \source {  { \source P }  } ; \source {  {\source { \Gamma_C } }  } ; \source {  {\source \Gamma}  }  \Rightarrow  \source {  {\source \tau }  } )  \mapsto  ( \source {  { \source P }  } ; \source {  {\source { \Gamma_C } }  } ; \source {  {\source \Gamma}'  }  \Rightarrow  \source {  {\source \tau }_{{\mathrm{2}}}  } )  \rightsquigarrow  \intermed {  {\intermed e}_{{\mathrm{12}}} \, {\intermed M}'_{{\mathrm{2}}}  } 
    \end{gather*}
    and we need to show that
    \begin{equation}
       \intermed {  {\intermed { \Gamma_C } }  } ; \intermed {  {\intermed \Gamma}'  }  \vdash  \intermed {  {\intermed \Sigma}_{{\mathrm{1}}}  } : \intermed {   {\intermed e}_{{\mathrm{11}}} \, {\intermed M}'_{{\mathrm{1}}}  [  {\intermed e}_{{\mathrm{1}}}  ]   }  \simeq_{log}  \intermed {  {\intermed \Sigma}_{{\mathrm{2}}}  } : \intermed {   {\intermed e}_{{\mathrm{12}}} \, {\intermed M}'_{{\mathrm{2}}}  [  {\intermed e}_{{\mathrm{2}}}  ]   } : \intermed {  {\intermed \sigma}_{{\mathrm{2}}}  } 
      \label{eqg:ctx_refl3}
    \end{equation}
    From the premises of the two \textsc{sM-inf-inf-appR} rules, we obtain:
    \begin{gather}
       \source {  {\source M}'  } : ( \source {  { \source P }  } ; \source {  {\source { \Gamma_C } }  } ; \source {  {\source \Gamma}  }  \Rightarrow  \source {  {\source \tau }  } )  \mapsto  ( \source {  { \source P }  } ; \source {  {\source { \Gamma_C } }  } ; \source {  {\source \Gamma}'  }  \Leftarrow  \source {  {\source \tau }_{{\mathrm{1}}}  } )  \rightsquigarrow  \intermed {  {\intermed M}'_{{\mathrm{1}}}  } 
      \label{eq:ctx_refl16} \\
       \source {  {\source M}'  } : ( \source {  { \source P }  } ; \source {  {\source { \Gamma_C } }  } ; \source {  {\source \Gamma}  }  \Rightarrow  \source {  {\source \tau }  } )  \mapsto  ( \source {  { \source P }  } ; \source {  {\source { \Gamma_C } }  } ; \source {  {\source \Gamma}'  }  \Leftarrow  \source {  {\source \tau }'_{{\mathrm{1}}}  } )  \rightsquigarrow  \intermed {  {\intermed M}'_{{\mathrm{2}}}  } 
      \label{eq:ctx_refl17} \\
       \source {  { \source P }  } ; \source {  {\source { \Gamma_C } }  } ; \source {  {\source \Gamma}'  }  \vdash_{tm}^{\intermed {M} }  \source {  {\source e}'_{{\mathrm{1}}}  } \Rightarrow \source {  {\source \tau }_{{\mathrm{1}}}  \rightarrow  {\source \tau }_{{\mathrm{2}}}  } \stoi{ \rightsquigarrow  \intermed {  {\intermed e}_{{\mathrm{11}}}  } } 
      \label{eq:ctx_refl18} \\
       \source {  { \source P }  } ; \source {  {\source { \Gamma_C } }  } ; \source {  {\source \Gamma}'  }  \vdash_{tm}^{\intermed {M} }  \source {  {\source e}'_{{\mathrm{1}}}  } \Rightarrow \source {  {\source \tau }'_{{\mathrm{1}}}  \rightarrow  {\source \tau }_{{\mathrm{2}}}  } \stoi{ \rightsquigarrow  \intermed {  {\intermed e}_{{\mathrm{12}}}  } } 
      \label{eq:ctx_refl19}
    \end{gather}
    By Lemma~\ref{lemma:ctx_type_unique}, we know that
    \( \source {  {\source \tau }'_{{\mathrm{1}}}  } = \source {  {\source \tau }_{{\mathrm{1}}}  } \).

    By applying Part 2 of this theorem to Equations~\ref{eq:ctx_refl16}
    and~\ref{eq:ctx_refl17}, we get:
    \begin{equation*}
       \intermed {  {\intermed \Sigma}_{{\mathrm{1}}}  } : \intermed {  {\intermed M}'_{{\mathrm{1}}}  }  \simeq_{log}  \intermed {  {\intermed \Sigma}_{{\mathrm{2}}}  } : \intermed {  {\intermed M}'_{{\mathrm{2}}}  } : ( \intermed {  {\intermed { \Gamma_C } }  } ; \intermed {  {\intermed \Gamma}  }  \Rightarrow  \intermed {  {\intermed \sigma}  } )  \mapsto  ( \intermed {  {\intermed { \Gamma_C } }  } ; \intermed {  {\intermed \Gamma}'  }  \Rightarrow  \intermed {  {\intermed \sigma}_{{\mathrm{1}}}  } ) 
    \end{equation*}
    By unfolding the definition of logical equivalence in the above and applying it on
    Equation~\ref{eq:ctx_refl0}, we get:
    \begin{equation}
       \intermed {  {\intermed { \Gamma_C } }  } ; \intermed {  {\intermed \Gamma}'  }  \vdash  \intermed {  {\intermed \Sigma}_{{\mathrm{1}}}  } : \intermed {   {\intermed M}'_{{\mathrm{1}}}  [  {\intermed e}_{{\mathrm{1}}}  ]   }  \simeq_{log}  \intermed {  {\intermed \Sigma}_{{\mathrm{2}}}  } : \intermed {   {\intermed M}'_{{\mathrm{2}}}  [  {\intermed e}_{{\mathrm{2}}}  ]   } : \intermed {  {\intermed \sigma}_{{\mathrm{1}}}  } 
      \label{eq:ctx_refl23}
    \end{equation}
    By applying Expression Coherence Theorem A (Theorem~\ref{thmA:coh_exprA}) to
    Equations~\ref{eq:ctx_refl18} and~\ref{eq:ctx_refl19} (and on the second and fourth
    hypotheses of the outer statement), we get:
    \begin{equation}
       \intermed {  {\intermed { \Gamma_C } }  } ; \intermed {  {\intermed \Gamma}'  }  \vdash  \intermed {  {\intermed \Sigma}_{{\mathrm{1}}}  } : \intermed {  {\intermed e}_{{\mathrm{11}}}  }  \simeq_{log}  \intermed {  {\intermed \Sigma}_{{\mathrm{2}}}  } : \intermed {  {\intermed e}_{{\mathrm{12}}}  } : \intermed {  {\intermed \sigma}_{{\mathrm{1}}}  \rightarrow  {\intermed \sigma}_{{\mathrm{2}}}  } 
      \label{eq:ctx_refl24}
    \end{equation}
    Goal~\ref{eqg:ctx_refl3} follows from compatibility of term applications
    (Lemma~\ref{lemma:compat_app}, together
    with Equations~\ref{eq:ctx_refl23} and~\ref{eq:ctx_refl24}). \\
  }
  \proofStep{\textsc{sM-inf-inf-letL}}
  {
     \source {  \textbf{\source {let} } \, {\source x}  \ottsym{:}  \forall  {\overline {\source a} }_{\ottmv{j}}  \ottsym{.}  {\overline {\source Q} }_{\ottmv{i}}  \Rightarrow  {\source \tau }_{{\mathrm{1}}}  \ottsym{=}  {\source M}' \, \textbf{\source {in} } \, {\source e}'_{{\mathrm{2}}}  } : ( \source {  { \source P }  } ; \source {  {\source { \Gamma_C } }  } ; \source {  {\source \Gamma}  }  \Rightarrow  \source {  {\source \tau }  } )  \mapsto  ( \source {  { \source P }  } ; \source {  {\source { \Gamma_C } }  } ; \source {  {\source \Gamma}'  }  \Rightarrow  \source {  {\source \tau }_{{\mathrm{2}}}  } )  \rightsquigarrow  \intermed {  {\intermed M}_{{\mathrm{1}}}  }  
  }
  {
    where \( \intermed {  {\intermed M}_{{\mathrm{1}}}  } = \intermed {  \textbf{\intermed {let} } \, {\intermed x}  \ottsym{:}  \forall  {\overline {\intermed a} }_{\ottmv{j}}  \ottsym{.}   {\overline {\intermed Q} }_{\ottmv{i}}  \Rightarrow  {\intermed \sigma}_{{\mathrm{1}}}   \ottsym{=}  \Lambda  {\overline {\intermed a} }_{\ottmv{j}}  \ottsym{.}  \lambda  {\overline {\intermed \delta} }_{\ottmv{i}}  \ottsym{:}  {\overline {\intermed Q} }_{\ottmv{i}}  \ottsym{.}  {\intermed M}'_{{\mathrm{1}}} \, \textbf{\intermed {in} } \, {\intermed e}_{{\mathrm{21}}}  } \).

    By case analysis on the second hypothesis of the nested statement, its last context
    typing rule
    must be \textsc{sM-inf-inf-letL} as well. Therefore, the first and second hypotheses
    of the nested statement become
    \begin{gather*}
       \source {  \textbf{\source {let} } \, {\source x}  \ottsym{:}  \forall  {\overline {\source a} }_{\ottmv{j}}  \ottsym{.}  {\overline {\source Q} }_{\ottmv{i}}  \Rightarrow  {\source \tau }_{{\mathrm{1}}}  \ottsym{=}  {\source M}' \, \textbf{\source {in} } \, {\source e}'_{{\mathrm{2}}}  } : ( \source {  { \source P }  } ; \source {  {\source { \Gamma_C } }  } ; \source {  {\source \Gamma}  }  \Rightarrow  \source {  {\source \tau }  } )  \mapsto  ( \source {  { \source P }  } ; \source {  {\source { \Gamma_C } }  } ; \source {  {\source \Gamma}'  }  \Rightarrow  \source {  {\source \tau }_{{\mathrm{2}}}  } )  \rightsquigarrow  \intermed {  {\intermed M}_{{\mathrm{1}}}  }  \\
       \source {  \textbf{\source {let} } \, {\source x}  \ottsym{:}  \forall  {\overline {\source a} }_{\ottmv{j}}  \ottsym{.}  {\overline {\source Q} }_{\ottmv{i}}  \Rightarrow  {\source \tau }_{{\mathrm{1}}}  \ottsym{=}  {\source M}' \, \textbf{\source {in} } \, {\source e}'_{{\mathrm{2}}}  } : ( \source {  { \source P }  } ; \source {  {\source { \Gamma_C } }  } ; \source {  {\source \Gamma}  }  \Rightarrow  \source {  {\source \tau }  } )  \mapsto  ( \source {  { \source P }  } ; \source {  {\source { \Gamma_C } }  } ; \source {  {\source \Gamma}'  }  \Rightarrow  \source {  {\source \tau }_{{\mathrm{2}}}  } )  \rightsquigarrow  \intermed {  {\intermed M}_{{\mathrm{2}}}  }  
    \end{gather*}
    where \( \intermed {  {\intermed M}_{{\mathrm{2}}}  } = \intermed {  \textbf{\intermed {let} } \, {\intermed x}  \ottsym{:}  \forall  {\overline {\intermed a} }_{\ottmv{j}}  \ottsym{.}   {\overline {\intermed Q} }_{\ottmv{i}}  \Rightarrow  {\intermed \sigma}_{{\mathrm{1}}}   \ottsym{=}  \Lambda  {\overline {\intermed a} }_{\ottmv{j}}  \ottsym{.}  \lambda  {\overline {\intermed \delta} }_{\ottmv{i}}  \ottsym{:}  {\overline {\intermed Q} }_{\ottmv{i}}  \ottsym{.}  {\intermed M}'_{{\mathrm{2}}} \, \textbf{\intermed {in} } \, {\intermed e}_{{\mathrm{22}}}  } \).

    Goal~\ref{eqg:ctx_refl0} becomes
    \begin{equation}
       \intermed {  {\intermed { \Gamma_C } }  } ; \intermed {  {\intermed \Gamma}'  }  \vdash  \intermed {  {\intermed \Sigma}_{{\mathrm{1}}}  } : \intermed {   {\intermed M}_{{\mathrm{1}}}  [  {\intermed e}_{{\mathrm{1}}}  ]   }  \simeq_{log}  \intermed {  {\intermed \Sigma}_{{\mathrm{2}}}  } : \intermed {   {\intermed M}_{{\mathrm{2}}}  [  {\intermed e}_{{\mathrm{2}}}  ]   } : \intermed {  {\intermed \sigma}_{{\mathrm{2}}}  } 
      \label{eqg:ctx_refl4}
    \end{equation}
    From the premises of the two \textsc{sM-inf-inf-letL} rules, we obtain:
    \begin{gather}
       \source {  {\source M}'  } : ( \source {  { \source P }  } ; \source {  {\source { \Gamma_C } }  } ; \source {  {\source \Gamma}  }  \Rightarrow  \source {  {\source \tau }  } )  \mapsto  ( \source {  { \source P }  } ; \source {  {\source { \Gamma_C } }  } ; \source {  {\source \Gamma}'  \ottsym{,}  {\overline {\source a} }_{\ottmv{j}}  \ottsym{,}  {\overline {\source \delta} }_{\ottmv{i}}  \ottsym{:}  {\overline {\source Q} }_{\ottmv{i}}  }  \Leftarrow  \source {  {\source \tau }_{{\mathrm{1}}}  } )  \rightsquigarrow  \intermed {  {\intermed M}'_{{\mathrm{1}}}  } 
      \label{eq:ctx_refl27} \\
       \source {  {\source M}'  } : ( \source {  { \source P }  } ; \source {  {\source { \Gamma_C } }  } ; \source {  {\source \Gamma}  }  \Rightarrow  \source {  {\source \tau }  } )  \mapsto  ( \source {  { \source P }  } ; \source {  {\source { \Gamma_C } }  } ; \source {  {\source \Gamma}'  \ottsym{,}  {\overline {\source a} }_{\ottmv{j}}  \ottsym{,}  {\overline {\source \delta} }_{\ottmv{i}}  \ottsym{:}  {\overline {\source Q} }_{\ottmv{i}}  }  \Leftarrow  \source {  {\source \tau }_{{\mathrm{1}}}  } )  \rightsquigarrow  \intermed {  {\intermed M}'_{{\mathrm{2}}}  } 
      \label{eq:ctx_refl28} \\
       \source {  { \source P }  } ; \source {  {\source { \Gamma_C } }  } ; \source {  {\source \Gamma}'  \ottsym{,}  {\source x}  \ottsym{:}  \forall  {\overline {\source a} }_{\ottmv{j}}  \ottsym{.}  {\overline {\source Q} }_{\ottmv{i}}  \Rightarrow  {\source \tau }_{{\mathrm{1}}}  }  \vdash_{tm}^{\intermed {M} }  \source {  {\source e}'_{{\mathrm{2}}}  } \Rightarrow \source {  {\source \tau }_{{\mathrm{2}}}  } \stoi{ \rightsquigarrow  \intermed {  {\intermed e}_{{\mathrm{21}}}  } } 
      \label{eq:ctx_refl29} \\
       \source {  { \source P }  } ; \source {  {\source { \Gamma_C } }  } ; \source {  {\source \Gamma}'  \ottsym{,}  {\source x}  \ottsym{:}  \forall  {\overline {\source a} }_{\ottmv{j}}  \ottsym{.}  {\overline {\source Q} }_{\ottmv{i}}  \Rightarrow  {\source \tau }_{{\mathrm{1}}}  }  \vdash_{tm}^{\intermed {M} }  \source {  {\source e}'_{{\mathrm{2}}}  } \Rightarrow \source {  {\source \tau }_{{\mathrm{2}}}  } \stoi{ \rightsquigarrow  \intermed {  {\intermed e}_{{\mathrm{22}}}  } } 
      \label{eq:ctx_refl30} \\
       \source {  {\source { \Gamma_C } }  } ; \source {  {\source \Gamma}'  }  \vdash_{ty}^{\intermed {M} }  \source {  \forall  {\overline {\source a} }_{\ottmv{j}}  \ottsym{.}  {\overline {\source Q} }_{\ottmv{i}}  \Rightarrow  {\source \tau }_{{\mathrm{1}}}  }\,{ \stoi{ \rightsquigarrow  \intermed {  \forall  {\overline {\intermed a} }_{\ottmv{j}}  \ottsym{.}   {\overline {\intermed Q} }_{\ottmv{i}}  \Rightarrow  {\intermed \sigma}_{{\mathrm{1}}}   } } } 
      \label{eq:ctx_refl31} \\
       \source {  {\overline {\source \delta} }_{\ottmv{i}}  }~ \textbf {fresh} 
      \label{eq:ctx_refl31b} \\
       \source {  {\source x}  }  \notin   \textbf {dom}  ( \source {  {\source \Gamma}'  } ) 
      \label{eq:ctx_refl31g}
    \end{gather}
    Through repeated case analysis on Equation~\ref{eq:ctx_refl31}
    (\textsc{sTy-scheme} and \textsc{sTy-qual}), we know that
    \begin{gather}
       \source {  {\overline {\source a} }_{\ottmv{j}}  }  \notin  \source {  {\source \Gamma}'  } 
      \nonumber \\
      \ottcomp{ \source {  {\source { \Gamma_C } }  } ; \source {  {\source \Gamma}'  \ottsym{,}  {\overline {\source a} }_{\ottmv{j}}  }  \vdash_{\mathit {Q} }^{\intermed {M} }  \source {  {\source Q}_{\ottmv{i}}  }  \rightsquigarrow  \intermed {  {\intermed Q}_{\ottmv{i}}  } }{\ottmv{i}}
      \nonumber \\
       \source {  {\source { \Gamma_C } }  } ; \source {  {\source \Gamma}'  }  \vdash_{ty}^{\intermed {M} }  \source {  {\source \tau }_{{\mathrm{1}}}  }\,{ \stoi{ \rightsquigarrow  \intermed {  {\intermed \sigma}_{{\mathrm{1}}}  } } } 
      \label{eq:ctx_refl31e}
    \end{gather}
    By repeated case analysis on the second hypothesis, we get that
    \begin{equation}
       \vdash_{ctx}^{\intermed {M} }  \source {  \bullet  } ; \source {  {\source { \Gamma_C } }  } ; \source {  \bullet  }  \rightsquigarrow  \intermed {  \bullet  } ; \intermed {  {\intermed { \Gamma_C } }  } ; \intermed {  \bullet  } 
      \label{eq:ctx_refl31f}
    \end{equation}
    By \textsc{sCtx-tyEnvTy} and \textsc{sCtx-tyEnvD}, in
    combination with these results and Equation~\ref{eq:ctx_refl31b}, we obtain
    \( \vdash_{ctx}^{\intermed {M} }  \source {  \bullet  } ; \source {  {\source { \Gamma_C } }  } ; \source {  \bullet  \ottsym{,}  {\overline {\source a} }_{\ottmv{j}}  \ottsym{,}  {\overline {\source \delta} }_{\ottmv{i}}  \ottsym{:}  {\overline {\source Q} }_{\ottmv{i}}  }  \rightsquigarrow  \intermed {  \bullet  } ; \intermed {  {\intermed { \Gamma_C } }  } ; \intermed {  \bullet  \ottsym{,}  {\overline {\intermed a} }_{\ottmv{j}}  \ottsym{,}  {\overline {\intermed \delta} }_{\ottmv{i}}  \ottsym{:}  {\overline {\intermed Q} }_{\ottmv{i}}  } \).
    Finally, Lemma~\ref{lemma:ctx_weakening_typing}, together with this
    result and the second and fourth hypothesis, teaches us that:
    \begin{gather}
       \vdash_{ctx}^{\intermed {M} }  \source {  { \source P }  } ; \source {  {\source { \Gamma_C } }  } ; \source {  {\source \Gamma}'  \ottsym{,}  {\overline {\source a} }_{\ottmv{j}}  \ottsym{,}  {\overline {\source \delta} }_{\ottmv{i}}  \ottsym{:}  {\overline {\source Q} }_{\ottmv{i}}  }  \rightsquigarrow  \intermed {  {\intermed \Sigma}_{{\mathrm{1}}}  } ; \intermed {  {\intermed { \Gamma_C } }  } ; \intermed {  {\intermed \Gamma}'  \ottsym{,}  {\overline {\intermed a} }_{\ottmv{j}}  \ottsym{,}  {\overline {\intermed \delta} }_{\ottmv{i}}  \ottsym{:}  {\overline {\intermed Q} }_{\ottmv{i}}  } 
      \label{eq:ctx_refl31c} \\
       \vdash_{ctx}^{\intermed {M} }  \source {  { \source P }  } ; \source {  {\source { \Gamma_C } }  } ; \source {  {\source \Gamma}'  \ottsym{,}  {\overline {\source a} }_{\ottmv{j}}  \ottsym{,}  {\overline {\source \delta} }_{\ottmv{i}}  \ottsym{:}  {\overline {\source Q} }_{\ottmv{i}}  }  \rightsquigarrow  \intermed {  {\intermed \Sigma}_{{\mathrm{2}}}  } ; \intermed {  {\intermed { \Gamma_C } }  } ; \intermed {  {\intermed \Gamma}'  \ottsym{,}  {\overline {\intermed a} }_{\ottmv{j}}  \ottsym{,}  {\overline {\intermed \delta} }_{\ottmv{i}}  \ottsym{:}  {\overline {\intermed Q} }_{\ottmv{i}}  } 
      \label{eq:ctx_refl31d}
    \end{gather}
    By applying Part 2 of this theorem to Equations~\ref{eq:ctx_refl27}
    and~\ref{eq:ctx_refl28}, in combination with Equations~\ref{eq:ctx_refl31e},
    \ref{eq:ctx_refl31c}
    and~\ref{eq:ctx_refl31d} and the first, third and fifth hypothesis, we get:
    \begin{equation*}
       \intermed {  {\intermed \Sigma}_{{\mathrm{1}}}  } : \intermed {  {\intermed M}'_{{\mathrm{1}}}  }  \simeq_{log}  \intermed {  {\intermed \Sigma}_{{\mathrm{2}}}  } : \intermed {  {\intermed M}'_{{\mathrm{2}}}  } : ( \intermed {  {\intermed { \Gamma_C } }  } ; \intermed {  {\intermed \Gamma}  }  \Rightarrow  \intermed {  {\intermed \sigma}  } )  \mapsto  ( \intermed {  {\intermed { \Gamma_C } }  } ; \intermed {  {\intermed \Gamma}'  \ottsym{,}  {\overline {\intermed a} }_{\ottmv{j}}  \ottsym{,}  {\overline {\intermed \delta} }_{\ottmv{i}}  \ottsym{:}  {\overline {\intermed Q} }_{\ottmv{i}}  }  \Rightarrow  \intermed {  {\intermed \sigma}_{{\mathrm{1}}}  } ) 
    \end{equation*}
    By unfolding the definition of logical equivalence in the above and applying it on
    Equation~\ref{eq:ctx_refl0}, we get:
    \begin{equation}
       \intermed {  {\intermed { \Gamma_C } }  } ; \intermed {  {\intermed \Gamma}'  \ottsym{,}  {\overline {\intermed a} }_{\ottmv{j}}  \ottsym{,}  {\overline {\intermed \delta} }_{\ottmv{i}}  \ottsym{:}  {\overline {\intermed Q} }_{\ottmv{i}}  }  \vdash  \intermed {  {\intermed \Sigma}_{{\mathrm{1}}}  } : \intermed {   {\intermed M}'_{{\mathrm{1}}}  [  {\intermed e}_{{\mathrm{1}}}  ]   }  \simeq_{log}  \intermed {  {\intermed \Sigma}_{{\mathrm{2}}}  } : \intermed {   {\intermed M}'_{{\mathrm{2}}}  [  {\intermed e}_{{\mathrm{2}}}  ]   } : \intermed {  {\intermed \sigma}_{{\mathrm{1}}}  } 
      \label{eq:ctx_refl35}
    \end{equation}
    By repeatedly applying compatibility Lemmas~\ref{lemma:compat_dabs}
    and~\ref{lemma:compat_tyabs}, we get:
    \begin{equation}
       \intermed {  {\intermed { \Gamma_C } }  } ; \intermed {  {\intermed \Gamma}'  }  \vdash  \intermed {  {\intermed \Sigma}_{{\mathrm{1}}}  } : \intermed {   \Lambda  {\overline {\intermed a} }_{\ottmv{j}}  \ottsym{.}  \lambda  {\overline {\intermed \delta} }_{\ottmv{i}}  \ottsym{:}  {\overline {\intermed Q} }_{\ottmv{i}}  \ottsym{.}  {\intermed M}'_{{\mathrm{1}}}  [  {\intermed e}_{{\mathrm{1}}}  ]   }  \simeq_{log}  \intermed {  {\intermed \Sigma}_{{\mathrm{2}}}  } : \intermed {   \Lambda  {\overline {\intermed a} }_{\ottmv{j}}  \ottsym{.}  \lambda  {\overline {\intermed \delta} }_{\ottmv{i}}  \ottsym{:}  {\overline {\intermed Q} }_{\ottmv{i}}  \ottsym{.}  {\intermed M}'_{{\mathrm{2}}}  [  {\intermed e}_{{\mathrm{2}}}  ]   } : \intermed {  \forall  {\overline {\intermed a} }_{\ottmv{j}}  \ottsym{.}   {\overline {\intermed Q} }_{\ottmv{i}}  \Rightarrow  {\intermed \sigma}_{{\mathrm{1}}}   } 
      \label{eq:ctx_refl35b}
    \end{equation}
    It follows from \textsc{sCtx-tyEnvTm}, in combination with
    Equations~\ref{eq:ctx_refl31f}, \ref{eq:ctx_refl31g}
    and~\ref{eq:ctx_refl31}, that \\
    \( \vdash_{ctx}^{\intermed {M} }  \source {  \bullet  } ; \source {  {\source { \Gamma_C } }  } ; \source {  \bullet  \ottsym{,}  {\source x}  \ottsym{:}  \forall  {\overline {\source a} }_{\ottmv{j}}  \ottsym{.}  {\overline {\source Q} }_{\ottmv{i}}  \Rightarrow  {\source \tau }_{{\mathrm{1}}}  }  \rightsquigarrow  \intermed {  \bullet  } ; \intermed {  {\intermed { \Gamma_C } }  } ; \intermed {  \bullet  \ottsym{,}  {\intermed x}  \ottsym{:}  \forall  {\overline {\intermed a} }_{\ottmv{j}}  \ottsym{.}   {\overline {\intermed Q} }_{\ottmv{i}}  \Rightarrow  {\intermed \sigma}_{{\mathrm{1}}}   } \).
    Similarly to before, by applying Lemma~\ref{lemma:ctx_weakening_typing} to
    this result, together with the second and fourth hypothesis, we get:
    \begin{gather}
       \vdash_{ctx}^{\intermed {M} }  \source {  { \source P }  } ; \source {  {\source { \Gamma_C } }  } ; \source {  {\source \Gamma}'  \ottsym{,}  {\source x}  \ottsym{:}  \forall  {\overline {\source a} }_{\ottmv{j}}  \ottsym{.}  {\overline {\source Q} }_{\ottmv{i}}  \Rightarrow  {\source \tau }_{{\mathrm{1}}}  }  \rightsquigarrow  \intermed {  {\intermed \Sigma}_{{\mathrm{1}}}  } ; \intermed {  {\intermed { \Gamma_C } }  } ; \intermed {  {\intermed \Gamma}'  \ottsym{,}  {\intermed x}  \ottsym{:}  \forall  {\overline {\intermed a} }_{\ottmv{j}}  \ottsym{.}   {\overline {\intermed Q} }_{\ottmv{i}}  \Rightarrow  {\intermed \sigma}_{{\mathrm{1}}}   } 
      \label{eq:ctx_refl36} \\
       \vdash_{ctx}^{\intermed {M} }  \source {  { \source P }  } ; \source {  {\source { \Gamma_C } }  } ; \source {  {\source \Gamma}'  \ottsym{,}  {\source x}  \ottsym{:}  \forall  {\overline {\source a} }_{\ottmv{j}}  \ottsym{.}  {\overline {\source Q} }_{\ottmv{i}}  \Rightarrow  {\source \tau }_{{\mathrm{1}}}  }  \rightsquigarrow  \intermed {  {\intermed \Sigma}_{{\mathrm{2}}}  } ; \intermed {  {\intermed { \Gamma_C } }  } ; \intermed {  {\intermed \Gamma}'  \ottsym{,}  {\intermed x}  \ottsym{:}  \forall  {\overline {\intermed a} }_{\ottmv{j}}  \ottsym{.}   {\overline {\intermed Q} }_{\ottmv{i}}  \Rightarrow  {\intermed \sigma}_{{\mathrm{1}}}   } 
      \label{eq:ctx_refl37}
    \end{gather}
    By applying Expression Coherence Theorem A (Theorem~\ref{thmA:coh_exprA}) to
    Equations~\ref{eq:ctx_refl29} and~\ref{eq:ctx_refl30}, together with
    Equations~\ref{eq:ctx_refl36} and~\ref{eq:ctx_refl37}, we get:
    \begin{equation}
       \intermed {  {\intermed { \Gamma_C } }  } ; \intermed {  {\intermed \Gamma}'  \ottsym{,}  {\intermed x}  \ottsym{:}  \forall  {\overline {\intermed a} }_{\ottmv{j}}  \ottsym{.}   {\overline {\intermed Q} }_{\ottmv{i}}  \Rightarrow  {\intermed \sigma}_{{\mathrm{1}}}   }  \vdash  \intermed {  {\intermed \Sigma}_{{\mathrm{1}}}  } : \intermed {  {\intermed e}_{{\mathrm{21}}}  }  \simeq_{log}  \intermed {  {\intermed \Sigma}_{{\mathrm{2}}}  } : \intermed {  {\intermed e}_{{\mathrm{22}}}  } : \intermed {  {\intermed \sigma}_{{\mathrm{2}}}  } 
      \label{eq:ctx_refl38}
    \end{equation}
    Goal~\ref{eqg:ctx_refl4} follows from compatibility of let expressions
    (Lemma~\ref{lemma:compat_let}, together
    with Equations~\ref{eq:ctx_refl35} and~\ref{eq:ctx_refl38}). \\
  }
  \proofStep{\textsc{sM-inf-inf-letR}}
  {
     \source {  \textbf{\source {let} } \, {\source x}  \ottsym{:}  \forall  {\overline {\source a} }_{\ottmv{j}}  \ottsym{.}  {\overline {\source Q} }_{\ottmv{i}}  \Rightarrow  {\source \tau }_{{\mathrm{1}}}  \ottsym{=}  {\source e}'_{{\mathrm{1}}} \, \textbf{\source {in} } \, {\source M}'  } : ( \source {  { \source P }  } ; \source {  {\source { \Gamma_C } }  } ; \source {  {\source \Gamma}  }  \Rightarrow  \source {  {\source \tau }  } )  \mapsto  ( \source {  { \source P }  } ; \source {  {\source { \Gamma_C } }  } ; \source {  {\source \Gamma}'  }  \Rightarrow  \source {  {\source \tau }_{{\mathrm{2}}}  } )  \rightsquigarrow  \intermed {  {\intermed M}_{{\mathrm{1}}}  } 
  }
  {
    where \( \intermed {  {\intermed M}_{{\mathrm{1}}}  } = \intermed {  \textbf{\intermed {let} } \, {\intermed x}  \ottsym{:}  \forall  {\overline {\intermed a} }_{\ottmv{j}}  \ottsym{.}   {\overline {\intermed Q} }_{\ottmv{i}}  \Rightarrow  {\intermed \sigma}_{{\mathrm{1}}}   \ottsym{=}  \Lambda  {\overline {\intermed a} }_{\ottmv{j}}  \ottsym{.}  \lambda  {\overline {\intermed \delta} }_{\ottmv{i}}  \ottsym{:}  {\overline {\intermed Q} }_{\ottmv{i}}  \ottsym{.}  {\intermed e}_{{\mathrm{11}}} \, \textbf{\intermed {in} } \, {\intermed M}'_{{\mathrm{1}}}  } \).

    By case analysis on the second hypothesis of the nested statement, its last context
    typing rule
    must be \textsc{sM-inf-inf-letR} as well. Therefore, the first and second hypotheses
    of the nested statement become
    \begin{gather*}
       \source {  \textbf{\source {let} } \, {\source x}  \ottsym{:}  \forall  {\overline {\source a} }_{\ottmv{j}}  \ottsym{.}  {\overline {\source Q} }_{\ottmv{i}}  \Rightarrow  {\source \tau }_{{\mathrm{1}}}  \ottsym{=}  {\source e}'_{{\mathrm{1}}} \, \textbf{\source {in} } \, {\source M}'  } : ( \source {  { \source P }  } ; \source {  {\source { \Gamma_C } }  } ; \source {  {\source \Gamma}  }  \Rightarrow  \source {  {\source \tau }  } )  \mapsto  ( \source {  { \source P }  } ; \source {  {\source { \Gamma_C } }  } ; \source {  {\source \Gamma}'  }  \Rightarrow  \source {  {\source \tau }_{{\mathrm{2}}}  } )  \rightsquigarrow  \intermed {  {\intermed M}_{{\mathrm{1}}}  } \\
       \source {  \textbf{\source {let} } \, {\source x}  \ottsym{:}  \forall  {\overline {\source a} }_{\ottmv{j}}  \ottsym{.}  {\overline {\source Q} }_{\ottmv{i}}  \Rightarrow  {\source \tau }_{{\mathrm{1}}}  \ottsym{=}  {\source e}'_{{\mathrm{1}}} \, \textbf{\source {in} } \, {\source M}'  } : ( \source {  { \source P }  } ; \source {  {\source { \Gamma_C } }  } ; \source {  {\source \Gamma}  }  \Rightarrow  \source {  {\source \tau }  } )  \mapsto  ( \source {  { \source P }  } ; \source {  {\source { \Gamma_C } }  } ; \source {  {\source \Gamma}'  }  \Rightarrow  \source {  {\source \tau }_{{\mathrm{2}}}  } )  \rightsquigarrow  \intermed {  {\intermed M}_{{\mathrm{2}}}  } 
    \end{gather*}
    where \( \intermed {  {\intermed M}_{{\mathrm{2}}}  } = \intermed {  \textbf{\intermed {let} } \, {\intermed x}  \ottsym{:}  \forall  {\overline {\intermed a} }_{\ottmv{j}}  \ottsym{.}   {\overline {\intermed Q} }_{\ottmv{i}}  \Rightarrow  {\intermed \sigma}_{{\mathrm{1}}}   \ottsym{=}  \Lambda  {\overline {\intermed a} }_{\ottmv{j}}  \ottsym{.}  \lambda  {\overline {\intermed \delta} }_{\ottmv{i}}  \ottsym{:}  {\overline {\intermed Q} }_{\ottmv{i}}  \ottsym{.}  {\intermed e}_{{\mathrm{12}}} \, \textbf{\intermed {in} } \, {\intermed M}'_{{\mathrm{2}}}  } \).
    
    Goal~\ref{eqg:ctx_refl0} becomes
    \begin{equation}
       \intermed {  {\intermed { \Gamma_C } }  } ; \intermed {  {\intermed \Gamma}'  }  \vdash  \intermed {  {\intermed \Sigma}_{{\mathrm{1}}}  } : \intermed {   {\intermed M}_{{\mathrm{1}}}  [  {\intermed e}_{{\mathrm{1}}}  ]   }  \simeq_{log}  \intermed {  {\intermed \Sigma}_{{\mathrm{2}}}  } : \intermed {   {\intermed M}_{{\mathrm{2}}}  [  {\intermed e}_{{\mathrm{2}}}  ]   } : \intermed {  {\intermed \sigma}_{{\mathrm{2}}}  } 
      \label{eqg:ctx_refl5}
    \end{equation}
    From the premises of the two \textsc{sM-inf-inf-letR} rules, we obtain:
    \begin{gather}
       \source {  {\source M}'  } : ( \source {  { \source P }  } ; \source {  {\source { \Gamma_C } }  } ; \source {  {\source \Gamma}  }  \Rightarrow  \source {  {\source \tau }  } )  \mapsto  ( \source {  { \source P }  } ; \source {  {\source { \Gamma_C } }  } ; \source {  {\source \Gamma}'  \ottsym{,}  {\source x}  \ottsym{:}  \forall  {\overline {\source a} }_{\ottmv{j}}  \ottsym{.}  {\overline {\source Q} }_{\ottmv{i}}  \Rightarrow  {\source \tau }_{{\mathrm{1}}}  }  \Rightarrow  \source {  {\source \tau }_{{\mathrm{2}}}  } )  \rightsquigarrow  \intermed {  {\intermed M}'_{{\mathrm{1}}}  } 
      \label{eq:ctx_refl41} \\
       \source {  {\source M}'  } : ( \source {  { \source P }  } ; \source {  {\source { \Gamma_C } }  } ; \source {  {\source \Gamma}  }  \Rightarrow  \source {  {\source \tau }  } )  \mapsto  ( \source {  { \source P }  } ; \source {  {\source { \Gamma_C } }  } ; \source {  {\source \Gamma}'  \ottsym{,}  {\source x}  \ottsym{:}  \forall  {\overline {\source a} }_{\ottmv{j}}  \ottsym{.}  {\overline {\source Q} }_{\ottmv{i}}  \Rightarrow  {\source \tau }_{{\mathrm{1}}}  }  \Rightarrow  \source {  {\source \tau }_{{\mathrm{2}}}  } )  \rightsquigarrow  \intermed {  {\intermed M}'_{{\mathrm{2}}}  } 
      \label{eq:ctx_refl42} \\
       \source {  { \source P }  } ; \source {  {\source { \Gamma_C } }  } ; \source {  {\source \Gamma}'  \ottsym{,}  {\overline {\source a} }_{\ottmv{j}}  \ottsym{,}  {\overline {\source \delta} }_{\ottmv{i}}  \ottsym{:}  {\overline {\source Q} }_{\ottmv{i}}  }  \vdash_{tm}^{\intermed {M} }  \source {  {\source e}_{{\mathrm{1}}}  } \Leftarrow \source {  {\source \tau }_{{\mathrm{1}}}  } \stoi{ \rightsquigarrow  \intermed {  {\intermed e}_{{\mathrm{11}}}  } } 
      \label{eq:ctx_refl43} \\
       \source {  { \source P }  } ; \source {  {\source { \Gamma_C } }  } ; \source {  {\source \Gamma}'  \ottsym{,}  {\overline {\source a} }_{\ottmv{j}}  \ottsym{,}  {\overline {\source \delta} }_{\ottmv{i}}  \ottsym{:}  {\overline {\source Q} }_{\ottmv{i}}  }  \vdash_{tm}^{\intermed {M} }  \source {  {\source e}_{{\mathrm{1}}}  } \Leftarrow \source {  {\source \tau }_{{\mathrm{1}}}  } \stoi{ \rightsquigarrow  \intermed {  {\intermed e}_{{\mathrm{12}}}  } } 
      \label{eq:ctx_refl44} \\
       \source {  {\source { \Gamma_C } }  } ; \source {  {\source \Gamma}'  }  \vdash_{ty}^{\intermed {M} }  \source {  \forall  {\overline {\source a} }_{\ottmv{j}}  \ottsym{.}  {\overline {\source Q} }_{\ottmv{i}}  \Rightarrow  {\source \tau }_{{\mathrm{1}}}  }\,{ \stoi{ \rightsquigarrow  \intermed {  \forall  {\overline {\intermed a} }_{\ottmv{j}}  \ottsym{.}   {\overline {\intermed Q} }_{\ottmv{i}}  \Rightarrow  {\intermed \sigma}_{{\mathrm{1}}}   } } } 
      \label{eq:ctx_refl45} \\
       \source {  {\overline {\source \delta} }_{\ottmv{i}}  }~ \textbf {fresh} 
      \label{eq:ctx_refl45b} \\
       \source {  {\source x}  }  \notin   \textbf {dom}  ( \source {  {\source \Gamma}'  } ) 
      \label{eq:ctx_refl45c}
    \end{gather}
    By repeated case analysis on the second hypothesis, we get that
    \begin{equation}
       \vdash_{ctx}^{\intermed {M} }  \source {  \bullet  } ; \source {  {\source { \Gamma_C } }  } ; \source {  \bullet  }  \rightsquigarrow  \intermed {  \bullet  } ; \intermed {  {\intermed { \Gamma_C } }  } ; \intermed {  \bullet  } 
      \label{eq:ctx_refl45d}
    \end{equation}
    By \textsc{sCtx-tyEnvTm}, in
    combination with this result and Equations~\ref{eq:ctx_refl45}
    and~\ref{eq:ctx_refl45c}, we obtain \\
    \( \vdash_{ctx}^{\intermed {M} }  \source {  \bullet  } ; \source {  {\source { \Gamma_C } }  } ; \source {  \bullet  \ottsym{,}  {\source x}  \ottsym{:}  \forall  {\overline {\source a} }_{\ottmv{j}}  \ottsym{.}  {\overline {\source Q} }_{\ottmv{i}}  \Rightarrow  {\source \tau }_{{\mathrm{1}}}  }  \rightsquigarrow  \intermed {  \bullet  } ; \intermed {  {\intermed { \Gamma_C } }  } ; \intermed {  \bullet  \ottsym{,}  {\intermed x}  \ottsym{:}  \forall  {\overline {\intermed a} }_{\ottmv{j}}  \ottsym{.}   {\overline {\intermed Q} }_{\ottmv{i}}  \Rightarrow  {\intermed \sigma}_{{\mathrm{1}}}   } \).
    Lemma~\ref{lemma:ctx_weakening_typing}, together with this
    result and the second and fourth hypothesis, teaches us that:
    \begin{gather}
       \vdash_{ctx}^{\intermed {M} }  \source {  { \source P }  } ; \source {  {\source { \Gamma_C } }  } ; \source {  {\source \Gamma}'  \ottsym{,}  {\source x}  \ottsym{:}  \forall  {\overline {\source a} }_{\ottmv{j}}  \ottsym{.}  {\overline {\source Q} }_{\ottmv{i}}  \Rightarrow  {\source \tau }_{{\mathrm{1}}}  }  \rightsquigarrow  \intermed {  {\intermed \Sigma}_{{\mathrm{1}}}  } ; \intermed {  {\intermed { \Gamma_C } }  } ; \intermed {  {\intermed \Gamma}'  \ottsym{,}  {\intermed x}  \ottsym{:}  \forall  {\overline {\intermed a} }_{\ottmv{j}}  \ottsym{.}   {\overline {\intermed Q} }_{\ottmv{i}}  \Rightarrow  {\intermed \sigma}_{{\mathrm{1}}}   } 
      \label{eq:ctx_refl46} \\
       \vdash_{ctx}^{\intermed {M} }  \source {  { \source P }  } ; \source {  {\source { \Gamma_C } }  } ; \source {  {\source \Gamma}'  \ottsym{,}  {\source x}  \ottsym{:}  \forall  {\overline {\source a} }_{\ottmv{j}}  \ottsym{.}  {\overline {\source Q} }_{\ottmv{i}}  \Rightarrow  {\source \tau }_{{\mathrm{1}}}  }  \rightsquigarrow  \intermed {  {\intermed \Sigma}_{{\mathrm{2}}}  } ; \intermed {  {\intermed { \Gamma_C } }  } ; \intermed {  {\intermed \Gamma}'  \ottsym{,}  {\intermed x}  \ottsym{:}  \forall  {\overline {\intermed a} }_{\ottmv{j}}  \ottsym{.}   {\overline {\intermed Q} }_{\ottmv{i}}  \Rightarrow  {\intermed \sigma}_{{\mathrm{1}}}   } 
      \label{eq:ctx_refl47}
    \end{gather}
    By applying the induction hypothesis to Equations~\ref{eq:ctx_refl41}
    and~\ref{eq:ctx_refl42}, in combination with Equations~\ref{eq:ctx_refl46}
    and~\ref{eq:ctx_refl47}, we get:
    \begin{equation*}
       \intermed {  {\intermed \Sigma}_{{\mathrm{1}}}  } : \intermed {  {\intermed M}'_{{\mathrm{1}}}  }  \simeq_{log}  \intermed {  {\intermed \Sigma}_{{\mathrm{2}}}  } : \intermed {  {\intermed M}'_{{\mathrm{2}}}  } : ( \intermed {  {\intermed { \Gamma_C } }  } ; \intermed {  {\intermed \Gamma}  }  \Rightarrow  \intermed {  {\intermed \sigma}  } )  \mapsto  ( \intermed {  {\intermed { \Gamma_C } }  } ; \intermed {  {\intermed \Gamma}'  \ottsym{,}  {\intermed x}  \ottsym{:}  \forall  {\overline {\intermed a} }_{\ottmv{j}}  \ottsym{.}   {\overline {\intermed Q} }_{\ottmv{i}}  \Rightarrow  {\intermed \sigma}_{{\mathrm{1}}}   }  \Rightarrow  \intermed {  {\intermed \sigma}_{{\mathrm{2}}}  } ) 
    \end{equation*}
    By unfolding the definition of logical equivalence in the above and then applying it on
    Equation~\ref{eq:ctx_refl0}, we get:
    \begin{equation}
       \intermed {  {\intermed { \Gamma_C } }  } ; \intermed {  {\intermed \Gamma}'  \ottsym{,}  {\intermed x}  \ottsym{:}  \forall  {\overline {\intermed a} }_{\ottmv{j}}  \ottsym{.}   {\overline {\intermed Q} }_{\ottmv{i}}  \Rightarrow  {\intermed \sigma}_{{\mathrm{1}}}   }  \vdash  \intermed {  {\intermed \Sigma}_{{\mathrm{1}}}  } : \intermed {   {\intermed M}'_{{\mathrm{1}}}  [  {\intermed e}_{{\mathrm{1}}}  ]   }  \simeq_{log}  \intermed {  {\intermed \Sigma}_{{\mathrm{2}}}  } : \intermed {   {\intermed M}'_{{\mathrm{2}}}  [  {\intermed e}_{{\mathrm{2}}}  ]   } : \intermed {  {\intermed \sigma}_{{\mathrm{2}}}  } 
      \label{eq:ctx_refl51}
    \end{equation}
    Through repeated case analysis on Equation~\ref{eq:ctx_refl45}
    (\textsc{sTy-scheme} and \textsc{sTy-qual}), we know that
    \begin{gather}
       \source {  {\overline {\source a} }_{\ottmv{j}}  }  \notin  \source {  {\source \Gamma}'  } 
      \nonumber \\
      \ottcomp{ \source {  {\source { \Gamma_C } }  } ; \source {  {\source \Gamma}'  \ottsym{,}  {\overline {\source a} }_{\ottmv{j}}  }  \vdash_{\mathit {Q} }^{\intermed {M} }  \source {  {\source Q}_{\ottmv{i}}  }  \rightsquigarrow  \intermed {  {\intermed Q}_{\ottmv{i}}  } }{\ottmv{i}}
      \nonumber \\
       \source {  {\source { \Gamma_C } }  } ; \source {  {\source \Gamma}'  }  \vdash_{ty}^{\intermed {M} }  \source {  {\source \tau }_{{\mathrm{1}}}  }\,{ \stoi{ \rightsquigarrow  \intermed {  {\intermed \sigma}_{{\mathrm{1}}}  } } } 
      \label{eq:ctx_refl45e}
    \end{gather}
    By \textsc{sCtx-tyEnvTy} and \textsc{sCtx-tyEnvD}, in
    combination with these results and Equations~\ref{eq:ctx_refl45b}
    and~\ref{eq:ctx_refl45d}, we obtain
    \( \vdash_{ctx}^{\intermed {M} }  \source {  \bullet  } ; \source {  {\source { \Gamma_C } }  } ; \source {  \bullet  \ottsym{,}  {\overline {\source a} }_{\ottmv{j}}  \ottsym{,}  {\overline {\source \delta} }_{\ottmv{i}}  \ottsym{:}  {\overline {\source Q} }_{\ottmv{i}}  }  \rightsquigarrow  \intermed {  \bullet  } ; \intermed {  {\intermed { \Gamma_C } }  } ; \intermed {  \bullet  \ottsym{,}  {\overline {\intermed a} }_{\ottmv{j}}  \ottsym{,}  {\overline {\intermed \delta} }_{\ottmv{i}}  \ottsym{:}  {\overline {\intermed Q} }_{\ottmv{i}}  } \).
    Lemma~\ref{lemma:ctx_weakening_typing}, together with this
    result and the second and fourth hypothesis, teaches us that:
    \begin{gather}
       \vdash_{ctx}^{\intermed {M} }  \source {  { \source P }  } ; \source {  {\source { \Gamma_C } }  } ; \source {  {\source \Gamma}'  \ottsym{,}  {\overline {\source a} }_{\ottmv{j}}  \ottsym{,}  {\overline {\source \delta} }_{\ottmv{i}}  \ottsym{:}  {\overline {\source Q} }_{\ottmv{i}}  }  \rightsquigarrow  \intermed {  {\intermed \Sigma}_{{\mathrm{1}}}  } ; \intermed {  {\intermed { \Gamma_C } }  } ; \intermed {  {\intermed \Gamma}'  \ottsym{,}  {\overline {\intermed a} }_{\ottmv{j}}  \ottsym{,}  {\overline {\intermed \delta} }_{\ottmv{i}}  \ottsym{:}  {\overline {\intermed Q} }_{\ottmv{i}}  } 
      \label{eq:ctx_refl45f} \\
       \vdash_{ctx}^{\intermed {M} }  \source {  { \source P }  } ; \source {  {\source { \Gamma_C } }  } ; \source {  {\source \Gamma}'  \ottsym{,}  {\overline {\source a} }_{\ottmv{j}}  \ottsym{,}  {\overline {\source \delta} }_{\ottmv{i}}  \ottsym{:}  {\overline {\source Q} }_{\ottmv{i}}  }  \rightsquigarrow  \intermed {  {\intermed \Sigma}_{{\mathrm{2}}}  } ; \intermed {  {\intermed { \Gamma_C } }  } ; \intermed {  {\intermed \Gamma}'  \ottsym{,}  {\overline {\intermed a} }_{\ottmv{j}}  \ottsym{,}  {\overline {\intermed \delta} }_{\ottmv{i}}  \ottsym{:}  {\overline {\intermed Q} }_{\ottmv{i}}  } 
      \label{eq:ctx_refl45g}
    \end{gather}
    By applying Expression Coherence Theorem A (Theorem~\ref{thmA:coh_exprA}) to
    Equations~\ref{eq:ctx_refl43} and~\ref{eq:ctx_refl44}, in combination with
    Equations~\ref{eq:ctx_refl45f} and~\ref{eq:ctx_refl45g}, we get:
    \begin{equation}
       \intermed {  {\intermed { \Gamma_C } }  } ; \intermed {  {\intermed \Gamma}'  \ottsym{,}  {\overline {\intermed a} }_{\ottmv{j}}  \ottsym{,}  {\overline {\intermed \delta} }_{\ottmv{i}}  \ottsym{:}  {\overline {\intermed Q} }_{\ottmv{i}}  }  \vdash  \intermed {  {\intermed \Sigma}_{{\mathrm{1}}}  } : \intermed {  {\intermed e}_{{\mathrm{11}}}  }  \simeq_{log}  \intermed {  {\intermed \Sigma}_{{\mathrm{2}}}  } : \intermed {  {\intermed e}_{{\mathrm{12}}}  } : \intermed {  {\intermed \sigma}_{{\mathrm{1}}}  } 
      \label{eq:ctx_refl52}
    \end{equation}
    By repeatedly applying compatibility Lemmas~\ref{lemma:compat_dabs}
    and~\ref{lemma:compat_tyabs}, we get:
    \begin{equation}
       \intermed {  {\intermed { \Gamma_C } }  } ; \intermed {  {\intermed \Gamma}'  }  \vdash  \intermed {  {\intermed \Sigma}_{{\mathrm{1}}}  } : \intermed {  \Lambda  {\overline {\intermed a} }_{\ottmv{j}}  \ottsym{.}  \lambda  {\overline {\intermed \delta} }_{\ottmv{i}}  \ottsym{:}  {\overline {\intermed Q} }_{\ottmv{i}}  \ottsym{.}  {\intermed e}_{{\mathrm{11}}}  }  \simeq_{log}  \intermed {  {\intermed \Sigma}_{{\mathrm{2}}}  } : \intermed {  \Lambda  {\overline {\intermed a} }_{\ottmv{j}}  \ottsym{.}  \lambda  {\overline {\intermed \delta} }_{\ottmv{i}}  \ottsym{:}  {\overline {\intermed Q} }_{\ottmv{i}}  \ottsym{.}  {\intermed e}_{{\mathrm{12}}}  } : \intermed {  \forall  {\overline {\intermed a} }_{\ottmv{j}}  \ottsym{.}   {\overline {\intermed Q} }_{\ottmv{i}}  \Rightarrow  {\intermed \sigma}_{{\mathrm{1}}}   } 
      \label{eq:ctx_refl52b}
    \end{equation}
    Goal~\ref{eqg:ctx_refl5} follows from Lemma~\ref{lemma:compat_let}, together
    with Equations~\ref{eq:ctx_refl51} and~\ref{eq:ctx_refl52b}. \\
  }
  \proofStep{\textsc{sM-inf-inf-ann}}
  {
     \source {  {\source M}'  ::  {\source \tau }'  } : ( \source {  { \source P }  } ; \source {  {\source { \Gamma_C } }  } ; \source {  {\source \Gamma}  }  \Rightarrow  \source {  {\source \tau }  } )  \mapsto  ( \source {  { \source P }  } ; \source {  {\source { \Gamma_C } }  } ; \source {  {\source \Gamma}'  }  \Rightarrow  \source {  {\source \tau }'  } )  \rightsquigarrow  \intermed {  {\intermed M}_{{\mathrm{1}}}  } 
  }
  {
    By case analysis, we know the final step in the second derivation has to be
    \textsc{sM-inf-inf-ann} as well. This means that:
    \begin{gather}
       \source {  {\source M}'  ::  {\source \tau }'  } : ( \source {  { \source P }  } ; \source {  {\source { \Gamma_C } }  } ; \source {  {\source \Gamma}  }  \Rightarrow  \source {  {\source \tau }  } )  \mapsto  ( \source {  { \source P }  } ; \source {  {\source { \Gamma_C } }  } ; \source {  {\source \Gamma}'  }  \Rightarrow  \source {  {\source \tau }'  } )  \rightsquigarrow  \intermed {  {\intermed M}_{{\mathrm{1}}}  } 
      \label{eq:ctx_refl53} \\
       \source {  {\source M}'  ::  {\source \tau }'  } : ( \source {  { \source P }  } ; \source {  {\source { \Gamma_C } }  } ; \source {  {\source \Gamma}  }  \Rightarrow  \source {  {\source \tau }  } )  \mapsto  ( \source {  { \source P }  } ; \source {  {\source { \Gamma_C } }  } ; \source {  {\source \Gamma}'  }  \Rightarrow  \source {  {\source \tau }'  } )  \rightsquigarrow  \intermed {  {\intermed M}_{{\mathrm{2}}}  } 
      \label{eq:ctx_refl54}
    \end{gather}
    The goal to be proven is the following:
    \begin{equation}
       \intermed {  {\intermed { \Gamma_C } }  } ; \intermed {  {\intermed \Gamma}'  }  \vdash  \intermed {  {\intermed \Sigma}_{{\mathrm{1}}}  } : \intermed {   {\intermed M}_{{\mathrm{1}}}  [  {\intermed e}_{{\mathrm{1}}}  ]   }  \simeq_{log}  \intermed {  {\intermed \Sigma}_{{\mathrm{2}}}  } : \intermed {   {\intermed M}_{{\mathrm{2}}}  [  {\intermed e}_{{\mathrm{2}}}  ]   } : \intermed {  {\intermed \sigma}'  } 
      \label{eqg:ctx_refl6}
    \end{equation}
    From the rule premise we know that:
    \begin{gather}
       \source {  {\source M}'  } : ( \source {  { \source P }  } ; \source {  {\source { \Gamma_C } }  } ; \source {  {\source \Gamma}  }  \Rightarrow  \source {  {\source \tau }  } )  \mapsto  ( \source {  { \source P }  } ; \source {  {\source { \Gamma_C } }  } ; \source {  {\source \Gamma}'  }  \Leftarrow  \source {  {\source \tau }'  } )  \rightsquigarrow  \intermed {  {\intermed M}_{{\mathrm{1}}}  } 
      \label{eq:ctx_refl55} \\
       \source {  {\source M}'  } : ( \source {  { \source P }  } ; \source {  {\source { \Gamma_C } }  } ; \source {  {\source \Gamma}  }  \Rightarrow  \source {  {\source \tau }  } )  \mapsto  ( \source {  { \source P }  } ; \source {  {\source { \Gamma_C } }  } ; \source {  {\source \Gamma}'  }  \Leftarrow  \source {  {\source \tau }'  } )  \rightsquigarrow  \intermed {  {\intermed M}_{{\mathrm{2}}}  } 
      \label{eq:ctx_refl56}
    \end{gather}
    Goal~\ref{eqg:ctx_refl6} follows directly from Part 2 of this theorem, in
    combination with Equations~\ref{eq:ctx_refl55} and~\ref{eq:ctx_refl56}. \\
  }
\item[Part 2] By case analysis on the first typing derivation. \\
  \proofStep{\textsc{sM-inf-check-abs}}
  {
     \source {  \lambda  {\source x}  \ottsym{.}  {\source M}'  } : ( \source {  { \source P }  } ; \source {  {\source { \Gamma_C } }  } ; \source {  {\source \Gamma}  }  \Rightarrow  \source {  {\source \tau }  } )  \mapsto  ( \source {  { \source P }  } ; \source {  {\source { \Gamma_C } }  } ; \source {  {\source \Gamma}'  }  \Leftarrow  \source {  {\source \tau }_{{\mathrm{1}}}  \rightarrow  {\source \tau }_{{\mathrm{2}}}  } )  \rightsquigarrow  \intermed {  \lambda  {\intermed x}  \ottsym{:}  {\intermed \sigma}_{{\mathrm{1}}}  \ottsym{.}  {\intermed M}'_{{\mathrm{1}}}  } 
  }
  {
    By case analysis, we know that the final step in the second derivation has
    to be either \textsc{sM-inf-check-abs} or \textsc{sM-inf-check-inf}. Note
    however that no matching inference rules exist. The final step in the second
    derivation thus has to be \textsc{sM-inf-check-abs} as well. This means that:
    \begin{gather}
       \source {  \lambda  {\source x}  \ottsym{.}  {\source M}'  } : ( \source {  { \source P }  } ; \source {  {\source { \Gamma_C } }  } ; \source {  {\source \Gamma}  }  \Rightarrow  \source {  {\source \tau }  } )  \mapsto  ( \source {  { \source P }  } ; \source {  {\source { \Gamma_C } }  } ; \source {  {\source \Gamma}'  }  \Leftarrow  \source {  {\source \tau }_{{\mathrm{1}}}  \rightarrow  {\source \tau }_{{\mathrm{2}}}  } )  \rightsquigarrow  \intermed {  \lambda  {\intermed x}  \ottsym{:}  {\intermed \sigma}_{{\mathrm{1}}}  \ottsym{.}  {\intermed M}'_{{\mathrm{1}}}  } 
      \label{eq:ctx_refl57} \\
       \source {  \lambda  {\source x}  \ottsym{.}  {\source M}'  } : ( \source {  { \source P }  } ; \source {  {\source { \Gamma_C } }  } ; \source {  {\source \Gamma}  }  \Rightarrow  \source {  {\source \tau }  } )  \mapsto  ( \source {  { \source P }  } ; \source {  {\source { \Gamma_C } }  } ; \source {  {\source \Gamma}'  }  \Leftarrow  \source {  {\source \tau }_{{\mathrm{1}}}  \rightarrow  {\source \tau }_{{\mathrm{2}}}  } )  \rightsquigarrow  \intermed {  \lambda  {\intermed x}  \ottsym{:}  {\intermed \sigma}_{{\mathrm{1}}}  \ottsym{.}  {\intermed M}'_{{\mathrm{2}}}  } 
      \label{eq:ctx_refl58}
    \end{gather}
    The goal to be proven is the following:
    \begin{equation}
       \intermed {  {\intermed { \Gamma_C } }  } ; \intermed {  {\intermed \Gamma}'  }  \vdash  \intermed {  {\intermed \Sigma}_{{\mathrm{1}}}  } : \intermed {   \lambda  {\intermed x}  \ottsym{:}  {\intermed \sigma}_{{\mathrm{1}}}  \ottsym{.}  {\intermed M}'_{{\mathrm{1}}}  [  {\intermed e}_{{\mathrm{1}}}  ]   }  \simeq_{log}  \intermed {  {\intermed \Sigma}_{{\mathrm{2}}}  } : \intermed {   \lambda  {\intermed x}  \ottsym{:}  {\intermed \sigma}_{{\mathrm{1}}}  \ottsym{.}  {\intermed M}'_{{\mathrm{2}}}  [  {\intermed e}_{{\mathrm{2}}}  ]   } : \intermed {  {\intermed \sigma}_{{\mathrm{1}}}  \rightarrow  {\intermed \sigma}_{{\mathrm{2}}}  } 
      \label{eqg:ctx_refl7}
    \end{equation}
    From the rule premise we know that:
    \begin{gather}
       \source {  {\source M}'  } : ( \source {  { \source P }  } ; \source {  {\source { \Gamma_C } }  } ; \source {  {\source \Gamma}  }  \Rightarrow  \source {  {\source \tau }  } )  \mapsto  ( \source {  { \source P }  } ; \source {  {\source { \Gamma_C } }  } ; \source {  {\source \Gamma}'  \ottsym{,}  {\source x}  \ottsym{:}  {\source \tau }_{{\mathrm{1}}}  }  \Leftarrow  \source {  {\source \tau }_{{\mathrm{2}}}  } )  \rightsquigarrow  \intermed {  {\intermed M}'_{{\mathrm{1}}}  } 
      \label{eq:ctx_refl59} \\
       \source {  {\source M}'  } : ( \source {  { \source P }  } ; \source {  {\source { \Gamma_C } }  } ; \source {  {\source \Gamma}  }  \Rightarrow  \source {  {\source \tau }  } )  \mapsto  ( \source {  { \source P }  } ; \source {  {\source { \Gamma_C } }  } ; \source {  {\source \Gamma}'  \ottsym{,}  {\source x}  \ottsym{:}  {\source \tau }_{{\mathrm{1}}}  }  \Leftarrow  \source {  {\source \tau }_{{\mathrm{2}}}  } )  \rightsquigarrow  \intermed {  {\intermed M}'_{{\mathrm{2}}}  } 
      \label{eq:ctx_refl60} \\
       \source {  {\source { \Gamma_C } }  } ; \source {  {\source \Gamma}'  }  \vdash_{ty}^{\intermed {M} }  \source {  {\source \tau }_{{\mathrm{1}}}  }\,{ \stoi{ \rightsquigarrow  \intermed {  {\intermed \sigma}_{{\mathrm{1}}}  } } } 
      \label{eq:ctx_refl61}
    \end{gather}
    We know from \textsc{sCtx-tyEnvTm}, in combination with
    Equation~\ref{eq:ctx_refl61} and the 4\textsuperscript{th} and
    6\textsuperscript{th} hypothesis that:
    \begin{gather}
       \vdash_{ctx}^{\intermed {M} }  \source {  { \source P }  } ; \source {  {\source { \Gamma_C } }  } ; \source {  {\source \Gamma}'  \ottsym{,}  {\source x}  \ottsym{:}  {\source \tau }_{{\mathrm{1}}}  }  \rightsquigarrow  \intermed {  {\intermed \Sigma}_{{\mathrm{1}}}  } ; \intermed {  {\intermed { \Gamma_C } }  } ; \intermed {  {\intermed \Gamma}'  \ottsym{,}  {\intermed x}  \ottsym{:}  {\intermed \sigma}_{{\mathrm{1}}}  } 
      \label{eq:ctx_refl62} \\
       \vdash_{ctx}^{\intermed {M} }  \source {  { \source P }  } ; \source {  {\source { \Gamma_C } }  } ; \source {  {\source \Gamma}'  \ottsym{,}  {\source x}  \ottsym{:}  {\source \tau }_{{\mathrm{1}}}  }  \rightsquigarrow  \intermed {  {\intermed \Sigma}_{{\mathrm{2}}}  } ; \intermed {  {\intermed { \Gamma_C } }  } ; \intermed {  {\intermed \Gamma}'  \ottsym{,}  {\intermed x}  \ottsym{:}  {\intermed \sigma}_{{\mathrm{1}}}  } 
      \label{eq:ctx_refl63}
    \end{gather}
    By applying the induction hypothesis to Equations~\ref{eq:ctx_refl59}
    and~\ref{eq:ctx_refl60}, in combination with Equations~\ref{eq:ctx_refl62}
    and~\ref{eq:ctx_refl63}, we get:
    \begin{equation}
       \intermed {  {\intermed \Sigma}_{{\mathrm{1}}}  } : \intermed {  {\intermed M}'_{{\mathrm{1}}}  }  \simeq_{log}  \intermed {  {\intermed \Sigma}_{{\mathrm{2}}}  } : \intermed {  {\intermed M}'_{{\mathrm{2}}}  } : ( \intermed {  {\intermed { \Gamma_C } }  } ; \intermed {  {\intermed \Gamma}  }  \Rightarrow  \intermed {  {\intermed \sigma}  } )  \mapsto  ( \intermed {  {\intermed { \Gamma_C } }  } ; \intermed {  {\intermed \Gamma}'  \ottsym{,}  {\intermed x}  \ottsym{:}  {\intermed \sigma}_{{\mathrm{1}}}  }  \Rightarrow  \intermed {  {\intermed \sigma}_{{\mathrm{2}}}  } ) 
      \label{eq:ctx_refl64}
    \end{equation}
    By unfolding the definition of logical equivalence in
    Equation~\ref{eq:ctx_refl64}, we get:
    \begin{gather}
      \forall {\intermed e}'_{{\mathrm{1}}}, {\intermed e}'_{{\mathrm{2}}} :  \intermed {  {\intermed { \Gamma_C } }  } ; \intermed {  {\intermed \Gamma}  }  \vdash  \intermed {  {\intermed \Sigma}_{{\mathrm{1}}}  } : \intermed {  {\intermed e}'_{{\mathrm{1}}}  }  \simeq_{log}  \intermed {  {\intermed \Sigma}_{{\mathrm{2}}}  } : \intermed {  {\intermed e}'_{{\mathrm{2}}}  } : \intermed {  {\intermed \sigma}  } 
      \label{eq:ctx_refl65} \\
      \Rightarrow  \intermed {  {\intermed { \Gamma_C } }  } ; \intermed {  {\intermed \Gamma}'  \ottsym{,}  {\intermed x}  \ottsym{:}  {\intermed \sigma}_{{\mathrm{1}}}  }  \vdash  \intermed {  {\intermed \Sigma}_{{\mathrm{1}}}  } : \intermed {   {\intermed M}'_{{\mathrm{1}}}  [  {\intermed e}'_{{\mathrm{1}}}  ]   }  \simeq_{log}  \intermed {  {\intermed \Sigma}_{{\mathrm{2}}}  } : \intermed {   {\intermed M}'_{{\mathrm{2}}}  [  {\intermed e}'_{{\mathrm{2}}}  ]   } : \intermed {  {\intermed \sigma}_{{\mathrm{2}}}  } 
      \label{eq:ctx_refl66}
    \end{gather}
    This result, together with Equation~\ref{eq:ctx_refl0}, tells us that:
    \begin{equation}
       \intermed {  {\intermed { \Gamma_C } }  } ; \intermed {  {\intermed \Gamma}'  \ottsym{,}  {\intermed x}  \ottsym{:}  {\intermed \sigma}_{{\mathrm{1}}}  }  \vdash  \intermed {  {\intermed \Sigma}_{{\mathrm{1}}}  } : \intermed {   {\intermed M}'_{{\mathrm{1}}}  [  {\intermed e}_{{\mathrm{1}}}  ]   }  \simeq_{log}  \intermed {  {\intermed \Sigma}_{{\mathrm{2}}}  } : \intermed {   {\intermed M}'_{{\mathrm{2}}}  [  {\intermed e}_{{\mathrm{2}}}  ]   } : \intermed {  {\intermed \sigma}_{{\mathrm{2}}}  } 
      \label{eq:ctx_refl67}
    \end{equation}
    Goal~\ref{eqg:ctx_refl7} follows from Lemma~\ref{lemma:compat_abs}, together
    with Equation~\ref{eq:ctx_refl67}. \\
  }
  \proofStep{\textsc{sM-inf-check-inf}}
  {
     \source {  {\source M}  } : ( \source {  { \source P }  } ; \source {  {\source { \Gamma_C } }  } ; \source {  {\source \Gamma}  }  \Rightarrow  \source {  {\source \tau }  } )  \mapsto  ( \source {  { \source P }  } ; \source {  {\source { \Gamma_C } }  } ; \source {  {\source \Gamma}'  }  \Leftarrow  \source {  {\source \tau }'  } )  \rightsquigarrow  \intermed {  {\intermed M}_{{\mathrm{1}}}  } 
  }
  {
    By case analysis, we know that the final step in the second derivation has
    to be either \textsc{sM-inf-check-abs} or \textsc{sM-inf-check-inf}. Note
    however that in the case of \textsc{sM-inf-check-abs}, \({\source M}\) would have
    to be of the form \( \source {  {\source M}  } = \source {  \lambda  {\source x}  \ottsym{.}  {\source M}'  } \). In this case, no matching
    inference rules exist, meaning that this is an impossible case.
    Consequently, the final step in the second derivation can only be
    \textsc{sM-inf-check-inf}. This means that:
    \begin{gather}
       \source {  {\source M}  } : ( \source {  { \source P }  } ; \source {  {\source { \Gamma_C } }  } ; \source {  {\source \Gamma}  }  \Rightarrow  \source {  {\source \tau }  } )  \mapsto  ( \source {  { \source P }  } ; \source {  {\source { \Gamma_C } }  } ; \source {  {\source \Gamma}'  }  \Leftarrow  \source {  {\source \tau }'  } )  \rightsquigarrow  \intermed {  {\intermed M}_{{\mathrm{1}}}  } 
      \label{eq:ctx_refl68} \\
       \source {  {\source M}  } : ( \source {  { \source P }  } ; \source {  {\source { \Gamma_C } }  } ; \source {  {\source \Gamma}  }  \Rightarrow  \source {  {\source \tau }  } )  \mapsto  ( \source {  { \source P }  } ; \source {  {\source { \Gamma_C } }  } ; \source {  {\source \Gamma}'  }  \Leftarrow  \source {  {\source \tau }'  } )  \rightsquigarrow  \intermed {  {\intermed M}_{{\mathrm{2}}}  } 
      \label{eq:ctx_refl69}
    \end{gather}
    The goal to be proven is the following:
    \begin{equation}
       \intermed {  {\intermed { \Gamma_C } }  } ; \intermed {  {\intermed \Gamma}'  }  \vdash  \intermed {  {\intermed \Sigma}_{{\mathrm{1}}}  } : \intermed {   {\intermed M}_{{\mathrm{1}}}  [  {\intermed e}_{{\mathrm{1}}}  ]   }  \simeq_{log}  \intermed {  {\intermed \Sigma}_{{\mathrm{2}}}  } : \intermed {   {\intermed M}_{{\mathrm{2}}}  [  {\intermed e}_{{\mathrm{2}}}  ]   } : \intermed {  {\intermed \sigma}'  } 
      \label{eqg:ctx_refl8}
    \end{equation}
    From the rule premise we know that:
    \begin{gather}
       \source {  {\source M}  } : ( \source {  { \source P }  } ; \source {  {\source { \Gamma_C } }  } ; \source {  {\source \Gamma}  }  \Rightarrow  \source {  {\source \tau }  } )  \mapsto  ( \source {  { \source P }  } ; \source {  {\source { \Gamma_C } }  } ; \source {  {\source \Gamma}'  }  \Rightarrow  \source {  {\source \tau }'  } )  \rightsquigarrow  \intermed {  {\intermed M}_{{\mathrm{1}}}  } 
      \label{eq:ctx_refl70} \\
       \source {  {\source M}  } : ( \source {  { \source P }  } ; \source {  {\source { \Gamma_C } }  } ; \source {  {\source \Gamma}  }  \Rightarrow  \source {  {\source \tau }  } )  \mapsto  ( \source {  { \source P }  } ; \source {  {\source { \Gamma_C } }  } ; \source {  {\source \Gamma}'  }  \Rightarrow  \source {  {\source \tau }'  } )  \rightsquigarrow  \intermed {  {\intermed M}_{{\mathrm{2}}}  } 
      \label{eq:ctx_refl71} 
    \end{gather}
    Goal~\ref{eqg:ctx_refl8} follows directly by applying Part 1 of this theorem
    to Equations~\ref{eq:ctx_refl70} and~\ref{eq:ctx_refl71}. \\
  }
\item[Part 3] By case analysis on the first typing derivation. \\
  Similar to Part 1. \\
\item[Part 4] By case analysis on the first typing derivation. \\
  Similar to Part 2. \\
\end{description}
\qed

\begin{theoremB}[Value Relation for Dictionary Values]\leavevmode
  If \( \intermed {  {\intermed \Sigma}_{{\mathrm{1}}}  } ; \intermed {  {\intermed { \Gamma_C } }  } ; \intermed {  \bullet  }  \vdash_{d}  \intermed {  {\intermed {dv} }_{{\mathrm{1}}}  } : \intermed {  {\intermed Q}  } \itot{\,  \rightsquigarrow  \target {  {\target e }_{{\mathrm{1}}}  } } \)
  and \( \intermed {  {\intermed \Sigma}_{{\mathrm{2}}}  } ; \intermed {  {\intermed { \Gamma_C } }  } ; \intermed {  \bullet  }  \vdash_{d}  \intermed {  {\intermed {dv} }_{{\mathrm{2}}}  } : \intermed {  {\intermed Q}  } \itot{\,  \rightsquigarrow  \target {  {\target e }_{{\mathrm{2}}}  } } \) \\
  and  \( \intermed {  {\intermed { \Gamma_C } }  }  \vdash  \intermed {  {\intermed \Sigma}_{{\mathrm{1}}}  }  \simeq_{log}  \intermed {  {\intermed \Sigma}_{{\mathrm{2}}}  } \) \\
  then \( ( \intermed {  {\intermed \Sigma}_{{\mathrm{1}}}  } : \intermed {  {\intermed {dv} }_{{\mathrm{1}}}  } , \intermed {  {\intermed \Sigma}_{{\mathrm{2}}}  } : \intermed {  {\intermed {dv} }_{{\mathrm{2}}}  } ) \in \mathcal{V} \llbracket \intermed {  {\intermed Q}  } \rrbracket ^{\intermed {  {\intermed { \Gamma_C } }  } }_{\intermed {  \bullet  } } \).
  \label{thmA:dval_rel}
\end{theoremB}

\proof
By induction on the size of \(\intermed{{\intermed {dv} }_{{\mathrm{1}}}}\) and \(\intermed{{\intermed {dv} }_{{\mathrm{2}}}}\). 

From the definition of dictionary values we know that
\begin{gather*}
   \intermed {  {\intermed {dv} }_{{\mathrm{1}}}  } = \intermed {  {\intermed D } \, \overline {\intermed \sigma}_{\ottmv{j}} \, {\overline {\intermed { dv } } }_{\ottmv{i}}  }  \\
   \intermed {  {\intermed {dv} }_{{\mathrm{2}}}  } = \intermed {  {\intermed D }' \, \overline {\intermed \sigma}'_{\ottmv{h}} \, {\overline {\intermed { dv } } }'_{\ottmv{k}}  } 
\end{gather*}
For some \(\intermed{{\intermed D }}\), \(\intermed{{\intermed D }'}\), \(\intermed{\overline {\intermed \sigma}_{\ottmv{j}}}\), \(\intermed{\overline {\intermed \sigma}'_{\ottmv{h}}}\),
\(\intermed{{\overline {\intermed { dv } } }_{\ottmv{i}}}\) and \(\intermed{{\overline {\intermed { dv } } }'_{\ottmv{k}}}\).

By case analysis on both the 1\textsuperscript{st} and 2\textsuperscript{nd}
hypothesis (\textsc{D-con}), we know that:
\begin{gather}
   ( \intermed {  {\intermed D }  } : \intermed {  \forall  {\overline {\intermed a} }_{\ottmv{j}}  \ottsym{.}  {\overline {\intermed Q} }_{\ottmv{i}}  \Rightarrow  {\intermed Q}'  } ) . \intermed {  {\intermed m}  }  \mapsto  \intermed {  {\intermed e}  }  \in  \intermed {  {\intermed \Sigma}_{{\mathrm{1}}}  } ~
  \textit{where}~ \intermed {  {\intermed Q}  } = \intermed {  [  \overline {\intermed \sigma}_{\ottmv{j}}  /  {\overline {\intermed a} }_{\ottmv{j}}  ]  {\intermed Q}'  } 
  \label{eq:dval1} \\
  \ottcomp{ \intermed {  {\intermed { \Gamma_C } }  } ; \intermed {  \bullet  }  \vdash_{ty}  \intermed {  {\intermed \sigma}_{\ottmv{j}}  } \itot{ \rightsquigarrow  \target {  {\target \sigma }_{\ottmv{j}}  } } }{\ottmv{j}}
  \label{eq:dval2} \\
  \ottcomp{ \intermed {  {\intermed \Sigma}_{{\mathrm{1}}}  } ; \intermed {  {\intermed { \Gamma_C } }  } ; \intermed {  \bullet  }  \vdash_{d}  \intermed {  {\intermed {dv} }_{\ottmv{i}}  } : \intermed {  [  \overline {\intermed \sigma}_{\ottmv{j}}  /  {\overline {\intermed a} }_{\ottmv{j}}  ]  {\intermed Q}_{\ottmv{i}}  } \itot{\,  \rightsquigarrow  \target {  {\target e }_{\ottmv{i}}  } } }{\ottmv{i}}
  \label{eq:dval3} \\
   \intermed {  {\intermed \Sigma}_{{\mathrm{11}}}  } ; \intermed {  {\intermed { \Gamma_C } }  } ; \intermed {  \bullet  }  \vdash_{tm}  \intermed {  {\intermed e}  } : \intermed {  \forall  {\overline {\intermed a} }_{\ottmv{j}}  \ottsym{.}   {\overline {\intermed Q} }_{\ottmv{i}}  \Rightarrow  {\intermed \sigma}   } \itot{ \rightsquigarrow  \target {  {\target e }  } } ~
  \textit{where}~ \intermed {  {\intermed \Sigma}_{{\mathrm{1}}}  } = \intermed {  {\intermed \Sigma}_{{\mathrm{11}}}  \ottsym{,}  \ottsym{(}  {\intermed D }  \ottsym{:}  \forall  {\overline {\intermed a} }_{\ottmv{j}}  \ottsym{.}  {\overline {\intermed Q} }_{\ottmv{i}}  \Rightarrow  {\intermed Q}'  \ottsym{)}  \ottsym{.}  {\intermed m}  \mapsto  {\intermed e}  \ottsym{,}  {\intermed \Sigma}_{{\mathrm{12}}}  } 
  \label{eq:dval4} \\
   \vdash_{ctx}  \intermed {  {\intermed \Sigma}_{{\mathrm{1}}}  } ; \intermed {  {\intermed { \Gamma_C } }  } ; \intermed {  \bullet  } 
  \label{eq:dval4b} \\
   ( \intermed {  {\intermed D }'  } : \intermed {  \forall  {\overline {\intermed b} }_{\ottmv{h}}  \ottsym{.}  {\overline {\intermed Q} }'_{\ottmv{k}}  \Rightarrow  {\intermed Q}''  } ) . \intermed {  {\intermed m}'  }  \mapsto  \intermed {  {\intermed e}'  }  \in  \intermed {  {\intermed \Sigma}_{{\mathrm{2}}}  } ~
  \textit{where}~ \intermed {  {\intermed Q}  } = \intermed {  [  \overline {\intermed \sigma}'_{\ottmv{h}}  /  {\overline {\intermed b} }_{\ottmv{h}}  ]  {\intermed Q}''  } 
  \label{eq:dval5} \\
  \ottcomp{ \intermed {  {\intermed { \Gamma_C } }  } ; \intermed {  \bullet  }  \vdash_{ty}  \intermed {  {\intermed \sigma}'_{\ottmv{h}}  } \itot{ \rightsquigarrow  \target {  {\target \sigma }'_{\ottmv{h}}  } } }{\ottmv{h}}
  \label{eq:dval6} \\
  \ottcomp{ \intermed {  {\intermed \Sigma}_{{\mathrm{2}}}  } ; \intermed {  {\intermed { \Gamma_C } }  } ; \intermed {  \bullet  }  \vdash_{d}  \intermed {  {\intermed {dv} }'_{\ottmv{k}}  } : \intermed {  [  \overline {\intermed \sigma}'_{\ottmv{h}}  /  {\overline {\intermed b} }_{\ottmv{h}}  ]  {\intermed Q}'_{\ottmv{k}}  } \itot{\,  \rightsquigarrow  \target {  {\target e }'_{\ottmv{k}}  } } }{\ottmv{k}}
  \label{eq:dval7} \\
   \intermed {  {\intermed \Sigma}_{{\mathrm{21}}}  } ; \intermed {  {\intermed { \Gamma_C } }  } ; \intermed {  \bullet  }  \vdash_{tm}  \intermed {  {\intermed e}'  } : \intermed {  \forall  {\overline {\intermed b} }_{\ottmv{h}}  \ottsym{.}   {\overline {\intermed Q} }'_{\ottmv{k}}  \Rightarrow  {\intermed \sigma}'   } \itot{ \rightsquigarrow  \target {  {\target e }'  } } ~
  \textit{where}~ \intermed {  {\intermed \Sigma}_{{\mathrm{2}}}  } = \intermed {  {\intermed \Sigma}_{{\mathrm{21}}}  \ottsym{,}  \ottsym{(}  {\intermed D }'  \ottsym{:}  \forall  {\overline {\intermed b} }_{\ottmv{h}}  \ottsym{.}  {\overline {\intermed Q} }'_{\ottmv{k}}  \Rightarrow  {\intermed Q}''  \ottsym{)}  \ottsym{.}  {\intermed m}'  \mapsto  {\intermed e}'  \ottsym{,}  {\intermed \Sigma}_{{\mathrm{22}}}  } 
  \label{eq:dval8} \\
   \vdash_{ctx}  \intermed {  {\intermed \Sigma}_{{\mathrm{2}}}  } ; \intermed {  {\intermed { \Gamma_C } }  } ; \intermed {  \bullet  } 
  \label{eq:dval8b}
\end{gather}
By case analysis on Equations~\ref{eq:dval4b} and~\ref{eq:dval8b} (\textsc{iCtx-MEnv})
and the definition of logical equivalence in the 3\textsuperscript{rd} hypothesis,
it follows from Equations~\ref{eq:dval1} and~\ref{eq:dval5} that:
\begin{gather}
   \intermed {  {\intermed D }  } = \intermed {  {\intermed D }'  } 
  \nonumber \\
   \intermed {  {\overline {\intermed a} }_{\ottmv{j}}  } = \intermed {  {\overline {\intermed b} }_{\ottmv{h}}  } 
  \nonumber \\
   \intermed {  {\overline {\intermed Q} }_{\ottmv{i}}  } = \intermed {  {\overline {\intermed Q} }'_{\ottmv{k}}  } 
  \nonumber \\
   \intermed {  {\intermed Q}'  } = \intermed {  {\intermed Q}''  } 
  \nonumber \\
   \intermed {  {\intermed m}  } = \intermed {  {\intermed m}'  } 
  \nonumber \\
   \intermed {  {\intermed { \Gamma_C } }  }  \vdash  \intermed {  {\intermed \Sigma}_{{\mathrm{11}}}  }  \simeq_{log}  \intermed {  {\intermed \Sigma}_{{\mathrm{21}}}  } 
  \nonumber \\
   \intermed {  {\intermed { \Gamma_C } }  }  \vdash  \intermed {  {\intermed \Sigma}_{{\mathrm{12}}}  }  \simeq_{log}  \intermed {  {\intermed \Sigma}_{{\mathrm{22}}}  } 
  \nonumber \\
   \intermed {  {\intermed { \Gamma_C } }  } ; \intermed {  \bullet  }  \vdash  \intermed {  {\intermed \Sigma}_{{\mathrm{11}}}  } : \intermed {  {\intermed e}  }  \simeq_{log}  \intermed {  {\intermed \Sigma}_{{\mathrm{21}}}  } : \intermed {  {\intermed e}'  } : \intermed {  \forall  {\overline {\intermed a} }_{\ottmv{j}}  \ottsym{.}   {\overline {\intermed Q} }_{\ottmv{i}}  \Rightarrow  {\intermed \sigma}   } 
  \nonumber
\end{gather}
Consequently, we also know that \(\intermed{\ottmv{j}} = \intermed{\ottmv{h}}\) and \(\intermed{\ottmv{i}} = \intermed{\ottmv{k}}\).

Furthermore, rule \textsc{iCtx-MEnv} also tells us that
\( \textbf {unambig}  ( \intermed {  \forall  {\overline {\intermed a} }_{\ottmv{j}}  \ottsym{.}  {\overline {\intermed Q} }_{\ottmv{i}}  \Rightarrow  {\intermed Q}'  } ) \).
The definition of unambiguity thus gives us
\( \intermed {  {\overline {\intermed a} }_{\ottmv{j}}  }  \in   \textbf {fv}  ( \intermed {  {\intermed Q}'  } ) \). 
This, in combination with Equations~\ref{eq:dval1} and \ref{eq:dval5}
tells us that \( \intermed {  \overline {\intermed \sigma}_{\ottmv{j}}  } = \intermed {  \overline {\intermed \sigma}'_{\ottmv{h}}  } \).

Unfolding the definition of the \(\mathcal{V}\) relation, reduces the goal to be
proven to:
\begin{gather}
  \ottcomp{ ( \intermed {  {\intermed \Sigma}_{{\mathrm{1}}}  } : \intermed {  {\intermed {dv} }_{\ottmv{i}}  } , \intermed {  {\intermed \Sigma}_{{\mathrm{2}}}  } : \intermed {  {\intermed {dv} }'_{\ottmv{k}}  } ) \in \mathcal{V} \llbracket \intermed {  [  \overline {\intermed \sigma}_{\ottmv{j}}  /  {\overline {\intermed a} }_{\ottmv{j}}  ]  {\intermed Q}_{\ottmv{i}}  } \rrbracket ^{\intermed {  {\intermed { \Gamma_C } }  } }_{\intermed {  \bullet  } } }{\ottmv{i}}
  \label{eq:dval9} \\
   \intermed {  {\intermed \Sigma}_{{\mathrm{1}}}  } ; \intermed {  {\intermed { \Gamma_C } }  } ; \intermed {  \bullet  }  \vdash_{d}  \intermed {  {\intermed D } \, \overline {\intermed \sigma}_{\ottmv{j}} \, {\overline {\intermed { dv } } }_{\ottmv{i}}  } : \intermed {  {\intermed Q}  } \itot{\,  \rightsquigarrow  \target {  {\target e }_{{\mathrm{1}}}  } } 
  \label{eq:dval10} \\
   \intermed {  {\intermed \Sigma}_{{\mathrm{2}}}  } ; \intermed {  {\intermed { \Gamma_C } }  } ; \intermed {  \bullet  }  \vdash_{d}  \intermed {  {\intermed D } \, \overline {\intermed \sigma}_{\ottmv{j}} \, {\overline {\intermed { dv } } }'_{\ottmv{k}}  } : \intermed {  {\intermed Q}  } \itot{\,  \rightsquigarrow  \target {  {\target e }_{{\mathrm{2}}}  } } 
  \label{eq:dval11} \\
   ( \intermed {  {\intermed D }  } : \intermed {  \forall  {\overline {\intermed a} }_{\ottmv{j}}  \ottsym{.}  {\overline {\intermed Q} }_{\ottmv{i}}  \Rightarrow  {\intermed Q}'  } ) . \intermed {  {\intermed m}  }  \mapsto  \intermed {  {\intermed e}  }  \in  \intermed {  {\intermed \Sigma}_{{\mathrm{1}}}  } ~
  \textit{where}~ \intermed {  {\intermed Q}  } = \intermed {  [  \overline {\intermed \sigma}_{\ottmv{j}}  /  {\overline {\intermed a} }_{\ottmv{j}}  ]  {\intermed Q}'  } 
  \label{eq:dval12}
\end{gather}
Goals~\ref{eq:dval10} and \ref{eq:dval11} are given by the hypothesis.
Goal~\ref{eq:dval12} follows directly from Equation~\ref{eq:dval1}.
Finally, Goal~\ref{eq:dval9} follows by applying the induction hypothesis on
Equations~\ref{eq:dval3} and \ref{eq:dval7}. \\
\qed

\begin{theoremB}[Environment Equivalence Preservation]
  If \( \vdash_{ctx}^{\intermed {M} }  \source {  { \source P }  } ; \source {  {\source { \Gamma_C } }  } ; \source {  {\source \Gamma}  }  \rightsquigarrow  \intermed {  {\intermed \Sigma}_{{\mathrm{1}}}  } ; \intermed {  {\intermed { \Gamma_C } }  } ; \intermed {  {\intermed \Gamma}  } \)
  and \( \vdash_{ctx}^{\intermed {M} }  \source {  { \source P }  } ; \source {  {\source { \Gamma_C } }  } ; \source {  {\source \Gamma}  }  \rightsquigarrow  \intermed {  {\intermed \Sigma}_{{\mathrm{2}}}  } ; \intermed {  {\intermed { \Gamma_C } }  } ; \intermed {  {\intermed \Gamma}  } \) \\
  then \( \intermed {  {\intermed { \Gamma_C } }  }  \vdash  \intermed {  {\intermed \Sigma}_{{\mathrm{1}}}  }  \simeq_{log}  \intermed {  {\intermed \Sigma}_{{\mathrm{2}}}  } \).
  \label{thmA:env_eq_pres}
\end{theoremB} 

\begin{figure}[htp]
  \centering

  \def\svgscale{0.75}
\begingroup%
  \makeatletter%
  \providecommand\color[2][]{%
    \errmessage{(Inkscape) Color is used for the text in Inkscape, but the package 'color.sty' is not loaded}%
    \renewcommand\color[2][]{}%
  }%
  \providecommand\transparent[1]{%
    \errmessage{(Inkscape) Transparency is used (non-zero) for the text in Inkscape, but the package 'transparent.sty' is not loaded}%
    \renewcommand\transparent[1]{}%
  }%
  \providecommand\rotatebox[2]{#2}%
  \newcommand*\fsize{\dimexpr\f@size pt\relax}%
  \newcommand*\lineheight[1]{\fontsize{\fsize}{#1\fsize}\selectfont}%
  \ifx\svgwidth\undefined%
    \setlength{\unitlength}{93.5922019bp}%
    \ifx\svgscale\undefined%
      \relax%
    \else%
      \setlength{\unitlength}{\unitlength * \real{\svgscale}}%
    \fi%
  \else%
    \setlength{\unitlength}{\svgwidth}%
  \fi%
  \global\let\svgwidth\undefined%
  \global\let\svgscale\undefined%
  \makeatother%
  \begin{picture}(1,2.0087144)%
    \lineheight{1}%
    \setlength\tabcolsep{0pt}%
    \put(0,0){\includegraphics[width=\unitlength,page=1]{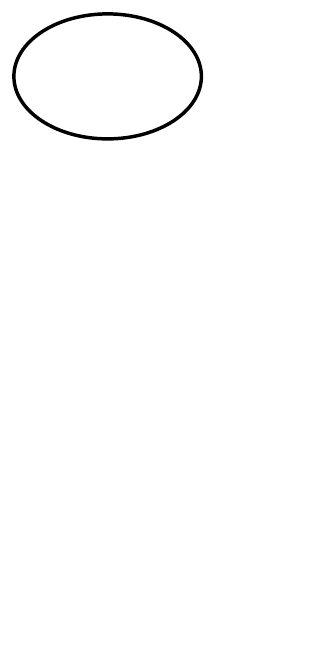}}%
    \put(0.33122418,1.7341189){\color[rgb]{0,0,0}\makebox(0,0)[t]{\lineheight{1.25}\smash{\begin{tabular}[t]{c}T25\end{tabular}}}}%
    \put(0,0){\includegraphics[width=\unitlength,page=2]{mut_dep_graph.pdf}}%
    \put(0.33122418,0.96482402){\color[rgb]{0,0,0}\makebox(0,0)[t]{\lineheight{1.25}\smash{\begin{tabular}[t]{c}T28\end{tabular}}}}%
    \put(0,0){\includegraphics[width=\unitlength,page=3]{mut_dep_graph.pdf}}%
    \put(0.33122418,0.19552912){\color[rgb]{0,0,0}\makebox(0,0)[t]{\lineheight{1.25}\smash{\begin{tabular}[t]{c}T27\end{tabular}}}}%
    \put(0,0){\includegraphics[width=\unitlength,page=4]{mut_dep_graph.pdf}}%
  \end{picture}%
\endgroup%

  \caption{Dependency graph for Theorems~\ref{thmA:env_eq_pres},
    \ref{thmA:dict_cohA} and
    \ref{thmA:coh_exprA}}
  \label{fig:dep_graph1}
\end{figure}

\proof 
By structural induction on \({ \source P }\) and mutually proven with
Theorems~\ref{thmA:dict_cohA} and~\ref{thmA:coh_exprA} (see
Figure~\ref{fig:dep_graph1}). 
Note that at the dependency between
Theorem~\ref{thmA:env_eq_pres} and~\ref{thmA:coh_exprA}, the size of \({ \source P }\) is
strictly decreasing, whereas \({ \source P }\) remains constant at every other
dependency. Because of this, the size of \({ \source P }\) is strictly decreasing in
every cycle. 
Consequently, the induction remains well-founded. \\
\proofStep{\( \source {  { \source P }  } = \source {  \bullet  } \)}{}
{
  By case analysis on the 1\textsuperscript{st} and 2\textsuperscript{nd}
  hypothesis:
  \begin{equation*}
    \intermed{{\intermed \Sigma}_{{\mathrm{1}}}} = \intermed{{\intermed \Sigma}_{{\mathrm{2}}}} = \intermed{\bullet}
  \end{equation*}
  The goal follows from \textsc{ctxLog-empty}. \\
}
\proofStep{\( \source {  { \source P }  } = \source {  { \source P }'  }, \source {  \ottsym{(}  {\source D }  \ottsym{:}  {\source C}  \ottsym{)}  \ottsym{.}  {\source m}  \mapsto  \bullet  \ottsym{,}  {\overline {\source b} }_{\ottmv{j}}  \ottsym{,}  {\overline {\source \delta} }_{\ottmv{i}}  \ottsym{:}  {\overline {\source Q} }_{\ottmv{i}}  \ottsym{,}  {\overline {\source a} }_{\ottmv{k}}  \ottsym{,}  {\overline {\source \delta} }_{\ottmv{h}}  \ottsym{:}  [  {\source \tau }  /  {\source a }  ]  {\overline {\source Q} }_{\ottmv{h}}  \ottsym{:}  {\source e}  } \)}{}
{
  By case analysis on the 1\textsuperscript{st} and 2\textsuperscript{nd}
  hypothesis (\textsc{sCtx-pgmInst}):
  \begin{gather}
     \intermed {  {\intermed \Sigma}_{{\mathrm{1}}}  } = \intermed {  {\intermed \Sigma}'_{{\mathrm{1}}}  \ottsym{,}  \ottsym{(}  {\intermed D }  \ottsym{:}  {\intermed C}  \ottsym{)}  \ottsym{.}  {\intermed m}  \mapsto  \Lambda  {\overline {\intermed b} }_{\ottmv{j}}  \ottsym{.}  \lambda  {\overline {\intermed \delta} }_{\ottmv{i}}  \ottsym{:}  {\overline {\intermed Q} }_{\ottmv{i}}  \ottsym{.}  \Lambda  {\overline {\intermed a} }_{\ottmv{k}}  \ottsym{.}  \lambda  {\overline {\intermed \delta} }_{\ottmv{h}}  \ottsym{:}  [  {\intermed \sigma}  /  {\intermed a }  ]  {\overline {\intermed Q} }_{\ottmv{h}}  \ottsym{.}  {\intermed e}_{{\mathrm{1}}}  } 
    \nonumber \\
     \intermed {  {\intermed \Sigma}_{{\mathrm{2}}}  } = \intermed {  {\intermed \Sigma}'_{{\mathrm{2}}}  \ottsym{,}  \ottsym{(}  {\intermed D }  \ottsym{:}  {\intermed C}'  \ottsym{)}  \ottsym{.}  {\intermed m}  \mapsto  \Lambda  {\overline {\intermed b} }_{\ottmv{j}}  \ottsym{.}  \lambda  {\overline {\intermed \delta} }_{\ottmv{i}}  \ottsym{:}  {\overline {\intermed Q} }'_{\ottmv{i}}  \ottsym{.}  \Lambda  {\overline {\intermed a} }_{\ottmv{k}}  \ottsym{.}  \lambda  {\overline {\intermed \delta} }_{\ottmv{h}}  \ottsym{:}  [  {\intermed \sigma}  /  {\intermed a }  ]  {\overline {\intermed Q} }'_{\ottmv{h}}  \ottsym{.}  {\intermed e}_{{\mathrm{2}}}  } 
    \nonumber \\
     \source {  { \source P }'  } ; \source {  {\source { \Gamma_C } }  } ; \source {  \bullet  \ottsym{,}  {\overline {\source b} }_{\ottmv{j}}  \ottsym{,}  {\overline {\source \delta} }_{\ottmv{i}}  \ottsym{:}  {\overline {\source Q} }_{\ottmv{i}}  \ottsym{,}  {\overline {\source a} }_{\ottmv{k}}  \ottsym{,}  {\overline {\source \delta} }_{\ottmv{h}}  \ottsym{:}  [  {\source \tau }  /  {\source a }  ]  {\overline {\source Q} }_{\ottmv{h}}  }  \vdash_{tm}^{\intermed {M} }  \source {  {\source e}  } \Leftarrow \source {  [  {\source \tau }  /  {\source a }  ]  {\source \tau }'  } \stoi{ \rightsquigarrow  \intermed {  {\intermed e}_{{\mathrm{1}}}  } } 
    \label{eq:env_pres3} \\
     \source {  { \source P }'  } ; \source {  {\source { \Gamma_C } }  } ; \source {  \bullet  \ottsym{,}  {\overline {\source b} }_{\ottmv{j}}  \ottsym{,}  {\overline {\source \delta} }_{\ottmv{i}}  \ottsym{:}  {\overline {\source Q} }_{\ottmv{i}}  \ottsym{,}  {\overline {\source a} }_{\ottmv{k}}  \ottsym{,}  {\overline {\source \delta} }_{\ottmv{h}}  \ottsym{:}  [  {\source \tau }  /  {\source a }  ]  {\overline {\source Q} }_{\ottmv{h}}  }  \vdash_{tm}^{\intermed {M} }  \source {  {\source e}  } \Leftarrow \source {  [  {\source \tau }  /  {\source a }  ]  {\source \tau }'  } \stoi{ \rightsquigarrow  \intermed {  {\intermed e}_{{\mathrm{2}}}  } } 
    \label{eq:env_pres4} \\
     \vdash_{ctx}^{\intermed {M} }  \source {  { \source P }'  } ; \source {  {\source { \Gamma_C } }  } ; \source {  {\source \Gamma}  }  \rightsquigarrow  \intermed {  {\intermed \Sigma}'_{{\mathrm{1}}}  } ; \intermed {  {\intermed { \Gamma_C } }  } ; \intermed {  {\intermed \Gamma}  } 
    \label{eq:env_pres5} \\
     \vdash_{ctx}^{\intermed {M} }  \source {  { \source P }'  } ; \source {  {\source { \Gamma_C } }  } ; \source {  {\source \Gamma}  }  \rightsquigarrow  \intermed {  {\intermed \Sigma}'_{{\mathrm{2}}}  } ; \intermed {  {\intermed { \Gamma_C } }  } ; \intermed {  {\intermed \Gamma}  } 
    \label{eq:env_pres6}
  \end{gather}
  Since the elaboration from \srcL{} constraints to \interL{} constraints is
  entirely deterministic (Lemma~\ref{lemma:c_elab_unique}), we know that 
  \( \intermed {  {\intermed C}'  } = \intermed {  {\intermed C}  } \), \( \intermed {  {\overline {\intermed Q} }'_{\ottmv{i}}  } = \intermed {  {\overline {\intermed Q} }_{\ottmv{i}}  } \), \( \intermed {  {\overline {\intermed Q} }'_{\ottmv{h}}  } = \intermed {  {\overline {\intermed Q} }_{\ottmv{h}}  } \) and
  consequently that \( \intermed {  [  {\intermed \sigma}  /  {\intermed a }  ]  {\overline {\intermed Q} }'_{\ottmv{h}}  } = \intermed {  [  {\intermed \sigma}  /  {\intermed a }  ]  {\overline {\intermed Q} }_{\ottmv{h}}  } \).

  From the induction hypothesis, together with Equations~\ref{eq:env_pres5}
  and~\ref{eq:env_pres6}, we get that:
  \begin{equation}
     \intermed {  {\intermed { \Gamma_C } }  }  \vdash  \intermed {  {\intermed \Sigma}'_{{\mathrm{1}}}  }  \simeq_{log}  \intermed {  {\intermed \Sigma}'_{{\mathrm{2}}}  } 
    \label{eq:env_pres7}
  \end{equation}
  From Expression Coherence Theorem A (Theorem~\ref{thmA:coh_exprA}), in
  combination with Equations~\ref{eq:env_pres3}, \ref{eq:env_pres4},
  \ref{eq:env_pres5} and~\ref{eq:env_pres6}, we know:
  \begin{equation}
     \intermed {  {\intermed { \Gamma_C } }  } ; \intermed {  {\intermed \Gamma}  }  \vdash  \intermed {  {\intermed \Sigma}'_{{\mathrm{1}}}  } : \intermed {  {\intermed e}_{{\mathrm{1}}}  }  \simeq_{log}  \intermed {  {\intermed \Sigma}'_{{\mathrm{2}}}  } : \intermed {  {\intermed e}_{{\mathrm{2}}}  } : \intermed {  [  {\intermed \sigma}  /  {\intermed a }  ]  {\intermed \sigma}'  } 
    \label{eq:env_pres8}
  \end{equation}
  where \( \source {  {\source { \Gamma_C } }  } ; \source {  {\source \Gamma}  }  \vdash_{ty}^{\intermed {M} }  \source {  [  {\source \tau }  /  {\source a }  ]  {\source \tau }'  }\,{ \stoi{ \rightsquigarrow  \intermed {  [  {\intermed \sigma}  /  {\intermed a }  ]  {\intermed \sigma}'  } } } \).

  The goal follows from \textsc{ctxLog-cons}, together with
  Equations~\ref{eq:env_pres7} and~\ref{eq:env_pres8}. \\
}
\qed

\renewcommand\itot[1]{#1}
\begin{theoremB}[Contextual Equivalence in \interL{} Implies Contextual
  Equivalence in \srcL{}]\ \\
  If \( \intermed {  {\intermed { \Gamma_C } }  } ; \intermed {  {\intermed \Gamma}  }  \vdash  \intermed {  {\intermed \Sigma}_{{\mathrm{1}}}  } : \target {  {\target e }_{{\mathrm{1}}}  }  \simeq_{ctx}  \intermed {  {\intermed \Sigma}_{{\mathrm{2}}}  } : \target {  {\target e }_{{\mathrm{2}}}  } : \intermed {  {\intermed \sigma}  } \) \\
  and \( \vdash_{ctx}^{\intermed {M} }  \source {  { \source P }  } ; \source {  {\source { \Gamma_C } }  } ; \source {  {\source \Gamma}  }  \rightsquigarrow  \intermed {  {\intermed \Sigma}_{{\mathrm{1}}}  } ; \intermed {  {\intermed { \Gamma_C } }  } ; \intermed {  {\intermed \Gamma}  } \)
  and \( \vdash_{ctx}^{\intermed {M} }  \source {  { \source P }  } ; \source {  {\source { \Gamma_C } }  } ; \source {  {\source \Gamma}  }  \rightsquigarrow  \intermed {  {\intermed \Sigma}_{{\mathrm{2}}}  } ; \intermed {  {\intermed { \Gamma_C } }  } ; \intermed {  {\intermed \Gamma}  } \) \\
  and \( \vdash_{ctx}  \source {  { \source P }  } ; \source {  {\source { \Gamma_C } }  } ; \source {  {\source \Gamma}  }  \rightsquigarrow  \target {  {\target \Gamma }  } \) \\
  and there exists an \({\source \tau }\) such that
  \( \source {  {\source { \Gamma_C } }  } ; \source {  {\source \Gamma}  }  \vdash_{ty}^{\intermed {M} }  \source {  {\source \tau }  }\,{ \stoi{ \rightsquigarrow  \intermed {  {\intermed \sigma}  } } } \) \\
  then \( \source {  { \source P }  } ; \source {  {\source { \Gamma_C } }  } ; \source {  {\source \Gamma}  }  \vdash  \target {  {\target e }_{{\mathrm{1}}}  }  \simeq_{ctx}  \target {  {\target e }_{{\mathrm{2}}}  } : \source {  {\source \tau }  } \).
  \label{thmA:ctx_equiv_pres}
\end{theoremB}

\proof
By unfolding the definition of contextual equivalence, the goal becomes:
\begin{gather}
  \forall  \source {  {\source M}  } : ( \source {  { \source P }  } ; \source {  {\source { \Gamma_C } }  } ; \source {  {\source \Gamma}  }  \Rightarrow  \source {  {\source \tau }  } )  \mapsto  ( \source {  { \source P }  } ; \source {  {\source { \Gamma_C } }  } ; \source {  \bullet  }  \Rightarrow  \source {  \mathit{\source {Bool} }  } )  \rightsquigarrow  \target {  {\target M }_{{\mathrm{1}}}  } 
  \label{eqg:ctx_equiv_pres0a} \\
   \source {  {\source M}  } : ( \source {  { \source P }  } ; \source {  {\source { \Gamma_C } }  } ; \source {  {\source \Gamma}  }  \Rightarrow  \source {  {\source \tau }  } )  \mapsto  ( \source {  { \source P }  } ; \source {  {\source { \Gamma_C } }  } ; \source {  \bullet  }  \Rightarrow  \source {  \mathit{\source {Bool} }  } )  \rightsquigarrow  \target {  {\target M }_{{\mathrm{2}}}  } 
  \label{eqg:ctx_equiv_pres0b} \\
  \textit{then}~ \target {   {\target M }_{{\mathrm{1}}}  [  {\target e }_{{\mathrm{1}}}  ]   }  \simeq  \target {   {\target M }_{{\mathrm{2}}}  [  {\target e }_{{\mathrm{2}}}  ]   } 
  \label{eqg:ctx_equiv_pres0}
\end{gather}
We thus assume Equations~\ref{eqg:ctx_equiv_pres0a}
and~\ref{eqg:ctx_equiv_pres0b} and prove Equation~\ref{eqg:ctx_equiv_pres0}.

By unfolding the definition of contextual equivalence in the first hypothesis, we
get that:
\begin{gather}
  \forall  \intermed {  {\intermed M}_{{\mathrm{1}}}  } : ( \intermed {  {\intermed \Sigma}_{{\mathrm{1}}}  } ; \intermed {  {\intermed { \Gamma_C } }  } ; \intermed {  {\intermed \Gamma}  }  \Rightarrow  \intermed {  {\intermed \sigma}  } )  \mapsto  ( \intermed {  {\intermed \Sigma}_{{\mathrm{1}}}  } ; \intermed {  {\intermed { \Gamma_C } }  } ; \intermed {  \bullet  }  \Rightarrow  \intermed {  \mathit{\intermed {Bool} }  } ) \itot{ \rightsquigarrow  \target {  {\target M }_{{\mathrm{1}}}  } } 
  \nonumber \\
  \forall  \intermed {  {\intermed M}_{{\mathrm{2}}}  } : ( \intermed {  {\intermed \Sigma}_{{\mathrm{2}}}  } ; \intermed {  {\intermed { \Gamma_C } }  } ; \intermed {  {\intermed \Gamma}  }  \Rightarrow  \intermed {  {\intermed \sigma}  } )  \mapsto  ( \intermed {  {\intermed \Sigma}_{{\mathrm{2}}}  } ; \intermed {  {\intermed { \Gamma_C } }  } ; \intermed {  \bullet  }  \Rightarrow  \intermed {  \mathit{\intermed {Bool} }  } ) \itot{ \rightsquigarrow  \target {  {\target M }_{{\mathrm{2}}}  } } 
  \nonumber \\
  \textit{if}~ \intermed {  {\intermed \Sigma}_{{\mathrm{1}}}  } : \intermed {  {\intermed M}_{{\mathrm{1}}}  }  \simeq_{log}  \intermed {  {\intermed \Sigma}_{{\mathrm{2}}}  } : \intermed {  {\intermed M}_{{\mathrm{2}}}  } : ( \intermed {  {\intermed { \Gamma_C } }  } ; \intermed {  {\intermed \Gamma}  }  \Rightarrow  \intermed {  {\intermed \sigma}  } )  \mapsto  ( \intermed {  {\intermed { \Gamma_C } }  } ; \intermed {  \bullet  }  \Rightarrow  \intermed {  \mathit{\intermed {Bool} }  } ) 
  \nonumber \\
  \textit{then}~ \target {   {\target M }_{{\mathrm{1}}}  [  {\target e }_{{\mathrm{1}}}  ]   }  \simeq  \target {   {\target M }_{{\mathrm{2}}}  [  {\target e }_{{\mathrm{2}}}  ]   } 
  \label{eq:ctx_equiv_pres1}
\end{gather}
By applying Lemma~\ref{lemma:env_strengthening} to the fourth hypothesis, we
know that:
\begin{equation*}
   \vdash_{ctx}  \source {  { \source P }  } ; \source {  {\source { \Gamma_C } }  } ; \source {  \bullet  }  \rightsquigarrow  \target {  \target {\bullet}  } 
\end{equation*}
By applying context equivalence (Theorem~\ref{thmA:equiv_ctx}) on
Equations~\ref{eqg:ctx_equiv_pres0a} and~\ref{eqg:ctx_equiv_pres0b}, together
with this result and hypotheses 2, 3 and 4, we know that:
\begin{gather}
   \source {  {\source M}  } : ( \source {  { \source P }  } ; \source {  {\source { \Gamma_C } }  } ; \source {  {\source \Gamma}  }  \Rightarrow  \source {  {\source \tau }  } )  \mapsto  ( \source {  { \source P }  } ; \source {  {\source { \Gamma_C } }  } ; \source {  \bullet  }  \Rightarrow  \source {  \mathit{\source {Bool} }  } )  \rightsquigarrow  \intermed {  {\intermed M}'_{{\mathrm{1}}}  } 
  \label{eq:ctx_equiv_pres2} \\
   \intermed {  {\intermed M}'_{{\mathrm{1}}}  } : ( \intermed {  {\intermed \Sigma}_{{\mathrm{1}}}  } ; \intermed {  {\intermed { \Gamma_C } }  } ; \intermed {  {\intermed \Gamma}  }  \Rightarrow  \intermed {  {\intermed \sigma}  } )  \mapsto  ( \intermed {  {\intermed \Sigma}_{{\mathrm{1}}}  } ; \intermed {  {\intermed { \Gamma_C } }  } ; \intermed {  \bullet  }  \Rightarrow  \intermed {  \mathit{\intermed {Bool} }  } ) \itot{ \rightsquigarrow  \target {  {\target M }'_{{\mathrm{1}}}  } } 
  \label{eq:ctx_equiv_pres3} \\
   \source {  {\source M}  } : ( \source {  { \source P }  } ; \source {  {\source { \Gamma_C } }  } ; \source {  {\source \Gamma}  }  \Rightarrow  \source {  {\source \tau }  } )  \mapsto  ( \source {  { \source P }  } ; \source {  {\source { \Gamma_C } }  } ; \source {  \bullet  }  \Rightarrow  \source {  \mathit{\source {Bool} }  } )  \rightsquigarrow  \intermed {  {\intermed M}'_{{\mathrm{2}}}  } 
  \label{eq:ctx_equiv_pres4} \\
   \intermed {  {\intermed M}'_{{\mathrm{2}}}  } : ( \intermed {  {\intermed \Sigma}_{{\mathrm{2}}}  } ; \intermed {  {\intermed { \Gamma_C } }  } ; \intermed {  {\intermed \Gamma}  }  \Rightarrow  \intermed {  {\intermed \sigma}  } )  \mapsto  ( \intermed {  {\intermed \Sigma}_{{\mathrm{2}}}  } ; \intermed {  {\intermed { \Gamma_C } }  } ; \intermed {  \bullet  }  \Rightarrow  \intermed {  \mathit{\intermed {Bool} }  } ) \itot{ \rightsquigarrow  \target {  {\target M }'_{{\mathrm{2}}}  } } 
  \label{eq:ctx_equiv_pres5}
\end{gather}
Similarly, by applying Lemma~\ref{lemma:env_mid_strengthening} to the first and
second hypothesis, we get:
\begin{gather}
   \vdash_{ctx}^{\intermed {M} }  \source {  { \source P }  } ; \source {  {\source { \Gamma_C } }  } ; \source {  \bullet  }  \rightsquigarrow  \intermed {  {\intermed \Sigma}_{{\mathrm{1}}}  } ; \intermed {  {\intermed { \Gamma_C } }  } ; \intermed {  \bullet  } 
  \nonumber \\
   \vdash_{ctx}^{\intermed {M} }  \source {  { \source P }  } ; \source {  {\source { \Gamma_C } }  } ; \source {  \bullet  }  \rightsquigarrow  \intermed {  {\intermed \Sigma}_{{\mathrm{1}}}  } ; \intermed {  {\intermed { \Gamma_C } }  } ; \intermed {  \bullet  } 
  \nonumber
\end{gather}
By applying Theorem~\ref{thmA:ctx_reflex} to
Equations~\ref{eq:ctx_equiv_pres2} and~\ref{eq:ctx_equiv_pres4},
together with this result, hypotheses 2, 3 and 5, and \textsc{sTy-bool}, we know
that:
\begin{equation}
   \intermed {  {\intermed \Sigma}_{{\mathrm{1}}}  } : \intermed {  {\intermed M}'_{{\mathrm{1}}}  }  \simeq_{log}  \intermed {  {\intermed \Sigma}_{{\mathrm{2}}}  } : \intermed {  {\intermed M}'_{{\mathrm{2}}}  } : ( \intermed {  {\intermed { \Gamma_C } }  } ; \intermed {  {\intermed \Gamma}  }  \Rightarrow  \intermed {  {\intermed \sigma}  } )  \mapsto  ( \intermed {  {\intermed { \Gamma_C } }  } ; \intermed {  \bullet  }  \Rightarrow  \intermed {  \mathit{\intermed {Bool} }  } ) 
  \label{eq:ctx_equiv_pres6}
\end{equation}
We take \( \intermed {  {\intermed M}_{{\mathrm{1}}}  } = \intermed {  {\intermed M}'_{{\mathrm{1}}}  } \) and \( \intermed {  {\intermed M}_{{\mathrm{2}}}  } = \intermed {  {\intermed M}'_{{\mathrm{2}}}  } \).
Consequently, since \interL{} context elaboration is deterministic
(Theorem~\ref{thmA:ctxelab:det}),
we get that \( \target {  {\target M }_{{\mathrm{1}}}  } = \target {  {\target M }'_{{\mathrm{1}}}  } \) and \( \target {  {\target M }_{{\mathrm{2}}}  } = \target {  {\target M }'_{{\mathrm{2}}}  } \).
Goal~\ref{eqg:ctx_equiv_pres0} follows from Equations~\ref{eq:ctx_equiv_pres1},
\ref{eq:ctx_equiv_pres3}, \ref{eq:ctx_equiv_pres5} and~\ref{eq:ctx_equiv_pres6}.
\\
\qed

\subsection{Partial Coherence Theorems}

\renewcommand\itot[1]{}
\begin{theoremB}[Coherence - Dictionaries - Part A]\leavevmode
  If \( \source {  { \source P }  } ; \source {  {\source { \Gamma_C } }  } ; \source {  {\source \Gamma}  }  \vDash^{\intermed {M} }  \source {  {\source Q}  } \stoi{ \rightsquigarrow  \intermed {  {\intermed d}_{{\mathrm{1}}}  } } \)
  and \( \source {  { \source P }  } ; \source {  {\source { \Gamma_C } }  } ; \source {  {\source \Gamma}  }  \vDash^{\intermed {M} }  \source {  {\source Q}  } \stoi{ \rightsquigarrow  \intermed {  {\intermed d}_{{\mathrm{2}}}  } } \) \\
  and \( \vdash_{ctx}^{\intermed {M} }  \source {  { \source P }  } ; \source {  {\source { \Gamma_C } }  } ; \source {  {\source \Gamma}  }  \rightsquigarrow  \intermed {  {\intermed \Sigma}_{{\mathrm{1}}}  } ; \intermed {  {\intermed { \Gamma_C } }  } ; \intermed {  {\intermed \Gamma}  } \)
  and \( \vdash_{ctx}^{\intermed {M} }  \source {  { \source P }  } ; \source {  {\source { \Gamma_C } }  } ; \source {  {\source \Gamma}  }  \rightsquigarrow  \intermed {  {\intermed \Sigma}_{{\mathrm{2}}}  } ; \intermed {  {\intermed { \Gamma_C } }  } ; \intermed {  {\intermed \Gamma}  } \) \\
  then \( \intermed {  {\intermed { \Gamma_C } }  } ; \intermed {  {\intermed \Gamma}  }  \vdash  \intermed {  {\intermed \Sigma}_{{\mathrm{1}}}  } : \intermed {  {\intermed d}_{{\mathrm{1}}}  }  \simeq_{log}  \intermed {  {\intermed \Sigma}_{{\mathrm{2}}}  } : \intermed {  {\intermed d}_{{\mathrm{2}}}  } : \intermed {  {\intermed Q}  } \) \\
  where \( \source {  {\source { \Gamma_C } }  } ; \source {  {\source \Gamma}  }  \vdash_{\mathit {Q} }^{\intermed {M} }  \source {  {\source Q}  }  \rightsquigarrow  \intermed {  {\intermed Q}  } \).
  \label{thmA:dict_cohA}
\end{theoremB}

\proof
By induction on the first constraint entailment derivation.
This theorem is mutually proven with Theorems~\ref{thmA:env_eq_pres}
and~\ref{thmA:coh_exprA} (see Figure~\ref{fig:dep_graph1}). 
Note that at the dependency between
Theorem~\ref{thmA:env_eq_pres} and~\ref{thmA:coh_exprA}, the size of \({ \source P }\) is
strictly decreasing, whereas \({ \source P }\) remains constant at every other
dependency. Because of this, the size of \({ \source P }\) is strictly decreasing in
every cycle.
Consequently, the induction remains well-founded. \\
\proofStepNL{\textsc{sEntail-inst}}
{
  \drule{sEntail-inst}
}
{
  The final step in the second derivation can be either \textsc{sEntail-inst} or
  \textsc{sEntail-local}: 
  \begin{itemize}
  \item \textsc{sEntail-inst}:
    This means that:
    \begin{gather}
       \source {  { \source P }  } ; \source {  {\source { \Gamma_C } }  } ; \source {  {\source \Gamma}  }  \vDash^{\intermed {M} }  \source {  {\source Q}  } \stoi{ \rightsquigarrow  \intermed {  {\intermed D } \, \overline {\intermed \sigma}_{\ottmv{j}} \, {\overline {\intermed d} }_{\ottmv{i}}  } } 
      \label{eq:dcoh_ii0a} \\
       \source {  { \source P }  } ; \source {  {\source { \Gamma_C } }  } ; \source {  {\source \Gamma}  }  \vDash^{\intermed {M} }  \source {  {\source Q}  } \stoi{ \rightsquigarrow  \intermed {  {\intermed D }' \, \overline {\intermed \sigma}'_{\ottmv{j}} \, {\overline {\intermed d} }'_{\ottmv{i}}  } } 
      \label{eq:dcoh_ii0b}
    \end{gather}
    The goal to be proven is the following:
    \begin{gather}
       \intermed {  {\intermed { \Gamma_C } }  } ; \intermed {  {\intermed \Gamma}  }  \vdash  \intermed {  {\intermed \Sigma}_{{\mathrm{1}}}  } : \intermed {  {\intermed D } \, \overline {\intermed \sigma}_{\ottmv{j}} \, {\overline {\intermed d} }_{\ottmv{i}}  }  \simeq_{log}  \intermed {  {\intermed \Sigma}_{{\mathrm{2}}}  } : \intermed {  {\intermed D }' \, \overline {\intermed \sigma}'_{\ottmv{j}} \, {\overline {\intermed d} }'_{\ottmv{i}}  } : \intermed {  {\intermed Q}  } 
      \label{eq:dcoh_ii1} \\
      \textit{where}~ \source {  {\source { \Gamma_C } }  } ; \source {  {\source \Gamma}  }  \vdash_{\mathit {Q} }^{\intermed {M} }  \source {  {\source Q}  }  \rightsquigarrow  \intermed {  {\intermed Q}  } 
      \label{eq:dcoh_ii1b}
    \end{gather}
    By repeated case analysis on the 3\textsuperscript{rd} hypothesis
    (\textsc{sCtx-pgmInst}), together with the fact that
    \begin{equation*}
     ( \source {  {\source D }  } : \source {  \forall  {\overline {\source a} }_{\ottmv{j}}  \ottsym{.}  {\overline {\source Q} }_{\ottmv{i}}  \Rightarrow  {\source Q}'  } ) . \source {  {\source m}  }  \mapsto  \source {  \bullet  \ottsym{,}  {\overline {\source a} }_{\ottmv{j}}  \ottsym{,}  {\overline {\source \delta} }_{\ottmv{i}}  \ottsym{:}  {\overline {\source Q} }_{\ottmv{i}}  \ottsym{,}  {\overline {\source b} }_{\ottmv{k}}  \ottsym{,}  {\overline {\source \delta} }_{\ottmv{h}}  \ottsym{:}  {\overline {\source Q} }_{\ottmv{h}}  } : \source {  {\source e}  }  \in  \source {  { \source P }  } 
    \end{equation*}
    (1\textsuperscript{st} rule premise (\textsc{sEntail-inst})), we know that:
    \begin{equation}
       \source {  {\source { \Gamma_C } }  } ; \source {  \bullet  }  \vdash_{\mathit {C} }^{\intermed {M} }  \source {  \forall  {\overline {\source a} }_{\ottmv{j}}  \ottsym{.}  {\overline {\source Q} }_{\ottmv{i}}  \Rightarrow  {\source Q}'  }  \rightsquigarrow  \intermed {  \forall  {\overline {\intermed a} }_{\ottmv{j}}  \ottsym{.}  {\overline {\intermed Q} }_{\ottmv{i}}  \Rightarrow  {\intermed Q}'  } 
    \end{equation}
    Consequently, by case analysis (\textsc{sQ-TC}) on this result, we get that:
    \begin{equation}
       \source {  {\source { \Gamma_C } }  } ; \source {  \bullet  \ottsym{,}  {\overline {\source a} }_{\ottmv{j}}  }  \vdash_{\mathit {Q} }^{\intermed {M} }  \source {  {\source Q}'  }  \rightsquigarrow  \intermed {  {\intermed Q}'  } 
    \end{equation}
    By applying Lemma~\ref{lemma:tvar_subst_ctr} to this result, in combination
    with the 4\textsuperscript{th} rule premise (\textsc{sEntail-inst}), we get
    that:
    \begin{equation}
       \source {  {\source { \Gamma_C } }  } ; \source {  \bullet  }  \vdash_{\mathit {Q} }^{\intermed {M} }  \source {  [  {\overline {\source \tau} }_{\ottmv{j}}  /  {\overline {\source a} }_{\ottmv{j}}  ]  {\source Q}'  }  \rightsquigarrow  \intermed {  [  \overline {\intermed \sigma}_{\ottmv{j}}  /  {\overline {\intermed a} }_{\ottmv{j}}  ]  {\intermed Q}'  } 
      \label{eq:dcoh_ii1c}
    \end{equation}
    Goal~\ref{eq:dcoh_ii1b} follows directly from Equation~\ref{eq:dcoh_ii1c},
    since we know that \( \source {  {\source Q}  } = \source {  [  {\overline {\source \tau} }_{\ottmv{j}}  /  {\overline {\source a} }_{\ottmv{j}}  ]  {\source Q}'  } \)
    (2\textsuperscript{nd} rule premise (\textsc{sEntail-inst})).
    
    By unfolding the definition of logical equivalence in
    Goal~\ref{eq:dcoh_ii1}, the goal reduces to:
    \begin{equation}
       ( \intermed {  {\intermed \Sigma}_{{\mathrm{1}}}  } : \intermed {   {\intermed \gamma }_{{\mathrm{1}}}  ( {\intermed \phi }_{{\mathrm{1}}}  ( {\intermed R }  ( {\intermed D } \, \overline {\intermed \sigma}_{\ottmv{j}} \, {\overline {\intermed d} }_{\ottmv{i}} )))   } , \intermed {  {\intermed \Sigma}_{{\mathrm{2}}}  } : \intermed {   {\intermed \gamma }_{{\mathrm{2}}}  ( {\intermed \phi }_{{\mathrm{2}}}  ( {\intermed R }  ( {\intermed D }' \, \overline {\intermed \sigma}'_{\ottmv{j}} \, {\overline {\intermed d} }'_{\ottmv{i}} )))   } ) \in \mathcal{V} \llbracket \intermed {  {\intermed Q}  } \rrbracket ^{\intermed {  {\intermed { \Gamma_C } }  } }_{\intermed {  {\intermed R }  } } 
      \label{eq:dcoh_ii13}
    \end{equation}
    for any \( \intermed {  {\intermed R }  } \in \mathcal{F} \llbracket \intermed {  {\intermed \Gamma}  } \rrbracket ^{\intermed {  {\intermed { \Gamma_C } }  } } \),
    \( \intermed {  {\intermed \phi }  } \in \mathcal{G} \llbracket \intermed {  {\intermed \Gamma}  } \rrbracket ^{\intermed {  {\intermed \Sigma}_{{\mathrm{1}}}  } , \intermed {  {\intermed \Sigma}_{{\mathrm{2}}}  } , \intermed {  {\intermed { \Gamma_C } }  } }_{\intermed {  {\intermed R }  } } \) and
    \( \intermed {  {\intermed \gamma }  } \in \mathcal{H} \llbracket \intermed {  {\intermed \Gamma}  } \rrbracket ^{\intermed {  {\intermed \Sigma}_{{\mathrm{1}}}  } , \intermed {  {\intermed \Sigma}_{{\mathrm{2}}}  } , \intermed {  {\intermed { \Gamma_C } }  } }_{\intermed {  {\intermed R }  } } \).

    By simplifying the substitutions (note that \({\intermed \phi }\) only substitutes
    term variables, which has no impact on the types or dictionaries),
    Goal~\ref{eq:dcoh_ii13} reduces to:
    \begin{equation}
       ( \intermed {  {\intermed \Sigma}_{{\mathrm{1}}}  } : \intermed {  {\intermed D } \, {\intermed R }  \ottsym{(}  \overline {\intermed \sigma}_{\ottmv{j}}  \ottsym{)} \, \ottsym{(}   {\intermed \gamma }_{{\mathrm{1}}}  ( \bullet  ( {\intermed R }  ( {\overline {\intermed d} }_{\ottmv{i}} )))   \ottsym{)}  } , \intermed {  {\intermed \Sigma}_{{\mathrm{2}}}  } : \intermed {  {\intermed D }' \, {\intermed R }  \ottsym{(}  \overline {\intermed \sigma}'_{\ottmv{j}}  \ottsym{)} \, \ottsym{(}   {\intermed \gamma }_{{\mathrm{2}}}  ( \bullet  ( {\intermed R }  ( {\overline {\intermed d} }'_{\ottmv{i}} )))   \ottsym{)}  } ) \in \mathcal{V} \llbracket \intermed {  {\intermed Q}  } \rrbracket ^{\intermed {  {\intermed { \Gamma_C } }  } }_{\intermed {  {\intermed R }  } } 
      \label{eq:dcoh_ii14}
    \end{equation}
    By applying preservation Theorem~\ref{thmA:typing_pres_constraint} on
    Equations~\ref{eq:dcoh_ii0a} and \ref{eq:dcoh_ii0b} respectively, we get:
    \begin{gather}
       \intermed {  {\intermed \Sigma}_{{\mathrm{1}}}  } ; \intermed {  {\intermed { \Gamma_C } }  } ; \intermed {  {\intermed \Gamma}  }  \vdash_{d}  \intermed {  {\intermed D } \, \overline {\intermed \sigma}_{\ottmv{j}} \, {\overline {\intermed d} }_{\ottmv{i}}  } : \intermed {  {\intermed Q}  } \itot{\,  \rightsquigarrow  \target {  {\target e }_{{\mathrm{1}}}  } } 
      \label{eq:dcoh_ii15} \\
       \intermed {  {\intermed \Sigma}_{{\mathrm{2}}}  } ; \intermed {  {\intermed { \Gamma_C } }  } ; \intermed {  {\intermed \Gamma}  }  \vdash_{d}  \intermed {  {\intermed D }' \, \overline {\intermed \sigma}'_{\ottmv{j}} \, {\overline {\intermed d} }'_{\ottmv{i}}  } : \intermed {  {\intermed Q}  } \itot{\,  \rightsquigarrow  \target {  {\target e }_{{\mathrm{2}}}  } } 
      \label{eq:dcoh_ii16}
    \end{gather}
    By repeatedly applying substitution lemmas~\ref{lemma:var_subst_dict},
    \ref{lemma:dvar_subst_dict} and \ref{lemma:tvar_subst_dict},
    using \({\intermed R }\) and \({\intermed \gamma }\), Equations~\ref{eq:dcoh_ii15} and
    \ref{eq:dcoh_ii16} simplify to:
    \begin{gather}
       \intermed {  {\intermed \Sigma}_{{\mathrm{1}}}  } ; \intermed {  {\intermed { \Gamma_C } }  } ; \intermed {  \bullet  }  \vdash_{d}  \intermed {  {\intermed D } \, {\intermed R }  \ottsym{(}  \overline {\intermed \sigma}_{\ottmv{j}}  \ottsym{)} \, \ottsym{(}   {\intermed \gamma }_{{\mathrm{1}}}  ( \bullet  ( {\intermed R }  ( {\overline {\intermed d} }_{\ottmv{i}} )))   \ottsym{)}  } : \intermed {  {\intermed R }  \ottsym{(}  {\intermed Q}  \ottsym{)}  } \itot{\,  \rightsquigarrow  \target {  {\target e }'_{{\mathrm{1}}}  } } 
      \label{eq:dcoh_ii17} \\
       \intermed {  {\intermed \Sigma}_{{\mathrm{2}}}  } ; \intermed {  {\intermed { \Gamma_C } }  } ; \intermed {  \bullet  }  \vdash_{d}  \intermed {  {\intermed D }' \, {\intermed R }  \ottsym{(}  \overline {\intermed \sigma}'_{\ottmv{j}}  \ottsym{)} \, \ottsym{(}   {\intermed \gamma }_{{\mathrm{2}}}  ( \bullet  ( {\intermed R }  ( {\overline {\intermed d} }'_{\ottmv{i}} )))   \ottsym{)}  } : \intermed {  {\intermed R }  \ottsym{(}  {\intermed Q}  \ottsym{)}  } \itot{\,  \rightsquigarrow  \target {  {\target e }'_{{\mathrm{2}}}  } } 
      \label{eq:dcoh_ii18}
    \end{gather}
    Furthermore, by applying preservation Theorem~\ref{thmA:typing_pres_env} on
    the 3\textsuperscript{rd} and 4\textsuperscript{th} hypothesis, we know
    that \( \vdash_{ctx}  \intermed {  {\intermed \Sigma}_{{\mathrm{1}}}  } ; \intermed {  {\intermed { \Gamma_C } }  } ; \intermed {  {\intermed \Gamma}  } \) and \( \vdash_{ctx}  \intermed {  {\intermed \Sigma}_{{\mathrm{2}}}  } ; \intermed {  {\intermed { \Gamma_C } }  } ; \intermed {  {\intermed \Gamma}  } \).
    By Theorem~\ref{thmA:env_eq_pres}, we know that \( \intermed {  {\intermed { \Gamma_C } }  }  \vdash  \intermed {  {\intermed \Sigma}_{{\mathrm{1}}}  }  \simeq_{log}  \intermed {  {\intermed \Sigma}_{{\mathrm{2}}}  } \).
    Consequently, by applying Theorem~\ref{thmA:dval_rel} on
    Equations~\ref{eq:dcoh_ii17} and \ref{eq:dcoh_ii18} we get:
    \begin{equation}
       ( \intermed {  {\intermed \Sigma}_{{\mathrm{1}}}  } : \intermed {  {\intermed D } \, {\intermed R }  \ottsym{(}  \overline {\intermed \sigma}_{\ottmv{j}}  \ottsym{)} \, \ottsym{(}   {\intermed \gamma }_{{\mathrm{1}}}  ( \bullet  ( {\intermed R }  ( {\overline {\intermed d} }_{\ottmv{i}} )))   \ottsym{)}  } , \intermed {  {\intermed \Sigma}_{{\mathrm{2}}}  } : \intermed {  {\intermed D }' \, {\intermed R }  \ottsym{(}  \overline {\intermed \sigma}'_{\ottmv{j}}  \ottsym{)} \, \ottsym{(}   {\intermed \gamma }_{{\mathrm{2}}}  ( \bullet  ( {\intermed R }  ( {\overline {\intermed d} }'_{\ottmv{i}} )))   \ottsym{)}  } ) \in \mathcal{V} \llbracket \intermed {  {\intermed R }  \ottsym{(}  {\intermed Q}  \ottsym{)}  } \rrbracket ^{\intermed {  {\intermed { \Gamma_C } }  } }_{\intermed {  \bullet  } } 
      \label{eq:dcoh_ii19}
    \end{equation}
    Goal~\ref{eq:dcoh_ii14} follows from Equation~\ref{eq:dcoh_ii19} by Lemma~\ref{lemma:dict_Rsubst}.

  \item \textsc{sEntail-local}:
    This means that:
    \begin{gather}
       \source {  { \source P }  } ; \source {  {\source { \Gamma_C } }  } ; \source {  {\source \Gamma}  }  \vDash^{\intermed {M} }  \source {  {\source Q}  } \stoi{ \rightsquigarrow  \intermed {  {\intermed D } \, \overline {\intermed \sigma}_{\ottmv{j}} \, {\overline {\intermed d} }_{\ottmv{i}}  } } 
      \label{eq:dcoh_iv0a} \\
       \source {  { \source P }  } ; \source {  {\source { \Gamma_C } }  } ; \source {  {\source \Gamma}  }  \vDash^{\intermed {M} }  \source {  {\source Q}  } \stoi{ \rightsquigarrow  \intermed {  {\intermed \delta }  } } 
      \label{eq:dcoh_iv0b}
    \end{gather}
    The goal to be proven is the following:
    \begin{gather}
       \intermed {  {\intermed { \Gamma_C } }  } ; \intermed {  {\intermed \Gamma}  }  \vdash  \intermed {  {\intermed \Sigma}_{{\mathrm{1}}}  } : \intermed {  {\intermed D } \, \overline {\intermed \sigma}_{\ottmv{j}} \, {\overline {\intermed d} }_{\ottmv{i}}  }  \simeq_{log}  \intermed {  {\intermed \Sigma}_{{\mathrm{2}}}  } : \intermed {  {\intermed \delta }  } : \intermed {  {\intermed Q}  } 
      \label{eq:dcoh_iv1} \\
      \textit{where}~ \source {  {\source { \Gamma_C } }  } ; \source {  {\source \Gamma}  }  \vdash_{\mathit {Q} }^{\intermed {M} }  \source {  {\source Q}  }  \rightsquigarrow  \intermed {  {\intermed Q}  } 
      \label{eq:dcoh_iv1b}
    \end{gather}
    Goal~\ref{eq:dcoh_iv1b} follows similarly to Goal~\ref{eq:dcoh_ii1b} in the
    previous part of this proof.
    
    The 1\textsuperscript{st} rule premise (\textsc{sEntail-local}) tells us that:
    \begin{equation}
       ( \source {  {\source \delta }  } : \source {  {\source Q}  } )  \in  \source {  {\source \Gamma}  } 
      \label{eq:dcoh_iv6}
    \end{equation}
    Unfolding the definition of logical equivalence reduces Goal~\ref{eq:dcoh_iv1} to:
    \begin{equation}
       ( \intermed {  {\intermed \Sigma}_{{\mathrm{1}}}  } : \intermed {   {\intermed \gamma }_{{\mathrm{1}}}  ( {\intermed \phi }_{{\mathrm{1}}}  ( {\intermed R }  ( {\intermed D } \, \overline {\intermed \sigma}_{\ottmv{j}} \, {\overline {\intermed d} }_{\ottmv{i}} )))   } , \intermed {  {\intermed \Sigma}_{{\mathrm{2}}}  } : \intermed {   {\intermed \gamma }_{{\mathrm{2}}}  ( {\intermed \phi }_{{\mathrm{2}}}  ( {\intermed R }  ( {\intermed \delta } )))   } ) \in \mathcal{V} \llbracket \intermed {  {\intermed Q}  } \rrbracket ^{\intermed {  {\intermed { \Gamma_C } }  } }_{\intermed {  {\intermed R }  } } 
      \label{eq:dcoh_iv7}
    \end{equation}
    for any \( \intermed {  {\intermed R }  } \in \mathcal{F} \llbracket \intermed {  {\intermed \Gamma}  } \rrbracket ^{\intermed {  {\intermed { \Gamma_C } }  } } \),
    \( \intermed {  {\intermed \phi }  } \in \mathcal{G} \llbracket \intermed {  {\intermed \Gamma}  } \rrbracket ^{\intermed {  {\intermed \Sigma}_{{\mathrm{1}}}  } , \intermed {  {\intermed \Sigma}_{{\mathrm{2}}}  } , \intermed {  {\intermed { \Gamma_C } }  } }_{\intermed {  {\intermed R }  } } \) and
    \( \intermed {  {\intermed \gamma }  } \in \mathcal{H} \llbracket \intermed {  {\intermed \Gamma}  } \rrbracket ^{\intermed {  {\intermed \Sigma}_{{\mathrm{1}}}  } , \intermed {  {\intermed \Sigma}_{{\mathrm{2}}}  } , \intermed {  {\intermed { \Gamma_C } }  } }_{\intermed {  {\intermed R }  } } \).

    By simplifying the substitutions (note that \({\intermed \phi }\) only substitutes
    term variables, which has no impact on the types or dictionaries),
    Goal~\ref{eq:dcoh_iv7} reduces to: 
    \begin{equation}
       ( \intermed {  {\intermed \Sigma}_{{\mathrm{1}}}  } : \intermed {  {\intermed D } \, {\intermed R }  \ottsym{(}  \overline {\intermed \sigma}_{\ottmv{j}}  \ottsym{)} \, \ottsym{(}   {\intermed \gamma }_{{\mathrm{1}}}  ( \bullet  ( {\intermed R }  ( {\overline {\intermed d} }_{\ottmv{i}} )))   \ottsym{)}  } , \intermed {  {\intermed \Sigma}_{{\mathrm{2}}}  } : \intermed {   {\intermed \gamma }_{{\mathrm{2}}}  ( {\intermed \phi }_{{\mathrm{2}}}  ( {\intermed R }  ( {\intermed \delta } )))   } ) \in \mathcal{V} \llbracket \intermed {  {\intermed Q}  } \rrbracket ^{\intermed {  {\intermed { \Gamma_C } }  } }_{\intermed {  {\intermed R }  } } 
      \label{eq:dcoh_iv13}
    \end{equation}
    The definition of \(\mathcal{H}\), together with Equation~\ref{eq:dcoh_iv6},
    combined with the fact that
    \( \vdash_{ctx}^{\intermed {M} }  \source {  { \source P }  } ; \source {  {\source { \Gamma_C } }  } ; \source {  {\source \Gamma}  }  \rightsquigarrow  \intermed {  {\intermed \Sigma}_{{\mathrm{1}}}  } ; \intermed {  {\intermed { \Gamma_C } }  } ; \intermed {  {\intermed \Gamma}  } \) and
    \( \vdash_{ctx}^{\intermed {M} }  \source {  { \source P }  } ; \source {  {\source { \Gamma_C } }  } ; \source {  {\source \Gamma}  }  \rightsquigarrow  \intermed {  {\intermed \Sigma}_{{\mathrm{2}}}  } ; \intermed {  {\intermed { \Gamma_C } }  } ; \intermed {  {\intermed \Gamma}  } \), tells us that:
    \begin{equation}
       \intermed {  {\intermed \delta }  }  \mapsto  ( \intermed {  {\intermed {dv} }_{{\mathrm{1}}}  } , \intermed {  {\intermed {dv} }_{{\mathrm{2}}}  } )  \in  \intermed {  {\intermed \gamma }  } ~
      \textit{where}~ ( \intermed {  {\intermed \Sigma}_{{\mathrm{1}}}  } : \intermed {  {\intermed {dv} }_{{\mathrm{1}}}  } , \intermed {  {\intermed \Sigma}_{{\mathrm{2}}}  } : \intermed {  {\intermed {dv} }_{{\mathrm{2}}}  } ) \in \mathcal{V} \llbracket \intermed {  {\intermed Q}  } \rrbracket ^{\intermed {  {\intermed { \Gamma_C } }  } }_{\intermed {  {\intermed R }  } } 
      \label{eq:dcoh_iv8}
    \end{equation}
    We thus know that:
    \begin{equation*}
       \intermed {   {\intermed \gamma }_{{\mathrm{2}}}  ( {\intermed \phi }_{{\mathrm{2}}}  ( {\intermed R }  ( {\intermed \delta } )))   } = \intermed {  {\intermed {dv} }_{{\mathrm{2}}}  } 
    \end{equation*}
    By unfolding the definition of the \(\mathcal{V}\) relation in
    Equation~\ref{eq:dcoh_iv8}, we get:
    \begin{equation}
       \intermed {  {\intermed \Sigma}_{{\mathrm{2}}}  } ; \intermed {  {\intermed { \Gamma_C } }  } ; \intermed {  \bullet  }  \vdash_{d}  \intermed {  {\intermed {dv} }_{{\mathrm{2}}}  } : \intermed {  {\intermed R }  \ottsym{(}  {\intermed Q}  \ottsym{)}  } \itot{\,  \rightsquigarrow  \target {  {\target e }'  } } 
      \label{eq:dcoh_iv10}
    \end{equation}
    Preservation Theorem~\ref{thmA:typing_pres_env}, applied to the
    3\textsuperscript{rd} and 4\textsuperscript{th} hypothesis, tells us that:
    \begin{gather*}
       \vdash_{ctx}  \intermed {  {\intermed \Sigma}_{{\mathrm{1}}}  } ; \intermed {  {\intermed { \Gamma_C } }  } ; \intermed {  {\intermed \Gamma}  }  \\
       \vdash_{ctx}  \intermed {  {\intermed \Sigma}_{{\mathrm{2}}}  } ; \intermed {  {\intermed { \Gamma_C } }  } ; \intermed {  {\intermed \Gamma}  } 
    \end{gather*}
    Applying preservation Theorem~\ref{thmA:typing_pres_constraint} to
    Equation~\ref{eq:dcoh_iv0a} results in:
    \begin{equation}
       \intermed {  {\intermed \Sigma}_{{\mathrm{1}}}  } ; \intermed {  {\intermed { \Gamma_C } }  } ; \intermed {  {\intermed \Gamma}  }  \vdash_{d}  \intermed {  {\intermed D } \, \overline {\intermed \sigma}_{\ottmv{j}} \, {\overline {\intermed d} }_{\ottmv{i}}  } : \intermed {  {\intermed Q}  } \itot{\,  \rightsquigarrow  \target {  {\target e }  } } 
      \label{eq:dcoh_iv11}
    \end{equation}
    By repeatedly applying substitution lemmas~\ref{lemma:var_subst_dict},
    \ref{lemma:dvar_subst_dict} and \ref{lemma:tvar_subst_dict},
    using \({\intermed R }\) and \({\intermed \gamma }\), Equation~\ref{eq:dcoh_iv11}
    simplifies to:
    \begin{equation}
       \intermed {  {\intermed \Sigma}_{{\mathrm{1}}}  } ; \intermed {  {\intermed { \Gamma_C } }  } ; \intermed {  \bullet  }  \vdash_{d}  \intermed {  {\intermed D } \, {\intermed R }  \ottsym{(}  \overline {\intermed \sigma}_{\ottmv{j}}  \ottsym{)} \, \ottsym{(}   {\intermed \gamma }_{{\mathrm{1}}}  ( \bullet  ( {\intermed R }  ( {\overline {\intermed d} }_{\ottmv{i}} )))   \ottsym{)}  } : \intermed {  {\intermed R }  \ottsym{(}  {\intermed Q}  \ottsym{)}  } \itot{\,  \rightsquigarrow  \target {  {\target e }''  } } 
      \label{eq:dcoh_iv12}
    \end{equation}
    From Theorem~\ref{thmA:env_eq_pres}, we know that \( \intermed {  {\intermed { \Gamma_C } }  }  \vdash  \intermed {  {\intermed \Sigma}_{{\mathrm{1}}}  }  \simeq_{log}  \intermed {  {\intermed \Sigma}_{{\mathrm{2}}}  } \).
    By applying Theorem~\ref{thmA:dval_rel} to Equations~\ref{eq:dcoh_iv10} and
    \ref{eq:dcoh_iv12} we get:
    \begin{equation}
       ( \intermed {  {\intermed \Sigma}_{{\mathrm{1}}}  } : \intermed {  {\intermed D } \, {\intermed R }  \ottsym{(}  \overline {\intermed \sigma}_{\ottmv{j}}  \ottsym{)} \, \ottsym{(}   {\intermed \gamma }_{{\mathrm{1}}}  ( \bullet  ( {\intermed R }  ( {\overline {\intermed d} }_{\ottmv{i}} )))   \ottsym{)}  } , \intermed {  {\intermed \Sigma}_{{\mathrm{2}}}  } : \intermed {  {\intermed {dv} }_{{\mathrm{2}}}  } ) \in \mathcal{V} \llbracket \intermed {  {\intermed R }  \ottsym{(}  {\intermed Q}  \ottsym{)}  } \rrbracket ^{\intermed {  {\intermed { \Gamma_C } }  } }_{\intermed {  \bullet  } } 
      \label{eq:dcoh_iv14}
    \end{equation}
    Goal~\ref{eq:dcoh_iv13} follows from Equation~\ref{eq:dcoh_iv14}, in
    combination with Lemma~\ref{lemma:dict_Rsubst}. \\

  \end{itemize}
}
\proofStep{\textsc{sEntail-local}}
{
  \drule{sEntail-local}
}
{
  The final step in the derivation can be either \textsc{sEntail-inst} or
  \textsc{sEntail-local}:
  \begin{itemize}
  \item \textsc{sEntail-inst}: This proof case is identical to the
    2\textsuperscript{nd} part of the previous case.
    
  \item \textsc{sEntail-local}: This means that:
    \begin{gather*}
       \source {  { \source P }  } ; \source {  {\source { \Gamma_C } }  } ; \source {  {\source \Gamma}  }  \vDash^{\intermed {M} }  \source {  {\source Q}  } \stoi{ \rightsquigarrow  \intermed {  {\intermed \delta }  } }  \\
       \source {  { \source P }  } ; \source {  {\source { \Gamma_C } }  } ; \source {  {\source \Gamma}  }  \vDash^{\intermed {M} }  \source {  {\source Q}  } \stoi{ \rightsquigarrow  \intermed {  {\intermed \delta }'  } } 
    \end{gather*}
    The goal to be proven is the following:
    \begin{gather}
       \intermed {  {\intermed { \Gamma_C } }  } ; \intermed {  {\intermed \Gamma}  }  \vdash  \intermed {  {\intermed \Sigma}_{{\mathrm{1}}}  } : \intermed {  {\intermed \delta }  }  \simeq_{log}  \intermed {  {\intermed \Sigma}_{{\mathrm{2}}}  } : \intermed {  {\intermed \delta }'  } : \intermed {  {\intermed Q}  } 
      \label{eq:dcoh_vv1} \\
      \textit{where}~ \source {  {\source { \Gamma_C } }  } ; \source {  {\source \Gamma}  }  \vdash_{\mathit {Q} }^{\intermed {M} }  \source {  {\source Q}  }  \rightsquigarrow  \intermed {  {\intermed Q}  } 
      \label{eq:dcoh_vv1b}
    \end{gather}
    The rule premise tells us that:
    \begin{gather}
       ( \source {  {\source \delta }  } : \source {  {\source Q}  } )  \in  \source {  {\source \Gamma}  } 
      \label{eq:dcoh_vv2} \\
       ( \source {  {\source \delta }'  } : \source {  {\source Q}  } )  \in  \source {  {\source \Gamma}  } 
      \label{eq:dcoh_vv3}
    \end{gather}
    Goal~\ref{eq:dcoh_vv1b} follows by repeated case analysis
    (\textsc{sCtx-tyEnvD}) on the 3\textsuperscript{rd} hypothesis, together
    with Equation~\ref{eq:dcoh_vv2}.
    
    Unfolding the definition of logical equivalence reduces
    Goal~\ref{eq:dcoh_vv1} to:
    \begin{equation}
       ( \intermed {  {\intermed \Sigma}_{{\mathrm{1}}}  } : \intermed {   {\intermed \gamma }_{{\mathrm{1}}}  ( {\intermed \phi }_{{\mathrm{1}}}  ( {\intermed R }  ( {\intermed \delta } )))   } , \intermed {  {\intermed \Sigma}_{{\mathrm{2}}}  } : \intermed {   {\intermed \gamma }_{{\mathrm{2}}}  ( {\intermed \phi }_{{\mathrm{2}}}  ( {\intermed R }  ( {\intermed \delta }' )))   } ) \in \mathcal{V} \llbracket \intermed {  {\intermed Q}  } \rrbracket ^{\intermed {  {\intermed { \Gamma_C } }  } }_{\intermed {  {\intermed R }  } } 
      \label{eq:dcoh_vv4}
    \end{equation}
    for any \( \intermed {  {\intermed R }  } \in \mathcal{F} \llbracket \intermed {  {\intermed \Gamma}  } \rrbracket ^{\intermed {  {\intermed { \Gamma_C } }  } } \),
    \( \intermed {  {\intermed \phi }  } \in \mathcal{G} \llbracket \intermed {  {\intermed \Gamma}  } \rrbracket ^{\intermed {  {\intermed \Sigma}_{{\mathrm{1}}}  } , \intermed {  {\intermed \Sigma}_{{\mathrm{2}}}  } , \intermed {  {\intermed { \Gamma_C } }  } }_{\intermed {  {\intermed R }  } } \) and
    \( \intermed {  {\intermed \gamma }  } \in \mathcal{H} \llbracket \intermed {  {\intermed \Gamma}  } \rrbracket ^{\intermed {  {\intermed \Sigma}_{{\mathrm{1}}}  } , \intermed {  {\intermed \Sigma}_{{\mathrm{2}}}  } , \intermed {  {\intermed { \Gamma_C } }  } }_{\intermed {  {\intermed R }  } } \).

    The definition of \(\mathcal{H}\), together with Equations~\ref{eq:dcoh_vv2}
    and \ref{eq:dcoh_vv3}, combined with the fact that
    \( \vdash_{ctx}^{\intermed {M} }  \source {  { \source P }  } ; \source {  {\source { \Gamma_C } }  } ; \source {  {\source \Gamma}  }  \rightsquigarrow  \intermed {  {\intermed \Sigma}_{{\mathrm{1}}}  } ; \intermed {  {\intermed { \Gamma_C } }  } ; \intermed {  {\intermed \Gamma}  } \) and
    \( \vdash_{ctx}^{\intermed {M} }  \source {  { \source P }  } ; \source {  {\source { \Gamma_C } }  } ; \source {  {\source \Gamma}  }  \rightsquigarrow  \intermed {  {\intermed \Sigma}_{{\mathrm{2}}}  } ; \intermed {  {\intermed { \Gamma_C } }  } ; \intermed {  {\intermed \Gamma}  } \), tells us that:
    \begin{gather}
       \intermed {  {\intermed \delta }  }  \mapsto  ( \intermed {  {\intermed {dv} }_{{\mathrm{1}}}  } , \intermed {  {\intermed {dv} }_{{\mathrm{2}}}  } )  \in  \intermed {  {\intermed \gamma }  } ~
      \textit{where}~ ( \intermed {  {\intermed \Sigma}_{{\mathrm{1}}}  } : \intermed {  {\intermed {dv} }_{{\mathrm{1}}}  } , \intermed {  {\intermed \Sigma}_{{\mathrm{2}}}  } : \intermed {  {\intermed {dv} }_{{\mathrm{2}}}  } ) \in \mathcal{V} \llbracket \intermed {  {\intermed Q}  } \rrbracket ^{\intermed {  {\intermed { \Gamma_C } }  } }_{\intermed {  {\intermed R }  } } 
      \label{eq:dcoh_vv5} \\
       \intermed {  {\intermed \delta }'  }  \mapsto  ( \intermed {  {\intermed {dv} }'_{{\mathrm{1}}}  } , \intermed {  {\intermed {dv} }'_{{\mathrm{2}}}  } )  \in  \intermed {  {\intermed \gamma }  } ~
      \textit{where}~ ( \intermed {  {\intermed \Sigma}_{{\mathrm{1}}}  } : \intermed {  {\intermed {dv} }'_{{\mathrm{1}}}  } , \intermed {  {\intermed \Sigma}_{{\mathrm{2}}}  } : \intermed {  {\intermed {dv} }'_{{\mathrm{2}}}  } ) \in \mathcal{V} \llbracket \intermed {  {\intermed Q}  } \rrbracket ^{\intermed {  {\intermed { \Gamma_C } }  } }_{\intermed {  {\intermed R }  } } 
      \label{eq:dcoh_vv6}
    \end{gather}
    We thus know that:
    \begin{gather*}
       \intermed {   {\intermed \gamma }_{{\mathrm{1}}}  ( {\intermed \phi }_{{\mathrm{1}}}  ( {\intermed R }  ( {\intermed \delta } )))   } = \intermed {  {\intermed {dv} }_{{\mathrm{1}}}  }  \\
       \intermed {   {\intermed \gamma }_{{\mathrm{2}}}  ( {\intermed \phi }_{{\mathrm{2}}}  ( {\intermed R }  ( {\intermed \delta }' )))   } = \intermed {  {\intermed {dv} }'_{{\mathrm{2}}}  } 
    \end{gather*}
    From the definition of the \(\mathcal{V}\) relation in
    Equations~\ref{eq:dcoh_vv5} and \ref{eq:dcoh_vv6} it follows that:
    \begin{gather}
       \intermed {  {\intermed \Sigma}_{{\mathrm{1}}}  } ; \intermed {  {\intermed { \Gamma_C } }  } ; \intermed {  \bullet  }  \vdash_{d}  \intermed {  {\intermed {dv} }_{{\mathrm{1}}}  } : \intermed {  {\intermed R }  \ottsym{(}  {\intermed Q}  \ottsym{)}  } \itot{\,  \rightsquigarrow  \target {  {\target e }  } } 
      \label{eq:dcoh_vv7} \\
       \intermed {  {\intermed \Sigma}_{{\mathrm{2}}}  } ; \intermed {  {\intermed { \Gamma_C } }  } ; \intermed {  \bullet  }  \vdash_{d}  \intermed {  {\intermed {dv} }'_{{\mathrm{2}}}  } : \intermed {  {\intermed R }  \ottsym{(}  {\intermed Q}  \ottsym{)}  } \itot{\,  \rightsquigarrow  \target {  {\target e }'  } } 
      \label{eq:dcoh_vv8}
    \end{gather}
    Applying preservation Theorem~\ref{thmA:typing_pres_env} to our hypothesis that
    \( \vdash_{ctx}^{\intermed {M} }  \source {  { \source P }  } ; \source {  {\source { \Gamma_C } }  } ; \source {  {\source \Gamma}  }  \rightsquigarrow  \intermed {  {\intermed \Sigma}_{{\mathrm{1}}}  } ; \intermed {  {\intermed { \Gamma_C } }  } ; \intermed {  {\intermed \Gamma}  } \) and
    \( \vdash_{ctx}^{\intermed {M} }  \source {  { \source P }  } ; \source {  {\source { \Gamma_C } }  } ; \source {  {\source \Gamma}  }  \rightsquigarrow  \intermed {  {\intermed \Sigma}_{{\mathrm{2}}}  } ; \intermed {  {\intermed { \Gamma_C } }  } ; \intermed {  {\intermed \Gamma}  } \) gives us
    \( \vdash_{ctx}  \intermed {  {\intermed \Sigma}_{{\mathrm{1}}}  } ; \intermed {  {\intermed { \Gamma_C } }  } ; \intermed {  {\intermed \Gamma}  } \) and
    \( \vdash_{ctx}  \intermed {  {\intermed \Sigma}_{{\mathrm{2}}}  } ; \intermed {  {\intermed { \Gamma_C } }  } ; \intermed {  {\intermed \Gamma}  } \), respectively.

    From Theorem~\ref{thmA:env_eq_pres}, we get that \( \intermed {  {\intermed { \Gamma_C } }  }  \vdash  \intermed {  {\intermed \Sigma}_{{\mathrm{1}}}  }  \simeq_{log}  \intermed {  {\intermed \Sigma}_{{\mathrm{2}}}  } \).
    Consequently, by applying Theorem~\ref{thmA:dval_rel} on Equations~\ref{eq:dcoh_vv7} and
    \ref{eq:dcoh_vv8} we get:
    \begin{equation}
       ( \intermed {  {\intermed \Sigma}_{{\mathrm{1}}}  } : \intermed {  {\intermed {dv} }_{{\mathrm{1}}}  } , \intermed {  {\intermed \Sigma}_{{\mathrm{2}}}  } : \intermed {  {\intermed {dv} }'_{{\mathrm{2}}}  } ) \in \mathcal{V} \llbracket \intermed {  {\intermed R }  \ottsym{(}  {\intermed Q}  \ottsym{)}  } \rrbracket ^{\intermed {  {\intermed { \Gamma_C } }  } }_{\intermed {  \bullet  } } 
      \label{eq:dcoh_vv9}
    \end{equation}
    Goal~\ref{eq:dcoh_vv4} follows from Equation~\ref{eq:dcoh_vv9} by Lemma~\ref{lemma:dict_Rsubst}.
    
  \end{itemize}
}
\qed

\begin{theoremB}[Coherence - Expressions - Part A]\leavevmode
  \begin{itemize}
  \item
    If \( \source {  { \source P }  } ; \source {  {\source { \Gamma_C } }  } ; \source {  {\source \Gamma}  }  \vdash_{tm}^{\intermed {M} }  \source {  {\source e}  } \Rightarrow \source {  {\source \tau }  } \stoi{ \rightsquigarrow  \intermed {  {\intermed e}_{{\mathrm{1}}}  } } \)
    and \( \source {  { \source P }  } ; \source {  {\source { \Gamma_C } }  } ; \source {  {\source \Gamma}  }  \vdash_{tm}^{\intermed {M} }  \source {  {\source e}  } \Rightarrow \source {  {\source \tau }  } \stoi{ \rightsquigarrow  \intermed {  {\intermed e}_{{\mathrm{2}}}  } } \) \\
    and \( \vdash_{ctx}^{\intermed {M} }  \source {  { \source P }  } ; \source {  {\source { \Gamma_C } }  } ; \source {  {\source \Gamma}  }  \rightsquigarrow  \intermed {  {\intermed \Sigma}_{{\mathrm{1}}}  } ; \intermed {  {\intermed { \Gamma_C } }  } ; \intermed {  {\intermed \Gamma}  } \)
    and \( \vdash_{ctx}^{\intermed {M} }  \source {  { \source P }  } ; \source {  {\source { \Gamma_C } }  } ; \source {  {\source \Gamma}  }  \rightsquigarrow  \intermed {  {\intermed \Sigma}_{{\mathrm{2}}}  } ; \intermed {  {\intermed { \Gamma_C } }  } ; \intermed {  {\intermed \Gamma}  } \) \\
    then \( \intermed {  {\intermed { \Gamma_C } }  } ; \intermed {  {\intermed \Gamma}  }  \vdash  \intermed {  {\intermed \Sigma}_{{\mathrm{1}}}  } : \intermed {  {\intermed e}_{{\mathrm{1}}}  }  \simeq_{log}  \intermed {  {\intermed \Sigma}_{{\mathrm{2}}}  } : \intermed {  {\intermed e}_{{\mathrm{2}}}  } : \intermed {  {\intermed \sigma}  } \)
    where \( \source {  {\source { \Gamma_C } }  } ; \source {  {\source \Gamma}  }  \vdash_{ty}^{\intermed {M} }  \source {  {\source \tau }  }\,{ \stoi{ \rightsquigarrow  \intermed {  {\intermed \sigma}  } } } \).
  \item
    If \( \source {  { \source P }  } ; \source {  {\source { \Gamma_C } }  } ; \source {  {\source \Gamma}  }  \vdash_{tm}^{\intermed {M} }  \source {  {\source e}  } \Leftarrow \source {  {\source \tau }  } \stoi{ \rightsquigarrow  \intermed {  {\intermed e}_{{\mathrm{1}}}  } } \)
    and \( \source {  { \source P }  } ; \source {  {\source { \Gamma_C } }  } ; \source {  {\source \Gamma}  }  \vdash_{tm}^{\intermed {M} }  \source {  {\source e}  } \Leftarrow \source {  {\source \tau }  } \stoi{ \rightsquigarrow  \intermed {  {\intermed e}_{{\mathrm{2}}}  } } \) \\
    and \( \vdash_{ctx}^{\intermed {M} }  \source {  { \source P }  } ; \source {  {\source { \Gamma_C } }  } ; \source {  {\source \Gamma}  }  \rightsquigarrow  \intermed {  {\intermed \Sigma}_{{\mathrm{1}}}  } ; \intermed {  {\intermed { \Gamma_C } }  } ; \intermed {  {\intermed \Gamma}  } \)
    and \( \vdash_{ctx}^{\intermed {M} }  \source {  { \source P }  } ; \source {  {\source { \Gamma_C } }  } ; \source {  {\source \Gamma}  }  \rightsquigarrow  \intermed {  {\intermed \Sigma}_{{\mathrm{2}}}  } ; \intermed {  {\intermed { \Gamma_C } }  } ; \intermed {  {\intermed \Gamma}  } \) \\
    then \( \intermed {  {\intermed { \Gamma_C } }  } ; \intermed {  {\intermed \Gamma}  }  \vdash  \intermed {  {\intermed \Sigma}_{{\mathrm{1}}}  } : \intermed {  {\intermed e}_{{\mathrm{1}}}  }  \simeq_{log}  \intermed {  {\intermed \Sigma}_{{\mathrm{2}}}  } : \intermed {  {\intermed e}_{{\mathrm{2}}}  } : \intermed {  {\intermed \sigma}  } \)
    where \( \source {  {\source { \Gamma_C } }  } ; \source {  {\source \Gamma}  }  \vdash_{ty}^{\intermed {M} }  \source {  {\source \tau }  }\,{ \stoi{ \rightsquigarrow  \intermed {  {\intermed \sigma}  } } } \).
  \end{itemize}
  \label{thmA:coh_exprA}
\end{theoremB}

\proof
By mutual induction on the first typing derivation.
This theorem is mutually proven with Theorems~\ref{thmA:env_eq_pres}
and~\ref{thmA:dict_cohA} (see Figure~\ref{fig:dep_graph1}).
Note that at the dependency between
Theorem~\ref{thmA:env_eq_pres} and~\ref{thmA:coh_exprA}, the size of \({ \source P }\) is
strictly decreasing, whereas \({ \source P }\) remains constant at every other
dependency. Because of this, the size of \({ \source P }\) is strictly decreasing in
every cycle. 
Consequently, the induction remains well-founded.

By applying Lemma~\ref{lemma:type_well_source_typing} to the
1\textsuperscript{st} hypothesis, we get that:
\begin{equation}
   \source {  {\source { \Gamma_C } }  } ; \source {  {\source \Gamma}  }  \vdash_{ty}^{\intermed {M} }  \source {  {\source \tau }  }\,{ \stoi{ \rightsquigarrow  \intermed {  {\intermed \sigma}  } } } 
  \label{eq:cohe0}
\end{equation}

\begin{description}
\item[Part 1]~\\
  \proofStep{\textsc{sTm-inf-true}}
  {
     \source {  { \source P }  } ; \source {  {\source { \Gamma_C } }  } ; \source {  {\source \Gamma}  }  \vdash_{tm}^{\intermed {M} }  \source {  \mathit{\source {True} }  } \Rightarrow \source {  \mathit{\source {Bool} }  } \stoi{ \rightsquigarrow  \intermed {  \mathit{\intermed {True} }  } } 
  }
  {
    Through case analysis, it is straightforward to see that the final step in
    the second derivation is \textsc{sTm-inf-true} as well.
    This means that \( \intermed {  {\intermed e}_{{\mathrm{1}}}  } = \intermed {  {\intermed e}_{{\mathrm{2}}}  } = \intermed {  \mathit{\intermed {True} }  } \).
    From Theorem~\ref{thmA:env_eq_pres}, we know that 
    \( \intermed {  {\intermed { \Gamma_C } }  }  \vdash  \intermed {  {\intermed \Sigma}_{{\mathrm{1}}}  }  \simeq_{log}  \intermed {  {\intermed \Sigma}_{{\mathrm{2}}}  } \).
    The goal follows from reflexivity Theorem~\ref{thmA:expr_reflex}. \\
  }
  \proofStep{\textsc{sTm-inf-false}}
  {
     \source {  { \source P }  } ; \source {  {\source { \Gamma_C } }  } ; \source {  {\source \Gamma}  }  \vdash_{tm}^{\intermed {M} }  \source {  \mathit{\source {True} }  } \Rightarrow \source {  \mathit{\source {Bool} }  } \stoi{ \rightsquigarrow  \intermed {  \mathit{\intermed {True} }  } } 
  }
  {
    The proof is identical to the \textsc{sTm-inf-true} case. \\
  }
  \proofStep{\textsc{sTm-inf-let}}
  {
     \source {  { \source P }  } ; \source {  {\source { \Gamma_C } }  } ; \source {  {\source \Gamma}  }  \vdash_{tm}^{\intermed {M} }  \source {  \textbf{\source {let} } \, {\source x}  \ottsym{:}  \forall  {\overline {\source a} }_{\ottmv{j}}  \ottsym{.}  {\overline {\source Q} }_{\ottmv{i}}  \Rightarrow  {\source \tau }_{{\mathrm{1}}}  \ottsym{=}  {\source e}_{{\mathrm{1}}} \, \textbf{\source {in} } \, {\source e}_{{\mathrm{2}}}  } \Rightarrow \source {  {\source \tau }_{{\mathrm{2}}}  } \stoi{ \rightsquigarrow  \intermed {  \textbf{\intermed {let} } \, {\intermed x}  \ottsym{:}  \forall  {\overline {\intermed a} }_{\ottmv{j}}  \ottsym{.}   {\overline {\intermed Q} }_{\ottmv{k}}  \Rightarrow  {\intermed \sigma}   \ottsym{=}  \Lambda  {\overline {\intermed a} }_{\ottmv{j}}  \ottsym{.}  \lambda  {\overline {\intermed \delta} }_{\ottmv{k}}  \ottsym{:}  {\overline {\intermed Q} }_{\ottmv{k}}  \ottsym{.}  {\intermed e}_{{\mathrm{1}}} \, \textbf{\intermed {in} } \, {\intermed e}_{{\mathrm{2}}}  } } 
  }
  {
    Through case analysis, we know that the final step in the second derivation
    has to be \textsc{sTm-inf-let} as well.
    This means that:
    \begin{gather*}
       \source {  { \source P }  } ; \source {  {\source { \Gamma_C } }  } ; \source {  {\source \Gamma}  }  \vdash_{tm}^{\intermed {M} }  \source {  \textbf{\source {let} } \, {\source x}  \ottsym{:}  \forall  {\overline {\source a} }_{\ottmv{j}}  \ottsym{.}  {\overline {\source Q} }_{\ottmv{i}}  \Rightarrow  {\source \tau }_{{\mathrm{1}}}  \ottsym{=}  {\source e}_{{\mathrm{1}}} \, \textbf{\source {in} } \, {\source e}_{{\mathrm{2}}}  } \Rightarrow \source {  {\source \tau }_{{\mathrm{2}}}  } \stoi{ \rightsquigarrow  \intermed {  {\intermed e}_{{\mathrm{3}}}  } }  \\
      \textit{where}~ \intermed {  {\intermed e}_{{\mathrm{3}}}  } = \intermed {  \textbf{\intermed {let} } \, {\intermed x}  \ottsym{:}  \forall  {\overline {\intermed a} }_{\ottmv{j}}  \ottsym{.}   {\overline {\intermed Q} }_{\ottmv{k}}  \Rightarrow  {\intermed \sigma}   \ottsym{=}  \Lambda  {\overline {\intermed a} }_{\ottmv{j}}  \ottsym{.}  \lambda  {\overline {\intermed \delta} }_{\ottmv{k}}  \ottsym{:}  {\overline {\intermed Q} }_{\ottmv{k}}  \ottsym{.}  {\intermed e}_{{\mathrm{1}}} \, \textbf{\intermed {in} } \, {\intermed e}_{{\mathrm{2}}}  }  \\
       \source {  { \source P }  } ; \source {  {\source { \Gamma_C } }  } ; \source {  {\source \Gamma}  }  \vdash_{tm}^{\intermed {M} }  \source {  \textbf{\source {let} } \, {\source x}  \ottsym{:}  \forall  {\overline {\source a} }_{\ottmv{j}}  \ottsym{.}  {\overline {\source Q} }_{\ottmv{i}}  \Rightarrow  {\source \tau }_{{\mathrm{1}}}  \ottsym{=}  {\source e}_{{\mathrm{1}}} \, \textbf{\source {in} } \, {\source e}_{{\mathrm{2}}}  } \Rightarrow \source {  {\source \tau }_{{\mathrm{2}}}  } \stoi{ \rightsquigarrow  \intermed {  {\intermed e}_{{\mathrm{4}}}  } }  \\
      \textit{where}~ \intermed {  {\intermed e}_{{\mathrm{4}}}  } = \intermed {  \textbf{\intermed {let} } \, {\intermed x}  \ottsym{:}  \forall  {\overline {\intermed a} }_{\ottmv{j}}  \ottsym{.}   {\overline {\intermed Q} }_{\ottmv{k}}  \Rightarrow  {\intermed \sigma}   \ottsym{=}  \Lambda  {\overline {\intermed a} }_{\ottmv{j}}  \ottsym{.}  \lambda  {\overline {\intermed \delta} }_{\ottmv{k}}  \ottsym{:}  {\overline {\intermed Q} }_{\ottmv{k}}  \ottsym{.}  {\intermed e}'_{{\mathrm{1}}} \, \textbf{\intermed {in} } \, {\intermed e}'_{{\mathrm{2}}}  } 
    \end{gather*}
    The goal to be proven is the following:
    \begin{equation}
       \intermed {  {\intermed { \Gamma_C } }  } ; \intermed {  {\intermed \Gamma}  }  \vdash  \intermed {  {\intermed \Sigma}_{{\mathrm{1}}}  } : \intermed {  {\intermed e}_{{\mathrm{3}}}  }  \simeq_{log}  \intermed {  {\intermed \Sigma}_{{\mathrm{2}}}  } : \intermed {  {\intermed e}_{{\mathrm{4}}}  } : \intermed {  {\intermed \sigma}_{{\mathrm{2}}}  } ~
      \textit{where}~ \source {  {\source { \Gamma_C } }  } ; \source {  {\source \Gamma}  }  \vdash_{ty}^{\intermed {M} }  \source {  {\source \sigma}_{{\mathrm{2}}}  }\,{ \stoi{ \rightsquigarrow  \intermed {  {\intermed \sigma}_{{\mathrm{2}}}  } } } 
      \label{eq:cohe_let1}
    \end{equation}
    The rule premise tells us that:
    \begin{gather}
       \source {  { \source P }  } ; \source {  {\source { \Gamma_C } }  } ; \source {  {\source \Gamma}  \ottsym{,}  {\overline {\source a} }_{\ottmv{j}}  \ottsym{,}  {\overline {\source \delta} }_{\ottmv{k}}  \ottsym{:}  {\overline {\source Q} }_{\ottmv{k}}  }  \vdash_{tm}^{\intermed {M} }  \source {  {\source e}_{{\mathrm{1}}}  } \Leftarrow \source {  {\source \tau }_{{\mathrm{1}}}  } \stoi{ \rightsquigarrow  \intermed {  {\intermed e}_{{\mathrm{1}}}  } } 
      \label{eq:cohe_let2} \\
       \source {  { \source P }  } ; \source {  {\source { \Gamma_C } }  } ; \source {  {\source \Gamma}  \ottsym{,}  {\source x}  \ottsym{:}  \forall  {\overline {\source a} }_{\ottmv{j}}  \ottsym{.}  {\overline {\source Q} }_{\ottmv{k}}  \Rightarrow  {\source \tau }_{{\mathrm{1}}}  }  \vdash_{tm}^{\intermed {M} }  \source {  {\source e}_{{\mathrm{2}}}  } \Rightarrow \source {  {\source \tau }_{{\mathrm{2}}}  } \stoi{ \rightsquigarrow  \intermed {  {\intermed e}_{{\mathrm{2}}}  } } 
      \label{eq:cohe_let3} \\
       \source {  { \source P }  } ; \source {  {\source { \Gamma_C } }  } ; \source {  {\source \Gamma}  \ottsym{,}  {\overline {\source a} }_{\ottmv{j}}  \ottsym{,}  {\overline {\source \delta} }_{\ottmv{k}}  \ottsym{:}  {\overline {\source Q} }_{\ottmv{k}}  }  \vdash_{tm}^{\intermed {M} }  \source {  {\source e}_{{\mathrm{1}}}  } \Leftarrow \source {  {\source \tau }_{{\mathrm{1}}}  } \stoi{ \rightsquigarrow  \intermed {  {\intermed e}'_{{\mathrm{1}}}  } } 
      \label{eq:cohe_let4} \\
       \source {  { \source P }  } ; \source {  {\source { \Gamma_C } }  } ; \source {  {\source \Gamma}  \ottsym{,}  {\source x}  \ottsym{:}  \forall  {\overline {\source a} }_{\ottmv{j}}  \ottsym{.}  {\overline {\source Q} }_{\ottmv{k}}  \Rightarrow  {\source \tau }_{{\mathrm{1}}}  }  \vdash_{tm}^{\intermed {M} }  \source {  {\source e}_{{\mathrm{2}}}  } \Rightarrow \source {  {\source \tau }_{{\mathrm{2}}}  } \stoi{ \rightsquigarrow  \intermed {  {\intermed e}'_{{\mathrm{2}}}  } } 
      \label{eq:cohe_let5} \\
      \textit{where}~ \textbf {closure}  ( \source {  {\source { \Gamma_C } }  } ; \source {  {\overline {\source Q} }_{\ottmv{i}}  } ) = \source {  {\overline {\source Q} }_{\ottmv{k}}  } 
      \nonumber
    \end{gather}
    From Lemma~\ref{lemma:ctx_well_source_mid_typing}, together with
    Equations~\ref{eq:cohe_let2}, \ref{eq:cohe_let3}, \ref{eq:cohe_let4}
    and~\ref{eq:cohe_let5}, and through repeated case analysis on the
    results to discover the contents of the elaborated environments, we get that:
    \begin{gather}
       \vdash_{ctx}^{\intermed {M} }  \source {  { \source P }  } ; \source {  {\source { \Gamma_C } }  } ; \source {  {\source \Gamma}  \ottsym{,}  {\overline {\source a} }_{\ottmv{j}}  \ottsym{,}  {\overline {\source \delta} }_{\ottmv{k}}  \ottsym{:}  {\overline {\source Q} }_{\ottmv{k}}  }  \rightsquigarrow  \intermed {  {\intermed \Sigma}_{{\mathrm{1}}}  } ; \intermed {  {\intermed { \Gamma_C } }  } ; \intermed {  {\intermed \Gamma}  \ottsym{,}  {\overline {\intermed a} }_{\ottmv{j}}  \ottsym{,}  {\overline {\intermed \delta} }_{\ottmv{k}}  \ottsym{:}  {\overline {\intermed Q} }_{\ottmv{k}}  } 
      \nonumber \\
       \vdash_{ctx}^{\intermed {M} }  \source {  { \source P }  } ; \source {  {\source { \Gamma_C } }  } ; \source {  {\source \Gamma}  \ottsym{,}  {\source x}  \ottsym{:}  \forall  {\overline {\source a} }_{\ottmv{j}}  \ottsym{.}  {\overline {\source Q} }_{\ottmv{k}}  \Rightarrow  {\source \tau }_{{\mathrm{1}}}  }  \rightsquigarrow  \intermed {  {\intermed \Sigma}_{{\mathrm{1}}}  } ; \intermed {  {\intermed { \Gamma_C } }  } ; \intermed {  {\intermed \Gamma}  \ottsym{,}  {\intermed x}  \ottsym{:}  \forall  {\overline {\intermed a} }_{\ottmv{j}}  \ottsym{.}   {\overline {\intermed Q} }_{\ottmv{k}}  \Rightarrow  {\intermed \sigma}   } 
      \nonumber \\
       \vdash_{ctx}^{\intermed {M} }  \source {  { \source P }  } ; \source {  {\source { \Gamma_C } }  } ; \source {  {\source \Gamma}  \ottsym{,}  {\overline {\source a} }_{\ottmv{j}}  \ottsym{,}  {\overline {\source \delta} }_{\ottmv{k}}  \ottsym{:}  {\overline {\source Q} }_{\ottmv{k}}  }  \rightsquigarrow  \intermed {  {\intermed \Sigma}_{{\mathrm{2}}}  } ; \intermed {  {\intermed { \Gamma_C } }  } ; \intermed {  {\intermed \Gamma}  \ottsym{,}  {\overline {\intermed a} }_{\ottmv{j}}  \ottsym{,}  {\overline {\intermed \delta} }_{\ottmv{k}}  \ottsym{:}  {\overline {\intermed Q} }_{\ottmv{k}}  } 
      \nonumber \\
       \vdash_{ctx}^{\intermed {M} }  \source {  { \source P }  } ; \source {  {\source { \Gamma_C } }  } ; \source {  {\source \Gamma}  \ottsym{,}  {\source x}  \ottsym{:}  \forall  {\overline {\source a} }_{\ottmv{j}}  \ottsym{.}  {\overline {\source Q} }_{\ottmv{k}}  \Rightarrow  {\source \tau }_{{\mathrm{1}}}  }  \rightsquigarrow  \intermed {  {\intermed \Sigma}_{{\mathrm{2}}}  } ; \intermed {  {\intermed { \Gamma_C } }  } ; \intermed {  {\intermed \Gamma}  \ottsym{,}  {\intermed x}  \ottsym{:}  \forall  {\overline {\intermed a} }_{\ottmv{j}}  \ottsym{.}   {\overline {\intermed Q} }_{\ottmv{k}}  \Rightarrow  {\intermed \sigma}   } 
      \nonumber
    \end{gather}
    By applying the induction hypothesis to Equations~\ref{eq:cohe_let3} and
    \ref{eq:cohe_let5}, we get:
    \begin{equation}
       \intermed {  {\intermed { \Gamma_C } }  } ; \intermed {  {\intermed \Gamma}  \ottsym{,}  {\intermed x}  \ottsym{:}  \forall  {\overline {\intermed a} }_{\ottmv{j}}  \ottsym{.}   {\overline {\intermed Q} }_{\ottmv{k}}  \Rightarrow  {\intermed \sigma}   }  \vdash  \intermed {  {\intermed \Sigma}_{{\mathrm{1}}}  } : \intermed {  {\intermed e}_{{\mathrm{2}}}  }  \simeq_{log}  \intermed {  {\intermed \Sigma}_{{\mathrm{2}}}  } : \intermed {  {\intermed e}'_{{\mathrm{2}}}  } : \intermed {  {\intermed \sigma}_{{\mathrm{2}}}  } 
      \label{eq:cohe_let6}
    \end{equation}
    Furthermore, applying Part 2 of this lemma to
    Equations~\ref{eq:cohe_let2} and \ref{eq:cohe_let4} results in:
    \begin{equation}
       \intermed {  {\intermed { \Gamma_C } }  } ; \intermed {  {\intermed \Gamma}  \ottsym{,}  {\overline {\intermed a} }_{\ottmv{j}}  \ottsym{,}  {\overline {\intermed \delta} }_{\ottmv{k}}  \ottsym{:}  {\overline {\intermed Q} }_{\ottmv{k}}  }  \vdash  \intermed {  {\intermed \Sigma}_{{\mathrm{1}}}  } : \intermed {  {\intermed e}_{{\mathrm{1}}}  }  \simeq_{log}  \intermed {  {\intermed \Sigma}_{{\mathrm{2}}}  } : \intermed {  {\intermed e}'_{{\mathrm{1}}}  } : \intermed {  {\intermed \sigma}  } 
      \label{eq:cohe_let7}
    \end{equation}
    Applying compatibility Lemma~\ref{lemma:compat_let}, together with
    Equation~\ref{eq:cohe_let6}, reduces Goal~\ref{eq:cohe_let1} to:
    \begin{equation}
       \intermed {  {\intermed { \Gamma_C } }  } ; \intermed {  {\intermed \Gamma}  }  \vdash  \intermed {  {\intermed \Sigma}_{{\mathrm{1}}}  } : \intermed {  \Lambda  {\overline {\intermed a} }_{\ottmv{j}}  \ottsym{.}  \lambda  {\overline {\intermed \delta} }_{\ottmv{k}}  \ottsym{:}  {\overline {\intermed Q} }_{\ottmv{k}}  \ottsym{.}  {\intermed e}_{{\mathrm{1}}}  }  \simeq_{log}  \intermed {  {\intermed \Sigma}_{{\mathrm{2}}}  } : \intermed {  \Lambda  {\overline {\intermed a} }_{\ottmv{j}}  \ottsym{.}  \lambda  {\overline {\intermed \delta} }_{\ottmv{k}}  \ottsym{:}  {\overline {\intermed Q} }_{\ottmv{k}}  \ottsym{.}  {\intermed e}'_{{\mathrm{1}}}  } : \intermed {  \forall  {\overline {\intermed a} }_{\ottmv{j}}  \ottsym{.}   {\overline {\intermed Q} }_{\ottmv{k}}  \Rightarrow  {\intermed \sigma}   } 
      \label{eq:cohe_let8}
    \end{equation}
    Combining compatibility Lemma~\ref{lemma:compat_dabs} with
    Equation~\ref{eq:cohe_let7} gives us:
    \begin{equation}
       \intermed {  {\intermed { \Gamma_C } }  } ; \intermed {  {\intermed \Gamma}  \ottsym{,}  {\overline {\intermed a} }_{\ottmv{j}}  }  \vdash  \intermed {  {\intermed \Sigma}_{{\mathrm{1}}}  } : \intermed {  \lambda  {\overline {\intermed \delta} }_{\ottmv{k}}  \ottsym{:}  {\overline {\intermed Q} }_{\ottmv{k}}  \ottsym{.}  {\intermed e}_{{\mathrm{1}}}  }  \simeq_{log}  \intermed {  {\intermed \Sigma}_{{\mathrm{2}}}  } : \intermed {  \lambda  {\overline {\intermed \delta} }_{\ottmv{k}}  \ottsym{:}  {\overline {\intermed Q} }_{\ottmv{k}}  \ottsym{.}  {\intermed e}'_{{\mathrm{1}}}  } : \intermed {   {\overline {\intermed Q} }_{\ottmv{k}}  \Rightarrow  {\intermed \sigma}   } 
      \label{eq:cohe_let9}
    \end{equation}
    Goal~\ref{eq:cohe_let8} follows directly by applying
    Equation~\ref{eq:cohe_let9} to compatibility Lemma~\ref{lemma:compat_tyabs}. \\
  }
  \proofStep{\textsc{sTm-inf-ArrE}}
  {
     \source {  { \source P }  } ; \source {  {\source { \Gamma_C } }  } ; \source {  {\source \Gamma}  }  \vdash_{tm}^{\intermed {M} }  \source {  {\source e}_{{\mathrm{1}}} \, {\source e}_{{\mathrm{2}}}  } \Rightarrow \source {  {\source \tau }_{{\mathrm{2}}}  } \stoi{ \rightsquigarrow  \intermed {  {\intermed e}_{{\mathrm{1}}} \, {\intermed e}_{{\mathrm{2}}}  } } 
  }
  {
    Through case analysis, we see that the final step in the second derivation
    can only be \textsc{sTm-inf-ArrE}.
    This means that:
    \begin{gather*}
       \source {  { \source P }  } ; \source {  {\source { \Gamma_C } }  } ; \source {  {\source \Gamma}  }  \vdash_{tm}^{\intermed {M} }  \source {  {\source e}_{{\mathrm{1}}} \, {\source e}_{{\mathrm{2}}}  } \Rightarrow \source {  {\source \tau }_{{\mathrm{2}}}  } \stoi{ \rightsquigarrow  \intermed {  {\intermed e}_{{\mathrm{1}}} \, {\intermed e}_{{\mathrm{2}}}  } }  \\
       \source {  { \source P }  } ; \source {  {\source { \Gamma_C } }  } ; \source {  {\source \Gamma}  }  \vdash_{tm}^{\intermed {M} }  \source {  {\source e}_{{\mathrm{1}}} \, {\source e}_{{\mathrm{2}}}  } \Rightarrow \source {  {\source \tau }_{{\mathrm{2}}}  } \stoi{ \rightsquigarrow  \intermed {  {\intermed e}'_{{\mathrm{1}}} \, {\intermed e}'_{{\mathrm{2}}}  } } 
    \end{gather*}
    The goal to be proven is the following:
    \begin{equation}
       \intermed {  {\intermed { \Gamma_C } }  } ; \intermed {  {\intermed \Gamma}  }  \vdash  \intermed {  {\intermed \Sigma}_{{\mathrm{1}}}  } : \intermed {  {\intermed e}_{{\mathrm{1}}} \, {\intermed e}_{{\mathrm{2}}}  }  \simeq_{log}  \intermed {  {\intermed \Sigma}_{{\mathrm{2}}}  } : \intermed {  {\intermed e}'_{{\mathrm{1}}} \, {\intermed e}'_{{\mathrm{2}}}  } : \intermed {  {\intermed \sigma}_{{\mathrm{2}}}  } ~
      \textit{where}~ \source {  {\source { \Gamma_C } }  } ; \source {  {\source \Gamma}  }  \vdash_{ty}^{\intermed {M} }  \source {  {\source \tau }_{{\mathrm{2}}}  }\,{ \stoi{ \rightsquigarrow  \intermed {  {\intermed \sigma}_{{\mathrm{2}}}  } } } 
      \label{eq:cohe_arrE7}
    \end{equation}
    The rule premise tells us that:
    \begin{gather}
       \source {  { \source P }  } ; \source {  {\source { \Gamma_C } }  } ; \source {  {\source \Gamma}  }  \vdash_{tm}^{\intermed {M} }  \source {  {\source e}_{{\mathrm{1}}}  } \Rightarrow \source {  {\source \tau }_{{\mathrm{1}}}  \rightarrow  {\source \tau }_{{\mathrm{2}}}  } \stoi{ \rightsquigarrow  \intermed {  {\intermed e}_{{\mathrm{1}}}  } } 
      \label{eq:cohe_arrE1} \\
       \source {  { \source P }  } ; \source {  {\source { \Gamma_C } }  } ; \source {  {\source \Gamma}  }  \vdash_{tm}^{\intermed {M} }  \source {  {\source e}_{{\mathrm{2}}}  } \Leftarrow \source {  {\source \tau }_{{\mathrm{1}}}  } \stoi{ \rightsquigarrow  \intermed {  {\intermed e}_{{\mathrm{2}}}  } } 
      \label{eq:cohe_arrE2} \\
       \source {  { \source P }  } ; \source {  {\source { \Gamma_C } }  } ; \source {  {\source \Gamma}  }  \vdash_{tm}^{\intermed {M} }  \source {  {\source e}_{{\mathrm{1}}}  } \Rightarrow \source {  {\source \tau }_{{\mathrm{1}}}  \rightarrow  {\source \tau }_{{\mathrm{2}}}  } \stoi{ \rightsquigarrow  \intermed {  {\intermed e}'_{{\mathrm{1}}}  } } 
      \label{eq:cohe_arrE3} \\
       \source {  { \source P }  } ; \source {  {\source { \Gamma_C } }  } ; \source {  {\source \Gamma}  }  \vdash_{tm}^{\intermed {M} }  \source {  {\source e}_{{\mathrm{2}}}  } \Leftarrow \source {  {\source \tau }_{{\mathrm{1}}}  } \stoi{ \rightsquigarrow  \intermed {  {\intermed e}'_{{\mathrm{2}}}  } } 
      \label{eq:cohe_arrE4}
    \end{gather}
    Applying the induction hypothesis on Equations~\ref{eq:cohe_arrE1} and
    \ref{eq:cohe_arrE3} results in:
    \begin{equation}
       \intermed {  {\intermed { \Gamma_C } }  } ; \intermed {  {\intermed \Gamma}  }  \vdash  \intermed {  {\intermed \Sigma}_{{\mathrm{1}}}  } : \intermed {  {\intermed e}_{{\mathrm{1}}}  }  \simeq_{log}  \intermed {  {\intermed \Sigma}_{{\mathrm{2}}}  } : \intermed {  {\intermed e}'_{{\mathrm{1}}}  } : \intermed {  {\intermed \sigma}_{{\mathrm{1}}}  \rightarrow  {\intermed \sigma}_{{\mathrm{2}}}  } ~
      \textit{where}~ \source {  {\source { \Gamma_C } }  } ; \source {  {\source \Gamma}  }  \vdash_{ty}^{\intermed {M} }  \source {  {\source \tau }_{{\mathrm{1}}}  \rightarrow  {\source \tau }_{{\mathrm{2}}}  }\,{ \stoi{ \rightsquigarrow  \intermed {  {\intermed \sigma}_{{\mathrm{1}}}  \rightarrow  {\intermed \sigma}_{{\mathrm{2}}}  } } } 
      \label{eq:cohe_arrE5}
    \end{equation}
    By applying Part 2 of this lemma to Equations~\ref{eq:cohe_arrE2} and
    \ref{eq:cohe_arrE4} we get:
    \begin{equation}
       \intermed {  {\intermed { \Gamma_C } }  } ; \intermed {  {\intermed \Gamma}  }  \vdash  \intermed {  {\intermed \Sigma}_{{\mathrm{1}}}  } : \intermed {  {\intermed e}_{{\mathrm{2}}}  }  \simeq_{log}  \intermed {  {\intermed \Sigma}_{{\mathrm{2}}}  } : \intermed {  {\intermed e}'_{{\mathrm{2}}}  } : \intermed {  {\intermed \sigma}_{{\mathrm{1}}}  } ~
      \textit{where}~ \source {  {\source { \Gamma_C } }  } ; \source {  {\source \Gamma}  }  \vdash_{ty}^{\intermed {M} }  \source {  {\source \tau }_{{\mathrm{1}}}  }\,{ \stoi{ \rightsquigarrow  \intermed {  {\intermed \sigma}_{{\mathrm{1}}}  } } } 
      \label{eq:cohe_arrE6}
    \end{equation}
    Goal~\ref{eq:cohe_arrE7} follows by applying Equations~\ref{eq:cohe_arrE5}
    and \ref{eq:cohe_arrE6} to compatibility Lemma~\ref{lemma:compat_app}. \\
  }
  \proofStep{\textsc{sTm-inf-Ann}}
  {
     \source {  { \source P }  } ; \source {  {\source { \Gamma_C } }  } ; \source {  {\source \Gamma}  }  \vdash_{tm}^{\intermed {M} }  \source {  {\source e}  ::  {\source \tau }  } \Rightarrow \source {  {\source \tau }  } \stoi{ \rightsquigarrow  \intermed {  {\intermed e}  } } 
  }
  {
    Through case analysis we know that the final step in the second derivation
    is \textsc{sTm-inf-Ann}. This means that:
    \begin{gather*}
       \source {  { \source P }  } ; \source {  {\source { \Gamma_C } }  } ; \source {  {\source \Gamma}  }  \vdash_{tm}^{\intermed {M} }  \source {  {\source e}  ::  {\source \tau }  } \Rightarrow \source {  {\source \tau }  } \stoi{ \rightsquigarrow  \intermed {  {\intermed e}  } }  \\
       \source {  { \source P }  } ; \source {  {\source { \Gamma_C } }  } ; \source {  {\source \Gamma}  }  \vdash_{tm}^{\intermed {M} }  \source {  {\source e}  ::  {\source \tau }  } \Rightarrow \source {  {\source \tau }  } \stoi{ \rightsquigarrow  \intermed {  {\intermed e}'  } } 
    \end{gather*}
    The goal to be proven is the following:
    \begin{equation}
       \intermed {  {\intermed { \Gamma_C } }  } ; \intermed {  {\intermed \Gamma}  }  \vdash  \intermed {  {\intermed \Sigma}_{{\mathrm{1}}}  } : \intermed {  {\intermed e}  }  \simeq_{log}  \intermed {  {\intermed \Sigma}_{{\mathrm{2}}}  } : \intermed {  {\intermed e}'  } : \intermed {  {\intermed \sigma}  } ~
      \textit{where}~ \source {  {\source { \Gamma_C } }  } ; \source {  {\source \Gamma}  }  \vdash_{ty}^{\intermed {M} }  \source {  {\source \tau }  }\,{ \stoi{ \rightsquigarrow  \intermed {  {\intermed \sigma}  } } } 
      \label{eq:cohe_ann1}
    \end{equation}
    From the rule premise we know:
    \begin{gather}
       \source {  { \source P }  } ; \source {  {\source { \Gamma_C } }  } ; \source {  {\source \Gamma}  }  \vdash_{tm}^{\intermed {M} }  \source {  {\source e}  } \Leftarrow \source {  {\source \tau }  } \stoi{ \rightsquigarrow  \intermed {  {\intermed e}  } } 
      \label{eq:cohe_ann2} \\
       \source {  { \source P }  } ; \source {  {\source { \Gamma_C } }  } ; \source {  {\source \Gamma}  }  \vdash_{tm}^{\intermed {M} }  \source {  {\source e}  } \Leftarrow \source {  {\source \tau }  } \stoi{ \rightsquigarrow  \intermed {  {\intermed e}'  } } 
      \label{eq:cohe_ann3}
    \end{gather}
    The goal follows directly from Part 2 of this lemma, applied to
    Equation~\ref{eq:cohe_ann2} and \ref{eq:cohe_ann3}.
  }
\item[Part 2]~\\
\proofStep{\textsc{sTm-check-var}}
{
   \source {  { \source P }  } ; \source {  {\source { \Gamma_C } }  } ; \source {  {\source \Gamma}  }  \vdash_{tm}^{\intermed {M} }  \source {  {\source x}  } \Leftarrow \source {  [  {\overline {\source \tau} }_{\ottmv{j}}  /  {\overline {\source a} }_{\ottmv{j}}  ]  {\source \tau }  } \stoi{ \rightsquigarrow  \intermed {  {\intermed x} \, \overline {\intermed \sigma}_{\ottmv{j}} \, {\overline {\intermed d} }_{\ottmv{i}}  } } 
}
{
  Through case analysis, we see that the final step in the second typing
  derivation can either be \textsc{sTm-check-var} or \textsc{sTm-check-Inf}.
  However, noting that no matching inference rules exist, we conclude that the
  final derivation step has to be \textsc{sTm-check-var}.
  This means that:
  \begin{gather*}
     \source {  { \source P }  } ; \source {  {\source { \Gamma_C } }  } ; \source {  {\source \Gamma}  }  \vdash_{tm}^{\intermed {M} }  \source {  {\source x}  } \Leftarrow \source {  [  {\overline {\source \tau} }_{\ottmv{j}}  /  {\overline {\source a} }_{\ottmv{j}}  ]  {\source \tau }  } \stoi{ \rightsquigarrow  \intermed {  {\intermed x} \, \overline {\intermed \sigma}_{\ottmv{j}} \, {\overline {\intermed d} }_{\ottmv{i}}  } }  \\
     \source {  { \source P }  } ; \source {  {\source { \Gamma_C } }  } ; \source {  {\source \Gamma}  }  \vdash_{tm}^{\intermed {M} }  \source {  {\source x}  } \Leftarrow \source {  [  {\overline {\source \tau} }_{\ottmv{j}}  /  {\overline {\source a} }_{\ottmv{j}}  ]  {\source \tau }  } \stoi{ \rightsquigarrow  \intermed {  {\intermed x} \, \overline {\intermed \sigma}'_{\ottmv{j}} \, {\overline {\intermed d} }'_{\ottmv{i}}  } } 
  \end{gather*}
  The goal to be proven is the following:
  \begin{equation}
     \intermed {  {\intermed { \Gamma_C } }  } ; \intermed {  {\intermed \Gamma}  }  \vdash  \intermed {  {\intermed \Sigma}_{{\mathrm{1}}}  } : \intermed {  {\intermed x} \, \overline {\intermed \sigma}_{\ottmv{j}} \, {\overline {\intermed d} }_{\ottmv{i}}  }  \simeq_{log}  \intermed {  {\intermed \Sigma}_{{\mathrm{2}}}  } : \intermed {  {\intermed x} \, \overline {\intermed \sigma}'_{\ottmv{j}} \, {\overline {\intermed d} }'_{\ottmv{i}}  } : \intermed {  {\intermed \sigma}'  } ~
    \textit{where}~ \source {  {\source { \Gamma_C } }  } ; \source {  {\source \Gamma}  }  \vdash_{ty}^{\intermed {M} }  \source {  [  {\overline {\source \tau} }_{\ottmv{j}}  /  {\overline {\source a} }_{\ottmv{j}}  ]  {\source \tau }  }\,{ \stoi{ \rightsquigarrow  \intermed {  {\intermed \sigma}'  } } } 
    \label{eq:cohe_var1}
  \end{equation}
  By inversion on Equation~\ref{eq:cohe0}, we know that \( \intermed {  {\intermed \sigma}'  } = \intermed {  [  \overline {\intermed \sigma}_{\ottmv{j}}  /  {\overline {\intermed a} }_{\ottmv{j}}  ]  {\intermed \sigma}  } \).
  
  The rule premise tells us that:
  \begin{gather}
     ( \source {  {\source x}  } : \source {  \forall  {\overline {\source a} }_{\ottmv{j}}  \ottsym{.}  {\overline {\source Q} }_{\ottmv{i}}  \Rightarrow  {\source \tau }  } )  \in  \source {  {\source \Gamma}  } 
    \label{eq:cohe_var2} \\
    \ottcomp{ \source {  { \source P }  } ; \source {  {\source { \Gamma_C } }  } ; \source {  {\source \Gamma}  }  \vDash^{\intermed {M} }  \source {  [  {\overline {\source \tau} }_{\ottmv{j}}  /  {\overline {\source a} }_{\ottmv{j}}  ]  {\source Q}_{\ottmv{i}}  } \stoi{ \rightsquigarrow  \intermed {  {\intermed d}_{\ottmv{i}}  } } }{\ottmv{i}}
    \label{eq:cohe_var3} \\
    \ottcomp{ \source {  {\source { \Gamma_C } }  } ; \source {  {\source \Gamma}  }  \vdash_{ty}^{\intermed {M} }  \source {  {\source \tau }_{\ottmv{j}}  }\,{ \stoi{ \rightsquigarrow  \intermed {  {\intermed \sigma}_{\ottmv{j}}  } } } }{\ottmv{j}}
    \label{eq:cohe_var4} \\
    \ottcomp{ \source {  { \source P }  } ; \source {  {\source { \Gamma_C } }  } ; \source {  {\source \Gamma}  }  \vDash^{\intermed {M} }  \source {  [  {\overline {\source \tau} }_{\ottmv{j}}  /  {\overline {\source a} }_{\ottmv{j}}  ]  {\source Q}_{\ottmv{i}}  } \stoi{ \rightsquigarrow  \intermed {  {\intermed d}'_{\ottmv{i}}  } } }{\ottmv{i}}
    \label{eq:cohe_var4b} \\
    \ottcomp{ \source {  {\source { \Gamma_C } }  } ; \source {  {\source \Gamma}  }  \vdash_{ty}^{\intermed {M} }  \source {  {\source \tau }_{\ottmv{j}}  }\,{ \stoi{ \rightsquigarrow  \intermed {  {\intermed \sigma}'_{\ottmv{j}}  } } } }{\ottmv{j}}
    \label{eq:cohe_var5}
  \end{gather}
  Since type elaboration is completely deterministic (Lemma~\ref{lemma:ty_elab_unique}),
  we know that \( \intermed {  \overline {\intermed \sigma}_{\ottmv{j}}  } = \intermed {  \overline {\intermed \sigma}'_{\ottmv{j}}  } \).

  From Lemma~\ref{lemma:pres_env_tm}, 
  combined with the 3\textsuperscript{rd} hypothesis and
  Equation~\ref{eq:cohe_var2}, we know that:
  \begin{equation}
     ( \intermed {  {\intermed x}  } : \intermed {  \forall  {\overline {\intermed a} }_{\ottmv{j}}  \ottsym{.}   {\overline {\intermed Q} }_{\ottmv{i}}  \Rightarrow  {\intermed \sigma}   } )  \in  \intermed {  {\intermed \Gamma}  } 
    \label{eq:cohe_var6}
  \end{equation}
  By applying Theorem~\ref{thmA:typing_pres_env} to the 3\textsuperscript{rd} and
  4\textsuperscript{th} hypothesis, we get:
  \begin{gather}
     \vdash_{ctx}  \intermed {  {\intermed \Sigma}_{{\mathrm{1}}}  } ; \intermed {  {\intermed { \Gamma_C } }  } ; \intermed {  {\intermed \Gamma}  } 
    \label{eq:cohe_var6b} \\
     \vdash_{ctx}  \intermed {  {\intermed \Sigma}_{{\mathrm{2}}}  } ; \intermed {  {\intermed { \Gamma_C } }  } ; \intermed {  {\intermed \Gamma}  } 
    \label{eq:cohe_var6c}
  \end{gather}
  Applying Equations~\ref{eq:cohe_var6}, \ref{eq:cohe_var6b} and~\ref{eq:cohe_var6c} to
  \textsc{iTm-var}, results in:
  \begin{gather}
     \intermed {  {\intermed \Sigma}_{{\mathrm{1}}}  } ; \intermed {  {\intermed { \Gamma_C } }  } ; \intermed {  {\intermed \Gamma}  }  \vdash_{tm}  \intermed {  {\intermed x}  } : \intermed {  \forall  {\overline {\intermed a} }_{\ottmv{j}}  \ottsym{.}   {\overline {\intermed Q} }_{\ottmv{i}}  \Rightarrow  {\intermed \sigma}   } \itot{ \rightsquigarrow  \target {  {\target x}  } } 
    \label{eq:cohe_var7} \\
     \intermed {  {\intermed \Sigma}_{{\mathrm{2}}}  } ; \intermed {  {\intermed { \Gamma_C } }  } ; \intermed {  {\intermed \Gamma}  }  \vdash_{tm}  \intermed {  {\intermed x}  } : \intermed {  \forall  {\overline {\intermed a} }_{\ottmv{j}}  \ottsym{.}   {\overline {\intermed Q} }_{\ottmv{i}}  \Rightarrow  {\intermed \sigma}   } \itot{ \rightsquigarrow  \target {  {\target x}  } } 
    \label{eq:cohe_var7b}
  \end{gather}
  From Theorem~\ref{thmA:env_eq_pres}, we know that:
  \begin{equation}
     \intermed {  {\intermed { \Gamma_C } }  }  \vdash  \intermed {  {\intermed \Sigma}_{{\mathrm{1}}}  }  \simeq_{log}  \intermed {  {\intermed \Sigma}_{{\mathrm{2}}}  } 
    \label{eq:cohe_var7c}
  \end{equation}
  From reflexivity Theorem~\ref{thmA:expr_reflex}, applied on
  Equations~\ref{eq:cohe_var7}, \ref{eq:cohe_var7b} and~\ref{eq:cohe_var7c}, we
  know that:
  \begin{equation}
     \intermed {  {\intermed { \Gamma_C } }  } ; \intermed {  {\intermed \Gamma}  }  \vdash  \intermed {  {\intermed \Sigma}_{{\mathrm{1}}}  } : \intermed {  {\intermed x}  }  \simeq_{log}  \intermed {  {\intermed \Sigma}_{{\mathrm{2}}}  } : \intermed {  {\intermed x}  } : \intermed {  \forall  {\overline {\intermed a} }_{\ottmv{j}}  \ottsym{.}   {\overline {\intermed Q} }_{\ottmv{i}}  \Rightarrow  {\intermed \sigma}   } 
    \label{eq:cohe_var8}
  \end{equation}
  From repeated case analysis on the 3\textsuperscript{rd} hypothesis
  (\textsc{sCtx-pgmInst}), we get:
  \begin{equation}
     \vdash_{ctx}^{\intermed {M} }  \source {  \bullet  } ; \source {  {\source { \Gamma_C } }  } ; \source {  {\source \Gamma}  }  \rightsquigarrow  \intermed {  \bullet  } ; \intermed {  {\intermed { \Gamma_C } }  } ; \intermed {  {\intermed \Gamma}  } 
    \label{eq:cohe_var8a}
  \end{equation}
  By applying Theorem~\ref{thmA:typing_pres_types} to Equations~\ref{eq:cohe_var4}
  and~\ref{eq:cohe_var8a}, we know that:
  \begin{equation}
    \ottcomp{ \intermed {  {\intermed { \Gamma_C } }  } ; \intermed {  {\intermed \Gamma}  }  \vdash_{ty}  \intermed {  {\intermed \sigma}_{\ottmv{j}}  } \itot{ \rightsquigarrow  \target {  {\target \sigma }'_{\ottmv{j}}  } } }{\ottmv{j}}
    \label{eq:cohe_var8b}
  \end{equation}
  Applying compatibility Lemma~\ref{lemma:compat_tyapp} \(j\) times to
  Equations~\ref{eq:cohe_var8} and~\ref{eq:cohe_var8b} results in:
  \begin{equation}
     \intermed {  {\intermed { \Gamma_C } }  } ; \intermed {  {\intermed \Gamma}  }  \vdash  \intermed {  {\intermed \Sigma}_{{\mathrm{1}}}  } : \intermed {  {\intermed x} \, \overline {\intermed \sigma}_{\ottmv{j}}  }  \simeq_{log}  \intermed {  {\intermed \Sigma}_{{\mathrm{2}}}  } : \intermed {  {\intermed x} \, \overline {\intermed \sigma}_{\ottmv{j}}  } : \intermed {  [  \overline {\intermed \sigma}_{\ottmv{j}}  /  {\overline {\intermed a} }_{\ottmv{j}}  ]   {\overline {\intermed Q} }_{\ottmv{i}}  \Rightarrow  [  \overline {\intermed \sigma}_{\ottmv{j}}  /  {\overline {\intermed a} }_{\ottmv{j}}  ]  {\intermed \sigma}   } 
    \label{eq:cohe_var9}
  \end{equation}
  Applying Theorem~\ref{thmA:dict_cohA} to Equations~\ref{eq:cohe_var3} and
  \ref{eq:cohe_var4b} gives us:
  \begin{gather}
    \ottcomp{ \intermed {  {\intermed { \Gamma_C } }  } ; \intermed {  {\intermed \Gamma}  }  \vdash  \intermed {  {\intermed \Sigma}_{{\mathrm{1}}}  } : \intermed {  {\intermed d}_{\ottmv{i}}  }  \simeq_{log}  \intermed {  {\intermed \Sigma}_{{\mathrm{2}}}  } : \intermed {  {\intermed d}'_{\ottmv{i}}  } : \intermed {  {\intermed Q}'_{\ottmv{i}}  } }{\ottmv{i}}
    \label{eq:cohe_var10} \\
    \textit{where}~\ottcomp{ \source {  {\source { \Gamma_C } }  } ; \source {  {\source \Gamma}  }  \vdash_{\mathit {Q} }^{\intermed {M} }  \source {  [  {\overline {\source \tau} }_{\ottmv{j}}  /  {\overline {\source a} }_{\ottmv{j}}  ]  {\source Q}_{\ottmv{i}}  }  \rightsquigarrow  \intermed {  {\intermed Q}'_{\ottmv{i}}  } }{\ottmv{i}}
    \label{eq:cohe_var10b}
  \end{gather}
  By inversion on Equation~\ref{eq:cohe_var10b}, we know that
  \(\ottcomp{ \intermed {  {\intermed Q}'_{\ottmv{i}}  } = \intermed {  [  \overline {\intermed \sigma}_{\ottmv{j}}  /  {\overline {\intermed a} }_{\ottmv{j}}  ]  {\intermed Q}_{\ottmv{i}}  } }{\ottmv{i}}\).
  
  Goal~\ref{eq:cohe_var1} follows by repeatedly applying compatibility
  Lemma~\ref{lemma:compat_dapp} to Equation~\ref{eq:cohe_var9}, combined with
  Equation~\ref{eq:cohe_var10}. \\
}
\proofStep{\textsc{sTm-check-meth}}
{
   \source {  { \source P }  } ; \source {  {\source { \Gamma_C } }  } ; \source {  {\source \Gamma}  }  \vdash_{tm}^{\intermed {M} }  \source {  {\source m}  } \Leftarrow \source {  [  {\overline {\source \tau} }_{\ottmv{j}}  /  {\overline {\source a} }_{\ottmv{j}}  ]  [  {\source \tau }  /  {\source a }  ]  {\source \tau }'  } \stoi{ \rightsquigarrow  \intermed {  {\intermed d}  \ottsym{.}  {\intermed m} \, \overline {\intermed \sigma}_{\ottmv{j}} \, {\overline {\intermed d} }_{\ottmv{i}}  } } 
}
{
  The proof is similar to the \textsc{sTm-check-var} case. \\
}
\proofStep{\textsc{sTm-check-ArrI}}
{
   \source {  { \source P }  } ; \source {  {\source { \Gamma_C } }  } ; \source {  {\source \Gamma}  }  \vdash_{tm}^{\intermed {M} }  \source {  \lambda  {\source x}  \ottsym{.}  {\source e}  } \Leftarrow \source {  {\source \tau }_{{\mathrm{1}}}  \rightarrow  {\source \tau }_{{\mathrm{2}}}  } \stoi{ \rightsquigarrow  \intermed {  \lambda  {\intermed x}  \ottsym{:}  {\intermed \sigma}_{{\mathrm{1}}}  \ottsym{.}  {\intermed e}  } } 
}
{
  Through case analysis, it is straightforward to note that the final step in
  the second typing derivation can be either \textsc{sTm-check-ArrI} or
  \textsc{sTm-check-Inf}. In the latter case however, no matching inference
  rules exist. The \textsc{sTm-check-ArrI} case is the only remaining
  possibility.
  This means that:
  \begin{gather*}
     \source {  { \source P }  } ; \source {  {\source { \Gamma_C } }  } ; \source {  {\source \Gamma}  }  \vdash_{tm}^{\intermed {M} }  \source {  \lambda  {\source x}  \ottsym{.}  {\source e}  } \Leftarrow \source {  {\source \tau }_{{\mathrm{1}}}  \rightarrow  {\source \tau }_{{\mathrm{2}}}  } \stoi{ \rightsquigarrow  \intermed {  \lambda  {\intermed x}  \ottsym{:}  {\intermed \sigma}_{{\mathrm{1}}}  \ottsym{.}  {\intermed e}  } }  \\
     \source {  { \source P }  } ; \source {  {\source { \Gamma_C } }  } ; \source {  {\source \Gamma}  }  \vdash_{tm}^{\intermed {M} }  \source {  \lambda  {\source x}  \ottsym{.}  {\source e}  } \Leftarrow \source {  {\source \tau }_{{\mathrm{1}}}  \rightarrow  {\source \tau }_{{\mathrm{2}}}  } \stoi{ \rightsquigarrow  \intermed {  \lambda  {\intermed x}  \ottsym{:}  {\intermed \sigma}'_{{\mathrm{1}}}  \ottsym{.}  {\intermed e}'  } } 
  \end{gather*}
  From the 3\textsuperscript{rd} rule premise we know that:
  \begin{gather}
     \source {  {\source { \Gamma_C } }  } ; \source {  {\source \Gamma}  }  \vdash_{ty}^{\intermed {M} }  \source {  {\source \tau }_{{\mathrm{1}}}  }\,{ \stoi{ \rightsquigarrow  \intermed {  {\intermed \sigma}_{{\mathrm{1}}}  } } } 
    \nonumber \\
     \source {  {\source { \Gamma_C } }  } ; \source {  {\source \Gamma}  }  \vdash_{ty}^{\intermed {M} }  \source {  {\source \tau }_{{\mathrm{1}}}  }\,{ \stoi{ \rightsquigarrow  \intermed {  {\intermed \sigma}'_{{\mathrm{1}}}  } } } 
    \nonumber
  \end{gather}
  Since type elaboration is entirely deterministic (Lemma~\ref{lemma:ty_elab_unique}), it is straightforward to
  note that \( \intermed {  {\intermed \sigma}_{{\mathrm{1}}}  } = \intermed {  {\intermed \sigma}'_{{\mathrm{1}}}  } \).
  
  The goal to be proven is the following:
  \begin{equation}
     \intermed {  {\intermed { \Gamma_C } }  } ; \intermed {  {\intermed \Gamma}  }  \vdash  \intermed {  {\intermed \Sigma}_{{\mathrm{1}}}  } : \intermed {  \lambda  {\intermed x}  \ottsym{:}  {\intermed \sigma}_{{\mathrm{1}}}  \ottsym{.}  {\intermed e}  }  \simeq_{log}  \intermed {  {\intermed \Sigma}_{{\mathrm{2}}}  } : \intermed {  \lambda  {\intermed x}  \ottsym{:}  {\intermed \sigma}_{{\mathrm{1}}}  \ottsym{.}  {\intermed e}'  } : \intermed {  {\intermed \sigma}'  } ~
    \textit{where}~ \source {  {\source { \Gamma_C } }  } ; \source {  {\source \Gamma}  }  \vdash_{ty}^{\intermed {M} }  \source {  {\source \tau }_{{\mathrm{1}}}  \rightarrow  {\source \tau }_{{\mathrm{2}}}  }\,{ \stoi{ \rightsquigarrow  \intermed {  {\intermed \sigma}'  } } } 
    \label{eq:cohe_arrI1}
  \end{equation}
  By inversion on Equation~\ref{eq:cohe0}, we know that \( \intermed {  {\intermed \sigma}'  } = \intermed {  {\intermed \sigma}_{{\mathrm{1}}}  \rightarrow  {\intermed \sigma}_{{\mathrm{2}}}  } \).

  From the rule premise we know:
  \begin{gather}
     \source {  { \source P }  } ; \source {  {\source { \Gamma_C } }  } ; \source {  {\source \Gamma}  \ottsym{,}  {\source x}  \ottsym{:}  {\source \tau }_{{\mathrm{1}}}  }  \vdash_{tm}^{\intermed {M} }  \source {  {\source e}  } \Leftarrow \source {  {\source \tau }_{{\mathrm{2}}}  } \stoi{ \rightsquigarrow  \intermed {  {\intermed e}  } } 
    \label{eq:cohe_arrI2} \\
     \source {  { \source P }  } ; \source {  {\source { \Gamma_C } }  } ; \source {  {\source \Gamma}  \ottsym{,}  {\source x}  \ottsym{:}  {\source \tau }_{{\mathrm{1}}}  }  \vdash_{tm}^{\intermed {M} }  \source {  {\source e}  } \Leftarrow \source {  {\source \tau }_{{\mathrm{2}}}  } \stoi{ \rightsquigarrow  \intermed {  {\intermed e}'  } } 
    \label{eq:cohe_arrI3}
  \end{gather}
  By applying Equation~\ref{eq:cohe_arrI2} to
  Lemma~\ref{lemma:ctx_well_source_mid_typing}, and through repeated case
  analysis on the result to discover the contents of the elaborated
  environments, we know that: 
  \begin{equation*}
     \vdash_{ctx}^{\intermed {M} }  \source {  { \source P }  } ; \source {  {\source { \Gamma_C } }  } ; \source {  {\source \Gamma}  \ottsym{,}  {\source x}  \ottsym{:}  {\source \tau }_{{\mathrm{1}}}  }  \rightsquigarrow  \intermed {  {\intermed \Sigma}_{{\mathrm{1}}}  } ; \intermed {  {\intermed { \Gamma_C } }  } ; \intermed {  {\intermed \Gamma}  \ottsym{,}  {\intermed x}  \ottsym{:}  {\intermed \sigma}_{{\mathrm{1}}}  } 
  \end{equation*}
  Applying the induction hypothesis on Equations~\ref{eq:cohe_arrI2} and
  \ref{eq:cohe_arrI3} results in:
  \begin{equation}
     \intermed {  {\intermed { \Gamma_C } }  } ; \intermed {  {\intermed \Gamma}  \ottsym{,}  {\intermed x}  \ottsym{:}  {\intermed \sigma}_{{\mathrm{1}}}  }  \vdash  \intermed {  {\intermed \Sigma}_{{\mathrm{1}}}  } : \intermed {  {\intermed e}  }  \simeq_{log}  \intermed {  {\intermed \Sigma}_{{\mathrm{2}}}  } : \intermed {  {\intermed e}'  } : \intermed {  {\intermed \sigma}_{{\mathrm{2}}}  } ~
    \textit{where}~ \source {  {\source { \Gamma_C } }  } ; \source {  {\source \Gamma}  \ottsym{,}  {\source x}  \ottsym{:}  {\source \tau }_{{\mathrm{1}}}  }  \vdash_{ty}^{\intermed {M} }  \source {  {\source \tau }_{{\mathrm{2}}}  }\,{ \stoi{ \rightsquigarrow  \intermed {  {\intermed \sigma}_{{\mathrm{2}}}  } } } 
    \label{eq:cohe_arrI4}
  \end{equation}
  Goal~\ref{eq:cohe_arrI1} follows directly from compatibility
  Lemma~\ref{lemma:compat_abs}. \\
}
\proofStep{\textsc{sTm-check-Inf}}
{
   \source {  { \source P }  } ; \source {  {\source { \Gamma_C } }  } ; \source {  {\source \Gamma}  }  \vdash_{tm}^{\intermed {M} }  \source {  {\source e}  } \Leftarrow \source {  {\source \tau }  } \stoi{ \rightsquigarrow  \intermed {  {\intermed e}  } } 
}
{
  Through case analysis, we note that the final step in the second typing
  derivation can either be \textsc{sTm-check-var}, \textsc{sTm-check-meth},
  \textsc{sTm-check-ArrI} or \textsc{sTm-check-Inf}. In the first 3 cases, the
  proof is symmetrical to the corresponding proof cases described above. We
  proceed with the last case:
  
  The goal to be proven is the following:
  \begin{equation}
     \intermed {  {\intermed { \Gamma_C } }  } ; \intermed {  {\intermed \Gamma}  }  \vdash  \intermed {  {\intermed \Sigma}_{{\mathrm{1}}}  } : \intermed {  {\intermed e}  }  \simeq_{log}  \intermed {  {\intermed \Sigma}_{{\mathrm{2}}}  } : \intermed {  {\intermed e}'  } : \intermed {  {\intermed \sigma}  } ~
    \textit{where}~ \source {  {\source { \Gamma_C } }  } ; \source {  {\source \Gamma}  }  \vdash_{ty}^{\intermed {M} }  \source {  {\source \tau }  }\,{ \stoi{ \rightsquigarrow  \intermed {  {\intermed \sigma}  } } } 
    \label{eq:cohe_inf1}
  \end{equation}
  where we know that:
  \begin{gather}
     \source {  { \source P }  } ; \source {  {\source { \Gamma_C } }  } ; \source {  {\source \Gamma}  }  \vdash_{tm}^{\intermed {M} }  \source {  {\source e}  } \Leftarrow \source {  {\source \tau }  } \stoi{ \rightsquigarrow  \intermed {  {\intermed e}  } } 
    \label{eq:cohe_inf2} \\
     \source {  { \source P }  } ; \source {  {\source { \Gamma_C } }  } ; \source {  {\source \Gamma}  }  \vdash_{tm}^{\intermed {M} }  \source {  {\source e}  } \Leftarrow \source {  {\source \tau }  } \stoi{ \rightsquigarrow  \intermed {  {\intermed e}'  } } 
    \label{eq:cohe_inf3} 
  \end{gather}
  The rule premise tells us that:
  \begin{gather}
     \source {  { \source P }  } ; \source {  {\source { \Gamma_C } }  } ; \source {  {\source \Gamma}  }  \vdash_{tm}^{\intermed {M} }  \source {  {\source e}  } \Rightarrow \source {  {\source \tau }  } \stoi{ \rightsquigarrow  \intermed {  {\intermed e}  } } 
    \label{eq:cohe_inf4} \\
     \source {  { \source P }  } ; \source {  {\source { \Gamma_C } }  } ; \source {  {\source \Gamma}  }  \vdash_{tm}^{\intermed {M} }  \source {  {\source e}  } \Rightarrow \source {  {\source \tau }  } \stoi{ \rightsquigarrow  \intermed {  {\intermed e}'  } } 
    \label{eq:cohe_inf5}
  \end{gather}
  The goal follows directly by applying Part 1 of this lemma to
  Equations~\ref{eq:cohe_inf4} and \ref{eq:cohe_inf5}. Part 1 can be applied on
  \({\source e}\), even though the term size did not decrease, because inference is
  defined to be smaller than type checking in our proof by induction.
}
\end{description}
\qed

\begin{theoremB}[Coherence - Expressions - Part B]\ \\
  If \( \intermed {  {\intermed { \Gamma_C } }  } ; \intermed {  {\intermed \Gamma}  }  \vdash  \intermed {  {\intermed \Sigma}_{{\mathrm{1}}}  } : \intermed {  {\intermed e}_{{\mathrm{1}}}  }  \simeq_{log}  \intermed {  {\intermed \Sigma}_{{\mathrm{2}}}  } : \intermed {  {\intermed e}_{{\mathrm{2}}}  } : \intermed {  {\intermed \sigma}  } \)
  then \( \intermed {  {\intermed { \Gamma_C } }  } ; \intermed {  {\intermed \Gamma}  }  \vdash  \intermed {  {\intermed \Sigma}_{{\mathrm{1}}}  } : \intermed {  {\intermed e}_{{\mathrm{1}}}  }  \simeq_{ctx}  \intermed {  {\intermed \Sigma}_{{\mathrm{2}}}  } : \intermed {  {\intermed e}_{{\mathrm{2}}}  } : \intermed {  {\intermed \sigma}  } \).
  \label{thmA:coh_exprB}
\end{theoremB}

\proof
By unfolding the definition of contextual equivalence, the goal becomes:
\begin{gather}
  \forall  \intermed {  {\intermed M}_{{\mathrm{1}}}  } : ( \intermed {  {\intermed \Sigma}_{{\mathrm{1}}}  } ; \intermed {  {\intermed { \Gamma_C } }  } ; \intermed {  {\intermed \Gamma}  }  \Rightarrow  \intermed {  {\intermed \sigma}  } )  \mapsto  ( \intermed {  {\intermed \Sigma}_{{\mathrm{1}}}  } ; \intermed {  {\intermed { \Gamma_C } }  } ; \intermed {  \bullet  }  \Rightarrow  \intermed {  \mathit{\intermed {Bool} }  } ) \itot{ \rightsquigarrow  \target {  {\target M }_{{\mathrm{1}}}  } } 
  \nonumber \\
  \forall  \intermed {  {\intermed M}_{{\mathrm{2}}}  } : ( \intermed {  {\intermed \Sigma}_{{\mathrm{2}}}  } ; \intermed {  {\intermed { \Gamma_C } }  } ; \intermed {  {\intermed \Gamma}  }  \Rightarrow  \intermed {  {\intermed \sigma}  } )  \mapsto  ( \intermed {  {\intermed \Sigma}_{{\mathrm{2}}}  } ; \intermed {  {\intermed { \Gamma_C } }  } ; \intermed {  \bullet  }  \Rightarrow  \intermed {  \mathit{\intermed {Bool} }  } ) \itot{ \rightsquigarrow  \target {  {\target M }_{{\mathrm{2}}}  } } 
  \nonumber \\
  \textit{if}~ \intermed {  {\intermed \Sigma}_{{\mathrm{1}}}  } : \intermed {  {\intermed M}_{{\mathrm{1}}}  }  \simeq_{log}  \intermed {  {\intermed \Sigma}_{{\mathrm{2}}}  } : \intermed {  {\intermed M}_{{\mathrm{2}}}  } : ( \intermed {  {\intermed { \Gamma_C } }  } ; \intermed {  {\intermed \Gamma}  }  \Rightarrow  \intermed {  {\intermed \sigma}  } )  \mapsto  ( \intermed {  {\intermed { \Gamma_C } }  } ; \intermed {  \bullet  }  \Rightarrow  \intermed {  \mathit{\intermed {Bool} }  } ) 
  \label{eq:cohexprB1a} \\
  \textit{then}~ \intermed {  {\intermed \Sigma}_{{\mathrm{1}}}  } : \intermed {   {\intermed M}_{{\mathrm{1}}}  [  {\intermed e}_{{\mathrm{1}}}  ]   }  \simeq  \intermed {  {\intermed \Sigma}_{{\mathrm{2}}}  } : \intermed {   {\intermed M}_{{\mathrm{2}}}  [  {\intermed e}_{{\mathrm{2}}}  ]   } 
  \label{eq:cohexprB1}
\end{gather}
We select any \(\intermed{{\intermed M}_{{\mathrm{1}}}}\) and \(\intermed{{\intermed M}_{{\mathrm{2}}}}\) such that
Equation~\ref{eq:cohexprB1a} holds, and thus need to prove Goal~\ref{eq:cohexprB1}.

From the congruence Theorem~\ref{thmA:congruence_expr} and the
1\textsuperscript{st} hypothesis, we know that:
\begin{equation*}
   \intermed {  {\intermed { \Gamma_C } }  } ; \intermed {  \bullet  }  \vdash  \intermed {  {\intermed \Sigma}_{{\mathrm{1}}}  } : \intermed {   {\intermed M}_{{\mathrm{1}}}  [  {\intermed e}_{{\mathrm{1}}}  ]   }  \simeq_{log}  \intermed {  {\intermed \Sigma}_{{\mathrm{2}}}  } : \intermed {   {\intermed M}_{{\mathrm{2}}}  [  {\intermed e}_{{\mathrm{2}}}  ]   } : \intermed {  \mathit{\intermed {Bool} }  } 
  \label{eq:cohexprB2}
\end{equation*}
By applying the definition of logical equivalence, we get:
\begin{equation}
   ( \intermed {  {\intermed \Sigma}_{{\mathrm{1}}}  } : \intermed {   {\intermed \gamma }_{{\mathrm{1}}}  ( {\intermed \phi }_{{\mathrm{1}}}  ( {\intermed R }  (  {\intermed M}_{{\mathrm{1}}}  [  {\intermed e}_{{\mathrm{1}}}  ]  )))   } , \intermed {  {\intermed \Sigma}_{{\mathrm{2}}}  } : \intermed {   {\intermed \gamma }_{{\mathrm{2}}}  ( {\intermed \phi }_{{\mathrm{2}}}  ( {\intermed R }  (  {\intermed M}_{{\mathrm{2}}}  [  {\intermed e}_{{\mathrm{2}}}  ]  )))   } ) \in \mathcal{E} \llbracket \intermed {  \mathit{\intermed {Bool} }  } \rrbracket ^{\intermed {  {\intermed { \Gamma_C } }  } }_{\intermed {  {\intermed R }  } } 
  \label{eq:cohexprB3}
\end{equation}
for any \( \intermed {  {\intermed R }  } \in \mathcal{F} \llbracket \intermed {  \bullet  } \rrbracket ^{\intermed {  {\intermed { \Gamma_C } }  } } \),
\( \intermed {  {\intermed \phi }  } \in \mathcal{G} \llbracket \intermed {  \bullet  } \rrbracket ^{\intermed {  {\intermed \Sigma}_{{\mathrm{1}}}  } , \intermed {  {\intermed \Sigma}_{{\mathrm{2}}}  } , \intermed {  {\intermed { \Gamma_C } }  } }_{\intermed {  {\intermed R }  } } \) and
\( \intermed {  {\intermed \gamma }  } \in \mathcal{H} \llbracket \intermed {  \bullet  } \rrbracket ^{\intermed {  {\intermed \Sigma}_{{\mathrm{1}}}  } , \intermed {  {\intermed \Sigma}_{{\mathrm{2}}}  } , \intermed {  {\intermed { \Gamma_C } }  } }_{\intermed {  {\intermed R }  } } \).

However, from the definition of \( \mathcal{F} \), \( \mathcal{G} \)
and \( \mathcal{H} \), it follows that \( \intermed {  {\intermed R }  } = \intermed {  \bullet  } \),
\( \intermed {  {\intermed \phi }  } = \intermed {  \bullet  } \) and
\( \intermed {  {\intermed \gamma }  } = \intermed {  \bullet  } \).

Equation~\ref{eq:cohexprB3} thus simplifies to:
\begin{equation*}
   ( \intermed {  {\intermed \Sigma}_{{\mathrm{1}}}  } : \intermed {   {\intermed M}_{{\mathrm{1}}}  [  {\intermed e}_{{\mathrm{1}}}  ]   } , \intermed {  {\intermed \Sigma}_{{\mathrm{2}}}  } : \intermed {   {\intermed M}_{{\mathrm{2}}}  [  {\intermed e}_{{\mathrm{2}}}  ]   } ) \in \mathcal{E} \llbracket \intermed {  \mathit{\intermed {Bool} }  } \rrbracket ^{\intermed {  {\intermed { \Gamma_C } }  } }_{\intermed {  \bullet  } } 
  \label{eq:cohexprB4}
\end{equation*}
Unfolding the definition of the \( \mathcal{E} \) relation, tells us that:
\begin{gather*}
   \intermed {  {\intermed \Sigma}_{{\mathrm{1}}}  } ; \intermed {  {\intermed { \Gamma_C } }  } ; \intermed {  \bullet  }  \vdash_{tm}  \intermed {   {\intermed M}  [  {\intermed e}_{{\mathrm{1}}}  ]   } : \intermed {  \mathit{\intermed {Bool} }  } \itot{ \rightsquigarrow  \target {  {\target e }_{{\mathrm{1}}}  } } 
  \label{eq:cohexprB5} \\
   \intermed {  {\intermed \Sigma}_{{\mathrm{2}}}  } ; \intermed {  {\intermed { \Gamma_C } }  } ; \intermed {  \bullet  }  \vdash_{tm}  \intermed {   {\intermed M}  [  {\intermed e}_{{\mathrm{2}}}  ]   } : \intermed {  \mathit{\intermed {Bool} }  } \itot{ \rightsquigarrow  \target {  {\target e }_{{\mathrm{2}}}  } } 
  \label{eq:cohexprB6} \\
  \begin{aligned}
    \exists {\intermed v}_{{\mathrm{1}}}, {\intermed v}_{{\mathrm{2}}} & :
     \intermed {  {\intermed \Sigma}_{{\mathrm{1}}}  }  \vdash  \intermed {   {\intermed M}  [  {\intermed e}_{{\mathrm{1}}}  ]   }  \longrightarrow^{*}  \intermed {  {\intermed v}_{{\mathrm{1}}}  } 
    \nonumber \\
    & \band  \intermed {  {\intermed \Sigma}_{{\mathrm{2}}}  }  \vdash  \intermed {   {\intermed M}  [  {\intermed e}_{{\mathrm{2}}}  ]   }  \longrightarrow^{*}  \intermed {  {\intermed v}_{{\mathrm{2}}}  } 
    \nonumber \\
    & \band  ( \intermed {  {\intermed \Sigma}_{{\mathrm{1}}}  } : \intermed {  {\intermed v}_{{\mathrm{1}}}  } , \intermed {  {\intermed \Sigma}_{{\mathrm{2}}}  } : \intermed {  {\intermed v}_{{\mathrm{2}}}  } ) \in \mathcal{V} \llbracket \intermed {  \mathit{\intermed {Bool} }  } \rrbracket ^{\intermed {  {\intermed { \Gamma_C } }  } }_{\intermed {  \bullet  } } 
    \nonumber
  \end{aligned}
\end{gather*}
From the definition of \( \mathcal{V} \), we know that either
\( \intermed {  {\intermed v}_{{\mathrm{1}}}  } = \intermed {  {\intermed v}_{{\mathrm{2}}}  } = \intermed {  \mathit{\intermed {True} }  } \) or \( \intermed {  {\intermed v}_{{\mathrm{1}}}  } = \intermed {  {\intermed v}_{{\mathrm{2}}}  } = \intermed {  \mathit{\intermed {False} }  } \).
The goal follows immediately. \\
\qed

\renewcommand\itot[1]{#1}

\begin{theoremB}[Coherence - Expressions - Part C]\ \\
  If \( \intermed {  {\intermed { \Gamma_C } }  } ; \intermed {  {\intermed \Gamma}  }  \vdash  \intermed {  {\intermed \Sigma}_{{\mathrm{1}}}  } : \intermed {  {\intermed e}_{{\mathrm{1}}}  }  \simeq_{ctx}  \intermed {  {\intermed \Sigma}_{{\mathrm{2}}}  } : \intermed {  {\intermed e}_{{\mathrm{2}}}  } : \intermed {  {\intermed \sigma}  } \) \\
  and \( \intermed {  {\intermed \Sigma}_{{\mathrm{1}}}  } ; \intermed {  {\intermed { \Gamma_C } }  } ; \intermed {  {\intermed \Gamma}  }  \vdash_{tm}  \intermed {  {\intermed e}_{{\mathrm{1}}}  } : \intermed {  {\intermed \sigma}  } \itot{ \rightsquigarrow  \target {  {\target e }_{{\mathrm{1}}}  } } \)
  and \( \intermed {  {\intermed \Sigma}_{{\mathrm{2}}}  } ; \intermed {  {\intermed { \Gamma_C } }  } ; \intermed {  {\intermed \Gamma}  }  \vdash_{tm}  \intermed {  {\intermed e}_{{\mathrm{2}}}  } : \intermed {  {\intermed \sigma}  } \itot{ \rightsquigarrow  \target {  {\target e }_{{\mathrm{2}}}  } } \) \\
  and \( \intermed {  {\intermed { \Gamma_C } }  } ; \intermed {  {\intermed \Gamma}  }  \rightsquigarrow  \target {  {\target \Gamma }  } \) \\
  then \( \intermed {  {\intermed { \Gamma_C } }  } ; \intermed {  {\intermed \Gamma}  }  \vdash  \intermed {  {\intermed \Sigma}_{{\mathrm{1}}}  } : \target {  {\target e }_{{\mathrm{1}}}  }  \simeq_{ctx}  \intermed {  {\intermed \Sigma}_{{\mathrm{2}}}  } : \target {  {\target e }_{{\mathrm{2}}}  } : \intermed {  {\intermed \sigma}  } \).
  \label{thmA:coh_exprC}
\end{theoremB}

\proof
By unfolding the definition of contextual equivalence, the goal becomes:
\begin{gather}
  \forall  \intermed {  {\intermed M}_{{\mathrm{1}}}  } : ( \intermed {  {\intermed \Sigma}_{{\mathrm{1}}}  } ; \intermed {  {\intermed { \Gamma_C } }  } ; \intermed {  {\intermed \Gamma}  }  \Rightarrow  \intermed {  {\intermed \sigma}  } )  \mapsto  ( \intermed {  {\intermed \Sigma}_{{\mathrm{1}}}  } ; \intermed {  {\intermed { \Gamma_C } }  } ; \intermed {  \bullet  }  \Rightarrow  \intermed {  \mathit{\intermed {Bool} }  } ) \itot{ \rightsquigarrow  \target {  {\target M }_{{\mathrm{1}}}  } } 
  \nonumber \\
  \forall  \intermed {  {\intermed M}_{{\mathrm{2}}}  } : ( \intermed {  {\intermed \Sigma}_{{\mathrm{2}}}  } ; \intermed {  {\intermed { \Gamma_C } }  } ; \intermed {  {\intermed \Gamma}  }  \Rightarrow  \intermed {  {\intermed \sigma}  } )  \mapsto  ( \intermed {  {\intermed \Sigma}_{{\mathrm{2}}}  } ; \intermed {  {\intermed { \Gamma_C } }  } ; \intermed {  \bullet  }  \Rightarrow  \intermed {  \mathit{\intermed {Bool} }  } ) \itot{ \rightsquigarrow  \target {  {\target M }_{{\mathrm{2}}}  } } 
  \nonumber \\
  \textit{if}~ \intermed {  {\intermed \Sigma}_{{\mathrm{1}}}  } : \intermed {  {\intermed M}_{{\mathrm{1}}}  }  \simeq_{log}  \intermed {  {\intermed \Sigma}_{{\mathrm{2}}}  } : \intermed {  {\intermed M}_{{\mathrm{2}}}  } : ( \intermed {  {\intermed { \Gamma_C } }  } ; \intermed {  {\intermed \Gamma}  }  \Rightarrow  \intermed {  {\intermed \sigma}  } )  \mapsto  ( \intermed {  {\intermed { \Gamma_C } }  } ; \intermed {  \bullet  }  \Rightarrow  \intermed {  \mathit{\intermed {Bool} }  } ) 
  \label{eqg:cohC0a} \\
  \textit{then}~ \target {   {\target M }_{{\mathrm{1}}}  [  {\target e }_{{\mathrm{1}}}  ]   }  \simeq  \target {   {\target M }_{{\mathrm{2}}}  [  {\target e }_{{\mathrm{2}}}  ]   } 
  \label{eqg:cohC0}
\end{gather}
We select any \(\intermed{{\intermed M}_{{\mathrm{1}}}}\) and \(\intermed{{\intermed M}_{{\mathrm{2}}}}\) such that
Equation~\ref{eqg:cohC0a} holds, and thus need to prove Goal~\ref{eqg:cohC0}.

By unfolding the definition of contextual equivalence in the
1\textsuperscript{st} hypothesis, we get that:
\begin{gather}
  \forall  \intermed {  {\intermed M}_{{\mathrm{1}}}  } : ( \intermed {  {\intermed \Sigma}_{{\mathrm{1}}}  } ; \intermed {  {\intermed { \Gamma_C } }  } ; \intermed {  {\intermed \Gamma}  }  \Rightarrow  \intermed {  {\intermed \sigma}  } )  \mapsto  ( \intermed {  {\intermed \Sigma}_{{\mathrm{2}}}  } ; \intermed {  {\intermed { \Gamma_C } }  } ; \intermed {  \bullet  }  \Rightarrow  \intermed {  \mathit{\intermed {Bool} }  } ) \itot{ \rightsquigarrow  \target {  {\target M }_{{\mathrm{1}}}  } } 
  \nonumber \\
  \forall  \intermed {  {\intermed M}_{{\mathrm{2}}}  } : ( \intermed {  {\intermed \Sigma}_{{\mathrm{2}}}  } ; \intermed {  {\intermed { \Gamma_C } }  } ; \intermed {  {\intermed \Gamma}  }  \Rightarrow  \intermed {  {\intermed \sigma}  } )  \mapsto  ( \intermed {  {\intermed \Sigma}_{{\mathrm{2}}}  } ; \intermed {  {\intermed { \Gamma_C } }  } ; \intermed {  \bullet  }  \Rightarrow  \intermed {  \mathit{\intermed {Bool} }  } ) \itot{ \rightsquigarrow  \target {  {\target M }_{{\mathrm{2}}}  } } 
  \nonumber \\
  \textit{if}~ \intermed {  {\intermed \Sigma}_{{\mathrm{1}}}  } : \intermed {  {\intermed M}_{{\mathrm{1}}}  }  \simeq_{log}  \intermed {  {\intermed \Sigma}_{{\mathrm{2}}}  } : \intermed {  {\intermed M}_{{\mathrm{2}}}  } : ( \intermed {  {\intermed { \Gamma_C } }  } ; \intermed {  {\intermed \Gamma}  }  \Rightarrow  \intermed {  {\intermed \sigma}  } )  \mapsto  ( \intermed {  {\intermed { \Gamma_C } }  } ; \intermed {  \bullet  }  \Rightarrow  \intermed {  \mathit{\intermed {Bool} }  } ) 
  \label{eq:cohC1a} \\
  \textit{then}~ \intermed {  {\intermed \Sigma}_{{\mathrm{1}}}  } : \intermed {   {\intermed M}_{{\mathrm{1}}}  [  {\intermed e}_{{\mathrm{1}}}  ]   }  \simeq  \intermed {  {\intermed \Sigma}_{{\mathrm{2}}}  } : \intermed {   {\intermed M}_{{\mathrm{2}}}  [  {\intermed e}_{{\mathrm{2}}}  ]   } 
  \label{eq:cohC1}
\end{gather}
By applying Theorem~\ref{thmA:tgt_ctx_compat} to \(\intermed{{\intermed M}_{{\mathrm{1}}}}\) and
\(\intermed{{\intermed M}_{{\mathrm{2}}}}\), together with the 2\textsuperscript{nd} and
3\textsuperscript{rd} hypothesis, we get:
\begin{gather}
   \intermed {  {\intermed \Sigma}_{{\mathrm{1}}}  } ; \intermed {  {\intermed { \Gamma_C } }  } ; \intermed {  \bullet  }  \vdash_{tm}  \intermed {   {\intermed M}_{{\mathrm{1}}}  [  {\intermed e}_{{\mathrm{1}}}  ]   } : \intermed {  \mathit{\intermed {Bool} }  } \itot{ \rightsquigarrow  \target {   {\target M }_{{\mathrm{1}}}  [  {\target e }_{{\mathrm{1}}}  ]   } } 
  \nonumber \\
   \intermed {  {\intermed \Sigma}_{{\mathrm{2}}}  } ; \intermed {  {\intermed { \Gamma_C } }  } ; \intermed {  \bullet  }  \vdash_{tm}  \intermed {   {\intermed M}_{{\mathrm{2}}}  [  {\intermed e}_{{\mathrm{2}}}  ]   } : \intermed {  \mathit{\intermed {Bool} }  } \itot{ \rightsquigarrow  \target {   {\target M }_{{\mathrm{2}}}  [  {\target e }_{{\mathrm{2}}}  ]   } } 
  \nonumber 
\end{gather}
From the definition of kleene equivalence, Equation~\ref{eq:cohC1} reduces to:
\begin{equation*}
  \exists {\intermed v} :  \intermed {  {\intermed \Sigma}_{{\mathrm{1}}}  }  \vdash  \intermed {   {\intermed M}_{{\mathrm{1}}}  [  {\intermed e}_{{\mathrm{1}}}  ]   }  \longrightarrow^{*}  \intermed {  {\intermed v}  }  \band  \intermed {  {\intermed \Sigma}_{{\mathrm{2}}}  }  \vdash  \intermed {   {\intermed M}_{{\mathrm{2}}}  [  {\intermed e}_{{\mathrm{2}}}  ]   }  \longrightarrow^{*}  \intermed {  {\intermed v}  } 
\end{equation*}
Finally, Lemma~\ref{lemma:val_sem_pres} applied to these results, tells us that:
\begin{gather}
   \intermed {  {\intermed \Sigma}_{{\mathrm{1}}}  } ; \intermed {  {\intermed { \Gamma_C } }  } ; \intermed {  \bullet  }  \vdash_{tm}  \intermed {  {\intermed v}  } : \intermed {  \mathit{\intermed {Bool} }  } \itot{ \rightsquigarrow  \target {  {\target v}_{{\mathrm{1}}}  } } 
  \nonumber \\
   \intermed {  {\intermed \Sigma}_{{\mathrm{2}}}  } ; \intermed {  {\intermed { \Gamma_C } }  } ; \intermed {  \bullet  }  \vdash_{tm}  \intermed {  {\intermed v}  } : \intermed {  \mathit{\intermed {Bool} }  } \itot{ \rightsquigarrow  \target {  {\target v}_{{\mathrm{2}}}  } } 
  \nonumber \\
   \target {   {\target M }_{{\mathrm{1}}}  [  {\target e }_{{\mathrm{1}}}  ]   }  \longrightarrow^{*}  \target {  {\target v}_{{\mathrm{1}}}  } 
  \nonumber \\
   \target {   {\target M }_{{\mathrm{2}}}  [  {\target e }_{{\mathrm{2}}}  ]   }  \longrightarrow^{*}  \target {  {\target v}_{{\mathrm{2}}}  } 
  \nonumber
\end{gather}
Goal~\ref{eqg:cohC0} follows from the definition of kleene equivalence since either
\( \target {  {\target v}_{{\mathrm{1}}}  } = \target {  {\target v}_{{\mathrm{2}}}  } = \target {  \mathit{\target {True} }  } \) or \( \target {  {\target v}_{{\mathrm{1}}}  } = \target {  {\target v}_{{\mathrm{2}}}  } = \target {  \mathit{\target {False} }  } \). \\
\qed

\subsection{Main Coherence Theorems}

\begin{theoremB}[Coherence]\leavevmode
  If \( \source {  \bullet  } ; \source {  \bullet  }  \vdash_{pgm}  \source {  {\source {pgm} }  } : \source {  {\source \tau }  } ; \source {  { \source P }_{{\mathrm{1}}}  } ; \source {  {\source { \Gamma_C } }_{{\mathrm{1}}}  }  \rightsquigarrow  \target {  {\target e }_{{\mathrm{1}}}  } \)
  and \( \source {  \bullet  } ; \source {  \bullet  }  \vdash_{pgm}  \source {  {\source {pgm} }  } : \source {  {\source \tau }  } ; \source {  { \source P }_{{\mathrm{2}}}  } ; \source {  {\source { \Gamma_C } }_{{\mathrm{2}}}  }  \rightsquigarrow  \target {  {\target e }_{{\mathrm{2}}}  } \) \\
  then \( \source {  {\source { \Gamma_C } }_{{\mathrm{1}}}  } = \source {  {\source { \Gamma_C } }_{{\mathrm{2}}}  } \), \( \source {  { \source P }_{{\mathrm{1}}}  } = \source {  { \source P }_{{\mathrm{2}}}  } \)
  and \( \source {  { \source P }_{{\mathrm{1}}}  } ; \source {  {\source { \Gamma_C } }_{{\mathrm{1}}}  } ; \source {  \bullet  }  \vdash  \target {  {\target e }_{{\mathrm{1}}}  }  \simeq_{ctx}  \target {  {\target e }_{{\mathrm{2}}}  } : \source {  {\source \tau }  } \).
  \label{thmA:coh}
\end{theoremB}

\proof
Since we know from \textsc{sCtxT-empty} that
\begin{equation*}
   \vdash_{ctx}  \source {  \bullet  } ; \source {  \bullet  } ; \source {  \bullet  }  \rightsquigarrow  \target {  \target {\bullet}  } 
\end{equation*}
the goal follows directly from the Program Coherence Theorem
(Theorem~\ref{thmA:ecoh_programs}). \\
\qed

\begin{theoremB}[Coherence - Expressions]\leavevmode
  \begin{itemize}
  \item If \( \source {  { \source P }  } ; \source {  {\source { \Gamma_C } }  } ; \source {  {\source \Gamma}  }  \vdash_{tm}  \source {  {\source e}  } \Rightarrow \source {  {\source \tau }  } \stot{ \rightsquigarrow  \target {  {\target e }_{{\mathrm{1}}}  } } \)
    and \( \source {  { \source P }  } ; \source {  {\source { \Gamma_C } }  } ; \source {  {\source \Gamma}  }  \vdash_{tm}  \source {  {\source e}  } \Rightarrow \source {  {\source \tau }  } \stot{ \rightsquigarrow  \target {  {\target e }_{{\mathrm{2}}}  } } \) \\
    then \( \source {  { \source P }  } ; \source {  {\source { \Gamma_C } }  } ; \source {  {\source \Gamma}  }  \vdash  \target {  {\target e }_{{\mathrm{1}}}  }  \simeq_{ctx}  \target {  {\target e }_{{\mathrm{2}}}  } : \source {  {\source \tau }  } \).
  \item If \( \source {  { \source P }  } ; \source {  {\source { \Gamma_C } }  } ; \source {  {\source \Gamma}  }  \vdash_{tm}  \source {  {\source e}  } \Leftarrow \source {  {\source \tau }  } \stot{ \rightsquigarrow  \target {  {\target e }_{{\mathrm{1}}}  } } \)
    and \( \source {  { \source P }  } ; \source {  {\source { \Gamma_C } }  } ; \source {  {\source \Gamma}  }  \vdash_{tm}  \source {  {\source e}  } \Leftarrow \source {  {\source \tau }  } \stot{ \rightsquigarrow  \target {  {\target e }_{{\mathrm{2}}}  } } \) \\
    then \( \source {  { \source P }  } ; \source {  {\source { \Gamma_C } }  } ; \source {  {\source \Gamma}  }  \vdash  \target {  {\target e }_{{\mathrm{1}}}  }  \simeq_{ctx}  \target {  {\target e }_{{\mathrm{2}}}  } : \source {  {\source \tau }  } \).
  \end{itemize}
  \label{thmA:ecoh}
\end{theoremB}

\proof

By applying Lemma~\ref{lemma:ctx_well_source_typing} to the
1\textsuperscript{st} hypothesis, we know that:
\begin{equation}
   \vdash_{ctx}  \source {  { \source P }  } ; \source {  {\source { \Gamma_C } }  } ; \source {  {\source \Gamma}  }  \rightsquigarrow  \target {  {\target \Gamma }  } 
  \label{eq:ecoh0}
\end{equation}
\begin{description}
\item[Part 1]
  From environment equivalence (Theorem~\ref{thmA:equiv_env}), we get that:
  \begin{gather}
     \vdash_{ctx}^{\intermed {M} }  \source {  { \source P }  } ; \source {  {\source { \Gamma_C } }  } ; \source {  {\source \Gamma}  }  \rightsquigarrow  \intermed {  {\intermed \Sigma}_{{\mathrm{1}}}  } ; \intermed {  {\intermed { \Gamma_C } }  } ; \intermed {  {\intermed \Gamma}  } 
    \label{eq:ecoh1} \\
     \vdash_{ctx}^{\intermed {M} }  \source {  { \source P }  } ; \source {  {\source { \Gamma_C } }  } ; \source {  {\source \Gamma}  }  \rightsquigarrow  \intermed {  {\intermed \Sigma}_{{\mathrm{2}}}  } ; \intermed {  {\intermed { \Gamma_C } }'  } ; \intermed {  {\intermed \Gamma}'  } 
    \label{eq:ecoh2} \\
     \intermed {  {\intermed { \Gamma_C } }  } ; \intermed {  {\intermed \Gamma}  }  \rightsquigarrow  \target {  {\target \Gamma }  } 
    \label{eq:ecoh3}
  \end{gather}
  Since class and typing environment elaboration is entirely deterministic (Lemma~\ref{lemma:env_elab_unique}),
  it is easy to see that \( \intermed {  {\intermed { \Gamma_C } }'  } = \intermed {  {\intermed { \Gamma_C } }  } \) and \( \intermed {  {\intermed \Gamma}'  } = \intermed {  {\intermed \Gamma}  } \).

  We know from expression equivalence (Theorem~\ref{thmA:equiv_expr}) that:
  \begin{gather}
     \source {  { \source P }  } ; \source {  {\source { \Gamma_C } }  } ; \source {  {\source \Gamma}  }  \vdash_{tm}^{\intermed {M} }  \source {  {\source e}  } \Leftarrow \source {  {\source \tau }  } \stoi{ \rightsquigarrow  \intermed {  {\intermed e}_{{\mathrm{1}}}  } } 
    \label{eq:ecoh4} \\
     \intermed {  {\intermed \Sigma}_{{\mathrm{1}}}  } ; \intermed {  {\intermed { \Gamma_C } }  } ; \intermed {  {\intermed \Gamma}  }  \vdash_{tm}  \intermed {  {\intermed e}_{{\mathrm{1}}}  } : \intermed {  {\intermed \sigma}_{{\mathrm{1}}}  } \itot{ \rightsquigarrow  \target {  {\target e }_{{\mathrm{1}}}  } } 
    \label{eq:ecoh5} \\
     \source {  { \source P }  } ; \source {  {\source { \Gamma_C } }  } ; \source {  {\source \Gamma}  }  \vdash_{tm}^{\intermed {M} }  \source {  {\source e}  } \Leftarrow \source {  {\source \tau }  } \stoi{ \rightsquigarrow  \intermed {  {\intermed e}_{{\mathrm{2}}}  } } 
    \label{eq:ecoh6} \\
     \intermed {  {\intermed \Sigma}_{{\mathrm{2}}}  } ; \intermed {  {\intermed { \Gamma_C } }  } ; \intermed {  {\intermed \Gamma}  }  \vdash_{tm}  \intermed {  {\intermed e}_{{\mathrm{2}}}  } : \intermed {  {\intermed \sigma}_{{\mathrm{2}}}  } \itot{ \rightsquigarrow  \target {  {\target e }_{{\mathrm{2}}}  } } 
    \label{eq:ecoh7} \\
    \textit{where}~ \source {  {\source { \Gamma_C } }  } ; \source {  {\source \Gamma}  }  \vdash_{ty}^{\intermed {M} }  \source {  {\source \tau }  }\,{ \stoi{ \rightsquigarrow  \intermed {  {\intermed \sigma}_{{\mathrm{1}}}  } } } 
    \label{eq:ecoh7b} \\
    \textit{and}~ \source {  {\source { \Gamma_C } }  } ; \source {  {\source \Gamma}  }  \vdash_{ty}^{\intermed {M} }  \source {  {\source \tau }  }\,{ \stoi{ \rightsquigarrow  \intermed {  {\intermed \sigma}_{{\mathrm{2}}}  } } } 
    \nonumber
  \end{gather}
  Since type elaboration is entirely deterministic
  (Lemma~\ref{lemma:ty_elab_unique}), it is easy to see that
  \( \intermed {  {\intermed \sigma}  } = \intermed {  {\intermed \sigma}_{{\mathrm{1}}}  } = \intermed {  {\intermed \sigma}_{{\mathrm{2}}}  } \). 

  By applying Expression Coherence Theorem A (Theorem~\ref{thmA:coh_exprA}) to Equations~\ref{eq:ecoh1},
  \ref{eq:ecoh2}, \ref{eq:ecoh4} and~\ref{eq:ecoh6}, we get:
  \begin{equation}
     \intermed {  {\intermed { \Gamma_C } }  } ; \intermed {  {\intermed \Gamma}  }  \vdash  \intermed {  {\intermed \Sigma}_{{\mathrm{1}}}  } : \intermed {  {\intermed e}_{{\mathrm{1}}}  }  \simeq_{log}  \intermed {  {\intermed \Sigma}_{{\mathrm{2}}}  } : \intermed {  {\intermed e}_{{\mathrm{2}}}  } : \intermed {  {\intermed \sigma}  } 
    \label{eq:ecoh8}
  \end{equation}
  By applying Expression Coherence Theorem B (Theorem~\ref{thmA:coh_exprB}) to Equation~\ref{eq:ecoh8},
  we get:
  \begin{equation}
     \intermed {  {\intermed { \Gamma_C } }  } ; \intermed {  {\intermed \Gamma}  }  \vdash  \intermed {  {\intermed \Sigma}_{{\mathrm{1}}}  } : \intermed {  {\intermed e}_{{\mathrm{1}}}  }  \simeq_{ctx}  \intermed {  {\intermed \Sigma}_{{\mathrm{2}}}  } : \intermed {  {\intermed e}_{{\mathrm{2}}}  } : \intermed {  {\intermed \sigma}  } 
    \label{eq:ecoh9}
  \end{equation}
  By applying Expression Coherence Theorem C (Theorem~\ref{thmA:coh_exprC}) to
  Equations~\ref{eq:ecoh3}, \ref{eq:ecoh5}, \ref{eq:ecoh7} and~\ref{eq:ecoh9},
  we get:
  \begin{equation}
     \intermed {  {\intermed { \Gamma_C } }  } ; \intermed {  {\intermed \Gamma}  }  \vdash  \intermed {  {\intermed \Sigma}_{{\mathrm{1}}}  } : \target {  {\target e }_{{\mathrm{1}}}  }  \simeq_{ctx}  \intermed {  {\intermed \Sigma}_{{\mathrm{2}}}  } : \target {  {\target e }_{{\mathrm{2}}}  } : \intermed {  {\intermed \sigma}  } 
    \label{eq:ecoh10}
  \end{equation}
  The goal follows directly from Theorem~\ref{thmA:ctx_equiv_pres}, together with
  Equations~\ref{eq:ecoh10}, \ref{eq:ecoh0}, \ref{eq:ecoh1}, \ref{eq:ecoh2}
  and~\ref{eq:ecoh7b}. \\
\item[Part 2]
  Similar to Part 1. \\
\end{description}
\qed

\renewcommand\itot[1]{}

\begin{theoremB}[Coherence - Programs]\ \\
  If \( \source {  { \source P }  } ; \source {  {\source { \Gamma_C } }  }  \vdash_{pgm}  \source {  {\source {pgm} }  } : \source {  {\source \tau }  } ; \source {  { \source P }_{{\mathrm{1}}}  } ; \source {  {\source { \Gamma_C } }_{{\mathrm{1}}}  }  \rightsquigarrow  \target {  {\target e }_{{\mathrm{1}}}  } \), \\
  \( \source {  { \source P }  } ; \source {  {\source { \Gamma_C } }  }  \vdash_{pgm}  \source {  {\source {pgm} }  } : \source {  {\source \tau }  } ; \source {  { \source P }_{{\mathrm{2}}}  } ; \source {  {\source { \Gamma_C } }_{{\mathrm{2}}}  }  \rightsquigarrow  \target {  {\target e }_{{\mathrm{2}}}  } \), \\
  \( \vdash_{ctx}  \source {  { \source P }  } ; \source {  {\source { \Gamma_C } }  } ; \source {  \bullet  }  \rightsquigarrow  \target {  \target {\bullet}  } \), \\
  then \( \source {  {\source { \Gamma_C } }_{{\mathrm{1}}}  } = \source {  {\source { \Gamma_C } }_{{\mathrm{2}}}  } \),
  \( \source {  { \source P }_{{\mathrm{1}}}  } = \source {  { \source P }_{{\mathrm{2}}}  } \) \\
  and \( \source {  { \source P }  \ottsym{,}  { \source P }_{{\mathrm{1}}}  } ; \source {  {\source { \Gamma_C } }  \ottsym{,}  {\source { \Gamma_C } }_{{\mathrm{1}}}  } ; \source {  \bullet  }  \vdash  \target {  {\target e }_{{\mathrm{1}}}  }  \simeq_{ctx}  \target {  {\target e }_{{\mathrm{2}}}  } : \source {  {\source \tau }  } \).
  \label{thmA:ecoh_programs}
\end{theoremB}

\proof
By structural induction on \({\source {pgm} }\). \\
\proofStep{\( \source {  {\source {pgm} }  } = \source {  {\source {cls} }  \ottsym{;}  {\source {pgm} }'  } \)}
{
}
{
  By case analysis on the program typing derivations (\textsc{sPgmT-cls}):
  \begin{gather}
     \source {  {\source { \Gamma_C } }  }  \vdash_{cls}  \source {  {\source {cls} }  } : \source {  {\source { \Gamma_C } }'_{{\mathrm{1}}}  } 
    \label{eq:pcoh:1} \\
     \source {  {\source { \Gamma_C } }  }  \vdash_{cls}  \source {  {\source {cls} }  } : \source {  {\source { \Gamma_C } }'_{{\mathrm{2}}}  } 
    \label{eq:pcoh:2} \\
     \source {  { \source P }  } ; \source {  {\source { \Gamma_C } }  \ottsym{,}  {\source { \Gamma_C } }'_{{\mathrm{1}}}  }  \vdash_{pgm}  \source {  {\source {pgm} }'  } : \source {  {\source \tau }  } ; \source {  { \source P }_{{\mathrm{1}}}  } ; \source {  {\source { \Gamma_C } }''_{{\mathrm{1}}}  }  \rightsquigarrow  \target {  {\target e }_{{\mathrm{1}}}  } 
    \nonumber \\
     \source {  { \source P }  } ; \source {  {\source { \Gamma_C } }  \ottsym{,}  {\source { \Gamma_C } }'_{{\mathrm{2}}}  }  \vdash_{pgm}  \source {  {\source {pgm} }'  } : \source {  {\source \tau }  } ; \source {  { \source P }_{{\mathrm{2}}}  } ; \source {  {\source { \Gamma_C } }''_{{\mathrm{2}}}  }  \rightsquigarrow  \target {  {\target e }_{{\mathrm{2}}}  } 
    \nonumber \\
     \source {  {\source { \Gamma_C } }_{{\mathrm{1}}}  } = \source {  {\source { \Gamma_C } }'_{{\mathrm{1}}}  \ottsym{,}  {\source { \Gamma_C } }''_{{\mathrm{1}}}  } 
    \nonumber \\
     \source {  {\source { \Gamma_C } }_{{\mathrm{2}}}  } = \source {  {\source { \Gamma_C } }'_{{\mathrm{2}}}  \ottsym{,}  {\source { \Gamma_C } }''_{{\mathrm{2}}}  } 
    \nonumber 
  \end{gather}
  Since class typing is entirely deterministic, we know that
  \( \source {  {\source { \Gamma_C } }'_{{\mathrm{1}}}  } = \source {  {\source { \Gamma_C } }'_{{\mathrm{2}}}  } \).
  From Theorem~\ref{thmA:typing_pres_class}, in combination with
  Equations~\ref{eq:pcoh:1} and~\ref{eq:pcoh:2}, we know that:
  \begin{gather}
     \vdash_{ctx}  \source {  { \source P }  } ; \source {  {\source { \Gamma_C } }  \ottsym{,}  {\source { \Gamma_C } }'_{{\mathrm{1}}}  } ; \source {  \bullet  }  \rightsquigarrow  \target {  \target {\bullet}  } 
    \nonumber \\
     \vdash_{ctx}  \source {  { \source P }  } ; \source {  {\source { \Gamma_C } }  \ottsym{,}  {\source { \Gamma_C } }'_{{\mathrm{2}}}  } ; \source {  \bullet  }  \rightsquigarrow  \target {  \target {\bullet}  } 
    \nonumber
  \end{gather}
  The goal follows from the induction hypothesis. \\
}
\proofStep{\( \source {  {\source {pgm} }  } = \source {  {\source {inst} }  \ottsym{;}  {\source {pgm} }'  } \)}
{
}
{
  By case analysis on the program typing derivations (\textsc{sPgmT-inst}):
  \begin{gather}
     \source {  { \source P }  } ; \source {  {\source { \Gamma_C } }  }  \vdash_{inst}  \source {  {\source {inst} }  } : \source {  { \source P }_{{\mathrm{11}}}  } 
    \label{eq:ecohp1} \\
     \source {  { \source P }  } ; \source {  {\source { \Gamma_C } }  }  \vdash_{inst}  \source {  {\source {inst} }  } : \source {  { \source P }_{{\mathrm{21}}}  } 
    \label{eq:ecohp2} \\
     \source {  { \source P }  \ottsym{,}  { \source P }_{{\mathrm{11}}}  } ; \source {  {\source { \Gamma_C } }  }  \vdash_{pgm}  \source {  {\source {pgm} }'  } : \source {  {\source \tau }  } ; \source {  { \source P }_{{\mathrm{12}}}  } ; \source {  {\source { \Gamma_C } }_{{\mathrm{1}}}  }  \rightsquigarrow  \target {  {\target e }_{{\mathrm{1}}}  } 
    \label{eq:ecohp2b} \\
     \source {  { \source P }  \ottsym{,}  { \source P }_{{\mathrm{21}}}  } ; \source {  {\source { \Gamma_C } }  }  \vdash_{pgm}  \source {  {\source {pgm} }'  } : \source {  {\source \tau }  } ; \source {  { \source P }_{{\mathrm{22}}}  } ; \source {  {\source { \Gamma_C } }_{{\mathrm{2}}}  }  \rightsquigarrow  \target {  {\target e }_{{\mathrm{2}}}  } 
    \label{eq:ecohp2c} \\
     \source {  { \source P }_{{\mathrm{1}}}  } = \source {  { \source P }_{{\mathrm{11}}}  \ottsym{,}  { \source P }_{{\mathrm{12}}}  } 
    \nonumber \\
     \source {  { \source P }_{{\mathrm{2}}}  } = \source {  { \source P }_{{\mathrm{21}}}  \ottsym{,}  { \source P }_{{\mathrm{22}}}  } 
    \nonumber
  \end{gather}
  The goal to be proven is the following:
  \begin{gather}
     \source {  {\source { \Gamma_C } }_{{\mathrm{1}}}  } = \source {  {\source { \Gamma_C } }_{{\mathrm{2}}}  } 
    \label{eqg:ecohp1} \\
     \source {  { \source P }_{{\mathrm{11}}}  \ottsym{,}  { \source P }_{{\mathrm{12}}}  } = \source {  { \source P }_{{\mathrm{21}}}  \ottsym{,}  { \source P }_{{\mathrm{22}}}  } 
    \label{eqg:ecohp2} \\
     \source {  { \source P }  \ottsym{,}  { \source P }_{{\mathrm{11}}}  \ottsym{,}  { \source P }_{{\mathrm{12}}}  } ; \source {  {\source { \Gamma_C } }  \ottsym{,}  {\source { \Gamma_C } }_{{\mathrm{1}}}  } ; \source {  \bullet  }  \vdash  \target {  {\target e }_{{\mathrm{1}}}  }  \simeq_{ctx}  \target {  {\target e }_{{\mathrm{2}}}  } : \source {  {\source \tau }  } 
    \label{eqg:ecohp3}
  \end{gather}
  By case analysis on Equations~\ref{eq:ecohp1} and~\ref{eq:ecohp2} (\textsc{sInstT-inst}),
  we know that:
  \begin{gather}
     \source {  {\source {inst} }  } = \source {  \textbf{\source {instance} } \, {\overline {\source Q} }_{\ottmv{p}}  \Rightarrow  {\source {\mathit{TC} } } \, {\source \tau } \, \textbf{\source {where} } \, \ottsym{\{}  {\source m}  \ottsym{=}  {\source e}  \ottsym{\}}  } 
    \nonumber \\
     ( \source {  {\source m}  } : \source {  {\overline {\source Q} }'_{\ottmv{i}}  }  \Rightarrow  \source {  {\source {\mathit{TC} } }  }~\source {  {\source a }  } : \source {  \forall  {\overline {\source a} }_{\ottmv{j}}  \ottsym{.}  {\overline {\source Q} }'_{\ottmv{h}}  \Rightarrow  {\source \tau }_{{\mathrm{1}}}  } )  \in  \source {  {\source { \Gamma_C } }  } 
    \label{eq:ecohpe2} \\
     \source {  {\overline {\source b} }_{\ottmv{k}}  } =  \textbf {fv}  ( \source {  {\source \tau }'  } ) 
    \nonumber \\
     \textbf {closure}  ( \source {  {\source { \Gamma_C } }  } ; \source {  {\overline {\source Q} }_{\ottmv{p}}  } ) = \source {  {\overline {\source Q} }_{\ottmv{q}\,{\mathrm{1}}}  } 
    \nonumber \\
     \textbf {closure}  ( \source {  {\source { \Gamma_C } }  } ; \source {  {\overline {\source Q} }_{\ottmv{p}}  } ) = \source {  {\overline {\source Q} }_{\ottmv{q}\,{\mathrm{2}}}  } 
    \nonumber \\
     \textbf {unambig}  ( \source {  \forall  {\overline {\source b} }_{\ottmv{k}}  \ottsym{.}  {\overline {\source Q} }_{\ottmv{q}\,{\mathrm{1}}}  \Rightarrow  {\source {\mathit{TC} } } \, {\source \tau }'  } ) 
    \label{eq:ecohpe1} \\
     \textbf {unambig}  ( \source {  \forall  {\overline {\source b} }_{\ottmv{k}}  \ottsym{.}  {\overline {\source Q} }_{\ottmv{q}\,{\mathrm{2}}}  \Rightarrow  {\source {\mathit{TC} } } \, {\source \tau }'  } ) 
    \nonumber \\
    \ottcomp{ \source {  { \source P }  } ; \source {  {\source { \Gamma_C } }  } ; \source {  \bullet  \ottsym{,}  {\overline {\source b} }_{\ottmv{k}}  \ottsym{,}  {\overline {\source \delta} }_{\ottmv{q}\,{\mathrm{1}}}  \ottsym{:}  {\overline {\source Q} }_{\ottmv{q}\,{\mathrm{1}}}  }  \vDash  \source {  [  {\source \tau }'  /  {\source a }  ]  {\source Q}'_{\ottmv{i}}  } \stot{ \rightsquigarrow  \target {  {\target e }_{{\mathrm{1}}\,\ottmv{i}}  } } }{\ottmv{i}}
    \nonumber \\
    \ottcomp{ \source {  { \source P }  } ; \source {  {\source { \Gamma_C } }  } ; \source {  \bullet  \ottsym{,}  {\overline {\source b} }_{\ottmv{k}}  \ottsym{,}  {\overline {\source \delta} }_{\ottmv{q}\,{\mathrm{2}}}  \ottsym{:}  {\overline {\source Q} }_{\ottmv{q}\,{\mathrm{2}}}  }  \vDash  \source {  [  {\source \tau }'  /  {\source a }  ]  {\source Q}'_{\ottmv{i}}  } \stot{ \rightsquigarrow  \target {  {\target e }_{{\mathrm{2}}\,\ottmv{i}}  } } }{\ottmv{i}}
    \nonumber \\
     \source {  {\source D }  }~ \textbf {fresh} 
    \label{eq:ecohpe3} \\
     \source {  {\overline {\source \delta} }_{\ottmv{q}\,{\mathrm{1}}}  }~ \textbf {fresh} 
    \nonumber \\
     \source {  {\overline {\source \delta} }_{\ottmv{q}\,{\mathrm{2}}}  }~ \textbf {fresh} 
    \nonumber \\
     \source {  {\overline {\source \delta} }'_{\ottmv{h}}  }~ \textbf {fresh} 
    \nonumber
  \end{gather}
  \begin{gather}
     \source {  { \source P }  } ; \source {  {\source { \Gamma_C } }  } ; \source {  \bullet  \ottsym{,}  {\overline {\source b} }_{\ottmv{k}}  \ottsym{,}  {\overline {\source \delta} }_{\ottmv{q}\,{\mathrm{1}}}  \ottsym{:}  {\overline {\source Q} }_{\ottmv{q}\,{\mathrm{1}}}  \ottsym{,}  {\overline {\source a} }_{\ottmv{j}}  \ottsym{,}  {\overline {\source \delta} }'_{\ottmv{h}}  \ottsym{:}  [  {\source \tau }'  /  {\source a }  ]  {\overline {\source Q} }'_{\ottmv{h}}  }  \vdash_{tm}  \source {  {\source e}  } \Leftarrow \source {  [  {\source \tau }'  /  {\source a }  ]  {\source \tau }_{{\mathrm{1}}}  } \stot{ \rightsquigarrow  \target {  {\target e }'_{{\mathrm{1}}}  } } 
    \label{eq:ecohp3} \\
     \source {  { \source P }  } ; \source {  {\source { \Gamma_C } }  } ; \source {  \bullet  \ottsym{,}  {\overline {\source b} }_{\ottmv{k}}  \ottsym{,}  {\overline {\source \delta} }_{\ottmv{q}\,{\mathrm{2}}}  \ottsym{:}  {\overline {\source Q} }_{\ottmv{q}\,{\mathrm{2}}}  \ottsym{,}  {\overline {\source a} }_{\ottmv{j}}  \ottsym{,}  {\overline {\source \delta} }'_{\ottmv{h}}  \ottsym{:}  [  {\source \tau }'  /  {\source a }  ]  {\overline {\source Q} }'_{\ottmv{h}}  }  \vdash_{tm}  \source {  {\source e}  } \Leftarrow \source {  [  {\source \tau }'  /  {\source a }  ]  {\source \tau }_{{\mathrm{1}}}  } \stot{ \rightsquigarrow  \target {  {\target e }'_{{\mathrm{2}}}  } } 
    \label{eq:ecohp4} \\
     \source {  { \source P }_{{\mathrm{11}}}  } = \source {  \ottsym{(}  {\source D }  \ottsym{:}  \forall  {\overline {\source b} }_{\ottmv{k}}  \ottsym{.}  {\overline {\source Q} }_{\ottmv{q}\,{\mathrm{1}}}  \Rightarrow  {\source {\mathit{TC} } } \, {\source \tau }'  \ottsym{)}  \ottsym{.}  {\source m}  \mapsto  \bullet  \ottsym{,}  {\overline {\source b} }_{\ottmv{k}}  \ottsym{,}  {\overline {\source \delta} }_{\ottmv{q}\,{\mathrm{1}}}  \ottsym{:}  {\overline {\source Q} }_{\ottmv{q}\,{\mathrm{1}}}  \ottsym{,}  {\overline {\source a} }_{\ottmv{j}}  \ottsym{,}  {\overline {\source \delta} }'_{\ottmv{h}}  \ottsym{:}  [  {\source \tau }'  /  {\source a }  ]  {\overline {\source Q} }'_{\ottmv{h}}  \ottsym{:}  {\source e}_{{\mathrm{1}}}  } 
    \label{eq:ecohp7} \\
     \source {  { \source P }_{{\mathrm{21}}}  } = \source {  \ottsym{(}  {\source D }  \ottsym{:}  \forall  {\overline {\source b} }_{\ottmv{k}}  \ottsym{.}  {\overline {\source Q} }_{\ottmv{q}\,{\mathrm{2}}}  \Rightarrow  {\source {\mathit{TC} } } \, {\source \tau }'  \ottsym{)}  \ottsym{.}  {\source m}  \mapsto  \bullet  \ottsym{,}  {\overline {\source b} }_{\ottmv{k}}  \ottsym{,}  {\overline {\source \delta} }_{\ottmv{q}\,{\mathrm{2}}}  \ottsym{:}  {\overline {\source Q} }_{\ottmv{q}\,{\mathrm{2}}}  \ottsym{,}  {\overline {\source a} }_{\ottmv{j}}  \ottsym{,}  {\overline {\source \delta} }'_{\ottmv{h}}  \ottsym{:}  [  {\source \tau }'  /  {\source a }  ]  {\overline {\source Q} }'_{\ottmv{h}}  \ottsym{:}  {\source e}_{{\mathrm{2}}}  } 
    \label{eq:ecohp8} \\
     \source {  {\source { \Gamma_C } }  } ; \source {  \bullet  \ottsym{,}  {\overline {\source b} }_{\ottmv{k}}  }  \vdash_{ty}  \source {  {\source \tau }'  }\,{ \stoi{ \rightsquigarrow  \target {  {\target \sigma }'  } } } 
    \label{eq:ecohp29} \\
    \ottcomp{ \source {  {\source { \Gamma_C } }  } ; \source {  \bullet  \ottsym{,}  {\overline {\source b} }_{\ottmv{k}}  }  \vdash_{\mathit {Q} }  \source {  {\source Q}_{\ottmv{q}\,{\mathrm{1}}}  }  \rightsquigarrow  \target {  {\target \sigma }_{\ottmv{q}\,{\mathrm{1}}}  } }{\ottmv{q}}
    \label{eq:ecohp30} \\
    \ottcomp{ \source {  {\source { \Gamma_C } }  } ; \source {  \bullet  \ottsym{,}  {\overline {\source b} }_{\ottmv{k}}  }  \vdash_{\mathit {Q} }  \source {  {\source Q}_{\ottmv{q}\,{\mathrm{2}}}  }  \rightsquigarrow  \target {  {\target \sigma }_{\ottmv{q}\,{\mathrm{2}}}  } }{\ottmv{q}}
    \label{eq:ecohp30b} \\
     ( \source {  {\source D }'  } : \source {  \forall  {\overline {\source b} }'_{\ottmv{s}}  \ottsym{.}  {\overline {\source Q} }'_{\ottmv{n}}  \Rightarrow  {\source {\mathit{TC} } } \, {\source \tau }_{{\mathrm{2}}}  } ) . \source {  {\source m}'  }  \mapsto  \source {  {\source \Gamma}'  } : \source {  {\source e}'  }  \notin  \source {  { \source P }  }~\textit{where}~\source {  [  {\overline {\source \tau} }'_{\ottmv{s}}  /  {\overline {\source b} }'_{\ottmv{s}}  ]  {\source \tau }_{{\mathrm{2}}}  } = \source {  [  {\overline {\source \tau} }'_{\ottmv{k}}  /  {\overline {\source b} }_{\ottmv{k}}  ]  {\source \tau }'  } 
    \label{eq:ecohpe4}
  \end{gather}
  Note that since instance typing is entirely deterministic,
  \( \source {  { \source P }_{{\mathrm{11}}}  } = \source {  { \source P }_{{\mathrm{21}}}  } \).
  Similarly, since the closure over the superclass relation is deterministic, we know that
  \( \source {  {\overline {\source Q} }_{\ottmv{q}\,{\mathrm{1}}}  } = \source {  {\overline {\source Q} }_{\ottmv{q}\,{\mathrm{2}}}  } \).
  Note that we assume that the fresh variables are identical in both program typing derivations.

  We can derive from \textsc{sCtxT-pgmInst} that
  \begin{gather}
     \vdash_{ctx}  \source {  { \source P }  \ottsym{,}  { \source P }_{{\mathrm{11}}}  } ; \source {  {\source { \Gamma_C } }  } ; \source {  \bullet  }  \rightsquigarrow  \target {  \target {\bullet}  } 
    \label{eq:ecohp14} \\
     \vdash_{ctx}  \source {  { \source P }  \ottsym{,}  { \source P }_{{\mathrm{21}}}  } ; \source {  {\source { \Gamma_C } }  } ; \source {  \bullet  }  \rightsquigarrow  \target {  \target {\bullet}  } 
    \label{eq:ecohp15}
  \end{gather}
  assuming we can show that:
  \begin{gather}
     \textbf {unambig}  ( \source {  \forall  {\overline {\source b} }_{\ottmv{k}}  \ottsym{.}  {\overline {\source Q} }_{\ottmv{q}\,{\mathrm{1}}}  \Rightarrow  {\source {\mathit{TC} } } \, {\source \tau }'  } ) 
    \label{eq:ecohp16} \\
     \source {  {\source { \Gamma_C } }  } ; \source {  \bullet  }  \vdash_{\mathit {C} }  \source {  \forall  {\overline {\source b} }_{\ottmv{k}}  \ottsym{.}  {\overline {\source Q} }_{\ottmv{q}\,{\mathrm{1}}}  \Rightarrow  {\source {\mathit{TC} } } \, {\source \tau }'  }  \rightsquigarrow  \target {  \forall  {\overline {\target b} }_{\ottmv{k}}  \ottsym{.}  {\overline {\target \sigma} }_{\ottmv{q}\,{\mathrm{1}}}  \rightarrow  [  {\target \sigma }'  /  {\target a }  ]  \ottsym{\{}  {\target m}  \ottsym{:}  \forall  {\overline {\target a} }_{\ottmv{j}}  \ottsym{.}  {\overline {\target \sigma} }'_{\ottmv{h}}  \rightarrow  {\target \sigma }_{{\mathrm{1}}}  \ottsym{\}}  } 
    \label{eq:ecohp17} \\
     ( \source {  {\source m}  } : \source {  {\overline {\source Q} }'_{\ottmv{i}}  }  \Rightarrow  \source {  {\source {\mathit{TC} } }  }~\source {  {\source a }  } : \source {  \forall  {\overline {\source a} }_{\ottmv{j}}  \ottsym{.}  {\overline {\source Q} }'_{\ottmv{h}}  \Rightarrow  {\source \tau }_{{\mathrm{1}}}  } )  \in  \source {  {\source { \Gamma_C } }  } 
    \label{eq:ecohp18} \\
     \source {  {\source { \Gamma_C } }  } ; \source {  \bullet  \ottsym{,}  {\source a }  }  \vdash_{ty}  \source {  \forall  {\overline {\source a} }_{\ottmv{j}}  \ottsym{.}  {\overline {\source Q} }'_{\ottmv{h}}  \Rightarrow  {\source \tau }_{{\mathrm{1}}}  }\,{ \stoi{ \rightsquigarrow  \target {  \forall  {\overline {\target a} }_{\ottmv{j}}  \ottsym{.}  {\overline {\target \sigma} }'_{\ottmv{h}}  \rightarrow  {\target \sigma }_{{\mathrm{1}}}  } } } 
    \label{eq:ecohp19} \\
     \source {  {\source { \Gamma_C } }  } ; \source {  \bullet  \ottsym{,}  {\overline {\source b} }_{\ottmv{k}}  }  \vdash_{ty}  \source {  {\source \tau }'  }\,{ \stoi{ \rightsquigarrow  \target {  {\target \sigma }'  } } } 
    \label{eq:ecohp20} \\
     \source {  { \source P }  } ; \source {  {\source { \Gamma_C } }  } ; \source {  \bullet  \ottsym{,}  {\overline {\source b} }_{\ottmv{k}}  \ottsym{,}  {\overline {\source \delta} }_{\ottmv{q}\,{\mathrm{1}}}  \ottsym{:}  {\overline {\source Q} }_{\ottmv{q}\,{\mathrm{1}}}  \ottsym{,}  {\overline {\source a} }_{\ottmv{j}}  \ottsym{,}  {\overline {\source \delta} }'_{\ottmv{h}}  \ottsym{:}  [  {\source \tau }'  /  {\source a }  ]  {\overline {\source Q} }'_{\ottmv{h}}  }  \vdash_{tm}  \source {  {\source e}  } \Leftarrow \source {  [  {\source \tau }'  /  {\source a }  ]  {\source \tau }_{{\mathrm{1}}}  } \stot{ \rightsquigarrow  \target {  {\target e }'_{{\mathrm{1}}}  } } 
    \label{eq:ecohp21} \\
     \source {  { \source P }  } ; \source {  {\source { \Gamma_C } }  } ; \source {  \bullet  \ottsym{,}  {\overline {\source b} }_{\ottmv{k}}  \ottsym{,}  {\overline {\source \delta} }_{\ottmv{q}\,{\mathrm{2}}}  \ottsym{:}  {\overline {\source Q} }_{\ottmv{q}\,{\mathrm{2}}}  \ottsym{,}  {\overline {\source a} }_{\ottmv{j}}  \ottsym{,}  {\overline {\source \delta} }'_{\ottmv{h}}  \ottsym{:}  [  {\source \tau }'  /  {\source a }  ]  {\overline {\source Q} }'_{\ottmv{h}}  }  \vdash_{tm}  \source {  {\source e}  } \Leftarrow \source {  [  {\source \tau }'  /  {\source a }  ]  {\source \tau }_{{\mathrm{1}}}  } \stot{ \rightsquigarrow  \target {  {\target e }'_{{\mathrm{2}}}  } } 
    \label{eq:ecohp22} \\
     \source {  {\source D }  }  \notin   \textbf {dom}  ( \source {  { \source P }  } ) 
    \label{eq:ecohp23} \\
     ( \source {  {\source D }'  } : \source {  \forall  {\overline {\source b} }'_{\ottmv{k}}  \ottsym{.}  {\overline {\source Q} }''_{\ottmv{h}}  \Rightarrow  {\source {\mathit{TC} } } \, {\source \tau }''  } ) . \source {  {\source m}'  }  \mapsto  \source {  {\source \Gamma}'  } : \source {  {\source e}'  }  \notin  \source {  { \source P }  } ~
    where~ \source {  [  {\overline {\source \tau} }_{\ottmv{k}}  /  {\overline {\source b} }_{\ottmv{k}}  ]  {\source \tau }'  } = \source {  [  {\overline {\source \tau} }'_{\ottmv{k}}  /  {\overline {\source b} }'_{\ottmv{k}}  ]  {\source \tau }''  } 
    \label{eq:ecohp25} \\
     \vdash_{ctx}  \source {  { \source P }  } ; \source {  {\source { \Gamma_C } }  } ; \source {  \bullet  }  \rightsquigarrow  \target {  \target {\bullet}  } 
    \label{eq:ecohp28} 
  \end{gather}
  Goals~\ref{eq:ecohp16}, \ref{eq:ecohp18}, \ref{eq:ecohp20}, \ref{eq:ecohp21},
  \ref{eq:ecohp22}, \ref{eq:ecohp23} and~\ref{eq:ecohp25} follow directly from
  Equations~\ref{eq:ecohpe1}, \ref{eq:ecohpe2}, \ref{eq:ecohp29}, \ref{eq:ecohp3},
  \ref{eq:ecohp4}, \ref{eq:ecohpe3} and~\ref{eq:ecohpe4} respectively. 

  Goal~\ref{eq:ecohp28}
  follows directly from the 3\textsuperscript{rd} hypothesis.
  Goal~\ref{eq:ecohp19} follows by applying case analysis on Equation~\ref{eq:ecohp28}
  (\textsc{sCtxT-clsEnv}), in combination with Equation~\ref{eq:ecohp18}.
  Goal~\ref{eq:ecohp17} follows from \textsc{sCT-abs} and \textsc{sQT-TC}, in
  combination with Equations~\ref{eq:ecohp18}, \ref{eq:ecohp19},
  \ref{eq:ecohp29} and~\ref{eq:ecohp30}.

  Finally, Goals~\ref{eqg:ecohp1}, \ref{eqg:ecohp2} and~\ref{eqg:ecohp3} follow
  by applying the induction hypothesis on Equations~\ref{eq:ecohp2b}
  and~\ref{eq:ecohp2c}, in combination with 
  Equations~\ref{eq:ecohp14} and~\ref{eq:ecohp15}. \\
}
\proofStep{\( \source {  {\source {pgm} }  } = \source {  {\source e}  } \)}
{
}
{
  The goal follows directly from coherence Theorem~\ref{thmA:ecoh}.
}
\qed

\renewcommand\itot[1]{}

\newpage
\newcommand{\pref}[1]{\ref{#1}}
\renewcommand\itot[1]{#1}

\section{\interL{}-to-\targetL{} Theorems}

\subsection{Lemmas}
\subsubsection{Determinism / Uniqueness}

\begin{lemmaB}[Dictionary Elaboration Uniqueness]\leavevmode
  \label{lem:det:Q}\ \\
  If \( \intermed {  {\intermed { \Gamma_C } }  } ; \intermed {  {\intermed \Gamma}  }  \vdash_{\mathit {Q} }  \intermed {  {\intermed Q}  }\, \itot{ \rightsquigarrow  \target {  {\target \sigma }_{{\mathrm{1}}}  } } \) and \( \intermed {  {\intermed { \Gamma_C } }  } ; \intermed {  {\intermed \Gamma}  }  \vdash_{\mathit {Q} }  \intermed {  {\intermed Q}  }\, \itot{ \rightsquigarrow  \target {  {\target \sigma }_{{\mathrm{2}}}  } } \),
  then \( \target {  {\target \sigma }_{{\mathrm{1}}}  } = \target {  {\target \sigma }_{{\mathrm{2}}}  } \).
\end{lemmaB}

\proof
By straightforward induction on the well-formedness derivation. \\
\qed

\begin{lemmaB}[Type Elaboration Uniqueness]
  \label{lem:det:ty}\ \\
  If \( \intermed {  {\intermed { \Gamma_C } }  } ; \intermed {  {\intermed \Gamma}  }  \vdash_{ty}  \intermed {  {\intermed \sigma}  } \itot{ \rightsquigarrow  \target {  {\target \sigma }_{{\mathrm{1}}}  } } \) and \( \intermed {  {\intermed { \Gamma_C } }  } ; \intermed {  {\intermed \Gamma}  }  \vdash_{ty}  \intermed {  {\intermed \sigma}  } \itot{ \rightsquigarrow  \target {  {\target \sigma }_{{\mathrm{2}}}  } } \),
  then \( \target {  {\target \sigma }_{{\mathrm{1}}}  } = \target {  {\target \sigma }_{{\mathrm{2}}}  } \).
\end{lemmaB}

\proof
By straightforward induction on the well-formedness derivation. \\
\qed

\begin{lemmaB}[Context Elaboration Uniqueness]
  \label{lem:det:GT}\ \\
  If $ \intermed {  {\intermed { \Gamma_C } }  } ; \intermed {  {\intermed \Gamma}  }  \rightsquigarrow  \target {  {\target \Gamma }_{{\mathrm{1}}}  } $
  and $ \intermed {  {\intermed { \Gamma_C } }  } ; \intermed {  {\intermed \Gamma}  }  \rightsquigarrow  \target {  {\target \Gamma }_{{\mathrm{2}}}  } $
  then $ \target {  {\target \Gamma }_{{\mathrm{1}}}  } = \target {  {\target \Gamma }_{{\mathrm{2}}}  } $.
\end{lemmaB}

\proof
By straightforward induction on the well-formedness derivation. \\
\qed

\begin{lemmaB}[Determinism of Evaluation]\ \\
  \label{lem:tgt_eval_unique}
  If \( \target {  {\target e }  }  \longrightarrow  \target {  {\target e }_{{\mathrm{1}}}  } \) and \( \target {  {\target e }  }  \longrightarrow  \target {  {\target e }_{{\mathrm{2}}}  } \)
  then \( \target {  {\target e }_{{\mathrm{1}}}  } = \target {  {\target e }_{{\mathrm{2}}}  } \).
\end{lemmaB}

\proof
By straightforward induction on both evaluation derivations. \\
\qed

\subsubsection{Soundness}

\begin{lemmaB}[Dictionary Variable Elaboration Soundness]
  \label{lem:qelab}\ \\
  If \( \intermed {  {\intermed { \Gamma_C } }  } ; \intermed {  {\intermed \Gamma}  }  \vdash_{\mathit {Q} }  \intermed {  {\intermed Q}  }\, \itot{ \rightsquigarrow  \target {  {\target \sigma }  } } \)
  and $ \intermed {  {\intermed { \Gamma_C } }  } ; \intermed {  {\intermed \Gamma}  }  \rightsquigarrow  \target {  {\target \Gamma }  } $
  then $ \target {  {\target \Gamma }  }  \vdash_{ty}  \target {  {\target \sigma }  } $.
\end{lemmaB}

\proof
By straightforward induction on the dictionary typing derivation. \\
\qed

\begin{lemmaB}[Type Elaboration Soundness]
  \label{lem:tyelab}\ \\
  If \( \intermed {  {\intermed { \Gamma_C } }  } ; \intermed {  {\intermed \Gamma}  }  \vdash_{ty}  \intermed {  {\intermed \sigma}  } \itot{ \rightsquigarrow  \target {  {\target \sigma }  } } \)
  and $ \intermed {  {\intermed { \Gamma_C } }  } ; \intermed {  {\intermed \Gamma}  }  \rightsquigarrow  \target {  {\target \Gamma }  } $
  then \( \target {  {\target \Gamma }  }  \vdash_{ty}  \target {  {\target \sigma }  } \).
\end{lemmaB}

\proof
By straightforward induction on the type well-formedness derivation. \\
\qed

\begin{lemmaB}[Term Variable Elaboration Soundness]
  \label{lem:tmvar:elab}\ \\
  If $ \vdash_{ctx}  \intermed {  {\intermed \Sigma}  } ; \intermed {  {\intermed { \Gamma_C } }  } ; \intermed {  {\intermed \Gamma}  } $
  and $ ( \intermed {  {\intermed x}  } : \intermed {  {\intermed \sigma}  } )  \in  \intermed {  {\intermed \Gamma}  } $,
  then there are unique ${\target \Gamma }$ and ${\target \sigma }$ \\
  such that $ \intermed {  {\intermed { \Gamma_C } }  } ; \intermed {  {\intermed \Gamma}  }  \rightsquigarrow  \target {  {\target \Gamma }  } $
  and $ \intermed {  {\intermed { \Gamma_C } }  } ; \intermed {  {\intermed \Gamma}  }  \vdash_{ty}  \intermed {  {\intermed \sigma}  } \itot{ \rightsquigarrow  \target {  {\target \sigma }  } } $
  and $ ( \target {  {\target x}  } : \target {  {\target \sigma }  } )  \in  \target {  {\target \Gamma }  } $.
\end{lemmaB}

\proof
By straightforward induction on the environment well-formedness derivation. \\
\qed

\begin{lemmaB}[Dictionary Variable in Environment Elaboration Soundness]
  \label{lem:elab:dvar}\ \\
  If $ \vdash_{ctx}  \intermed {  {\intermed \Sigma}  } ; \intermed {  {\intermed { \Gamma_C } }  } ; \intermed {  {\intermed \Gamma}  } $
  and $ ( \intermed {  {\intermed \delta }  } : \intermed {  {\intermed {\mathit{TC} } } \, {\intermed \sigma}  } )  \in  \intermed {  {\intermed \Gamma}  } $,
  then there are unique ${\target \Gamma }$ and ${\target \sigma }$ \\
  such that $ \intermed {  {\intermed { \Gamma_C } }  } ; \intermed {  {\intermed \Gamma}  }  \rightsquigarrow  \target {  {\target \Gamma }  } $
  and $ \vdash_{ctx}  \target {  {\target \Gamma }  } $
  and $ \intermed {  {\intermed { \Gamma_C } }  } ; \intermed {  {\intermed \Gamma}  }  \vdash_{\mathit {Q} }  \intermed {  {\intermed {\mathit{TC} } } \, {\intermed \sigma}  }\, \itot{ \rightsquigarrow  \target {  {\target \sigma }  } } $
  and $ ( \target {  {\target \delta }  } : \target {  {\target \sigma }  } )  \in  \target {  {\target \Gamma }  } $.
\end{lemmaB}

\proof
By straightforward induction on the environment well-formedness derivation. \\
\qed

\begin{lemmaB}[Environment Elaboration Soundness]
  \label{lem:ctx_elab}\ \\
  If $ \vdash_{ctx}  \intermed {  {\intermed \Sigma}  } ; \intermed {  {\intermed { \Gamma_C } }  } ; \intermed {  {\intermed \Gamma}  } $,
  then there is a unique ${\target \Gamma }$ such
  that $ \intermed {  {\intermed { \Gamma_C } }  } ; \intermed {  {\intermed \Gamma}  }  \rightsquigarrow  \target {  {\target \Gamma }  } $
  and $ \vdash_{ctx}  \target {  {\target \Gamma }  } $.
\end{lemmaB}

\proof
By straightforward induction on the environment well-formedness derivation. \\
\qed

\subsubsection{Canonical Forms Lemmas}

\begin{lemmaB}[Canonical Forms for Functions]
  \label{lem:canon_func}\ \\
  If $ \target {  {\target \Gamma }  }  \vdash_{tm}  \target {  {\target v}  } : \target {  {\target \sigma }_{{\mathrm{1}}}  \rightarrow  {\target \sigma }_{{\mathrm{2}}}  } $ for some value ${\target v}$,
  then ${\target v}$ is of the form $\target{\lambda  {\target x}  \ottsym{:}  {\target \sigma }_{{\mathrm{1}}}  \ottsym{.}  {\target e }}$,
  for some ${\target x}$ and ${\target e }$.
\end{lemmaB}

\proof
By straightforward induction on the typing derivation. \\
\qed

\begin{lemmaB}[Canonical Forms for Type Abstractions]
  \label{lem:canon_tyabs}\ \\
  If $ \target {  {\target \Gamma }  }  \vdash_{tm}  \target {  {\target v}  } : \target {  \forall  {\target a }  \ottsym{.}  {\target \sigma }  } $ for some value ${\target v}$,
  then ${\target v}$ is of the form $\target{\Lambda  {\target a }  \ottsym{.}  {\target e }}$,
  for some ${\target e }$.
\end{lemmaB}

\proof
By straightforward induction on the typing derivation. \\
\qed

\subsubsection{Evaluation Lemmas}

\begin{lemmaB}[Distribution of \textsc{tEval-app}]
  \label{lem:tEval_app_dist}\ \\
  If $ \target {  \target {\bullet}  }  \vdash_{tm}  \target {  {\target e }  } : \target {  {\target \sigma }_{{\mathrm{1}}}  \rightarrow  {\target \sigma }_{{\mathrm{2}}}  } $
  and $ \target {  \target {\bullet}  }  \vdash_{tm}  \target {  {\target e }'  } : \target {  {\target \sigma }_{{\mathrm{1}}}  \rightarrow  {\target \sigma }_{{\mathrm{2}}}  } $
  and $ \target {  \target {\bullet}  }  \vdash_{tm}  \target {  {\target e }_{{\mathrm{1}}}  } : \target {  {\target \sigma }_{{\mathrm{1}}}  } $
  and $ \target {  {\target e }  }  \longrightarrow^{*}  \target {  {\target e }'  } $,
  then $ \target {  {\target e } \, {\target e }_{{\mathrm{1}}}  }  \longrightarrow^{*}  \target {  {\target e }' \, {\target e }_{{\mathrm{1}}}  } $.
\end{lemmaB}

\proof
The goal follows from Canonical Forms Lemma~\ref{lem:canon_func},
together with the well-known strong normalization
of System F with records. \\
\qed

\begin{lemmaB}[Distribution of \textsc{tEval-Tapp}]
  \label{lem:tEval_tapp_dist}\ \\
  If $ \target {  \target {\bullet}  }  \vdash_{tm}  \target {  {\target e }  } : \target {  \forall  {\target a }  \ottsym{.}  {\target \sigma }'  } $
  and $ \target {  \target {\bullet}  }  \vdash_{tm}  \target {  {\target e }'  } : \target {  \forall  {\target a }  \ottsym{.}  {\target \sigma }'  } $
  and $ \target {  \target {\bullet}  }  \vdash_{ty}  \target {  {\target \sigma }  } $
  and $ \target {  {\target e }  }  \longrightarrow^{*}  \target {  {\target e }'  } $,
  then $ \target {  {\target e } \, {\target \sigma }  }  \longrightarrow^{*}  \target {  {\target e }' \, {\target \sigma }  } $.
\end{lemmaB}

\proof
The goal follows from Canonical Forms Lemma~\ref{lem:canon_tyabs},
together with the well-known strong normalization
of System F with records. \\
\qed

\subsection{Soundness}
\begin{theoremB}[Term Elaboration Soundness]
  \label{thmA:tmelab}
  \begin{flalign}
  &\text{If}~ \intermed {  {\intermed \Sigma}  } ; \intermed {  {\intermed { \Gamma_C } }  } ; \intermed {  {\intermed \Gamma}  }  \vdash_{tm}  \intermed {  {\intermed e}  } : \intermed {  {\intermed \sigma}  } \itot{ \rightsquigarrow  \target {  {\target e }  } } \label{tmelab:hyp} &\\
  &\text{then, there are unique}~{\target \Gamma }~\text{and}~{\target \sigma }~\text{such
    that}\nonumber &\\
  & \intermed {  {\intermed { \Gamma_C } }  } ; \intermed {  {\intermed \Gamma}  }  \rightsquigarrow  \target {  {\target \Gamma }  } \label{tmelab:res1} &\\
  &\text{and}~ \intermed {  {\intermed { \Gamma_C } }  } ; \intermed {  {\intermed \Gamma}  }  \vdash_{ty}  \intermed {  {\intermed \sigma}  } \itot{ \rightsquigarrow  \target {  {\target \sigma }  } } \label{tmelab:res2} &\\
  &\text{and}~ \target {  {\target \Gamma }  }  \vdash_{tm}  \target {  {\target e }  } : \target {  {\target \sigma }  } \label{tmelab:res3}
  \end{flalign}
\end{theoremB}

\proof This theorem is proved mutually with Theorem~\ref{thmA:delab}. The proof follows
structural induction on Hypothesis~\ref{tmelab:hyp} of the theorem.\\ \\
\proofStep{Case \textsc{iTm-true}}
          {\ottdruleiTmXXtrue{}}
          {We need to show that there are unique ${\target \Gamma }$ and ${\target \sigma }$ such that
            $ \intermed {  {\intermed { \Gamma_C } }  } ; \intermed {  {\intermed \Gamma}  }  \rightsquigarrow  \target {  {\target \Gamma }  } $ and $ \intermed {  {\intermed { \Gamma_C } }  } ; \intermed {  {\intermed \Gamma}  }  \vdash_{ty}  \intermed {  \mathit{\intermed {Bool} }  } \itot{ \rightsquigarrow  \target {  {\target \sigma }  } } $ and
            $ \target {  {\target \Gamma }  }  \vdash_{tm}  \target {  \mathit{\target {True} }  } : \target {  {\target \sigma }  } $.
            
            Obviously, ${\target \sigma }$ can only be equal to $\mathit{\target {Bool} }$.
            By Lemma~\ref{lem:ctx_elab} applied on the premise of rule
            \textsc{iTm-true}, there is a unique ${\target \Gamma }$ such that $ \intermed {  {\intermed { \Gamma_C } }  } ; \intermed {  {\intermed \Gamma}  }  \rightsquigarrow  \target {  {\target \Gamma }  } $ and
            $ \vdash_{ctx}  \target {  {\target \Gamma }  } $. We use the latter result to instantiate rule \textsc{tTm-True},
            which concludes with $ \target {  {\target \Gamma }  }  \vdash_{tm}  \target {  \mathit{\target {True} }  } : \target {  \mathit{\target {Bool} }  } $.\\
          }
\proofStep{Case \textsc{iTm-false}}
          {\ottdruleiTmXXfalse{}}
          {Similar to case \textsc{iTm-true}.\\}
\proofStep{Case \textsc{iTm-var}}
          {\ottdruleiTmXXvar{}}
          {
            By Lemma~\ref{lem:ctx_elab}, there is a unique ${\target \Gamma }$ such that
            \begin{gather}
              \nonumber
               \intermed {  {\intermed { \Gamma_C } }  } ; \intermed {  {\intermed \Gamma}  }  \rightsquigarrow  \target {  {\target \Gamma }  } \\
              \label{tmelab:var:2}
              \text{and}~ \vdash_{ctx}  \target {  {\target \Gamma }  } 
            \end{gather}
            With Lemma~\ref{lem:tmvar:elab} applied on the two premises of rule \textsc{iTm-var},
            we also find a unique ${\target \sigma }$ such that $ \intermed {  {\intermed { \Gamma_C } }  } ; \intermed {  {\intermed \Gamma}  }  \vdash_{ty}  \intermed {  {\intermed \sigma}  } \itot{ \rightsquigarrow  \target {  {\target \sigma }  } } $ and
            $ ( \target {  {\target x}  } : \target {  {\target \sigma }  } )  \in  \target {  {\target \Gamma }  } $. Then, by rule \textsc{tTm-Var} applied on the latter
            result and on Equation \pref{tmelab:var:2}, we reach the goal.\\
          }
\proofStep{Case \textsc{iTm-let}}
          {\ottdruleiTmXXlet{}}
          {The induction hypothesis for the first premise of rule \textsc{iTm-let} is
           \begin{gather}
             \text{There are unique}~{\target \Gamma },~\target{{\target \sigma }_{{\mathrm{1}}}},~\text{such}\nonumber\\
             \label{tmelab:let:IH1:GT}
             \text{that}~ \intermed {  {\intermed { \Gamma_C } }  } ; \intermed {  {\intermed \Gamma}  }  \rightsquigarrow  \target {  {\target \Gamma }  } \\
             \label{tmelab:let:IH1:ty}
             \text{and}~ \intermed {  {\intermed { \Gamma_C } }  } ; \intermed {  {\intermed \Gamma}  }  \vdash_{ty}  \intermed {  {\intermed \sigma}_{{\mathrm{1}}}  } \itot{ \rightsquigarrow  \target {  {\target \sigma }_{{\mathrm{1}}}  } } \\
             \label{tmelab:let:IH1:wftm}
             \text{and}~ \target {  {\target \Gamma }  }  \vdash_{tm}  \target {  {\target e }_{{\mathrm{1}}}  } : \target {  {\target \sigma }_{{\mathrm{1}}}  } 
           \end{gather}           
           The induction hypothesis for the second premise of rule \textsc{iTm-let} is
           \begin{gather}
             \text{There are unique}~\target{{\target \Gamma }'},~\target{{\target \sigma }_{{\mathrm{2}}}},~\text{such}\nonumber\\
             \label{tmelab:let:IH2:GT}
             \text{that}~ \intermed {  {\intermed { \Gamma_C } }  } ; \intermed {  {\intermed \Gamma}  \ottsym{,}  {\intermed x}  \ottsym{:}  {\intermed \sigma}_{{\mathrm{1}}}  }  \rightsquigarrow  \target {  {\target \Gamma }'  } \\
             \label{tmelab:let:IH2:ty}
             \text{and}~ \intermed {  {\intermed { \Gamma_C } }  } ; \intermed {  {\intermed \Gamma}  \ottsym{,}  {\intermed x}  \ottsym{:}  {\intermed \sigma}_{{\mathrm{1}}}  }  \vdash_{ty}  \intermed {  {\intermed \sigma}_{{\mathrm{2}}}  } \itot{ \rightsquigarrow  \target {  {\target \sigma }_{{\mathrm{2}}}  } } \\
             \label{tmelan:let:IH2:wftm}
             \text{and}~ \target {  {\target \Gamma }'  }  \vdash_{tm}  \target {  {\target e }_{{\mathrm{2}}}  } : \target {  {\target \sigma }_{{\mathrm{2}}}  } 
           \end{gather}
           By inversion on Equation~\pref{tmelab:let:IH2:GT}, there are $\target{{\target \Gamma }_{{\mathrm{0}}}}$ and $\target{{\target \sigma }'_{{\mathrm{1}}}}$ such
           that
           \begin{gather}
             \label{tmelab:let:1}
              \intermed {  {\intermed { \Gamma_C } }  } ; \intermed {  {\intermed \Gamma}  }  \rightsquigarrow  \target {  {\target \Gamma }_{{\mathrm{0}}}  }  \\
             \label{tmelab:let:2}
             \text{and}~ \intermed {  {\intermed { \Gamma_C } }  } ; \intermed {  {\intermed \Gamma}  }  \vdash_{ty}  \intermed {  {\intermed \sigma}_{{\mathrm{1}}}  } \itot{ \rightsquigarrow  \target {  {\target \sigma }'_{{\mathrm{1}}}  } }  \\
             \text{and}~ \target {  {\target \Gamma }'  } = \target {  {\target \Gamma }_{{\mathrm{0}}}  \ottsym{,}  {\target x}  \ottsym{:}  {\target \sigma }'_{{\mathrm{1}}}  } \nonumber
           \end{gather}
           By uniqueness (Lemma~\ref{lem:det:GT} on Equations~\pref{tmelab:let:IH1:GT} and
           \pref{tmelab:let:1} and Lemma~\ref{lem:det:ty} on Equations~\pref{tmelab:let:IH1:ty} and
           \pref{tmelab:let:2}), we get $ \target {  {\target \Gamma }_{{\mathrm{0}}}  } = \target {  {\target \Gamma }  } $ and $ \target {  {\target \sigma }'_{{\mathrm{1}}}  } = \target {  {\target \sigma }_{{\mathrm{1}}}  } $.
           This refines Equation \pref{tmelan:let:IH2:wftm} into
           \begin{equation}
             \label{tmelab:let:3}
              \target {  {\target \Gamma }  \ottsym{,}  {\target x}  \ottsym{:}  {\target \sigma }_{{\mathrm{1}}}  }  \vdash_{tm}  \target {  {\target e }_{{\mathrm{2}}}  } : \target {  {\target \sigma }_{{\mathrm{2}}}  } 
           \end{equation}
           By applying Lemma~\ref{lem:tyelab} on Equations \pref{tmelab:let:IH1:ty} and
           \pref{tmelab:let:IH1:GT}, we get
           \begin{equation}
             \label{tmelab:let:4}
              \target {  {\target \Gamma }  }  \vdash_{ty}  \target {  {\target \sigma }_{{\mathrm{1}}}  } 
           \end{equation}
           By applying rule \textsc{tTm-Let} on Equations \pref{tmelab:let:IH1:wftm},
           \pref{tmelab:let:3} and \pref{tmelab:let:4}, we get
           \begin{equation}
              \target {  {\target \Gamma }  }  \vdash_{tm}  \target {  \textbf{\target {let} } \, {\target x}  \ottsym{:}  {\target \sigma }_{{\mathrm{1}}}  \ottsym{=}  {\target e }_{{\mathrm{1}}} \, \textbf{\target {in} } \, {\target e }_{{\mathrm{2}}}  } : \target {  {\target \sigma }_{{\mathrm{2}}}  } 
           \end{equation}
           It remains to show that $ \intermed {  {\intermed { \Gamma_C } }  } ; \intermed {  {\intermed \Gamma}  }  \vdash_{ty}  \intermed {  {\intermed \sigma}_{{\mathrm{2}}}  } \itot{ \rightsquigarrow  \target {  {\target \sigma }_{{\mathrm{2}}}  } } $. This is easily derived from
           Equation \pref{tmelab:let:IH2:ty}, since the existence of variable ${\target x}$ in the
           context does not affect the well-formedness nor the translation of $\intermed{{\intermed \sigma}_{{\mathrm{2}}}}$.\\
          }
\proofStep{Case \textsc{iTm-method}}
          {\ottdruleiTmXXmethod{}}
          {
          Applying Theorem~\ref{thmA:delab} to the first premise of rule
          \textsc{iTm-method} results in
          \begin{gather}
            \text{There are unique}~{\target \Gamma }~\text{and}~\target{{\target \sigma }_{{\mathrm{1}}}}~\text{such}~\nonumber\\
            \label{tmelab:meth:IH1:GT}
            \text{that}~ \intermed {  {\intermed { \Gamma_C } }  } ; \intermed {  {\intermed \Gamma}  }  \rightsquigarrow  \target {  {\target \Gamma }  } \\
            \label{tmelab:meth:IH1:q}
            \text{and}~ \intermed {  {\intermed { \Gamma_C } }  } ; \intermed {  {\intermed \Gamma}  }  \vdash_{\mathit {Q} }  \intermed {  {\intermed {\mathit{TC} } } \, {\intermed \sigma}  }\, \itot{ \rightsquigarrow  \target {  {\target \sigma }_{{\mathrm{1}}}  } } \\
            \label{tmelab:meth:IH1:wftm}
            \text{and}~ \target {  {\target \Gamma }  }  \vdash_{tm}  \target {  {\target e }  } : \target {  {\target \sigma }_{{\mathrm{1}}}  } 
          \end{gather}
          By inversion on Equation \pref{tmelab:meth:IH1:q}, we get
          \begin{gather}
            \label{tmelan:meth:1}
             \intermed {  {\intermed { \Gamma_C } }  } = \intermed {  {\intermed { \Gamma_C } }_{{\mathrm{1}}}  }, \intermed {  {\intermed m}'  \ottsym{:}  {\intermed {\mathit{TC} } } \, {\intermed a }'  \ottsym{:}  {\intermed \sigma}_{\ottmv{m}}  \ottsym{,}  {\intermed { \Gamma_C } }_{{\mathrm{2}}}  } \\ 
            \label{tmelan:meth:2}
             \intermed {  {\intermed { \Gamma_C } }  } ; \intermed {  {\intermed \Gamma}  }  \vdash_{ty}  \intermed {  {\intermed \sigma}  } \itot{ \rightsquigarrow  \target {  {\target \sigma }  } } \\
            \label{tmelan:meth:3}
             \intermed {  {\intermed { \Gamma_C } }_{{\mathrm{1}}}  } ; \intermed {  \bullet  \ottsym{,}  {\intermed a }'  }  \vdash_{ty}  \intermed {  {\intermed \sigma}_{\ottmv{m}}  } \itot{ \rightsquigarrow  \target {  {\target \sigma }_{\ottmv{m}}  } } 
          \end{gather}
          for some $\intermed{{\intermed m}'}$, $\intermed{{\intermed a }'}$, $\intermed{{\intermed \sigma}_{\ottmv{m}}}$, ${\target \sigma }$ and $\target{{\target \sigma }_{\ottmv{m}}}$.
          However, each dictionary ${\intermed {\mathit{TC} } }$ corresponds to a unique entry in the class
          environment ${\intermed { \Gamma_C } }$. By this uniqueness, we get $ \intermed {  {\intermed m}'  } = \intermed {  {\intermed m}  } $,
          $ \intermed {  {\intermed a }  } = \intermed {  {\intermed a }'  } $, $ \intermed {  {\intermed \sigma}_{\ottmv{m}}  } = \intermed {  {\intermed \sigma}'  } $. 
          Then,
          $ \target {  {\target \sigma }_{{\mathrm{1}}}  } = \target {  [  {\target \sigma }  /  {\target a }  ]  \ottsym{\{}  {\target m}  \ottsym{:}  {\target \sigma }_{\ottmv{m}}  \ottsym{\}}  } $, and Equations \pref{tmelab:meth:IH1:q}
          and \pref{tmelab:meth:IH1:wftm} become
          \begin{gather}
             \intermed {  {\intermed { \Gamma_C } }  } ; \intermed {  {\intermed \Gamma}  }  \vdash_{\mathit {Q} }  \intermed {  {\intermed {\mathit{TC} } } \, {\intermed \sigma}  }\, \itot{ \rightsquigarrow  \target {  [  {\target \sigma }  /  {\target a }  ]  \ottsym{\{}  {\target m}  \ottsym{:}  {\target \sigma }_{\ottmv{m}}  \ottsym{\}}  } } \nonumber\\
            \label{tmelab:meth:5}
            \text{and}~ \target {  {\target \Gamma }  }  \vdash_{tm}  \target {  {\target e }  } : \target {  [  {\target \sigma }  /  {\target a }  ]  \ottsym{\{}  {\target m}  \ottsym{:}  {\target \sigma }_{\ottmv{m}}  \ottsym{\}}  } 
          \end{gather}
          Lemma~\ref{lemma:tvar_tsubst} applied on Equations~\pref{tmelan:meth:2} and
          \pref{tmelan:meth:3}, results in
          \begin{equation*}
             \intermed {  {\intermed { \Gamma_C } }  } ; \intermed {  {\intermed \Gamma}  }  \vdash_{ty}  \intermed {  [  {\intermed \sigma}  /  {\intermed a }  ]  {\intermed \sigma}'  } \itot{ \rightsquigarrow  \target {  [  {\target \sigma }  /  {\target a }  ]  {\target \sigma }_{\ottmv{m}}  } } 
          \end{equation*}
          and rule \textsc{tTm-Proj} instantiated with Equation~\pref{tmelab:meth:5} gives
          $ \target {  {\target \Gamma }  }  \vdash_{tm}  \target {  {\target e }  \ottsym{.}  {\target m}  } : \target {  [  {\target \sigma }  /  {\target a }  ]  {\target \sigma }_{\ottmv{m}}  } $.\\
          }
\proofStep{Case \textsc{iTm-arrI}}
          {\ottdruleiTmXXarrI{}}
          {
          The induction hypothesis from the first premise of rule \textsc{iTm-arrI} is
          \begin{gather}
            \text{There are unique}~\target{{\target \Gamma }'}~\text{and}~\target{{\target \sigma }_{{\mathrm{1}}}}~\text{such}~\nonumber\\
            \text{that}~ \intermed {  {\intermed { \Gamma_C } }  } ; \intermed {  {\intermed \Gamma}  \ottsym{,}  {\intermed x}  \ottsym{:}  {\intermed \sigma}_{{\mathrm{1}}}  }  \rightsquigarrow  \target {  {\target \Gamma }'  } \nonumber\\
            \label{tmelab:arri:IH:ty}
            \text{and}~ \intermed {  {\intermed { \Gamma_C } }  } ; \intermed {  {\intermed \Gamma}  \ottsym{,}  {\intermed x}  \ottsym{:}  {\intermed \sigma}_{{\mathrm{1}}}  }  \vdash_{ty}  \intermed {  {\intermed \sigma}_{{\mathrm{2}}}  } \itot{ \rightsquigarrow  \target {  {\target \sigma }_{{\mathrm{2}}}  } } \\
            \label{tmelab:arri:IH:wftm}
            \text{and}~ \target {  {\target \Gamma }'  }  \vdash_{tm}  \target {  {\target e }  } : \target {  {\target \sigma }_{{\mathrm{2}}}  } 
          \end{gather}
          Similarly to the \textsc{iTm-Let} case, $\target{{\target \Gamma }'}$ is of the form ${\target \Gamma }  \ottsym{,}  {\target x}  \ottsym{:}  {\target \sigma }_{{\mathrm{1}}}$,
          where
          \begin{gather}
            \label{tmelab:arri:1}
             \intermed {  {\intermed { \Gamma_C } }  } ; \intermed {  {\intermed \Gamma}  }  \rightsquigarrow  \target {  {\target \Gamma }  } \\
            \label{tmelab:arri:2}
            \text{and}~ \intermed {  {\intermed { \Gamma_C } }  } ; \intermed {  {\intermed \Gamma}  }  \vdash_{ty}  \intermed {  {\intermed \sigma}_{{\mathrm{1}}}  } \itot{ \rightsquigarrow  \target {  {\target \sigma }_{{\mathrm{1}}}  } } 
          \end{gather}
          In addition, from Lemma~\ref{lem:tyelab} applied on Equation~\pref{tmelab:arri:1} and the
          second premise of rule \textsc{iTm-arrI}, we obtain \( \target {  {\target \Gamma }  }  \vdash_{ty}  \target {  {\target \sigma }_{{\mathrm{1}}}  } \).

          By instantiating the premises of rule \textsc{tTm-Abs} with the above result and with
          Equation \pref{tmelab:arri:IH:wftm}, we get
          $ \target {  {\target \Gamma }  }  \vdash_{tm}  \target {  \lambda  {\target x}  \ottsym{:}  {\target \sigma }_{{\mathrm{1}}}  \ottsym{.}  {\target e }  } : \target {  {\target \sigma }_{{\mathrm{1}}}  \rightarrow  {\target \sigma }_{{\mathrm{2}}}  } $, where $ \intermed {  {\intermed { \Gamma_C } }  } ; \intermed {  {\intermed \Gamma}  }  \vdash_{ty}  \intermed {  {\intermed \sigma}_{{\mathrm{2}}}  } \itot{ \rightsquigarrow  \target {  {\target \sigma }_{{\mathrm{2}}}  } } $. The
          latter holds from Equation \pref{tmelab:arri:IH:ty}, where ${\target x}$ has been removed
          from the context (it does not affect the well-formedness of type $\intermed{{\intermed \sigma}_{{\mathrm{2}}}}$). This, in
          combination with Equation \pref{tmelab:arri:2} in rule \textsc{iTy-Arr}, gives
          $ \intermed {  {\intermed { \Gamma_C } }  } ; \intermed {  {\intermed \Gamma}  }  \vdash_{ty}  \intermed {  {\intermed \sigma}_{{\mathrm{1}}}  \rightarrow  {\intermed \sigma}_{{\mathrm{2}}}  } \itot{ \rightsquigarrow  \target {  {\target \sigma }_{{\mathrm{2}}}  \rightarrow  {\target \sigma }_{{\mathrm{2}}}  } } $.\\
          }
\proofStep{Case \textsc{iTm-arrE}}
          {\ottdruleiTmXXarrE{}}
          {
            The induction hypothesis from the first premise of rule \textsc{iTm-arrE} is
            \begin{gather}
              \text{There are unique}~{\target \Gamma }~\text{and}~{\target \sigma }~\text{such}~\nonumber\\
              \label{tmelab:arre:IH1:GT}
              \text{that}~ \intermed {  {\intermed { \Gamma_C } }  } ; \intermed {  {\intermed \Gamma}  }  \rightsquigarrow  \target {  {\target \Gamma }  } \\
              \label{tmelab:arre:IH1:ty}
              \text{and}~ \intermed {  {\intermed { \Gamma_C } }  } ; \intermed {  {\intermed \Gamma}  }  \vdash_{ty}  \intermed {  {\intermed \sigma}_{{\mathrm{1}}}  \rightarrow  {\intermed \sigma}_{{\mathrm{2}}}  } \itot{ \rightsquigarrow  \target {  {\target \sigma }  } } \\
              \label{tmelab:arre:IH1:wftm}
             \text{and}~ \target {  {\target \Gamma }  }  \vdash_{tm}  \target {  {\target e }_{{\mathrm{1}}}  } : \target {  {\target \sigma }  } 
            \end{gather}
            By inversion on Equation~\pref{tmelab:arre:IH1:ty}, ${\target \sigma }$ can only be of the form
            $\target{{\target \sigma }_{{\mathrm{1}}}  \rightarrow  {\target \sigma }_{{\mathrm{2}}}}$ for the $\target{{\target \sigma }_{{\mathrm{1}}}}$ and $\target{{\target \sigma }_{{\mathrm{2}}}}$, uniquely determined by equations
            \begin{gather}
              \label{tmelab:arre:1}
               \intermed {  {\intermed { \Gamma_C } }  } ; \intermed {  {\intermed \Gamma}  }  \vdash_{ty}  \intermed {  {\intermed \sigma}_{{\mathrm{1}}}  } \itot{ \rightsquigarrow  \target {  {\target \sigma }_{{\mathrm{1}}}  } } \\
              \text{and}~ \intermed {  {\intermed { \Gamma_C } }  } ; \intermed {  {\intermed \Gamma}  }  \vdash_{ty}  \intermed {  {\intermed \sigma}_{{\mathrm{2}}}  } \itot{ \rightsquigarrow  \target {  {\target \sigma }_{{\mathrm{2}}}  } } \nonumber
            \end{gather}.
            The induction hypothesis from the second premise of rule \textsc{iTm-arrE} is
            \begin{gather}
              \text{There are unique}~\target{{\target \Gamma }'}~\text{and}~\target{{\target \sigma }'_{{\mathrm{1}}}}~\text{such}~\nonumber\\
              \label{tmelab:arre:IH2:GT}
              \text{that}~ \intermed {  {\intermed { \Gamma_C } }  } ; \intermed {  {\intermed \Gamma}  }  \rightsquigarrow  \target {  {\target \Gamma }'  } \\
              \label{tmelab:arre:IH2:ty}
              \text{and}~ \intermed {  {\intermed { \Gamma_C } }  } ; \intermed {  {\intermed \Gamma}  }  \vdash_{ty}  \intermed {  {\intermed \sigma}_{{\mathrm{1}}}  } \itot{ \rightsquigarrow  \target {  {\target \sigma }'_{{\mathrm{1}}}  } } \\
              \label{tmelab:arre:IH2:wftm}
              \text{and}~ \target {  {\target \Gamma }  }  \vdash_{tm}  \target {  {\target e }_{{\mathrm{2}}}  } : \target {  {\target \sigma }'_{{\mathrm{1}}}  } 
            \end{gather}
            By uniqueness (Lemma~\ref{lem:det:GT}) on Equations \pref{tmelab:arre:IH1:GT} and \pref{tmelab:arre:IH2:GT},
            it must hold that $ \target {  {\target \Gamma }'  } = \target {  {\target \Gamma }  } $. Also, uniqueness (Lemma~\ref{lem:det:ty}) on Equations
            \pref{tmelab:arre:1} and \pref{tmelab:arre:IH2:ty}, gives $ \target {  {\target \sigma }'_{{\mathrm{1}}}  } = \target {  {\target \sigma }_{{\mathrm{1}}}  } $.

            Combining Equations \pref{tmelab:arre:IH1:wftm} and \pref{tmelab:arre:IH2:wftm},
            (rewritten with the equalities holding so far) in rule \textsc{tTm-App}, we get
            $ \target {  {\target \Gamma }  }  \vdash_{tm}  \target {  {\target e }_{{\mathrm{1}}} \, {\target e }_{{\mathrm{2}}}  } : \target {  {\target \sigma }_{{\mathrm{2}}}  } $.\\
          }
\proofStep{Case \textsc{iTm-constrI}}
          {\ottdruleiTmXXconstrI{}}
          {
          The induction hypothesis from the first premise of rule \textsc{iTm-constrI} is
          \begin{gather}
            \text{There are unique}~\target{{\target \Gamma }'}~\text{and}~{\target \sigma }~\text{such}~\nonumber\\
            \text{that}~ \intermed {  {\intermed { \Gamma_C } }  } ; \intermed {  {\intermed \Gamma}  \ottsym{,}  {\intermed \delta }  \ottsym{:}  {\intermed Q}  }  \rightsquigarrow  \target {  {\target \Gamma }'  } \nonumber\\
            \label{tmelab:constrI:IH:ty}
            \text{and}~ \intermed {  {\intermed { \Gamma_C } }  } ; \intermed {  {\intermed \Gamma}  \ottsym{,}  {\intermed \delta }  \ottsym{:}  {\intermed Q}  }  \vdash_{ty}  \intermed {  {\intermed \sigma}  } \itot{ \rightsquigarrow  \target {  {\target \sigma }  } } \\
            \label{tmelab:constrI:IH:wftm}
            \text{and}~ \target {  {\target \Gamma }'  }  \vdash_{tm}  \target {  {\target e }  } : \target {  {\target \sigma }  } 
          \end{gather}
          Similarly to the \textsc{iTm-Let} case, $\target{{\target \Gamma }'}$ is of the form $\target{{\target \Gamma }  \ottsym{,}  {\target \delta }  \ottsym{:}  {\target \sigma }_{\ottmv{q}}}$,
          where
          \begin{gather}
            \label{tmelab:constrI:1}
             \intermed {  {\intermed { \Gamma_C } }  } ; \intermed {  {\intermed \Gamma}  }  \rightsquigarrow  \target {  {\target \Gamma }  } \\
            \label{tmelab:constrI:2}
            \text{and}~ \intermed {  {\intermed { \Gamma_C } }  } ; \intermed {  {\intermed \Gamma}  }  \vdash_{\mathit {Q} }  \intermed {  {\intermed Q}  }\, \itot{ \rightsquigarrow  \target {  {\target \sigma }_{\ottmv{q}}  } } 
          \end{gather}
          In addition, from Lemma~\ref{lem:qelab} applied on Equation~\pref{tmelab:constrI:1} and the
          second premise of rule \textsc{iTm-constrI}, we obtain \( \target {  {\target \Gamma }  }  \vdash_{ty}  \target {  {\target \sigma }_{{\mathrm{1}}}  } \).

          By instantiating the premises of rule \textsc{tTm-Abs} with the above result and with
          Equation \pref{tmelab:constrI:IH:wftm}, we get
          $ \target {  {\target \Gamma }  }  \vdash_{tm}  \target {  \lambda  {\target x}  \ottsym{:}  {\target \sigma }_{\ottmv{q}}  \ottsym{.}  {\target e }  } : \target {  {\target \sigma }_{\ottmv{q}}  \rightarrow  {\target \sigma }  } $, where $ \intermed {  {\intermed { \Gamma_C } }  } ; \intermed {  {\intermed \Gamma}  }  \vdash_{ty}  \intermed {  {\intermed \sigma}  } \itot{ \rightsquigarrow  \target {  {\target \sigma }  } } $. The
          latter holds from Equation \pref{tmelab:constrI:IH:ty}, where ${\target x}$ has been
          removed from the context (it does not affect the well-formedness of type $\intermed{{\intermed \sigma}_{{\mathrm{2}}}}$).
          This, in combination with Equation \pref{tmelab:constrI:2} in rule \textsc{iTy-Qual},
          gives $ \intermed {  {\intermed { \Gamma_C } }  } ; \intermed {  {\intermed \Gamma}  }  \vdash_{ty}  \intermed {   {\intermed Q}  \Rightarrow  {\intermed \sigma}   } \itot{ \rightsquigarrow  \target {  {\target \sigma }_{\ottmv{q}}  \rightarrow  {\target \sigma }  } } $.\\
          }
\proofStep{Case \textsc{iTm-constrE}}
          {\ottdruleiTmXXconstrE{}}
          {
            The induction hypothesis from the first premise of rule \textsc{iTm-constrE} is
            \begin{gather}
              \text{There are unique}~{\target \Gamma }~\text{and}~\target{{\target \sigma }'}~\text{such}~\nonumber\\
              \label{tmelab:constrE:IH1:GT}
              \text{that}~ \intermed {  {\intermed { \Gamma_C } }  } ; \intermed {  {\intermed \Gamma}  }  \rightsquigarrow  \target {  {\target \Gamma }  } \\
              \label{tmelab:constrE:IH1:ty}
              \text{and}~ \intermed {  {\intermed { \Gamma_C } }  } ; \intermed {  {\intermed \Gamma}  }  \vdash_{ty}  \intermed {   {\intermed Q}  \Rightarrow  {\intermed \sigma}   } \itot{ \rightsquigarrow  \target {  {\target \sigma }'  } }  \\
              \label{tmelab:constrE:IH1:wftm}
             \text{and}~ \target {  {\target \Gamma }  }  \vdash_{tm}  \target {  {\target e }_{{\mathrm{1}}}  } : \target {  {\target \sigma }  } 
            \end{gather}
            By inversion on Equation~\pref{tmelab:constrE:IH1:ty}, $\target{{\target \sigma }'}$ can only be of the form
            $\target{{\target \sigma }_{\ottmv{q}}  \rightarrow  {\target \sigma }}$ for the $\target{{\target \sigma }_{\ottmv{q}}}$ and ${\target \sigma }$, uniquely determined by equations
            \begin{gather}
              \label{tmelab:constrE:1}
               \intermed {  {\intermed { \Gamma_C } }  } ; \intermed {  {\intermed \Gamma}  }  \vdash_{\mathit {Q} }  \intermed {  {\intermed Q}  }\, \itot{ \rightsquigarrow  \target {  {\target \sigma }_{\ottmv{q}}  } } \\
              \text{and}~ \intermed {  {\intermed { \Gamma_C } }  } ; \intermed {  {\intermed \Gamma}  }  \vdash_{ty}  \intermed {  {\intermed \sigma}  } \itot{ \rightsquigarrow  \target {  {\target \sigma }  } } \nonumber
            \end{gather}
            Theorem~\ref{thmA:delab}, applied to the second premise of rule
            \textsc{iTm-constrE}, is
            \begin{gather}
              \text{There are unique}~\target{{\target \Gamma }'}~\text{and}~\target{{\target \sigma }_{{\mathrm{0}}}}~\text{such}~\nonumber\\
              \label{tmelab:constrE:IH2:GT}
              \text{that}~ \intermed {  {\intermed { \Gamma_C } }  } ; \intermed {  {\intermed \Gamma}  }  \rightsquigarrow  \target {  {\target \Gamma }'  } \\
              \label{tmelab:constrE:IH2:ty}
              \text{and}~ \intermed {  {\intermed { \Gamma_C } }  } ; \intermed {  {\intermed \Gamma}  }  \vdash_{\mathit {Q} }  \intermed {  {\intermed Q}  }\, \itot{ \rightsquigarrow  \target {  {\target \sigma }_{{\mathrm{0}}}  } } \\
              \label{tmelab:constrE:IH2:wftm}
              \text{and}~ \target {  {\target \Gamma }'  }  \vdash_{tm}  \target {  {\target e }_{{\mathrm{2}}}  } : \target {  {\target \sigma }_{{\mathrm{0}}}  } 
            \end{gather}
            By uniqueness (Lemma~\ref{lem:det:GT}) on Equations \pref{tmelab:constrE:IH1:GT} and
            \pref{tmelab:constrE:IH2:GT},
            it must hold that $ \target {  {\target \Gamma }'  } = \target {  {\target \Gamma }  } $. Also, uniqueness (Lemma~\ref{lem:det:Q}) on Equations
            \pref{tmelab:constrE:1} and \pref{tmelab:constrE:IH2:ty}, gives
            $ \target {  {\target \sigma }_{{\mathrm{0}}}  } = \target {  {\target \sigma }_{\ottmv{q}}  } $.

            Combining Equations \pref{tmelab:constrE:IH1:wftm} and
            \pref{tmelab:constrE:IH2:wftm}, (rewritten with the above-mentioned equalities) in
            rule \textsc{tTm-App}, we get $ \target {  {\target \Gamma }  }  \vdash_{tm}  \target {  {\target e }_{{\mathrm{1}}} \, {\target e }_{{\mathrm{2}}}  } : \target {  {\target \sigma }  } $.\\
          }
\proofStep{Case \textsc{iTm-forallI}}
          {\ottdruleiTmXXforallI{}}
          {The induction hypothesis for the premise of rule \textsc{iTm-forallI} is the
            following.
            \begin{gather}
              \text{There are unique}~\target{{\target \Gamma }'}~\text{and}~{\target \sigma }~\text{such}~\nonumber\\
              \label{tmelab:foralli:IH:GT}
              \text{that}~ \intermed {  {\intermed { \Gamma_C } }  } ; \intermed {  {\intermed \Gamma}  \ottsym{,}  {\intermed a }  }  \rightsquigarrow  \target {  {\target \Gamma }'  } \\
              \label{tmelab:foralli:IH:ty}
              \text{and}~ \intermed {  {\intermed { \Gamma_C } }  } ; \intermed {  {\intermed \Gamma}  \ottsym{,}  {\intermed a }  }  \vdash_{ty}  \intermed {  {\intermed \sigma}  } \itot{ \rightsquigarrow  \target {  {\target \sigma }  } } \\
              \label{tmelab:foralli:IH:wftm}
              \text{and}~ \target {  {\target \Gamma }'  }  \vdash_{tm}  \target {  {\target e }  } : \target {  {\target \sigma }  } 
            \end{gather}
            By inversion on Equation~\pref{tmelab:foralli:IH:GT}, $\target{{\target \Gamma }'}$ can only be of the form
            $\target{{\target \Gamma }  \ottsym{,}  {\target a }}$, where ${\target \Gamma }$ uniquely determined by
            \begin{equation*}
               \intermed {  {\intermed { \Gamma_C } }  } ; \intermed {  {\intermed \Gamma}  }  \rightsquigarrow  \target {  {\target \Gamma }  } 
            \end{equation*}
            Then, Equation \pref{tmelab:foralli:IH:wftm} becomes
            \begin{equation*}
               \target {  {\target \Gamma }  \ottsym{,}  {\target a }  }  \vdash_{tm}  \target {  {\target e }  } : \target {  {\target \sigma }  } 
            \end{equation*}
            Using this to instantiate rule \textsc{tTm-Tabs}, we get
            \begin{equation*}
               \target {  {\target \Gamma }  }  \vdash_{tm}  \target {  \Lambda  {\target a }  \ottsym{.}  {\target e }  } : \target {  \forall  {\target a }  \ottsym{.}  {\target \sigma }  } 
            \end{equation*}
            where, by rule \textsc{iTy-scheme} on Equation~\pref{tmelab:foralli:IH:ty}, it follows that
            $ \intermed {  {\intermed { \Gamma_C } }  } ; \intermed {  {\intermed \Gamma}  }  \vdash_{ty}  \intermed {  \forall  {\intermed a }  \ottsym{.}  {\intermed \sigma}  } \itot{ \rightsquigarrow  \target {  \forall  {\target a }  \ottsym{.}  {\target \sigma }  } } $.\\
          }
\proofStep{Case \textsc{iTm-forallE}}
          {\ottdruleiTmXXforallE{}}
          {The induction hypothesis from the first premise of rule \textsc{iTm-forallE} is as
            follows.
            \begin{gather}
              \text{There are unique}~{\target \Gamma }~\text{and}~\target{{\target \sigma }_{{\mathrm{0}}}}~\text{such}~\nonumber\\
              \label{tmelab:forallE:IH2:GT}
              \text{that}~ \intermed {  {\intermed { \Gamma_C } }  } ; \intermed {  {\intermed \Gamma}  }  \rightsquigarrow  \target {  {\target \Gamma }  } \\
              \label{tmelab:forallE:IH2:ty}
              \text{and}~ \intermed {  {\intermed { \Gamma_C } }  } ; \intermed {  {\intermed \Gamma}  }  \vdash_{ty}  \intermed {  \forall  {\intermed a }  \ottsym{.}  {\intermed \sigma}'  } \itot{ \rightsquigarrow  \target {  {\target \sigma }_{{\mathrm{0}}}  } } \\
              \label{tmelab:forallE:IH2:wftm}
              \text{and}~ \target {  {\target \Gamma }  }  \vdash_{tm}  \target {  {\target e }  } : \target {  {\target \sigma }_{{\mathrm{0}}}  } 
            \end{gather}
            By inversion on Equation~\pref{tmelab:forallE:IH2:ty}, $\target{{\target \sigma }_{{\mathrm{0}}}}$ can only be of the form
            $\target{\forall  {\target a }  \ottsym{.}  {\target \sigma }'}$, where $\target{{\target \sigma }'}$ is uniquely determined by equation
            \begin{equation}
              \label{tmelab:forallE:1}
               \intermed {  {\intermed { \Gamma_C } }  } ; \intermed {  {\intermed \Gamma}  \ottsym{,}  {\intermed a }  }  \vdash_{ty}  \intermed {  {\intermed \sigma}'  } \itot{ \rightsquigarrow  \target {  {\target \sigma }'  } } 
            \end{equation}
            Lemma~\ref{lemma:tvar_tsubst}, applied on the second premise of rule
            \textsc{iTm-forallE} and on Equation~\pref{tmelab:forallE:1}, results in
            $ \intermed {  {\intermed { \Gamma_C } }  } ; \intermed {  {\intermed \Gamma}  }  \vdash_{ty}  \intermed {  [  {\intermed \sigma}  /  {\intermed a }  ]  {\intermed \sigma}'  } \itot{ \rightsquigarrow  \target {  [  {\target \sigma }  /  {\target a }  ]  {\target \sigma }'  } } $.

            Also, Lemma~\ref{lem:tyelab} applied on the second premise of rule
            \textsc{iTm-forallE} and on Equation~\pref{tmelab:forallE:IH2:GT} results in
            $ \target {  {\target \Gamma }  }  \vdash_{ty}  \target {  {\target \sigma }  } $.
            We use the latter, together with Equation \pref{tmelab:forallE:IH2:wftm}, to
            instantiate rule \textsc{tTm-Tapp}, which concludes
            $ \target {  {\target \Gamma }  }  \vdash_{tm}  \target {  {\target e }  } : \target {  [  {\target \sigma }  /  {\target a }  ]  {\target \sigma }'  } $.
          }
\qed

\begin{theoremB}[Dictionary Elaboration Soundness]
\label{thmA:delab}
  \begin{flalign}
  &\text{If}~ \intermed {  {\intermed \Sigma}  } ; \intermed {  {\intermed { \Gamma_C } }  } ; \intermed {  {\intermed \Gamma}  }  \vdash_{d}  \intermed {  {\intermed \delta }  } : \intermed {  {\intermed Q}  } \itot{\,  \rightsquigarrow  \target {  {\target e }  } } \label{delab:hyp} &\\
  &\text{then, there are unique}~{\target \Gamma }~\text{and}~{\target \sigma }~\text{such that}\nonumber &\\
  & \intermed {  {\intermed { \Gamma_C } }  } ; \intermed {  {\intermed \Gamma}  }  \rightsquigarrow  \target {  {\target \Gamma }  } \label{delab:res1} &\\
  &\text{and}~ \intermed {  {\intermed { \Gamma_C } }  } ; \intermed {  {\intermed \Gamma}  }  \vdash_{\mathit {Q} }  \intermed {  {\intermed Q}  }\, \itot{ \rightsquigarrow  \target {  {\target \sigma }  } } \label{delab:res2} &\\
  &\text{and}~ \target {  {\target \Gamma }  }  \vdash_{tm}  \target {  {\target e }  } : \target {  {\target \sigma }  } \label{delab:res3}
  \end{flalign}
\end{theoremB}

\proof
This theorem is proved mutually with Theorem~\ref{thmA:tmelab}. The proof follows
structural induction on the first hypothesis.\\
\proofStep{Case \textsc{D-var}}
          {\ottdruleDXXvar{}}
          {By Lemma~\ref{lem:elab:dvar}, there exist ${\target \Gamma }$ and ${\target \sigma }$ such that
           $ \intermed {  {\intermed { \Gamma_C } }  } ; \intermed {  {\intermed \Gamma}  }  \rightsquigarrow  \target {  {\target \Gamma }  } $ and
            $ \intermed {  {\intermed { \Gamma_C } }  } ; \intermed {  {\intermed \Gamma}  }  \vdash_{\mathit {Q} }  \intermed {  {\intermed {\mathit{TC} } } \, {\intermed \sigma}  }\, \itot{ \rightsquigarrow  \target {  {\target \sigma }  } } $ and $ ( \target {  {\target \delta }  } : \target {  {\target \sigma }  } )  \in  \target {  {\target \Gamma }  } $. This satisfies
            Equations \pref{delab:res1} and \pref{delab:res2} of the theorem.

            By instantiating rule \textsc{tTm-Var} on $ ( \target {  {\target \delta }  } : \target {  {\target \sigma }  } )  \in  \target {  {\target \Gamma }  } $, we get
            $ \target {  {\target \Gamma }  }  \vdash_{tm}  \target {  {\target \delta }  } : \target {  {\target \sigma }  } $, which satisfies Equation \pref{delab:res3} of the
            theorem.\\
          }
\proofStep{Case \textsc{D-con}}
          {\ottdruleDXXcon{}}
          { We need to show that there exist $\target{{\target \Gamma }_{{\mathrm{0}}}}$ and ${\target \sigma }$ such that
            $ \intermed {  {\intermed { \Gamma_C } }  } ; \intermed {  {\intermed \Gamma}  }  \rightsquigarrow  \target {  {\target \Gamma }_{{\mathrm{0}}}  } $ and
            \begin{gather}
              \label{delab:dcon:goal:ty}
               \intermed {  {\intermed { \Gamma_C } }  } ; \intermed {  {\intermed \Gamma}  }  \vdash_{\mathit {Q} }  \intermed {  {\intermed {\mathit{TC} } } \, [  \overline {\intermed \sigma}_{\ottmv{j}}  /  {\overline {\intermed a} }_{\ottmv{j}}  ]  {\intermed \sigma}_{\ottmv{q}}  }\, \itot{ \rightsquigarrow  \target {  {\target \sigma }  } } \\
              \label{delab:dcon:goal:te_rec}
               \target {  {\target \Gamma }_{{\mathrm{0}}}  }  \vdash_{tm}  \target {  \ottsym{(}  \Lambda  {\target a }_{\ottmv{j}}  \ottsym{.}  \lambda \, \ottcomp{{\target \delta }_{\ottmv{i}}  \ottsym{:}  {\target \sigma }'_{\ottmv{i}}}{\ottmv{i}} \, \ottsym{.}  \ottsym{\{}  {\target m}  \ottsym{=}  {\target e }  \ottsym{\}}  \ottsym{)} \, {\overline {\target \sigma} }_{\ottmv{j}} \, {\overline {\target e} }_{\ottmv{i}}  } : \target {  {\target \sigma }  } 
            \end{gather}
            From the 2\textsuperscript{nd} premise of rule \textsc{D-con}, applied on Lemma
            \ref{lem:ctx_elab}, there is a unique ${\target \Gamma }$ such that
            \begin{gather}
              \label{delab:dcon:GT}
               \intermed {  {\intermed { \Gamma_C } }  } ; \intermed {  {\intermed \Gamma}  }  \rightsquigarrow  \target {  {\target \Gamma }  } \\
              \nonumber
               \vdash_{ctx}  \target {  {\target \Gamma }  } 
            \end{gather}
            We thus take $ \target {  {\target \Gamma }_{{\mathrm{0}}}  } = \target {  {\target \Gamma }  } $.
            
            Next, we will try to refine type ${\target \sigma }$.
            By Lemma
            \ref{lemma:inSS2C} applied on the first premise of rule \textsc{D-con}, there are
            ${\intermed a }$, $\target{{\target \sigma }_{{\mathrm{0}}}}$ and $\target{{\overline {\target \sigma} }_{\ottmv{i}}}$ such that
            \begin{gather}
              \label{delab:dcon:im}
               ( \intermed {  {\intermed m}  } : \intermed {  {\intermed {\mathit{TC} } } \, {\intermed a }  } : \intermed {  {\intermed \sigma}_{\ottmv{m}}  } )  \in  \intermed {  {\intermed { \Gamma_C } }  } \\
              \label{delab:dcon:iqli}
              \text{and}~\ottcomp{ \intermed {  {\intermed { \Gamma_C } }  } ; \intermed {  \bullet  \ottsym{,}  {\overline {\intermed a} }_{\ottmv{j}}  }  \vdash_{\mathit {Q} }  \intermed {  {\intermed Q}_{\ottmv{i}}  }\, \itot{ \rightsquigarrow  \target {  {\target \sigma }''_{\ottmv{i}}  } } }{\ottmv{i}}\\
              \label{delab:dcon:iq}
              \text{and}~ \intermed {  {\intermed { \Gamma_C } }  } ; \intermed {  \bullet  \ottsym{,}  {\overline {\intermed a} }_{\ottmv{j}}  }  \vdash_{\mathit {Q} }  \intermed {  {\intermed {\mathit{TC} } } \, {\intermed \sigma}_{\ottmv{q}}  }\, \itot{ \rightsquigarrow  \target {  {\target \sigma }_{{\mathrm{0}}}  } } 
            \end{gather}
            Since Equation \pref{delab:dcon:iq} holds, the premises of the only rule that
            applies, namely rule \textsc{iQ-TC}, must hold as well.
            \begin{gather}
              \label{delab:dcon:isq}
               \intermed {  {\intermed { \Gamma_C } }  } ; \intermed {  \bullet  \ottsym{,}  {\overline {\intermed a} }_{\ottmv{j}}  }  \vdash_{ty}  \intermed {  {\intermed \sigma}_{\ottmv{q}}  } \itot{ \rightsquigarrow  \target {  {\target \sigma }_{\ottmv{q}}  } } \\
              \label{delab:dcon:im'}
               \intermed {  {\intermed { \Gamma_C } }  } = \intermed {  {\intermed { \Gamma_C } }_{{\mathrm{1}}}  }, \intermed {  {\intermed m}'  \ottsym{:}  {\intermed {\mathit{TC} } } \, {\intermed a }'  \ottsym{:}  {\intermed \sigma}'_{\ottmv{m}}  \ottsym{,}  {\intermed { \Gamma_C } }_{{\mathrm{2}}}  } \\
              \label{delab:dcon:ism}
               \intermed {  {\intermed { \Gamma_C } }_{{\mathrm{1}}}  } ; \intermed {  \bullet  \ottsym{,}  {\intermed a }  }  \vdash_{ty}  \intermed {  {\intermed \sigma}'_{\ottmv{m}}  } \itot{ \rightsquigarrow  \target {  {\target \sigma }_{\ottmv{m}}  } } 
            \end{gather}
            for some $\intermed{{\intermed m}'}$, $\intermed{{\intermed a }'}$, $\intermed{{\intermed \sigma}'_{\ottmv{m}}}$, $\target{{\target \sigma }_{\ottmv{q}}}$ and $\target{{\target \sigma }_{\ottmv{m}}}$, where
            $ \target {  {\target \sigma }_{{\mathrm{0}}}  } = \target {  [  {\target \sigma }_{\ottmv{q}}  /  {\target a }'  ]  \ottsym{\{}  {\target m}'  \ottsym{:}  {\target \sigma }_{\ottmv{m}}  \ottsym{\}}  } $.
            By uniqueness on Equations \pref{delab:dcon:im} and \pref{delab:dcon:im'}, we get
            $ \intermed {  {\intermed m}'  } = \intermed {  {\intermed m}  } $, $ \intermed {  {\intermed a }  } = \intermed {  {\intermed a }'  } $,  $ \intermed {  {\intermed \sigma}'_{\ottmv{m}}  } = \intermed {  {\intermed \sigma}_{\ottmv{m}}  } $. Then,
            $ \target {  {\target \sigma }_{{\mathrm{0}}}  } = \target {  [  {\target \sigma }_{\ottmv{q}}  /  {\target a }  ]  \ottsym{\{}  {\target m}  \ottsym{:}  {\target \sigma }_{\ottmv{m}}  \ottsym{\}}  } = \target {  \ottsym{\{}  {\target m}  \ottsym{:}  [  {\target \sigma }_{\ottmv{q}}  /  {\target a }  ]  {\target \sigma }_{\ottmv{m}}  \ottsym{\}}  } $.
            By Lemma~\ref{lemma:tvar_tsubst_iq} applied on Equation~\pref{delab:dcon:isq} and
            the 4\textsuperscript{th} set of premises of rule \textsc{D-con}, Goal
            \pref{delab:dcon:goal:ty} is satisfied with
            $ \target {  {\target \sigma }  } = \target {  \ottsym{\{}  {\target m}  \ottsym{:}  [ \, \ottcomp{{\target \sigma }_{\ottmv{j}}  /  {\target a }_{\ottmv{j}}}{\ottmv{j}} \, ]  [  {\target \sigma }_{\ottmv{q}}  /  {\target a }  ]  {\target \sigma }_{\ottmv{m}}  \ottsym{\}}  } $.

            Then, Goal \pref{delab:dcon:goal:te_rec} becomes
            \begin{equation}
              \label{delab:dcon:goal:te_ref}
               \target {  {\target \Gamma }  }  \vdash_{tm}  \target {  \ottsym{(}  \Lambda  {\target a }_{\ottmv{j}}  \ottsym{.}  \lambda \, \ottcomp{{\target \delta }_{\ottmv{i}}  \ottsym{:}  {\target \sigma }'_{\ottmv{i}}}{\ottmv{i}} \, \ottsym{.}  \ottsym{\{}  {\target m}  \ottsym{=}  {\target e }  \ottsym{\}}  \ottsym{)} \, {\overline {\target \sigma} }_{\ottmv{j}} \, {\overline {\target e} }_{\ottmv{i}}  } : \target {  \ottsym{\{}  {\target m}  \ottsym{:}  [ \, \ottcomp{{\target \sigma }_{\ottmv{j}}  /  {\target a }_{\ottmv{j}}}{\ottmv{j}} \, ]  [  {\target \sigma }_{\ottmv{q}}  /  {\target a }  ]  {\target \sigma }_{\ottmv{m}}  \ottsym{\}}  } 
            \end{equation}
            For that, it suffices to show that
            \begin{gather}
              \label{delab:dcon:goal:te}
               \target {  {\target \Gamma }  \ottsym{,}  {\overline {\target a} }_{\ottmv{j}}  \ottsym{,} \, \ottcomp{{\target \delta }_{\ottmv{i}}  \ottsym{:}  {\target \sigma }'_{\ottmv{i}}}{\ottmv{i}}  }  \vdash_{tm}  \target {  {\target e }  } : \target {  [  {\target \sigma }_{\ottmv{q}}  /  {\target a }  ]  {\target \sigma }_{\ottmv{m}}  } \\
              \label{delab:dcon:goal:ttjs}
              \ottcomp{ \target {  {\target \Gamma }  }  \vdash_{ty}  \target {  {\target \sigma }_{\ottmv{j}}  } }{\ottmv{j}}\\
              \label{delab:dcon:goal:teis}
              \text{and}~\ottcomp{ \target {  {\target \Gamma }  }  \vdash_{tm}  \target {  {\target e }_{\ottmv{i}}  } : \target {  [ \, \ottcomp{{\target \sigma }_{\ottmv{j}}  /  {\target a }_{\ottmv{j}}}{\ottmv{j}} \, ]  {\target \sigma }'_{\ottmv{i}}  } }{\ottmv{i}}
            \end{gather}
            because, then we can use rules \textsc{tTm-Rec}, \textsc{tTm-Tapp} and
            \textsc{tTm-App} to reach Equation~\pref{delab:dcon:goal:te_ref}.
            
            \paragraph{For Goal~\pref{delab:dcon:goal:ttjs}}
            Passing the 4\textsuperscript{th} set of premises of rule \textsc{D-con} in Lemma
            \ref{lem:qelab}, we get $\ottcomp{ \target {  {\target \Gamma }  }  \vdash_{ty}  \target {  {\target \sigma }_{\ottmv{j}}  } }{\ottmv{j}}$, satisfying
            Equation~\pref{delab:dcon:goal:ttjs}. 

            \paragraph{For Goal~\pref{delab:dcon:goal:te}}
            Theorem~\ref{thmA:tmelab}, applied to the 5\textsuperscript{th} premise of rule
            \textsc{D-con}, is:
            \begin{gather}
              \text{There are}~\target{{\target \Gamma }'}~\text{and}~\target{{\target \sigma }_{{\mathrm{0}}}}~\text{such that}\nonumber\\
               \intermed {  {\intermed { \Gamma_C } }  } ; \intermed {  \bullet  \ottsym{,}  {\overline {\intermed a} }_{\ottmv{j}}  \ottsym{,}  {\overline {\intermed \delta} }_{\ottmv{i}}  \ottsym{:}  {\overline {\intermed Q} }_{\ottmv{i}}  }  \rightsquigarrow  \target {  {\target \Gamma }'  } \nonumber\\
              \text{and}~ \intermed {  {\intermed { \Gamma_C } }  } ; \intermed {  \bullet  \ottsym{,}  {\overline {\intermed a} }_{\ottmv{j}}  \ottsym{,}  {\overline {\intermed \delta} }_{\ottmv{i}}  \ottsym{:}  {\overline {\intermed Q} }_{\ottmv{i}}  }  \vdash_{ty}  \intermed {  [  {\intermed \sigma}_{\ottmv{q}}  /  {\intermed a }  ]  {\intermed \sigma}_{\ottmv{m}}  } \itot{ \rightsquigarrow  \target {  {\target \sigma }_{{\mathrm{0}}}  } } \label{delab:dcon:IH5:tt}\\
              \text{and}~ \target {  {\target \Gamma }'  }  \vdash_{tm}  \target {  {\target e }  } : \target {  {\target \sigma }_{{\mathrm{0}}}  } .\label{delab:dcon:IH1:wftm}
            \end{gather}
            It is easy to verify that $ \target {  {\target \Gamma }'  } = \target {  \target {\bullet}  \ottsym{,}  {\overline {\target a} }_{\ottmv{j}}  \ottsym{,} \, \ottcomp{{\target \delta }_{\ottmv{i}}  \ottsym{:}  {\target \sigma }'_{\ottmv{i}}}{\ottmv{i}}  } $.
            From Lemma~\ref{lemma:ty_weakening} on Equation \pref{delab:dcon:ism}, we get
            the weakened equation
            \begin{equation*}
               \intermed {  {\intermed { \Gamma_C } }  } ; \intermed {  \bullet  \ottsym{,}  {\intermed a }  }  \vdash_{ty}  \intermed {  {\intermed \sigma}_{\ottmv{m}}  } \itot{ \rightsquigarrow  \target {  {\target \sigma }_{\ottmv{m}}  } } 
            \end{equation*}
            where ${\intermed { \Gamma_C } }$ is specified in Equation~\pref{delab:dcon:im'}.
            Using this result together with Equation \pref{delab:dcon:isq} in Lemma
            \ref{lemma:tvar_tsubst}, we get
            $ \intermed {  {\intermed { \Gamma_C } }  } ; \intermed {  \bullet  \ottsym{,}  {\overline {\intermed a} }_{\ottmv{j}}  }  \vdash_{ty}  \intermed {  [  {\intermed \sigma}_{\ottmv{q}}  /  {\intermed a }  ]  {\intermed \sigma}_{\ottmv{m}}  } \itot{ \rightsquigarrow  \target {  [  {\target \sigma }_{\ottmv{q}}  /  {\target a }  ]  {\target \sigma }_{\ottmv{m}}  } } $, which we weaken
            as 
            \begin{equation*}
               \intermed {  {\intermed { \Gamma_C } }  } ; \intermed {  \bullet  \ottsym{,}  {\overline {\intermed a} }_{\ottmv{j}}  \ottsym{,}  {\overline {\intermed \delta} }_{\ottmv{i}}  \ottsym{:}  {\overline {\intermed Q} }_{\ottmv{i}}  }  \vdash_{ty}  \intermed {  [  {\intermed \sigma}_{\ottmv{q}}  /  {\intermed a }  ]  {\intermed \sigma}_{\ottmv{m}}  } \itot{ \rightsquigarrow  \target {  [  {\target \sigma }_{\ottmv{q}}  /  {\target a }  ]  {\target \sigma }_{\ottmv{m}}  } } 
            \end{equation*}
            Then, by uniqueness on the latter and on Equation \pref{delab:dcon:IH5:tt}, we have
            $ \target {  {\target \sigma }_{{\mathrm{0}}}  } = \target {  [  {\target \sigma }_{\ottmv{q}}  /  {\target a }  ]  {\target \sigma }_{\ottmv{m}}  } $, and Equation \pref{delab:dcon:IH1:wftm}
            becomes
            \begin{equation*}
               \target {  \target {\bullet}  \ottsym{,}  {\overline {\target a} }_{\ottmv{j}}  \ottsym{,} \, \ottcomp{{\target \delta }_{\ottmv{i}}  \ottsym{:}  {\target \sigma }'_{\ottmv{i}}}{\ottmv{i}}  }  \vdash_{tm}  \target {  {\target e }  } : \target {  [  {\target \sigma }_{\ottmv{q}}  /  {\target a }  ]  {\target \sigma }_{\ottmv{m}}  } 
            \end{equation*}
            Prefixing the typing environment of the above with ${\target \Gamma }$, Goal
            \pref{delab:dcon:goal:te} is satisfied.

            \paragraph{For Goal~\pref{delab:dcon:goal:teis}} The 6\textsuperscript{th} set of
            premises of rule \textsc{D-con} induces, for each $i$, the following induction
            hypothesis.
            \begin{gather}
              \text{There are}~\target{{\target \Gamma }_{\ottmv{i}}}~\text{and}~\target{{\target \sigma }''_{\ottmv{i}}}~\text{such that}\nonumber\\
              \label{delab:dcon:IHi:GT}
               \intermed {  {\intermed { \Gamma_C } }  } ; \intermed {  {\intermed \Gamma}  }  \rightsquigarrow  \target {  {\target \Gamma }_{\ottmv{i}}  } \\
              \label{delab:dcon:IHi:q}
               \intermed {  {\intermed { \Gamma_C } }  } ; \intermed {  {\intermed \Gamma}  }  \vdash_{\mathit {Q} }  \intermed {  [  \overline {\intermed \sigma}_{\ottmv{j}}  /  {\overline {\intermed a} }_{\ottmv{j}}  ]  {\intermed Q}_{\ottmv{i}}  }\, \itot{ \rightsquigarrow  \target {  {\target \sigma }''_{\ottmv{i}}  } }  \\
              \label{delab:dcon:IHi:wftm}
              \text{and}~ \target {  {\target \Gamma }_{\ottmv{i}}  }  \vdash_{tm}  \target {  {\target e }_{\ottmv{i}}  } : \target {  {\target \sigma }''_{\ottmv{i}}  } 
            \end{gather}
            From Equations \pref{delab:dcon:GT} and \pref{delab:dcon:IHi:GT} in Lemma
            \ref{lem:det:GT}, we have $ \target {  {\target \Gamma }_{\ottmv{i}}  } = \target {  {\target \Gamma }  } $.
            For each equation in the 3\textsuperscript{rd} set of premises of rule
            \textsc{D-con}, we appply Lemma~\ref{lemma:tvar_tsubst_iq} multiple times with the
            4\textsuperscript{th} set of premises of rule \textsc{D-con}, to get
            \begin{equation*}
              \ottcomp{ \intermed {  {\intermed { \Gamma_C } }  } ; \intermed {  {\intermed \Gamma}  }  \vdash_{\mathit {Q} }  \intermed {  [  \overline {\intermed \sigma}_{\ottmv{j}}  /  {\overline {\intermed a} }_{\ottmv{j}}  ]  {\intermed Q}_{\ottmv{i}}  }\, \itot{ \rightsquigarrow  \target {  [ \, \ottcomp{{\target \sigma }_{\ottmv{j}}  /  {\target a }_{\ottmv{j}}}{\ottmv{j}} \, ]  {\target \sigma }'_{\ottmv{i}}  } } }{\ottmv{i}}
            \end{equation*}
            Then $ \target {  {\target \sigma }''_{\ottmv{i}}  } = \target {  [ \, \ottcomp{{\target \sigma }_{\ottmv{j}}  /  {\target a }_{\ottmv{j}}}{\ottmv{j}} \, ]  {\target \sigma }'_{\ottmv{i}}  } $, for all $i$, and Equation
            \pref{delab:dcon:IHi:wftm} satisfies Goal \pref{delab:dcon:goal:teis}.
            }
\qed

\subsection{Determinism}
\begin{theoremB}[Deterministic Dictionary Elaboration]\ \\
  \label{thmA:delab:det}
  If \( \intermed {  {\intermed \Sigma}  } ; \intermed {  {\intermed { \Gamma_C } }  } ; \intermed {  {\intermed \Gamma}  }  \vdash_{d}  \intermed {  {\intermed d}  } : \intermed {  {\intermed Q}  } \itot{\,  \rightsquigarrow  \target {  {\target e }_{{\mathrm{1}}}  } } \)
  and \( \intermed {  {\intermed \Sigma}  } ; \intermed {  {\intermed { \Gamma_C } }  } ; \intermed {  {\intermed \Gamma}  }  \vdash_{d}  \intermed {  {\intermed d}  } : \intermed {  {\intermed Q}  } \itot{\,  \rightsquigarrow  \target {  {\target e }_{{\mathrm{2}}}  } } \),
  then \( \target {  {\target e }_{{\mathrm{1}}}  } = \target {  {\target e }_{{\mathrm{2}}}  } \).
\end{theoremB}
\proof This theorem is proved mutually with Theorem~\ref{thmA:tmelab:det}. The proof follows
structural induction on both hypotheses.\\\\
\proofStep{Case \textsc{D-var}}
          {}
          {The first and second hypotheses of the theorem are:
            \begin{gather*}
               \intermed {  {\intermed \Sigma}  } ; \intermed {  {\intermed { \Gamma_C } }  } ; \intermed {  {\intermed \Gamma}  }  \vdash_{d}  \intermed {  {\intermed \delta }  } : \intermed {  {\intermed Q}  } \itot{\,  \rightsquigarrow  \target {  {\target \delta }_{{\mathrm{1}}}  } } \\
              \text{and}~ \intermed {  {\intermed \Sigma}  } ; \intermed {  {\intermed { \Gamma_C } }  } ; \intermed {  {\intermed \Gamma}  }  \vdash_{d}  \intermed {  {\intermed \delta }  } : \intermed {  {\intermed Q}  } \itot{\,  \rightsquigarrow  \target {  {\target \delta }_{{\mathrm{2}}}  } } 
            \end{gather*}
            From the convention regarding namespace translations, explained
            in Section~\ref{sec:namespace_translation}, it follows directly that
            $ \target {  {\target \delta }_{{\mathrm{1}}}  } = \target {  {\target \delta }_{{\mathrm{2}}}  } = \target {  {\target \delta }  } $.\\
          }
\proofStep{Case \textsc{D-con}}
          {}
          {The first and second hypotheses of the theorem are:
            \begin{gather}
              \label{delab:det:dcon:1}
               \intermed {  {\intermed \Sigma}  } ; \intermed {  {\intermed { \Gamma_C } }  } ; \intermed {  {\intermed \Gamma}  }  \vdash_{d}  \intermed {  {\intermed D } \, \overline {\intermed \sigma}_{\ottmv{j}} \, {\overline {\intermed d} }_{\ottmv{i}}  } : \intermed {  {\intermed {\mathit{TC} } } \, [  \overline {\intermed \sigma}_{\ottmv{j}}  /  {\overline {\intermed a} }_{\ottmv{j}}  ]  {\intermed \sigma}_{\ottmv{q}}  } \itot{\,  \rightsquigarrow  \target {  \ottsym{(}  \Lambda  {\overline {\target a} }_{\ottmv{j}}  \ottsym{.}  \lambda \, \ottcomp{{\target \delta }_{\ottmv{i}}  \ottsym{:}  {\target \sigma }'_{\ottmv{i}}}{\ottmv{i}} \, \ottsym{.}  \ottsym{\{}  {\target m}  \ottsym{=}  {\target e }  \ottsym{\}}  \ottsym{)} \, {\overline {\target \sigma} }_{\ottmv{j}} \, {\overline {\target e} }_{\ottmv{i}}  } } \\
              \label{delab:det:dcon:2}
              \text{and}~
               \intermed {  {\intermed \Sigma}  } ; \intermed {  {\intermed { \Gamma_C } }  } ; \intermed {  {\intermed \Gamma}  }  \vdash_{d}  \intermed {  {\intermed D } \, \overline {\intermed \sigma}_{\ottmv{j}} \, {\overline {\intermed d} }_{\ottmv{i}}  } : \intermed {  {\intermed {\mathit{TC} } } \, [  \overline {\intermed \sigma}_{\ottmv{j}}  /  {\overline {\intermed a} }_{\ottmv{j}}  ]  {\intermed \sigma}_{\ottmv{q}}  } \itot{\,  \rightsquigarrow  \target {  \ottsym{(}  \Lambda  {\overline {\target a} }_{\ottmv{j}}  \ottsym{.}  \lambda \, \ottcomp{{\target \delta }'_{\ottmv{i}}  \ottsym{:}  {\target \sigma }''_{\ottmv{i}}}{\ottmv{i}} \, \ottsym{.}  \ottsym{\{}  {\target m}'  \ottsym{=}  {\target e }'  \ottsym{\}}  \ottsym{)} \, {\overline {\target \sigma} }'_{\ottmv{j}} \, {\overline {\target e} }'_{\ottmv{i}}  } } 
            \end{gather}
            The 5\textsuperscript{th} premise of the two instantiations of rule \textsc{D-con},
            are
            \begin{gather}
              \label{delab:det:dcon:3}
               \intermed {  {\intermed \Sigma}_{{\mathrm{1}}}  } ; \intermed {  {\intermed { \Gamma_C } }  } ; \intermed {  \bullet  \ottsym{,}  {\overline {\intermed a} }_{\ottmv{j}}  \ottsym{,}  {\overline {\intermed \delta} }_{\ottmv{i}}  \ottsym{:}  {\overline {\intermed Q} }_{\ottmv{i}}  }  \vdash_{tm}  \intermed {  {\intermed e}  } : \intermed {  [  {\intermed \sigma}_{\ottmv{q}}  /  {\intermed a }  ]  {\intermed \sigma}_{\ottmv{m}}  } \itot{ \rightsquigarrow  \target {  {\target e }  } } \\
              \label{delab:det:dcon:4}
              \text{and}~
               \intermed {  {\intermed \Sigma}_{{\mathrm{1}}}  } ; \intermed {  {\intermed { \Gamma_C } }  } ; \intermed {  \bullet  \ottsym{,}  {\overline {\intermed a} }_{\ottmv{j}}  \ottsym{,}  {\overline {\intermed \delta} }'_{\ottmv{i}}  \ottsym{:}  {\overline {\intermed Q} }'_{\ottmv{i}}  }  \vdash_{tm}  \intermed {  {\intermed e}  } : \intermed {  [  {\intermed \sigma}_{\ottmv{q}}  /  {\intermed a }  ]  {\intermed \sigma}_{\ottmv{m}}  } \itot{ \rightsquigarrow  \target {  {\target e }'  } } 
            \end{gather}
            and the 1\textsuperscript{st} premise of the two \textsc{D-con} rules are
            \begin{gather*}
               ( \intermed {  {\intermed D }  } : \intermed {  \forall  {\overline {\intermed a} }_{\ottmv{j}}  \ottsym{.}  {\overline {\intermed Q} }_{\ottmv{i}}  \Rightarrow  {\intermed {\mathit{TC} } } \, {\intermed \sigma}_{\ottmv{q}}  } ) . \intermed {  {\intermed m}  }  \mapsto  \intermed {  \Lambda  {\overline {\intermed a} }_{\ottmv{j}}  \ottsym{.}  \lambda  {\overline {\intermed \delta} }_{\ottmv{i}}  \ottsym{:}  {\overline {\intermed Q} }_{\ottmv{i}}  \ottsym{.}  {\intermed e}  }  \in  \intermed {  {\intermed \Sigma}  } \\
               ( \intermed {  {\intermed D }  } : \intermed {  \forall  {\overline {\intermed a} }_{\ottmv{j}}  \ottsym{.}  {\overline {\intermed Q} }'_{\ottmv{i}}  \Rightarrow  {\intermed {\mathit{TC} } } \, {\intermed \sigma}_{\ottmv{q}}  } ) . \intermed {  {\intermed m}'  }  \mapsto  \intermed {  \Lambda  {\overline {\intermed a} }_{\ottmv{j}}  \ottsym{.}  \lambda  {\overline {\intermed \delta} }'_{\ottmv{i}}  \ottsym{:}  {\overline {\intermed Q} }'_{\ottmv{i}}  \ottsym{.}  {\intermed e}  }  \in  \intermed {  {\intermed \Sigma}  } 
            \end{gather*}
            However, a valid method-implementations environment, like ${\intermed \Sigma}$ in this case,
            contains a unique entry for each constructor ${\intermed D }$. From the two above premises
            and this uniqueness property, we have that
            \begin{equation}
              \label{delab:det:dcon:nm_eqs}
            \begin{aligned}
               \intermed {  {\overline {\intermed Q} }'_{\ottmv{i}}  } = \intermed {  {\overline {\intermed Q} }_{\ottmv{i}}  } ~~\text{and}~~ \intermed {  {\overline {\intermed \delta} }'_{\ottmv{i}}  } = \intermed {  {\overline {\intermed \delta} }_{\ottmv{i}}  }  &,~~\text{for all}~i\\
              \text{and}~ \intermed {  {\intermed m}'  } = \intermed {  {\intermed m}  } &
            \end{aligned}
            \end{equation}
            By applying these equations to Equations
            \pref{delab:det:dcon:3} and \pref{delab:det:dcon:4}, their typing environment
            becomes identical. We can, now, use Theorem~\ref{thmA:tmelab:det} on these two equations,
            from which we get $ \target {  {\target e }  } = \target {  {\target e }'  } $.
            Also, from the namespace-translation convention, we get $ \target {  {\target m}'  } = \target {  {\target m}  } $ and
            $ \target {  {\overline {\target \delta} }'_{\ottmv{i}}  } = \target {  {\overline {\target \delta} }_{\ottmv{i}}  } $.

            If we rewrite Equations \pref{delab:det:dcon:1} and \pref{delab:det:dcon:2} with the
            equations obtained so far, we have
            \begin{gather}
              \label{delab:det:dcon:5}
               \intermed {  {\intermed \Sigma}  } ; \intermed {  {\intermed { \Gamma_C } }  } ; \intermed {  {\intermed \Gamma}  }  \vdash_{d}  \intermed {  {\intermed D } \, \overline {\intermed \sigma}_{\ottmv{j}} \, {\overline {\intermed d} }_{\ottmv{i}}  } : \intermed {  {\intermed {\mathit{TC} } } \, [  \overline {\intermed \sigma}_{\ottmv{j}}  /  {\overline {\intermed a} }_{\ottmv{j}}  ]  {\intermed \sigma}_{\ottmv{q}}  } \itot{\,  \rightsquigarrow  \target {  \ottsym{(}  \Lambda  {\overline {\target a} }_{\ottmv{j}}  \ottsym{.}  \lambda \, \ottcomp{{\target \delta }_{\ottmv{i}}  \ottsym{:}  {\target \sigma }'_{\ottmv{i}}}{\ottmv{i}} \, \ottsym{.}  \ottsym{\{}  {\target m}  \ottsym{=}  {\target e }  \ottsym{\}}  \ottsym{)} \, {\overline {\target \sigma} }_{\ottmv{j}} \, {\overline {\target e} }_{\ottmv{i}}  } } \\
              \label{delab:det:dcon:6}
              \text{and}~
               \intermed {  {\intermed \Sigma}  } ; \intermed {  {\intermed { \Gamma_C } }  } ; \intermed {  {\intermed \Gamma}  }  \vdash_{d}  \intermed {  {\intermed D } \, \overline {\intermed \sigma}_{\ottmv{j}} \, {\overline {\intermed d} }_{\ottmv{i}}  } : \intermed {  {\intermed {\mathit{TC} } } \, [  \overline {\intermed \sigma}_{\ottmv{j}}  /  {\overline {\intermed a} }_{\ottmv{j}}  ]  {\intermed \sigma}_{\ottmv{q}}  } \itot{\,  \rightsquigarrow  \target {  \ottsym{(}  \Lambda  {\overline {\target a} }_{\ottmv{j}}  \ottsym{.}  \lambda \, \ottcomp{{\target \delta }_{\ottmv{i}}  \ottsym{:}  {\target \sigma }''_{\ottmv{i}}}{\ottmv{i}} \, \ottsym{.}  \ottsym{\{}  {\target m}  \ottsym{=}  {\target e }  \ottsym{\}}  \ottsym{)} \, {\overline {\target \sigma} }'_{\ottmv{j}} \, {\overline {\target e} }'_{\ottmv{i}}  } } 
            \end{gather}
            From Equations~\pref{delab:det:dcon:nm_eqs} and the 3\textsuperscript{rd} set
            of premises of the two \textsc{D-con} instantiations in Equations~\pref{delab:det:dcon:5} and
            \pref{delab:det:dcon:6}, we get
            \begin{gather*}
              \ottcomp{ \intermed {  {\intermed { \Gamma_C } }  } ; \intermed {  \bullet  \ottsym{,}  {\overline {\intermed a} }_{\ottmv{j}}  }  \vdash_{\mathit {Q} }  \intermed {  {\intermed Q}_{\ottmv{i}}  }\, \itot{ \rightsquigarrow  \target {  {\target \sigma }'_{\ottmv{i}}  } } }{\ottmv{i}}\\
              \text{and}~\ottcomp{ \intermed {  {\intermed { \Gamma_C } }  } ; \intermed {  \bullet  \ottsym{,}  {\overline {\intermed a} }_{\ottmv{j}}  }  \vdash_{\mathit {Q} }  \intermed {  {\intermed Q}_{\ottmv{i}}  }\, \itot{ \rightsquigarrow  \target {  {\target \sigma }''_{\ottmv{i}}  } } }{\ottmv{i}}
            \end{gather*}
            By passing the two above in Lemma~\ref{lem:det:Q}, we get
            $ \target {  {\target \sigma }''_{\ottmv{i}}  } = \target {  {\target \sigma }'_{\ottmv{i}}  } $, for all $i$.

            From the 4\textsuperscript{th} set of premises of the two \textsc{D-con}
            instantiations in Equations~\pref{delab:det:dcon:5} and \pref{delab:det:dcon:6}, we get
            \begin{gather}
              \ottcomp{ \intermed {  {\intermed { \Gamma_C } }  } ; \intermed {  {\intermed \Gamma}  }  \vdash_{ty}  \intermed {  {\intermed \sigma}_{\ottmv{j}}  } \itot{ \rightsquigarrow  \target {  {\target \sigma }_{\ottmv{j}}  } } }{\ottmv{j}}\\
              \text{and}~\ottcomp{ \intermed {  {\intermed { \Gamma_C } }  } ; \intermed {  {\intermed \Gamma}  }  \vdash_{ty}  \intermed {  {\intermed \sigma}_{\ottmv{j}}  } \itot{ \rightsquigarrow  \target {  {\target \sigma }'_{\ottmv{j}}  } } }{\ottmv{j}}
            \end{gather}
            By passing the two above in Lemma~\ref{lem:det:ty}, we get
            $ \target {  {\target \sigma }'_{\ottmv{j}}  } = \target {  {\target \sigma }_{\ottmv{j}}  } $, for all $j$.

            From the 6\textsuperscript{th} set of premises of the two \textsc{D-con}
            instantiations in Equations~\pref{delab:det:dcon:5} and \pref{delab:det:dcon:6}, we get
            \begin{gather}
              \ottcomp{ \intermed {  {\intermed \Sigma}  } ; \intermed {  {\intermed { \Gamma_C } }  } ; \intermed {  {\intermed \Gamma}  }  \vdash_{d}  \intermed {  {\intermed d}_{\ottmv{i}}  } : \intermed {  [  \overline {\intermed \sigma}_{\ottmv{j}}  /  {\overline {\intermed a} }_{\ottmv{j}}  ]  {\intermed Q}_{\ottmv{i}}  } \itot{\,  \rightsquigarrow  \target {  {\target e }_{\ottmv{i}}  } } }{\ottmv{i}}\\
              \text{and}~\ottcomp{ \intermed {  {\intermed \Sigma}  } ; \intermed {  {\intermed { \Gamma_C } }  } ; \intermed {  {\intermed \Gamma}  }  \vdash_{d}  \intermed {  {\intermed d}_{\ottmv{i}}  } : \intermed {  [  \overline {\intermed \sigma}_{\ottmv{j}}  /  {\overline {\intermed a} }_{\ottmv{j}}  ]  {\intermed Q}_{\ottmv{i}}  } \itot{\,  \rightsquigarrow  \target {  {\target e }'_{\ottmv{i}}  } } }{\ottmv{i}}
            \end{gather}
            From the induction hypothesis for the two above equations, we have
            $ \target {  {\target e }'_{\ottmv{i}}  } = \target {  {\target e }_{\ottmv{i}}  } $, for all $i$.

            Rewritting the derivations in Equations~\pref{delab:det:dcon:5} and \pref{delab:det:dcon:6}
            with the new equations obtained, we have
            \begin{equation*}
               \target {  \ottsym{(}  \Lambda  {\overline {\target a} }_{\ottmv{j}}  \ottsym{.}  \lambda \, \ottcomp{{\target \delta }_{\ottmv{i}}  \ottsym{:}  {\target \sigma }''_{\ottmv{i}}}{\ottmv{i}} \, \ottsym{.}  \ottsym{\{}  {\target m}  \ottsym{=}  {\target e }  \ottsym{\}}  \ottsym{)} \, {\overline {\target \sigma} }'_{\ottmv{j}} \, {\overline {\target e} }'_{\ottmv{i}}  } = \target {  \ottsym{(}  \Lambda  {\overline {\target a} }_{\ottmv{j}}  \ottsym{.}  \lambda \, \ottcomp{{\target \delta }_{\ottmv{i}}  \ottsym{:}  {\target \sigma }'_{\ottmv{i}}}{\ottmv{i}} \, \ottsym{.}  \ottsym{\{}  {\target m}  \ottsym{=}  {\target e }  \ottsym{\}}  \ottsym{)} \, {\overline {\target \sigma} }_{\ottmv{j}} \, {\overline {\target e} }_{\ottmv{i}}  } 
            \end{equation*}
         }
\qed

\begin{theoremB}[Deterministic Term Elaboration]\ \\
  \label{thmA:tmelab:det}
  If \( \intermed {  {\intermed \Sigma}  } ; \intermed {  {\intermed { \Gamma_C } }  } ; \intermed {  {\intermed \Gamma}  }  \vdash_{tm}  \intermed {  {\intermed e}  } : \intermed {  {\intermed \sigma}  } \itot{ \rightsquigarrow  \target {  {\target e }_{{\mathrm{1}}}  } } \)
  and \( \intermed {  {\intermed \Sigma}  } ; \intermed {  {\intermed { \Gamma_C } }  } ; \intermed {  {\intermed \Gamma}  }  \vdash_{tm}  \intermed {  {\intermed e}  } : \intermed {  {\intermed \sigma}  } \itot{ \rightsquigarrow  \target {  {\target e }_{{\mathrm{2}}}  } } \),
  then \( \target {  {\target e }_{{\mathrm{1}}}  } = \target {  {\target e }_{{\mathrm{2}}}  } \).
\end{theoremB}
\proof This theorem is proved mutually with Theorem~\ref{thmA:delab:det}. The proof follows
structural induction on both hypotheses.\\\\
\proofStep{Case \textsc{iTm-true}}{}
          {The two hypotheses of the theorem are:
            \begin{align*}
              & \intermed {  {\intermed \Sigma}  } ; \intermed {  {\intermed { \Gamma_C } }  } ; \intermed {  {\intermed \Gamma}  }  \vdash_{tm}  \intermed {  \mathit{\intermed {True} }  } : \intermed {  \mathit{\intermed {Bool} }  } \itot{ \rightsquigarrow  \target {  {\target e }_{{\mathrm{1}}}  } } \\
              \text{and}~& \intermed {  {\intermed \Sigma}  } ; \intermed {  {\intermed { \Gamma_C } }  } ; \intermed {  {\intermed \Gamma}  }  \vdash_{tm}  \intermed {  \mathit{\intermed {True} }  } : \intermed {  \mathit{\intermed {Bool} }  } \itot{ \rightsquigarrow  \target {  {\target e }_{{\mathrm{2}}}  } } 
            \end{align*}
            From rule \textsc{iTm-true}, it must hold that $ \target {  {\target e }_{{\mathrm{1}}}  } = \target {  {\target e }_{{\mathrm{2}}}  } = \target {  \mathit{\target {True} }  } $.\\
          }
\proofStep{Case \textsc{iTm-false}}{}
          {The two hypotheses of the theorem are:
            \begin{align*}
              & \intermed {  {\intermed \Sigma}  } ; \intermed {  {\intermed { \Gamma_C } }  } ; \intermed {  {\intermed \Gamma}  }  \vdash_{tm}  \intermed {  \mathit{\intermed {False} }  } : \intermed {  \mathit{\intermed {Bool} }  } \itot{ \rightsquigarrow  \target {  {\target e }_{{\mathrm{1}}}  } } \\
              \text{and}~& \intermed {  {\intermed \Sigma}  } ; \intermed {  {\intermed { \Gamma_C } }  } ; \intermed {  {\intermed \Gamma}  }  \vdash_{tm}  \intermed {  \mathit{\intermed {False} }  } : \intermed {  \mathit{\intermed {Bool} }  } \itot{ \rightsquigarrow  \target {  {\target e }_{{\mathrm{2}}}  } } 
            \end{align*}
            From rule \textsc{iTm-false}, it must hold that $ \target {  {\target e }_{{\mathrm{1}}}  } = \target {  {\target e }_{{\mathrm{2}}}  } = \target {  \mathit{\target {False} }  } $.\\
          }
\proofStep{Case \textsc{iTm-var}}{}
          {The two hypotheses of the theorem are:
            \begin{align*}
               \intermed {  {\intermed \Sigma}  } ; \intermed {  {\intermed { \Gamma_C } }  } ; \intermed {  {\intermed \Gamma}  }  \vdash_{tm}  \intermed {  {\intermed x}  } : \intermed {  {\intermed \sigma}  } \itot{ \rightsquigarrow  \target {  {\target x}_{{\mathrm{1}}}  } } \\
              \text{and}~ \intermed {  {\intermed \Sigma}  } ; \intermed {  {\intermed { \Gamma_C } }  } ; \intermed {  {\intermed \Gamma}  }  \vdash_{tm}  \intermed {  {\intermed x}  } : \intermed {  {\intermed \sigma}  } \itot{ \rightsquigarrow  \target {  {\target x}_{{\mathrm{2}}}  } } 
            \end{align*}
            From rule \textsc{iTm-var}, it must hold that $ \target {  {\target x}_{{\mathrm{1}}}  } = \target {  {\target x}_{{\mathrm{2}}}  } = \target {  {\target x}  } $, where
            ${\target x}$ is a target-term-variable with the same identifier as ${\intermed x}$.\\
          }
\proofStep{Case \textsc{iTm-let}}{}
          {The two hypotheses of the theorem are:
            \begin{align}
              \label{tmelab:det:let:hyp1}
              & \intermed {  {\intermed \Sigma}  } ; \intermed {  {\intermed { \Gamma_C } }  } ; \intermed {  {\intermed \Gamma}  }  \vdash_{tm}  \intermed {  \textbf{\intermed {let} } \, {\intermed x}  \ottsym{:}  {\intermed \sigma}_{{\mathrm{1}}}  \ottsym{=}  {\intermed e}_{{\mathrm{1}}} \, \textbf{\intermed {in} } \, {\intermed e}_{{\mathrm{2}}}  } : \intermed {  {\intermed \sigma}_{{\mathrm{2}}}  } \itot{ \rightsquigarrow  \target {  \textbf{\target {let} } \, {\target x}_{{\mathrm{0}}}  \ottsym{:}  {\target \sigma }_{{\mathrm{1}}}  \ottsym{=}  {\target e }_{{\mathrm{1}}} \, \textbf{\target {in} } \, {\target e }_{{\mathrm{2}}}  } } \\
              \label{tmelab:det:let:hyp2}
              \text{and}~& \intermed {  {\intermed \Sigma}  } ; \intermed {  {\intermed { \Gamma_C } }  } ; \intermed {  {\intermed \Gamma}  }  \vdash_{tm}  \intermed {  \textbf{\intermed {let} } \, {\intermed x}  \ottsym{:}  {\intermed \sigma}_{{\mathrm{1}}}  \ottsym{=}  {\intermed e}_{{\mathrm{1}}} \, \textbf{\intermed {in} } \, {\intermed e}_{{\mathrm{2}}}  } : \intermed {  {\intermed \sigma}_{{\mathrm{2}}}  } \itot{ \rightsquigarrow  \target {  \textbf{\target {let} } \, {\target x}'_{{\mathrm{0}}}  \ottsym{:}  {\target \sigma }'_{{\mathrm{1}}}  \ottsym{=}  {\target e }'_{{\mathrm{1}}} \, \textbf{\target {in} } \, {\target e }'_{{\mathrm{2}}}  } } 
            \end{align}
            From our convention regarding translation of identifiers, we have
            \begin{equation}
              \label{tmelab:det:let:tx}
               \target {  {\target x}_{{\mathrm{0}}}  } = \target {  {\target x}'_{{\mathrm{0}}}  } = \target {  {\target x}  } 
            \end{equation}
            where ${\target x}$ is a target-term variable with the same identifier as ${\intermed x}$.
            
            The first premise of the two \textsc{iTm-let} instantiations above, are
            \begin{align*}
               \intermed {  {\intermed \Sigma}  } ; \intermed {  {\intermed { \Gamma_C } }  } ; \intermed {  {\intermed \Gamma}  }  \vdash_{tm}  \intermed {  {\intermed e}_{{\mathrm{1}}}  } : \intermed {  {\intermed \sigma}_{{\mathrm{1}}}  } \itot{ \rightsquigarrow  \target {  {\target e }_{{\mathrm{1}}}  } } \\
              \text{and}~ \intermed {  {\intermed \Sigma}  } ; \intermed {  {\intermed { \Gamma_C } }  } ; \intermed {  {\intermed \Gamma}  }  \vdash_{tm}  \intermed {  {\intermed e}_{{\mathrm{1}}}  } : \intermed {  {\intermed \sigma}_{{\mathrm{1}}}  } \itot{ \rightsquigarrow  \target {  {\target e }'_{{\mathrm{1}}}  } } 
            \end{align*}
            By induction hypothesis, we get
            \begin{equation}
              \label{tmelab:det:let:te1}
               \target {  {\target e }_{{\mathrm{1}}}  } = \target {  {\target e }'_{{\mathrm{1}}}  } 
            \end{equation}
            The 3\textsuperscript{rd} premise of the two instantiations in Equations
            \pref{tmelab:det:let:hyp1} and \pref{tmelab:det:let:hyp2} of rule \text{iTm-let}
            are:
            \begin{align*}
               \intermed {  {\intermed { \Gamma_C } }  } ; \intermed {  {\intermed \Gamma}  }  \vdash_{ty}  \intermed {  {\intermed \sigma}_{{\mathrm{1}}}  } \itot{ \rightsquigarrow  \target {  {\target \sigma }_{{\mathrm{1}}}  } } \\
              \text{and}~ \intermed {  {\intermed { \Gamma_C } }  } ; \intermed {  {\intermed \Gamma}  }  \vdash_{ty}  \intermed {  {\intermed \sigma}_{{\mathrm{1}}}  } \itot{ \rightsquigarrow  \target {  {\target \sigma }'_{{\mathrm{1}}}  } } 
            \end{align*}
            By uniqueness (Lemma~\ref{lem:det:ty}), we get
            \begin{equation}
              \label{tmelab:det:let:tt1}
               \target {  {\target \sigma }_{{\mathrm{1}}}  } = \target {  {\target \sigma }'_{{\mathrm{1}}}  } 
            \end{equation}
            The 2\textsuperscript{nd} premise of the two instantiations in Equations
            \pref{tmelab:det:let:hyp1} and \pref{tmelab:det:let:hyp2} of rule \text{iTm-let}
            are:
            \begin{align*}
               \intermed {  {\intermed \Sigma}  } ; \intermed {  {\intermed { \Gamma_C } }  } ; \intermed {  {\intermed \Gamma}  \ottsym{,}  {\intermed x}  \ottsym{:}  {\intermed \sigma}_{{\mathrm{1}}}  }  \vdash_{tm}  \intermed {  {\intermed e}_{{\mathrm{2}}}  } : \intermed {  {\intermed \sigma}_{{\mathrm{2}}}  } \itot{ \rightsquigarrow  \target {  {\target e }_{{\mathrm{2}}}  } } \\
              \text{and}~ \intermed {  {\intermed \Sigma}  } ; \intermed {  {\intermed { \Gamma_C } }  } ; \intermed {  {\intermed \Gamma}  \ottsym{,}  {\intermed x}  \ottsym{:}  {\intermed \sigma}_{{\mathrm{1}}}  }  \vdash_{tm}  \intermed {  {\intermed e}_{{\mathrm{2}}}  } : \intermed {  {\intermed \sigma}_{{\mathrm{2}}}  } \itot{ \rightsquigarrow  \target {  {\target e }'_{{\mathrm{2}}}  } } 
            \end{align*}
            By induction hypothesis, we get
            \begin{equation}
              \label{tmelab:det:let:te2}
               \target {  {\target e }_{{\mathrm{2}}}  } = \target {  {\target e }'_{{\mathrm{2}}}  } 
            \end{equation}
            From Equations~\pref{tmelab:det:let:tx}, \pref{tmelab:det:let:te1},
            \pref{tmelab:det:let:tt1} and \pref{tmelab:det:let:te2}, we obtain
            \begin{equation*}
               \target {  \textbf{\target {let} } \, {\target x}_{{\mathrm{0}}}  \ottsym{:}  {\target \sigma }_{{\mathrm{1}}}  \ottsym{=}  {\target e }_{{\mathrm{1}}} \, \textbf{\target {in} } \, {\target e }_{{\mathrm{2}}}  } = \target {  \textbf{\target {let} } \, {\target x}'_{{\mathrm{0}}}  \ottsym{:}  {\target \sigma }'_{{\mathrm{1}}}  \ottsym{=}  {\target e }'_{{\mathrm{1}}} \, \textbf{\target {in} } \, {\target e }'_{{\mathrm{2}}}  } 
            \end{equation*}
          }
\proofStep{Case \textsc{iTm-method}}{}
          {The two hypotheses of the theorem are:
            \begin{align}
              \label{tmelab:det:meth:hyp1}
              & \intermed {  {\intermed \Sigma}  } ; \intermed {  {\intermed { \Gamma_C } }  } ; \intermed {  {\intermed \Gamma}  }  \vdash_{tm}  \intermed {  {\intermed d}  \ottsym{.}  {\intermed m}  } : \intermed {  [  {\intermed \sigma}_{\ottmv{q}}  /  {\intermed a }  ]  {\intermed \sigma}_{\ottmv{m}}  } \itot{ \rightsquigarrow  \target {  {\target e }  \ottsym{.}  {\target m}_{{\mathrm{0}}}  } } \\
              \label{tmelab:det:meth:hyp2}
              \text{and}~& \intermed {  {\intermed \Sigma}  } ; \intermed {  {\intermed { \Gamma_C } }  } ; \intermed {  {\intermed \Gamma}  }  \vdash_{tm}  \intermed {  {\intermed d}  \ottsym{.}  {\intermed m}  } : \intermed {  [  {\intermed \sigma}_{\ottmv{q}}  /  {\intermed a }  ]  {\intermed \sigma}_{\ottmv{m}}  } \itot{ \rightsquigarrow  \target {  {\target e }'  \ottsym{.}  {\target m}'_{{\mathrm{0}}}  } } 
            \end{align}
            By our convention for dictionary labels, we have
            \begin{equation*}
               \target {  {\target m}_{{\mathrm{0}}}  } = \target {  {\target m}'_{{\mathrm{0}}}  } = \target {  {\target m}  } 
            \end{equation*}
            where ${\target m}$ is a record field with the same identifier as class method ${\intermed m}$.

            The first premise of the two \textsc{iTm-method} instantiations in Equations
            \pref{tmelab:det:meth:hyp1} and \pref{tmelab:det:meth:hyp2} are:
            \begin{align*}
                         & \intermed {  {\intermed \Sigma}  } ; \intermed {  {\intermed { \Gamma_C } }  } ; \intermed {  {\intermed \Gamma}  }  \vdash_{d}  \intermed {  {\intermed d}  } : \intermed {  {\intermed {\mathit{TC} } } \, {\intermed \sigma}_{\ottmv{q}}  } \itot{\,  \rightsquigarrow  \target {  {\target e }  } } \\
              \text{and}~& \intermed {  {\intermed \Sigma}  } ; \intermed {  {\intermed { \Gamma_C } }  } ; \intermed {  {\intermed \Gamma}  }  \vdash_{d}  \intermed {  {\intermed d}  } : \intermed {  {\intermed {\mathit{TC} } } \, {\intermed \sigma}_{\ottmv{q}}  } \itot{\,  \rightsquigarrow  \target {  {\target e }'  } } 
            \end{align*}
            By Theorem~\ref{thmA:delab:det}, we have
            \begin{equation*}
               \target {  {\target e }  } = \target {  {\target e }'  } 
            \end{equation*}
            Then, $ \target {  {\target e }  \ottsym{.}  {\target m}_{{\mathrm{0}}}  } = \target {  {\target e }'  \ottsym{.}  {\target m}'_{{\mathrm{0}}}  } $.\\
          }
\proofStep{Case \textsc{iTm-arrI}}{}
          {The two hypotheses of the theorem are:
            \begin{align}
              \label{tmelab:det:arri:hyp1}
              & \intermed {  {\intermed \Sigma}  } ; \intermed {  {\intermed { \Gamma_C } }  } ; \intermed {  {\intermed \Gamma}  }  \vdash_{tm}  \intermed {  \lambda  {\intermed x}  \ottsym{:}  {\intermed \sigma}_{{\mathrm{1}}}  \ottsym{.}  {\intermed e}  } : \intermed {  {\intermed \sigma}_{{\mathrm{1}}}  \rightarrow  {\intermed \sigma}_{{\mathrm{2}}}  } \itot{ \rightsquigarrow  \target {  \lambda  {\target x}_{{\mathrm{0}}}  \ottsym{:}  {\target \sigma }_{{\mathrm{1}}}  \ottsym{.}  {\target e }  } } \\
              \label{tmelab:det:arri:hyp2}
              \text{and}~& \intermed {  {\intermed \Sigma}  } ; \intermed {  {\intermed { \Gamma_C } }  } ; \intermed {  {\intermed \Gamma}  }  \vdash_{tm}  \intermed {  \lambda  {\intermed x}  \ottsym{:}  {\intermed \sigma}_{{\mathrm{1}}}  \ottsym{.}  {\intermed e}  } : \intermed {  {\intermed \sigma}_{{\mathrm{1}}}  \rightarrow  {\intermed \sigma}_{{\mathrm{2}}}  } \itot{ \rightsquigarrow  \target {  \lambda  {\target x}'_{{\mathrm{0}}}  \ottsym{:}  {\target \sigma }'_{{\mathrm{1}}}  \ottsym{.}  {\target e }'  } } 
          \end{align}
          From our identifiers' translation, it is implied that
          \begin{equation}
            \label{tmelab:det:arri:eq_tx}
             \target {  {\target x}_{{\mathrm{0}}}  } = \target {  {\target x}'_{{\mathrm{0}}}  } = {\target x}
          \end{equation}
          where ${\target x}$ is the target-term variable with the same identifier as ${\intermed x}$.

          The 2\textsuperscript{nd} premise of the two \textsc{iTm-arrI} instantiations above
          are:
          \begin{align*}
                       & \intermed {  {\intermed { \Gamma_C } }  } ; \intermed {  {\intermed \Gamma}  }  \vdash_{ty}  \intermed {  {\intermed \sigma}_{{\mathrm{1}}}  } \itot{ \rightsquigarrow  \target {  {\target \sigma }_{{\mathrm{1}}}  } } \\
            \text{and}~& \intermed {  {\intermed { \Gamma_C } }  } ; \intermed {  {\intermed \Gamma}  }  \vdash_{ty}  \intermed {  {\intermed \sigma}_{{\mathrm{1}}}  } \itot{ \rightsquigarrow  \target {  {\target \sigma }'_{{\mathrm{1}}}  } } 
          \end{align*}
          By uniqueness (Lemma~\ref{lem:det:ty}), we have
          \begin{equation}
            \label{tmelab:det:arri:eq_tt1}
             \target {  {\target \sigma }_{{\mathrm{1}}}  } = \target {  {\target \sigma }'_{{\mathrm{1}}}  } 
          \end{equation}
          The first premise of the two \textsc{iTm-arrI} instantiations in Equations
          \pref{tmelab:det:arri:hyp1} and \pref{tmelab:det:arri:hyp2} are:
          \begin{align*}
                       & \intermed {  {\intermed \Sigma}  } ; \intermed {  {\intermed { \Gamma_C } }  } ; \intermed {  {\intermed \Gamma}  \ottsym{,}  {\intermed x}  \ottsym{:}  {\intermed \sigma}_{{\mathrm{1}}}  }  \vdash_{tm}  \intermed {  {\intermed e}  } : \intermed {  {\intermed \sigma}_{{\mathrm{2}}}  } \itot{ \rightsquigarrow  \target {  {\target e }  } } \\
            \text{and}~& \intermed {  {\intermed \Sigma}  } ; \intermed {  {\intermed { \Gamma_C } }  } ; \intermed {  {\intermed \Gamma}  \ottsym{,}  {\intermed x}  \ottsym{:}  {\intermed \sigma}_{{\mathrm{1}}}  }  \vdash_{tm}  \intermed {  {\intermed e}  } : \intermed {  {\intermed \sigma}_{{\mathrm{2}}}  } \itot{ \rightsquigarrow  \target {  {\target e }'  } } 
          \end{align*}
          By the induction hypothesis, we get
          \begin{equation}
            \label{tmelab:det:arri:eq_te}
             \target {  {\target e }  } = \target {  {\target e }'  } 
          \end{equation}
          From Equations~\pref{tmelab:det:arri:eq_tx}, \pref{tmelab:det:arri:eq_tt1} and
          \pref{tmelab:det:arri:eq_te}, we obtain
          \begin{equation*}
             \target {  \lambda  {\target x}_{{\mathrm{0}}}  \ottsym{:}  {\target \sigma }_{{\mathrm{1}}}  \ottsym{.}  {\target e }  } = \target {  \lambda  {\target x}'_{{\mathrm{0}}}  \ottsym{:}  {\target \sigma }'_{{\mathrm{1}}}  \ottsym{.}  {\target e }'  } 
          \end{equation*}
          }
\proofStep{Case \textsc{iTm-arrE}}{}
          {The two hypotheses of the theorem are:
            \begin{align}
              \label{tmelab:det:arre:hyp1}
                         & \intermed {  {\intermed \Sigma}  } ; \intermed {  {\intermed { \Gamma_C } }  } ; \intermed {  {\intermed \Gamma}  }  \vdash_{tm}  \intermed {  {\intermed e}_{{\mathrm{1}}} \, {\intermed e}_{{\mathrm{2}}}  } : \intermed {  {\intermed \sigma}_{{\mathrm{2}}}  } \itot{ \rightsquigarrow  \target {  {\target e }_{{\mathrm{1}}} \, {\target e }_{{\mathrm{2}}}  } } \\
              \label{tmelab:det:arre:hyp2}
              \text{and}~& \intermed {  {\intermed \Sigma}  } ; \intermed {  {\intermed { \Gamma_C } }  } ; \intermed {  {\intermed \Gamma}  }  \vdash_{tm}  \intermed {  {\intermed e}_{{\mathrm{1}}} \, {\intermed e}_{{\mathrm{2}}}  } : \intermed {  {\intermed \sigma}_{{\mathrm{2}}}  } \itot{ \rightsquigarrow  \target {  {\target e }'_{{\mathrm{1}}} \, {\target e }'_{{\mathrm{2}}}  } } 
            \end{align}
            The first premise of the above two instantiated \textsc{iTm-arrE} rules are:
            \begin{align*}
                         & \intermed {  {\intermed \Sigma}  } ; \intermed {  {\intermed { \Gamma_C } }  } ; \intermed {  {\intermed \Gamma}  }  \vdash_{tm}  \intermed {  {\intermed e}_{{\mathrm{1}}}  } : \intermed {  {\intermed \sigma}_{{\mathrm{1}}}  \rightarrow  {\intermed \sigma}_{{\mathrm{2}}}  } \itot{ \rightsquigarrow  \target {  {\target e }_{{\mathrm{1}}}  } } \\
              \text{and}~& \intermed {  {\intermed \Sigma}  } ; \intermed {  {\intermed { \Gamma_C } }  } ; \intermed {  {\intermed \Gamma}  }  \vdash_{tm}  \intermed {  {\intermed e}_{{\mathrm{1}}}  } : \intermed {  {\intermed \sigma}_{{\mathrm{1}}}  \rightarrow  {\intermed \sigma}_{{\mathrm{2}}}  } \itot{ \rightsquigarrow  \target {  {\target e }'_{{\mathrm{1}}}  } } 
            \end{align*}
            By induction hypothesis, we have
            \begin{equation}
              \label{tmelab:det:arre:eq_te1}
               \target {  {\target e }_{{\mathrm{1}}}  } = \target {  {\target e }'_{{\mathrm{1}}}  } 
            \end{equation}
            Similarly, from the second premise of the two \textsc{iTm-arrE} rules, and the
            induction hypothesis, we get
            \begin{equation}
              \label{tmelab:det:arre:eq_te2}
               \target {  {\target e }_{{\mathrm{2}}}  } = \target {  {\target e }'_{{\mathrm{2}}}  } 
            \end{equation}
            Then, we obtain
            \begin{equation*}
               \target {  {\target e }_{{\mathrm{1}}} \, {\target e }_{{\mathrm{2}}}  } = \target {  {\target e }'_{{\mathrm{1}}} \, {\target e }'_{{\mathrm{2}}}  } 
            \end{equation*}
            by Equations~\pref{tmelab:det:arre:eq_te1} and \pref{tmelab:det:arre:eq_te2}.\\
          }
\proofStep{Case \textsc{iTm-constrI}}{}
          {The two hypotheses of the theorem are:
            \begin{align}
              \label{tmelab:det:constri:hyp1}
              & \intermed {  {\intermed \Sigma}  } ; \intermed {  {\intermed { \Gamma_C } }  } ; \intermed {  {\intermed \Gamma}  }  \vdash_{tm}  \intermed {  \lambda  {\intermed \delta }  \ottsym{:}  {\intermed Q}  \ottsym{.}  {\intermed e}  } : \intermed {   {\intermed Q}  \Rightarrow  {\intermed \sigma}   } \itot{ \rightsquigarrow  \target {  \lambda  {\target \delta }_{{\mathrm{0}}}  \ottsym{:}  {\target \sigma }  \ottsym{.}  {\target e }  } } \\
              \label{tmelab:det:constri:hyp2}
              \text{and}~& \intermed {  {\intermed \Sigma}  } ; \intermed {  {\intermed { \Gamma_C } }  } ; \intermed {  {\intermed \Gamma}  }  \vdash_{tm}  \intermed {  \lambda  {\intermed \delta }  \ottsym{:}  {\intermed Q}  \ottsym{.}  {\intermed e}  } : \intermed {   {\intermed Q}  \Rightarrow  {\intermed \sigma}   } \itot{ \rightsquigarrow  \target {  \lambda  {\target \delta }'_{{\mathrm{0}}}  \ottsym{:}  {\target \sigma }'  \ottsym{.}  {\target e }'  } } 
            \end{align}
            From our identifiers' translation, it is implied that
          \begin{equation}
            \label{tmelab:det:constri:eq_txd}
             \target {  {\target \delta }_{{\mathrm{0}}}  } = \target {  {\target \delta }'_{{\mathrm{0}}}  } = {\target \delta }
          \end{equation}
          where ${\target \delta }$ is the target-term variable with the same identifier as the
          dictionary variable ${\intermed \delta }$.

          The 2\textsuperscript{nd} premise of the two \textsc{iTm-constrI} instantiations above
          are:
          \begin{align*}
                       & \intermed {  {\intermed { \Gamma_C } }  } ; \intermed {  {\intermed \Gamma}  }  \vdash_{\mathit {Q} }  \intermed {  {\intermed Q}  }\, \itot{ \rightsquigarrow  \target {  {\target \sigma }  } } \\
            \text{and}~& \intermed {  {\intermed { \Gamma_C } }  } ; \intermed {  {\intermed \Gamma}  }  \vdash_{\mathit {Q} }  \intermed {  {\intermed Q}  }\, \itot{ \rightsquigarrow  \target {  {\target \sigma }'  } } 
          \end{align*}
          By uniqueness (Lemma~\ref{lem:det:Q}), we have
          \begin{equation}
            \label{tmelab:det:constri:eq_tt}
             \target {  {\target \sigma }  } = \target {  {\target \sigma }'  } 
          \end{equation}
          The first premise of the two \textsc{iTm-constrI} instantiations in Equations
          \pref{tmelab:det:constri:hyp1} and \pref{tmelab:det:constri:hyp2} are:
          \begin{align*}
                       & \intermed {  {\intermed \Sigma}  } ; \intermed {  {\intermed { \Gamma_C } }  } ; \intermed {  {\intermed \Gamma}  \ottsym{,}  {\intermed \delta }  \ottsym{:}  {\intermed Q}  }  \vdash_{tm}  \intermed {  {\intermed e}  } : \intermed {  {\intermed \sigma}  } \itot{ \rightsquigarrow  \target {  {\target e }  } } \\
            \text{and}~& \intermed {  {\intermed \Sigma}  } ; \intermed {  {\intermed { \Gamma_C } }  } ; \intermed {  {\intermed \Gamma}  \ottsym{,}  {\intermed \delta }  \ottsym{:}  {\intermed Q}  }  \vdash_{tm}  \intermed {  {\intermed e}  } : \intermed {  {\intermed \sigma}  } \itot{ \rightsquigarrow  \target {  {\target e }'  } } 
          \end{align*}
          By the induction hypothesis, we get
          \begin{equation}
            \label{tmelab:det:constri:eq_te}
             \target {  {\target e }  } = \target {  {\target e }'  } 
          \end{equation}
          From Equations~\pref{tmelab:det:constri:eq_txd}, \pref{tmelab:det:constri:eq_tt} and
          \pref{tmelab:det:constri:eq_te}, we obtain
          \begin{equation*}
             \target {  \lambda  {\target \delta }_{{\mathrm{0}}}  \ottsym{:}  {\target \sigma }  \ottsym{.}  {\target e }  } = \target {  \lambda  {\target \delta }'_{{\mathrm{0}}}  \ottsym{:}  {\target \sigma }'  \ottsym{.}  {\target e }'  } 
          \end{equation*}
          }
\proofStep{Case \textsc{iTm-constrE}}{}
          {The two hypotheses of the theorem are:
            \begin{align}
              \label{tmelab:det:constre:hyp1}
                         & \intermed {  {\intermed \Sigma}  } ; \intermed {  {\intermed { \Gamma_C } }  } ; \intermed {  {\intermed \Gamma}  }  \vdash_{tm}  \intermed {  {\intermed e} \, {\intermed d}  } : \intermed {  {\intermed \sigma}  } \itot{ \rightsquigarrow  \target {  {\target e }_{{\mathrm{1}}} \, {\target e }_{{\mathrm{2}}}  } } \\
              \label{tmelab:det:constre:hyp2}
              \text{and}~& \intermed {  {\intermed \Sigma}  } ; \intermed {  {\intermed { \Gamma_C } }  } ; \intermed {  {\intermed \Gamma}  }  \vdash_{tm}  \intermed {  {\intermed e} \, {\intermed d}  } : \intermed {  {\intermed \sigma}  } \itot{ \rightsquigarrow  \target {  {\target e }'_{{\mathrm{1}}} \, {\target e }'_{{\mathrm{2}}}  } } 
            \end{align}
            The first premise of the above two instantiated \textsc{iTm-constrE} rules are:
            \begin{align*}
                         & \intermed {  {\intermed \Sigma}  } ; \intermed {  {\intermed { \Gamma_C } }  } ; \intermed {  {\intermed \Gamma}  }  \vdash_{tm}  \intermed {  {\intermed e}  } : \intermed {   {\intermed Q}  \Rightarrow  {\intermed \sigma}   } \itot{ \rightsquigarrow  \target {  {\target e }_{{\mathrm{1}}}  } } \\
              \text{and}~& \intermed {  {\intermed \Sigma}  } ; \intermed {  {\intermed { \Gamma_C } }  } ; \intermed {  {\intermed \Gamma}  }  \vdash_{tm}  \intermed {  {\intermed e}  } : \intermed {   {\intermed Q}  \Rightarrow  {\intermed \sigma}   } \itot{ \rightsquigarrow  \target {  {\target e }'_{{\mathrm{1}}}  } } 
            \end{align*}
            By induction hypothesis, we have
            \begin{equation}
              \label{tmelab:det:constre:eq_te1}
               \target {  {\target e }_{{\mathrm{1}}}  } = \target {  {\target e }'_{{\mathrm{1}}}  } 
            \end{equation}
            Similarly, from the second premise of the two \textsc{iTm-constrE} rules, and the
            induction hypothesis, we get
            \begin{equation}
              \label{tmelab:det:constre:eq_te2}
               \target {  {\target e }_{{\mathrm{2}}}  } = \target {  {\target e }'_{{\mathrm{2}}}  } 
            \end{equation}
            Then, we obtain
            \begin{equation*}
               \target {  {\target e }_{{\mathrm{1}}} \, {\target e }_{{\mathrm{2}}}  } = \target {  {\target e }'_{{\mathrm{1}}} \, {\target e }'_{{\mathrm{2}}}  } 
            \end{equation*}
            by Equations~\pref{tmelab:det:constre:eq_te1} and \pref{tmelab:det:constre:eq_te2}.\\}
\proofStep{Case \textsc{iTm-forallI}}{}
          {The two hypotheses of the theorem are:
            \begin{align*}
                         & \intermed {  {\intermed \Sigma}  } ; \intermed {  {\intermed { \Gamma_C } }  } ; \intermed {  {\intermed \Gamma}  }  \vdash_{tm}  \intermed {  \Lambda  {\intermed a }  \ottsym{.}  {\intermed e}  } : \intermed {  \forall  {\intermed a }  \ottsym{.}  {\intermed \sigma}  } \itot{ \rightsquigarrow  \target {  \Lambda  {\target a }_{{\mathrm{0}}}  \ottsym{.}  {\target e }  } } \\
              \text{and}~& \intermed {  {\intermed \Sigma}  } ; \intermed {  {\intermed { \Gamma_C } }  } ; \intermed {  {\intermed \Gamma}  }  \vdash_{tm}  \intermed {  \Lambda  {\intermed a }  \ottsym{.}  {\intermed e}  } : \intermed {  \forall  {\intermed a }  \ottsym{.}  {\intermed \sigma}  } \itot{ \rightsquigarrow  \target {  \Lambda  {\target a }'_{{\mathrm{0}}}  \ottsym{.}  {\target e }'  } } 
            \end{align*}
            It is implied by our identifiers' translation convention, that
            \begin{equation}
              \label{tmelab:det:foralli:eq_ta}
               \target {  {\target a }_{{\mathrm{0}}}  } = \target {  {\target a }'_{{\mathrm{0}}}  } = \target {  {\target a }  } 
            \end{equation}
            where ${\target a }$ is the target-type variable with the same identifier as the ${\intermed a }$.

            The premise of the two \textsc{iTm-forallI} instantiations above are:
            \begin{align*}
                         & \intermed {  {\intermed \Sigma}  } ; \intermed {  {\intermed { \Gamma_C } }  } ; \intermed {  {\intermed \Gamma}  \ottsym{,}  {\intermed a }  }  \vdash_{tm}  \intermed {  {\intermed e}  } : \intermed {  {\intermed \sigma}  } \itot{ \rightsquigarrow  \target {  {\target e }  } } \\
              \text{and}~& \intermed {  {\intermed \Sigma}  } ; \intermed {  {\intermed { \Gamma_C } }  } ; \intermed {  {\intermed \Gamma}  \ottsym{,}  {\intermed a }  }  \vdash_{tm}  \intermed {  {\intermed e}  } : \intermed {  {\intermed \sigma}  } \itot{ \rightsquigarrow  \target {  {\target e }'  } } 
            \end{align*}
            By the induction hypothesis, we have
            \begin{equation}
              \label{tmelab:det:foralli:eq_te}
               \target {  {\target e }  } = \target {  {\target e }'  } 
            \end{equation}
            Equations~\pref{tmelab:det:foralli:eq_ta} and \pref{tmelab:det:foralli:eq_te}
            result in
            \begin{equation*}
               \target {  \Lambda  {\target a }_{{\mathrm{0}}}  \ottsym{.}  {\target e }  } = \target {  \Lambda  {\target a }'_{{\mathrm{0}}}  \ottsym{.}  {\target e }'  } 
            \end{equation*}
          }
\proofStep{Case \textsc{iTm-forallE}}{}
          {The two hypotheses of the theorem are:
            \begin{align}
              \label{tmleba:det:foralle:prem1}
                         & \intermed {  {\intermed \Sigma}  } ; \intermed {  {\intermed { \Gamma_C } }  } ; \intermed {  {\intermed \Gamma}  }  \vdash_{tm}  \intermed {  {\intermed e} \, {\intermed \sigma}  } : \intermed {  [  {\intermed \sigma}  /  {\intermed a }  ]  {\intermed \sigma}'  } \itot{ \rightsquigarrow  \target {  {\target e } \, {\target \sigma }  } } \\
              \label{tmleba:det:foralle:prem1'}
              \text{and}~& \intermed {  {\intermed \Sigma}  } ; \intermed {  {\intermed { \Gamma_C } }  } ; \intermed {  {\intermed \Gamma}  }  \vdash_{tm}  \intermed {  {\intermed e} \, {\intermed \sigma}  } : \intermed {  [  {\intermed \sigma}  /  {\intermed a }  ]  {\intermed \sigma}'  } \itot{ \rightsquigarrow  \target {  {\target e }' \, {\target \sigma }'  } } 
            \end{align}
            The first premise of the above two instantiations of rule \textsc{iTm-forallE} are:
            \begin{align*}
                         & \intermed {  {\intermed \Sigma}  } ; \intermed {  {\intermed { \Gamma_C } }  } ; \intermed {  {\intermed \Gamma}  }  \vdash_{tm}  \intermed {  {\intermed e}  } : \intermed {  \forall  {\intermed a }  \ottsym{.}  {\intermed \sigma}'  } \itot{ \rightsquigarrow  \target {  {\target e }  } } \\
              \text{and}~& \intermed {  {\intermed \Sigma}  } ; \intermed {  {\intermed { \Gamma_C } }  } ; \intermed {  {\intermed \Gamma}  }  \vdash_{tm}  \intermed {  {\intermed e}  } : \intermed {  \forall  {\intermed a }  \ottsym{.}  {\intermed \sigma}'  } \itot{ \rightsquigarrow  \target {  {\target e }'  } } 
            \end{align*}
            From the induction hypothesis, we get
            \begin{equation}
              \label{tmelab:det:foralle:eq_te}
               \target {  {\target e }  } = \target {  {\target e }'  } 
            \end{equation}
            The second premise of the two instantiations of rule
            \textsc{iTm-forallE} in Equations
            \pref{tmleba:det:foralle:prem1} and \pref{tmleba:det:foralle:prem1'} are:
            \begin{align*}
                           \intermed {  {\intermed { \Gamma_C } }  } ; \intermed {  {\intermed \Gamma}  }  \vdash_{ty}  \intermed {  {\intermed \sigma}  } \itot{ \rightsquigarrow  \target {  {\target \sigma }  } } 
              \text{and}~& \intermed {  {\intermed { \Gamma_C } }  } ; \intermed {  {\intermed \Gamma}  }  \vdash_{ty}  \intermed {  {\intermed \sigma}  } \itot{ \rightsquigarrow  \target {  {\target \sigma }'  } } 
            \end{align*}
            Applying Lemma~\ref{lem:det:ty} on these equations gives
            \begin{equation}
              \label{tmelab:det:foralle:eq_tt}
               \target {  {\target \sigma }  } = \target {  {\target \sigma }'  } 
            \end{equation}
            From Equations~\pref{tmelab:det:foralle:eq_te} and \pref{tmelab:det:foralle:eq_tt}
            we have
            \begin{equation*}
               \target {  {\target e } \, {\target \sigma }  } = \target {  {\target e }' \, {\target \sigma }'  } 
            \end{equation*}
          }
\qed

\begin{theoremB}[Deterministic Context Elaboration]\ \\
  \label{thmA:ctxelab:det}
  If \( \intermed {  {\intermed M}  } : ( \intermed {  {\intermed \Sigma}  } ; \intermed {  {\intermed { \Gamma_C } }  } ; \intermed {  {\intermed \Gamma}  }  \Rightarrow  \intermed {  {\intermed \sigma}  } )  \mapsto  ( \intermed {  {\intermed \Sigma}  } ; \intermed {  {\intermed { \Gamma_C } }  } ; \intermed {  {\intermed \Gamma}'  }  \Rightarrow  \intermed {  {\intermed \sigma}'  } ) \itot{ \rightsquigarrow  \target {  {\target M }_{{\mathrm{1}}}  } } \)
  and \( \intermed {  {\intermed M}  } : ( \intermed {  {\intermed \Sigma}  } ; \intermed {  {\intermed { \Gamma_C } }  } ; \intermed {  {\intermed \Gamma}  }  \Rightarrow  \intermed {  {\intermed \sigma}  } )  \mapsto  ( \intermed {  {\intermed \Sigma}  } ; \intermed {  {\intermed { \Gamma_C } }  } ; \intermed {  {\intermed \Gamma}'  }  \Rightarrow  \intermed {  {\intermed \sigma}'  } ) \itot{ \rightsquigarrow  \target {  {\target M }_{{\mathrm{2}}}  } } \), \\
  then \( \target {  {\target M }_{{\mathrm{1}}}  } = \target {  {\target M }_{{\mathrm{2}}}  } \).
\end{theoremB}

\proof
This proof proceeds by straightforward structural induction on both hypotheses,
in combination with Lemmas~\ref{lem:det:Q} and~\ref{lem:det:ty} and
Theorems~\ref{thmA:delab:det} and~\ref{thmA:tmelab:det}. \\
\qed

\subsection{Semantic Preservation}
\begin{theoremB}[Semantic Preservation]
  \label{thmA:sem_pres}
  \begin{flalign}
    &\text{If}~ \intermed {  {\intermed \Sigma}  }  \vdash  \intermed {  {\intermed e}  }  \longrightarrow  \intermed {  {\intermed e}'  } \nonumber &\\
    \label{semelab:hyp1}
    &\text{and}~ \intermed {  {\intermed \Sigma}  } ; \intermed {  {\intermed { \Gamma_C } }  } ; \intermed {  \bullet  }  \vdash_{tm}  \intermed {  {\intermed e}  } : \intermed {  {\intermed \sigma}  } \itot{ \rightsquigarrow  \target {  {\target e }  } }  &\\
    \label{semelab:hyp2}
    &\text{and}~ \intermed {  {\intermed \Sigma}  } ; \intermed {  {\intermed { \Gamma_C } }  } ; \intermed {  \bullet  }  \vdash_{tm}  \intermed {  {\intermed e}'  } : \intermed {  {\intermed \sigma}  } \itot{ \rightsquigarrow  \target {  {\target e }'  } }  &\\
    &\text{for some}~{\intermed \sigma},~{\target e }~\text{and}~\target{{\target e }'},\nonumber &\\
    &\text{then, there is an}~\target{{\target e }_{\ottmv{h}}}~\text{such that}\nonumber &\\
    & \target {  {\target e }  }  \longrightarrow^{*}  \target {  {\target e }_{\ottmv{h}}  } ~\text{and}~ \target {  {\target e }'  }  \longrightarrow^{*}  \target {  {\target e }_{\ottmv{h}}  } .\nonumber
  \end{flalign}
\end{theoremB}

\proof This proof proceeds by induction on the first hypothesis.

\proofStep{Case \textsc{iEval-app}}{\drule{iEval-app}}
          {
            Hypotheses \pref{semelab:hyp1} and \pref{semelab:hyp2} of the theorem, adapted to
            this case, are:
            \begin{align*}
                         & \intermed {  {\intermed \Sigma}  } ; \intermed {  {\intermed { \Gamma_C } }  } ; \intermed {  \bullet  }  \vdash_{tm}  \intermed {  {\intermed e}_{{\mathrm{1}}} \, {\intermed e}_{{\mathrm{2}}}  } : \intermed {  {\intermed \sigma}_{{\mathrm{2}}}  } \itot{ \rightsquigarrow  \target {  {\target e }_{{\mathrm{1}}} \, {\target e }_{{\mathrm{2}}}  } } \\
              \text{and}~& \intermed {  {\intermed \Sigma}  } ; \intermed {  {\intermed { \Gamma_C } }  } ; \intermed {  \bullet  }  \vdash_{tm}  \intermed {  {\intermed e}'_{{\mathrm{1}}} \, {\intermed e}_{{\mathrm{2}}}  } : \intermed {  {\intermed \sigma}_{{\mathrm{2}}}  } \itot{ \rightsquigarrow  \target {  {\target e }'_{{\mathrm{1}}} \, {\target e }_{{\mathrm{2}}}  } } 
            \end{align*}
            for some $\intermed{{\intermed \sigma}_{{\mathrm{2}}}}$, $\target{{\target e }_{{\mathrm{1}}}}$, $\target{{\target e }'_{{\mathrm{1}}}}$ and $\target{{\target e }_{{\mathrm{2}}}}$.
            
            The last rule of these derivations must be instances of \textsc{iTm-arrE}. For
            Hypothesis \pref{semelab:hyp1}, we have:
            \begin{equation}
              \label{semelab:app:der1}
            \ottdrule['']{
              \ottpremise{ \intermed {  {\intermed \Sigma}  } ; \intermed {  {\intermed { \Gamma_C } }  } ; \intermed {  \bullet  }  \vdash_{tm}  \intermed {  {\intermed e}_{{\mathrm{1}}}  } : \intermed {  {\intermed \sigma}_{{\mathrm{1}}}  \rightarrow  {\intermed \sigma}_{{\mathrm{2}}}  } \itot{ \rightsquigarrow  \target {  {\target e }_{{\mathrm{1}}}  } } }
              \ottpremise{ \intermed {  {\intermed \Sigma}  } ; \intermed {  {\intermed { \Gamma_C } }  } ; \intermed {  \bullet  }  \vdash_{tm}  \intermed {  {\intermed e}_{{\mathrm{2}}}  } : \intermed {  {\intermed \sigma}_{{\mathrm{1}}}  } \itot{ \rightsquigarrow  \target {  {\target e }_{{\mathrm{2}}}  } } }
            }{
               \intermed {  {\intermed \Sigma}  } ; \intermed {  {\intermed { \Gamma_C } }  } ; \intermed {  \bullet  }  \vdash_{tm}  \intermed {  {\intermed e}_{{\mathrm{1}}} \, {\intermed e}_{{\mathrm{2}}}  } : \intermed {  {\intermed \sigma}_{{\mathrm{2}}}  } \itot{ \rightsquigarrow  \target {  {\target e }_{{\mathrm{1}}} \, {\target e }_{{\mathrm{2}}}  } } 
            }{\textsc{iTm-arrE}
            }
            \end{equation}
            and for Hypothesis \pref{semelab:hyp2}, we have:
            \begin{equation}
              \label{semelab:app:der2}
            \ottdrule['']{
              \ottpremise{ \intermed {  {\intermed \Sigma}  } ; \intermed {  {\intermed { \Gamma_C } }  } ; \intermed {  \bullet  }  \vdash_{tm}  \intermed {  {\intermed e}'_{{\mathrm{1}}}  } : \intermed {  {\intermed \sigma}'_{{\mathrm{1}}}  \rightarrow  {\intermed \sigma}_{{\mathrm{2}}}  } \itot{ \rightsquigarrow  \target {  {\target e }'_{{\mathrm{1}}}  } } }
              \ottpremise{ \intermed {  {\intermed \Sigma}  } ; \intermed {  {\intermed { \Gamma_C } }  } ; \intermed {  \bullet  }  \vdash_{tm}  \intermed {  {\intermed e}_{{\mathrm{2}}}  } : \intermed {  {\intermed \sigma}'_{{\mathrm{1}}}  } \itot{ \rightsquigarrow  \target {  {\target e }_{{\mathrm{2}}}  } } }
            }{
               \intermed {  {\intermed \Sigma}  } ; \intermed {  {\intermed { \Gamma_C } }  } ; \intermed {  \bullet  }  \vdash_{tm}  \intermed {  {\intermed e}'_{{\mathrm{1}}} \, {\intermed e}_{{\mathrm{2}}}  } : \intermed {  {\intermed \sigma}_{{\mathrm{2}}}  } \itot{ \rightsquigarrow  \target {  {\target e }'_{{\mathrm{1}}} \, {\target e }_{{\mathrm{2}}}  } } 
            }{\textsc{iTm-arrE}
            }
            \end{equation}
            From the second premise of the two above rules and by uniqueness (Lemma~\ref{lem:det:ty}), we get
            $ \intermed {  {\intermed \sigma}_{{\mathrm{1}}}  } = \intermed {  {\intermed \sigma}'_{{\mathrm{1}}}  } $.

            The induction hypothesis is:
            \begin{align*}
              \text{If}~ \intermed {  {\intermed \Sigma}  } ; \intermed {  {\intermed { \Gamma_C } }  } ; \intermed {  \bullet  }  \vdash_{tm}  \intermed {  {\intermed e}_{{\mathrm{1}}}  } : \intermed {  {\intermed \sigma}_{\ottmv{h}}  } \itot{ \rightsquigarrow  \target {  {\target e }_{\ottmv{p}}  } } \\
              \text{and}~ \intermed {  {\intermed \Sigma}  } ; \intermed {  {\intermed { \Gamma_C } }  } ; \intermed {  \bullet  }  \vdash_{tm}  \intermed {  {\intermed e}'_{{\mathrm{1}}}  } : \intermed {  {\intermed \sigma}_{\ottmv{h}}  } \itot{ \rightsquigarrow  \target {  {\target e }_{\ottmv{q}}  } } \\
              \text{for some}~\intermed{{\intermed \sigma}_{\ottmv{h}}},~\target{{\target e }_{\ottmv{p}}}~\text{and}~\target{{\target e }_{\ottmv{q}}},\nonumber\\
              \text{then, there is an}~\target{{\target e }'}~\text{such that}\nonumber\\
               \target {  {\target e }_{\ottmv{p}}  }  \longrightarrow^{*}  \target {  {\target e }'  } ~\text{and}~ \target {  {\target e }_{\ottmv{q}}  }  \longrightarrow  \target {  {\target e }'  } 
            \end{align*}
            Then, an appropriate choice for $\target{{\target e }_{\ottmv{h}}}$ is $\target{{\target e }' \, {\target e }_{{\mathrm{2}}}}$, since from Lemma~
            \pref{lem:tEval_app_dist}, we have $ \target {  {\target e }_{{\mathrm{1}}} \, {\target e }_{{\mathrm{2}}}  }  \longrightarrow^{*}  \target {  {\target e }' \, {\target e }_{{\mathrm{2}}}  } $ and
            $ \target {  {\target e }'_{{\mathrm{1}}} \, {\target e }_{{\mathrm{2}}}  }  \longrightarrow^{*}  \target {  {\target e }' \, {\target e }_{{\mathrm{2}}}  } $.\\
          }
\proofStep{Case \textsc{iEval-appAbs}}{\drule{iEval-appAbs}}
          {
            Hypotheses \pref{semelab:hyp1} and \pref{semelab:hyp2} of the theorem, adapted to
            this case, are:
            \begin{align}
              \label{semelab:appabs:der1}
                         & \intermed {  {\intermed \Sigma}  } ; \intermed {  {\intermed { \Gamma_C } }  } ; \intermed {  \bullet  }  \vdash_{tm}  \intermed {  \ottsym{(}  \lambda  {\intermed x}  \ottsym{:}  {\intermed \sigma}  \ottsym{.}  {\intermed e}_{{\mathrm{1}}}  \ottsym{)} \, {\intermed e}_{{\mathrm{2}}}  } : \intermed {  {\intermed \sigma}'  } \itot{ \rightsquigarrow  \target {  {\target e }_{{\mathrm{0}}}  } } \\
              \label{semelab:appabs:der2}
              \text{and}~& \intermed {  {\intermed \Sigma}  } ; \intermed {  {\intermed { \Gamma_C } }  } ; \intermed {  \bullet  }  \vdash_{tm}  \intermed {  [  {\intermed e}_{{\mathrm{2}}}  /  {\intermed x}  ]  {\intermed e}_{{\mathrm{1}}}  } : \intermed {  {\intermed \sigma}'  } \itot{ \rightsquigarrow  \target {  {\target e }'_{{\mathrm{0}}}  } } 
            \end{align}
            for some $\intermed{{\intermed \sigma}'}$, $\target{{\target e }_{{\mathrm{0}}}}$ and $\target{{\target e }'_{{\mathrm{0}}}}$. We need to show that
            there exists an \(\target{{\target e }_{\ottmv{h}}}\) such that \( \target {  {\target e }_{{\mathrm{0}}}  }  \longrightarrow^{*}  \target {  {\target e }_{\ottmv{h}}  } \) and
            \( \target {  {\target e }'_{{\mathrm{0}}}  }  \longrightarrow^{*}  \target {  {\target e }_{\ottmv{h}}  } \). We do this by showing that $ \target {  {\target e }_{{\mathrm{0}}}  }  \longrightarrow^{*}  \target {  {\target e }'_{{\mathrm{0}}}  } $.
            
            By inversion, the last part of Derivation \pref{semelab:appabs:der1} must be an
            instance of \textsc{iTm-arrI} directly followed by \textsc{iTm-arrE}, as shown
            below.
            \begin{equation*}
            \ottdrule{
              \ottpremise{
                \ottdrule{
                  \ottpremise{ \intermed {  {\intermed \Sigma}  } ; \intermed {  {\intermed { \Gamma_C } }  } ; \intermed {  \bullet  \ottsym{,}  {\intermed x}  \ottsym{:}  {\intermed \sigma}  }  \vdash_{tm}  \intermed {  {\intermed e}_{{\mathrm{1}}}  } : \intermed {  {\intermed \sigma}'  } \itot{ \rightsquigarrow  \target {  {\target e }_{{\mathrm{1}}}  } } }
                  \ottpremise{ \intermed {  {\intermed { \Gamma_C } }  } ; \intermed {  \bullet  }  \vdash_{ty}  \intermed {  {\intermed \sigma}  } \itot{ \rightsquigarrow  \target {  {\target \sigma }  } } }
                }{
                   \intermed {  {\intermed \Sigma}  } ; \intermed {  {\intermed { \Gamma_C } }  } ; \intermed {  \bullet  }  \vdash_{tm}  \intermed {  \lambda  {\intermed x}  \ottsym{:}  {\intermed \sigma}  \ottsym{.}  {\intermed e}_{{\mathrm{1}}}  } : \intermed {  {\intermed \sigma}  \rightarrow  {\intermed \sigma}'  } \itot{ \rightsquigarrow  \target {  \lambda  {\target x}  \ottsym{:}  {\target \sigma }  \ottsym{.}  {\target e }_{{\mathrm{1}}}  } } 
                }{\textsc{iTm-arrI}}}
              \ottpremise{ \intermed {  {\intermed \Sigma}  } ; \intermed {  {\intermed { \Gamma_C } }  } ; \intermed {  \bullet  }  \vdash_{tm}  \intermed {  {\intermed e}_{{\mathrm{2}}}  } : \intermed {  {\intermed \sigma}  } \itot{ \rightsquigarrow  \target {  {\target e }_{{\mathrm{2}}}  } } }
            }{
               \intermed {  {\intermed \Sigma}  } ; \intermed {  {\intermed { \Gamma_C } }  } ; \intermed {  \bullet  }  \vdash_{tm}  \intermed {  \ottsym{(}  \lambda  {\intermed x}  \ottsym{:}  {\intermed \sigma}  \ottsym{.}  {\intermed e}_{{\mathrm{1}}}  \ottsym{)} \, {\intermed e}_{{\mathrm{2}}}  } : \intermed {  {\intermed \sigma}'  } \itot{ \rightsquigarrow  \target {  \ottsym{(}  \lambda  {\target x}  \ottsym{:}  {\target \sigma }  \ottsym{.}  {\target e }_{{\mathrm{1}}}  \ottsym{)} \, {\target e }_{{\mathrm{2}}}  } } 
            }{\textsc{iTm-arrE}
            }
            \end{equation*}
            where $ \target {  {\target e }_{{\mathrm{0}}}  } = \target {  \ottsym{(}  \lambda  {\target x}  \ottsym{:}  {\target \sigma }  \ottsym{.}  {\target e }_{{\mathrm{1}}}  \ottsym{)} \, {\target e }_{{\mathrm{2}}}  } $. From the above equation, we can use
            premises
            \begin{align*}
                         & \intermed {  {\intermed \Sigma}  } ; \intermed {  {\intermed { \Gamma_C } }  } ; \intermed {  \bullet  \ottsym{,}  {\intermed x}  \ottsym{:}  {\intermed \sigma}  }  \vdash_{tm}  \intermed {  {\intermed e}_{{\mathrm{1}}}  } : \intermed {  {\intermed \sigma}'  } \itot{ \rightsquigarrow  \target {  {\target e }_{{\mathrm{1}}}  } } \\
              \text{and}~& \intermed {  {\intermed \Sigma}  } ; \intermed {  {\intermed { \Gamma_C } }  } ; \intermed {  \bullet  }  \vdash_{tm}  \intermed {  {\intermed e}_{{\mathrm{2}}}  } : \intermed {  {\intermed \sigma}  } \itot{ \rightsquigarrow  \target {  {\target e }_{{\mathrm{2}}}  } } 
            \end{align*}
            in Lemma~\ref{lemma:var_subst}, to obtain
            \begin{equation*}
               \intermed {  {\intermed \Sigma}  } ; \intermed {  {\intermed { \Gamma_C } }  } ; \intermed {  \bullet  }  \vdash_{tm}  \intermed {  [  {\intermed e}_{{\mathrm{2}}}  /  {\intermed x}  ]  {\intermed e}_{{\mathrm{1}}}  } : \intermed {  {\intermed \sigma}'  } \itot{ \rightsquigarrow  \target {  [  {\target e }_{{\mathrm{1}}}  /  {\target x}  ]  {\target e }_{{\mathrm{2}}}  } } 
            \end{equation*}
            Then, by uniqueness (Theorem~\ref{thmA:tmelab:det} on the latter and on
            Equation~\pref{semelab:appabs:der2}), we have $ \target {  {\target e }'_{{\mathrm{0}}}  } = \target {  [  {\target e }_{{\mathrm{1}}}  /  {\target x}  ]  {\target e }_{{\mathrm{2}}}  } $.

            We set $ \target {  {\target e }_{\ottmv{h}}  } = \target {  [  {\target e }_{{\mathrm{1}}}  /  {\target x}  ]  {\target e }_{{\mathrm{2}}}  } $, since
            $ \target {  \ottsym{(}  \lambda  {\target x}  \ottsym{:}  {\target \sigma }  \ottsym{.}  {\target e }_{{\mathrm{1}}}  \ottsym{)} \, {\target e }_{{\mathrm{2}}}  }  \longrightarrow^{*}  \target {  [  {\target e }_{{\mathrm{1}}}  /  {\target x}  ]  {\target e }_{{\mathrm{2}}}  } $, by evaluation rule
            \textsc{tEval-AppAbs}, and $ \target {  [  {\target e }_{{\mathrm{1}}}  /  {\target x}  ]  {\target e }_{{\mathrm{2}}}  }  \longrightarrow^{*}  \target {  [  {\target e }_{{\mathrm{1}}}  /  {\target x}  ]  {\target e }_{{\mathrm{2}}}  } $, by
            reflexivity of $ \longrightarrow^{*} $.\\
          }
\proofStep{Case \textsc{iEval-tyApp}}{\drule{iEval-tyApp}}
          {Hypotheses \pref{semelab:hyp1} and \pref{semelab:hyp2} of the theorem, adapted to
            this case, are:
            \begin{align*}
                         & \intermed {  {\intermed \Sigma}  } ; \intermed {  {\intermed { \Gamma_C } }  } ; \intermed {  \bullet  }  \vdash_{tm}  \intermed {  {\intermed e} \, {\intermed \sigma}  } : \intermed {  [  {\intermed \sigma}  /  {\intermed a }  ]  {\intermed \sigma}_{\ottmv{p}}  } \itot{ \rightsquigarrow  \target {  {\target e } \, {\target \sigma }  } } \\
              \text{and}~& \intermed {  {\intermed \Sigma}  } ; \intermed {  {\intermed { \Gamma_C } }  } ; \intermed {  \bullet  }  \vdash_{tm}  \intermed {  {\intermed e}' \, {\intermed \sigma}  } : \intermed {  [  {\intermed \sigma}  /  {\intermed a }'  ]  {\intermed \sigma}_{\ottmv{q}}  } \itot{ \rightsquigarrow  \target {  {\target e }' \, {\target \sigma }  } } 
            \end{align*}
            for some $\intermed{{\intermed a }}$, $\intermed{{\intermed a }'}$, $\intermed{{\intermed \sigma}_{\ottmv{p}}}$, $\intermed{{\intermed \sigma}_{\ottmv{q}}}$, ${\target e }$, $\target{{\target e }'}$ and ${\target \sigma }$.

            The last rule of both derivations above must be instances of \textsc{iTm-forallE}.
            For the first, we have:
            \begin{equation}
              \label{semelab:tyapp:der1}
              \ottdrule{
                \ottpremise{ \intermed {  {\intermed \Sigma}  } ; \intermed {  {\intermed { \Gamma_C } }  } ; \intermed {  \bullet  }  \vdash_{tm}  \intermed {  {\intermed e}  } : \intermed {  \forall  {\intermed a }  \ottsym{.}  {\intermed \sigma}_{\ottmv{p}}  } \itot{ \rightsquigarrow  \target {  {\target e }  } } }
                \ottpremise{ \intermed {  {\intermed { \Gamma_C } }  } ; \intermed {  \bullet  }  \vdash_{ty}  \intermed {  {\intermed \sigma}  } \itot{ \rightsquigarrow  \target {  {\target \sigma }  } } }
              }{
                 \intermed {  {\intermed \Sigma}  } ; \intermed {  {\intermed { \Gamma_C } }  } ; \intermed {  \bullet  }  \vdash_{tm}  \intermed {  {\intermed e} \, {\intermed \sigma}  } : \intermed {  [  {\intermed \sigma}  /  {\intermed a }  ]  {\intermed \sigma}_{\ottmv{p}}  } \itot{ \rightsquigarrow  \target {  {\target e } \, {\target \sigma }  } } 
              }{\textsc{iTm-forallE}}
            \end{equation}
            and for the second:
            \begin{equation}
              \label{semelab:tyapp:der2}
              \ottdrule{
                \ottpremise{ \intermed {  {\intermed \Sigma}  } ; \intermed {  {\intermed { \Gamma_C } }  } ; \intermed {  \bullet  }  \vdash_{tm}  \intermed {  {\intermed e}'  } : \intermed {  \forall  {\intermed a }'  \ottsym{.}  {\intermed \sigma}_{\ottmv{q}}  } \itot{ \rightsquigarrow  \target {  {\target e }'  } } }
                \ottpremise{ \intermed {  {\intermed { \Gamma_C } }  } ; \intermed {  \bullet  }  \vdash_{ty}  \intermed {  {\intermed \sigma}  } \itot{ \rightsquigarrow  \target {  {\target \sigma }  } } }
              }{
                 \intermed {  {\intermed \Sigma}  } ; \intermed {  {\intermed { \Gamma_C } }  } ; \intermed {  \bullet  }  \vdash_{tm}  \intermed {  {\intermed e} \, {\intermed \sigma}  } : \intermed {  [  {\intermed \sigma}  /  {\intermed a }'  ]  {\intermed \sigma}_{\ottmv{q}}  } \itot{ \rightsquigarrow  \target {  {\target e }' \, {\target \sigma }  } } 
              }{\textsc{iTm-forallE}}
            \end{equation}
            By applying Theorem~\ref{thmA:intermediate_preservation} on the first premise of
            Equation \pref{semelab:tyapp:der1} and on the premise of rule \textsc{iEval-tyApp},
            we have that $ \intermed {  \forall  {\intermed a }  \ottsym{.}  {\intermed \sigma}_{\ottmv{p}}  } = \intermed {  \forall  {\intermed a }'  \ottsym{.}  {\intermed \sigma}_{\ottmv{q}}  } $.

            The induction hypothesis is:
            \begin{align*}
              \text{If}~ \intermed {  {\intermed \Sigma}  } ; \intermed {  {\intermed { \Gamma_C } }  } ; \intermed {  \bullet  }  \vdash_{tm}  \intermed {  {\intermed e}_{{\mathrm{1}}}  } : \intermed {  {\intermed \sigma}_{\ottmv{h}}  } \itot{ \rightsquigarrow  \target {  {\target e }_{\ottmv{p}}  } } \\
              \text{and}~ \intermed {  {\intermed \Sigma}  } ; \intermed {  {\intermed { \Gamma_C } }  } ; \intermed {  \bullet  }  \vdash_{tm}  \intermed {  {\intermed e}'_{{\mathrm{1}}}  } : \intermed {  {\intermed \sigma}_{\ottmv{h}}  } \itot{ \rightsquigarrow  \target {  {\target e }_{\ottmv{q}}  } } \\
              \text{for some}~\intermed{{\intermed \sigma}_{\ottmv{h}}},~\target{{\target e }_{\ottmv{p}}}~\text{and}~\target{{\target e }_{\ottmv{q}}},\nonumber\\
              \text{then, there is an}~\target{{\target e }'_{\ottmv{h}}}~\text{such that}\nonumber\\
               \target {  {\target e }_{\ottmv{p}}  }  \longrightarrow^{*}  \target {  {\target e }'_{\ottmv{h}}  } ~\text{and}~ \target {  {\target e }_{\ottmv{q}}  }  \longrightarrow  \target {  {\target e }'_{\ottmv{h}}  } 
            \end{align*}
            For $ \intermed {  {\intermed \sigma}_{\ottmv{h}}  } = \intermed {  \forall  {\intermed a }  \ottsym{.}  {\intermed \sigma}_{\ottmv{p}}  } $, $ \target {  {\target e }_{\ottmv{p}}  } = \target {  {\target e }  } $ and $ \target {  {\target e }_{\ottmv{q}}  } = \target {  {\target e }'  } $, the two
            conditions are fulfilled by the first premise of Equations
            \pref{semelab:tyapp:der1} and \pref{semelab:tyapp:der2}.

            We choose $ \target {  {\target e }_{\ottmv{h}}  } = \target {  {\target e }'_{\ottmv{h}} \, {\target \sigma }  } $, because from Lemma~\pref{lem:tEval_tapp_dist},
            we have $ \target {  {\target e } \, {\target \sigma }  }  \longrightarrow^{*}  \target {  {\target e }'_{\ottmv{h}} \, {\target \sigma }  } $ and $ \target {  {\target e }' \, {\target \sigma }  }  \longrightarrow^{*}  \target {  {\target e }'_{\ottmv{h}} \, {\target \sigma }  } $.\\
          }
\proofStep{Case \textsc{iEval-tyAppAbs}}{\drule{iEval-tyAppAbs}}
          {
            Hypotheses \pref{semelab:hyp1} and \pref{semelab:hyp2} of the theorem, adapted to
            this case, are:
            \begin{align}
              \label{semelab:tyappabs:der1}
                         & \intermed {  {\intermed \Sigma}  } ; \intermed {  {\intermed { \Gamma_C } }  } ; \intermed {  \bullet  }  \vdash_{tm}  \intermed {  \ottsym{(}  \Lambda  {\intermed a }  \ottsym{.}  {\intermed e}  \ottsym{)} \, {\intermed \sigma}  } : \intermed {  {\intermed \sigma}_{{\mathrm{0}}}  } \itot{ \rightsquigarrow  \target {  {\target e }_{{\mathrm{0}}}  } } \\
              \label{semelab:tyappabs:der2}
              \text{and}~& \intermed {  {\intermed \Sigma}  } ; \intermed {  {\intermed { \Gamma_C } }  } ; \intermed {  \bullet  }  \vdash_{tm}  \intermed {  [  {\intermed \sigma}  /  {\intermed a }  ]  {\intermed e}  } : \intermed {  {\intermed \sigma}_{{\mathrm{0}}}  } \itot{ \rightsquigarrow  \target {  {\target e }'_{{\mathrm{0}}}  } } 
            \end{align}
            for some $\intermed{{\intermed \sigma}_{{\mathrm{0}}}}$, $\target{{\target e }_{{\mathrm{0}}}}$ and $\target{{\target e }'_{{\mathrm{0}}}}$.
            We need to show that
            there exists an \(\target{{\target e }_{\ottmv{h}}}\) such that \( \target {  {\target e }_{{\mathrm{0}}}  }  \longrightarrow^{*}  \target {  {\target e }_{\ottmv{h}}  } \) and
            \( \target {  {\target e }'_{{\mathrm{0}}}  }  \longrightarrow^{*}  \target {  {\target e }_{\ottmv{h}}  } \). We do this by showing that $ \target {  {\target e }_{{\mathrm{0}}}  }  \longrightarrow^{*}  \target {  {\target e }'_{{\mathrm{0}}}  } $.
            
            By inversion, the last part of Derivation \pref{semelab:tyappabs:der1} must be an
            instance of \textsc{iTm-forallI} directly followed by \textsc{iTm-forallE}, as shown
            below.
            \begin{equation*}
            \ottdrule{
              \ottpremise{
                \ottdrule{
                  \ottpremise{ \intermed {  {\intermed \Sigma}  } ; \intermed {  {\intermed { \Gamma_C } }  } ; \intermed {  \bullet  \ottsym{,}  {\intermed a }  }  \vdash_{tm}  \intermed {  {\intermed e}  } : \intermed {  {\intermed \sigma}'  } \itot{ \rightsquigarrow  \target {  {\target e }  } } }
                }{
                   \intermed {  {\intermed \Sigma}  } ; \intermed {  {\intermed { \Gamma_C } }  } ; \intermed {  \bullet  }  \vdash_{tm}  \intermed {  \Lambda  {\intermed a }  \ottsym{.}  {\intermed e}  } : \intermed {  \forall  {\intermed a }  \ottsym{.}  {\intermed \sigma}'  } \itot{ \rightsquigarrow  \target {  \Lambda  {\target a }  \ottsym{.}  {\target e }  } } 
                }{\textsc{iTm-forallI}}}
              \ottpremise{ \intermed {  {\intermed { \Gamma_C } }  } ; \intermed {  \bullet  }  \vdash_{ty}  \intermed {  {\intermed \sigma}  } \itot{ \rightsquigarrow  \target {  {\target \sigma }  } } }
            }{
               \intermed {  {\intermed \Sigma}  } ; \intermed {  {\intermed { \Gamma_C } }  } ; \intermed {  \bullet  }  \vdash_{tm}  \intermed {  \ottsym{(}  \Lambda  {\intermed a }  \ottsym{.}  {\intermed e}  \ottsym{)} \, {\intermed \sigma}  } : \intermed {  [  {\intermed \sigma}  /  {\intermed a }  ]  {\intermed \sigma}'  } \itot{ \rightsquigarrow  \target {  \ottsym{(}  \Lambda  {\target a }  \ottsym{.}  {\target e }  \ottsym{)} \, {\target \sigma }  } } 
            }{\textsc{iTm-forallE}
            }
            \end{equation*}
            where $ \intermed {  {\intermed \sigma}_{{\mathrm{0}}}  } = \intermed {  [  {\intermed \sigma}  /  {\intermed a }  ]  {\intermed \sigma}'  } $ and $ \target {  {\target e }_{{\mathrm{0}}}  } = \target {  \ottsym{(}  \Lambda  {\target a }  \ottsym{.}  {\target e }  \ottsym{)} \, {\target \sigma }  } $. From
            the above equation, we can use premises
            \begin{align*}
                         & \intermed {  {\intermed \Sigma}  } ; \intermed {  {\intermed { \Gamma_C } }  } ; \intermed {  \bullet  \ottsym{,}  {\intermed a }  }  \vdash_{tm}  \intermed {  {\intermed e}  } : \intermed {  {\intermed \sigma}'  } \itot{ \rightsquigarrow  \target {  {\target e }  } } \\
              \text{and}~& \intermed {  {\intermed { \Gamma_C } }  } ; \intermed {  \bullet  }  \vdash_{ty}  \intermed {  {\intermed \sigma}  } \itot{ \rightsquigarrow  \target {  {\target \sigma }  } } 
            \end{align*}
            in Lemma~\ref{lemma:tvar_subst}, to obtain
            \begin{equation*}
               \intermed {  {\intermed \Sigma}  } ; \intermed {  {\intermed { \Gamma_C } }  } ; \intermed {  \bullet  }  \vdash_{tm}  \intermed {  [  {\intermed \sigma}  /  {\intermed a }  ]  {\intermed e}  } : \intermed {  [  {\intermed \sigma}  /  {\intermed a }  ]  {\intermed \sigma}'  } \itot{ \rightsquigarrow  \target {  [  {\target \sigma }  /  {\target a }  ]  {\target e }  } } 
            \end{equation*}
            Then, by uniqueness (Theorem~\ref{thmA:tmelab:det} on the latter and on
            Equation~\pref{semelab:tyappabs:der2}), we have $ \target {  {\target e }'_{{\mathrm{0}}}  } = \target {  [  {\target \sigma }  /  {\target a }  ]  {\target e }  } $.

            We set $ \target {  {\target e }_{\ottmv{h}}  } = \target {  [  {\target \sigma }  /  {\target a }  ]  {\target e }  } $, since
            $ \target {  \ottsym{(}  \Lambda  {\target a }  \ottsym{.}  {\target e }  \ottsym{)} \, {\target \sigma }  }  \longrightarrow^{*}  \target {  [  {\target \sigma }  /  {\target a }  ]  {\target e }  } $, by evaluation rule
            \textsc{tEval-TAppAbs}, and $ \target {  [  {\target \sigma }  /  {\target a }  ]  {\target e }  }  \longrightarrow^{*}  \target {  [  {\target \sigma }  /  {\target a }  ]  {\target e }  } $, by
            reflexivity of $ \longrightarrow^{*} $.\\
          }
\proofStep{Case \textsc{iEval-DApp}}{\drule{iEval-DApp}}
          {
            Hypotheses \pref{semelab:hyp1} and \pref{semelab:hyp2} of the theorem, adapted to
            this case, are:
            \begin{align*}
                         & \intermed {  {\intermed \Sigma}  } ; \intermed {  {\intermed { \Gamma_C } }  } ; \intermed {  \bullet  }  \vdash_{tm}  \intermed {  {\intermed e} \, {\intermed d}  } : \intermed {  {\intermed \sigma}  } \itot{ \rightsquigarrow  \target {  {\target e }_{{\mathrm{1}}} \, {\target e }_{{\mathrm{2}}}  } } \\
              \text{and}~& \intermed {  {\intermed \Sigma}  } ; \intermed {  {\intermed { \Gamma_C } }  } ; \intermed {  \bullet  }  \vdash_{tm}  \intermed {  {\intermed e}' \, {\intermed d}  } : \intermed {  {\intermed \sigma}  } \itot{ \rightsquigarrow  \target {  {\target e }'_{{\mathrm{1}}} \, {\target e }_{{\mathrm{2}}}  } } 
            \end{align*}
            for some $\target{{\target e }_{{\mathrm{1}}}}$, $\target{{\target e }'_{{\mathrm{1}}}}$ and $\target{{\target e }_{{\mathrm{2}}}}$.
            
            The last rule of these derivations must be instances of \textsc{iTm-constrE}. For
            Hypothesis \pref{semelab:hyp1}, we have:
            \begin{equation}
              \label{semelab:dapp:der1}
            \ottdrule{
              \ottpremise{ \intermed {  {\intermed \Sigma}  } ; \intermed {  {\intermed { \Gamma_C } }  } ; \intermed {  \bullet  }  \vdash_{tm}  \intermed {  {\intermed e}  } : \intermed {   {\intermed Q}  \Rightarrow  {\intermed \sigma}   } \itot{ \rightsquigarrow  \target {  {\target e }_{{\mathrm{1}}}  } } }
              \ottpremise{ \intermed {  {\intermed \Sigma}  } ; \intermed {  {\intermed { \Gamma_C } }  } ; \intermed {  \bullet  }  \vdash_{d}  \intermed {  {\intermed d}  } : \intermed {  {\intermed Q}  } \itot{\,  \rightsquigarrow  \target {  {\target e }_{{\mathrm{2}}}  } } }
            }{
               \intermed {  {\intermed \Sigma}  } ; \intermed {  {\intermed { \Gamma_C } }  } ; \intermed {  \bullet  }  \vdash_{tm}  \intermed {  {\intermed e} \, {\intermed d}  } : \intermed {  {\intermed \sigma}  } \itot{ \rightsquigarrow  \target {  {\target e }_{{\mathrm{1}}} \, {\target e }_{{\mathrm{2}}}  } } 
            }{\textsc{iTm-constrE}
            }
            \end{equation}
            and for Hypothesis \pref{semelab:hyp2}, we have:
            \begin{equation}
              \label{semelab:dapp:der2}
            \ottdrule['']{
              \ottpremise{ \intermed {  {\intermed \Sigma}  } ; \intermed {  {\intermed { \Gamma_C } }  } ; \intermed {  \bullet  }  \vdash_{tm}  \intermed {  {\intermed e}'  } : \intermed {   {\intermed Q}'  \Rightarrow  {\intermed \sigma}   } \itot{ \rightsquigarrow  \target {  {\target e }'_{{\mathrm{1}}}  } } }
              \ottpremise{ \intermed {  {\intermed \Sigma}  } ; \intermed {  {\intermed { \Gamma_C } }  } ; \intermed {  \bullet  }  \vdash_{d}  \intermed {  {\intermed d}  } : \intermed {  {\intermed Q}'  } \itot{\,  \rightsquigarrow  \target {  {\target e }_{{\mathrm{2}}}  } } }
            }{
               \intermed {  {\intermed \Sigma}  } ; \intermed {  {\intermed { \Gamma_C } }  } ; \intermed {  \bullet  }  \vdash_{tm}  \intermed {  {\intermed e}'_{{\mathrm{1}}} \, {\intermed d}  } : \intermed {  {\intermed \sigma}  } \itot{ \rightsquigarrow  \target {  {\target e }'_{{\mathrm{1}}} \, {\target e }_{{\mathrm{2}}}  } } 
            }{\textsc{iTm-constrE}
            }
            \end{equation}
            From the second premise of the two above rules and by uniqueness, we get
            $ \intermed {  {\intermed Q}  } = \intermed {  {\intermed Q}'  } $.

            The induction hypothesis is:
            \begin{align*}
              \text{If}~ \intermed {  {\intermed \Sigma}  } ; \intermed {  {\intermed { \Gamma_C } }  } ; \intermed {  \bullet  }  \vdash_{tm}  \intermed {  {\intermed e}  } : \intermed {  {\intermed \sigma}_{\ottmv{h}}  } \itot{ \rightsquigarrow  \target {  {\target e }_{\ottmv{p}}  } } \\
              \text{and}~ \intermed {  {\intermed \Sigma}  } ; \intermed {  {\intermed { \Gamma_C } }  } ; \intermed {  \bullet  }  \vdash_{tm}  \intermed {  {\intermed e}'  } : \intermed {  {\intermed \sigma}_{\ottmv{h}}  } \itot{ \rightsquigarrow  \target {  {\target e }_{\ottmv{q}}  } } \\
              \text{for some}~\intermed{{\intermed \sigma}_{\ottmv{h}}},~\target{{\target e }_{\ottmv{p}}}~\text{and}~\target{{\target e }_{\ottmv{q}}},\nonumber\\
              \text{then, there is an}~\target{{\target e }'_{\ottmv{h}}}~\text{such that}\nonumber\\
               \target {  {\target e }_{\ottmv{p}}  }  \longrightarrow^{*}  \target {  {\target e }'_{\ottmv{h}}  } ~\text{and}~ \target {  {\target e }_{\ottmv{q}}  }  \longrightarrow  \target {  {\target e }'_{\ottmv{h}}  } 
            \end{align*}
            Then, the conditions of the above hold from the first premise of Derivations
            \pref{semelab:dapp:der1} and \pref{semelab:dapp:der2}, where
            $ \intermed {  {\intermed \sigma}_{\ottmv{h}}  } = \intermed {   {\intermed Q}  \Rightarrow  {\intermed \sigma}   } $, $ \target {  {\target e }_{\ottmv{p}}  } = \target {  {\target e }_{{\mathrm{1}}}  } $ and $ \target {  {\target e }_{\ottmv{q}}  } = \target {  {\target e }'_{{\mathrm{1}}}  } $.

            Then, an appropriate choice for $\target{{\target e }_{\ottmv{h}}}$ is $\target{{\target e }'_{\ottmv{h}} \, {\target e }_{{\mathrm{2}}}}$, since from Lemma~
            \pref{lem:tEval_app_dist}, we have $ \target {  {\target e }_{{\mathrm{1}}} \, {\target e }_{{\mathrm{2}}}  }  \longrightarrow^{*}  \target {  {\target e }'_{\ottmv{h}} \, {\target e }_{{\mathrm{2}}}  } $ and
            $ \target {  {\target e }'_{{\mathrm{1}}} \, {\target e }_{{\mathrm{2}}}  }  \longrightarrow^{*}  \target {  {\target e }'_{\ottmv{h}} \, {\target e }_{{\mathrm{2}}}  } $.\\
          }
\proofStep{Case \textsc{iEval-DAppAbs}}{\drule{iEval-DAppAbs}}
          {
            Hypotheses \pref{semelab:hyp1} and \pref{semelab:hyp2} of the theorem, adapted to
            this case, are:
            \begin{align}
              \label{semelab:dappabs:der1}
                         & \intermed {  {\intermed \Sigma}  } ; \intermed {  {\intermed { \Gamma_C } }  } ; \intermed {  \bullet  }  \vdash_{tm}  \intermed {  \ottsym{(}  \lambda  {\intermed \delta }  \ottsym{:}  {\intermed Q}  \ottsym{.}  {\intermed e}  \ottsym{)} \, {\intermed d}  } : \intermed {  {\intermed \sigma}  } \itot{ \rightsquigarrow  \target {  {\target e }_{{\mathrm{0}}}  } } \\
              \label{semelab:dappabs:der2}
              \text{and}~& \intermed {  {\intermed \Sigma}  } ; \intermed {  {\intermed { \Gamma_C } }  } ; \intermed {  \bullet  }  \vdash_{tm}  \intermed {  [  {\intermed d}  /  {\intermed \delta }  ]  {\intermed e}  } : \intermed {  {\intermed \sigma}  } \itot{ \rightsquigarrow  \target {  {\target e }'_{{\mathrm{0}}}  } } 
            \end{align}
            for some $\intermed{{\intermed \sigma}'}$, $\target{{\target e }_{{\mathrm{0}}}}$ and $\target{{\target e }'_{{\mathrm{0}}}}$.
            We need to show that
            there exists an \(\target{{\target e }_{\ottmv{h}}}\) such that \( \target {  {\target e }_{{\mathrm{0}}}  }  \longrightarrow^{*}  \target {  {\target e }_{\ottmv{h}}  } \) and
            \( \target {  {\target e }'_{{\mathrm{0}}}  }  \longrightarrow^{*}  \target {  {\target e }_{\ottmv{h}}  } \). We do this by showing that $ \target {  {\target e }_{{\mathrm{0}}}  }  \longrightarrow^{*}  \target {  {\target e }'_{{\mathrm{0}}}  } $.
            
            By inversion, the last part of Derivation \pref{semelab:dappabs:der1} must be an
            instance of \textsc{iTm-constrI} directly followed by \textsc{iTm-constrE}, as shown
            below.
            \begin{equation*}
            \ottdrule{
              \ottpremise{
                \ottdrule{
                  \ottpremise{ \intermed {  {\intermed \Sigma}  } ; \intermed {  {\intermed { \Gamma_C } }  } ; \intermed {  \bullet  \ottsym{,}  {\intermed \delta }  \ottsym{:}  {\intermed Q}  }  \vdash_{tm}  \intermed {  {\intermed e}  } : \intermed {  {\intermed \sigma}  } \itot{ \rightsquigarrow  \target {  {\target e }_{{\mathrm{1}}}  } } }
                  \ottpremise{ \intermed {  {\intermed { \Gamma_C } }  } ; \intermed {  \bullet  }  \vdash_{ty}  \intermed {  {\intermed \sigma}  } \itot{ \rightsquigarrow  \target {  {\target \sigma }  } } }
                }{
                   \intermed {  {\intermed \Sigma}  } ; \intermed {  {\intermed { \Gamma_C } }  } ; \intermed {  \bullet  }  \vdash_{tm}  \intermed {  \lambda  {\intermed \delta }  \ottsym{:}  {\intermed Q}  \ottsym{.}  {\intermed e}  } : \intermed {   {\intermed Q}  \Rightarrow  {\intermed \sigma}   } \itot{ \rightsquigarrow  \target {  \lambda  {\target \delta }  \ottsym{:}  {\target \sigma }  \ottsym{.}  {\target e }_{{\mathrm{1}}}  } } 
                }{\textsc{iTm-constrI}}}
              \ottpremise{ \intermed {  {\intermed \Sigma}  } ; \intermed {  {\intermed { \Gamma_C } }  } ; \intermed {  \bullet  }  \vdash_{d}  \intermed {  {\intermed d}  } : \intermed {  {\intermed Q}  } \itot{\,  \rightsquigarrow  \target {  {\target e }_{{\mathrm{2}}}  } } }
            }{
               \intermed {  {\intermed \Sigma}  } ; \intermed {  {\intermed { \Gamma_C } }  } ; \intermed {  \bullet  }  \vdash_{tm}  \intermed {  \ottsym{(}  \lambda  {\intermed \delta }  \ottsym{:}  {\intermed Q}  \ottsym{.}  {\intermed e}  \ottsym{)} \, {\intermed d}  } : \intermed {  {\intermed \sigma}  } \itot{ \rightsquigarrow  \target {  \ottsym{(}  \lambda  {\target \delta }  \ottsym{:}  {\target \sigma }  \ottsym{.}  {\target e }_{{\mathrm{1}}}  \ottsym{)} \, {\target e }_{{\mathrm{2}}}  } } 
            }{\textsc{iTm-constrE}
            }
            \end{equation*}
            where $ \target {  {\target e }_{{\mathrm{0}}}  } = \target {  \ottsym{(}  \lambda  {\target \delta }  \ottsym{:}  {\target \sigma }  \ottsym{.}  {\target e }_{{\mathrm{1}}}  \ottsym{)} \, {\target e }_{{\mathrm{2}}}  } $. From the above equation, we can use
            premises
            \begin{align*}
                         & \intermed {  {\intermed \Sigma}  } ; \intermed {  {\intermed { \Gamma_C } }  } ; \intermed {  \bullet  \ottsym{,}  {\intermed \delta }  \ottsym{:}  {\intermed Q}  }  \vdash_{tm}  \intermed {  {\intermed e}  } : \intermed {  {\intermed \sigma}  } \itot{ \rightsquigarrow  \target {  {\target e }_{{\mathrm{1}}}  } } \\
              \text{and}~& \intermed {  {\intermed \Sigma}  } ; \intermed {  {\intermed { \Gamma_C } }  } ; \intermed {  \bullet  }  \vdash_{d}  \intermed {  {\intermed d}  } : \intermed {  {\intermed Q}  } \itot{\,  \rightsquigarrow  \target {  {\target e }_{{\mathrm{2}}}  } } 
            \end{align*}
            in Lemma~\ref{lemma:dvar_subst}, to obtain
            \begin{equation*}
               \intermed {  {\intermed \Sigma}  } ; \intermed {  {\intermed { \Gamma_C } }  } ; \intermed {  \bullet  }  \vdash_{tm}  \intermed {  [  {\intermed d}  /  {\intermed \delta }  ]  {\intermed e}  } : \intermed {  {\intermed \sigma}  } \itot{ \rightsquigarrow  \target {  [  {\target e }_{{\mathrm{1}}}  /  {\target \delta }  ]  {\target e }_{{\mathrm{2}}}  } } 
            \end{equation*}
            Then, by uniqueness (Theorem~\ref{thmA:delab:det} on the latter and on
            Equation~\pref{semelab:dappabs:der2}), we have $ \target {  {\target e }'_{{\mathrm{0}}}  } = \target {  [  {\target e }_{{\mathrm{1}}}  /  {\target \delta }  ]  {\target e }_{{\mathrm{2}}}  } $.

            We set $ \target {  {\target e }_{\ottmv{h}}  } = \target {  [  {\target e }_{{\mathrm{1}}}  /  {\target \delta }  ]  {\target e }_{{\mathrm{2}}}  } $, since
            $ \target {  \ottsym{(}  \lambda  {\target \delta }  \ottsym{:}  {\target \sigma }  \ottsym{.}  {\target e }_{{\mathrm{1}}}  \ottsym{)} \, {\target e }_{{\mathrm{2}}}  }  \longrightarrow^{*}  \target {  [  {\target e }_{{\mathrm{1}}}  /  {\target \delta }  ]  {\target e }_{{\mathrm{2}}}  } $, by evaluation rule
            \textsc{tEval-AppAbs}, and $ \target {  [  {\target e }_{{\mathrm{1}}}  /  {\target \delta }  ]  {\target e }_{{\mathrm{2}}}  }  \longrightarrow^{*}  \target {  [  {\target e }_{{\mathrm{1}}}  /  {\target \delta }  ]  {\target e }_{{\mathrm{2}}}  } $, by
            reflexivity of $ \longrightarrow^{*} $.\\
          }
\proofStep{Case \textsc{iEval-method}}
          {
            \ottdrule{
              \ottpremise{ ( \intermed {  {\intermed D }  } : \intermed {  \forall  {\overline {\intermed a} }_{\ottmv{j}}  \ottsym{.}  {\overline {\intermed Q} }_{\ottmv{i}}  \Rightarrow  {\intermed {\mathit{TC} } } \, {\intermed \sigma}_{\ottmv{q}}  } ) . \intermed {  {\intermed m}  }  \mapsto  \intermed {  {\intermed e}  }  \in  \intermed {  {\intermed \Sigma}  } }
            }{ \intermed {  {\intermed \Sigma}  }  \vdash  \intermed {  \ottsym{(}  {\intermed D } \, \overline {\intermed \sigma}_{\ottmv{j}} \, {\overline {\intermed d} }_{\ottmv{i}}  \ottsym{)}  \ottsym{.}  {\intermed m}  }  \longrightarrow  \intermed {  {\intermed e} \, \overline {\intermed \sigma}_{\ottmv{j}} \, {\overline {\intermed d} }_{\ottmv{i}}  } 
            }{\textsc{iEval-method}}
          }
          {
            Hypotheses \pref{semelab:hyp1} and \pref{semelab:hyp2} of the theorem, adapted to
            this case, are:
            \begin{align}
              \label{semelab:meth:der1}
                         & \intermed {  {\intermed \Sigma}  } ; \intermed {  {\intermed { \Gamma_C } }  } ; \intermed {  \bullet  }  \vdash_{tm}  \intermed {  \ottsym{(}  {\intermed D } \, \overline {\intermed \sigma}_{\ottmv{j}} \, {\overline {\intermed d} }_{\ottmv{i}}  \ottsym{)}  \ottsym{.}  {\intermed m}  } : \intermed {  {\intermed \sigma}_{{\mathrm{0}}}  } \itot{ \rightsquigarrow  \target {  {\target e }_{{\mathrm{0}}}  } } \\
              \label{semelab:meth:der2}
              \text{and}~& \intermed {  {\intermed \Sigma}  } ; \intermed {  {\intermed { \Gamma_C } }  } ; \intermed {  \bullet  }  \vdash_{tm}  \intermed {  {\intermed e} \, \overline {\intermed \sigma}_{\ottmv{j}} \, {\overline {\intermed d} }_{\ottmv{i}}  } : \intermed {  {\intermed \sigma}_{{\mathrm{0}}}  } \itot{ \rightsquigarrow  \target {  {\target e }'_{{\mathrm{0}}}  } } 
            \end{align}
            for some $\intermed{{\intermed \sigma}_{{\mathrm{0}}}}$, $\target{{\target e }_{{\mathrm{0}}}}$ and $\target{{\target e }'_{{\mathrm{0}}}}$. We need to show that there is a
            $\target{{\target e }_{\ottmv{h}}}$ such that $ \target {  {\target e }_{{\mathrm{0}}}  }  \longrightarrow^{*}  \target {  {\target e }_{\ottmv{h}}  } $ and $ \target {  {\target e }'_{{\mathrm{0}}}  }  \longrightarrow^{*}  \target {  {\target e }_{\ottmv{h}}  } $.
            
            By inversion on Equation~\pref{semelab:meth:der1}, we have
            $ \target {  {\target e }_{{\mathrm{0}}}  } = \target {  \ottsym{(}  \ottsym{(}  \Lambda  {\overline {\target a} }_{\ottmv{j}}  \ottsym{.}  \lambda \, \ottcomp{{\target \delta }_{\ottmv{i}}  \ottsym{:}  {\target \sigma }'_{\ottmv{i}}}{\ottmv{i}} \, \ottsym{.}  \ottsym{\{}  {\target m}  \ottsym{=}  {\target e }_{\ottmv{m}}  \ottsym{\}}  \ottsym{)} \, {\overline {\target \sigma} }_{\ottmv{j}} \, {\overline {\target e} }_{\ottmv{i}}  \ottsym{)}  \ottsym{.}  {\target m}  } $
            and $ \intermed {  {\intermed \sigma}_{{\mathrm{0}}}  } = \intermed {  [  [  \overline {\intermed \sigma}_{\ottmv{j}}  /  {\overline {\intermed a} }_{\ottmv{j}}  ]  {\intermed \sigma}_{\ottmv{q}}  /  {\intermed a }  ]  {\intermed \sigma}_{\ottmv{m}}  } $, for some $\intermed{{\intermed \sigma}_{\ottmv{m}}}$,
            $\intermed{{\intermed e}_{\ottmv{m}}}$, $\intermed{{\intermed a }}$, $\target{{\target \sigma }'_{\ottmv{i}}}$, $\target{{\target e }_{\ottmv{m}}}$, $\target{{\target \sigma }_{\ottmv{j}}}$ and $\target{{\target e }_{\ottmv{i}}}$,
            such that 
            \begin{gather*}
               ( \intermed {  {\intermed m}  } : \intermed {  {\intermed {\mathit{TC} } } \, {\intermed a }  } : \intermed {  {\intermed \sigma}_{\ottmv{m}}  } )  \in  \intermed {  {\intermed { \Gamma_C } }  } \\
               \intermed {  {\intermed { \Gamma_C } }  } ; \intermed {  \bullet  \ottsym{,}  {\intermed a }  }  \vdash_{ty}  \intermed {  {\intermed \sigma}_{\ottmv{m}}  } \itot{ \rightsquigarrow  \target {  {\target \sigma }_{\ottmv{m}}  } } \\
               ( \intermed {  {\intermed D }  } : \intermed {  \forall  {\overline {\intermed a} }_{\ottmv{j}}  \ottsym{.}  {\overline {\intermed Q} }_{\ottmv{i}}  \Rightarrow  {\intermed {\mathit{TC} } } \, {\intermed \sigma}_{\ottmv{q}}  } ) . \intermed {  {\intermed m}  }  \mapsto  \intermed {  \Lambda  {\overline {\intermed a} }_{\ottmv{j}}  \ottsym{.}  \lambda  {\overline {\intermed \delta} }_{\ottmv{i}}  \ottsym{:}  {\overline {\intermed Q} }_{\ottmv{i}}  \ottsym{.}  {\intermed e}_{\ottmv{m}}  }  \in  \intermed {  {\intermed \Sigma}  } \\
               \intermed {  {\intermed \Sigma}_{{\mathrm{1}}}  } ; \intermed {  {\intermed { \Gamma_C } }  } ; \intermed {  \bullet  \ottsym{,}  {\overline {\intermed a} }_{\ottmv{j}}  \ottsym{,}  {\overline {\intermed \delta} }_{\ottmv{i}}  \ottsym{:}  {\overline {\intermed Q} }_{\ottmv{i}}  }  \vdash_{tm}  \intermed {  {\intermed e}_{\ottmv{m}}  } : \intermed {  [  {\intermed \sigma}_{\ottmv{q}}  /  {\intermed a }  ]  {\intermed \sigma}_{\ottmv{m}}  } \itot{ \rightsquigarrow  \target {  {\target e }_{\ottmv{m}}  } } \\
              \ottcomp{ \intermed {  {\intermed { \Gamma_C } }  } ; \intermed {  \bullet  \ottsym{,}  {\overline {\intermed a} }_{\ottmv{j}}  }  \vdash_{\mathit {Q} }  \intermed {  {\intermed Q}_{\ottmv{i}}  }\, \itot{ \rightsquigarrow  \target {  {\target \sigma }'_{\ottmv{i}}  } } }{\ottmv{i}}\\
              \ottcomp{ \intermed {  {\intermed { \Gamma_C } }  } ; \intermed {  \bullet  }  \vdash_{ty}  \intermed {  {\intermed \sigma}_{\ottmv{j}}  } \itot{ \rightsquigarrow  \target {  {\target \sigma }_{\ottmv{j}}  } } }{\ottmv{j}}\\
              \ottcomp{ \intermed {  {\intermed \Sigma}  } ; \intermed {  {\intermed { \Gamma_C } }  } ; \intermed {  \bullet  }  \vdash_{d}  \intermed {  {\intermed d}_{\ottmv{i}}  } : \intermed {  [  \overline {\intermed \sigma}_{\ottmv{j}}  /  {\overline {\intermed a} }_{\ottmv{j}}  ]  {\intermed Q}_{\ottmv{i}}  } \itot{\,  \rightsquigarrow  \target {  {\target e }_{\ottmv{i}}  } } }{\ottmv{i}}
            \end{gather*}
            Because ${\intermed \Sigma}$ contains a unique method implementation per class instance, we
            also have
            \begin{equation}
              \label{semelab:meth:ie}
               \intermed {  {\intermed e}  } = \intermed {  \Lambda  {\overline {\intermed a} }_{\ottmv{j}}  \ottsym{.}  \lambda  {\overline {\intermed \delta} }_{\ottmv{i}}  \ottsym{:}  {\overline {\intermed Q} }_{\ottmv{i}}  \ottsym{.}  {\intermed e}_{\ottmv{m}}  } 
            \end{equation}
            Then, term $\intermed{{\intermed e} \, \overline {\intermed \sigma}_{\ottmv{j}} \, {\overline {\intermed d} }_{\ottmv{i}}}$, equal to
            $\intermed{\ottsym{(}  \Lambda  {\overline {\intermed a} }_{\ottmv{j}}  \ottsym{.}  \lambda  {\overline {\intermed \delta} }_{\ottmv{i}}  \ottsym{:}  {\overline {\intermed Q} }_{\ottmv{i}}  \ottsym{.}  {\intermed e}_{\ottmv{m}}  \ottsym{)} \, \overline {\intermed \sigma}_{\ottmv{j}} \, {\overline {\intermed d} }_{\ottmv{i}}}$, is deterministcally elaborated to
            $\target{\ottsym{(}  \Lambda  {\overline {\target a} }_{\ottmv{j}}  \ottsym{.}  \lambda \, \ottcomp{{\target \delta }_{\ottmv{i}}  \ottsym{:}  {\target \sigma }'_{\ottmv{i}}}{\ottmv{i}} \, \ottsym{.}  {\target e }_{\ottmv{m}}  \ottsym{)} \, {\overline {\target \sigma} }_{\ottmv{j}} \, {\overline {\target e} }_{\ottmv{i}}}$.

            Indeed, $ \target {  {\target e }_{\ottmv{h}}  } = \target {  [ \, \ottcomp{{\target e }_{\ottmv{i}}  /  {\target x}_{\ottmv{i}}}{\ottmv{i}} \, ]  [ \, \ottcomp{{\target \sigma }_{\ottmv{j}}  /  {\target a }_{\ottmv{j}}}{\ottmv{j}} \, ]  {\target e }_{\ottmv{m}}  } $
            is an appropriate choice, since
            \begin{equation*}
            \begin{aligned}
              \target{{\target e }_{{\mathrm{0}}}} &= \target{\ottsym{(}  \ottsym{(}  \Lambda  {\overline {\target a} }_{\ottmv{j}}  \ottsym{.}  \lambda \, \ottcomp{{\target \delta }_{\ottmv{i}}  \ottsym{:}  {\target \sigma }'_{\ottmv{i}}}{\ottmv{i}} \, \ottsym{.}  \ottsym{\{}  {\target m}  \ottsym{=}  {\target e }_{\ottmv{m}}  \ottsym{\}}  \ottsym{)} \, {\overline {\target \sigma} }_{\ottmv{j}} \, {\overline {\target e} }_{\ottmv{i}}  \ottsym{)}  \ottsym{.}  {\target m}}\\
              &\to \target{\ottsym{(}  \lambda \, \ottcomp{{\target \delta }_{\ottmv{i}}  \ottsym{:}  [ \, \ottcomp{{\target \sigma }_{\ottmv{j}}  /  {\target a }_{\ottmv{j}}}{\ottmv{j}} \, ]  {\target \sigma }'_{\ottmv{i}}}{\ottmv{i}} \, \ottsym{.}  \ottsym{\{}  {\target m}  \ottsym{=}  [ \, \ottcomp{{\target \sigma }_{\ottmv{j}}  /  {\target a }_{\ottmv{j}}}{\ottmv{j}} \, ]  {\target e }_{\ottmv{m}}  \ottsym{\}}  \ottsym{)}  \ottsym{.}  {\target m}}\\
              &\to \target{\ottsym{\{}  {\target m}  \ottsym{=}  [ \, \ottcomp{{\target e }_{\ottmv{i}}  /  {\target \delta }_{\ottmv{i}}}{\ottmv{i}} \, ]  [ \, \ottcomp{{\target \sigma }_{\ottmv{j}}  /  {\target a }_{\ottmv{j}}}{\ottmv{j}} \, ]  {\target e }_{\ottmv{m}}  \ottsym{\}}  \ottsym{.}  {\target m}}\\
              &\to \target{[ \, \ottcomp{{\target e }_{\ottmv{i}}  /  {\target \delta }_{\ottmv{i}}}{\ottmv{i}} \, ]  [ \, \ottcomp{{\target \sigma }_{\ottmv{j}}  /  {\target a }_{\ottmv{j}}}{\ottmv{j}} \, ]  {\target e }_{\ottmv{m}}}
            \end{aligned}
            \end{equation*}
            and
            \begin{equation*}
              \begin{aligned}
                \target{{\target e }'_{{\mathrm{0}}}} &= \target{\ottsym{(}  \Lambda  {\overline {\target a} }_{\ottmv{j}}  \ottsym{.}  \lambda \, \ottcomp{{\target \delta }_{\ottmv{i}}  \ottsym{:}  {\target \sigma }'_{\ottmv{i}}}{\ottmv{i}} \, \ottsym{.}  {\target e }_{\ottmv{m}}  \ottsym{)} \, {\overline {\target \sigma} }_{\ottmv{j}} \, {\overline {\target e} }_{\ottmv{i}}}\\
                &\to \target{\lambda \, \ottcomp{{\target \delta }_{\ottmv{i}}  \ottsym{:}  [ \, \ottcomp{{\target \sigma }_{\ottmv{j}}  /  {\target a }_{\ottmv{j}}}{\ottmv{j}} \, ]  {\target \sigma }'_{\ottmv{i}}}{\ottmv{i}} \, \ottsym{.}  [ \, \ottcomp{{\target \sigma }_{\ottmv{j}}  /  {\target a }_{\ottmv{j}}}{\ottmv{j}} \, ]  {\target e }_{\ottmv{m}}}\\
                &\to \target{[ \, \ottcomp{{\target e }_{\ottmv{i}}  /  {\target \delta }_{\ottmv{i}}}{\ottmv{i}} \, ]  [ \, \ottcomp{{\target \sigma }_{\ottmv{j}}  /  {\target a }_{\ottmv{j}}}{\ottmv{j}} \, ]  {\target e }_{\ottmv{m}}}
              \end{aligned}
            \end{equation*}
          }
\proofStep{Case \textsc{iEval-let}}{\drule{iEval-let}}
          {
            Hypotheses \pref{semelab:hyp1} and \pref{semelab:hyp2} of the theorem, adapted to
            this case, are:
            \begin{align}
              \label{semelab:let:der1}
                         & \intermed {  {\intermed \Sigma}  } ; \intermed {  {\intermed { \Gamma_C } }  } ; \intermed {  \bullet  }  \vdash_{tm}  \intermed {  \textbf{\intermed {let} } \, {\intermed x}  \ottsym{:}  {\intermed \sigma}  \ottsym{=}  {\intermed e}_{{\mathrm{1}}} \, \textbf{\intermed {in} } \, {\intermed e}  } : \intermed {  {\intermed \sigma}'  } \itot{ \rightsquigarrow  \target {  {\target e }_{{\mathrm{0}}}  } } \\
              \label{semelab:let:der2}
              \text{and}~& \intermed {  {\intermed \Sigma}  } ; \intermed {  {\intermed { \Gamma_C } }  } ; \intermed {  \bullet  }  \vdash_{tm}  \intermed {  [  {\intermed e}_{{\mathrm{1}}}  /  {\intermed x}  ]  {\intermed e}_{{\mathrm{2}}}  } : \intermed {  {\intermed \sigma}'  } \itot{ \rightsquigarrow  \target {  {\target e }'_{{\mathrm{0}}}  } } 
            \end{align}
            for some $\intermed{{\intermed \sigma}'}$, $\target{{\target e }_{{\mathrm{0}}}}$ and $\target{{\target e }'_{{\mathrm{0}}}}$. 
            We need to show that
            there exists an \(\target{{\target e }_{\ottmv{h}}}\) such that \( \target {  {\target e }_{{\mathrm{0}}}  }  \longrightarrow^{*}  \target {  {\target e }_{\ottmv{h}}  } \) and
            \( \target {  {\target e }'_{{\mathrm{0}}}  }  \longrightarrow^{*}  \target {  {\target e }_{\ottmv{h}}  } \). We do this by showing that $ \target {  {\target e }_{{\mathrm{0}}}  }  \longrightarrow^{*}  \target {  {\target e }'_{{\mathrm{0}}}  } $.
            
            By inversion, the last rule used for Derivation \pref{semelab:let:der1} must be an
            instance of \textsc{iTm-let}.
            \begin{equation*}
                \ottdrule{
                  \ottpremise{ \intermed {  {\intermed \Sigma}  } ; \intermed {  {\intermed { \Gamma_C } }  } ; \intermed {  \bullet  }  \vdash_{tm}  \intermed {  {\intermed e}_{{\mathrm{1}}}  } : \intermed {  {\intermed \sigma}  } \itot{ \rightsquigarrow  \target {  {\target e }_{{\mathrm{1}}}  } } }
                  \ottpremise{ \intermed {  {\intermed \Sigma}  } ; \intermed {  {\intermed { \Gamma_C } }  } ; \intermed {  \bullet  \ottsym{,}  {\intermed x}  \ottsym{:}  {\intermed \sigma}  }  \vdash_{tm}  \intermed {  {\intermed e}_{{\mathrm{2}}}  } : \intermed {  {\intermed \sigma}'  } \itot{ \rightsquigarrow  \target {  {\target e }_{{\mathrm{2}}}  } } }
                  \ottpremise{ \intermed {  {\intermed { \Gamma_C } }  } ; \intermed {  \bullet  }  \vdash_{ty}  \intermed {  {\intermed \sigma}  } \itot{ \rightsquigarrow  \target {  {\target \sigma }  } } }
                }{
                   \intermed {  {\intermed \Sigma}  } ; \intermed {  {\intermed { \Gamma_C } }  } ; \intermed {  \bullet  }  \vdash_{tm}  \intermed {  \textbf{\intermed {let} } \, {\intermed x}  \ottsym{:}  {\intermed \sigma}  \ottsym{=}  {\intermed e}_{{\mathrm{1}}} \, \textbf{\intermed {in} } \, {\intermed e}  } : \intermed {  {\intermed \sigma}'  } \itot{ \rightsquigarrow  \target {  \textbf{\target {let} } \, {\target x}  \ottsym{:}  {\target \sigma }  \ottsym{=}  {\target e }_{{\mathrm{1}}} \, \textbf{\target {in} } \, {\target e }_{{\mathrm{2}}}  } } 
                }{\textsc{iTm-constrI}}
            \end{equation*}
            where $ \target {  {\target e }_{{\mathrm{0}}}  } = \target {  \textbf{\target {let} } \, {\target x}  \ottsym{:}  {\target \sigma }  \ottsym{=}  {\target e }_{{\mathrm{1}}} \, \textbf{\target {in} } \, {\target e }_{{\mathrm{2}}}  } $. From the above equation, we can use
            the first two premises in Lemma~\ref{lemma:var_subst}, to obtain
            \begin{equation*}
               \intermed {  {\intermed \Sigma}  } ; \intermed {  {\intermed { \Gamma_C } }  } ; \intermed {  \bullet  }  \vdash_{tm}  \intermed {  [  {\intermed e}_{{\mathrm{1}}}  /  {\intermed x}  ]  {\intermed e}_{{\mathrm{2}}}  } : \intermed {  {\intermed \sigma}'  } \itot{ \rightsquigarrow  \target {  [  {\target e }_{{\mathrm{1}}}  /  {\target x}  ]  {\target e }_{{\mathrm{2}}}  } } 
            \end{equation*}
            Then, by uniqueness (Theorem~\ref{thmA:tmelab:det} on the latter and on
            Equation~\pref{semelab:let:der2}), we have $ \target {  {\target e }'_{{\mathrm{0}}}  } = \target {  [  {\target e }_{{\mathrm{1}}}  /  {\target x}  ]  {\target e }_{{\mathrm{2}}}  } $.

            We set $ \target {  {\target e }_{\ottmv{h}}  } = \target {  [  {\target e }_{{\mathrm{1}}}  /  {\target x}  ]  {\target e }_{{\mathrm{2}}}  } $, since
            $ \target {  \ottsym{(}  \lambda  {\target x}  \ottsym{:}  {\target \sigma }  \ottsym{.}  {\target e }_{{\mathrm{1}}}  \ottsym{)} \, {\target e }_{{\mathrm{2}}}  }  \longrightarrow^{*}  \target {  [  {\target e }_{{\mathrm{1}}}  /  {\target x}  ]  {\target e }_{{\mathrm{2}}}  } $, by evaluation rule
            \textsc{tEval-AppAbs}, and $ \target {  [  {\target e }_{{\mathrm{1}}}  /  {\target x}  ]  {\target e }_{{\mathrm{2}}}  }  \longrightarrow^{*}  \target {  [  {\target e }_{{\mathrm{1}}}  /  {\target x}  ]  {\target e }_{{\mathrm{2}}}  } $, by
            reflexivity of $ \longrightarrow^{*} $.\\
          }
\qed

\begin{lemmaB}[\targetL{} Preservation of Values]\ \\
  \label{lemma:val_pres}
  If \( \intermed {  {\intermed \Sigma}  } ; \intermed {  {\intermed { \Gamma_C } }  } ; \intermed {  \bullet  }  \vdash_{tm}  \intermed {  {\intermed v}  } : \intermed {  {\intermed \sigma}  } \itot{ \rightsquigarrow  \target {  {\target e }  } } \)
  then \({\target e }\) is a value.
\end{lemmaB}
\proof By straightforward case analysis on the typing derivation.

\begin{theoremB}[Value Semantic Preservation]\ \\
  \label{lemma:val_sem_pres}
  If \( \intermed {  {\intermed \Sigma}  } ; \intermed {  {\intermed { \Gamma_C } }  } ; \intermed {  \bullet  }  \vdash_{tm}  \intermed {  {\intermed e}  } : \intermed {  {\intermed \sigma}  } \itot{ \rightsquigarrow  \target {  {\target e }  } } \)
  and \( \intermed {  {\intermed \Sigma}  }  \vdash  \intermed {  {\intermed e}  }  \longrightarrow^{*}  \intermed {  {\intermed v}  } \)
  then \( \intermed {  {\intermed \Sigma}  } ; \intermed {  {\intermed { \Gamma_C } }  } ; \intermed {  \bullet  }  \vdash_{tm}  \intermed {  {\intermed v}  } : \intermed {  {\intermed \sigma}  } \itot{ \rightsquigarrow  \target {  {\target v}  } } \)
  and \( \target {  {\target e }  }  \simeq  \target {  {\target v}  } \).
\end{theoremB}
\proof
From the Preservation Theorem~\ref{thmA:intermediate_preservation} and Progress
Theorem~\ref{thmA:progress_expressions}, in combination with the hypothesis,
we know that:
\begin{equation*}
   \intermed {  {\intermed \Sigma}  } ; \intermed {  {\intermed { \Gamma_C } }  } ; \intermed {  \bullet  }  \vdash_{tm}  \intermed {  {\intermed v}  } : \intermed {  {\intermed \sigma}  } \itot{ \rightsquigarrow  \target {  {\target e }'  } } 
\end{equation*}
Lemma~\ref{lemma:val_pres} teaches us that \({\target e }'\) is some value \({\target v}\).

The goal follows by repeatedly applying Theorem~\ref{thmA:sem_pres}, in
combination with the fact that evaluation in both \interL{} and \targetL{} is
deterministic (Lemmas~\ref{lemma:eval_unique} and~\ref{lem:tgt_eval_unique}). \\
\qed





\end{document}